\documentclass[twocolumn,amsmath,amssymb,10pt,floatfix,superscriptaddress,a4paper,letterpaper,fleqn,nofootinbib]{revtex4-1}

\usepackage{amssymb}
\usepackage{graphicx}
\usepackage{subfigure}
\usepackage{epsfig}
\usepackage{dcolumn}
\usepackage{array}
\usepackage{bm}
\usepackage{amssymb}
\usepackage{fancyhdr}
\usepackage{floatflt}
\usepackage{tabularx}
\usepackage{amsmath}
\usepackage{amssymb}
\usepackage[table]{xcolor}
\usepackage{longtable}
\usepackage{threeparttable}
\usepackage{multirow}
\usepackage{float}
\usepackage{afterpage}
\usepackage{color}
\usepackage{textcomp}
\usepackage{multirow}
\usepackage{mathdesign}
\usepackage{booktabs}
\usepackage{paralist}
\usepackage{makecell}

\usepackage{tabularray}
\UseTblrLibrary{booktabs}

\newcolumntype{P}[1]{>{\centering\arraybackslash}p{#1}}

\usepackage{units}

\newcommand\T{\rule{0pt}{2.6ex}}       
\def\ql{``}
\def\qr{''\hspace{0.5mm}}

\usepackage[warnundef]{jabbrv}
\DefineJournalAbbreviation{Test}{Tes} 
\setlongtables
\definecolor{Darkgreen}{rgb}{0,0.4,0}
\usepackage[breaklinks=true,linkbordercolor={1 1 1}]{hyperref}
\hypersetup{
colorlinks = true,    
linktoc    = all,     
linkcolor  = blue,    
citecolor  = Darkgreen
}
\def\etal{\emph{et~al.}\ }

\extrafloats{1000}
\newcommand{\nuc}[2]{$^{{\mathrm{#1}}}${#2}}

\newcommand{\CoH}{\texttt{CoH}}
\newcommand{\CGMF}{\texttt{CGMF}}
\newcommand{\NJOY}{\texttt{NJOY}}
\newcommand{\AMPX}{\texttt{AMPX}}
\newcommand{\SCALE}{\texttt{SCALE}}
\newcommand{\FLASSH}{\texttt{FLASSH}}
\newcommand{\LEAPR}{\texttt{LEAPR}}
\newcommand{\FUDGE}{\texttt{FUDGE}}
\newcommand{\GANDR}{\texttt{GANDR}}
\newcommand{\DICEBOX}{\texttt{DICEBOX}}
\newcommand{\OpenMC}{\texttt{OpenMC}}
\newcommand{\SERPENT}{\texttt{Serpent}}
\newcommand{\Serpent}{\texttt{Serpent}}
\newcommand{\Shift}{\texttt{Shift}}
\newcommand{\CASMO}{\texttt{CASMO5}}
\newcommand{\CUPIDO}{\texttt{CUPIDO}}
\newcommand{\TEDCA}{\texttt{TEDCA}}
\newcommand{\EMPIRE}{\texttt{EMPIRE}}
\newcommand{\GIDI}{\texttt{GIDI+}}
\newcommand{\SAMMY}{\texttt{SAMMY}}

\newcommand{\cah}{CaH$_2$}
\newcommand{\beo}{BeO}
\newcommand{\sic}{SiC}
\newcommand{\csic}{3C-SiC}
\newcommand{\sio}{SiO$_2$-$\alpha$}
\newcommand{\bem}{Be-metal}
\newcommand{\uc}{UC}
\newcommand{\gr}{reactor graphite}
\newcommand{\besd}{Be+S$_d$}
\newcommand{\sd}{S$_d$}
\newcommand{\flibe}{FLiBe}
\newcommand{\grsd}{graph+S$_d$}
\newcommand{\hf}{HF}

\newcommand{\puo}{PuO$_2$}
\newcommand{\ninePu}{$^{239}$Pu}

\newcommand{\fiveU}{$^{235}$U}
\newcommand{\sixCa}{$^{46}$Ca}
\newcommand{\oneH}{$^{1}$H}
\newcommand{\uo}{UO$_2$}
\newcommand{\un}{UN}
\newcommand{\alo}{Al$_2$O$_3$}

\newcommand{\mgft}{MgF$_2$}
\newcommand{\beft}{BeF$_2$}

\newcommand{\ENDF}{ENDF/B-VIII.1}
\newcommand{\prENDF}{ENDF/B-VIII.0}
\newcommand{\keff}{$k_\mathrm{eff}$}
\newcommand{\ab}{\emph{ab initio}}
\newcommand{\Ab}{\emph{Ab initio}}
\newcommand{\Ss}{S$_s$($\alpha$,$\beta$)}
\newcommand{\s}{S($\alpha$,$\beta$)}
\newcommand{\AILD}{\cite{Hawari2014,Hawari2004}}
\newcommand{\VASP}{\cite{Kresse1996_1,Kresse1996_2,Kresse1999}}

\newcommand{\ps}{(C$_8$H$_8$)$_n$}
\newcommand{\pmma}{(C$_5$O$_2$H$_8$)$_n$}

\newcommand{\be}{\begin{equation}}
\newcommand{\ee}{\end{equation}}

\newcommand{\bea}{\begin{eqnarray}}
\newcommand{\eea}{\end{eqnarray}}

\newcommand{\ANL}{Argonne National Laboratory, Argonne, IL 60439-4842 USA}
\newcommand{\AWE}{AWE.plc Aldermaston, Reading, Berkshire, RG7 4PR, United Kingdom}
\newcommand{\BNL}{Brookhaven National Laboratory, Upton, NY 11973-5000, USA}
\newcommand{\CEA}{CEA, DEN, DER, SPRC, Cadarache, 13108 Saint-Paul-l\`{e}z-Durance, France}
\newcommand{\CNL}{Canadian Nuclear Laboratories, Chalk River, Ontario, Canada}
\newcommand{\Ecole}{Ecole Polytechnique F\'{e}d\'{e}rale de Lausanne,  1015 Lausanne, Switzerland}
\newcommand{\ESS}{European Spallation Source ERIC, Lund, Sweden}
\newcommand{\HZDR}{Helmholtz-Zentrum Dresden - Rossendorf e.V., Dresden, Germany}
\newcommand{\IAEA}{International Atomic Energy Agency, Vienna A-1400, PO Box 100, Austria}
\newcommand{\IAEAconsultant}{International Atomic Energy Agency (consultant), Vienna A-1400, PO Box 100, Austria}
\newcommand{\IRSN}{Institut de Radioprotection et de S\^{u}ret\'{e} Nucl\'{e}aire, 92262 Fontenay aux Roses, Cedex, France}
\newcommand{\JSI}{Jo\v{z}ef Stefan Institute, Jamova 39, SI-1000, Ljubljana, Slovenia}
\newcommand{\KAERI}{Korea Atomic Energy Research Institute, Daejeon, Republic of Korea}

\newcommand{\LANL}{Los Alamos National Laboratory, Los Alamos, NM 87545, USA}
\newcommand{\LLNL}{Lawrence Livermore National Laboratory, Livermore, CA 94551-0808, USA}
\newcommand{\NCSU}{North Carolina State University, Department of Nuclear Engineering, Raleigh, NC 27695}
\newcommand{\NIST}{National Institute of Standards and Technology, Gaithersburg, MD 20899-8463, USA}
\newcommand{\NNLb}{Naval Nuclear Laboratory, West Mifflin, PA 15122-0079, USA}
\newcommand{\NNLk}{Naval Nuclear Laboratory, Schenectady, NY 12301-1072, USA}
\newcommand{\NRG}{NRG Westerduinweg 3, 1755 LE Petten, Netherlands}
\newcommand{\ORNL}{Oak Ridge National Laboratory, Oak Ridge, TN 37831-6171, USA}
\newcommand{\PSI}{Laboratory for Reactor Physics Systems Behaviour, Paul Scherrer Institut, Villigen, Switzerland}
\newcommand{\Rez}{Research Centre \v{R}e\v{z} Ltd, Husinec-\v{R}e\v{z}, Czech Republic}
\newcommand{\RPI}{Rensselaer Polytechnic Institute, Troy, NY 12180, USA}
\newcommand{\Sharjah}{Department of Nuclear, University of Sharjah, Sharjah, United Arab Emirates}
\newcommand{\Spectra}{Spectra Tech, Inc., Oak Ridge, TN 37830, USA}
\newcommand{\Studvisk}{Studsvik Scandpower, Inc., 1070 Riverwalk Dr., Idaho Falls, ID 83401, USA}
\newcommand{\Surrey}{University of Surrey, Guildford, Surrey, GU2 7XH, United Kingdom}
\newcommand{\TAMU}{Department of Nuclear Engineering, Texas A\&M University, College Station, TX 77843, USA}
\newcommand{\UPM}{Universidad Polit\'{e}cnica de Madrid, Jos\'{e} Guti\'{e}rrez Abascal, 2 28006, Madrid, Spain}
\newcommand{\Yarmouk}{Physics Department, Yarmouk University, Irbid, Jordan}
 \newcommand{\IGFAE}{IGFAE-Universidad de Santiago de Compostela, 1782 Spain}

\begin{document}
\setcounter{page}{1}

\title{
     \qquad \\ \qquad \\ \qquad \\  \qquad \\  \qquad \\ \qquad \\
ENDF/B-VIII.1: Updated Nuclear Reaction Data Library for Science and Applications
}


%
%



\author{G.P.A. Nobre}
      \email[Corresponding author: ]{gnobre@bnl.gov}
      \affiliation{\BNL}


\author{R.~Capote}
      \affiliation{\IAEA}

\author{M.T.~Pigni}
      \affiliation{\ORNL}

\author{A.~Trkov}
      \affiliation{\JSI}

\author{C.M.~Mattoon}
      \affiliation{\LLNL}
      
\author{D.~Neudecker}
      \affiliation{\LANL}
      
\author{D.A.~Brown}
      \affiliation{\BNL}
      
\author{M.B.~Chadwick}
      \affiliation{\LANL}
      
\author{A.C.~Kahler}
      \affiliation{\LANL}
      
\author{N.A.~Kleedtke}
      \affiliation{\LANL}
      
% ...

\date{\today}
   \received{December 23, 2024; revised October 21, 2025}

\begin{abstract}{
\newpage
The ENDF/B-VIII.1 library is the newest recommended evaluated nuclear data file by the 
Cross Section Evaluation Working Group (CSEWG) for use in nuclear science and technology applications, and incorporates advances made in the six years since the release of ENDF/B-VIII.0. 
Among key advances made are that the $^{239}$Pu file was reevaluated by a joint international effort and that updated $^{16,18}$O, $^{19}$F, $^{28-30}$Si, $^{50-54}$Cr, $^{55}$Mn, $^{54,56,57}$Fe, $^{63,65}$Cu, $^{139}$La, $^{233,235,238}$U, and $^{240,241}$Pu neutron nuclear data from the IAEA coordinated INDEN collaboration were adopted. Over 60 neutron dosimetry cross sections were adopted from the IAEA's IRDFF-II library. In addition, the new library includes significant changes for $^3$He, $^6$Li, $^9$Be, $^{51}$V, $^{88}$Sr, $^{103}$Rh, $^{140,142}$Ce, Dy, $^{181}$Ta, Pt, $^{206-208}$Pb, and $^{234,236}$U neutron data, and new nuclear data for the photonuclear, charged-particle and atomic sublibraries. 
Numerous thermal neutron scattering kernels were reevaluated or provided for the very first time.
On the covariance side, work was undertaken to introduce better uncertainty quantification standards and testing for nuclear data covariances.
The significant effort to reevaluate important nuclides has reduced bias in the simulations of many integral experiments with particular progress noted for fluorine, copper, and stainless steel containing benchmarks. Data issues hindered the successful deployment of the previous ENDF/B-VIII.0 for commercial nuclear power applications in high burnup situations.  
These issues 
were addressed by improving the $^{238}$U and $^{239,240,241}$Pu evaluated data in the resonance region. The new library performance as a function of burnup is similar to the reference ENDF/B-VII.1 library. The ENDF/B-VIII.1 data are available in ENDF-6 and GNDS format at \url{https://doi.org/10.11578/endf/2571019}. 

}
\end{abstract}
\maketitle


\lhead{ENDF/B-VIII.1: Updated Nuclear$\dots$}     
\chead{NUCLEAR DATA SHEETS}                  
\rhead{G.P.A. Nobre \textit{et al.}}        
\lfoot{}
\rfoot{}
\renewcommand{\footrulewidth}{0.4pt}
\tableofcontents{}



\section{INTRODUCTION}
\label{sec:introduction}


This paper describes the contents of the next generation of the United States' premier nuclear data library, ENDF/B-VIII.1.  The ENDF library is the product of the Cross Section Evaluation Working Group (CSEWG), a long running collaboration formed in 1968.  The version numbering for the ENDF library indicates that this release is a ``minor'' one in that it does not include an update to the standards observables published by the International Atomic Energy Agency (IAEA) Neutron Data Standard Committee \cite{carlson2018}. However, the version numbering does not tell the whole story: we have updated many neutron sublibrary evaluations including fissionable materials $^{233,234,235,236,238}$U, $^{239,240,241}$Pu and structural materials, such as F, Si, Cr, Fe, Cu, Ta, and Pb. Additionally, we have developed several thermal neutron scattering evaluations and introduced other smaller but impactful changes. We have also addressed issues with actinide resolved-resonance-region data in high burnup situations. Finally, we have adopted many of the high-quality reactor dosimetry evaluations from the IRDFF-II library~\cite{IRDFF}. These changes will be outlined in the next sections and expanded on throughout this paper.

Our main goal while developing ENDF/B-VIII.1 is maintaining and/or improving the overall quality and performance of the library while ensuring both the transparency and reproducibility of methods and decisions within the CSEWG collaboration\footnote{Due to the nature of the collaboration, in-development data usually cannot be made fully open to the public.}. 
Updating an ENDF evaluation requires a detailed study to confirm that a new proposed evaluation is indeed superior. This cautious approach helps prevent an unintended step backwards.  The key principles we have followed are:
\label{sec:Principles}
\begin{itemize}
\item{Do no harm}
\item{Be aware of historical and previous evaluation decisions}
\item{Ensure proposed changes are demonstrably better}
\item{Anticipate potential problems and conduct validation testing}
\end{itemize}

For the most part, each ENDF/B-VIII.1 evaluation is ``hand crafted'' by nuclear data experts merging the best available experiment and theory,  and in some cases, with limited Artificial Intelligence and/or Machine Learning (AI/ML) assistance.  This process is unlike TENDL (which is arguably purely ML-generated) and more like the JENDL and JEFF evaluation efforts.  
Once the evaluator has completed their work, they upload their files to CSEWG's git repository (see Sec.~\ref{sec:GitLab}) where a battery of automated ``phase 1'' tests are run on the data.  The results of these tests are used in a peer review process by nuclear data subject matter experts  (see Sec.~\ref{sec:reviews}).  This review takes place directly in the git repository's web page, preserving both the file modification and review history and comments.  Once the files pass this peer review step, the library is made available to data validators coordinated by the CSEWG Validation Committee.  This fully online review process is new for ENDF/B-VIII.1 and is expected to smooth the development of the follow-on ENDF/B-IX.0 library.



We are in a rare moment of optimism in the field of nuclear data.  While nuclear science is being
recognized for its important historical impacts \cite{doi:10.1080/15361055.2023.2297128,doi:10.1080/15361055.2024.2346868,Chadwick_2024,doi:10.1080/00295450.2021.1903301,doi:10.1080/00295450.2021.1901002}, 
in a broad sense, we are also experiencing a ``nuclear renaissance'' and not just for traditional nuclear reactors.  
Recently, the Nuclear Ignition Facility (NIF) at Lawrence Livermore National Laboratory (LLNL) achieved Lawson's criteria for ignition \cite{2022:PRL:lawsonCriteria} sparking
discussions about nuclear data for fusion applications.  In addition, we are eagerly
anticipating the possibility of human-crewed missions to the moon and Mars, all of which spark interest
in space reactors \cite{Rearden:2021} and nuclear propulsion \cite{Wall:2023}.  Planetary science uses
methods, such as active neutron interrogation which uses techniques pioneered by the petroleum industry
\cite{Mauborgne:2017} but eagerly adopted for nuclear non-proliferation \cite{McConchie:2021}.  Put simply, nuclear
science and specifically nuclear data are being viewed as part of the solution to the world's problems in
combating climate change, ensuring nuclear security, enabling space exploration and providing new radionuclides for medical applications (see recent nuclear data needs for medical applications compiled at the IAEA meeting \cite{INDC(NDS)-0884}). 
ENDF/B-VIII.1 changes to actinides and thermal neutron scattering evaluations will aid all reactor application, both on Earth and in space.  The changes to the structural materials will benefit nearly all applications.  The addition/correction of outgoing particle spectra will improve energy balance in all applications.  Improvements to outgoing gamma data will have a substantial impact by enabling spectroscopy applications in exploration and non-proliferation applications.

Our ability to take advantage of these opportunities is enhanced by the Nuclear Data Working Group's
efforts to organize the Workshop on Applied Nuclear Data Activities (WANDA) series of workshops which
helps us identify cross-cutting needs for both current and emerging nuclear data users. This has also
expanded the communities' funding and brought in new members into the nuclear data workforce as reflected
in several new evaluations in ENDF/B-VIII.1.  These new entrants have helped preserve the expert knowledge that might have otherwise been lost when the senior members of our community have passed away.



\section{OVERVIEW OF THE CURRENT ENDF/B RELEASE}
\label{sec:overview}


Table \ref{table:libraryOverview} provides an overview of the ENDF/B-VIII.1 release and compares it to the last four ENDF/B versions. In this release, there are significant changes to most sublibraries, with only the standards and decay sublibraries remaining untouched.  The sublibraries in which new evaluations were added, either by replacing existing evaluations or creating new ones are highlighted in bold font in  Table \ref{table:libraryOverview}.  Below, we summarize the largest changes to various sublibraries and the largest changes to the ENDF workflow.

\begin{table}
\caption{\label{table:libraryOverview}Overview of the ENDF/B library releases and the 15 sublibraries in ENDF/B-VIII.1.
Shown in the columns are the number of materials present in each sublibrary in each release.  Here Spontaneous Fission Yields is abbreviated as SFY and Neutron-induced Fission Yields as NFY.}
\begin{tabular}{lrrrrr}
\toprule\toprule
Sublibrary                 & VIII.1  & VIII.0 & VII.1 & VII.0 & VI.8 \\
\midrule
Neutron                    & 558     & 557   & 423   & 393   & 328  \\
Thermal n-scattering       & 114     & 33    & 21    & 20    & 15   \\
\midrule
Proton                     & 49      & 49     & 48    & 48    & 35   \\
Deuteron                   &  6     & 5     & 5     & 5     & 2   \\
Triton                     &  5      & 5      & 3     & 3     & 1    \\
Helium3                    &  4      & 3     & 2     & 2     & 1    \\
Alpha                      &  5      & 1     & n/a   & n/a   & n/a  \\
Photonuclear               & 222     & 163    & 163   & 163   & n/a  \\
\midrule
Atomic relaxation          & 100     & 100    & 100   & 100   & 100  \\
Electron                   & 100     & 100    & 100   & 100   & 100  \\
Photoatomic                & 100     & 100    & 100   & 100   & 100  \\
\midrule
Decay data                 & 3821    & 3821   & 3817  & 3838  & 979 \\
\midrule
SFY                        & 9       & 9      & 9     & 9     & 9    \\
NFY                        & 31      & 31     & 31    & 31    & 31   \\
\midrule
Standards                  & 10      & 10     & 8     & 8     & 8    \\
\bottomrule\bottomrule
\end{tabular}
\end{table}

\subsection{Neutron sublibrary}

In this release, many neutron files were reevaluated or received major changes.  Below, we highlight these changes and, where appropriate, attempt to relate the evaluations to an application need, at least in broad terms.  Roughly one third of the new neutron evaluations were performed as part of the International Nuclear Data Evaluation Network (INDEN) collaboration:
\begin{itemize}
\item \textbf{$^{239}$Pu:} As a product of a joint evaluation from IAEA, Los Alamos National Laboratory (LANL), LLNL and Oak Ridge National Laboratory (ORNL), a new $^{239}$Pu was developed.  Changes to the fission neutron multiplicity $\bar{\nu}$, Prompt Fission Neutron Spectrum, resonance region and fast region were made.  The changes helped address the depletion issues noted elsewhere.

\item \textbf{INDEN light isotopes:} $^{16}$O, $^{18}$O, $^{19}$F

\item \textbf{INDEN structural materials:}
	$^{28,29,30}$Si, $^{63,65}$Cu, and components of stainless steel ($^{50,51,52,53,54}$Cr, $^{55}$Mn and $^{54,56,57}$Fe)
	
\item \textbf{INDEN detector materials:} $^{139}$La is used in scintillators and other detector materials

\item \textbf{INDEN actinides:} $^{233}$U, $^{235}$U, $^{238}$U, and $^{240,241}$Pu

\item \textbf{Non-INDEN light isotopes:} $^{3}$He, $^{6}$Li, and $^{9}$Be

\item \textbf{Non-INDEN structural materials:} $^{51}$V and $^{181}$Ta

\item \textbf{Dosimetry materials:} in addition to adopting IRDFF-II neutron dosimetry reactions in many materials, full evaluations were made for other dosimetry materials: $^{88}$Sr, $^{103}$Rh, and $^{190,191,192,193,194,195,196,197,198}$Pt 

\item \textbf{Rare earths:} $^{140,142}$Ce is often alloyed in fuel and refractory materials and $^{156,158,160,161,162,163,164}$Dy is used in reactor control rods 

\item \textbf{Lead:} is used in shielding and in lead cooled fast reactors, so $^{206,207,208}$Pb were reevaluated 

\item \textbf{Non-INDEN actinides:} $^{234}$U and $^{236}$U
\end{itemize}

In addition, there were many fixes or improvements to existing evaluations, including $^{106,108,110,111,112,114,116}$Cd, 
gamma-spectra fixes, addition of previously missing outgoing-particle distributions, updates to the unresolved resonance region (URR) for fission products and prompt $\bar{\nu}$ evaluations of minor actinides.

Finally, we note that there was much activity addressing neutron reaction covariances, including improved covariance testing, adoption of Templates of Expected Measurement Uncertainties to identify missing components in covariance data and a major decision to not provide mathematically adjusted neutron libraries for ENDF/B-VIII.1.
Covariances were corrected (some are very minor changes), updated or newly included, for many nuclei, such as $^{1,2}$H, $^{6,7}$Li, $^9$Be, $^{10}$B, $^{13}$C, 
$^{23}$Na, $^{24}$Mg, $^{27}$Al, $^{35,36}$Cl, $^{29,30,32}$Si, $^{31}$P, $^{32}$S, $^{39,41}$K, $^{46-48}$Ti
$^{49,51}$V, $^{50-53}$Cr, $^{55}$Mn, $^{54,56}$Fe, $^{59}$Co, $^{58,60,63}$Ni, $^{63,65}$Cu, $^{64,67,68}$Zn, $^{75}$As, $^{86}$Kr, $^{86}$Kr, 
$^{89}$Y, $^{90,91,95}$Zr, $^{92,97,98,100}$Mo, $^{102}$Ru, $^{103}$Rh, $^{113,115}$In, $^{127}$I, $^{132}$Xe, $^{139}$La, $^{140,142}$Ce, $^{139}$La, $^{141}$Pr, 
$^{143}$Nd, $^{147}$Pm, $^{155}$Eu, $^{152,160}$Gd, $^{156,158,160-164}$Dy, $^{169,170}$Tm, $^{181}$Ta, $^{199}$Hg, 
$^{190-198}$Pt,  $^{182-184,186}$W, $^{197}$Au, $^{204,206-208}$Pb, $^{209}$Bi, $^{233-236}$U and $^{239,240,242}$Pu.

\subsection{Thermal neutron scattering sublibrary}

CSEWG has dramatically increased the number of evaluations in the thermal scattering neutron law (TSL) sublibrary:
\begin{itemize} 
  \item \textbf{Traditional moderators:}
 light water (H$_2$O),
 Beryllium metal,
 Beryllium Oxide (BeO),
 Calcium Hydride (CaH$_2$),
 plastics (Polystyrene ((C$_8$H$_8$)$_n$) and Lucite ((C$_5$O$_2$H$_8$)$_n$)),
 graphite (reactor-grade graphite (20\%) and crystalline graphite (graph+S$_d$)),
 anhydrous Hydrogen Fluoride (HF), and
 heavy paraffinic oil.
 
 \item \textbf{Exotic moderators:}
 Beryllium Carbide (Be\textsubscript {2}C),
 Zirconium Hydride (ZrH$_x$ and ZrH\textsubscript {2}),
 Yttrium Hydride (YH\textsubscript {2}),
 Lithium-7 Hydride (\textsuperscript {7}LiH), and Deuteride (\textsuperscript {7}LiD)

\item \textbf{FLiBe molten salt}

\item \textbf{Structural materials and cladding:}
 Silicon Carbide (SiC),
 Silicon Dioxide (SiO$_2$), and
 Zirconium Carbide (ZrC)

\item \textbf{Fuels:}
Plutonium Dioxide (PuO$_2$),
Uranium Carbide (UC),
Uranium metal,
Uranium Nitride (UN),
Uranium Dioxide (UO$_2$),  and
Uranium Hydride (UH\textsubscript {3})

\item \textbf{Special purpose materials:}
liquid hydrogen and deuterium (l-H$_2$, l-D$_2$), and 
neutron filters (
Sapphire (Al$_2$O$_3$),
Magnesium Oxide (MgO),
Magnesium Fluoride (MgF$_2$), and
Beryllium Fluoride (BeF$_2$))

\end{itemize}


These additions meant that number of evaluated files in the ENDF/B library is larger than 
the previous number of available material (MAT) numbers as allotted by the ENDF-6 manual.
Therefore, CSEWG approved a format change that allows the use of all numbers from 1 to 9999. 
To organize and manage the administrative task of assigning unique MAT numbers to individual materials in a consistent manner within a given ENDF/B release and to ensure the user is effectively informed, we distribute alongside the TSL sublibrary of \ENDF\ a comma-separated file (CSV), named \texttt{TSL\_MAT\_numbers.csv}, which lists all evaluated files in the current release and their corresponding MAT number. In the future, as more materials evaluations are added and thus more MAT numbers are assigned, this file will be updated accordingly.

\subsection{Photonuclear sublibrary}

Around 200 evaluated files originating from the IAEA Coordinated Research Project (CRP) on photonuclear data~\cite{Kawano2020} were adopted into \ENDF. For a few select mission-critical materials, evaluations from \prENDF, which can be traced back to an earlier CRP~\cite{IAEAPhoto1999}, were adopted instead. Additionally,  fixes addressing format and other small issues were implemented. Lastly, \ENDF\ contains one new photonuclear file when compared to \prENDF\ -- that is \nuc{242}{Pu} which was adopted from JENDL-5~\cite{jendl5}.

\subsection{Charged particle sublibraries}

There were major and minor updates to evaluated files from the alphas, helions (\nuc{3}{He}), deuterons, protons and tritons sublibraries in \ENDF. Among these, we can highlight:

\begin{itemize}

\item For the alphas sublibrary, we adopted: the LLNL   Evaluated  Charged-Partilce Library  (ECPL)~\cite{white:1991ecp} evaluation for \nuc{6}{Li}; the NNL-modified JENDL evaluation for \nuc{9}{Be} and \nuc{16,17}{O}.

\item For the helions sublibrary we adopted: new \nuc{4}{He} evaluation from the INDEN collaboration; the LLNL ECPL~\cite{white:1991ecp} evaluation for \nuc{7}{Li}.

\item For the deuterons sublibrary we adopted: additions and corrections to the \prENDF\  d+t file; new LANL evaluation updates for \nuc{3}{He} and \nuc{6}{Li}.

\item For the protons sublibrary we adopted: evaluation updates for \nuc{4}{He}.

\item For the tritons sublibrary we adopted: evaluation updates for \nuc{4}{He}.

\end{itemize}

\subsection{Fission product yield sublibraries}

There were no new complete evaluations for either spontaneous (SFY) or neutron-induced fission yields (NFY) sublibraries in \ENDF. However, an important fix was implemented into the \nuc{241}{Pu} NFY file, solving a long-standing anomalous discontinuity issue introduced in ENDF/B-VI.2, decades ago. Additionally, many erroneously large uncertainties, present in both NFY and SFY from \prENDF, were corrected.

\subsection{Migration to GitLab repository}
\label{sec:GitLab}

Shortly after the \prENDF~release, the development version of the ENDF library was migrated from an Apache Subversion (SVN) repository to a GitLab server
hosted by the National Nuclear Data Center (NNDC)~\cite{NNDC-GitLab}. This migration provided several benefits leading up to the ENDF/B-VIII.1 release. The Git versioning system has by far surpassed 
SVN and competitors to become the main tool for developing and maintaining versioned repositories. Git provides more flexibility than SVN in creating, editing and merging branches.
This capability proved especially useful when several competing candidate evaluations were being developed and reviewed simultaneously.
In addition to hosting the repository, GitLab provides a web interface with integrated issue trackers and merge requests that helped
collaborators identify, document, track and resolve issues.  This allowed the implementation of continuous-integrated, continuous-development tools that would automatically trigger a series of format and processing tests for every update submitted. Although access to the ENDF GitLab repository cannot be made public due to the nature of the sponsors of much of its development, any member of the CSEWG collaboration can request access to it.

\subsection{Review process} 
\label{sec:reviews}

Even though evaluation reviews have always existed in the ENDF/B library, from series of paper forms a few decades ago to something more dynamic but less structured in the recent years, for \ENDF, this was the first time that a formal,  streamlined,  electronic/digital peer-review was completely integrated into the evaluation submission process. This was accomplished by leveraging the GitLab ENDF/B repository~\cite{NNDC-GitLab}, where contributions would be directly uploaded to one branch (\texttt{phase1}) by evaluators, but could only move to a second branch (\texttt{phase2}) after being reviewed by one or more peers. 

The user-friendliness of the GitLab platform, which allows for back-and-forth interactions with mark-down text, figures, file attachments, hyperlinks and GitLab cross-links, greatly encouraged iterations between evaluators, reviewers and other collaborators. The whole review discussion history, in the form of review branch merge requests, is completely preserved in GitLab, allowing for future consultation of the documentation and rationale that went into a particular decision. 
Although one could argue this process would slow down prompt access to the evaluations, quite often the evaluators would praise the process as it gave them a chance to find and fix errors, while users would benefit from this quality-assurance step. 

This integrated process often led to issues with evaluated files being found and addressed well before beta and final releases, minimizing chances of unwanted surprises after the releases, therefore, increasing the quality and reliability of the final product. Although the details of the review process are restricted to the CSEWG collaboration,  important information about the development of the files can be openly found in the change logs (\texttt{CHANGELOG.md} files) distributed within each \ENDF\ released sublibrary~\cite{ENDF-release-page,ENDF-release-alphas,ENDF-release-atom-relax,ENDF-release-decay,ENDF-release-deuterons,ENDF-release-electrons,ENDF-release-gammas,ENDF-release-he3,ENDF-release-neutrons,ENDF-release-nfy,ENDF-release-photoatomic,ENDF-release-protons,ENDF-release-sfy,ENDF-std,ENDF-tsl,ENDF-tritons}, as well as with the whole \ENDF\ release package \cite{ENDF-release-page}).





\section{NEUTRON CROSS SECTIONS}
\label{sec:neutron-sublib}

\subsection{Jointly-assembled \nuc{239}{Pu}}
\label{subsec:n:239Pu}



\subsubsection{Background \& Previous Evaluations}
A joint international effort on \nuc{239}{Pu} has been adopted for ENDF/B-VIII.1. This work involved substantial contributions from LANL, LLNL, ORNL, and the IAEA-coordinated INDEN collaboration. A completely new evaluation has been developed in the whole energy range. A concerted effort has been made by the international team to include the latest measured differential data, improve the agreement with existing measurements and neutron standards, use the IAEA evaluated prompt fission neutron spectrum (PFNS) at the thermal point, make use of new and decisive LANL-LLNL Chi-Nu PFNS and CEA measurements~\cite{NNSACEA,Marini2020,Kelly2021,Neudecker:2023Pu9fissionsourceterm} above the thermal point, and increase the evaluation range up to 30~MeV incident neutron energy. Additional work on resonance parameters and resulting fission, capture and total cross sections in the resolved resonance range below 2~eV was required to increase the criticality of power reactors at high burnup, a known ENDF/B-VIII.0 deficiency~\cite{kim2021-318}.

\subsubsection{(n,f) fission neutron multiplicity $\overline\nu$}
\begin{figure}[!thb]
\centering
\includegraphics[width=\columnwidth]{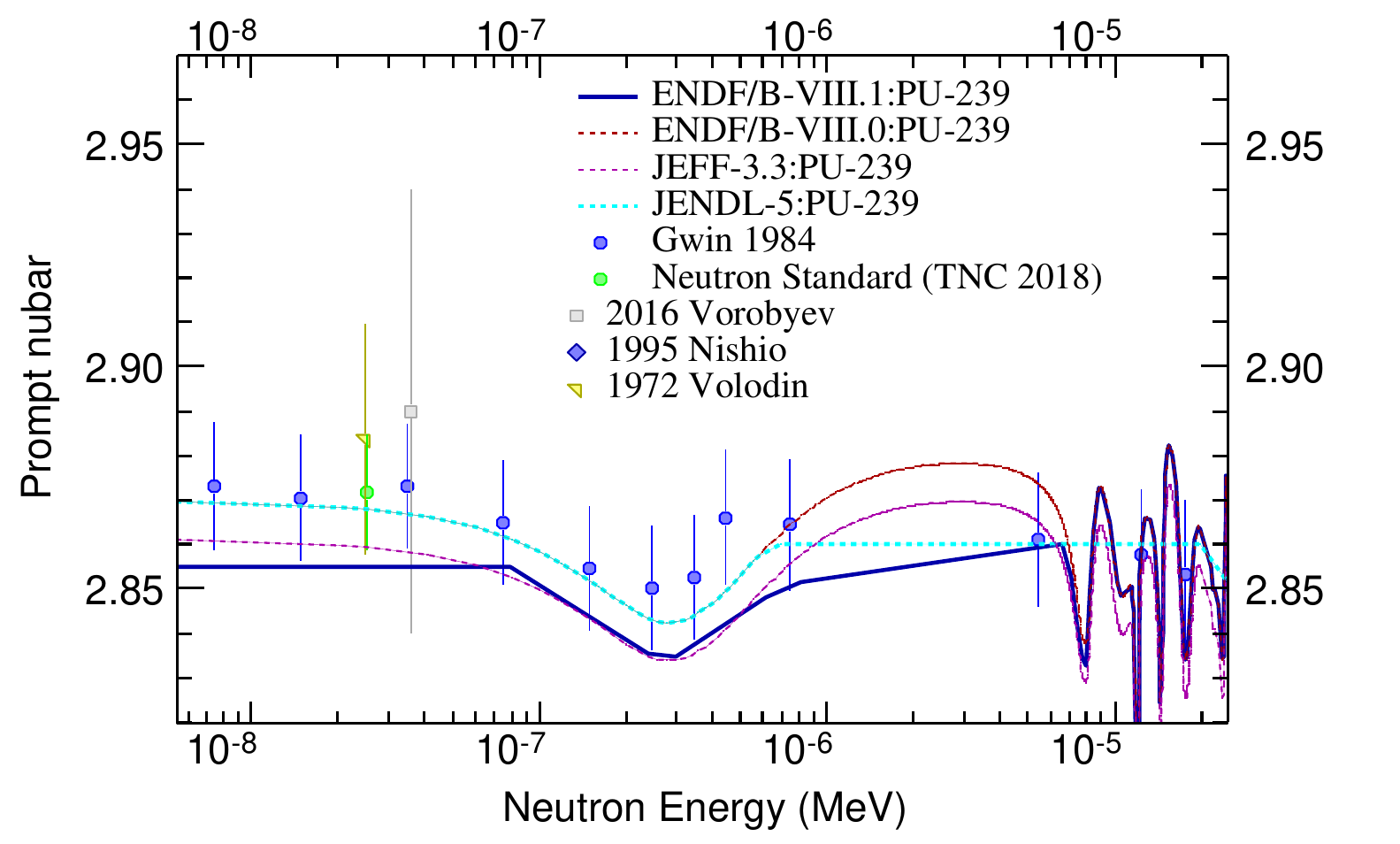}
\caption{Evaluated $^{239}$Pu $\overline\nu$ in ENDF/B-VIII.1 and previous evaluations are compared with selected experimental data 
\cite{Nishio1995,Volodin1972,capote:2016} from EXFOR \cite{EXFOR}. Gwin data \cite{Gwin:1984-nubar} were derived using the recommended $^{252}$Cf(sf) $\overline\nu$ \cite{carlson2018}.}
\label{fig:Pu239nu-LE}
\end{figure}

The neutron-induced average prompt fission neutron multiplicity, $\overline\nu$, for n+$^{239}$Pu in the resonance region was reevaluated to consider the Thermal Neutron Constant (TNC) values \cite{carlson2018} as well as experimental values measured by Gwin \cite{Gwin:1984-nubar,Gwin:1986-nubar}. The $\overline\nu$ evaluation shape up to 0.8 eV is very similar to \prENDF{} and largely driven by Gwin 1984 data \cite{Gwin:1984-nubar}. The comparison with Gwin data and other evaluations is shown in Fig.~\ref{fig:Pu239nu-LE}. The thermal $\overline\nu_{pr}$=2.855 is 0.6\% lower than the recommended standard TNC value $\overline\nu_{pr}$ of $2.872\pm0.012$. The evaluated prompt $\overline\nu$ value is driven by lower criticality required for Pu solutions benchmarks\footnote{Private communication by A.C. ``Skip'' Kahler.} and the need to increase the criticality at high burnup, see the discussion in Section~\ref{sec:depletion}.

Note that Kahler calculated 158 plutonium thermal solutions (PST1, PST2, PST3, PST4, PST5, PST6, PST9, PST10, PST11, PST12, PST18, PST22, PST28, PST32, PST34, and PST38): the average $\Delta k_\mathrm{eff}$ for ENDF/B-VII.1 was 1.00451(455), for ENDF/B-VIII.0 was 0.99996(468), and for the latest ENDF/B-VIII.1 evaluation became 1.00236(493). The excellent ENDF/B-VIII.0 results were driven by the adoption of the Working Party on International Nuclear Data Evaluation Co-operation (WPEC) Sub-Group-34 (SG-34) resonance parameters, but the thermal-neutron-induced PFNS was not changed from the ENDF/B-VII.0 evaluation. The dispersion of calculated results remain at ~500~pcm level for all three evaluations, which is certainly too high. Further work is needed to improve this performance considering criticality, depletion, and temperature reactivity coefficients.

A bump present in ENDF/B-VIII.0 $\overline\nu$ evaluation between 0.8~eV and 5~eV was reduced in this work as previously done in the JENDL-5 evaluation \cite{jendl5} as shown in Fig.~\ref{fig:Pu239nu-LE}. There is contradictory information from integral experiments: to increase power-reactor criticality at high burnup ~\cite{kim2021-318}, a higher Pu thermal average neutron multiplicity is required. However, the average criticality of many well-thermalized Pu thermal solutions (as listed above) indicates a need for lower thermal average neutron multiplicity as suggested by A.C.~Kahler. This contradiction required a compromise solution.

\begin{figure}[!thb]
\centering
\includegraphics[width=\columnwidth]{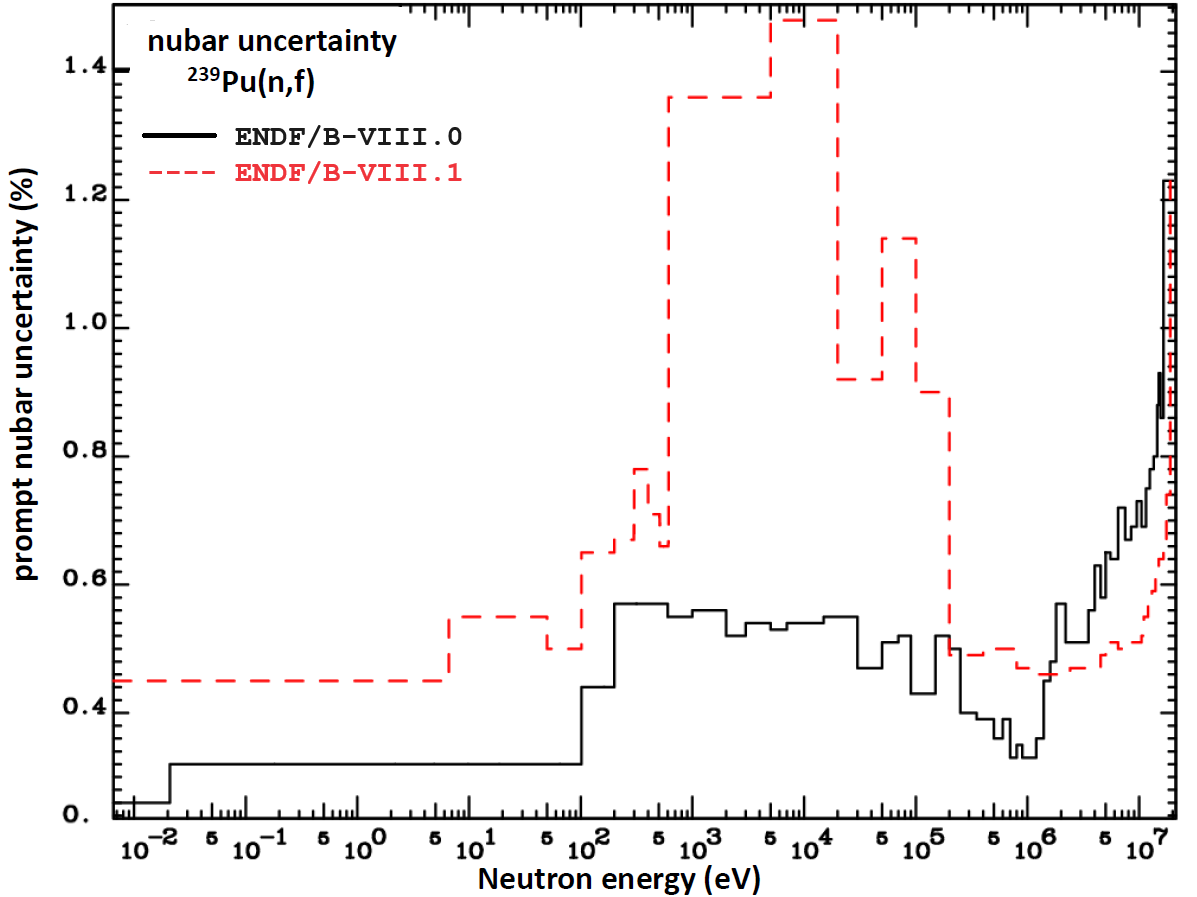}
\caption{Evaluated $^{239}$Pu $\overline\nu$ uncertainty in ENDF/B-VIII.1 versus ENDF/B-VIII.0.}
\label{fig:Pu239nu-LE-unc}
\end{figure}

The evaluated $\overline\nu$ uncertainty for n+$^{239}$Pu is shown in Fig.~\ref{fig:Pu239nu-LE-unc} in a logarithmic energy scale to emphasize the resonance range. The current evaluated $\overline\nu$ uncertainty has increased significantly up to 1.2~MeV of incident neutron energy (minimum uncertainty around 0.44\%) due to the increase in the evaluated $\overline\nu$ uncertainty of the $^{252}$Cf(sf) standard from 0.13\% up to 0.42\% \cite{carlson2018}, and its impact on experimental data measured as a ratio to the $^{252}$Cf(sf) $\overline\nu$ standard.

Additionally, the $\overline\nu$ uncertainty from 500~eV up to 80~keV has increased up to 1.5\% to consider the Gwin 1986 experimental data \cite{Gwin:1986-nubar} (also measured relative to the $^{252}$Cf(sf) standard). No other measurements are available from about 10~eV up to ~80~keV of neutron incident energy.

\begin{figure}[htb!]
\centering
\includegraphics[width=\columnwidth]{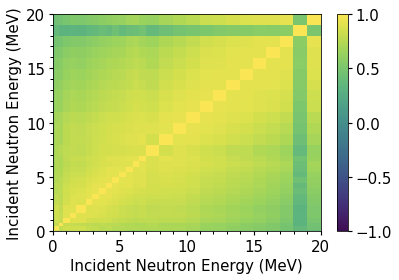}
\caption{Evaluated $^{239}$Pu $\overline\nu$ correlation matrix.}
\label{fig:Pu239nucor}
\end{figure}

The ENDF/B-VIII.1 $\overline\nu$ evaluated uncertainties in the fast-neutron range in Fig.~\ref{fig:Pu239nu-LE-unc} are larger than in ENDF/B-VIII.0 from 80~keV to approximately 2~MeV due to the increased $^{252}$Cf(sf) $\overline\nu$ standard uncertainty (0.42\%). The evaluated uncertainties are smaller than ENDF/B-VIII.0 above 2~MeV due to high-precision data from Marini \etal \cite{Marini2021,Marini2024} as discussed below. The correlation matrix in Fig.~\ref{fig:Pu239nucor} is strongly and positively correlated due to the full correlation stemming from the dominant $^{252}$Cf(sf) $\overline\nu$ standard uncertainty. A reevaluation of the $^{252}$Cf(sf) $\overline\nu$ standard is planned for ENDF/B-IX.0. 


The neutron-induced average prompt fission neutron multiplicity, $\overline\nu$, for n+$^{239}$Pu in the fast-neutron range (above 80~keV of incident neutron energy) was reevaluated from scratch for ENDF/B-VIII.0~\cite{Neudecker:2023Pu9fissionsourceterm,Neudecker2021nu}.
Experimental-data uncertainties were reevaluated using templates of expected $\overline\nu$ measurement uncertainties~\cite{nu} and information provided in EXFOR and published papers.
The very recent high-precision data by Marini \etal~\cite{Marini2021,Marini2024} for incident energies above 700~keV were included in the new evaluation. Unfortunately, no measurement was carried out at the thermal incident energy. New Marini~\etal measurements validated the evaluated ENDF/B-VIII.0 prompt fission neutron multiplicity from 700~keV up to about 5~MeV~\cite{Marini2024}.

\begin{figure}[!thb]
\vspace{-3mm}
\centering
\includegraphics[width=\columnwidth]{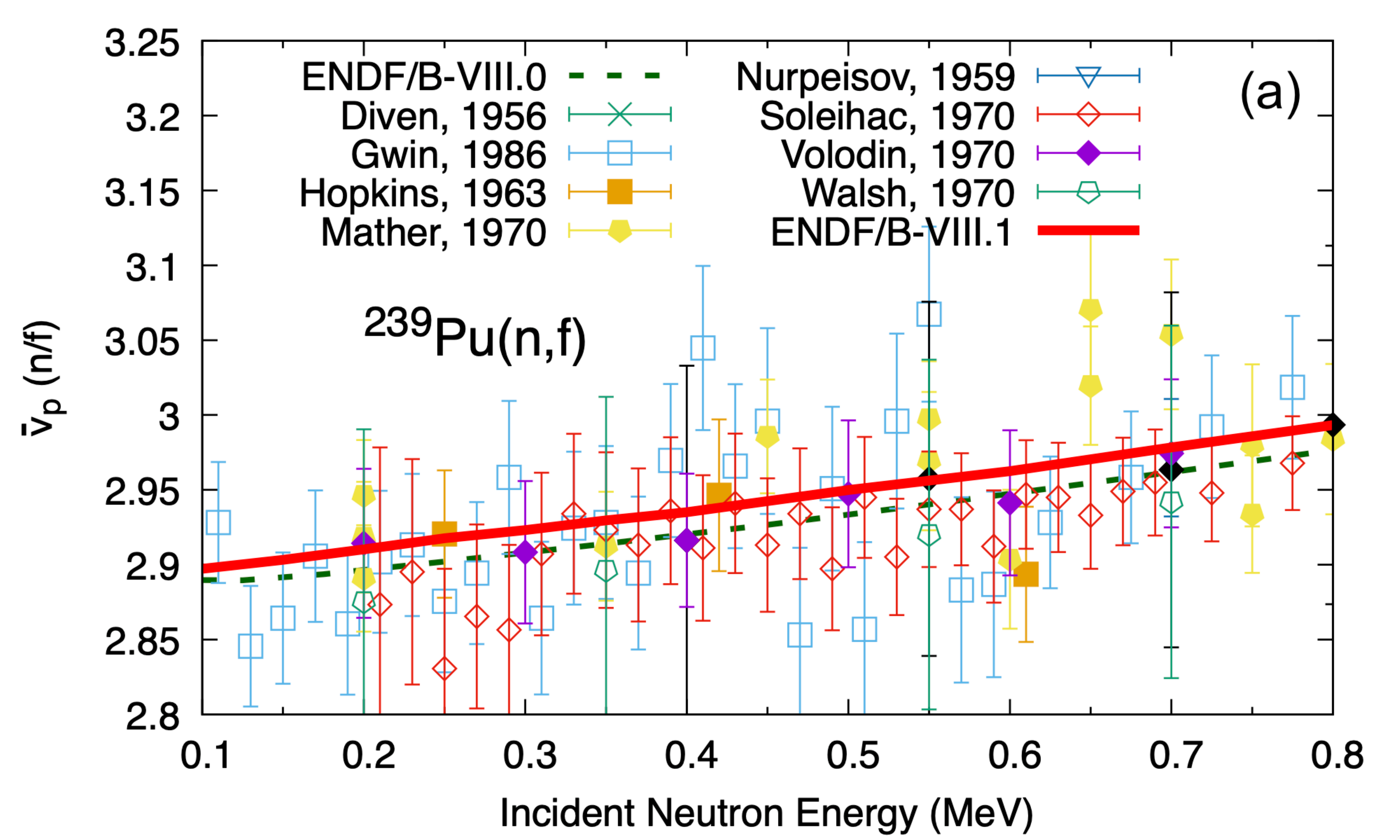}
\includegraphics[width=\columnwidth]{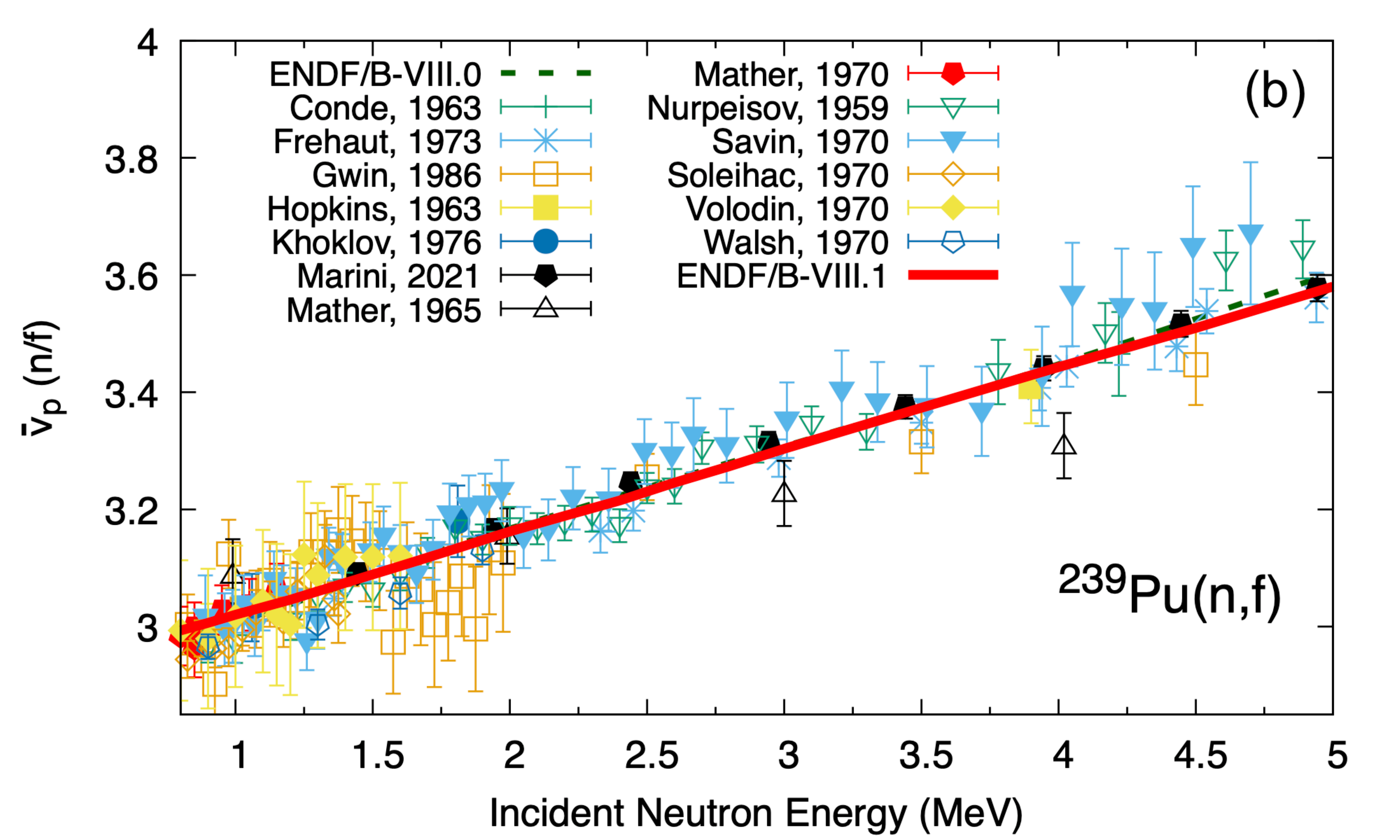}
\includegraphics[width=\columnwidth]{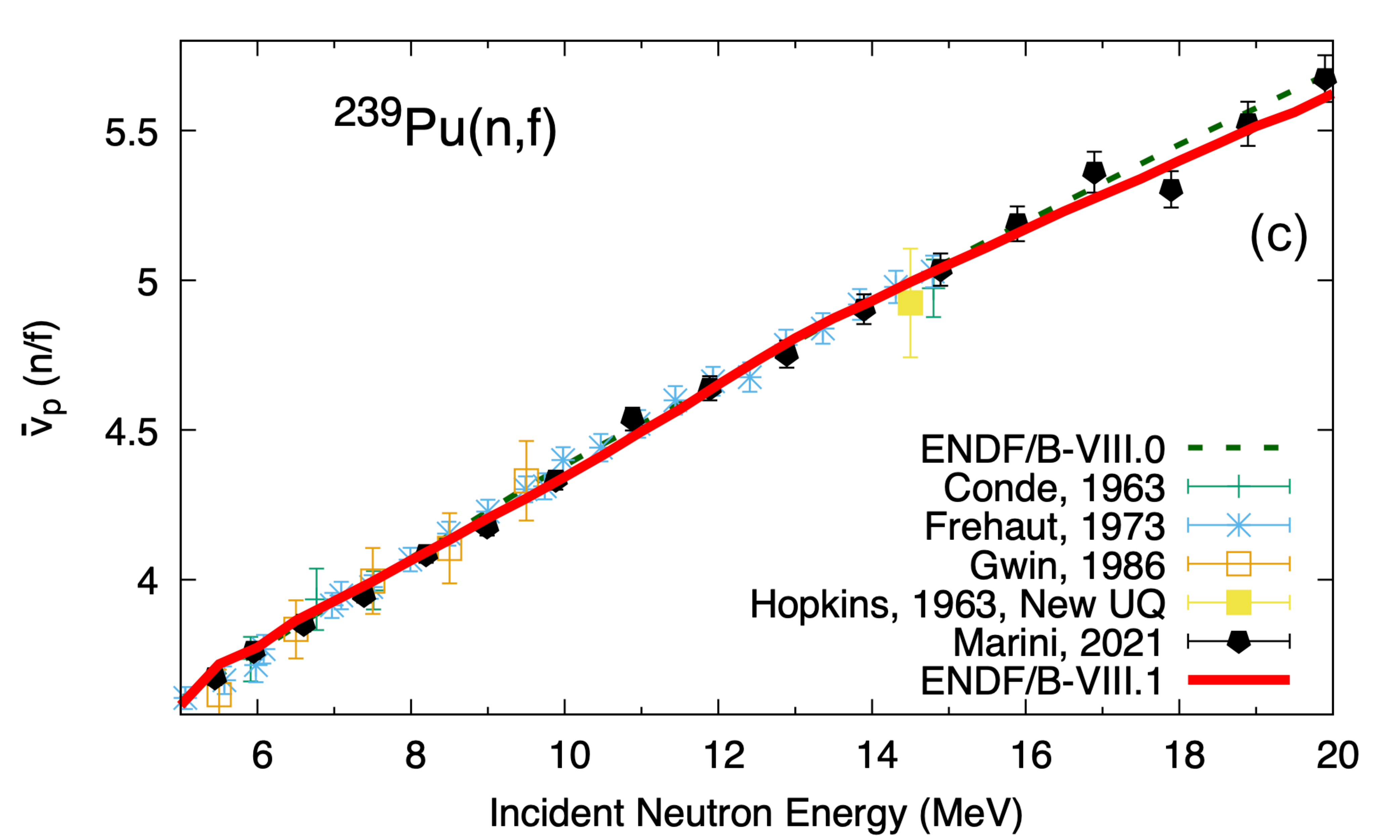}
\caption{Selected experimental data \cite{Diven:1955,Hopkins:1963,Gwin:1986,MARINI2022137513,Mather:1965,Nurpeisov:1975,Volodin:1972,Savin:1970,Soleihac:1970,Walsh:1971,Conde:1968,Frehaut:1973,SavinKhokhlov:1973} taken from EXFOR \cite{EXFOR} and evaluated $^{239}$Pu $\overline\nu$ in the fast neutron range  compared to \prENDF\ (dashed black line) and \ENDF\ (solid line) for 0.1~MeV to  0.8~MeV (a), 0.9~MeV to 5~MeV (b) and 5~MeV to 20~MeV (c) neutron incident energies.}
\label{fig:Pu239nu}
\vspace{-2mm}
\end{figure}

\begin{figure}[!thb]
\vspace{-2mm}
\centering
\includegraphics[width=\columnwidth]{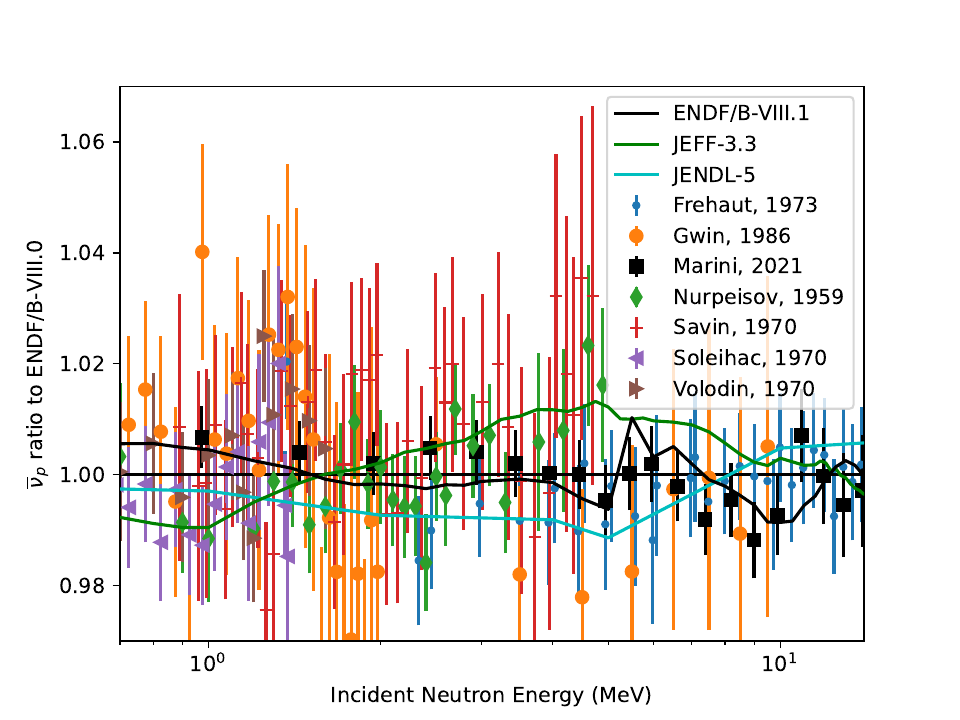}
\vspace{-4mm}
\caption{Evaluated $^{239}$Pu $\overline\nu$ ratio to \prENDF{} for selected experimental data \cite{Frehaut:1973,Gwin:1986,
MARINI2022137513,Nurpeisov:1975,Savin:1970,Soleihac:1970,Volodin:1972} and for JENDL-5 \cite{jendl5}, JEFF-3.3 \cite{JEFF33}, and the 
new ENDF/B-VIII.1 evaluations.}
\label{fig:pu239-nubar-ratio-B8}
\vspace{-2mm}
\end{figure}

For the very first time, the \CGMF\ fission-event generator~\cite{Talou2021} was used as a model to evaluate $\overline\nu$.
Including that model in the evaluation made it possible to employ evaluated model parameters to compute other fission quantities, such as yields as a function of mass, the average total kinetic energy of fission fragments, \textit{etc.} It was shown in Ref.~\cite{Neudecker:2023Pu9fissionsourceterm} that most of these predicted fission quantities agree well with experimental data except for the PFNS -- a known model deficiency that is being investigated. This combined new information led to distinct $\overline\nu$ changes in the fast energy range from ENDF/B-VIII.0 as shown in Figs.~\ref{fig:Pu239nu} and \ref{fig:pu239-nubar-ratio-B8}.

A new $\overline\nu$ evaluation is compared to JENDL-5, JEFF-3.3 evaluations and selected experimental data and plotted as a ratio to \prENDF{} in Fig.~\ref{fig:pu239-nubar-ratio-B8}. The new evaluation is up to 0.5\% higher than the \prENDF{} below 1.5~MeV (the curves cross around 1.5~MeV) and is in reasonable agreement with Marini data~\cite{Marini2021,Marini2024} within the experiment's uncertainties except at 2.2, 5.5, and 12~MeV.

From 5~MeV up to 5.5~MeV, the $\overline\nu$ jump predicted by the model at the opening of the second fission chance is barely (if at all) seen in the data. The new evaluation overestimates the Marini data around 5~MeV by about 0.5\%, but the agreement with Marini data improves above 6~MeV. 

The changed $\overline\nu$ resulted in a slight increase in Jezebel $k_\mathrm{eff}$ (about 100--150 pcm) that was counter-balanced by the criticality reduction induced by a new and softer n+$^{239}$Pu PFNS. However, the disagreement of model-predicted PFNS versus the experimental PFNS data implies that strong correlations brought by the model into the evaluation should be reviewed and models improved. 

The evaluated $\overline\nu$ described above was increased at the IAEA by up to 0.1\%\footnote{representing about +20~pcm of reactivity worth} in the energy range 1.5--4~MeV (scaling factors: 1.0 at 1.5~MeV, 1.001 at 2.5--3~MeV, and 1.0 at 4.0~MeV, linear interpolation) to calculate the Jezebel $k_\mathrm{eff}\equiv 1$ for the Jezebel v5 of the benchmark computational model. It is worth noting that such increase improves the agreement of the evaluated $\overline\nu$ with Marini data~\cite{Marini2021,Marini2024}. The Jezebel v5 \ql calibration\qr was recommended by the expert CSEWG group led by R. Little. The $\overline\nu$-tweak was well within the assessed relative uncertainty of about 0.5\%. No changes in the $\overline\nu$-covariance matrix were introduced.
\begin{figure}
\centering
\includegraphics[width=\columnwidth]{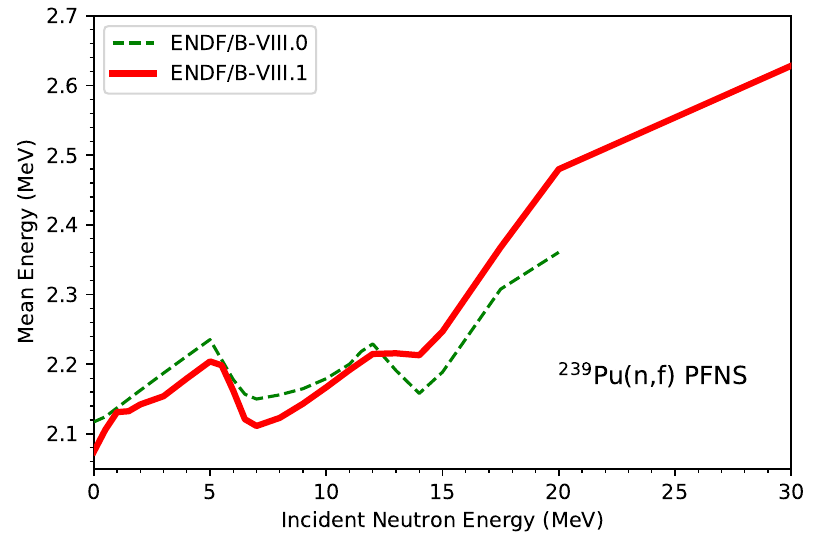}
\caption{Evaluated $^{239}$Pu PFNS mean energy.}
\label{fig:Pu239PFNSmeanenergy}
\vspace{-2mm}
\end{figure}

\begin{figure}[!thb]
\vspace{-1mm}
\centering
\includegraphics[width=\columnwidth]{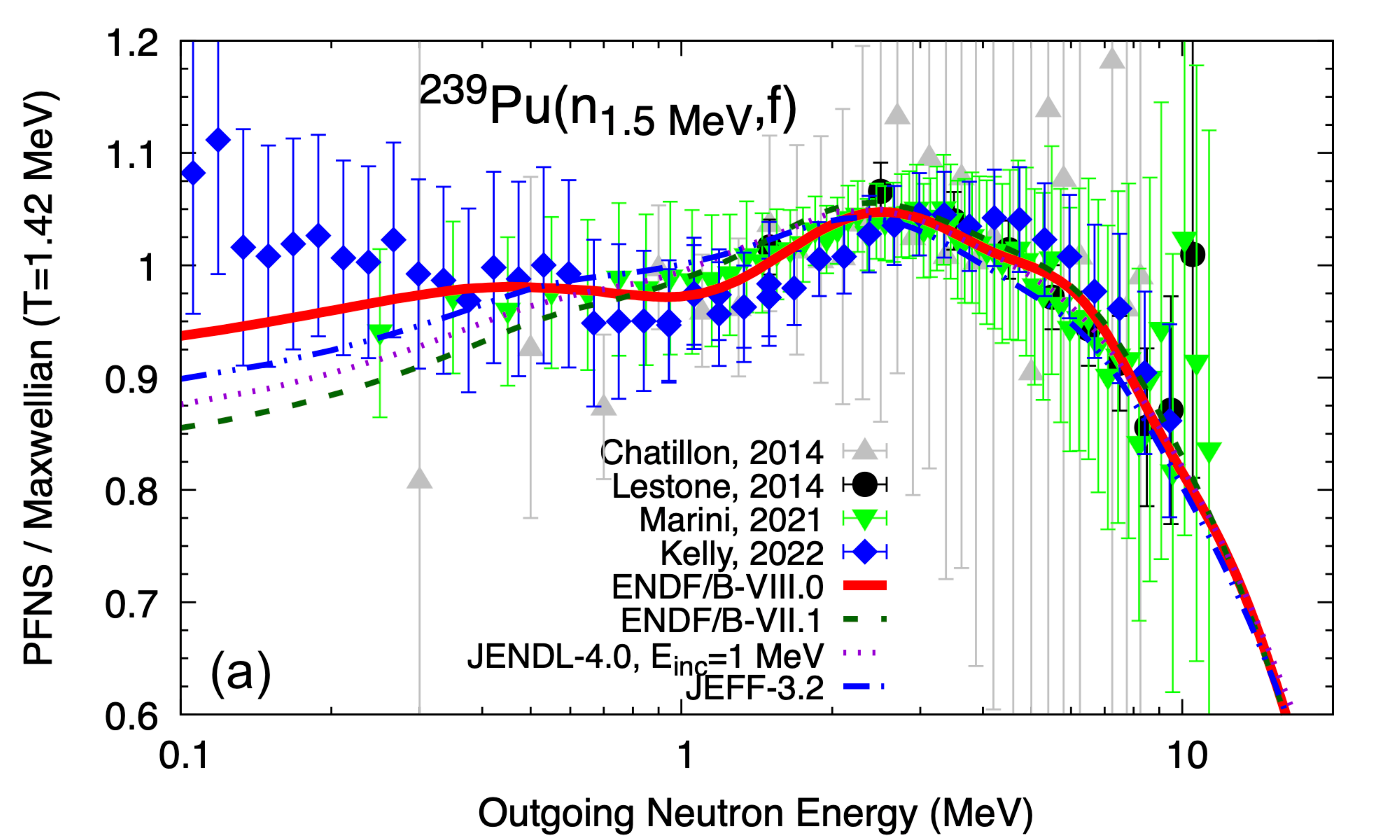}
\includegraphics[width=\columnwidth]{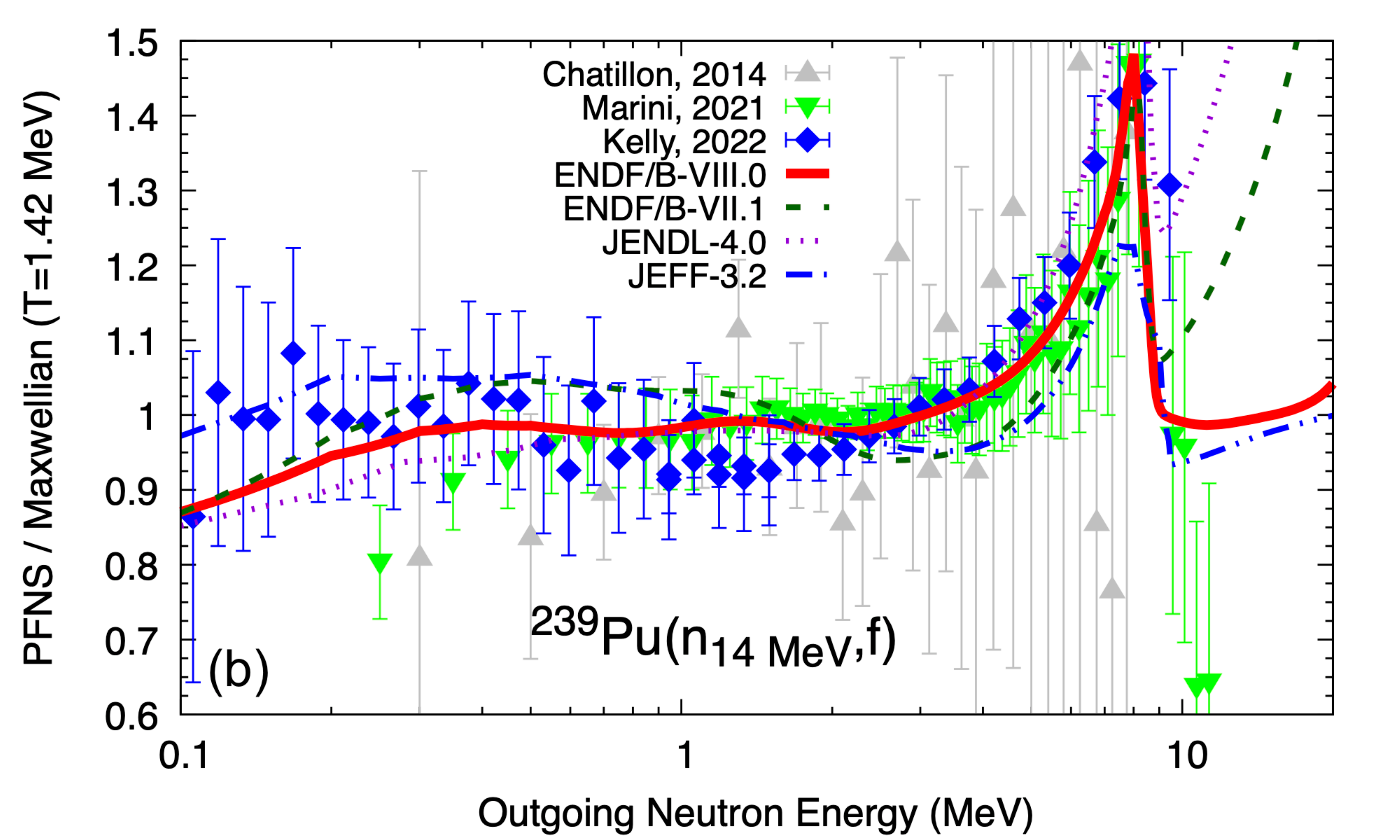}
\caption{Selected experimental data \cite{Marini2021,Lestone:2014,Kelly2021,Chatillon2014} from EXFOR \cite{EXFOR} and 
evaluated $^{239}$Pu PFNS for $E_\mathrm{inc}$=1.5 (a) and 14~MeV (b).}
\label{fig:Pu239PFNS}
\vspace{-3mm}
\end{figure}

\subsubsection{Prompt Fission Neutron Spectrum}
The n+$^{239}$Pu PFNS at thermal energies was adopted from the IAEA CRP work \cite{capote:2016} which used a generalized least-square fit to available PFNS experimental data for the three major fissile actinides: $^{233,235}$U and $^{239}$Pu. A common feature of the IAEA-evaluated PFNS at the thermal point was the reduction of the PFNS average energy by about 30~keV compared to the ENDF/B-VII.1 values for all three fissile nuclei. Such reduction required important changes in the evaluated resonance parameters and neutron multiplicity below 5~eV to compensate for the increased criticality induced by softer PFNS. Those changes were introduced for $^{235}$U in the ENDF/B-VIII.0 library \cite{capote2018,Brown2018}, but the IAEA-evaluated PFNS at the thermal point were only adopted for $^{233}$U and $^{239}$Pu targets in the current ENDF/B-VIII.1 library release.

PFNS at all other energies were reevaluated as described in Refs.~\cite{Neudecker:2023Pu9fissionsourceterm,Neudecker2022PFNS} and built on PFNS work included in ENDF/B-VIII.0~\cite{Neudecker2018PFNS}.
The main difference is the inclusion of high-precision $^{239}$Pu PFNS measured by the CEA and Chi-Nu teams~\cite{Kelly2021,NNSACEA,Marini2020,PhysRevLett.122.072503} and Lestone data at $E_\mathrm{inc}$=2~MeV~\cite{Lestone:2014}.
The evaluated PFNS differ distinctly at all incident-neutron energies from ENDF/B-VIII.0 as can be seen in the change of mean energy in Fig.~\ref{fig:Pu239PFNSmeanenergy}.
The new PFNS evaluation follows more closely the new experimental data sets as can be seen for $E_\mathrm{inc}$=1.5 and 14~MeV in Fig.~\ref{fig:Pu239PFNS}.
In Fig.~\ref{fig:Pu239PFNSmeanenergy}, the average mean energy shows a non-linear behavior from thermal to 1~MeV, where we would expect a linear one from a physics point of view.
The reason for that is we prioritized following the high-precision experiments starting at 0.5~MeV rather than  obtaining a smooth trend in the mean energy for the first-chance fission. New PFNS experiments would be needed in the low energy range, especially at the thermal up to 1~MeV, to better evaluate the PFNS in this energy range and confirm the IAEA PFNS evaluation at the thermal point.

This distinct change in the PFNS led to a minor decrease in Jezebel and Flattop-Pu $k_\mathrm{eff}$ values by 128(1) and 114(1)~pcm~\cite{Neudecker:2023Pu9fissionsourceterm,Neudecker2022PFNS}.
It was counter-balanced by other changes in $^{239}$Pu, such as in the average prompt fission neutron multiplicity. The changes in the prediction of the Pu LLNL pulsed-sphere neutron-leakage spectra were modest~\cite{Neudecker:2023Pu9fissionsourceterm,Neudecker2022PFNS}.

\begin{figure}[!tphb]
\vspace{-4mm}
\centering
\includegraphics[width=\columnwidth, clip, trim = 0mm 0mm 0mm 5mm]{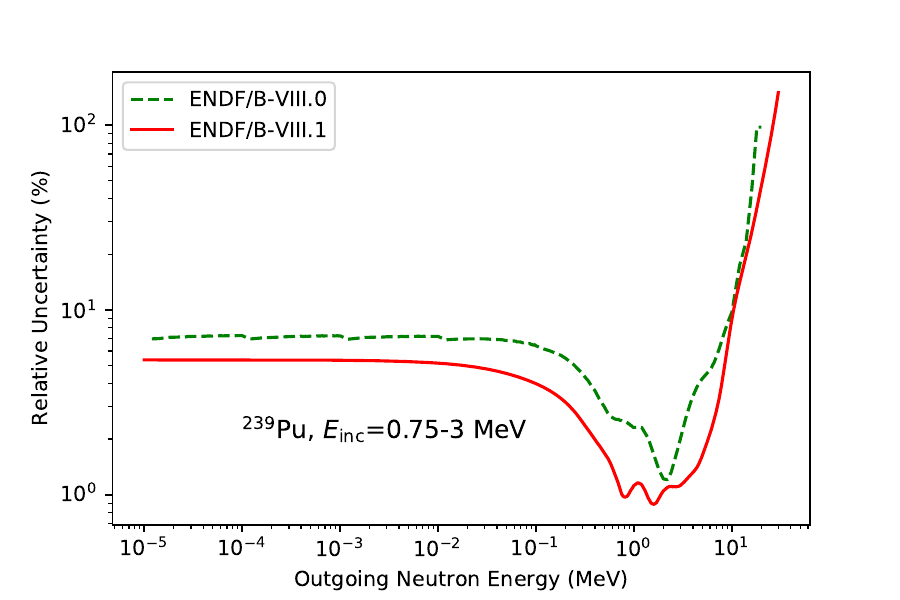}
\caption{Evaluated $^{239}$Pu PFNS uncertainties for $E_\mathrm{inc}$=0.75--3~MeV.}
\label{fig:Pu239PFNSUnc}
\vspace{-2mm}
\end{figure}

Evaluated PFNS covariances are provided for seven $E_\mathrm{inc}$ bins.
Their mean energy uncertainties, $\delta \langle E \rangle$, are listed in Table~\ref{tab:Pu239PFNSmeanenergyunc} at thermal and from 0.75--3~MeV, where most experimental data are available.
In the $E_\mathrm{inc}$ bin spanning 0.75--3~MeV  (Fig.~\ref{fig:Pu239PFNSUnc}), the low uncertainties are driven by Marini, Chi-Nu and Lestone data at 1.5~MeV~\cite{Kelly2021,NNSACEA,Marini2020,Lestone:2014}.

\begin{table}
\vspace{-2mm}
\centering
\caption{\label{tab:Pu239PFNSmeanenergyunc} The mean energy uncertainties, $\delta \langle E \rangle$, for the $^{239}$Pu PFNS are listed per incident-neutron energy, $E_\mathrm{inc}$.  For ENDF/B-VIII.0, the mean energy uncertainty was 37~keV for thermal--5~MeV.}
\begin{tabular}{lc}
\toprule \toprule
$E_\mathrm{inc}$ & $\delta \langle E \rangle$ (keV) \\
\midrule
Thermal & 16 \\
0.25--0.75 & 37 \\
0.75--3 & 15 \\
3--5 & 33 \\
5--6.5 & 33 \\
6.5--13 & 34 \\
13--30 & 40 \\
\bottomrule \bottomrule
\end{tabular}
\vspace{-3mm}
\end{table}

\subsubsection{Resonance Region}
The resolved resonance region (RRR) and unresolved resonance region (URR) parameters have been updated by M. Pigni \etal \cite{Pigni2022} within the INDEN collaboration since \prENDF{}.

Similar to the \nuc{235}{U} evaluation, the R-matrix analysis of the n+\nuc{239}{Pu} system is an ongoing milestone of the Nuclear Criticality Safety Program (NCSP), and, in the ENDF/B-VIII.1 nuclear data library release, the major updates in chronological order to this evaluation work were the extension of the RRR evaluation up to 5~keV and the updates in the thermal and low-energy region up to a few eVs. The extension to 5~keV of the RRR improved benchmark performance for integral experiments that are not fully thermalized. The updates below 5~eV were to guarantee an excellent performance in criticality benchmarks affected by the changes in the PFNS evaluation, together with the adoption of the thermal constants recommended by the standard evaluation \cite{carlson2018}, and to address issues in depletion calculations at high burnup, a known ENDF/B-VIII.0 deficiency~\cite{kim2021-318}.

\begin{figure}[!htpb]
\centering
\includegraphics[width=0.98\columnwidth,clip, trim = 115mm 95mm 80mm 95mm]{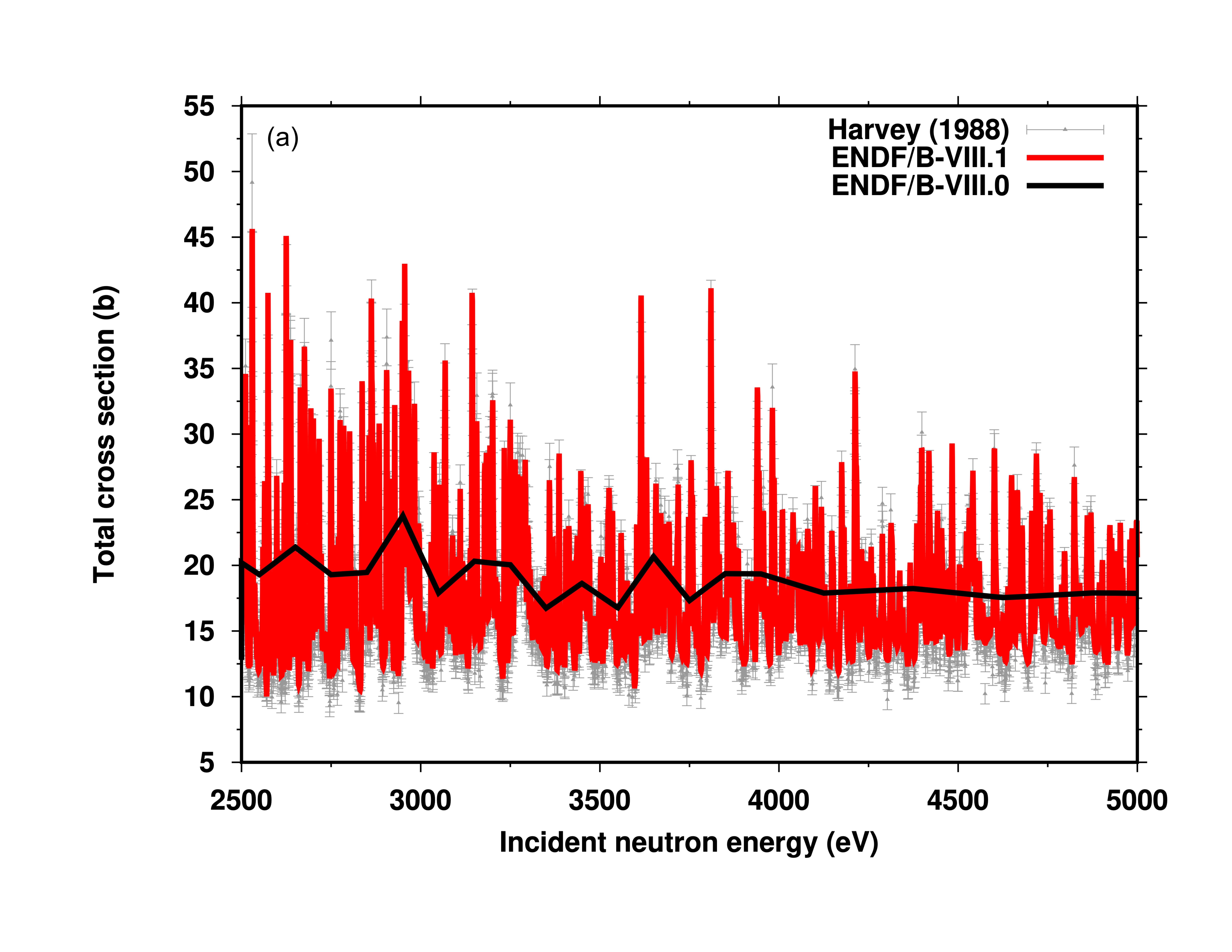}
\includegraphics[width=0.98\columnwidth,clip, trim = 115mm 95mm 80mm 95mm]{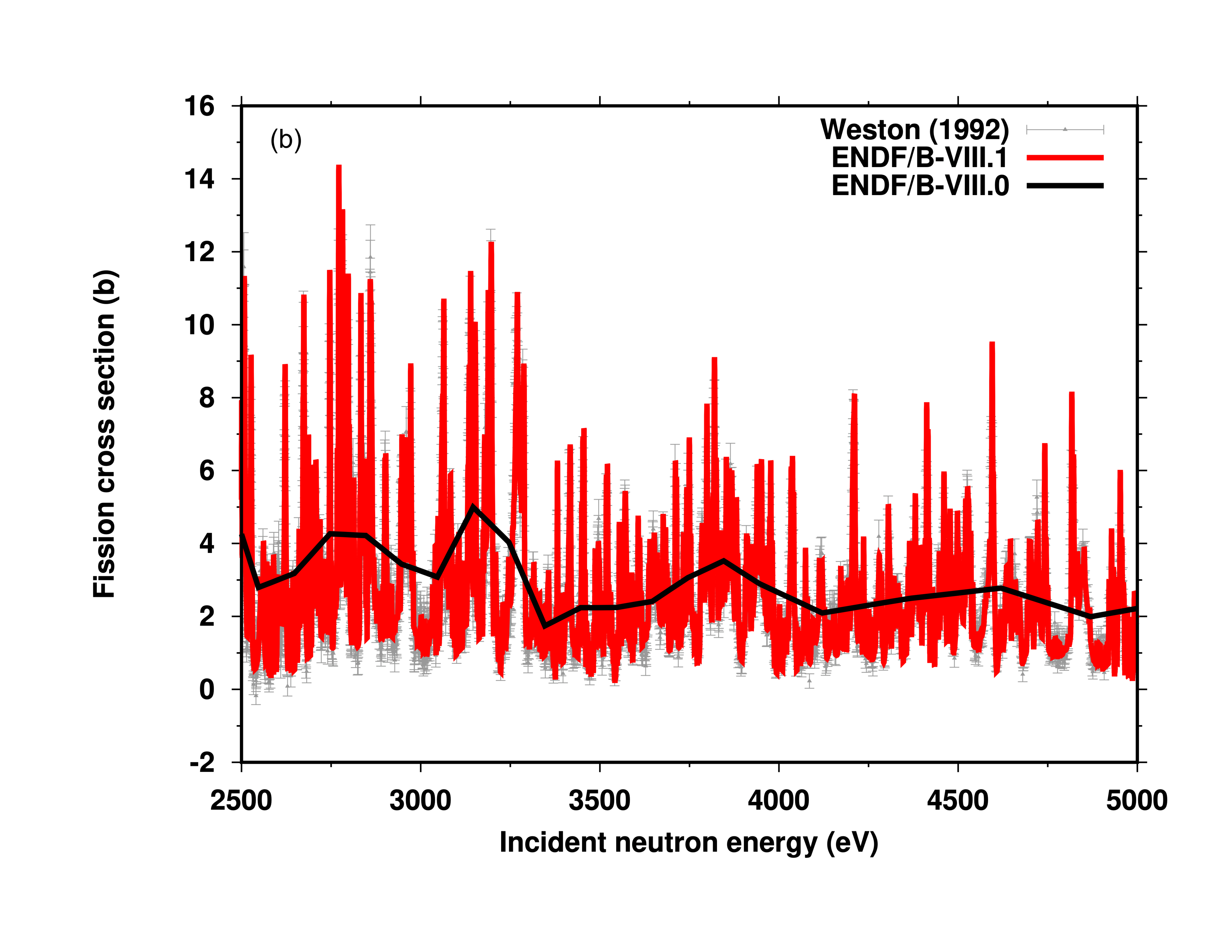}
\vspace{-.1in}
\caption{n+\nuc{239}{Pu} total and fission neutron cross section calculated for two nuclear data libraries together with Harvey's (a) and Weston's (b) measured data~\cite{harvey:1988, weston:1992}.}\label{fig:pu239_extension}
\end{figure}

Starting from energy levels and resonance widths randomly distributed from Wigner and Porter-Thomas distributions, respectively, the extension of the RRR evaluation from 2.5~keV up to 5~keV used both Harvey's thick-target data transmission and Weston's fission cross sections\footnote{Weston's fission data were aligned to the Harvey's thick-target transmission data.}. Due to the lack of measured capture data in this energy range, the cross sections of this reaction channel were obtained by using an average value of 40~meV assigned to the capture widths consistently with the Reich-Moore (RM) approximation of the R-matrix theory. In this regard, Fig.~\ref{fig:pu239_extension} highlights the neutron-energy region between 2.5--5~keV comparing the ENDF/B-VIII.0 (in black) and ENDF/B-VIII.1 (in red) evaluation together with Harvey's total cross sections and Weston's fission cross section. The Bayesian sequential fit of the measured data generated reasonable agreement. The fidelity of the newly released ENDF/B-VIII.1 library has been enormously improved in comparison to the ENDF/B-VIII.0 evaluation, particularly in view of the fluctuating behavior of the measured data. Although the $\chi^{2}$ value for both cases is satisfactory (also most likely due to the large number of degrees of freedom), the fit of the fission cross section was easier to achieve than the fit of the transmission data.

\begin{figure}[!htbp]
\vspace{-2mm}
\centering
\includegraphics[width=\columnwidth]{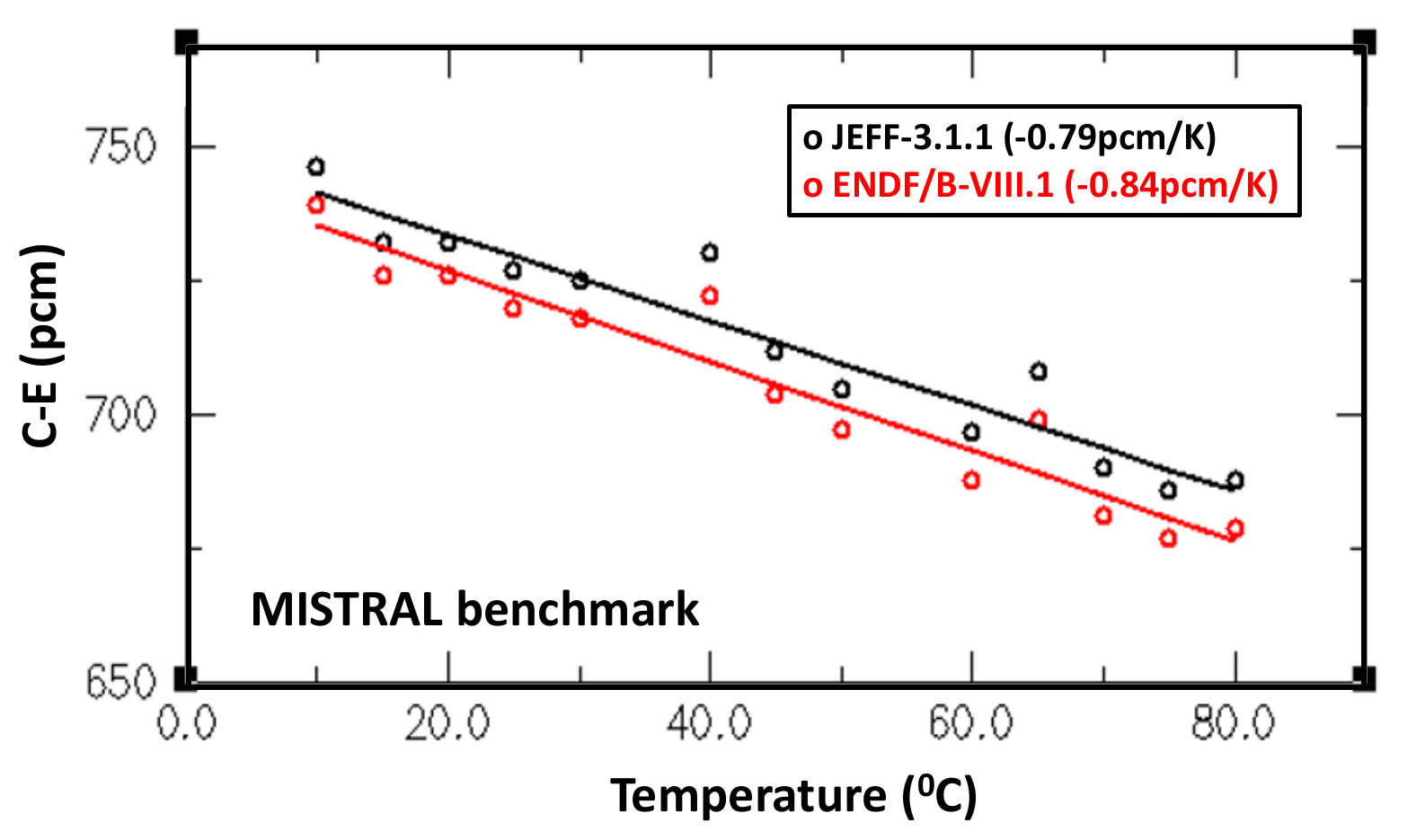}
\caption{MISTRAL benchmark reactivity temperature coefficient $C-E$ (pcm) for the ENDF/B-VIII.1 evaluation versus the JEFF-3.1.1 reference result.}\label{fig:pu239-mistral}
\vspace{-2mm}
\end{figure}

In the low neutron-energy region below around 5~eV, the performance of the ENDF/B-VIII.1 evaluation was tested on the basis of multiple integral constraints, i.e., depletion calculations coupled to criticality safety benchmarks and the CEA Mistral-2 experiment that are particularly sensitive to the thermal and sub-thermal energy region, respectively. MISTRAL $C-E$ criticality differences as a function of temperature are compared versus the JEFF-3.1.1 reference result in Fig.~\ref{fig:pu239-mistral}. There is a known bias in the calculated results, but the important quantity is the slope which defines the reactivity temperature coefficient (RTC). The RTC calculated for the ENDF/B-VIII.1 evaluations is similar to the JEFF-3.1.1 RTC reference value and agrees with data within the estimated uncertainty of around 1~pcm/K.

\begin{figure}[!tbph]
\vspace{-2mm}
\centering
\includegraphics[width=0.98\columnwidth,clip,trim = 21mm 21mm 13mm 21mm]{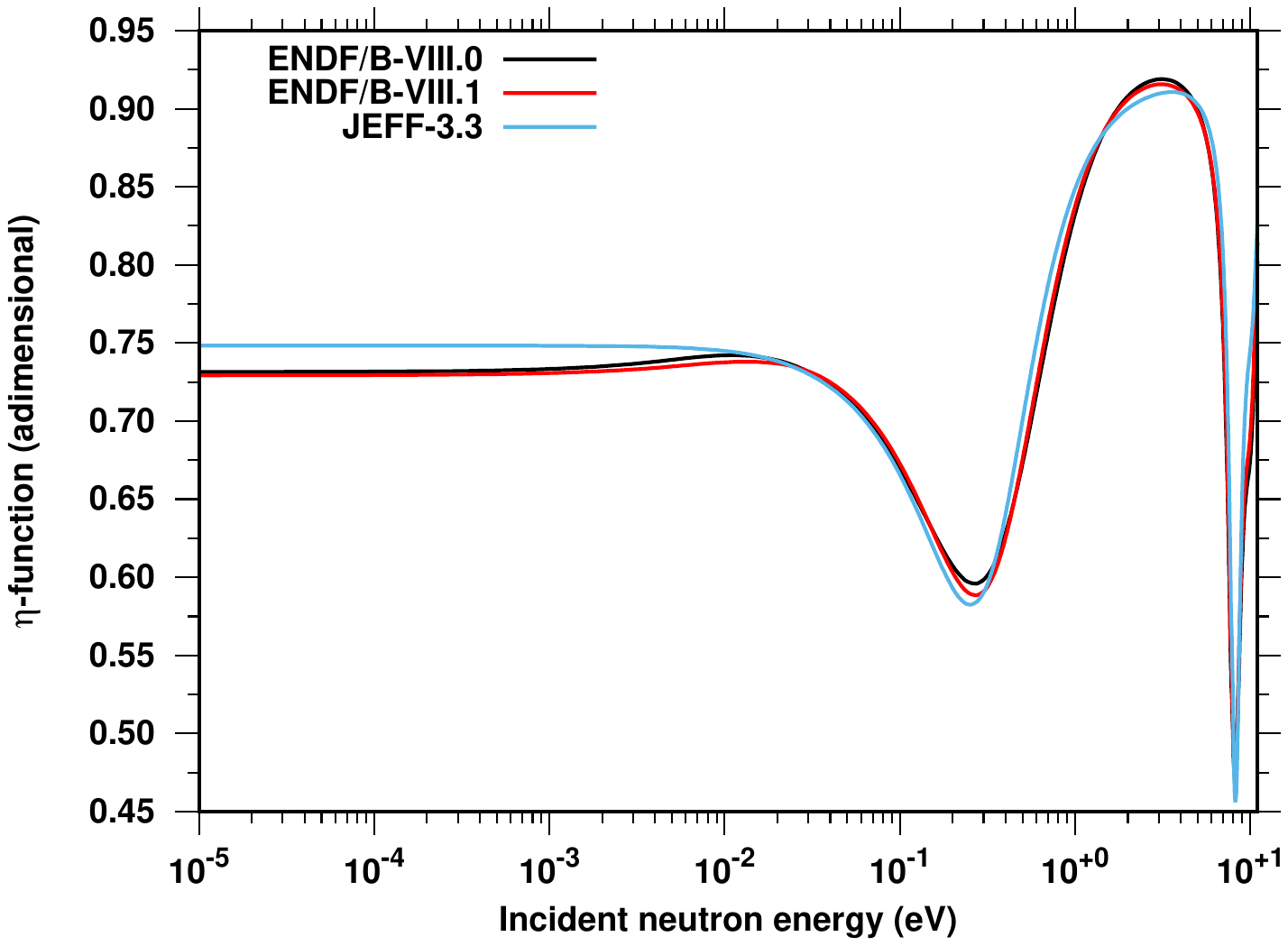}
\vspace{-.1in}
\caption{\nuc{239}{Pu} energy dependent $\eta$-function for three nuclear data libraries.}\label{fig:pu239_eta}
\end{figure}

Starting the discussion from the sub-thermal energy region, Fig.~\ref{fig:pu239_eta} shows the energy-dependent reproduction factor $\eta$-function ($\eta=\overline\nu \sigma_f / (\sigma_f + \sigma_\gamma)$) for three nuclear data libraries up to 10~eV. Besides non-negligible differences in both magnitude and energy shift of the resonance level at 0.3~eV, the energy behavior of the $\eta$-function for neutron incident energies below the thermal point $E_{n}$=0.0253~eV is significantly different, particularly with respect to the JEFF-3.3 library. In fact, sensitivity analyses of integral benchmarks showed that the non-linear sub-thermal energy trend of the $\eta$-function, together with its increased value above the thermal point up to about the 0.3~eV resonance level, are of fundamental importance to achieve a reasonable agreement with the Mistral-2 integral experiment. 
Overall, the energy behavior of the $\eta$-function for ENDF/B-VIII.0 and \ENDF\ is very similarly driven by the adopted bound states; it shows a slowly increasing trend in $\eta$ up to the thermal point where all three libraries converge. 

\begin{figure}[!htbp]
\vspace{+2mm}
\centering
\includegraphics[width=0.98\columnwidth,clip,trim = 105mm 95mm 80mm 90mm]{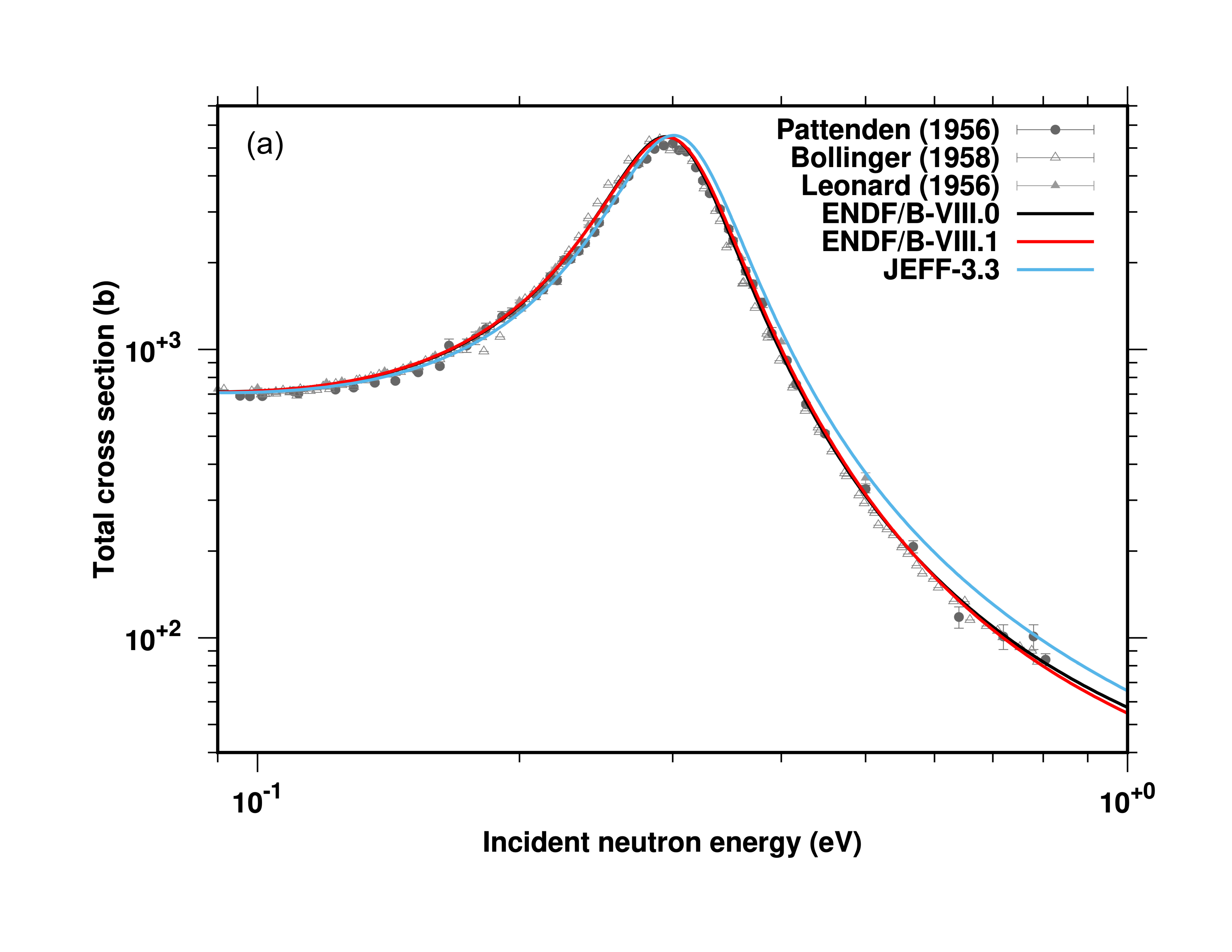}
\includegraphics[width=0.98\columnwidth,clip,trim = 105mm 95mm 80mm 90mm]{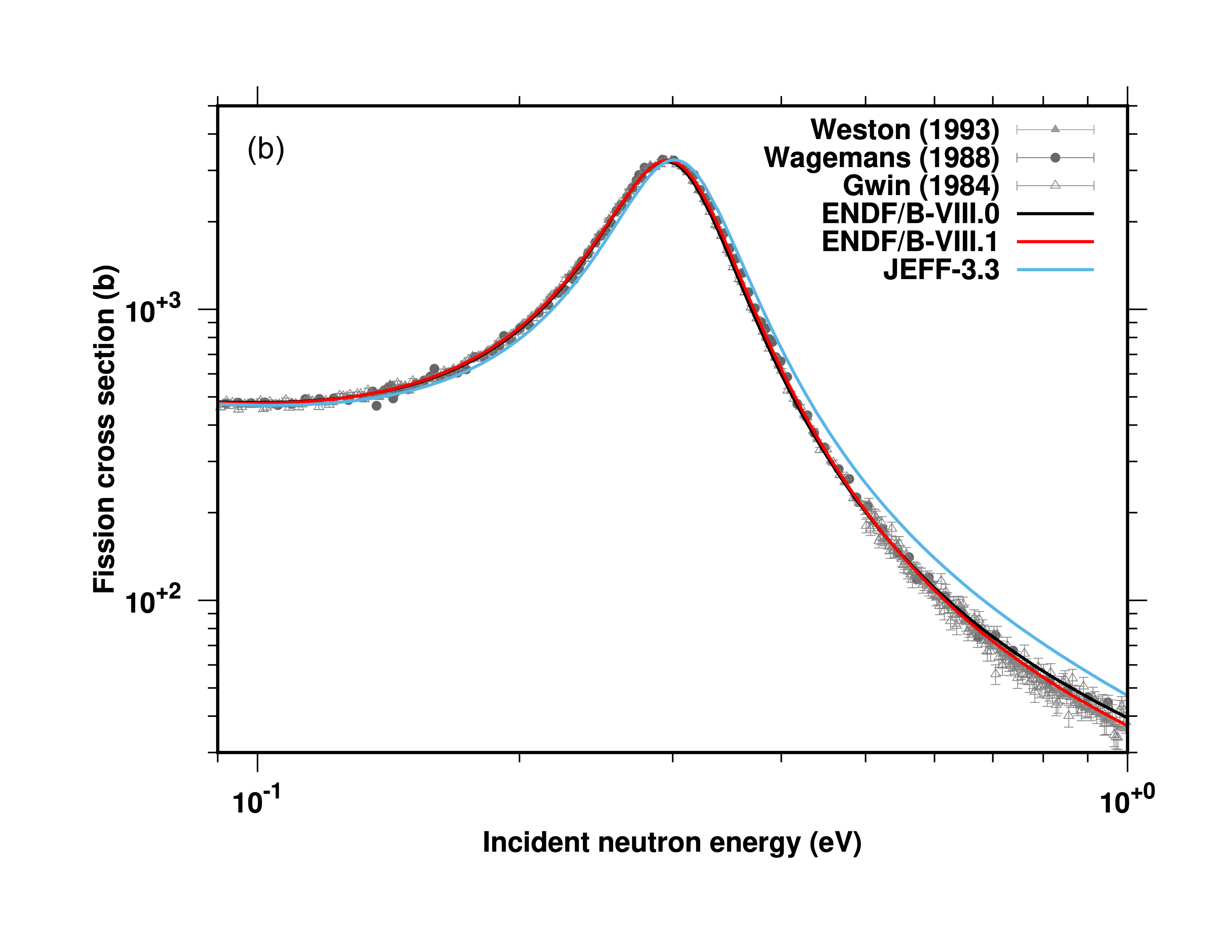}
\vspace{-.1in}
\caption{n+\nuc{239}{Pu} total and fission cross sections calculated for three nuclear data libraries together with selected measured data~\cite{wagemans:1988,gwin:1984,leonard:1956,pattenden:1956,bollinger:1958,weston:1993}. Impurity effects due to \nuc{240,242}{Pu} isotopes in the sample are not plotted.}\label{fig:pu239_xcs}
\vspace{-2mm}
\end{figure}

Above the thermal point (in the epithermal region below 2~eV), it is important that the fission-to-capture cross-section ratio is well calibrated to achieve a local minimum of the $\eta$-function at about 0.3~eV to consistently guarantee a higher \nuc{239}{Pu} production at high burnup in depletion calculations. The correct ratio of the 0.3~eV resonance level was suggested by the fit of measured data including their scaling factors as shown in Fig.~\ref{fig:pu239_xcs}, while the energy trend in the sub-thermal energy region was achieved by the energy recalibration of the bound level (negative-energy level) which is very close to zero energy. As clearly shown in these figures for the JEFF-3.3 library, an energy shift of the 0.3~eV energy level would help in having a flat gradient as a function of Energy of Average Lethargy causing Fission (EALF) in criticality benchmarks. However, this choice poorly describes the measured differential data for all reaction channels, as seen in Fig.~\ref{fig:pu239_xcs}. We extensively discussed this issue at the INDEN meeting in Vienna in May 2024 and came to the final compromise solution used in the ENDF/B-VIII.1 evaluation.  
In this regard, additional sensitivity analyses suggested that the achievement of reasonable depletion calculations is anti-correlated with having a flat gradient as a function of EALF in criticality benchmarks.
\begin{figure}[!htbp]
\vspace{-2mm}
\centering
\includegraphics[width=0.98\columnwidth,clip,trim = 21mm 21mm 13mm 21mm]{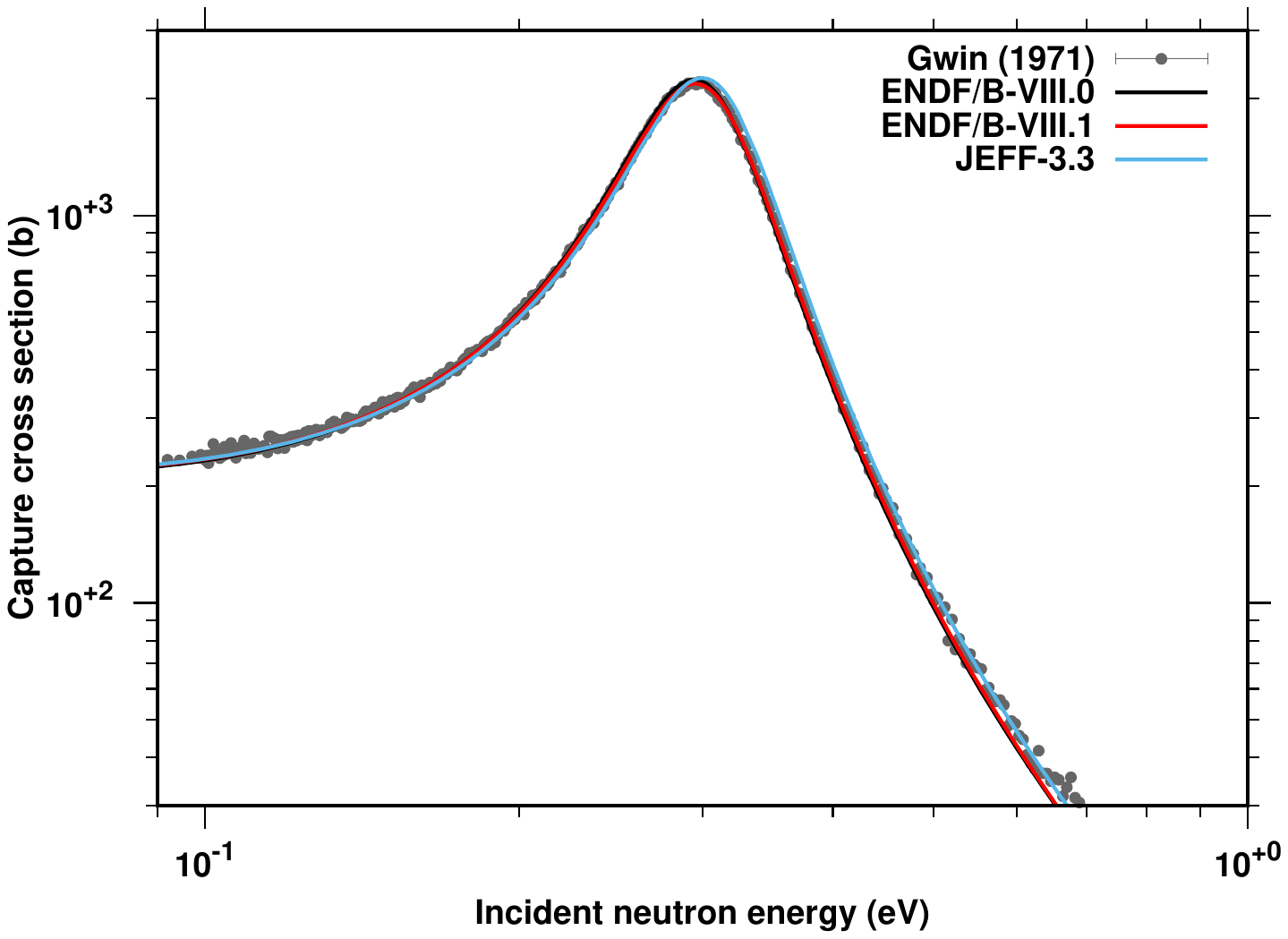}
\vspace{-.1in}
\caption{n+\nuc{239}{Pu} capture cross sections calculated for three nuclear data libraries together with Gwin's measured data~\cite{Gwin:1971}. Impurity effects due to  \nuc{240,242}{Pu} isotopes in the sample are not plotted.}\label{fig:pu239_xcs2}
\vspace{-2mm}
\end{figure}
 In Table~\ref{tab:thermal_xs_p9}, the thermal constants for scattering, fission, and capture cross sections 
 are reported for three nuclear data libraries: JEFF-3.3, ENDF/B-VIII.0 and the new ENDF/B-VIII.1 library. The thermal constant evaluated data show deviations within 1\% for fission and capture cross sections. For the elastic scattering channel, there are deviations up to -6\% with the JEFF-3.3 library.  In the current ENDF/B-VIII.1 evaluation, the thermal-elastic scattering value was estimated to be about 8~b, similar to the ENDF/B-VIII.0 release. However, the thermal value for this reaction channel is in need of a verification measurement. 

\begin{table}
\vspace{-3mm}
\caption{n+\nuc{239}{Pu} 2200 m/s thermal cross sections 
for three nuclear data libraries compared with the Thermal Neutron Constants derived by the Standard group \cite{carlson2018} and Duran \etal \cite{duran2024,duran2023} estimates.
Available experimental data by Lounsbury \etal~\cite{Lounsbury1970,Beer1972,Beer1975} are also listed. 
}\label{tab:thermal_xs_p9}
\vspace{-4mm}
\footnotesize
\begin{center}
\begin{threeparttable}
\begin{tabular}{l  c c c c}
\toprule \toprule
   Source                                 &  $\alpha_{\mathrm{therm}}$    &  $\sigma_{el}$ (b) &  $\sigma_{f}$  (b) &  $\sigma_{c}$  (b) \\
\midrule
ENDF/B-VIII.1                             & 0.3600(17)             &  8.08(8)    & 751.1(22)    & 270.4(10)    \\                   
ENDF/B-VIII.0                             & 0.3614(535)            &  8.07(7)    & 747.4(22)    & 270.1(400)    \\
JEFF-3.3                                  & 0.3622(145)            &  7.76(31)    & 749.4(64)    & 271.4(68)    \\
Standards \cite{carlson2018}              & 0.3586(35)             &  7.8(10)      & 752.4(22)    & 269.8(25)    \\
Duran \cite{duran2024,duran2023}          & 0.3588(34)             &  8.0(10)      & 751.0(19)    & 269.5(24)    \\
Lounsbury \cite{Lounsbury1970,Beer1972,Beer1975}& 0.3555(57)       &  --            & --            & --            \\
\bottomrule \bottomrule
\end{tabular}
\end{threeparttable}
\end{center}
\vspace{-4mm}
\end{table}

 In fact, among transmission data measured at ORELA in the mid- to late-eighties on Pu samples of thin, medium, and thick size~\cite{harvey:1988}, particularly those measured at the 18-m flight path extend to low energies just below the large Pu resonance at 0.25~eV. By fitting the thick-target sample, a relatively large increase of the thermal scattering cross section was derived. The increase was estimated to be up to +19\% higher than the value reported in ENDF/B-VIII.0, which stands at about 8~b. A thermal scattering cross section of 9 -- 9.5 b is largely overestimating the value of the scattering cross section reported by National Institute of Standards and Technology (NIST) based on the Sears'  compilation~\cite{Sears1992}, which is about the value adopted by the JEFF-3.3 library and, therefore, already 3\% lower than ENDF/B evaluations. A similar R-matrix analysis for the \nuc{238}{U} evaluation with transmission data extending to low energies and measured at ORELA showed similar inconsistencies. Therefore, waiting for additional experimental evidence, ORELA thick-target transmission data measured at the 18-m flight path was excluded in the current evaluation and the thermal scattering value of 8~b was adopted from the ENDF/B-VIII.0 evaluation.

The new ENDF/B-VIII.1 evaluation shows the best agreement with standard values \cite{carlson2018} as well as with thermal cross sections recommended by Duran \etal \cite{duran2024,duran2023}. An excellent agreement of evaluated ENDF/B-VIII.1 $\alpha_{therm}$ with measurements of Lounsbury \etal \cite{Lounsbury1970,Beer1972,Beer1975} is observed; evaluated values of ENDF/B-VIII.0 and JEFF-3.3 $\alpha_{therm}$ are slightly larger than Lounsbury 1-sigma uncertainty interval.

It is worth noting that the first iteration of the INDEN $^{239}$Pu RRR evaluation (ENDF/B-VIII.1$\beta$1) resulted in a MISTRAL reactivity temperature coefficient of -1.5~pcm/K, significantly worse than the JEFF-3.1.1 reference value equal to -0.79~pcm/K. This poor result was traceable to an overley large value of $\alpha_2\equiv I^c_2/I^f_2=0.653$ in the energy region from 0.1 up to 1~eV, being $I^c_2$ and $I^f_2$ -- the corresponding integrals of the capture and fission cross section in the mentioned energy interval. Calculated $\alpha_2$ values are listed in Table~\ref{tab:alpha2ratio} for ENDF/B-VII.1, ENDF/B-VIII.0, ENDF/B-VIII.1$\beta$1, and ENDF/B-VIII.1 evaluations. The $\alpha_2$ integral value determines the MISTRAL reactivity temperature coefficient and, to a large extent, the reactivity at high burnup for power reactors. If we wanted to reproduce the ENDF/B-VII.1 depletion trend, then we have to reproduce the ENDF/B-VII.1 value of $\alpha_2=0.624$. The ENDF/B-VIII.0 value was 1.6\% larger, the ENDF/B-VIII.1$\beta$1 value was an unacceptable 4.8\% larger than the ENDF/B-VII.1 $\alpha_2$ value. 

We kept the newly fit resonance parameters in ENDF/B-VIII.1$\beta$1, but adopted the ENDF/B-VIII.0 negative resonances as discussed at the INDEN May 2024 meeting at the IAEA. As a result, we achieved the resulting ENDF/B-VIII.1 value of $\alpha_2=0.623$, which is in excellent agreement with the target ENDF/B-VII.1 $\alpha_2$ value. Additional fitting to satisfy differential data constraints resulted in the final resonance parameters of the new ENDF/B-VIII.1 $^{239}$Pu evaluation.

\begin{table}[!tbh]
\vspace{-2mm}
\caption{n+\nuc{239}{Pu} capture and fission cross-section integrals $I^c_2$ and $I^f_2$ (in eVb) in the energy interval [0.1--1] eV covering the first resonance
are compared with the ratio $\alpha_2\equiv I^c_2/I^f_2$ for ENDF/B-VII.1, ENDF/B-VIII.0, ENDF/B-VIII.1$\beta1$, and ENDF/B-VIII.1 libraries}\label{tab:alpha2ratio}
\vspace{-2mm}
\small
\begin{center}
\begin{threeparttable}
\begin{tabular}{l | c c c }
\toprule \toprule
    \multirow{2}{*}{Source}                                   & $\alpha_2$ &  $I^f_2$(b.eV)  &  $I^c_2$(b.eV)\\
                    & [0.1 -- 1eV] &  [0.1 -- 1eV]     &  [0.1 -- 1eV]   \\
\midrule
ENDF/B-VIII.1                          & 0.623         & 564.3        &   351.6       \\
ENDF/B-VIII.1$\beta$1                  & 0.653         & 542.6        &   354.3       \\
ENDF/B-VIII.0                          & 0.633         & 553.6        &   350.7       \\
ENDF/B-VII.1                           & 0.624         & 558.4        &   348.6       \\
\bottomrule \bottomrule
\end{tabular}
\end{threeparttable}
\end{center}
\vspace{-4mm}
\end{table}

%
\subsubsection{Fast Region}
New data have been incorporated in the fast energy range since the last ENDF/B release.
Notable differential data include radiative capture from Mosby \cite{Mosby2018} (already used in the ENDF/B-VIII.0), fissionTPC data from Snyder \cite{Snyder2019} and (n,2n) data measured at CEA just above the threshold by M\'{e}ot and colleagues \cite{Meot2021}.

Continual progress has been made in the fast energy region with respect to integral benchmarks.
However, compensating errors in individual reaction channel data may still play a consequential role in the prediction of criticality.
Capture and (n,2n) evaluations were derived by the INDEN collaboration using a least-squares fit of experimental data as described below.
Elastic and inelastic data were calculated using nuclear models \cite{Mumpower2023a}.

\begin{figure}
 \begin{center}
 \includegraphics[width=\columnwidth]{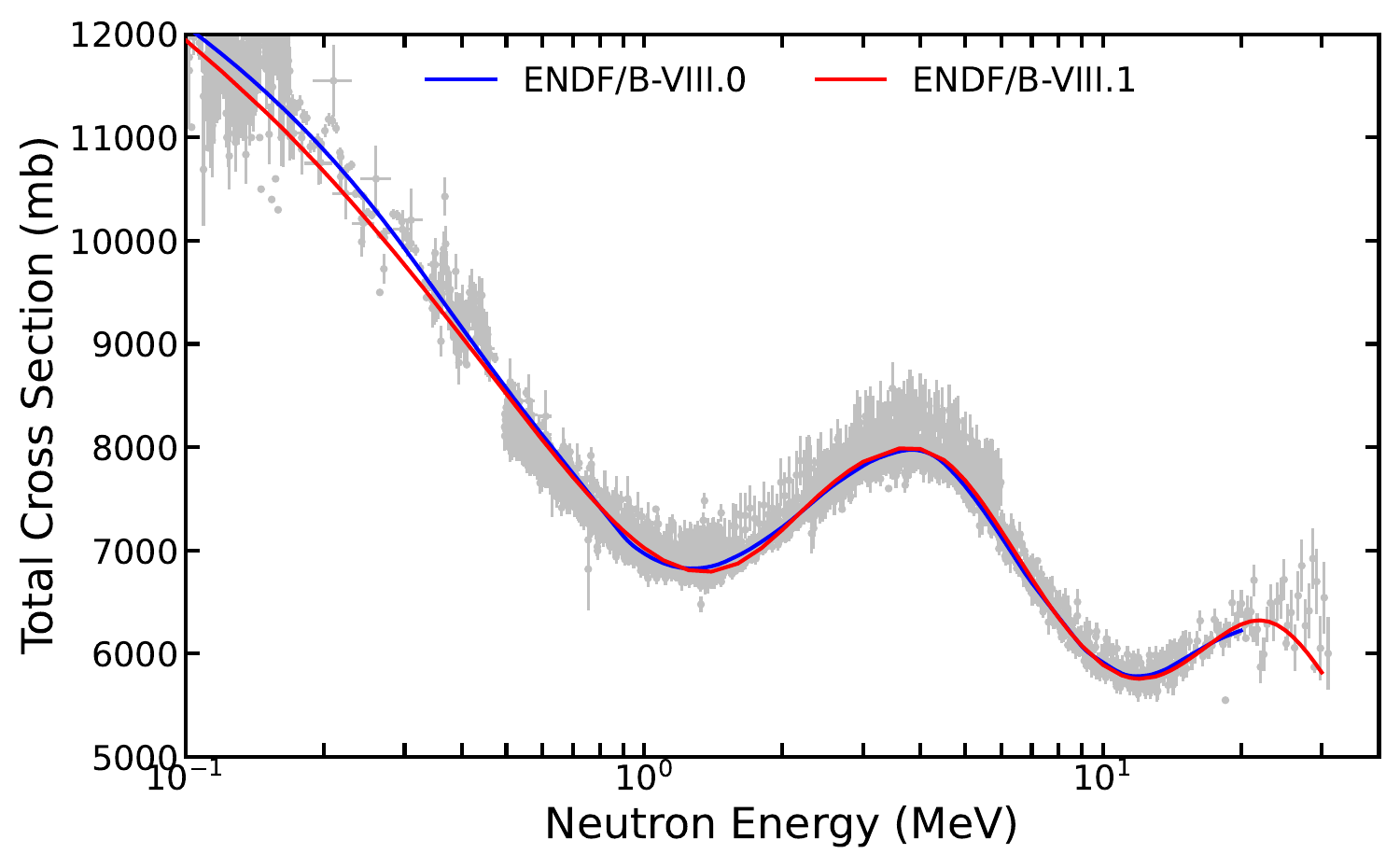}
 \caption{The total cross section in the fast energy range for \nuc{239}{Pu} along with comparison to \prENDF{}. The spread in experimental data in the EXFOR database is shown in grey \cite{EXFOR}.}
 \label{fig:cs_239pu_tot}
 \end{center}
\end{figure}

\paragraph{($n$,tot) total cross section\newline}
The new total cross section in the fast neutron range was calculated using the modern reaction code \CoH{}~\cite{Kawano2019, Kawano2021a, Kawano2021b} with input parameters carefully adjusted to reproduce the total cross-section experimental data, see Ref.~\cite{Mumpower2023a} for details. The evaluation of the total cross section is essentially equal to \prENDF{} except in the region between 100 and 400~keV where it is marginally lower.
Fig.~\ref{fig:cs_239pu_tot} highlights the present evaluation in comparison with \prENDF{}.

\begin{figure}[htb!]
\centering
\includegraphics[width=\columnwidth]{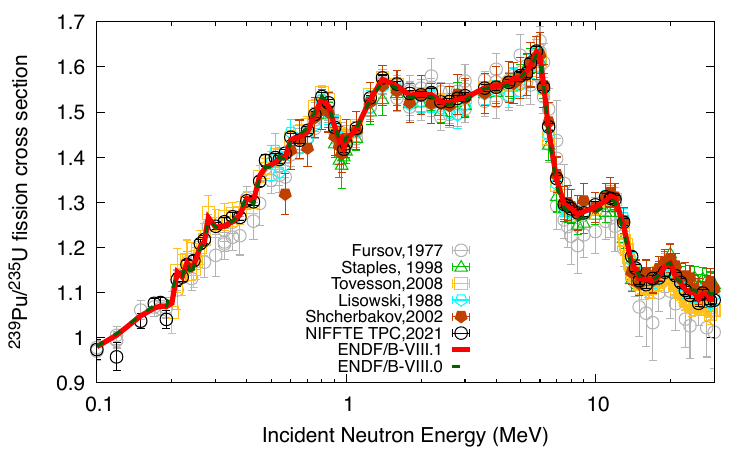}
\caption{Evaluated $^{239}$Pu/$^{235}$U cross-section ratios from ENDF/B-VIII.1 and ENDF/B-VIII.0 libraries versus
 selected experimental data \cite{Snyder2022,Tovesson2010,Staples1998,Lisowski1988,Fursov1977,Shcherbakov2002} taken from EXFOR \cite{EXFOR}.}
\label{fig:Pu239nfcs}
\end{figure}

\begin{figure}[htb!]
\centering
\includegraphics[width=\columnwidth]{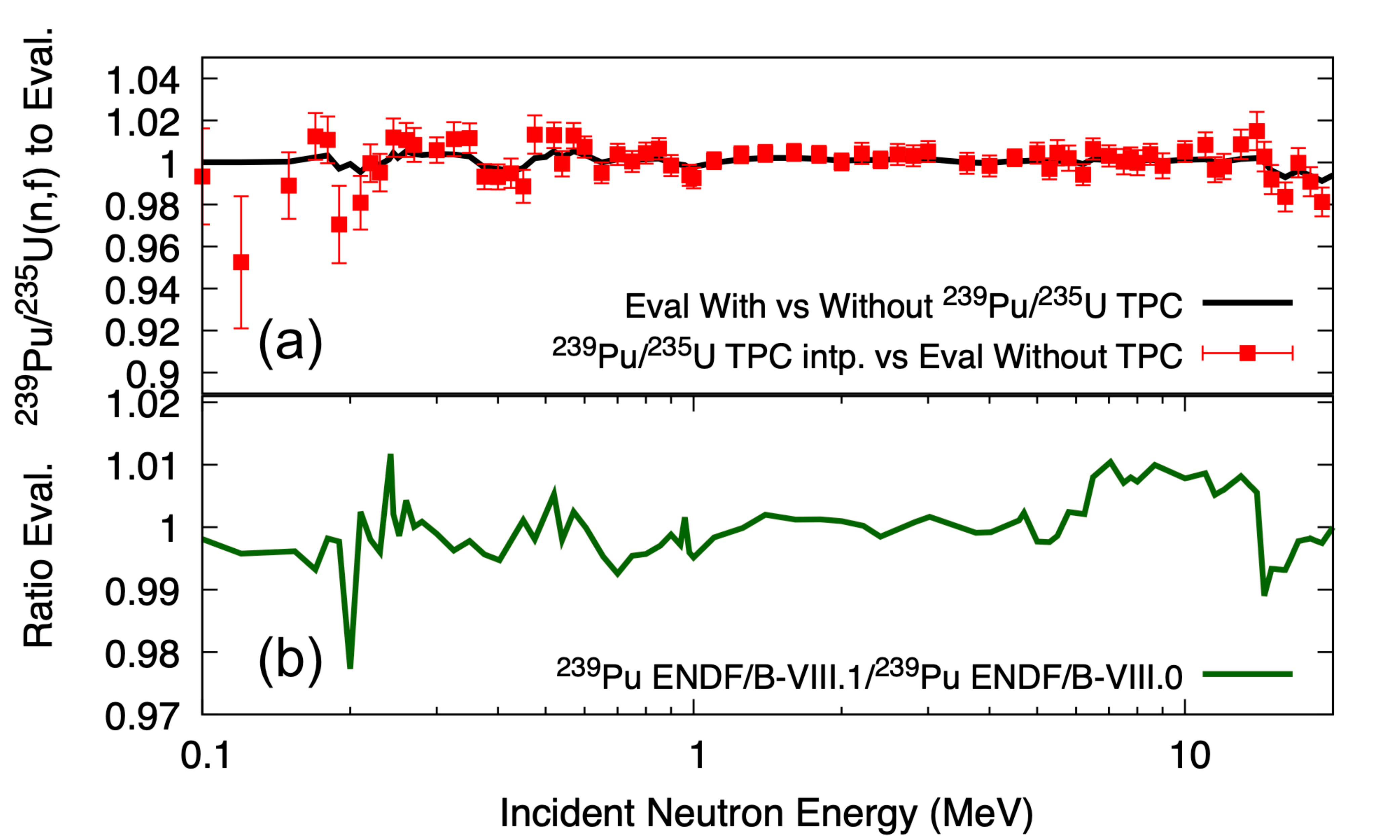}
\caption{$^{239}$Pu/$^{235}$U cross-section ratios (from fission TPC \cite{Snyder2022} or ENDF/B-VIII.1) are divided by the evaluation
 without fissionTPC data (a) or by the ENDF/B-VIII.0 $^{239}$Pu/$^{235}$U cross-section ratio (b).}
\label{fig:Pu239nfcsratio}
\end{figure}

\paragraph{($n$,$f$) fission cross section\newline}
For ENDF/B-VIII.0, neutron-induced fission cross sections for $^{239}$Pu were reported by the Neutron Data Standards committee~\cite{carlson2018} with covariances enlarged by a fully correlated USU (unrecognized sources of uncertainties~\cite{USU}) component equal to 1.2\%.
This USU  should account for known but missing uncertainty sources and correlations in the input to the evaluation as well as unknown uncertainties.
While the latter can only be assessed by the spread of data among experiments employing different measurement techniques, the former were tackled by creating templates of expected fission cross-section uncertainties and applying them to $^{239}$Pu(n,f) cross-section data in the database~\cite{Neudecker2020, Neudecker:2019tempnf}.
In addition to that, fission TPC $^{238}$U/$^{235}$U and $^{239}$Pu/$^{235}$U~\cite{Snyder2022} neutron-induced fission cross-section ratios were included into the evaluation undertaken by the Standards  group~\cite{Neudecker2021}.
These data were treated as shape data as their normalization is about 2\% off from the current standard~\cite{Snyder2022}. 
The final $^{239}$Pu(n,f) cross section implemented in ENDF/B-VIII.1 was presented in Ref.~\cite{Neudecker:2023Pu9fissionsourceterm} and is compared to fission TPC data in Fig.~\ref{fig:Pu239nfcs} as a ratio to $^{235}$U(n,f) cross sections.
Inclusion of the high-precision fission TPC data reduced the cross section above 10~MeV (see Fig.~\ref{fig:Pu239nfcsratio}), while changes from ENDF/B-VIII.0 to ENDF/B-VIII.1 cross sections stem from applying templates of expected measurement uncertainties leading to an improved uncertainty estimate.
In the energy region above 10~MeV, discrepant data sets led to questions on the shape of the cross section~\cite{carlson2018}; potential biases were suspected in quantifying the detector efficiency, especially the effect of angular distribution on the efficiency.

Fission TPC data were measured with a time-projection chamber, a new and different type of fission detector than previously used. This detector has a more fine-grained view of the angular distribution of fission fragments~\cite{Snyder2022}. Hence, these new data lead to more confidence in the new shape of the fission cross section at higher incident-neutron energies.

\paragraph{($n,f$) fission cross-section covariances\newline}
Finally, work was done to update the USU  component of the fission cross-section uncertainties to have energy-dependent correlations rather than being fully correlated.
This work is motivated by the belief that ENDF/B-VIII.0 estimated USU fission cross-section uncertainty of 1.2\% was probably overestimated in the very important energy region from 0.1~MeV up to 5~MeV (the first-chance fission plateau). Energy-independent USU also introduced overly strong, long-range, cross-energy correlations which were probably unrealistic. Such correlations present a challenge to data adjustment based on critical assemblies which feature sensitivities to narrow energy regions (e.g., typically fast critical assemblies -- PMF -- are sensitive to neutron energies from 10~keV up to 5~MeV, mostly).

\begin{figure}[!thb]
\vspace{-3mm}
 \begin{center}
 \includegraphics[width=\columnwidth]{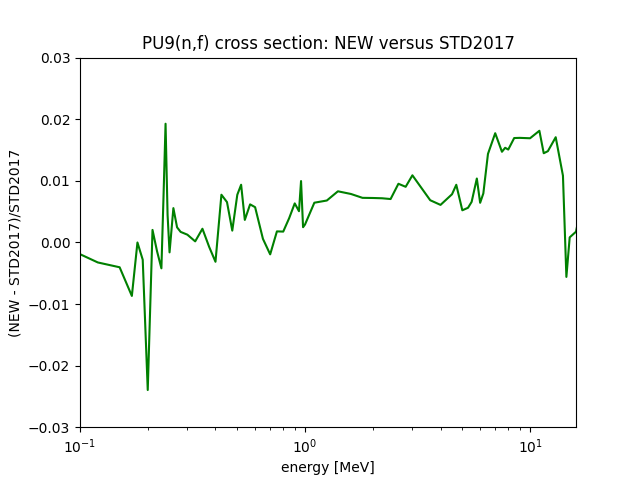}
\vspace{-3mm}
 \caption{Newly evaluated \nuc{239}{Pu}($n$,$f$) cross section with latest \texttt{gmapy} code and inputs including a new USU treatment as a ratio to the \prENDF{} cross sections. These cross sections were not used in the ENDF/B-VIII.1 file.}
 \label{fig:pu239-nf-new-to-std2017}
 \end{center}
\vspace{-4mm}
\end{figure}

The USU energy-dependence was accounted for by the introduction of additional energy-dependent bias functions for each dataset, which are parameterized as a piecewise linear functions. The same energy mesh is assumed for all datasets of the same reaction channel, but the bias function of each dataset is allowed to vary independently of bias functions of other datasets. The energy-dependent USU  was then also defined as piecewise linear function using the same energy mesh as for the bias functions. At a conceptual level, the spread of the bias function at each energy mesh point informed the magnitude of the uncertainty associated with USU  at the same energy. 

Technically, the generalized least squares (GLS) method was further generalized to a Bayesian hierarchical model in order to also incorporate the bias functions and the associated USU uncertainties. The posterior expectation of the cross sections and associated  covariance matrix was then determined by the Hamilton Monte Carlo method. A more detailed description of the procedure will be the subject of a separate publication. All changes were implemented within the open-source IAEA \texttt{gmapy} code developed by G. Schnabel, which was used for calculations. The \texttt{gmapy} code employed as input the revised and updated GMA input database already used for Neutron Data Standards \cite{carlson2018}.

Energy-dependent USU components were added to the revised GMA database used for the evaluation undertaken by the Standards group~\cite{Neudecker2021}, which includes new relative measurements of $^{238}$U(n,f) and $^{239}$Pu(n,f) from the NIFFTE collaboration~\cite{Snyder2022}, revised uncertainty quantification of $^{239}$Pu(n,f) datasets based on templates~\cite{Neudecker2020, Neudecker:2019tempnf}, and revised spectrum-averaged cross sections (SACS) and SACS ratio data \cite{capote2023}. Additionally, datasets or parts of datasets have been removed when they were deemed inconsistent with other datasets or exhibited problematic energy behavior. In particular, datapoints from the Pankratov $^{235}$U(n,f) and $^{238}$U(n,f) shape datasets (GMA ID 722 and 874) above 23~MeV were removed; the absolute $^{238}$U(n,f) measurement of Nolte between 34 and 200~MeV (GMA ID 8008) as well as datapoints above 27~MeV of the Carlson $^{235}$U(n,f) measurement (GMA ID 524) were also discarded. At an IAEA Experts Meeting in 2023, an agreement was reached to remove these data from the analysis.

\begin{figure}[!thb]
\vspace{-3mm}
 \begin{center}
 \includegraphics[width=\columnwidth]{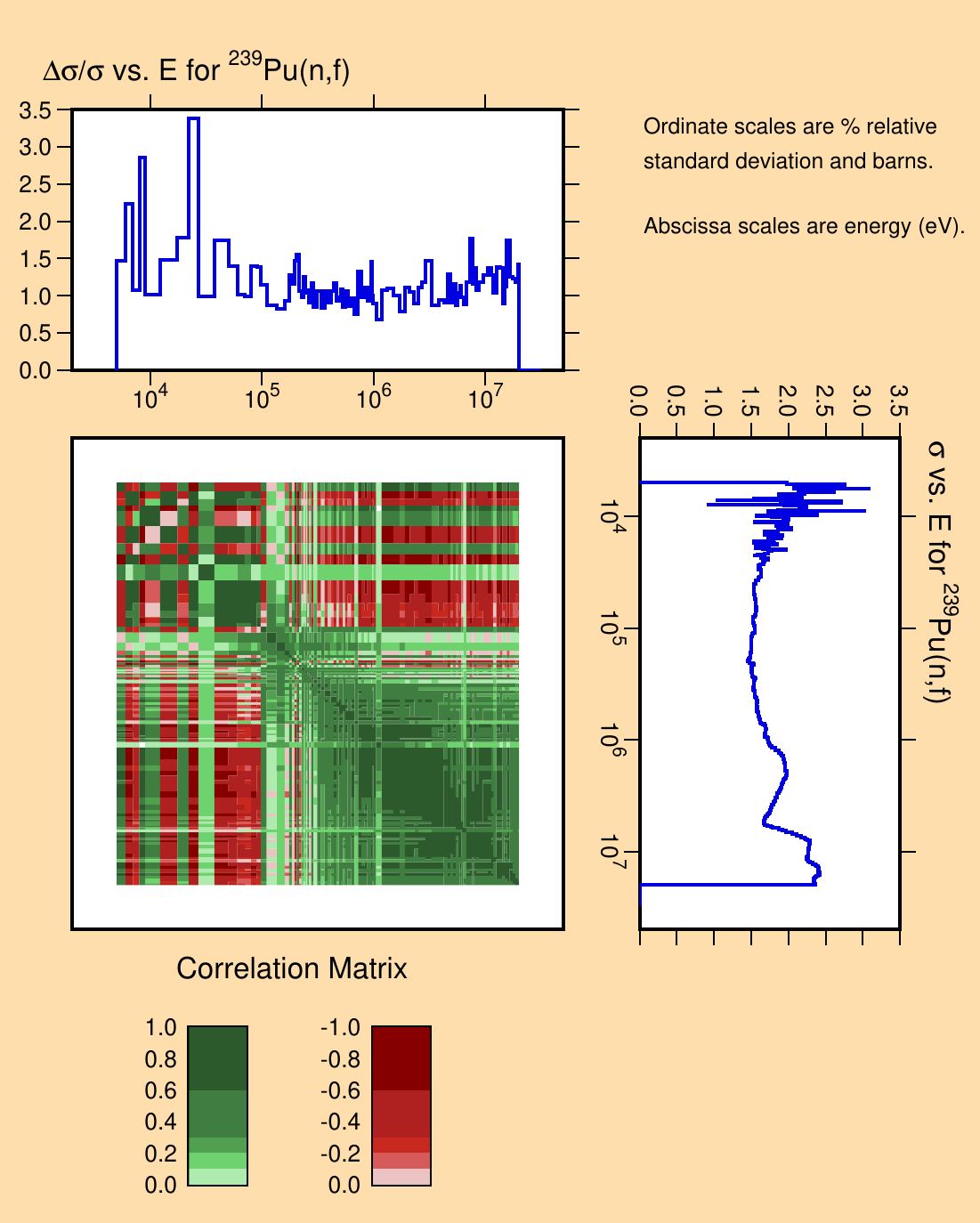}
\vspace{-3mm}
 \caption{The ENDF/B-VIII.1 evaluated \nuc{239}{Pu}($n$,$f$) cross-section uncertainties and correlation matrix for incident neutron energies from 5~keV up to 30~MeV.}
 \label{fig:pu239-nf-cov}
 \end{center}
\vspace{-4mm}
\end{figure}
The energy mesh for USU components was manually defined taking into account the availability of data and features of the cross-section curve, with the lowest energy mesh point starting as low as 1~keV and going up to at least 30~MeV. USU components were defined for absolute and relative measurements as well as shape measurements of absolute and relative cross sections involving $^{235,238}$U and $^{239}$Pu targets and their combinations. 30000 samples were obtained from the posterior which proved to be sufficiently large for the convergence of posterior cross-section estimates and the associated covariance matrix. Note that data from the entire GMA input database was used for the fitting, not only the $^{239}$Pu(n,f) data.

Due to the fact that, in this ENDF/B release, the $^{239}$Pu(n,f) covariance matrix coming from a new USU treatment was introduced after the mean-value evaluation had been concluded, special care must be taken to ensure consistency between the employed mean values~\cite{Neudecker2021} and the updated covariance matrix using the energy-dependent USU. This was achieved by a post-processing of the covariance matrix that inflates elements in the covariance matrix based on a cross-entropy criterion.

The difference between the STD2017 evaluation \cite{carlson2018} (part of the ENDF/B-VIII.0 release) and the newly evaluated (\texttt{gmapy}) $^{239}$Pu(n,f) cross sections due to updated USU treatment and vetted data is about +0.7\% larger than the standards in the first-chance plateau from 1~MeV up to 6~MeV. The fission cross-section increase reaches almost 1.8\% from 7~MeV up to 13~MeV, but it vanishes at 14~MeV as shown in Fig.~\ref{fig:pu239-nf-new-to-std2017}.

The final correlation plot and the associated relative uncertainties with respect to ENDF/B-VIII.1 evaluated mean values are shown in Fig.~\ref{fig:pu239-nf-cov}.

\paragraph{($n$,$\gamma$) capture cross section\newline} \label{ssec:pu9capture}
The INDEN evaluation of the neutron-capture channel was adopted for the ENDF/B-VIII.1 library. The INDEN evaluation is a Bayesian Generalized Least-Square Evaluation using the \GANDR\ code \cite{GANDR} with the model prior generated using the \EMPIRE\ code system~\cite{Herman:2007} as described in Ref.~\cite{trkov:2011}.

To measure neutron capture on fissile elements, it is imperative to measure the fission-to-capture ratio (usually called the $\alpha$-ratio). This implies that the systematic uncertainty from fission measurements will correlate all capture datasets derived from the measured $\alpha$-ratio. Experimental data used as input for the \GANDR\  system are listed in Tables~\ref{tab:pu239ng} and ~\ref{tab:pu239alpha}.

The newest $^{239}$Pu($n$,$\gamma$) dataset is the one measured by Mosby and colleagues \cite{Mosby2018} at LANL which used the ENDF/B-VIII.0 fission cross section (standard) to derive the capture cross section from the measured $\alpha$-ratio. This dataset also features the lowest uncertainties of about 4\%, which defines the ENDF/B-VIII.1 evaluated mean values and uncertainties. There is also an extensive set of older $\alpha$-ratio measurements which are listed in Table~\ref{tab:pu239alpha}. We used evaluated $^{239}$Pu fission cross sections for ENDF/B-VIII.1 undertaken by the Standards group~\cite{Neudecker2021} to derive the measured capture cross sections with uncertainties from the corresponding $\alpha$-ratio listed in Table~\ref{tab:pu239alpha}. Data derived in this way are automatically corrected to the latest fission standard and represent a much larger and consistent experimental database than those data listed in Table~\ref{tab:pu239ng} (with the exception of Mosby 2018 which were also considered).

\begin{table}[tbp]
\caption{Selection of radiative capture cross-section measurements of $^{239}$Pu for incident neutron energy above 5~keV
from the EXFOR database~\cite{EXFOR}. Experiments selected for fitting are marked by a ``+'' sign in the first column.}
\label{tab:pu239ng}
\par
\begin{center}
\begin{tabular}{clll}
\toprule \toprule
  & Author                                        & Year   & EXFOR No. \\ \midrule
+ & Mosby \textit{et al.} \cite{Mosby2018}        & (2018) & 14794002  \\
- & Belyaev \textit{et al.} \cite{Belyaev1970}    & (1970) & 40087006  \\
- & Hopkins \textit{et al.} \cite{Hopkins1962}    & (1962) & 12331008  \\
- & Andreev \cite{Andreev1958}                    & (1958) & 40385007  \\
- & Spivak \textit{et al.} \cite{Spivak1956}      & (1956) & 40350006  \\ \midrule
\multicolumn{4}{l}{Systematic uncertainties:} \\
3\% & \multicolumn{3}{l}{for each EXFOR set} \\
3\% & \multicolumn{3}{l}{common to all measurements} \\ \bottomrule \bottomrule
\end{tabular}%
\end{center}
\vspace{-3mm}
\end{table}

\begin{table}[tbp]
\caption{Selection of absorption (alpha ratio $\equiv$ capture/fission) cross-section measurements of $^{239}$Pu for
incident neutron energy above 5~keV from the EXFOR database \cite{EXFOR}. Experiments selected for fitting are marked by a ``+'' sign in the first column.}
\label{tab:pu239alpha}
\par
\begin{center}
\begin{tabular}{clll}
\toprule \toprule
  & Author                                            & Year   & EXFOR No.  \\ \midrule
+ & Bolotskii \textit{et al.} \cite{Bolotskii1977}    & (1977) & 40548002   \\
+ & Ryabov \cite{Ryabov1976,Ryabov1976e}              & (1976) & 40312003   \\
+ & Gwin \textit{et al.} \cite{Gwin1976}              & (1976) & 10267029   \\
+ & Kononov \textit{et al.} \cite{Kononov1975}        & (1975) & 40412005   \\
\multicolumn{4}{l}{~~~Kononov data discarded for $E_n>60$~keV:} \\
+ & Bandl \textit{et al.} \cite{Bandl1972}            & (1972) & 20158003   \\
+ & Kurov \textit{et al.} \cite{Kurov1972}            & (1972) & 40024003   \\
+ & Weston \textit{et al.} \cite{Weston1972}          & (1972) & 10301003   \\
- & Farrell \textit{et al.} \cite{Farrell1970}        & (1970) & 10326002   \\
+ & Belyaev \textit{et al.} \cite{Belyaev1970}        & (1970) & 40087008   \\
+ & Schomberg \textit{et al.} \cite{Schomberg1970}    & (1970) & 20476005   \\
+ & de Saussure \textit{et al.} \cite{DeSaussure1967} & (1966) & 12409004   \\
+ & de Saussure \textit{et al.} \cite{DeSaussure1967} & (1966) & 12409005   \\
+ & Hopkins \textit{et al.} \cite{Hopkins1962}        & (1962) & 12331007   \\
- & Andreev \cite{Andreev1958}                        & (1958) & 40385008   \\
- & Spivak \textit{et al.} \cite{Spivak1956}          & (1956) & 40350016   \\
\midrule
\multicolumn{4}{l}{Systematic uncertainties:} \\
3\% & \multicolumn{3}{l}{for each EXFOR set} \\
3\% & \multicolumn{3}{l}{common to all measurements} \\
\bottomrule \bottomrule
\end{tabular}%
\end{center}
\vspace{-3mm}
\end{table}

\begin{figure}[!thb]
 \begin{center}
 \includegraphics[clip,width=\columnwidth]{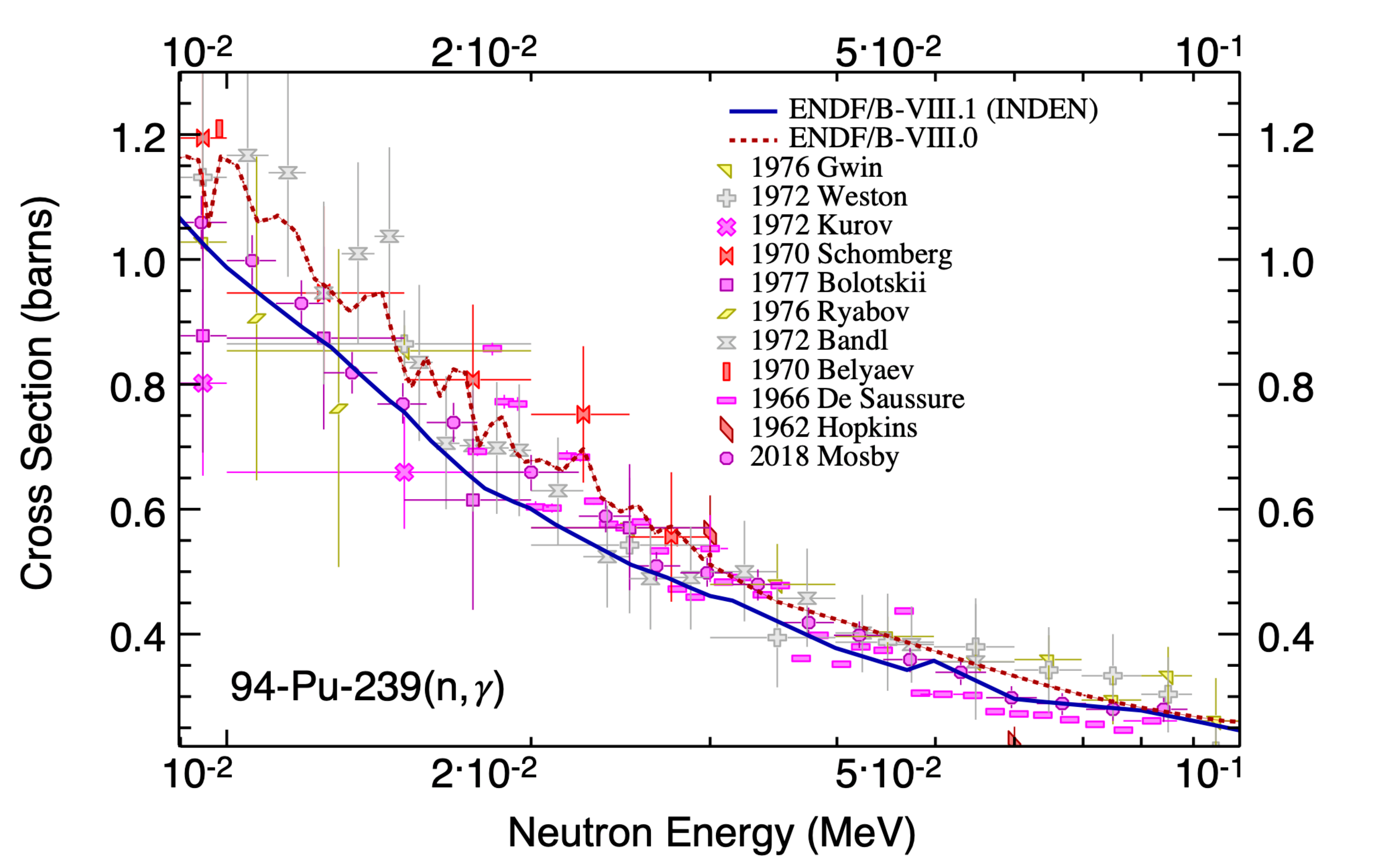}
 \caption{The ENDF/B-VIII.1 evaluated \nuc{239}{Pu}($n$,$\gamma$) cross section for incident neutron energies from 10~keV up to 100~keV along with comparison to experimental data from EXFOR \cite{EXFOR} as listed in Tables \ref{tab:pu239ng} and \ref{tab:pu239alpha} and \prENDF{}. Data from Refs.~\cite{Gwin1976,Weston1972,Kurov1972,Schomberg1970,Bolotskii1977,Ryabov1976,Ryabov1976e,Bandl1972,Belyaev1970,DeSaussure1967,Hopkins1962,Mosby2018}.}
 \label{fig:pu239ng}
 \end{center}
\end{figure}

\begin{figure}[!thb]
 \begin{center}
 \includegraphics[clip,width=\columnwidth]{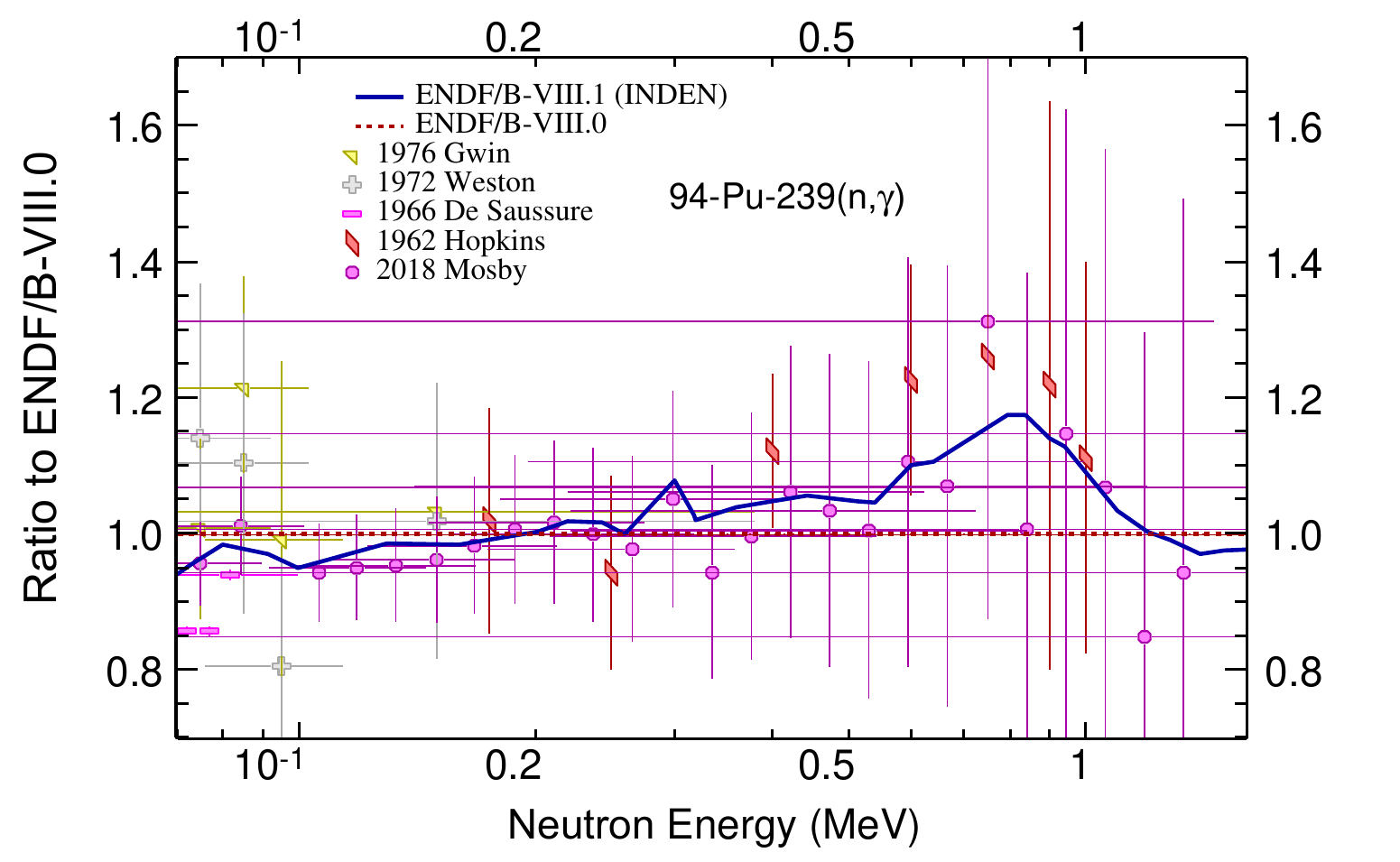}
 \caption{The ratio of the ENDF/B-VIII.1 evaluated \nuc{239}{Pu}($n$,$\gamma$) cross section to \prENDF{} for incident neutron energies from 70~keV up to 1.5~MeV is compared versus the same ratio using  selected experimental data from EXFOR \cite{EXFOR}. Data from Refs.~\cite{Gwin1976,Weston1972,DeSaussure1967,Hopkins1962,Mosby2018}.}
 \label{fig:pu239ng-ratioB8}
 \end{center}
\end{figure}

The ENDF/B-VIII.1 evaluated \nuc{239}{Pu}($n$,$\gamma$) cross section for incident neutron energies from 9~keV up to 100~keV is compared with the data described above and \prENDF\ in Fig.~\ref{fig:pu239ng}.
A significant decrease compared to \prENDF\ is observed in the whole URR up to 50~keV mainly driven by Mosby data. Fluctuations observed in the \prENDF\ file were related to the URR evaluation. In the region from 20~keV up to 50~keV, older measured data led to evaluated results slightly lower than the Mosby measurements. The \nuc{239}{Pu}($n$,$\gamma$) ratio to \prENDF\ is shown in Fig.~\ref{fig:pu239ng-ratioB8} from 70~keV up to 1.3~MeV. A significant increase in the capture cross section compared to \prENDF{} is noted in the fast neutron region from 300~keV up to 1~MeV, reaching almost 20\% around 800~keV. This capture increase is driven by Hopkins and Mosby data and the impact on reducing the criticality of fast Pu assemblies is significant.

\begin{figure}[!thb]
 \begin{center}
 \includegraphics[clip,width=\columnwidth]{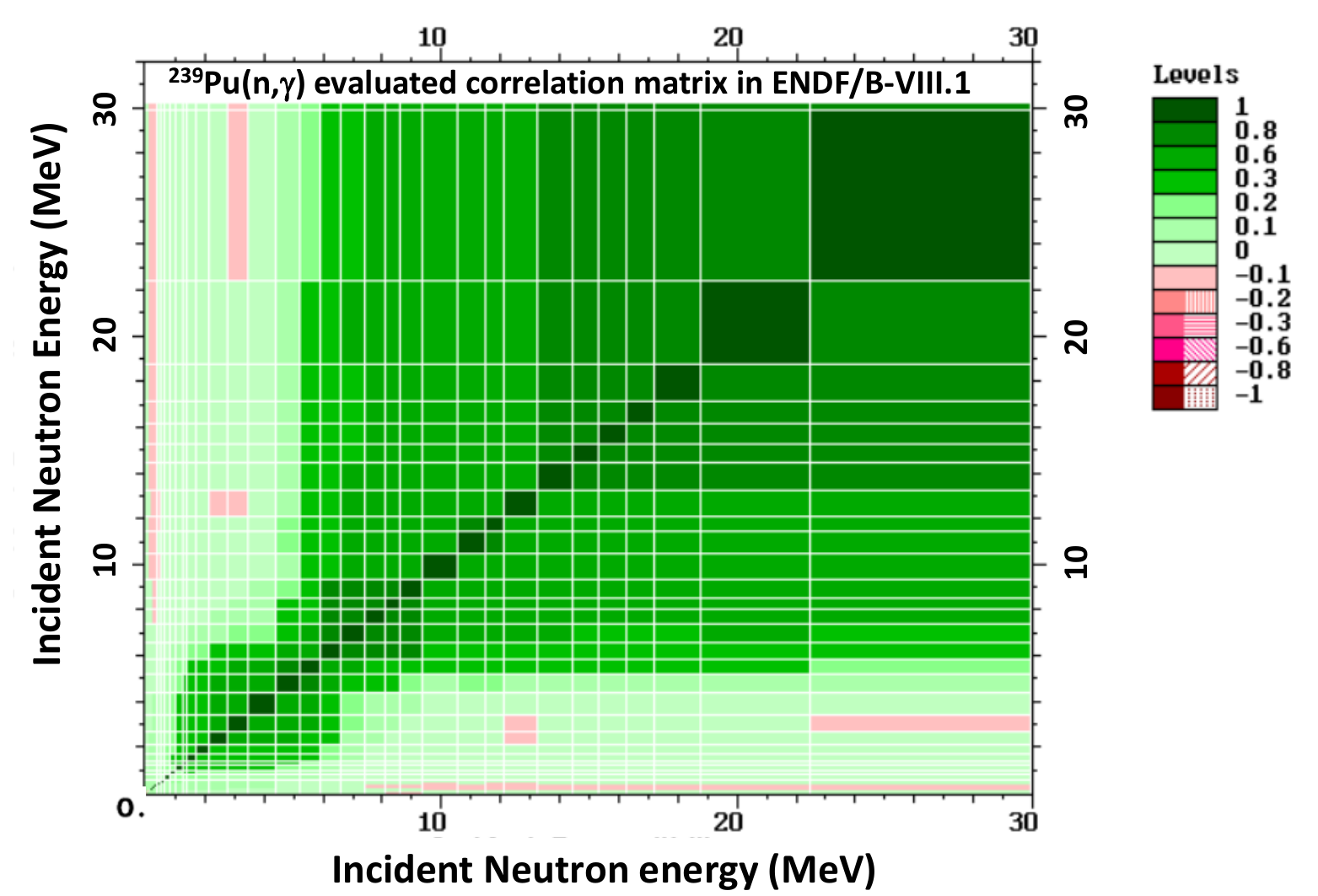}
 \caption{ENDF/B-VIII.1 evaluated \nuc{239}{Pu}($n$,$\gamma$) cross-section correlations for incident neutron energies from 5~keV up to 30~MeV.}
 \label{fig:pu239-ng-corr}
 \end{center}
\end{figure}

\begin{figure}[!thb]
 \begin{center}
 \includegraphics[clip,width=\columnwidth]{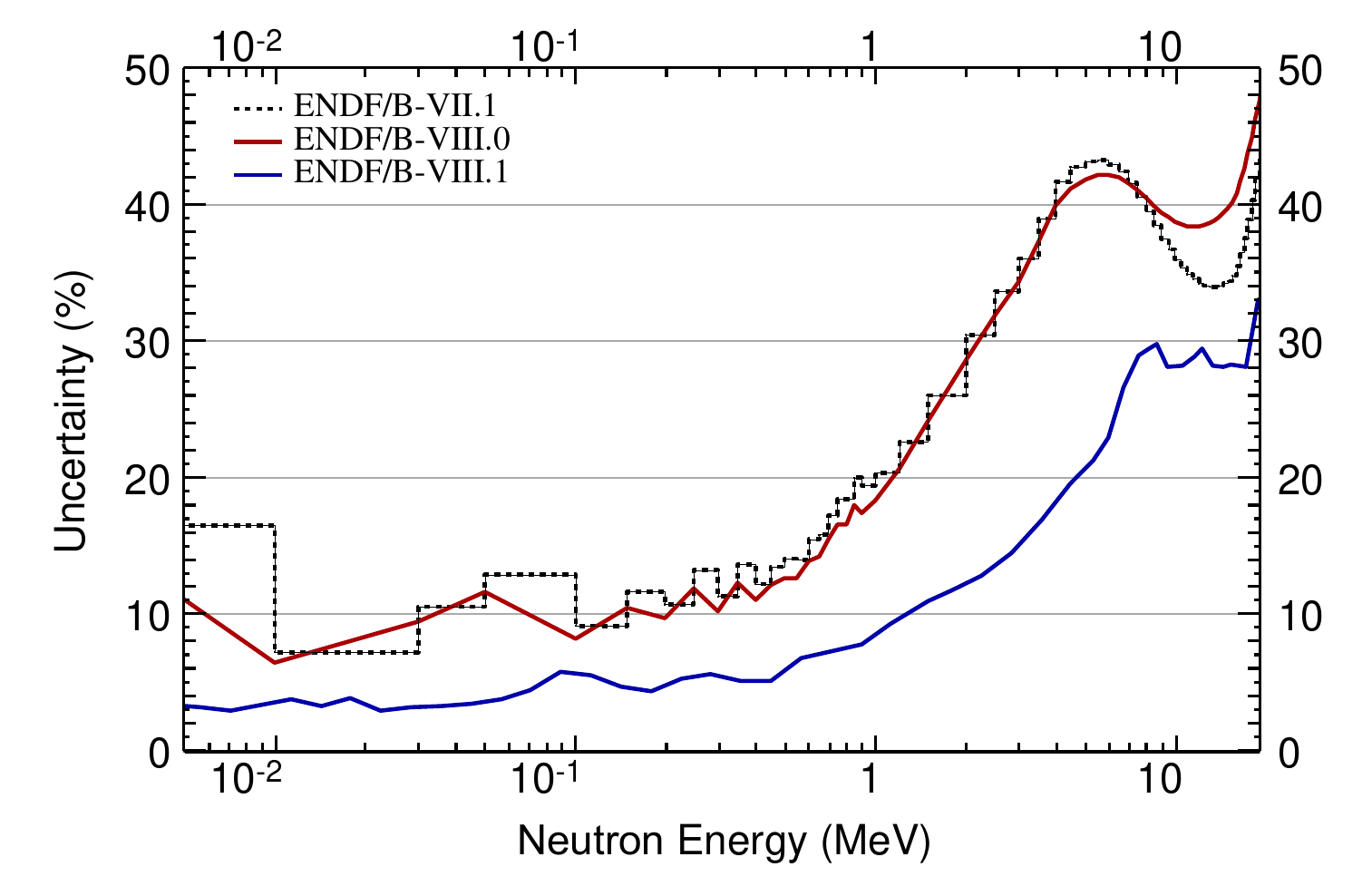}
 \caption{ENDF/B-VIII.1 evaluated \nuc{239}{Pu}($n$,$\gamma$) cross-section uncertainties for incident neutron energies from 5~keV up to 20~MeV are compared for the ENDF/B-VII.1 and ENDF/B-VIII.0 libraries.}
 \label{fig:pu239-ng-unc}
 \end{center}
\end{figure}

\paragraph{($n$,$\gamma$) covariances\newline}
The \EMPIRE\ code system is a well-tested system package for nuclear model calculations and nuclear data evaluation \cite{Herman:2007}. Its
capabilities include random sampling of model parameters, which can be utilized to generate a full model covariance matrix of all calculated cross
sections, including cross-reaction correlations. The \EMPIRE\ system was employed to calculate the prior covariance matrices of reaction cross sections of neutron-induced reactions on a $^{239}$Pu target for incident neutron energies up to 30~MeV, including both the capture and (n,2n) cross sections. The model parameter uncertainties were conservatively estimated to get capture cross-section uncertainties covering the spread of experimental data. 
The resulting model prior covariance for the capture reaction features an uncertainty estimate of around 30\% up to 2~MeV, and much larger for higher energies.
The model prior correlations are strong and positive with near-energy correlations reaching 95\% and correlations between 10~keV and 100~keV of about 50\%. A typical model correlation matrix is shown in Fig.~1 of Ref.~\cite{trkov:2011}. The derived modeling prior covariance matrix was input to the \GANDR\ system \cite{GANDR,GANDR-std}, which is a software package based on the GLS method that makes use of the experimental data retrieved from EXFOR in the C4/C5 format \cite{EXFOR}.

By introducing the vetted experimental data from the EXFOR database listed in Tables \ref{tab:pu239ng} and \ref{tab:pu239alpha} above into \GANDR, the constrained covariance matrices
and evaluated cross section were obtained using the method outlined in Ref.~\cite{trkov:2011}. Posterior evaluated values were subsequently formatted in ENDF-6 format.
A typical evaluated correlation matrix shows negative and positive correlations due to the experimental data constraints as shown in Fig.~\ref{fig:pu239-ng-corr}.

The evaluated uncertainty of the ENDF/B-VIII.1 capture cross sections is about 4\% from 5~keV up to 100~keV, being reduced by a factor of about 2 compared to \prENDF{} and ENDF/B-VII.1 uncertainties as shown in Fig.~\ref{fig:pu239-ng-unc}. The uncertainty increases monotonically reaching 10\% at 1~MeV and 20\% at 5~MeV. The uncertainty reduction is driven mostly by the lower uncertainty of the data from Mosby \etal~\cite{Mosby2018}, which is about 4\% below 500~keV. The lower uncertainty of Mosby data were not properly considered in the ENDF/B-VIII.0 evaluation, therefore the evaluated uncertainty remained much larger. Below 500~keV,  there is also some uncertainty reduction driven by the availability of several independent experiments derived from the measured $\alpha-$ratio data as listed in Table~\ref{tab:pu239alpha}.

\paragraph{($n$,$2n$) cross section\newline}
The INDEN evaluation of the (n,2n) channel was adopted for the ENDF/B-VIII.1 library. The INDEN evaluation is a Bayesian Generalized Least-Square evaluation using the \GANDR\ code \cite{GANDR} with the model prior generated using the \EMPIRE\ code system~\cite{Herman:2007} as described in Ref.~\cite{trkov:2011} and Section~\ref{ssec:pu9capture}. The evaluation started from the McNabb analysis \cite{McNabb2001} which was adopted for the ENDF/B-VII.1 library. The most precise experiment considered at that time was the activation measurement done by Lougheed \etal and carefully analysed and corrected by McNabb and colleagues \cite{McNabb2001}. A joint LANL GEANIE experiment was coupled to a model simulation to derive the most comprehensive set of the (n,2n) cross sections over a broad energy range by Bernstein and collaborators \cite{Bernstein2002}. We also used a new experiment undertaken by M\'{e}ot \etal \cite{Meot2021} which was normalized to the \prENDF{} at 9~MeV. We discarded Mather \cite{Mather1972} and Frehaut \cite{Frehaut1986} data as discussed in Ref.~\cite{McNabb2001}. We have also discarded a new surrogate experiment by Ma \etal \cite{Ma2020} as having very large uncertainties below 10~MeV and a shape not corresponding to the theoretical expectations. Selected experimental data for the (n,2n) evaluations are listed in Table \ref{tab:pu239n2n}.

\begin{table}[tbp]
\footnotesize
\caption{Selection of (n,2n) cross-section measurements of $^{239}$Pu from the EXFOR database \cite{EXFOR}. Selected experiments are marked by a ``+'' sign in the first column.}
\label{tab:pu239n2n}
\par
\begin{center}
\begin{tabular}{cllll}
\toprule \toprule
& Author                                                     & Year   & EXFOR No. & Comment         \\ \midrule
+ & M\'{e}ot \textit{et al.} \cite{Meot2021}                 & (2021) & 23777002  & New measurement \\
+ & Bernstein \textit{et al.} \cite{Bernstein2002}           & (2002) & 13787006  & Meas + Model    \\
+ & Lougheed \textit{et al.} \cite{Lougheed2002,McNabb2001}  & (2002) & 13787006  & Meas + Model  \\
- & Mather \textit{et al.} \cite{Mather1972}                 & (1972) & 32815003  & An outlier      \\
- & Frehaut \textit{et al.} \cite{Frehaut1986}               & (1986) & 32815003  & Wrong shape     \\
- & Nanru Ma \textit{et al.} \cite{Ma2020}                   & (2020) & 32815003  & Surrogate       \\
\midrule
\multicolumn{5}{l}{Systematic uncertainties:} \\
1\% & \multicolumn{4}{l}{for each EXFOR  set} \\
4\% & \multicolumn{4}{l}{common to all measurements (altogether 4.1 \%)} \\
\bottomrule \bottomrule
\end{tabular}%
\end{center}
\end{table}

\begin{figure}[!thb]
 \begin{center}
 \includegraphics[clip,width=\columnwidth]{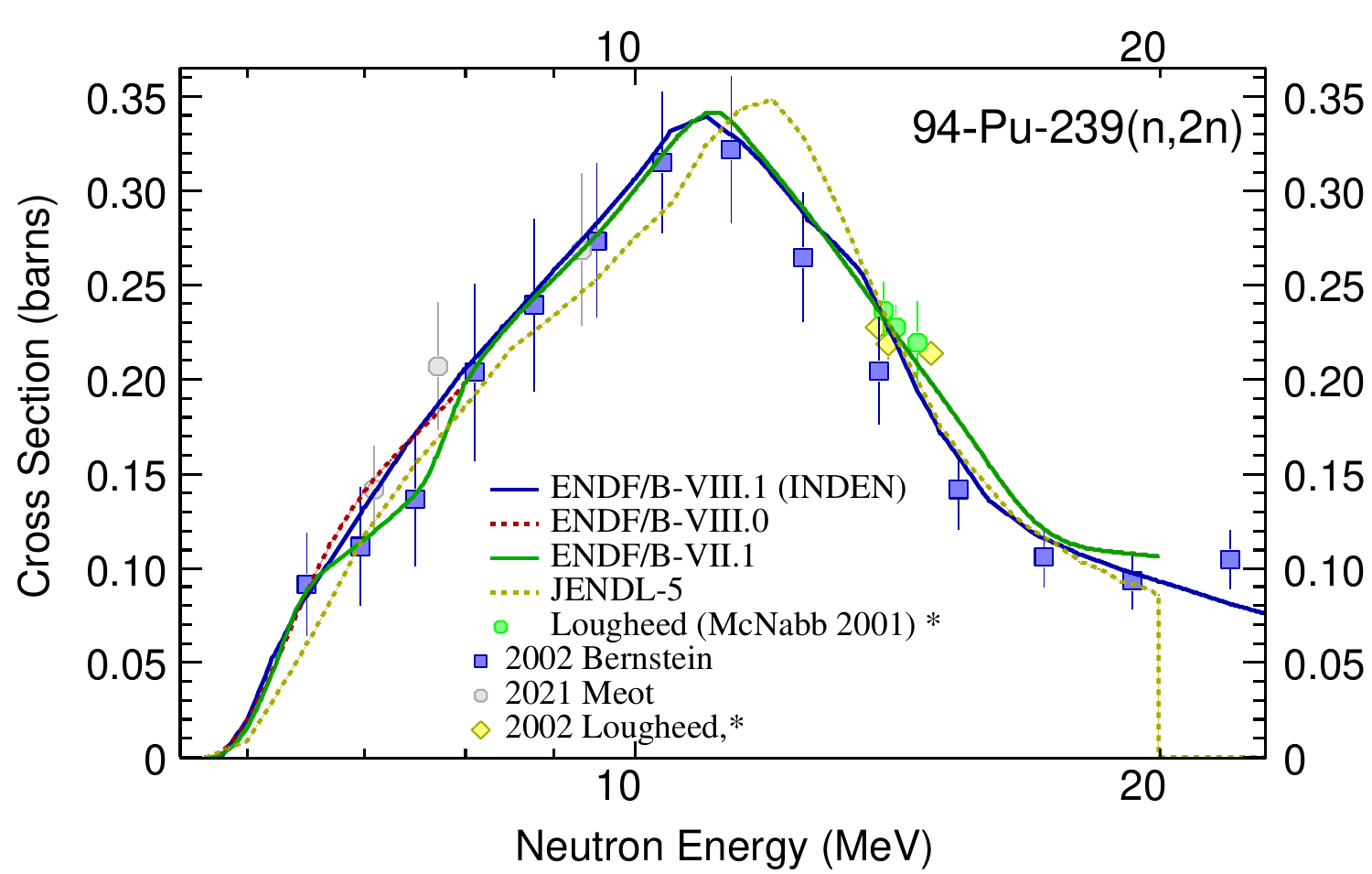}
 \caption{ENDF/B-VIII.1 evaluated \nuc{239}{Pu}($n$,2$n$) cross sections versus selected experimental data~\cite{Meot2021,Bernstein2002,Lougheed2002,McNabb2001} and JENDL-5, ENDF/B-VII.1, and \prENDF{} evaluations up to 25~MeV.}
 \label{fig:pu239-n2n}
 \end{center}
\end{figure}

The ENDF/B-VIII.1 evaluated ($n$,2$n$) cross sections versus experimental data and ENDF/B-VIII.0, ENDF/B-VII.1, and JENDL-5 evaluations are shown in Fig.~\ref{fig:pu239-n2n}.
Note that both the original Lougheed data \cite{Lougheed2002} (yellow rhombi)  and the revised data by McNabb \cite{McNabb2001} (green circles) are plotted. The highest energy point of McNabb revised data was considered an outlier and its uncertainty was doubled. Below 9~MeV, there is an excellent agreement between the \prENDF{} and new ENDF/B-VIII.1 evaluation which will lead to almost the same calculated $^{239}$Pu(n,2n) reaction rate in a fission spectrum. The energy-dependence of the new evaluation below 10~MeV is driven by the shape of the \EMPIRE\ model calculations used as a prior in the least-squares evaluation. This shape validates the changes in this energy region undertaken in the ENDF/B-VIII.0 evaluation. The JENDL-5 evaluation results in a much lower calculated $^{239}$Pu(n,2n) reaction rate in a fission spectrum. Note that the new M\'{e}ot \etal \cite{Meot2021} measured shape is in excellent agreement with the new evaluation from 7~MeV up to 9~MeV. The ENDF/B-VIII.1 and JENDL-5 evaluations are in excellent agreement above 13~MeV and shows the best agreement with GEANIE-derived data \cite{Bernstein2002} and a very good agreement with McNabb derived data~\cite{McNabb2001}. There is a disagreement between JENDL-5 and other evaluations in the energy corresponding to the maximum of the $^{239}$Pu(n,2n) excitation function. The ENDF/B evaluated maximum energy around 11~MeV relies on the measured data while the JENDL-5 maximum of about 12~MeV aligns better with model calculations; further investigation is needed.

\paragraph{($n$,$2n$) covariances\newline}
The (n,2n) covariance matrix and uncertainties were derived as described for the capture cross-section covariances in Section~\ref{ssec:pu9capture}.
A derived modeling prior covariance matrix was input to the \GANDR\ system \cite{GANDR,GANDR-std}. 
By introducing the vetted (marked with + sign) experimental data from the EXFOR database listed in Table \ref{tab:pu239n2n} into \GANDR, the constrained covariance matrix and corresponding evaluated cross section were obtained using a method described in Ref.~\cite{trkov:2011} and formatted in ENDF-6 format.

\begin{figure}[!thb]
\vspace{-2mm}
 \begin{center}
 \includegraphics[clip,width=\columnwidth]{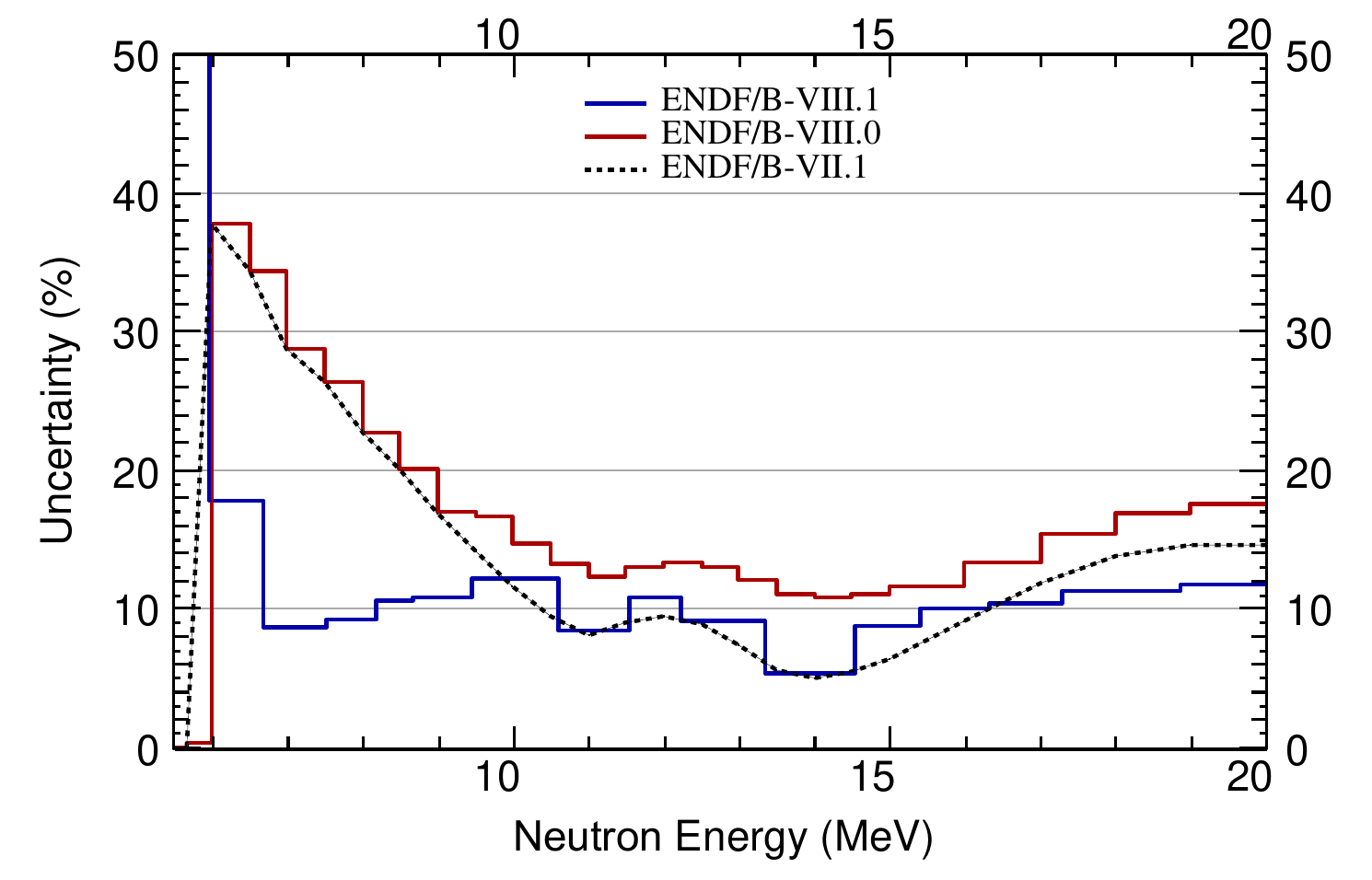}
\vspace{-2mm}
 \caption{ENDF/B-VIII.1 evaluated \nuc{239}{Pu}($n$,2$n$) cross-section uncertainties for incident neutron energies from 5~MeV up to 20~MeV are compared for the ENDF/B-VII.1 and ENDF/B-VIII.0 libraries.}
 \label{fig:pu239-n2n-unc}
 \end{center}
\vspace{-2mm}
\end{figure}
The evaluated uncertainty of the ENDF/B-VIII.1 (n,2n) cross sections is about 10\% from 7~MeV up to 13~MeV, being reduced significantly compared to the uncertainties in the ENDF/B-VII.1 and \prENDF{} libraries in the region below 9~MeV due to the use of newly available M\'{e}ot data \cite{Meot2021} as shown in Fig.~\ref{fig:pu239-n2n-unc}. The uncertainty reaches the minimum around 14~MeV in excellent agreement with McNabb 2001 evaluation \cite{McNabb2001} used in the ENDF/B-VII.1 library. This uncertainty reduction is driven by the lower uncertainty estimated by McNabb \etal~\cite{McNabb2001} from the Lougheed experimental data \cite{Lougheed2002}. Note that a much larger uncertainty (about 12\%) defined in the ENDF/B-VIII.0 evaluation  is inconsistent with the  McNabb analysis~\cite{McNabb2001}.
The uncertainty of the current evaluation is lower at energies above 15~MeV due to the strong cross-energy correlations in the model prior and available GEANIE experimental data.
Evaluated uncertainties are driven by the overall uncertainty of the best experiments and modeling as evidenced in the discussion above.

\begin{figure}[!thb]
\vspace{-2mm}
 \begin{center}
 \includegraphics[clip,width=\columnwidth]{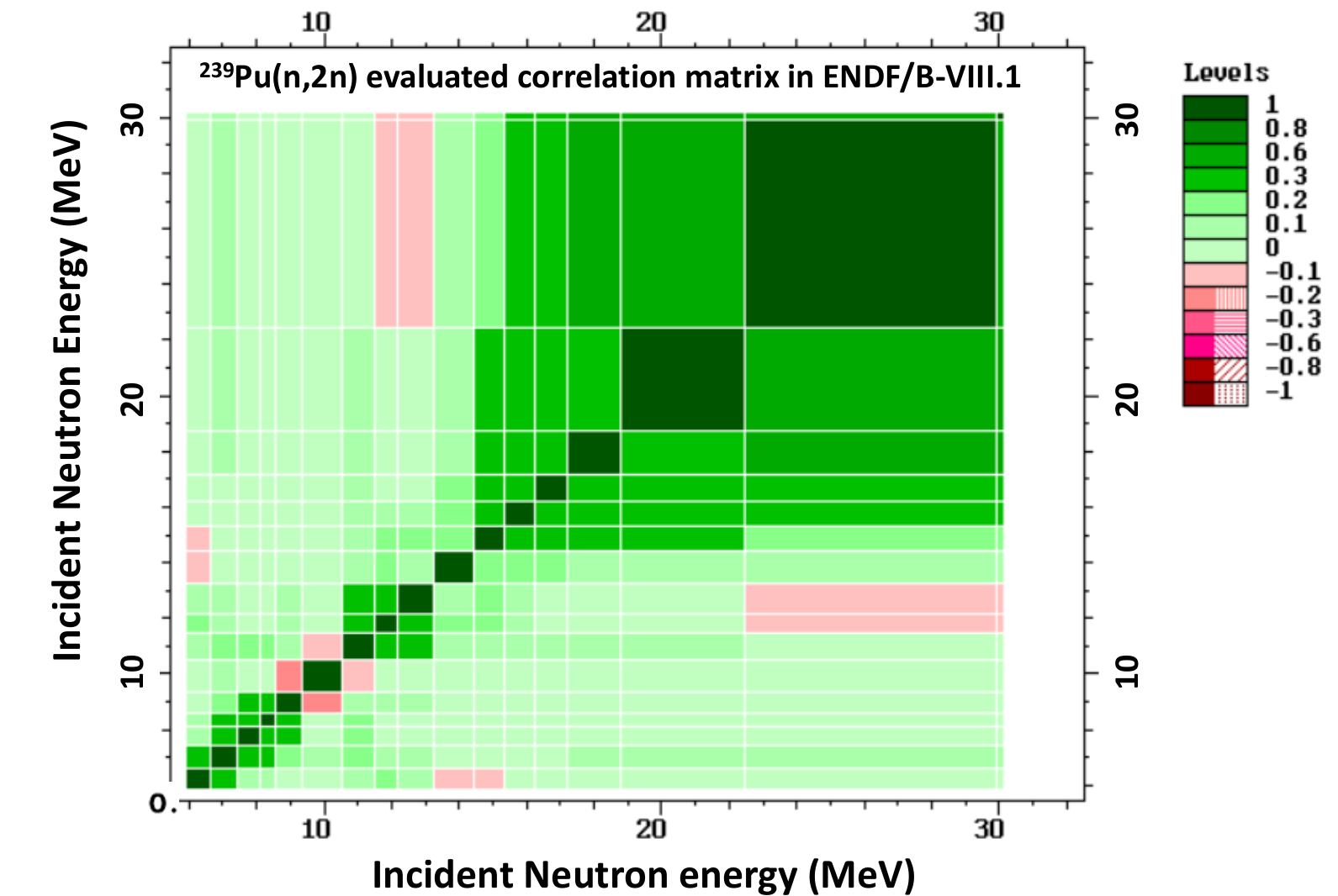}
\vspace{-2mm}
 \caption{ENDF/B-VIII.1 evaluated \nuc{239}{Pu}($n$,2$n$) cross-section correlations for incident neutron energies from 5~MeV up to 30~MeV.}
 \label{fig:pu239-n2n-corr}
 \end{center}
\vspace{-2mm}
\end{figure}

A typical evaluated correlation matrix shows negative and positive correlations due to the experimental data constraints as shown in Fig.~\ref{fig:pu239-n2n-corr}.

\paragraph{Elastic \& inelastic cross sections\newline}
A theoretical improvement in the treatment of the direct reaction channels in a coupled-channel calculation is the diagonalization of the S-matrix using the Engelbrecht-Weidenm\"{u}ller (EW) transformation \cite{Engelbrecht1973}.
\begin{figure}[!thb]
\vspace{-3mm}
 \begin{center}
 \includegraphics[clip,width=\columnwidth]{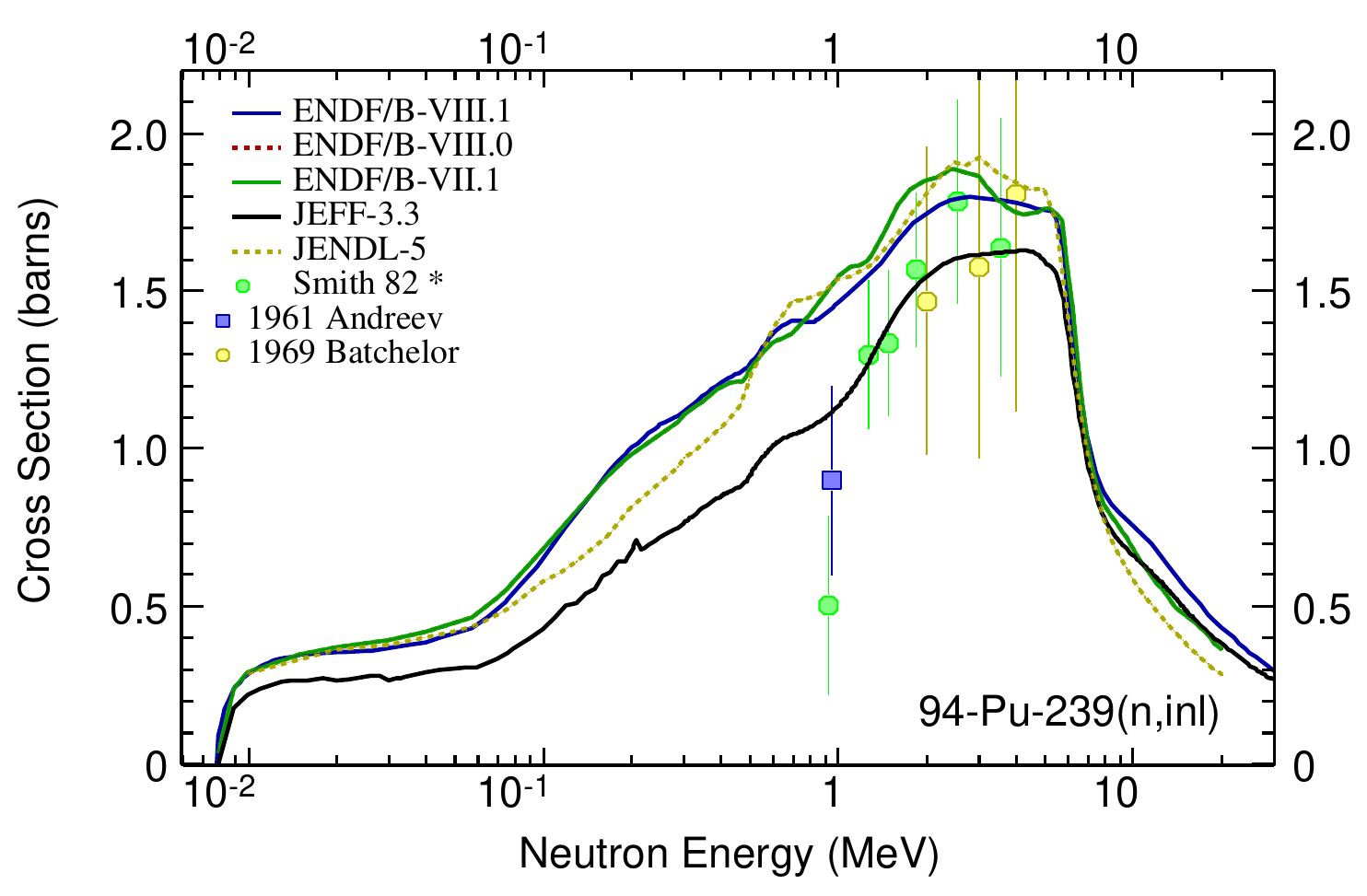}
 \caption{The inelastic cross section in the fast energy range for \nuc{239}{Pu} along with comparison to \prENDF{}, ENDF/B-VII.1, JENDL-5, JEFF-3.3 and selected experimental data from EXFOR \cite{EXFOR}.
 Note that Smith 1982 data~\cite{Smith1982,Smith1982a} are weakly model dependent (model calculations are needed to estimate the inelastic cross section below the experimental threshold).}
 \label{fig:cs_239pu_inel}
 \end{center}
\vspace{-3mm}
\end{figure}
This development was implemented in the \CoH\ code~\cite{Kawano2016}, resulting in a more realistic treatment of the inelastic cross sections when the direct channels are strongly coupled, which is the case in the study of low-energy neutron-induced reactions on $^{239}$Pu. However, this important theoretical improvement has a small impact on the calculated total inelastic cross section as shown in Fig.~\ref{fig:cs_239pu_inel}. Such small impact was expected: there are many open inelastic channels even at very low incident-neutron energy. This is due to the high level density in odd targets and reduces the predicted inelastic enhancement~\cite{Kawano2016} of the EW transformation. The ENDF/B-VIII.1 evaluated inelastic cross section is very close to the ENDF/B-VII.1 evaluation (adopted for \prENDF{}) up to 1~MeV incident neutron energy.

\begin{figure}[!thb]
\vspace{-3mm}
 \begin{center}
 \includegraphics[clip,width=\columnwidth]{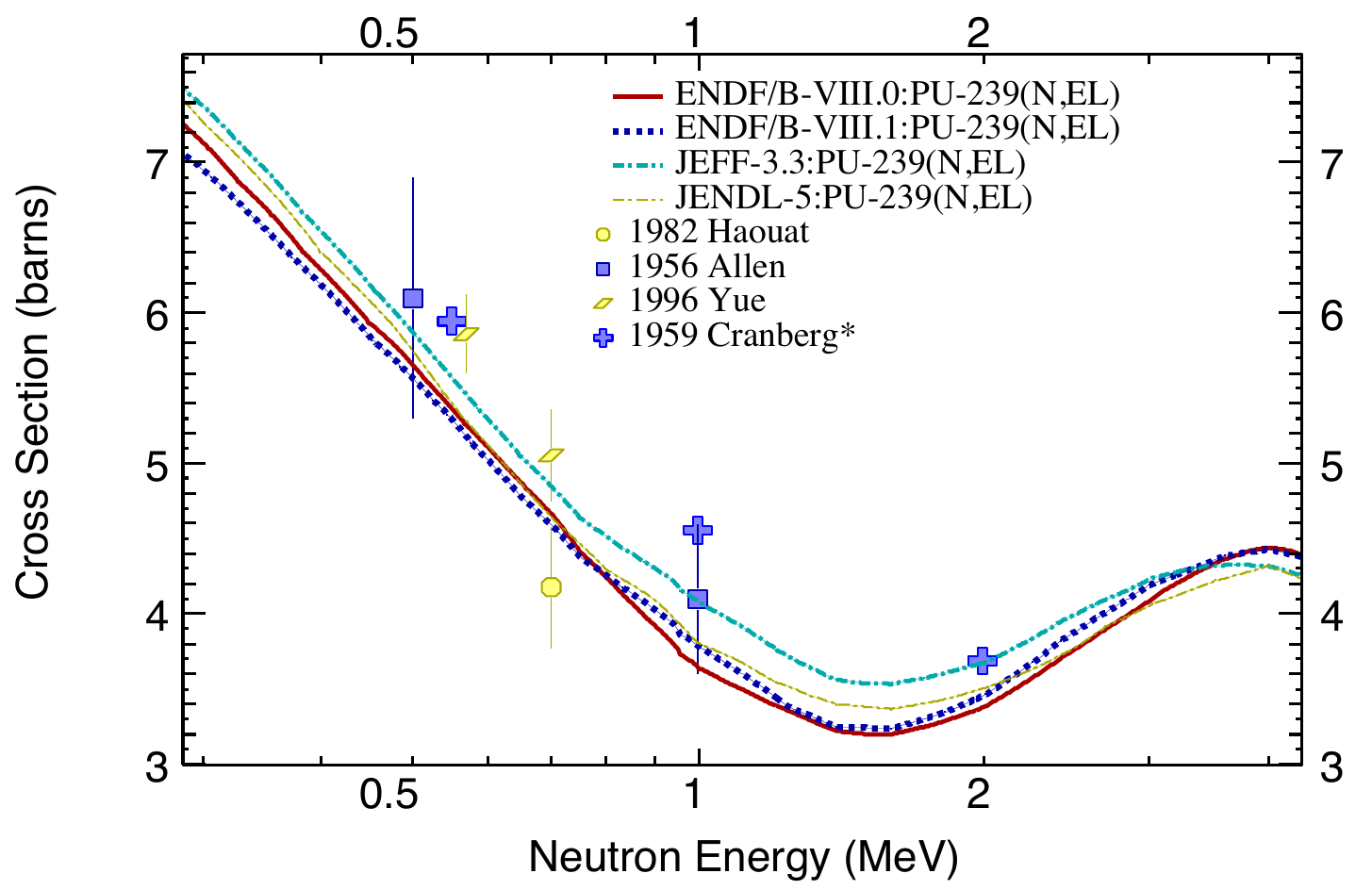}
 \caption{The elastic cross section in the fast energy range for \nuc{239}{Pu} along with comparison to \prENDF{}, ENDF/B-VII.1, JENDL-5, JEFF-3.3 and selected experimental data \cite{yue1994,cranberg1959,knitter1969,coppola1970,Smith1973} from EXFOR \cite{EXFOR}. Note that Cranberg 59 \cite{cranberg1959} data were reduced by 500 mb to consider the inelastic scattering to the first excited level. 
 }
 \label{fig:cs_239pu_el}
 \end{center}
\vspace{-3mm}
\end{figure}
The ENDF/B-VIII.1 total inelastic cross section reaches a maximum of 1.8 b from 2~MeV up to 3~MeV being about 100 mb lower than \prENDF{} and JENDL-5 evaluations. The JEFF-3.3 evaluation shows the best agreement with experimental data below 2~MeV, being about 200~mb lower from 2 up to 5~MeV than the ENDF/B-VIII.1 evaluation. This decrease in the inelastic channel is related to the increased JEFF-3.3 elastic cross section from 1~MeV up to 5~MeV as shown in Fig.~\ref{fig:cs_239pu_el}. These evaluation differences deserve further study, as this energy region corresponds to the maximum number of fission neutrons; therefore, it is the most important energy region for fast reactors.

It was noted that there is the need to increase the inelastic scattering spectrum to the continuum to describe both the differential 14~MeV scattering data by Kammerdiener \cite{Kammerdiener1972} as well as the LLNL pulsed-sphere neutron-leakage spectra \cite{Wong1972}. Description of inelastic neutron emission scattering cross sections and spectra above 5~MeV is a long-standing issue for neutron scattering on actinide nuclei (e.g., see Ref.~\cite{Marcinkowski1993}). Different theoretical approaches have been proposed.  A common empirical method used in ENDF/B-VII and ENDF/B-VIII.0 major actinide evaluations \cite{young2007,capote2018} relies on the combination of direct excitation to the continuum treated by Distorted Wave Born Approximation (DWBA) of collective levels of the even-even core whose properties (excitation energy, spin, parity, deformation lengths) are adjusted to reproduce the 14~MeV scattering neutron spectra and time-of-flight neutron distributions from pulsed-sphere experiments \cite{Wong1972}. The exciton model is used to describe the smooth continuum pre-equilibrium emission. A semi-microscopic approach to inelastic scattering relies on the Tamura Udagawa Lenske pre-equilibrium QRPA structure model \cite{wienke2008}. A new empirical approach was used for calculations of the inelastic scattering in this evaluation that also relies on the exciton model with an effective 1p-1h level density at low energy that emulates the excitation of collective levels \cite{Mumpower2023b}. This approach results in higher total inelastic scattering cross section in ENDF/B-VIII.1 \cite{Mumpower2023a} compared to other evaluations above 10~MeV (including the \prENDF{}) as shown in Fig.~\ref{fig:cs_239pu_inel}. A sound theoretical treatment of the direct inelastic neutron scattering requires a deformed QRPA calculation as discussed in Ref.~\cite{dupuis2024}. Unfortunately, this is technically very challenging and is still work in progress.

\begin{figure}[!thb]
\vspace{-3mm}
 \begin{center}
 \includegraphics[width=80mm]{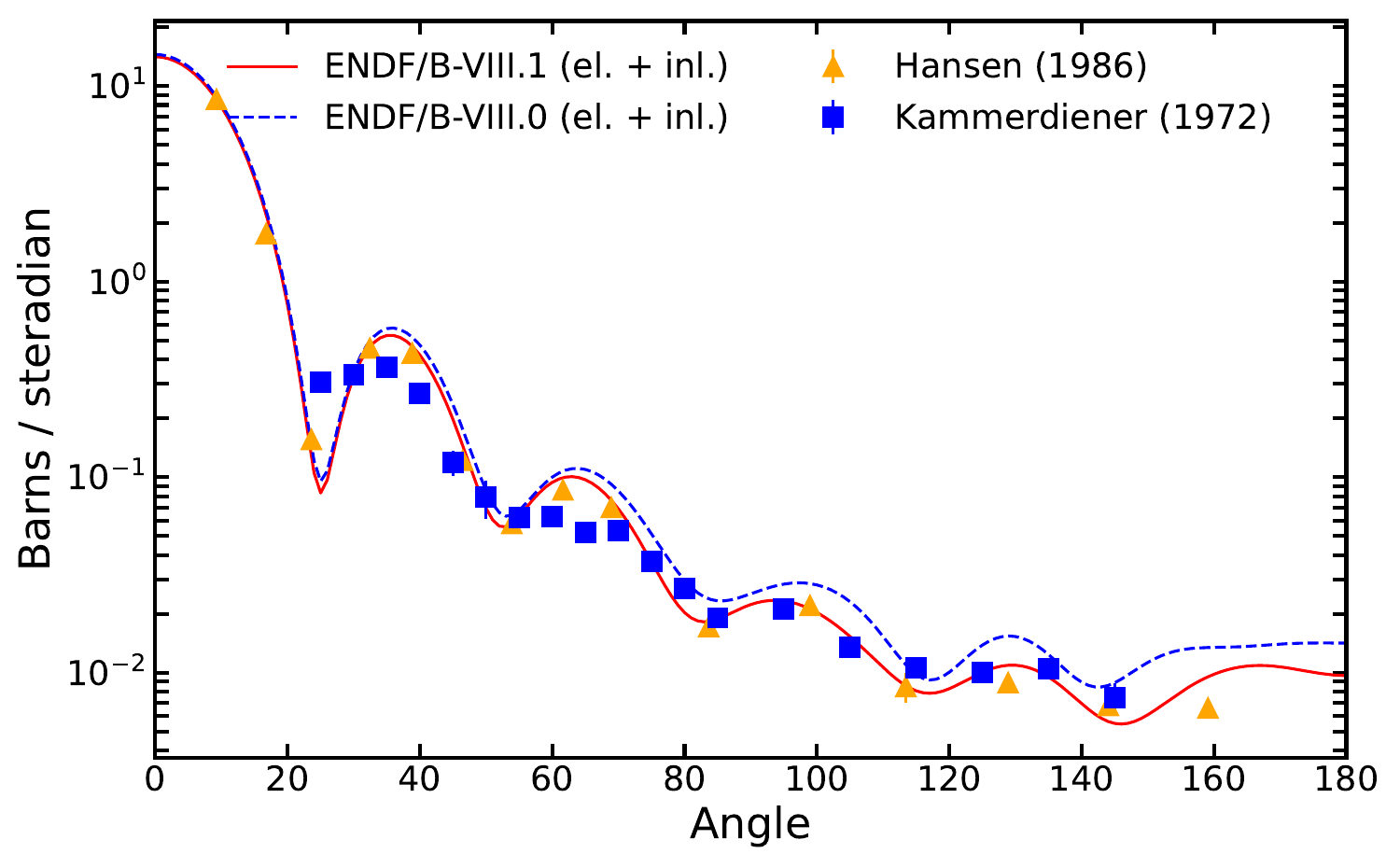}
 \caption{The angular distribution of \nuc{239}{Pu} at incident neutron energy of 14~MeV along with comparison to \prENDF{} and data \cite{Hansen1986,Kammerdiener1972}.}
 \label{fig:cs_239pu_angdist}
 \end{center}
\vspace{-3mm}
\end{figure}

The summed elastic and inelastic angular distribution at 14~MeV is shown in Fig.~\ref{fig:cs_239pu_angdist}.
A marked improvement over \prENDF{} is found with respect to the Hansen \cite{Hansen1986} and Kammerdiener data \cite{Kammerdiener1972}, which is due to the improved dispersive coupled-channel optical model potential (OMP) -- Reference Input Parameter Library (RIPL) 2408\footnote{The OMP was slightly refitted to change the calculated total cross section \cite{Mumpower2023a}, but those changes are not expected to effect the calculated angular distributions.} -- used in the current evaluation \cite{Capote2005,Soukhovitskii2005,Capote2008} where those 14~MeV data were used in the fit.

Cross-section covariances for the elastic, inelastic, and total channels were estimated using a Kalman filter~\cite{Kalman1960}. 
For these channels, the Kalman uncertainties were scaled to match the magnitude used in the ENDF/B-VIII.0 approach. 
For \nuc{239}{Pu}, the fission channel is evaluated independently which causes problems in developing self-consistent cross-channel covariances. 
Cross-channel covariances were, therefore, not included in the present effort and studies are in progress for inclusion in future releases. 

\subsubsection{Integral benchmarks}
Having a new evaluation of neutron-induced reactions on a \nuc{239}{Pu} target in this release, it was essential to perform extensive testing and validation. This comprehensive validation is documented in Section~\ref{sec:integral}.

\begin{table}[!tbh]
\vspace{-2mm}
    \centering
    \caption{Calculated and experimental values of effective delayed neutron fraction for selected fast plutonium benchmark experiments. The notation of * was used to signify that the uncertainty of the calculated result is less than 1 pcm. Only experimental uncertainties were provided for both assemblies, which were unrealistically small. Hence, 8\% uncertainties were added to experimental ones in quadrature following IRPhEP.}
    \label{tab:beta_eff}
    \begin{tabular}{cccc}
        \toprule\toprule
        Benchmark & Measured & ENDF/B-VIII.0 & ENDF/B-VIII.1 \\ \midrule
        A & 0.00195(19) & 0.00183(0)* & 0.00185(0)* \\
        B & 0.00276(23) & 0.00284(4) & 0.00275(4) \\ \bottomrule \bottomrule
    \end{tabular}
\end{table}
Also included in our validation was the calculation of the effective delayed neutron fraction, which describes the fraction of delayed neutrons that contribute to the criticality of the system.
Two systems were considered: (A)~Jezebel (PU-MET-FAST-001-001) and (B) Flattop with plutonium core (PU-MET-FAST-006-001). The results in Table \ref{tab:beta_eff} show that the simulated values from the present evaluation are in better agreement with measured experimental values than ENDF/B-VIII.0.

\paragraph{Future work\newline}
The competition between the elastic and inelastic channels as well as the elastic angular distributions 
across a wide range of incident neutron energies remain open issues in the evaluation of \nuc{239}{Pu}.
Additional targeted measurements may be required to ascertain information on these channels.
Additional information may be needed to confirm the ground-breaking Lougheed $(n,2n)$ experiment at 14~MeV \cite{Lougheed2002,McNabb2001} that drove the (n,2n) evaluations and defined uncertainties since the ENDF/B-VII.0 release.
In the meantime, as we work towards more self-consistent evaluations, it is important to continue to incorporate the latest experimental efforts along with advancements in theoretical modeling.


\subsection{INDEN evaluations}
\label{subsec:n:INDEN}

\subsubsection{\nuc{16}{O}}
\label{subsec:n:16O}


Much work has been done on the $^{17}$O R-matrix analysis at LANL in the last few years: extending the data to higher
energies for some reactions, and adding $^{16}$O$^*$ excited states to the channel configuration. However, it was decided after preliminary data
testing to carry over the ENDF/B-VIII.0 evaluation \cite{Brown2018} to ENDF/B-VIII.1. This decision was taken because much of the new data added to the
analysis was from $(\alpha,n)$ measurements on $^{13}$C targets, and although many of these measurements are clear improvements over the
earlier ones, they still suffer from difficulties in characterizing their absolute normalizations and energy calibrations to the extent that
strong disagreements remain among them.  Most of the experimental groups are diligently working to improve their data, and hopefully the situation
concerning the $(\alpha,n)$ data will become clearer in the near future.  We will not repeat here the discussion about the $^{16}$O
evaluation in Ref.~\cite[Section 10]{Brown2018}, but rather summarize the few changes made for ENDF/B-VIII.1.

Some of the newer $(\alpha,n_0)$ measurements, especially those at lower energies, confirm the normalization scale that was determined in
the ENDF/B-VIII.0 analysis, which was 0.94 times the original cross-section scale of Bair and Haas \cite{BH73}.  This was also consistent
with the $^{16}$O$(n,\alpha_0)$ cross section measured by Giorginis \cite{Giorginis17} up to $E_n=9$ MeV.

Therefore, no changes were made in the $(n,\alpha_0)$ cross section for ENDF/B-VIII.1.  However, the $(n,\alpha_1)$, $(n,\alpha_2)$, and
$(n,\alpha_3)$ cross sections were reduced by a factor of 2 relative to ENDF/B-VIII.0, as suggested by R. Capote, to agree better with
experimental data from Refs.~\cite{Davis63} and \cite{Parnell68}.  The total cross section remains identical to the ENDF/B-VIII.0 evaluation above 5 MeV.

Simakov and Fischer \cite{Simakov2022} made a comprehensive analysis of the ORNL oxygen 60'' broomstick experiment \cite{Maerker72} and demonstrated a much better performance of the JEFF-3.3 (= ENDF/B-VII.1) $^{16}$O evaluation compared to ENDF/B-VIII.0 from 5.7~MeV up to 7.2~MeV \cite{Simakov2022}. Simakov and Fischer estimated the need for a 3\% reduction of the ENDF/B-VIII.0 total cross section from 5.5~MeV up to 7~MeV and a 20\% reduction of the $(n,\alpha_0)$ cross section. Capote and Trkov have found that a proposed reduction in the total cross section, especially in the region from 6.2~MeV up to 6.7~MeV is consistent with measured data by Finlay (1993), Cierjacks (1980), Larson (1980), Schracks (1972), Perey (1972) and renormalized Cierjacks (1968). This is one indication that further work needs to be done on the R-matrix analysis to accommodate experimental information about the $^{17}$O system in a unitary framework that describes all the reactions well with special focus on improving the total neutron cross section and alpha emission channels.

\subsubsection{\nuc{18}{O}}
\label{subsec:n:18O}

The evaluation was adopted from the ENDF/B-VIII.0 library, except for the radiative capture, taken from EAF-2010 as proposed by A. Trkov to improve the agreement with existing experimental data.


\subsubsection{\nuc{19}{F}}
\label{subsec:n:19F}


LiF--NaF (FLiNa) or LiF--BeF$_2$ (FLiBe) salts with highly enriched $^7$Li isotope are strong candidates for Generation IV reactor designs, such as the High Temperature Graphite Reactor (HTGR) and Fluoride-cooled High-temperature Reactor (FHR) or Molten Salt Reactor  (MSR). Some of those advanced reactor designs are being commercially developed, e.g., by Kairos Power and Terrapower. Fluorine is also an integral component of uranium manufacturing/enrichment process as uranium hexafluoride (UF$_6$) and is also used directly as nuclear fuel. Finally, fluorine is a major component of Teflon$^{\circledR}$ which is commonly used to reduce the average neutron energy in critical assemblies.

\begin{figure}[!thb]
\vspace{-2mm}
\centering
\includegraphics[clip,width=\columnwidth]{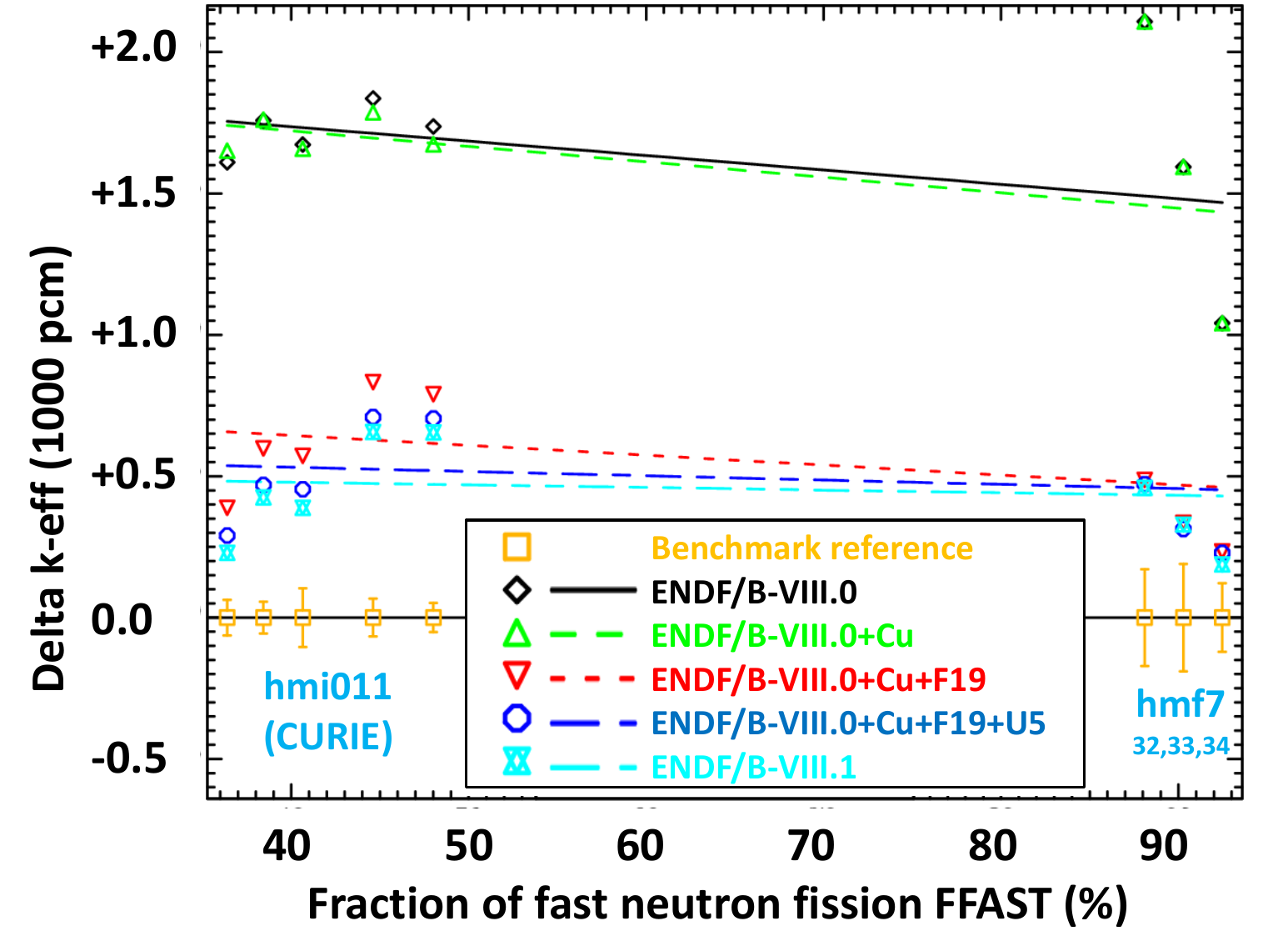}
\vspace{-3mm}
\caption{Criticality differences to the experimental benchmark values $\Delta~k_{\mathrm{eff}}$ of $^{19}$F-sensitive fast ICBESP benchmarks (HMI011/1,2,3,4,5 and HMF007/32,33,34~\cite{ICSBEP}). Calculated values are given for the reference ENDF/B-VIII.0 library, then subsequently adding $^{63,65}$Cu evaluations from ENDF/B-VIII.1 (+Cu), $^{63,65}$Cu and $^{19}$F evaluations from ENDF/B-VIII.1 (+Cu+F19), and $^{63,65}$Cu, $^{19}$F, and $^{235}$U evaluations from ENDF/B-VIII.1 (+Cu+F19+U5) library. Finally, the full ENDF/B-VIII.1 results are also shown.
}
\label{fig:f19-icsbep-Teflon}
\end{figure}

Therefore, $^{19}$F nuclear data are very important for criticality safety, design, and operation of many nuclear facilities including some advanced reactor designs. $^{19}$F nuclear data deficiencies have been shown in the validation of the ENDF/B-VIII.0 library \cite{Brown2018} (e.g., for the HMF007/ cases 32,33,34 \cite{ICSBEP}) which were calculated high by 1500 pcm on average.

\begin{figure}
\vspace{-5mm}
\centering
\includegraphics[width=0.70\columnwidth]{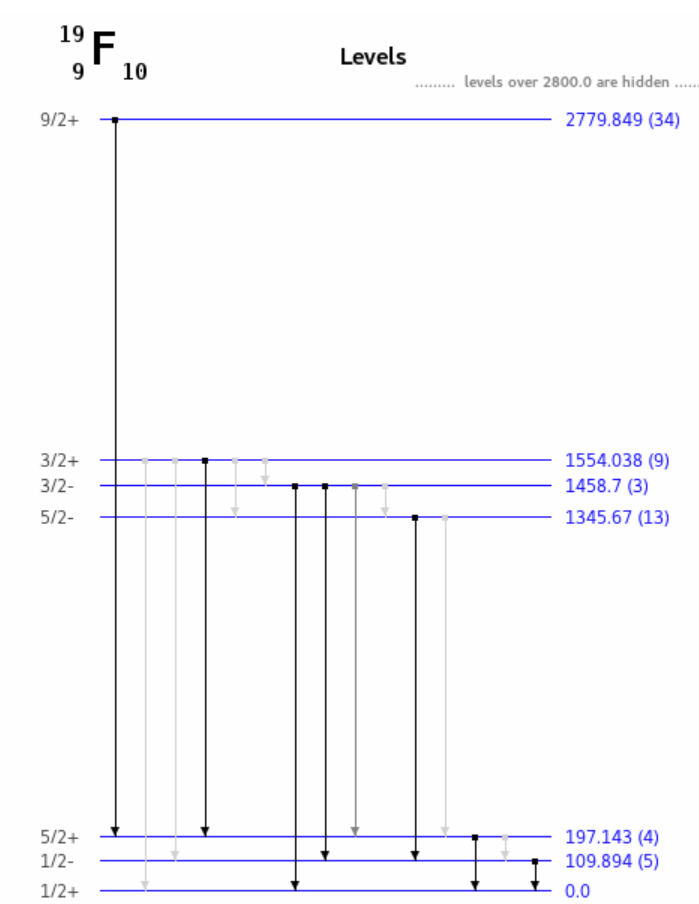}
\vspace{-3mm}
\caption{Low lying nuclear levels of $^{19}$F  below 3 MeV of excitation energy~\cite{TILLEY1995}.}
\label{fig:f19-levels}
\vspace{-1mm}
\end{figure}

\begin{figure}[!thbp]
\vspace{-2mm}
\includegraphics[width=\columnwidth]{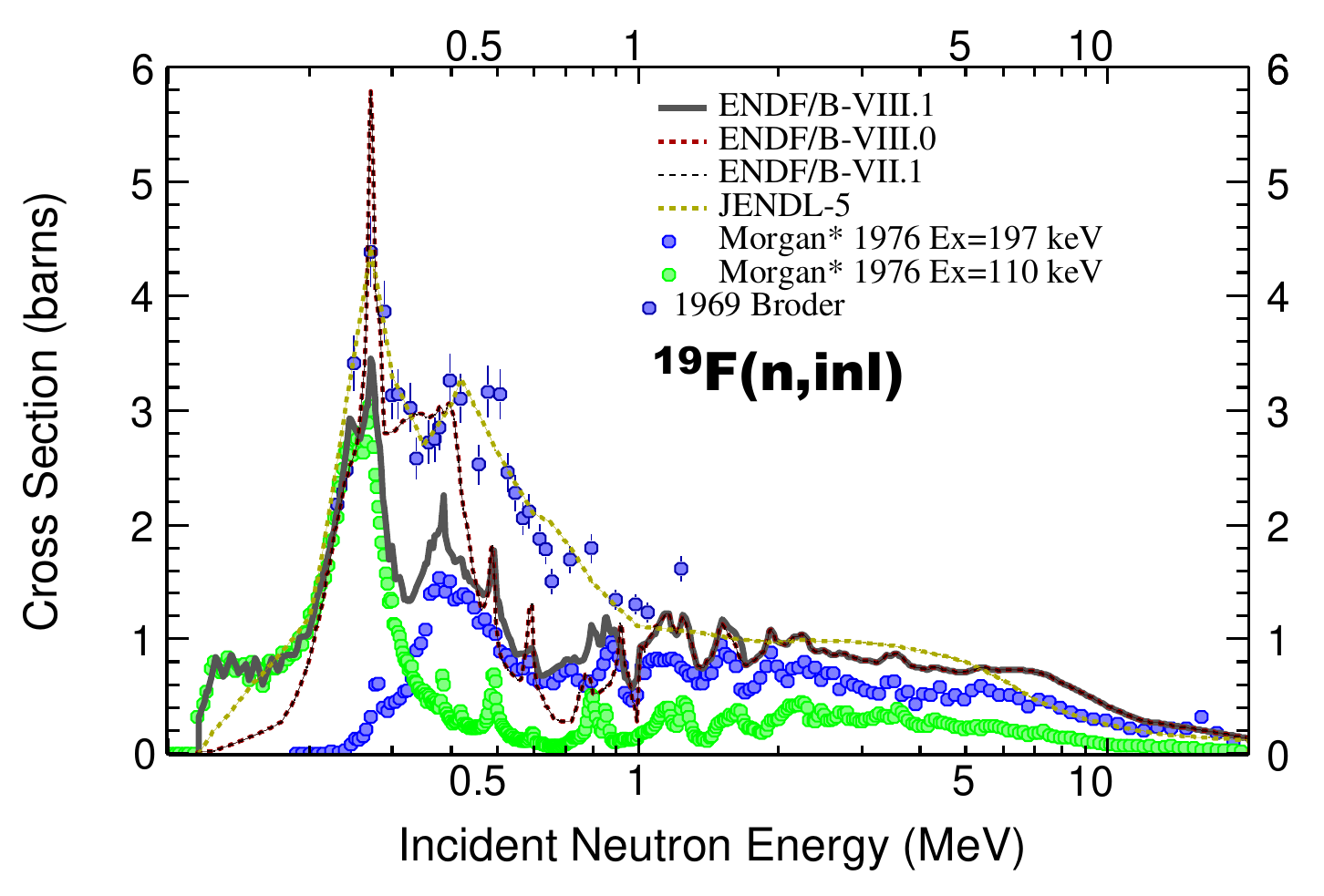}
\vspace{-3mm}
\caption{Neutron induced inelastic scattering cross sections of $^{19}$F up to 20~MeV. Experimental data from Broder {\it et al.}~\cite{Broder:1969} and total inelastic data derived from Morgan {\it et al.}~\cite{Morgan:1976} are compared to JENDL-5 \cite{jendl5}, ENDF/B-VIII.0=ENDF/B-VII.1 \cite{ENDF-VII.1,Brown2018} and ENDF/B-VIII.1 (INDEN) evaluations.}
\label{fig:f19-INL}
\end{figure}

Fluorine also constitutes the first case of a nuclear data deficiency identified through machine learning (ML) techniques by Neudecker and collaborators~\cite{Neudecker2020f19}: the $^{19}$F(n,inl) cross section from $0.4-0.9$~MeV was shown to be problematic by a study of biases in simulating approximately 1,000 ISCBEP $k_\mathrm{eff}$ values. After these ML results pointed to a potential shortcoming in this cross section,  evaluated ENDF/B-VIII.0 $^{19}$F(n,inl) cross section were compared to associated differential experimental data and large discrepancies were found. This problematic behavior was attributed to the resonance range of the evaluation having been extended to too high incident neutron energies. 

\begin{figure}[!thbp]
\vspace{-4mm}
\includegraphics[width=\columnwidth]{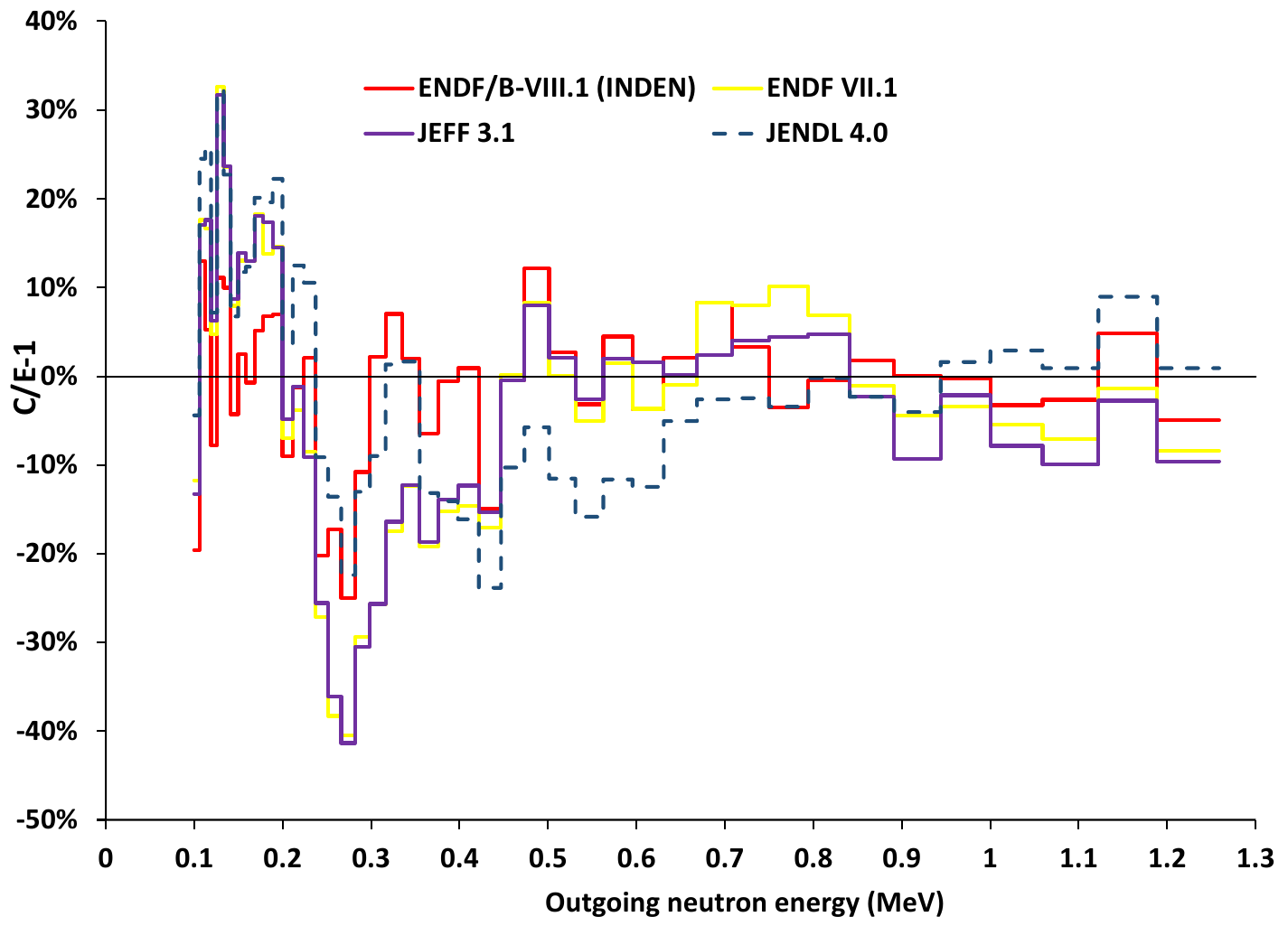}
\vspace{-2mm}
    \caption{LR0 reactor spectrum neutron leakage of FLINA solution from 100~keV up to 1.2~MeV of leaked neutrons measured with HPD detectors~\cite{Kostal:2015,Losa:2018}. Experimental benchmark values (E) are compared to calculations (C) using ENDF/B-VII.0, JEFF-3.1, JENDL-4, and the new INDEN evaluation adopted for the ENDF/B-VIII.1 library\footnote{private communication by M. Kostal, December 2023.}.}
\label{fig:f19-FLINA-Rez2015}
\vspace{-1mm}
\end{figure}

Of particular concern were the recently developed CURIE criticality experiments, which are intermediate-spectrum high-enriched uranium (HEU) teflon-reflected assemblies, which also contain a large amount of copper. They are listed in the International Criticality Safety Benchmark Evaluation Project (ICSBEP) Handbook \cite{ICSBEP} as HEU-MET-INTER-011 (HMI011) and showed an overestimation of criticality from reference benchmark values by about 1700 pcm for \prENDF{} as shown in Fig.~\ref{fig:f19-icsbep-Teflon}. The impact of latest improvements in $^{63,65}$Cu evaluations within the INDEN/ORNL collaboration adopted for the ENDF/B-VIII.1 library is negligible (compare green triangles versus black rhombi in the figure), and the performance remains pretty poor. $^{63,65}$Cu  ENDF/B-VIII.1 evaluation induced small criticality changes in HEU intermediate-spectrum assemblies (e.g., Zeus benchmarks), which is confirmed in HMI011 cases. The criticality overestimation of HMF007/32,33,34 cases remain between 1100 and 2000 pcm as shown in Fig.~\ref{fig:f19-icsbep-Teflon}.
The largest improvements in calculated criticality of more than 1000 pcm is due to the improvements in $^{19}$F data (red triangles). Additional 100 pcm are due to $^{235}$U changes (blue circles). Other materials have a very minor impact.

\paragraph{$^{19}$F evaluation changes: \newline}
\label{subsec:19F_validation}

\begin{figure*}[!t]
    \centering
    \subfigure[~Carbon measured and simulated quasi-differential neutron scattering at 45 degrees.]
    {\includegraphics[width=0.49\textwidth]{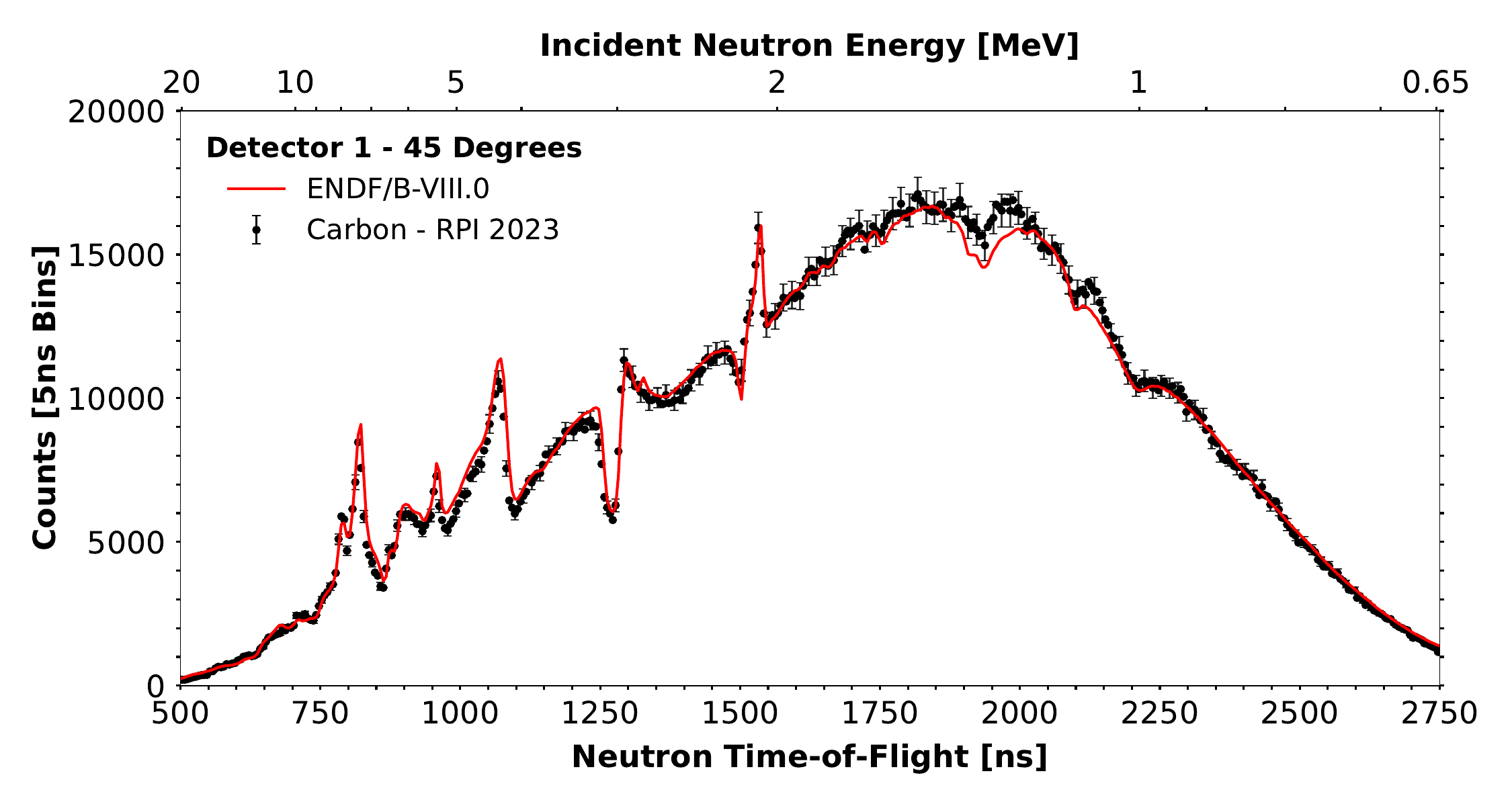}
    \label{fig:carbon-RPI-det1-45deg-Teflon}
    }
    \hfill
    \subfigure[~Carbon measured and simulated quasi-differential neutron scattering at 150 degrees.]{%
        \includegraphics[width=0.49\textwidth]{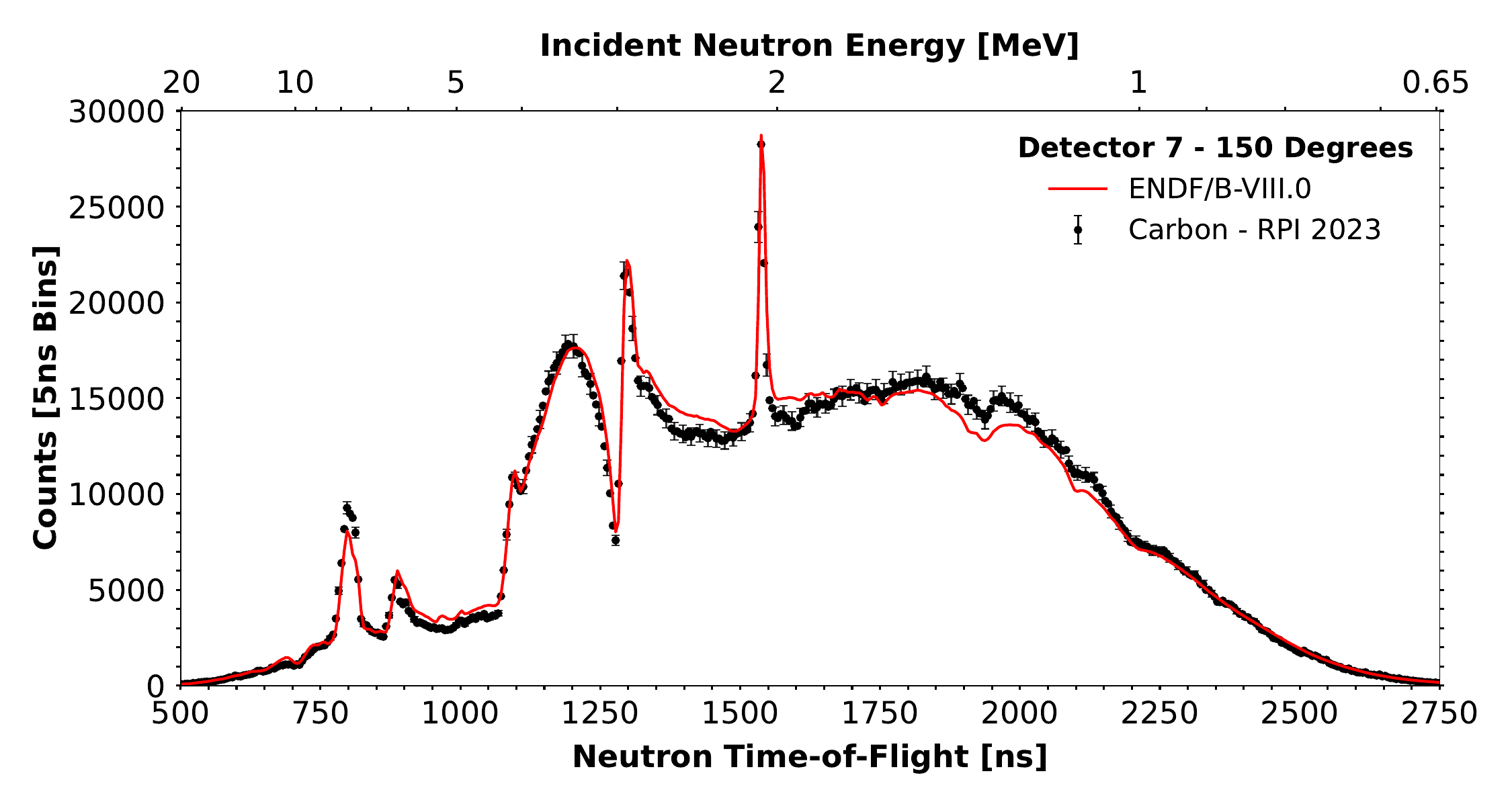}
        \label{fig:carbon-RPI-det7-150deg-Teflon}
    }

    \vspace{0.25cm}
    \subfigure[~Teflon$^{\circledR}$ measured (E) and simulated (C) quasi-differential neutron scattering at 45 degrees.]{%
    \includegraphics[width=0.49\textwidth]{neutrons/figs/teflon-RPI-det1-45deg.pdf}
    \label{fig:teflon-RPI-det1-45deg}
    }
    \hfill
    \subfigure[~Teflon$^{\circledR}$ measured (E) and simulated (C) quasi-differential neutron scattering at 150 degrees.]{%
        \includegraphics[width=0.49\textwidth]{neutrons/figs/teflon-RPI-det7-150deg.pdf}
        \label{fig:teflon-RPI-det7-150deg}   
    }
    \vspace{-2mm}
    \caption{Results from the Teflon$^{\circledR}$ quasi-differential neutron scattering experiment at RPI \cite{SiemersCSEWG2023,SiemersWINS2023} in the incident neutron energy range of 0.65 to 20 MeV. Carbon neutron scattering yields, measured during the Teflon$^{\circledR}$ experiment, at a forward and backward detector compared to MCNP simulations of the ENDF/B-VIII.0 carbon evaluation are presented in Figs.~\ref{fig:carbon-RPI-det1-45deg-Teflon} and \ref{fig:carbon-RPI-det7-150deg-Teflon}, respectively. Measured Teflon$^{\circledR}$ neutron scattering yields at corresponding forward and backward detectors compared to MCNP simulations of the ENDF/B-VIII.0, JEFF-3.3, JENDL-5.0, and ENDF/B-VIII.1 ${}^{19}$F evaluations are presented in Figs.~\ref{fig:teflon-RPI-det1-45deg} and \ref{fig:teflon-RPI-det7-150deg}, respectively. $C/E$ ratios are provided to compare the experimental and simulated results along with the experimental uncertainty band.}
    \label{fig:teflon-rpi-qd-scattering}
\end{figure*}

Starting from the \mbox{ENDF/B-VIII.0} evaluation, the following cross-section changes were made:
\begin{itemize}

\item Japanese colleagues pointed out deficiencies of existing $^{19}$F evaluations in the elastic angular distributions below 1 MeV of the neutron incident energy in the RRR~\cite{jendl5}. In this evaluation, elastic angular distributions were derived by fitting Elwyn {\it et al.} data up to 2 MeV~\cite{Elwyn:1964}; above 2 MeV, the elastic angular distributions were adopted from the JENDL-5 evaluation \cite{jendl5}.

\item The total and capture cross sections were adopted from JENDL-5 \cite{jendl5}.

\item The first two discrete level inelastic cross sections (levels at 110 and 197 keV as shown in Fig.~\ref{fig:f19-levels}) were replaced by the (n,n'$\gamma$) data of Morgan~\cite{Morgan:1976}. A comparison of the total inelastic experimental data versus evaluated cross sections is shown in Fig.~\ref{fig:f19-INL}. Many evaluations, including ENDF/B-VIII.0 \cite{Brown2018} and  JENDL-5 \cite{jendl5}, followed the (n,n'$\gamma$) data by Broder {\it et al.}~\cite{Broder:1969}, which were measured relative to the $^{56}$Fe(n,n'$\gamma$) data. Note that this monitor reaction was not recommended as a reference (n,n'$\gamma$) cross section by the IAEA Neutron Standard committee~\cite{carlson2018}.
Therefore, Fe was considered a poor reference for (n,n'$\gamma$), which makes Broder measurement~\cite{Broder:1969} questionable.

    Morgan and coworkers measured the $^7$Li(n,n'$\gamma$) data at a given detection angle together with the $^{19}$F(n,n'$\gamma$) data. Morgan measurements for $^7$Li(n,n'$\gamma$) are in good agreement with the proposed $^7$Li(n,n'$\gamma$) reference cross section \cite{carlson2018}. The angular distributions of the first two $^{19}$F inelastic levels are known to be almost isotropic (the first level is isotropic by definition having spin=1/2). This allows us to calculate the discrete level excitation function shown in Fig.~\ref{fig:f19-INL} from Morgan measured data by simply multiplying the measured angular distribution by $4\pi$. Derived inelastic cross sections for the first two levels from Morgan \etal ~\cite{Morgan:1976} were used in the newly evaluated file as shown in Fig.~\ref{fig:f19-INL} and reduced the evaluated total inelastic cross section below 1~MeV by up to 30\% compared to the evaluations, which followed Broder data \cite{Broder:1969} (e.g., the ENDF/B-VIII.0 and JENDL-5 files).
\item The $^{19}$F(n,2n) cross sections were adopted from the IRDFF-II evaluation \cite{IRDFF}.
\end{itemize}

\paragraph{$^{19}$F Validation\newline}

Improvements in $^{19}$F evaluation, in particular the reduction of the inelastic cross section and the elastic angular distribution changes, decreased the excess criticality in fast HEU benchmarks to below 500~pcm on average as shown in Fig.~\ref{fig:f19-icsbep-Teflon}. The average excess-criticality dependence on the fast fission neutron fraction (FFAST\%) also flattened significantly. Further improvement is needed and is probably related to remaining issues in cross sections and angular distributions of excited levels above 1.3~MeV as discussed below.

Neutron spectrum in FLINA salt for energies between 0.1~MeV and 1.3~MeV was measured using hydrogen proportional counters by Kostal {\it et al.}~\cite{Kostal:2015,Losa:2018}. The direct comparison between experimental and calculated values is realized by means of a $C/E-1$ comparison as shown in Fig.~\ref{fig:f19-FLINA-Rez2015}: a significant improvement in the $C/E$ is observed for the new ENDF/B-VIII.1 evaluation.

An additional validation of the proposed ${}^{19}$F changes is provided by a recent quasi-differential measurement of the scattered neutron yield from a Teflon$^{\circledR}$ (C$_2$F$_4$)$_n$ sample, undertaken by Siemers and colleagues at the Rensselaer Polytechnic Institute (RPI) and presented at the WINS 2023 workshop \cite{SiemersWINS2023} and CSEWG 2023 \cite{SiemersCSEWG2023} as shown in Fig.~\ref{fig:teflon-rpi-qd-scattering}. Neutron scattering and emissions were measured in time-of-flight from a 1.95'' thick 3'' diameter right cylinder of virgin Teflon$^{\circledR}$ with an estimated 6\% systematic uncertainty in the incident neutron energies range of 0.65~MeV up to 20~MeV. A 2.75'' thick 3'' diameter right cylinder sample of carbon was also measured as part of the Teflon$^{\circledR}$ experiment to validate the experimental findings. MCNP6 was used to perform the radiation transport calculations of the experiment to compare evaluated nuclear data with the experimental data. 

First, the measured neutron scattering yield from the carbon sample was compared to evaluated carbon nuclear data from ENDF/B-VIII.0 (which is equal to ENDF/B-VIII.1). Results for a typical forward scattering angle (45 degrees) are shown in Fig.~\ref{fig:carbon-RPI-det1-45deg-Teflon} and for a typical backward scattering angle (150 degrees) in Fig.~\ref{fig:carbon-RPI-det7-150deg-Teflon}. Excellent agreement is observed between experiment and evaluation throughout most of the experimental energy region. Minor disagreements are noted around 2.5 MeV and 7.5 MeV at the backward angle and around 1.3 MeV and 5 MeV at both the forward and backward neutron scattering angles. Note that in another, independent, quasi-differential neutron scattering experiment from RPI (undertaken to measure the scattered neutron yield from tantalum), similarly good agreement between the measured carbon sample (same sample as measured in the Teflon$^{\circledR}$ experiment) and the simulation was observed throughout the experimental incident neutron energy region at both detection angles provided (see Figs.~\ref{fig:carbon-RPI-det1-45deg-Ta} and ~\ref{fig:carbon-RPI-det6-110deg-Ta} in a later section). Therefore, the minor discrepancies observed for carbon in Teflon$^{\circledR}$ experiments may show deficiencies in evaluated carbon nuclear data, especially at backward angles. Note that the carbon elastic angular distributions is the only standard angular distribution up to 1.8~MeV of neutron incident energy~\cite{carlson2018}.

Evaluated ${}^{19}$F nuclear data from the ENDF/B-VIII.0, JEFF-3.3, JENDL-5, and ENDF/B-VIII.1 libraries were used to reproduce the measured Teflon$^{\circledR}$ neutron scattering yield in simulation and compared with the measured data. Results for a typical forward scattering angle (45 deg) are shown in Fig.~\ref{fig:teflon-RPI-det1-45deg} and for a typical backward scattering angle (150 deg) in Fig.~\ref{fig:teflon-RPI-det7-150deg}. Note that these scattering angles correspond to the angles of the aforementioned carbon validation measurement. The new ENDF/B-VIII.1 evaluation adopted from INDEN is observed to agree best with the experimental data. However, some underestimation of the measured scattered neutron yield remains between 1.3~MeV and 2~MeV at forward angles. Despite this, evaluation and experimental agreement for forward angles is acceptable.

On the other hand, significant differences are observed between the evaluation and experiment at backward angles. Particularly in the low-lying resonances corresponding to inelastic scattering to the excited ${}^{19}$F nuclear levels from 1.3~MeV up to 3~MeV. These discrete levels of the $^{19}$F nucleus are shown in Fig.~\ref{fig:f19-levels}. The third, fourth, fifth (grouped between 1.3--1.5~MeV), and sixth (2.78~MeV) excited levels are present in this region of disagreement. Although the ENDF/B-VIII.1 evaluation shows better performance at backward angles, no evaluation is able to reproduce the resonance structure observed in the experiment between 1300 and 2100 ns (roughly from 1.2~MeV up to 3.0~ MeV). The results from this experiment demonstrate that we have advanced our understanding. However, new neutron inelastic scattering measurements, including angular distributions, of the third to sixth excited levels with excitation energies from 1.34~MeV up to 2.78~MeV are needed to further improve our knowledge of the $^{19}$F inelastic scattering cross sections. ENDF/B-VIII.0 covariances were unchanged.


\subsubsection{\nuc{28,29,30}{Si}}
\label{subsec:n:28-29-30Si}


The evaluation work on three major silicon isotopes was an INDEN collaboration led by ORNL/IAEA as described in Ref.~\cite{pigni:2018}. In that work, resonance parameter evaluations performed at ORNL in 2002~\cite{Derrien:2002} were revised in their thermal constant values following the reference values recommended by the IAEA project Evaluated Gamma-ray Activation File (EGAF) \cite{EGAF,EGAF-report}, especially the $^{28}$Si(n,$\gamma$) cross section of $\sigma_{th}=186$~mb. The energy-dependent contribution of the \TEDCA~code for direct capture cross-section calculations in the 2002 ORNL evaluation was replaced by new \CUPIDO\footnote{F. Dietrich, unpublished code.} direct capture calculations.

\begin{figure}[tbp]
\vspace{-2mm}
\centering
\includegraphics[scale=0.70]{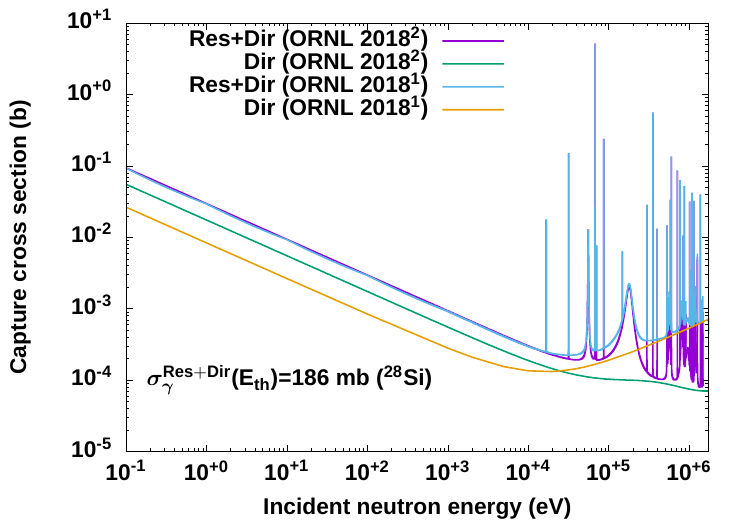}
\vspace{-2mm}
\caption{\nuc{28}{Si}(n,$\gamma$) cross sections reconstructed from the
  ORNL~(2002) evaluation and the ORNL-1/ORNL-2 (2018) evaluations in the neutron
  energy range up to 1.75~MeV. Figure taken from Ref.~\cite{pigni:2018}.}\label{fig:xcscap28}
\vspace{-2mm}
\end{figure}
\begin{figure}[tbp]
\vspace{-2mm}
\centering
\includegraphics[scale = 0.70]{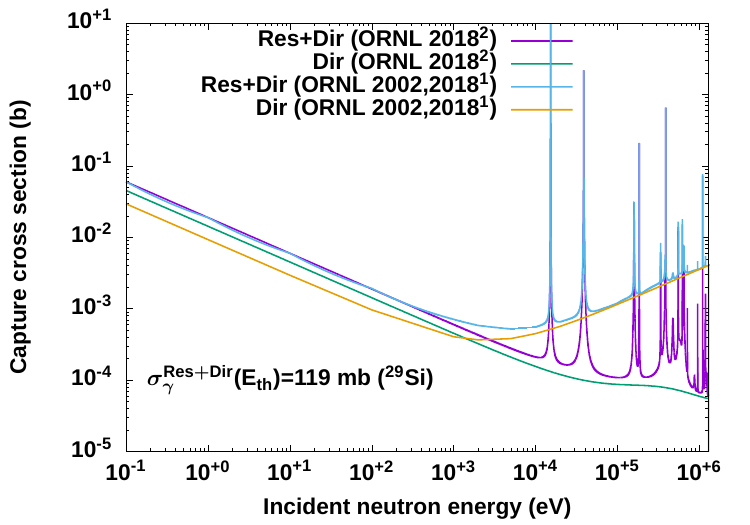}
\vspace{-2mm}
\caption{\nuc{29}{Si}(n,$\gamma$) cross sections reconstructed from the
  ORNL~(2002) evaluation and the ORNL-1/ORNL-2 (2018) evaluations in the neutron
  energy range up to 1.75~MeV. Figure taken from Ref.~\cite{pigni:2018}.}\label{fig:xcscap29}
\vspace{-2mm}
\end{figure}
\begin{figure}[tbp]
\vspace{-2mm}
\centering
\includegraphics[scale = 0.70]{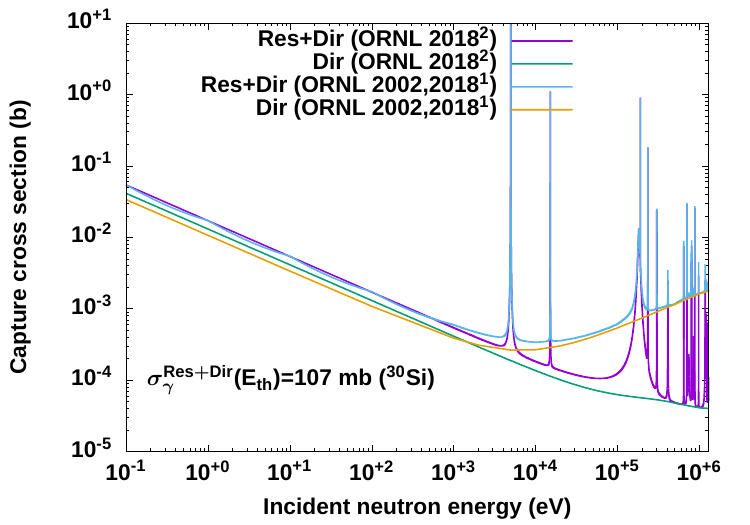}
\vspace{-2mm}
\caption{\nuc{30}{Si}(n,$\gamma$) cross sections reconstructed from the
  ORNL~(2002) evaluation and the ORNL-1/ORNL-2 (2018) evaluations in the neutron
  energy range up to 1.75~MeV. Figure taken from Ref.~\cite{pigni:2018}.}\label{fig:xcscap30}
\vspace{-2mm}
\end{figure}

\begin{figure}[tbp]
\vspace{-2mm}
\centering
\hspace{-7mm}\includegraphics[clip,scale=0.42, trim = 33mm 22mm 42mm 15mm]{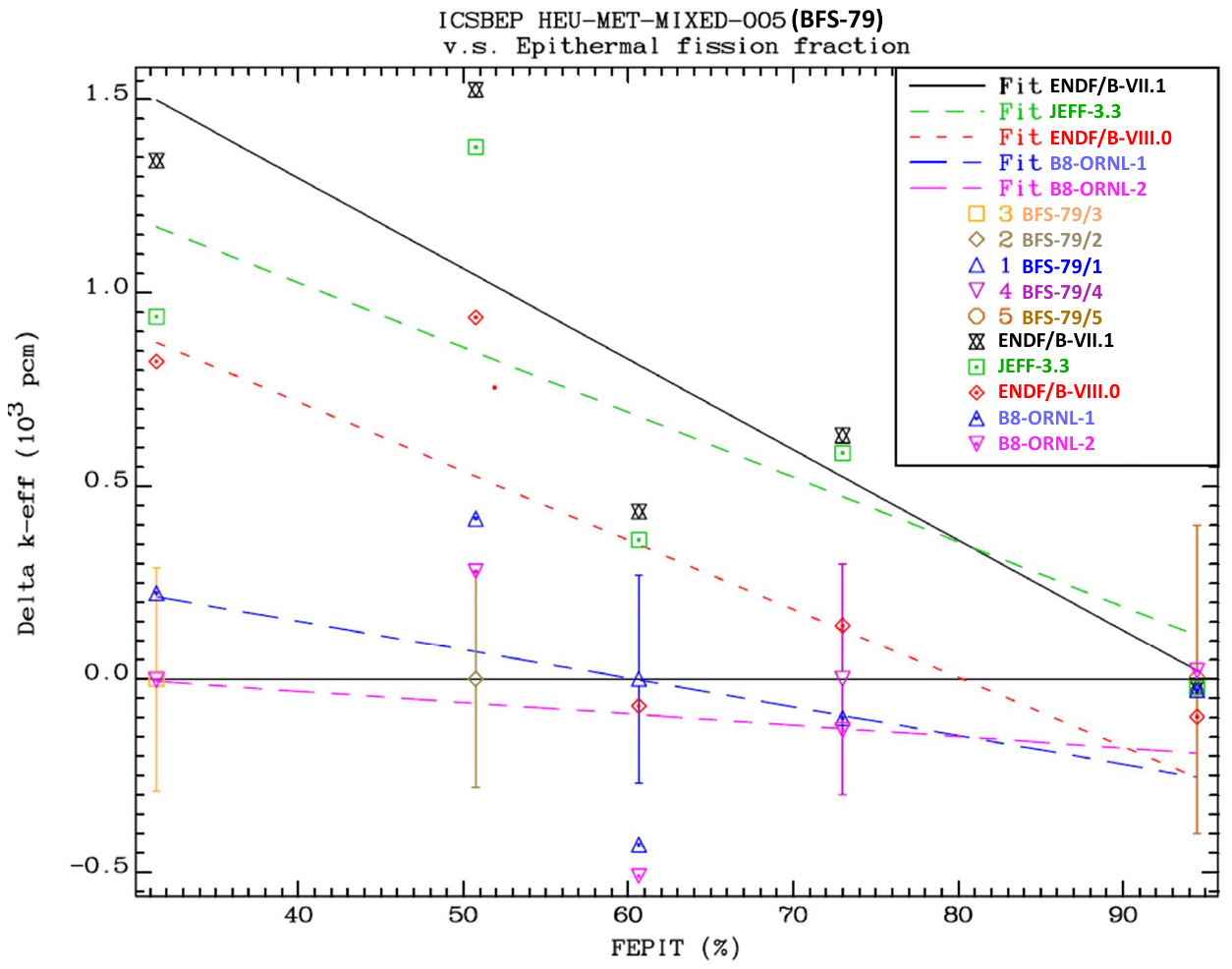}
\vspace{-3mm}
\caption{Results for the HEU-MET-MIXED THERM (HMM005) experiments from the ICSBEP compilation \cite{ICSBEP} for different evaluated libraries: ENDF/B-VIII.1 (e80-ORNL-2), ENDF/B-VIII.0,
JEFF-3.3, and ENDF/B-VII.1. Comparison between ORNL-1 and ORNL-2 evaluations. Figure taken from Ref.~\cite{pigni:2018}.}\label{fig:si-hmm005}
\vspace{-2mm}
\end{figure}

Figs.~\ref{fig:xcscap28}--\ref{fig:xcscap30}, taken from Ref.~\cite{pigni:2018}, graphically summarize the changes in the direct capture contribution and corresponding adopted thermal capture values. It should be noted that \CUPIDO\ calculations showed a better physical behavior of the capture increase in the fast neutron region. As shown in Ref.~\cite{pigni:2018}, results of benchmark calculations performed at the IAEA evidenced an excellent agreement with the IPPE HMM005 benchmark (BFS-079), especially designed to test silicon cross sections as shown in Fig.~\ref{fig:si-hmm005}. Benchmark testing was performed for two different evaluated datafiles: ORNL-1 and ORNL-2. The first included changes only for the \nuc{28}{Si} isotope and the latter for all three isotopes -- \nuc{28,29,30}{Si}. This was motivated by the unphysical behavior of the \TEDCA\ calculation of the direct capture cross sections reported in the 2002 ORNL evaluation. Due to these changes, a decreased reactivity of about 800~pcm for thermal assemblies was expected by the overall increased capture cross sections in the low-energy region.

Calculated results for the HMM005 benchmark experiments from the ICSBEP compilation \cite{ICSBEP} are plotted in Fig.~\ref{fig:si-hmm005} versus measured benchmark values. Calculations were undertaken for different evaluated libraries: ENDF/B-VIII.1 (e80-ORNL-2), ENDF/B-VIII.0, JEFF-3.3, and ENDF/B-VII.1. It is clearly seen that the new INDEN Si evaluation adopted for the ENDF/B-VIII.1 library shows a perfectly flat trend well within experimental benchmark uncertainties as a function of the fraction of epithermal neutrons (FEPIT). This represents a big performance improvement compared to previous evaluations. ENDF/B-VIII.0 covariances were unchanged.

\subsubsection{\nuc{50,51,52,53,54}{Cr}}
\label{subsec:n:50-51-52-53-54Cr}


Chromium is present in structural materials in  many nuclear systems. It is an important component in alloys with iron and other materials to form stainless steel~\cite{Klueh:2001}.
The ENDF/B-VIII.1 evaluated files for \nuc{50,52,53,54}{Cr}, which form the natural occurrence of chromium,  were mostly based on the recent evaluation by Nobre \etal \cite{chromium}, with a few minor modifications which are discussed below.

One of the main motivations behind the new set of evaluations detailed in Ref.~\cite{chromium} was the poor description of a cluster of  large capture resonances centered around neutron-incident energy of 5 keV and going up to around 10 keV. This resonance cluster, which drives capture by elemental chromium in reactor systems, is formed by resonances from \nuc{50}{Cr} and \nuc{53}{Cr} and have been historically poorly described. Careful investigation of previous data from Refs.~\cite{Stieglitz:1971,Guber:2011,Stieglitz:1970} allowed to fit capture more reliably in that important energy region. This directly improved the performance in criticality benchmarks sensitive to chromium, as discussed in Ref.~\cite{chromium}. Additionally, Ref.~\cite{chromium} presented complete new fast-region isotopic evaluations using modern theoretical models (e.g., soft-rotor optical model potentials~\cite{Li:2013}, tuning of microscopic level densities~\cite{Nobre:2020}) with newer reaction isotopic data for many reaction channels, in particular inelastic and (n,p). The resulting evaluated files for \nuc{50,52,53,54}{Cr} represented a significant improvement in the description of experimental data for angle-integrated cross sections, angular distributions, energy spectra, and neutron and gamma double-differential spectra. 

Updates in the evaluated files for chromium isotopes present in ENDF/B-VIII.1 done after Ref.~\cite{chromium} are:
\begin{itemize}

\item Extended model calculations in \nuc{50,52,53,54}{Cr} from 20~MeV to 65~MeV to allow its use in fusion applications. This led to the adoption of these files by the Fusion Evaluated Nuclear Data Library (FENDL)~\cite{FENDL}.

\item Adoption of selected capture resonance widths from BROND-3.1 \cite{BROND-3.1} for certain resonances in \nuc{52}{Cr}; namely, the resonances at neutron energies of -30.06818~keV (artificial resonance background), 1.625867~keV, 22.95014~keV, 27.59859~keV, 33.91804~keV, 47.94530~keV, 78.82184~keV, 94.94674~keV, 106.4305~keV, and 122.9100~keV. The main goal of this particular change was to increase the \nuc{52}{Cr} Maxwellian-averaged cross section (MACS) for capture, which led to improvement both in low-enriched uranium (LEU) lattices and in pmi002 criticality.

\item Extrapolation of model calculations to generate a higher-fidelity evaluated file for the unstable \nuc{51}{Cr} isotope which now replaces the TENDL-based version from ENDF/B-VIII.0, taking advantage of the  complete new data-driven model parametrization from the stable isotopes done in Ref.~\cite{chromium}. In addition to that, exit distributions for outgoing particles were added as part of the effort described in Section~\ref{subsec:n:exit-dist}.

\item Covariances from ENDF/B-VIII.0 were adopted for ENDF/B-VIII.1 for chromium isotopes whenever they were available; that is, for \nuc{50,52,53}{Cr}. Unfortunately, new fast-region covariances consistent with the new fast-region evaluations were not generated in time. Also, there were no new complete resonance evaluations to justify new resonance covariances. Hence, the pragmatic choice was made of adopting ENDF/B-VIII.0 covariances when possible. A few consistency tweaks were necessary, such as removing MF=33 MT=22 in \nuc{50,53}{Cr} since MF=3 MT=22 is now not present in the current file; updating the resonance widths in MF=32 in \nuc{52}{Cr} for the resonances with widths taken from BROND-3.1 to match the new values; and other minor fixes.

\end{itemize}

Although these new evaluations for chromium isotopes incorporated into \ENDF~represent a major improvement relative to \prENDF, corresponding to one of the highlights of the current release, there are clear pathways for further improvements:

\begin{itemize}

\item Most of the performance improvement for the current set of  \ENDF~chromium files came from a proper fit of the 0 -- 10~keV cluster of resonances from the minor isotopes \nuc{50,53}{Cr}. However, these were fits of older or corrected data. Ideally, new capture and/or transmission measurements on enriched \nuc{50}{Cr} and \nuc{53}{Cr} should be performed to guide new evaluations.

\item As detailed above, the resonance evaluations for chromium isotopes in \ENDF~were restricted to initial cluster of resonances in \nuc{50,53}{Cr}. New complete reevaluations of resonances in \nuc{50,53}{Cr} as well as in \nuc{52}{Cr}, preferably with the new measurements available as mentioned in the bullet above, would bring an additional round of improvements.

\item New consistent covariances in the fast region and, if there are new resonance evaluations, in the resonance region as well.

\item Neutron leakage experiments using natural chromium may help in improving the elastic and inelastic scattering cross sections and elastic angular distributions, which remain poorly constrained below 4 MeV of neutron incident energy. Such experiments are planned in Rez, CZ within the INDEN collaboration.

\item The PETALE experimental programme \cite{PETALE2021} in the CROCUS zero power reactor at the  Federal Polytechnic School of Lausanne (EPFL), Switzerland aims to measure reaction rates inside Cr plate used as reactor reflector. This experiment may provide key data to improve Cr evaluations. Results are being analyzed; preliminary data were first presented at the JEFF 2024 meeting showing some issues for the adopted ENDF/B-VIII.1 Cr evaluation. Additional work is warranted.
\end{itemize}

\subsubsection{\nuc{55}{Mn}}
\label{subsec:n:55-Mn}

Two updates were done to the original ENDF/B-VIII.0 evaluation \cite{Brown2018}: dosimetry cross sections and thermal capture prompt gammas.
$^{55}$Mn(n,2n) dosimetry cross sections were replaced by those evaluated within the IRDFF-II library \cite{IRDFF} as described in Ref.~\cite{INDC(NDS)-0900}. Thermal capture gammas were updated as described in Ref.~\cite{INDC(NDS)-0810}. The gamma emission changes are summarized below.

At the CSEWG meeting at BNL in November 2019, Marie-Laure Mauborgne presented results of calculations relevant to oil-well logging\footnote{see CSWEG 2019 Mauborgne presentation at \url{https://indico.bnl.gov/event/6642/contributions/31815/attachments/25323/37862/BNL_-_Mauborgne.pptx}}, where some evaluations showed degraded performance compared to ENDF/B-VI.3. Experimental prompt-gamma data for the thermal-capture gamma spectra were used in the ENDF/B-VI.3 evaluation. 

The case of $^{55}$Mn was investigated because this is an IAEA evaluation that was adopted for ENDF/B-VII.1 \cite{ENDF-VII.1} and transferred without change into ENDF/B-VIII.0 \cite{Brown2018}.  The IAEA $^{55}$Mn evaluation adopted for the ENDF/B-VII.1 library was based on \EMPIRE\ modeling \cite{Herman:2007} with discrete levels adopted from the RIPL-3 (2009 release) database \cite{RIPL3}. Unfortunately, the gamma spectra were stored on a coarse outgoing photon energy grid with more than 100 keV energy bins. That makes the resolution of the primary gammas too poor for radionuclide identification. Additionally, the probability of emission of high-energy photons (e.g., higher than 4 MeV), which are crucial for many applications, can be accurately derived from the EGAF library \cite{EGAF,EGAF-report}, but it is not accurate to describe them with state-of-art theoretical modeling.

The EGAF library \cite{EGAF,EGAF-report} was used to update the evaluated gamma emission spectrum corresponding to the thermal capture for incident neutron energies below 100 eV. It is worth noticing that a more accurate update of the evaluation was recently demonstrated by RPI colleagues by combining a \DICEBOX\ code simulation\footnote{see \DICEBOX\ distribution at \url{https://nds.iaea.org/dicebox/}} of the thermal-capture gamma cascade with an updated MCNP gamma transport \cite{Cook2023}. Further investigations are warranted within the GRIN project \cite{GRIN-GAP:2023}.

\subsubsection{\nuc{54,56,57}{Fe}}
\label{subsec:n:54-56-57Fe}


A new suite of evaluations for $^{54,56,57,58}$Fe isotopes was developed in the framework of the CIELO international collaboration \cite{Herman2018,CIELO-res} and adopted for the ENDF/B-VIII.0 library \cite{Brown2018}. New RRR were evaluated for $^{54}$Fe and $^{57}$Fe, while modifications were applied to resonances in $^{56}$Fe. However, during the ENDF/B-VIII.0 library validation, a 30\% underestimation of the fast neutron transmission through thick iron shells (for neutrons with energies between 1 and 10 MeV) was identified, see Fig.~32 of Ref.~\cite{Herman2018}.

While the CIELO iron evaluation was adopted for the ENDF/B-VIII.0 library, it was recognized that these identified deficiencies need to be addressed. A new evaluation was developed by Trkov and colleagues \cite{Trkov:2023} within the IAEA INDEN project \cite{Capote2022}; for a detailed description, see the mentioned Ref.~\cite{Trkov:2023}. Covariances from ENDF/B-VIII.0 were adopted for ENDF/B-VIII.1 for iron isotopes which is a pragmatic choice.

\begin{figure*}[!thbp]
\centering
\includegraphics[width=\textwidth, clip, trim = 50mm 14mm 35mm 15mm]{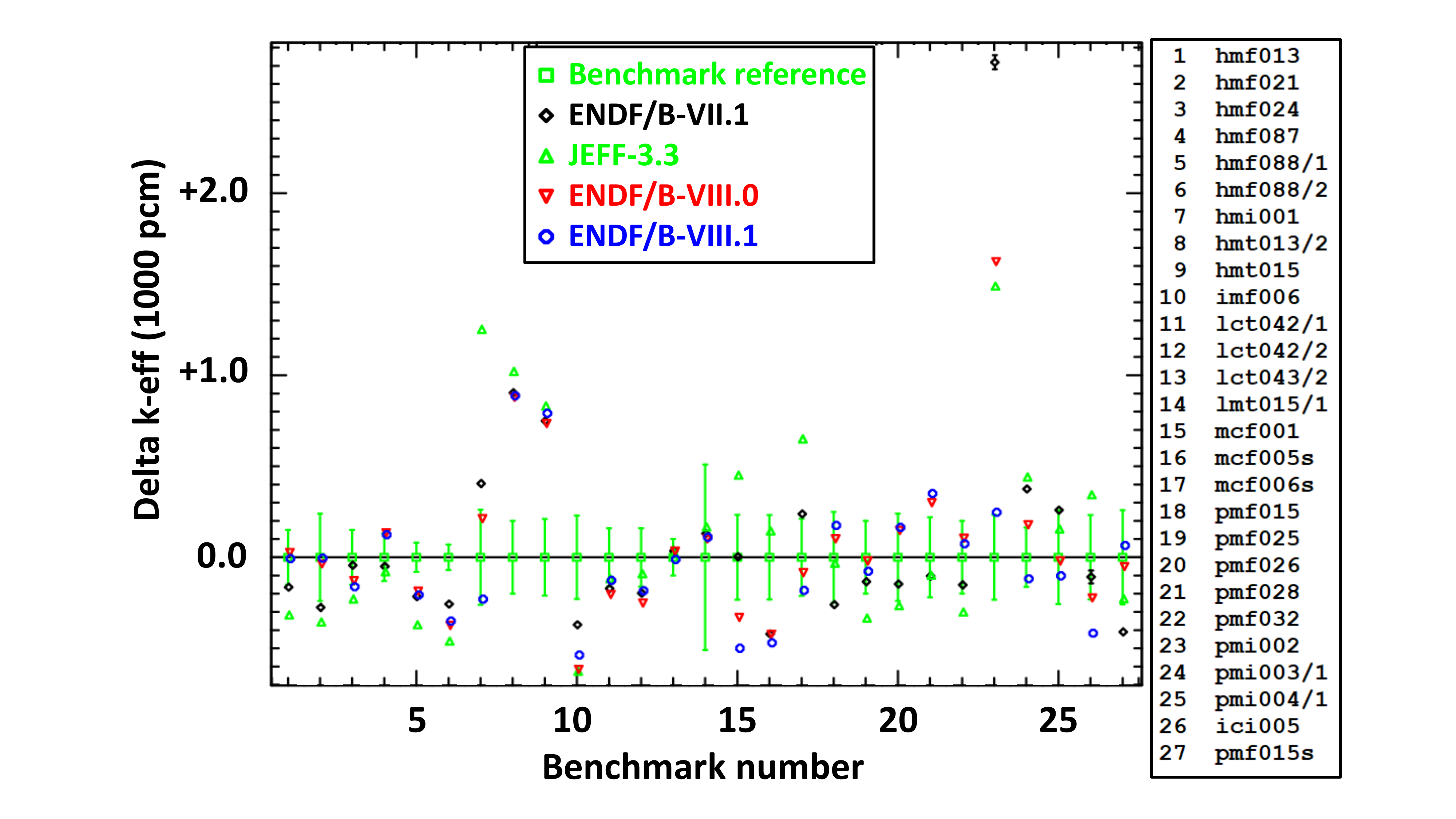}
\vspace{-.1in}
\caption{Criticality differences to the experimental benchmark values C/E of selected stainless steel ICSBEP benchmarks. Experimental benchmark values are compared  to JEFF-3.3, ENDF/B-VII.1, ENDF/B-VIII.0 and the new INDEN evaluation of iron (and chromium) isotopes adopted for the ENDF/B-VIII.1 library (INDEN r61). 
}
\label{fig:SS-benchmarks}
\end{figure*}

Following discussions with Plompen, this new evaluation adopted the CIELO resolved resonance parameters for $^{56}$Fe with one important change in the $\Gamma_{\gamma}$ of the 27.7~keV resonance: the ENDF/B-VIII.0 value, $\Gamma_{\gamma}=1.288$~eV, was found to overestimate Spencer ORNL data and the recommended $\Gamma_{\gamma}=1.0005$~eV value used in the JEFF-3.1 evaluation was adopted. These resonance parameters were essentially those evaluated by Perey and Perey \cite{Perey:1980} for the ENDF/B-V.2 material 1326. The ENDF/B-V.2 resonance parameters were adopted by Fr\"ohner's evaluation for the JEF-2.2 library \cite{Frohner:1989}. Some typos in the original evaluation were corrected (one resonance energy was changed from 767.240 keV to 766.724 keV and the spurious resonance at 59.5 keV was deleted \cite{Herman2018}).

\paragraph{Iron validation\newline}
In the ICSBEP Handbook \cite{ICSBEP}, there are many benchmarks sensitive either to iron or stainless steel cross sections. A selection of the most sensitive benchmarks was made (see Table 1 of Ref.~\cite{Trkov:2023}). The comparison between the calculated values for the ENDF/B-VIII.0, JEFF-3.3 and the ENDF/B-VIII.1 evaluations is shown in Fig.~\ref{fig:SS-benchmarks}.
Current criticality performance is as good as for ENDF/B-VIII.0 (compare blue triangles versus red circles). The largest improvement is observed for the pmi002 benchmark (Case 23) mainly due to the improvement in Cr cross sections in ENDF/B-VIII.1.

\begin{figure}[!htbp]
\centering
\includegraphics[clip,width=\columnwidth]{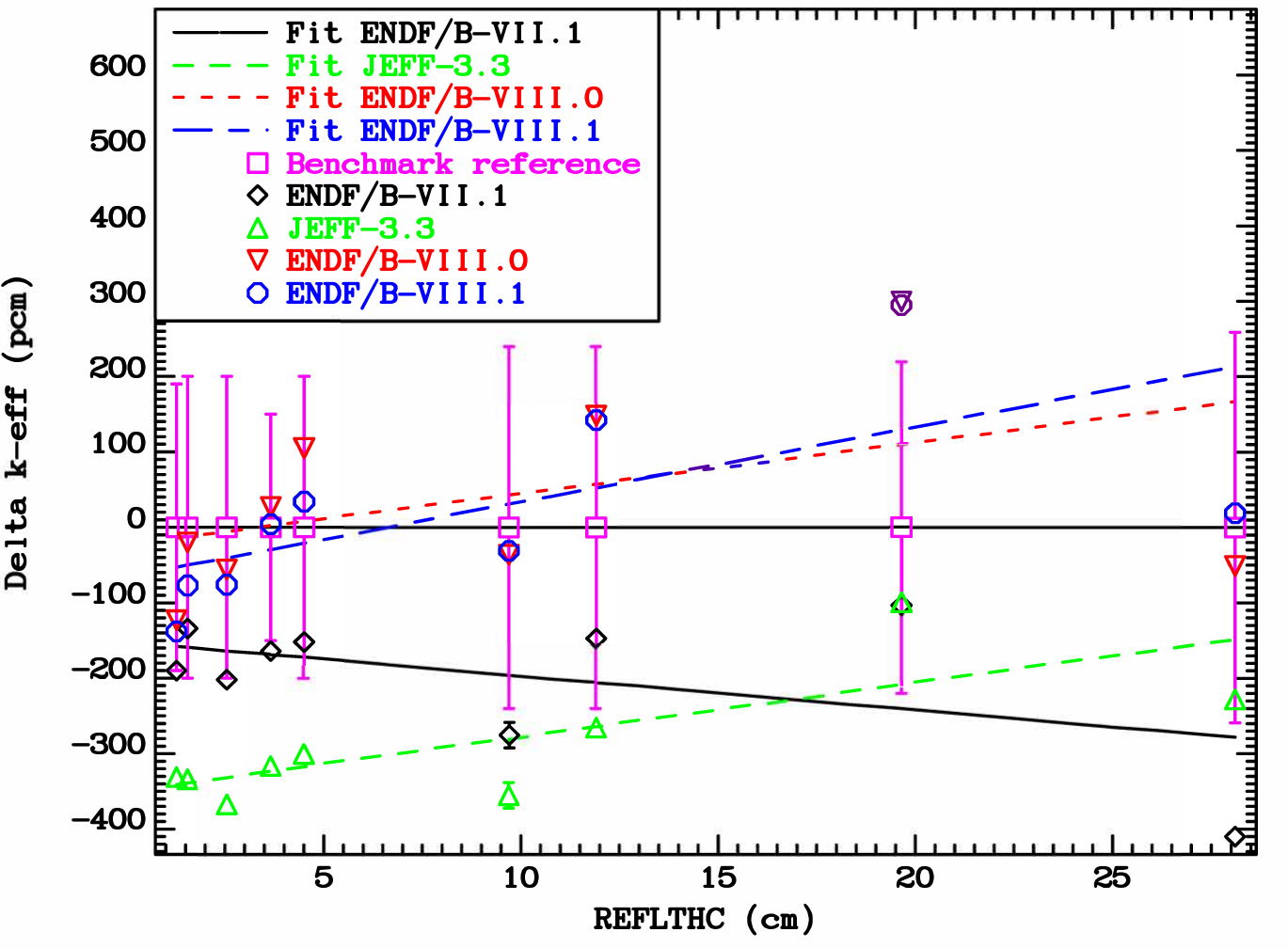}
\vspace{-3mm}
\caption{Criticality differences to the experimental benchmark values C/E of stainless-steel reflected ICSBEP benchmarks \cite{ICSBEP} as a function of the reflector thickness in cm. Experimental benchmark values are compared  to JEFF-3.3, ENDF/B-VIII.0 and the new INDEN evaluation of iron and chromium isotopes adopted for the ENDF/B-VIII.1 library. See Table 2 of Ref.~\cite{Trkov:2023} for the list of 9 benchmarks.}
\label{fig:SS-refl}
\end{figure}

We also checked the reactivity prediction of fast assemblies as a function of the stainless steel reflector thickness for values from 1.27~cm up to 28~cm thick (see Table 2 of Ref.~\cite{Trkov:2023}).
The results are shown in Fig.~\ref{fig:SS-refl}: the ENDF/B-VIII.1 performance is very similar to the \prENDF\ and significantly better than the ENDF/B-VII.1 and JEFF-3.3 libraries.

For the joint validation of both Cr and Fe ENDF/B-VIII.1 evaluations in deep penetration problems, we relied on a recent
experiment at Rez, CZ that uses a well-validated neutron spectrometer \cite{schulc2022,kostal2023}. The experiment used a 50.2$\times$50.2$\times$50.4~cm$^3$ stainless steel (SS) cube with a $^{252}$Cf(sf) source located inside. The neutron leakage spectrum was measured at one meter for incident neutron energies above 1~MeV.

\begin{figure}[!htbp]
\centering
\includegraphics[clip,scale = 0.34, trim = 2mm 1mm 2mm 2mm]{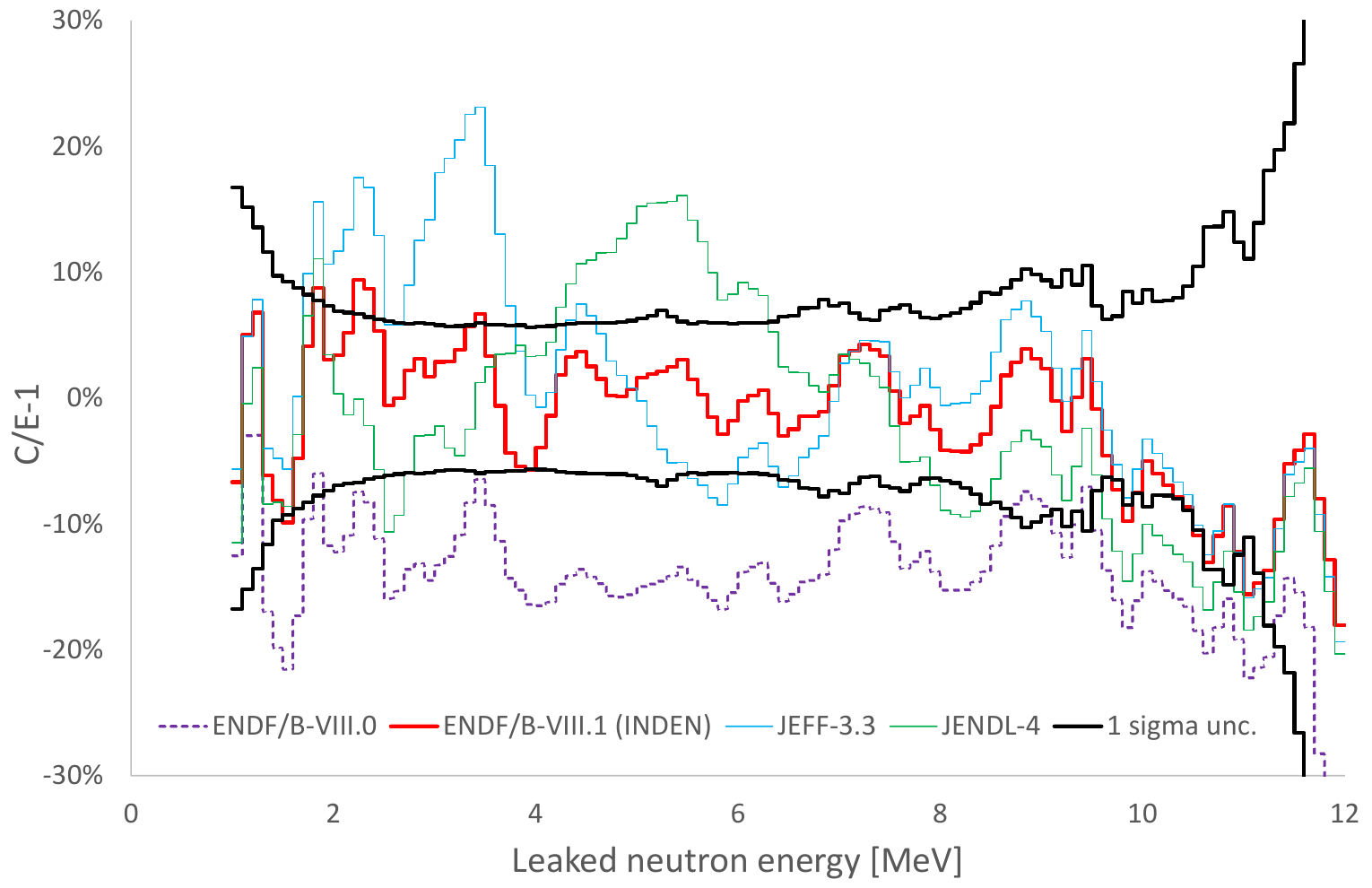}
\vspace{-4mm}
\caption{Measured neutron leakage of the $^{252}$Cf(sf) neutron source measured at 1~m distance from a 50.2x50.2x50.4~cm$^3$ stainless steel block \cite{schulc2022,kostal2023}. Experimental benchmark values $C/E-1$ are compared with transport calculations using JEFF-3.3, JENDL-4.0, ENDF/B-VIII.0 and the new ENDF/B-VIII.1 library that adopted the INDEN Fe and Cr evaluations.}
\label{fig:SS-Cf-leakage}
\vspace{-3mm}
\end{figure}

Calculated leaked neutron spectra through the SS cube with different libraries are compared to the measured spectrum using the $C/E-1$ criteria for outgoing neutron energies from 1~MeV up to 10~MeV. The best results of all participating libraries were obtained with the INDEN Cr and Fe evaluations, adopted for the ENDF/B-VIII.1 library as shown in Fig.~\ref{fig:SS-Cf-leakage}. The 15\% underestimation of the neutron leakage through stainless steel observed in Refs.~\cite{Herman2018,CIELO-res} for the ENDF/B-VIII.0 library is clearly seen in Fig.~\ref{fig:SS-Cf-leakage} in the whole energy range. The JENDL-4 evaluation slightly overestimates the measured data around 5 MeV whilst the JEFF-3.3 evaluation overestimates the measured neutron leakage from 2~MeV up to 4~MeV of neutron leaked energy.

\subsubsection{\nuc{63,65}{Cu}}
\label{subsec:n:63-65Cu}


Copper is an important structural material in various nuclear energy applications, including criticality safety. A significant number of ICBESP criticality benchmarks show a large sensitivity to copper data due to the extensive use of copper reflectors in various critical assemblies (e.g., the Zeus, Comet and CURIE benchmarks in the USA and the IMF020, IMF022 series in Sweden). Significant efforts have been devoted to improving the experimental database leading to improved copper evaluations.

New experimental neutron capture data on Cu isotopes have been measured by activation and time-of-flight methods by Newsome \cite{Newsome:2018}, Weigand \cite{Weigand:2017}, and Prokop \cite{Prokop:2019}. The evaluations of neutron reactions on \nuc{63,65}{Cu} have been updated within the INDEN collaboration. Cross sections for \nuc{63}{Cu}(n,$\alpha$), \nuc{63}{Cu}(n,2n),  and \nuc{65}{Cu}(n,2n) reactions were adopted from the IRDFF-II library \cite{IRDFF,INDC(NDS)-0900}.

\paragraph{Angular distributions\newline}
Discussions within the INDEN collaboration and our previous work \cite{mcdonnell_updates_2021} established that the angular distributions (represented by the Legendre coefficients) and capture cross sections strongly affect integral benchmark calculations; this work continued that investigation. 
The angular distribution parameterization in the RRR incorporates measured data from literature that have become available since the previous \nuc{63,65}{Cu} evaluation for ENDF/B-VIII.0 \cite{Sobes:2014,Brown2018}. 

\begin{figure}[!htbp]
\centering
\includegraphics[width=0.95\columnwidth]{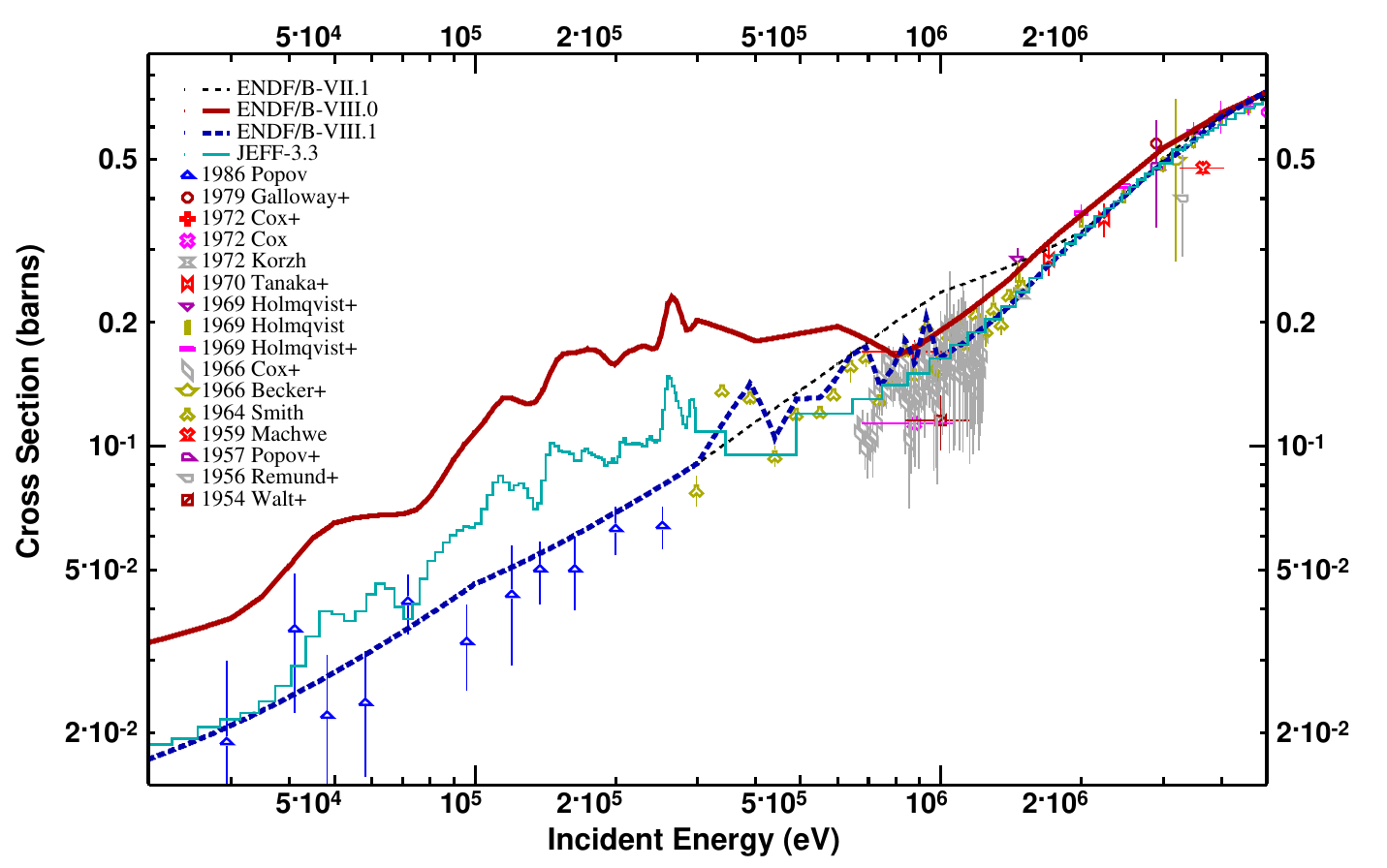}
\vspace{-.1in}
\caption{$^{63}$Cu $\mu$-bar ($P_1$ Legendre coefficient in the LAB system) as a function of the neutron incident energy for different evaluated libraries versus selected experimental data \cite{Popov:1986,Smith:1964,Tanaka:1970,Becker:1966,Korzh:1968,Popov:1957} from EXFOR \cite{EXFOR} up to 4~MeV.} \label{fig:Cu-mubar}
\vspace{-.1in}
\end{figure}

\begin{figure*}[!bthp]
\centering
\subfigure[~Detector at 35 deg]{\includegraphics[width=\columnwidth]{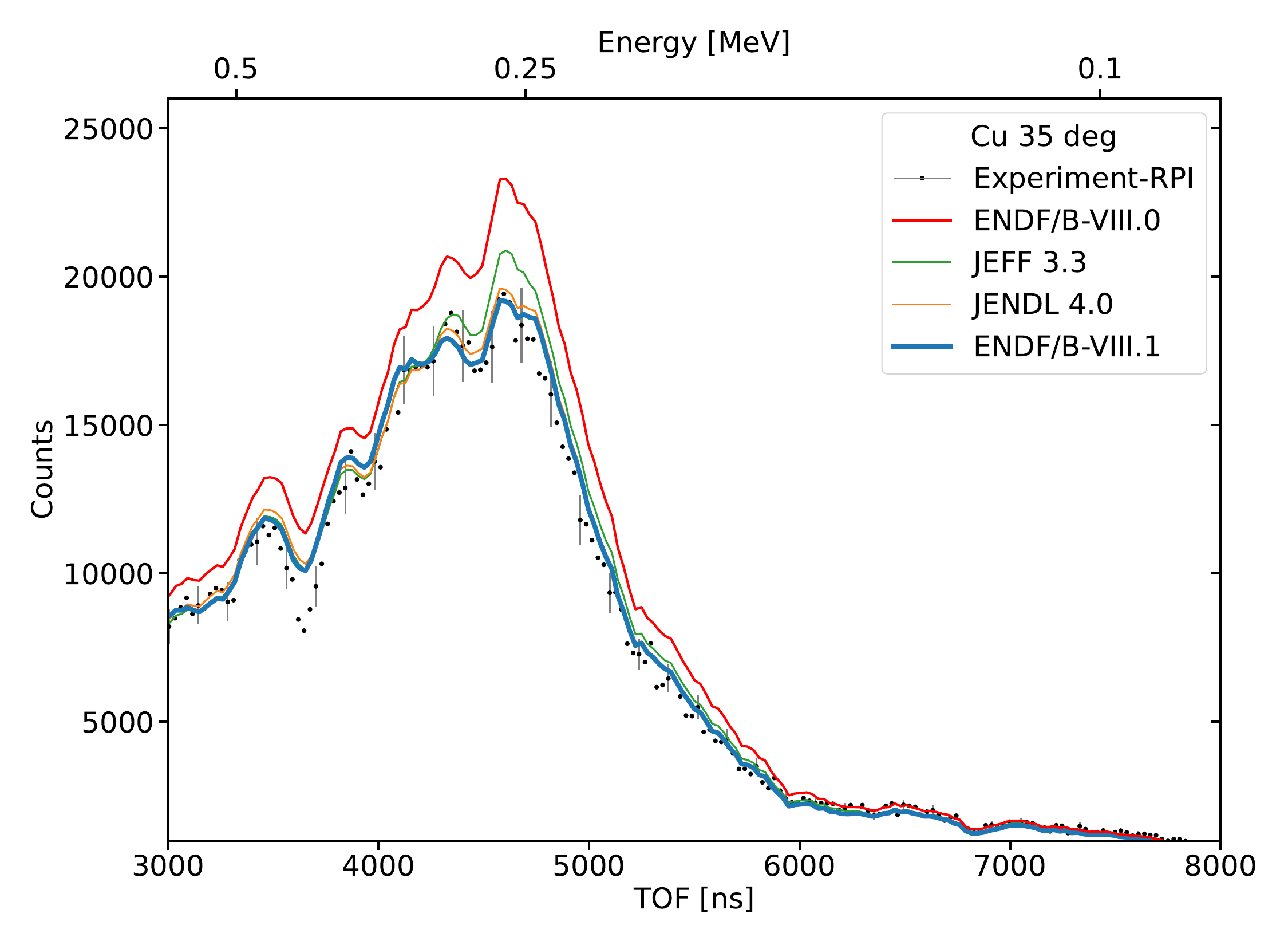}}
\subfigure[~Detector at 150deg]{\includegraphics[width=\columnwidth]{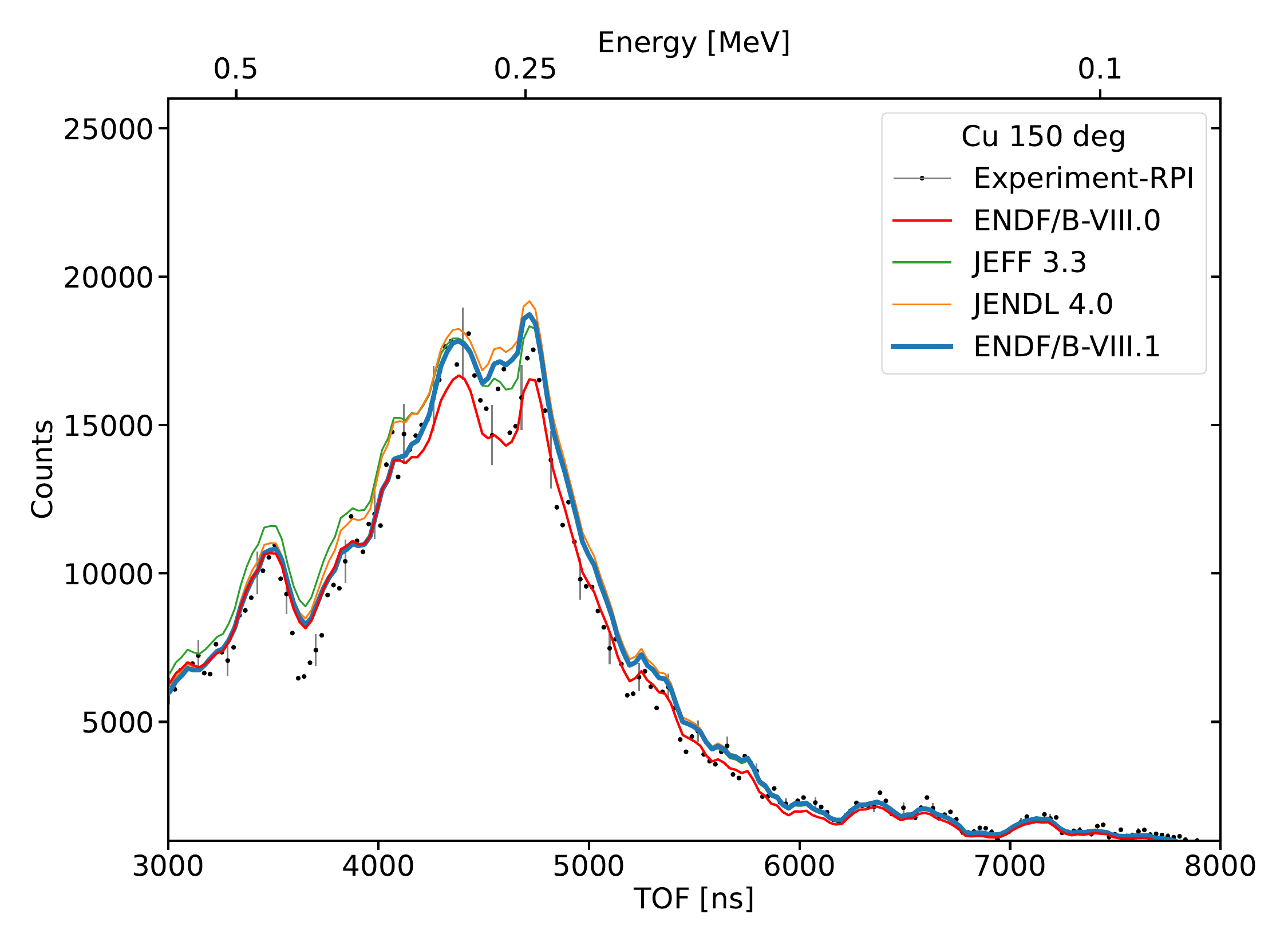}}
\centering
\vspace{-2mm}
\caption{Quasi differential copper scattering data measured with 35 deg and 150 deg detectors by Blain \etal \cite{Blain:2022} below 600~keV is compared with simulations using different evaluated libraries.}
\label{fig:Cu-RPI}
\vspace{-2mm}
\end{figure*}

The $\mu$-bar (the Legendre coefficient $P_1$ in the LAB system) dependence as a function of the neutron incident energy is shown in Fig.~\ref{fig:Cu-mubar}. A clear overestimation of the evaluated $\mu$-bar compared to the measured data is observed in the ENDF/B-VIII.0 evaluation up to about 700~keV of neutron incident energy. The evaluated angular distributions below 300~keV in the ENDF/B-VIII.0 library, which were reconstructed from the resonance parameters \cite{Sobes:2014}, were found to disagree with measured data by Popov and Samosvat~\cite{Popov:1986}. The evaluated angular distributions from 300~keV up to 700~keV in the ENDF/B-VIII.0 library also disagree with Smith {\it et al.}~\cite{Smith:1964} data. The ENDF/B-VII.1 evaluation agrees with Popov data \cite{Popov:1986} on average, so a significant deviation from the experimental data happened during the ENDF/B-VIII.0 evaluation.

A similar disagreement between ENDF/B-VIII.0 evaluated and measured angular distribution data was shown by Blain and colleagues in RPI quasi-differential experiments \cite{Blain:2022}, see Fig.~\ref{fig:Cu-RPI} at forward (35 degrees) and backward (150 degrees) angles. The largest disagreement between the ENDF/B-VIII.0 simulation and measured RPI data is observed in the region near 250 keV of incident neutron energy for a 35-degree detector. Much better agreement is shown for JENDL-4.0 and JEFF-3.3 libraries. The INDEN evaluation adopted for ENDF/B-VIII.1 fitted the Legendre coefficients representing angular distributions to Popov \cite{Popov:1986} and Smith \cite{Smith:1964} data up to 1.1 MeV. The new ENDF/B-VIII.1 evaluation is in very good agreement with RPI quasi-differential data on copper which confirmed our evaluation choices.

\begin{table*}[!tbh]
\vspace{-2mm}
  \caption[Experimental data sets measured for several copper sample configurations]{Experimental data sets measured for several copper sample configurations, such as thickness $n$ and enrichment in the RRR. \textmd{These data represent the experimental database used to derive the set of resonance parameters in the RRR for \nuc{63,65}{Cu}.} \\
  \textbf{Note}: The Weigand \cite{Weigand:2017} and Prokop \cite{Prokop:2019} references do not communicate sample thickness.  As such, neither data set was included in the \SAMMY\ R-matrix
  analysis.}
  \label{table:Datasets}
\begin{center}
\begin{tabular}{l|cccccc}
\toprule \toprule
    Author~(year)&  Facility&   Sample&  Type& $n$~($10^{-3}$~a/b)&  Energy Range & Data Points\\
 \midrule
    \multirow{2}{*}{Pandey~(1977)~\cite{pandey_neutron_1977}}     &  \multirow{2}{*}{ORELA}   & \multirow{2}{*}{Enriched} & \multirow{2}{*}{Trans} &\multirow{2}{*}{78.964}       &  1~keV--1.4~MeV   & 24086\\
          &     &  &  &      &  33~eV--185~keV   & 31393\\
    Kauwen.~(2013)~\cite{kauwenberghs_2013} & GELINA   & Enriched & Trans &8.804        & 150~eV--90~keV  & 21528  \\
    \multirow{2}{*}{Tsuchiya~(2014)~\cite{tsuchiya_impact_2014}}  & \multirow{2}{*}{GELINA}   & \multirow{2}{*}{Natural}  &Trans& 0.989 & \multirow{2}{*}{150~eV--90~keV} & \multirow{2}{*}{37673} \\
                                                                  &    &   & Trans&164.63 & & \\
    Guber~(2014)~\cite{guber_2014}               & GELINA   & Enriched & Capture Yield&8.804        &  90~eV--300~keV & 31296  \\
    Weigand~(2017)~\cite{Weigand:2017}               & LANSCE   & \textsuperscript{63}Cu & Capture & --        &  74~eV--690~keV & 398  \\
    Prokop~(2019)~\cite{Prokop:2019} & LANSCE   & \textsuperscript{65}Cu & Capture &
    -- &  1~eV--933~keV & 264  \\
\bottomrule \bottomrule
\end{tabular}
\end{center}
\vspace{-3mm}
\end{table*}

\paragraph{Resolved Resonance parameters\newline}
The RRR in the ENDF/B-VIII.0 evaluation was limited to 100 keV \cite{Brown2018} due to the poor integral performance traced to the decreased average capture cross section if the RRR was extended to 300 keV \cite{Sobes:2014,Brown2018}. The experimental data considered in the current R-matrix analysis were discussed in a previous work by McDonnell and Pigni~\cite{mcdonnell_updates_2021}.
Table~\ref{table:Datasets} summarizes (reproduced from the aforementioned work by the authors~\cite{mcdonnell_updates_2021}) the experimental data sets measured for several sample configurations and list experimental properties, such as thickness $n$, target enrichment, reaction type, and energy range.

The Pandey~\cite{pandey_neutron_1977} measurements are valuable for constraining the
R-matrix parameters in the middle and the upper end of the RRR. At the lower end of the RRR, however, the Kauwenberghs data~\cite{kauwenberghs_2013}
are of a higher resolution and lower uncertainty and, thus, provide a more accurate constraint for R-matrix analysis.

The Tsuchiya~\cite{tsuchiya_impact_2014} measurements on natural copper samples were primarily used for validation
rather than fitting.  The exception is that it proved useful to incorporate the
data into the fit in the vicinity of the 579~eV resonance in \nuc{63}{Cu}, since this is the only
measurement in which that resonance is not blacked out; thus, this measurement
provides valuable information for the true shape and magnitude of that resonance.

The Guber~\cite{guber_2014} measurements provide necessary constraints for the capture widths.
As discussed in our previous work \cite{mcdonnell_updates_2021},
this experiment's measured data are estimated to be lower than the true capture yield
above 100~keV.  Therefore, we  augmented the information with the more recent
(but lower resolution) measurement by Weigand \emph{et al.}~\cite{Weigand:2017}.
Although the resolution of the Weigand data is too low for fitting resonance parameters,
the data were used to normalize the capture cross section for incident neutron energies
above 50~keV.

The additional recent measurement of the capture cross section for \nuc{65}{Cu}
by Prokop~\cite{Prokop:2019} was found to agree well with the
existing ENDF/B-VIII.0 evaluated data for \nuc{65}{Cu}, so it was used as validation
rather than a fitting constraint.

The R-matrix analysis is described in detail in Ref.~\cite{mcdonnell_updates_2021}.
We summarize the discussion and update with final parameter values here.

One feature of the ENDF/B-VIII.0 evaluations for \nuc{63,65}{Cu} is that average values were used for each isotope's capture widths.
Resonance parameters for \nuc{63}{Cu} from a previous beta release of ENDF/B-VIII.0 (``revision 900'')
were found to describe the shapes of the resonances in the data better, and so we chose these parameters for our prior.
However, the ENDF/B-VIII.0 resonance parameters themselves served as the prior parameters for \nuc{65}{Cu} as they give good agreement with recently measured Prokop capture data \cite{Prokop:2019}.

The comments about the underestimation of the capture yield data in Guber experiments discussed during the ENDF/B-VIII.0 evaluation, which led to cutting the RRR at 100 keV, were confirmed \cite{mcdonnell_updates_2021}. The measured data by Guber~\etal~\cite{guber_2014} are thought to be too small for incident neutron energies above 100~keV due to an issue with the flux
monitor during the experiment.  A more recent experiment by Weigand \emph{et al.}~\cite{Weigand:2017} corroborates a higher capture cross section in \nuc{63}{Cu} as suggested within the INDEN collaboration.
For energies below 50~keV, the R-matrix parameters describe the increased capture cross section sufficiently well. This was accomplished by scaling the capture yield data from Guber~\etal by a factor 
of 20\% for energies above 19~keV, so that the magnitude of the capture cross section is similar to that of the
data from Weigand~\emph{et al}. However, from 50~keV up to 100~keV, the cross section is modeled with an additional background term to reproduce Weigand data.

In this work, a refit of the capture widths below 100~keV was performed, which allows an improved description of several resonances. In contrast, the ENDF/B-VIII.0 evaluation files for \nuc{63,65}{Cu} used identical average values for the capture widths for each resonance.

The thermal values for the two isotopes, as obtained by this work and ENDF/B-VIII.0, are compared with the National Institute of Standards and Technology (NIST) recommended values by Sears~\cite{Sears1992} in Table~\ref{tab:cu_thermal_values}, showing a marginal increase of the elastic cross section and overall good agreement with NIST recommendations.
\begin{table}[!htb]
\small
\centering
\caption{Thermal cross section values in barn for \nuc{63,65}{Cu} in ENDF/B-VIII.0 (labeled B-VIII.0) and ENDF/B-VIII.1 (labeled B-VIII.1) evaluations are compared with NIST \cite{Sears1992} and Firestone recommendations \cite{firestone2021}. 
ENDF/B evaluations do not feature uncertainty components. If uncertainties and covariances are needed they may be adopted from JEFF-3.3 \cite{JEFF33}.}
\label{tab:cu_thermal_values}
\begin{tabular}{lcccc}
        \toprule \toprule
        Quantity                  & B-VIII.0  & B-VIII.1  & NIST \cite{Sears1992}   &   Firestone \cite{firestone2021}\\
        \midrule
        \nuc{63}{Cu} (n,$\gamma$) & 4.47               & 4.48           &     4.52  &  4.3(3)    \\
        \nuc{63}{Cu} (n,el)       & 5.10               & 5.22           &     5.2   &  --        \\
        \nuc{65}{Cu} (n,$\gamma$) & 2.15               & 2.15           &     2.17  &  2.11(17)  \\
        \nuc{65}{Cu} (n,el)       & 13.8               & 13.8           &     14.1  &  --        \\
        \bottomrule \bottomrule
\end{tabular}
\vspace{-3mm}
\end{table}

In the analysis with the \SAMMY\ code, both isotopes, \nuc{63,65}{Cu}, were analyzed together,
accounting for the isotopic enrichment in each experimental data set.
Most of the experimental data sets considered for capture used highly enriched isotopic samples
($>$99\%), which minimized the correlation between the isotopes in the analysis.
The exception was in the vicinity of the 579~eV resonance in \nuc{63}{Cu},
where the Tsuchiya data set, measured on a natural copper sample, was incorporated.
This introduced a small correlation between the two isotopes in that region.

\begin{figure}[!htbp]
\centering
\includegraphics[width=0.99\columnwidth]{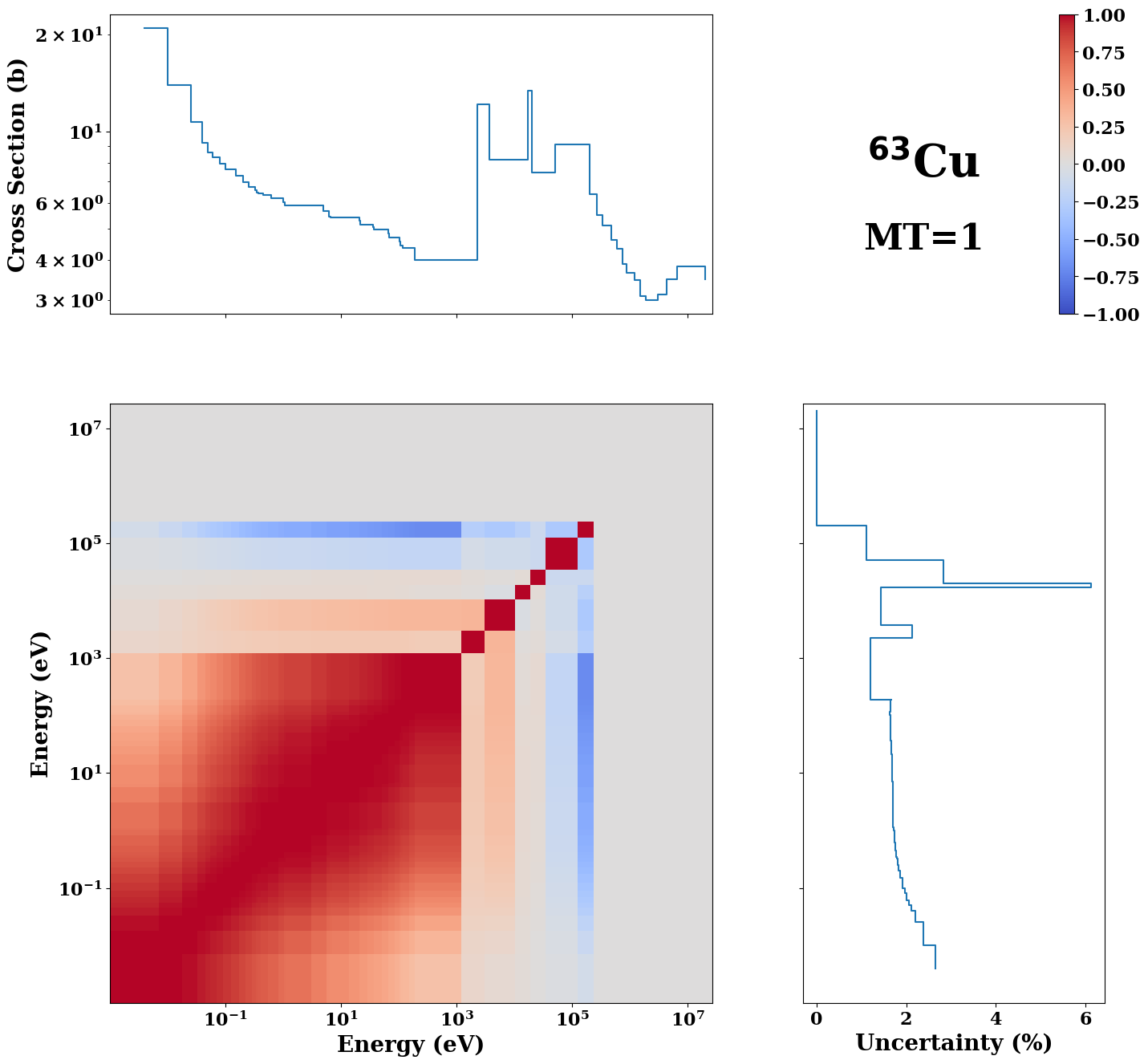}
\caption{Correlation matrix and uncertainties for the total cross section of n+\nuc{63}{Cu} are shown.}
\label{fig:Cu63_mt1_cov}
\end{figure}

The resonance parameter covariance matrix (RPCM) was obtained in the course of the
R-matrix analysis. The RPCM is reported in the ENDF File 32.  
The correlation matrix and uncertainties in the cross section for n+\nuc{63}{Cu} 
are shown in Fig.~\ref{fig:Cu63_mt1_cov}. 

\paragraph{Fast neutron region\newline}

The elastic scattering angular distributions for both \nuc{63,65}{Cu} isotopes up to 1100 keV were updated using the Legendre polynomials derived from the fit to Popov and Samosvat~\cite{Popov:1986} and Smith {\it et al.}~\cite{Smith:1964} data taken on natural copper. The same angular distributions were used for both copper isotopes. Smith and Popov angular distribution data are consistent at 300 keV. Popov data were used below 300 keV; Smith data are used from 300 keV up to 1100 keV.

From 1.1~MeV up to 3~MeV, the angular distributions were adopted from the JEFF-3.3 library, which is in better agreement with experimental data by Tanaka \etal~\cite{Tanaka:1970}, Becker \etal~\cite{Becker:1966}, Korzh \emph{et al.}~\cite{Korzh:1968}, and Popov \cite{Popov:1957}, as shown in the Fig.~\ref{fig:Cu-mubar}. We found a huge sensitivity of the calculated criticality in fast assemblies to small differences in the evaluated $\mu$-bar up to 3~MeV. Above 3~MeV, the JEFF-3.3 evaluation smoothly matches the \prENDF{} as shown in Fig.~\ref{fig:Cu-mubar}.

\begin{figure}[!htbp]
\centering
\includegraphics[width=0.99\columnwidth]{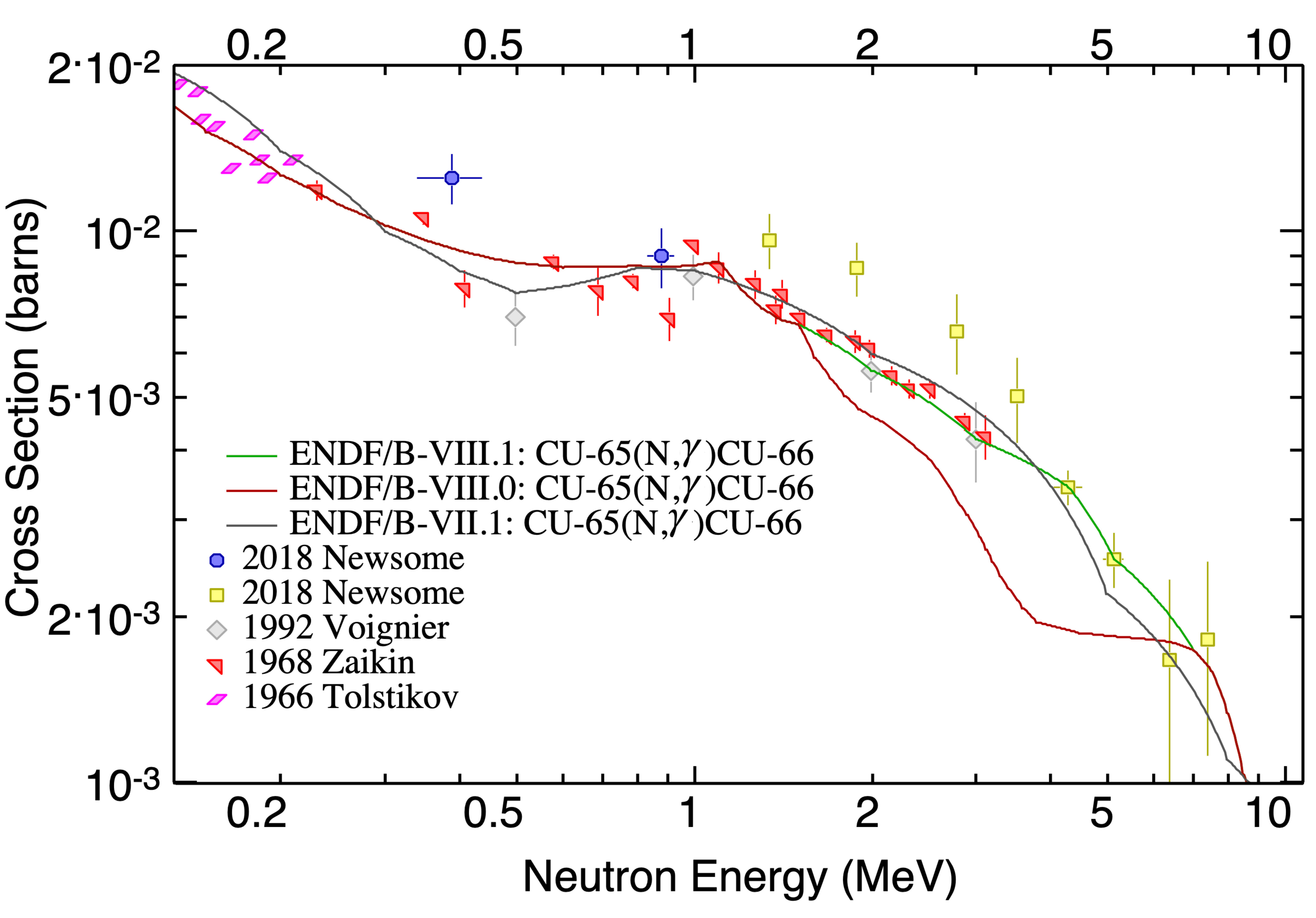}
\caption{n+\nuc{65}{Cu} capture cross sections calculated for ENDF/B-VII.1, ENDF/B-VIII.0, and ENDF/B-VIII.1 nuclear data libraries are compared with selected experimental data \cite{Newsome:2018,Voignier:1992,Zaikin:1968,Tolstikov:1966}.}
\label{fig:Cu65-ng}
\end{figure}

Capture cross  section on $^{65}$Cu have been recently measured by Newsome \etal \cite{Newsome:2018} and Prokop \etal \cite{Prokop:2019}. Prokop data are not yet in EXFOR, but authors showed excellent agreement in the fast neutron region with the ENDF/B-VIII.0 evaluation up to 1~MeV (see Fig.~8 of Ref.~\cite{Prokop:2019}). Older measurements by Voignier \cite{Voignier:1992}, Zaikin \cite{Zaikin:1968}, and Tolstikov \cite{Tolstikov:1966} are shown in Fig.~\ref{fig:Cu65-ng} and agree with the ENDF/B-VIII.0 evaluation below ~1.5~MeV. Unfortunately, Newsome activation measurements \cite{Newsome:2018} above 1~MeV seem to be too high. Our new evaluation followed ENDF/B-VIII.0 up to 1.5~MeV in agreement with Prokop and Zaikin data. Above 1.5~MeV, the new evaluation is closer to the ENDF/B-VII.1 evaluation, being significantly higher than the ENDF/B-VIII.0 evaluation. The new evaluation agrees much better with experimental data than \prENDF{} as shown in Fig.~\ref{fig:Cu65-ng}.

\begin{figure}[!htbp]
\centering
\includegraphics[width=0.99\columnwidth]{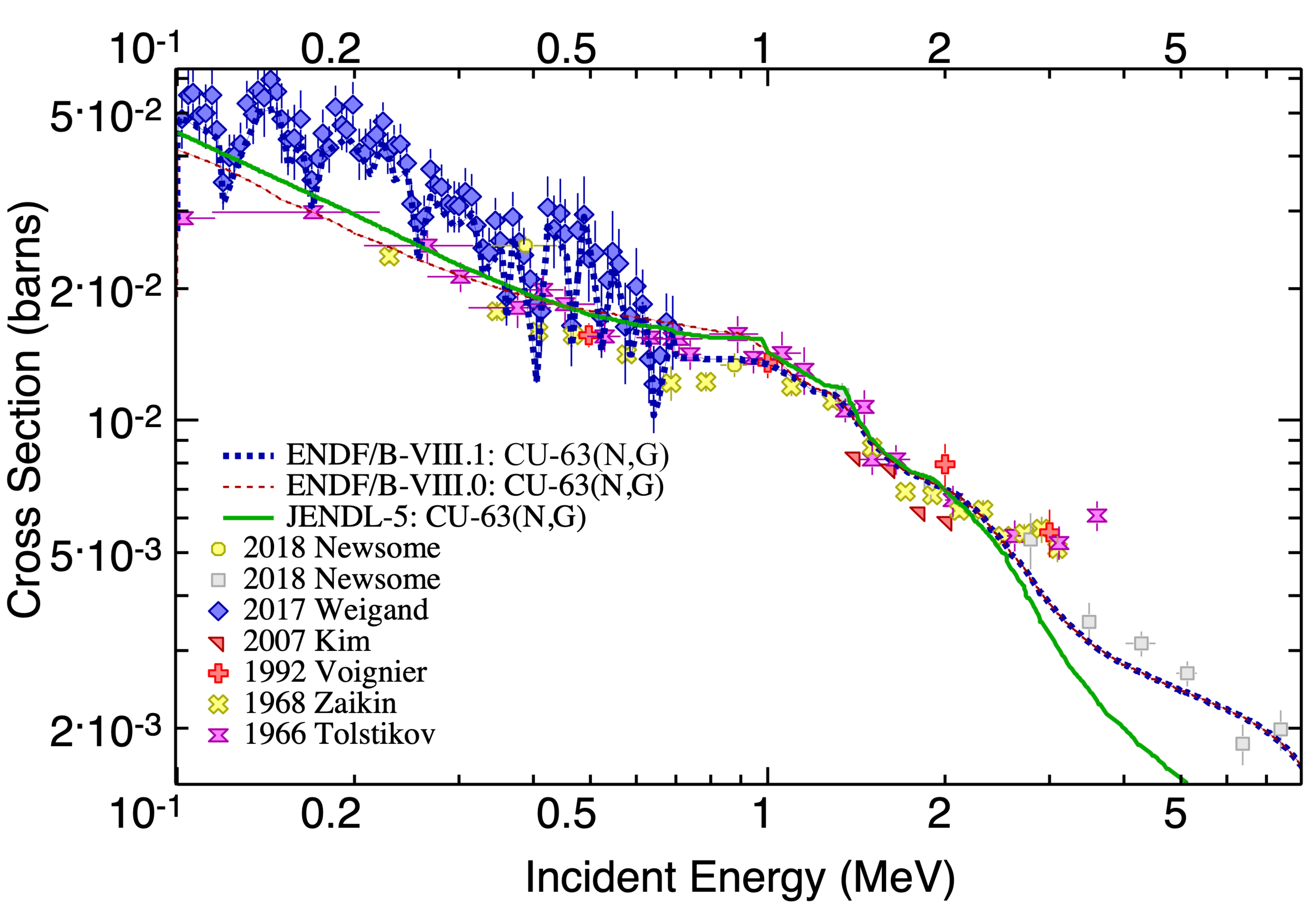}
\caption{n+\nuc{63}{Cu} capture cross sections calculated for JENDL-5, ENDF/B-VIII.0, and ENDF/B-VIII.1 nuclear data libraries are compared with selected experimental data \cite{Newsome:2018,Weigand:2017,Voignier:1992,Zaikin:1968,Tolstikov:1966}.}
\label{fig:Cu63-ng}
\end{figure}

Capture cross  section on $^{63}$Cu have been recently measured by Newsome \etal \cite{Newsome:2018} and Weigand \etal \cite{Weigand:2017}. 
Weigand data below 0.6~MeV and Newsome measured point at 0.4~MeV are significantly higher than older measurements by Voignier \cite{Voignier:1992}, Zaikin \cite{Zaikin:1968}, and Tolstikov \cite{Tolstikov:1966} in this energy region as shown in Fig.~\ref{fig:Cu63-ng}. 
For this isotope, Voignier, Tolstikov, and Zaikin data were also in good agreement and were the basis of the ENDF/B-VIII.0 and JENDL-5 evaluations. In this work we choose to follow Weigand data in the fast neutron region starting at ~100 keV (and below as discussed in the RRR evaluation above). Implicitly, that means that the new evaluation is higher than older evaluations from 100 keV up to 600~keV as driving experimental data are discrepant.
We significantly reduced capture cross section from ENDF/B-VIII.0 and JENDL-5 evaluations from 600~keV up to 1~MeV to better agree with Newsome and older data (especially Zaikin) in that region. 
From 1~MeV up to 2~MeV there is good agreement with data and the other evaluations, and the new evaluation is close to the ENDF/B-VIII.0 and JENDL-5 evaluations there. 
It should be noted that from 2~MeV up to 3~MeV the JENDL-5 evaluation is lower than measured datasets, but we followed newer Newsome data up to 7~MeV in agreement with the ENDF/B-VIII.0 library.

\begin{figure}[!htbp]
\centering
\includegraphics[width=\columnwidth]{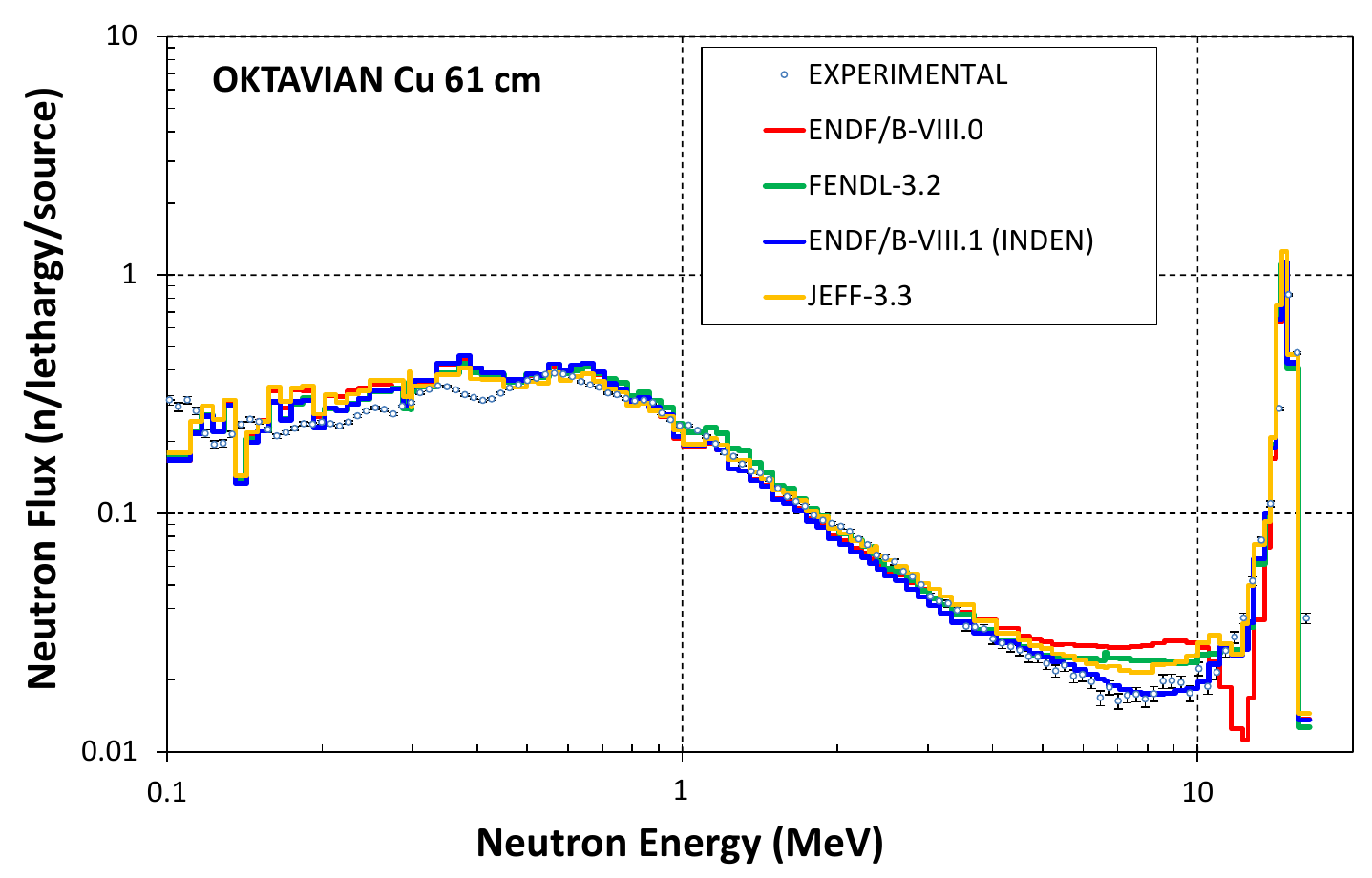}
\vspace{-.1in}
\caption{The measured neutron leakage from a copper assembly using pulsed 14-MeV DT neutron beam at the Oktavian facility is compared with simulations using different nuclear data libraries. Experimental data taken from Ref.~\cite{TAKAHASHI1989}.}
\label{fig:Cu-Oktavian-14MeV}
\end{figure}

Inelastic cross section and scattering spectra (including continuum double differential cross sections and discrete level angular distributions) were adopted from the JENDL-4 evaluation. This was very important to correct deficiencies of the ENDF/B-VIII.0 evaluation in 14~MeV fusion benchmarks as shown, e.g., for the Oktavian benchmark in Fig.~\ref{fig:Cu-Oktavian-14MeV}. Data issues in previous evaluations (except JENDL-4) are clearly seen from 4~MeV up to 12~MeV of the outgoing neutron energy in the above mentioned figure leading to an overestimated response compared to measured data.

\paragraph{Validation\newline}
We have selected two benchmark suites: one thermal, the second one including the intermediate and fast neutron assemblies. Criticality calculations were done by A.~Trkov. A similar set of benchmarks were also calculated at ORNL in previous work~\cite{mcdonnell_updates_2021}.

\begin{figure}[!thpb]
    \centering
    \includegraphics[width=\columnwidth]{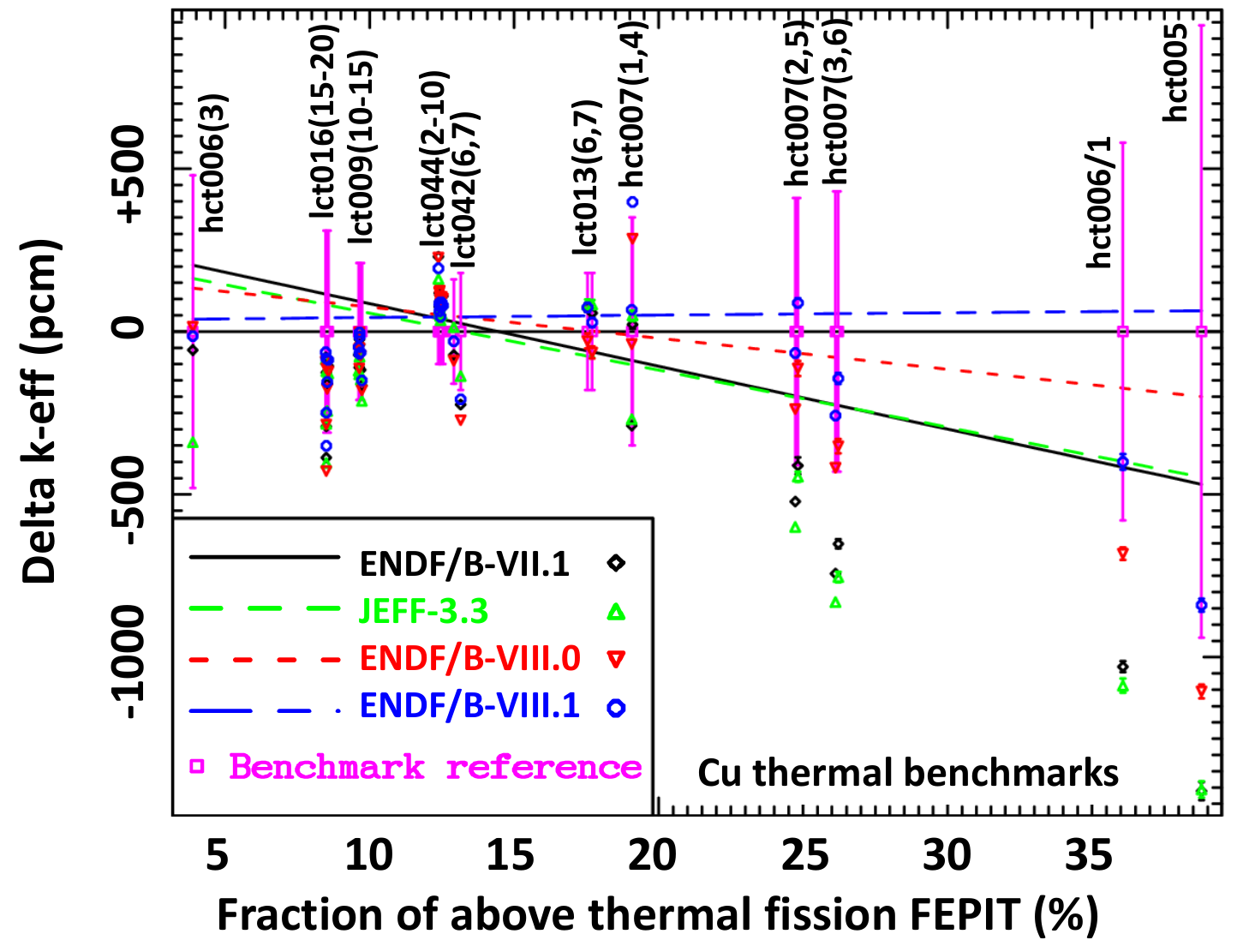}
\vspace{-2mm}
    \caption{The plot of $\Delta k_{\mathrm{eff}}$ comparing calculated to experimental values for
    the suite of thermal ICSBEP benchmarks \cite{ICSBEP} that show sensitivity to copper.
    The $x$-axis corresponds to the percentage fraction of above-thermal fissions---FEPIT.}
    \label{fig:delta_keff-thermal}
\vspace{-2mm}
\end{figure}

Results for the thermal set of benchmarks are shown in Fig.~\ref{fig:delta_keff-thermal}. The R-matrix parameters derived and updated in this work does improve agreement between calculated
and measured criticality benchmark values compared to previous libraries -- ENDF/B-VII.1, JEFF-3.3, and ENDF/B-VIII.0.

The agreement for well-thermalized assemblies with FEPIT$<20$\% was similar for all libraries as shown in Fig.~\ref{fig:delta_keff-thermal}. A perfectly flat trend was achieved for the ENDF/B-VIII.1 copper evaluation covering from very well-thermalized assemblies (HCT006/3 with FEPIT$\approx 0$) to epithermal assemblies with FEPIT$\approx 40$\% (HCT005), clearly an improved performance for lattices and thermal benchmarks.

All previous evaluations systematically underestimated the criticality of epithermal assemblies, being ENDF/B-VIII.0 slightly better (showing smaller gradient) than ENDF/B-VII.1 and JEFF-3.3, especially for HCT007(cases 1,2,3,5,6) and HCT005, but still worse than the new evaluation.

\begin{figure}[!thpb]
    \centering
    \includegraphics[width=\columnwidth]{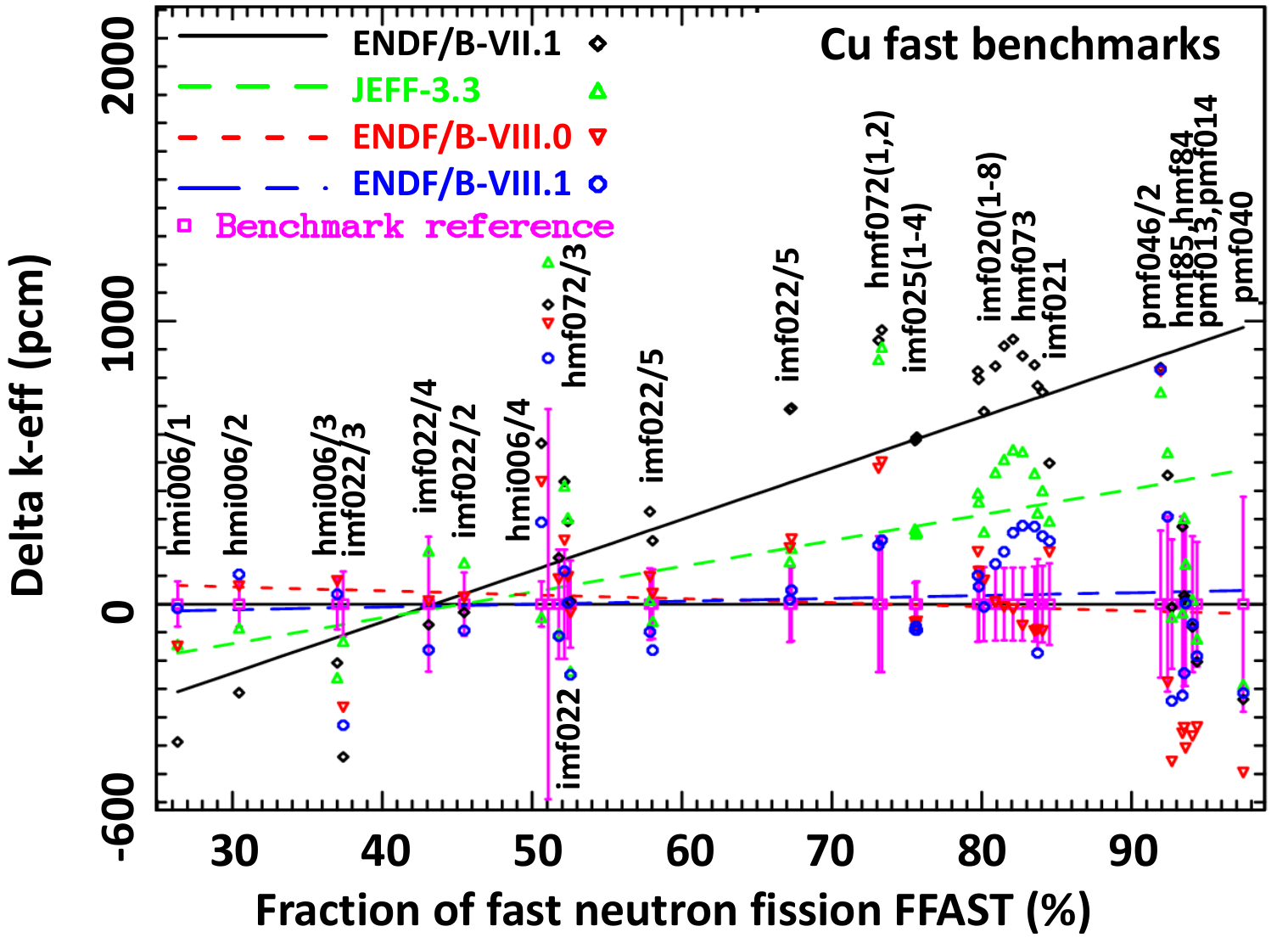}
\vspace{-2mm}
    \caption{The plot of $\Delta k_{\mathrm{eff}}$ comparing calculated to experimental values for
    the suite of fast and intermediate ICSBEP benchmarks \cite{ICSBEP} that show sensitivity to copper.
    The $x$-axis corresponds to the percentage fraction of fast-neutron fissions -- FFAST.}
    \label{fig:delta_keff-fast}
\vspace{-2mm}
\end{figure}
Results for the intermediate and fast neutron spectrum set of benchmarks are shown in Fig.~\ref{fig:delta_keff-fast}. This benchmark set covers a broad range of the percentage fraction of fast-neutron fissions  FFAST from ~25\% up to close to 100\%. Both ENDF/B-VIII.0 and the new ENDF/B-VIII.1 (INDEN) evaluations show a very flat trend as a function of the percentage of fast-neutron fissions FFAST. The new evaluation shows worse performance for the Swedish set of benchmarks IMF20,21,22, which will be discussed in detail below. However, the new evaluation improved the ENDF/B-VIII.0 performance both for ZEUS intermediate benchmarks (e.g., HMI006) as well as for the fastest studied benchmarks (FFAST>90\%) in this set (e.g., HMF084, HMF085 and PMF013, PMF014, PMF040, and PMF046/2).

\begin{figure}[!htbp]
\centering
\includegraphics[width=\columnwidth]{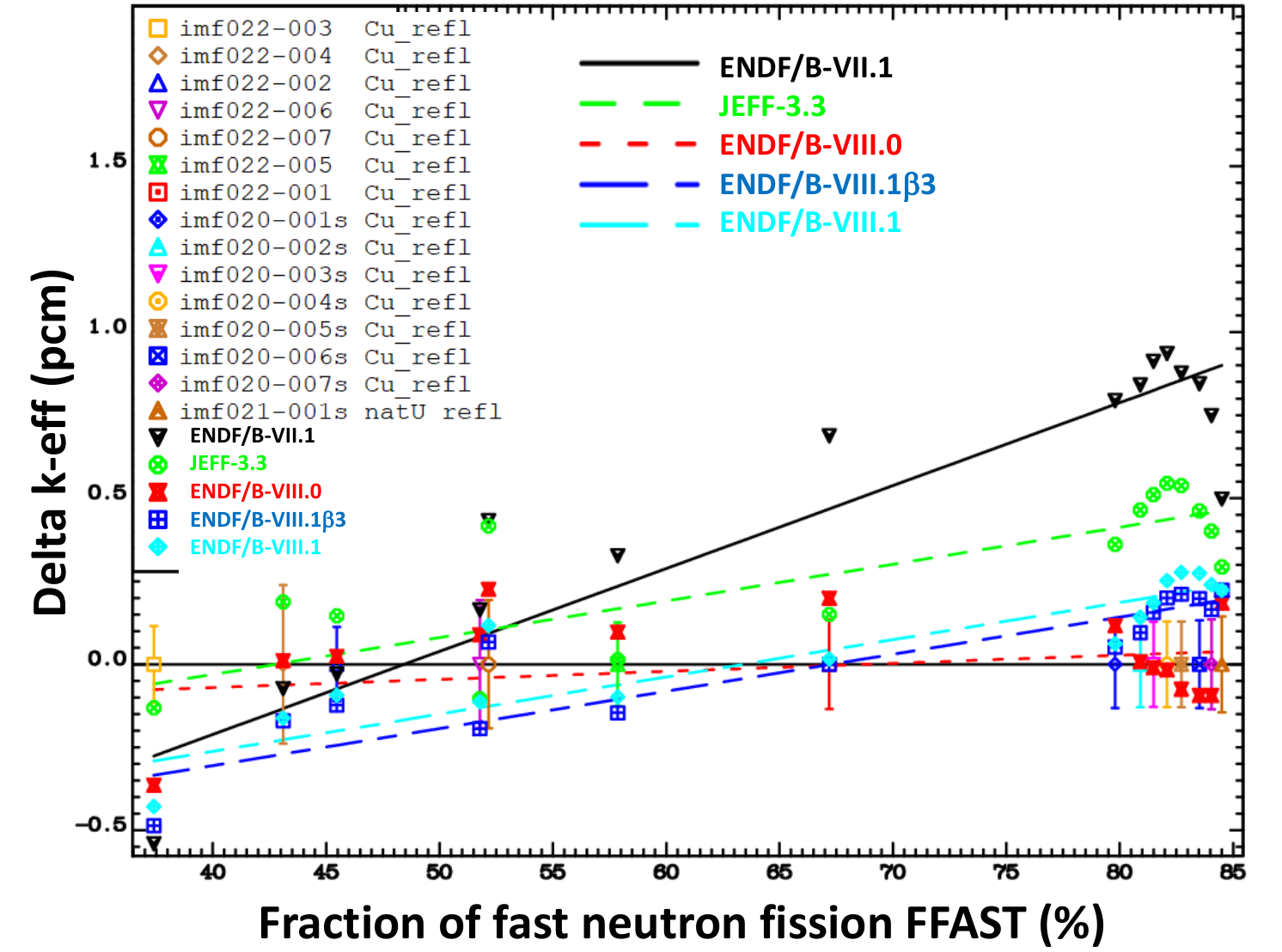}
\vspace{-.1in}
\caption{Criticality of Cu-sensitive intermediate Swedish ICSBEP benchmarks (IMF020,21,22) relative to the measured benchmark values as a function of the percentage fraction of fast-neutron fission (FFAST).}
\label{fig:IMF20-21-22}
\vspace{-2mm}
\end{figure}

As previously discussed, the $\Delta k_{\mathrm{eff}}$ calculated for the ENDF/B-VIII.1 evaluation increased for the Swedish IMF020 and IMF022 series of benchmarks compared to the ENDF/B-VIII.0 as shown in Fig.~\ref{fig:IMF20-21-22}. In fact, the new ENDF/B-VIII.1 evaluation (dashed cyan line) shows a larger positive gradient along this benchmark series as a function of the fraction of fast-neutron fissions (FFAST). The ENDF/B-VIII.0 (dashed red line) better agreement was driven by a too large $\mu$-bar in the RRR up to 700 keV, which was in disagreement with Popov \cite{Popov:1986} and Smith \cite{Smith:1964} data, as previously discussed.
However, we noticed that the IMF021 benchmark (brown upper triangle in Fig.~\ref{fig:IMF20-21-22}) uses exactly the same fuel as IMF020 and IMF022 benchmarks, but features a natural uranium reflector instead of the copper one. Considering how well we know uranium cross sections, the fact that the IMF021 calculates high by about +200 pcm both for ENDF/B-VIII.0 and ENDF/B-VIII.1 indicates an unidentified +200~pcm benchmark bias in the IMF020,21,22 benchmark series. If we consider this bias by subtracting 200 pcm from the $C$ values, the ENDF/B-VIII.1 evaluation will perform perfectly for this benchmark set as well, better than the ENDF/B-VIII.0.

The preliminary calculations of the new CERBERUS integral experiments~\cite{amundson_critical_2021} also show a much improved agreement for the ENDF/B-VIII.1 evaluation.

A very important shielding benchmark was the measurement of the $^{252}$Cf(sf) neutron leakage from a 48x50x50 cm$^3$ copper cube undertaken at the CVR Rez near Prague using stilbene scintillator detectors~\cite{Schulc:2021,Schulc:2022}.
This type of benchmark is very sensitive to the elastic and inelastic cross sections and elastic angular distributions up to 10~MeV of neutron incident energy, but it is practically insensitive to capture. It does perfectly complement the criticality benchmarks. The success of this benchmark to validate evaluated data is also driven by our excellent knowledge of the standard $^{252}$Cf(sf) neutron spectrum \cite{IRDFF2}.

\begin{figure}[!htbp]
\centering
\includegraphics[width=\columnwidth]{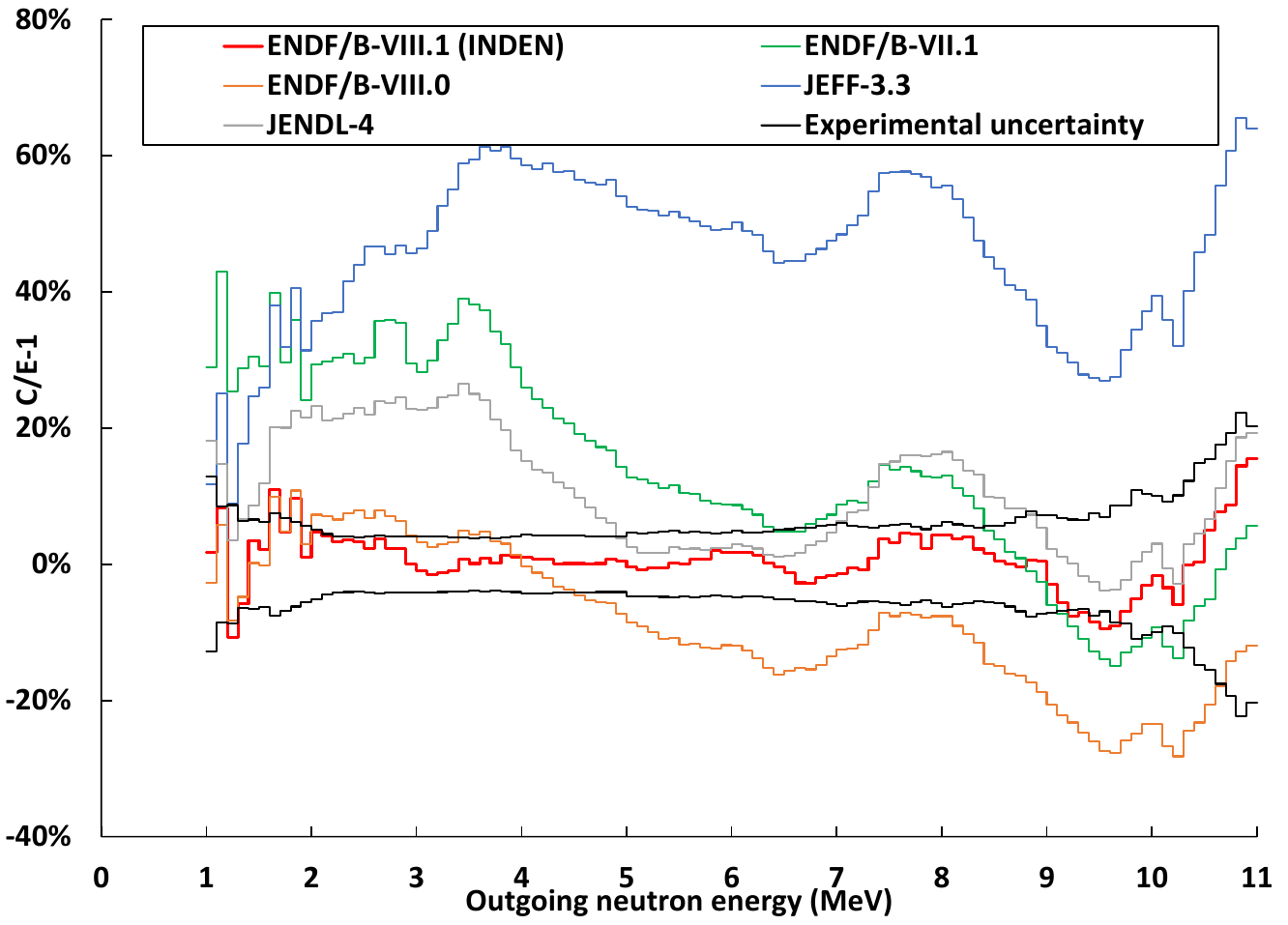}
\vspace{-.1in}
\caption{Measured neutron leakage of the \nuc{252}{Cf}(sf) neutron source measured at 1m distance from a 48x50x50~cm$^3$ copper block \cite{Schulc:2021,Schulc:2022}. Experimental benchmark values are compared with transport calculations using JEFF-3.3, JENDL-4.0, ENDF/B-VII.1, ENDF/B-VIII.0 and the new ENDF/B-VIII.1 library that adopted the INDEN Cu evaluations by means of C/E-1 ratio.}
\label{fig:Rez-shielding-Cu}
\vspace{-2mm}
\end{figure}

Calculated leaked neutron spectrum through the copper cube with different libraries is compared to the measured spectrum using the $C/E-1$ criteria for outgoing neutron energies from 1~MeV up to 10~MeV as shown in Fig.~\ref{fig:Rez-shielding-Cu}. The best results of all participating libraries were obtained with the INDEN Cu evaluation adopted for the ENDF/B-VIII.1 library.
JEFF-3.3 overestimated the leakage by up to +40\% due to disagreement of evaluated total cross section with available experimental data. The ENDF/B-VIII.0 evaluation underestimated the measured neutron leakage above 5~MeV by up to -20 to -30\% due to inappropriate description of the isotopic elastic cross section data in that energy region. ENDF/B-VII.1 and JENDL-4 showed similar performance overestimating the measured neutron leakage below 5~MeV of the outgoing neutron energy, and slightly overestimating the measured leakage around 8~MeV.

JEFF-3.3 covariances \cite{JEFF33} are recommended to be used, if needed, as ENDF/B-VIII.0 did not include covariances for copper. 

\subsubsection{\nuc{139}{La}}
\label{subsec:n:139La}

%
%
%

The central values for \nuc{139}{La} were adopted from TENDL-2014, with RRR and URR resonance parameters adopted from \prENDF. As for any isotopic TENDL evaluation, the fast neutron range was optimized by adjusting TALYS~\cite{TALYS} nuclear model parameters. In the case of \nuc{139}{La}, we adjusted the photon strength function for the neutron capture cross section, the proton optical model and particle-hole densities for the (n,n'), (n,2n) and (n,p) channels, and the alpha optical model parameters for the alpha-emitting channels.
The estimation of covariance data for \nuc{139}{La} was performed following an approach similar to that of the TENDL library~\cite{TENDL}. First, the cross sections and angular distributions are obtained by adjusting model parameters in order to be close to selected experimental data. In the fast neutron range, these parameters concern the models included in TALYS, and in the resonance range, resonance parameters are adjusted with the code TARES~\cite{TARES}. Covariances are then obtained by randomly modifying all model parameters. In the resonance range, multi-level Breit Wigner parameters are sampled given specific widths and standard deviations, in order to reproduce recommended values for the thermal cross sections and resonance integrals. In the fast range, a Bayesian Monte Carlo approach is followed~\cite{koning_bayesian_2015,rochman_monte_2018} with the code TASMAN, based on large prior distributions for parameters. In this case, the final sampled parameters are obtained from posterior distributions updated with experimental data. Finally,  covariance data from sampled cross sections and angular distributions are formatted into the ENDF-6 format with the code TEFAL.

It is worth noting that special nuclear data processing was needed when the \nuc{139}{La} evaluation was adopted for the FENDL library \cite{FENDL}. The analysis of a numerical benchmark for a neutron transport through a 1-meter sphere of \nuc{139}{La} proposed by C. Konno using ACE- and MATXS-formatted cross section files from the FENDL-3.1c library showed problems in the URR of \nuc{139}{La} evaluation adopted from TENDL-2014 \cite{aldama2021}. The total neutron flux computed using MCNP/ACE differed up to 50\% from the one calculated applying ANISN/MATXS deterministic transport codes. The problem arose when too small total cross section values were sampled by the PURR module of \NJOY2016 mainly due to the limitations of the processing method \cite{aldama2021}, but also due to the quality of the URR evaluated data. An \NJOY2016 (PURR) patch to correct this issue was proposed and applied, and the TENDL-2014 resonance data was replaced by the ENDF/B-VIII.0 resonance evaluation. The ENDF/B-VIII.0 data replacement and the \NJOY\ patch increased the neutron flux in the URR for Monte Carlo calculations by 4\%. The corresponding increase of the calculated neutron flux for deterministic codes is about 15\% for the ENDF/B-VIII.0 data. The agreement between deterministic and Monte Carlo benchmark results was significantly improved.

\subsubsection{\nuc{233}{U}}
\label{subsec:n:233U}

$^{233}$U evaluated data have poor performance in critical solutions, which has been well documented \cite{Brown2018}. Performance in fast critical assemblies is much better. No change has been done to the ENDF/B-VIII.0 evaluation in the fast energy range except the capture cross section that was changed from 3~keV up to 200~keV to agree on average with new experimental data by Leal-Cidoncha \etal \cite{Leal-Cidoncha2023}, and the neutron multiplicity change which is described below. A joint effort has been made by the international team led by ORNL within the INDEN collaboration to include the latest measured differential data and use the IAEA evaluated prompt fission neutron spectrum at the thermal point. Whilst performance problems remain, this new evaluation represents a step forward compared to \prENDF{}.

\paragraph{Thermal and resonance region\newline}
Similarly to R-matrix analyses on the two major fissile actinides \nuc{235}U and \nuc{239}Pu, the resonance parameter evaluation of n+\nuc{233}{U} reactions is an ongoing milestone of the US NCSP. In the ENDF/B-VIII.1 nuclear data library release, the major updates, in chronological order, to this evaluation work were the extension of the RRR evaluation up to 2.5~keV featuring an average increased fission cross section above 50 eV and the updates in the thermal and low-energy region up to a few eVs.

The extension up to 2.5~keV used transmission data and fission cross sections measured at the ORELA facility in the early 2000s~\cite{Guber:2000,Guber:2001}.  Due to the lack of capture measured data in the extended energy range, the cross sections of this reaction channel were obtained by using an average value of 39 meV assigned to the capture widths consistently with the RM approximation of the $R$-matrix theory. The second update was to guarantee a reasonable performance in criticality solution benchmarks affected by the changes in the thermal PFNS evaluation together with the adoption of the thermal constants recommended by the standard evaluations as discussed in the preliminary benchmark analysis found in Ref.~\cite{pigni:2021}.

\begin{table}[!thb]
\vspace{-3mm}
\caption{n+\nuc{233}{U} 2200 m/s thermal cross sections 
for three nuclear data libraries compared with the Thermal Neutron Constants derived by the Standard group \cite{carlson2018} and Duran \etal \cite{duran2024} estimates.
Available experimental data by Lounsbury \etal~\cite{Lounsbury1970,Beer1972,Beer1975} are also listed.}\label{tab:thermal_xs_u3}
\small
\vspace{-4mm}
\begin{center}
\begin{threeparttable}
\begin{tabular}{l | c c c c}
\toprule \toprule
               Source                     &  $\alpha_{therm}$ (b)      &  $\sigma_{el}$ (b) &  $\sigma_{f}$ (b) &  $\sigma_{c}$ (b) \\
\midrule
ENDF/B-VIII.1                             & 0.0827(28)             & 12.08(24)      & 533.4(91)    & 44.1(13) \\
relat. uncert.                            & 3.4\%                  & 2.0\%          & 1.7\%         & 2.9\%     \\
ENDF/B-VIII.0                             & 0.0792                 & 12.18          & 534.1         & 42.3      \\
relat. uncert.                            & --                     & --             & --            & --        \\
ENDF/B-VII.1                              & 0.0852(19)             & 12.18(75)      & 531.4(32)    & 45.3(10) \\
relat. uncert.                            & 2.3\%                  & 6.1\%          & 0.6\%         & 2.2\%     \\
Standards \cite{carlson2018}                    & 0.0842(17)       & 12.2(7)      & 533.0(22)    & 44.9(9) \\
relat. uncert.                                  & 2.0\%            & 5.7\%          & 0.4\%         & 2.0\%     \\
Duran \cite{duran2024,duran2023}                & 0.0845(41)       & 12.4(5)      & 533.0(7)    & 44.8(22) \\
relat. uncert.                                  & 4.9\%            & 4.0\%          & 1.3\%         & 4.9\%     \\
Lounsbury \cite{Lounsbury1970,Beer1972,Beer1975}& 0.0861(21)       &  --            &               &           \\
relat. uncert.                                  & 2.4\%            & --             & --            & -         \\
\bottomrule \bottomrule
\end{tabular}
\end{threeparttable}
\vspace{-3mm}
\end{center}
\end{table}
In Table~\ref{tab:thermal_xs_u3}, the fitted cross sections calculated at the thermal energy $E$=0.0253~eV 
are reported for three libraries. A very good agreement with standard values \cite{carlson2018} as well with thermal cross section recommended by Duran \etal \cite{duran2024,duran2023} is observed.
An excellent agreement of evaluated $\alpha_{therm}$ with Lounsbury \etal measurement \cite{Lounsbury1970,Beer1972,Beer1975} is also seen for all libraries except ENDF/B-VIII.0, which featured too low capture cross section.

\begin{figure}[!htbp]
\centering
\includegraphics[scale=.062,clip, trim = 115mm 15mm 115mm 115mm]{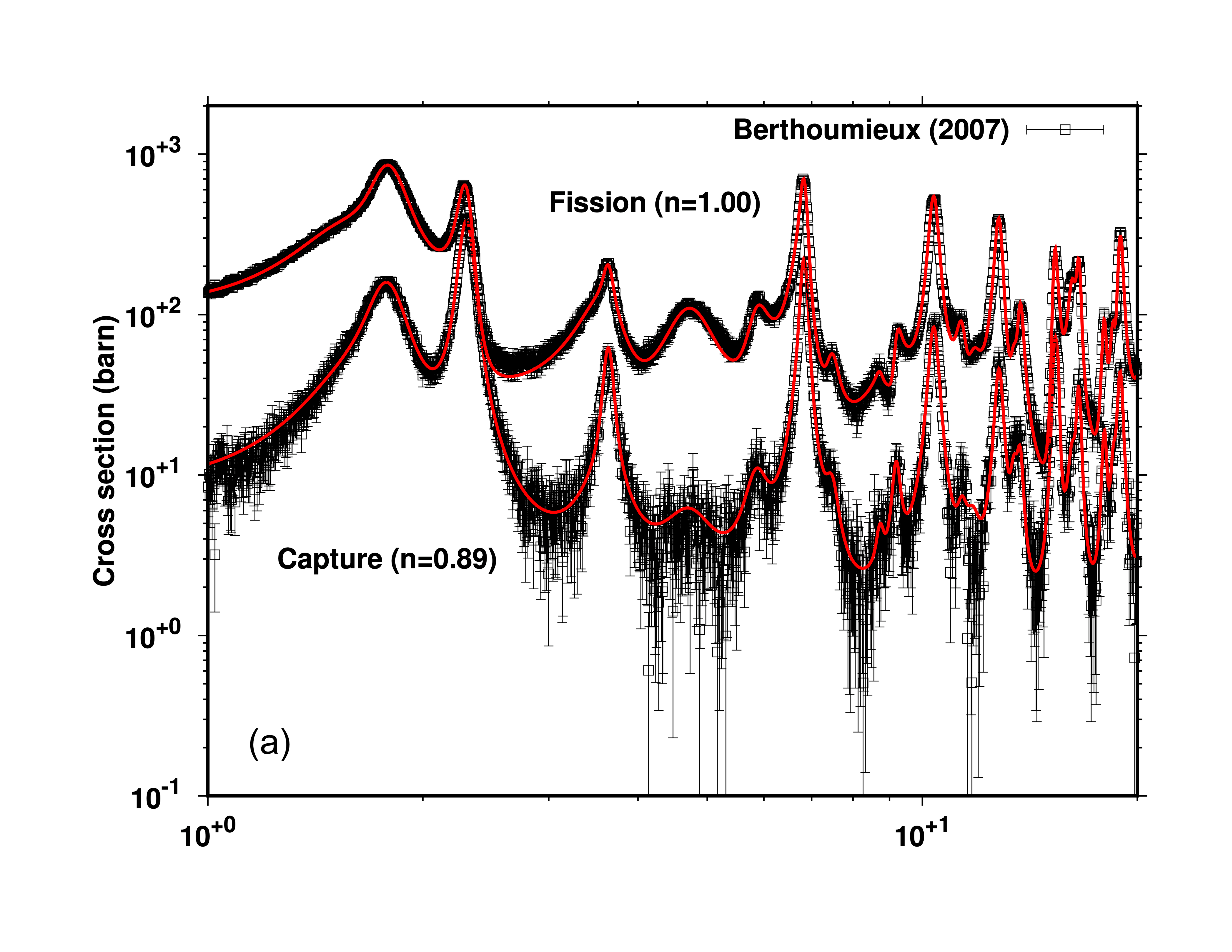} \\%
\includegraphics[scale=.062,clip, trim = 115mm 15mm 115mm 115mm]{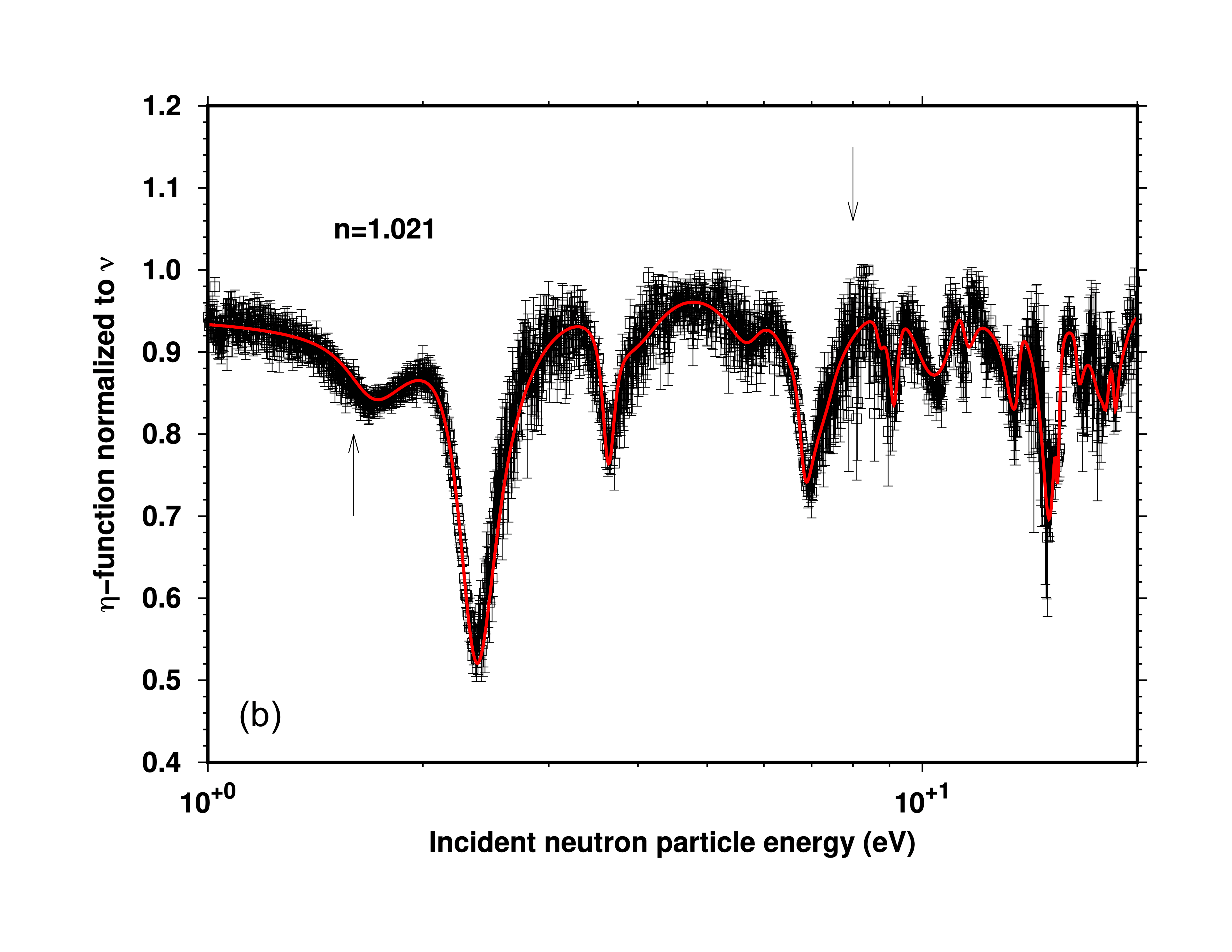}%
\caption[$n$+\nuc{233}{U} comparison with Berthomieux's measured data~\cite{Berthoumieux:2007}.]{
n+\nuc{233}{U} comparison of calculated (red) and measured (black)
cross sections (a) and $\eta$-function normalized to the number of
neutrons $\nu$ (b). Normalization factors to the calculated data
are also reported. Measured are data from Ref.~\cite{Berthoumieux:2007}.}
\label{fig:u233_cap_fiss_eta}
\end{figure}
In the energy region above thermal, the evaluation procedure consisted
on fitting sequentially fission and capture measured data. Among
these, Berthomieux's and Weston's
measurements~\cite{Berthoumieux:2007, Weston:1968} are particularly relevant because the fission and capture reaction channels were measured simultaneously with anti-coincidence techniques. Although the capture channel is associated with a relatively large scaling factor of about 11\%, Berthomieux's fission measured data~\cite{Berthoumieux:2007} (that are consistent to other fission cross sections measured by Calviani~\cite{Calviani:2009}) were taken as main reference for this evaluation because of their improved resolution and statistics over Weston's measured data. Instead, Weston's data were used for estimating the average magnitude of the capture reaction channel.

Plots in Fig.~\ref{fig:u233_cap_fiss_eta} show the calculated fission and capture cross sections (a) together with the $\eta$-function normalized to the average number of neutrons $\bar{\nu}$ (b). The quantity $\eta$ is a measure of the average number of neutron released per neutron absorption and related to the ratio capture-to-fission $\alpha$ as $\eta=\bar{\nu}/(1+\alpha)$. Compared to corresponding measured data (in black), the calculated fission and capture cross sections (in red) largely differ in their average magnitude. In fact, except for the resonance levels at 2.3~eV and 7.4~eV, the fission cross section is about an order of magnitude larger than the capture cross section. This results in a capture reaction channel associated with large uncertainties and highly correlated to small variations in the fission reaction channel. This can significantly impact the magnitude of the capture cross section and, in turn, greatly affect the reactivity in criticality benchmarks.

The bottom plot of Fig.~\ref{fig:u233_cap_fiss_eta} is about the energy dependence of the $\eta$-function and can be used to highlight critical points of incompatibility between measured data as well as the difficulty of the $R$-matrix model to describe the shape of the measured data. For instance, although the relatively good fit is achieved for the fission and capture channel, the energy region marked by the first arrow at about 1.5~eV could not be perfectly optimized to describe the measured data as their energy tail augmented by not negligible deviations between Weston's and Berthomieux's measured data in the same energy region. The second arrow marking the energy region at 8~eV is also about large deviations in the small resonance levels between Berthomieux's and Weston's measured data. This represents one of the cases where the magnitude of both fission and capture cross section is small and fission widths are much larger than the average capture width.

\begin{figure}[!htbp]
\centering
\includegraphics[scale=0.74,clip, trim = 5mm 0mm 5mm 0mm]{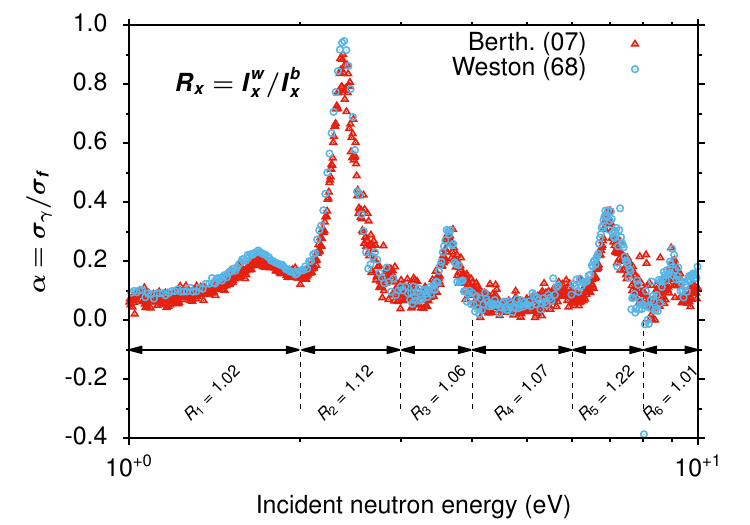}%
\caption[$n$+\nuc{233}{U} comparison between Berthomieux's and
Weston's ratio capture-to-fission $\alpha$ measured
data.]{n+\nuc{233}{U} comparison between Berthomieux's and Weston's
ratio capture-to-fission $\alpha$ measured
data~\cite{Berthoumieux:2007, Weston:1968}. The ratio $R_{x}$ of Weston to Berthomieux integrals of several energy ranges up to 10 eV is also shown.}\label{fig:u233_alpha}
\vspace{-3mm}
\end{figure}
In addition to the $\eta$-function, average differences in the Weston's and Berthomieux's measured data are shown in Fig.~\ref{fig:u233_alpha} where the integral comparison of the ratio-capture-to-fission $\alpha$ is shown up to 10~eV. Not negligible deviations (up to 22\%) between the reported measured data can be seen confirming the difficulties in the characterization of the capture channel among different measurements.

\begin{figure}[htb!]
\centering
\includegraphics[width=\columnwidth]{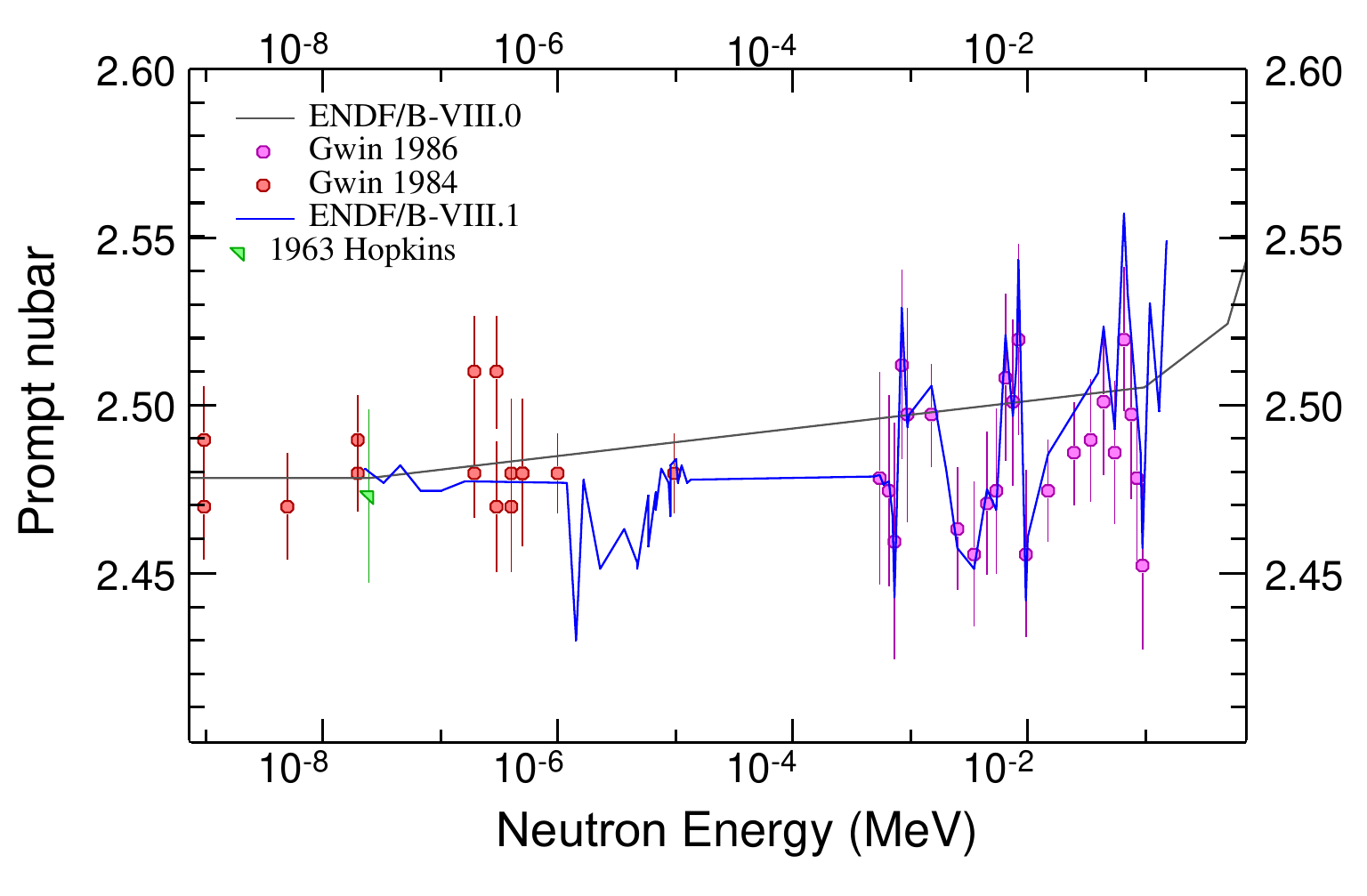}
\vspace{-4mm}
\caption{Experimental data from Gwin \etal~\cite{Gwin:1984-nubar,Gwin:1986-nubar} and Hopkins \etal~\cite{Hopkins1963} are compared to evaluated ENDF/B-VIII.0 and ENDF/B-VIII.1 $^{233}$U $\overline\nu$ below 100~keV.}
\label{fig:U233nu-RR}
\end{figure}

As in other cases of fissile actinides, an accurate and precise characterization of the capture cross sections remain elusive from the experimental point of view. For these reasons, some guidance was taken from the reactivity of criticality benchmarks as a function of the epithermal fission fraction. Evaluation of n+\nuc{233}{U} reactions is one of the most challenging cases due to the large magnitude of the fission reaction channel compared to the capture cross section.
%
%
%
\paragraph{(n,f) Fission $\overline\nu$\newline}
The thermal $\overline\nu_{tot}$ in the new evaluation was slightly increased to 2.4869 in perfect agreement with the thermal neutron constant value of 2.487(0.011)~\cite{carlson2018}. The \prENDF{} value was slightly smaller  2.48524 but agrees well within estimated uncertainties of 0.42\%. Gwin data \cite{Gwin:1986-nubar}  renormalized to the above-discussed thermal value was used to define the $\overline\nu$ energy dependence below 10 eV. Gwin data \cite{Gwin:1984-nubar} was also employed to define the neutron multiplicity from 500~eV up to 500 keV as shown in Fig.~\ref{fig:U233nu-RR}.

\begin{figure}[htb!]
\centering
\includegraphics[width=\columnwidth]{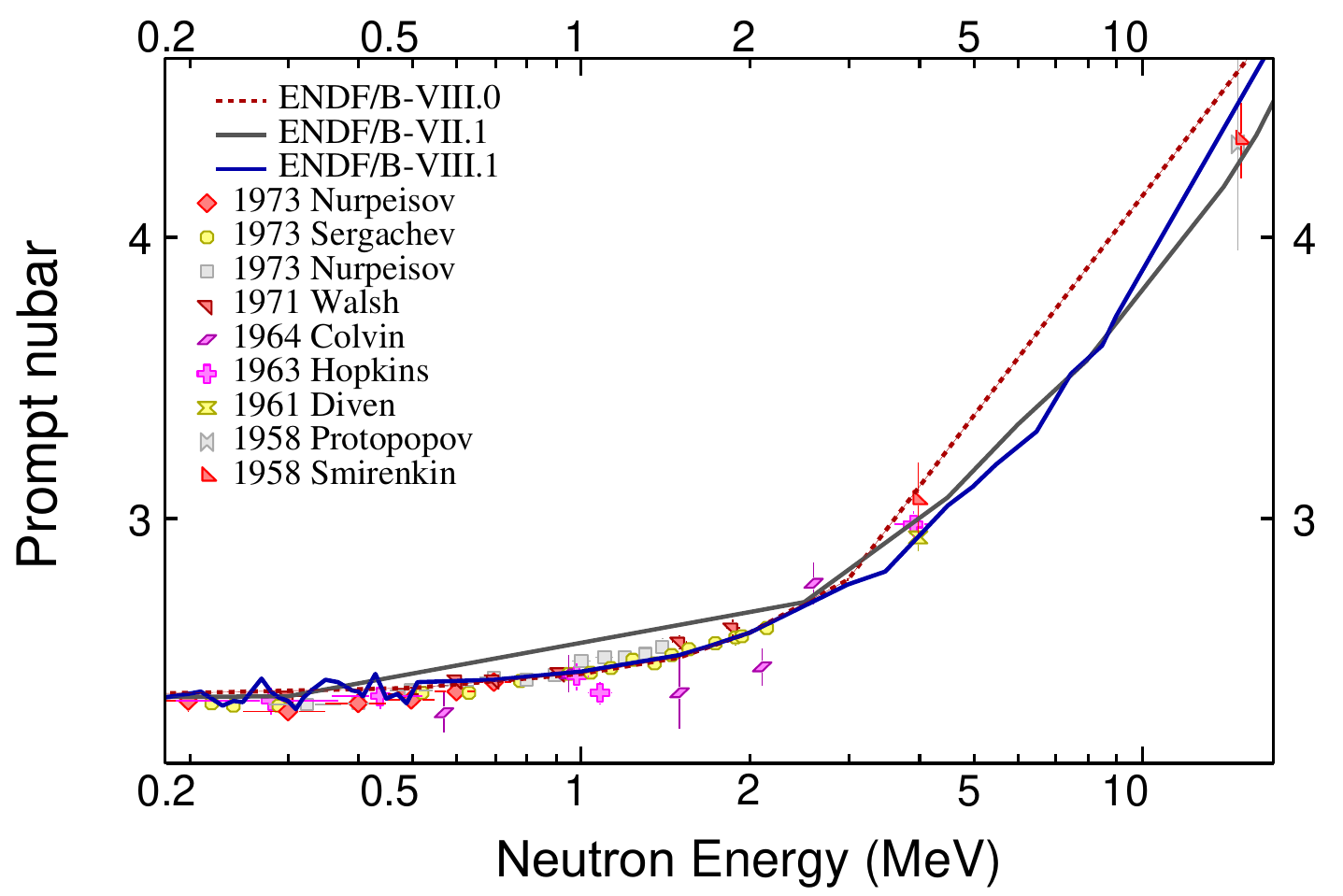}
\caption{Selected experimental data \cite{Diven1961,Hopkins1963,Nurpeisov1973,Sergachev1973,Walsh1971,Colvin1964,Smirenkin1958,Protopopov1958} from EXFOR \cite{EXFOR} versus evaluated $^{233}$U $\overline\nu$ from 100~keV up to 20~MeV.}
\label{fig:U233nu-fast}
\vspace{-3mm}
\end{figure}

ENDF/B-VIII.1 neutron multiplicity was changed to optimize the agreement with Jezebel-23 benchmark and is compared to \prENDF{} in Fig.~\ref{fig:U233nu-fast}.

\begin{figure}[!thb]
\centering
\includegraphics[width=\columnwidth]{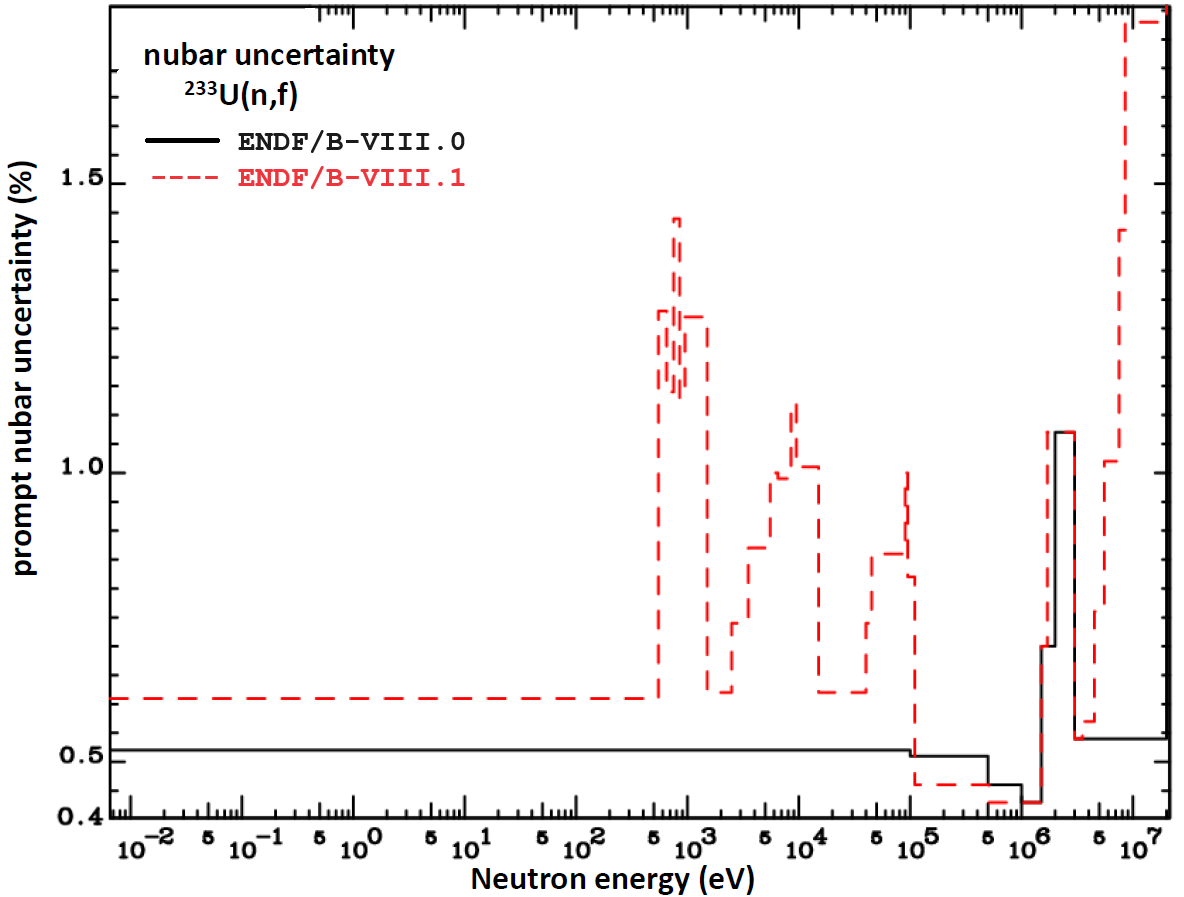}
\caption{Evaluated $^{233}$U $\overline\nu$ uncertainty. 
}
\label{fig:U233nurelunc}
\end{figure}

The ENDF/B-VIII.1 evaluated $\overline\nu$ uncertainties in Fig.~\ref{fig:U233nurelunc} are distinctly larger up to 100 keV due to the increased $\overline\nu$($^{252}$Cf(sf)) standard uncertainties (0.42\%). The use of Gwin nubar data \cite{Gwin:1984-nubar} from 500~eV up to 100 keV led to a significant increase of the nubar uncertainty to values up to 1.4\% in the new evaluation (from about 0.5\% in ENDF/B-VIII.0 to roughly the uncertainty of Gwin's experiment) as shown in Fig.~\ref{fig:U233nurelunc}. The evaluated $^{233}$U $\overline\nu$ uncertainty was similar to \prENDF{} up to 3~MeV and significantly increased above.

\subsubsection{\nuc{235}{U}}
\label{subsec:n:235U}



A focused international effort on $^{235}$U evaluations adopted for the ENDF/B-VIII.0 library was coordinated by the IAEA within the CIELO project \cite{IAEA-CIELO} and comprehensively documented
by Capote \etal\ \cite{capote2018} and Chadwick \etal \cite{CIELO-res}. Cross-section evaluation in the fast neutron range remains unchanged for ENDF/B-VIII.1. PFNS and $\overline\nu$ were revisited as described below. Minor changes were also undertaken in the resonance region driven by newly available experimental data~\cite{danon:2017}.

\begin{figure}[!htbp]
\centering
\includegraphics[scale=.35, trim = 21mm 21mm 13mm 21mm]{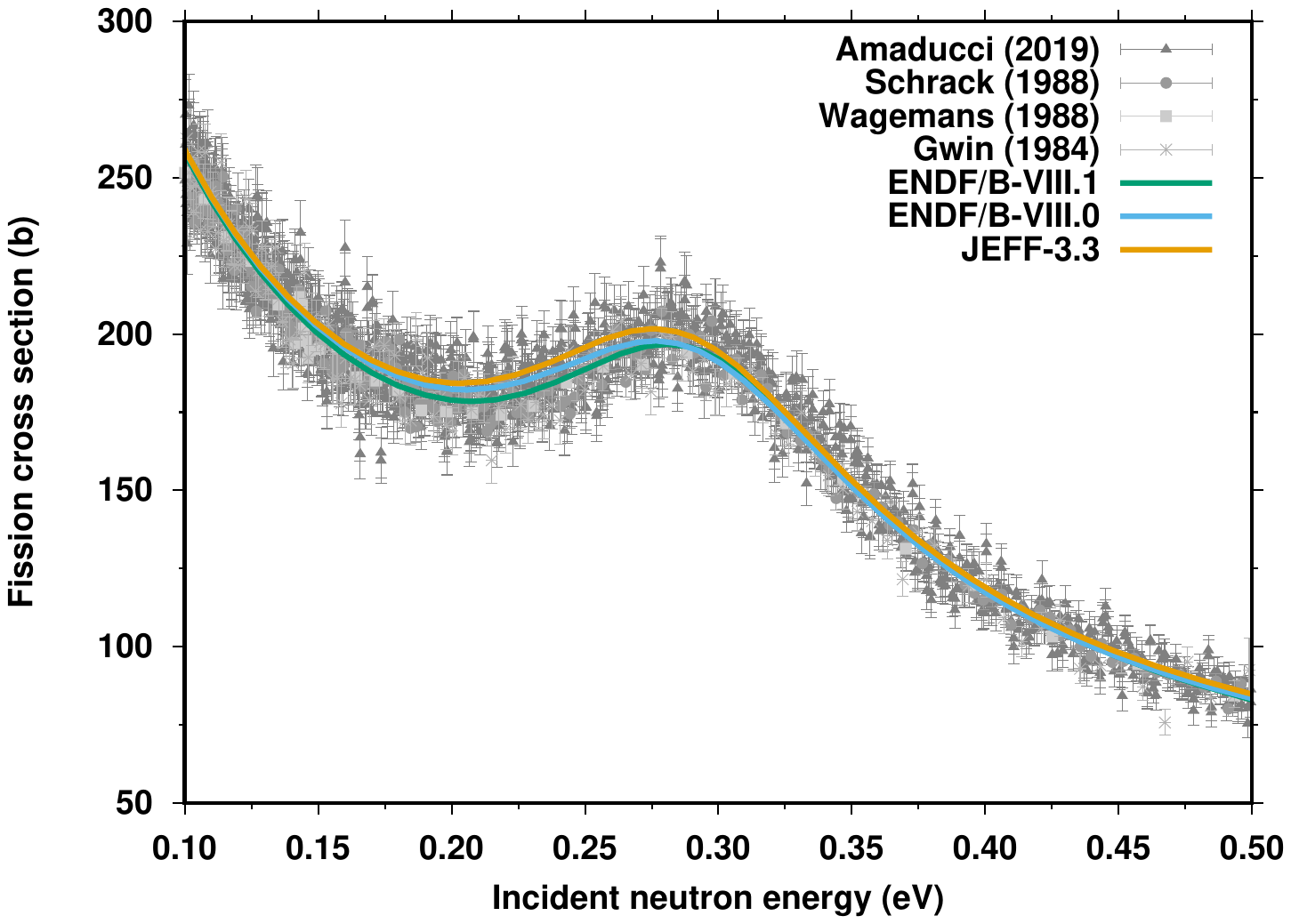}
\vspace{-.1in}
\caption{n+\nuc{235}{U} fission neutron cross section calculated for three nuclear data libraries together with selected measured data below 0.5~eV of neutron incident energy~\cite{amaducci:2019,schrack:1988,wagemans:1988,gwin:1984}.}\label{fig:u235fis}
\end{figure}

\paragraph{Thermal and Resonance regions\newline}
The R-matrix analysis of the n+\nuc{235}{U} system is an ongoing milestone of the NCSP. In the ENDF/B-VIII.1 nuclear data library release, the major update to this evaluation work has been the inclusion of the neutron capture and fission yield data measured at the RPI Gaerttner LINear ACcelerator (LINAC) facility below 2.5~keV~\cite{danon:2017}.
Two measurements were done---thermal and epithermal---that together cover the energy range from 0.01~eV to 2.5~keV. The measurements recorded gammas from $^{235}$U reactions using the RPI multiplicity detectors. Fission and capture events were separated by event multiplicity and total energy deposition~\cite{danon:2017}. The importance of these yield data relies in the simultaneous measurement of both capture and fission reaction channels including the low-lying resonance at 0.3~eV. In fact, besides being extremely important for the correct quantification of reactivity coefficients in criticality benchmarks, the first $^{235}$U resonance located at 0.3~eV features a very small neutron width and its magnitude is determined by the fit of the fission and capture channels. Although there is a large amount of measured data for the fission reaction channel as shown in Fig.~\ref{fig:u235fis}, the capture channel has been particularly difficult to estimate due to the very limited amount of experimental data below 1~eV.

\begin{figure}[!htbp]
\centering
\includegraphics[scale=.062, trim = 115mm 21mm 115mm 80mm]{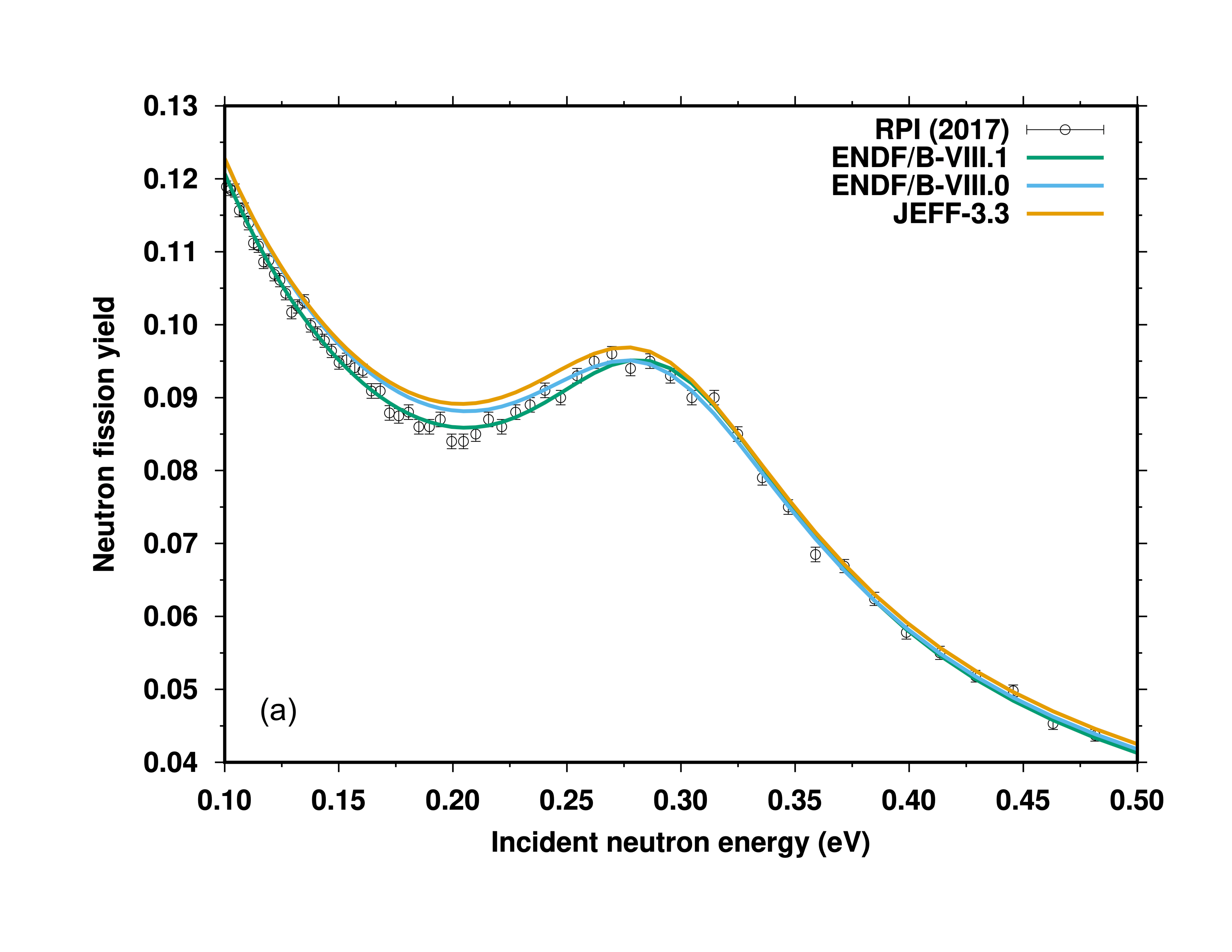}
\includegraphics[scale=.062, trim = 115mm 21mm 115mm 80mm]{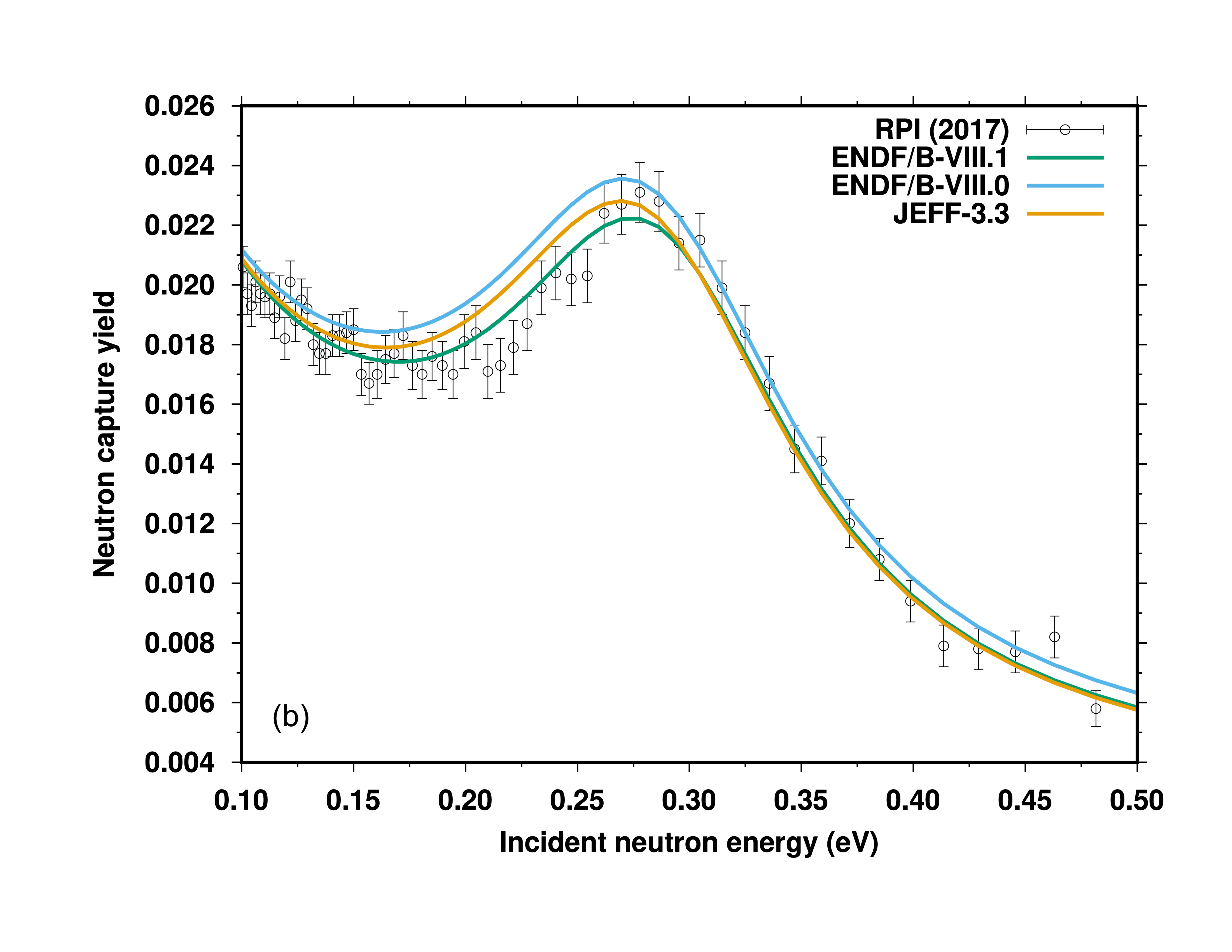}
\vspace{-.1in}
\caption{n+\nuc{235}{U} fission and capture neutron yield data calculated for three nuclear data libraries are compared with the RPI measured data~\cite{danon:2017} below 0.5~eV.}
\label{fig:u235yield}
\end{figure}

Since the evaluation of the capture reaction channel has a straightforward impact on the criticality of a fissile nuclear system, the increase of the capture cross section, as shown in Fig.~\ref{fig:u235yield} for the ENDF/B-VIII.0 release (cyan curve in (b) panel), was suggested by updates in the PFNS~\cite{trkov:2015,trkov:2015a,capote:2016} that featured a softer PFNS with lower average neutron energy than in the previous ENDF/B-VII.1 nuclear data library. By keeping the thermal values consistent with the thermal neutron constants recommended by the Standards group \cite{carlson2018}, the recent measured data induced a reduction of both fission and capture yields for the 0.3~eV level (green curve in the top and bottom panels). The energy region below 0.5~eV  was fitted with particular emphasis in evaluating the interference effect at the minimum-yield neutron energy of 0.2~eV that was not very well reproduced in ENDF/B-VIII.0 or JEFF-3.3 evaluations as shown in the figure. This has the consequence of slightly underestimating the peak of the 0.3~eV level for the capture channel although still within the lower limit of the experimental uncertainty.
The fission and capture yields measured at RPI~\cite{danon:2017} were compared to ENDF/B-VII.1 evaluation \cite{ENDF-VII.1}, 
ENDF/B-VIII.0 and the new resonance evaluation (note: labelled IAEA 2020) 
by Danon \etal\footnote{see Danon presentation at the INDEN meeting at \url{https://www-nds.iaea.org/index-meeting-crp/CM-INDEN-2020-res/docs/Danon-INDEN-RR-III.pdf}} as shown in the Table~\ref{fig:u235-low-en-vs-RPI}. Systematic uncertainties are estimated to be 2\% for fission and 3\% for measured capture yields. It is observed that the capture yield of the 8.76~eV resonance \cite{danon:2017} was overestimated by 10\% in the ENDF/B-VIII.0 evaluation; the capture yield was 5\% higher than the RPI measured already above the 0.6~eV valley. 
The new ENDF/B-VIII.1 resonance evaluation agrees within quoted uncertainty with the measured RPI data \cite{danon:2017} below 250 eV. 
Small differences between the ENDF/B-VIII.1 evaluation and measured data remain from 250 up to 350 eV and 
from 850 up to 950 eV for the capture yield. 

\begin{table}[!htbp]
\vspace{-3mm}
\caption{C/E of energy-grouped fission and capture yields for both RPI experiments compared to ENDF/B-VII.1 (B7.1), ENDF/B-VIII.0 (B8.0) and the ENDF/B-VIII.1 (B8.1) $^{235}$U resonance evaluation. Values highlighted in \textcolor{red}{red} show evaluated deviations in the capture yield larger than the estimated experimental uncertainty  of 3\%.}\label{fig:u235-low-en-vs-RPI}
\vspace{-6pt}
\begin{center}
\begin{threeparttable}
\begin{tabular}{r r |c c c | c c c}
\toprule \toprule
\multicolumn{8}{l}{From Thermal experiment:} \\
    E1 &  E2    &  \multicolumn{3}{l}{Fission yield} & \multicolumn{3}{l}{Capture yield}\\
   (eV)& (eV)   & B7.1 & B8.0 & B8.1 & B7.1 &   B8.0           & B8.1 \\
\midrule
  0.01 & 0.0206 & 1.01 & 1.00 & 1.00 & 1.02 &   1.01           & 1.01 \\
  0.02 & 0.03   & 1.00 & 1.00 & 1.00 & 0.99 &   0.99           & 0.99 \\
0.0206 & 0.0623 & 1.00 & 1.00 & 1.00 & 0.98 &   0.99           & 0.99 \\
0.0623 & 0.6    & 0.99 & 1.01 & 1.01 & 0.99 & {\color{red}1.05}& 1.03 \\
   0.6 & 7.8    & 0.97 & 0.99 & 1.00 & 0.99 & {\color{red}1.05}& 1.02 \\
0.0253 & 9.4    & 0.98 & 0.99 & 1.01 & 0.99 & {\color{red}1.06}& 1.02 \\
   7.8 & 11.    & 0.98 & 0.98 & 1.01 & 1.00 & {\color{red}1.10}& 1.02 \\
\midrule
\multicolumn{8}{l}{From Epithermal experiment:} \\
    E1 &  E2    &  \multicolumn{3}{l}{Fission yield} & \multicolumn{3}{l}{Capture yield}\\
   (eV)& (eV)   &  B7.1 & B8.0 & B8.1 & B7.1 &   B8.0          & B8.1 \\
\midrule
  9.4  & 150    & 1.02 & 1.01 & 1.01 & {\color{red}1.04} &  1.03            & 1.03 \\
  150  & 250    & 1.02 & 1.00 & 1.00 & {\color{red}1.07} &  1.03            & 1.03 \\
  250  & 350    & 1.04 & 1.02 & 1.02 & {\color{red}1.06} & {\color{red}0.96}&{\color{red}0.96}\\
  350  & 450    & 1.03 & 1.04 & 1.04 & {\color{red}1.12} &  0.99            & 0.99            \\
  450  & 550    & 1.02 & 1.02 & 1.02 & {\color{red}1.17} &  1.00            & 1.00            \\
  550  & 650    & 1.03 & 1.01 & 1.01 & {\color{red}1.18} &  1.00            & 1.00            \\
  650  & 750    & 1.03 & 1.02 & 1.02 & {\color{red}1.17} &  1.02            & 1.02            \\
  750  & 850    & 1.03 & 1.03 & 1.03 & {\color{red}1.17} &  1.03            & 1.03            \\
  850  & 950    & 1.00 & 1.01 & 1.01 & {\color{red}1.17} & {\color{red}1.07}&{\color{red}1.07}\\
  950  & 1500   & 1.02 & 1.00 & 1.00 & {\color{red}1.25} &  1.02            & 1.02            \\
\bottomrule \bottomrule
\end{tabular}
\end{threeparttable}
\end{center}
\vspace{-2mm}
\end{table}

%
It was independently tested at ORNL that the current resonance-parameter updates implemented in the ENDF/B-VIII.1 had negligible impact on the loss of reactivity of the ENDF/B-VIII.0 library in depletion calculations as a function of the burnup. This has been an indication of the little correlation between the loss of reactivity in depletion calculations and the excellent performance of the $^{235}$U ENDF/B-VIII.0 library in criticality benchmarks particularly for HEU configurations. In fact, additional tests demonstrated that the impact of \nuc{238}U and \nuc{239,240,241}Pu was directly responsible for the loss of reactivity in depletion calculations~\cite{kim2021-318}.

In Table~\ref{tab:thermal_xs_u5}, the thermal constants for scattering, fission, and capture cross sections 
are reported for three nuclear data libraries showing very consistent values; values for the new ENDF/B-VIII.1 evaluation agree with the recommended standard Thermal Neutron Constants \cite{carlson2018} as well as with thermal cross sections recommended by Duran \etal \cite{duran2024,duran2023}.
An excellent agreement of evaluated $\alpha_{therm}$ with Lounsbury \etal measurement \cite{Lounsbury1970,Beer1972,Beer1975} is observed for all libraries well within quoted uncertainties.

\begin{table}[!tbh]
\vspace{-3mm}
\caption{n+\nuc{235}{U} 2200 m/s thermal cross sections 
for three nuclear data libraries compared with the Thermal Neutron Constants derived by the Standard group \cite{carlson2018} and Duran \etal \cite{duran2024,duran2023} estimates. Available experimental data by Adamchuk \etal~\cite{Adamchuk1988} and Lounsbury \etal~\cite{Lounsbury1970,Beer1972,Beer1975} are also listed.}\label{tab:thermal_xs_u5}
\vspace{-3mm}
\begin{center}
\begin{threeparttable}
\begin{tabular}{l | c c c c}
\toprule \toprule
   Source                                   &  $\alpha_{therm}$ (b) &  $\sigma_{el}$ (b) &  $\sigma_{f}$ (b) &  $\sigma_{c}$ (b) \\
\midrule
 ENDF/B-VIII.1                           & 0.1696(8)         & 14.07(35)      & 586.2(29)    & 99.4(2)  \\
 relat. uncert.                          & 0.5\%             & 2.5\%          & 0.5\%         & 0.2\%    \\
 ENDF/B-VIII.0                           & 0.1694(10)        & 14.11(37)      & 586.7(37)    & 99.4(2)  \\
 relat. uncert.                          & 0.6\%             & 2.6\%          & 0.6\%         & 0.2\%    \\
 ENDF/B-VII.1                            & 0.1687(27)        & 15.11(77)      & 585.0(20)    & 98.7(16)\\
 relat. uncert.                          & 1.6\%             & 5.1\%          & 0.3\%         & 1.6\%    \\
 Standards \cite{carlson2018}            & 0.1694(22)        & 14.1(2)        & 587.3(14)    & 99.5(13)\\
 relat. uncert.                          & 1.3\%             & 1.4\%          & 0.2\%         & 1.3\%    \\
 Duran \cite{duran2024,duran2023}        & 0.1711(43)        & 14.3(5)        & 586.1(26)    & 100.3(25)\\
 relat. uncert.                          & 2.5\%             & 3.5\%          & 0.4\%         & 2.5\%    \\
 Lounsbury \cite{Lounsbury1970,Beer1972,Beer1975} & 0.1697(29)  &  --            & --            & --        \\
 relat. uncert.                          & 1.7\%                &  --            & --            & --        \\
 Adamchuk  \cite{Adamchuk1988}           & 0.1690(35)           &  --            & --            & --        \\
 relat. uncert.                          & 2.1\%                &  --            & --            & --        \\
 \bottomrule \bottomrule
\end{tabular}
\end{threeparttable}
\end{center}
\vspace{-2mm}
\end{table}

\paragraph{(n,f) Prompt Fission Neutron Spectrum\newline}
The $^{235}$U PFNS at thermal neutron energies was carried over from ENDF/B-VIII.0. This PFNS evaluation by the IAEA~\cite{trkov:2015,trkov:2015a,capote:2016} was obtained as part of the IAEA standard effort by a least-squares analysis of experimental data and their covariances and adjusting the high-energy tail above 10 MeV to reproduce the evaluated $^{90}$Zr($n$,$2n$) spectrum averaged cross section in the standard $^{235}$U($n_{th}$,f) neutron field. The physical origin of the PFNS high-energy neutrons above 10 MeV is still being clarified. It seems very unlikely that those neutrons are emitted during the deexcitation of fission fragments. An excellent agreement was found between evaluated PFNS and measured spectrum averaged cross section data of very high threshold dosimetry reactions \cite{schulc2024}, which explicitly supports the IAEA PFNS thermal evaluation.  Further investigations are warranted.

PFNS at all other energies were reevaluated as described in Ref.~\cite{Neudecker:2022U235PFNS}. This evaluation builds on PFNS work included in ENDF/B-VIII.0 and documented in detail in Ref.~\cite{Neudecker2018PFNS}. The main difference is the inclusion of final high-precision $^{235}$U PFNS measured by the Chi-Nu team of LANL and LLNL~\cite{Kelly2022}.
For ENDF/B-VIII.0, preliminary Chi-Nu PFNS were included for PFNS above incident neutron energies, $E_\mathrm{inc}$, of 5 MeV; these preliminary data were only going up to outgoing neutron energies, $E_\mathrm{out}$, of 2 MeV. Now, the final Chi-Nu data are included for $E_\mathrm{inc}=$ 2--10 MeV and up to $E_\mathrm{out}$ of 10 MeV.
Previously available data sets were measured at a limited incident neutron energy range or had larger uncertainties.

\begin{figure}[htb!]
\centering
\includegraphics[width=0.45\textwidth]{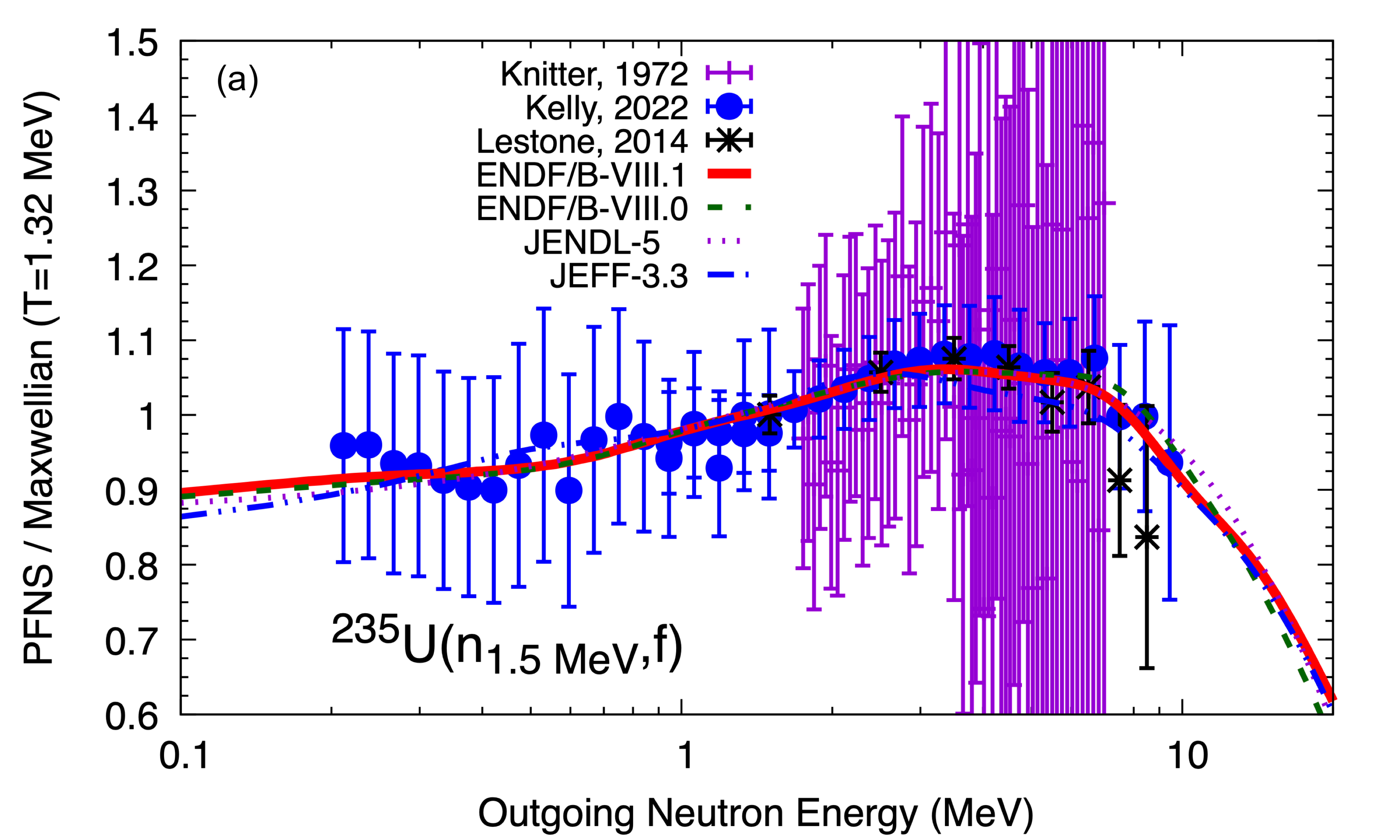}
\includegraphics[width=0.45\textwidth]{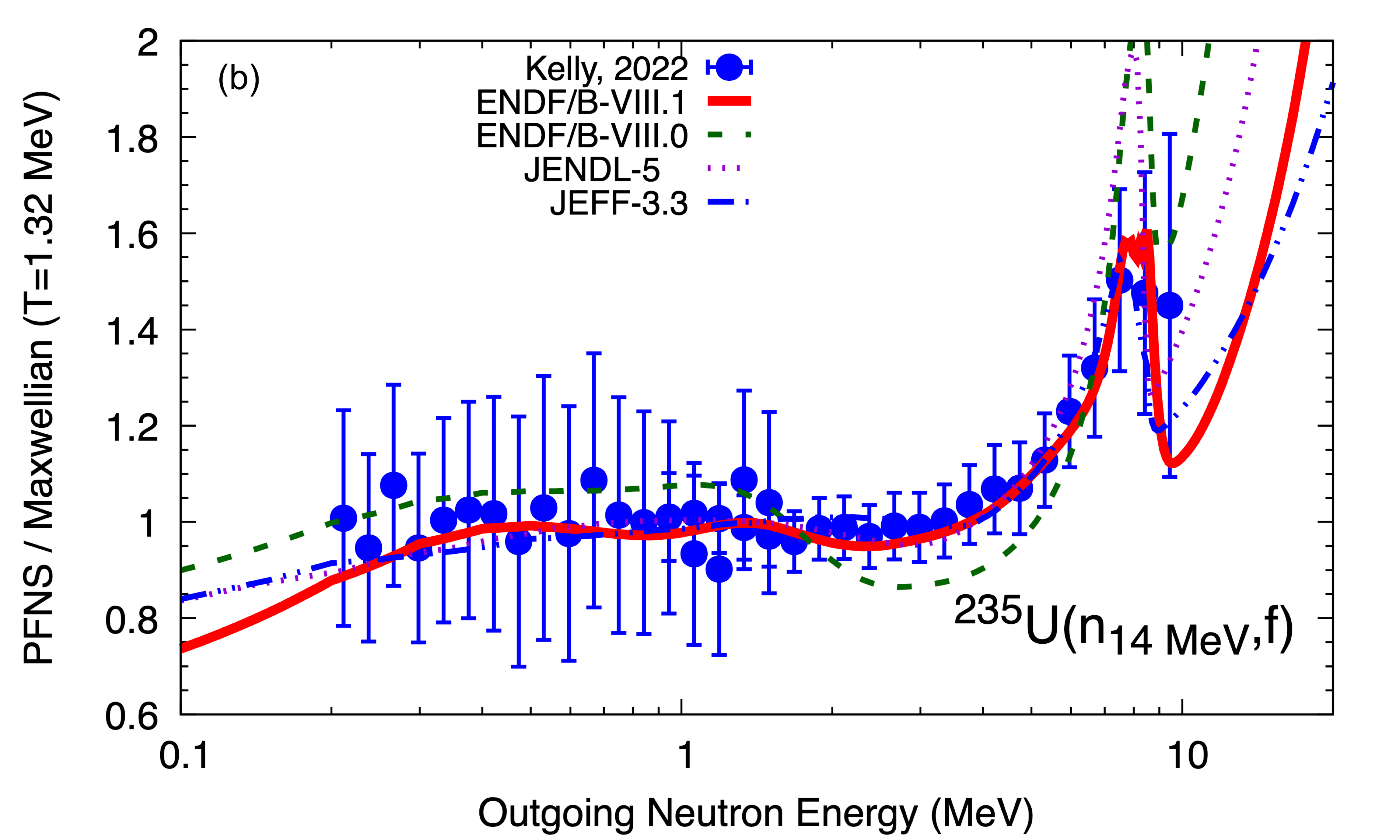}
\caption{Experimental data~\cite{Kelly2022,Lestone:2014,Knitter1972} and evaluated $^{235}$U PFNS for $E_\mathrm{inc}$=1.5 (a) and 14 MeV (b).}
\label{fig:U235PFNS}
\end{figure}

The newly evaluated PFNS shown in Fig.~\ref{fig:U235PFNS}(a) follow closely ENDF/B-VIII.0 for $E_\mathrm{inc} <$ 6 MeV except for a small decrease of the PFNS for $E_\mathrm{out}>$ 7 MeV which reflects Chi-Nu data's shape. This decrease leads to a modest criticality reduction of Godiva, Flattop and BigTen $k_\mathrm{eff}$ values by 52(11), 40(13) and 31(13)~pcm~\cite{Neudecker:2022U235PFNS}.

\begin{figure}[htb!]
\centering
\includegraphics[width=0.49\textwidth, clip, trim=0mm 0mm 0mm 10mm]{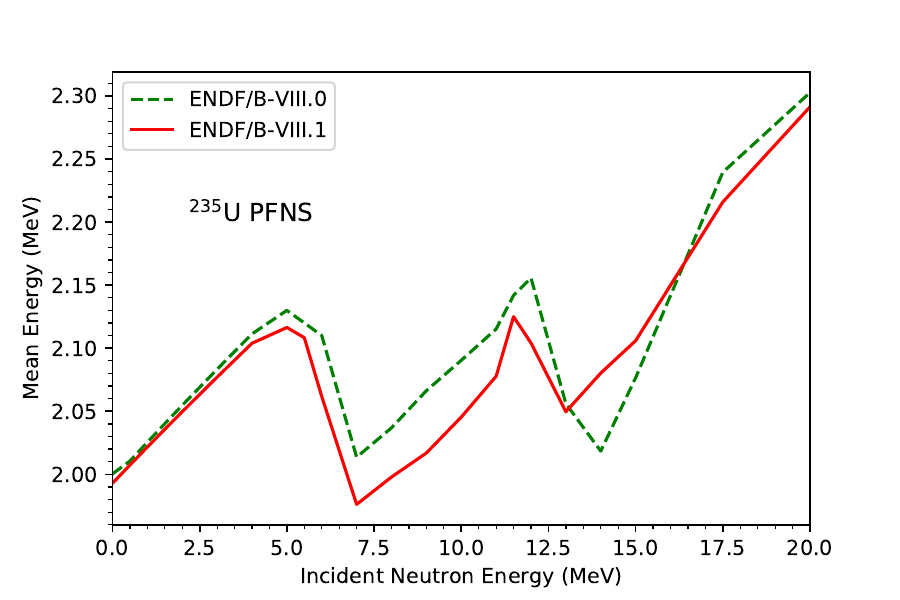}
\caption{Evaluated $^{235}$U PFNS mean energy.}
\label{fig:U235PFNSmeanenergy}
\end{figure}

The changes from ENDF/B-VIII.0 at higher $E_\mathrm{inc}$ are more pronounced as can be seen in the mean energy of the PFNS (Fig.~\ref{fig:U235PFNSmeanenergy}) and individual PFNS (Fig.~\ref{fig:U235PFNS}(b)). The distinct change around $E_\mathrm{inc}$ of 14 MeV led to improvements in predicting $^{235}$U LLNL 14-MeV pulsed sphere neutron leakage spectra in the valley after the peak,  in Ref.~\cite[Figs.~10 and 11]{Neudecker:2022U235PFNS}, and shown later in Fig.~\ref{fig:LLNLpulsedspheres235U}.

\begin{figure}[htb!]
\centering
\includegraphics[width=0.45\textwidth]{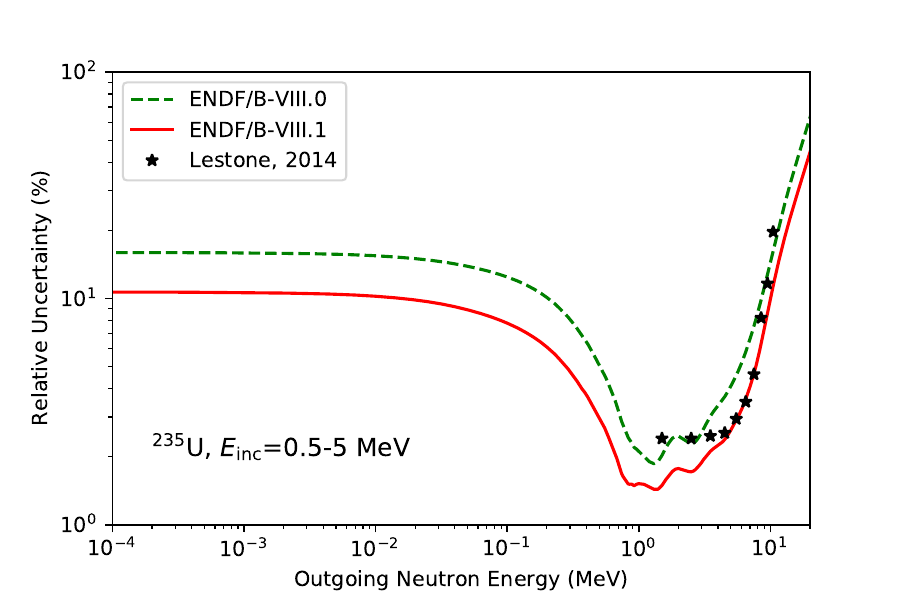}
\caption{Evaluated $^{235}$U PFNS uncertainties for $E_\mathrm{inc}$=0.5--5 MeV compared to experimental data of Ref.~\cite{Lestone:2014}.}
\label{fig:U235PFNSRelUnc}
\end{figure}

Evaluated PFNS covariances are provided for five $E_\mathrm{inc}$ bins.
Their mean energy uncertainties, $\delta \langle E \rangle$, are lowest in Table~\ref{tab:U235PFNSmeanenergyunc} at thermal and from 0.5--5 MeV, where most experimental data are available.
In the $E_\mathrm{inc}$ bin spanning 0.5--5 MeV, the low uncertainties are driven by Chi-Nu (K.J. Kelly) and Lestone data at 1.5~MeV~\cite{Kelly2022,Lestone:2014}.
In fact, the too-low uncertainties resulting from an evaluation with the Los Alamos model were increased to match Lestone uncertainties from $E_\mathrm{out}=$ 5--8 MeV, which have the lowest $^{235}$U PFNS uncertainties (Fig.~\ref{fig:U235PFNSRelUnc}).

\begin{table}
\caption{\label{tab:U235PFNSmeanenergyunc} The mean energy uncertainties, $\delta \langle E \rangle$, for the $^{235}$U PFNS are listed per incident-neutron energy, $E_\mathrm{inc}$. }
\centering
\begin{tabular}{lc}
\toprule \toprule
$E_\mathrm{inc}$ (MeV) & $\delta \langle E \rangle$ (keV) \\
\midrule
Thermal & 10 \\
0.5--5 & 24 \\
5--7 & 57 \\
7--12 & 39 \\
12--30 & 43 \\
\bottomrule \bottomrule
\end{tabular}
\end{table}

\begin{figure}[htb!]
\centering
\includegraphics[width=0.45\textwidth]{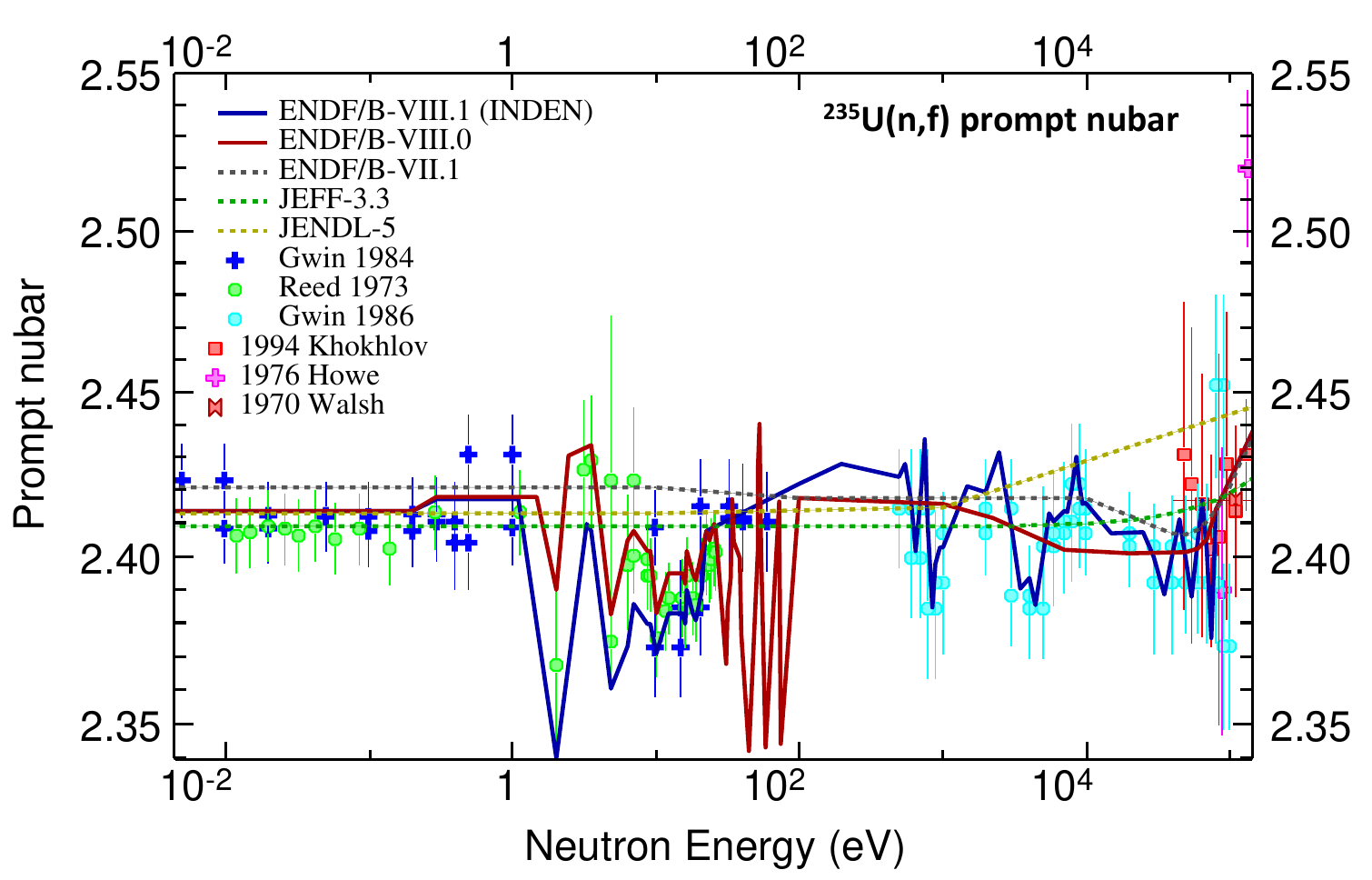}
\caption{Reed \cite{Reed73a,Reed73b} and Gwin \etal~\cite{Gwin:1984-nubar} experimental data versus evaluated $^{235}$U $\overline\nu$ below 100~keV.}
\label{fig:U235nu-RR}
\end{figure}

\paragraph{(n,f) Fission $\overline\nu$\newline}
The thermal $\overline\nu_{tot}$ in the new evaluation was slightly reduced to 2.42685, which is 0.13\% lower than the ENDF/B-VIII.0 value. The newly adopted value is in agreement with the thermal neutron constant value of 2.425(0.011) within uncertainties \cite{carlson2018}. Following the ENDF/B-VIII.0 evaluation, the Reed experimental data \cite{Reed73a,Reed73b} renormalized to the above-discussed thermal value is used to define the $\overline\nu$ energy dependence below 20~eV. ENDF/B-VIII.0 nubar fluctuations from about 20~eV up to 500~eV were replaced by linear interpolation as shown in Fig.~\ref{fig:U235nu-RR}. The tweak used in ENDF/B-VIII.0 evaluation for the negative resonance at 2~eV was eliminated.

\begin{figure}[!thb]
\centering
\includegraphics[width=\columnwidth]{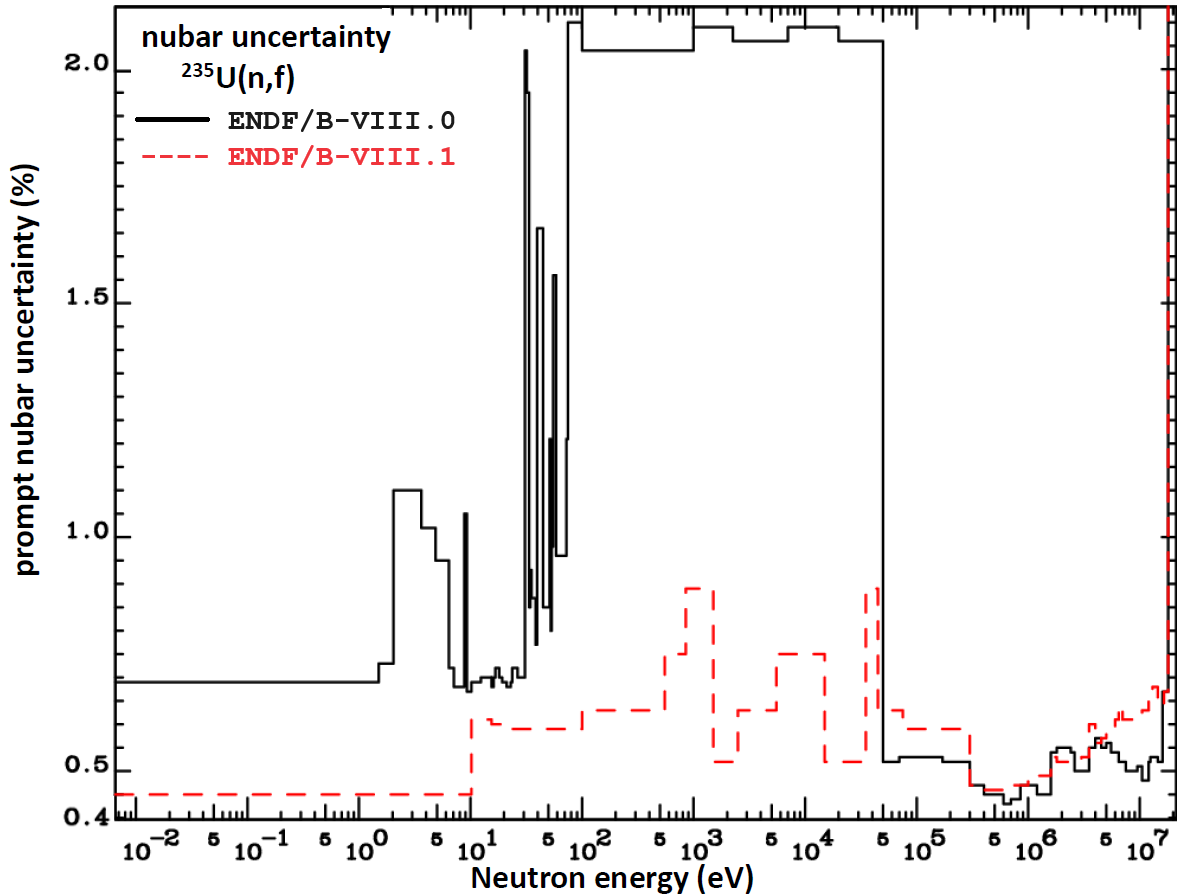}
\caption{Evaluated $^{235}$U $\overline\nu$ uncertainty.
}
\label{fig:U235nurelunc}
\end{figure}

Gwin nubar data \cite{Gwin:1984-nubar} were directly used from 500~eV up to 80~keV, leading to a significant reduction of the nubar uncertainty to values from 0.6\% up to 0.8\% in that region in the new evaluation (from about 2\% in ENDF/B-VIII.0 to roughly the uncertainty of Gwin's experiment) as shown in Fig.~\ref{fig:U235nurelunc}. Overall, the evaluated $^{235}$U $\overline\nu$ uncertainty was reduced compared to \prENDF{} except for the fast neutron range above 80~keV. The ENDF/B-VIII.1 evaluated $\overline\nu$ uncertainties in Fig.~\ref{fig:U235nurelunc} are distinctly larger above 80 keV due to the increased $\overline\nu$($^{252}$Cf(sf)) standard uncertainties (0.42\%). The correlation matrix is strongly correlated due to the full correlation stemming from the dominant $^{252}$Cf(sf) standard uncertainties.

\begin{figure}[htb!]
\centering
\includegraphics[scale=0.25, clip,trim=2mm 0mm 3mm 2mm]{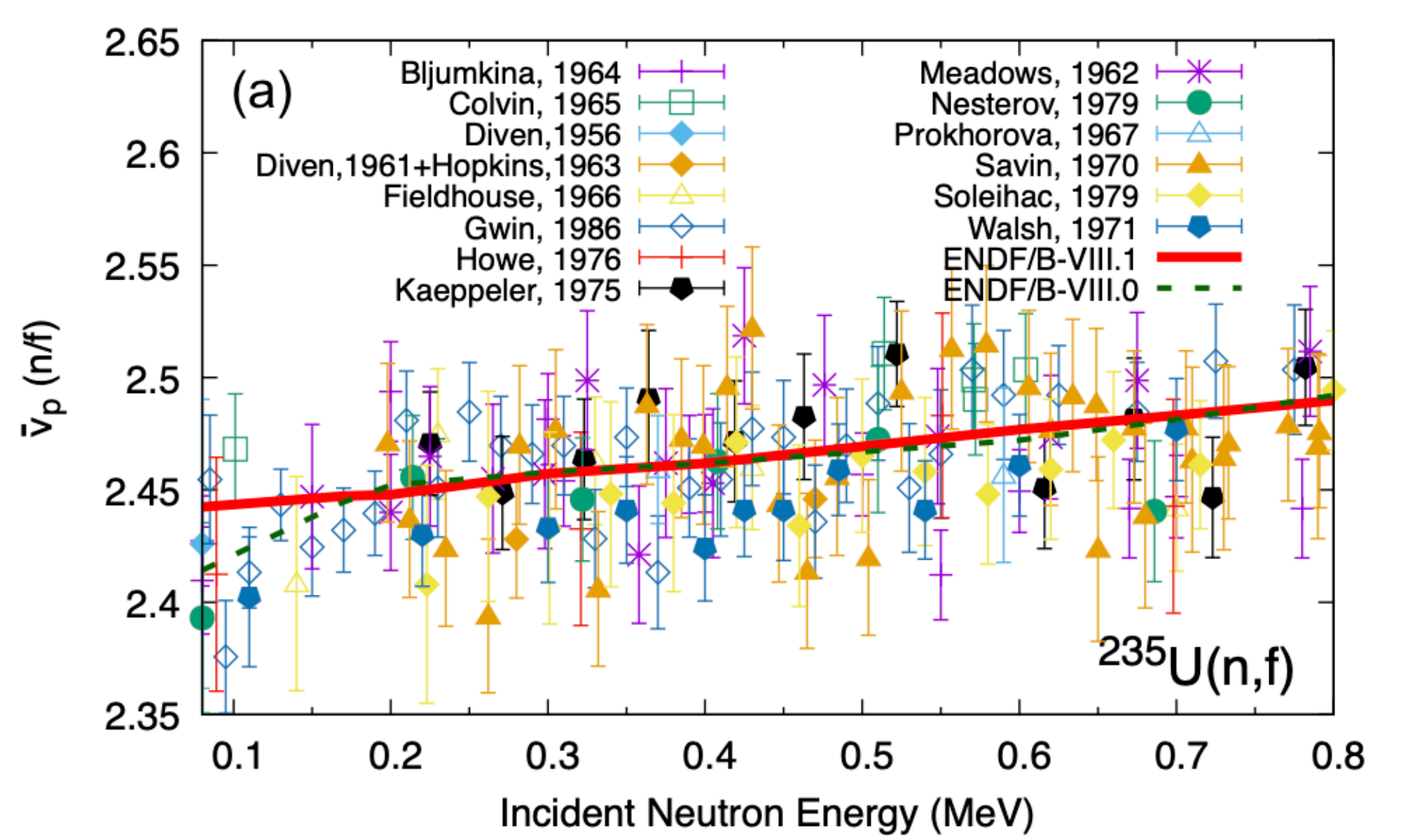}
\includegraphics[scale=0.25, clip,trim=2mm 0mm 3mm 2mm]{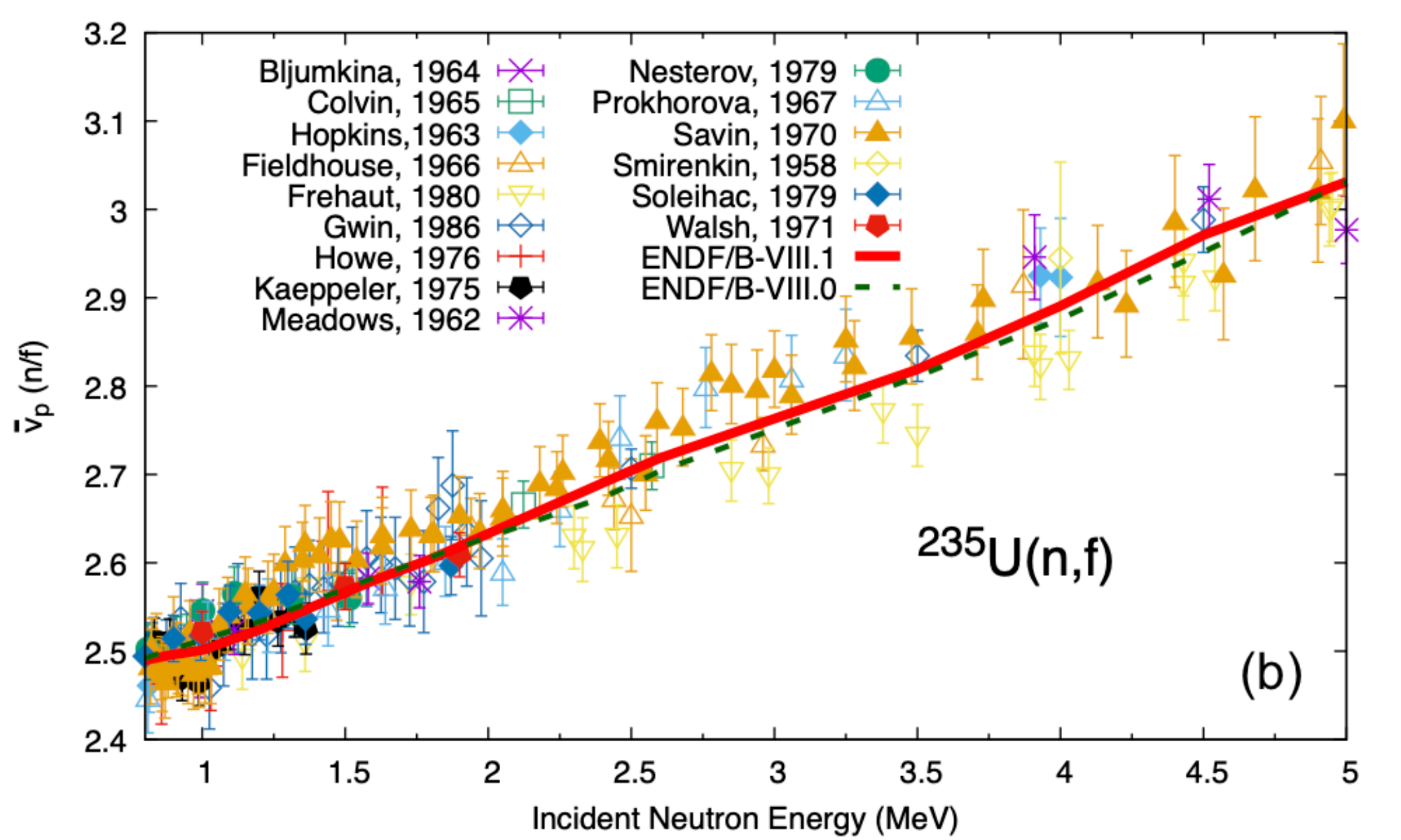}
\includegraphics[scale=0.25, clip,trim=2mm 0mm 3mm 2mm]{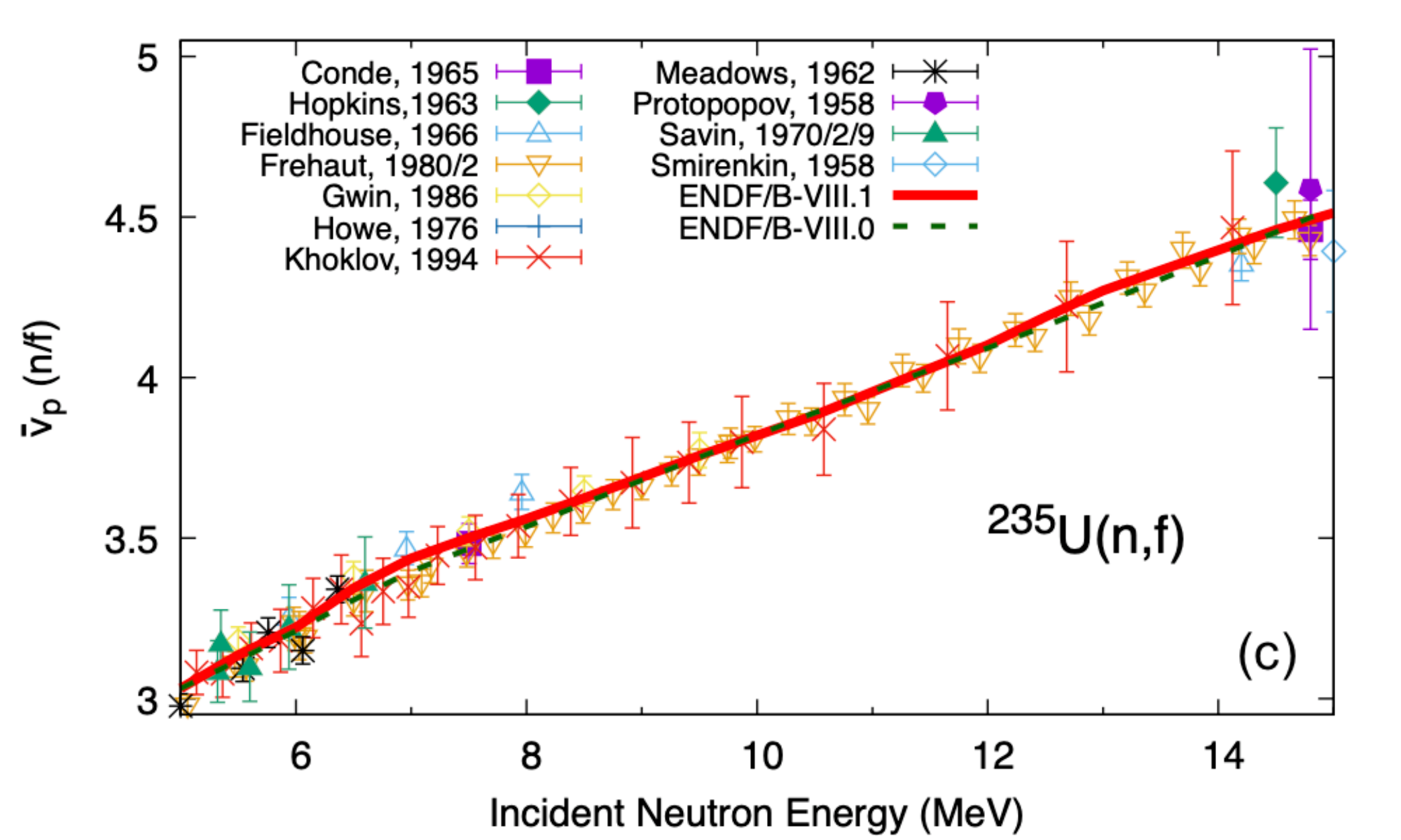}
\caption{Experimental and evaluated $^{235}$U $\overline\nu$ shown for three relevant incident neutron energy ranges: $\sim$100~keV to 800~keV in panel (a); $\sim$1~MeV to 5~MeV in panel (b); and $\sim$5~MeV to $\sim$15~MeV in panel (c). Experimental data from Refs.~\cite{Bljumkina1964,Colvin1966,Conde1965,Diven1956,Diven1961,Fieldhouse1966,Frehaut1969,Frehaut1981,Gwin1986,Hopkins1963,Howe1976,Kaeppeler1975,Meadows1962,Prokhorova1971,Prokhorova1968,Khokhlov1994,Protopopov1958-b,Savin1970,Smirenkin1958,Soleihac1970,Walsh1971}}
\label{fig:U235nu}
\end{figure}

The neutron-induced average prompt fission neutron multiplicity, $\overline\nu$, for $^{235}$U was reevaluated from scratch for ENDF/B-VIII.1~\cite{Neudecker2021nu,Lovell2022nu}.
Uncertainties for all experimental data were reevaluated in detail using the literature of data sets as well as templates of expected $\overline\nu$ measurement uncertainties~\cite{nu}.
It was unclear what data were used for the previous ENDF/B-VII.1 evaluation, but all data currently in EXFOR were re-visited \cite{EXFOR}.
Also, similar to the $^{239}$Pu $\overline\nu$ evaluation, the \CGMF\ fission-event generator~\cite{Talou2021} was used as model to evaluate the $\overline\nu$.
Including that model into the evaluation made it possible to employ evaluated model parameters to compute other fission quantities, such as yields as a function of mass, the average total kinetic energy of fission fragments, etc.
It was shown in Ref.~\cite{Lovell2022nu} that most of these predicted fission quantities agree well with experimental data except for the PFNS -- a known model deficiency that is being investigated.
This combined new information led to changes from ENDF/B-VIII.0 (Fig.~\ref{fig:U235nu})---on the order of 0.5\% up to 5 MeV.
The changed $\overline\nu$ resulted originally in an increase in Godiva $k_\mathrm{eff}$ by 114(10)~pcm.
However, $\overline\nu$ was tweaked from 2.5--5.5 MeV where the model stiffness led to a departure from the evaluation with only experimental data~\cite{Lovell2022nu}.
The tweaked $\overline\nu$ changes Godiva $k_\mathrm{eff}$ by 31(10)~pcm. The finally adopted $\overline\nu$ was further reduced at the IAEA by about 0.1\% from 2.5--3~MeV to reduce Godiva $k_\mathrm{eff}$ by 30~pcm to 1.00009(10).


\subsubsection{\nuc{238}{U}}
\label{subsec:n:238U}


A focused international effort on the evaluation of neutron induced reactions on $^{238}$U target adopted for the ENDF/B-VIII.0 library was coordinated by the IAEA within the CIELO project \cite{IAEA-CIELO}. The IAEA CIELO evaluation was described by Capote \etal~\cite{capote2018} and Chadwick \etal~\cite{CIELO-res}. $^{238}$U cross section evaluation in the fast neutron range remains unchanged for ENDF/B-VIII.1. PFNS and $\overline\nu$ were revisited as described below. Changes were also undertaken in the resonance region driven by identified deficiencies in ENDF/B-VIII.0 evaluations for depletion in power reactors~\cite{kim2021-318}.

\paragraph{Thermal and Resonance regions\newline}
One of the identified deficiencies of the $^{238}$U CIELO evaluation was the -500~pcm criticality swing at high burnup (exposures higher than 60 MWd/kgU) relative to the ENDF/B-VII.1 reference value~\cite{kim2021-318}. It was clearly stated by Kim \etal that such swing precluded the use of the ENDF/B-VIII.0 library for power reactor calculations~\cite{kim2021-318}. A solution of the \textit{burnup} problem became a high priority for next generation nuclear data libraries, and special attention was paid to this problem within the INDEN collaboration with inputs from both JEFF and ENDF research groups.
The -500~pcm reactivity swing was partially tracked down to the insufficient $^{239}$Pu production via neutron capture on the $^{238}$U. This reduced $^{239}$Pu production was due to the reduced capture from 100~eV up to 20~keV in the CIELO evaluation of n+$^{238}$U resonance parameters. Two solutions were discussed within the INDEN collaboration: changing the resonance gamma widths or changing the resonance parameters above the 100~eV.

\begin{figure}[phtb!]
\vspace{-2mm}
\centering
\includegraphics[width=0.5\textwidth]{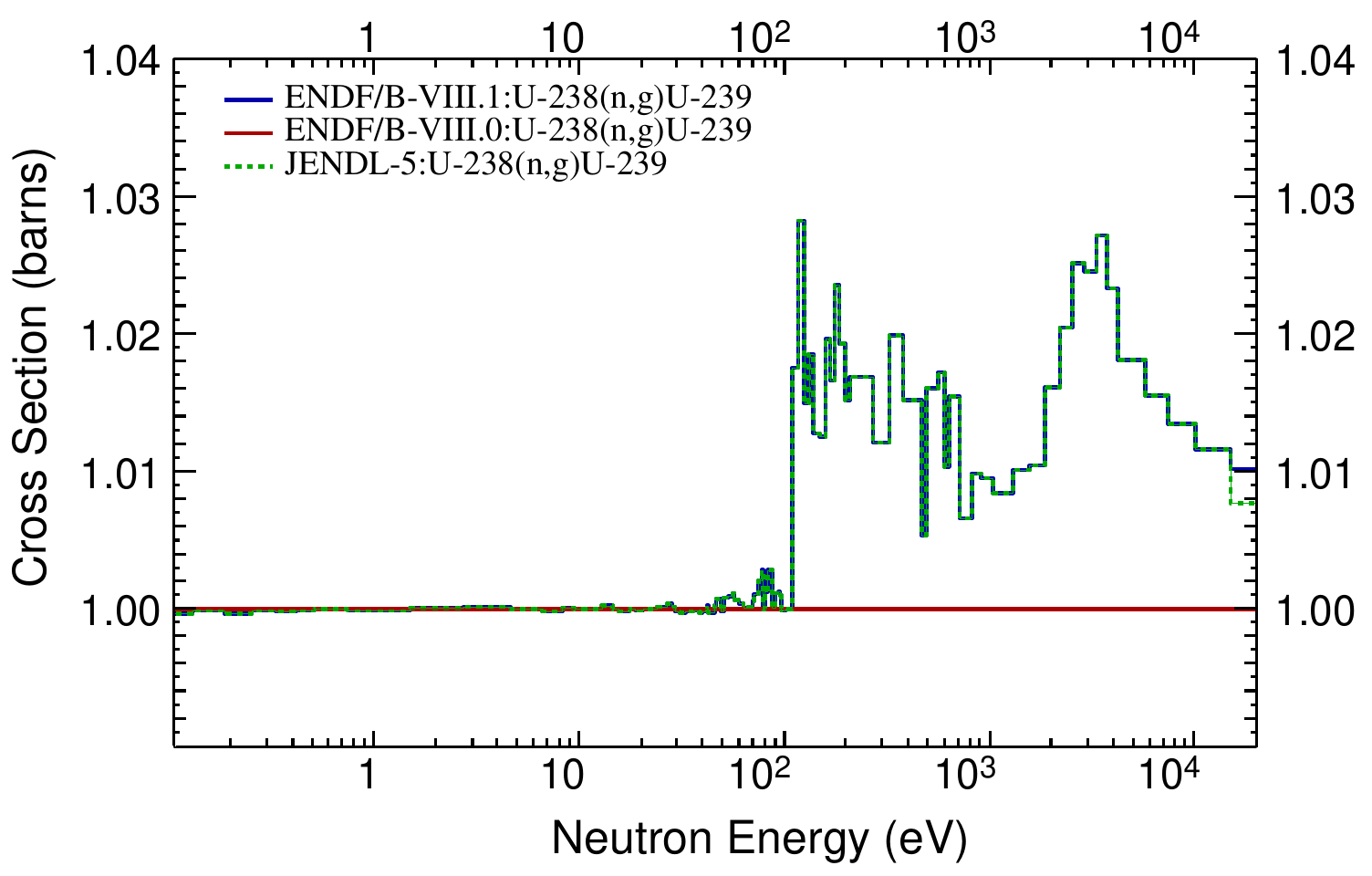}
\caption{Group averaged capture cross sections of the JENDL-5 library \cite{jendl5} and the ENDF/B-VIII.1 new library in the resolved resonance region of $^{238}$U as a ratio to the ENDF/B-VIII.0 cross sections.}
\label{fig:U238-ng}
\vspace{-2mm}
\end{figure}

For the ENDF/B-VIII.1 library, the n+$^{238}$U resonance parameters proposed by our Japanese colleagues for the JENDL-5 library \cite{jendl5} were adopted.
Reconstructed $^{238}$U capture cross sections of the JENDL-5 evaluation are shown in Fig.~\ref{fig:U238-ng} as a ratio to the ENDF/B-VIII.0 CIELO cross sections. No changes were made to resonance parameters below 100~eV, and reconstructed cross sections are practically identical to ENDF/B-VIII.0 below 100~eV preserving the criticality at zero power (BOL).
However, a 1\% to 3\% increase in the average capture cross sections is observed in the figure from 100~eV up to 20~keV, which increases significantly the $^{239}$Pu production in reactors via neutron capture on the $^{238}$U. This change of capture cross sections (resonance parameters), combined with additional changes in the $^{239,240,241}$Pu cross sections, flattens the calculated depletion (e.g., see Figs.~\ref{fig:post} and \ref{fig:casmo-depl1} in Section~\ref{sec:depletion}), largely eliminating the \textit{burnup} problem.

\begin{figure}[phtb!]
\vspace{-2mm}
\centering
\includegraphics[width=0.5\textwidth]{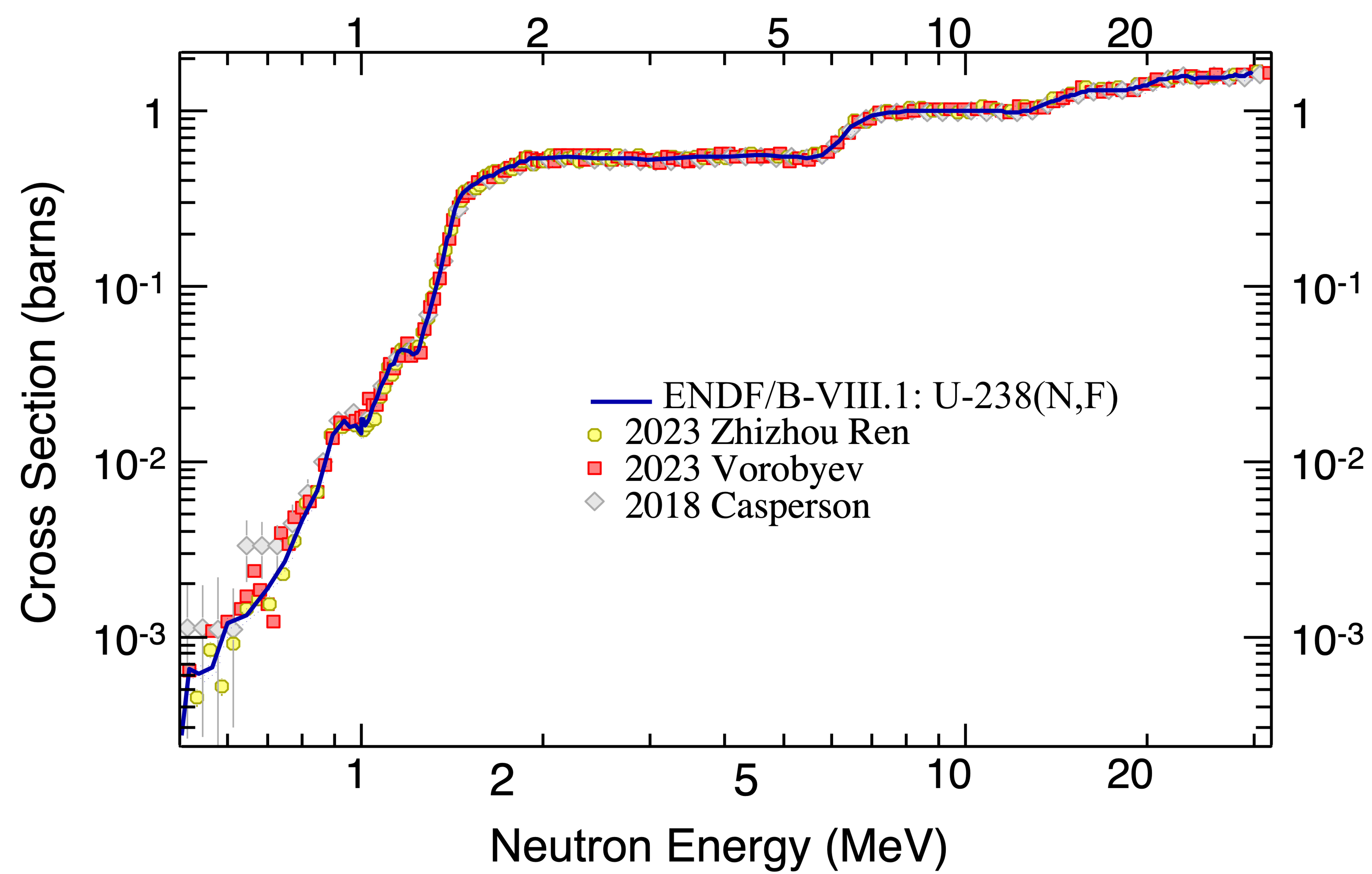}
\caption{New experimental data since 2018 \cite{Casperson:2018,Ren:2023,Vorobyev:2023} are compared to the ENDF/B-VIII.1 $^{238}$U(n,f) cross section, which is identical to the evaluated ENDF/B-VIII.0 and includes the standard energy range from 2~MeV up to 200~MeV.}
\label{fig:U238-nf}
\vspace{-2mm}
\end{figure}

\paragraph{Fission cross section\newline}
$^{238}$U neutron induced fission cross section is an standard reaction \cite{carlson2018}. Three new time-of-flight measurements of the $^{238}$U(n,f)/$^{235}$U(n,f) fission cross-section ratio in three different laboratories have been published since the ENDF/B-VIII.0 release: a measurement using a newly developed TPC detector by Casperson \etal \cite{Casperson:2018} at WNR neutron source in LANL, USA; a measurement by Ren \etal \cite{Ren:2023} using the Back\_n neutron flight path at the China Spallation Neutron Source; and a measurement by Vorobyev \etal  \cite{Vorobyev:2023} at the GNEIS neutron complex based on the 1~GeV proton synchrocyclotron in Russia. The three measured $^{238}$U(n,f) cross sections have been derived by multiplying the measured ratio by the standard $^{235}$U(n,f) cross section and are compared with the $^{238}$U(n,f) ENDF/B-VIII.0 cross section in Fig.~\ref{fig:U238-nf} from 0.5~MeV up to 30~MeV. An excellent agreement is observed inspiring confidence in the neutron standard evaluation \cite{carlson2018}. Note that the $^{238}$U(n,f) cross section is standard above 2~MeV \cite{carlson2018}.
\ENDF\ evaluated cross sections are identical to the \prENDF. Only the neutron multiplicity changed as described below.

\paragraph{(n,f) Prompt Fission Neutron Spectrum\newline}
New Chi-Nu PFNS experimental data measured at LANL in collaboration with LLNL were recently published~\cite{Kelly:2024U238PFNS}. A new LANL evaluation~\cite{U238PFNS_Neudecker} was finalized in spring 2024 and will be reviewed for inclusion into ENDF/B-IX.0.
For ENDF/B-VIII.1, the $^{238}$U PFNS evaluation was carried over from ENDF/B-VIII.0 with the exception of adopting the JENDL-5 PFNS evaluation \cite{jendl5} from 5~MeV up to 8~MeV to consistently follow the JENDL-4 (JENDL-5) evaluation above the first chance fission.
Note that the JENDL-4 (JENDL-5) PFNS was already adopted in ENDF/B-VIII.0 above 8~MeV. This change improves the agreement with measured Livermore Pulsed Sphere response.

\begin{figure}[htb!]
\vspace{-2mm}
    \centering
    \subfigure[]{\includegraphics[width=0.45\textwidth]{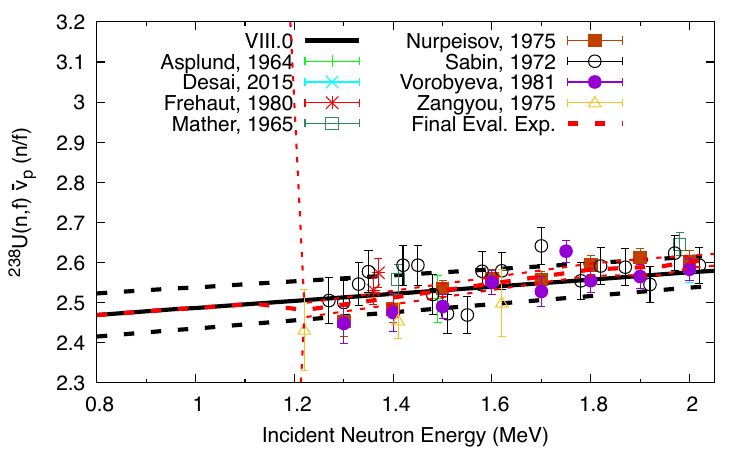}}
    \subfigure[]{\includegraphics[width=0.45\textwidth]{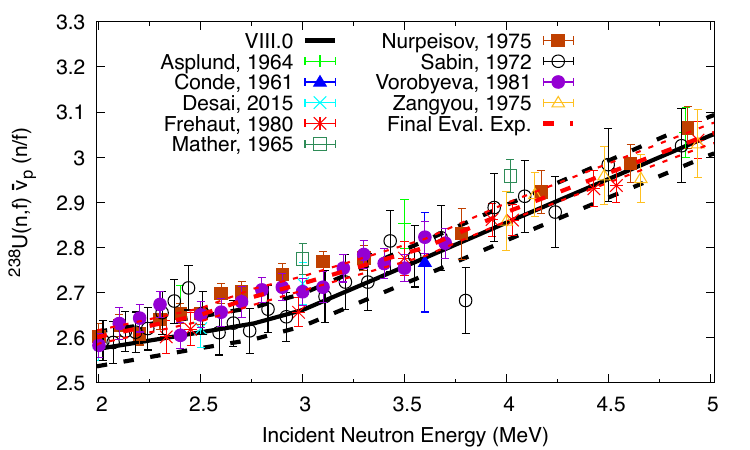}}
    \caption{Selected experimental data \cite{Asplund1964,Conde1961,Desai2015,Frehaut1969,Frehaut1973,Mather1965,Nurpeisov1975,Sabin1972,Vorobyeva1981,Zongyu1975} are compared to ENDF/B-VIII.0 and a LANL $^{238}$U $\overline\nu$ evaluation based on only experimental data shown. Panel (a) spans from 0.8~MeV to 2.5~MeV  and (b) from 2~MeV to 5~MeV. Plots were taken from Ref.~\cite{Neudecker:2022U8nubar}.}
    \label{fig:U238nu}
\end{figure}

\paragraph{(n,f) fission $\overline\nu$\newline}
Three different evaluations were considered for ENDF/B-VIII.1 $^{238}$U $\overline\nu$ in the fast range: ENDF/B-VIII.0, JENDL-5 and a new LANL evaluation~\cite{Neudecker:2022U8nubar}. Open questions remain; namely, there is a large spread in trustworthy experimental data below 5~MeV, and no new experimental data are available. 
ENDF/B-VIII.0 follows Frehaut \textit{et al.} data around 2 MeV as can be seen in Fig.~\ref{fig:U238nu}.

\begin{figure}[htb!]
\vspace{-2mm}
    \centering
    \includegraphics[width=\columnwidth]{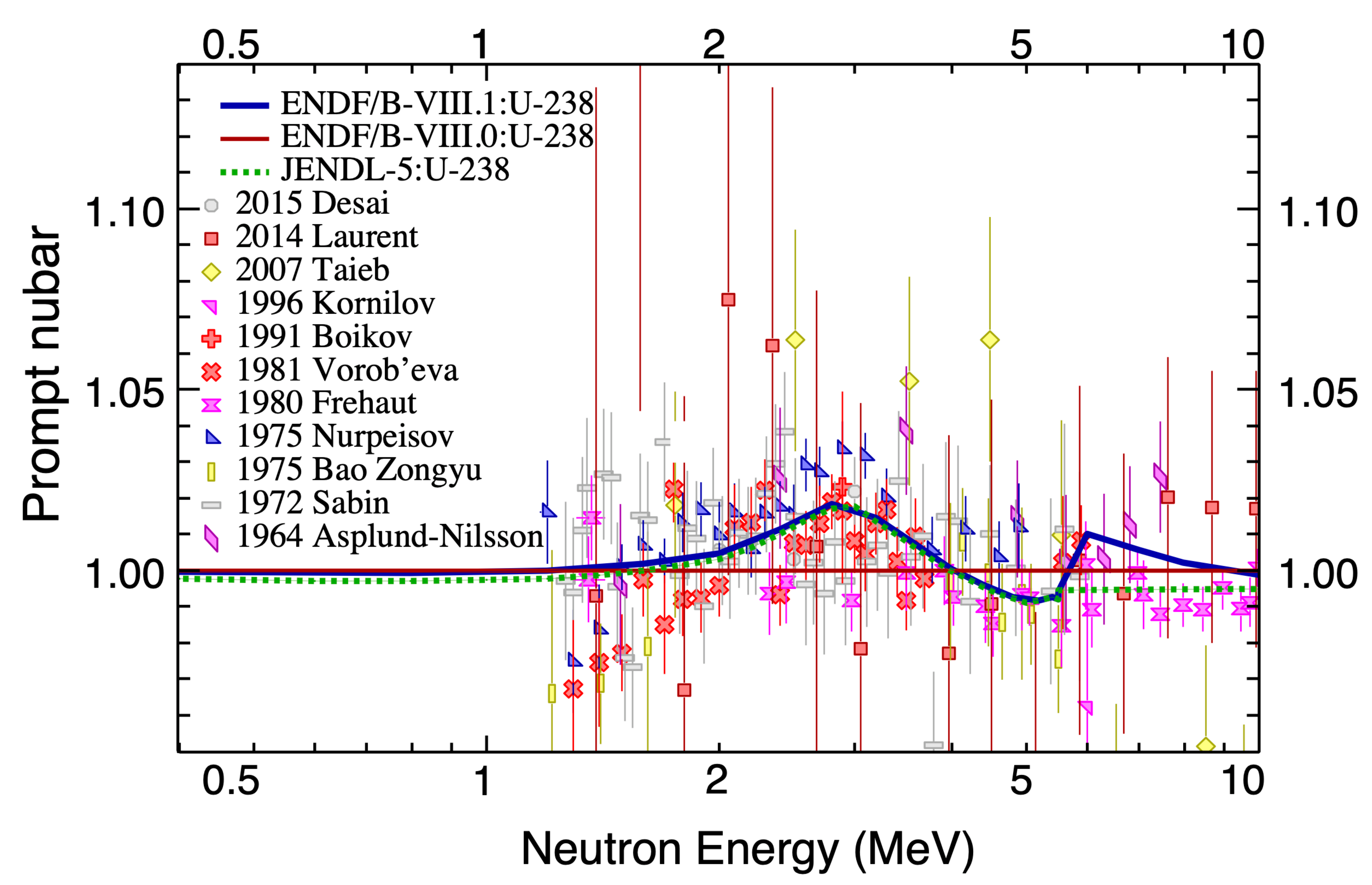}
\vspace{-4mm}
    \caption{Selected experimental data \cite{Asplund1964,Conde1961,Desai2015,Frehaut1969,Frehaut1973,Mather1965,Nurpeisov1975,Sabin1972,Vorobyeva1981,Zongyu1975,Laurent2014,Taieb2007,Kornilov1980,boikov1991} and evaluated $^{238}$U $\overline\nu$ as a ratio to the ENDF/B-VIII.0 $^{238}$U $\overline\nu$ evaluation.}
    \label{fig:U238nu-ratio}
\vspace{-2mm}
\end{figure}

The new LANL evaluation~\cite{Neudecker:2022U8nubar} based on the experimental data shown (including Frehaut data but also the 2\% higher data by Nurpeisov \textit{et al.} and Vorobyeva \textit{et al.}) would indicate a significantly (2\%) higher $\overline\nu$ value around 2~MeV as does JENDL-5 data in Fig.~\ref{fig:U238nu-ratio}.
LANL evaluators took the cautious approach of not recommending their new evaluation as long as we have no new experimental data.
Simulation studies by the IAEA indicated that a change in $^{238}$U $\overline\nu$ would be beneficial for benchmark performance, especially considering the criticality decrease associated with the new $^{234}$U evaluation. So, a pragmatic approach was adopted by the IAEA team that used trends seen across all three evaluations for different energy regions. This approach is described below.


The LANL evaluation~\cite{Neudecker2021nu,Lovell2022nu,Neudecker:2022U8nubar} was done from scratch;
uncertainties for all experimental data were re-evaluated in detail using the literature of data sets as well as templates of expected $\overline\nu$ measurement uncertainties~\cite{nu}.
All data currently in EXFOR were re-visited.
Despite discussions with experimenters, no clear conclusion could be reached on the spread in data below 5 MeV.
Similar to the $^{239}$Pu $\overline\nu$ evaluation, the CGMF fission-event generator~\cite{Talou2021} was used as a model to evaluate $\overline\nu$.
The model information combined with experimental data led to significant changes from the ENDF/B-VIII.0 evaluation as shown in Fig.~\ref{fig:U238nu-ratio} -- over 2\% at 3~MeV.
As can be seen from Fig.~\ref{fig:U238nu}, this large change is mostly driven by experimental data. 
A sudden increase of the evaluated neutron multiplicity is observed relative to the ENDF/B-VIII.0 evaluation going from -1\% at 5 MeV to +1\% at 6~MeV -- a 2\% increase. The observed increase washes out around 9~MeV. 

The $\overline\nu$ was tweaked by the IAEA team.
The changed values are well within evaluated uncertainties.
The goal of this modification was to improve the BigTen C/E (calculation/experiment ratio) benchmark performance whilst keeping the good ENDF/B-VIII.0 performance for Jemima benchmarks.
We checked that the region below 1.5~MeV of neutron incident energy has no impact on BigTen criticality as the $^{238}$U(n,f) cross section is too small.
Therefore, we adopted ENDF/B-VIII.0 $\overline\nu$ evaluation below 1.4~MeV, which is consistent with the new LANL evaluation.
Above, we started from the JENDL-5 evaluation \cite{jendl5} 
as its performance was closer to our target.
Tweaked values are within 0.1\% of the JENDL-5 evaluation from 1.4~MeV up to 5~MeV. At 5~MeV, the JENDL-5 evaluation is almost 1\% lower than the ENDF/B-VIII.0 evaluation as shown in Fig.~\ref{fig:U238nu-ratio}. Following the LANL evaluation and theoretical insights, we have increased $\overline\nu$ above 5~MeV to about 0.3\% larger than ENDF/B-VIII.0 (almost 1\% higher than the JENDL-5 at 6~MeV).
There is an increase in the LANL evaluation around the second chance fission threshold in Fig.~\ref{fig:U238nu-ratio} that, while lying within the spread of shown data, would benefit from additional (future) experimental information.

Summarizing, the tweaked $\overline\nu$ agrees with ENDF/B-VIII.0 below 1.4~MeV, and it is within 0.1\% of the JENDL-5 evaluation from 1.4~MeV up to 5~MeV, also following a trend seen in the LANL evaluation.
The ENDF/B-VIII.1 evaluation is similar in shape to the increase observed in the LANL evaluation above 5~MeV but with a lower amplitude at 6~MeV and slightly larger value above 8~MeV.

The 2\% increase of the $^{238}$U $\overline\nu$ at 3~MeV  was partially compensated in fast assemblies' criticality by the significant increase in newly evaluated $^{234}$U capture in the fast region. The $^{238}$U $\overline\nu$ increase also played a positive role in the reactivity increase of critical lattices (LCT benchmarks) by about 70~pcm compared to the ENDF/B-VIII.0 performance. Lattice criticality will be discussed in the validation section.

\subsubsection{\nuc{240,241}{Pu}}
\label{subsec:n:240-241Pu}



Deficiencies found in the resonance region evaluation of $^{240,241}$Pu isotopes led to changes described below. These changes were proposed for the JEFF-4 library and were also adopted here. From the application's point of view, the changes were important to increase criticality of Pu thermal systems at high burnup.

\paragraph{Resolved resonance range of $^{240}$Pu\newline}

\begin{figure}[!tbph]
\centering
\includegraphics[width=0.97\columnwidth]{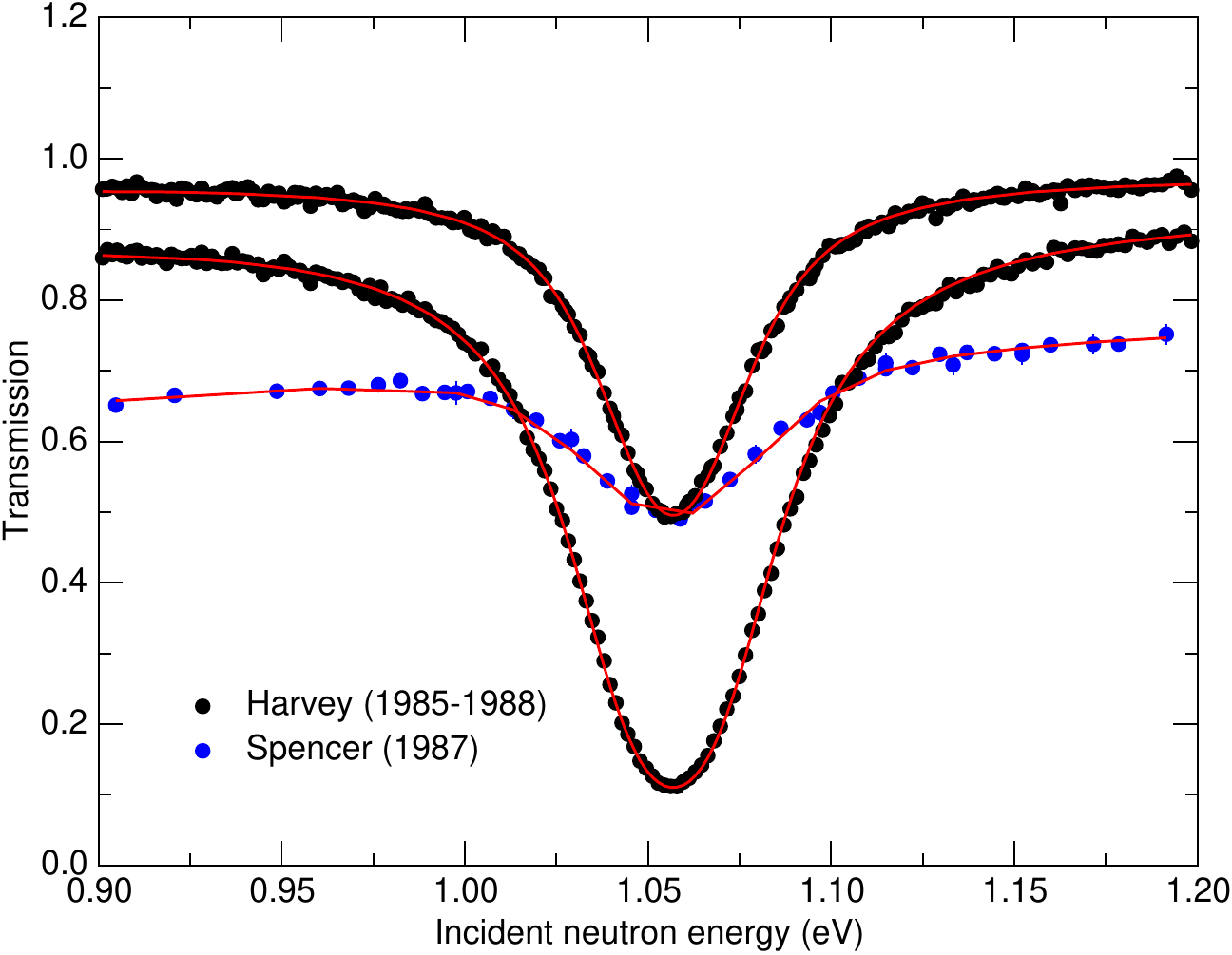}
\caption{CONRAD analysis of three $^{240}$Pu transmission experiments (1$^{\mathrm{st}}$ resonance at 1.06~eV) taken from the EXFOR database.
}
\label{fig_pu240_crd}
\end{figure}

An update to the $^{240}$Pu resonance parameters up to 5.7~keV was a necessity for reducing the calculated reactivity of some thermal critical benchmarks, such as PST018 from the ICSBEP database,  and improving the prediction of $^{244}$Cm buildup in spent nuclear fuel inventory. The resonance analysis was performed with the CONRAD code using data available in the EXFOR database \cite{EXFOR}.
For the first resonance of $^{240}$Pu, close to 1.06~eV, we have used $^{239}$Pu transmission data measured by Spencer (L=18~m, 1987) \cite{Spencer1987} and two sets measured by Harvey (L=18~m, 1985 and 1988) \cite{harvey:1988} in which $^{240}$Pu is an impurity (Fig.~\ref{fig_pu240_crd}). Two transmission data from Kolar (L= 100~m, 1968) \cite{Kolar:1968} and a fission cross section from Weston (1984) \cite{Weston:1984}  were also included in the fitting procedure. Prior resonance parameters were taken from the JEFF-3.3 library \cite{JEFF33}. The results leads to a thermal capture cross section $\sigma_\gamma$ of 285.6~barns. The capture resonance integral ($I_\gamma$=8829~barns) increases by $+4.1\%$ compared to ENDF/B-VIII.0 value.

\paragraph{Resolved resonance range of $^{241}$Pu\newline}
\begin{figure}[!tbhp]
\vspace{-3mm}
\centering
\includegraphics[width=0.97\columnwidth]{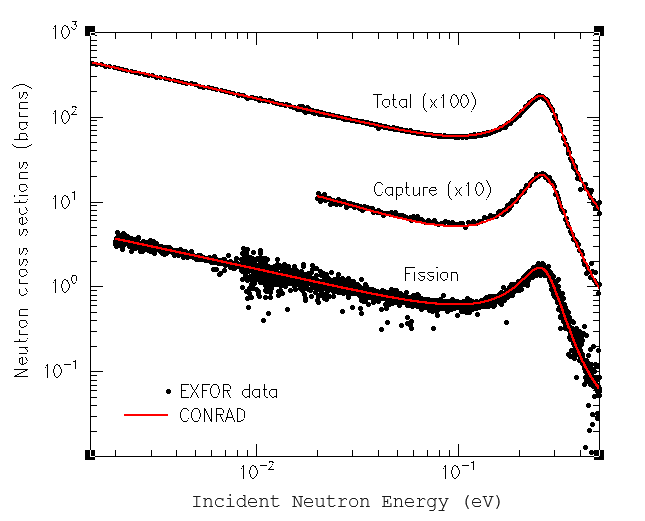}
\caption{Comparison of the CONRAD results with experimental data available in the EXFOR database for total, fission and capture neutron cross sections of $^{241}$Pu.}
\label{fig_pu241_crd}
\vspace{-3mm}
\end{figure}

The need for revisiting the $^{241}$Pu resonance range up to 300~eV emerged mainly from the underestimation of the calculated reactivity in power reactors at high burn up. At the same time, the Thermal Neutron Constants were updated according to works done in the framework of the neutron Standard group of the IAEA \cite{carlson2018,duran2024}. Most of the EXFOR data reported since the 1950s were included in the CONRAD analysis. The transmission data come from Harvey (1972), Kolar (1971), Pattenden (1963) and Simpson (1961). A few total cross sections from Smith (1970), Young (1968) and Craig (1964) were also included for the $1^{st}$ resonance. For fission, we used data sets from Wagemans (1976, 1991), Weston (1978), Blons (1971), Migneco (1970), James (1965), Watanabe (1964), Raffle (1959), Leonard (1959), Seppi (1958), Richmond (1956) and Adamchuk (1955). In contrast, a single capture data set was reported by Weston (1978).

\begin{table}[!tbph]
\vspace{-3mm}
\caption{n+\nuc{241}{Pu} 2200 m/s thermal cross sections 
for three nuclear data libraries compared with the Thermal Neutron Constants derived by the Standard group \cite{carlson2018} and Duran \etal estimates \cite{duran2024,duran2023}. Note that the ENDF/B-VIII.0 evaluation was adopted from ENDF/B-VII.1.
}\label{tab:thermal_xs_pu1}
\small
\vspace{-4mm}
\begin{center}
\begin{threeparttable}
\begin{tabular}{l | c c c c}
\toprule \toprule
    Source                                   &  $\alpha_{therm}$  (b)    &  $\sigma_{el}$ (b) &  $\sigma_{f}$ (b) &  $\sigma_{c}$ (b)\\
\midrule
ENDF/B-VIII.1                             & 0.3554                 & 12.0           & 1023.7        & 363.8        \\
relat. uncert.                            & 2.5\%                  & 10\%           & 1.5\%         & 2.0\%        \\
ENDF/B-VIII.0                             & 0.3586                 & 11.23          & 1012.3        & 363.0        \\
relat. uncert.                            & 2.5\%                  & 10\%           & 1.5\%         & 2.0\%        \\
JEFF-3.3                                  & 0.3586                 & 11.26          & 1012.3        & 363.0        \\
relat. uncert.                            & --                     & 5\%            & 2.0\%         &  --          \\
Standards \cite{carlson2018}              & 0.3539(79)             & 11.9(26)      & 1023.6(108)  & 362.3(61)   \\
Duran \cite{duran2024,duran2023}          & 0.3550(36)             & 11.5(15)      & 1018.9(25)   & 361.7(36)   \\
\bottomrule \bottomrule
\end{tabular}
\end{threeparttable}
\end{center}
\vspace{-4mm}
\end{table}
Evaluated cross section for all libraries show good agreement with standard values \cite{carlson2018} as well as with thermal cross sections recommended by Duran \etal \cite{duran2024,duran2023} as shown in Table~\ref{tab:thermal_xs_pu1}. For the resonance integrals, the one for fission ($I_f=588.8$~barns) increases by $+3.3\%$ compared to the ENDF/B-VIII.0 evaluation, while the one for capture ($I_\gamma=176$~barns) decreases by $-2.3\%$, accordingly. Such changes increase significantly the resonance $\alpha$ value, and therefore, the reactivity in reactors at high-burnup.

\subsection{Non-INDEN evaluations}
\label{subsec:n:non-INDEN}

\subsubsection{\nuc{3}{He}}
\label{subsec:n:3He}


An updated evaluation for the $n+^3$He reactions was submitted for ENDF/B-VIII.1. The primary purpose of the update was to add angular
distributions for the reactions for which only integrated cross sections had been given in ENDF/B-VIII.0.  However, in the process, small
changes were made in the integrated cross section for the $^3$He$(n,p)^3$H reaction above about 8 MeV and in the $^3$He$(n,d)^2$H reaction
above about 12 MeV neutron energy.  This occurred because we adopted the evaluated differential cross sections of Drosg and Otuka
\cite{Drosg15} for those reactions in the MeV region, including the integrated cross sections.  No changes were made in the $^3$He$(n,p)^3$H
cross section in the region where it is a standard ($E_n \le 50$ keV).

In addition, we calculated angular distributions for the $^3$He$(n,\gamma)^4$He reaction from the parameters of the $^4$He R-matrix analysis
that had been used to fit the capture cross section data.  The mass of the $^3$He target was changed to the nuclear mass, and the energy of
the outgoing gamma-ray at zero incident neutron energy was adjusted accordingly. These changes made the evaluation more useful for tracking
charged particles that are produced by neutrons incident on $^3$He.

\subsubsection{\nuc{6}{Li}}
\label{subsec:n:6Li}


The new evaluation for $n+^6$Li reactions is based on an R-matrix analysis of reactions in the $^7$Li system at excitation energies up to
$E_x=14.1$ MeV, which corresponds to $E_n=8$ MeV.  The analysis included six arrangement channels, as shown in the top part of Table
\ref{7Lisumm}.  The lower part of the table summarizes the data analyzed for each of the nine reactions included.  The fit to more than 6300
data points has a $\chi^2$ per datum of 1.39, or $\chi^2$ per degree-of-freedom is 1.43 for 170 parameters.

\begin{table}[!htp]
\vspace{-3mm}
\caption{Channel configuration (top) and data summary (bottom) for each reaction in the $^7$Li system R-matrix analysis. The projectile
   energy $E_{proj}$ range given in the second column of the lower panel corresponds to the laboratory bombarding energy of the projectile
   corresponding to either trition $E_t$ or the neutron $E_n$ energies.} \centering
\begin{center}
\begin{tabular}{ccc}
\toprule \toprule
Channel      & $l_{\rm max}$ & $a_c$ (fm) \\
\midrule
$t+^4$He        &  5  &    4.0  \\
$n+^6$Li        &  3  &    5.0  \\
$d+^5$He        &  1  &    7.5  \\
$n+^6$Li$^*$    &  1  &    5.0  \\
$p+^6$He        &  1  &    5.0  \\
$n+^6$Li$^{**}$ &  1  &    5.5  \\
\bottomrule 
\end{tabular}
\vspace{+4mm}
\begin{tabular}{cccrr}
\toprule 
    Reaction              & $E_{proj}$ (MeV) &   Observables                                                             & \# data & $\chi^2$/pt. \\
\midrule
$^4$He$(t,t)^4$He         &  $3-17$      &  $\sigma(\theta),A_y(\theta)$                                                 & 1689    & 1.03  \\
$^4$He$(t,n_0)^6$Li       &  $8.75-14.4$ &  $\sigma(\theta)$                                                             & 39      & 1.14  \\
$^4$He$(t,n_1)^6$Li$^*$   &  $8.75-14.4$ &  $\sigma(\theta)$                                                             & 3       & 0.42  \\
$^6$Li$(n,t)^4$He         &  $0-8$       &  $\sigma_{\rm int},\sigma(\theta)$                                            & 2840    & 1.44  \\
$^6$Li$(n,n_0)^6$Li       &  $0-8$       &  \makecell{$\sigma_{\rm int}$,\\ $\sigma_{\rm T},\sigma(\theta),P_y(\theta)$} & 1451    & 1.36  \\
$^6$Li$(n,d)^5$He         &  $3.4-8$     &  $\sigma_{\rm int},\sigma(\theta)$                                            & 28      & 11.9  \\
$^6$Li$(n,n_1)^6$Li$^*$   &  $3.4-8$     &  $\sigma_{\rm int},\sigma(\theta)$                                            & 175     & 2.11  \\
$^6$Li$(n,p)^6$He         &  $3.2-9$     &  $\sigma_{\rm int},\sigma(\theta)$                                            & 92      & 1.58  \\
$^6$Li$(n,n_2)^6$Li$^{**}$&  $4.2-7$     &  $\sigma_{\rm int}$                                                           & 41      & 0.30  \\
\midrule
Totals                   & \hspace{2em} 2268           &             17                              & 6358              & 1.39 \\
\bottomrule \bottomrule
\end{tabular}
\end{center}
\label{7Lisumm}
\vspace{-6mm}
\end{table}%

We show in the following only the integrated neutron cross sections that changed significantly in ENDF/B-VIII.1.  However, it should be
noted that a very good fit was maintained to the high-precision $t+\alpha$ elastic scattering data included, except at the highest energies
($\sim 17$ MeV).  Generally speaking, the fits to the angular distributions for the neutron-induced reactions were also quite good,
especially for the $^6$Li($n,t)^4$He reaction measured by Bai~\etal~\cite{Bai20} in the energy range $0<E_n\le 3$ MeV.  Some
measurements of the neutron angular distributions at higher energies were in significant disagreement, however.

\begin{figure}[htpb!]
   \centering
   \includegraphics[width=0.99\columnwidth, clip]{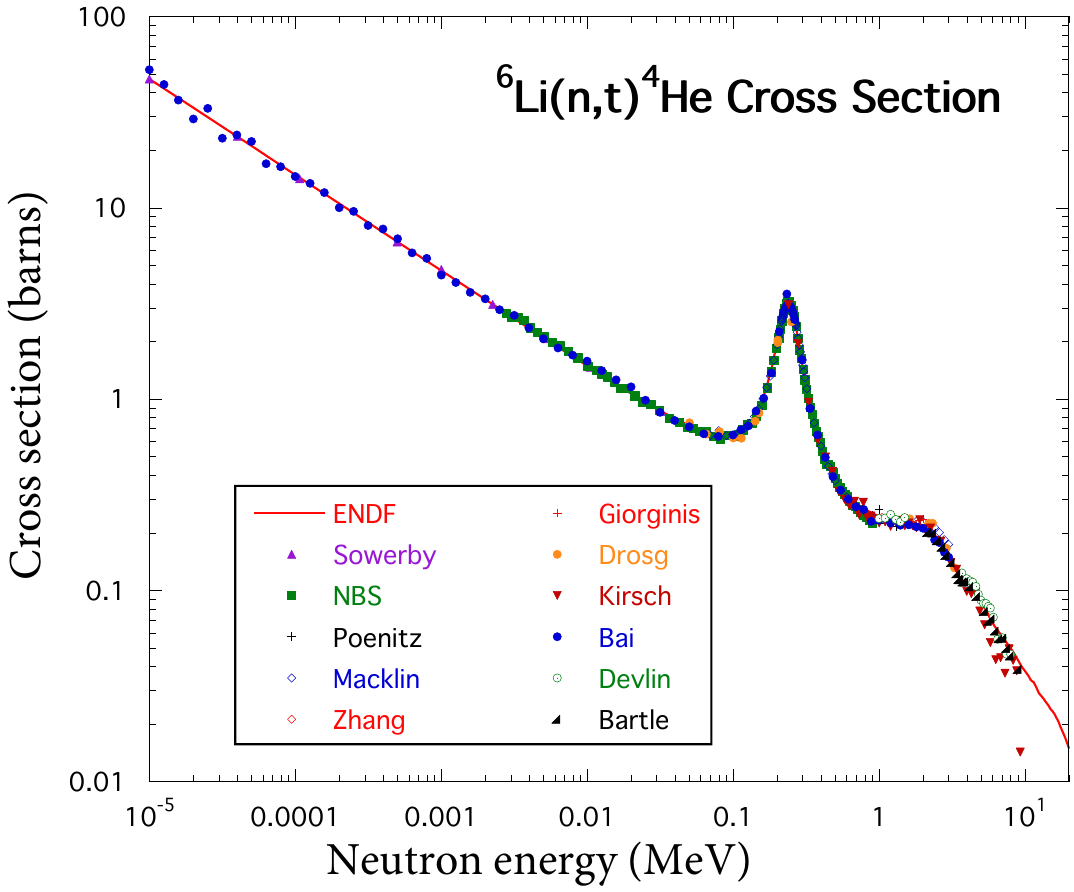}
   \caption{Measurements of the $^6$Li$(n,t)^4$He cross section compared to the ENDF/B-VIII.1 evaluation at incident  neutron energies
   between 10 eV and 20 MeV.  The experimental data are from Refs.~\cite{Sowerby70a,Lamaze78,Poenitz74,Macklin79,Zhang06a,Giorginis17,Drosg94,Kirsch17,Bai20,Devlin08,Bartle83}.}
   \label{6Lint1}
\end{figure}

\begin{figure}[htpb!]
   \centering
   \includegraphics[width=\columnwidth, clip, trim = 34mm 16mm 42mm 31mm]{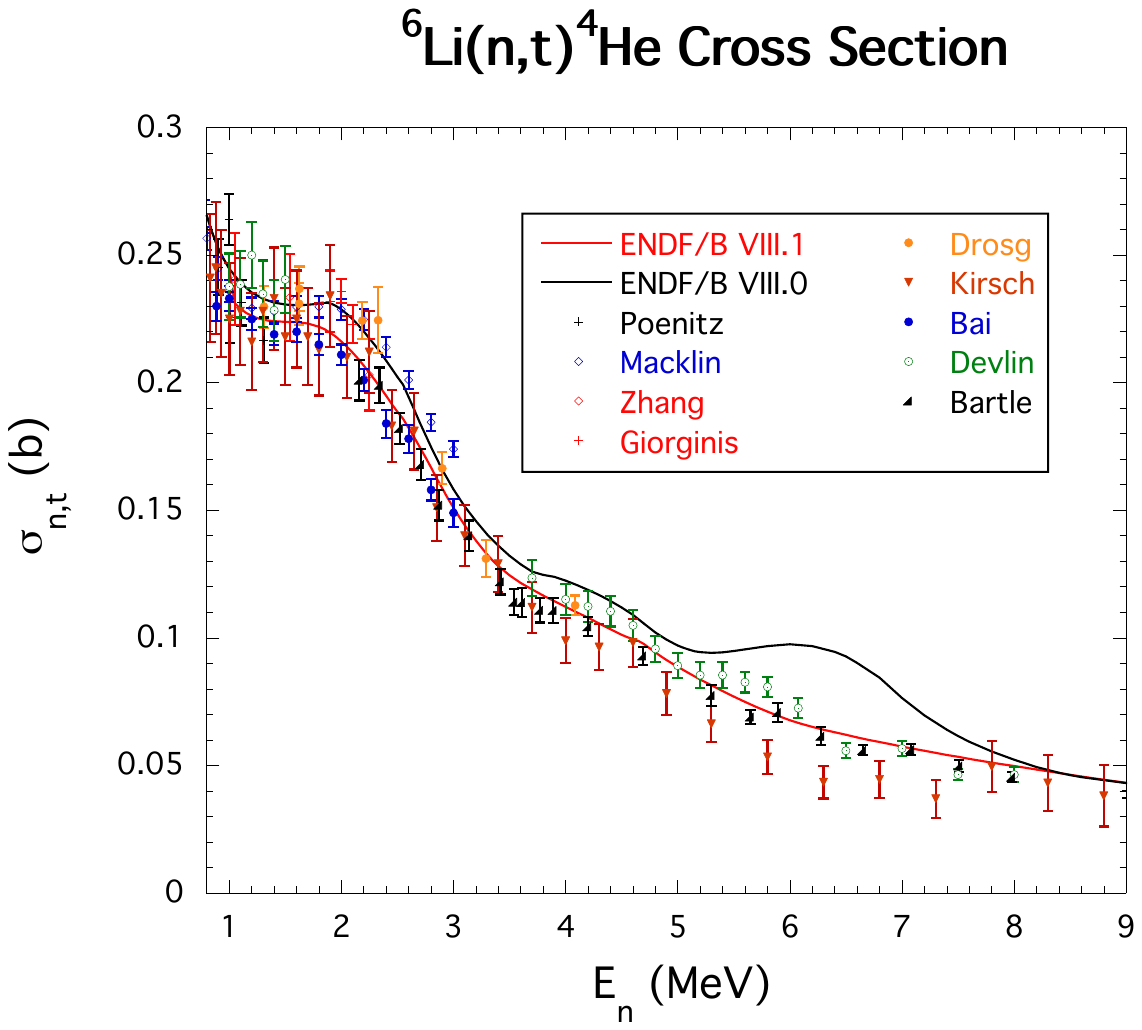}
   \caption{
   Comparison of measurements of the $^6$Li$(n,t)^4$He cross section to the ENDF/B-VIII.1 evaluation and to \prENDF\ at incident  neutron energies between 1 and 9 MeV.}
   \label{6Lint2}
\end{figure}

The $^6$Li$(n,t)^4$He cross section is shown in Fig.~\ref{6Lint1} at energies up to 20 MeV.  Below 1 MeV, it is unchanged from the standards
cross section determined for ENDF/B-VIII.0 \cite{15Car}.  It becomes substantially different above 1 MeV, especially in the 4--8 MeV region,
which is above the range of the previous R-matrix analysis.  This is illustrated in Fig.~\ref{6Lint2}, where it can be seen that a prominent
peak in the ENDF/B-VIII.0 cross section at around 6~MeV is no longer present.
This makes the new evaluation consistent with older ENDF/B evaluations and with the IRDFF-II evaluation~\cite{IRDFF} (see Ref.~\cite[Fig.75]{IRDFF}).
The level structure of the decreasing cross section in the MeV
region is rather subtle and not clearly defined because of disagreements among the different data sets.  The shoulder at around 2 MeV and
the broad rise at around 4.5~MeV come from $J^\pi=3/2^-$ levels that are more evident in other cross sections.

\begin{figure}[htpb!]
   \centering
   \includegraphics[width=\columnwidth, clip]{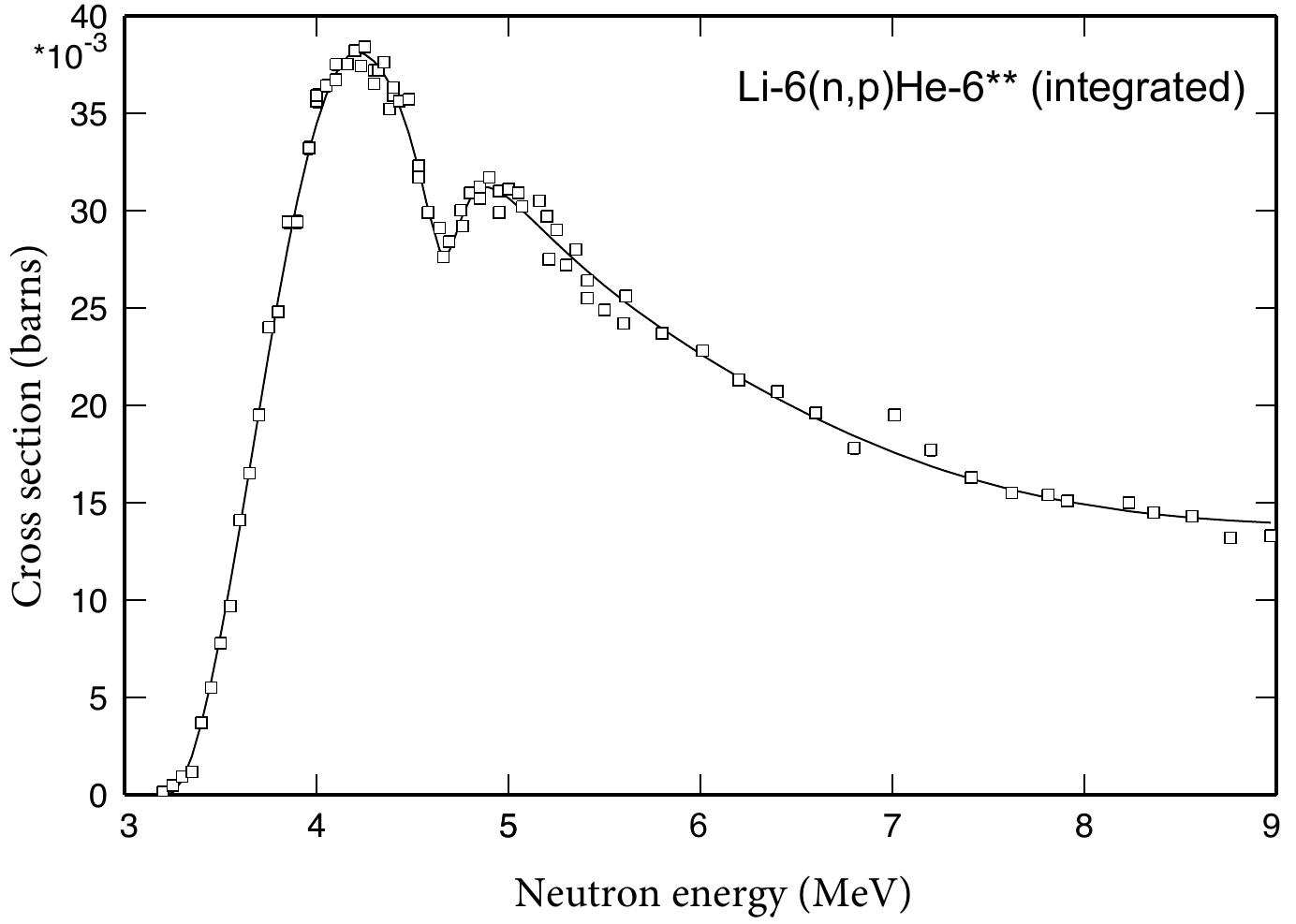}
   \caption{Measurements of the $^6$Li$(n,p)^6$He cross section compared to the ENDF/B-VIII.1 evaluation at incident  neutron energies
   between 3 and 9 MeV.  The experimental data are from Presser~\etal \cite{Presser69}.}
   \label{6Linp}
\end{figure}

\begin{figure}[hptb!]
   \centering
   \includegraphics[width=\columnwidth, clip]{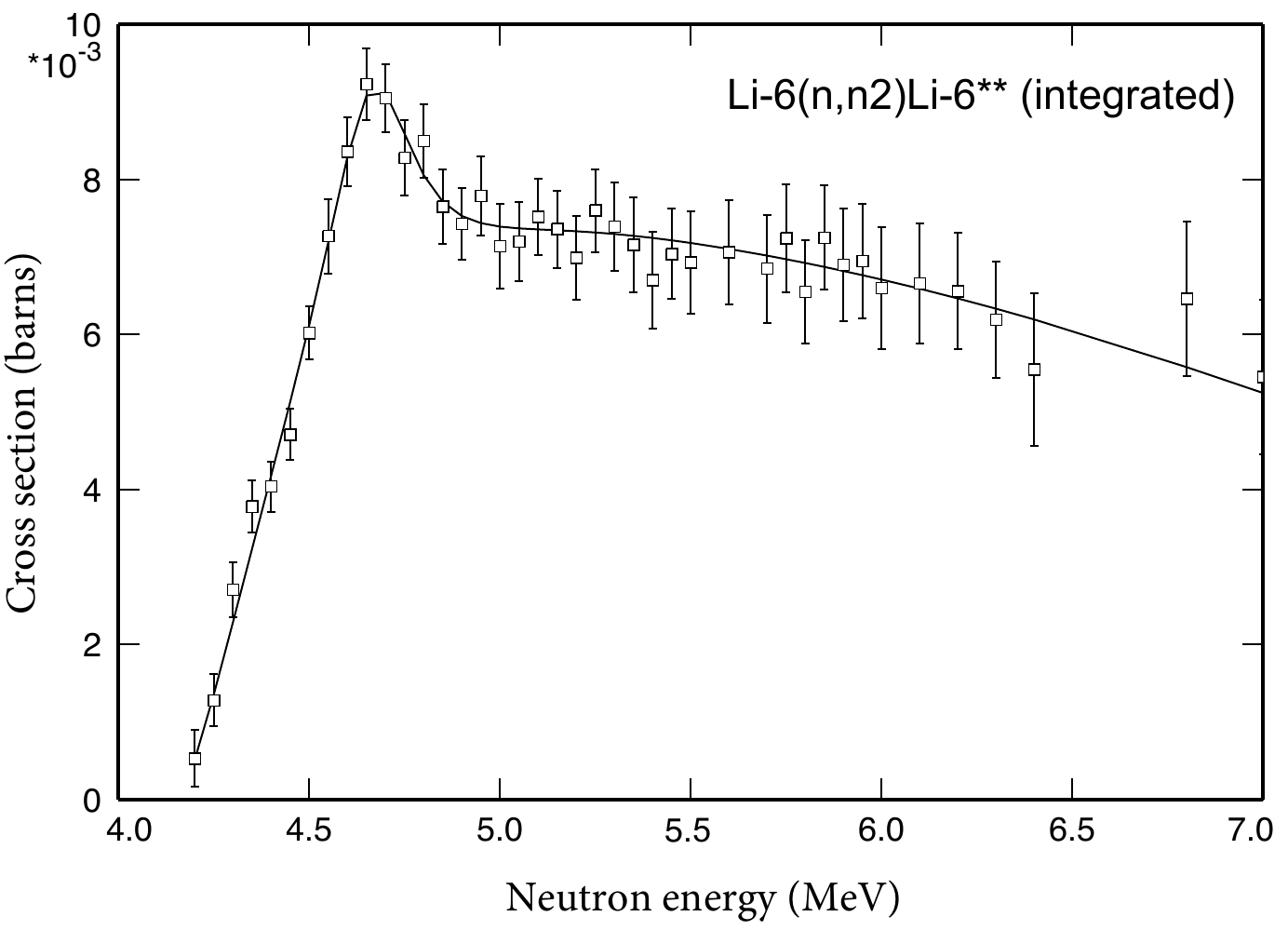}
   \caption{Measurements of the $^6$Li$(n,n_2)^6$Li$^{**}$ cross section compared to the ENDF/B-VIII.1 evaluation at incident neutron
   energies between 4 and 7 MeV.  The experimental data are from Presser~\etal \cite{Presser69}.}   \label{6Linn2}
\end{figure}

Figs. \ref{6Linp} and \ref{6Linn2} show the cross sections for the $^6$Li$(n,p)^6$He and $^6$Li$(n,n_2)^6$Li$^{**}$ reactions in which the
effects of the higher-energy $3/2^-$ levels in the 4--5 MeV region are clearly seen.  The double-peak structure of the interfering $3/2^-$
levels does not show up in the $(n,n_2)$ reaction because its threshold is too high.

\begin{figure}[!phtb]
   \centering
  \includegraphics[width=\columnwidth, clip]{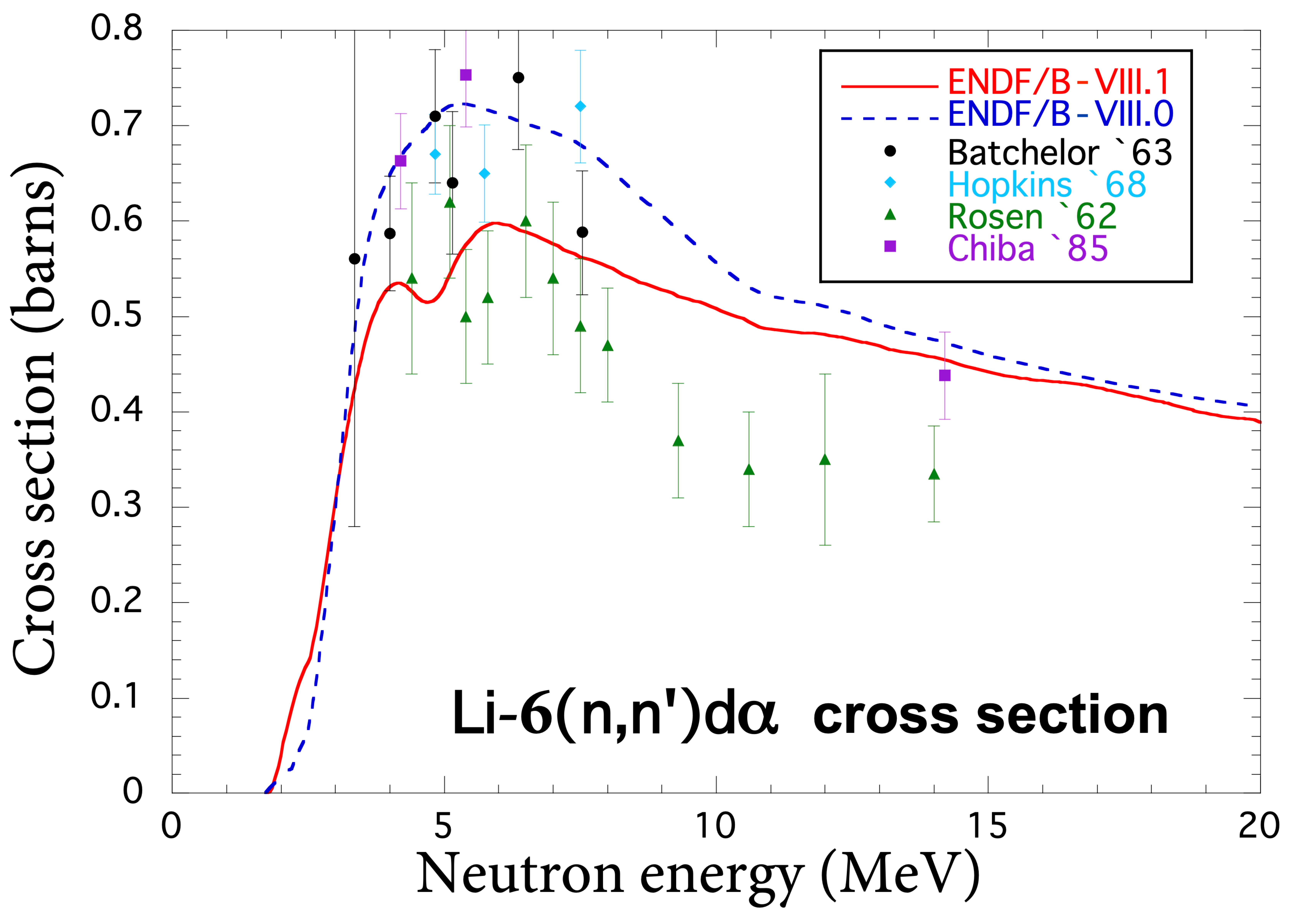}
   \caption{Measurements of the $^6$Li(n,n$^\prime$)d$\alpha$ cross section compared to the \ENDF\  and \prENDF\ evaluations at incident neutron
   energies between 0 and 20 MeV.  The experimental data are from Refs.~\cite{rosen:1962nid,batchelor:1963inl,hopkins:1968eis,chiba:1985ddn}.}
  \label{6linnprimeda}
\end{figure}

Finally, we show in Fig.~\ref{6linnprimeda} the three-body breakup cross section, $^6$Li$(n,n^\prime)d\alpha$, which in this evaluation is
represented by the sum of a continuum contribution from the $d+^5$He channel and one from the $n+^6$Li$^*$ channel in the $n+d+\alpha$ final
state.  The cross section is somewhat lower than before (blue dashed curve) and shows the double-peak structure of the interfering $3/2^-$
levels in the 4--6 MeV region.

Covariances were not updated from those in ENDF/B-VIII.0 library.

\subsubsection{\nuc{9}{Be}}
\label{subsec:n:9Be}

The current ENDF/B-VIII.1 file is identical to the previous ENDF/B-VIII.0 with two exceptions: \textit{i)} the neutron capture cross section
(MF=3, MT=102) has been updated to include more recent data in the resonance region; this has resulted in a smaller cross section in the
thermal-capture, $1/v$-region, with energies $E \lesssim 10^{-2}$ MeV, than the previous evaluation; and \textit{ii)} the total cross section
(MF=3, MT=1) has been updated in accordance with the changes to $(n,\gamma)$ capture, which are small relative to the total for most of the energy region.

The current evaluation is based on the previous $R$-matrix evaluation, to which the $\gamma + ^{10}$Be partition was added, with channels
sufficient to cover multipole transitions $E_\ell$ and $M_\ell$ with $\ell = 1,2,3$. This single added channel treats the $\gamma$
production as an inclusive (or ``lumped'') channel, as the sum over all possible $^{10}$Be excited-state transitions. Observed data has been
included in the evaluation from Refs.~\cite{conneely:1986,shibata:1992,firestone:2016,wallner:2019,marinlambarri:2020}.
The data point labeled 25~keV MACS is from Ref.~\cite{wallner:2019},
with its energy transformed from the center of mass value of 25 keV to the laboratory frame value of 27.78 keV, as shown in
Table \ref{tab:n9be-cap}.

\begin{figure}[htb!]
   \centering
   \includegraphics[width=\columnwidth, clip, trim = 15mm 22mm 22mm 22mm]{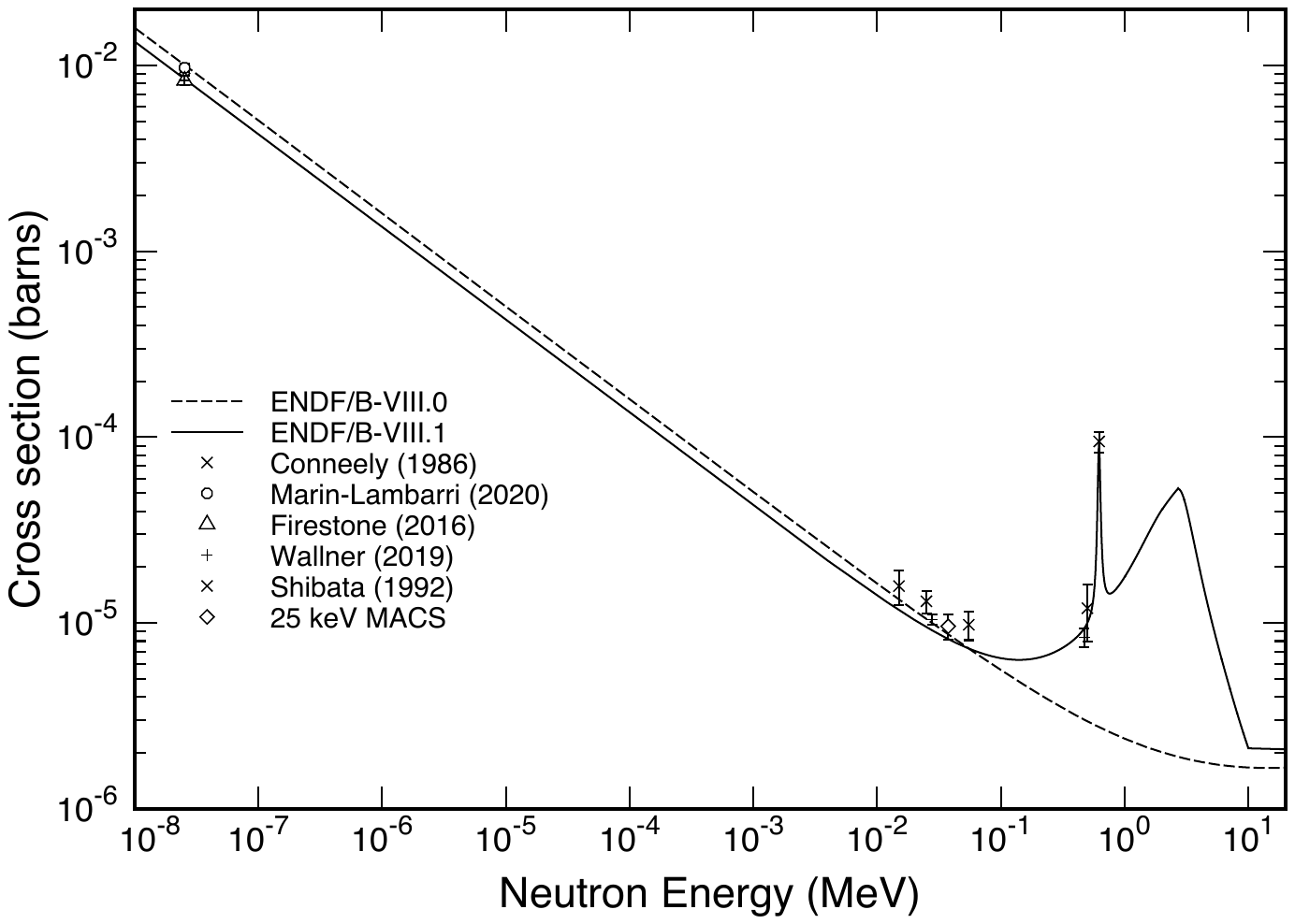}
  \vspace{-2mm}
   \caption{Comparison of the integrated capture cross section (in barns), $^9$Be$(n,\gamma)^{10}$Be of the updated evaluation for ENDF/B-VIII.1 to that of
   ENDF/B-VIII.0 as a function of incident neutron energy $E$ in MeV, compared to the available data. The observed data points in the legend
   are marked by references given in the text; the legend marked ``25 keV MACS'' is from Ref.~\cite{wallner:2019}.}
   \label{fig:n9be-mf3mt102}
\end{figure}

\begin{table}[htb!]
   \caption{\label{tab:n9be-cap}Observed $^9$Be$(n,\gamma)^{10}$Be capture cross section $\sigma_{(n,\gamma)}$ (in mb and $\mu$b) and
   uncertainty $\delta\sigma_{(n,\gamma)}$. The 30.0 keV data point from the original document of Ref.~\cite{shibata:1992} differs from that
   given in Ref.~\cite{wallner:2019}.}
   \begin{tabular*}{0.35\textwidth}{@{\extracolsep{\fill}}rlrlrlc}
   \toprule \toprule
      \multicolumn{2}{c}{$E$}  & \multicolumn{2}{c}{$\sigma_{(n,\gamma)}$}  & \multicolumn{2}{c}{$\delta\sigma_{(n,\gamma)}$}  & Ref.\\
\midrule
      0.0253  &  eV   &  8.49   &  mb      &  0.34  &  mb      &  \cite{conneely:1986}       \\
      0.0253  &  eV   &  9.70   &  mb      &  0.53  &  mb      &  \cite{marinlambarri:2020}  \\
      0.0253  &  eV   &  8.27   &  mb      &  0.13  &  mb      &  \cite{firestone:2016}      \\
      0.0253  &  eV   &  8.31   &  mb      &  0.52  &  mb      &  \cite{wallner:2019}        \\
      15.0    &  keV  &  15.81  &  $\mu$b  &  3.3   &  $\mu$b  &  \cite{shibata:1992}        \\
      30.0    &  keV  &  13.00  &  $\mu$b  &  1.8   &  $\mu$b  &  \cite{shibata:1992}        \\
      55.0    &  keV  &  9.79   &  $\mu$b  &  1.7   &  $\mu$b  &  \cite{shibata:1992}        \\
      27.8    &  keV  &  10.44  &  $\mu$b  &  0.63  &  $\mu$b  &  \cite{wallner:2019}        \\
      473     &  keV  &  8.4    &  $\mu$b  &  1.0   &  $\mu$b  &  \cite{wallner:2019}        \\
      500     &  keV  &  12.02  &  $\mu$b  &  4.1   &  $\mu$b  &  \cite{shibata:1992}        \\
      622     &  keV  &  94.80  &  $\mu$b  &  12.1  &  $\mu$b  &  \cite{shibata:1992}        \\
      \bottomrule \bottomrule
   \end{tabular*}
\end{table}

\subsubsection{\nuc{51}{V}}
\label{subsec:n:51V}

A $^{51}$V resolved resonance evaluation was carried out up to $200$ keV based on the reduced R-matrix RM formalism. The $^{51}$V resonance evaluation included in the ENDF/B-VIII.0 is listed up to $100$ keV. The new evaluation represents an extension of the RRR of $100$ keV. The evaluation uses  the $LRF=7$ resonance parameter format, referred to as the R-matrix Limited (RML) format. $^{51}$V is an isotope with ground-state spin and parity of $I^{\pi}=7/2^{-}$ which leads to channel spins and parities, $s^{\pi}=3^{-}$ and $s^{\pi}=4^{-}$, respectively. Below $200$ keV, neutron penetration factor shows that angular momentum $l=0$ and $l=1$; that is, $s-$, and $p-$ waves contribute the most. The normalized neutron penetration factor for $l=0$, $l=1$, and $l=2$; that is, $s-$, $p-$, and $d-$ waves, at $200$ keV are $1$, $0.21$, and $7.23\times10^{-3}$, respectively. Clearly, $d-$wave contributions are negligible below $200$ keV. For $p-$wave, the total angular momentum; that is, the resonance spin $J^\pi$,  shows identical values for the spin channels $s^\pi=3^-$  and $s^\pi=4^-$ which are $J^\pi=3^+$, and $J^\pi=4^+$, respectively. In the ENDF format, this characteristic can only be described by using the format option LRF=$7$ of the resolved resonance parameters. The evaluation includes resonance parameters covariance using the LCOMP=$2$ (compact format). Above $200$ keV, covariance information was taken from the TENDL library~\cite{TENDL}.

\begin{figure}[bt] 
  \centering
\includegraphics[width=1.02\columnwidth,keepaspectratio]{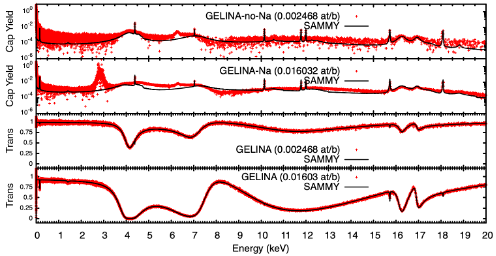}
\caption{GELINA transmission and capture yield data up to $20$ keV.}
  \label{figure1}
\end{figure}

\paragraph{Resonance evaluation\newline}
The $^{51}$V resolved resonance parameter evaluation was done in  the energy range from $10^{-5}$ eV to $200$ keV with the RM methodology included in the code \SAMMY~\cite{SAMMY}.
The experimental data used in the evaluation were those carried out at the Geel Electron
LINear Accelerator Facility (GELINA) LINAC \cite{Bensussan:1978} by Guber \etal~\cite{guber_neutron_2017}. They are capture and transmission measurements for which detailed information is available in Ref.~\cite{guber_neutron_2017}. In the energy range below $0.1$ eV, measurements covering thermal energy, $0.0253$ eV, cross-section data available in the EXFOR~\cite{Vertebnyi73,Egelstaff53,Pavlenko76,EXFOR} database were used in the evaluation. The GELINA measurements consisted of two transmissions and two capture yields for natural vanadium. The transmissions data were taken at a flight-path length of $47.64$ meters with thickness of $0.016032$ at/b and $0.002468$ at/b, respectively. Capture yield measurements were performed on a $58.586$ meter flight-path with sample thicknesses as mentioned previously. The \SAMMY\ fitting, solid dark line, to the GELINA transmission and capture cross section data, in red, is displayed in Fig. \ref{figure1} up to $20$ keV.

\begin{figure}[bt] 
  \centering \includegraphics[width=1.02\columnwidth,keepaspectratio]{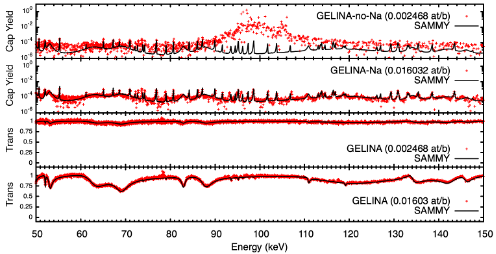}
\caption{GELINA transmission and capture yield data from $50$ keV to $150$ keV.}
  \label{figure2}
\end{figure}

Capture yield data show some structures around the energies $3$ keV and 6--7 keV that are not resonances, since they are not shown in the transmission data, but rather are artifacts due to the presence of background filters in the beam and background effects due to sample scattered neutrons which are capture events in the detectors environment. Multiple scattering effects in the capture data are reflected in the \SAMMY\ fit of the data, as can be seen around energy $4.5$ keV.
Fitting of the transmission and capture yield data in the energy range $50$ keV to $150$ keV is shown in Fig.~\ref{figure2}.
In the energy range 90~keV to 110~keV, anomalies are also observed for the $0.002468$ at/b capture yield data. This is due to the sulphur background filter.

The $^{51}$V resonance parameters obtained in this work were used to calculate some quantities of interest. Thermal capture, scattering and total  cross section  and resonance integral with uncertainties in connection to the Resonance Parameter Covariance (RPC) derived in this work are presented in Table~\ref{table4_therm}. 

\begin{table}[bt]

  \begin{center}
    \caption{Cross section at thermal ($0.0253$ eV) and resonance integral for \nuc{51}{V} in \ENDF. }
    \label{table4_therm}
    \begin{tabular}{ccc}
     \toprule \toprule
Cross Section (b) & \ENDF\ & Atlas~\cite{Mugh2006} \\ \midrule
$\sigma_t$ & $9.80(21)$ & $-$ \\
$\sigma_{\gamma}$    & $4.88(21)$ & $4.94(4)$ \\
$\sigma_s$    & $4.92(6)$ & $4.90(6)$\\
$I_\gamma$    & $2.57(23)$ & $-$ \\
\bottomrule \bottomrule
    \end{tabular}

  \end{center}

\end{table}

\paragraph{Benchmark results\newline}
Very few experiments for which $k_\mathrm{eff}$ is sensitive to the cross sections of $^{51}$V are available in the literature. However, a benchmark evaluation with five different experimental cases from the ICSBEP Handbook (\textbf{HMF-025}) \cite{ICSBEP} shows $k_\mathrm{eff}$ sensitive to the cross sections of $^{51}$V in the fast energy range of neutrons. As a result, various evaluations of $^{51}$V cross sections were tested on this benchmark evaluation, using the JEFF-3.3 evaluation as a reference. The calculations were performed with the MORET $5$ code (Monte Carlo standard deviation of 0.00010) \cite{moret} using cross sections processed with the \NJOY\ code \cite{NJOY}. The results of the testing together with the benchmark $k_\mathrm{eff}$ and uncertainties are gathered in Table~\ref{tabletwo}.

\begin{table}[bt]
\begin{center}
  \caption{$k_{\mathrm{eff}}$ values of the \textbf{HEU-MET-FAST-025} benchmark for various evaluations of $^{51}$V. }
    \label{tabletwo}
\scalebox{0.9}{
\begin{tabular}{ccc}
     \toprule \toprule
Case & Benchmark & $k_{\mathrm{eff}}$ \\ \midrule
\multirow{ 5}{*}{1} &
Benchmark $\pm$ Uncertainty & 0.99870 $\pm$ 0.00140  \\
&  JEFF-3.3 &  1.00322  \\
& JEFF-3.3  with $^{51}$V from JEFF-4T3 &  0.99918  \\
& JEFF-3.3  with $^{51}$V from ENDF/B-VIII.0 & 0.99913   \\
& JEFF-3.3  with $^{51}$V from \ENDF\ & 0.99909   \\ \midrule

\multirow{ 5}{*}{2}  &
Benchmark $\pm$ Uncertainty & 0.99900 $\pm$ 0.00160 \\
& JEFF-3.3 &  1.00739  \\
& JEFF-3.3  with  $^{51}$V from JEFF-4T3 & 1.00106   \\
& JEFF-3.3  with  $^{51}$V from ENDF/B-VIII.0 & 1.00128   \\
& JEFF-3.3  with  $^{51}$V from \ENDF\  & 1.00129   \\ \midrule
\multirow{ 5}{*}{3}  &
Benchmark $\pm$ Uncertainty & 0.99910 $\pm$ 0.00160 \\
& JEFF-3.3 &  1.01184  \\
& JEFF-3.3  with  $^{51}$V from JEFF-4T3 &  1.00237  \\
& JEFF-3.3  with  $^{51}$V from ENDF/B-VIII.0 & 1.00371   \\
& JEFF-3.3  with  $^{51}$V from \ENDF\  &  1.00386  \\ \midrule
\multirow{ 5}{*}{4} &
Benchmark $\pm$ Uncertainty & 0.99950 $\pm$ 0.00160 \\
& JEFF-3.3 &  1.01285  \\
& JEFF-3.3  with  $^{51}$V from JEFF-4T3 &  1.00250  \\
& JEFF-3.3  with  $^{51}$V from ENDF/B-VIII.0 & 1.00497   \\
& JEFF-3.3  with  $^{51}$V from \ENDF\  & 1.00558   \\ \midrule
\multirow{ 5}{*}{5}  &
Benchmark $\pm$ Uncertainty & 0.99910 $\pm$ 0.00160 \\
& JEFF-3.3 &  1.01329  \\
& JEFF-3.3  with  $^{51}$V from JEFF-4T3 &  1.00279  \\
& JEFF-3.3  with  $^{51}$V from ENDF/B-VIII.0 & 1.00549   \\
& JEFF-3.3  with  $^{51}$V from \ENDF\  & 1.00553   \\
\bottomrule \bottomrule

 \end{tabular}}

\end{center}

\end{table}

It can be concluded from the table that the JEFF-4T3, ENDF/B-VIII.0, and the new $^{51}$V evaluation contribute to improve quite significantly the $k_\mathrm{eff}$ results leading to a good agreement (within experimental uncertainties at $3$-$\sigma$ level) between the calculated $k_\mathrm{eff}$ and the benchmark $k_\mathrm{eff}$ at least for cases 1 to 3 and with the three libraries. The agreement is still good for cases 4 and 5 with the JEFF-4T3 library. However, it is a little worse with the ENDF/B-VIII.0 and the new   $^{51}$V evaluation. The reason for this difference remains unclear since the neutron spectrum and the Energy corresponding to the Average Lethargy of neutrons causing Fission (EALF) values are quite the same for the five benchmark cases. Finally, it can be pointed out that the JEFF-4T3, ENDF/B-VIII.0 and the new evaluation of $^{51}$V perform quite similarly.

\paragraph{Conclusions\newline}
A resonance region evaluation of $^{51}$V in the energy range $10^{-5}$ eV to $200$ keV was performed using the computer code \SAMMY. Measurements taken at the GELINA LINAC were included in the evaluation. Uncertainty and covariance were generated. Issues with the GELINA capture yield measurements below $20$~keV precluded a complete analysis and evaluation of the $^{51}$V cross sections leading to a higher uncertainty on the capture cross section. However, the use of a more complete representation of the ENDF resonance parameters based on the LRF=$7$ option makes the evaluation more attractive. It is recommended that new capture cross section data below $20$~keV be re-measured with very thin target samples.
Above the resonance region, the new evaluation borrow information from the TENDL library.
Benchmark results show improvement with the evaluation.



\subsubsection{\nuc{88}{Sr}}
\label{subsec:n:88Sr}



\begin{figure}[tbp]
\centering
\includegraphics[scale=.35]{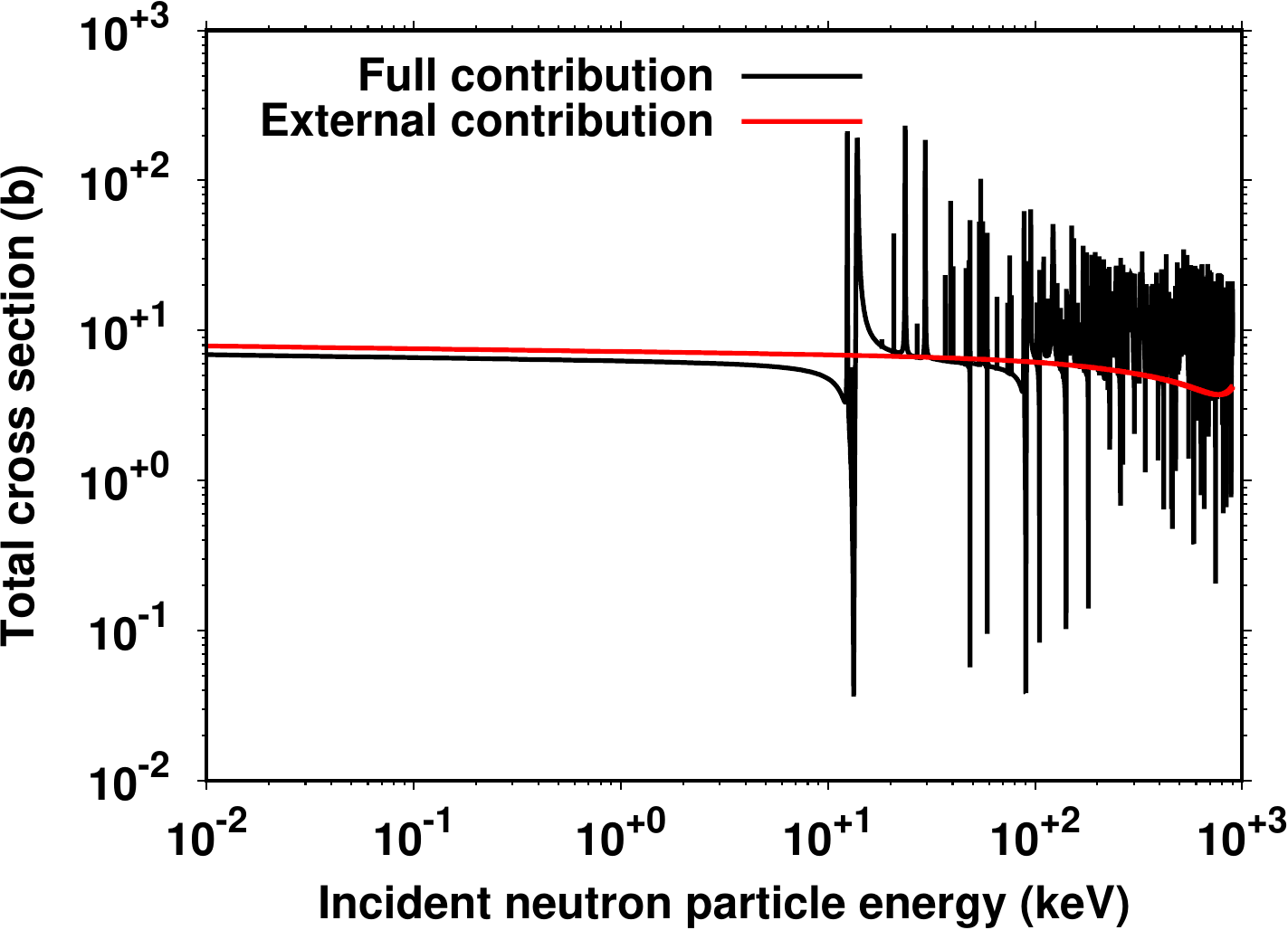}
\vspace{-.1in}
\caption[\nuc{88}{Sr}$(n,\textrm{tot})$ cross sections up to
  950~keV.]{\nuc{88}{Sr}(n,$\textrm{tot}$) cross sections up to
  950~keV, including both resonance and external parameters (in
  black). The energy dependent contribution to the total cross
  sections from the $R$ external parameterization is shown in
  red. Figure taken from Ref.~\cite{Pigni_88Sr_report}.}\label{fig:total}
\end{figure}

The resonance parameter evaluation of \nuc{88}{Sr} for incident neutron energy up to 950~keV was compiled from a publicly available set of resonance parameters~\cite{koehler:2000,koehler:2001} and used to generate an Evaluated Nuclear Data File (ENDF) submitted for inclusion to the Evaluated Nuclear Data File ENDF/B-VIII.1 release. Details of the evaluation work are discussed in the letter report~\cite{Pigni_88Sr_report} and only key features of the evaluation work are briefly described in this journal paper. The novelty of this evaluation work is an
external function representation defined by pole strengths and
$R^{\infty}$ parameters over the entire RRR
extended thrice with respect to the
current ENDF/B-VIII.0 resolved resonance evaluation. As an example, the external function is shown in red for the total cross section in Fig.~\ref{fig:total}. Moreover, the RM formalism used for the newly assembled ENDF file represents an improvement over the multi-level Breit-Wigner approximation.

\begin{figure}[tbp]
\centering
\includegraphics[scale=.35]{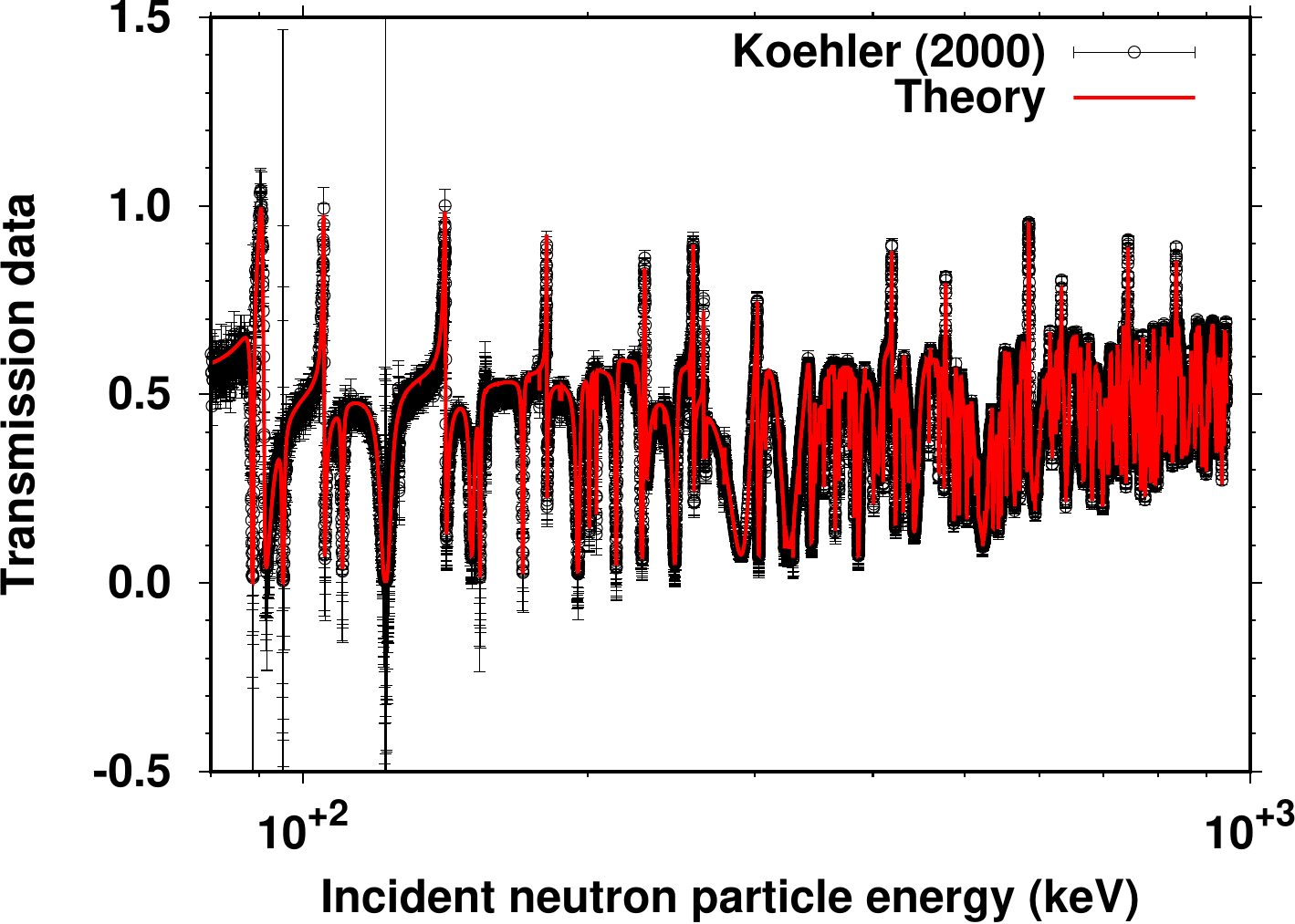}
\vspace{-.1in}
\caption[$n$+\nuc{88}{Sr} transmission data measured at 200~m flight path (FP).]{n+\nuc{88}{Sr}
  transmission data measured at 200~m flight path (FP) from 80~keV up to 1~MeV at the
  ORELA facility~(black dots) compared to theoretical data~(solid red
  line). The obtained $\chi^{2}$/dof value is 0.72 over the specified
  energy range. Figure taken from Ref.~\cite{Pigni_88Sr_report}.}\label{fig:trs200m}
\end{figure}

\begin{figure}[tbp]
\centering
\includegraphics[scale=.35]{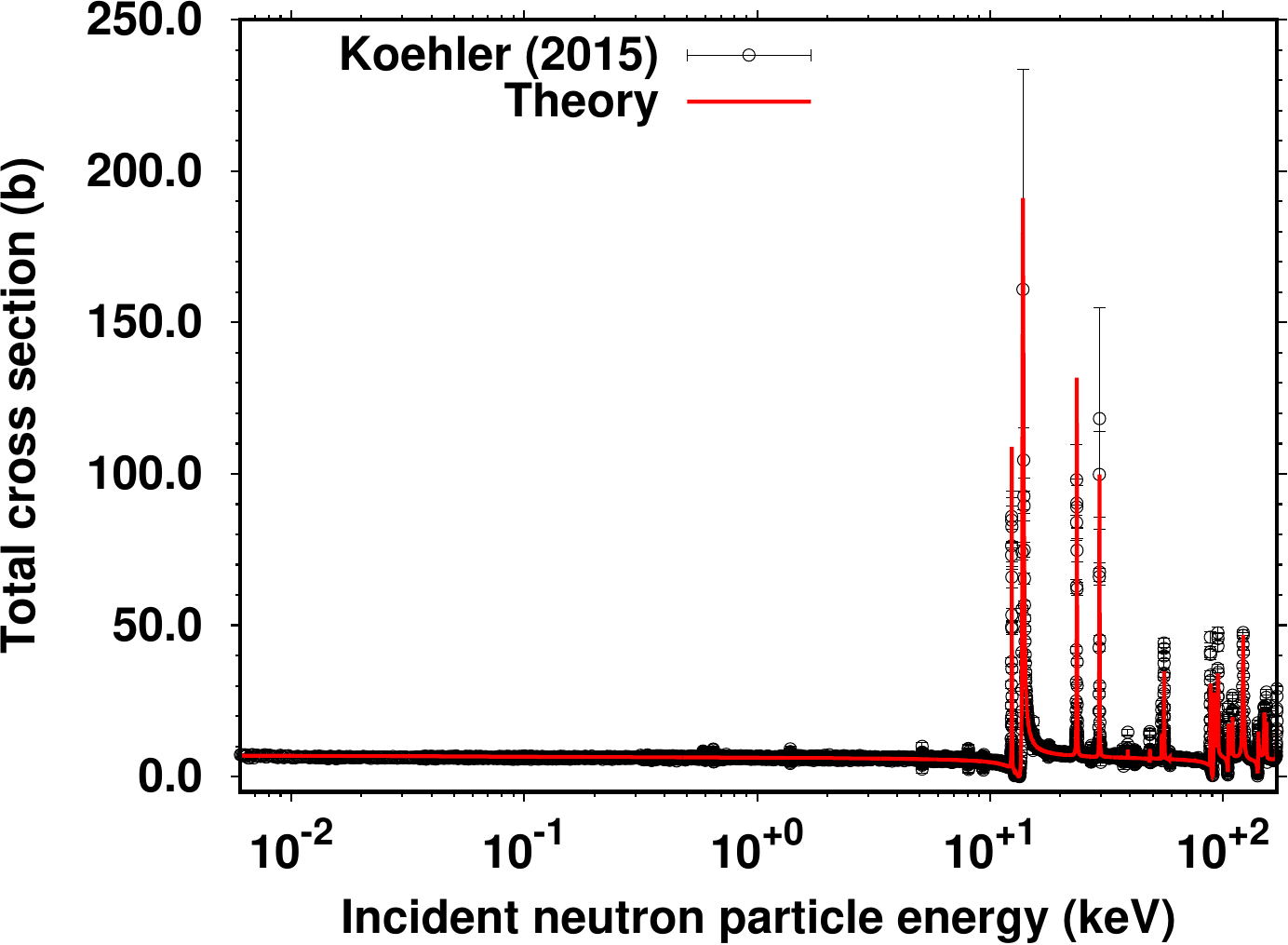}
\vspace{-.1in}
\caption[$n$+\nuc{88}{Sr} total data measured at 80~m flight path (FP).]{n+\nuc{88}{Sr}
 total cross-section data measured at 80~m flight path (FP) from 6 up to 170~keV at the
  ORELA facility~(black dots) compared to theoretical data~(solid red
  line). The obtained $\chi^{2}$/dof value is 3.34 over the specified
  energy range. Figure taken from Ref.~\cite{Pigni_88Sr_report}.}\label{fig:tot80m}
\end{figure}

\begin{figure}[tbp]
\centering
\includegraphics[scale=.35]{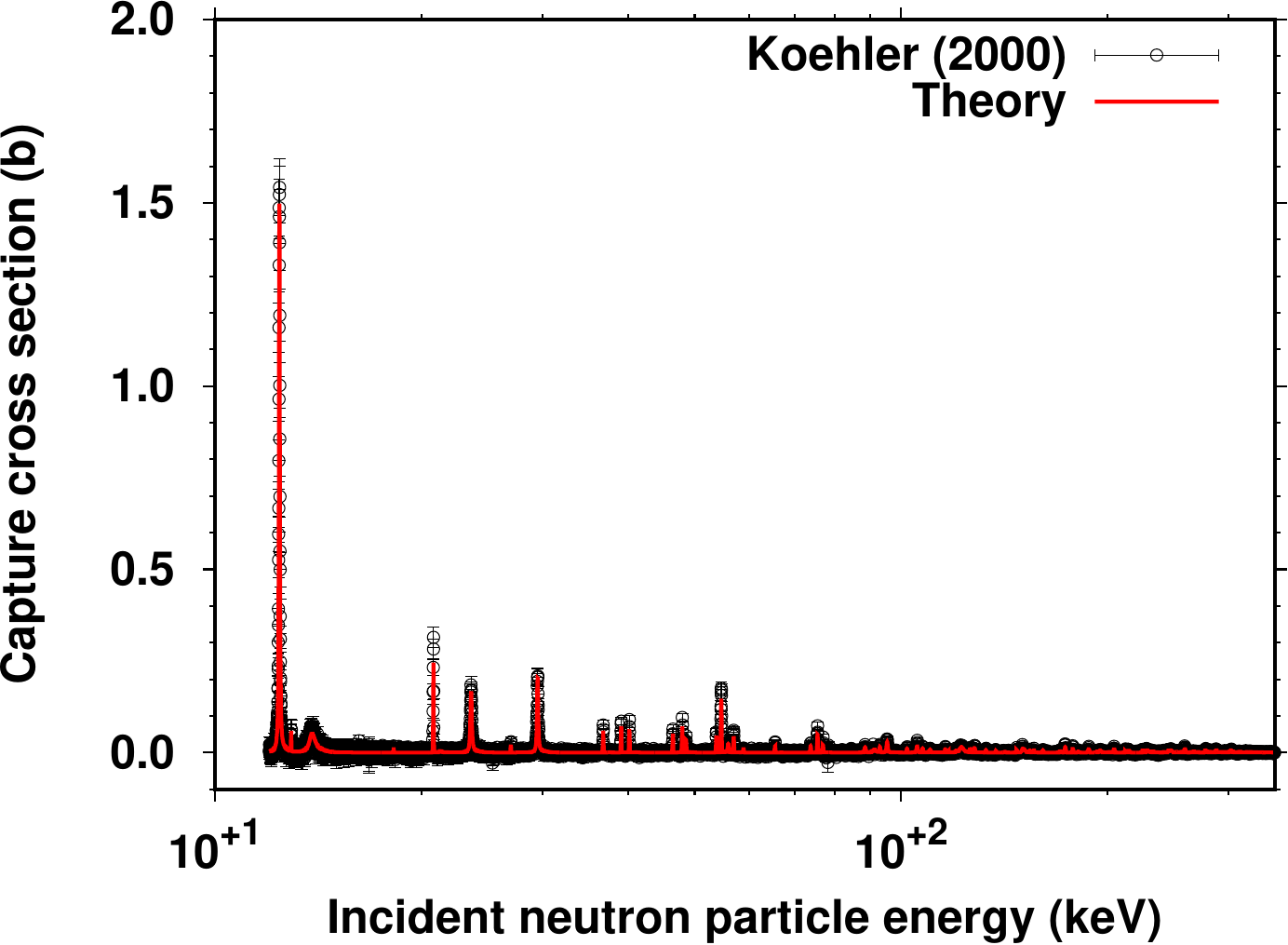}
\vspace{-.1in}
\caption[n+\nuc{88}{Sr} capture data measured at 40~m flight path (FP).]{n+\nuc{88}{Sr}
  capture data measured at 40~m flight path (FP) from 12 up to 350~keV at the
  ORELA facility~(black dots) compared to theoretical data~(solid red
  line). The obtained $\chi^{2}$/dof value is 0.57 over the specified
  energy range. Figure taken from Ref.~\cite{Pigni_88Sr_report}.}\label{fig:cap40m}
\end{figure}

Selected experimental data measured at the
ORELA facility~\cite{koehler:2000,koehler:2001} were chosen to test the quality of the resonance parameters as plotted in Figs.~\ref{fig:trs200m}--\ref{fig:cap40m} showing the overall good agreement between calculated and experimental (cross section and transmission) data. This also resulted in generating \SAMMY\ inputs, including experimental configurations for those selected measurements.


In the thermal energy region, the present evaluation work
features an increased elastic cross section of about 5\% and a
reduction of about 60\% in the capture cross section; recent
activation measurements~\cite{krane:2021} support these
changes in the capture channel. In view of this reduction, the $s$-wave direct capture contribution at the thermal energy was reduced to 0.45~mb and considered negligible at 5~keV. Table~\ref{tab:thermal_sr88} shows the thermal cross sections for both \ENDF\ and \prENDF. Despite the thermal scattering cross section of 6.4(1)~b reported by the National Institute of Standards and Technology that is based on the coherent scattering lengths measured by Koester~\cite{koester:1981}, the value of 7.7~b was adopted in this work since it is preserving the quality of the ORELA 80~m data. In this regard, new measurements may be needed to evaluate accurately the thermal scattering cross section.
\begin{table}[t]
\caption[$n$+\nuc{88}{Sr} thermal cross
  sections]{n+\nuc{88}{Sr} thermal cross sections (in barns)
  calculated at $T$=0~K.}\label{tab:thermal_sr88} 
\small
\vspace{-6pt}
\begin{center}
\begin{threeparttable}
\begin{tabular}{l c c c r }
\toprule \toprule
Isotope &  Source\tnote{*} &  Total & Scattering & Capture\tnote{**}\\
\midrule 
\multirow{2}{*}{\nuc{88}{Sr}} &ENDF/B-VIII.0 & 7.30710  & 7.29841  & 0.00869\\
&ENDF/B-VIII.1          & 7.71347  & 7.70983  & 0.00364\\
\bottomrule \bottomrule
\end{tabular}
\begin{tablenotes} \footnotesize
\item[*] Cross section uncertainties are missing in the ENDF/B-VIII.0
library as well as in the ENDF/B-VIII.1 release as the contribution of the external function uncertainty is not yet implemented in processing codes.
\item[**] Calculated from resonance parameters only. The estimated contribution of the direct capture cross section is 0.45~mb, which added to the value above gives about 41~mb.
\end{tablenotes}
\end{threeparttable}
\end{center}
\end{table}
Furthermore, the set of resonance parameters was merged in
the existing $n$+\nuc{88}{Sr} evaluation of the ENDF/B-VIII.0 library
by replacing the MF=2 section (resonance parameter) and consistently
adjusting the beginning of the MF=3 section (cross section) at
950~keV. The ENDF section related to MF=3, MT=102 was also updated to
account for $s$-wave contribution of the direct capture cross section
up to 5~keV. Moreover, due to the incompatibility with Koester's measured data, new measurements may be needed to evaluate
accurately the thermal scattering cross section. This work was one of the
milestones of the NCSP (Appendix B), and
future work will consist of measuring and evaluating minor Sr isotopes
and related uncertainty quantification. No covariance information was
added in this release as the contribution of the external function
uncertainty is not yet implemented in processing codes.

\subsubsection{\nuc{103}{Rh}}
\label{subsec:n:103Rh}


A $^{103}$Rh resolved resonance evaluation was carried out up to $8000$ eV based on the reduced R-matrix RM formalism in the energy range from $10^{-5}$ eV to $8000$ eV. The $^{103}$Rh resonance evaluation included in ENDF/B-VIII.0 is listed up to $4170$ eV. The new evaluation represents an extension of the RRR of $\sim 4000$ eV. The evaluation uses  the LRF=7 resonance parameter format, referred to as the RML format. $^{103}$Rh is an isotope with ground-state spin and parity of $I^{\pi}=1/2^{-}$ which leads to channel spins and parities, $s^{\pi}=0^{-}$ and $s^{\pi}=1^{-}$, respectively. Below $8000$ eV, neutron penetration factor shows that angular momentum $l=0$ and $l=1$, that is, $s$-, and $p$-waves contribute the most. For $p$-wave, the resonance spin $J^\pi$,  shows identical values for the spin channels $s^\pi=0^-$ and $s^\pi=1^-$ which is $J^\pi=1^+$. In the ENDF format, this characteristic can only be described by using the format option LRF=$7$ of the resolved resonance parameters. The evaluation includes resonance parameters covariance using the LCOMP=$2$ (compact format).

\begin{figure}[bt]
  \centering
\includegraphics[scale=0.172,clip,trim= 12mm 75mm 26mm 3mm]{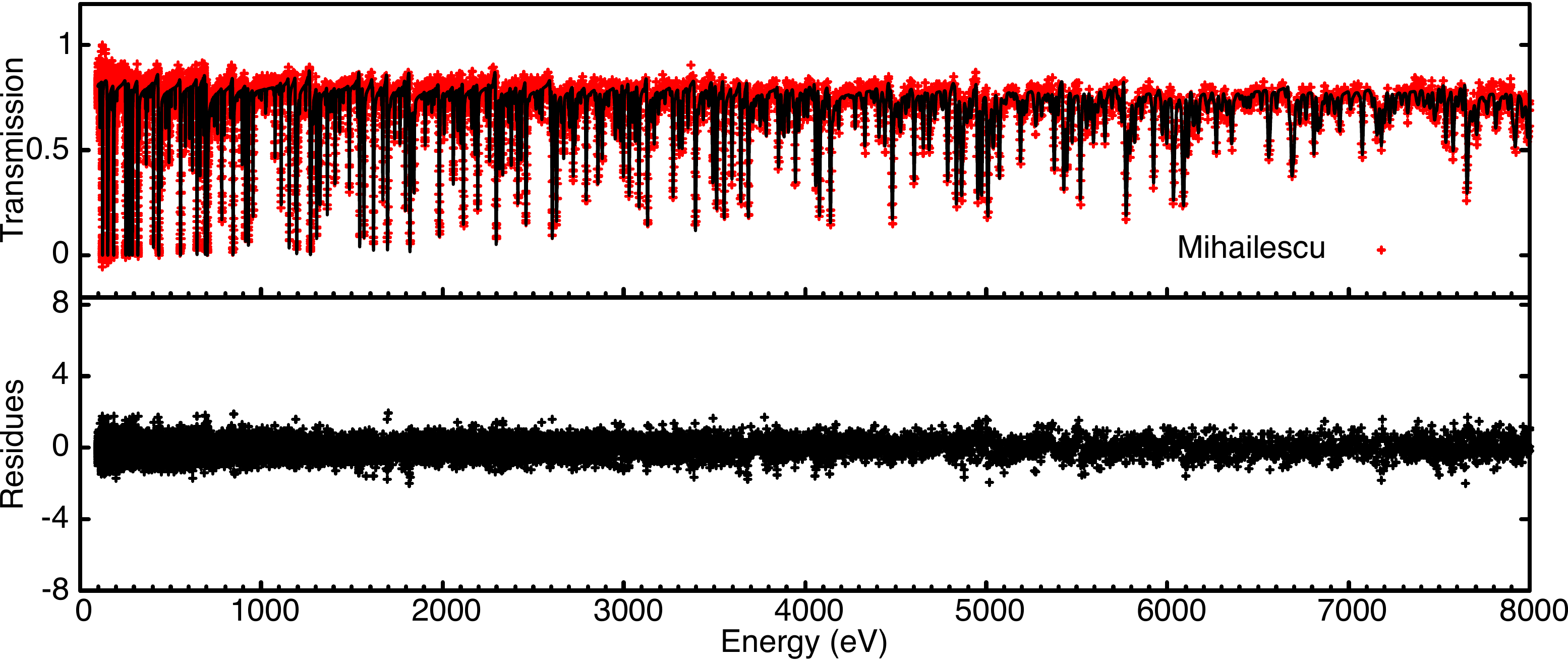}
  \caption{\SAMMY\ fitting of Mihailescu~\cite{Mih} \nuc{103}{Rh} transmission data and residues.}
  \label{Mih_trans_res}
\end{figure}

\begin{figure}[bt]
  \centering
\includegraphics[scale=0.172,clip,trim= 11mm 75mm 26mm 3mm]{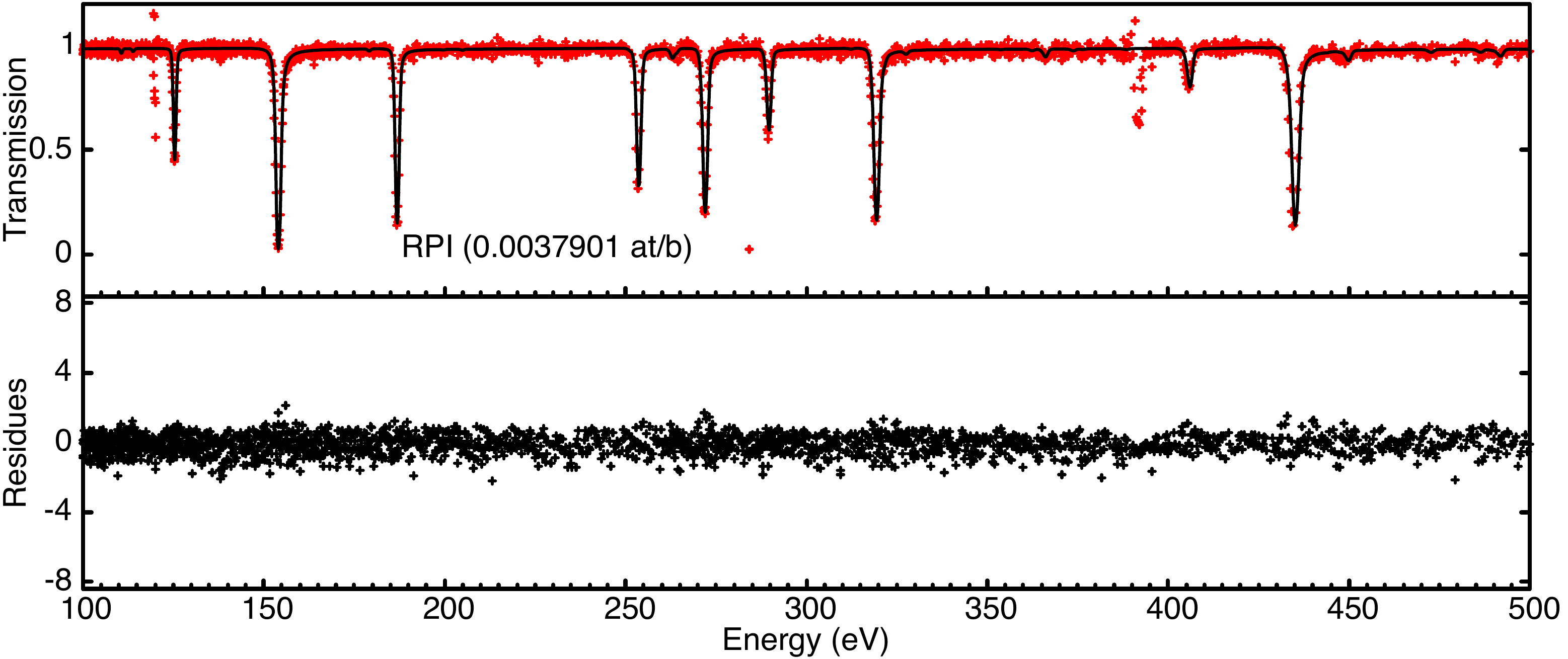}
  \caption{\SAMMY\ fitting of RPI \nuc{103}{Rh}  transmission data and residues.}
  \label{rpi_trans_residues}
\end{figure}

\begin{figure}[bt]
  \centering
\includegraphics[scale=0.170,clip,trim= 6mm 75mm 26mm 3mm]{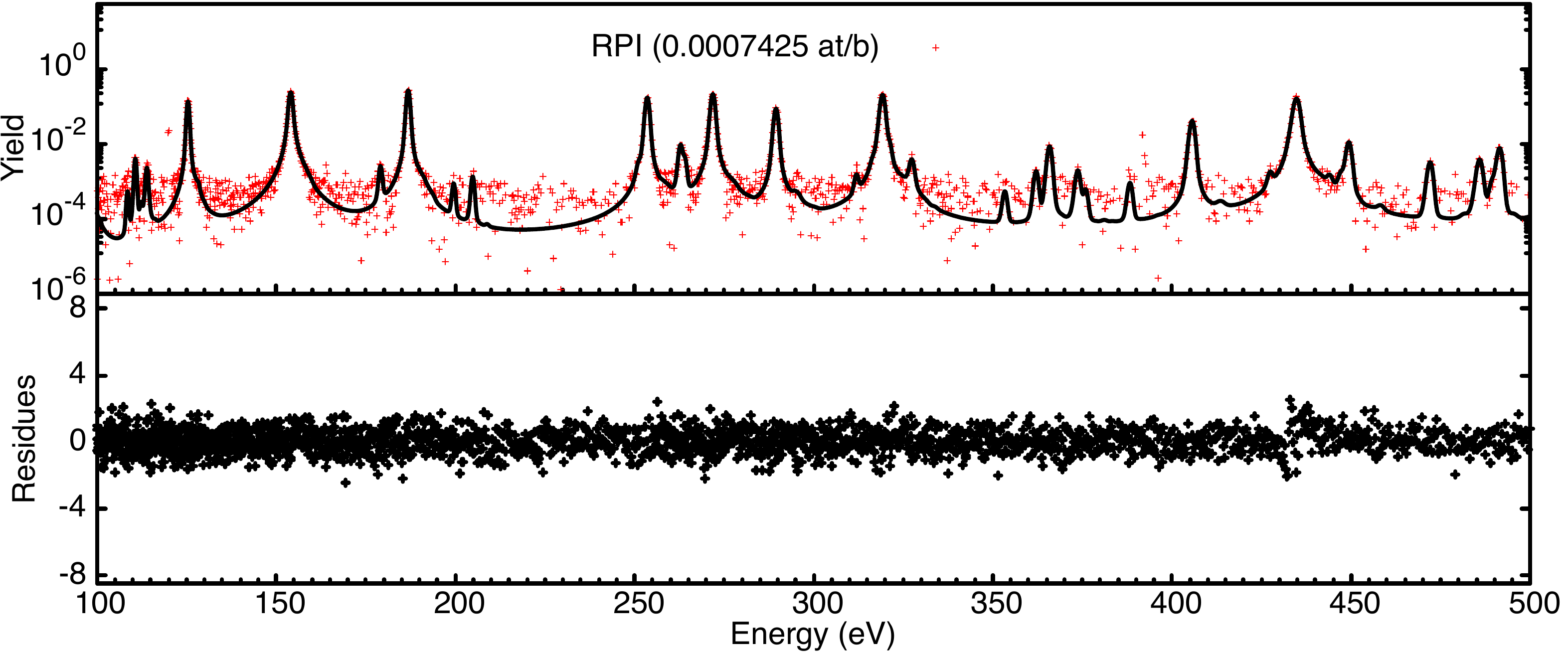}
  \caption{\SAMMY\ fitting of RPI \nuc{103}{Rh} capture yield data and residues.}
  \label{rpi_cap_residues}
\end{figure}

\paragraph{Resonance evaluation\newline}
Resonance parameters and resonance parameter covariance
for $^{103}$Rh were determined by a self-consistent analysis
of relevant experimental data; namely, high-resolution transmission
and capture yield data. The multi-level R-matrix code
\SAMMY~\cite{SAMMY} was used for the analysis. In addition to  existing experimental data
available in the EXFOR~\cite{EXFOR} system, neutron transmission and
capture yield measurements performed at the GELINA~\cite{Bensussan:1978} of the Institute for Reference Materials and Measurements (IRMM) of the
Joint Research Centre (JRC) Belgium and from the Gaerttner LINAC Center at RPI~\cite{Gaert} were used to extend
the upper energy range to 8000 eV. Three energy bound levels
(negative energies) and three energy levels above $8000$
eV were used to mock up the interference effects of external
levels in the energy range $10^{-5}$ eV to $8000$ eV. The GELINA experimental data used in the evaluation are displayed in Table~\ref{table_GELINA}. The \SAMMY\ analysis of the experimental data identified $549$ resonances; namely $197$ $s$-wave and $352$ $p$-wave. Excellent fit of the experimental data with the \SAMMY\ code were obtained for all experimental data up to $8000$ eV. As an example, a fit of Mihailescu~\cite{Mih} transmission data from $1.85$ eV to $8000$ eV is displayed in Fig.~\ref{Mih_trans_res} together with residues. Likewise, Fig.~\ref{rpi_trans_residues} and Fig.~\ref{rpi_cap_residues} display the fit of the RPI transmission and capture cross section and respective residues in the energy range $100$ eV to $500$ eV. Values of the thermal capture cross section compared with the experimental data for Brusegan~\cite{Brus} is listed in Table~\ref{table4_therm-rh}. Additionally, listed in Table~\ref{table4_therm-rh} are the resonance integral values calculated with the evaluated resonance parameters and the values listed in the Atlas of Neutron Resonances~\cite{Mugh2006}.
The uncertainty in the  calculated values are from the resonance parameter covariance derived in the evaluation.
\begin{table}[bt]
\begin{center}
\caption{GELINA experimental data}
   \label{table_GELINA}
\scalebox{0.9}{\begin{tabular}{ ccc }
\toprule \toprule
 Reference & Energy Range (eV) & Data \\
\midrule
 \multicolumn{3}{c}{Transmission} \\
\midrule
 Brusegan~\cite{Brus} & $0.4-1000$& 49.3 m, 0.002207 at/b\\
 Brusegan~\cite{Brus} & $0.4-1000$& 49.3 m, 0.000337 at/b\\
 Ribon~\cite{Rib} & $18.0-95.0$& 53.7 m, 0.00608 at/b\\
 Ribon~\cite{Rib} & $84.0-503.0$& 53.7 m, 0.0001487 at/b\\
 Ribon~\cite{Rib} & $178.0-757.0$& 53.7 m, 0.05 at/b\\
 Ribon~\cite{Rib} & $600.0-4000.0$& 103.7 m, 0.02435 at/b\\
 Mihailescu~\cite{Mih} & $1.85-8000.0$& 49.343 m, 0.0458 at/b\\
\midrule
 \multicolumn{3}{c}{Capture} \\
\midrule
Brusegan~\cite{Brus} & $0.01-1000$& 14.3624 m, 0.00187 at/b\\
Mihailescu~\cite{Mih} & $1.72-8000.0$& 28.814 m, 0.000337 at/b\\
Mihailescu~\cite{Mih} & $1.72-8000.0$& 28.814 m, 0.00187 at/b\\
\bottomrule \bottomrule
\end{tabular}}
  \end{center}
\end{table}

\begin{table}[bt]

  \begin{center}
    \caption{Capture cross section at thermal energy  ($0.0253$~eV) and resonance integral for \nuc{103}{Rh}.}
    \label{table4_therm-rh}
\scalebox{0.9}{    \begin{tabular}{cccc}
\toprule \toprule
Cross Section (b) & This Work & Atlas~\cite{Mugh2006,12Pri}&Brusegam \cite{Brus} \\ \midrule

$\sigma_{\gamma}$ & $142.8(23)$ & $143.5(15)$ &$142.5(15)$\\

$I_\gamma$    & $1007(41)$ & $1012(50)$&$-$ \\
\bottomrule \bottomrule
    \end{tabular}}

  \end{center}

\end{table}

\paragraph{Unresolved resonance evaluation\newline}
A new $^{103}$Rh URR evaluation was performed concurrently with the aforementioned RRR evaluation. The URR evaluation procedure and corresponding results were previously published~\cite{BARRY2023Rh103URR}. The \SAMMY\ code~\cite{SAMMY} was used to simultaneously fit experimental total and capture cross section data and extract energy dependent average URR parameters in the energy region from 8 keV up to the first inelastic threshold at 40.146 keV. The \SAMMY\ code was also used to generate covariance information for the URR parameters.

\paragraph{Benchmark results\newline}
Benchmark calculations were performed using the present evaluation. The resonance parameters, converted to ENDF format, were used instead of those from ENDF/B-VIII.0. The evaluation is referred to as ENDF/B-VIII.0mod. The Institute for Radiological Protection and Nuclear Safety from the French name ``Institut de Radioprotection et de S\^{u}ret\'{e} Nucl\'{e}aire (IRSN)'' participated on critical experiments aimed at nuclear data measurements for code and data validations, in particular for testing the $^{103}$Rh cross section data and uncertainties. The critical benchmark program named ``Mat\'{e}riaux Interaction R\'{e}flexion Toutes Epaisseurs'' (MIRTE) was designed by IRSN and carried out at the ``Commissariat \`{a} l'Energie Atomique (CEA)'' Valduc Center in the Apparatus B assembly. The benchmark is listed in the  ICSBEP handbook \cite{ICSBEP}
as LEU-COMP-THERM-106. Benchmark calculations were performed with the MORET $5$ code (Monte Carlo standard deviation of 0.00010) \cite{moret} using cross sections processed with the \NJOY\ code \cite{NJOY}. The results of the testing together with the benchmark $k_{\mathrm{eff}}$  and uncertainties are gathered in Table \ref{tabletwo-rh}.

\begin{table}[tb]
  \begin{center}
    \caption{Comparisons of $k_{\mathrm{eff}}$ results for the LEU-COMP-THERM-106 \cite{ICSBEP} benchmark.}
    \label{tabletwo-rh}

    \begin{tabular}{lc} 
\toprule \toprule
      Evaluation&
      $k_{\mathrm{eff}}$  results	\\   \midrule
     Benchmark&1.00040(64)\\

     ENDF/B-VIII.0& 1.00296\\
     ENDF/B-VIII.0mod&1.00216\\
     \bottomrule \bottomrule
    \end{tabular}
  \end{center}
\end{table}

Comparisons of the ENDF/B-VIII.0 results with calculations using the  ENDF/B-VIII.0mod library indicates a decrease on $\Delta k$ corresponding to about 80 pcm. The new evaluation provides a good indication that the criticality safety prediction for thermal system including $^{103}$Rh has improved.

Resolved resonance evaluation of the $^{103}$Rh cross sections included several experimental data sets. Resonance parameter covariance and uncertainties have  been generated together with the resonance parameters evaluation. The new $^{103}$Rh resonance evaluation extends the upper energy limit of existing evaluation of about $4000$~eV. Thermal benchmark results indicate that the new $^{103}$Rh resolved resonance evaluation is as good as those of existing evaluations or even potentially slightly better. Presently, no benchmark in the epithermal energy region exists to test the efficiency of the new evaluation.


\subsubsection{\nuc{140,142}{Ce}}
\label{subsec:n:140-142Ce}

\begin{figure*}[tbp]
\centering
\subfigure[\nuc{140}Ce cross section uncertainties.]{\includegraphics[scale=.35]{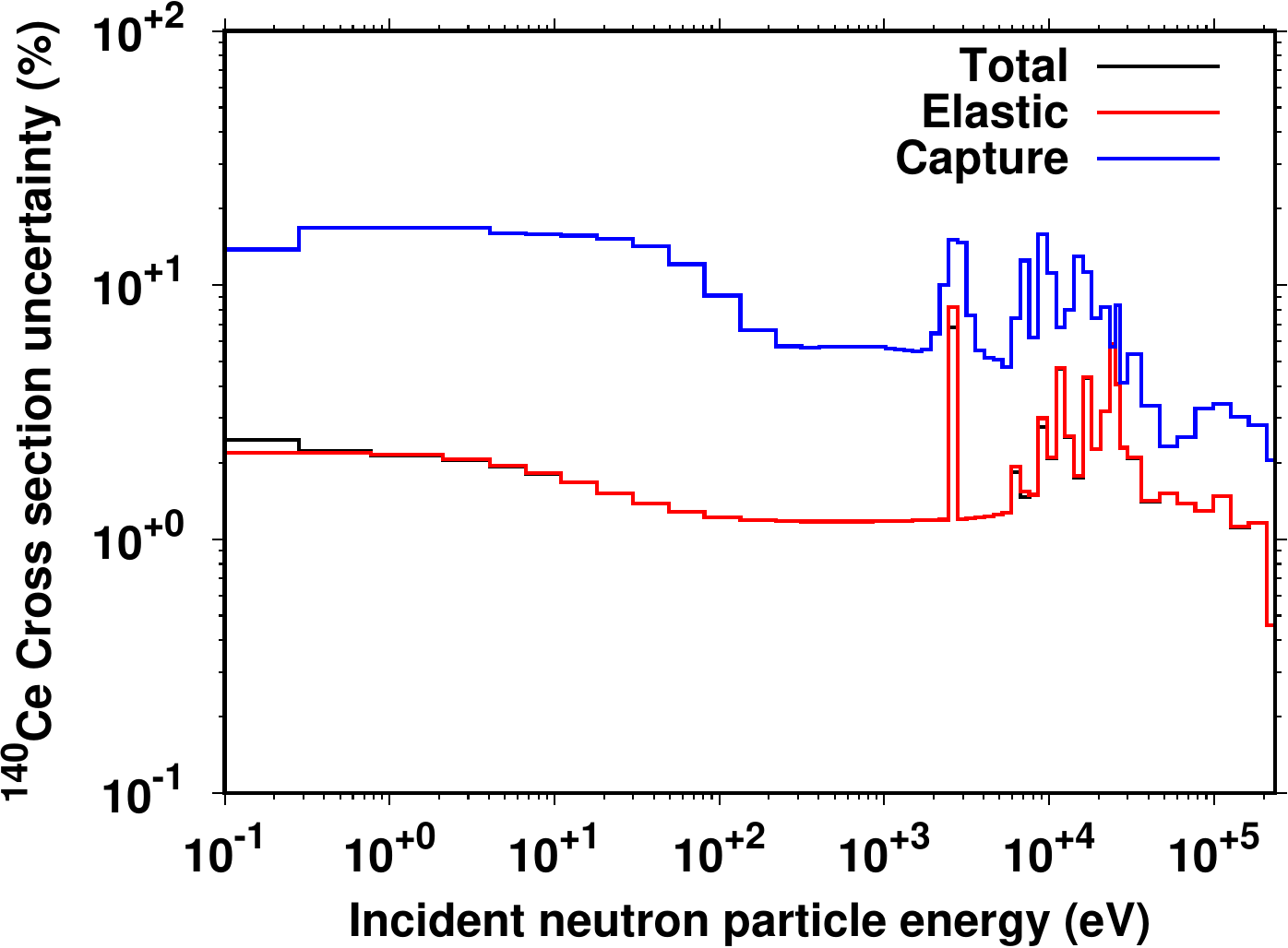}}
\subfigure[\nuc{142}Ce cross section uncertainties.]{\includegraphics[scale=.35]{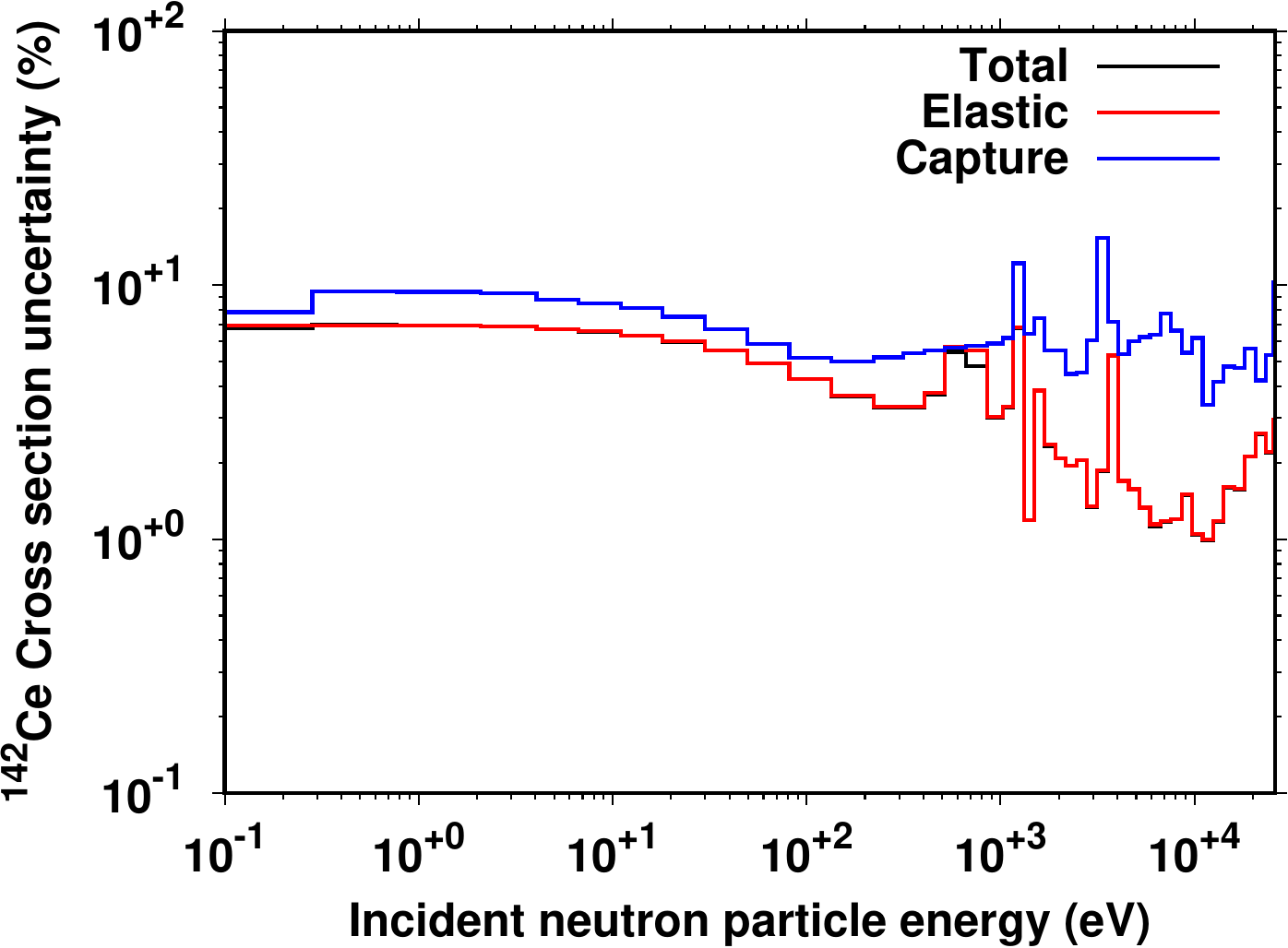}}
\caption{80-group representation of \nuc{140,142}{Ce} cross section uncertainties for total, elastic, and capture reaction channels.}\label{fig:cerium_uncer}
\end{figure*}

Details of the evaluation work in the RRR on even-$A$ cerium isotopes \nuc{140,142}{Ce} are reported in Ref.~\cite{chapman:2023}. Constrained by measured thermal constants available from the EXFOR database, these evaluations are based on recent high-resolution transmission and capture measurements performed on \nuc{nat}{Ce} and highly enriched \nuc{142}{Ce} samples at the JRC-GELINA facility.  Due to their poor quality and documentation, other available transmission data were disregarded and not included in the evaluation work. Improved with the RM $R$-matrix approximation over the Multi-level Breit Wigner approximation, the resonance parameterization performed in this work also resolves the long standing issue of the cerium evaluations of the ENDF/B nuclear data releases reporting incorrect resonance widths, particularly, for the 1.3~keV energy level. 
The updated and corrected set of resonance parameters included their uncertainty quantification.
The uncertainty quantification was performed with a retroactive
methodology to generate covariance matrices compatible with
experimental uncertainty guidelines, although these can be questionable
particularly for theoretical observables, such as resonance
parameters. The generated uncertainty and related correlations for
both isotopes and all available reaction channel in the RRR are reported in
Fig.~\ref{fig:cerium_uncer} for a defined 80-group
energy-averaged representation. Following the resonance structure of
the reconstructed cross sections, the related fluctuating uncertainty
of the total reaction channel is ranging between 1 and 7\%.
Uncertainty between 7 and 14\% in the thermal energy region for the
capture reaction channel seems consistent with the small magnitude of
the cross sections for both isotopes. For this reason, the
contribution of the capture channels to the total uncertainty is small,
and the total and elastic reaction channel are comparable in both
magnitude and structure.  
%

\subsubsection{\nuc{156,158,160,161,162,163,164}{Dy}}
\label{subsec:n:156-158-160-161-162-163-164Dy}


In support of the NCSP, the evaluation work in the RRR for the set of seven Dy isotopes, \nuc{156,158,160-164}{Dy}, is described in detail in Ref.~\cite{pigni:2023}. This is a brief summary of the evaluation work about the generated resonance parameters (File MF=2) and related covariance information (File MF=32).
By adopting the RM approximation implemented in the $R$-matrix code \SAMMY, the high-resolution capture and transmission measurements performed at the RPI Gaerttner LINAC facility~\cite{Block:2017,Shin:2017} were included in the Bayesian fitting procedure. Additional transmission data measured on
enriched samples by Liou~\cite{Liou:1975} at the Columbia University Nevis
synchrocyclotron in the mid-seventies were used to gauge the neutron
widths above 15~eV. The experimental conditions, such as resolution function, finite size sample, detector efficiencies, and
nuclide abundances of sample, multiple scattering, self-shielding,
normalization, background, and Doppler broadening, were taken into
account as implemented in the \SAMMY\ input files reported in Ref.~\cite{pigni:2023}. Thermal constants, such as absorption and
(in)coherent scattering cross sections and corresponding scattering
lengths, were calibrated to the National Institute of Standards and
Technology's compilation~\cite{Sears1992} except for \nuc{161,164}{Dy} isotopes. The guidance provided by Sears' (or NIST)
compilation~\cite{Sears1992} for the thermal constants was augmented by high-resolution capture yields and transmission
data~\cite{Block:2017} due to their extension to thermal energies. In the thermal region, the evaluated thermal scattering and
capture constant for the \nuc{164}{Dy} isotope shows differences with the 
NIST compilation up to 8\%. Also, for \nuc{161}{Dy} isotope, this work
reports a value of the scattering cross section 10\% higher than the
NIST's reported value. Particular emphasis should be devoted in future
experimental campaigns to measuring scattering lengths for these two
isotopes, particularly for the most abundant isotope, \nuc{164}{Dy},
with the largest total thermal cross section among the seven Dy
isotopes. In this regard, the capability to include scattering lengths in the fitting procedure, among other possible measured quantities and as an additional constraint, should be developed. This work included uncertainty quantification
of the total, elastic scattering and capture cross section for all
isotopes. The uncertainty in the
thermal energy region were about 1\% constrained by the transmission,
and capture measurements on natural samples~\cite{Block:2017} were
reported to have a similar uncertainty. In the energy region above 15~eV, although high-resolution capture
yield data measured by Shin~\cite{Shin:2017} are available, high-resolution
transmission data are still needed for isotopically enriched
samples. The currently available Liou's transmission
data~\cite{Liou:1975} are poorly documented in the uncertainty
quantification analysis, and it is very likely that the reported
data communicated to the EXFOR database are not measured data but
total cross section data reconstructed from resonance parameters
fitted to the measured transmission data. Moreover, Liou's
measurements campaign included capture measurements that are not
available in the EXFOR library. The lack of modern transmission data
measured on isotopically enriched samples in the neutron energy above
15~eV also precludes any attempt to extend the RRR to the currently
reported upper energy ranges.

In the specific case of Dy isotopes, there is a very limited number of modern
benchmarks that can be used to perform a conclusive validation over a
comprehensive suite of cases. Therefore, no particular validation test
was performed on the evaluated data, although Block's transmission and Shin's capture data represents a perfect example of
nuclear data ready for validation purposes due to their extension to thermal energies.

Finally, this work represents one of the first attempts to generate a fully
reproducible evaluation in the RRR as this is the primary goal of the
Working Party on International Nuclear Data Evaluation Co-operation Subgroup~49~\cite{mwherman:2021}.

\subsubsection{\nuc{181}{Ta}}
\label{subsec:n:181Ta}


\paragraph{Resonance range\newline}
The previous $^{181}$Ta evaluation was identified as one of the oldest evaluations within ENDF/B-VIII.0 \cite{NCSP5YP}, containing some resonance parameters that appear to have originated in ENDF/B-IV. The $^{181}$Ta ENDF/B-VIII.0 evaluation uses a currently outdated multi-level Breit-Wigner (MLBW) formalism to represent the RRR and no covariance data are provided. The previous RRR energy range only extends to 330 eV.  The URR in ENDF/B-VIII.0 extended from 330 eV to 5 keV. The energy region from 5 keV and above contained no resonance self-shielding, causing major discrepancies with experimental data. There was no evaluated uncertainty or covariance data reported with the URR cross sections.

The new $^{181}$Ta RRR parameters are a result of fitting several experimental transmission and capture yield data sets using the modern \SAMMY\ code~\cite{SAMMY}. The \SAMMY\ fitting process used a RM approximation to the R-matrix formalism. The starting RRR parameters were based on JEFF-3.3~\cite{JEFF33}. Spin assignments were adopted from the neutron ATLAS \cite{ATLAS2018}. For spins not given in the ATLAS, the values were randomly generated using a Monte Carlo process. This Monte Carlo process was extended to randomly add small fictitious resonances above $\approx$ 777 eV that improve RRR parameter statistics. All relevant experimental information, such as the resolution function, were included in the \SAMMY\ fitting process. The uncertainty on these experimental parameters were propagated into the final RRR parameter uncertainty. The new evaluation extends the RRR region up to 2.554 keV and provides covariance information for all RRR parameters. As part of validation, the new RRR parameters produce thermal coherent and incoherent scattering cross section values that agree with the NIST~\cite{Sears1992} values.  The details of $^{181}$Ta RRR evaluation were previously published \cite{BARRY2024Ta181RRR}. 

The new URR parameters for \nuc{181}{Ta} are the posterior values of a \SAMMY~\cite{SAMMY} Bayesian evaluation of multiple datasets for total cross section \cite{harvey1988,brown-exp-2023,Poenitz1981}, capture cross section \cite{mcdermott_physrev,wisshak_ta_capture,wisshak2004stellar,brzosko1969,bokhovko1991,kononov1977,yamamuro_1980}, and elastic cross section \cite{zo1985} found in EXFOR \cite{EXFOR}. A coordinated inter-institution effort was made to ensure consistency between the RRR, URR, and fast neutron evaluation efforts. To achieve self-consistency for RRR--URR evaluation, a subset of the posterior resonance parameters (in the energy range $<$~330 eV) from the RRR evaluation were directly used to calculate prior parameters for the URR. The URR evaluation was extended from the end of the RRR (2.554 keV) to 100 keV where the fast neutron evaluation began. The URR and fast neutron evaluators agreed on several high-accuracy datasets (\emph{e.g.}, Poenitz \etal \cite{Poenitz1981} and McDermott \etal \cite{mcdermott_physrev}) that would constrain the evaluations to similar final mean values at the 100 keV matching point. The posterior average resonance parameters included a full covariance matrix from the generalized least squares fit. The posterior covariance on the average resonance parameters was used to calculate a full covariance matrix for the total, capture, elastic, and 1$^{\text{st}}$ excited state inelastic cross sections to conform to the current ENDF/B format requirements. The new URR evaluation shows better agreement with energy-differential measurements (\emph{e.g.}, Brown \etal~\cite{brown2020} and Byoun \cite{byoun-thesis-1973}) and integral benchmark experiments. Reasonable covariance estimates are now available to estimate the impact of nuclear data uncertainties on modeled transport applications. Due to an unfortunate clerical error, however, the URR covariances for \nuc{181}{Ta} in the final ENDF/B-VIII.1 release were unintentionally overwritten in the process of file assembly with higher values that are incorrect but more conservative. The incorrect covariances were added from 2 keV to 100 keV. The released standard deviations on total cross section ranged from 4-10\%, elastic cross section from 4-9\%, capture cross section from 2-4\%, and inelastic cross section from 20-60\%. The correct values will be released in an updated file that will be available from the NNDC. The finer details of the URR evaluation were published by Brown \etal~\cite{brown-2024}.

\paragraph{Fast neutron range\newline}
The new  $^{181}$Ta  evaluation in the fast neutron range has been performed by combining selected differential measurements with modern nuclear reaction physics into a consistent set of nuclear data and at 100 keV merged with the above-described evaluation in the unresolved resonance range.

Tantalum is a nearly mono-isotopic metal with \nuc{181}{Ta} being 99.988\% of the natural mixture.  There is, therefore, an exceptional wealth of experimental data
available for total, capture, (n,p), (n,$\alpha$), (n,t), (n,$^3$He)
and (n,2n) cross sections, elastic and inelastic
angular distributions, as well as neutron- and $\gamma$-spectra (including
double-differential cross sections) in the high energy region. The total number of EXFOR datasets is over 340
covering nearly 140 quantities.
Because of the space limitation, we do not review here such a vast data set,
but Table~\ref{Tab:Ta181exp} lists most trusted and consistent experiments that were considered in the development of the evaluation.
We used the EXFOR-interface feature to renormalize
old experimental data to the current standards. In spite of this, more than 30\% of
experimental data were considered incompatible with the consistency condition
and dropped out from the analysis.

\begin{table}[!htbp]
\vspace{-3mm}
\caption{Experiments adopted for guiding $^{181}$Ta evaluation. }
\vspace{-6pt}
\begin{center}
 \begin{threeparttable}
\begin{tabular}{ l l }
\toprule \toprule
  Reaction/Quantity &  Adopted references    \\
\midrule
  Total x-sect.              	& \cite{Poenitz1981, Poenitz1983, Tsubone1984TS08, Rapp2019RA19, Smith1968SM03, Hannaske2013HA37, Byoun1973BYZY, 
                                    		Carlson1967CA23, Finlay1993FI01, Foster1971FO24}  \\
  Elastic x-sect.           	 & \cite{Smith2005SMZY, Zo-In-Ok1985_exfor40937.1}    \\
  Inelastic x-sect.          	& \cite{Rogers1971RO26,  exfor40680.2, Owens1968OW02,  exfor40603.2}  \\
  Capture x-sect.            	& \cite{mcdermott_physrev, wisshak2004stellar, wisshak_ta_capture, Voignier:1992, Poenitz1975}    \\
  (n,p) x-sect.              	& \cite{Filatenkov2016_exfor41614.1, Kasugai1994_exfor23011.1, Semkova2008SEZT, Begun2002_exfor32205.1,
                                        Wolfle1988WO05,Mukherjee1963_exfor31330.1}  \\
  (n,2n) x-sect.             	& \cite{Frehaut1974FRZG, Veeser1977VE10} \\
  (n,3n) x-sect.             	& \cite{Veeser1977VE10} \\
  (n,$\alpha$) x-sect.      	& \cite{Filatenkov2016_exfor41614.1, Mukherjee1963_exfor31330.1, Majdeddin1997_exfor31481.1} \\
  Angular distr.            	& \cite{Smith2005SMZY, Holmqvist1970HO18, Beyster1956BE32, Buccino1966_exfor11877.1, exfor40603.2, 
                                     Ferrer1977FE01, Takahashi1992_exfor22136.1, Hansen1985HA02, Cross1960CR10, Hudson1962HU11} \\
  Neutron energy spectra 	& \cite{Matsuyama1993_exfor22352.1, Marcinkowski1993, Vonach1980_exfor21599.1, Takahashi1992_exfor22136.1} \\
  Neutron double-diff. dist. & \cite{Matsuyama1992, Owens1968OW02, exfor40603.2, Soda1996_exfor22343.1, Takahashi1992_exfor22136.1,
                                     		Vonach1980_exfor21599.1, Marcinkowski1993} \\
\bottomrule \bottomrule
\end{tabular}
\label{Tab:Ta181exp}
 \end{threeparttable}
\end{center}
\vspace{-2mm}
\end{table}

\paragraph{Evaluation concept and model\newline}
The overall methodology employed in the current evaluation is based on the concept of limited trust, consistency, and hierarchy. The principal notions of such approach are:
\begin{itemize}

\item The limited trust applies to experimental data  which can be subjectively adopted, included after renormalization, or ignored. It applies  also to the model calculations which, while using default parameters, are usually not capable of reproducing experimental data within the required accuracy. In the extreme cases, model calculations can be tuned by energy dependent factors to reproduce particular shapes of the well-trusted experimental data. Such tuning, however, preserves all the constraints imposed by the reaction physics.

\item The consistency implies selecting nuclear reaction models (along with their respective parameters) and the experimental data that agree with each other creating a consistent picture for all reaction observables (e.g., cross sections, spectra, angular distributions and isomeric cross sections). To some extent, the same may apply to integral experiments, but this was not the case in the present evaluation.

\item The hierarchy concept is applied to adjusting the model parameters. Rather than fitting all model parameters to all selected experimental data, we opt for a step-wise gradual approach, in which a set of model parameters is adjusted to the observables which predominantly depend of this set, e.g., spherical optical model parameters are fitted only to total and elastic cross sections and elastic angular distributions (in case of coupled-channel, calculations experimental database includes also inelastic observables).  The idea is to link certain model parameters to a set of  physical quantities that predominantly depend on these parameters and are not (or weakly) sensitive to the others. In the following evaluation phases, we are going to keep this subset of parameters fixed.


\item The evaluation is entirely given by the model calculations, which are adjusted to the selected set of  differential experimental data that are considered reliable. Thus, the evaluation is effectively reduced to the set of model parameters and the exact version of the modeling code and is perfectly reproducible.

\end{itemize}

The  $^{181}$Ta calculations were performed with \EMPIRE-3.2.3~\cite{Herman:2007, EmpireManual}  reaction model code. In view of multiple
choices for selecting specific models available in \EMPIRE, various combinations of
these models were first explored in an attempt to reproduce the large set of available experimental
data, while refraining from  adjusting individual model parameters.  The  adopted models are:

\begin{itemize}
\item Coupled-Channels (CC) with dispersive Optical Model potential.
\item Quantum-mechanical Multistep  Direct~\cite{Tamura:82, ORTRI} (MSD)  model for pre-equilibrium neutron emission.
\item Heidelberg formulation of the Multistep Compound (MSC) model for pre-equilibrium neutron~\cite{Nishioka:86, Herman:92a}  and gamma emission~\cite{GammaMSC, GammaMSCapp}.
\item Exciton model with Iwamoto-Harada~\cite{Iwamoto:82} extension and  Kalbach~\cite{Kalbach1988} systematics for angular distributions for pre-equilibrium emission of protons and clusters.
\item Hauser-Feshbach with Moldauer~\cite{Moldauer:76} width fluctuation correction, anisotropy calculated using Blatt-Biedenharn coefficients,   full gamma cascade using E1 $\gamma$-strength function set to RIPL-3  MLO1~\cite{ripl2}.
\item Gilbert-Cameron~\cite{Gilbert65} level densities.
\end{itemize}

Selection of the models and their parameters is discussed briefly below. For more details, we refer readers to the descriptive part of the ENDF file.

\begin{figure}[!htbp]
\centering
\includegraphics[scale=.65]{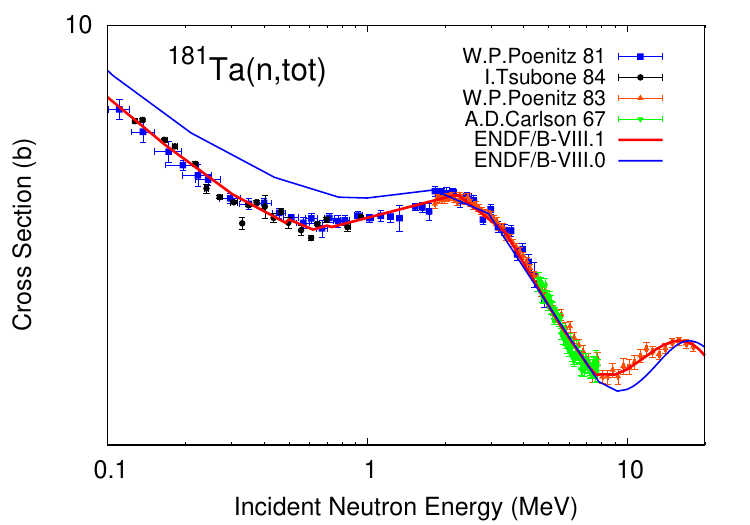}
\vspace{-.1in}
\caption{Evaluated total cross sections compared with the experimental data selected to drive the evaluation.}
\label{Ta181fig-tot}
\end{figure}

\paragraph{Optical model - total and elastic\newline}
Strongly deformed   $^{181}$Ta requires CC
modeling of the incident channel. The RIPL-3~\cite{RIPL3}  regional CC potential  (\#610)
that covers \nuc{181}{Ta} and extends up to 200 MeV  was not designed specifically
for tantalum and needed adjustments to optimally describe the incident energy region
from 50 keV up to 20 MeV.  The two very reliable and consistent measurements of
total by Poenitz \etal~\cite{Poenitz1981, Poenitz1983}, which cover entirely this energy range and overlap each other between 2 and 4 MeV, were used to refine the RIPL-3 potential. The fitting has been
performed using Kalman filter~\cite{Kalman1960} and resulted in rather minor (a few percent)  corrections to the three
geometry and two deformation parameters while the third deformation parameter has been reduced by 51\%.
Fig.~\ref{Ta181fig-tot}  shows that  experiments by Tsubone~\cite{Tsubone1984TS08}  and Carlson~\cite{Carlson1967CA23} perfectly agree with
the evaluated total calculated using updated optical potential. Results by Rapp~\cite{Rapp2019RA19} and Smith~\cite{Smith1968SM03} also agree with the evaluation but were not included in the plot for clarity. Also not shown on the plot are  data by Hannaske~\cite{Hannaske2013HA37} and Byoun~\cite{Byoun1973BYZY} that reveal the same shape as current evaluation and agree after ~3\% and ~4.5\% reduction, respectively. Similarly,  a 1.5\% upscaling brings into agreement Foster~\cite{Foster1971FO24} data, while Finlay's~\cite{Finlay1993FI01} shape is slightly different but still within 2\% of the new evaluation.

\begin{figure}[!htbp]
\centering
\includegraphics[scale=.131,clip,trim= 65mm 9mm 14mm 14mm]{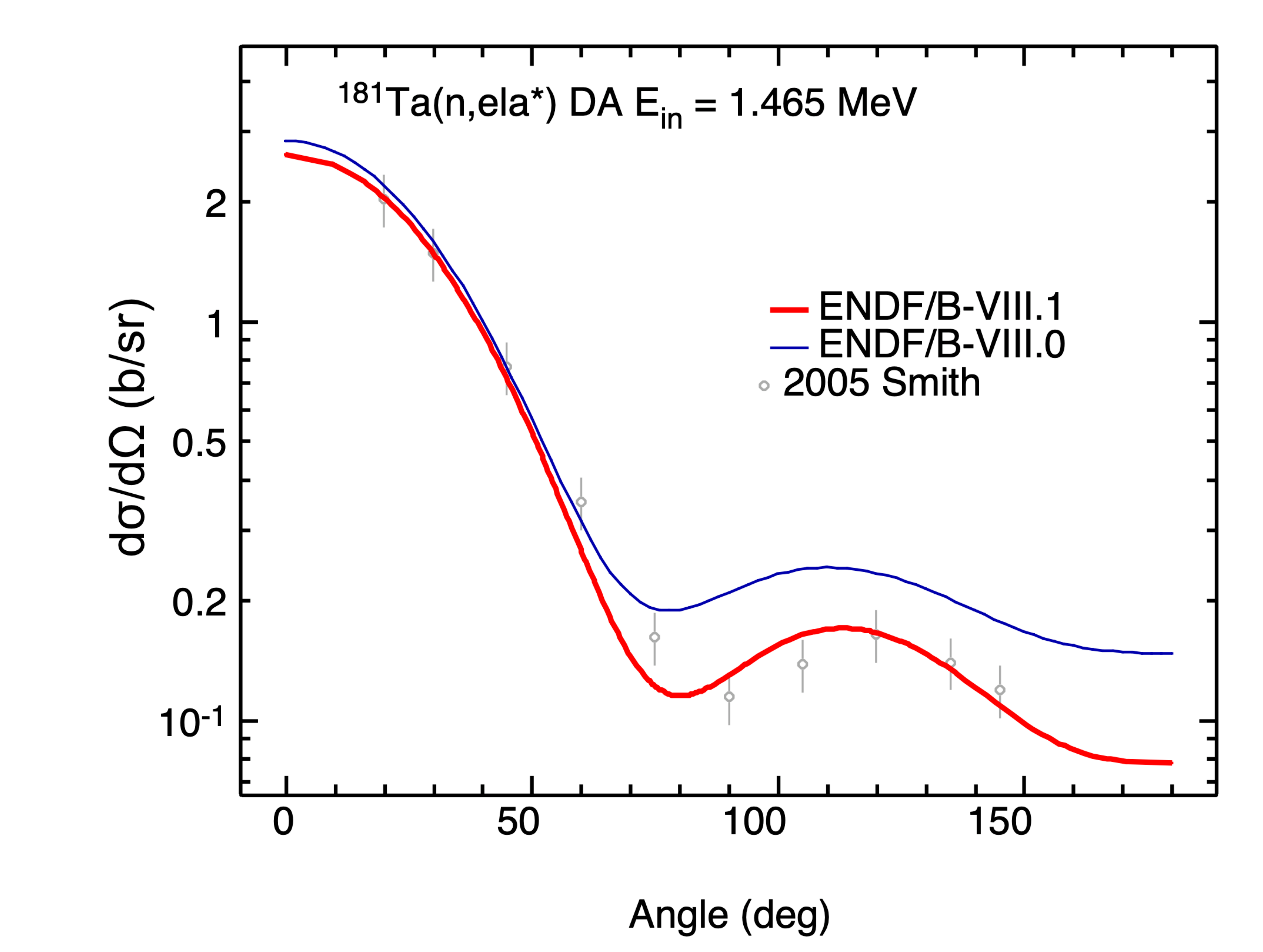}
\vspace{-.1in}
\caption{An example of semi-elastic angular distribution.  Agreement  between Smith data~\cite{Smith2005SMZY} and the new evaluation is good, but 
similar comparisons at lower incident energies are often much better. The ENDF/B-VIII.0 overestimates experimental results in the backward hemisphere, but it should be noted that
if the first inelastic were the only inelastic supposed to contribute to the semi-elastic, the agreement with the experimental data would be comparable to ENDF/B-VIII.1.}
\label{Ta18-elaDA1.465}
\end{figure}

\begin{figure}[!htbp]
\centering
\includegraphics[scale=.131,clip,trim= 15mm 9mm 23mm 14mm]{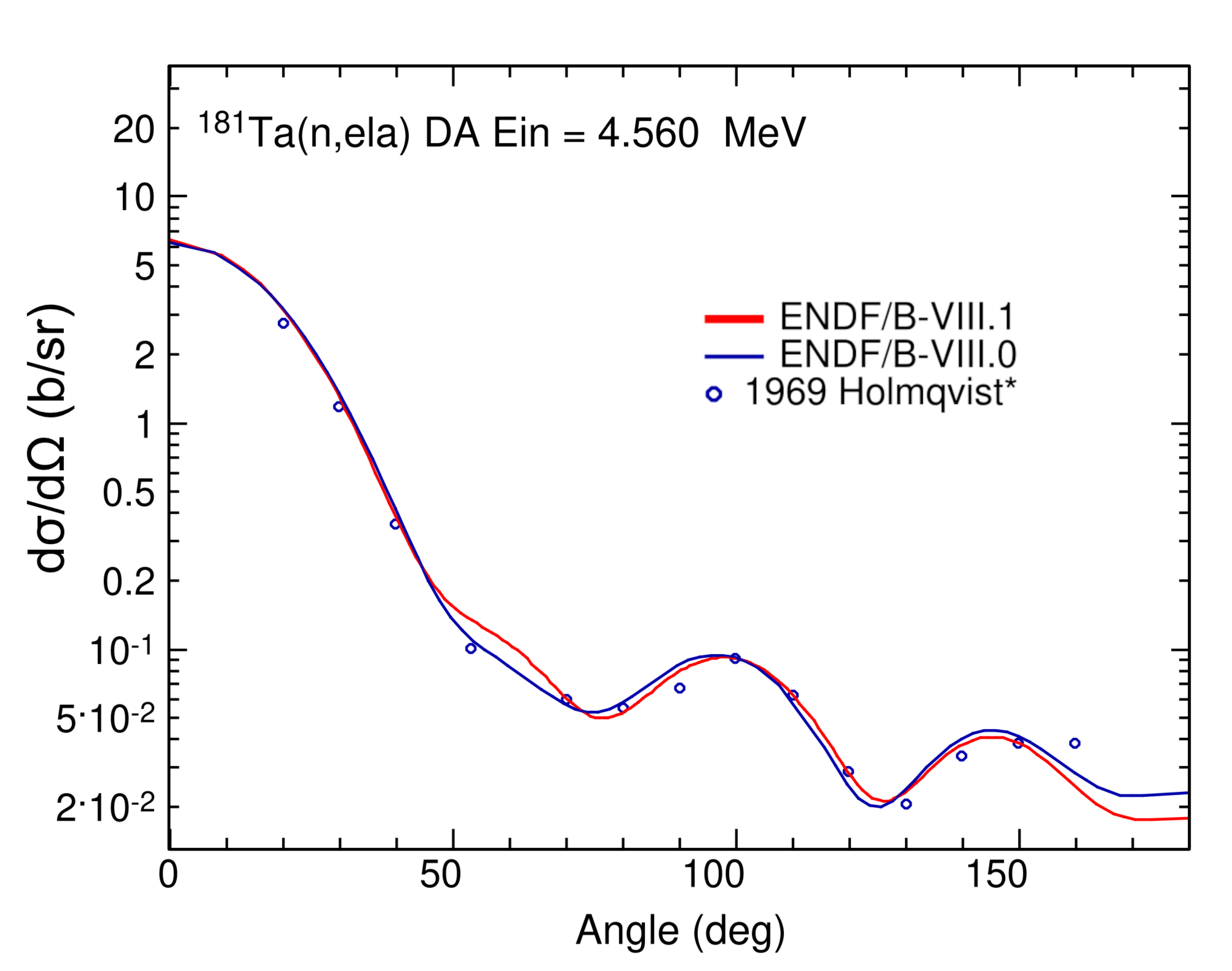}
\vspace{-.1in}
\caption{Semi-elastic angular distribution at 4.56 MeV compared with the current evaluation and ENDF/B-VIII.0.  Both evaluations agree 
very well with the experimental data, and consequently with each other,  in spite of very different optical potential employed in the 
calculations and large differences in total cross sections shown in Fig.~\ref{Ta181fig-tot}.}
\label{Ta181-elaDA4.56}
\end{figure}

The  performance of the adjusted potential on semi-elastic\footnote{Because of the experimental energy resolution, the 1$^{st}$, 2$^{nd}$ and eventually also the 3$^{rd}$ inelastics are summed with the elastic.} 
cross section data is good considering the wide spread of experimental data. Comparison with elastic angular distributions demonstrate good or excellent agreement with A. Smith~\cite{Smith2005SMZY} 
data up to 1.5 MeV (see ``good'' example at 1.465 MeV on Fig.~\ref{Ta18-elaDA1.465}). At higher incident energies, abundant and smooth  Smith data tend to be higher than the evaluation. 
The overall shape is very similar, but the difference is forward peaked suggesting that either more inelastics should be added to the semi-elastic or the three added inelastics might be underestimated.  
On the other hand, other experiments in majority support the current evaluation. An example is shown in Fig.~\ref{Ta181-elaDA4.56}.

\begin{figure}[!htbp]
\centering
\includegraphics[scale=.45]{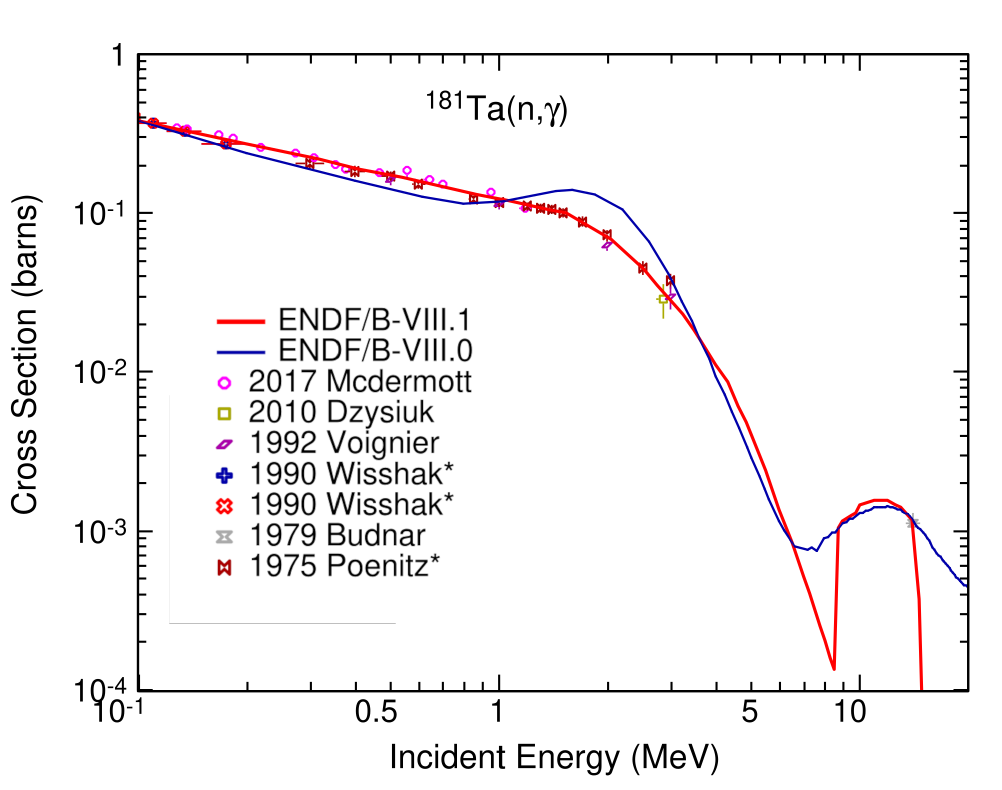}
\vspace{-.1in}
\caption{Selected set of experimental data compared with the new and previous evaluation.  The GDR bump at  high energies comes in the new evaluation from the $\gamma$ emission through the MSC mechanism. }
\label{Ta181-capture}
\end{figure}

\paragraph{Capture evaluation\newline}
There are over 30 capture experiments which spread about 20\% above and below the average. Default model 
calculations were well within these limits, but we decided to follow a few, most trusted and statistically 
consistent sets. These were measured by McDermott~\cite{mcdermott_physrev}, Wisshak~\cite{wisshak2004stellar, 
wisshak_ta_capture}, Voignier~\cite{Voignier:1992}, and Poenitz~\cite{Poenitz1975}. 	We used manually adjusted 
energy-dependent scaling factors to reproduce these data up to 2.75 MeV (see Fig.~\ref{Ta181-capture}). We note that 
this choice is supported by a number of other experiments which are very close to the evaluation but are either 
consistently off by a couple of percent, or  affected by larger spread of points, or agree with absolute value but 
cover too small an energy range to provide additional information.  For the first time, the preequilibrium emission of 
$\gamma$-rays was analyzed using the MSC mechanism~\cite{GammaMSCapp}. This mechanism, without any manual intervention, 
reflects Giant Dipole Resonance (GDR) with a hump between 9 and 15 MeV. However, between 6 and 9 MeV, the MSC contribution to the compound nucleus 
capture cross sections is relatively small, resulting in a discrepancy with the semi-direct results apparently used in 
the ENDF/B-VIII.0 evaluation. While applying larger smoothing to the MSC results could potentially alleviate this difference, 
there is no experimental evidence to support either of the two theoretical approaches. Furthermore, the capture cross sections 
in this energy range lack practical significance. Therefore, we have decided to retain the original MSC results without any 
artificial modifications.

\begin{figure}[!htbp]
\centering
\includegraphics[scale=.19,clip,trim= 39mm 0mm 0mm 0mm]{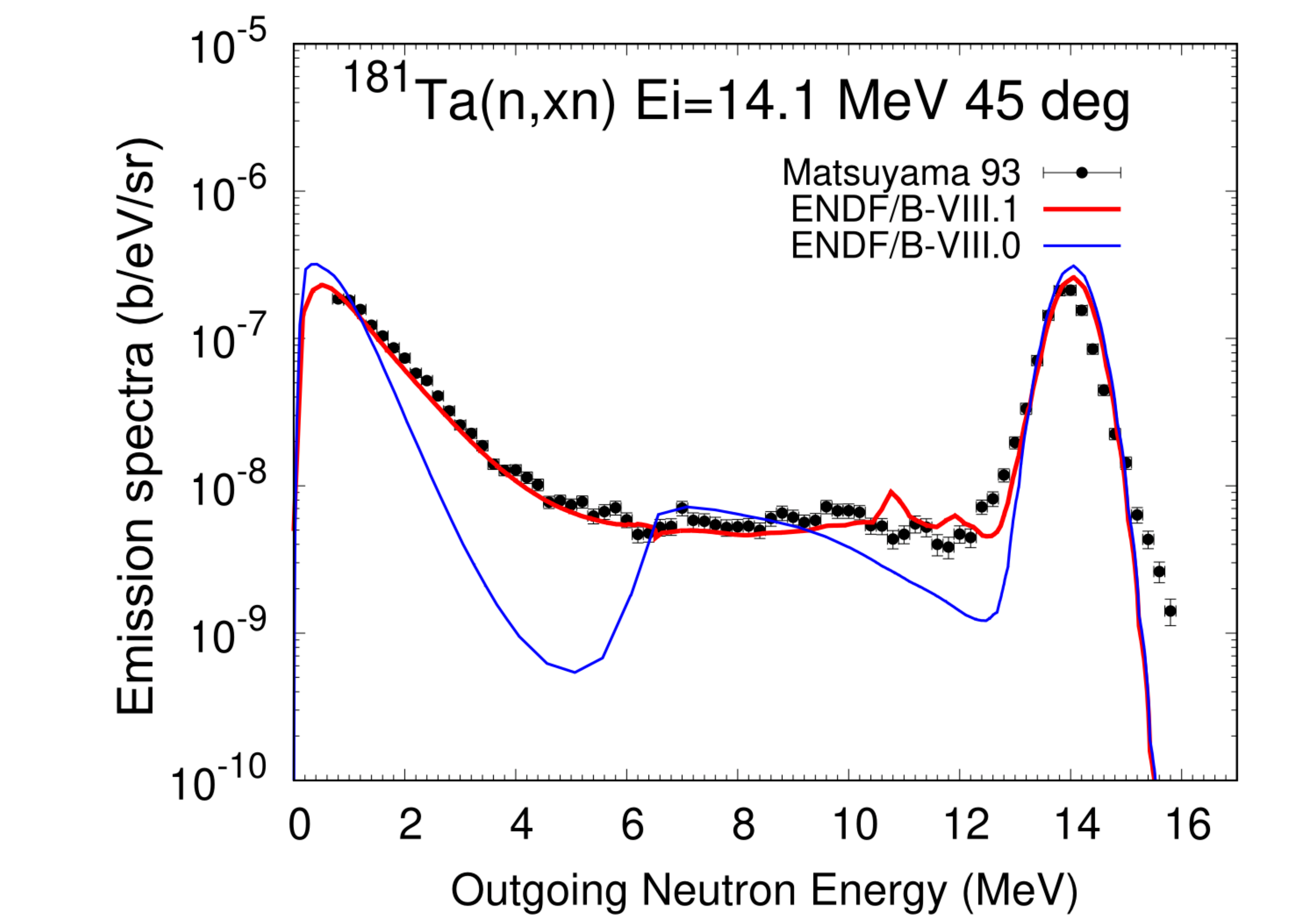}
\vspace{-.1in}
\caption{Double-differential neutron emission spectrum compared with the experimental data by Matsuyama~\cite{Matsuyama1992} showing essential improvement compared to ENDF/B-VIII.0.}
\label{Ta181-nSpec}
\end{figure}

\paragraph{Preequilibrium emission\newline}
The Multistep Direct (MSD) model provides an excellent tool for description of the high energy tail of
neutron spectra, which is impossible for the classical exciton model.   Only one
parameter, $l$=0 transfer response function,  was slightly tuned to improve neutron spectra
(see Fig.~\ref{Ta181-nSpec}) and at the same time ameliorate agreement for (n,2n) and (n,p) reactions.
According to the hierarchical concept, this parameter was kept unchanged in the remaining evaluation process.

The MSC mechanism is responsible for the middle-energy range of neutron spectra. None of the MSC
parameters was adjusted but the gradual absorption concept has been turned off.

Using MSD and MSC for neutron emission leaves more flexibility for reproducing charged particle
channels with the exciton model. The mean free path in the model has been set at 81\% of the theoretical
value (default is 150\%) to reproduce experimental proton and cluster emission data.

\paragraph{Compound Nucleus\newline}
Three components are critical for the compound nucleus calculations: level densities, discrete levels
and their decay schemes, and transmission coefficients.

Level densities affect cross sections and  have strong impact on the spectra.
Gilbert-Cameron (GC) level densities were chosen over Enhanced Generalized Superfluid Model (EGSM) and microscopic Hartree-Fock-Bogolyubov
because GC produce better capture cross sections between 1 and 3 MeV and slightly better
$\gamma$-spectra.  The effect on the neutron spectra is mixed with GC working better at incident
energies around 14 MeV while EGSM has an advantage at 20 MeV.
Following standard routine, level density parameters and numbers of discrete levels were adjusted to ensure smooth transition between discrete levels and continuum. Level density '$a$'-parameters were globally adjusted with the Kalman filter to improve overall agreement with reaction cross sections.

Discrete levels decay schemes are often incomplete. For the purpose of evaluating isomeric cross sections, the non-decaying levels were assigned transition(s) to the lowest energy level of the same parity and closest spin.
Isomeric cross sections are provided for $^{181}$Ta(n,2n)$^{g,m}$,  $^{181}$Ta(n,n$\alpha$)$^{g,m}$,
$^{181}$Ta(n,2n$\alpha$)$^{m}$, $^{181}$Ta(n,$\gamma$)$^{g,m}$, and $^{181}$Ta(n,$\alpha$)$^{g,m}$.

A spherical optical model is used  to calculate transmission coefficients for the outgoing particles. Usually these coefficients are not adjusted, but we realized that increasing real potential depth and radius by ~2\% for the second neutron in (n,2n) is an efficient way of lowering cross sections just above the threshold.


\begin{figure*}[t]
    \centering
    \subfigure[~Carbon measured and simulated quasi-differential neutron scattering at 45 degrees.]{%
    \includegraphics[width=0.49\textwidth]{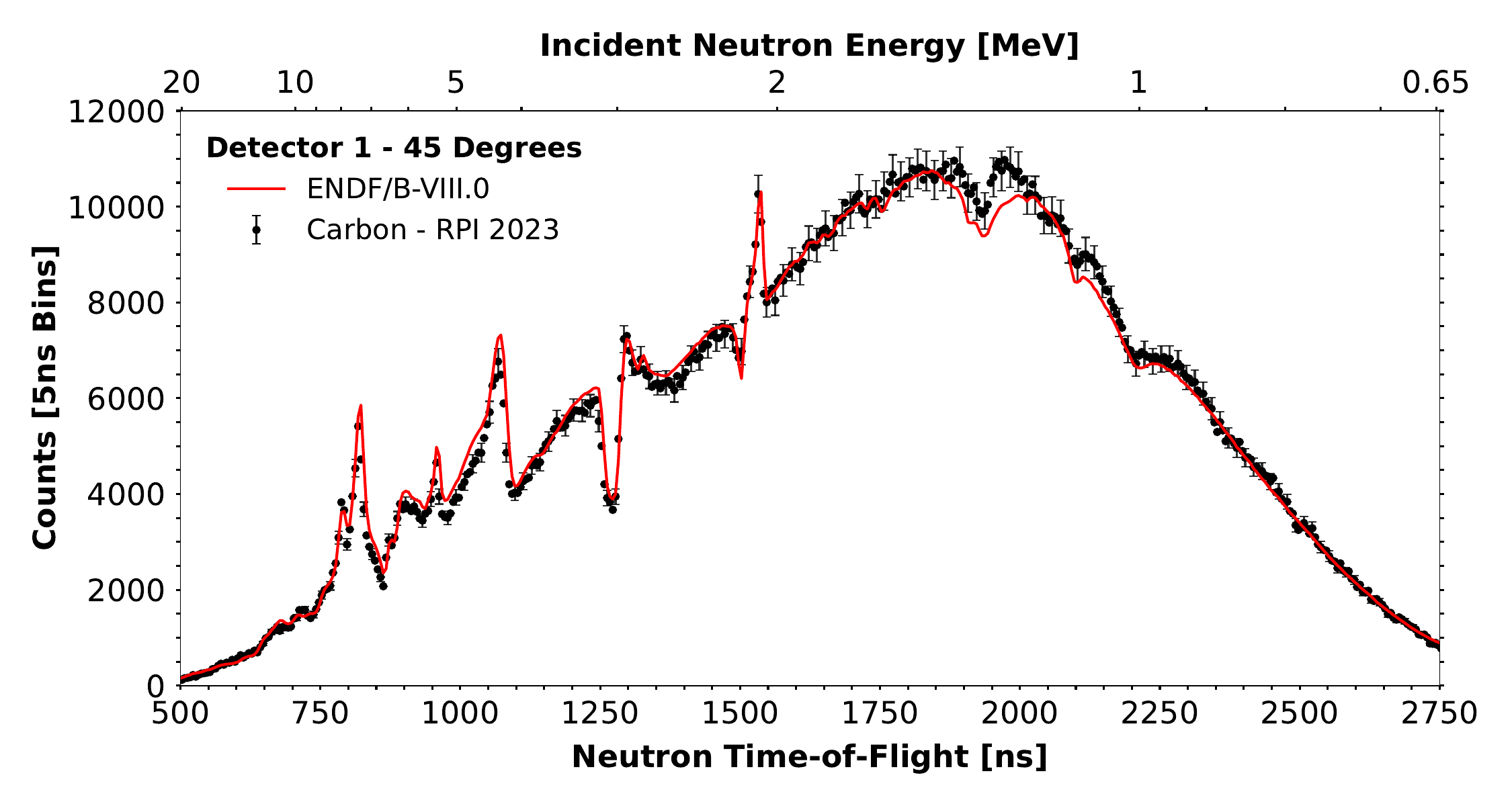}
    \label{fig:carbon-RPI-det1-45deg-Ta}
    }
    \hfill
    \subfigure[~Carbon measured and simulated quasi-differential neutron scattering at 110 degrees.]{%
        \includegraphics[width=0.49\textwidth]{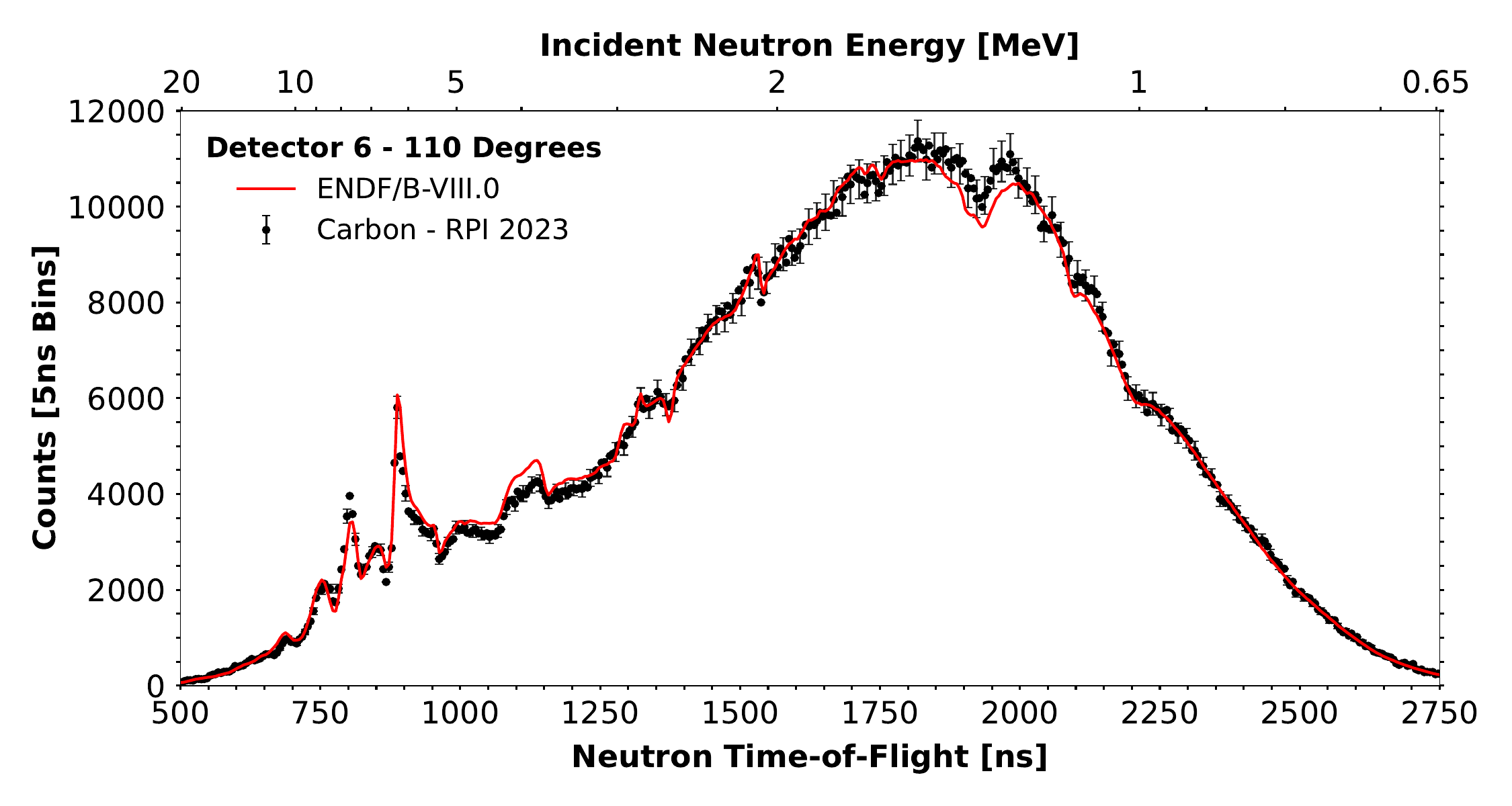}
        \label{fig:carbon-RPI-det6-110deg-Ta}
    }

    \vspace{0.25cm}

    \subfigure[~Ta measured (E) and simulated (C) quasi-differential neutron scattering at 45 degrees.]{%
    \includegraphics[width=0.49\textwidth]{neutrons/figs/Ta-RPI-det1-45deg.pdf}
    \label{fig:Ta-RPI-det1-45deg}
    }
    \hfill
    \subfigure[~Ta measured (E) and simulated (C) quasi-differential neutron scattering at 110 degrees.]{%
        \includegraphics[width=0.49\textwidth]{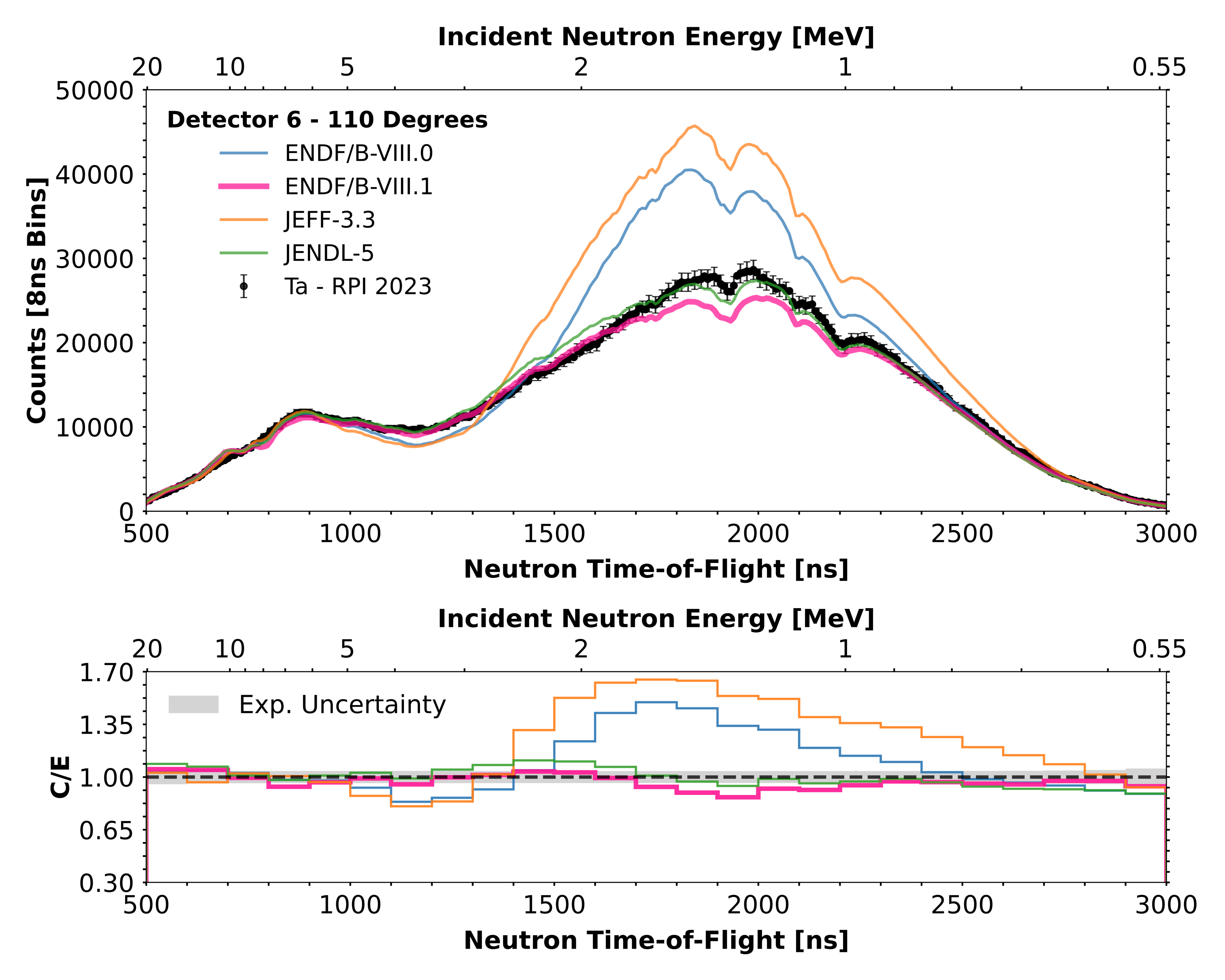}
        \label{fig:Ta-RPI-det6-110deg}   
    }
    
    \caption{Results from the elemental tantalum quasi-differential neutron scattering experiment at RPI \cite{SIEMERS2025111676} in the incident neutron energy range of 0.55 to 20 MeV. Carbon neutron scattering yields, measured during the Ta experiment, at a forward and backward detector compared to MCNP simulations of the ENDF/B-VIII.0 carbon evaluation are presented in Fig.~\ref{fig:carbon-RPI-det1-45deg-Ta} and Fig.~\ref{fig:carbon-RPI-det6-110deg-Ta}, respectively. Measured tantalum neutron scattering yields at corresponding forward and backward detectors compared to MCNP simulations of the ENDF/B-VIII.0, JEFF-3.3, JENDL-5.0, and ENDF/B-VIII.1 evaluations are presented in Fig.~\ref{fig:Ta-RPI-det1-45deg} and Fig.~\ref{fig:Ta-RPI-det6-110deg}, respectively. $C/E$ ratios are provided to compare the experimental and simulated results along with the experimental uncertainty band.}
    \label{fig:ta-rpi-qd-scattering}

\end{figure*}

The essential difference between current evaluation and ENDF/B-VIII.0 (as well as JENDL-4.0) is the inelastic cross section to the first excited level in $^{181}$Ta. We adopted model calculations that are 
up to 300 mb lower than EXFOR values by Rogers~\cite{Rogers1971RO26}. Other evaluations follow Rogers' experimental data. 
However, the Rogers' values for the first inelastic of 6.237 keV are actually not an experimental result. Due to its low excitation energy, the first inelastic is practically impossible to separate 
from the elastic, and Rogers' values are actually coming out of their calculations. This is explicitly stated in Ref.~\cite{Rogers1971RO26}, but it was inadvertently overlooked in the EXFOR 
compilation process, resulting in its listing as experimental and causing confusion in previous evaluations.

\paragraph{Covariances\newline}
Covariances were calculated with the Kalman filter using cross section sensitivities to 53 model parameters (mostly level density and optical model parameters involved in the incident and major outgoing channels). Experimental data sets used in covariance calculations were additionally curated in order to: eliminate outliers, thin overly abundant data sets, smooth statistical fluctuations, and correct missing or statistical only uncertainties.

The Kalman filter has a tendency to yield uncertainties that decrease with the total number of experimental points reaching unacceptably small values. To counteract this undesirable feature, we applied, in addition to thinning big datasets,
an $1/\sqrt{n}$  weight  to each experiment ($n$ being the number of points in the experiment). This procedure allows to keep uncertainties within acceptable limits.

Relative covariances, including cross-reaction terms, were put into the ENDF-6 format using \EMPIRE's IO modules.

\paragraph{Validation \newline}

Robust validation of the $^{181}$Ta evaluation is provided by a recent quasi-differential measurement of the scattered neutron yield from a tantalum sample, undertaken by Siemers and colleagues at RPI \cite{SIEMERS2025111676}. Neutron scattering and emissions were measured by time-of-flight from a 2.165'' thick 3'' diameter right metallic cylinder of tantalum with an estimated 6\% systematic uncertainty in the incident neutron energies range of 0.55~MeV to 20~MeV. A 2.75'' thick 3'' diameter right cylindrical sample of carbon was also measured as part of the Ta experiment to validate the experimental findings. MCNP6 was used to perform the radiation transport calculations of the experiment to compare evaluated nuclear data with the experimental data. The methodology describing the carbon validation was previously described in the \hyperref[subsec:19F_validation]{$^{19}$F validation} section. The carbon results for a typical forward scattering angle (45 degrees) are shown in Fig.~\ref{fig:carbon-RPI-det1-45deg-Ta} and for a typical perpendicular to backward scattering angle (110 degrees) in Fig.~\ref{fig:carbon-RPI-det6-110deg-Ta}. 

Similarly to the Teflon$^{\circledR}$ carbon validation measurement, excellent agreement between the measured scattering yield and the MCNP6 simulated scattering yield is observed in most of the experimental energy region. Minor disagreements are noted around 1.3 MeV and 5~MeV at both the forward and backward neutron scattering angles. Carbon results from this measurement and the Teflon$^{\circledR}$ carbon measurement suggest deficiencies may exist in evaluated carbon nuclear data, especially at backward angles. Note that the carbon elastic angular distributions is the only standard angular distribution up to 1.8~MeV of neutron incident energy~\cite{carlson2018}.

Evaluated ${}^{181}$Ta nuclear data from the ENDF/B-VIII.1, ENDF/B-VIII.0, JEFF-3.3, and JENDL-5 libraries were used to reproduce the measured tantalum neutron scattering yield in simulation and compared with the measured data. Results for a typical forward scattering angle (45 degrees) are shown in Fig.~\ref{fig:Ta-RPI-det1-45deg} and for a perpendicular to backward scattering angle (110 degrees) in Fig.~\ref{fig:Ta-RPI-det6-110deg}. Note that these scattering angles correspond to the angles of the aforementioned carbon validation measurement. At forward angles from 1.5~MeV up to 3~MeV, ENDF/B-VIII.0 and JEFF-3.3 slightly overestimate the RPI data, whilst JENDL-5 and ENDF/B-VIII.1 slightly underestimate the data. At perpendicular to backward angles (110 degrees), the JENDL-5 evaluation shows the best agreement while the new ENDF/B-VIII.1 evaluation remains low from 1~MeV up to 2~MeV. ENDF/B-VIII.0 and JEFF-3.3 significantly overestimate the measured RPI data. The JENDL-5 and ENDF/B-VIII.1 evaluations were observed to best reproduce the experimental data. Although the ENDF/B-VIII.1 evaluation shows large improvements over ENDF/B-VIII.0, especially at 110 degrees, an underestimation of the measured scattered neutron yield remains between 1.2 and 2 MeV at both detection angles.  This underestimation suggests that some deficiencies in the elastic and inelastic cross sections and angular distributions still remain between 1~MeV and 2~MeV in the current  evaluation. Despite this, the results from this experiment demonstrate that we have advanced our understanding of high energy incident neutron interactions with ${}^{181}$Ta.

\subsubsection{\nuc{190,191,192,193,194,195,196,197,198}{Pt}}
\label{subsec:n:190-198Pt}


ENDF/B-VIII.0 contains nine Pt isotopes adopted from TENDL-15 -- a large scale evaluation project using the TALYS code~\cite{TALYS}. It is safe to assume that all these evaluations are at least nine years old and, therefore, a more focused and more up-to-date approach would be beneficial.  There are 22 isomers in the residues of  neutron interaction with Pt isotopes. Some of them might be of practical interest but are not included in the current evaluations. The new evaluation is trying to fix these gaps by providing cross sections for isomeric or unstable ground states if the cross sections are of a measurable size. 

A fair amount of experimental data, mostly for  abundant isotopes and natural element, are available. They are not enough, however, to fully pinpoint evaluations, especially as many of the available experiments are not very useful.  Thus, the evaluation procedure must, to a large extent, rely on the model calculations. 

\paragraph{Evaluation in the fast neutron energy range\newline}
The chain of isotopes must be evaluated in parallel in order to make use of experimental data taken on the natural element, 
which often are the most reliable source of information. In the case of Pt, the evaluation methodology was the same as the one 
used for $^{181}$Ta  detailed in the previous section. Dealing with a chain of several isotopes implies, however,  that the 
consistency requirement must be extended. We need to ensure that the same model parameters are used when a given isotope appears 
in different calculations. Technically, this condition was imposed by defining the common input file containing all input parameters, 
except those pertinent to the particular target, such as target A and Z, or CC coupling scheme. This file was included 
while reading input files for the individual isotopes. 
In contrast to $^{181}$Ta, we employed the Hofmann, Richert, Tepel, and Weidenm\"uller (HRTW)~\cite{HRTW} width fluctuation correction and EGSM level 
densities. The consistency was compromised only for $^{198}$Pt, where GC level densities were utilized. This deviation 
from the rule was attributed to the fact that GC level densities provided a more accurate representation of the experimental 
data, and we prioritized agreement with data over consistency of the models. From a physics perspective, the ability to reproduce data 
with a consistent set of parameters enhances our confidence in the accuracy of our reaction modeling. However, the necessity to 
compromise consistency for one of the nine isotopes suggests that our understanding, likely of level densities, may not be comprehensive.

\begin{figure}[!htbp]
\centering
\includegraphics[scale=.60]{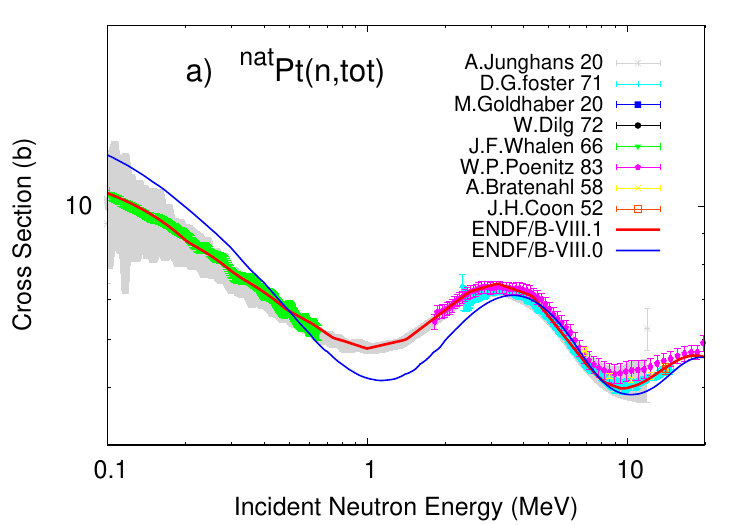}
\includegraphics[scale=.60]{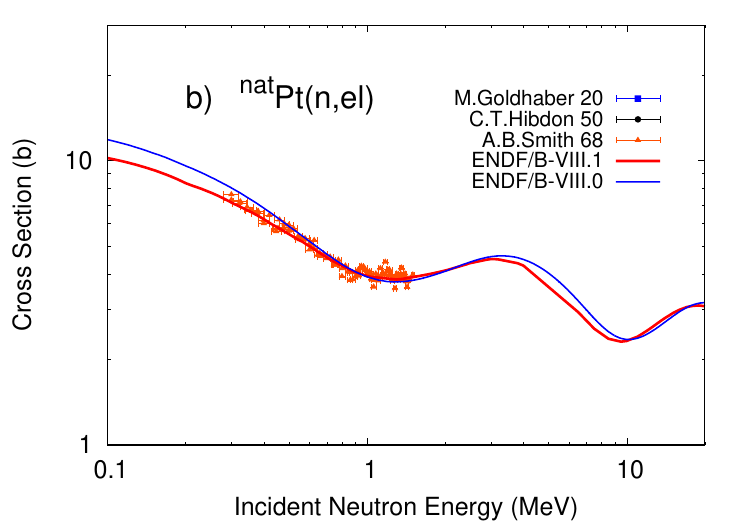}
\includegraphics[scale=.23]{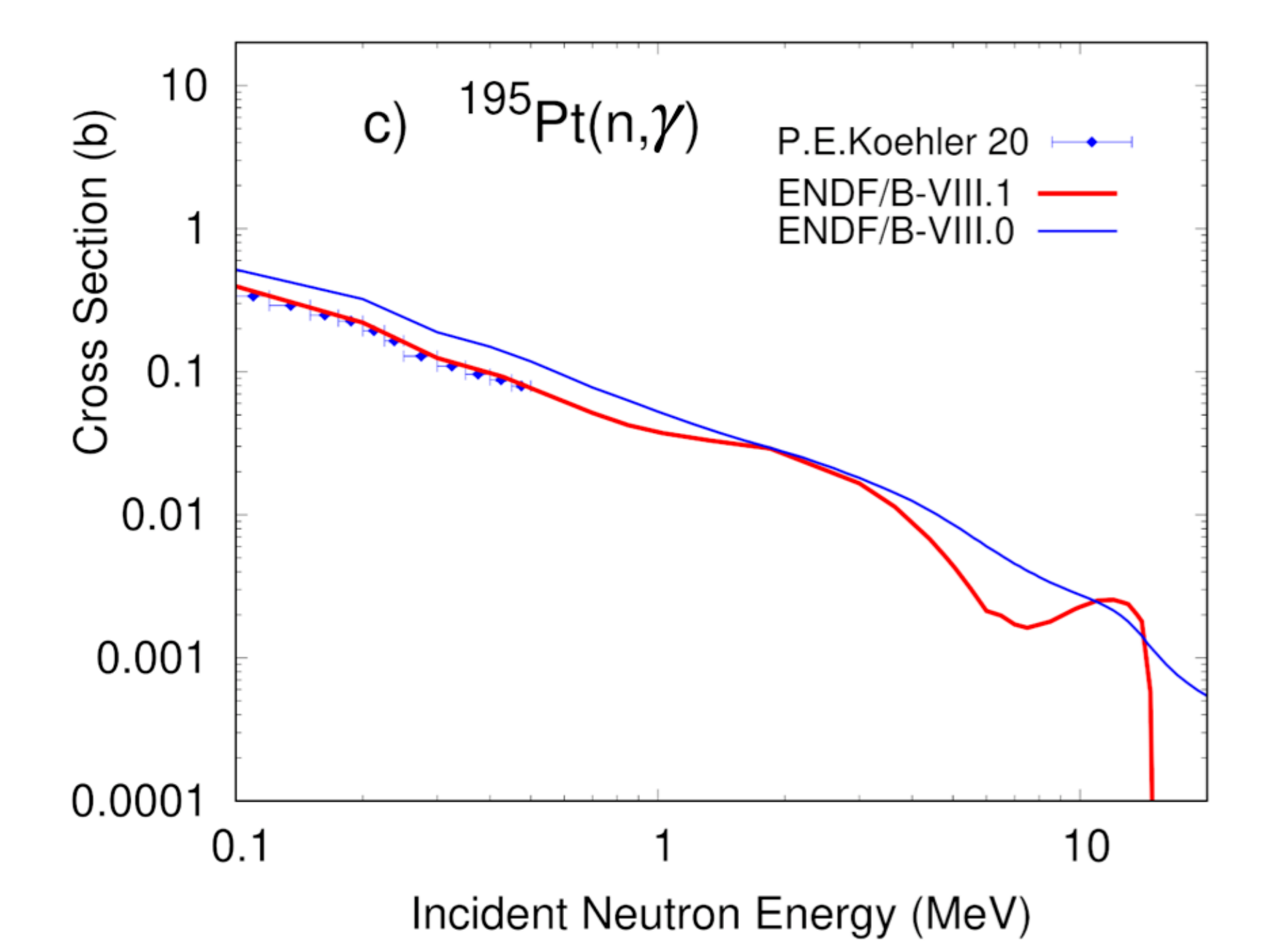}
\includegraphics[scale=.60, clip, trim= 0mm 0mm 5mm 0mm]{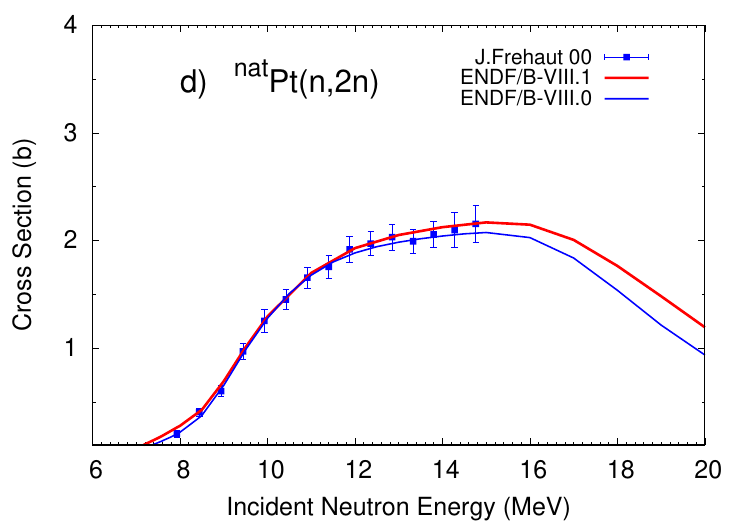}
\vspace{-.1in}
\caption{Evaluated cross sections for total, elastic, capture and (n,2n) compared with the experimental data and ENDF/B-VIII.0.  Except capture, all reactions are for the natural target and their cross sections are weighted sum over all stable isotopes. Data taken from Refs.\cite{Junghans:2020,Foster:1971,Goldhaber:1937,Dilg:1972,Whalen:1966,Poenitz:1983,Bratenahl:1958,Coon:1952,Hibdon:1950,Smith:1968,Koehler:2008,Koehler:2013,Frehaut:2016}.  }
\label{Pt00fig}
\end{figure}

Soukhovitskii-Capote CC potential (RIPL \#1483~\cite{RIPL3}) was adopted for all Pt isotopes with their respective level-scheme couplings. This potential, originally developed for gold, turns out to work so well on platinum that we labeled it the ``Jewels Potential.''  It calculates total cross sections for elemental Pt much better than the potential used in TENDL evaluations (see Fig.~\ref{Pt00fig}).
The same is confirmed also for $^{194}$Pt.  Elastic cross sections on the natural Pt are very good as well. Excellent agreement is obtained for elastic angular distributions  on $^{194}$Pt at 2.5~MeV and 4.55 MeV.
Inelastic cross sections for  $^{194,196,198}$Pt  are considerably higher than those in ENDF/B-VIII.0 although still slightly below the experimental values. 
The agreement for the inelastic angular distributions is fair, which is quite typical for these quantities. 

The shape of the calculated capture cross sections is very similar to the Koehler~\cite{Koehler:2013} measurements (see 3rd plot of Fig.~\ref{Pt00fig}).  We only scaled down calculated cross sections with a single, energy-independent factor per each isotope. These factors, which multiply the original MLO1 $\gamma$-strength, are closer to 1 than the default factors resulting from normalization to the thermal $\gamma$ width.  Choosing to follow isotopic measurements by Koehler, we underestimate by about 5\% the respectable experiments on the natural platinum that were very well reproduced with the default (thermal normalization)  calculations. There is a disagreement between these experiments and those by Koehler, and we hope that the new LANSCE results will help to make a final decision.

Elemental (n,2n) cross sections by Frehaut~\cite{Frehaut}, corrected by a factor of 1.078 as suggested by Vonach \cite{Vonach:1990},  are  well reproduced by the new evaluation. The ENDF/B-VIII.0 evaluation is 
a bit lower but equally good. The two evaluations differ mostly in the high energy tail where no experimental data exist, as shown in the bottom plot of Fig.~\ref{Pt00fig}.

In the case of (n,p) and (n,$\alpha$) reactions, we followed primarily data by Filatenkov~\cite{Filatenkov2016_exfor41614.1}. Ground-state cross sections agree with ENDF/B-VIII.0 within uncertainties. No isomeric meta state cross sections are given in the current ENDF/B-VIII.0 evaluations. Our (n,p) isomeric and ground-state cross sections reproduce measurements but at the price of modifying particle-hole spin distribution in the preequilibrium emission of protons.
Typically, this distribution is assumed to be $\sigma^2 = f \times n \times A^{1/3}
$
where $n$ is a number of excitons (in this case $n=2$) and A stands for the isotope mass number. Factor $f$ is commonly set to 0.24. We had to reduce this factor to 0.12 for $^{194}$Pt(n,p), 0.14 for $^{195}$Pt(n,p), and 0.04 for $^{196}$Pt(n,p). 
This rather  significant reduction is likely related to the fact that Multistep Direct mechanism, that is here simulated by the exciton model, favors low angular momentum transfer populating states with the spin close to the ground state.

Fast neutron evaluations for nine Pt isotopes were merged with ENDF/B-VIII.0 resonance regions.

\paragraph{Covariances\newline}
The covariances were calculated using a Kalman filter that combined the above model calculations with available experimental data on all of the Pt isotopes. Each experimental data set was weighted
by $1/ n$, where $n$ is the number of data points in each data set to eliminate the need for a posteriori scaling on any channel uncertainty. For each isotope, the same set of 53 model parameters were varied. We note that the optical model parameters included not only the incident channel but also  outgoing neutrons, protons, $\alpha$s,  deuterons, tritons, and $^3$He. The level density parameters  for 11 residuals, and four parameters in  the preequilibrium emission model were varied.
Using the Kalman filter, the prior parameter uncertainties will be reduced where constraining experimental data are available. For the unstable isotopes, Kalman propagates the default parameter variations, unless the parameters are constrained by other targets. These additional constraints are one benefit of performing such a simultaneous evaluation, as they provide realistic cross sections and uncertainties even though no experimental data are available.

 In addition to providing the correlations between incident energies in each channel, we also provide cross correlations between different reaction channels.  As expected, the cross correlations between elastic and total cross sections show similar structure to their in-channel correlations (which are similar in structure to one another). There are also significant correlations between the total cross section and the inelastic and capture channels. On the other hand, the correlations between the total cross section and both (n,p) and (n,$\alpha$) cross sections are close to zero.

One additional benefit of performing this type of simultaneous evaluation is that cross-isotope covariances could be provided between the various Pt isotopes. These are currently not included in ENDF/B-VIII.1 but could be constructed and formatted if they were of interest.

%
%
%

\subsubsection{\nuc{206,207,208}{Pb}}
\label{subsec:n:206-208Pb}


The major isotopes of lead, $^{206,207,208}$Pb, constitute 98.6\% of the natural element. Full reevaluation of these isotopes was undertaken as part of the US's Nuclear Energy University Program (DOE-NEUP) endeavors \cite{Proposal} towards lead fast reactors and accompanying technologies \cite{LFRRoad}. A comprehensive examination of all benchmark experiments containing lead were used to prioritize reactions channels and isotopes for the evaluation \cite{ALARM_CF,HMF27,PMF35,HMF64}. Across the board, $^{208}$Pb elastic scattering has the largest impact to system performance as $^{208}$Pb comprises over 52\% of natural lead and has an uncharacteristically high inelastic threshold for its mass. This characteristic derives from the fact that $^{208}$Pb is a double magic nucleus. Other outcomes of the double magic nucleus are a large level spacing and resonances that can be resolved well past 1 MeV. Similarly, as $^{206}$Pb and $^{207}$Pb are near the closed shell, these isotopes also have sufficiently large level spacing to resolve resonances higher than 0.5 MeV. Since the evaluation focused on cross sections relevant for fast spectra systems, neutron energies above 100 keV were prioritized. However, the RM formalism of R-matrix used in the evaluation exhibits strong dependency on the channel radius, distant, and bound levels. This means it was then necessary to perform evaluations down to $10^{-5}$ eV for consistency within the RRR. The RRR evaluations were performed using the R-matrix code \SAMMY~\cite{SAMMY} and the fast region  evaluations were modeled with the Hauser-Feshbach code  CoH3 \cite{COH3}.

\begin{figure}[!h]
      \centering
      \includegraphics[width =\columnwidth]{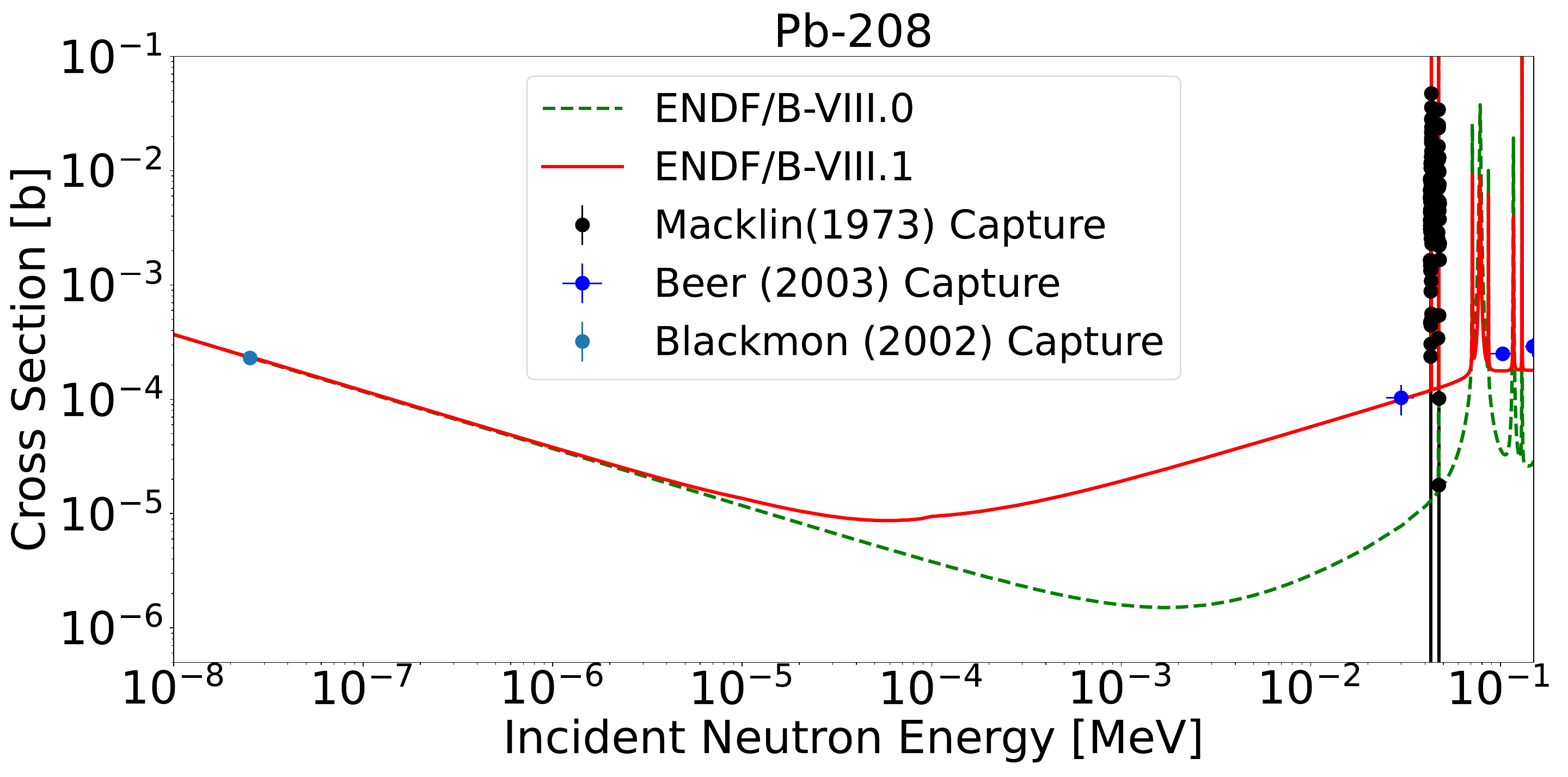}
      \caption{New capture cross section of $^{208}$Pb compared to ENDF/B-VIII.0 \cite{Brown2018} and data from Macklin \cite{Macklin208}, Blackmon \cite{Blackmon}, and Beer \cite{BEER}.}
      \label{fig:Pb208}
\end{figure}

Comparison of the $^{208}$Pb resonance parameters from ENDF/B-VIII.0 to the ORELA transmission measurement by Carlton \cite{Carlton} yields good agreement. Little was done in terms of adjustments to neutron widths ($\Gamma_n$) below 1 MeV as the total (and by extension the elastic) cross sections are well constrained. Time was spent on the capture cross section of $^{208}$Pb around 30 keV, relevant for the 30 keV MACS. Here, the capture cross section is dominated by compound reactions of two p-wave resonances near 40 keV. However, several measurements make note of evidence of direct capture in the region below 40 keV where no resonances have been reported \cite{BEER,Macklin208}. An energy-dependent background was provided in MF-3 to account for this direct capture component which adds an order of magnitude to the capture cross section in between resonances, Fig.~\ref{fig:Pb208}. The thermal capture cross section is constrained by Blackmon \cite{Blackmon} with $\sigma_\gamma = 230\; \mathrm{\mu b}$. The evaluator acknowledges conflicting accounts of this value from D.J. Hughes \cite{Hughes} and J.F. Emery \cite{Emery} who both claim $\sigma_\gamma = 470\; \mathrm{\mu b}$; however, documentation for either of these values cannot be obtained. Extension of the RRR from 1.0 to 1.5 MeV was deemed necessary to account for scattering anisotropy observed in RPI quasi-differential high energy scattering experiments \cite{Youmans} that was attributed solely to $^{208}$Pb. A novel methodology of employing quasi-differential scattering data simulations in MCNPv6.2 \cite{MCNP6.2}, \NJOY2016 \cite{NJOY}, and \SAMMY\ to validate $(\ell,J)$ states of resonances was developed in order to provide additional information of doublets observed while fitting transmission data. The new RRR evaluation of $^{208}$Pb was joined with a LANL evaluation above 1.5 MeV
, resulting in new cross sections from 10$^{-5}$ to 20 MeV.

The $^{206}$Pb RRR evaluation focused on updating the radiation widths, $\Gamma_\gamma$, from two differential capture yields by Domingo-Pardo \cite{Pb206Domingo} and Borella \cite{Pb206Borella}. These capture yield data provide the basis for adding 50 minor p- and d-wave resonances below 900 keV. However, the effects of these resonance are only observable in statistical plots as they contribute little to overall system performance. The fast region evaluation was undertaken to fit inelastic gamma production data from Negret \cite{Pb206Negret}. The purpose of this step is to add resonance structure to both the elastic and inelastic channels where previous evaluations had resonance structure only in the elastic channel. Post model fitting, the computed inelastic and elastic cross sections from CoH3 were normalized by the calculated total cross section and multiplied by total cross section measurements \cite{Pb206Horen}. The overall outcome is that now the \ENDF\ inelastic cross section  is larger than that of the ENDF/B-VIII.0.

\begin{figure}[!h]
      \centering
      \includegraphics[width =\columnwidth]{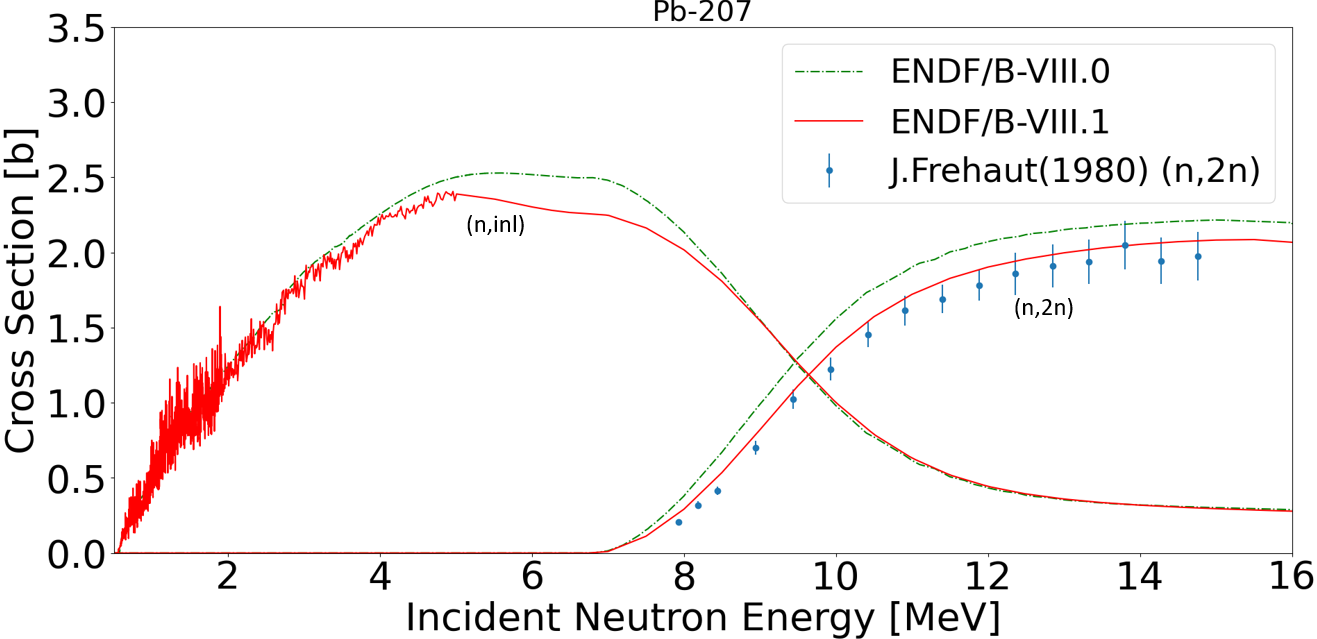}
      \caption{New $^{207}$Pb inelastic and (n,2n) cross sections compared to ENDF/B-VIII.0 and Frehaut \cite{Frehaut}. }
      \label{fig:Pb207}
\end{figure}

Much like $^{206}$Pb, the $^{207}$Pb evaluation encompassed updates to the inelastic scattering and (n,2n) cross sections from measurements by Mihailescu \cite{Pb207Mihailescu} as well as Frehaut (2016 version  \cite{Frehaut} including Vonach 1.078 normalization correction \cite{Vonach:1990}), as seen in Fig.~\ref{fig:Pb207}. In addition, updates to $\Gamma_\gamma$ were performed using Domingo-Pardo \cite{Pb207Domingo} data and resonance parameters adopted from JENDL-4.0 \cite{JENDL-4.0} as the library supports a higher upper energy of the RRR. Besides refitting to Domingo-Pardo capture yield data and Horen transmission \cite{Pb207Horen}, special care was taken in the bound levels for this isotope. Thermal scattering lengths and thermal capture ratios calculated from resonance parameters in the previous evaluation pointed to less than ideal prediction of incoherent scattering \cite{Sears1992} and the capture from the $(\ell=0, J=1)$ state which dominates thermal capture. Changes to the energies and strengths of the bound s-wave resonances have since fixed this issue, giving thermal capture cross sections that are consistent with the corresponding measured gamma spectra.

\begin{figure}[!h]
      \centering
      \includegraphics[width =\columnwidth]{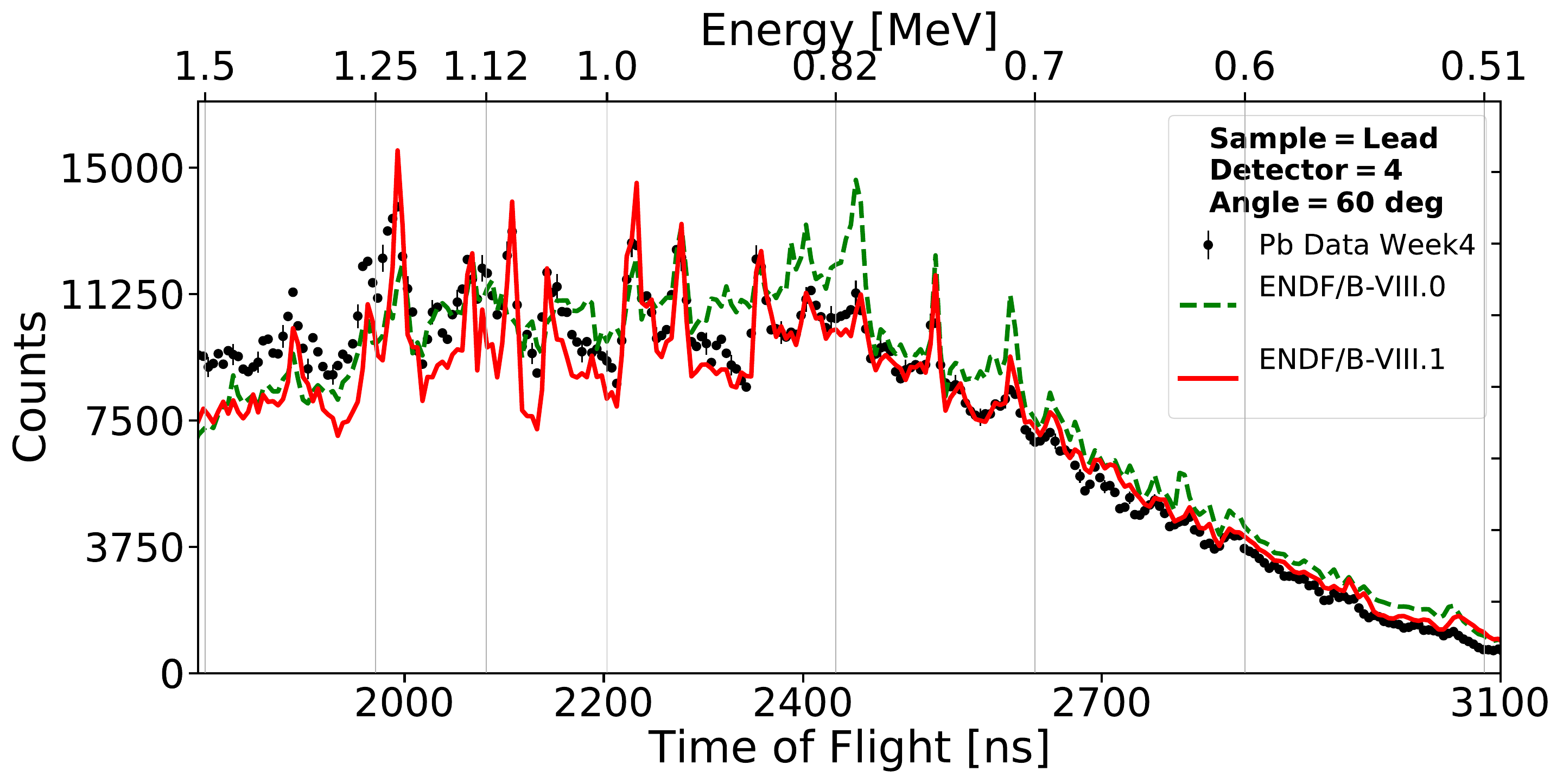}
      \caption{Comparison of simulation and experiment of high energy quasi-differential scattering experiment of natural lead \cite{Youmans} from ENDF/B-VIII.0 and ENDF/B-VIII.1.   }
      \label{fig:Scatter}
\end{figure}

New elastic scattering angular distributions were calculated for all isotopes within their respective RRR using the Blatt-Biedenharn formalism in \NJOY2016. Fast region scattering distributions are likewise updated from the CoH3 calculations. Nuclear data covariance for total, elastic, inelastic, capture, and (n,2n) are provided for all isotopes in the evaluation, the representations of which are split between MF-32 and MF-33. For MF-32, \SAMMY\ generated covariance in addition to scattering radii uncertainty provide sufficient quantification of uncertainties. In the fast region, a CoH3-KALMAN approach \cite{Lovell} was undertaken in Python. Uncertainties for elastic scattering angular distributions, MF-34, are provided for $^{208}$Pb as this is the only isotope able to be quantified in natural quasi-differential scattering data. Validation of the lead isotopes was done with critical, shielding, and scattering benchmarks. There is great improvement over ENDF/B-VIII.0 for fast critical benchmarks and scattering experiments resulting largely from $^{208}$Pb scattering updates observable in scattering data, Fig.~\ref{fig:Scatter}.

Specifically, fast critical benchmarks go from an average bias of -350 pcm with ENDF/B-VIII.0 to +70 pcm with ENDF/B-VIII.1. These benchmarks include PU-MET-INTER-004 \cite{PMI4}, HEU-MET-FAST-064 \cite{HMF64}, HEU-MET-FAST-027 \cite{HMF27}, PU-MET-FAST-035 \cite{PMF35}, and MIXED-MET-FAST-006 \cite{MMF6}.  Shielding benchmarks prove to be rather insensitive to the lead library chosen except for rather thick cases, 40 cm or more of lead \cite{ALARM_TRAN,ALARM_CF}. The effect of improving the fast assemblies resulted in a higher bias on thermal benchmarks when compared to ENDF/B-VIII.0. While a portion of the ``worsening deviation'' from C/E $=$ 1 is expected and supported from changes in the high energy scattering distributions, an equal portion was a result from the RRR analysis. Upon consensus of the CSEWG committee, a small increase in the $^{207}$Pb capture cross section at thermal energies was included as well as lowering the thermal elastic scattering cross sections of all three isotopes. Both adjustments are supported by quantities from NIST \cite{Sears1992} and integral measurements. Still, lead-reflected water moderated LEU assemblies trend high, and effort should be expended to verify the benchmark geometries. The evaluator points to high energy capture measurements on $^{207}$Pb and quasi-differential scattering of enriched samples for new measurements to be performed. In addition, the current ensemble of fast critical benchmarks \cite{ICSBEP} are not sufficient in terms of uncertainty analysis and geometry description to be considered adequate for nuclear data validation. New sub-critical lead-reflected plutonium experiments have been proposed as a remedy for this lack of well-validated integral experiments \cite{PbSubcrit}.

\subsubsection{\nuc{234}{U}}
\label{subsec:n:234U}

\begin{figure}
\includegraphics[scale=0.45,clip=true,trim=0mm 0mm 0mm 0mm]{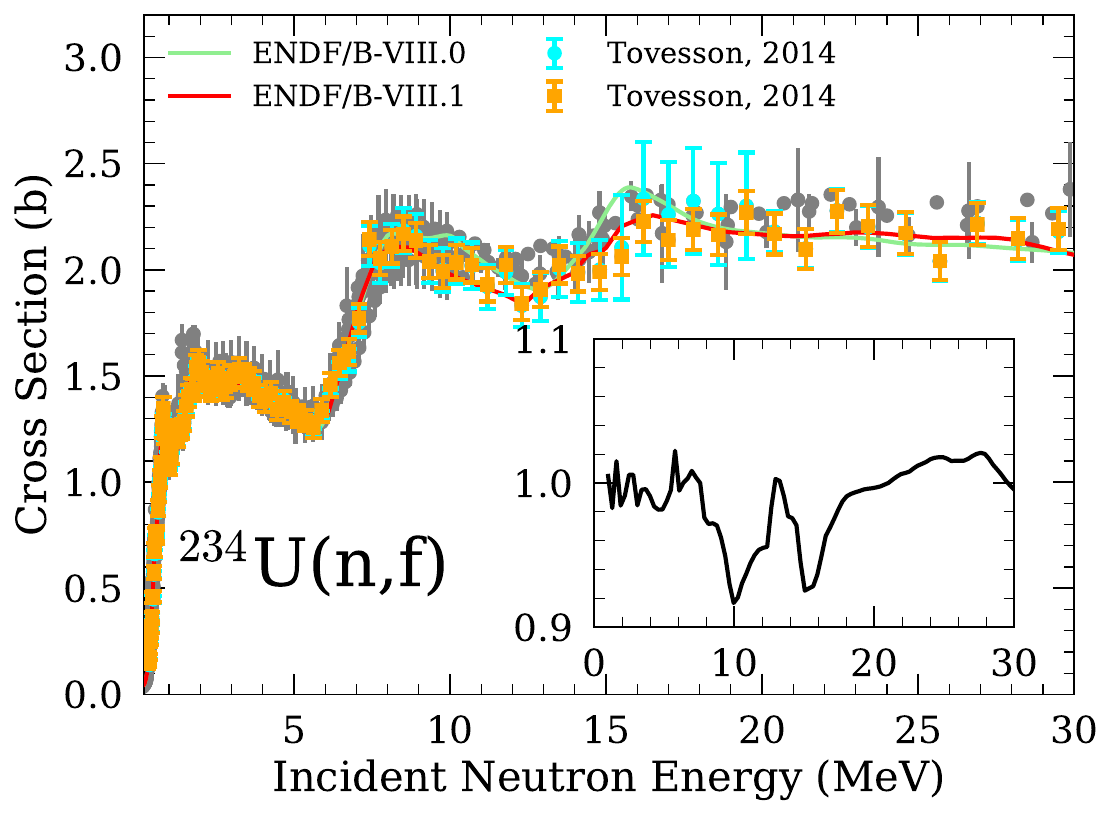}
\caption{
\label{fig:u4nf}
Cross section for the $^{234}$U(n,f) reaction. In the insert we show the ratio of the ENDF/B-VIII.1 fission cross section to ENDF/B-VIII.0. The gray data points are other available experiments except for Tovesson \textit{et al.} \cite{doi:10.13182/NSE13-56}. }
\end{figure}

In this update, we have performed a new GLS fit of the fission cross section data, including the newest Tovesson data \cite{doi:10.13182/NSE13-56}, which were not considered in the previous evaluations and have smaller reported uncertainties. The energy range in which the fit was performed was 0.8 MeV to 30 MeV. Below 0.8 MeV, we have kept the previous evaluation. As can be seen from Fig.~\ref{fig:u4nf}, the largest changes are between 9~MeV and 17~MeV, where the fit was strongly influenced by the new Tovesson data due to their smaller uncertainties. 

The main theoretical tool was the code CoH$_3$, which was used to generate the total cross section and the individual channel cross sections. Since the model is not flexible enough for the fission channel, we have used an energy-dependent adjustment of the transmission in the fission channel, so that the calculated fission cross section reproduces exactly the evaluation based on a direct fit to the data. 

\begin{figure}
\includegraphics[scale=0.45,clip=true,trim=0mm 0mm 0mm 0mm]{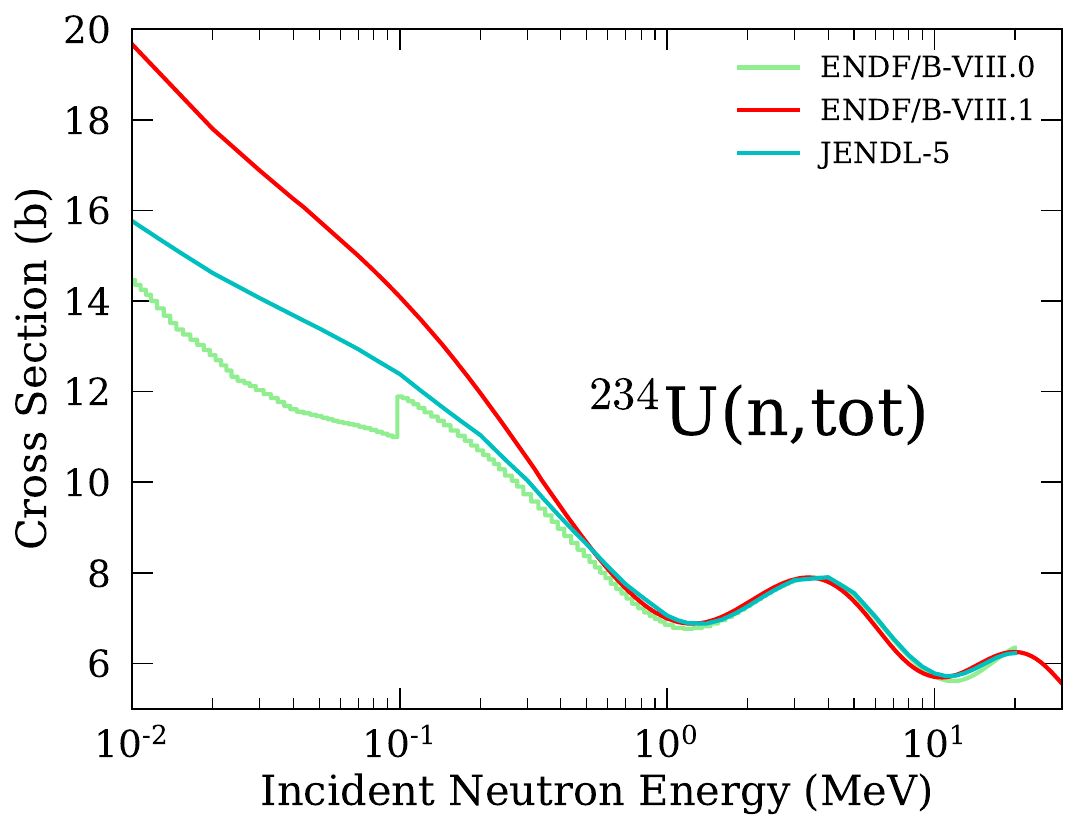}
\caption{
\label{fig:u4tot}
Cross section for the $^{234}$U(n,tot) reaction.}
\end{figure}

For the total cross section plotted in Fig.~\ref{fig:u4tot}, we have employed the Soukhovitski\~{\i} 2005 optical model potential \cite{Soukhovitskii2005}. The calculated total cross sections on $^{234}$U target below 500 keV is much larger than in ENDF/B-VIII.0 and JENDL-5 libraries despite the same optical model potential being used. This points to inconsistencies in the undertaken CC calculations as well as used deformation parameters ($\beta_2=0.215$, $\beta_4=0.109$ and $\beta_6= 0.00039$). Optical model calculations will have to be carefully revisited in a future reevaluation effort. Observed total cross section differences have a negative impact on calculated neutron strength functions at 1 keV, which is a well-defined quantity \cite{RIPL3}. The main impact of the new evaluation on criticality benchmarks is due to the changed capture and fission cross sections in the fast range. However, the elastic cross section is also increased by the too-high total cross section below 500 keV.

\begin{figure}
\includegraphics[scale=0.45,clip=true,trim=0mm 0mm 0mm 0mm]{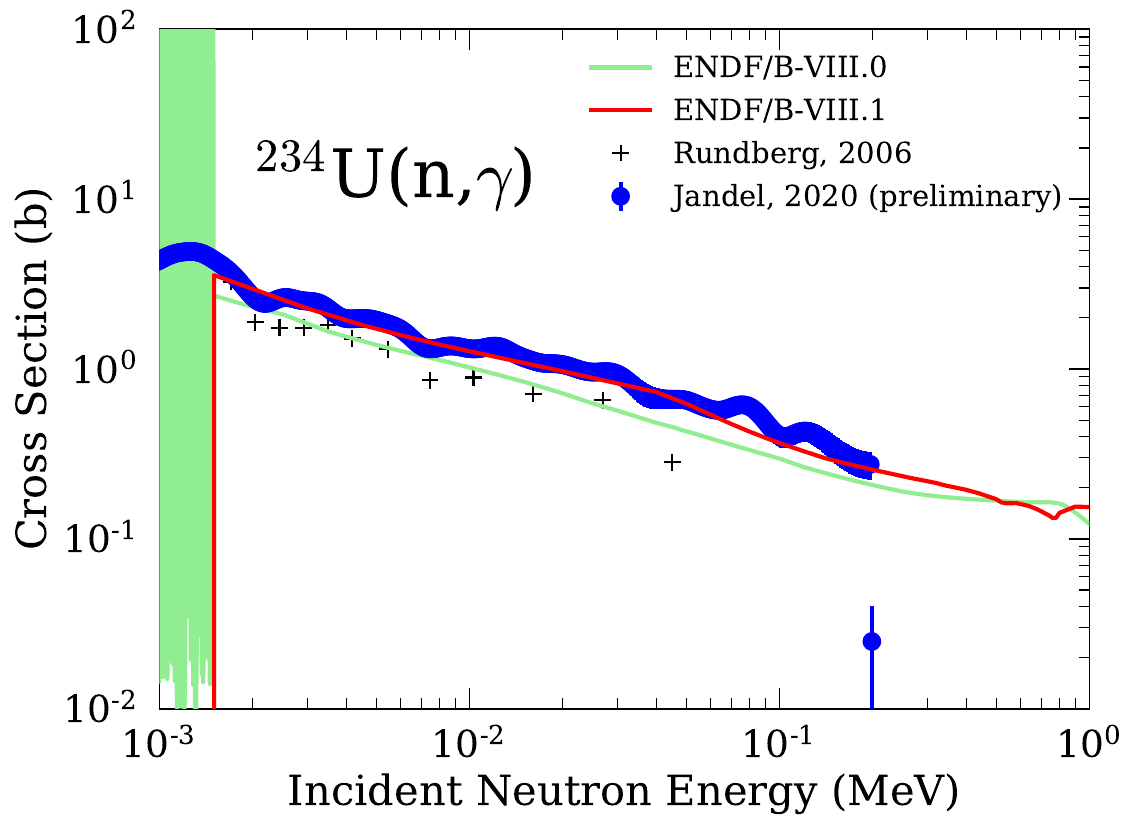}
\caption{
\label{fig:u4ng}
Cross section for the $^{234}$U(n,$\gamma$) reaction.}
\end{figure}

The capture cross section is shown in Fig.~\ref{fig:u4ng}, and it is based on CoH$_3$ calculations with enhanced $M1$ transitions. As  can be observed in Fig.~\ref{fig:u4ng}, the new evaluation is in good agreement with the preliminary data from an earlier DANCE measurement and higher than in ENDF/B-VIII.0. The $^{234}$U intentionally matches the new Los Alamos Jandel data and not the old Los Alamos (unpublished) Rundberg 2006 data. Note, however, that a final analysis of the DANCE data is not available yet.

All the other channels, except elastic, were taken from CoH$_3$ calculations. The elastic cross section was calculated as the difference between total and non-elastic. Significant deviations from the previous evaluations can be observed for (n,n'), (n,2n), and (n,3n), but there is no data to invalidate the newest evaluation. However, we note that the peak of the (n,2n) cross section follows an expected trend when looking at $^{236}$U (new evaluation) and $^{238}$U evaluation in ENDF/B-VIII.0. The newest (n,3n) peak is in better agreement with the JENDL-5.

\begin{figure}
\includegraphics[scale=0.45,clip=true,trim=0mm 0mm 0mm 0mm]{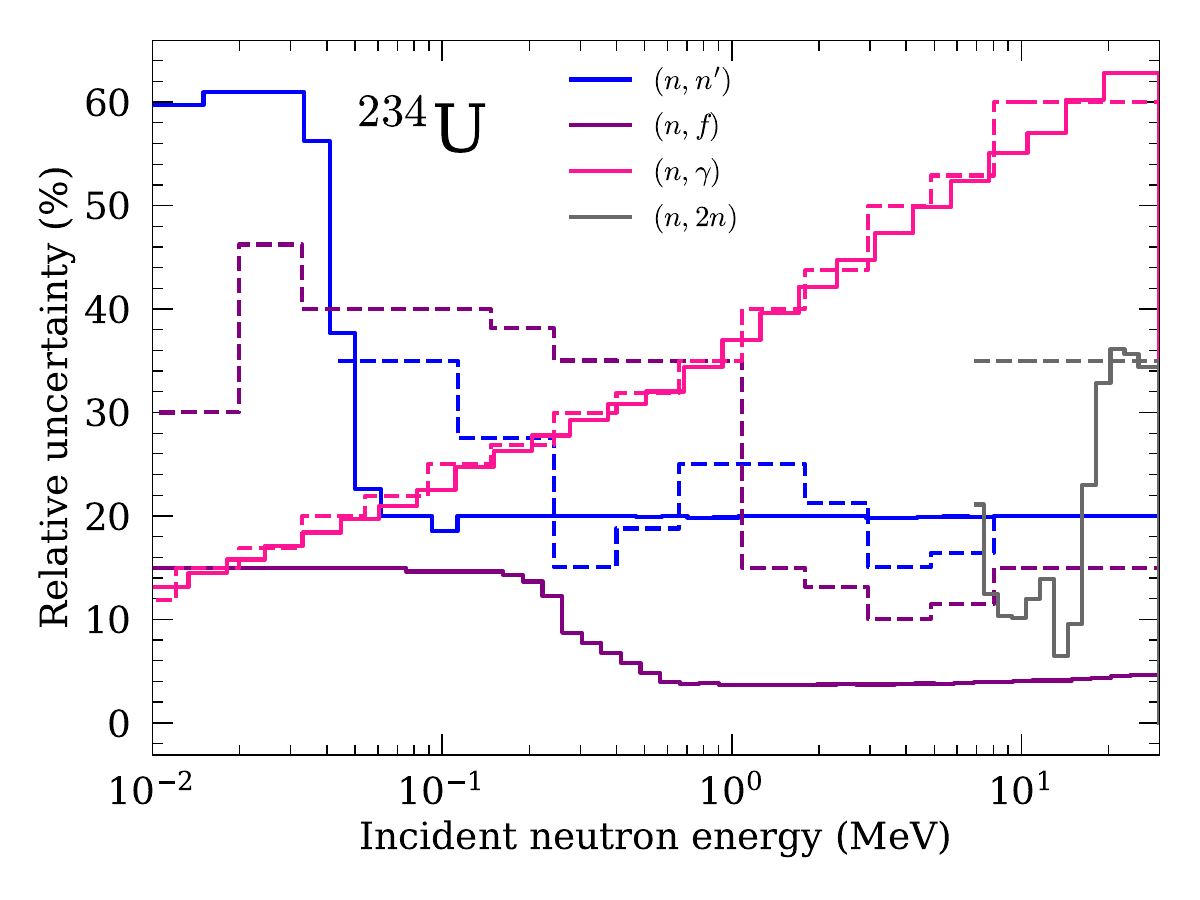}
\caption{
\label{fig:u4uncert}
Comparison between the relative uncertainties for $^{234}$U from ENDF/B-VIII.0 (dashed) and ENDF/B-VIII.1 (solid) for $(n,n^\prime)$, $(n,f)$, $(n,\gamma)$, and $(n,2n)$.}
\end{figure}

The correlation matrices in all channels have been updated using a Kalman filter, including available experimental data in all channels with sensitivities calculated to CoH$_3$ parameters.  For fission, the shape of the uncertainties from Kalman was used below 300 keV and matched to the magnitude of the ENDF/B-VIII.0 uncertainties, and above 300 keV they have been adjusted to match the Tovesson data.  The capture uncertainties follow the shape and magnitude of ENDF/B-VIII.0 uncertainties.  For the total cross section, the uncertainties come from Kalman below 10 keV; above 10 keV, the uncertainties remain at the value at 10 keV (about 3.5\%).  For the inelastic channel, the shape of the Kalman uncertainties were scaled so that we reproduce the average magnitude from ENDF/B-VIII.0 with a minimum set to 20\%, as recommended by templates of expected uncertainties \cite{nxn}.  For all other channels, the uncertainties from Kalman were scaled so that the maximum values were close to the ENDF/B-VIII.0 values at about 30\%, while keeping the shape of the energy-dependence extracted from Kalman.  Fig.~\ref{fig:u4uncert} shows the comparison between the relative uncertainties of ENDF/B-VIII.0 and ENDF/B-VIII.1 for the inelastic, fission, capture, and $(n,2n)$ channels.

\subsubsection{\nuc{236}{U}}
\label{subsec:n:236U}

\begin{figure}
\includegraphics[scale=0.45,clip=true,trim=0mm 0mm 0mm 0mm]{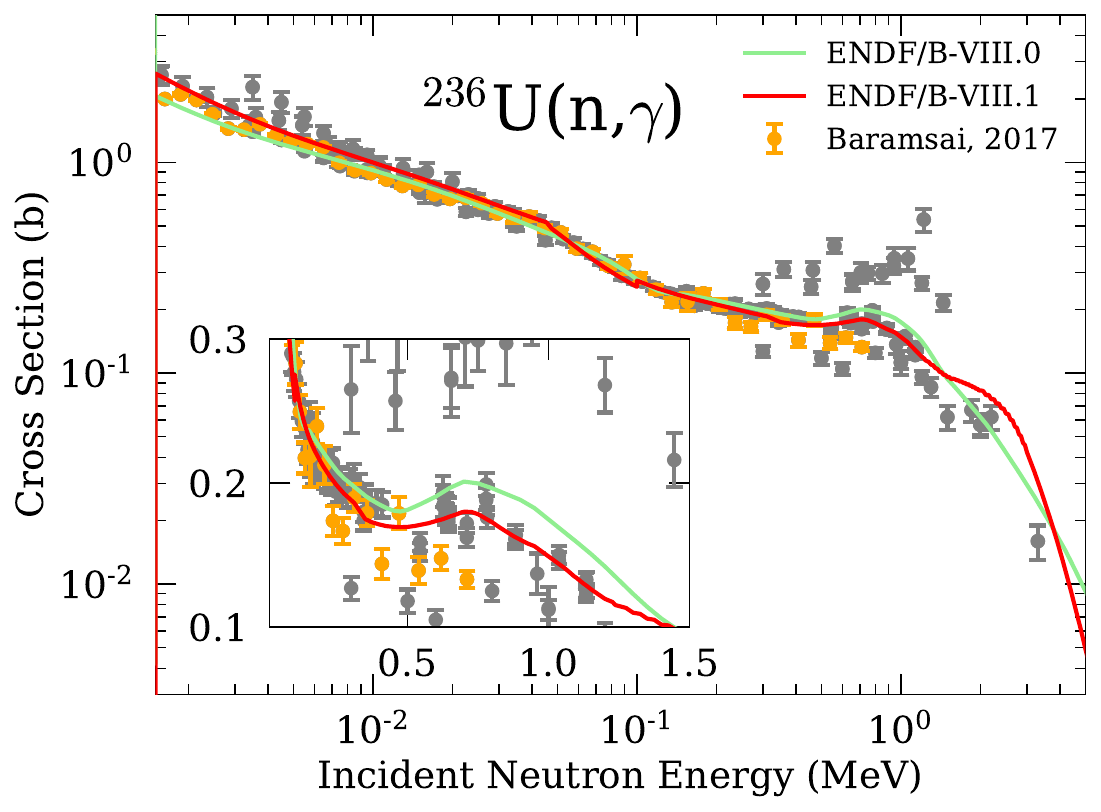}
\caption{
\label{fig:u6ng}
Cross section for the $^{236}$U(n,$\gamma$) reaction. In the insert we show the same quantity in linear scale up to 1.5 MeV incident energy. The gray data points are other available experiments except for Baramsai \textit{et al.}  \cite{PhysRevC.96.024619}.}
\end{figure}

We have used a very similar methodology to update the evaluation of neutron-induced reactions on $^{236}$U with the one for neutron-induced reactions on $^{234}$U targets presented in Section~\ref{subsec:n:234U}. The only change in methodology versus $^{234}$U was that we have used Kalman to refit the capture cross section above 100 keV to the available data. The CoH$_3$ calculation reproduce well some of the data points from LANSCE \cite{PhysRevC.96.024619}, but it tends to be low with respect to other data sets in this region. Hence, we have performed a parametric Kalman fit in which we also increased the uncertainties of the LANSCE measurement. This brought the evaluation of the capture cross section in better agreement with the available measurements and lowered it with respect to the ENDF/B-VIII.0 evaluation, as showed in Fig. \ref{fig:u6ng}.

\begin{figure}
\includegraphics[scale=0.16,clip=true,trim=0mm 0mm 0mm 0mm]{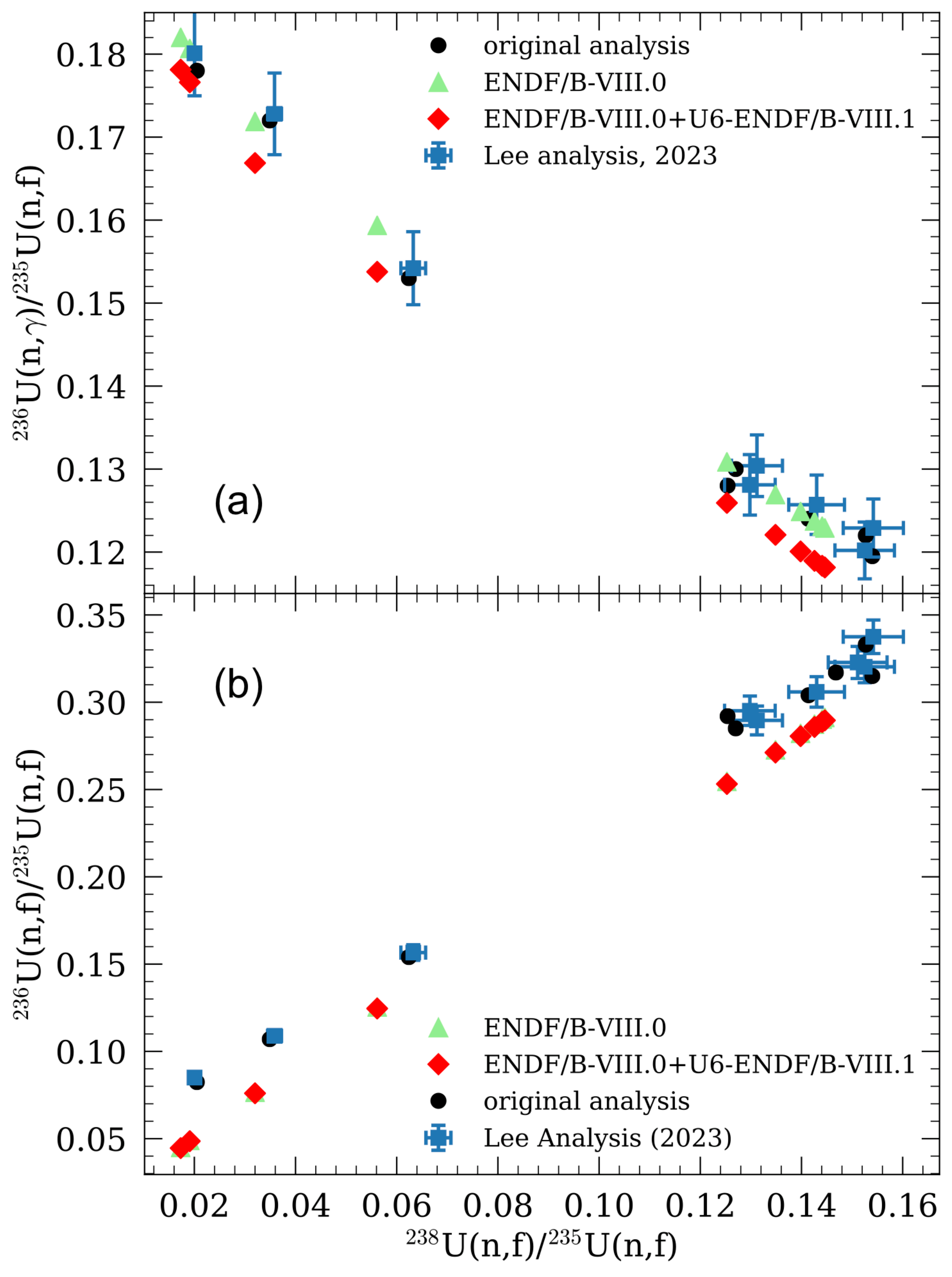}
\caption{
\label{fig:u6rr}
Comparison of the measured and experimental values of the ratio of the $^{236}$U capture rates to $^{235}$U fission rates (a) as a function of spectral index at different locations in the Flattop-25 critical assembly, and the ratio of the $^{236}$U fission rates to the $^{235}$U fission rates (b) as a function of the same spectral index.}
\end{figure}

We have also updated the total cross section and the fission cross section in the fast region. In both cases, the change from the ENDF/B-VIII.0 evaluation is of a few percent, with a maximum of about 5\% change. This relatively small change is unlikely to impact any benchmark that includes $^{236}$U. An example is shown in Fig.~\ref{fig:u6rr}(a), where we show the reaction rates in the Flattop 25 critical assembly.

$^{236}$U capture reaction rates now in ENDF/B-VIII.1 are lower than for ENDF/B-VIII.0, as shown in the upper panel of Fig.~\ref{fig:u6rr}, and are below the measurements (but are generally within or just outside the $\sigma$ uncertainties). This reflects our intentional decision to lower the capture cross section based on our analysis of the cross section data.  

\begin{figure}
\includegraphics[scale=0.45,clip=true,trim=0mm 0mm 0mm 0mm]{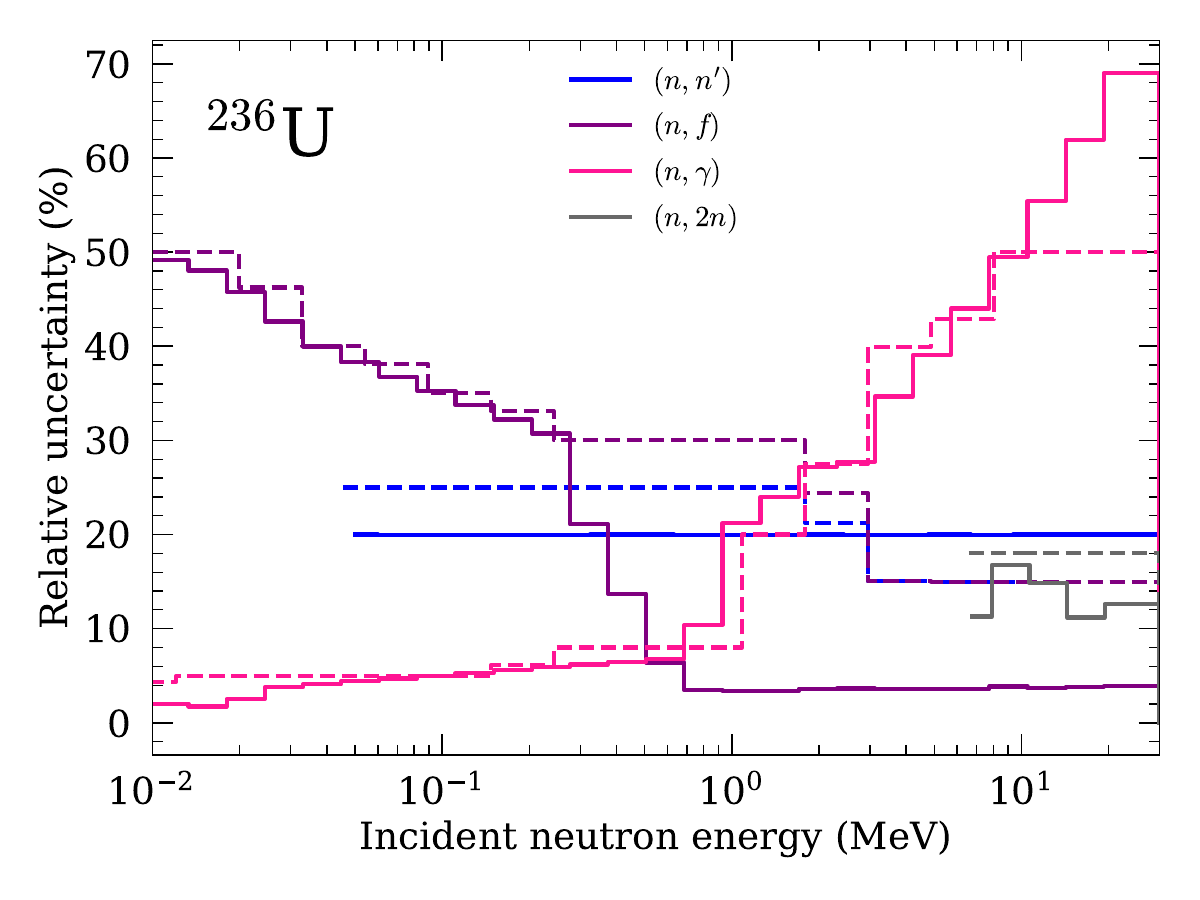}
\caption{
\label{fig:u6uncert}
Comparison between the relative uncertainties for $^{236}$U from ENDF/B-VIII.0 (dashed) and ENDF/B-VIII.1 (solid) for $(n,n^\prime)$, $(n,f)$, $(n,\gamma)$, and $(n,2n)$.}
\end{figure}

The correlation matrices in all channels have been updated using a similar procedure as for $^{234}$U. The capture uncertainties have been evaluated based on DANCE data below 1 MeV, and above we use a logarithmic function that follows the ENDF/B-VIII.0 uncertainties.  For fission, the ENDF/B-VIII.0 uncertainties have been used below 300 keV, and above 800 keV they have been adjusted to match Tovesson. Between 300 and 800 keV, the uncertainties linearly decrease. Inelastic uncertainties have been set to 20\%, as recommended by templates of expected uncertainties \cite{nxn}. For all other channels, the uncertainties from Kalman were scaled so that the maximum values were close to the ENDF/B-VIII.0 values at about 17\%, while keeping the shape of the energy-dependence extracted from Kalman.  Figure~\ref{fig:u6uncert} shows the comparison between the relative uncertainties of ENDF/B-VIII.0 and ENDF/B-VIII.1 for the inelastic, fission, capture, and $(n,2n)$ channels.

\subsubsection{Adoption of IRDFF-II neutron dosimetry evaluations}
\label{subsec:n:IRDFF}


The \mbox{IRDFF-II} library~\cite{IRDFF} contains 119 reaction entries and addresses incident neutron energies from 0 to 60~MeV. IRDFF evaluations include consistent mean values and covariances~\cite{IRDFF} which were adopted where possible in the ENDF/B-VIII.1 library. Only the reactions on isotopic materials were considered to update the ENDF/B-VIII.1 library. The IRDFF evaluations have been extremely well-validated and generally show at least equal or better performance than the corresponding \prENDF{} evaluations. This is illustrated in Table~\ref{tab:SACS-LR0} with a comparison of measured SACS in the LR-0 reactor near Prague \cite{kostal2024a} versus calculated SACS.
\begin{table}
\small
\vspace{-2mm}
\centering
\caption{$C/E-1$ of measured SACS averaged in LR-0 reactor spectrum \cite{kostal2024a} compared with SACS calculations using IRDFF-II or \prENDF{} evaluations with MCNP-6.2(R) code. Reactions with $C/E$ outside the 1-sigma experimental uncertainty are highlighted in red. Cd denotes measurements with Cd cover.}
\begin{tabular}{l ccc}
\toprule \toprule
Dosimeters &\multirow{ 2}{*}{ Exp.unc. }& \multicolumn{2}{c}{C/E-1} \\
Reaction                   &        &IRDFF-II  & ENDF/B-VIII.0 \\
\midrule
$^{58}$Fe(n,$\gamma$)      & 4.2\% &              2.9\% &              3.2\% \\
$^{59}$Co(n,$\gamma$)      & 4.1\% &              4.6\% &              4.6\% \\
Cd $^{23}$Na(n,$\gamma$)   & 3.3\% & {\color{red} 5.0\%}& {\color{red} 5.4\%}\\
Cd $^{58}$Fe(n,$\gamma$)   & 3.5\% & {\color{red} 4.4\%}& {\color{red} 6.1\%}\\
$^{47}$Ti(n,p)             & 2.0\% &              1.6\% & {\color{red} 7.3\%}\\
$^{64}$Zn(n,p)             & 5.3\% &             -0.1\% & {\color{red}-9.5\%}\\
$^{58}$Ni(n,p)             & 2.2\% &              0.1\% &             -1.0\% \\
$^{54}$Fe(n,p)             & 2.4\% &             -1.9\% &             -1.9\% \\
$^{46}$Ti(n,p)             & 2.2\% &              2.3\% & {\color{red}-6.2\%}\\
$^{60}$Ni(n,p)             & 8.5\% &             -0.1\% &             -7.1\% \\
$^{63}$Cu(n,$\alpha$)      & 2.9\% &              1.6\% & {\color{red}45.4\%}\\
$^{54}$Fe(n,$\alpha$)      & 10.5\% &             5.7\% & {\color{red}-16.2\%}\\
$^{56}$Fe(n,p)             & 2.6\% &             -0.3\% &             -0.3\%\\
$^{48}$Ti(n,p)             & 2.6\% &             -0.2\% & {\color{red}17.1\%}\\
$^{24}$Mg(n,p)             & 4.6\% &              1.6\% & {\color{red}10.6\%}\\
$^{27}$Al(n,$\alpha$)      & 2.3\% &              0.4\% &             2.3\%\\
$^{51}$V(n,$\alpha$)       & 3.5\% &              1.4\% & {\color{red}8.7\%}\\
$^{197}$Au(n,2n)           & 4.0\% &             -2.9\% &             -7.6\%\\
$^{127}$I(n,2n)            & 4.0\% &              0.1\% &             1.5\%\\
$^{55}$Mn(n,2n)            & 4.5\% & {\color{red} 6.7\%}&             4.5\%\\
$^{75}$As(n,2n)            & 4.3\% &              0.6\% &             3.5\%\\
$^{89}$Y(n,2n)             & 3.2\% &              2.1\% &             3.2\%\\
$^{19}$F(n,2n)             & 4.0\% & {\color{red} 5.6\%}&{\color{red}26.6\%}\\
$^{90}$Zr(n,2n)            & 4.0\% &              0.9\% &             2.9\%\\
$^{23}$Na(n,2n)            & 4.8\% &             -1.5\% & {\color{red}55.2\%}\\
\bottomrule \bottomrule
\end{tabular}
\vspace{-2mm}
\label{tab:SACS-LR0}
\end{table}
Note that capture cross sections cannot be replaced as the IRDFF-II library contains point-wise evaluations without resonance parameters, making the replacement impossible.
However, we can see in Table~\ref{tab:SACS-LR0} that the agreement of calculated SACS capture cross sections from both evaluations (\prENDF{} and IRDFF-II) with experimental data is pretty good for the thermal and epithermal regions.

Only for the $^{55}$Mn(n,2n) reaction, the $C/E$ of the IRDFF-II SACS is slightly worse than the \prENDF{}: 6.7\% for the IRDFF-II evaluation versus 4.5\% for the \prENDF{} with experimental uncertainty of 4.5\%. However, there are 12 dosimetry reactions (highlighted in red in Table~\ref{tab:SACS-LR0}) where the $C/E$ of the IRDFF-II SACS is  much better than the one calculated using \prENDF{}.
Worse $C/E$ cases beyond 2-sigma experimental uncertainty are $^{46}$Ti(n,p), $^{47}$Ti(n,p), $^{63}$Cu(n,$\alpha$), $^{48}$Ti(n,p), $^{24}$Mg(n,p), $^{51}$V(n,$\alpha$), $^{19}$F(n,2n), and $^{23}$Na(n,2n).

Threshold dosimetry reaction cross sections and associated covariances in the ENDF/B-VIII.1 library were replaced by those enumerated in Table~\ref{Table_Long_I} from the IRDFF-II library as described by Trkov and Capote in Ref.~\cite{INDC(NDS)-0900}, and noted in Comment's column in the Table. All discrepant reactions discussed above were replaced, which resulted in a significant improvement in ENDF/B-VIII.1 neutron dosimetry performance compared to \prENDF{}.

After the cross-section and covariance replacements by IRDFF, consistency fixes were implemented, such as reconstruction of the elastic cross section to preserve unitarity with the total cross section and rescaling of partial threshold reactions so their sum corresponds to the adopted IRDFF cross section, whenever appropriate.

Also noteworthy is that other partial or complete evaluations included in the \ENDF\ release and discussed in the present paper may have already chosen to include IRDFF dosimetry cross sections and associated covariances (e.g., \nuc{19}{F}(n,2n), \nuc{63}{Cu}(n,2n), etc.). These cases are discussed in their corresponding sections but are also listed here.

\begin{figure*}[p]
\centering
\subfigure[]{\includegraphics[width=\columnwidth]{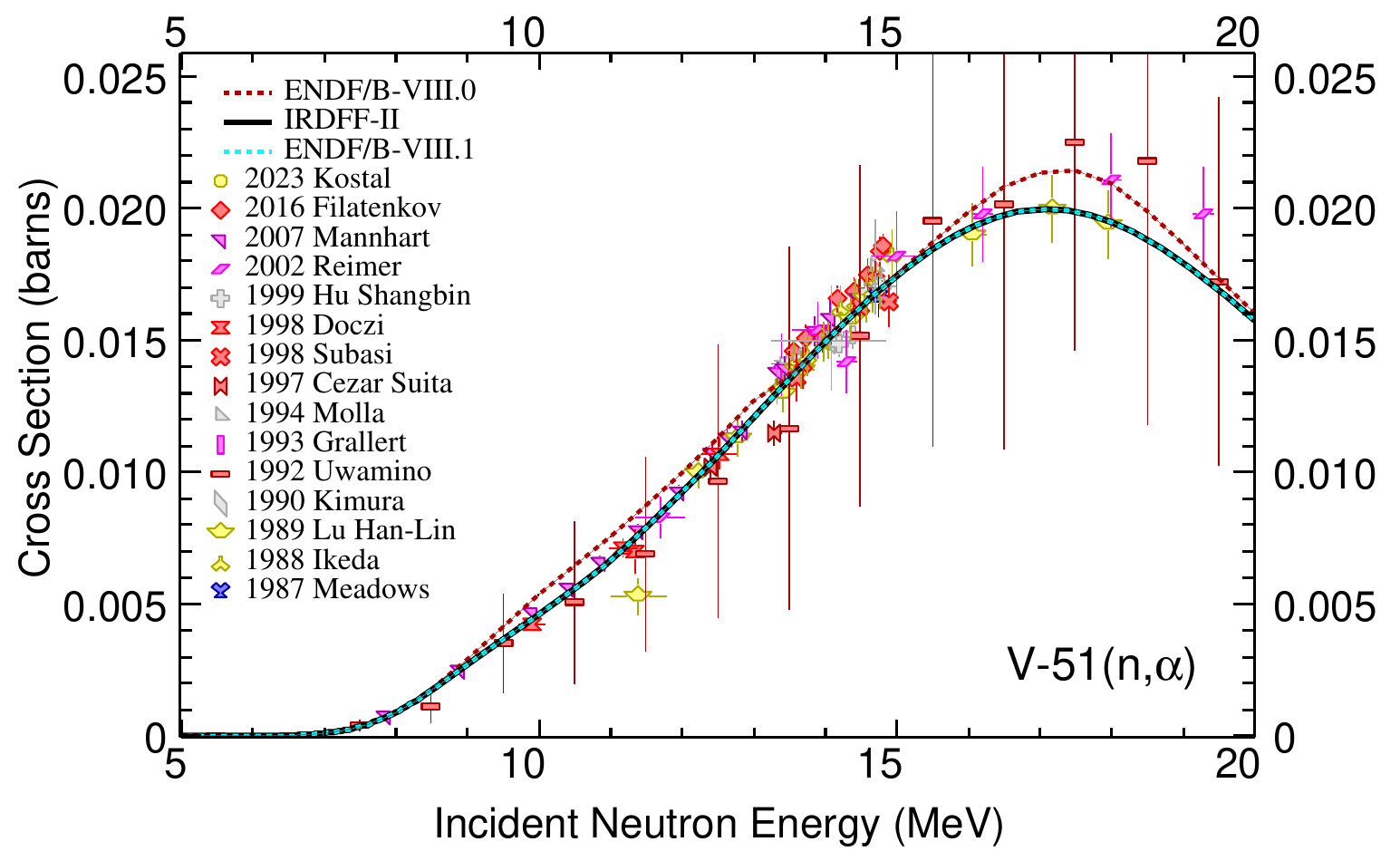}}
\subfigure[]{\includegraphics[width=\columnwidth]{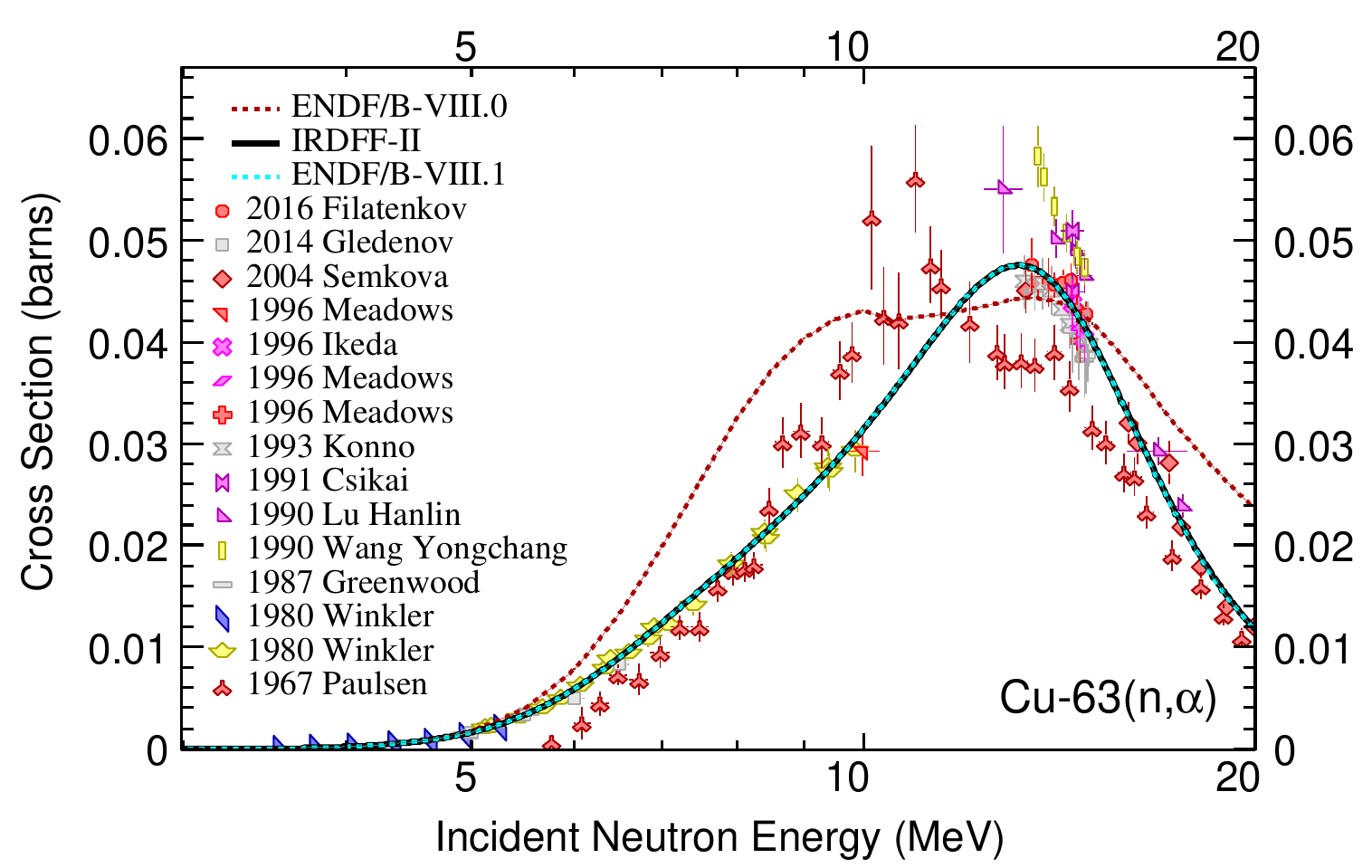}}
\subfigure[]{\includegraphics[width=1.0\columnwidth]{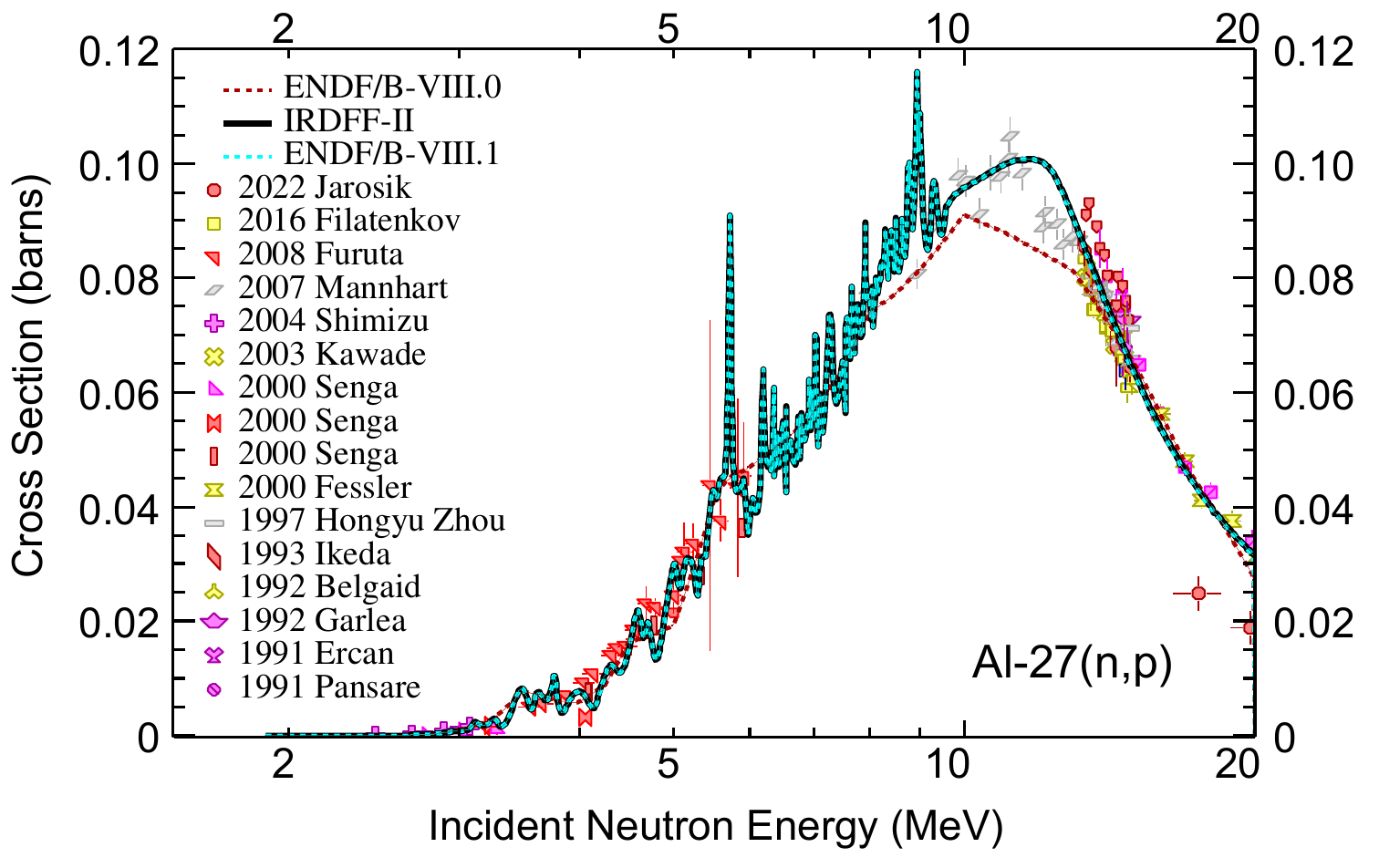}}
\subfigure[]{\includegraphics[width=1.0\columnwidth]{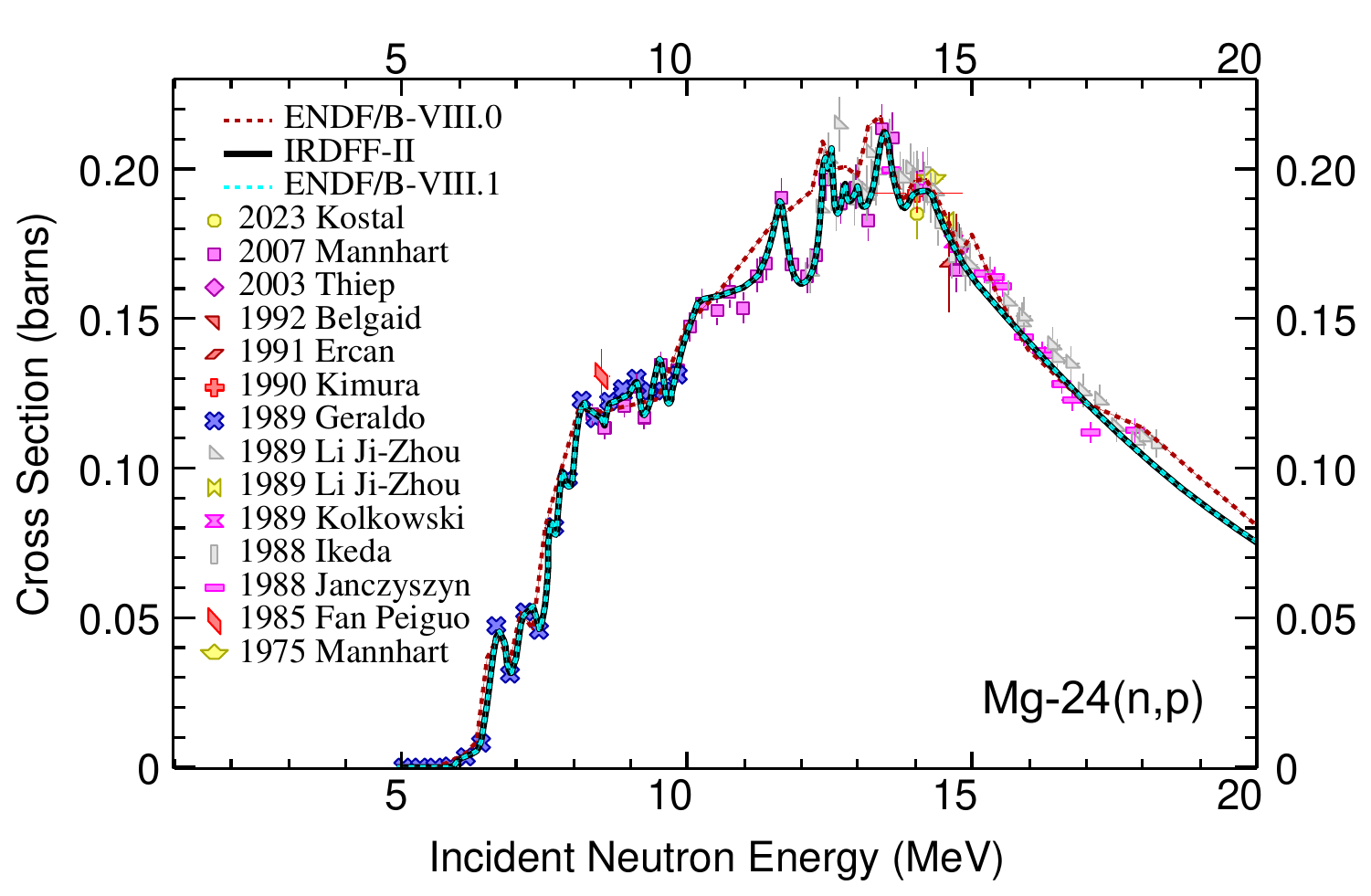}}
\subfigure[]{\includegraphics[width=1.0\columnwidth]{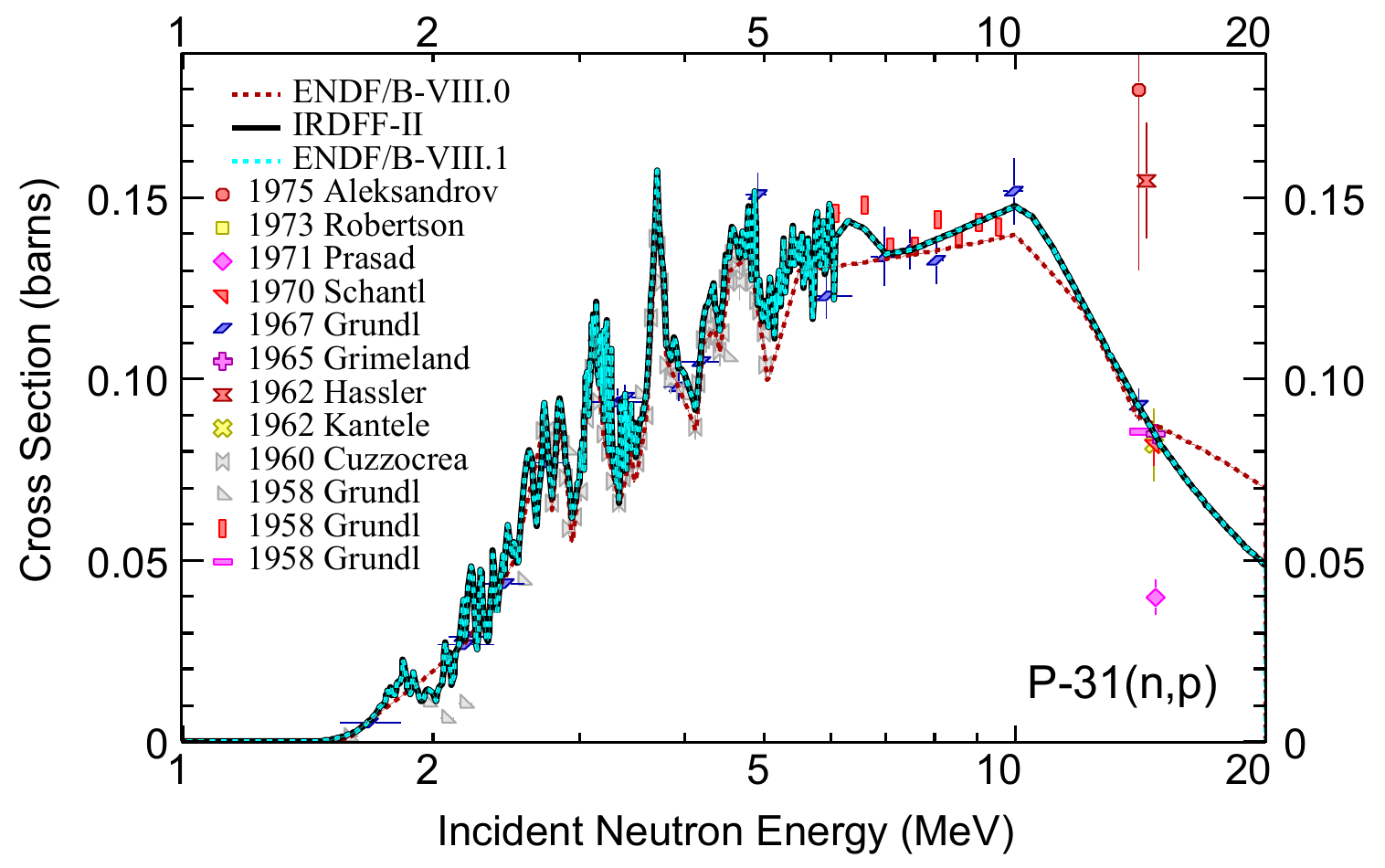}}
\subfigure[]{\includegraphics[width=1.0\columnwidth]{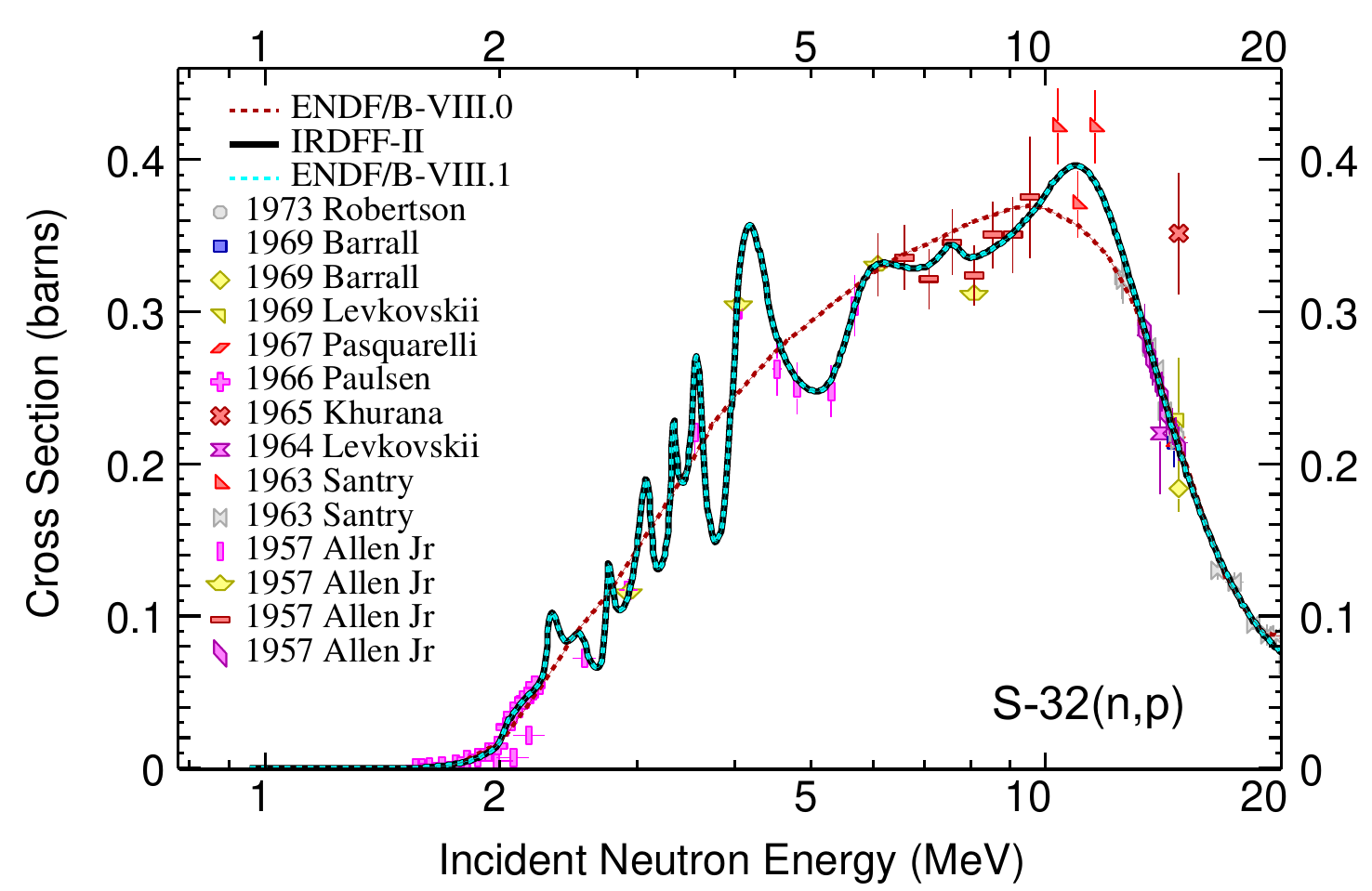}}
\centering
\vspace{-2mm}
\caption{(Color online) Selected IRDFF-II~\cite{IRDFF} (n,$\alpha$) and (n,p) dosimetry evaluations adopted for ENDF/B-VIII.1 are compared with selected data from EXFOR~\cite{EXFOR} and \prENDF{}. On the plots, \ENDF\ (cyan dashed lines) curves are exactly on top of IRDFF-II (solid black lines).}
\label{fig:IRDFF-2}
\vspace{-2mm}
\end{figure*}

\begin{figure*}[p]
\centering
\subfigure[]{\includegraphics[width=\columnwidth]{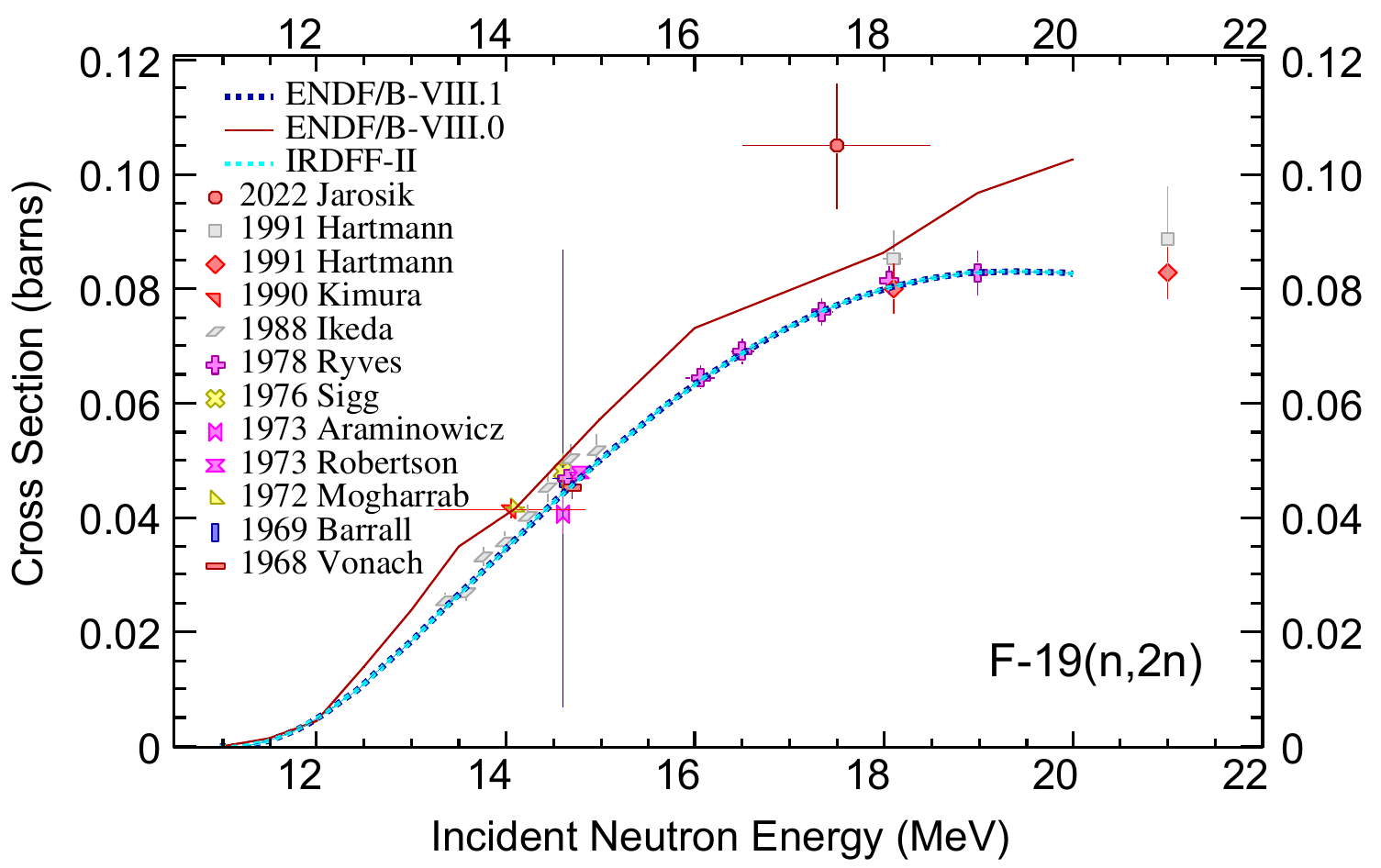}}
\subfigure[]{\includegraphics[width=\columnwidth]{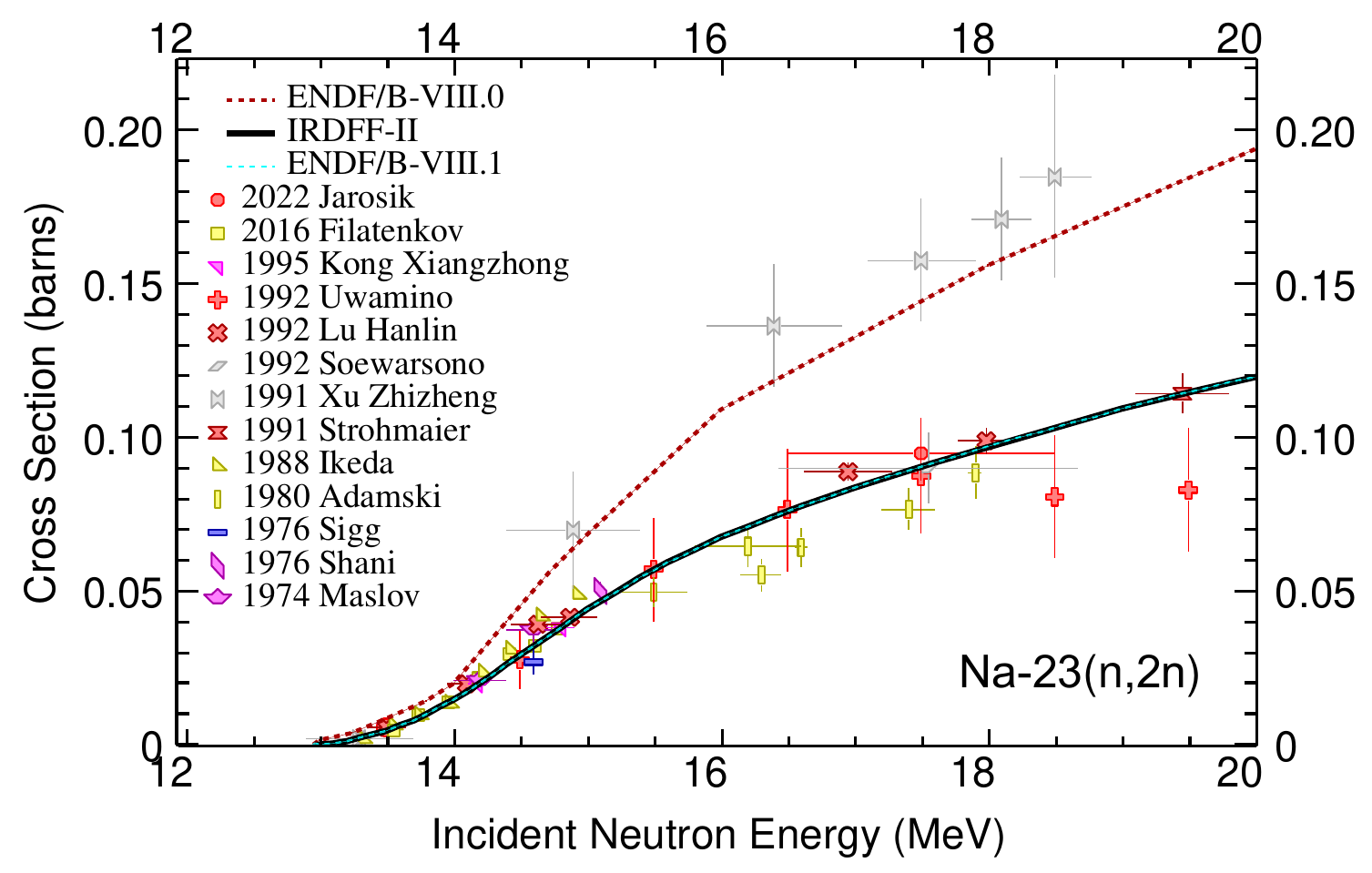}}
\subfigure[]{\includegraphics[width=\columnwidth]{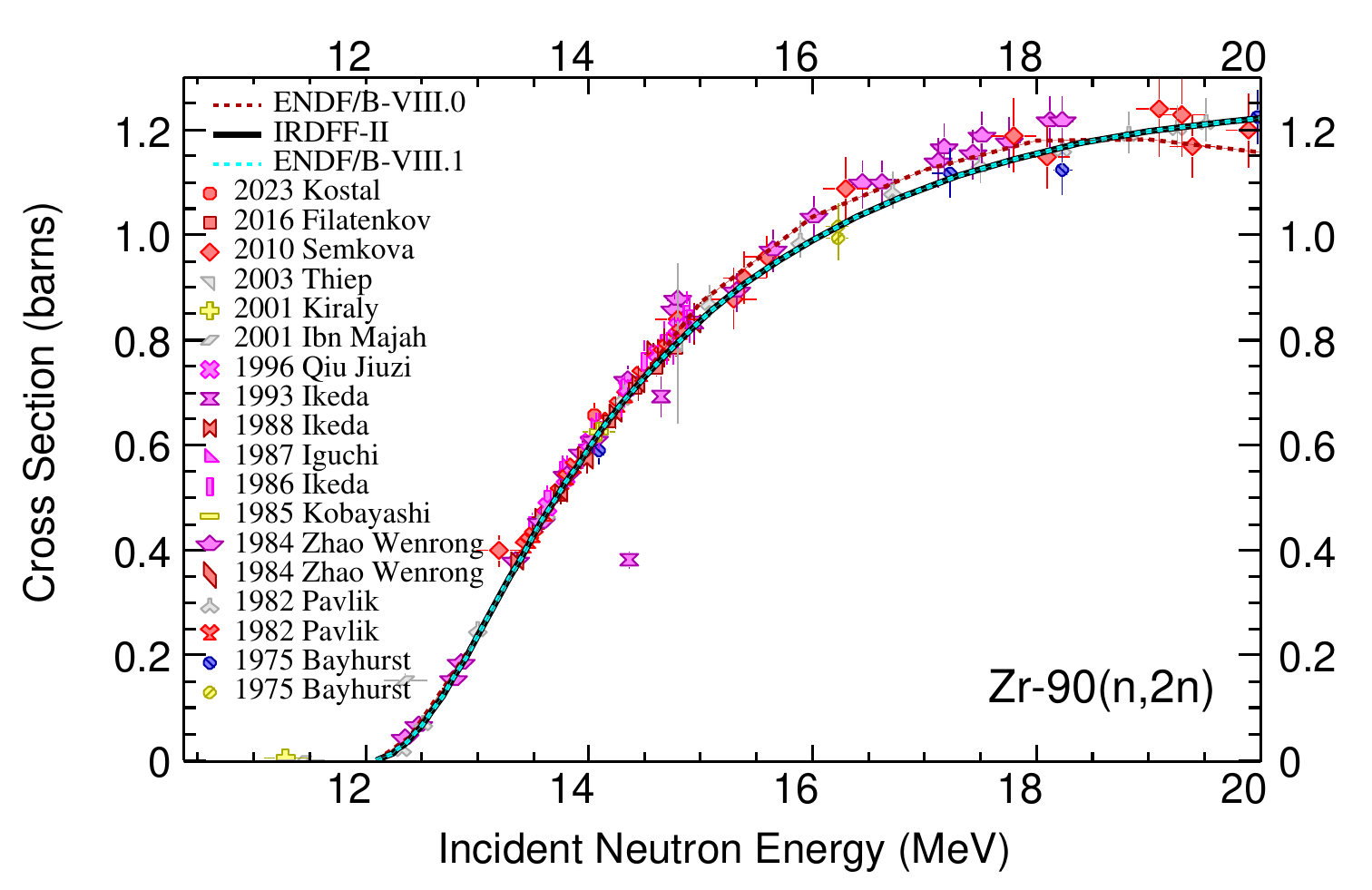}}
\subfigure[]{\includegraphics[width=\columnwidth]{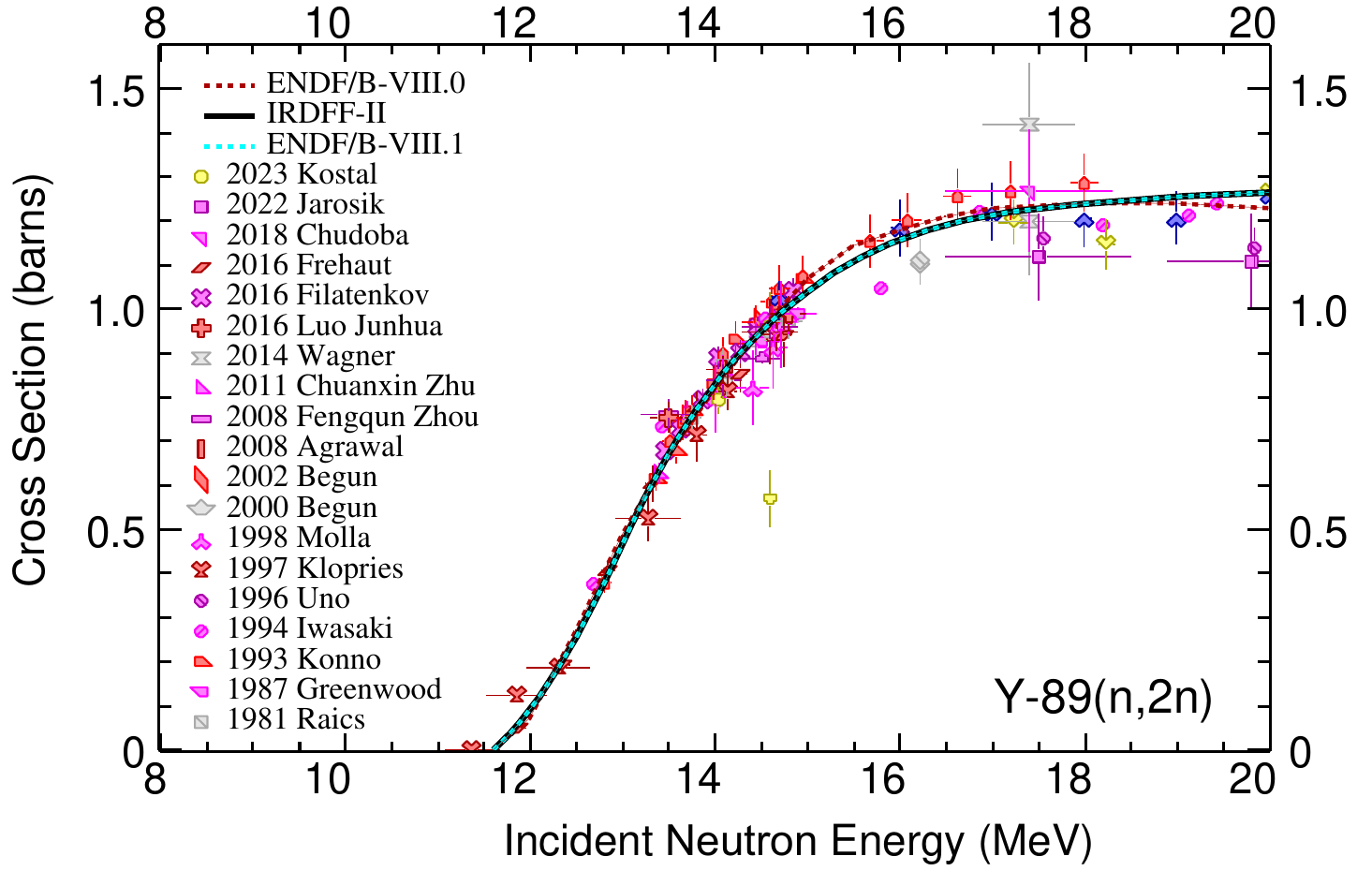}}
\subfigure[]{\includegraphics[width=\columnwidth]{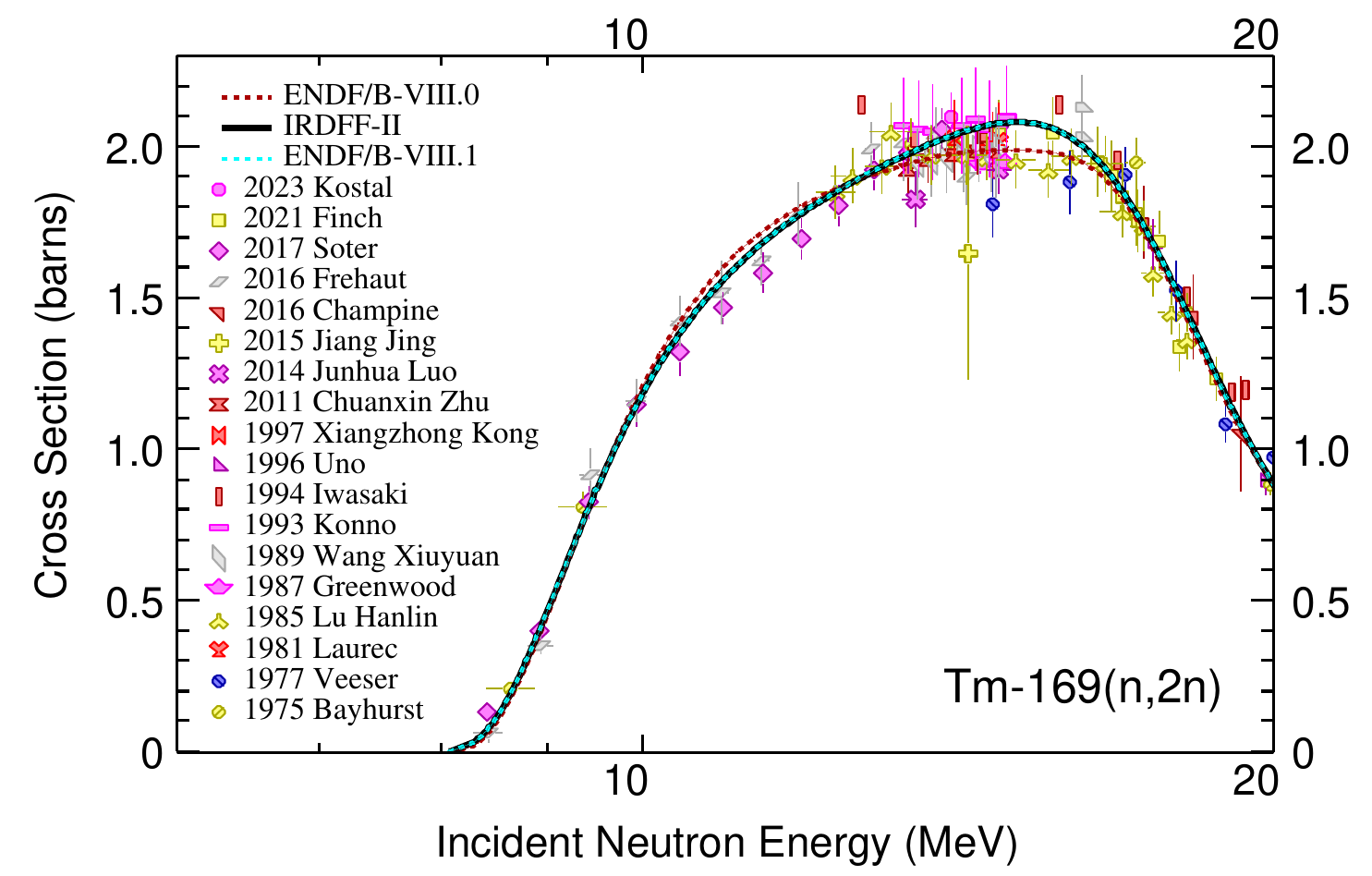}}
\subfigure[]{\includegraphics[width=\columnwidth]{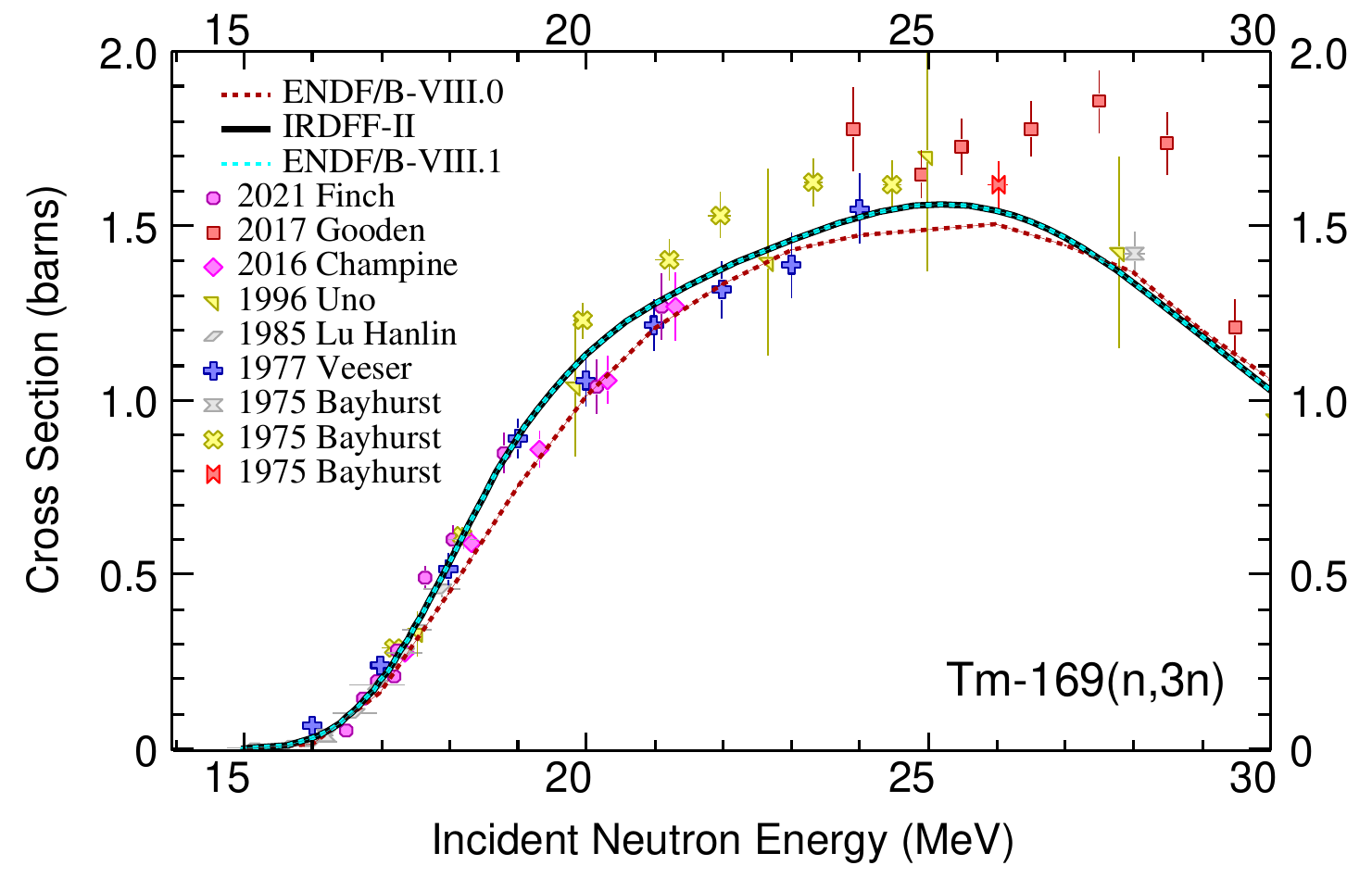}}
\vspace{-2mm}
\caption{(Color online) Selected IRDFF-II~\cite{IRDFF} (n,2n) and (n,3n) dosimetry evaluations adopted for ENDF/B-VIII.1 are compared with selected data from EXFOR~\cite{EXFOR} and \prENDF{}. On the plots, \ENDF\ (cyan dashed lines) curves are exactly on top of IRDFF-II (solid black lines).}
\label{fig:IRDFF-3}
\end{figure*}

The differences seen in the $C/E$ listed in Table~\ref{tab:SACS-LR0} are also shown in Figs.~\ref{fig:IRDFF-2} and \ref{fig:IRDFF-3}. For example, the $^{24}$Mg(n,p), $^{51}$V(n,$\alpha$), $^{63}$Cu(n,$\alpha$), $^{19}$F(n,2n), and $^{23}$Na(n,2n) plots show that the \prENDF{} evaluations clearly overestimate measured differential cross-section data leading to the large positive $C/E$ shown in Table~\ref{tab:SACS-LR0}. Light-element evaluations generally show IRDFF-II evaluations to be in better agreement with experimental data than \prENDF{}. Several cases like $^{89}$Y(n,2n) and $^{169}$Tm(n,xn) show very good agreement of IRDFF-II evaluations with the \prENDF{}, especially within the first few MeV above the reaction threshold which is the most important region for neutron dosimetry. 

\LTcapwidth=\textwidth
\newpage
\LTcapwidth=\textwidth
\begin{longtable*}{l  c  c  c  c c c}
\caption{IRDFF-II reaction list and evaluation sources \cite{IRDFF} adopted for the ENDF/B-VIII.1 library. Major sources of data were K.I. Zolotarev and IPPE colleagues evaluations \cite{INDC(CCP)-0360,INDC(CCP)-0431,INDC(CCP)-0438,INDC(NDS)-0526,INDC(NDS)-0546,INDC(NDS)-0584,INDC(NDS)-0657,INDC(NDS)-0668,INDC(NDS)-0705,INDC(NDS)-0796}. Selected energy regions and/or evaluations were also adopted from IRDF-2002~\cite{IRDF2002}, IRDF-2005~\cite{IRDFF1,IRDFF2}, EAF-2010~\cite{EAF2010}, JENDL-4~\cite{JENDL-4.0}, 
ENDF/B-VIII.0~\cite{Brown2018}, and JEFF-3.3~\cite{JEFF33} libraries. All IRDFF-II evaluations were cut to comply with the energy limit of the corresponding ENDF/B-VIII.1 neutron-transport evaluations. That means IRDFF-II extensions beyond 20~MeV that were mostly based on TENDL evaluations~\cite{TENDL-2015,TENDL-2017} were not used. In the ``Energy Interval'' column, $E_{\mathrm{thr.}}$ corresponds to the energy threshold of that particular reaction. The ``Reaction ID'' column identifies the corresponding reaction in the IRDFF-II evaluation~\cite[Table I]{IRDFF} that was adopted in \ENDF.  
\label{Table_Long_I}} \\
\toprule \toprule
 \multirow{ 2}{*}{No.}   &   \multirow{ 2}{*}{Reaction}           &  \multirow{ 2}{*}{Reaction ID} & \multirow{ 2}{*}{MAT}    &  Energy        &  \multirow{ 2}{*}{Evaluation  source}&           \multirow{ 2}{*}{Comments} \\
   &     &       &        &  Interval      &       &   \\
\midrule
\endfirsthead
\caption{(continued). IRDFF-II reaction list and evaluation sources \cite{IRDFF} adopted for the ENDF/B-VIII.1 library.} \\
\toprule
 \multirow{ 2}{*}{No.}   &   \multirow{ 2}{*}{Reaction}           &  \multirow{ 2}{*}{Reaction ID} & \multirow{ 2}{*}{MAT}    &  Energy        &  \multirow{ 2}{*}{Evaluation  source}&           \multirow{ 2}{*}{Comments} \\
   &     &       &        &  Interval      &       &   \\
\midrule
\endhead
\bottomrule \multicolumn{7}{r}{{Continued on next page}} \\ 
\endfoot
\bottomrule \bottomrule
\endlastfoot

\T
  \multirow{ 4}{*}{6} & \multirow{ 4}{*}{$^{7}$Li(n,X)$^{4}$He}                                  & \multirow{ 4}{*}{Li7He4}     &   \multirow{ 4}{*}{328}  & 0 -- 20~MeV      &  IRDFF-II Li7t                             & adopted     \\
     &                                                        &            &      & 0 -- 20~MeV      &  MT25 -- ENDF/B-VIII.0                       & unchanged   \\
     &                                                        &            &      & 0 -- 20~MeV      &  MT33=ENDF/B-VIII.0 (n,n')                 & unchanged   \\
     &                                                        &            &      & 0 -- 20~MeV      &  MT102 -- ENDF/B-VIII.0                      & unchanged   \\[1.25mm]

   7 & $^{7}$Li(n,X)$^{3}$H                                   & Li7H3      &   328  & 0 -- 20~MeV      &  MT33 -- ENDF/B-VIII.0 (n,n')                & adopted     \\[1.25mm]

  15 & $^{19}$F(n,2n)$^{18}$F                                 & F192       &   925  &$E_{\mathrm{thr.}}$ -- 20 MeV & RRDF-2002                                  & adopted     \\[1.25mm]

  16 & $^{23}$Na(n,2n)$^{22}$Na                               & Na232      &  1125  &$E_{\mathrm{thr.}}$ -- 20 MeV & INDC(NDS)-0705                             & adopted     \\[1.25mm]

  19 & $^{24}$Mg(n,p)$^{24}$Na                                & Mg24p      &  1225  &$E_{\mathrm{thr.}}$ -- 20 MeV & INDC(NDS)-0526                             & adopted  \\[1.25mm]

  20 & $^{27}$Al(n,p)$^{27}$Mg                                & Al27p      &  1325  &$E_{\mathrm{thr.}}$ -- 20 MeV & INDC(CCP)-0438                             & adopted  \\[1.25mm]

  21 & $^{27}$Al(n,$\alpha$)$^{24}$Na                         & Al27a      &  1325  &$E_{\mathrm{thr.}}$ -- 20 MeV & INDC(NDS)-0546                             & adopted  \\[1.25mm]

  23 & $^{27}$Al(n,2n)$^{26g}$Al                              & Al272g     &  1325  &$E_{\mathrm{thr.}}$ -- 20 MeV & INDC(NDS)-0705                             & adopted  \\[1.25mm]

  24 & $^{28}$Si(n,p)$^{28}$Al                                & Si28p      &  1425  &$E_{\mathrm{thr.}}$ -- 20 MeV & INDC(NDS)-0668                             & adopted  \\[1.25mm]

  25 & $^{29}$Si(n,X)$^{28}$Al                                & Si29Al28   &  1428  &$E_{\mathrm{thr.}}$ -- 60 MeV & IRDFF-II Si29p                             & adopted (MF10)\\[1.25mm]

  27 & $^{31}$P(n,p)$^{31}$Si                                 & P31p       &  1525  &$E_{\mathrm{thr.}}$ -- 20 MeV & INDC(NDS)-0668                             & adopted \\[1.25mm]

  29 & $^{32}$S(n,p)$^{32}$P                                  & S32p       &  1625  &$E_{\mathrm{thr.}}$ -- 20 MeV & INDC(NDS)-0526                             & adopted \\[1.25mm]

  31 & $^{46}$Ti(n,2n)$^{45}$Ti                               & Ti462      &  2225  &$E_{\mathrm{thr.}}$ -- 20 MeV & INDC(CCP)-0360                             & adopted \\[1.25mm]

  32 & $^{46}$Ti(n,p)$^{46}$Sc                                & Ti46p      &  2225  &$E_{\mathrm{thr.}}$ -- 20 MeV & IRDF-2002                                  & adopted \\[1.25mm]

  33 & $^{47}$Ti(n,p)$^{47}$Sc                                & Ti47p      &  2228  &$E_{\mathrm{thr.}}$ -- 20 MeV & RRDF-2002                                  & adopted \\[1.25mm]

  34 & $^{48}$Ti(n,p)$^{48}$Sc                                & Ti48p      &  2231  &$E_{\mathrm{thr.}}$ -- 20 MeV & IRDF-2002                                  & adopted \\[1.25mm]

  39 & $^{51}$V(n,$\alpha$)$^{48}$Sc                          & V51a       &  2328  &$E_{\mathrm{thr.}}$ -- 20 MeV & IRDF-2002                                  & adopted \\[1.25mm]

  42 & $^{55}$Mn(n,2n)$^{54}$Mn                               & Mn552      &  2525  &$E_{\mathrm{thr.}}$ -- 60 MeV & K.I.Zolotarev                              & adopted  \\[1.25mm]

  43 & $^{55}$Mn(n,$\gamma$)$^{56}$Mn                         & Mn55g      &  2525  &$E_{\mathrm{thr.}}$ -- 60 MeV & ENDF/B-VII.1~\cite{trkov:2011,ENDF-VII.1}  & unchanged\\[1.25mm]

  48 & $^{54}$Fe(n,2n)$^{53}$Fe                               & Fe542      &  2625  &$E_{\mathrm{thr.}}$ -- 20 MeV & INDC(CCP)-0360                             &adopted  \\[1.25mm]

  51 & $^{56}$Fe(n,p)$^{56}$Mn                                & Fe56p      &  2631  &$E_{\mathrm{thr.}}$ -- 20 MeV & INDC(CCP)-0438                             &unchanged\\[1.25mm]

  53 & $^{59}$Co(n,2n)$^{58}$Co                               & Co592      &  2725  &$E_{\mathrm{thr.}}$ -- 20 MeV & INDC(NDS)-0546                             &adopted  \\[1.25mm]

  54 & $^{59}$Co(n,3n)$^{57}$Co                               & Co593      &  2725  &$E_{\mathrm{thr.}}$ -- 20 MeV & INDC(NDS)-0584                             &adopted  \\[1.25mm]

  55 & $^{59}$Co(n,$\gamma$)$^{60}$Co                         & Co59g      &  2725  &$E_{\mathrm{thr.}}$ -- 20 MeV & IRDF-2002                                  &unchanged\\[1.25mm]

  56 & $^{59}$Co(n,p)$^{59}$Fe                                & Co59p      &  2725  &$E_{\mathrm{thr.}}$ -- 20 MeV & INDC(NDS)-0546                             &adopted  \\[1.25mm]

  57 & $^{59}$Co(n,$\alpha$)$^{56}$Mn                         & Co59a      &  2725  &$E_{\mathrm{thr.}}$ -- 20 MeV & IRDF-2002                                  &adopted  \\[1.25mm]

  62 & $^{58}$Ni(n,2n)$^{57}$Ni                               & Ni582      &  2825  &$E_{\mathrm{thr.}}$ -- 20 MeV & INDC(NDS)-0657                             &adopted  \\[1.25mm]

  63 & $^{58}$Ni(n,p)$^{58}$Co                                & Ni58p      &  2825  & 0 -- 20 MeV      & K.I.Zolotarev (RRDF-2002)                  &adopted  \\[1.25mm]

  64 & $^{60}$Ni(n,p)$^{60}$Co                                & Ni60p      &  2831  &$E_{\mathrm{thr.}}$ -- 20 MeV & INDC(NDS)-0526                             &adopted  \\[1.25mm]

  68 & $^{63}$Cu(n,2n)$^{62}$Cu                               & Cu632      &  2925  &$E_{\mathrm{thr.}}$ -- 20 MeV & INDC(NDS)-0526                             &adopted  \\[1.25mm]

  70 & $^{63}$Cu(n,$\alpha$)$^{60}$Co                         & Cu63a      &  2925  &$E_{\mathrm{thr.}}$ -- 20 MeV & IRDF-2002                                  &adopted  \\[1.25mm]

  71 & $^{65}$Cu(n,2n)$^{64}$Cu                               & Cu652      &  2931  &$E_{\mathrm{thr.}}$ -- 20 MeV & INDC(NDS)-0526                             &adopted  \\[1.25mm]

  74 & $^{64}$Zn(n,p)$^{64}$Cu                                & Zn64p      &  3025  & 0 -- 20 MeV      & INDC(NDS)-0526                             &adopted  \\[1.25mm]

  75 & $^{67}$Zn(n,p)$^{67}$Cu                                & Zn67p      &  3034  &$E_{\mathrm{thr.}}$ -- 20 MeV & INDC(NDS)-0526                             &adopted  \\[1.25mm]

  76 & $^{68}$Zn(n,X)$^{67}$Cu                                & Zn68Cu67   &  3034  &$E_{\mathrm{thr.}}$ -- 20 MeV & INDC(NDS)-0796                             & adopted (MF10)\\[1.25mm]

  77 & $^{75}$As(n,2n)$^{74}$As                               & As752      &  3325  &$E_{\mathrm{thr.}}$ -- 20 MeV & IRDF-2002                                  &adopted  \\[1.25mm]

  78 & $^{89}$Y(n,2n)$^{88}$Y                                 & Y892       &  3925  &$E_{\mathrm{thr.}}$ -- 20 MeV & INDC(NDS)-0584                             &adopted  \\[1.25mm]
\T
  80 & $^{90}$Zr(n,2n)$^{89}$Zr                               & Zr902      &  4025  &$E_{\mathrm{thr.}}$ -- 20 MeV & INDC(NDS)-0546                             &adopted  \\[1.25mm]

  86 & $^{92}$Mo(n,p)$^{92m}$Nb                               & Mo92pm     &  4225  &$E_{\mathrm{thr.}}$ -- 20 MeV & INDC(NDS)-0657                             &adopted  \\[1.25mm]

  87 & $^{103}$Rh(n,n')$^{103m}$Rh                            & Rh103nm    &  4525  &$E_{\mathrm{thr.}}$ -- 20 MeV & IRDF-2002                                  &adopted (MF10)\\[1.25mm]

  88 & $^{109}$Ag(n,$\gamma$)$^{110m}$Ag                      & Ag109gm    &  4731  & 0 -- 60 MeV      & JENDL-3.2                                  &adopted (MF10)\\[1.25mm]

  91 & $^{113}$In(n,n')$^{113m}$In                            & In113nm    &  4925  &$E_{\mathrm{thr.}}$ -- 20 MeV & INDC(NDS)-0657                             &adopted (MF10)\\[1.25mm]

  92 & $^{115}$In(n,2n)$^{114m}$In                            & In1152m    &  4931  &$E_{\mathrm{thr.}}$ -- 20 MeV & INDC(NDS)-0526                             &adopted (MF10)\\[1.25mm]

  93 & $^{115}$In(n,n')$^{115m}$In                            & In115nm    &  4931  &$E_{\mathrm{thr.}}$ -- 20 MeV & IRDF-2002                                  &adopted (MF10)\\[1.25mm]

  95 & $^{127}$I(n,2n)$^{126}$I                               & I1272      &  5325  &$E_{\mathrm{thr.}}$ -- 20 MeV & INDC(NDS)-0526                             &adopted\\[1.25mm]

  97 & $^{141}$Pr(n,2n)$^{140}$Pr                             & Pr1412     &  5925  &$E_{\mathrm{thr.}}$ -- 20 MeV & IRDF-2002                                  &adopted  \\[1.25mm]

  98 & $^{169}$Tm(n,2n)$^{168}$Tm                             & Tm1692     &  6925  &$E_{\mathrm{thr.}}$ -- 20 MeV & INDC(NDS)-0584                             &adopted  \\[1.25mm]

  99 & $^{169}$Tm(n,3n)$^{167}$Tm                             & Tm1693     &  6925  &$E_{\mathrm{thr.}}$ -- 20 MeV & INDC(NDS)-0657                             &adopted  \\[1.25mm]

 101 & $^{186}$W(n,$\gamma$)$^{187}$W                         & W186g      &  7443  & 0 -- 20 MeV      & ENDF/B-VII.1~\cite{trkov:2011,ENDF-VII.1}  &unchanged\\[1.25mm]

 102 & $^{197}$Au(n,2n)$^{196}$Au                             & Au1972     &  7925  &$E_{\mathrm{thr.}}$ -- 20 MeV & INDC(NDS)-0526                             &adopted  \\[1.25mm]

 103 & $^{197}$Au(n,$\gamma$)$^{198}$Au                       & Au197g     &  7925  & 0 -- 20 MeV      & IAEA STD 2017~\cite{carlson2018}           &unchanged\\[1.25mm]

 104 & $^{199}$Hg(n,n')$^{199m}$Hg                            & Hg199nm    &  8034  &$E_{\mathrm{thr.}}$ -- 20 MeV & INDC(NDS)-0526                             &adopted (MF10)\\[1.25mm]

 105 & $^{204}$Pb(n,n')$^{204m}$Pb                            & Pb204nm    &  8225  &$E_{\mathrm{thr.}}$ -- 20 MeV & INDC(CCP)-0431                             &adopted (MF10)\\[1.25mm]

 106 & $^{209}$Bi(n,2n)$^{208}$Bi                             & Bi2092     &  8325  &$E_{\mathrm{thr.}}$ -- 20MeV & V.G.Pronyaev                                &adopted  \\[1.25mm]

 107 & $^{209}$Bi(n,3n)$^{207}$Bi                             & Bi2093     &  8325  &$E_{\mathrm{thr.}}$ -- 20MeV & V.G.Pronyaev                                &adopted  \\[1.25mm]

 113 & $^{235}$U(n,f)                                         & U235f      &  9228  & 0 -- 20 MeV      & ENDF/B-VIII.0,CIELO~\cite{capote2018,CIELO,CIELO-res}&unchanged\\[1.25mm]

 114 & $^{238}$U(n,2n)$^{237}$U                               & U2382      &  9237  &$E_{\mathrm{thr.}}$ -- 20 MeV & ENDF/B-VIII.0,CIELO~\cite{capote2018,CIELO,CIELO-res}&unchanged\\[1.25mm] 

 115 & $^{238}$U(n,f)                                         & U238f      &  9237  &~0 -- 20 MeV      & ENDF/B-VIII.0,CIELO~\cite{capote2018,CIELO,CIELO-res}&unchanged\\[1.25mm]

 116 & $^{238}$U(n,$\gamma$)$^{239}$U                         & U238g      &  9237  &~0 -- 20 MeV      & ENDF/B-VIII.0,CIELO~\cite{capote2018,CIELO,CIELO-res}&unchanged\\[1.25mm]
\end{longtable*}

\clearpage
\newpage


\subsection{Fixes or improvements to existing evaluations}
\label{subsec:n:fixes-improvements}

\subsubsection{\nuc{106,108,110,111,112,114,116}{Cd}}
\label{subsec:n:106-108-110-111-112-114-116Cd}




The cadmium isotopes \nuc{106,108,110,111,112,114,116}{Cd} had one fix implemented related to the scattering radius,  AP in the ENDF-6 format. The ENDF-6 format manual~\cite{ENDF6-Format-2024}  states that if the $\ell$-dependent scattering radius APL is provided for all $\ell$-values  NLS levels, a value of AP is still required. For the definitions of the aforementioned parameters, refer to the ENDF-6 format manual~\cite{ENDF6-Format-2024}.



\subsubsection{Gamma-spectra fixes}
\label{subsec:n:GRIN}


To improve the accuracy and utility of the gamma-ray production data library, the GRIN project \cite{GRIN-GAP:2023} has performed systematic improvements in the evaluations of capture and inelastic gamma-ray production for neutron interactions. The isotopes \nuc{13}{C}, \nuc{16}{O}, \nuc{19}{F}, \nuc{28}{Si} have revised data in the MT=102 reaction channel where missing emissions were included, and the proper flagging primary gammas now ensures the representation of incident neutron energy dependence. Additionally, the dataset for inelastic gamma-ray production for these isotopes (and \nuc{32,34}{S}) have been updated to fill in missing entries and to update energies and branching ratios, in accordance with the latest values from RIPL~\cite{RIPL3,RIPL}. These changes not only resolve previous discrepancies but help to reconcile the structure and reaction information of these files.

\subsubsection{Previously missing outgoing-particle distributions}
\label{subsec:n:exit-dist}


There were several missing exit distributions in many of the neutron evaluations in ENDF/B-VIII.0. Not only there were no distributions given for outgoing charged particles, but about half of these evaluations were missing energy distributions for gamma production after capture, which made it difficult for transport codes.

\paragraph{Distributions adopted from TENDL\newline}
\label{subsec:n:exit-dist-LLNL}



A first round of patched exit distributions for charged particle and gamma production were, therefore, imported from TENDL2019 \cite{Koning2019}. 
The imported reactions and exit products are described in Table \ref{table:endf8.1_patch}. 
A second round of improvements are described in the next subsection, so this list excludes those nuclides with additional proton and alpha distributions supplied by LANL also for this release.

{\small
\begin{longtable*}{llll}
\caption{Patched exit distributions for missing gammas and charged particles  by importing energy and angular distributions from TENDL2019.
Labels give MT number and the exit particle which now has an exit distribution.}\label{table:endf8.1_patch} \\
\toprule \toprule
Nuclide & MT :  outgoing particle\\
\midrule
\endfirsthead
\caption{Patched exit distributions for missing gammas and charged particles  by importing energy and angular distributions from TENDL2019.
Labels give MT number and the exit particle which now has an exit distribution.}
\endhead
\\
\multicolumn{2}{c}%
{{\bfseries \tablename\ \thetable{} -- continued from previous page}} \\
\toprule
Nuclide & MT :  outgoing particle \\ \midrule
\endhead
\bottomrule \multicolumn{2}{r}{{Continued on next page}} \\ 
\endfoot
\bottomrule \bottomrule
\endlastfoot
\small

\nuc{22}{Na}  &  28:\nuc{1}{H}, 22:\nuc{4}{He}, 103:\nuc{1}{H}, 104:\nuc{2}{H}, 107:\nuc{4}{He}, 108:\nuc{4}{He}, 102:photon \\   
\nuc{23}{Na}  &  103:\nuc{1}{H}, 107:\nuc{4}{He} \\   
\nuc{24}{Mg}  &  28:\nuc{1}{H}, 22:\nuc{4}{He}, 103:\nuc{1}{H}, 107:\nuc{4}{He} \\   
\nuc{25}{Mg}  &  28:\nuc{1}{H}, 22:\nuc{4}{He}, 103:\nuc{1}{H}, 107:\nuc{4}{He} \\   
\nuc{26}{Mg}  &  28:\nuc{1}{H}, 22:\nuc{4}{He}, 103:\nuc{1}{H}, 107:\nuc{4}{He} \\   
\nuc{28}{Si}  &  104:\nuc{2}{H} \\   
\nuc{31}{P}  &  28:\nuc{1}{H}, 103:\nuc{1}{H}, 107:\nuc{4}{He} \\   
\nuc{32}{S}  &  28:\nuc{1}{H}, 22:\nuc{4}{He}, 103:\nuc{1}{H}, 107:\nuc{4}{He} \\
\end{longtable*}
}

\paragraph{Missing distributions adopted from reaction models\newline}
\label{subsec:n:exit-dist-LANL-KAERI}




To complete missing angular distributions and energy spectra 
of secondary particles in the ENDF/B-VIII.0 library, a statistical Hauser-Feshbach code, 
CoH$_3$~\cite{Kawano2021b}, was used for calculating angular
distributions for neutron-induced charged particle reactions including (n,p), (n,$\alpha$), (n,d), (n,t) and (n,$^3$He). 

In the CoH$_3$ calculations, we employ the 
OMP parameters by Koning and Delaroche~\cite{Koning2003} for proton and neutron,
 Avrigeanu {\it et al.}~\cite{Avrigeanu94} for $\alpha$-particle, 
Bojowald {\it et al.}~\cite{bojowald88} for deuteron, and Becchetti {\it el al.}~\cite{becchetti69} for triton and $^3$He.
The Gilbert-Cameron level density~\cite{Gilbert65} is adopted for all nuclei. 
$Q$-values are updated using the
mass data in Audi's 2012 mass table~\cite{Audi2012} and FRDM2012~\cite{Moller2016}. 
Isotropic angular distributions in ENDF/B-VIII.0 are replaced with explicit angular distributions from
 CoH$_3$-calculated results. Energy distributions of the emitted particles
are described in terms of the preequilibrium process and the compound
nuclear reaction. 
In the preequilibrium process, the two-component exciton
model~\cite{Kalbach1986} is employed with the angular distribution described by Kalbach's systematics~\cite{Kalbach1988}.

 Depending on the reaction in ENDF/B-VIII.0, differential cross sections are presented in two ways: 
 (1) only the total (n,$X$) reaction cross section is given, where $X$=p, d, $\alpha$, and so on, or 
 (2) the (n,$X$) cross section is divided into the discrete level population $\sigma_0$, $\sigma_1$,
$\ldots$, $\sigma_i$ and the continuum part $\sigma_{\rm cont}$. 
For the case of (1), missing discrete level cross sections were newly calculated and renormalized to make 
the total cross section ($\sum_i \sigma_i + \sigma_{\rm cont}$) equal to the cross section given in ENDF/B-VIII.0. 
For example, $^{58}$Ni in the ENDF/B-VIII.0 library has no partial (n,p) cross sections for discrete levels. 
We included 10 discrete levels of these residual nuclei to calculate cross
sections as well as angular distributions of those levels.

\begin{figure}
\subfigure[\nuc{58}{Ni}(n,p) cross sections.\label{subfig:58Ni_np}]{\includegraphics[width=.89\columnwidth, clip, trim= 00mm 00mm 00mm 13.7mm]{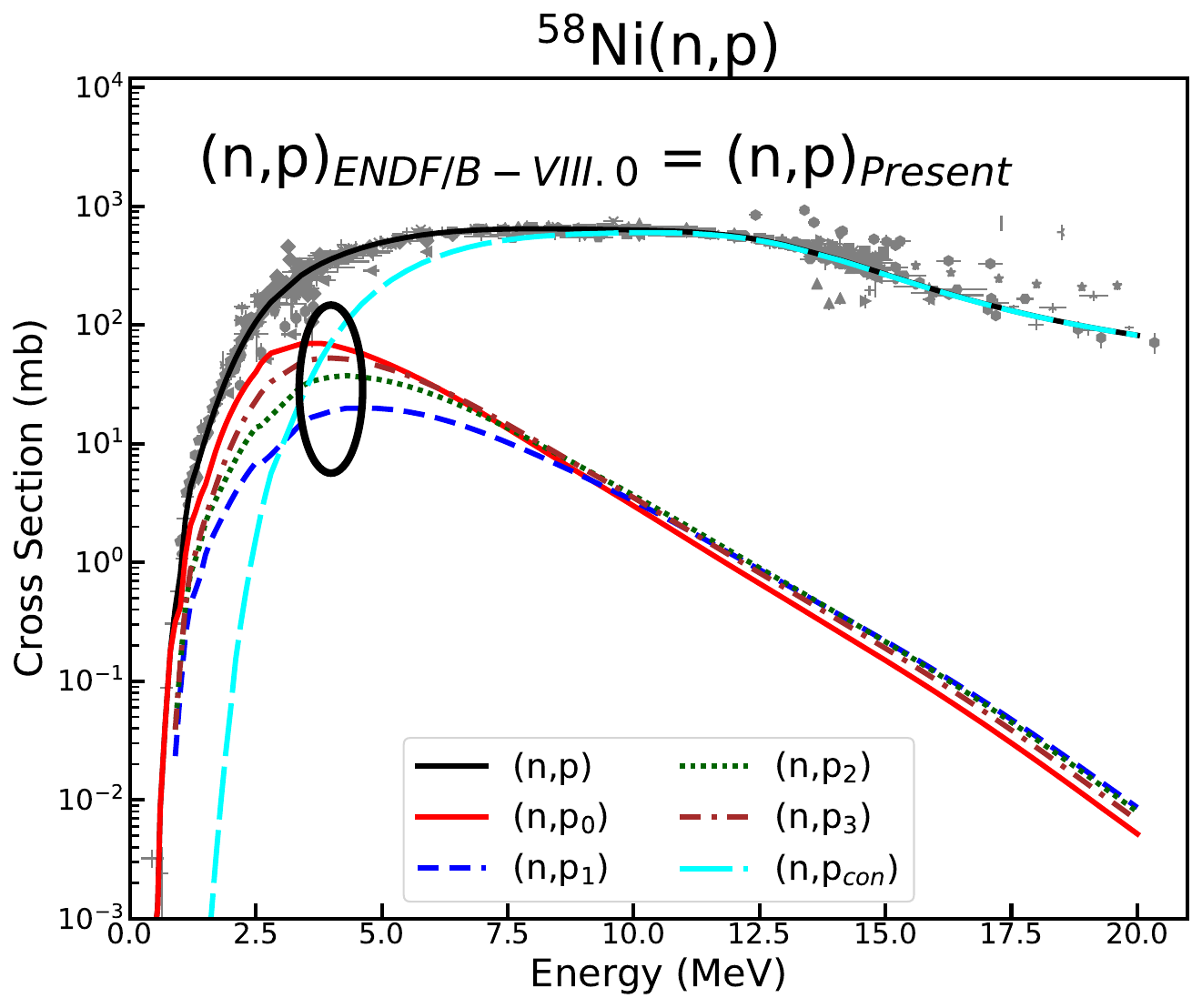}}
\subfigure[\nuc{58}{Ni}(n,p) angular distributions at neutron incident energy of 4.0MeV.\label{subfig:58Ni_angdist}]{\includegraphics[width=.90\columnwidth, clip, trim= 00mm 00mm 00mm 11.5mm]{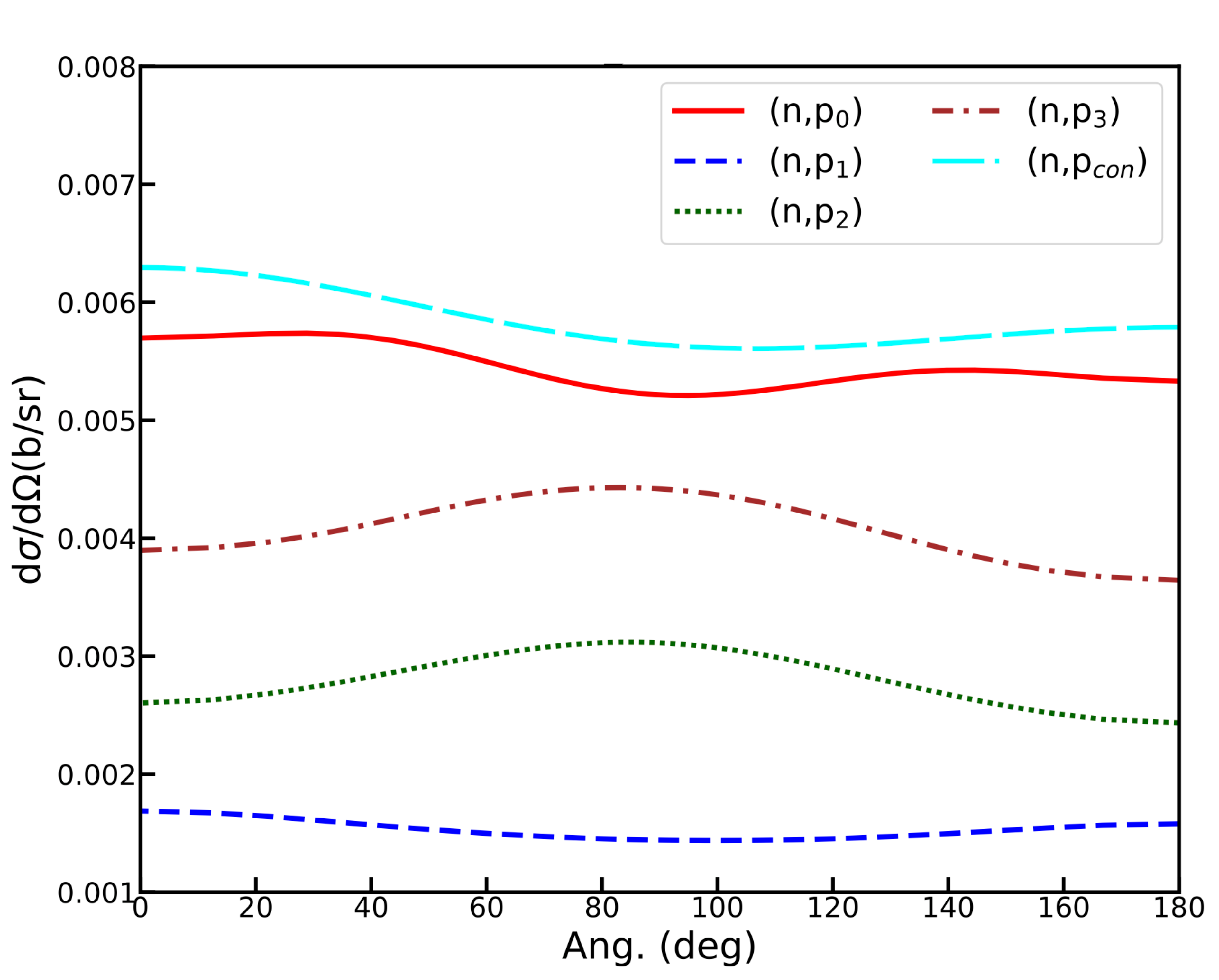}}
\vspace{-2mm}
\caption{Evaluated $^{58}$Ni(n,p) cross sections are exactly the same as those of ENDF/B-VIII.0.
The total (n,p) cross sections, along with those to the ground state, the first three excited states and to the
continuum, are shown in the panel (a).
Angular distributions of emitted protons induced by 4.0 MeV neutrons (black ellipse in the top plot)
are shown in the panel (b) after converting calculations to the Lab system.}
\label{ni58xsangcomp}
\vspace{-2mm}
\end{figure}

\begin{figure}
\subfigure[\nuc{90}{Zr}(n,p) angular distributions.\label{subfig:90zr)np_angdist}]{\includegraphics[width=.90\columnwidth, clip, trim= 00mm 00mm 00mm 12.2mm]{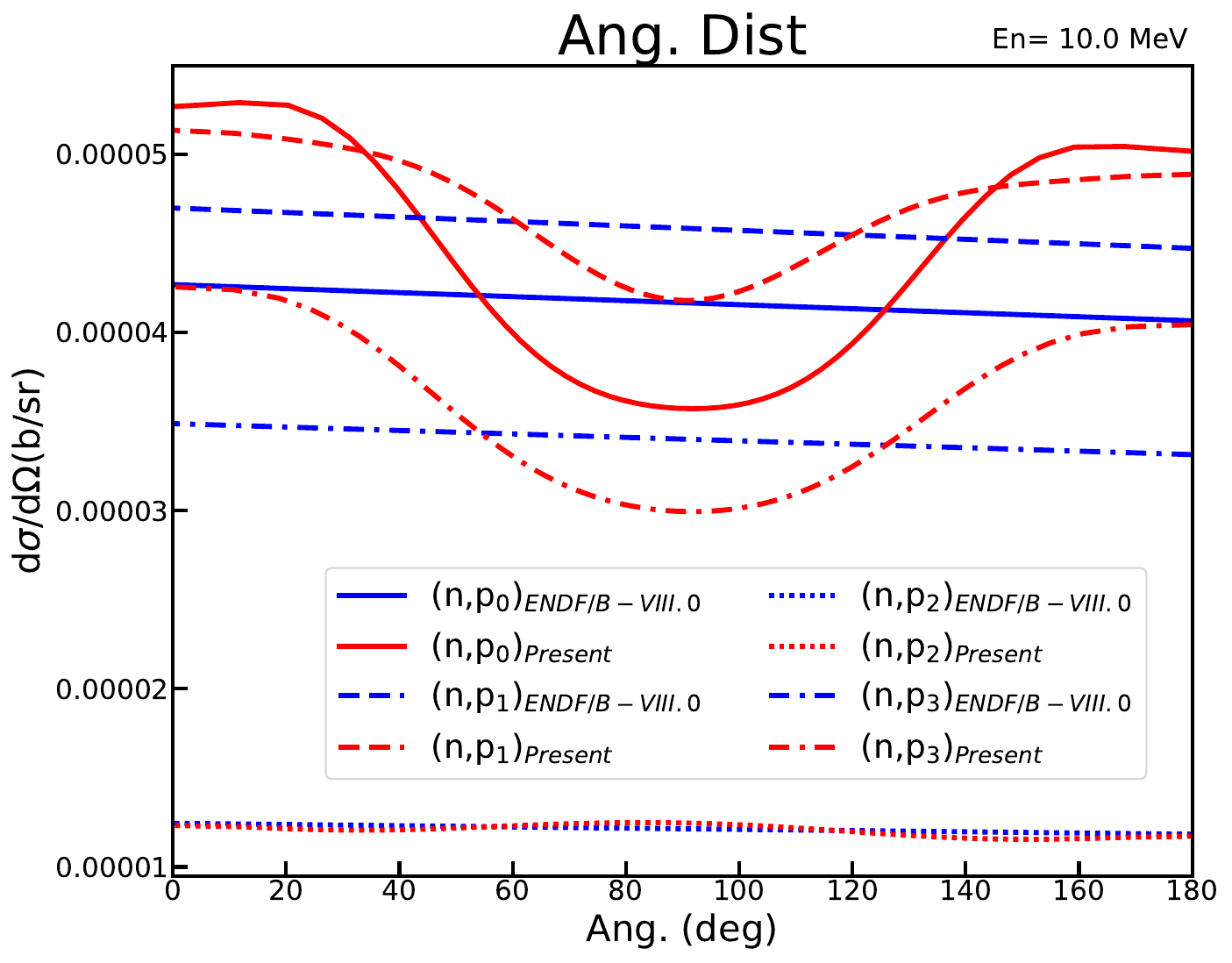}} \\
\subfigure[\nuc{90}{Zr}(n,$\alpha$) angular distributions.\label{subfig:90zr)na_angdist}]{\includegraphics[width=.90\columnwidth, clip, trim= 00mm 00mm 00mm 12.2mm]{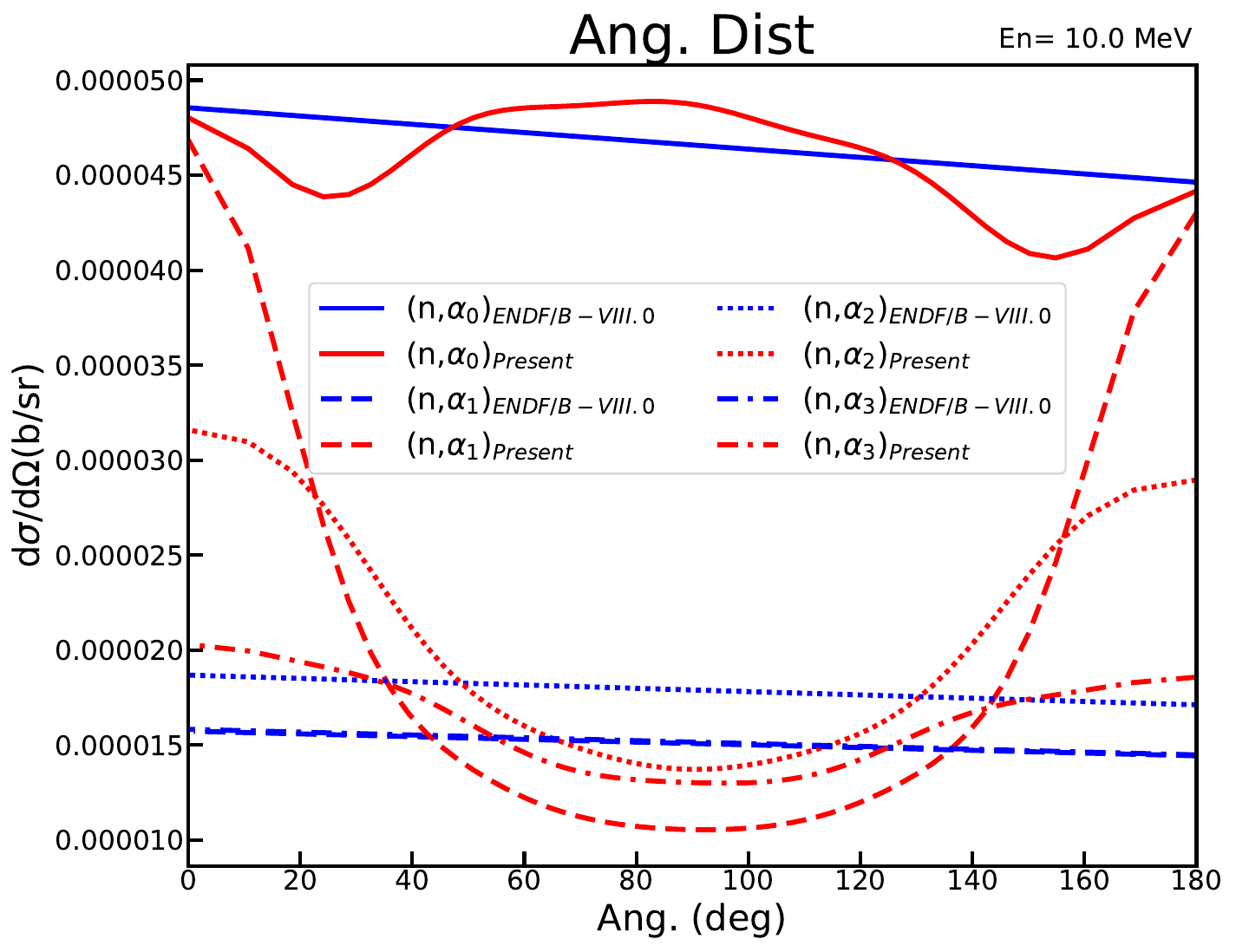}}
\vspace{-2mm}
\caption{Angular distributions at neutron incident energy of 10~MeV for discrete levels of (n,p)  and (n,$\alpha$) in  reactions in panels (a) and (b), respectively,
in  the Lab system are compared to those for $^{90}$Zr.
Most of the distributions of new evaluations are 90-degree symmetric in the CM system
while those of ENDF/B-VIII.0 are isotropic.}
\label{zr90angdist}
\vspace{-2mm}
\end{figure}

Fig.~\ref{subfig:58Ni_np} shows that the total (n,p) cross section is
exactly the same as that given in ENDF/B-VIII.0, along with four newly decomposed
cross sections of the discrete levels and the continuum part. The additional angular distributions
for these levels in the laboratory system are calculated and shown Fig.~\ref{subfig:58Ni_angdist} at the neutron incident energy of
4.0~MeV, which corresponds to the black circle in the top panel.
For the case of (2), we simply adopted partial and continuum cross sections as given in ENDF/B-VIII.0 
and replaced the isotropic angular distributions in ENDF/B-VIII.0 with 
the new calculations by CoH$_3$ for discrete levels of charged particle emissions.
Fig.~\ref{zr90angdist} shows the angular distributions for discrete levels of proton (Fig.~\ref{subfig:90zr)np_angdist}) and alpha (Fig.~\ref{subfig:90zr)na_angdist}) emissions 
induced by 10~MeV neutrons for $^{90}$Zr.

Numbers of discrete levels to calculate the (n,p), (n,$\alpha$),
(n,d), (n,t) and (n,$^3$He) reaction cross sections are listed in
Table~\ref{tab:discretelevelMore} for 53 nuclides ~\cite{Kim20}.

The
numbers in parenthesis are those given in ENDF/B-VIII.0, which
show that many evaluations in ENDF/B-VIII.0 do not contain any
information on discrete-level productions.  As a consistent approach, we 
calculated discrete-level cross sections up to
the maximum ten discrete levels for these.
Photon productions both in discrete transitions and in the
continuum are also replaced by CoH$_3$ results to preserve proper energy
balance. Details of this work can be found in Ref.~\cite{Kim20}.

\begin{center}
\begin{table}
\vspace{-4mm}
\caption{\label{tab:discretelevelMore} {\footnotesize Numbers of discrete levels in the
	residual nuclei included to calculate (n,p), (n,$\alpha$), (n,d), (n,t) and (n,$^3$He)
  reaction cross sections. The numbers in parenthesis
  present the number of discrete levels given in ENDF/B-VIII.0, where ``0''
  stands for no partial cross section given in the evaluation.}}
\begin{tblr}{
        colspec={clllll}, columns={font=\footnotesize}, column{2-6} = {0.95cm},
         rows={rowsep=0.4pt}, 
         row{1} = {c} 
}
\toprule \toprule
   Target      &     p         & $\alpha$   &    d        &      t       &    $^3$He     \\
\midrule
   $^{27}$Al   &   20  (20)    &   20 (20)  &   20 (20)   &   11 (11)    &     -         \\       
   $^{29}$Si   &   16  (16)    &   20 (20)  &     -       &     -        &     -         \\   
   $^{30}$Si   &    6   (6)    &   12 (12)  &     -       &     -        &     -         \\     
   $^{31}$Si   &    1   (1)    &   15 (15)  &     -       &     -        &     -         \\    
   $^{32}$Si   &    1   (1)    &    1  (1)  &     -       &     -        &     -         \\   
   $^{35}$Cl   &   30  (30)    &   21 (21)  &   31 (31)   &   31 (31)    &     -         \\    
   $^{36}$Cl   &   16  (16)    &   32 (32)  &   16 (16)   &   32 (32)    &     -         \\    
   $^{37}$Cl   &   10   (0)    &    6  (6)  &   12 (12)   &   16 (16)    &     -         \\     
   $^{39}$K    &   10   (0)    &   10  (0)  &     -       &     -        &     -         \\     
   $^{40}$K    &   10   (0)    &   10  (0)  &     -       &     -        &     -         \\     
   $^{41}$K    &   10   (0)    &   10  (0)  &     -       &     -        &     -         \\   
   $^{46}$Ti   &   10   (0)    &   10  (0)  &   10 (0)    &   10 (0)     &   10 (0)      \\     
   $^{47}$Ti   &   10   (0)    &   10  (0)  &   10 (0)    &   10 (0)     &   10 (0)      \\     
   $^{48}$Ti   &   10   (0)    &   10  (0)  &   10 (0)    &   10 (0)     &   10 (0)      \\   
   $^{49}$Ti   &   10   (0)    &   10  (0)  &   10 (0)    &   10 (0)     &   10 (0)      \\       
   $^{50}$Ti   &    9   (0)    &   10  (0)  &   10 (0)    &   10 (0)     &   10 (0)      \\      
   $^{49}$V    &   40  (40)    &   40 (40)  &     -       &     -        &     -         \\   
   $^{50}$V    &   10   (0)    &   10  (0)  &   10 (0)    &   10 (0)     &   10 (0)      \\   
   $^{51}$V    &   10   (0)    &   10  (0)  &   10 (0)    &   10 (0)     &   10 (0)      \\   
   $^{51}$Cr   &   10   (0)    &   10  (0)  &   10 (0)    &   10 (0)     &   10 (0)      \\   
   $^{58}$Co   &   40  (40)    &   40 (40)  &   10 (0)    &   10 (0)     &     -         \\         
   $^{59}$Co   &   10   (0)    &   10  (0)  &   10 (0)    &   10 (0)     &   10 (0)      \\   
   $^{58}$Ni   &   10   (0)    &   10  (0)  &   10 (0)    &   10 (0)     &     -         \\   
   $^{59}$Ni   &   10   (0)    &   10  (0)  &   10 (0)    &   10 (0)     &   10  (0)     \\   
   $^{60}$Ni   &   10   (0)    &   10  (0)  &   10 (0)    &   10 (0)     &     -         \\   
   $^{61}$Ni   &   10   (0)    &   10  (0)  &   10 (0)    &   10 (0)     &     -         \\   
   $^{62}$Ni   &   10   (0)    &   10  (0)  &   10 (0)    &   10 (0)     &     -         \\
   $^{63}$Ni   &   26  (26)    &   28 (28)  &     -       &     -        &     -         \\  
   $^{64}$Ni   &   10   (0)    &    1  (0)  &   10 (0)    &   10 (0)     &     -         \\  
   $^{63}$Cu   &   10   (0)    &   10  (0)  &   10 (0)    &   10 (0)     &     -         \\  
   $^{64}$Cu   &   40  (40)    &   40 (40)  &     -       &     -        &     -         \\  
   $^{65}$Cu   &   10   (0)    &   10  (0)  &   10 (0)    &   10 (0)     &     -         \\  
   $^{64}$Zn   &   10   (0)    &   10  (0)  &   10 (0)    &   10 (0)     &   10 (0)      \\  
   $^{65}$Zn   &   10   (0)    &   10  (0)  &   10 (0)    &   10 (0)     &   10 (0)      \\  
   $^{66}$Zn   &   10   (0)    &   10  (0)  &   10 (0)    &   10 (0)     &   10 (0)      \\  
   $^{67}$Zn   &   10   (0)    &   10  (0)  &   10 (0)    &   10 (0)     &   10 (0)      \\  
   $^{68}$Zn   &    8   (0)    &   10  (0)  &   10 (0)    &   10 (0)     &   10 (0)      \\  
   $^{69}$Zn   &   17  (17)    &   18 (18)  &     -       &     -        &     -         \\  
   $^{70}$Zn   &    1   (0)    &    1  (0)  &   10 (0)    &   10 (0)     &     -         \\  
   $^{73}$As   &   10   (0)    &   10  (0)  &     -       &     -        &     -         \\  
   $^{74}$As   &   10   (0)    &   10  (0)  &     -       &     -        &     -         \\  
   $^{90}$Zr   &   12  (12)    &    9  (9)  &   10 (0)    &   10 (0)     &     -         \\  
   $^{91}$Zr   &    6   (6)    &   40 (40)  &   10 (0)    &   10 (0)     &     -         \\  
   $^{92}$Zr   &    1   (1)    &   40 (40)  &   10 (0)    &   10 (0)     &     -         \\  
   $^{93}$Zr   &   17  (17)    &   27 (27)  &    9 (0)    &   10 (0)     &     -         \\  
   $^{94}$Zr   &   10  (10)    &   40 (40)  &   10 (0)    &    9 (0)     &     -         \\  
   $^{95}$Zr   &   16  (16)    &    9  (9)  &   10 (0)    &   10 (0)     &     -         \\  
   $^{96}$Zr   &    3   (3)    &   10 (10)  &   10 (0)    &   10 (0)     &     -         \\  
   $^{107}$Ag  &   10   (0)    &   10  (0)  &     -       &     -        &     -         \\  
   $^{109}$Ag  &   31  (31)    &    2  (2)  &     -       &     -        &     -         \\  
   $^{180}$Ta  &   10   (0)    &   10  (0)  &   10 (0)    &   10 (0)     &     -         \\  
   $^{181}$Ta  &   10   (0)    &   10  (0)  &     -       &     -        &   10 (0)      \\  
   $^{197}$Au  &   10   (0)    &   10  (0)  &     -       &     -        &     -         \\   
\bottomrule \bottomrule
\end{tblr}
\end{table}
\end{center}

\subsubsection{Fission-product updates in the URR}
\label{subsec:n:fission-products}



\begin{figure}
\subfigure[ n+\nuc{99}{Tc}(n,$\gamma$). Data taken from Refs.~ \cite{Chou1973,Little1977,Macklin1982,Matsumoto2003,Kobayashi2004}. \label{fig:Tc99-ng}]{\includegraphics[width=\columnwidth,clip,trim = 0mm 0mm 82mm 42mm]{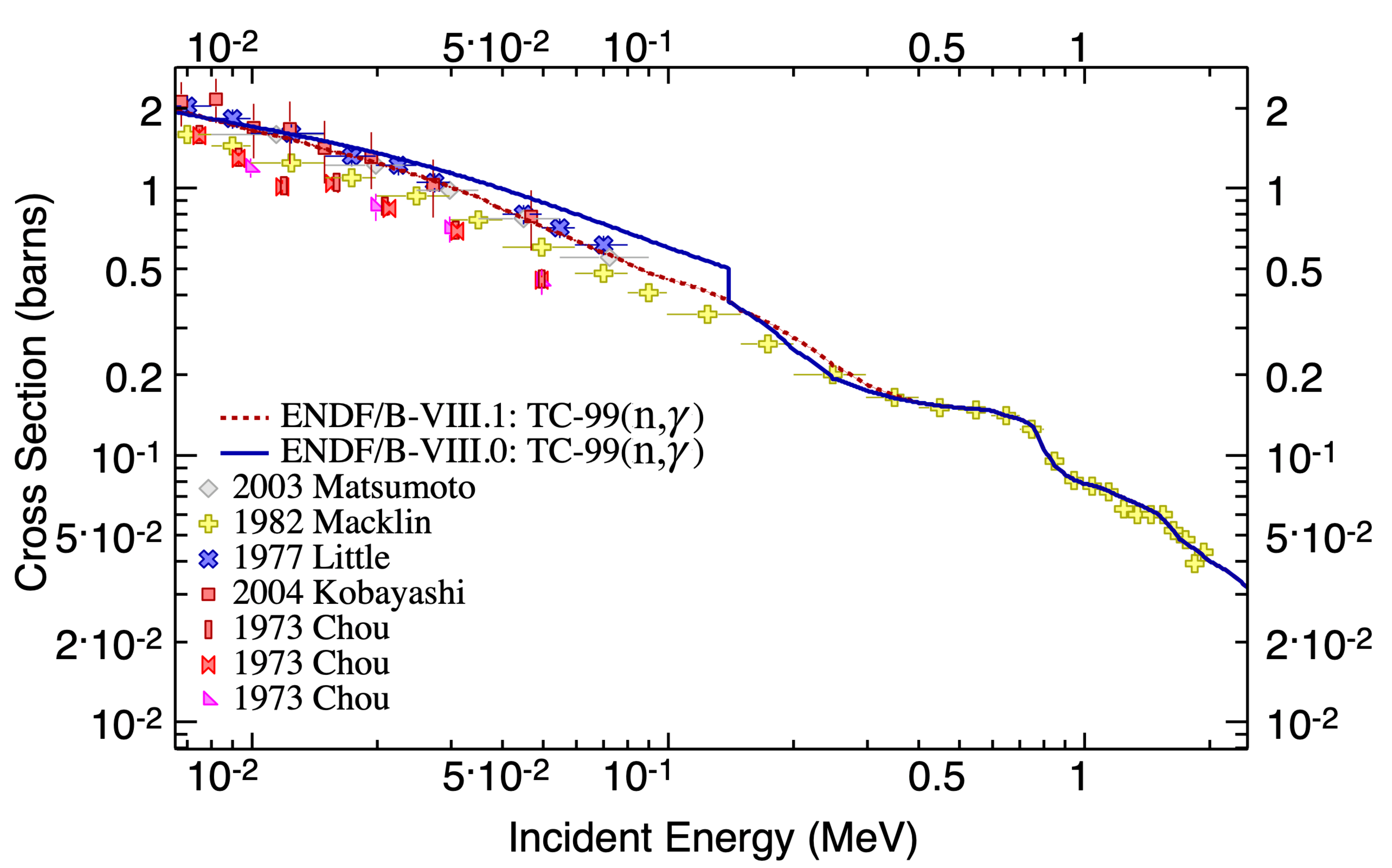}}\\
\subfigure[ n+\nuc{137}{Ba}(n,$\gamma$). Data taken from Refs.~ \cite{Musgrove1976,Voss1994,Koehler1998}. \label{fig:Ba137-ng}]{\includegraphics[width=\columnwidth,clip,trim = 0mm 0mm 83mm 42mm]{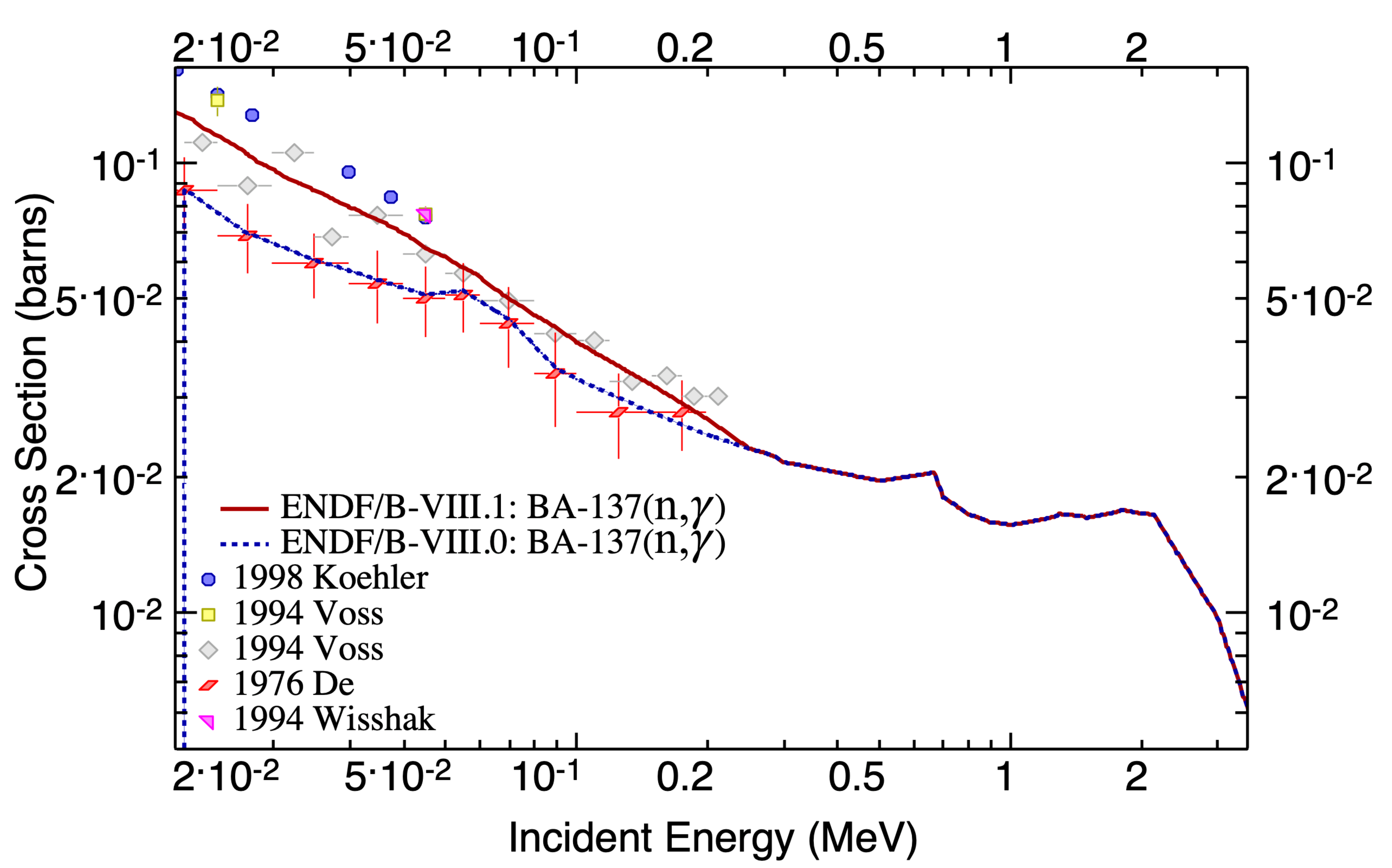}}
\caption{Comparison of the capture cross sections in the URR between \ENDF\ and \prENDF. Selected experimental data retrieved from EXFOR~\cite{EXFOR}.}
\label{fig:Tc99-ng:Ba137-ng}
\end{figure}

The capture, elastic and total cross sections in the URR were revised and updated for many fission products. Experimental data and other evaluations, in particular \prENDF\ and JENDL-4.0~\cite{JENDL-4.0}, were considered when revising outdated values. Table~\ref{table:URR-FP} summarizes all these updates implemented in \ENDF. Fig.~\ref{fig:Tc99-ng:Ba137-ng} shows a couple of illustrative and representative  examples of the changes made in the files described in this section, in this case for \nuc{99}{Tc} (Fig.~\ref{fig:Tc99-ng}) and \nuc{137}{Ba} (Fig.~\ref{fig:Ba137-ng}) in the URR.


\newcommand{\spacebetweenrows}{3mm}

\begin{center}
\begin{longtable*}{cc}

\caption{\label{table:URR-FP}Description of the URR updates for fission products. They were based in either \prENDF\ and/or JENDL-4.0~\cite{JENDL-4.0} with revisions made to the specified reactions and energy ranges.}\label{tab:fission_prod}\\

\toprule \toprule
Isotope & Description \\
\midrule
\endfirsthead

\multicolumn{2}{c}%
{{\bfseries \tablename\ \thetable{} -- continued from previous page}} \\
\toprule Isotope & Description \\
\midrule
\endhead

\bottomrule \multicolumn{2}{r}{{Continued on next page}} \\ 
\endfoot

\bottomrule \bottomrule
\endlastfoot

\parbox[t][][c]{2cm}{\nuc{78}{Se}} & \parbox[t][][t]{15cm}{\raggedright Total, elastic, and capture were revised between 12 to 600 keV.  Adopted JENDL-4 evaluation above 600~keV. } \vspace{\spacebetweenrows}
\\
\parbox[t][][c]{2cm}{\nuc{84}{Kr}} & \parbox[t][][t]{15cm}{\raggedright  URR parameters from JENDL-4  from 20 to 500 keV. Elastic and capture cross sections were revised between 20 to 100 keV; above 100 keV is unchanged from JENDL-4.} \vspace{\spacebetweenrows}
\\
\parbox[t][][c]{2cm}{\nuc{85}{Rb}} & \parbox[t][][t]{15cm}{\raggedright URR is from 8.468 to 100 keV. Elastic and total are revised between 8.468 to 100 keV. Revised capture is in good agreement with Beer \& Macklin data~\cite{Beer:1989}.} \vspace{\spacebetweenrows}
\\
\parbox[t][][c]{2cm}{\nuc{97}{Mo}} & \parbox[t][][t]{15cm}{\raggedright URR  parameters from JENDL-4 from 2 to 100 keV. Total, elastic, and capture revised between 2 to 300~keV. Same as \prENDF\ between 0.30 to 20 MeV.} \vspace{\spacebetweenrows}
\\
\parbox[t][][c]{2cm}{\nuc{99}{Tc}} & \parbox[t][][t]{15cm}{\raggedright Adopted initially \prENDF\ URR, which is defined from 6.38 to 140.34 keV. Elastic and capture were revised from 6.38 to 450 keV. Cross sections above 450~keV were maintained from \prENDF.
} \vspace{\spacebetweenrows}
\\
\parbox[t][][c]{2cm}{\nuc{102}{Pd}} & \parbox[t][][t]{15cm}{\raggedright In \prENDF, resolved parameters are up to 820 eV, without URR. Elastic and capture revised from 820 eV to 4 MeV. Same as \prENDF\ above 4~MeV.} \vspace{\spacebetweenrows}
\\
\parbox[t][][c]{2cm}{\nuc{109}{Ag}} & \parbox[t][][t]{15cm}{\raggedright URR parameters taken from JENDL-4
  	for the energy range 7.0095 to 88.915 keV. Total, elastic, and capture are revised for 7.0095 to 100 keV. Capture is lower than \prENDF\ by 3.6 -- 21.8\% from 10 to 80 keV.
} \vspace{\spacebetweenrows}
\\
\parbox[t][][c]{2cm}{\nuc{113}{In}} & \parbox[t][][t]{15cm}{\raggedright Capture and elastic were revised from JENDL-4 between 0.83 and 250 keV. Capture is 13\% higher from 0.83 to 100 keV.} \vspace{\spacebetweenrows}
\\
\parbox[t][][c]{2cm}{\nuc{115}{In}} & \parbox[t][][t]{15cm}{\raggedright Total, elastic and capture changed from \prENDF\ between 2 -- 300 keV. Revised capture is in good agreement with Kononov \etal \cite{kononov1977,kononov:1977}.
} \vspace{\spacebetweenrows}
\\
\parbox[t][][c]{2cm}{\nuc{115}{Sn}} & \parbox[t][][t]{15cm}{\raggedright Elastic and total are changed from \prENDF\ between 0.9 -- 300 keV. Revised capture is in good agreement with Wisshak \etal~\cite{Wisshak:1996}.
}\vspace{\spacebetweenrows}
\\
\parbox[t][][c]{2cm}{\nuc{119}{Sn}} & \parbox[t][][t]{15cm}{\raggedright Adopted  JENDL-4 with elastic and capture revised between 1.3 -- 100~keV. Revised capture is in good agreement with Timokhov \etal~\cite{Timokhov:1988}.
} \vspace{\spacebetweenrows}
\\
\parbox[t][][c]{2cm}{\nuc{127}{I}} & \parbox[t][][t]{15cm}{\raggedright Adopted URR from JENDL-4, with cross section (total, elastic, and capture) revisions between 5.2 to 200 keV. Capture is in very close agreement with Macklin~\cite{Macklin:1983} between 10 to 100 keV.  
} \vspace{\spacebetweenrows}
\\
\parbox[t][][c]{2cm}{\nuc{122}{Te}} & \parbox[t][][t]{15cm}{\raggedright Adopted JENDL-4 with elastic and total changed between 11 - 200 keV. Revised capture is in good agreement with Wisshak~\etal~\cite{Wisshak:1992}.
} \vspace{\spacebetweenrows}
\\
\parbox[t][][c]{2cm}{\nuc{124}{Te}} & \parbox[t][][t]{15cm}{\raggedright Elastic and total cross sections changed between  15.0 - 150 keV. 	Revised capture is in good agreement with Wisshak~\etal~\cite{Wisshak:1992}.
} \vspace{\spacebetweenrows}
\\
\parbox[t][][c]{2cm}{\nuc{133}{Cs}} & \parbox[t][][t]{15cm}{\raggedright Adopted unresolved parameters from ENDF/B-VII.0 between 4 and 81.607 keV. Revised elastic and capture cross sections were revised from 4 to 100 keV. Capture is excellent agreement with Bokhovko~\etal~\cite{bokhovko:1991}.
} \vspace{\spacebetweenrows}
\\
\parbox[t][][c]{2cm}{\nuc{134}{Cs}} & \parbox[t][][t]{15cm}{\raggedright Adopted JENDL-4, which has URR defined from 270 eV to 100 keV.  Elastic and capture were revised between 270 eV and 170 keV, resulting in capture being 15\% higher than JENDL-4 between 4 and 100 keV.
} \vspace{\spacebetweenrows}
\\
\parbox[t][][c]{2cm}{\nuc{130}{Ba}} & \parbox[t][][t]{15cm}{\raggedright URR from \prENDF\ (2.8 keV - 100 keV), with revised capture and elastic cross sections from  2.8 to 200 keV.
} \vspace{\spacebetweenrows}
\\
\parbox[t][][c]{2cm}{\nuc{134}{Ba}} & \parbox[t][][t]{15cm}{\raggedright Adopted JENDL-4, which has URR defined from 10.575 to 200 keV. Capture and elastic cross sections revised between 10.575 keV and 1 MeV, leading to a good capture agreement with Voss~\etal~\cite{Voss:1994}.
} \vspace{\spacebetweenrows}
\\
\parbox[t][][c]{2cm}{\nuc{137}{Ba}} & \parbox[t][][t]{15cm}{\raggedright Adopted JENDL-4, which has URR from 11.885 to 100 keV, with revised capture and elastic cross sections between 11.885 and 250 keV. Revised capture is in good agreement with Voss~\etal~\cite{Voss:1994}.
} \vspace{\spacebetweenrows}
\\
\parbox[t][][c]{2cm}{\nuc{138}{La}} & \parbox[t][][t]{15cm}{\raggedright Revised capture, total, and elastic cross sections from \prENDF\ from  330 eV to 150 keV, leading to capture being 15\% higher than \prENDF\ between 330 eV and 80 keV.
} \vspace{\spacebetweenrows}
\\
\parbox[t][][c]{2cm}{\nuc{147}{Pm}} & \parbox[t][][t]{15cm}{\raggedright Adopted JENDL-4 (URR from  102 eV to 100 keV) without changes. The half-life of \nuc{147}{Pm} is 2.62 years.
} \vspace{\spacebetweenrows}
\\
\parbox[t][][c]{2cm}{\nuc{148}{Nd}} & \parbox[t][][t]{15cm}{\raggedright Revised capture and elastic cross sections from \prENDF\ between 8.15 and 200 keV. URR is now defined between 8.15 and 300 keV.
} \vspace{\spacebetweenrows}
\\
\parbox[t][][c]{2cm}{\nuc{150}{Nd}} & \parbox[t][][t]{15cm}{\raggedright Revised capture and elastic cross sections from \prENDF\ between 13.85 and 200 keV. URR defined from 13.85 to 130.97 keV.} \vspace{\spacebetweenrows}
\\
\parbox[t][][c]{2cm}{\nuc{153}{Sm}} & \parbox[t][][t]{15cm}{\raggedright Resolved parameters adopted from \prENDF\ up to 25 eV. Adopted JENDL-4 evaluation above 25~eV.
} \vspace{\spacebetweenrows}
\\
\parbox[t][][c]{2cm}{\nuc{155}{Eu}} & \parbox[t][][t]{15cm}{\raggedright Adopted revised JENDL-4 for elastic and capture from 1 to 200 keV, leading to capture being 15\% higher than JENDL-4 between 4 and 80 keV. Same as JENDL-4 above 200~keV.
} \vspace{\spacebetweenrows}
\\
\parbox[t][][c]{2cm}{\nuc{160}{Gd}} & \parbox[t][][t]{15cm}{\raggedright Adopted \prENDF\ resolved parameters up to 9.7 keV. Adopted URR from 9.7 to 100 keV from JENDL-4, with revisions made to total, elastic, and capture cross sections from 9.7 to 200 keV. In the range between 0.2 and 20 MeV, cross sections were kept from \prENDF.
} \vspace{\spacebetweenrows}
\\
\parbox[t][][c]{2cm}{\nuc{159}{Tb}} & \parbox[t][][t]{15cm}{\raggedright URR parameters taken from JENDL-4 (1.2 to 100 keV). Total, elastic, and capture cross sections were revised between 1.2 and 200 keV.
} \vspace{\spacebetweenrows}
\\
\parbox[t][][c]{2cm}{\nuc{166}{Er}} & \parbox[t][][t]{15cm}{\raggedright URR parameters taken fom JENDL-4. Elastic and capture cross sections were revised between 5 and 200 keV, leading to good capture agreement with Kononov~\etal~\cite{kononov:1977,kononov:1977b}.
} \vspace{\spacebetweenrows}
\\
\parbox[t][][c]{2cm}{\nuc{168}{Er}} & \parbox[t][][t]{15cm}{\raggedright URR parameters were initially adopted from JENDL-4. Capture and elastic cross sections were revised from 9.96  to 100 keV.
} \vspace{\spacebetweenrows}
\\
\parbox[t][][c]{2cm}{\nuc{170}{Er}} & \parbox[t][][t]{15cm}{\raggedright URR parameters adopted from JENDL-4 (5.0 to 79.147 keV). Capture and elastic cross sections were revised from 5 to 110 keV.
} \vspace{\spacebetweenrows}
\\
\parbox[t][][c]{2cm}{\nuc{175}{Lu}} & \parbox[t][][t]{15cm}{\raggedright Capture and elastic cross sections from \prENDF\ were revised between 10 and 200 keV. Revised capture is in good agreement with Wisshak~\etal~\cite{Wisshak:2006:Lu}.
} \vspace{\spacebetweenrows}
\\
\parbox[t][][c]{2cm}{\nuc{176}{Lu}} & \parbox[t][][t]{15cm}{\raggedright URR initially adopted from \prENDF\ (102 eV  to 10 keV). Total, capture, and elastic cross sections were revised from 102 eV to 150 keV.
} \vspace{\spacebetweenrows}
\\
\parbox[t][][c]{2cm}{\nuc{168}{Yb}} & \parbox[t][][t]{15cm}{\raggedright URR initially adopted from \prENDF\ (190 eV to 250 keV). Capture and elastic cross sections were revised from 0.19 to 100 keV.
} \vspace{\spacebetweenrows}
\\
\parbox[t][][c]{2cm}{\nuc{176}{Yb}} & \parbox[t][][t]{15cm}{\raggedright URR initially adopted from \prENDF\ (5 to 250 keV). Capture and elastic cross sections were revised from 5 to 100 keV.
} \vspace{\spacebetweenrows}
\\
\parbox[t][][c]{2cm}{\nuc{174}{Hf}} & \parbox[t][][t]{15cm}{\raggedright Adopted JENDL-4 without changes in the URR.
} \vspace{\spacebetweenrows}
\\
\parbox[t][][c]{2cm}{\nuc{176}{Hf}} & \parbox[t][][t]{15cm}{\raggedright Adopted JENDL-4 without changes in the URR.
} \vspace{\spacebetweenrows}
\\
\parbox[t][][c]{2cm}{\nuc{177}{Hf}} & \parbox[t][][t]{15cm}{\raggedright URR parameters initially adopted from JENDL-4. Capture cross section was revised from 60 to 600 keV based on the measurements of Beer~\etal~\cite{Beer:1984} and  Bokhovko~\etal~\cite{bokhovko1991}. Non-elastic channel consistently recalculated.
} \vspace{\spacebetweenrows}
\\
\parbox[t][][c]{2cm}{\nuc{178}{Hf}} & \parbox[t][][t]{15cm}{\raggedright URR parameters initially adopted from JENDL-4. The first three positive resonances were then revised based on Trbovich~\etal~\cite{Trbovich:2009}. Bound level (-49.1 eV), total and elastic cross section between 0.1 to 4.5 MeV, and capture cross section between  0.1 to 2.0 MeV, were also revised.
} \vspace{\spacebetweenrows}
\\
\parbox[t][][c]{2cm}{\nuc{179}{Hf}} & \parbox[t][][t]{15cm}{\raggedright URR parameters initially adopted from JENDL-4. Total and elastic cross section between 0.1 to 4.5 MeV and capture cross section between  0.1 to 2.0 MeV were then revised.
} \vspace{\spacebetweenrows}
\\
\parbox[t][][c]{2cm}{\nuc{180}{Hf}} & \parbox[t][][t]{15cm}{\raggedright URR parameters initially adopted from JENDL-4. Total and elastic cross section between 0.1 to 4.5 MeV and capture cross section between  0.1 to 2.0 MeV were then revised.
} \vspace{\spacebetweenrows}
\\
\end{longtable*}
\end{center}



\subsubsection{Prompt nubar evaluations of minor actinides}
\label{subsec:n:nubar}



Values of prompt nubar for 18 minor actinide ground state files, plus one isomeric state file, were updated based on previous nubar evaluation efforts by Maslov  \etal found in a series of reports \cite{Maslov-BLR-002,Maslov-BLR-006,Maslov-BLR-015,Maslov-BLR-020} and model calculations. Table~\ref{tab:prompt_nubar} shows the nuclei for which prompt nubar were updated in \ENDF\ and the corresponding source of nubar values. Total nubar was modified to account for the changes in prompt nubar.



\begin{table}[!ht]
\caption{Files with prompt nubar updated in \ENDF\ and their respective source.\label{tab:prompt_nubar}} 
\begin{center}
\begin{tabular}{l c}
\toprule \toprule
Nuclide & Prompt nubar source \\
\midrule

\nuc{230}{Pa}     &  Ref.~\cite{Maslov-BLR-015}  \\
\nuc{232}{Pa}     &  Ref.~\cite{Maslov-BLR-020}  \\
\nuc{230}{U}      &  Ref.~\cite{Maslov-BLR-015}   \\
\nuc{231}{U}      &  Ref.~\cite{Maslov-BLR-015}   \\
\nuc{232}{U}      &  Ref.~\cite{Maslov-BLR-015}   \\
\nuc{237}{Pu}     &  Madland-Nix calculations   \\
\nuc{240}{Am}     &  Ref.~\cite{Maslov-BLR-015}  \\
\nuc{244}{Am}     &  Ref.~\cite{Maslov-BLR-006}  \\
\nuc{240}{Cm}     &  Ref.~\cite{Maslov-BLR-002}  \\
\nuc{246}{Cf}     &  Madland-Nix calculations  \\
\nuc{248}{Cf}     &  Madland-Nix calculations  \\
\nuc{249}{Cf}     &  Madland-Nix calculations  \\
\nuc{250}{Cf}     &  Madland-Nix calculations  \\
\nuc{251}{Cf}     &  Madland-Nix calculations  \\
\nuc{252}{Cf}     &  Madland-Nix calculations  \\
\nuc{253}{Cf}     &  Madland-Nix calculations  \\
\nuc{254}{Cf}     &  Madland-Nix calculations  \\
\nuc{254m1}{Es}   &  Same as \nuc{254}{Es}  \\

\bottomrule \bottomrule
\end{tabular}

\end{center}
\end{table}

\subsection{File updates due to only minor fixes}
\label{subsec:n:minor-fixes}



In addition to all the major updates from \prENDF\ to evaluated files in the neutron sublibrary described in the previous sections, there were also updates to more than 200 files due solely to minor to small fixes. Those correspond mostly to minor format, processing or cosmetic fixes. Due to their large number and limited importance, we chose not to list and describe them individually. More details about each individual update can be found in the library change logs (\texttt{CHANGELOG.md} files) distributed within each \ENDF\ released sublibrary~\cite{ENDF-release-page,ENDF-release-alphas,ENDF-release-atom-relax,ENDF-release-decay,ENDF-release-deuterons,ENDF-release-electrons,ENDF-release-gammas,ENDF-release-he3,ENDF-release-neutrons,ENDF-release-nfy,ENDF-release-photoatomic,ENDF-release-protons,ENDF-release-sfy,ENDF-std,ENDF-tsl,ENDF-tritons}, as well as with the whole \ENDF\ release package \cite{ENDF-release-page}), or through contact with corresponding author.

\section{NEUTRON REACTION COVARIANCES}
\label{sec:covariance}


The covariance session ran three big, distinctive initiatives for this library release cycle. Since the last release of the library:
\begin{itemize}
\item A concerted effort was made to deliver covariances for ENDF/B-VIII.1 earlier in the $\beta$-release phase than previously to allow for more stringent verification and validation. Also, CSEWG member institutions developed new codes for better testing of covariances.
\item Templates of expected measurement uncertainties were developed to aid evaluators and experimenters in estimating or providing more complete experimental uncertainties.
\item A discussion and vote was held on whether CSEWG wants to provide a library of formally adjusted nuclear data mean values and covariances.
\end{itemize}

\subsection{Improved Testing of Nuclear Data Covariances}

It was discussed at WANDA 2020~\cite{WANDA2020:summaryreport} that some issues were found by the community and users in ENDF/B-VIII.0 covariances after their release.
One issue that was raised regarding testing ENDF/B-VIII.0 covariances was that evaluators provided covariances too late in the $\beta$-release phases, resulting in insufficient time for thorough testing. 
Therefore, the executive committee requested that preliminary nuclear data covariances should already be delivered for ENDF/B-VIII.1$\beta$1 in February 2023 -- about half a year earlier than was the case for ENDF/B-VIII.0.
This additional time allowed for more stringent testing with LANL's ENDFtk~\cite{ENDFtk} and CovVal codes~\cite{CovVal} and ORNL's \AMPX\ code~\cite{AMPX} during the release phases.
The LLNL covariance testing employed a combination of \NJOY~\cite{NJOY}, \texttt{empy} tools from \EMPIRE\ \cite{Herman:2007}, and \texttt{x4i}~\cite{Bro11}.

The following verification and validation tests were performed on covariances:
\begin{enumerate}
\item It was tested whether the covariance formatting was correct by processing ENDF-6 formatted files with codes, such as \NJOY\ and \AMPX~\cite{NJOY,AMPX}. This test requires processing dedicated specifically to covariances which is sometimes omitted when processing mean values for quick testing of beta libraries.
\item Are mathematical properties satisfied?
\item Are physics properties satisfied?
\end{enumerate}

Below examples are given of succesfully identifying and correcting issues given increased testing. 
For instance, $^{50,52,53}$Cr covariances in \ENDF$\beta3$ could not be processed because of an inconsistency in mean values and covariances.
$^{50,52,53}$Cr mean values were updated for ENDF/B-VIII.1 and no longer supplied MT=22, but ENDF/B-VIII.0 covariances were carried over, including those for MT=22.
This processing issue was resolved by removing MF=33, MT=22.
Covariances obtained from the same evaluation as the mean values will be supplied for ENDF/B-IX.0 as mentioned in Section~\ref{subsec:n:50-51-52-53-54Cr}.

On item 2., we tested for symmetry, positive semi-definiteness, if correlations were within $\pm$ 1, and if all covariance constraints were satisfied.  The checks for semi-definiteness and correlations often implicitly test for the same thing--negative eigenvalues.  Differences between the checks arise since different energy ranges are emphasized in each of the tests (absolute covariances versus relative covariances, and correlations).  For example, a negative eigenvalue found in the relative covariances at energies where the associated cross section is very small will not
be very significant in the absolute covariances. The covariance constraint tests check whether the zero-sum rule for MF=35 data is enforced~\cite{ENDF6-Format-2012}. A related constraint (though not explicitly documented  in Ref.~\cite{ENDF6-Format-2012}) is whether the total MF=33 covariances are a sum of the partials covariances.
For instance, MF=31, MT= 456 (covariances for the average prompt fission neutron multiplicity, $\overline\nu$ ) for $^{239}$Pu for ENDF/B-VIII.1$\beta$1 failed the test for positive definiteness.
This covariance matrix was assembled from several pieces from three different institutes, and a mistake had happened in the non-trivial assembly.
This issue was found through testing for positive semi-definiteness as it led to medium-sized negative eigenvalues.
It was corrected for ENDF/B-VIII.1$\beta$2.

\begin{figure}
\centering
\includegraphics[width=\columnwidth]{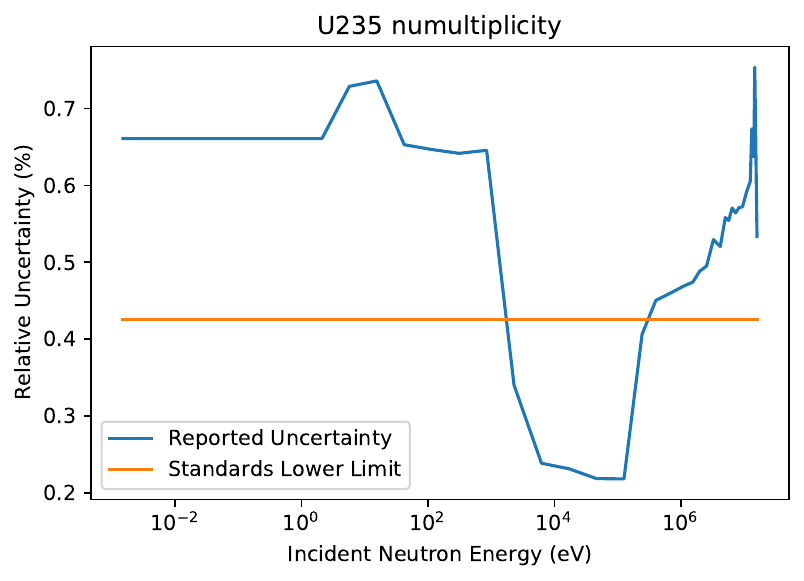}
\vspace{-2mm}
\caption{\label{fig:u235nubaruncVIII1b1} $^{235}$U(n,f) $\overline\nu$ uncertainties for ENDF/B-VIII.1$\beta$1 are compared to standard, $^{252}$Cf(sf) $\overline\nu_\mathrm{tot}$, uncertainties.}
\vspace{-2mm}
\end{figure}

On item 3, we test  for several issues. One test is whether uncertainties are within reasonable limits (e.g., no $>$100\% uncertainties for mubar). Another important test is whether the evaluated uncertainties are larger than those of the Neutron Standard data, which are frequently used as monitor for experiments supporting the evaluated uncertainties. For example, the $^{237}$Np(n,f) cross section uncertainties are expected to be larger than those of $^{235}$U(n,f) Standards cross sections,  as $^{237}$Np(n,f) cross section measurement use in most cases the  $^{235}$U(n,f) cross sections as a monitor. Finally, the uncertainties are counter-checked with those of templates~\cite{Neudecker:CovTesting2021,Smith:CovTesting,tempintro,ngamma_cp,LewisPhD,nxn,PFNS,nu,MatthewPhD,Neudecker2020, Neudecker:2019tempnf} to verify  if uncertainties are realistic.
 Also, the previous covariance session chair, D.L.~Smith, provided in  Ref.~\cite[Table 1]{Smith:CovTesting} (repeated in Ref.~\cite[Table I]{Neudecker:CovTesting2021}) reasonable lower bounds for evaluated uncertainties of specific nuclear data observables.
Another test was employing upper and lower bounds obtained from the ``Physical Uncertainty Bounds''~\cite{Neudecker2020PUBs} method for fission cross-section; PFNS and $\overline\nu$ uncertainties to cross-check whether evaluated uncertainties are realistic.
While testing mathematical properties is often unambiguous in that the mathematical property is either satisfied or not, the tests of item 3. are less clear-cut but aim to provide some clues that uncertainties are either over or underestimated.
It is then the job of the evaluator to judge, after being pointed to potential issues by these clues, whether uncertainties are indeed unrealistic given the evaluation input.
One example is given with Fig.~\ref{fig:u235nubaruncVIII1b1} that drew attention to the fact that ENDF/B-VIII.1$\beta$1 $^{235}$U(n,f) $\overline\nu$ uncertainties were smaller than the current standard, $^{252}$Cf(sf) $\overline\nu_\mathrm{tot}$, uncertainties from approximately 1--100 keV.
This uncertainty lower than the standard is implausible as many $\overline\nu$ experimental data are measured relative to $^{252}$Cf(sf) $\overline\nu_\mathrm{tot}$ and few data sets are available in that energy range.
In fact, $^{235}$U(n,f) $\overline\nu$ covariances were updated for thermal, resonance, and fast ranges but were carried over unchanged from ENDF/B-VIII.0 for 1--100 keV.
However, the previous, much lower, $^{252}$Cf(sf) $\overline\nu_\mathrm{tot}$ were applied for ENDF/B-VIII.0 $^{235}$U(n,f) $\overline\nu$ covariances and it was missed to update the 1--100~keV range for  ENDF/B-VIII.1$\beta$1 with the new standard uncertainties.
This issue was succesfully resolved for ENDF/B-VIII.1$\beta$4 by Capote and Trkov. 
The updated covariance matrix is described in Section~\ref{subsec:n:235U}.

For ENDF/B-VIII.1, all those covariances were tested with the CovVal code~\cite{CovVal} where at least a single digit changed in the covariance file. Due to that, on some cases, covariances were checked that were not new in ENDF/B-VIII.1 and could not be corrected as no one was working on them.
Below we note covariance issues that could not be resolved in ENDF/B-VIII.1 and will be addressed in later releases:
\begin{itemize}
    \item Covariances are missing in the RRR for $^{51}$Cr, $^{58}$Ni, $^{54}$Fe, $^{86}$Kr, and $^{236}$U(n,$\gamma$) cross section.
    \item Covariances are missing in the fast range for $^{35,36}$Cl, $^{39}$K, $^{51}$Cr, $^{156,158,160-164}$Dy,$^{170}$Tm, $^{35,36}$Cl, $^{63,65}$Cu, $^{140}$Ce, and $^{169}$Er.
    \item Some covariances in the RRR for $^{181}$Ta, $^{182,183}$W and  $^{206,207,208}$Pb were flagged as unrealistically low despite the best efforts of the evaluators to provide realistic uncertainties. These low uncertainties might be the result of an ENDF-6 format formalism that does not allow to store all relevant RRR uncertainties; namely, those coming from combined experiment information. This problem might affect several RRR evaluation. Solving it will require detailed discussions on potentially missed uncertainties in the evaluation and required extensions to the RRR format.
    \item $^{234}$U, and $^{242}$Pu PFNS ENDF/B-VIII.1 covariances have lower experimental uncertainties than the standard ($^{252}$Cf(sf) PFNS) without supporting experiments. These covariances are underestimated and need to be updated.
    \item Fast $^{240}$Pu $\overline\nu$ and (n,f) cross section uncertainties were not updated with the newest standard covariances ($^{252}$Cf(sf) $\overline\nu$ and $^{235}$U(n,f) cross sections) and are now  below standard uncertainties. The same is true for fast $^{233}$U(n,f) cross section covariances. These $^{233}$U and $^{240}$Pu covariances are underestimated and need to be updated.
\end{itemize}

To mention another example of a successful fix: Updated $^{233}$U PFNS covariances in the 0--5 MeV incident-neutron energy range were identified as needed, as new PFNS from Ref.~\cite{capote:2016} at thermal and from Ref.~\cite{Rising2013} from 0.5--5 MeV were adopted.
One covariance matrix was supplied for ENDF/B-VIII.1 covering 0--5 MeV based on work of Ref.~\cite{Rising2013}.

It should be emphasized that the CovVal testing focused on covariances that were new in ENDF/B-VIII.1.
Issues in covariances that did not change might exist and need to be studied for the next release now that the code was developed.

\begin{figure}
\centering
\includegraphics[width=\columnwidth]{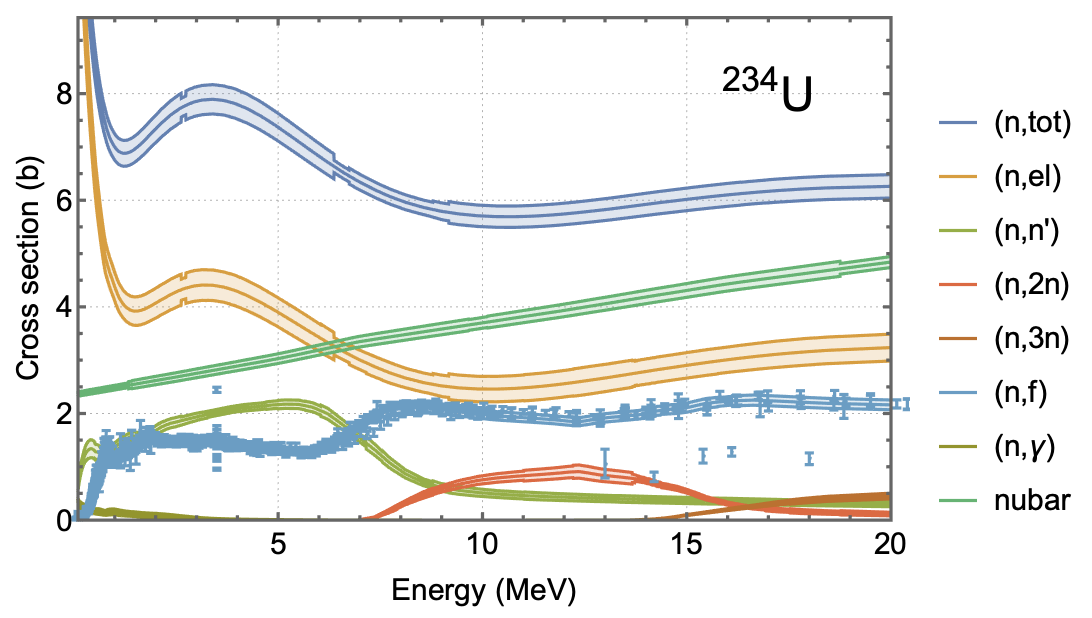}
\vspace{-2mm}
\caption{\label{fig:234Uunc81b1} $^{234}$U cross-section uncertainties for ENDF/B-VIII.1$\beta$1 are compared to EXFOR data to assess whether the uncertainties are reasonable given the spread in the experimental data.}
\vspace{-2mm}
\end{figure}

\begin{figure}
\centering
\includegraphics[width=\columnwidth]{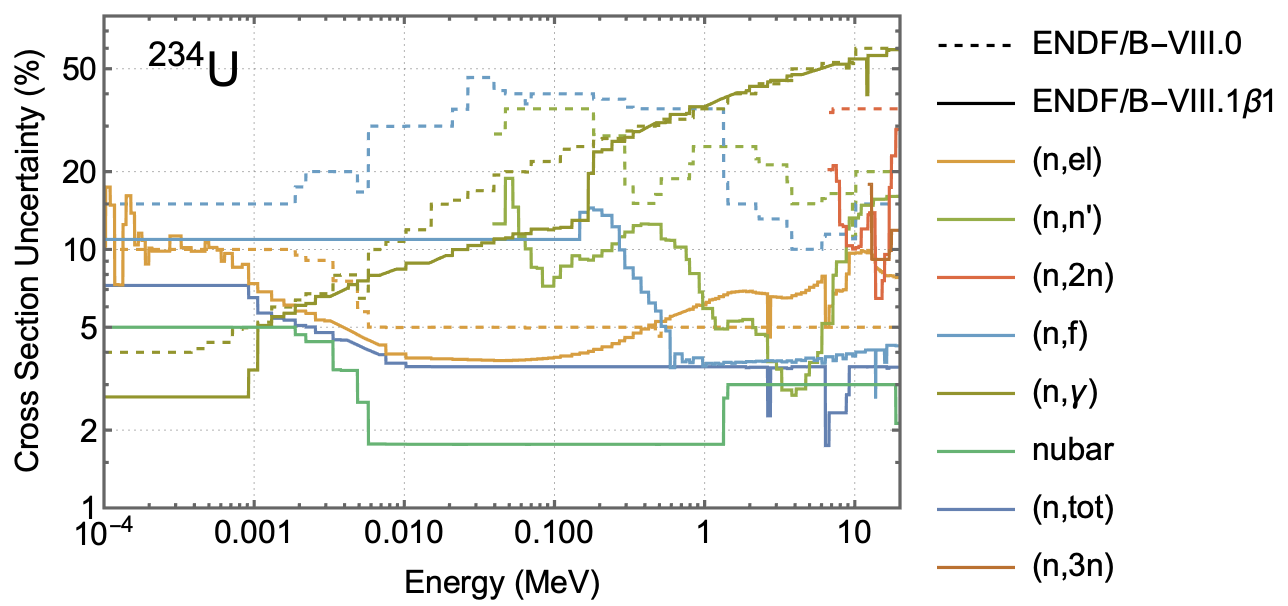}
\vspace{-2mm}
\caption{\label{fig:234Uunccomp8081b1} $^{234}$U cross-section uncertainties for ENDF/B-VIII.0 and ENDF/B-VIII.1$\beta$1.}
\vspace{-2mm}
\end{figure}

Additionally, visual comparisons were made between cross section confidence bands and EXFOR-data uncertainties \cite{EXFOR}, primarily looking for unjustifiably small or large uncertainties. For instance, $^{234}$U(n,f) cross-section uncertainties improved significantly from ENDF/B-VIII.0 to ENDF/B-VIII.1$\beta$1: the evaluated uncertainties at 2 MeV incident neutron energy represent better now the existing experimental data.  $^{234}$U(n,n') cross-section uncertainties, on the other hand, were found to have a smaller uncertainty at 4 MeV than any of the major actinides despite an absence of data as can be seen from Fig.~\ref{fig:234Uunc81b1}.  Fig.~\ref{fig:234Uunccomp8081b1} shows a comparison of $^{234}$U cross section uncertainties for ENDF/B-VIII.0 and ENDF/B-VIII.1$\beta$1.

Also, the covariances provided in ENDF/B-VIII.1$\beta$1 were processed through \AMPX\ and combined with the \SCALE\ 6.2 covariance library for any nuclides with no covariance data to generate a complete library for testing. This combined library was used to determine the estimated uncertainty in $k_\mathrm{eff}$ propagated from the nuclear data covariances for benchmarks in the VALID library maintained at ORNL~\cite{VALID} and for spent fuel in a hypothetical storage cask \cite{NUREGCR-6747}. The majority of the data-induced uncertainty in these systems comes from the major actinides, but the spent fuel system allows for testing the covariance data changes in major fission product nuclides typically credited in spent nuclear fuel (SNF) storage and transportation applications licensed by the US Nuclear Regulatory Commission \cite{NUREG-2216}.

The results of the data-induced uncertainty for the benchmarks are summarized below in Table~\ref{tab:keffbounds}. The standard deviation of the C/E values within each benchmark category is compared to the average data-induced uncertainty using the \SCALE\ covariance library based on ENDF/B-VIII.0 and the equivalent library based on ENDF/B-VIII.1 covariance data. In all nine categories, including different uranium enrichments, plutonium, fast spectrum benchmarks, and thermal spectrum systems, the observed variability is significantly lower than the predicted variation based on the nuclear data covariance. The ENDF/B-VIII.1-based data has significantly lower overpredictions compared to the ENDF/B-VIII.0 data for thermal spectrum Pu-fueled systems. The overall uncertainty in these systems is reduced by 20--30\%, with the change being driven by significant reductions in the $^{239}$Pu PFNS, fission, and (n,$\gamma$) cross sections. The (n,$\gamma$) uncertainty dropped approximately 95\%, which may require further evaluation in future releases. Unfortunately, the data-induced uncertainty increased for thermal spectrum U-fueled systems. This change was driven by increased uncertainty in $^{235}$U $\overline\nu$.

The reduction in $^{239}$Pu uncertainty and increase in $^{235}$U uncertainty for thermal systems will induce a shift in similarity metrics, such as c\textsubscript{k} for thermal SNF storage systems. c\textsubscript{k} values will be increased for LEU experiments and reduced for mixed systems. Most major fission product evaluations remained unchanged, though $^{103}$Rh uncertainty increased almost 30\%. This represents the $k_\mathrm{eff}$ uncertainty from $^{103}$Rh increasing from 21 pcm to 27 pcm in pressurized water reactor (PWR) fuel with a burnup of 40 GWd/MTU, so it is only a minor impact.

\begin{table}
\caption{\label{tab:keffbounds}  Average simulated uncertainties, $\sigma$, on $k_\mathrm{eff}$ values of ICSBEP due to ENDF/B-VIII.0 and ENDF/B-VIII.1 covariances. All values except for \# cases are in pcm. $\delta E$ are average experiment uncertainties and $\Sigma$ the standard devaiation of C/E.}
\centering
\footnotesize
\begin{tabular}{lcccccc}
\toprule \toprule
 &    \# Cases &  $\langle C/E \rangle$&  $\delta E$ & $\Sigma_{\mathrm{VIII.0}}$  & $\sigma$ & $\sigma_{\mathrm{VIII.1}}$   \\
\midrule
HMF  &      50     &1.00002    &193  &467 &     979 & 950 \\
HST    & 52  &0.999  &  494 & 615 & 652 & 792 \\
IMF    & 13 &     1.00132 &   269 & 362    &1027 &    1003 \\
LCT    & 140       &0.99874 & 195 & 162    &603 &     737 \\
LST    & 19 &     0.9992 &    318 & 283    &824  &944 \\
MCT    &49  & 0.99244 & 400    &313 &     973 & 758 \\
MST    & 10 & 0.99177 & 452    &384  &1323 &    1019 \\
PMF    & 12  & 0.99902   &207  &133  &1022       &1038 \\
PST    &81  & 0.99927    &497  &429  &1344 &    937 \\
\bottomrule \bottomrule
\end{tabular}

\end{table}

In Tables~\ref{tab:keffuncHMF001} and \ref{tab:keffuncPMF001}, it is shown that Godiva and Jezebel $k_\mathrm{eff}$ uncertainties due to PFNS covariances dropped significantly from ENDF/B-VIII.0 to ENDF/B-VIII.1.
This drop was expected and is supported by inclusion of high-precision experimental data that were recently released.
Jezebel $k_\mathrm{eff}$ uncertainties due to $^{239}$Pu $\overline\nu$ increased due to inclusion of somewhat larger standard uncertainties, while Godiva $k_\mathrm{eff}$ uncertainties due to $^{235}$U $\overline\nu$ are fairly similar.
A significant drop in Jezebel $k_\mathrm{eff}$ uncertainties was observed due to $^{239}$Pu capture uncertainties.

\begin{table}
\caption{\label{tab:keffuncHMF001} Impact of various $^{235}$U ENDF/B-VIII.0 and ENDF/B-VIII.1 covariances on simulated HMF-001 (``Godiva'' simple sphere) $k_\mathrm{eff}$ uncertainties. All values are in pcm.}
\centering
\begin{tabular}{lcc}
\toprule \toprule
&  ENDF/B-VIII.0 &  ENDF/B-VIII.1 \\
\midrule
(n,f) cs & 787 & 787\\
$\overline\nu$ & 399 & 382\\
PFNS & 124 & 33 \\
(n,el) cs &  224 & 224\\
(n,inl) cs & 239 & 239 \\
(n,$\gamma$) cs & 277 & 277 \\
\midrule
Sum & 982 & 975 \\
Exp. Unc. & 100 & 100 \\
$C-E$ &  14 & 0 \\
\bottomrule \bottomrule
\end{tabular}
\end{table}

\begin{table}
\caption{\label{tab:keffuncPMF001} Impact of various $^{239}$Pu ENDF/B-VIII.0 and ENDF/B-VIII.1 covariances on simulated PMF-001 (``Jezebel'' simple sphere) $k_\mathrm{eff}$ uncertainties. All values are in pcm. The asterisk (*) highlights that the adjustment for the cross correlation between elastic and inelastic scattering is larger than the inelastic scattering contribution. The total scattering uncertainty (elastic and inelastic) is 469 pcm with ENDF/B-VIII.1$\beta$1 covariances compared to 396 pcm with ENDF/B-VIII.0 covariances.}
\centering
\begin{tabular}{lcc}
\toprule \toprule
&  ENDF/B-VIII.0 &  ENDF/B-VIII.1 \\
\midrule
(n,f) cs & 877 & 920 \\
$\overline\nu$ & 317 & 416\\
PFNS & 179 & 37\\
(n,el) cs &  484 &360 \\
(n,inl) cs & -119* &165 \\
(n,$\gamma$) cs &64 &19 \\
\midrule
Sum & 1061 & 1085 \\
Exp. Unc. & 130 & 130 \\
$C-E$ & -28 & -116\\
\bottomrule \bottomrule
\end{tabular}
\end{table}

Finally, O.~Cabellos processed ENDF/B-VIII.0, ENDF/B-VIII.1$\beta$2 and ENDF/B-VIII.1$\beta$4 covariances with \NJOY2016.74~\cite{NJOY}.
Then, covariances were sandwiched with ICSBEP benchmark sensitivities using the NDaST tool (www.oecd-nea.org/ndast) using the BOXER format. 
Average simulated $k_\mathrm{eff}$ uncertainties for Pu, HEU and $^{233}$U due to the processed covariances are tabulated in Tables~\ref{tab:keffboundsPu}--\ref{tab:keffboundsU3}.
As mentioned before, PFNS uncertainties for Pu assemblies dropped due to the inclusion of Chi-nu, CEA and Lestone experimental data, while $\overline\nu$ uncertainties increased because of considering the new standard uncertainties ($^{252}$Cf(sf)) in the evaluation process.
The lower cross section uncertainties in the RRR triggered some discussions as mentioned above and is an active field of investigation.

\begin{table}
\caption{\label{tab:keffboundsPu}  Average simulated uncertainties on $k_\mathrm{eff}$ values of Pu ICSBEP assemblies due to ENDF/B-VIII.0 and ENDF/B-VIII.1 covariances for specific observables. All values except for \# cases are in pcm. $\delta E$ are average experiment uncertainties. Cross-section uncertainties consider cross correlations. ENDF/B-VIII.1$\beta$4 was used.}
\centering
\begin{tabular}{lcccc}
\toprule \toprule
Spectrum & FAST & INTER & MIXED & THERM  \\
\midrule
\# benchmarks & 152 & 4 &9 & 624\\
$\delta E$ & 334 & 710 & 587 & 427\\
\midrule
Total VIII.1  & 931 & 551 & 536 & 645\\
Total VIII.0  & 921 & 1403 & 1055 & 1099\\
\midrule
Cross sections VIII.1  & 841 & 360 & 224& 106 \\
Cross sections VIII.0  & 857 & 1368 & 983 & 998 \\
\midrule
P1-elastic VIII.1  & 66 &-&-&-\\
P1-elastic VIII.0  & 66 & -&-&-\\
\midrule
$\overline\nu$ VIII.1  & 390 & 389 & 479 & 632\\
$\overline\nu$ VIII.0  & 300 & 271 & 275 & 308 \\
\midrule
PFNS VIII.1  & 27 & 25 & 53& 59\\
PFNS VIII.0  & 117 & 106 & 265 & 297\\
\bottomrule \bottomrule
\end{tabular}
\end{table}
    
\begin{table}
\caption{\label{tab:keffboundsHEU}  Average simulated uncertainties on $k_\mathrm{eff}$ values of HEU ICSBEP assemblies due to ENDF/B-VIII.0 and ENDF/B-VIII.1 covariances for specific observables. All values except for \# cases are in pcm. $\delta E$ are average experiment uncertainties. Cross-section uncertainties consider cross correlations. ENDF/B-VIII.1$\beta$2 was used.}
\centering
\begin{tabular}{lcccc}
\toprule \toprule
Spectrum & FAST & INTER & MIXED & THERM  \\
\midrule
\# benchmarks & 477 & 21 & 78 & 798\\
$\delta E$ & 225 & 310 & 415 & 473\\
\midrule
Total VIII.1  & 883 & 649 & 588 & 667 \\
 Total VIII.0  & 911 & 758 & 649 & 497\\
\midrule
Cross sections VIII.1  & 811 & 514 & 442 & 143 \\
Cross sections VIII.0  & 816 & 545 & 470 & 155\\
\midrule
P1-elastic VIII.1  & - & - & - & - \\
P1-elastic VIII.0  & - & - & - & -\\
\midrule
$\overline\nu$ VIII.1  & 346 & 333 & 364 & 640 \\
$\overline\nu$ VIII.0  & 402 & 501 & 435 & 457\\
\midrule
PFNS VIII.1  & 29 & 60 & 61 & 98 \\
PFNS VIII.0  & 29 & 60 & 61 & 98\\
\bottomrule \bottomrule
\end{tabular}
\end{table}
    
\begin{table}
\caption{\label{tab:keffboundsU3}  Average simulated uncertainties on $k_\mathrm{eff}$ values of $^{233}$U ICSBEP assemblies due to ENDF/B-VIII.0 and ENDF/B-VIII.1 covariances for specific observables. All values except for \# cases are in pcm. $\delta E$ are average experiment uncertainties. Cross-section uncertainties consider cross correlations. ENDF/B-VIII.1$\beta$4 was used.}
\centering
\begin{tabular}{lcccc}
\toprule \toprule
Spectrum & FAST & INTER & MIXED & THERM  \\
\midrule
\# benchmarks & 8 & 29 & 8 & 194\\
$\delta E$ & 174 & 670 & 601 & 600\\
\midrule
Total VIII.1  & 810 & 1001 & 1026& 1018\\
Total VIII.0  & 810 & 1132 & 1155 & 1131\\
\midrule
Cross sections VIII.1  & 640 & 145 & 161 & 347 \\
Cross sections VIII.0  & 640 & 315 & 329  & 514 \\
\midrule
P1-elastic VIII.1  & - & - & - & -\\
P1-elastic VIII.0  & - & - & - & - \\
\midrule
$\overline\nu$ VIII.1  & 475 & 479 & 490 & 539 \\
$\overline\nu$ VIII.0  & 484 & 496 & 501 & 508 \\
\midrule
PFNS VIII.1  & 137 & 867 & 887 & 758 \\
PFNS VIII.0  & 95 & 967 & 987 & 837\\
\bottomrule \bottomrule
\end{tabular}
\end{table}

\subsection{Templates of Expected Measurement Uncertainties}
Templates of expected measurement uncertainties were established in Refs.~\cite{tempintro,ngamma_cp,tot_template,LewisPhD,nxn,PFNS,nu,MatthewPhD}.
This effort was started after fission cross section templates~\cite{Neudecker2020, Neudecker:2019tempnf} and initial versions of total and capture templates were discussed at the fall CSEWG meeting of 2018.
In spring 2019, a mini-CSEWG on that topic was held, and the scope was extended to encompass prompt fission neutron spectra, average prompt and total fission neutron multiplicities, charged-particle cross sections, (n,xn) cross sections, and fission yield.

These provide listings of expected uncertainties for a specific measurement type related to one nuclear-data observables; for instance, transmission for total cross sections or absolute liquid-scintillator measurements for the average prompt fission neutron multiplicity. These listings can be used by experimenters as a check-list before the release of their data, or by evaluators to counter-check whether all pertinent uncertainties were provided for experiments chosen for their evaluation input database.
Also, recommended values of many uncertainty sources are provided based on literature, EXFOR entries \cite{EXFOR}, and expert judgment from experimenters. These uncertainty values can be used as a last resort if they are not provided for past experiments. These uncertainty values should only be used for evaluations in the absence of other information to estimate the uncertainty source.
Estimates of correlation coefficient between uncertainties of the same and different experiments for evaluation purposes are also given.
The long-term goal of these templates is to provide better experimental uncertainties for evaluation purposes, be it either that experimenters use them as check-lists to provide data, or by evaluators to fill in missing uncertainties for their evaluation database.
Ultimately, having better quantified experimental uncertainties for evaluation purpose will lead to more realistic evaluated uncertainties for nuclear data libraries.

Most of the templates were published in 2023, at the end of the library release cycle. 
Despite that, they were already applied for some evaluations:
The fission cross section template was applied in Ref.~\cite{Neudecker:2019tempnf} for the experimental uncertainty quantification of the fast ENDF/B-VIII.1 $^{239}$Pu(n,f) cross section, while the templates for $\overline\nu_p$~\cite{nu} were applied to inform part of ENDF/B-VIII.1 $^{235,238}$U and $^{239}$Pu $\overline\nu_p$ evaluations.
The experimental uncertainty quantification for fast ENDF/B-VIII.1 $^{235}$U and $^{239}$Pu PFNS evaluations also made use of PFNS template uncertainties~\cite{PFNS}, where individual uncertainty sources were not provided by the author.
Finally, information from all the templates described above were used for covariances testing with the  CovVal codes~\cite{CovVal} as described above. 

\subsection{Decision Process of CSEWG to not Provide Mathematically Adjusted Libraries for ENDF/B-VIII.1 }
Meeting records as far back as  the eighties of the last century show that CSEWG regularly had discussions on whether evaluated mean values and covariances should be formally adjusted by the CSEWG community to trusted integral data.
Formally adjusted means in this context that nuclear data mean values and covariances are updated with integral mean values and uncertainties via a mathematical algorithm, such as generalized least squares.
So far, this step has not been undertaken but been left for the application user to do with integral experiments of their choice.
However, ENDF/B nuclear data libraries do use information coming from integral data.

For instance, ENDF/B-VIII.0 nuclear data mean values were partially tweaked by hand or favorably selected to reproduce $k_\mathrm{eff}$ values of selected ICBSEP critical assemblies~\cite{ICSBEP}, LLNL pulsed spheres~\cite{Wong1972}, reaction rates, etc.~\cite{Brown2018}.
Such changes were undertaken by computing integral data first with nuclear data obtained without integral information.
Then, a limited set of nuclear data, usually one or two observables, were changed by hand until one reached the desired agreement with integral data.
Examples of such changes were documented for the thermal $^{239}$Pu prompt fission neutron spectrum or for the $^{235}$U average prompt fission neutron multiplicity in the resonance range~\cite{Brown2018}.
Such changes are usually undertaken within the bounds of differential experimental data.
In formal adjustment procedures, on the contrary, usually all nuclear data that are sensitive to the particular integral data are adjusted simultaneously.

Another important difference between by-hand tweaking and formal adjustment is that in the former case ENDF/B covariances remain unchanged.
This leads to seemingly over predicted $k_\mathrm{eff}$ uncertainties when one forward-propagates nuclear data covariances to simulated $k_\mathrm{eff}$ bounds and compares them to the spread in calculated over experimental $k_\mathrm{eff}$ values C/E~\cite{WANDA2020:summaryreport}.
As a consequence, some users do not use ENDF/B covariances because this tweaking renders covariances inconsistent with mean values.
Some users have expressed concerns about this approach and the lack of documentation of it, because many of these changes and choices on nuclear data are not fully documented and neither are the benchmarks used for this by-hand tweaking and the reasoning for the choices.
This lack of documentation also constitutes a problem for some user communities as it is unclear which $k_\mathrm{eff}$ benchmarks can be used for formal adjustment~\cite{WANDA2020:summaryreport} while others are already incorporated in ENDF/B data.
Of course, many of these tweaks are built into ENDF/B libraries through the work of generations of scientists. Removing them means re-doing major parts of ENDF/B.
On the other hand, a formal adjustment may lead to producing a library which will be only valid to describe specific assemblies, geometries, etc., which may not be acceptable for a general purpose library.

Given this situation and frequent criticism by users~\cite{WANDA2020:summaryreport}, discussions on whether CSEWG should provide one true general-purpose library (i.e., without by-hand tweaking), and additional adjusted nuclear data libraries using formal mathematical algorithms came up regularly at CSEWG meetings.
To reach a decision for what to do for ENDF/B-VIII.1, a detailed discussion was held on the topic of adjustment as part of a mini-CSEWG covariance session on August 19, 2021 culminating in a vote whether adjustment should be undertaken for ENDF/B-VIIII.1.
This vote came out with more than two-thirds of attendees preferring not to adjust, reflecting the reality that there are no tools applicable to all integral responses catering to many subject areas, and that CSEWG is not yet ready to distribute and vet several adjusted libraries.

However, many people showed interest in learning about potential weaknesses in the library through adjustment results. 
Also, the following recommendations were given for the future:
\begin{itemize}
\item Results from adjustment and tool development should be discussed in the covariance session. The aim is to give recommendations to user how adjustment could be done and warn them about potential caveats. Also, evaluators could learn from adjustment results where nuclear data could be improved.
\item If possible, evaluators should document some of their by-hand tweaks in ENDF/B-VIII.1 nuclear data along with the integral experiments used,
\item Work should be undertaken in the future with the validation session to agree on what benchmarks/integral responses we use for tweaking. Integral responses used for tweaking should be carefully vetted.
\end{itemize}

\section{THERMAL NEUTRON SCATTERING SUBLIBRARY}
\label{sec:tsl-sublib}


Accurate description of thermal neutron scattering on different materials is required in modeling neutron transport in many different circumstances, from nuclear power reactors to cryogenic setups at neutron beam facilities. 
Nuclear data files (File 7), which allow the double differential scattering cross sections in the thermal energy region (and up to $E \sim $1--5 eV) to be constructed
for the specified materials, are distributed in a special sublibrary called the thermal neutron scattering law (TSL) sublibrary, or $S(\alpha, \beta)$ data. 
Materials are classified as moderators, fuel, or special purpose -- usage restrictions defined below.

The TSL sublibrary of ENDF/B-VIII.1 contains 114 evaluations for 69 materials. Nine materials (light water, beryllium oxide, beryllium-metal, yttrium hydride, silicon carbide, silicon dioxide, Lucite, liquid hydrogen, and liquid deuterium) were reevaluated, 
and 18 new materials were added (calcium hydride, uranium hydride, lithium-7 hydride, lithium-7 deuteride, paraffinic oil, polystyrene, Teflon, FLiBe, hydrogen fluoride, sapphire, uranium-metal, plutonium dioxide, magnesium oxide, beryllium fluoride, magnesium fluoride, beryllium carbide, uranium carbide, and zirconium carbide). 
Two materials were evaluated with added physics (crystalline-graphite+\sd and beryllium-metal+\sd). Four material evaluation sets were expanded to include new phases, material conditions, or enrichments (zirconium hydride, 20\% porous graphite, uranium dioxide, and uranium nitride). The remaining evaluations were imported from the ENDF/B-VIII.0 TSL sublibrary. Due to the increase in the number of materials in the current release, a new database of MAT numbers for File 7 has been established.

New File 7 evaluations and reevaluations for ENDF/B-VIII.1 were generated in ENDF-6 format \cite{ENDF6-Format-2024} by Naval Nuclear Laboratory (NNL), the Low Energy Interaction Physics (LEIP) group from Texas A\&M University (TAMU) and North Carolina State University (NCSU), ORNL, and European Spallation Source ERIC (ESS).
The efforts were coordinated through NEA/WPEC subgroups 42 and 48 \cite{report42_2020}. The new evaluations and reevaluations were performed using the \FLASSH~\cite{Fleming2023}, \NJOY~\cite{NJOY}, and \NJOY+\texttt{NCrystal}~\cite{RAMIC2022166227} code systems. The evaluations were developed with the use of computational material science methods (density functional theory, lattice dynamics, and molecular dynamics). 
Most evaluations were developed in the incoherent approximation with semi-empirical force-field molecular dynamics (MD) or \ab\ lattice dynamics (AILD) techniques \cite{Hawari2004,Hawari2014} introduced in ENDF/B-VIII.0. 
However, several physics and methodology advances are introduced in this release. An \ab\ molecular dynamics (AIMD) methodology \cite{Wormald2017,Wormald2017-epj} has been introduced to treat temperature effects on vibration in ZrH$_x$ and ZrH\textsubscript{2}. A technique to adjust computed molecular vibration spectra with inelastic neutron measurements \cite{RAMIREZCUESTA2004226,RAMIC2019425} has been introduced for polystyrene and Lucite. Coherent inelastic 1-phonon corrections have been introduced to Be-metal+\sd and crystalline-graphite+\sd as a relaxation of the incoherent and cubic approximations used in other evaluations.  The ENDF-6 format was extended to support thermal scattering evaluations for materials, which include of both coherent and incoherent elastic scattering effects \cite{mixed-elastic}. 
 
Unlike prior ENDF/B releases, uranium compounds (UC, UN, UO\textsubscript{2}, U-metal) included in this release are evaluated for several enrichments of \textsuperscript{235}U: natural, 5\%, 10\%, High-Assay Low-Enriched Uranium (HALEU) (19.75\%), HEU (93\%), and 100\%. 
Although it is the same material for each value of enrichment, each evaluation (e.g., U(UO\textsubscript{2}-HEU) and O(UO\textsubscript{2}-HEU)) is assigned a unique MAT numbers. 
The UN and UO\textsubscript{2} reevaluations use the same material models as ENDF/B-VIII.0 with updated lattice parameter; whereas UC and U-metal are considered new evaluations.

Several materials (BeO, SiC, SiO\textsubscript{2}, Be-metal, graphite, UO\textsubscript{2}, UN) were reevaluated to use room temperature experimental data rather than computational ones for the crystal lattice parameters. 
The material models and evaluation methodologies of these material evaluations are otherwise the same as or highly similar to ENDF/B-VIII.0.

Details of the evaluation and validation methodologies for each new and updated material in this release are given below in three subsections: moderators, fuels, and special purpose. Moderator and fuel evaluations support criticality analyses. Fuel evaluations can also aid fundamental investigations of chemical binding effects on Doppler broadening. Special purpose evaluations, such as liquid H$_2$ and D$_2$, sapphire single-crystal, etc., 
are optimized for neutron beamline applications and are \textbf{not
suitable} for use in nuclear reactor design and criticality
safety applications.

As the ENDF-6/GNDS formats for the TSL covariances (File 37) are not finalized yet, and additional research is needed on methods to generate consistent covariances, 
the TSL evaluations in ENDF/B-VIII.1 are given without estimations of their uncertainties. 
The consistent evaluations of the TSL data with their covariances are 
left for the future releases of the ENDF/B family of nuclear data libraries.

\subsection{Moderators}
\label{subsec:tsl:moderators}

\subsubsection{Light water (H$_2$O)} 
\label{sec:H2Otsl}

Similarly to ENDF/B-VIII.0, the proposed evaluation for light water is based on the CAB Model for neutron scattering on water 
\cite{marquez2014cab}. This model reproduces very well the properties of water and the evolution of the total cross section with
 temperature \cite{marquez2020experimental}. Nevertheless, the discrete nature of the MD simulations used in this 
 model made it a collection of evaluations for different temperatures instead of a single continuous evaluation.

\begin{figure}
    \centering
    \includegraphics[width=1.0\columnwidth]{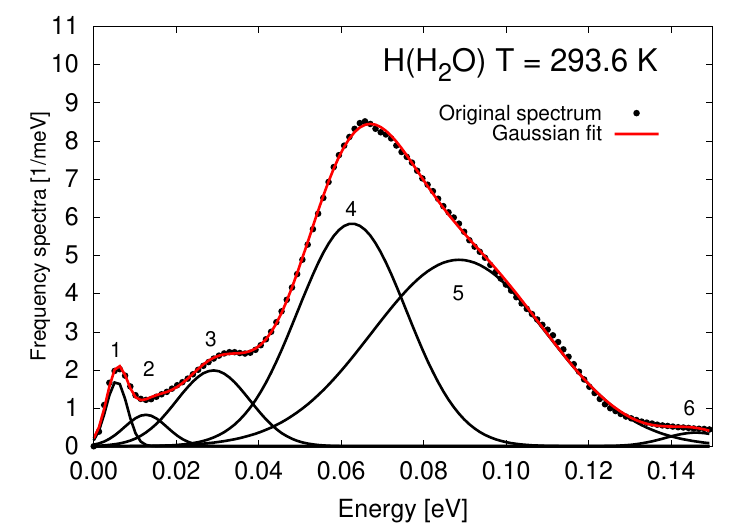}
    \caption{(Color online) Frequency spectrum of hydrogen in light water at $T = 293.6$ K and its decomposition in Gaussian 
    distributions, which are numbered from low to high energy. Although this description is purely mathematical, there are certain 
    features that can be associated with each distribution: 1, 2 and 3 roughly correspond to stretching and bending of the hydrogen
     bond network, and 4, 5 and 6 correspond to hindered molecular rotations (librations)  \cite{lisichkin2005temperature}.}
    \label{fig:rho_HH2O}
\end{figure}

To overcome this deficiency, the evaluation was updated to include a continuous temperature behavior. In this new methodology, the 
computed vibrational spectra for all temperatures is decomposed as a sum of Gaussian contributions, following the work by Esch \cite
{esch1971temperature}, Lisichkin \cite{lisichkin2005temperature} and Maul \cite{maul2018perturbation}. In Fig. \ref{fig:rho_HH2O}, we 
show the Gaussian decomposition of the frequency spectrum at $T = 293.6$~K. The parameters for the Gaussian distributions and the 
translational weight are then fitted using a fourth-order polynomial in inverse temperature, using the values at $T_0 = 293.6$ K as 
anchor points. The diffusion coefficient is interpolated using a logarithmic fifth-order polynomial in inverse temperature, following
 the work by Yoshida \cite{yoshida2010scaled} but modified to reproduce the value of the diffusion coefficient used in ENDF/B-VIII.0 at
  room temperature. The data from the CAB Model was complemented with a new MD simulation at $T = 275$~K to improve the
   performance below room temperature. Further details on the analysis, and the source code used to perform it, can be found in 
   Ref.~\cite{h2oparam}.

Using this methodology, the thermal scattering library for hydrogen in light water was computed on a $5$ K grid from $273.15$ K to 
$647.1$ K, and a extrapolated $50$ K grid above the critical temperature from $650$ K to $1000$ K. As in previous evaluations, oxygen 
in light water should be included using a free gas model.

\begin{figure}
    \centering
    \includegraphics[width=1.0\columnwidth]{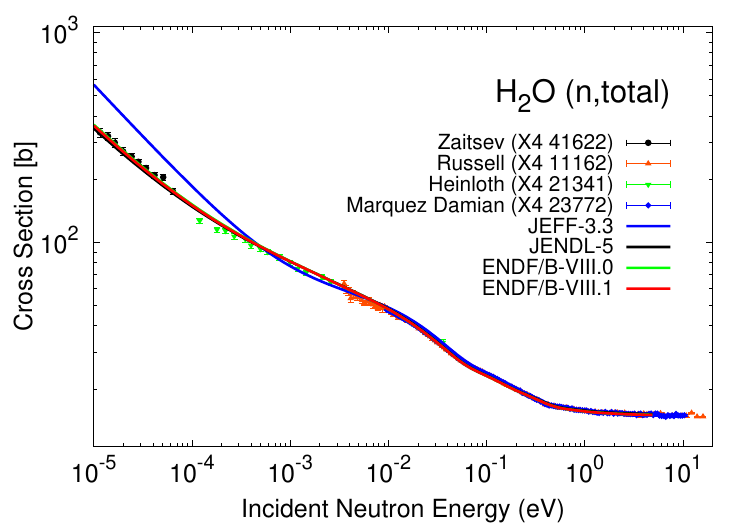}
    \caption{(Color online) Total cross section for the interaction of neutrons with light water at room temperature as a function of
     energy.}
    \label{fig:xs_H2O_RT}
\end{figure}

\begin{figure}
    \centering
    \includegraphics[width=1.0\columnwidth]{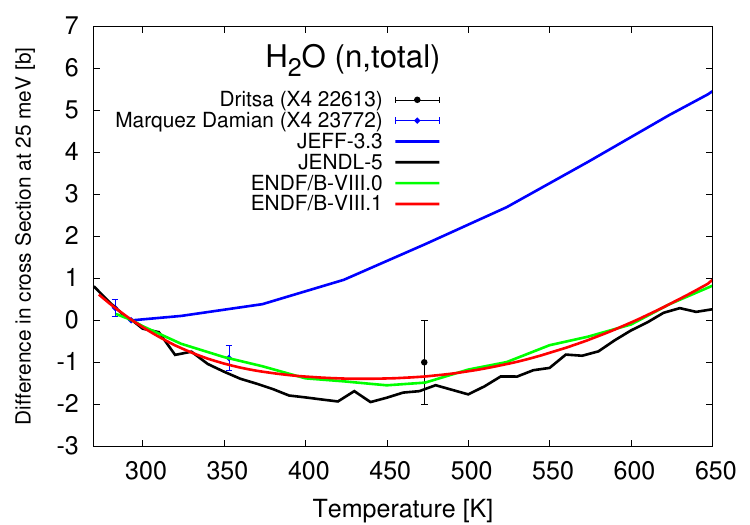}
    \caption{(Color online) Difference in total cross section for the interaction of neutrons with light water at $E = 0.0253$ eV as a 
    function of water temperature against the reference value at $T = 293.6$ K.}
    \label{fig:xs_24meV}
\end{figure}

Overall, the good agreement found in the ENDF/B-VIII.0 evaluation at room temperature is preserved in the updated evaluation of H$_2$O 
in ENDF/B-VIII.1 (Fig.~\ref{fig:xs_H2O_RT}). In addition, the temperature dependence of the neutron scattering cross section is now 
smoother (see Fig.~\ref{fig:xs_24meV}). Compared with the JENDL-5 \cite{jendl5} evaluation, a general agreement is observed, which 
might be caused from both evaluations being derived from MD simulations using the TIP4P/2005f model. However, the 
JENDL-5 evaluation shows oscillations similar to the ones observed in the ENDF/B-VIII.0 evaluation.
These oscillations were suppressed in update-16 (January 2025) of JENDL-5 $S(\alpha,\beta)$ for H in H$_2$O and O in  H$_2$O \cite{update_16} 
by smoothing the temperature-dependent vibrational density of states (frequency spectra)  
based on multiple MD simulations of liquid H$_2$O at a given temperature \cite{New_h2o_JENDL5}.
Bigger differences are seen when 
the computed cross sections are compared with the JEFF-3.3 \cite{JEFF33} light water evaluation, which is based on the model by Keinert
 and Mattes \cite{mattes2005thermal}.

\subsubsection{Beryllium (\bem)}
\label{sec:be}

This contribution updates the \bem\ evaluation of \prENDF\ with average experimental lattice parameters \cite{Larsen1984,Gordon1949} in the calculation of the coherent elastic data instead of \ab\ values as well as additional temperatures for potential utilization as a cold neutron moderator/reflector. 

Be-metal was evaluated using modern AILD methods \AILD\ for its hexagonal close-packed crystal structure \cite{Wormald2017}. This \ab\ model is the same as that used in the \prENDF\ evaluation. Electronic and bulk structure minimization was conducted using density functional theory (DFT) to validate thermophysical properties predicted in the \texttt{VASP} code~\VASP. Subsequently, Hellmann-Feynman forces were calculated using the finite displacement method and used in PHONON \cite{Parlinski1997} for predicting phonon dispersion relations and sampling phonon density of states (PDOS). Computed dispersion relations map phonon wave vectors to energy along high-symmetry lattice directions and demonstrate good agreement with experimental inelastic neutron scattering measurements as shown in Ref.~\cite[Fig. 3]{Wormald2017}. 

\begin{figure}
    \centering
    \includegraphics[width=1.0\columnwidth,clip,trim=  20mm 10mm 30mm 15mm]{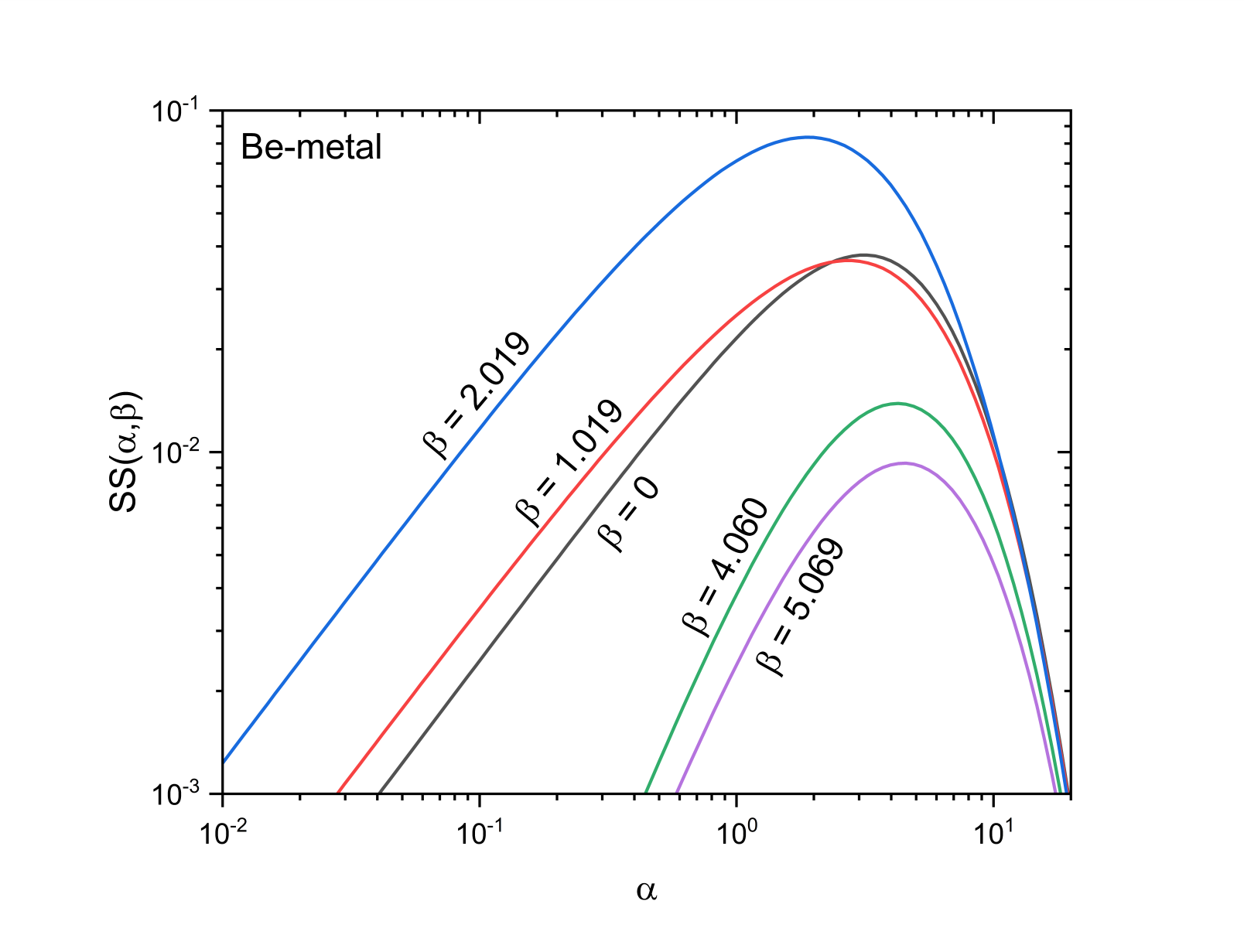}
    \caption{(Color online) The symmetric TSL of \bem\ as a function of momentum transfer, $\alpha$, for various values of $\beta$ at 296 K. $SS(\alpha,\beta)$  for each $\beta$ is labeled with the corresponding line.}
    \label{fig:TSL_Be}
\end{figure}

The ENDF TSL for \bem\ (File 7, MAT26) was generated using the AILD density of states (DOS) in the Full Law Analysis Scattering System Hub (\FLASSH) \cite{Fleming2023} for cryogenic and standard temperatures of 77, 100, 293.6, 296, 400, 500, 600, 700, 800, 1000, and 1200 K. Mass and free atom cross sections \cite{Brown2018} for $^9$Be 
are used as input. The symmetric TSL, \emph{i.e.}, \Ss, is tabulated as ENDF File 7 (MAT26), MT4 in the incoherent approximation \cite{NJOY91, ENDF6-Format-2012} and is illustrated at 296 K versus $\alpha$ for select $\beta$ in Fig.~\ref{fig:TSL_Be}. Coherent elastic scattering is evaluated in \FLASSH\ within the cubic approximation and stored as ENDF File 7, MT2.  

\begin{figure}
    \centering
    \includegraphics[width=1.1\columnwidth,clip,trim=  15mm 5mm 15mm 10mm]{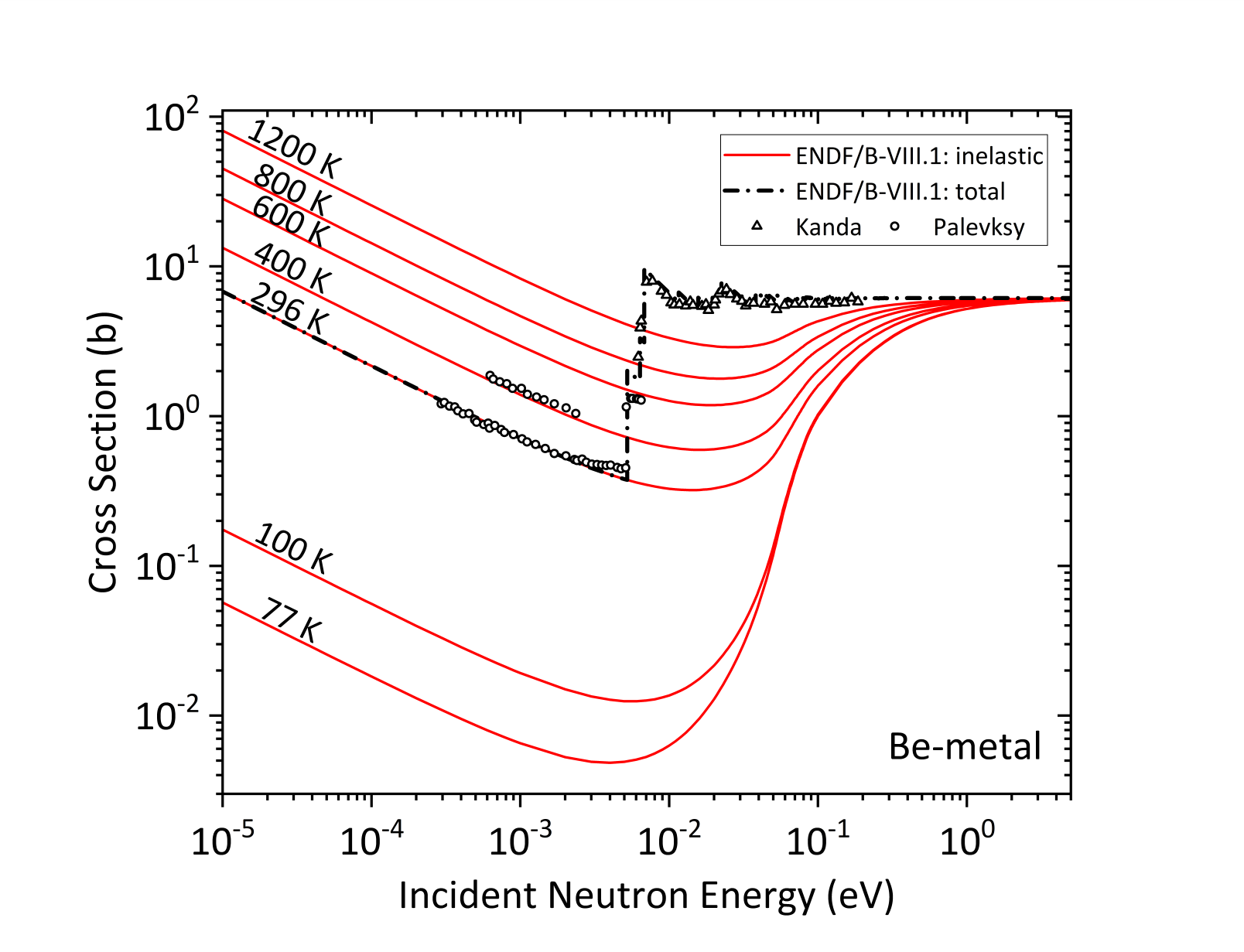}
    \caption{(Color online) The inelastic scattering cross section of \bem\ shown at selected temperatures. The total cross section of \bem, experimental measurements at room temperature \cite{Palevsky1952,Kanda1975} and measurements at 440 K \cite{Palevsky1952} are shown.}
    \label{fig:Xsec_Be}
\end{figure}

 The \ENDF\ total cross section of \bem\ is shown in Fig.~\ref{fig:Xsec_Be} \cite{Palevsky1952, Kanda1975}. Experimental cross sections have been used for validation at room temperature and higher temperatures and show excellent agreement with the \ENDF\ data. Lower temperature cross section measurements have been discarded for validation as competing effects from absorption dominate the total cross section. The inelastic scattering cross sections are also shown at select temperatures including the extension to cryogenic temperatures. 

\subsubsection{Beryllium-Metal with Distinct Effects (\besd)}
\label{sec:besd}

Beryllium-metal was evaluated using AILD techniques \AILD\ to generate the phonon dispersion relationships and resulting PDOS. The validated \prENDF\ material model \cite{Wormald2017} was used as the primary input to generate a novel evaluation which removes traditional approximations. The incoherent, cubic, and atom-site approximations to the TSL evaluation are removed; instead, this evaluation provides a non-cubic formulation of the TSL with additional distinct effects (\sd), which include coherent interactions in the inelastic cross section \cite{Hawari2008}. The \sd\ libraries were generated in a consistent fashion with the standard \bem\ evaluation. The same inputs and codes are used for both evaluations. Therefore, users would be able to recognize directly the impact of the isolated \sd\ effect on their calculations. Consequently, including the \ENDF\ \sd\ evaluations as presented here is highly informative to the community. 
 
\begin{figure}
    \centering
    \includegraphics[width=1.0\columnwidth,clip,trim=  0mm 0mm 0mm 0mm]{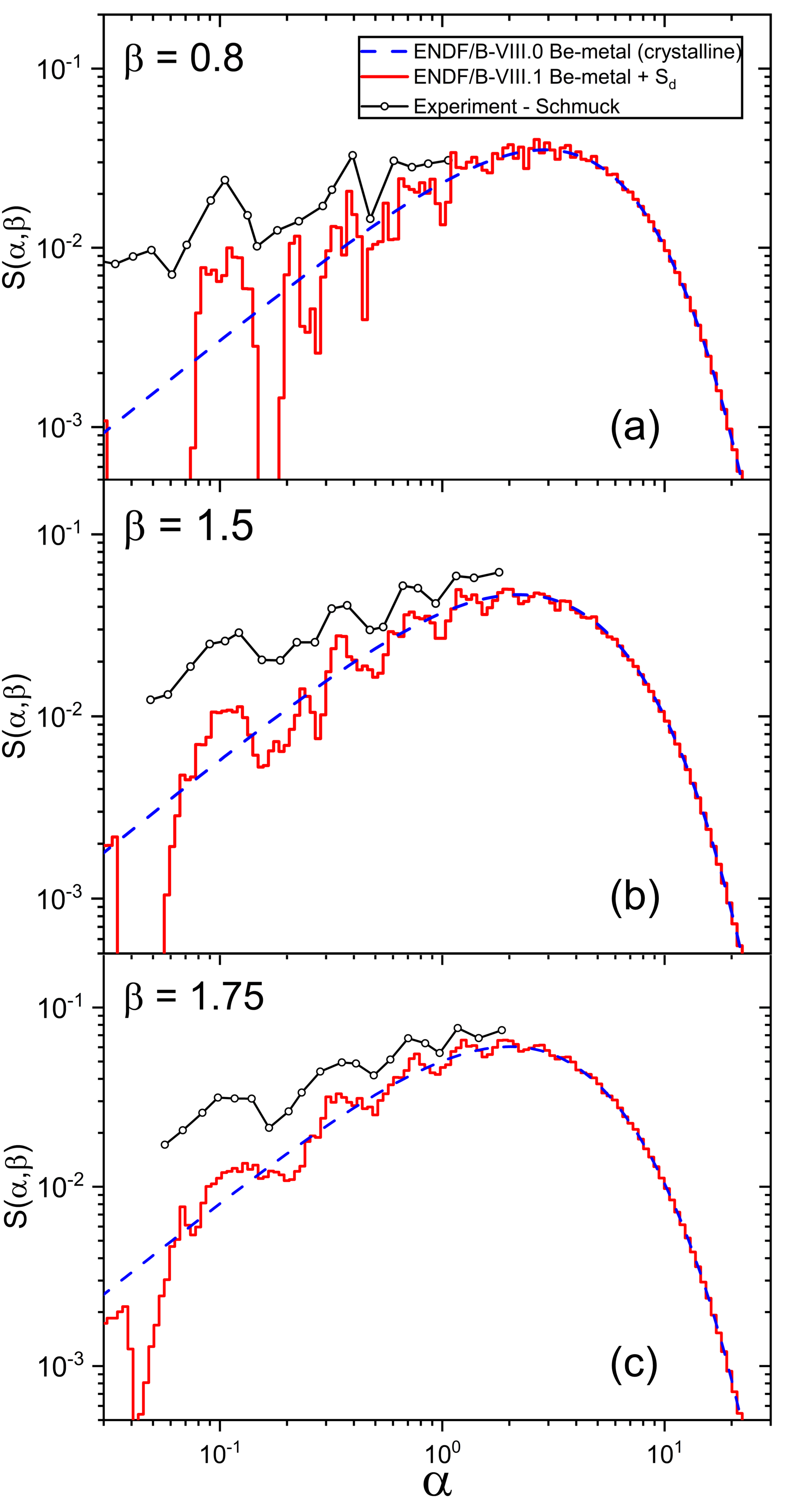}
    \caption{(Color online)The symmetric TSL for beryllium metal with \sd\ effects at 296 K as a function of momentum transfer $\alpha$ compared to experimental data \cite{Schmunk1964}. }
    \label{fig:TSL_BeSd}
\end{figure}

The dispersion relations, which were developed in this model, demonstrated reasonable agreement with the measurement \cite{Wormald2017}. The corresponding phonon DOS is fundamental input for the generation of the TSL with mass and free atom cross sections \cite{Brown2018} for $^9$Be. Using the \FLASSH\ code's advanced physics modules, the TSL for beryllium-metal with \sd\ effects (File 7, MAT204) was evaluated at temperatures of 77, 100, 296, 400, 500, 600, 700, 800, 1000, and 1200 K without the incoherent approximation and using the non-cubic formulation \cite{Fleming2023}. The symmetric TSL at 296 K tabulated in File 7, MT4 of the ENDF file is illustrated in Fig. \ref{fig:TSL_BeSd} and compared with experimental measurements \cite{Schmunk1964}.

\begin{figure}
    \centering
    \includegraphics[width=1.0\columnwidth,clip,trim=  10mm 8mm 10mm 3mm]{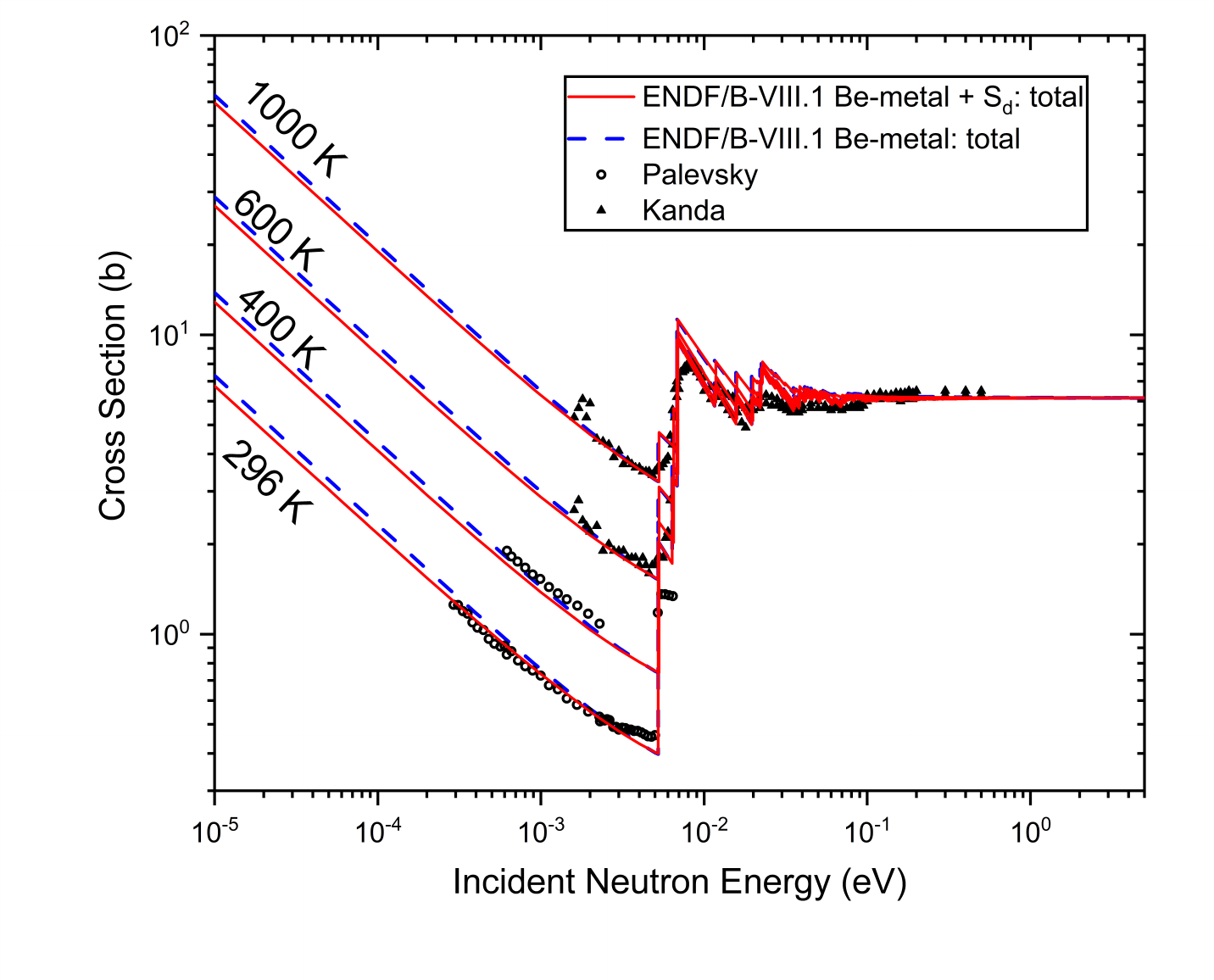}
    \caption{(Color online) Total cross sections of beryllium-metal with and without \sd\ effects at select temperatures are compared to experimental data at 300 \cite{Palevsky1952}, 440 \cite{Palevsky1952}, 573 \cite{Kanda1975}, and 973 K \cite{Kanda1975}.}
    \label{fig:Xsec_BeSd}
\end{figure}

Coherent elastic scattering was evaluated using the generalized non-cubic routine in \FLASSH\ with a Debye-Waller matrix capturing asymmetric lattice effects \cite{Fleming2023} with average experimental lattice parameters \cite{Larsen1984,Gordon1949}. This allows a consistent evaluation method between the TSL and resulting inelastic cross sections and the coherent elastic cross sections. The resulting total integrated scattering cross section for the \ENDF\ Be-metal with \sd\ effects, corrected for absorption, is shown in Fig. \ref{fig:Xsec_BeSd}. The comparison with experimental data shows excellent agreement of the \sd\ data.

This addition of distinct (\sd) effects provides the ability to directly benchmark the TSL evaluation against experimentally measured TSL data for beryllium-metal and improves the agreement with experimentally measured cross sections \cite{Palevsky1952,Kanda1975}. The beryllium-metal with \sd\ evaluation represents a new TSL that is included for the first time in the ENDF/B database.

\subsubsection{Beryllium Oxide (\beo)}
\label{sec:beo}

This contribution updates the \beo\ evaluation of \prENDF\ with the atomic mass and cross sections \cite{Brown2018} of natural  oxygen \cite{Sears1992}, as opposed to $^{16}$O, and uses experimental lattice parameters \cite{Hazen1986} in the calculation of the coherent elastic data instead of \ab\ values.

\begin{figure}
    \centering
    \includegraphics[width=1.0\columnwidth,clip,trim=  10mm 10mm 10mm 15mm]{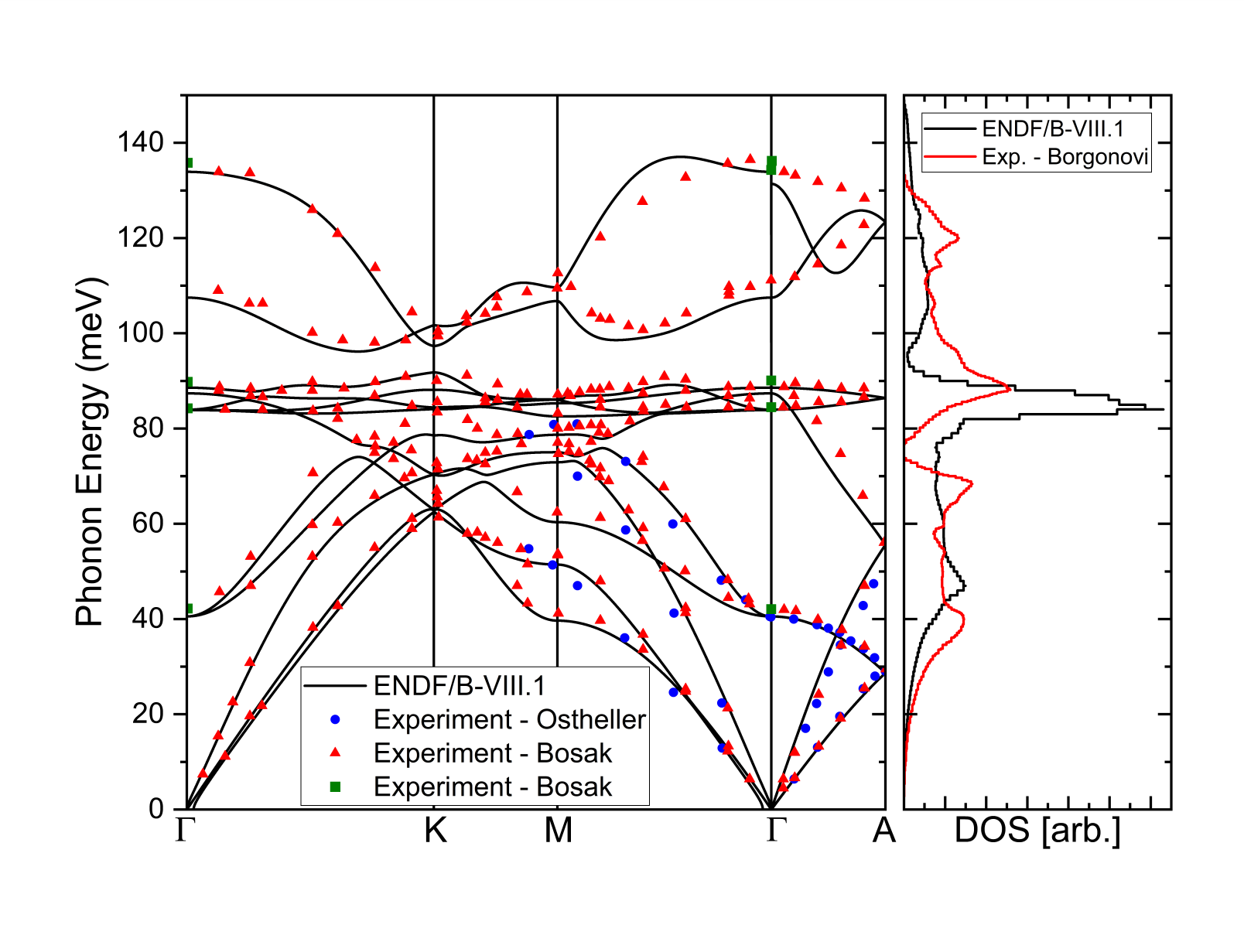}
    \caption{(Color online) The phonon spectrum of BeO used as input for the TSL evaluation compared with experimental data \cite{Borgonovi1968,Ostheller1968,Bosak2008}.} 
    \label{fig:Phonon_BeO}
\end{figure}

\beo\ was evaluated using AILD methods \AILD. \beo\ has a Wurtzite crystal structure. The \texttt{VASP} code \VASP\ was used to generate the Hellmann-Feynman forces for this structure, and the PHONON code \cite{Parlinski1997} was used to calculate the phonon dispersion curves and DOS for each element from the Hellmann-Feynman forces using the dynamical matrix method. This DOS was used in both the \prENDF\ and current evaluation. The phonon dispersions calculated using this method reasonably agree with experiment as shown in Fig.~\ref{fig:Phonon_BeO} \cite{Borgonovi1968,Ostheller1968,Bosak2008}.

\begin{figure}
    \centering
    \includegraphics[width=1.0\columnwidth,clip,trim=  20mm 10mm 30mm 15mm]{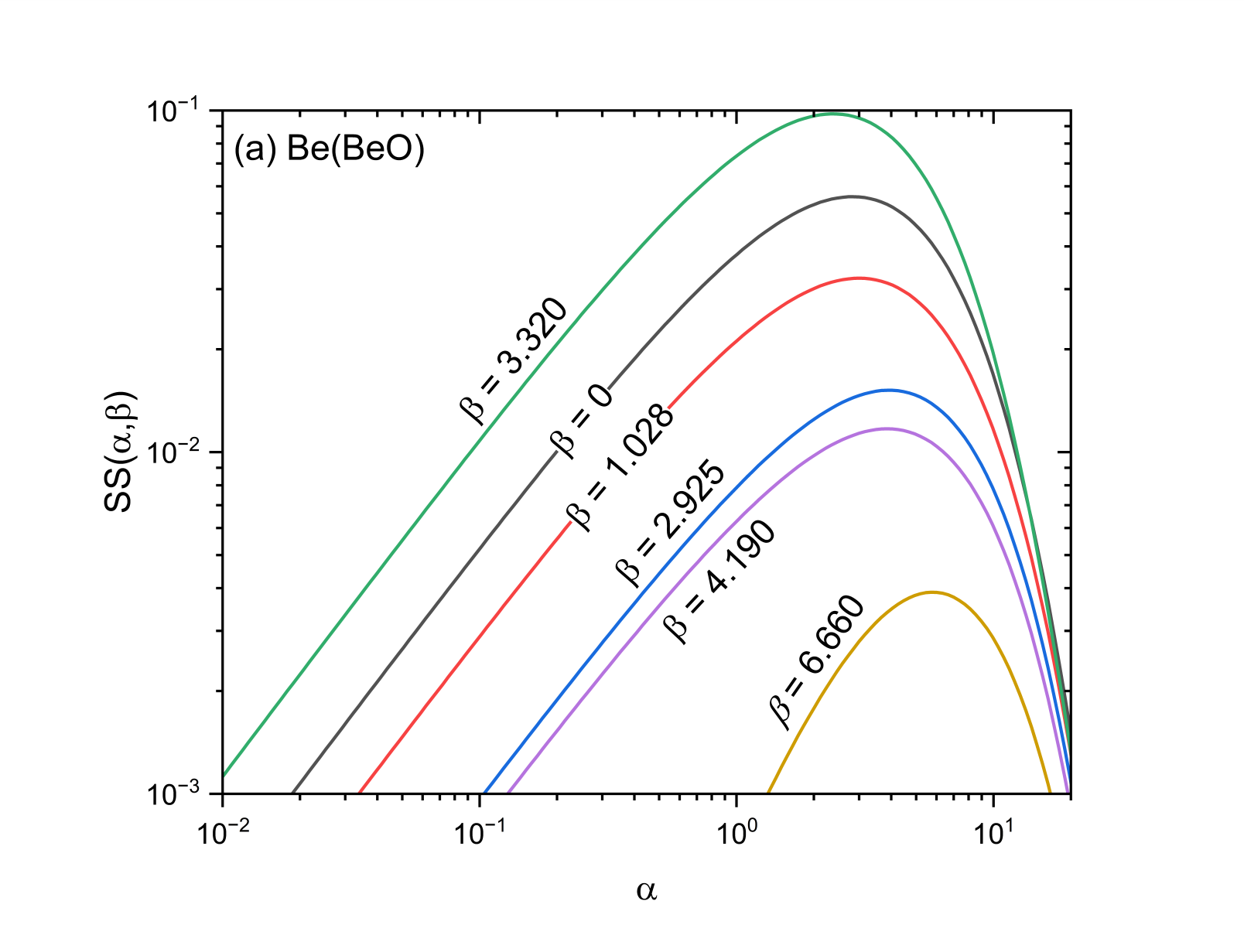}
    \includegraphics[width=1.0\columnwidth,clip,trim=  20mm 10mm 30mm 15mm]{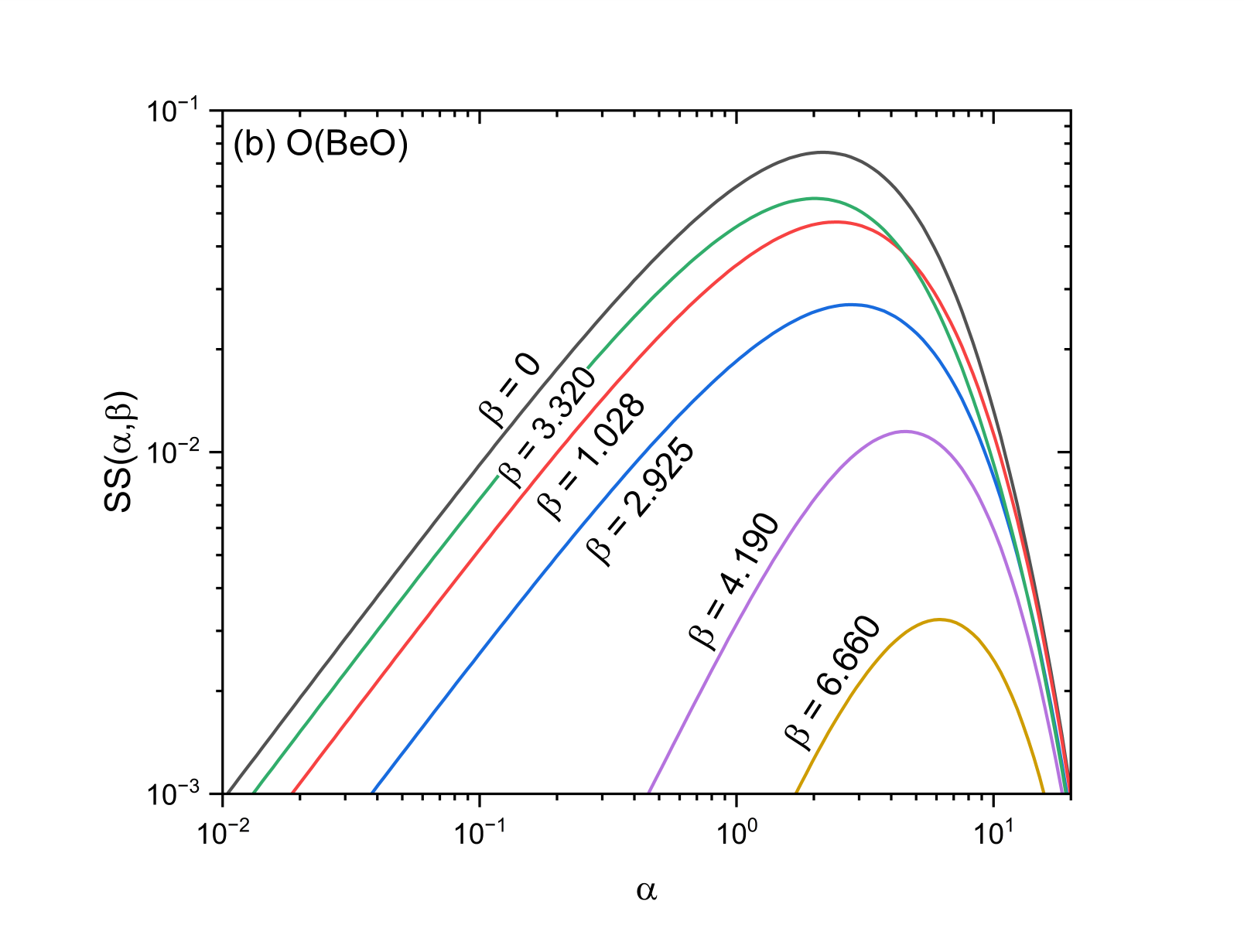}
    \caption{(Color online) The symmetric TSL of (a) Be(\beo) and (b) O(\beo) as a function of momentum transfer, $\alpha$, for various values of $\beta$ at 293.6 K. $SS(\alpha,\beta)$ for each $\beta$ is labeled with the corresponding line.}
    \label{fig:TSL_BeO}
\end{figure}

The TSL (File 7) was calculated from the partial PDOS for Be(\beo) (MAT27) and O(\beo) (MAT46) at 293.6, 400, 500, 600, 700, 800, 1000, and 1200 K using \FLASSH~\cite{Fleming2023}. The incoherent approximation was applied in the evaluation of the TSL; Fig.~\ref{fig:TSL_BeO} shows the symmetric TSL of Be(\beo) and O(\beo) at 293.6 K.

\begin{figure}
    \centering
    \includegraphics[width=1.0\columnwidth,clip,trim=  20mm 10mm 30mm 15mm]{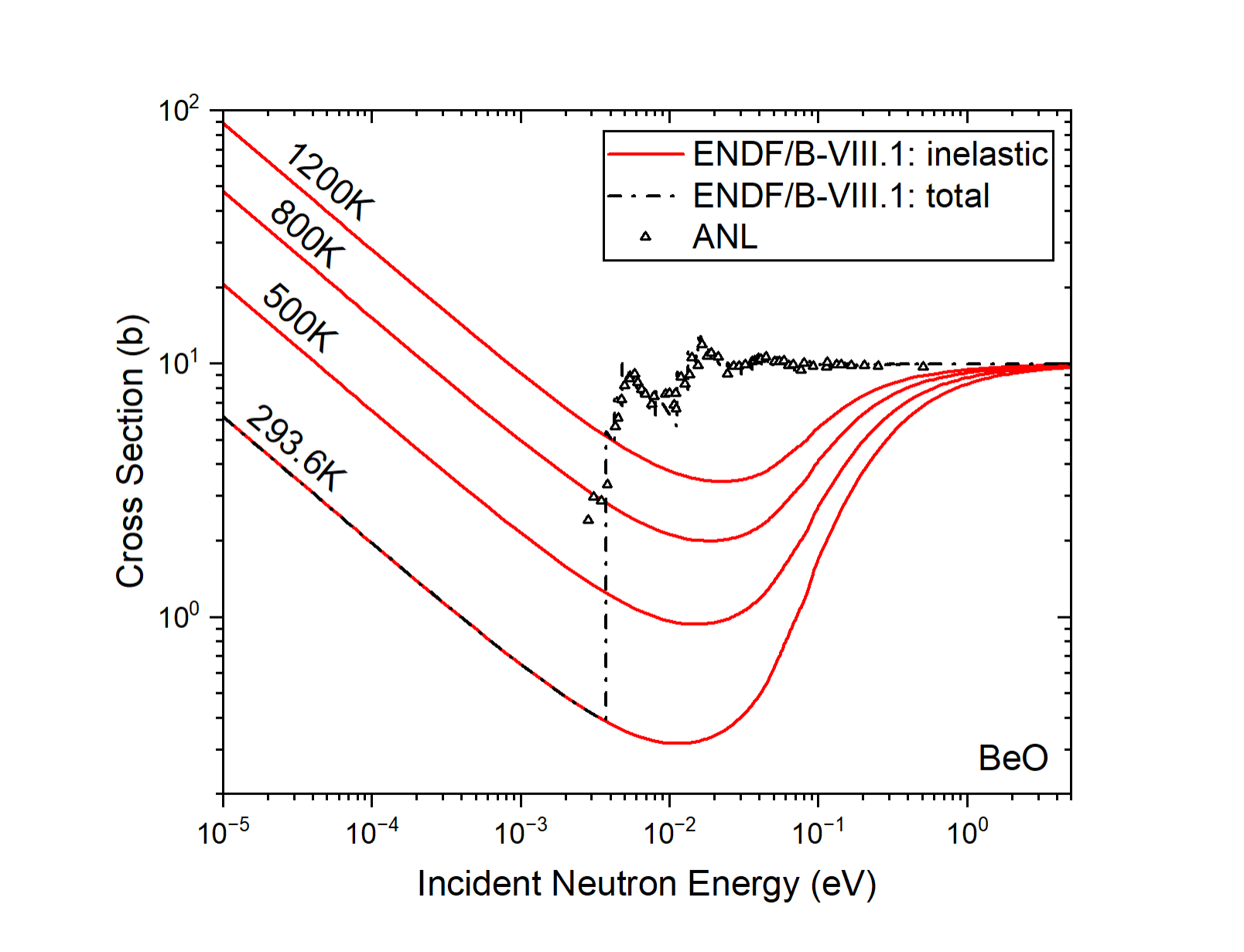}
    \caption{(Color online) The total scattering cross section of \beo\ for \ENDF\ at 293.6 K; measurement is at room temperature from Argonne National Laboratory (ANL) \cite{Hughes1958}. The inelastic scattering cross section of \beo\ is also shown at selected temperatures.}
    \label{fig:Xsec_BeO}
\end{figure}

The cubic approximation was used to calculate the coherent elastic scattering cross section of the compound, which is split evenly between the File 7 for each of Be(\beo) and O(\beo) (MT2). The \ENDF\ total scattering cross section of \beo\ is shown in Fig.~\ref{fig:Xsec_BeO} and found to be in good agreement with measured total cross sections \cite{Hughes1958}. The inelastic scattering cross sections are also compared at select temperatures.

\subsubsection{Calcium Hydride (\cah)}
\label{sec:cah2}

Calcium hydride (\cah) was evaluated using standard AILD methods \AILD. \cah\ has an orthorhombic crystal structure belonging to the \emph{Pnma} space group. An optimized structure with lattice parameters close to experimental values \cite{Wu2007} was created using MedeA-\texttt{VASP} \cite{MedeA, Kresse1996_1, Kresse1996_2, Kresse1999}. The Hellmann-Feynman forces were calculated using \texttt{VASP} \VASP\ and used in PHONON \cite{Parlinski1997} to calculate the phonon dispersion curves and DOS using the dynamical matrix method \cite{Laramee2022}. The generated partial DOS for each nonequivalent atom site matches experimental data well, as shown in  Ref.~\cite[Fig.~2]{Laramee2022}.

\begin{figure}
    \centering
    \includegraphics[width=1.0\columnwidth,clip,trim=  20mm 10mm 30mm 15mm]{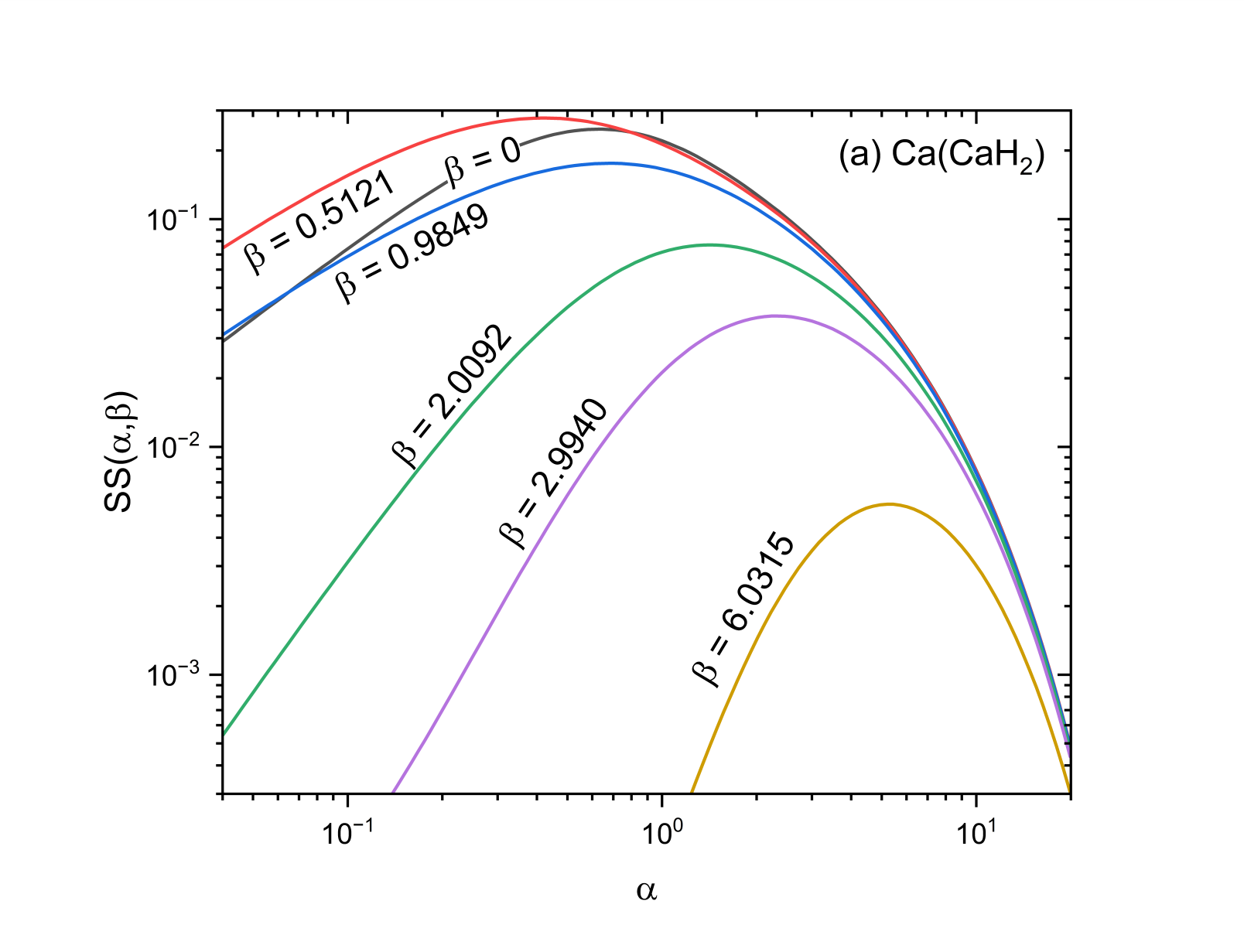}
    \includegraphics[width=1.0\columnwidth,clip,trim=  20mm 10mm 30mm 15mm]{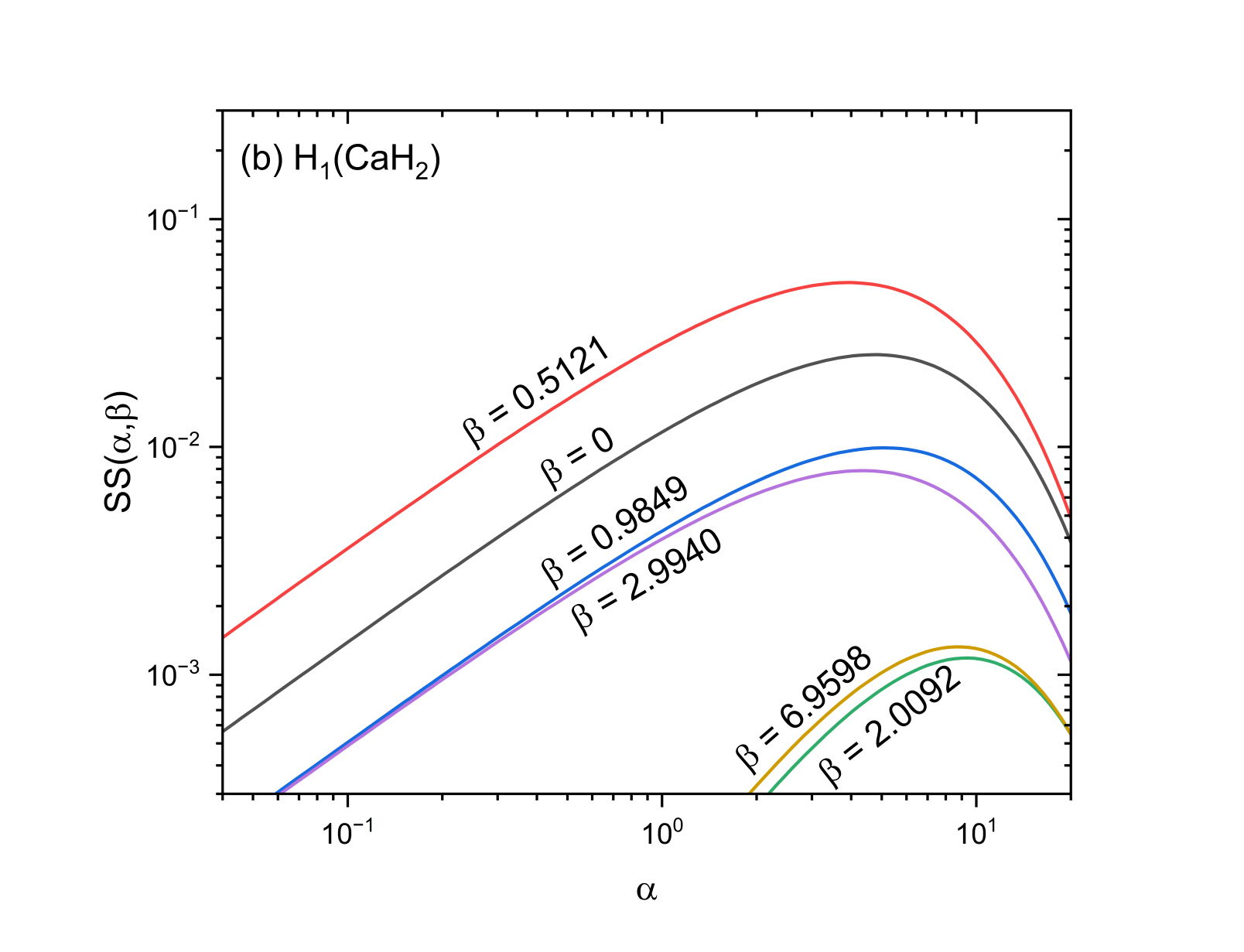}
    \includegraphics[width=1.0\columnwidth,clip,trim=  20mm 10mm 30mm 15mm]{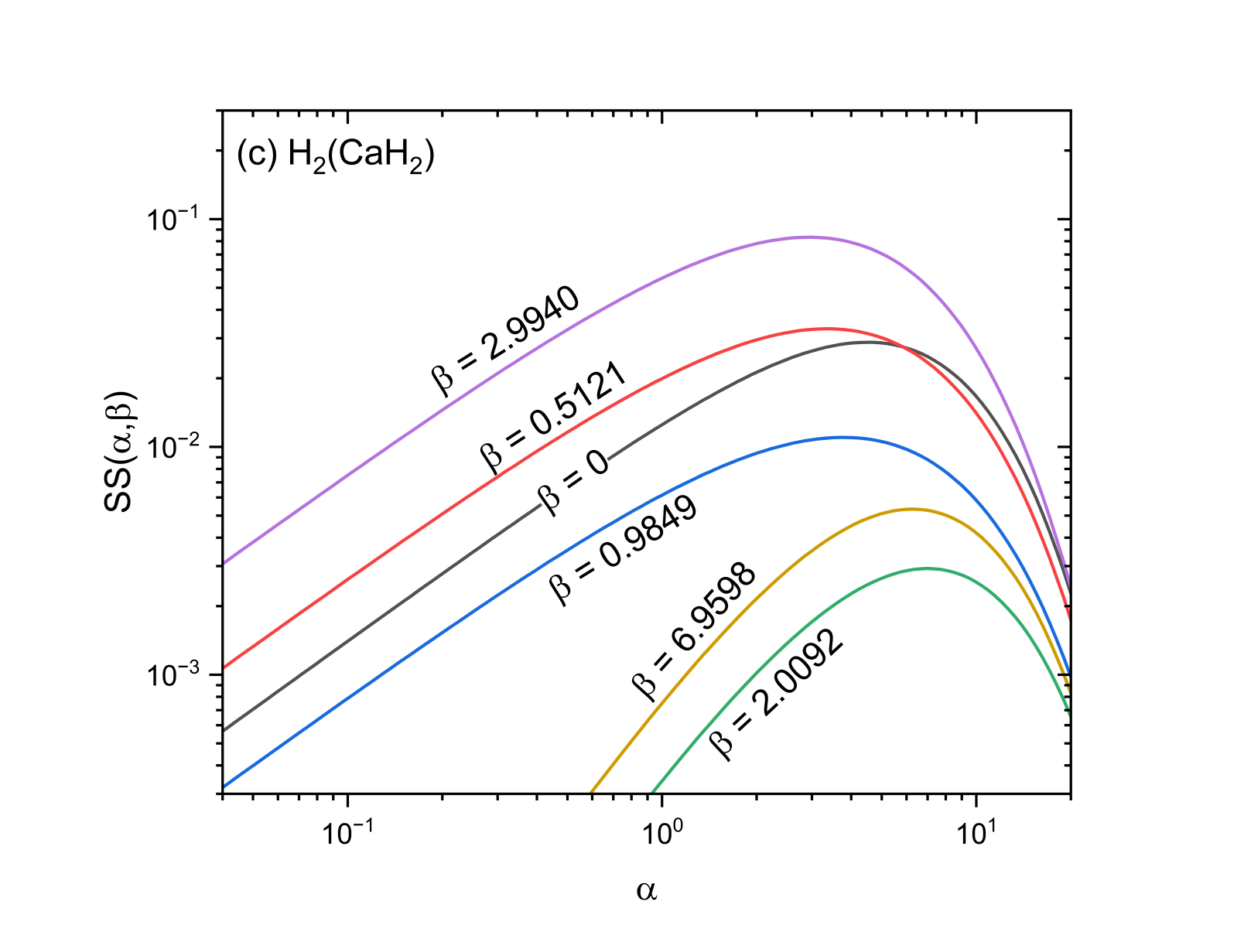}
    \caption{(Color online) The symmetric TSL of (a) Ca(\cah), (b) H$_1$(\cah), and (c) H$_2$(\cah) as a function of momentum transfer, $\alpha$, for various values of $\beta$ at 293.6 K. $SS(\alpha,\beta)$ for each $\beta$ is labeled with the corresponding line.}
    \label{fig:TSL_CaH2}
\end{figure}

The TSL (File 7) was calculated from the partial phonon DOS for Ca(\cah) (MAT3011) and both non-equivalent hydrogen atom site locations H$_1$(\cah) (MAT3013) and H$_2$(\cah) (MAT3014) at 293.6, 400, 500, 600, 700, 800, 1000, and 1200 K using the \FLASSH\ code \cite{Fleming2023}. Mass and total isotopic free atom cross sections were derived from the ENDF/B-VIII.0 nuclide evaluations \cite{Brown2018} at 1.0E-05 eV for all isotopes except \sixCa. For the \sixCa\ isotope, the total free atom cross section was obtained from the NIST database \cite{Sears1992}. The incoherent free atom cross sections for all isotopes were likewise obtained from the NIST database \cite{Sears1992}. For calcium, natural isotopic abundances were assumed \cite{Sears1992}, and \oneH\ was used as opposed to naturally weighted hydrogen. The incoherent approximation was applied in the evaluation of the TSL; Fig.~\ref{fig:TSL_CaH2} shows the symmetric TSL of Ca(\cah), H$_1$(\cah), and H$_2$(\cah) at 293.6 K.

\begin{figure}
    \centering
    \includegraphics[width=1.0\columnwidth,clip,trim=  10mm 10mm 30mm 15mm]{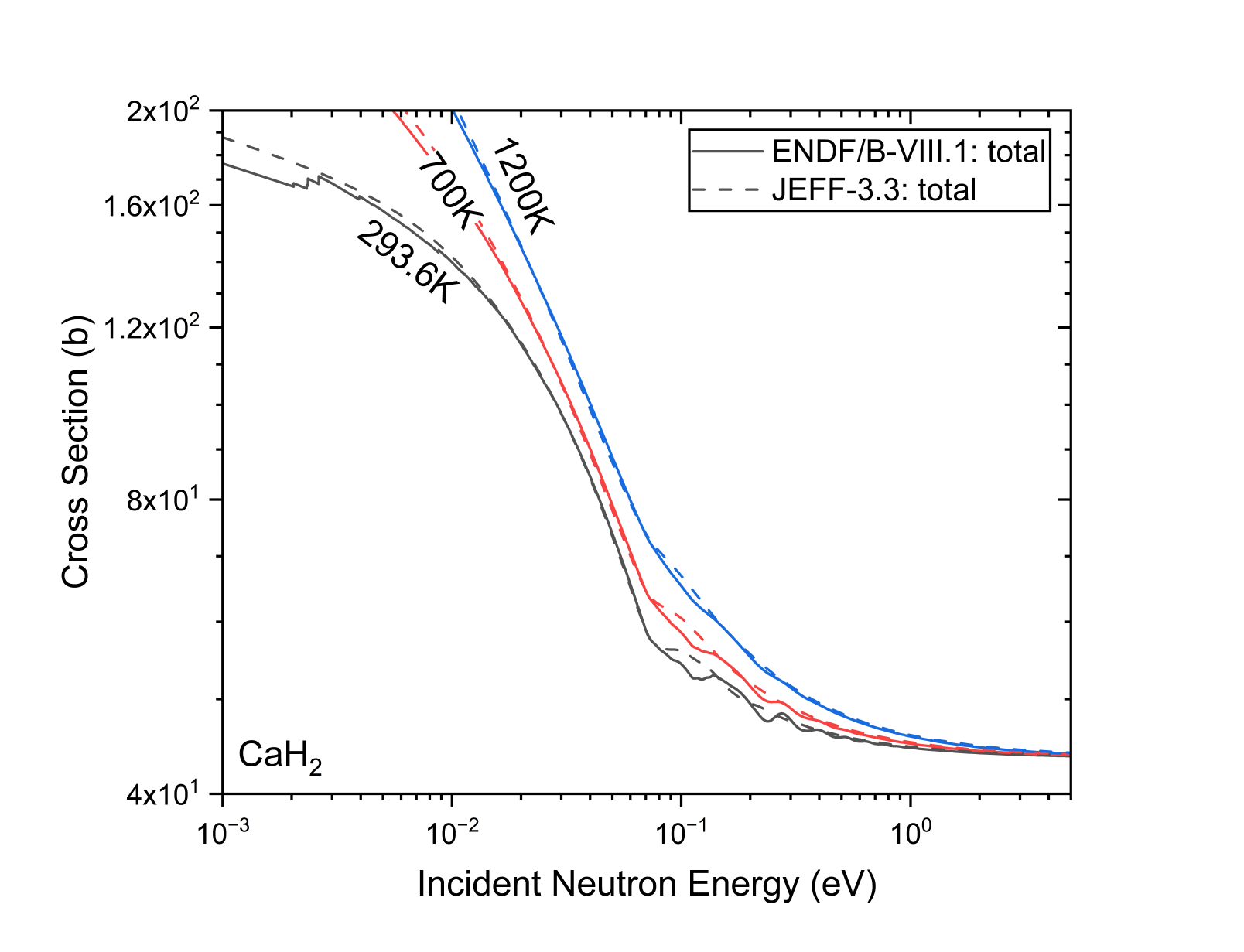}
    \caption{(Color online) The total scattering cross section of \cah\ shown at select temperatures. The total cross sections of \cah\ for 296, 700, and 1200 K from the JEFF-3.3 library are also shown for comparison \cite{JEFF33, Serot2005}. The energy range is limited to highlight the differences in the coherent elastic and inelastic behavior of the cross sections.}
    \label{fig:Xsec_CaH2_Trimmed}
\end{figure}

The cubic approximation was used to calculate the coherent elastic scattering cross section of the full compound, which is stored on the Ca(\cah) File 7 (MT2). The coherent elastic cross sections are calculated to account for the negative scattering length of hydrogen. Hydrogen has a large incoherent scattering length, so the incoherent elastic scattering cross section is stored on the File 7 for each of H$_1$(\cah) and H$_2$(\cah) (MT2). Fig.~\ref{fig:Xsec_CaH2_Trimmed} compares the \ENDF\ total scattering cross section of \cah\ with JEFF-3.3 \cite{JEFF33, Serot2005} at select temperatures. The robust ability of \FLASSH\ to process the coherent scattering behavior of any crystal structure enabled the inclusion of both the coherent elastic scattering of the full compound and the inelastic scattering of the distinct H sites. 

\begin{figure}
    \centering
    \includegraphics[width=1.0\columnwidth,clip,trim=  15mm 10mm 30mm 15mm]{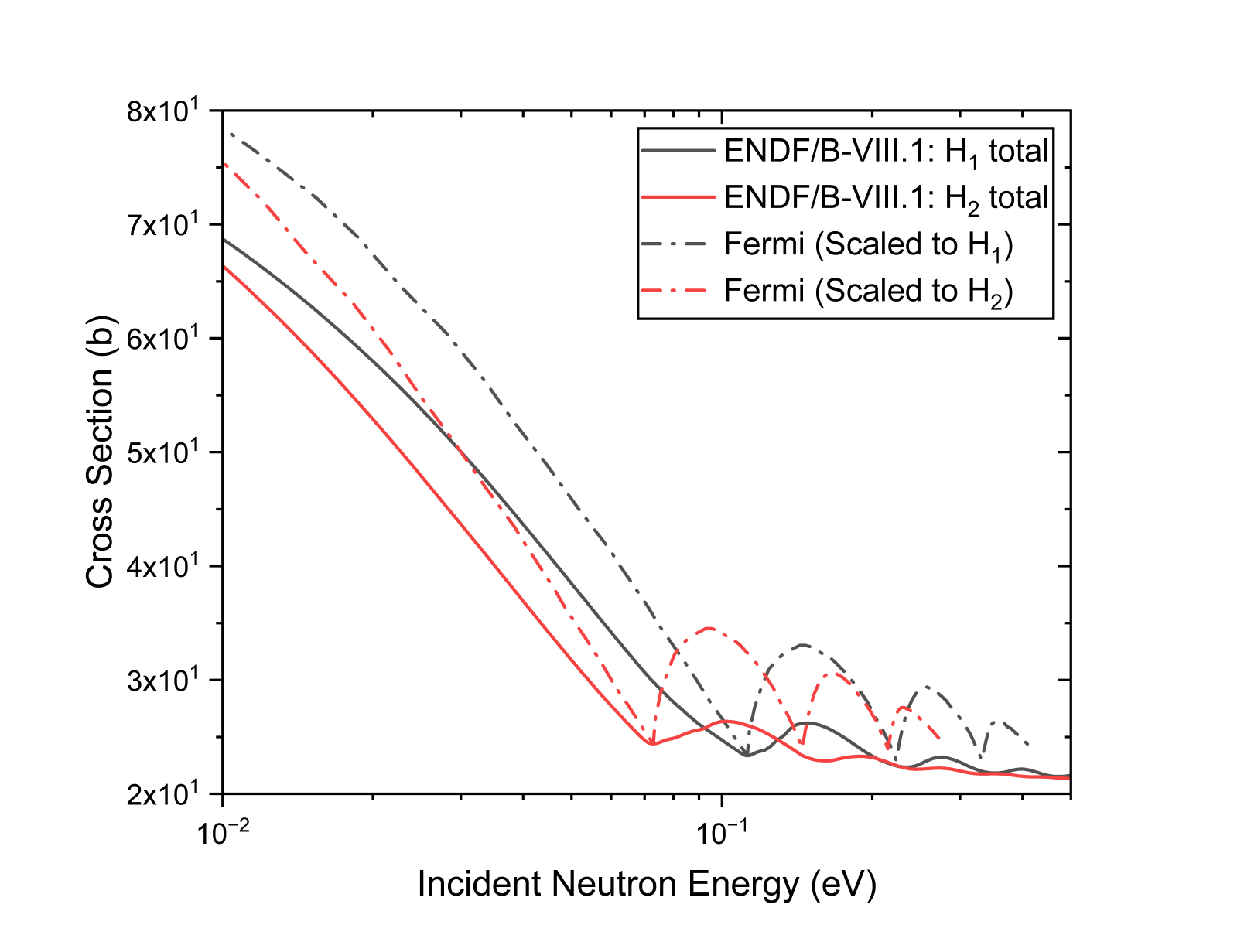}
    \caption{(Color online) The \ENDF\ total cross sections of H$_1$ and H$_2$ in \cah\ compared to the scattering cross section of a neutron in a hydrogenous substance as predicted by Fermi \cite{Fermi1936}, scaled to the \ENDF\ data. The predicted behavior can be seen in the \ENDF\ cross sections, with diminished magnitudes due to the finite mass of Ca in \cah.}
    \label{fig:Xsec_CaH2_Fermi}
\end{figure}

Fig.~\ref{fig:Xsec_CaH2_Fermi} compares the total H cross sections of this evaluation with the behavior of neutrons in a hydrogenous substance as predicted by Fermi \cite{Fermi1936}. The Fermi data has been digitized and scaled such that the start of the first oscillation aligns with those of the calculated H$_1$ and H$_2$ data. The behavior predicted by Fermi can be seen in the calculated cross sections for H$_1$ and H$_2$ albeit with discrepancy in the magnitudes of individual oscillations likely due to the difference between Ca's finite mass and Fermi's assumed infinite binding mass.

\subsubsection{FLiBe Molten Salt}
\label{sec:flibe}

To support design and operational analysis of \flibe\ molten salt reactors, the TSL for \flibe\ molten salt was evaluated using fundamental data generated using classical MD simulations \cite{Zhu2020, ZhuPhD}. The LAMMPS MD code \cite{Thompson2022} was used to construct an atomistic supercell of FLiBe molecules assuming a Born-Mayer type force field \cite{Zhu2020}. As described in Ref. \cite{ZhuPhD}, FLiBe density, viscosity, diffusion coefficients, heat capacity, and thermal conductivity are compared with experimental data and used to benchmark the FLiBe MD model which was used to generate the inputs for the \ENDF\ TSL evaluations. Using this model, the velocity auto-correlation function of F, Li, and Be was calculated and converted by Fourier transform into the excitation DOS of each species. 

\begin{figure}
    \centering
    \includegraphics[width=1.0\columnwidth,clip,trim=  15mm 10mm 30mm 15mm]{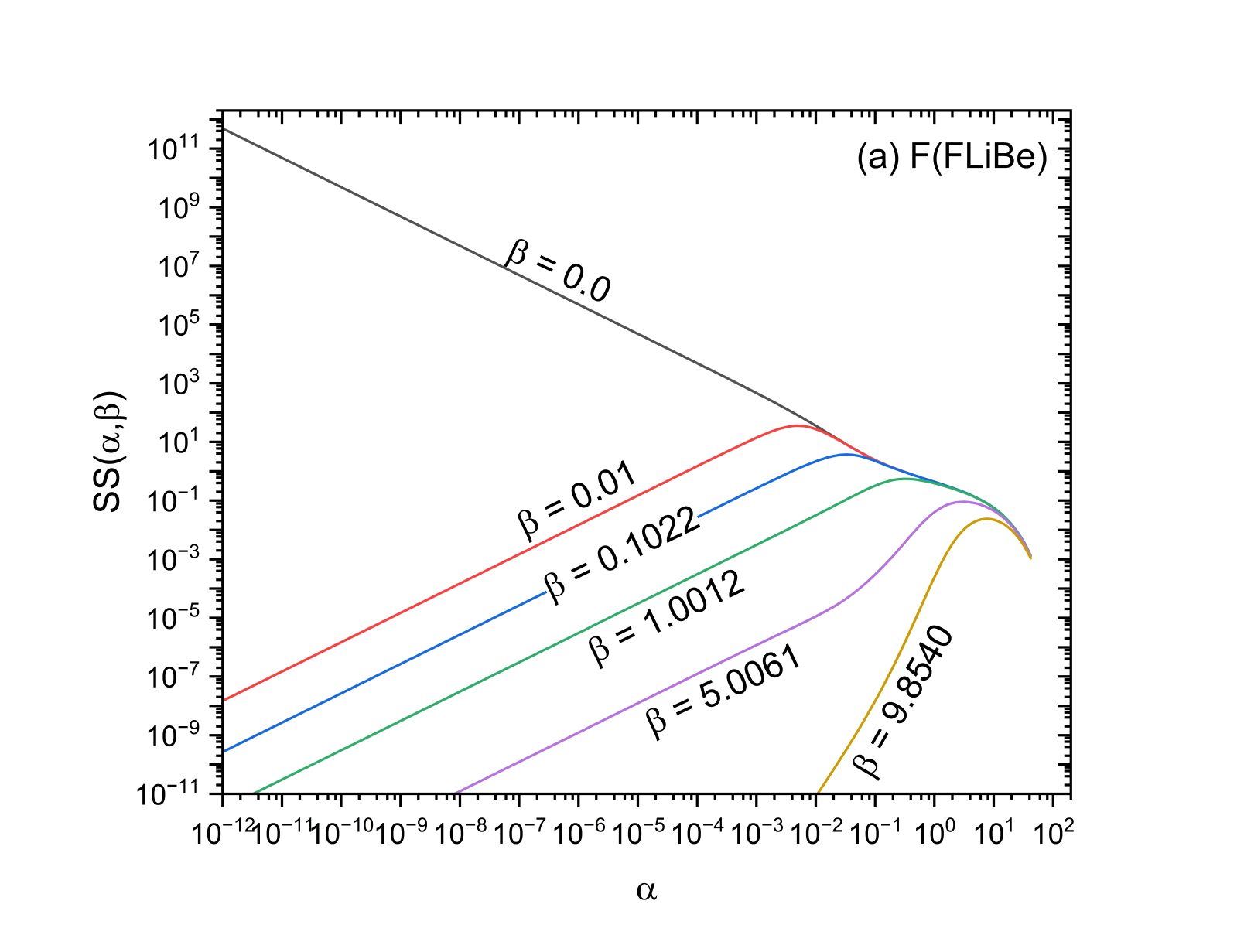}
    \includegraphics[width=1.0\columnwidth,clip,trim=  15mm 10mm 30mm 15mm]{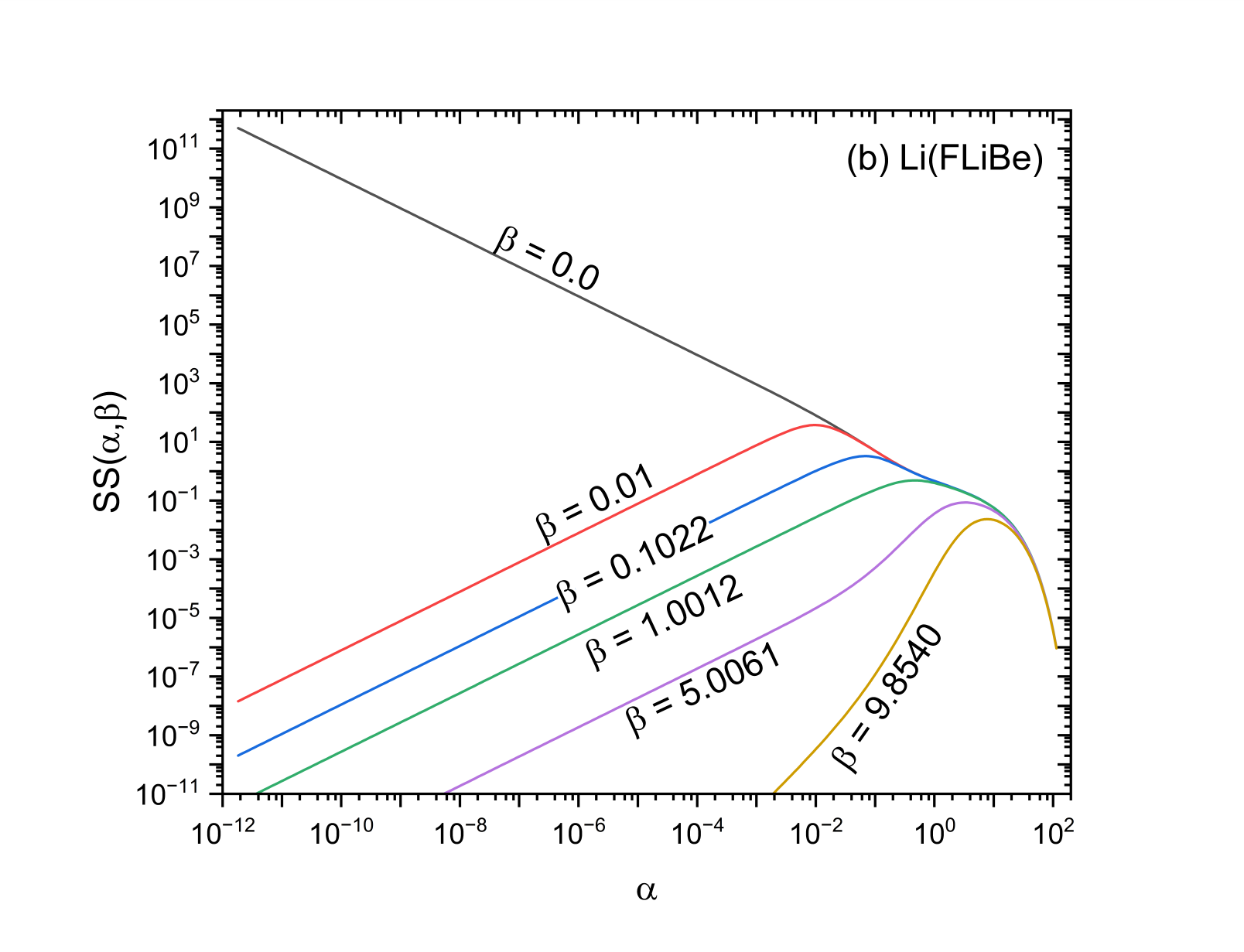}
    \includegraphics[width=1.0\columnwidth,clip,trim=  15mm 10mm 30mm 15mm]{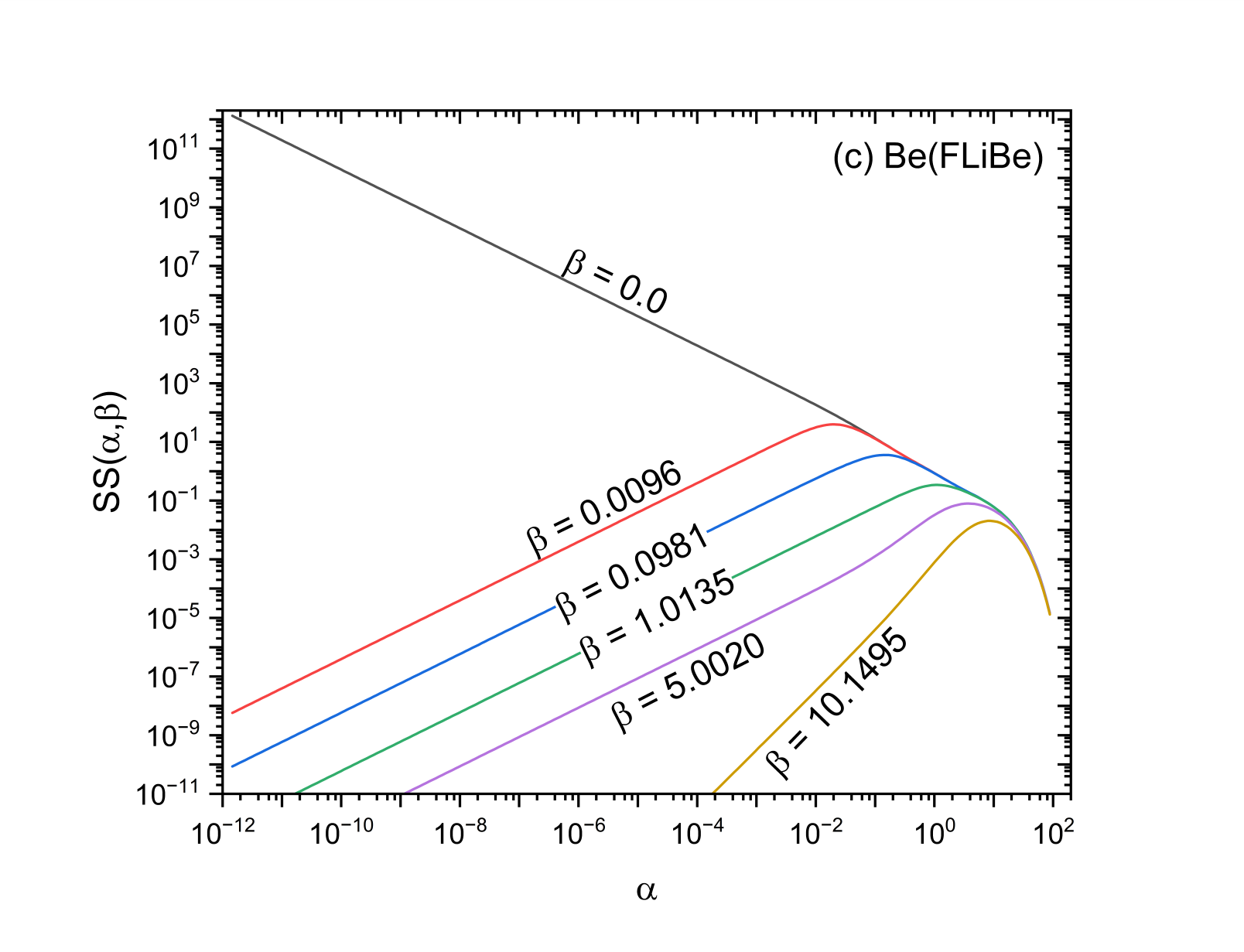}
    \caption{(Color online) The symmetric TSL of (a) F(\flibe), (b) Li(\flibe), and (c) Be(\flibe) as a function of momentum transfer, $\alpha$, for various values of $\beta$ at 773 K. $SS(\alpha,\beta)$ for each $\beta$ is labeled with the corresponding line.}
    \label{fig:TSL_FLiBe}
\end{figure}

\begin{figure}
    \centering
    \includegraphics[width=1.0\columnwidth,clip,trim=  20mm 10mm 30mm 15mm]{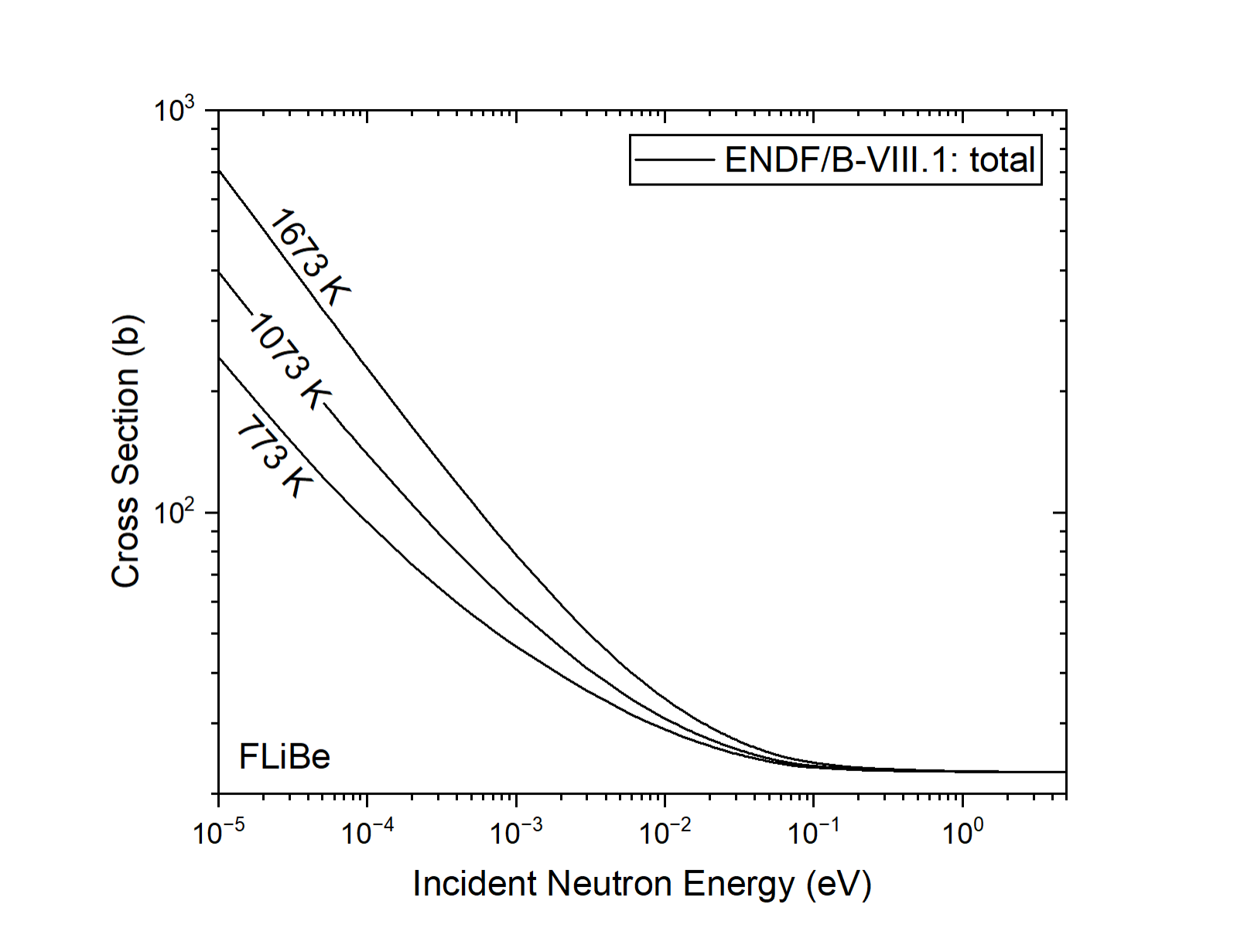}
    \caption{Total thermal neutron scattering cross section of liquid \flibe\ at select temperatures.}
    \label{fig:Xsec_FLiBe}
\end{figure}

Subsequently, the DOS of each species was used in the \FLASSH\ \cite{Fleming2023} code to generate the TSL (File 7, MAT4001, MAT4002, and MAT4003), under the incoherent approximation, and assuming the convolution of the solid component of the TSL with the diffusive component, to produce the total TSL.  Fig.~\ref{fig:TSL_FLiBe} shows the symmetric TSL for F, Li, and Be in \flibe. The total thermal neutron scattering cross section of \flibe\ is shown in Fig.~\ref{fig:Xsec_FLiBe}. The evaluation was completed at 773, 873, 923, 973, 1073, 1173, 1273, 1473, and 1673~K, which corresponds to relevant operation and potential safety conditions of molten salt reactors. Mass and free atom cross sections \cite{Brown2018} for $^{19}$F, $^7$Li, and $^9$Be are used for the evaluations. The \flibe\ evaluations represent new TSL contributions that are included for the first time in the ENDF/B database.

\subsubsection{Nuclear/Reactor Graphite (20\%)}
\label{sec:gr}

\begin{figure}
    \centering
    \includegraphics[width=1.0\columnwidth,clip,trim=  20mm 8mm 30mm 15mm]{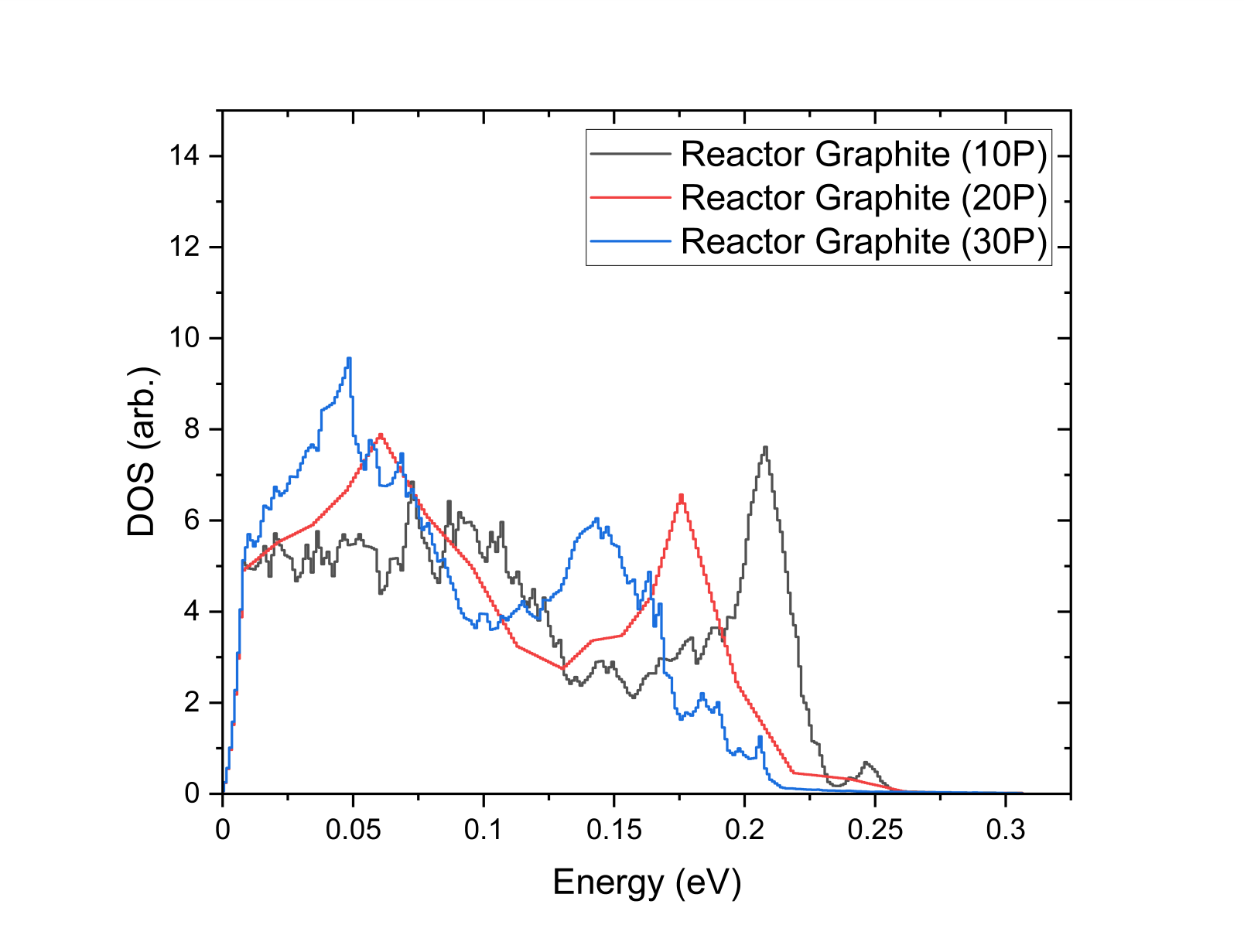}
    \caption{(Color online) The DOS used for \gr\ (20\% porosity) as compared with 10\% and 30\%.}
    \label{fig:DOS_C20P}
\end{figure}

Nuclear/reactor graphite with 20\% porosity (reactor-graphite-20P) was evaluated using a PDOS that is constructed using the systematics observed based in the classical MD models of 10\% and 30\% porous graphite. Nuclear graphite is a multi-phase material whose micro-structure differs from that of ideal, crystalline graphite. Specifically, localized graphite microcrystals are located in a carbon binder matrix, which impacts characteristic lattice vibrations and distinguishes its thermal scattering cross sections from crystalline graphite. Initially, a MD model for crystalline graphite was validated in LAMMPS \cite{Thompson2022} against experimental thermophysical properties. Next, porosity was introduced by stochastically removing atoms from the crystalline structure. For the 20\% porous nuclear graphite, the porosity would correspond to a graphite density of approximately 1.7 g/cm\textsuperscript{3}. Time-dependent atomic velocities are calculated and used to generate the PDOS as the Fourier transform of the velocity auto-correlation function \cite{Hawari2014_2}. The DOS used for the evaluation is shown in Fig.~\ref{fig:DOS_C20P}. 

\begin{figure}
    \centering
    \includegraphics[width=1.0\columnwidth,clip,trim=  20mm 8mm 30mm 15mm]{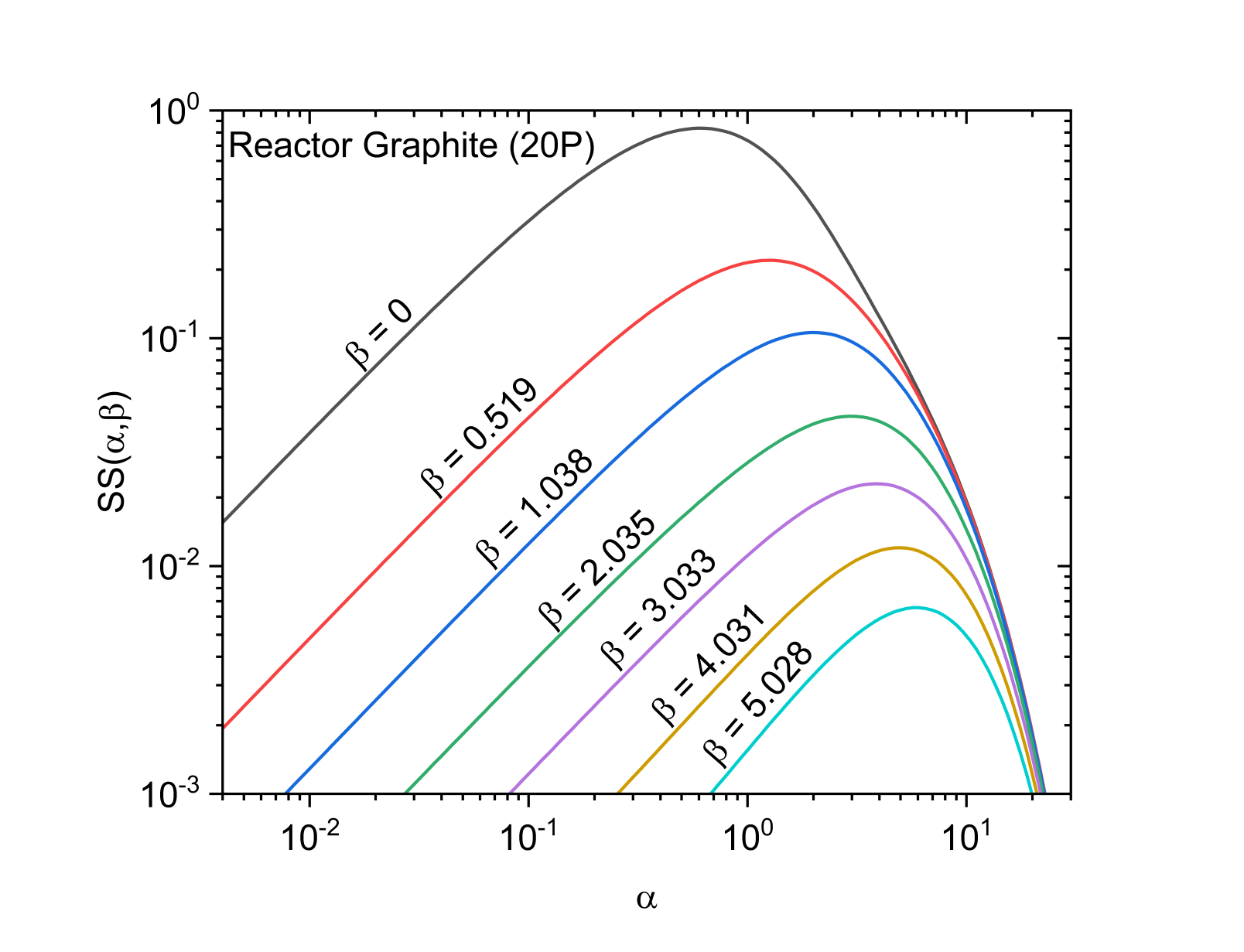}
    \caption{(Color online) The symmetric TSL of \gr\ (20\% porosity) as a function of momentum transfer, $\alpha$, for various values of $\beta$ at 296 K. $SS(\alpha,\beta)$ for each $\beta$ is labeled with the corresponding line.}
    \label{fig:TSL_C20P}
\end{figure}

\begin{figure}
    \centering
    \includegraphics[width=1.0\columnwidth,clip,trim=  20mm 8mm 30mm 15mm]{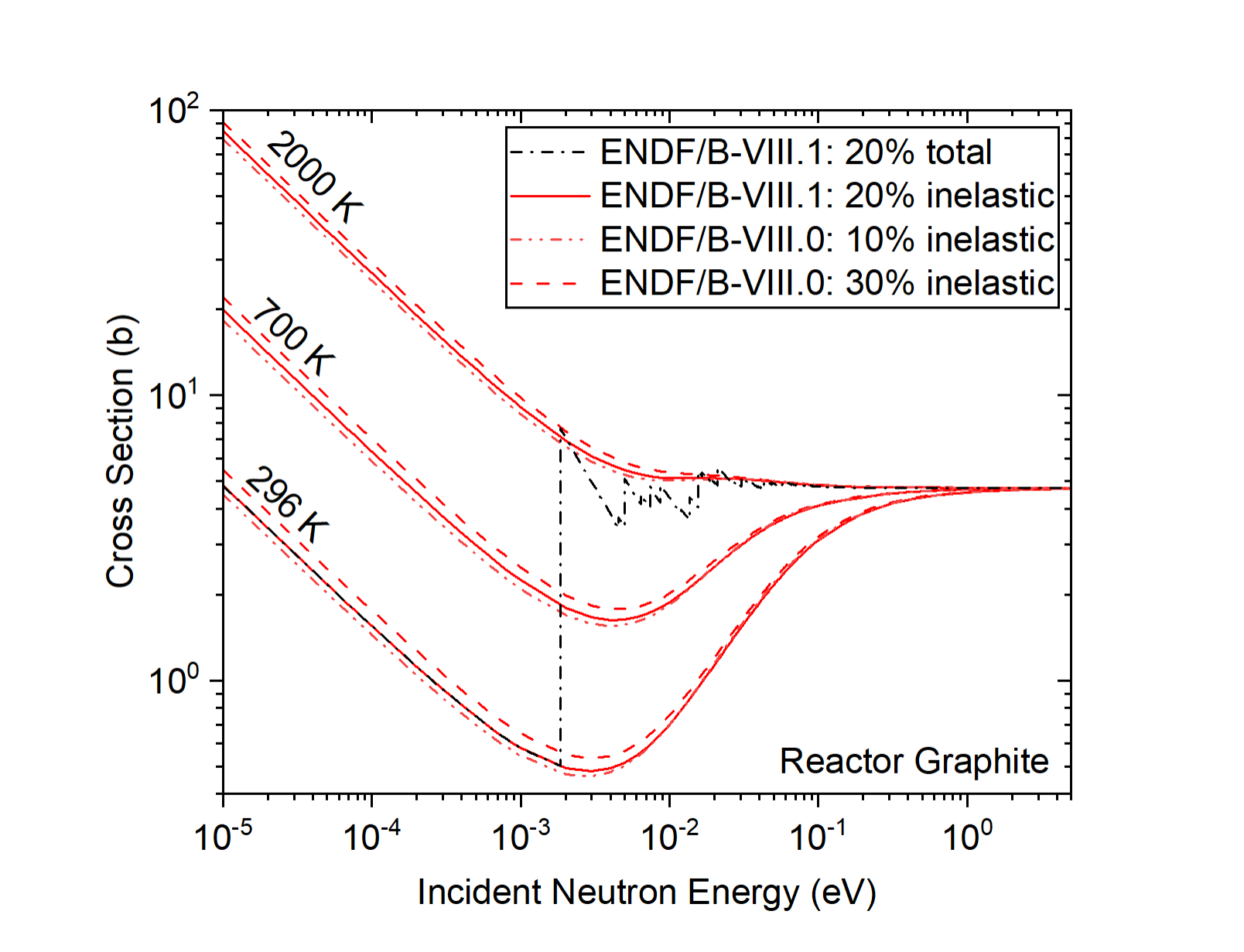}
    \caption{(Color online) The inelastic scattering cross section of 20\% porosity \gr\ (\ENDF) compared with 10\% and 30\% \gr\ from \prENDF\ at selected temperatures. The total scattering cross section of 20\% \gr\ at 296 K is also shown.}
    \label{fig:Xsec_C20P}
\end{figure}

The ENDF TSL for reactor-graphite-20P (File 7, MAT320) was generated in \FLASSH\ \cite{Fleming2023} with mass and free atom cross sections for natural carbon \cite{ENDF-VII.1} at standard temperatures of 296, 400, 500, 600, 700, 800, 1000, 1200, 1600, and 2000 K. In the incoherent approximation  \cite{NJOY91, ENDF6-Format-2012}, the symmetric TSL is stored as ENDF File 7, MT4 and presented in Fig.~\ref{fig:TSL_C20P} at 296 K versus $\alpha$ for select $\beta$. Coherent elastic scattering describes zero-phonon interactions, is evaluated with the cubic approximation in \FLASSH, and tabulated in File 7, MT2. In this case, crystalline graphite\textquotesingle{}s coherent elastic behavior is assumed to be representative of nuclear/reactor graphite as well. Thermal scattering cross sections generated in \FLASSH\ are depicted in Fig.~\ref{fig:Xsec_C20P}. This is a novel TSL evaluation to be included for the first time in the \ENDF\ database. These cross sections have additionally been benchmarked against thermal scattering benchmarks, including the graphite slowing-down-time spectroscopy experiment that was performed at the ORELA facility \cite{Lee2023}. As seen in Ref.~\cite{FUND-ORELA-ACC-GRAPH-PNSDT-001}, the use of the nuclear/reactor graphite with 20\% porosity reduces the discrepancy between the measured and calculated thermal slowing-down-time spectrum (mean absolute deviation) from 4.14\% with crystalline graphite to 2.13\% with 20\% porosity. 

\subsubsection{Crystalline Graphite with Distinct Effects (\grsd)}
\label{sec:grsd}

\begin{figure}
    \centering
    \includegraphics[width=1.0\columnwidth,clip,trim=  10mm 8mm 10mm 8mm]{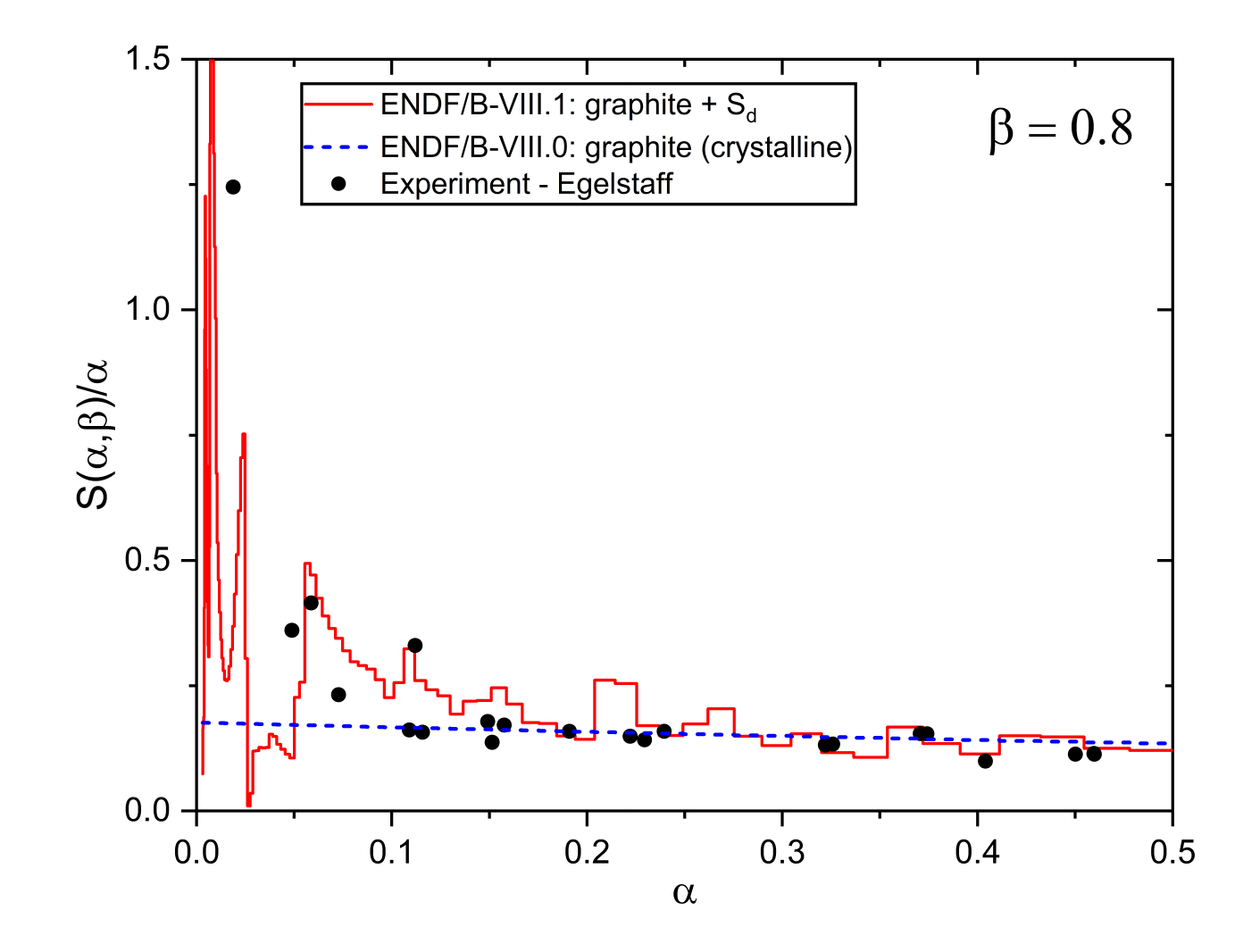}
    \caption{(Color online) The symmetric TSL for graphite with \sd\ effects at 296 K as a function of momentum transfer $\alpha$ compared to experimental data \cite{Egelstaff1962, Wikner1964}. }
    \label{fig:TSL_GrSd}
\end{figure}

Crystalline graphite was evaluated as part of the \prENDF\ database using AILD techniques \AILD\ to generate the phonon dispersion relationships and PDOS. Graphite is a crystalline material with tightly bound basal planes separated by Van der Waals forces. Its asymmetric crystalline structure results in directionally dependent thermal scattering effects. Using the validated \prENDF\ crystalline model of graphite \cite{Wormald2017}, a novel evaluation was created, which accounts for the directional dependence of the graphite structure and removes traditional incoherent, cubic, and atom-site approximations to the TSL \cite{Fleming2021}. This evaluation utilizes a non-cubic formulation of the TSL with additional distinct effects (\sd) which include coherent interactions in the inelastic cross section \cite{Hawari2008}. The dispersion relations, which were developed in the AILD model demonstrated reasonable agreement with measurement \cite{Wormald2017}. These relationships then form the fundamental input for the generation of the TSL with mass and free atom cross section for natural carbon \cite{ENDF-VII.1}. Using the \FLASSH\ code and the advanced physics modules, the TSL for crystalline graphite with \sd\ effects (File 7, MAT301) was evaluated at temperatures of 296, 400, 500, 600, 700, 800, 1000, and 1200 K without the incoherent approximation and using the non-cubic formulation \cite{Fleming2023}. The symmetric TSL at 296 K tabulated in File 7, MT4 of the ENDF file is illustrated in Fig. \ref{fig:TSL_GrSd}, and an example comparison with experimental measurements is shown \cite{Egelstaff1962, Wikner1964}. Additional TSL and experimental data comparisons of this data can be found in Ref.~\cite{Fleming2021}.

\begin{figure}
    \centering
    \includegraphics[width=1.0\columnwidth,clip,trim=  10mm 8mm 10mm 3mm]{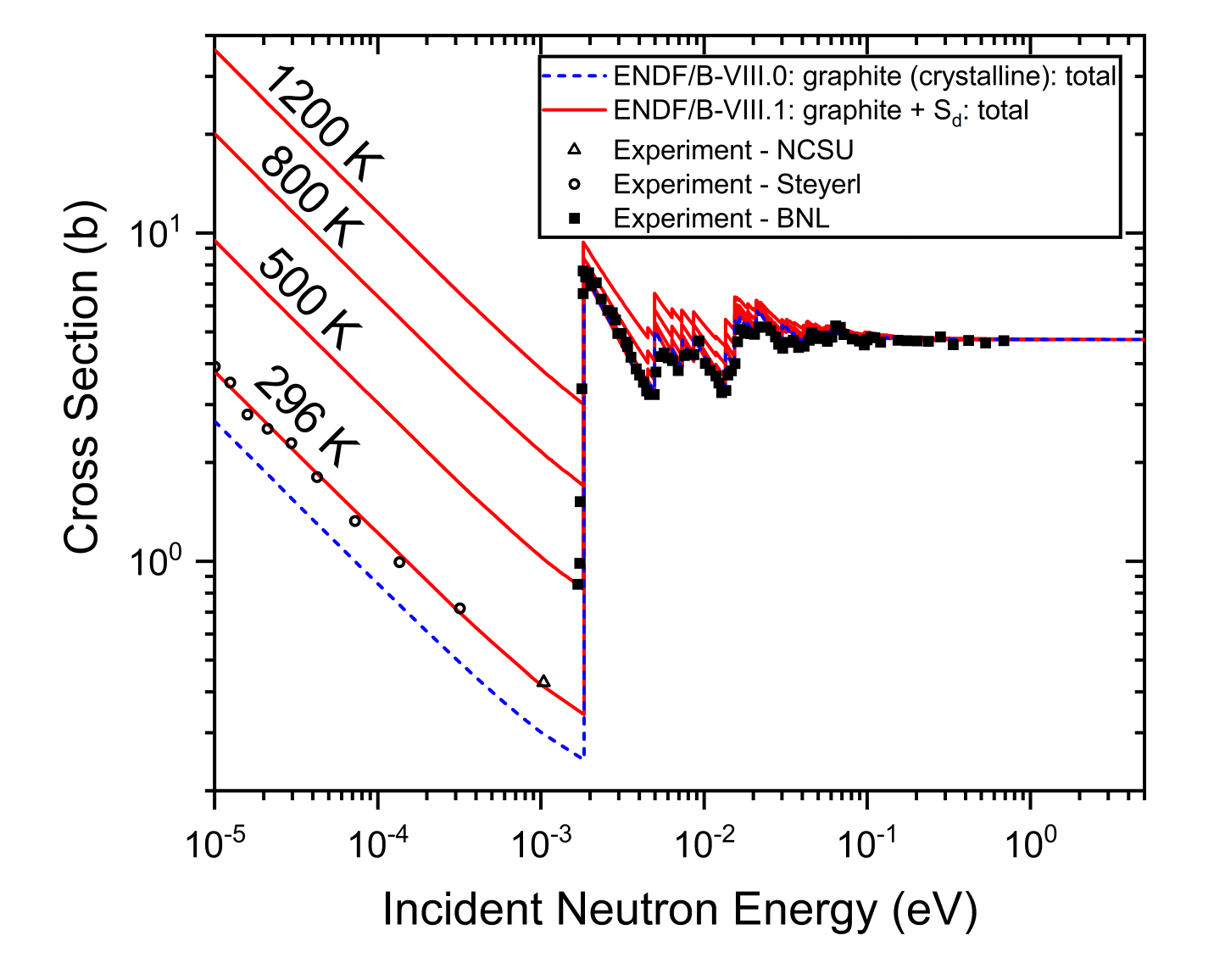}
    \caption{(Color online) Total thermal neutron scattering cross section of crystalline graphite with \sd\ effects at 296 K compared to experimental data \cite{Hawari2009, Steyerl1974, Hughes1958}. The total cross section with \sd\ effects is also shown at select temperatures.}
    \label{fig:Xsec_GrSd}
\end{figure}

Coherent elastic scattering was evaluated using the generalized non-cubic routine in \FLASSH\ through the use of the directionally dependent Debye-Waller matrix that explicitly treats asymmetric vibrational behavior in different crystallographic directions \cite{Fleming2023}. This allows a consistent evaluation method between the TSL and resulting inelastic cross sections and the coherent elastic cross sections. The resulting total integrated scattering cross section for the \ENDF\ crystalline graphite with \sd\ effects is shown in Fig. \ref{fig:Xsec_GrSd}. The comparison with experimental data shows improved agreement over the similar evaluation in \prENDF\ without \sd\ effects.

This addition of distinct (\sd) effects provides the ability to directly benchmark the TSL evaluation against experimentally measured TSL data \cite{Fleming2021} for crystalline graphite and improves the agreement with experimentally measured inelastic cross sections \cite{Hawari2009, Steyerl1974, Hughes1958}. The crystalline graphite with \sd\ evaluation represents a new TSL evaluation that is included for the first time in the ENDF/B database.

The users of the crystalline graphite of ENDF/B-VIII.0 are encouraged 
to try the graphite with S$_d$ of ENDF/B-VIII.1 and provide feedback to the authors 
if feasible. 
Similarly, the authors encourage the users of the crystalline graphite 
to test the porosity effect in the systems with ``real'' graphite 
by changing the ideal graphite $S(\alpha, \beta)$ 
to one of the porous graphite $S(\alpha, \beta)$ available in ENDF/B-VIII.1 
and provide feedback to the authors.

\subsubsection{Anhydrous Hydrogen Fluoride (\hf)}
\label{sec:hf}

Hydrogen Fluoride (\hf) is a liquid material which is used at elevated temperatures and pressures to process and manufacture nuclear fuel. It is comprised of long chains of  bonded hydrogen and fluorine molecules. The evaluation was based on a classical MD model composed of a supercell of 1000 HF molecules \cite{Ahmed2025}. The model was executed using the GROMACS code \cite{Abraham2015} to produce temperature-dependent DOS of intra- and inter-molecular excitations as the power spectrum of relevant atomic trajectory autocorrelation functions for thermodynamic state-points between 343 K (3 atmospheres) and 383 K (17 atmospheres). The MD model validated using density, diffusion coefficients, potential energy, and bond length experimental data, and the resulting DOS is shown in Ref.~\cite[Fig.~9]{Ahmed2025}.   

\begin{figure}
    \centering
    \includegraphics[width=1.0\columnwidth,clip,trim=  15mm 10mm 30mm 15mm]{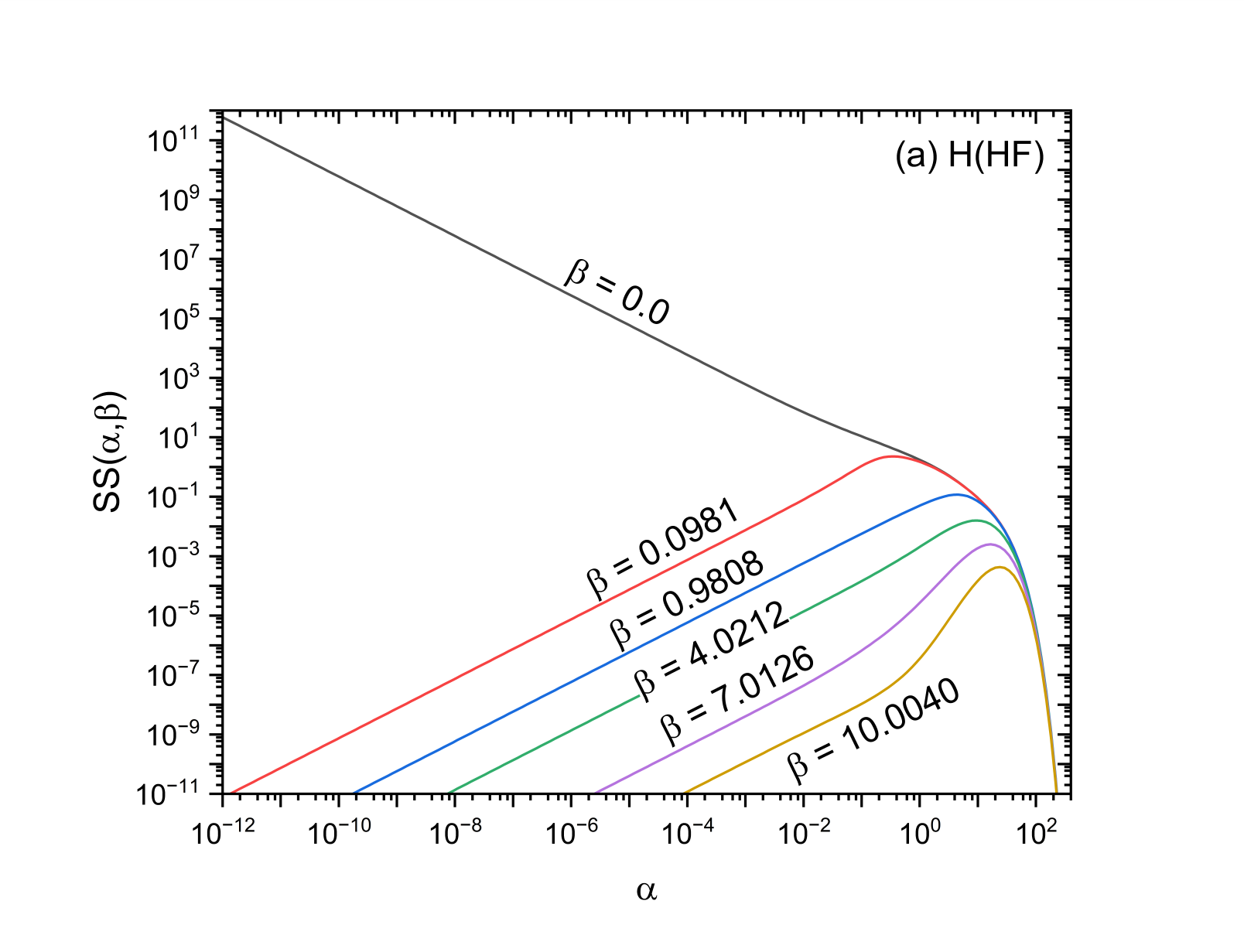}
    \includegraphics[width=1.0\columnwidth,clip,trim=  15mm 10mm 30mm 15mm]{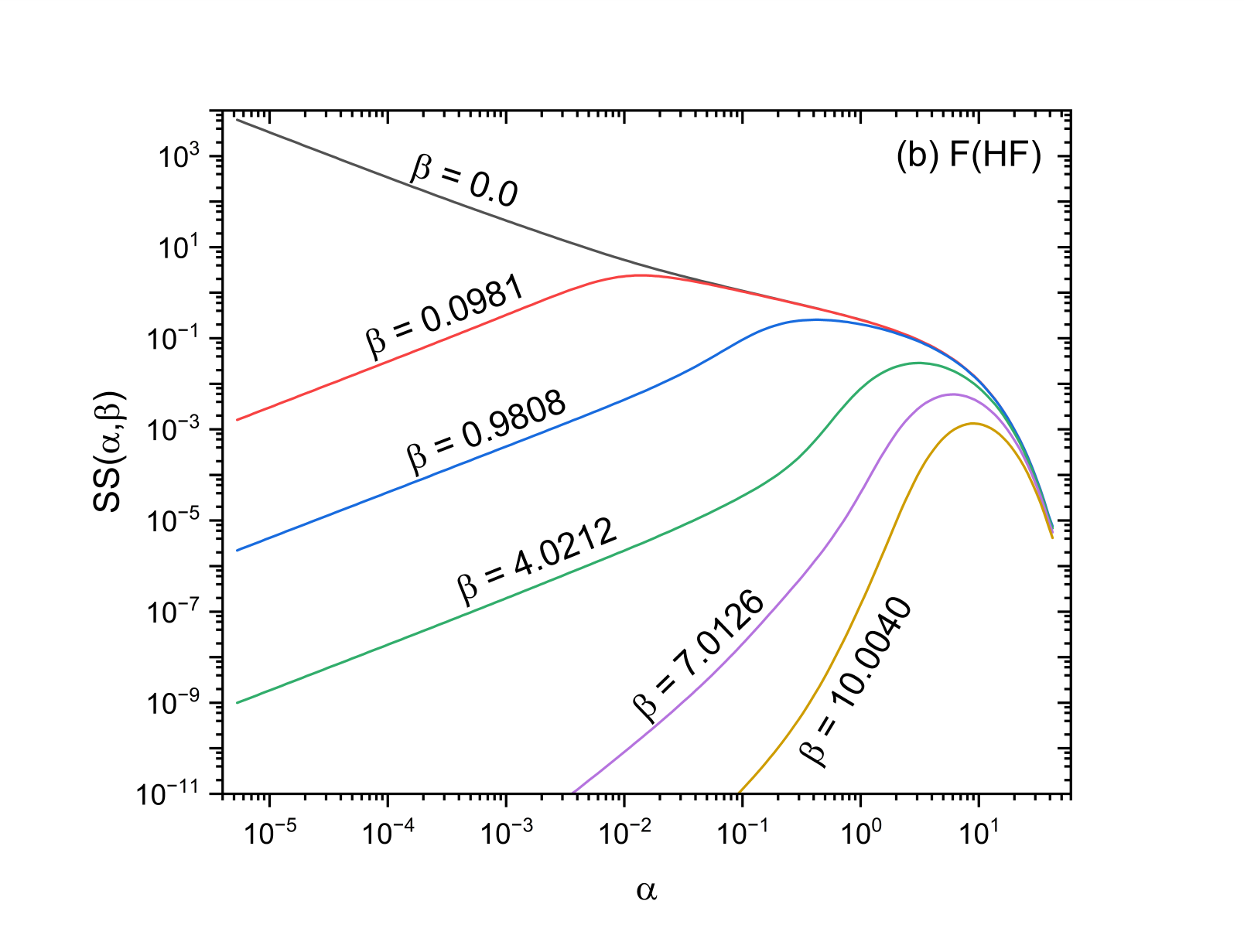}
    \caption{(Color online) The symmetric TSL for (a) H(\hf) and (b) F(\hf) 343 K as a function of momentum transfer $\alpha$, for a range of neutron energy transfers, $\beta$. \Ss\ for each $\beta$ is labeled with the corresponding line.}
    \label{fig:TSL_HF}
\end{figure}

\begin{figure}
    \centering
    \includegraphics[width=1.0\columnwidth,clip,trim=  20mm 10mm 30mm 15mm]{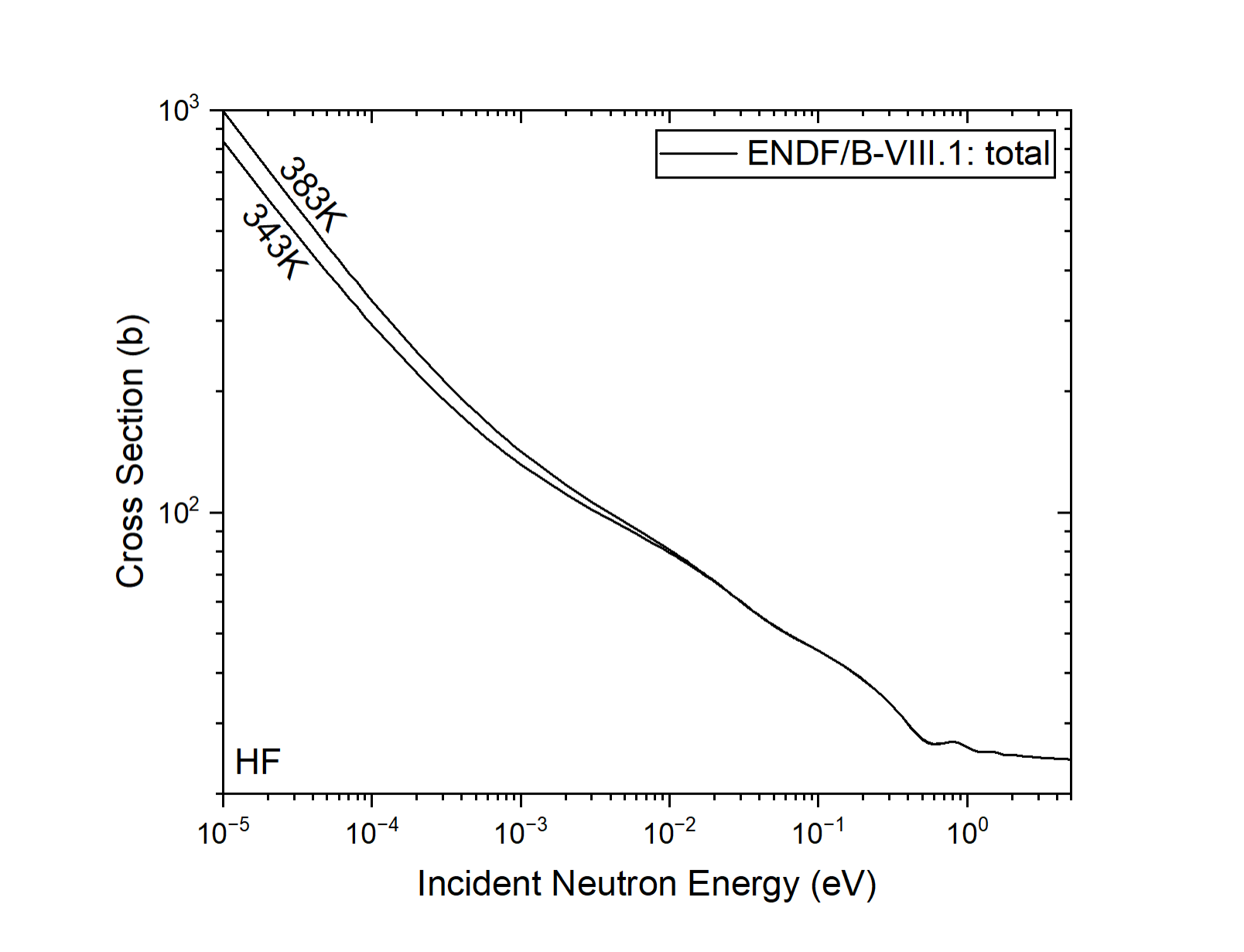}
    \caption{Total thermal neutron scattering cross section of \hf\ at select temperatures.}
    \label{fig:Xsec_HF}
\end{figure}

The TSLs of H(\hf) (File 7, MAT3048) and F(\hf) (File 7, MAT3047) were generated in \FLASSH\ \cite{Fleming2023} under the incoherent approximation with a Schofield diffusional model \cite{Schofield1962}. Diffusional parameters were derived from the MD model based on the model's effective cluster size of eight molecules. The TSL was generated for temperatures of 343, 348, 353, 363, 373, and 383 K with mass and free atom cross sections \cite{Brown2018} for $^1$H and $^{19}$F. Fig.~\ref{fig:TSL_HF} shows the scattering law at various $\alpha$ and $\beta$ values. Fig.~\ref{fig:Xsec_HF} shows the total scattering cross sections for H and F in \hf. 

These evaluations represent new additions to the ENDF/B database. Benchmark testing of these evaluations in ICSBEP HEU-SOL-THERM-039 \cite{HEU-SOL-THERM-039} have been completed and showed improvement on the order of 1000 pcm in the criticality calculations in comparison to experimental data \cite{Ahmed2025}. The \hf\ evaluations represent new TSL contributions that are included for the first time in the ENDF/B database.

\subsubsection{Heavy Paraffinic Oil}
\label{sec:oil}

Paraffinic-based oils (a hydrocarbon material) appear in various industrial applications, including those involving nuclear reactor and neutronic systems. In such cases, the description of neutron interactions and thermalization in paraffinic oil environments requires the use of TSL data and related cross sections. Therefore, to evaluate the TSL of paraffinic oil, classical MD methods, with the COMPASS force field, were used to establish the atomistic models for generating the velocity auto-correlation function and DOS of hydrogen in oil \cite{Manring2019}. The MD models were designed to capture a highly viscous oil and are described in Ref. \cite{Manring2019}. 

\begin{figure}
    \centering
    \includegraphics[width=1.0\columnwidth,clip,trim=  15mm 10mm 30mm 15mm]{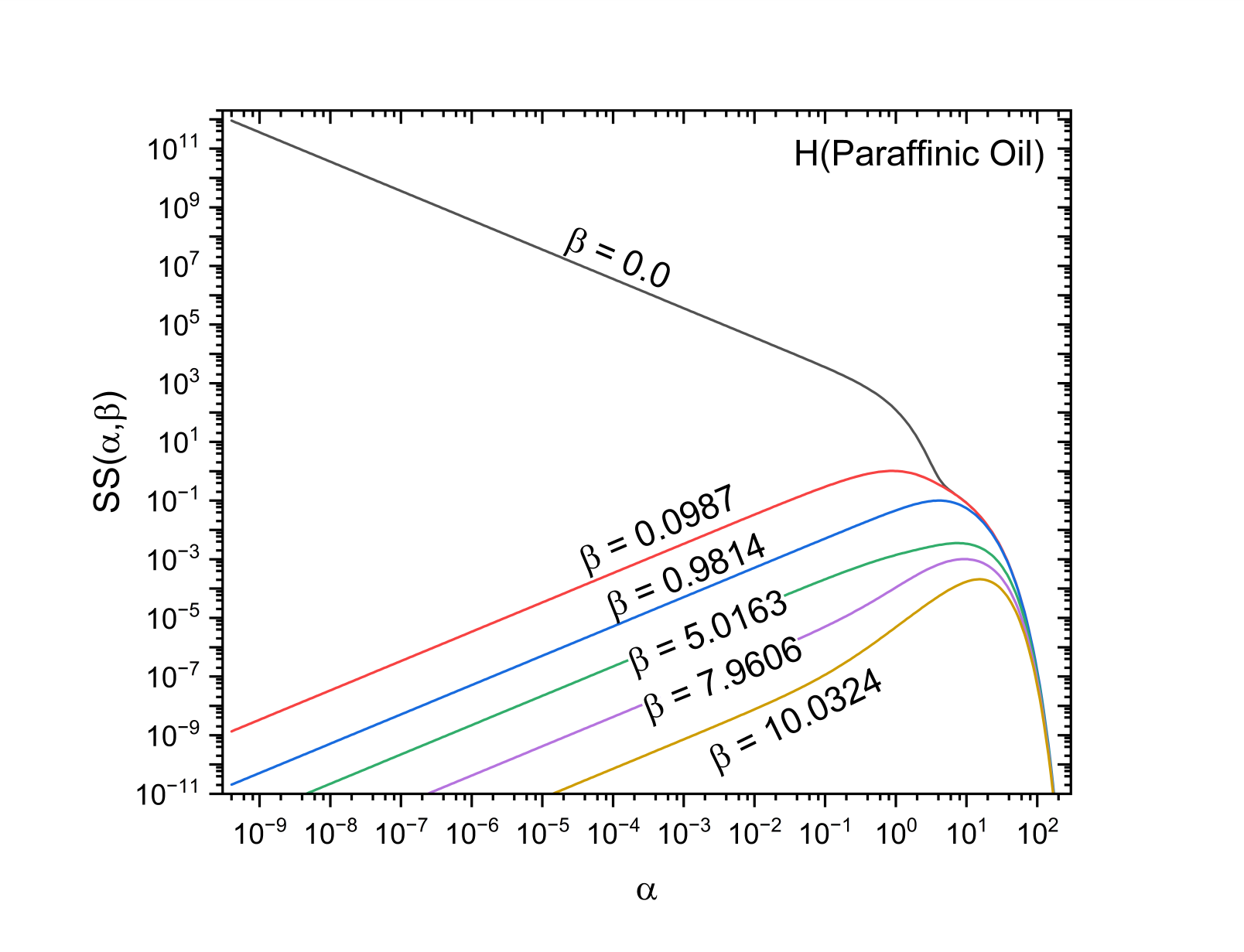}
    \caption{(Color online) The symmetric TSL of H(ParaffinicOil) as a function of momentum transfer, $\alpha$, for various values of $\beta$ at 293.6 K. $SS(\alpha,\beta)$ for each $\beta$ is labeled with the corresponding line.}
    \label{fig:TSL_ParaffinicOil}
\end{figure}

\begin{figure}
    \centering
    \includegraphics[width=1.0\columnwidth,clip,trim=  20mm 10mm 30mm 15mm]{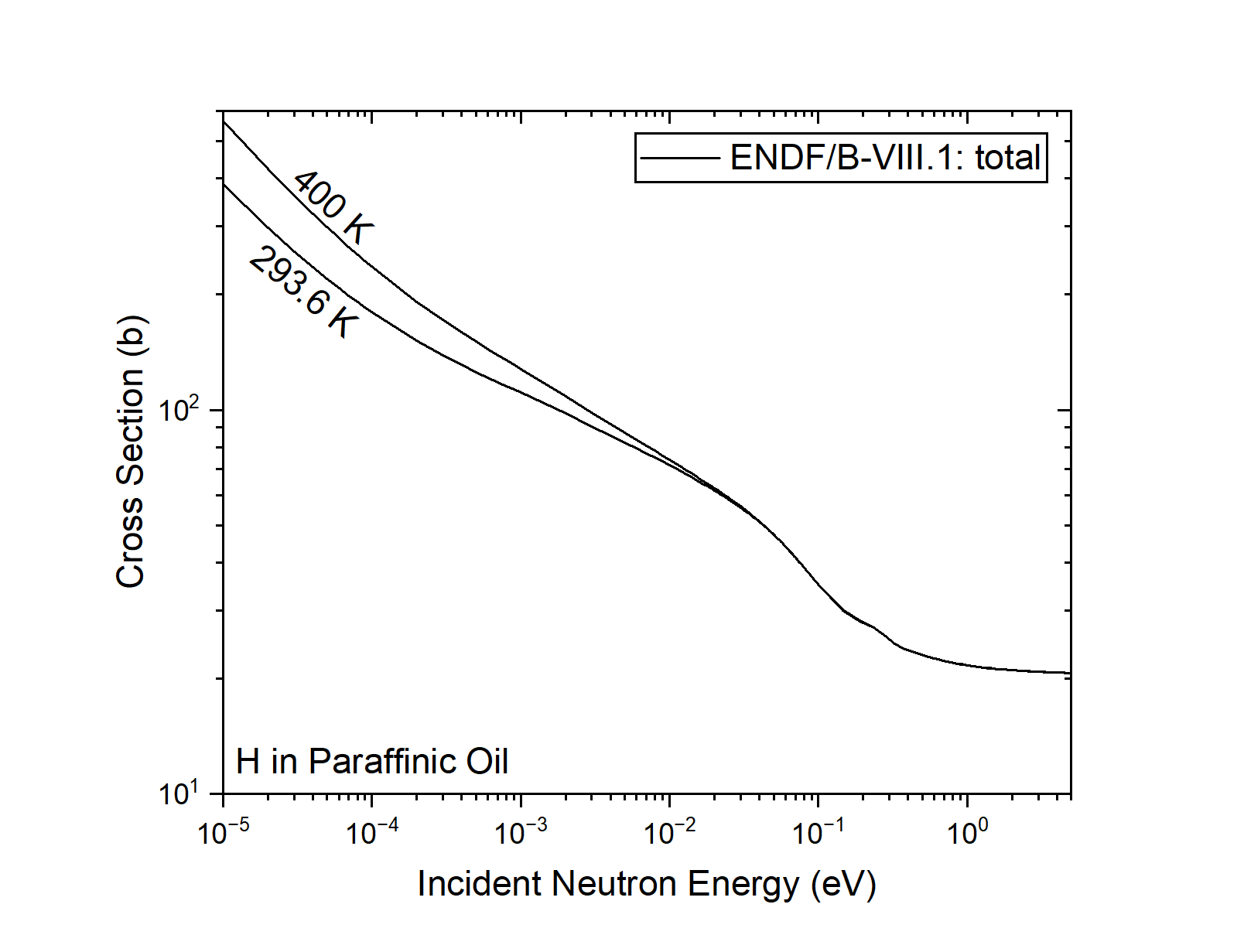}
    \caption{Total thermal neutron scattering cross section of H in paraffinic oil at select temperatures.}
    \label{fig:Xsec_ParaffinicOil}
\end{figure}

The \FLASSH\ \cite{Fleming2023} code was used to perform the TSL evaluation and produce the H(ParaffinicOil) data in the ENDF File 7 format (MAT41). The total TSL, SS($\alpha$,$\beta$), was calculated assuming the incoherent approximation and the convolution of the solid TSL component with a diffusive component based on a Langevin model with mass and free atom cross sections \cite{Brown2018} for $^1$H. Fig.~\ref{fig:TSL_ParaffinicOil} shows the symmetric TSL of hydrogen in paraffinic oil as a function of $\alpha$ for various values of $\beta$. The total thermal scattering cross section of hydrogen is shown in Fig.~\ref{fig:Xsec_ParaffinicOil}. Benchmark comparison of this evaluation against transmission data shows excellent agreement \cite{Daskalakis2025}. The evaluation was performed for 293.6, 300, 325, 350, 375, 400 K. The paraffinic oil evaluation represents a new TSL contribution that is included for the first time in the ENDF/B database.

\subsubsection{Silicon Carbide (\sic)}
\label{sec:sic}

The evaluation for 3C-SiC (\sic) is based on the previous \prENDF\ evaluation with key updates to improve the TSL and resulting cross section. The \prENDF\ AILD simulations of the lattice structure were maintained \cite{Zhu2013,ZhuMS}; however, updates to the evaluation of the PDOS were performed to provide needed energy resolution. In addition, this contribution updates the evaluation to use the natural abundances \cite{Sears1992} of the atomic mass and cross sections of C and Si \cite{Brown2018}, as opposed to $^{12}$C and $^{28}$Si, and to use experimental lattice parameters \cite{Madelung1982} in the calculation of the coherent elastic data instead of \ab\ values.

3C-SiC was evaluated using typical AILD methods \AILD. \Ab\ simulations of \csic\ were performed using the \texttt{VASP} code \VASP, and the resultant Hellmann-Feynman forces were used in the PHONON code \cite{Parlinski1997} to calculate the phonon dispersion relations and PDOS for each element using the dynamical matrix method \cite{ZhuMS}. Ref.~\cite[Fig.~9]{ZhuMS} shows a comparison of the calculated and experimental phonon dispersion curves.

\begin{figure}
    \centering
    \includegraphics[width=1.0\columnwidth,clip,trim=  20mm 10mm 30mm 15mm]{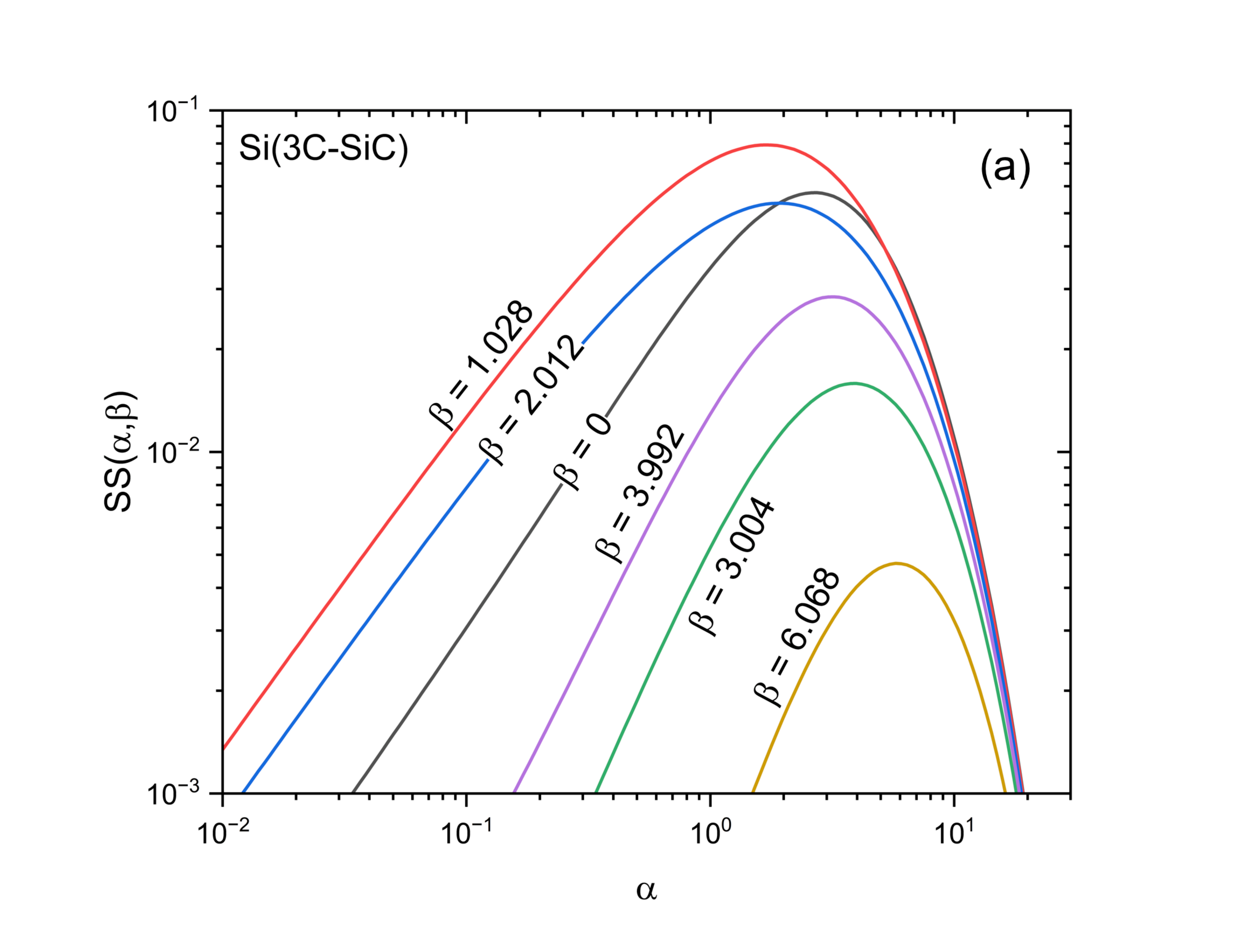}
    \includegraphics[width=1.0\columnwidth,clip,trim=  20mm 10mm 30mm 15mm]{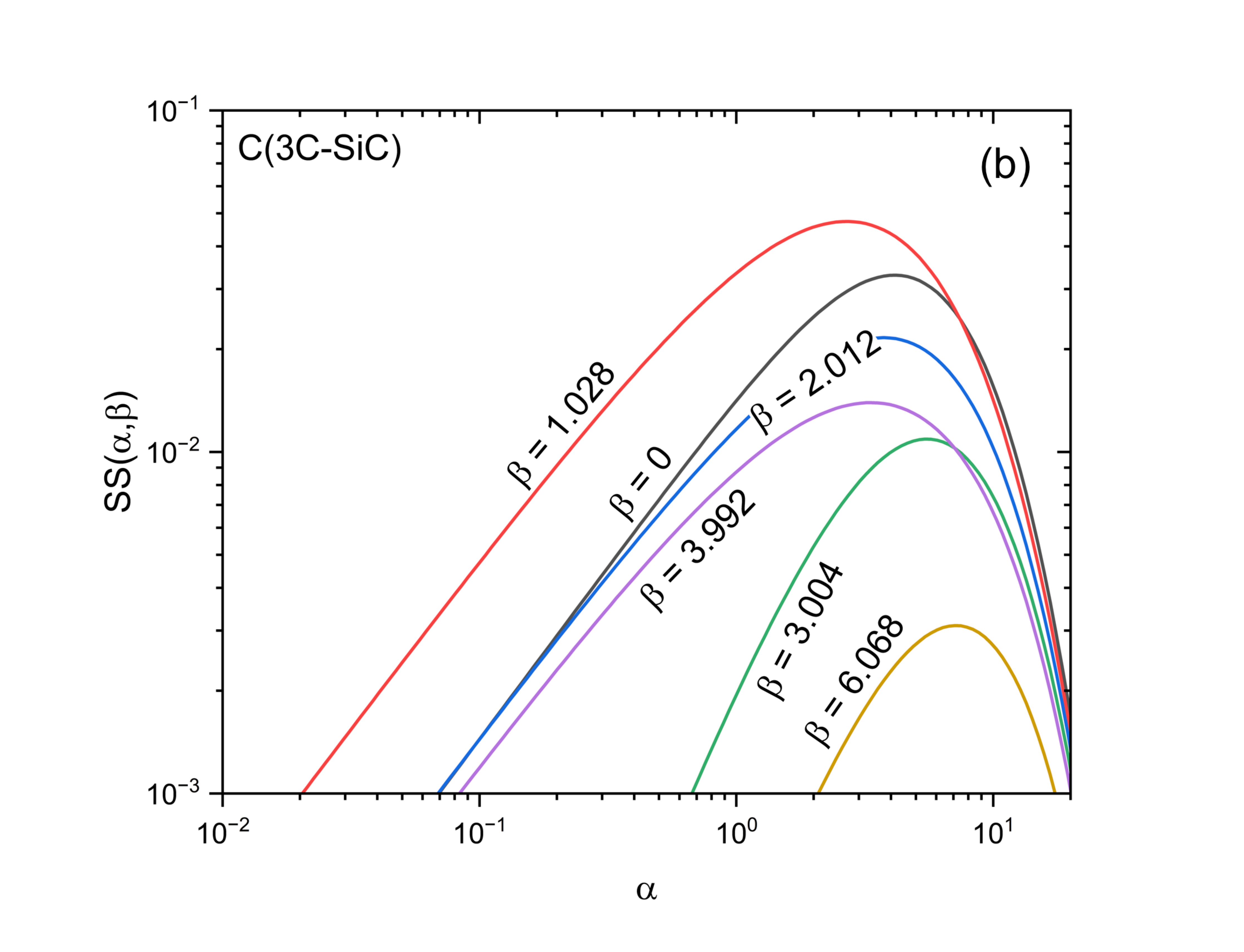}
    \caption{(Color online) The symmetric TSL of (a) Si(\csic), and (b) C(\csic) as a function of momentum transfer, $\alpha$, for various values of $\beta$ at 296 K. $SS(\alpha,\beta)$ for each $\beta$ is labeled with the corresponding line.}
    \label{fig:TSL_SiC}
\end{figure}

The partial PDOS for Si and C in \sic\ were used to calculate the TSL (File 7) at 296, 300, 400, 500, 600, 700, 800, 1000, and 1200 K under the incoherent approximation for Si(\csic) (MAT43) and C(\csic) (MAT44) using \FLASSH\ \cite{Fleming2023}. The symmetric TSL of Si(\csic) and C(\csic) at 296 K are shown in Fig.~\ref{fig:TSL_SiC}.

\begin{figure}
    \centering
    \includegraphics[width=1.0\columnwidth,clip,trim=  20mm 10mm 30mm 15mm]{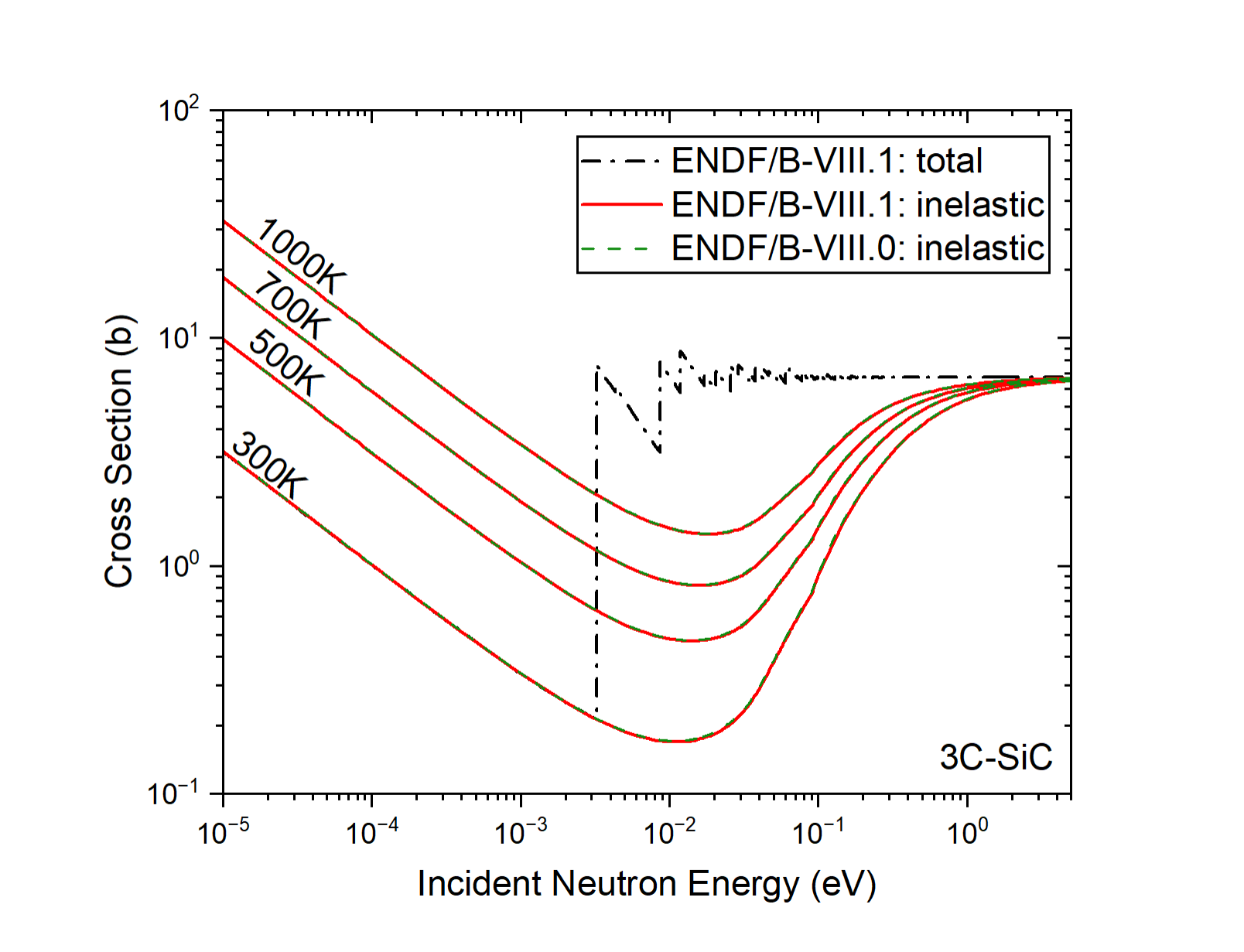}
    \caption{(Color online) The inelastic scattering cross section of \csic\ shown at selected temperatures. The total scattering cross section of \csic\ at 300 K is also shown. The inelastic scattering cross section from \prENDF\ is shown for comparison.}
    \label{fig:Xsec_SiC}
\end{figure}

The coherent elastic scattering cross section of \csic\ was generated using the cubic approximation in \FLASSH\ and is split evenly between the File 7 for each of Si(\csic) and C(\csic) (MT2). The \ENDF\ total scattering cross section of \sic\ at 296 K is shown in Fig.~\ref{fig:Xsec_SiC}, along with the inelastic scattering cross section at select temperatures.

\subsubsection{Silicon Dioxide (SiO$_2$, $\alpha$ Phase)}
\label{sec:sio2}

\sio\ was reevaluated using standard AILD methods \AILD. Also called $\alpha$-quartz, \sio\ is a common material and a major component of rock and earth. The $\alpha$-phase is the low-temperature phase, existing in this form until temperatures exceed 836 K, at which point it transitions to $\beta$-quartz. \sio\ has a trigonal crystal structure belonging to the \emph{P}3$_2$21 (\#154) or \emph{P}3$_1$21 (\#152) space group depending on chirality. The \texttt{VASP} code \VASP\ was used to perform the \ab\ modeling and calculate the Hellmann-Feynman forces of the crystal. These were used by the PHONON code \cite{Parlinski1997} to calculate the phonon dispersion relations and PDOS via the dynamical matrix method. 

\begin{figure}
    \centering
    \includegraphics[width=1.0\columnwidth,clip,trim=  20mm 10mm 30mm 15mm]{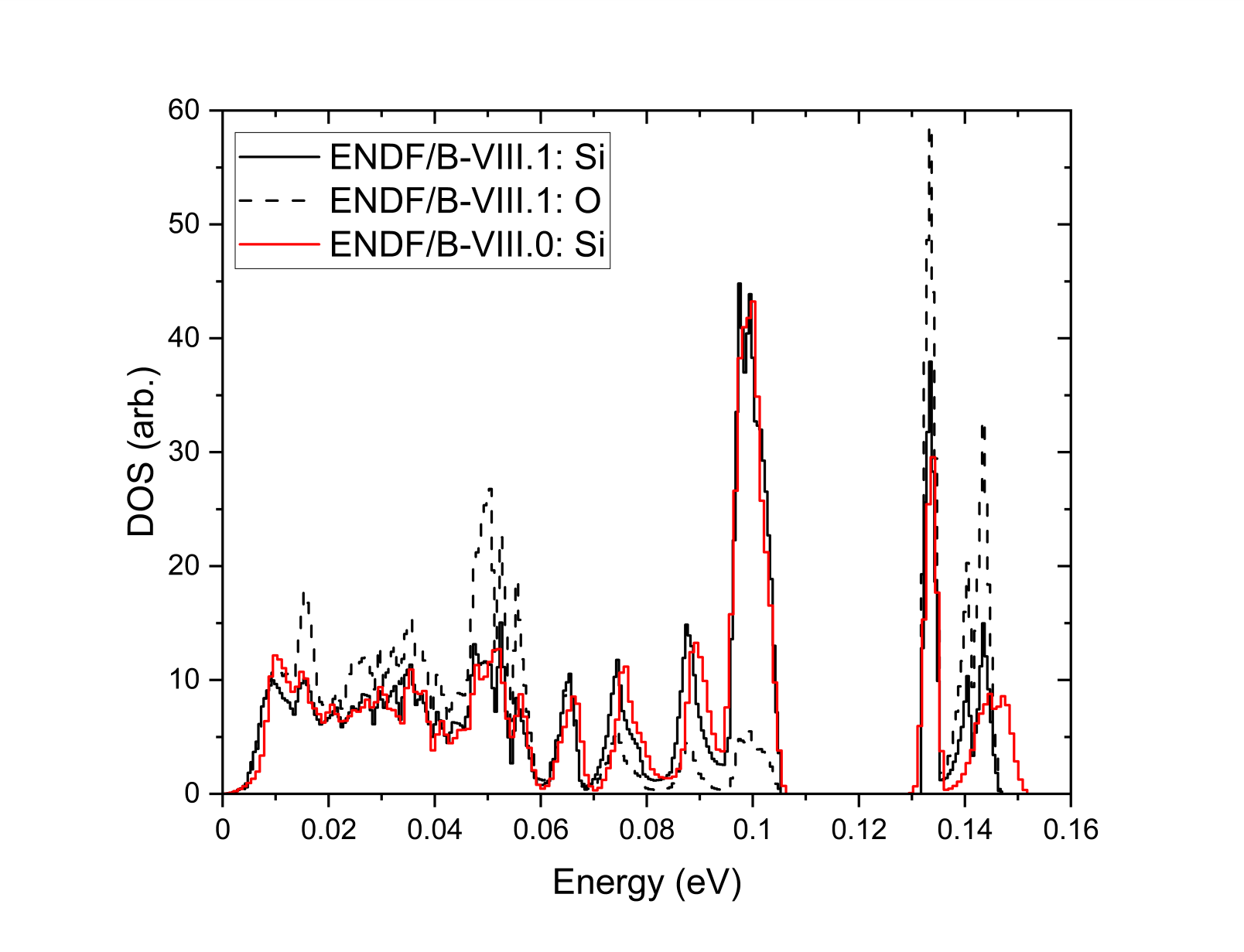}
    \caption{(Color online) The Si(\sio) and O(\sio) DOSs used in the \prENDF\ and \ENDF\ evaluations.}
    \label{fig:DOS_SiO2}
\end{figure}

The \sio\ evaluation has been improved for \ENDF\ through the use of a new PDOS for Si and O in \sio\ and the separation of the single \sio\ evaluation from \prENDF\ into distinct evaluations of Si(\sio) and O(\sio). A comparison of the \prENDF\ and \ENDF\ DOSs are shown in Fig.~\ref{fig:DOS_SiO2}. The calculation of the coherent elastic scattering cross section was updated to use synthesized experimental lattice parameters and atomic positions \cite{Carpenter1998, Cohen1958, Page1976}, and naturally weighted \cite{Sears1992} atomic masses and cross sections \cite{Brown2018} were used throughout the evaluation.

\begin{figure}
    \centering
    \includegraphics[width=1.0\columnwidth,clip,trim=  20mm 10mm 30mm 15mm]{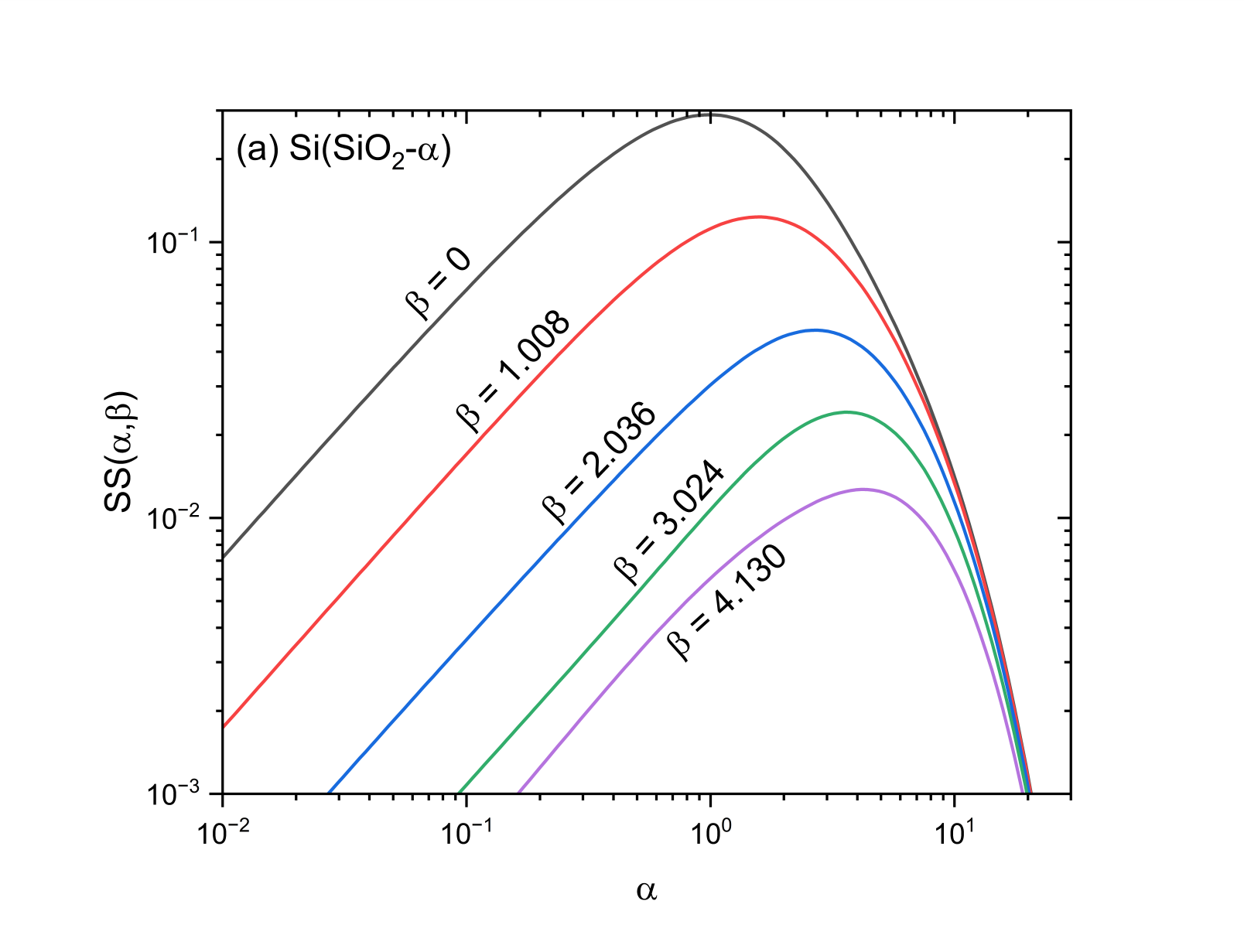}
    \includegraphics[width=1.0\columnwidth,clip,trim=  20mm 10mm 30mm 15mm]{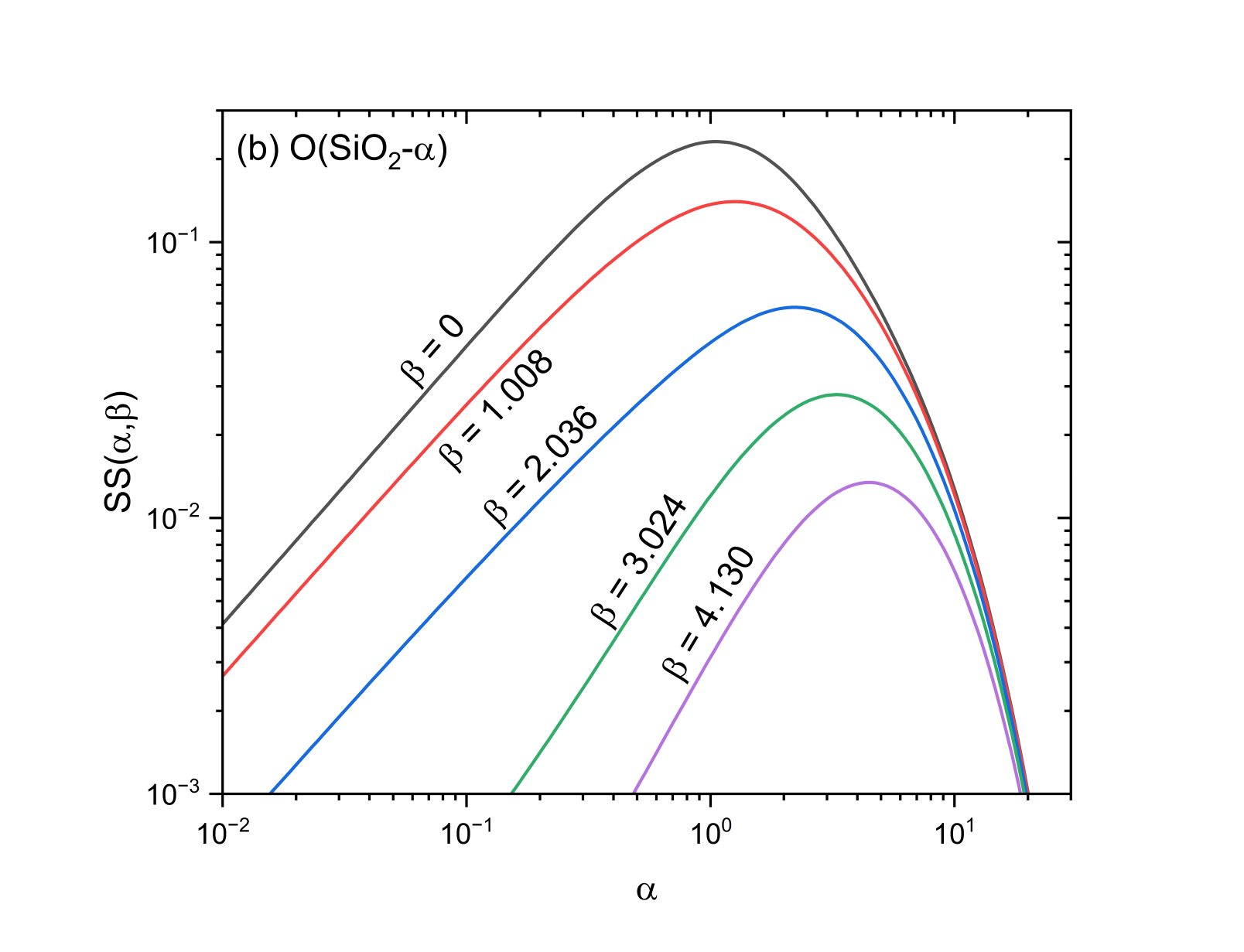}
    \caption{(Color online) The symmetric TSL of (a) Si(\sio) and (b) O(\sio) as a function of momentum transfer, $\alpha$, for various values of $\beta$ at 293.6 K. $SS(\alpha,\beta)$ for each $\beta$ is labeled with the corresponding line.}
    \label{fig:TSL_SiO2}
\end{figure}

The partial PDOS were used to calculate the TSL (File 7) for Si(\sio) (MAT3016) and O(\sio) (MAT3017) at 293.6, 350, 400, 500, 650, 800, and 846 K. The calculation of the TSL and thermal neutron scattering cross sections were performed using \FLASSH\ \cite{Fleming2023}, using the incoherent approximation for the former. Fig.~\ref{fig:TSL_SiO2} presents the symmetric TSL of Si(\sio) and O(\sio) at 293.6~K.

\begin{figure}
    \centering
    \includegraphics[width=1.0\columnwidth,clip,trim=  20mm 10mm 30mm 15mm]{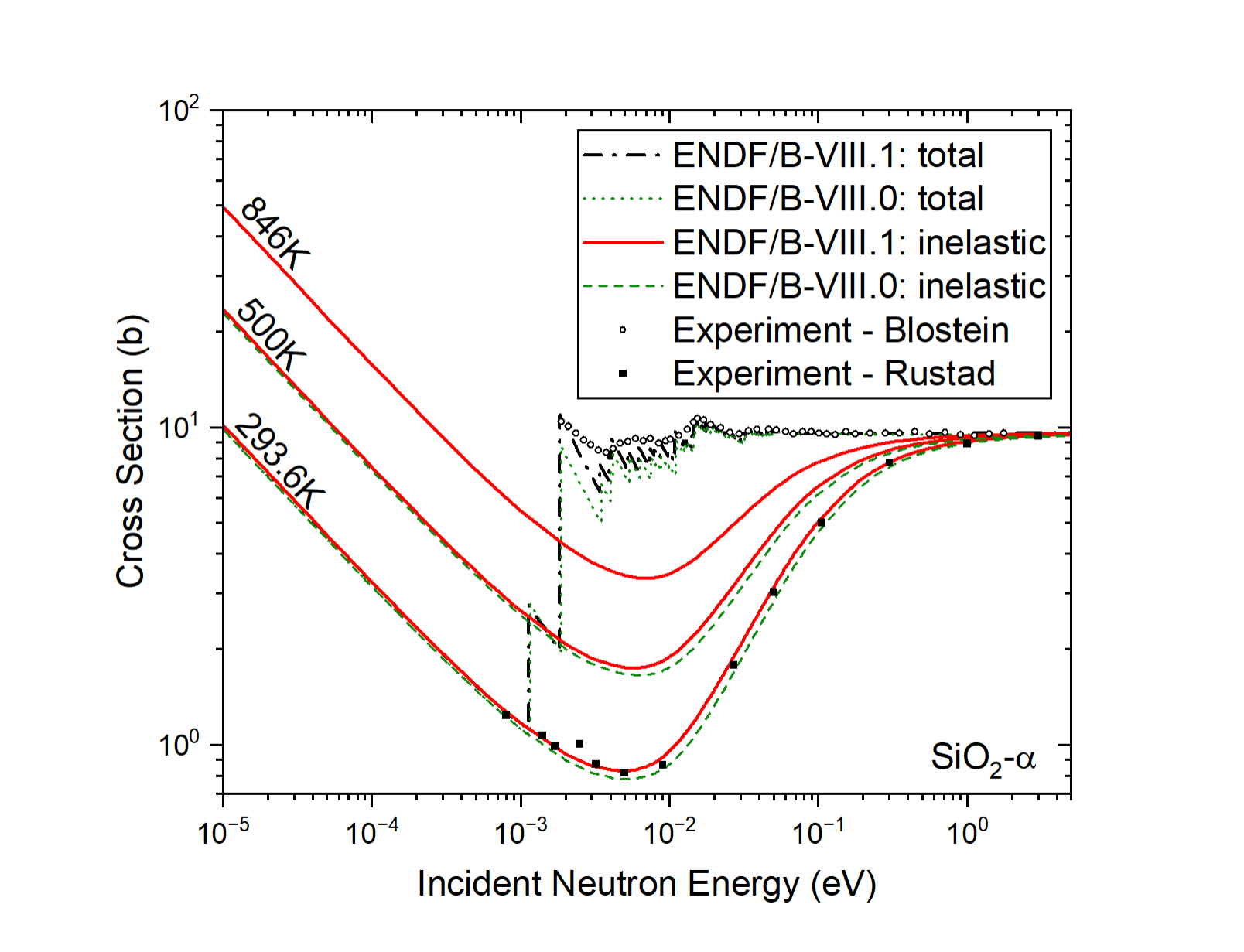}
    \caption{(Color online) The inelastic scattering cross section of \sio\ shown at selected temperatures. The total scattering cross section of \sio\ at 293.6 K is also shown. The total and inelastic cross sections are compared to experimental data at room temperature \cite{Blostein1999, Rustad1965}. The \prENDF\ \sio\ inelastic scattering cross sections below 500 K are shown for comparison.}
    \label{fig:Xsec_SiO2}
\end{figure}

The cubic approximation was invoked in the calculation of the coherent elastic scattering contribution of the lattice, and the coherent elastic data is split between the File 7 of Si(\sio) and O(\sio) according to stoichiometry, with a weighting factor of $1/3$. Fig.~\ref{fig:Xsec_SiO2} shows the \ENDF\ total scattering cross section of \sio\ at 293.6 K, as well as the inelastic scattering cross section of the compound at selected temperatures. The total and inelastic cross sections are compared to experimental data at room temperature \cite{Blostein1999, Rustad1965}, where the experimental data for inelastic scattering has been corrected for absorption.

\subsubsection{Polystyrene (\ps)} 
\label{sec:PStsl}

The neutronic interest in polystyrene stemmed from a question surrounding several benchmarks in the ICSBEP benchmark suite. Specifically, MCT-002 and PCM-034 both contain appreciable quantities of polystyrene. Historically, the TSL file for polyethylene was used as a surrogate for calculating integral eigenvalue. This file was created to try and quantify the accuracy of this assumption.

\begin{figure}[h] 
  \centering
  \includegraphics[width=1.0\columnwidth]{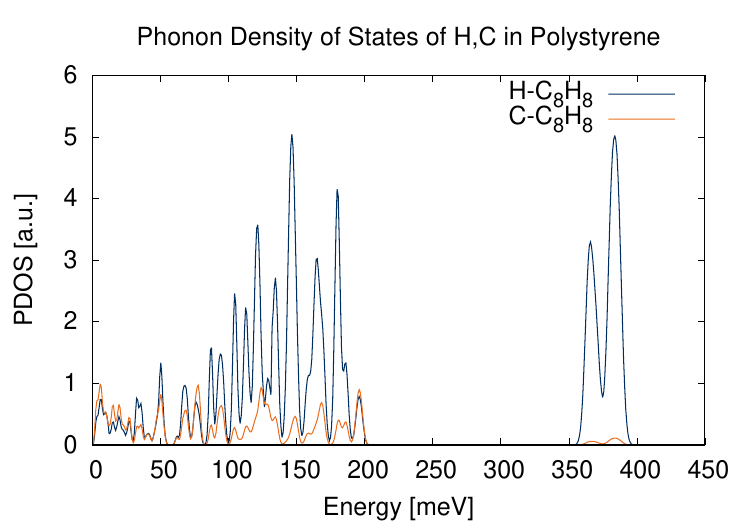}
  \caption{Phonon density of states for H \& C in polystyrene.}
  \label{fig:PS_PDOS}
\end{figure}

The PDOS for hydrogen and carbon in polystyrene was calculated using CASTEP \cite{ref_castep}. The starting structure was taken from Ref.~\cite{ref_ps_structure}, with a unit cell containing 48 carbon and hydrogen atoms. The structure was first relaxed using the Broyden-Fletcher-Goldfarb-Shanno (BFGS) minimization algorithm to identify an energetically stable atomic configuration and unit cell shape. A 750 eV energy cutoff was used in the Perdew-Burke-Ernzerhof formulation of the generalized gradient approximation. The system was relaxed until a tolerance of 5e-9 eV/atom for the ground state energy and 5e-4 \AA~ for the geometric displacement of each atom was met. This relaxed unit cell was then used to calculate phonon properties using CASTEP's built-in density functional perturbation theory (DFPT) method. The advantage of CASTEP's DFPT implementation is that the calculations can be performed on the unit cell instead of the (often much larger) supercell.  Because of the large size of the unit cell, a k-point grid of 1~$\times$~1~$\times$~3 was found to be sufficient for calculating the phonon properties. The resulting phonon file was then fed into OCLIMAX \cite{ref_oclimax_1,ref_oclimax_2}, which was used to calculate the partial PDOS for carbon and hydrogen. Phonon spectra were then used to calculate the resulting ENDF TSL files using \NJOY2016~\cite{NJOY}. A plot of the constituent phonon spectra is shown in Fig..~\ref{fig:PS_PDOS}.

The basis of ORNL-developed TSLs are inelastic neutron scattering (INS) and transmission (i.e., total scattering) measurements. INS measurements are performed using the VISION instrument \cite{SEEGER2009719} at the Spallation Neutron Source at ORNL to measure \textit{S(Q},$\omega$), which is directly comparable to $S(\alpha, \beta)$: $S(\alpha,\beta,T)=k_BT \textrm{ }exp\left(\frac{-\hbar\omega}{2k_BT}\right) S(Q,\omega,T)$, where \textit{T} is the temperature, and $k_B$ is the Boltzmann constant; $S(\alpha, \beta)$ is the quantity  stored in ENDF File 7, MT4 of the TSL file. This means that INS measurements are explicit tools for validation of TSLs. Transmission measurements were performed at the Gaerttner LINAC at RPI~\cite{FRITZ2023109651}.

\begin{figure}[h] 
  \centering
  \includegraphics[width=1.0\columnwidth]{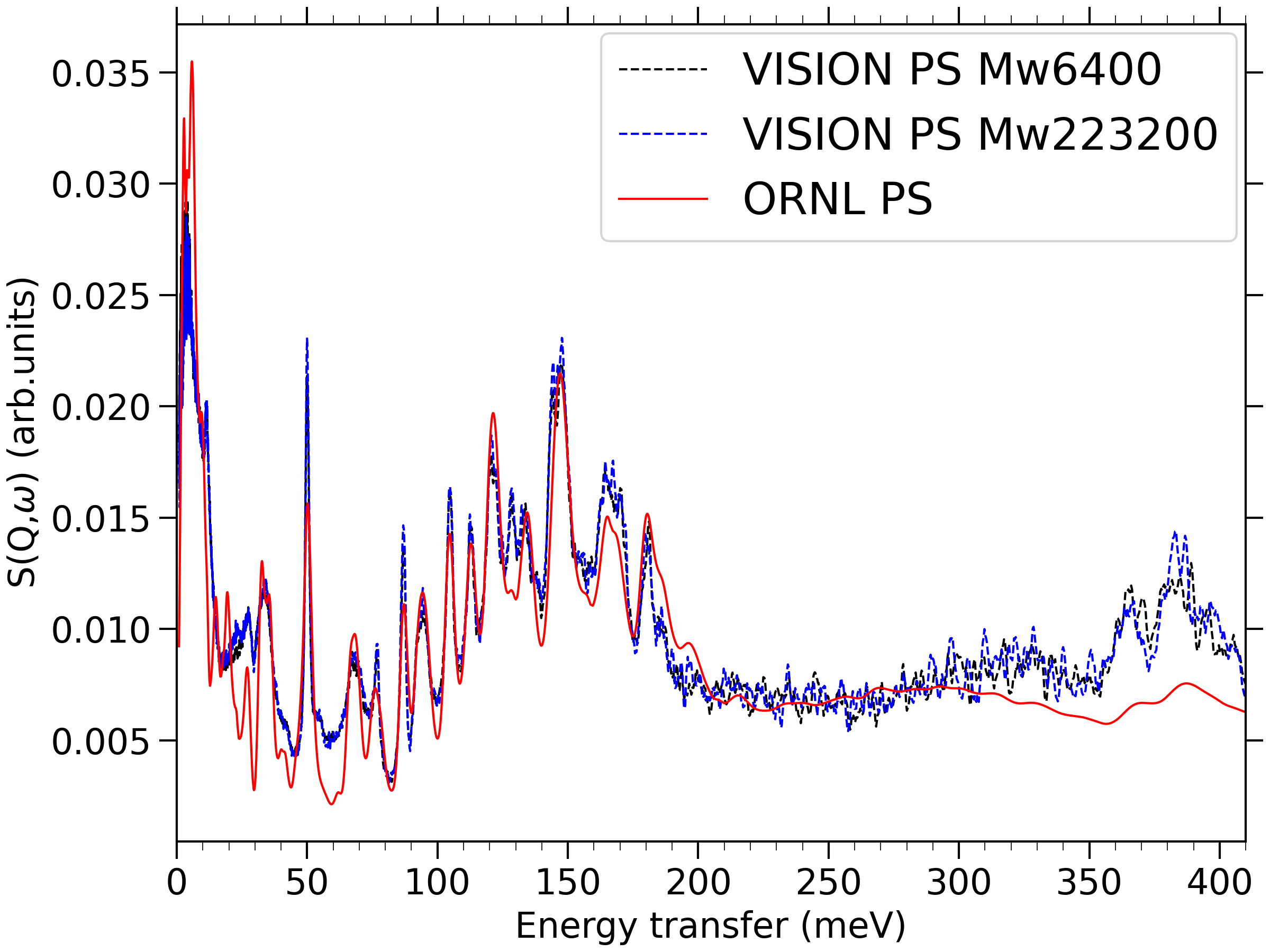}
  \caption{Comparison of calculated S(Q,$\omega$) spectra versus measured for PS.}
  \label{fig:PS_vision_comparison}
\end{figure}

For polystyrene, two samples were measured using the VISION instrument: one with a molecular weight of 223200 g/mol and the other with a molecular weight of 6400 g/mol. The difference in the molecular weight implies that the samples were different structurally, where a higher molecular weight polymer tends to have a more complex and larger structure compared to a polymer with a lower molecular weight. To compare with the VISION measurement, we converted $S(\alpha,\beta)$ from an ENDF file to \textit{S(}\textit{Q},$\omega$), extracted the values along the same \textit{Q} and $\omega$ trajectory as in the VISION experiment, and applied the VISION instrumental resolution. The comparison with the experimentally measured \textit{S(}\textit{Q},$\omega$) is shown in Fig.~\ref{fig:PS_vision_comparison}, where the agreement between measured and calculated spectra is excellent. Most of the vibrational modes can be observed, although some small inconsistencies were seen at lower energies.

\begin{figure}[h] 
  \centering
  \includegraphics[width=1.0\columnwidth]{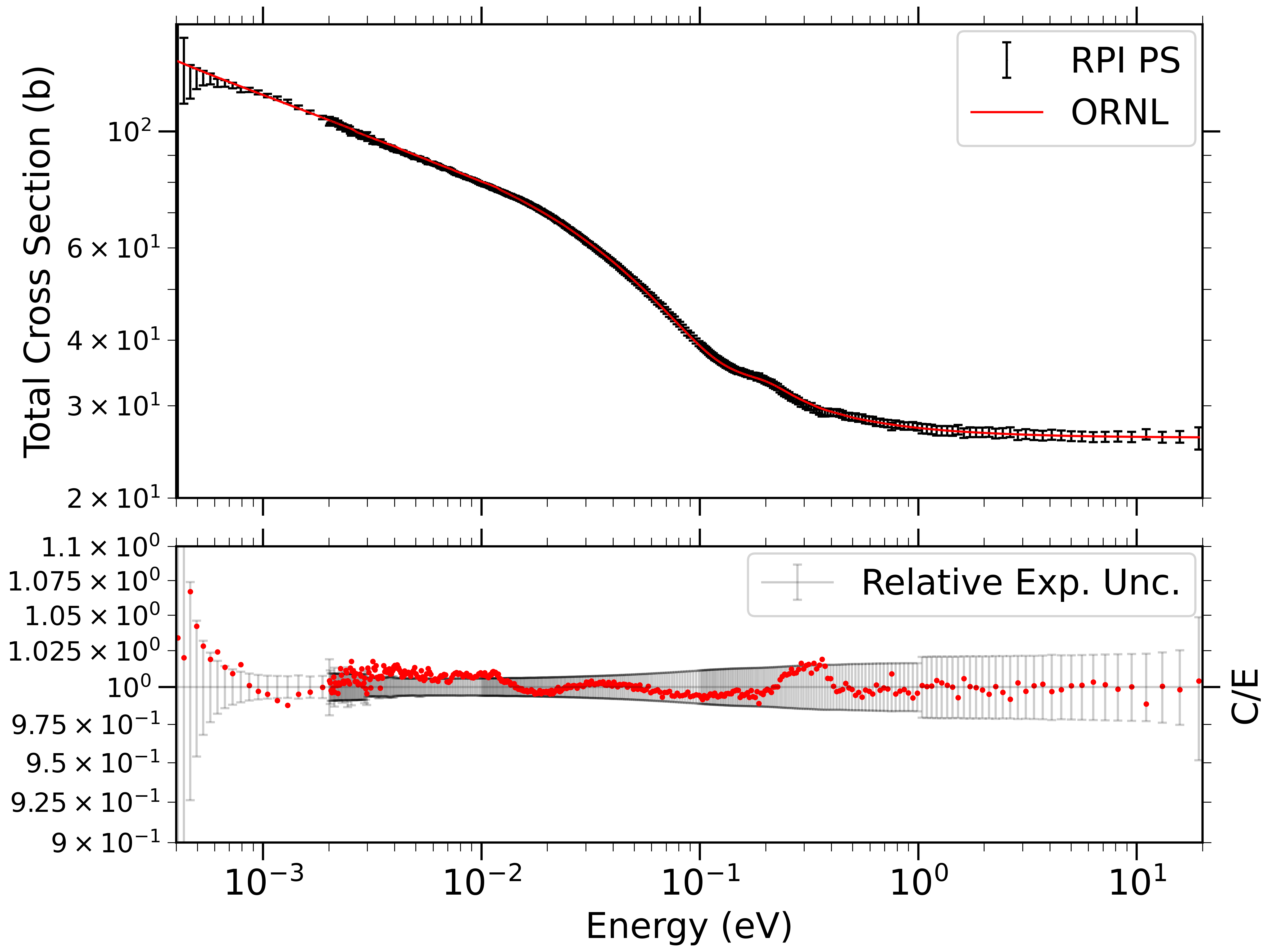}
  \caption{Comparison of calculated total cross section versus that measured for PS (top) and C/E plot (bottom).}
  \label{fig:PS_tsx_comparison}
\end{figure}

Note that the TSL evaluation used an optimization method developed at ORNL. The basis for any TSL evaluation is the phonon spectrum, and from it, the whole scattering kernel---that is, $S(\alpha,\beta)$---as well as the total cross section can be calculated. Thus, because the phonon spectra have distinct vibrational modes (i.e., peaks), the developed method assigns weights to each peak and uses an optimization procedure using Dakota \cite{ref_dakota} to change the weights while minimizing the $\chi^2$ square value between the calculated and measured \textit{S(Q},$\omega$), as well as the calculated and measured total cross section, as measured at the RPI LINAC. The detailed optimization procedure details will be provided in a companion article. The comparison between measured and calculated total cross section is shown in Fig.~\ref{fig:PS_tsx_comparison}, as well as the C/E plot. More information about this fitting procedure can be found in \cite{ref_ornl_nd4_report}.
Fig.~\ref{fig:PS_tsx_comparison} shows that the agreement between the measured and calculated total cross section is excellent. This is in part because the evaluation was optimized using our methodology to match the RPI experimental transmission data. Ideally, for validation purposes, additional polystyrene transmission data are necessary.

For both carbon and hydrogen in polystyrene, TSLs were calculated using \NJOY2016 in the incoherent approximation, and the symmetric $S(\alpha,\beta)$ is stored as ENDF File 7, MT 4, with incoherent elastic stored in ENDF File 7, MT 2.

\subsubsection{Poly(methyl methacrylate) (\pmma)} 
\label{sec:luciteTSL}

\begin{figure}[h] 
  \centering
  \includegraphics[width=1.0\columnwidth]{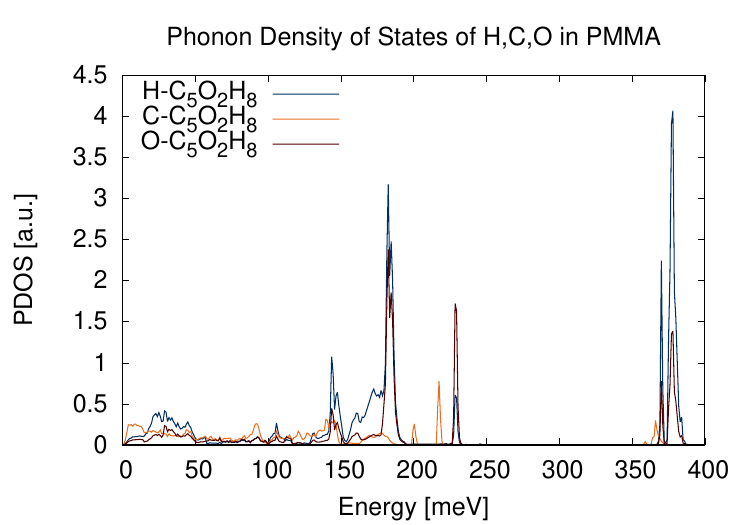}
  \caption{Phonon density of states for H, C, \& O in Poly(methyl methacrylate).}
  \label{fig:PMMA_PDOS}
\end{figure}

Polymethyl methacrylate (\pmma, or PMMA), also known as Lucite or Plexiglass, was evaluated using LAMMPS \cite{Thompson2022}. Phonon spectra for all atoms in lucite (i.e., hydrogen, carbon, and oxygen) were calculated. Lucite is a polymer with an amorphous structure. The initial structure for MD calculations was obtained using Polymer Builder \cite{doi:10.1021/acs.jctc.1c00169}, as implemented in CHARMM-GUI \cite{https://doi.org/10.1002/jcc.20945}. The structure created by the Polymer Builder consisted of 55 chains and 20 monomers of lucite at each chain. The specific computational details will be provided in the follow-up article to this paper. The partial DOS for each atom were calculated using velocity auto-correlation function, as implemented in MDANSE \cite{doi:10.1021/acs.jcim.6b00571}. A plot of the constituent phonon spectra is shown in Fig..~\ref{fig:PMMA_PDOS}.

\begin{figure}[h] 
  \centering
  \includegraphics[width=1.0\columnwidth]{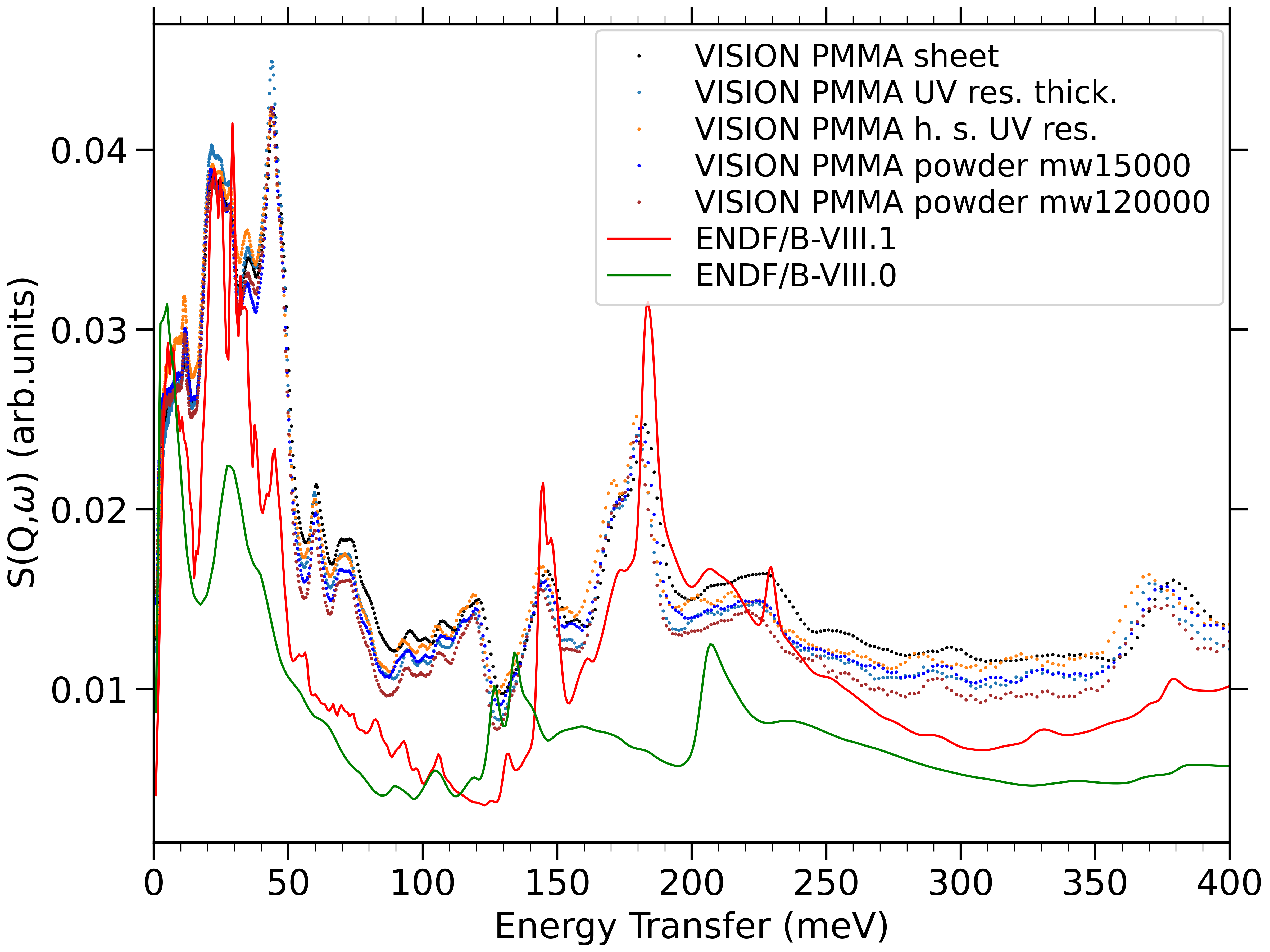}
  \caption{Comparison of calculated S(Q,$\omega$) spectra from different TSLs versus that measured for PMMA.}
  \label{fig:PMMA_vision_comparison}
\end{figure}

Five different samples of lucite were measured with the VISION instrument: two lucite powders (with a molecular weight of 15,000 g/mol and 120,000 g/mol), a sheet of regular store-bought lucite, a clear scratch and UV-resistant sheet of lucite, and a clear high-strength UV-resistant sheet of lucite. The main difference between all these samples was in the additives used to achieve different properties; lucite powders had no additives. Additionally, samples were of different thicknesses. The comparison of different TSLs with the experimentally measured \textit{S(Q},$\omega$) can be seen in Fig.~\ref{fig:PMMA_vision_comparison}. Very little variation in the shape of the measured spectra was observed even though the samples were of significantly different compositions. The variations in the intensity of the spectra are due to differences in the thickness due to multiple neutron scattering. Additionally, it is noticeable that none of the TSLs do a great job of reproducing the measured spectra, but the ORNL (ENDF/B-VIII.1) TSL reproduces the shape better than ENDF/B-VIII.0 TSL. This can be seen in the calculated $\chi^2$ values for each TSL in Table \ref{table:chi_pmma_sqw}.

\begin{table}[ht]
    \centering
    \caption{Calculated $\chi^2$ values for ENDF/B-VIII.0 and ENDF/B-VIII.1 TSLs with respect to \textit{S(Q},$\omega$) measurements with the VISION instrument.}
    \label{table:chi_pmma_sqw}
    \begin{tabular}{cc}
    \toprule \toprule
                TSL & $\chi^2$  \\ \midrule
                ENDF/B-VIII.0 & 1.24E+6  \\ 
                ENDF/B-VIII.1 & 4.94E+5 \\
                \bottomrule \bottomrule
    \end{tabular}
\end{table}

\begin{figure}[h] 
  \centering
  \includegraphics[width=1.0\columnwidth]{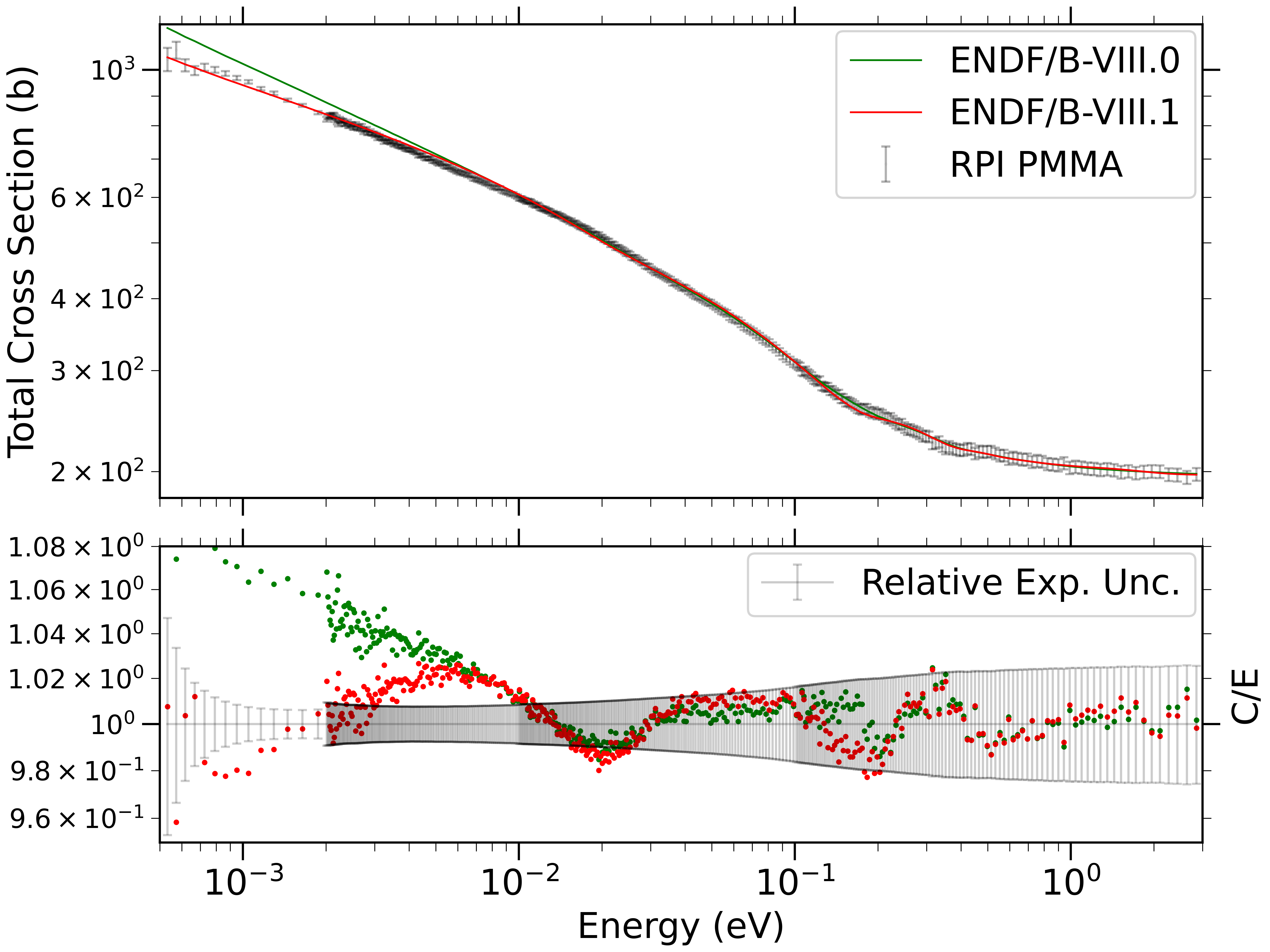}
  \caption{Comparison of calculated total cross sections versus those measured for PMMA (top) and C/E plot (bottom).}
  \label{fig:PMMA_tsx_comparison}
\end{figure}

The comparison between measured and calculated total cross sections is given in Fig.~\ref{fig:PMMA_tsx_comparison}, as well as the C/E plot. The ENDF/B-VIII.0 TSL overpredicts the total cross section below 3 meV, whereas ENDF/B-VIII.1 TSLs does significantly better. Both TSLs overcalculate the total cross section between 5 meV and 10 meV. Looking at the $\chi^2$ values for each TSL in Table \ref{table:chi_pmma_tsx}, we can see that ORNL TSL has slightly lower value.
\begin{table}[ht]
    \centering
    \caption{Calculated $\chi^2$ values for ENDF/B-VIII.0 and ENDF/B-VIII.1 TSLs with respect to total cross section measurements at RPI.}
    \label{table:chi_pmma_tsx}
    \begin{tabular}{cc}
    \toprule \toprule
                TSL & $\chi^2$  \\ \midrule
                ENDF/B-VIII.0 & 294.4  \\ 
                ENDF/B-VIII.1 & 293.1 \\ 
\bottomrule \bottomrule               
    \end{tabular}

\end{table}

\begin{figure}[h] 
  \centering
  \includegraphics[width=1.0\columnwidth]{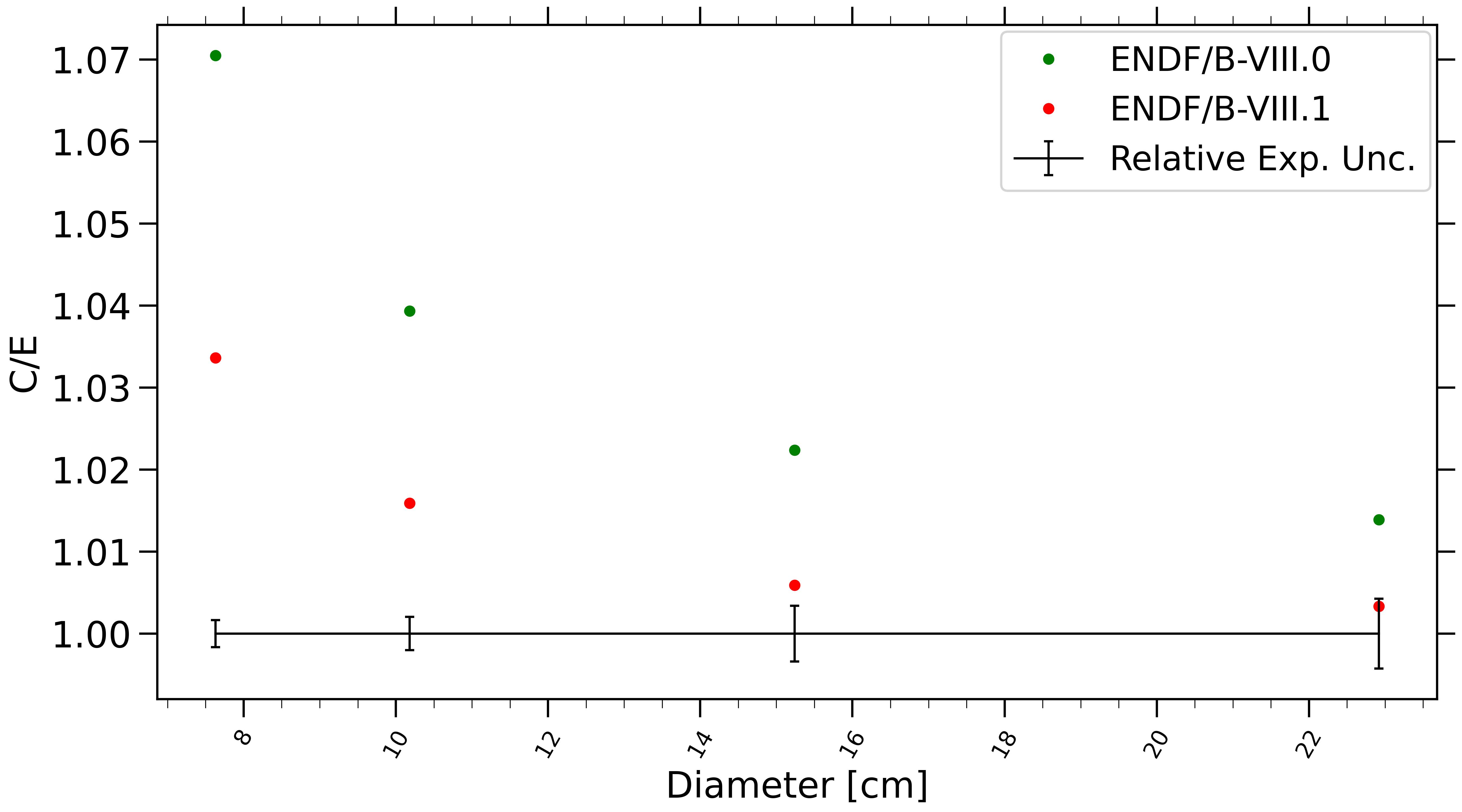}
  \caption{Comparison of calculated decay constant eigenvalues from ENDF/B-VIII.0 and ENDF/B-VIII.1 TSLs versus measured for PMMA.}
  \label{fig:PMMA_pnda_comparison}
\end{figure}

The TSLs for PMMA were compared against the new pulsed-neutron die-away (PNDA) measurements from LLNL~\cite{PMMA_PNDA}. The comparison between the calculated and measured decay constant eigenvalue, $\alpha$, is shown in Fig.~\ref{fig:PMMA_pnda_comparison}. As shown in comparison with the measured total cross section, the ENDF/B-VIII.0 TSL overpredicts the measured decay constant eigenvalues, whereas the ORNL TSL performs significantly better. The calculated cumulative $\chi^2$ values for each TSL can be seen in Table \ref{table:chi_pmma_pnda}.

\begin{table}[ht]
\centering
    \caption{Calculated cumulative $\chi^2$ values for ENDF/B-VIII.0 and ENDF/B-VIII.1 TSLs with respect to PNDA measurements.}
    \label{table:chi_pmma_pnda}
    \begin{tabular}{cc}
    \toprule \toprule
        TSL & $\chi^2$  \\ \midrule
        ENDF/B-VIII.0 & 2.24E+3 \\
        ENDF/B-VIII.1 & 4.77E+2  \\
\bottomrule \bottomrule
    \end{tabular}
\end{table}

\begin{figure}[h] 
  \centering
  \includegraphics[width=1.0\columnwidth]{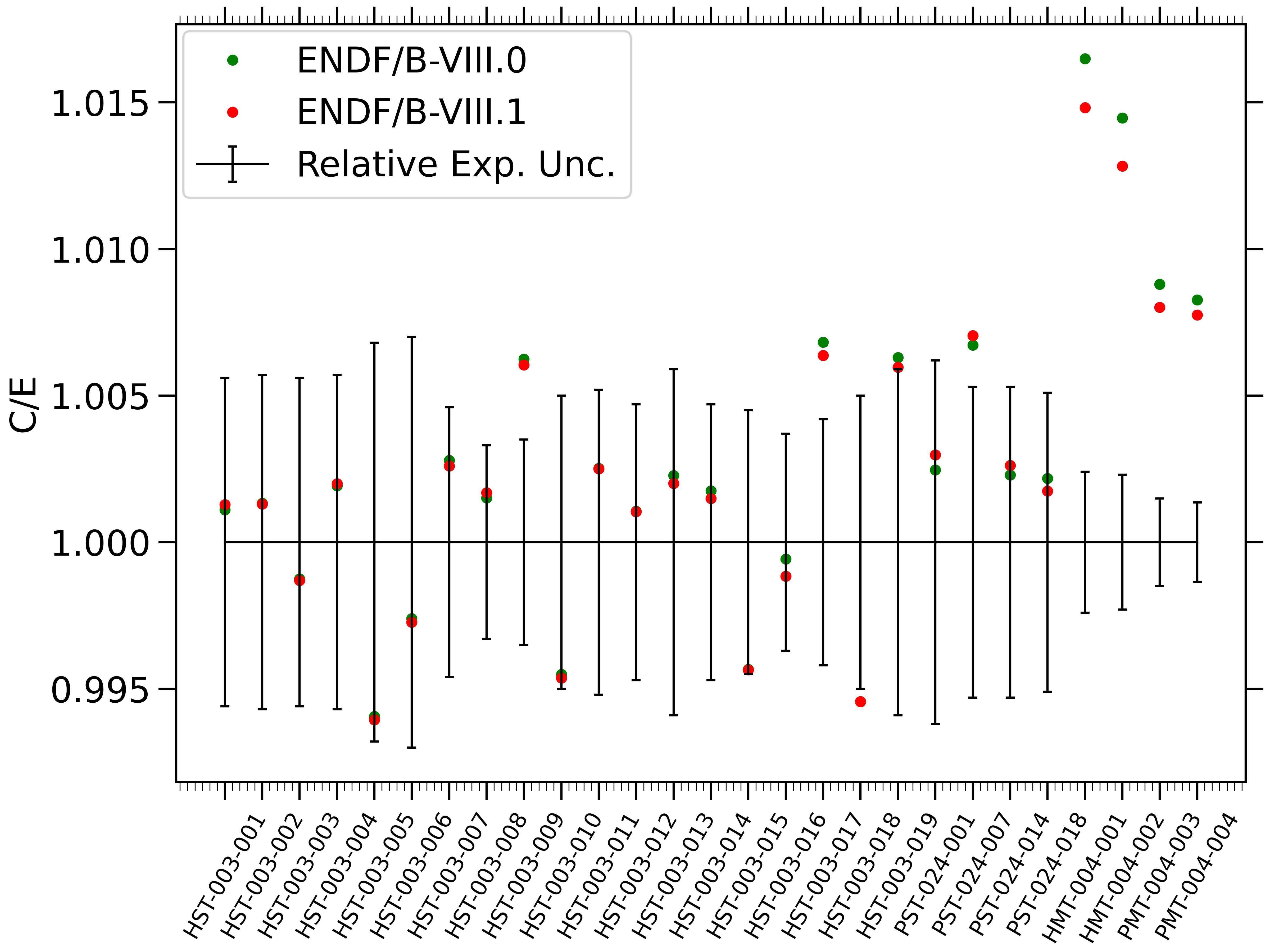}
  \caption{Comparison of calculated k$_\mathrm{eff}$ values from ENDF/B-VIII.0 and ENDF/B-VIII.1 TSLs versus those measured for different thermal criticality benchmarks, taken from the ICSBEP 2021 Handbook.}
  \label{fig:PMMA_benchmarks_comparison}
\end{figure}

The available PMMA TSLs were compared against most of the thermal benchmarks that contain PMMA in the ICSBEP Handbook, 2021 edition. The comparison of calculated k$_\mathrm{eff}$ C/E values versus various benchmarks can be seen in Fig.~\ref{fig:PMMA_benchmarks_comparison}. Similar to what was observed in the PNDA and total cross section comparisons,  the ENDF/B-VIII.0 TSL overpredicts k$_\mathrm{eff}$ values, whereas the ORNL TSL performs better, including new HMT-004 benchmarks from LLNL. The calculated cumulative $\chi^2$ values for each TSL can be seen in Table \ref{table:chi_pmma_keff}.

\begin{table}[ht]
\centering
    \caption{Calculated cumulative $\chi^2$ values for ENDF/B-VIII.0 and ENDF/B-VIII.1 TSLs with respect to k$_\mathrm{eff}$ measurements.}
    \label{table:chi_pmma_keff}
    \begin{tabular}{cc}
    \toprule \toprule
        TSL & $\chi^2$  \\ \midrule
        ENDF/B-VIII.0 & 1.33E+2 \\
        ENDF/B-VIII.1 & 9.34E+1  \\
\bottomrule \bottomrule
    \end{tabular}
\end{table}

Hydrogen, carbon, and oxygen in PMMA TSLs were calculated using \NJOY2016 in the incoherent approximation, and the symmetric $S(\alpha,\beta)$ is stored as ENDF File 7, MT 4, with incoherent elastic stored in ENDF File 7, MT 2.

\subsubsection{Zirconium Carbide (ZrC)}
\label{sec:zrc}

The TSL for C and Zr bound in ZrC have been evaluated by AILD techniques with \FLASSH~\cite{Fleming2023} and constitute a new material evaluation, included in the ENDF/B database for the first time. ZrC has a rock-salt crystal structure. Partial PDOS for Zr(ZrC) and C(ZrC) were generated with density functional theory and lattice dynamics; the calculations are detailed in Ref. \cite{WORMALD2023}. Lattice parameters, phonon dispersion relations and Debye-Waller coefficients are found to be in good agreement with experiment \cite{WORMALD2023}.

TSL File 7 evaluations are available for Zr(ZrC) and C(ZrC) at temperatures of 77, 293.6, 400, 500, 600, 700, 800, 1000, 1200, 1400, 1600, 1800, and 2000~K. This temperature range supports cryogenic through high temperature simulations. Each element was evaluated as the natural isotopic composition recommended by NIST \cite{NIST_AWIC4.1} with total scattering cross sections and atomic weight ratios (AWR) extracted from the ENDF/B-VIII.1 nuclide evaluations; NIST cross sections were used to supplement incoherent elastic \cite{Sears1992}. The \textsuperscript{12}C AWR is set to exactly 12~amu. The mixed elastic scattering format \cite{mixed-elastic,ENDF6-Format-2024} was used to capture both coherent and incoherent contributions to elastic scattering (MT2). Coherent elastic scattering was treated with random alloy theory to account for isotopic variation in neutron scattering lengths of each element \cite{WORMALD2023,SCHWEIKA2007},  although, the natural C scattering length differs slightly from Ref. \cite{WORMALD2023} due to the updated definition of \textsuperscript{12}C mass.  Inelastic scattering (MT4) was evaluated in the incoherent approximation with the phonon expansion method and automatic $\left(\alpha,\beta\right)$ gridding in \FLASSH.

\begin{figure}
    \centering
    \subfigure[~Zr(ZrC).]{\includegraphics[width=0.92\columnwidth]{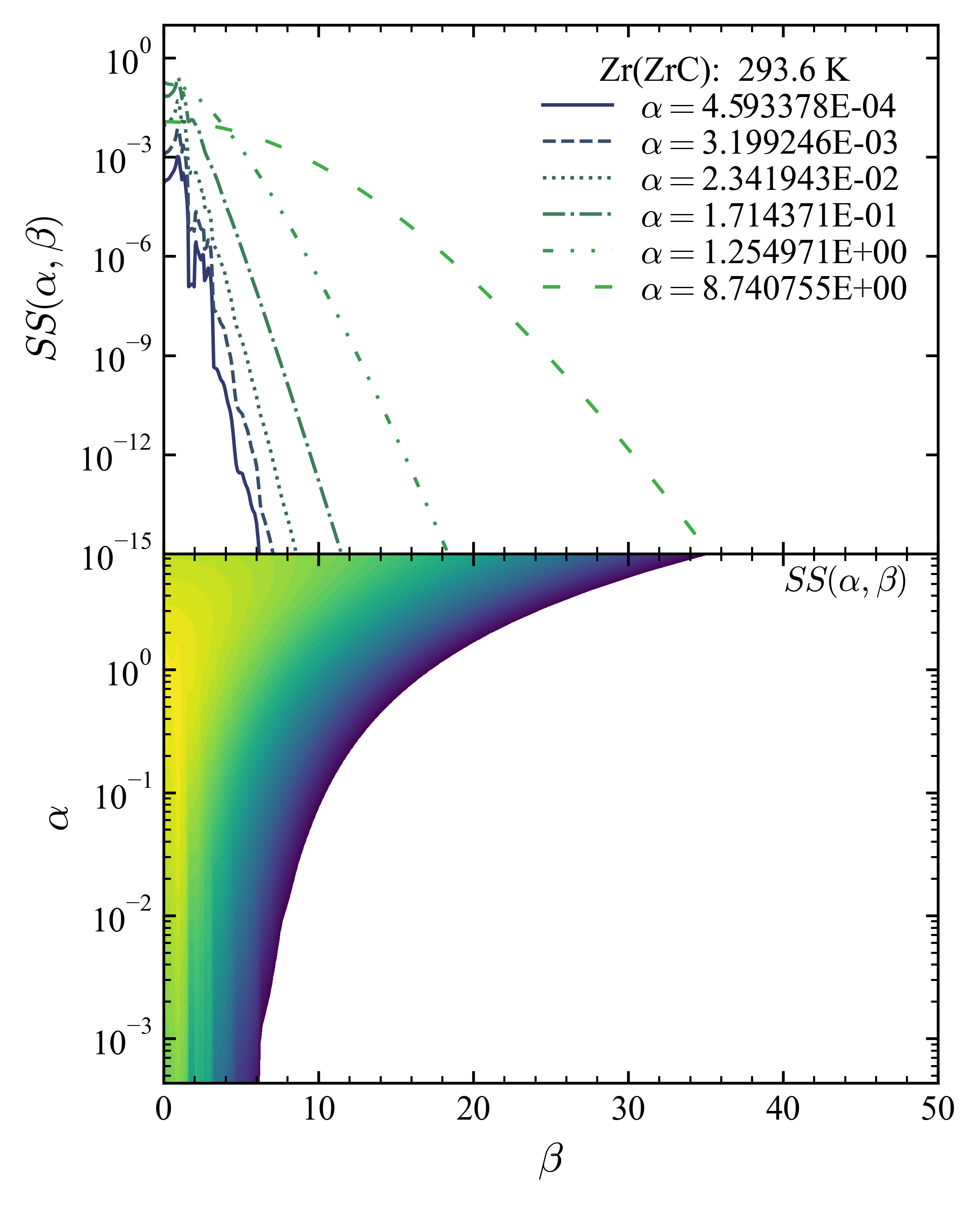}}
        \subfigure[~C(ZrC).]{\includegraphics[width=0.92\columnwidth]{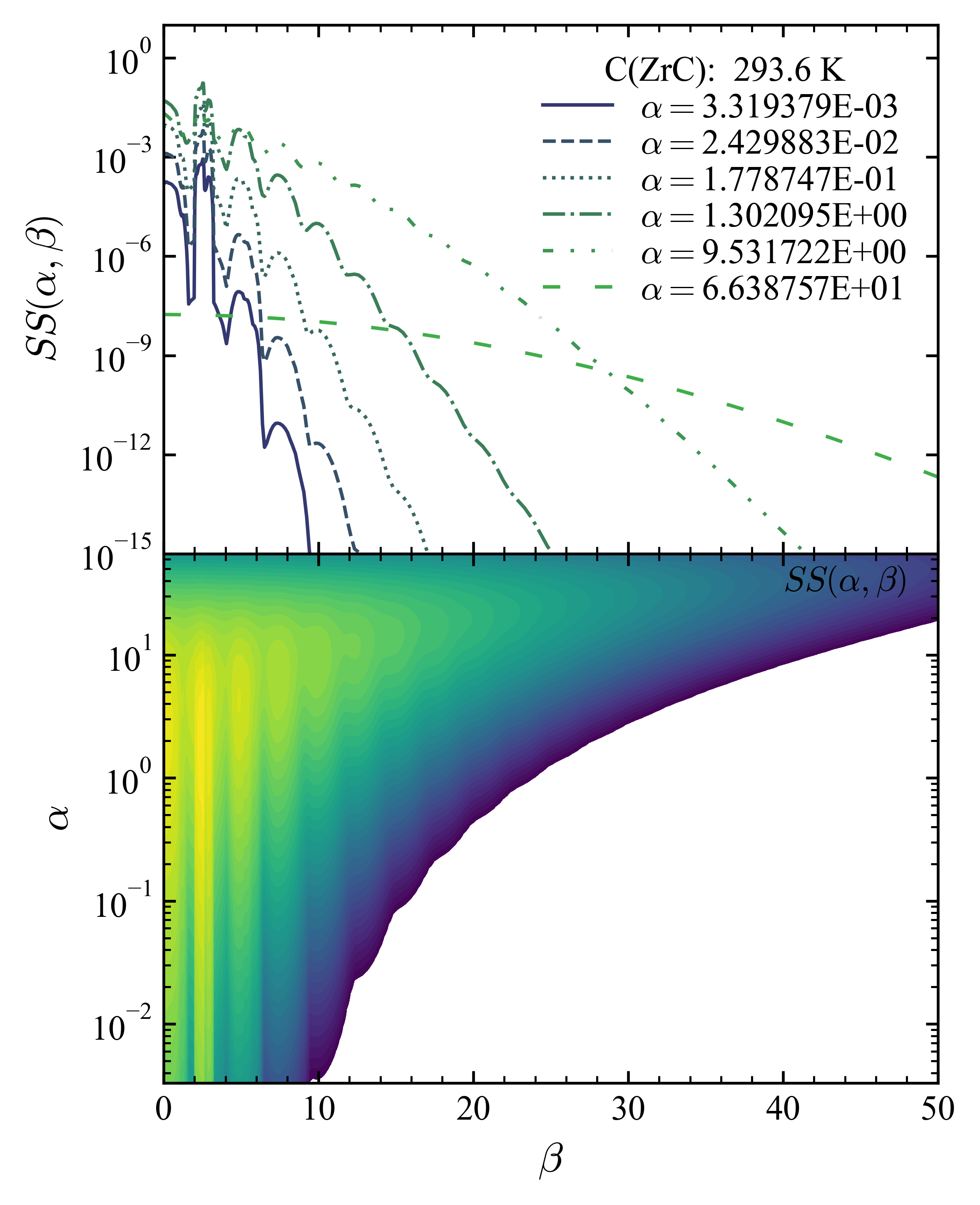}}
    \caption{(Color online) Zr(ZrC) and C(ZrC) symmetric TSL, $SS\left(\alpha,\beta\right)$, at 293.6~K. The $\left(\alpha,\beta\right)$ grid is referenced to ${k_B}T=0.0253$~eV. The contour plot illustrates only $SS\left(\alpha,\beta\right)>10^{-15}$; other non-zero values are not shown. Quantized oscillations in the C(ZrC) occur at integer values of $k_B T\beta\approx0.065$~eV. }
    \label{fig:TSL_ZrC}
\end{figure}

The symmetric TSL at 293.6~K tabulated in File 7, MT4 is illustrated in Fig.~\ref{fig:TSL_ZrC} for Zr(ZrC) and C(ZrC). Quantum oscillations in the C(ZrC) TSL, like those of metal hydrides, occur at fixed $\beta$ intervals corresponding to localization of the partial phonon spectra around 0.065~eV, which is accompanied by a phonon band gap between acoustic and optical phonon modes. Monte Carlo transport simulations indicate that the quantized energy exchange can influence reactivity and shape the thermal neutron spectra from ambient to high temperatures \cite{WORMALD2023}. Further, the gapped, optical phonon spectrum imparts quantum driven neutron thermalization behaviors similar to metal hydrides with analogs to electronic semiconductor physics, as described in Ref. \cite{WORMALD2023}.

\begin{figure}
    \centering
    \includegraphics[width=1.0\columnwidth,clip,trim=  0mm 5mm 0mm 0mm]{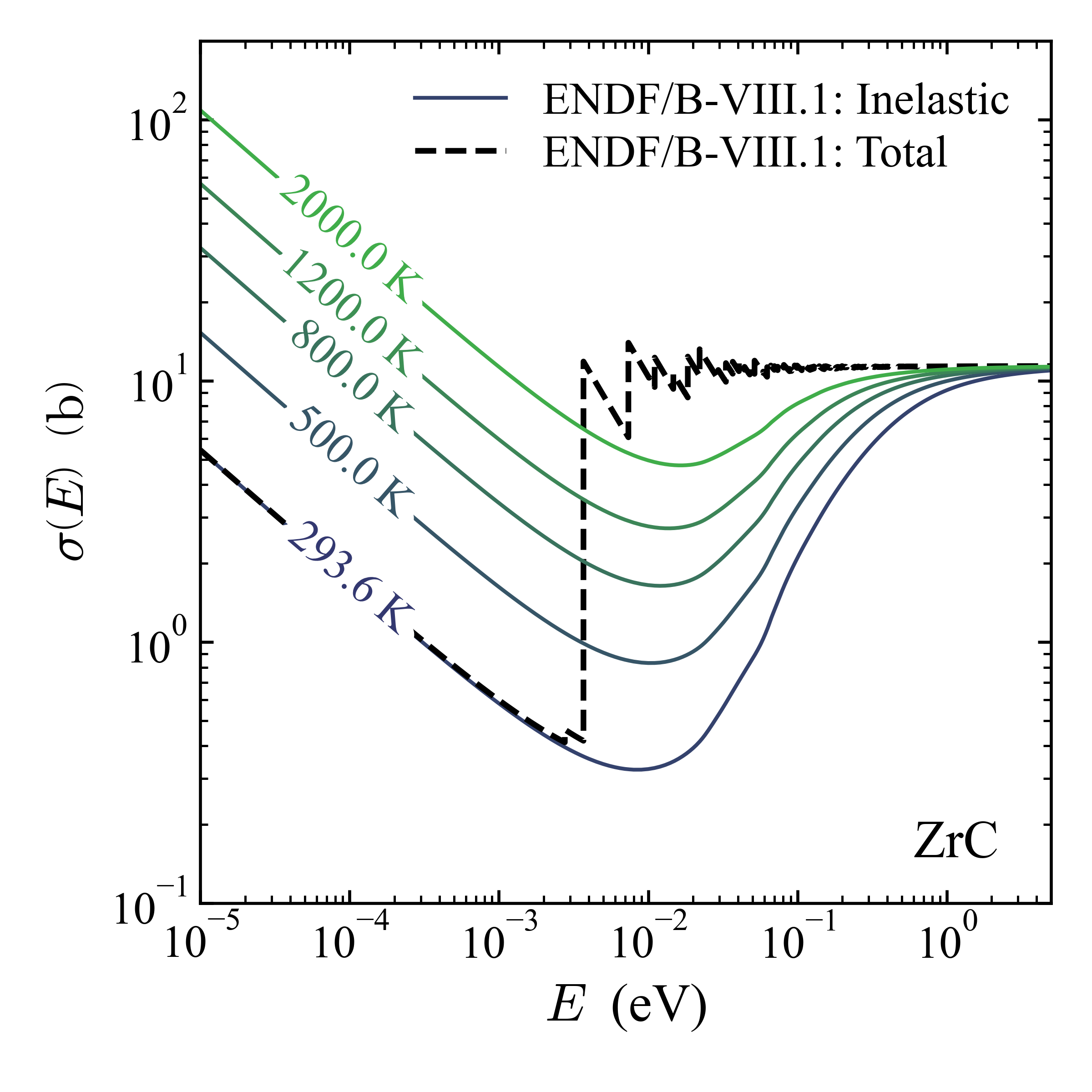}
    \caption{(Color online) Total scattering cross section of ZrC at 293.6~K. The inelastic cross section between 293.6~K and 2000~K is also shown. The abrupt change in slope of the inelastic cross section near 0.065~eV is due to the gapped phonon spectra. }
    \label{fig:xsec_ZrC}
\end{figure}

The ZrC integrated inelastic and total scattering cross sections computed with the NDEX nuclear data processing code \cite{NDEX_TSL1,NDEX_TSL2} are illustrated in Fig.~\ref{fig:xsec_ZrC}. Inelastic scattering for C(ZrC) has an apparent discontinuity in slope near 0.065~ eV due to the localized optical phonon peak.

\subsubsection{Beryllium Carbide (Be\textsubscript{2}C)}
\label{sec:be2c}

TSL evaluations Be and C bound in Be\textsubscript{2}C were generated with \FLASSH~\cite{Fleming2023} and AILD techniques as a new material evaluation to the ENDF/B database. Be\textsubscript{2}C has an anti-fluorite crystal structure. The partial phonon spectra were generated with density functional theory and lattice dynamics. The AILD methodology is similar to Ref. \cite{Hehr2010} and the PDOS demonstrate reasonable agreement with heat capacity measurements \cite{Zerkle2020-csewg}.

TSL File 7 evaluations are available for Be(Be\textsubscript{2}C) and C(Be\textsubscript{2}C) at 293.6, 400, 500, 600, 700, 800, 1000, 1200, 1600, and 2000~K. The material TSL evaluation considers \textsuperscript{9}Be but uses an elemental composition for C from NIST \cite{NIST_AWIC4.1}. For both elements, the total scattering cross sections and AWR were extracted from the ENDF/B-VIII.1 nuclide evaluation. Coherent elastic scattering (MT2) treatment was selected and evaluated using the experimental lattice parameter at room temperature \cite{STARITZKY1956-be2c}. Inelastic scattering was evaluated in the incoherent approximation with the phonon expansion and automatic $\left(\alpha,\beta\right)$ gridding in \FLASSH.

\begin{figure}
    \centering
    \subfigure[~Be(Be$_2$C).]{\includegraphics[width=0.92\columnwidth]{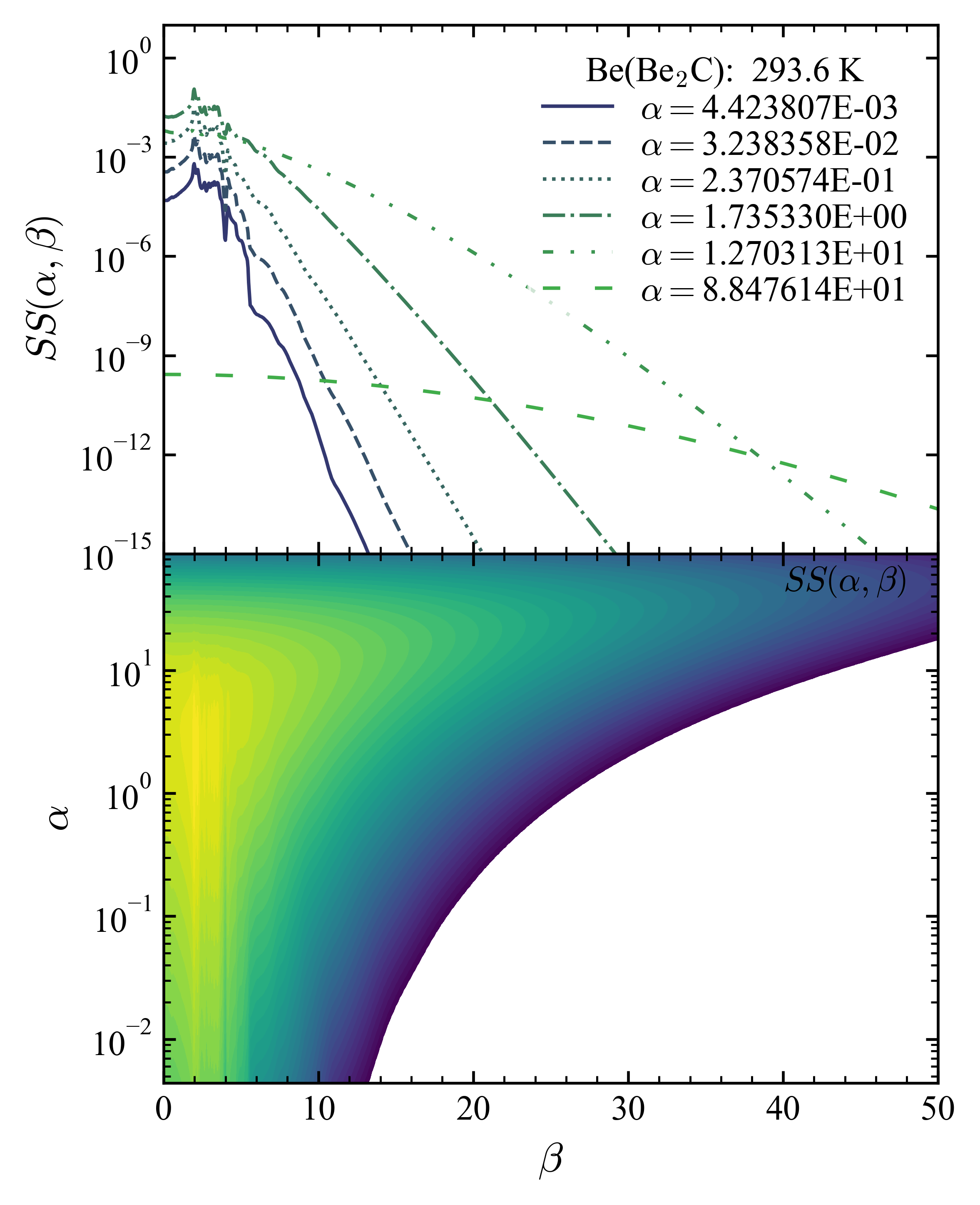}}
    \subfigure[~C(Be$_2$C).]{\includegraphics[width=0.92\columnwidth]{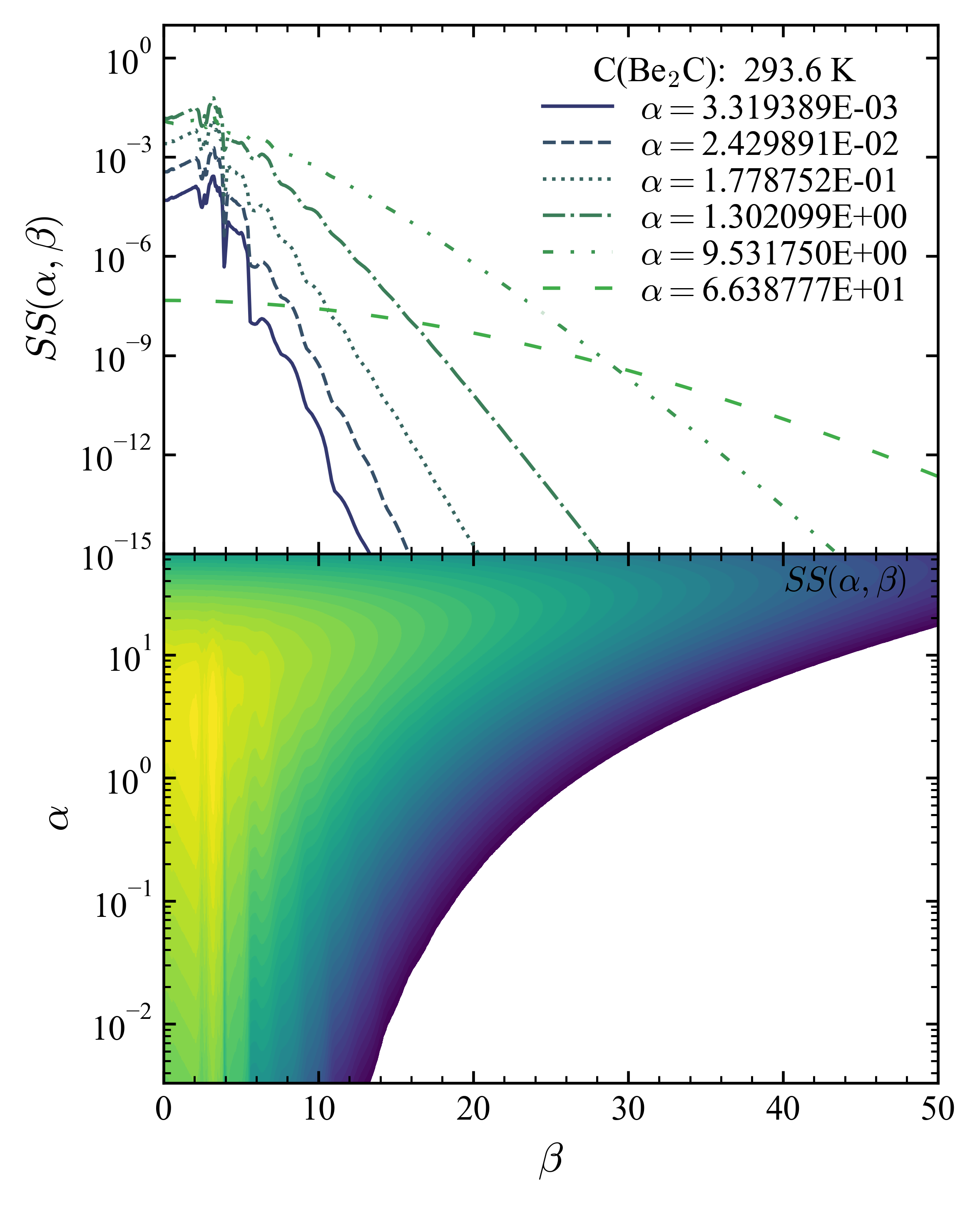}}
    \caption{(Color online) Be(Be\textsubscript{2}C) and C(Be\textsubscript{2}C) symmetric TSL, $SS\left(\alpha,\beta\right)$, at 293.6~K. The $\left(\alpha,\beta\right)$ grid is referenced to ${k_B}T=0.0253$~eV. The contour plot illustrates only $SS\left(\alpha,\beta\right)>10^{-15}$; other non-zero values are not shown.}
    \label{fig:TSL_Be2C}
\end{figure}

\begin{figure}
    \centering
    \includegraphics[width=1.0\columnwidth,clip,trim=  0mm 5mm 0mm 0mm]{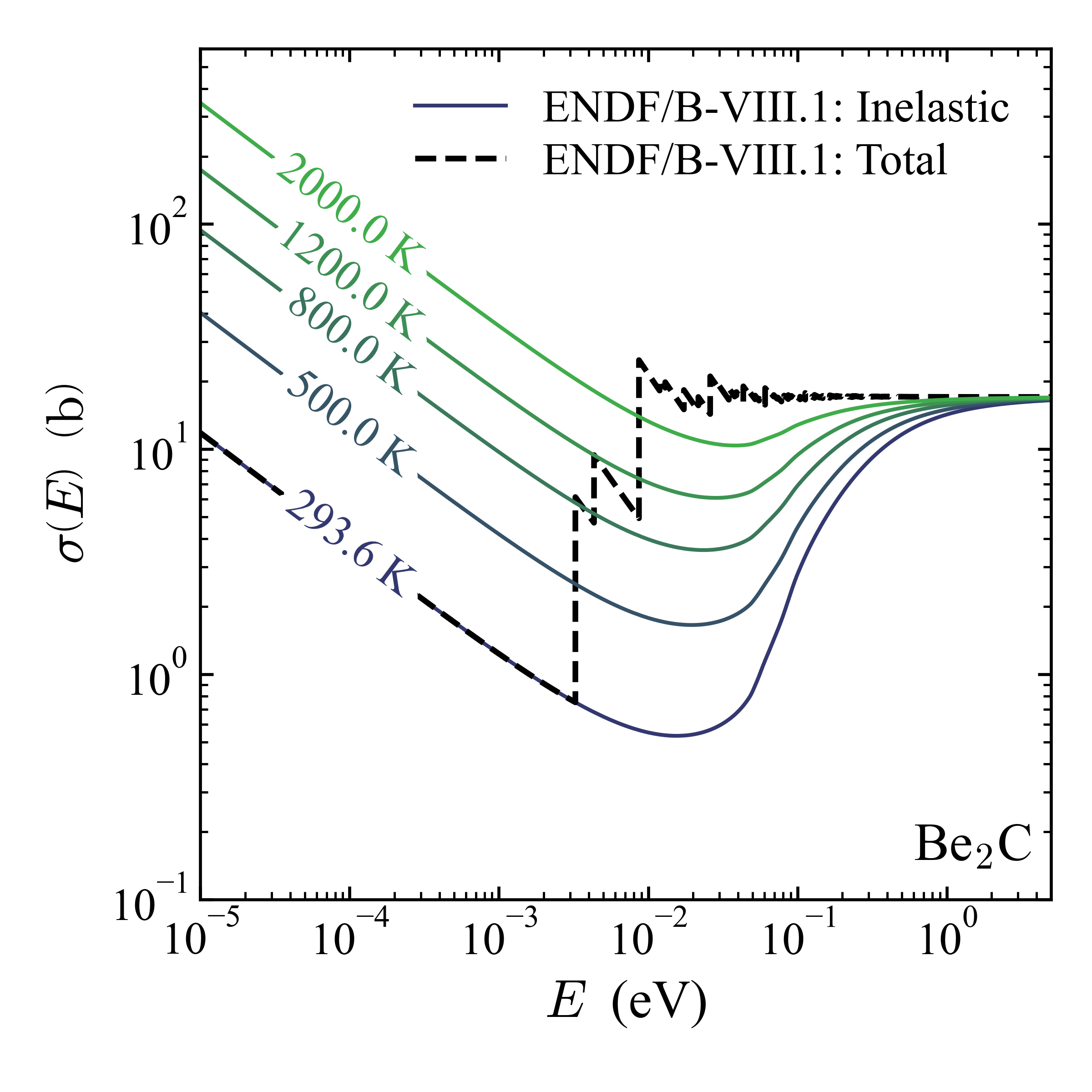}
    \caption{(Color online) Total scattering cross section of Be\textsubscript{2}C at 293.6~K. The inelastic cross section between 293.6~K and 2000~K is also shown. }
    \label{fig:xsec_Be2C}
\end{figure}

Symmetric TSLs at 293.6~K tabulated in File 7, MT4, are illustrated for Be(Be\textsubscript{2}C) and C(Be\textsubscript{2}C) in Fig.~\ref{fig:TSL_Be2C}, respectively. The TSL for both elements lack significant structure as a function of $\beta$ beyond two phonon orders ($\approx$~0.16~eV). The integrated inelastic and total cross sections computed with NDEX nuclear data processing code \cite{NDEX_TSL1,NDEX_TSL2} is illustrated in Fig.~\ref{fig:xsec_Be2C}.

\subsubsection{Zirconium Hydride (ZrH$_x$ and ZrH\textsubscript{2})}
\label{sec:zrhx}

Zirconium hydride exists in two stable phases: the cubic $\delta$-ZrH$_x$ (Fluorite structure) and tetragonal $\epsilon$-ZrH\textsubscript{2} (distorted-Fluorite structure). TSL evaluations were generated with \FLASSH~\cite{Fleming2023} using a combination of AILD and AIMD techniques for H and Zr for each phase \cite{WORMALD2020}. The inclusion of both phases is an expansion of the ENDF database, where up through ENDF/B-VIII.0 a single evaluation was used to approximate both material phases.

The partial phonon spectra for Zr(ZrH$_x$) and Zr(ZrH\textsubscript{2}) were generated with AILD methods, whereas H(ZrH$_x$) and H(ZrH\textsubscript{2}) were generated with AIMD methods to capture anharmonicity \cite{WORMALD2020}. The hydrogen phonon spectra capture experimentally observed differences in phases and have improved agreement with experiment when compared to the General Atomics (GA) spectrum used in prior ENDF evaluations \cite{SLAGGIE1968}. TSL evaluations are likewise demonstrated to be in reasonable agreement with neutron scattering measurements \cite{WORMALD2020}.

TSL File 7 evaluations are available for H(ZrH$_x$), Zr(ZrH$_x$), H(ZrH\textsubscript{2}), and Zr(ZrH\textsubscript{2}) at temperatures of 77, 293.6, 400, 500, 600, 700, 800, 1000, and 1200~K. Hydrogen was evaluated as \textsuperscript{1}H and Zr with the natural isotopic abundances recommended by NIST \cite{NIST_AWIC4.1}. For both elements, the total scattering cross sections and AWR were extracted from the ENDF/B-VIII.1 nuclide evaluations. Coherent elastic scattering (MT2), which was omitted in prior ENDF releases, has been added for the zirconium sublattice of both material phases and is treated with random alloy theory \cite{SCHWEIKA2007} to account for the variation in zirconium nuclide scattering lengths. In this case, the zirconium scattering lengths and cross sections are the same those used in ZrC \cite{WORMALD2023}. For completeness, mixed elastic scattering is used to add incoherent elastic for Zr(ZrH$_x$) and Zr(ZrH\textsubscript{2}). H(ZrH$_x$) and H(ZrH\textsubscript{2}) are evaluated with incoherent elastic scattering only such that the evaluations can be used with variable H stoichiometry defined by the phase diagram (e.g., the Zr-H system proposed by Zuzek~\etal~\cite{Zuzek1990}). Inelastic scattering (MT4) contributions were evaluated in the incoherent approximation with the phonon expansion method and automatic $\left(\alpha,\beta\right)$ gridding in \FLASSH.

\begin{figure}
    \centering
    \subfigure[~H(ZrH$_x$).]{\includegraphics[width=0.92\columnwidth]{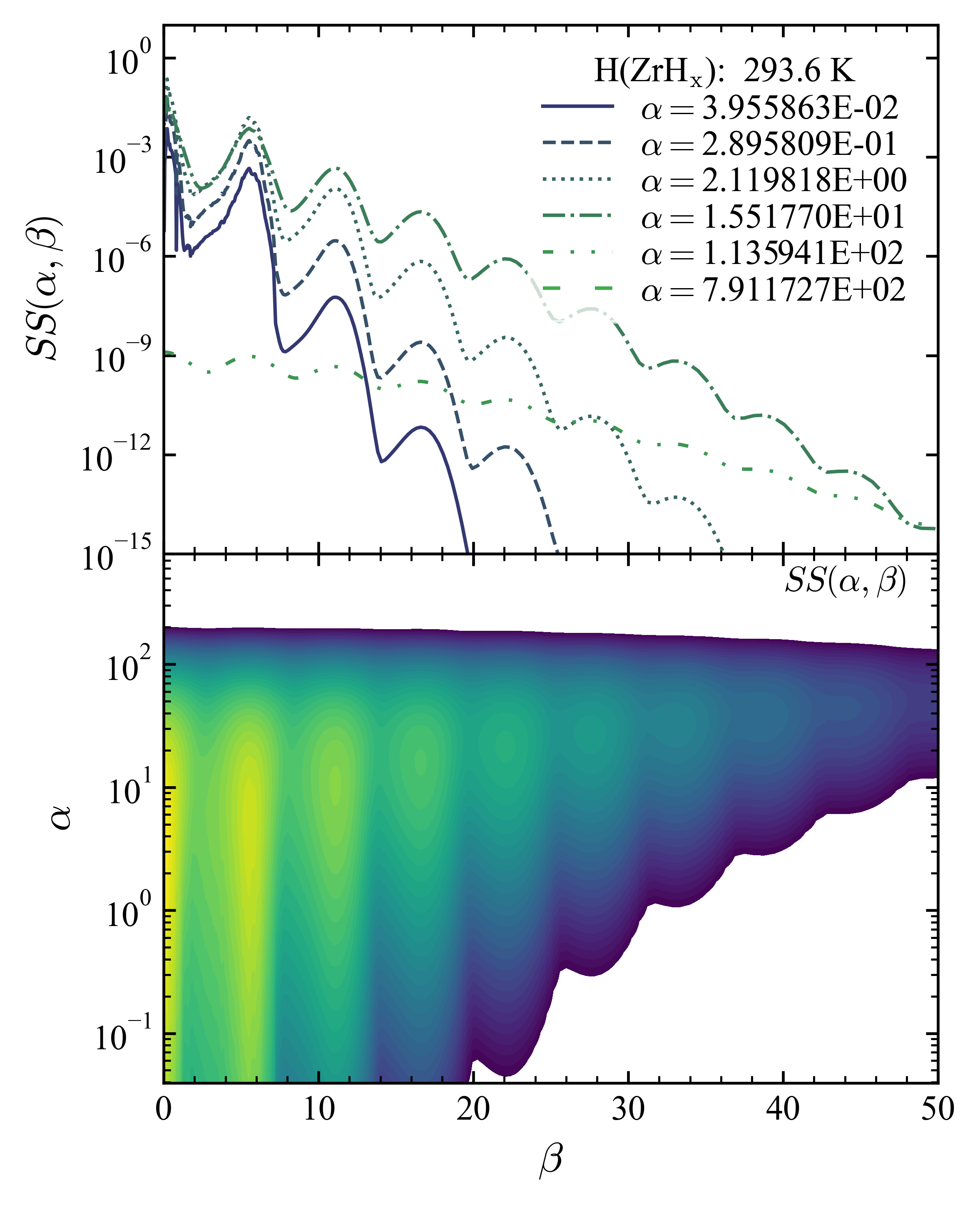}}
    \subfigure[~H(ZrH$_2$).]{\includegraphics[width=0.92\columnwidth]{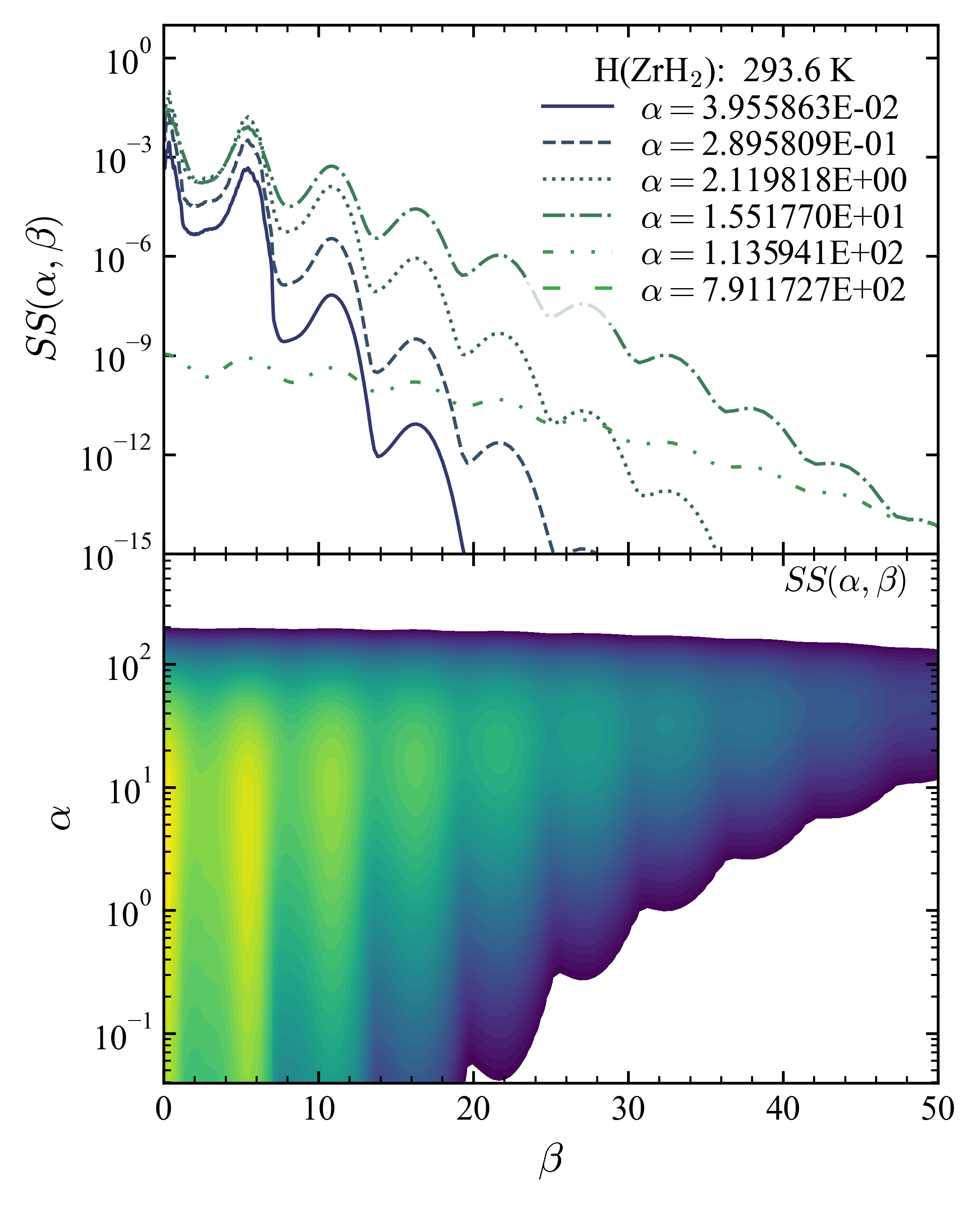}}
    \caption{(Color online) H(ZrH$_x$) and H(ZrH\textsubscript{2}) symmetric TSL, $SS\left(\alpha,\beta\right)$, at 293.6~K. The $\left(\alpha,\beta\right)$ grid is referenced to ${k_B}T=0.0253$~eV. The contour plot illustrates only $SS\left(\alpha,\beta\right)>10^{-15}$; other non-zero values are not shown. Quantized oscillations in each TSL occur at integer values of $k_B T\beta\approx0.13$~eV.}
    \label{fig:TSL_ZrHx}
\end{figure}

\begin{figure}
    \centering
    \subfigure[~H(ZrH$_x$).]{\includegraphics[width=0.98\columnwidth,clip,trim=  0mm 5mm 0mm 0mm]{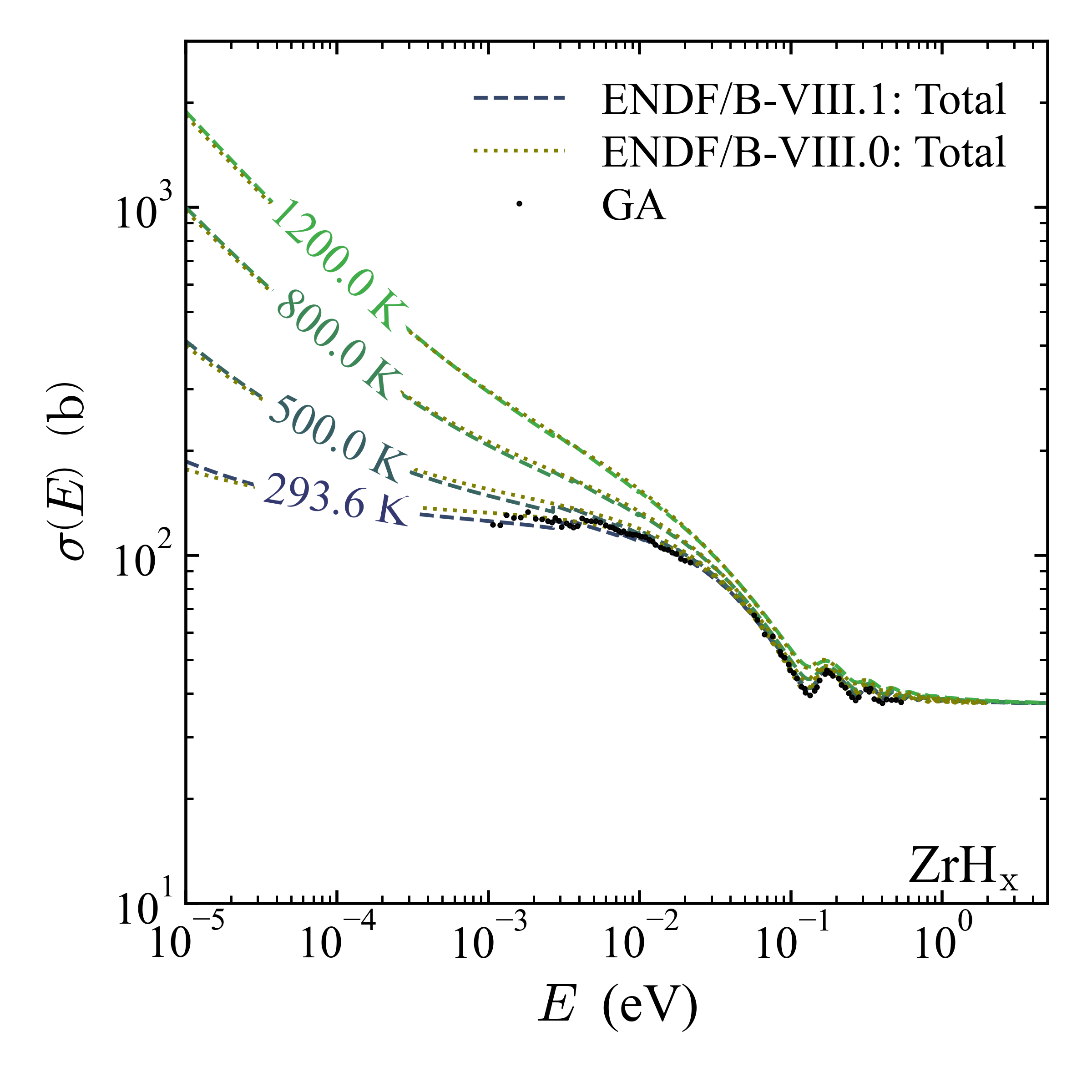}}
    \subfigure[~H(ZrH$_2$).]{\includegraphics[width=0.98\columnwidth,clip,trim=  0mm 5mm 0mm 0mm]{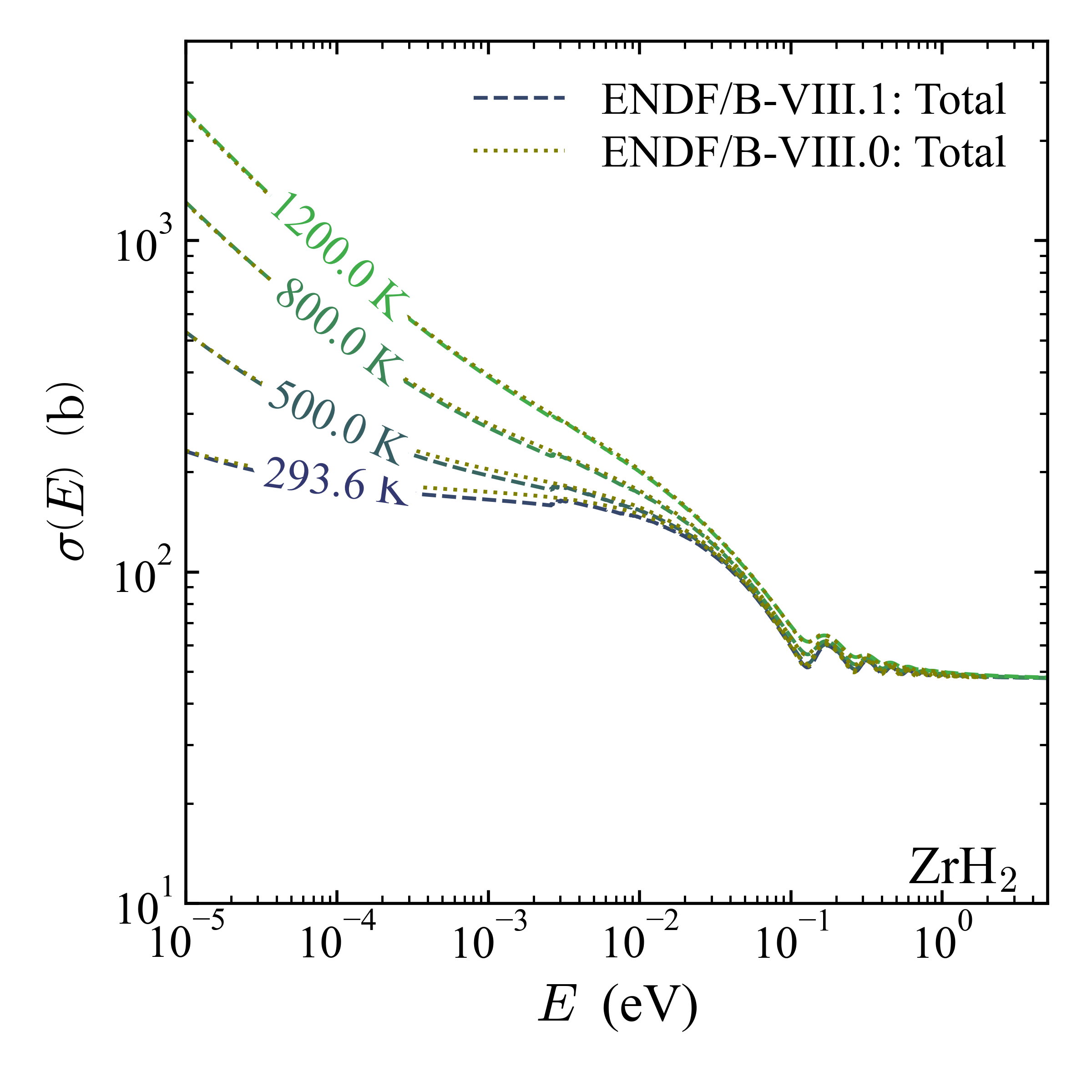}}
    \caption{(Color online) Total scattering cross section of ZrH$_x$ ($x=1.5$) and ZrH\textsubscript{2} ($x=2$) between 293.6--1200~K as compared to the ENDF/B-VIII.0 ZrH material evaluation. Oscillations in the cross section correspond to the H oscillator energy of  $\approx0.13$~eV. Neutron transmission measurements from GA \cite{GA4490} at room temperature are also compared to the ZrH$_x$ material evaluation. GA cross sections have been adjusted by the ENDF/B-VIII.1 elemental zirconium free atom scattering cross section above 0.004~eV per Ref. \cite{GA4490}. The ENDF/B-VIII.1 ZrH$_x$ material evaluation captures the Bragg edges neglected in the ENDF/B-VIII.0 ZrH material evaluation.}
    \label{fig:xsec_ZrHx}
\end{figure}

Symmetric TSLs at 293.6~K tabulated in File 7, MT4, are illustrated in Fig.~\ref{fig:TSL_ZrHx} for H(ZrH$_x$) and H(ZrH\textsubscript{2}). Quantized oscillations in each hydrogen TSL are present at fixed intervals of $k_B T\beta\approx0.13$~eV. The integrated inelastic and total cross sections for ZrH$_x$ and ZrH\textsubscript{2} computed with NDEX nuclear data processing code \cite{NDEX_TSL1,NDEX_TSL2} are illustrated in Fig.~\ref{fig:xsec_ZrHx}. Notably, the adaptive energy meshing in NDEX is necessary to fully capture the oscillator behavior in the cross section. The ENDF/B-VIII.1 evaluations are found to capture the Bragg edge features observed in neutron transmission measurements \cite{GA4490} that are neglected from the ENDF/B-VIII.0 ZrH material evaluation; however, oscillations in the inelastic cross section at higher energies are similar between ENDF/B-VIII.0 and ENDF/B-VIII.1 evaluations.

Initial testing of ZrH$_x$ in the ICT-003 ICSBEP model of the Jo\v zef Stefan Institute (JSI) TRIGA \cite{ICSBEP} using NDEX/MC21 \cite{MC21v10} indicate a 300~pcm bias in $k_\mathrm{eff}$ between ZrH$_x$ and the historic ZrH, which is well within the 1120~pcm benchmark uncertainty that results from fuel enrichment and hydrogen content. Additional TSL testing at JSI indicates a small variation in $k_\mathrm{eff}$ ($<100$~pcm) between the ZrH$_x$ evaluation and the historic ZrH evaluations \cite{Svajger2023}. A variation in $ k_\mathrm{eff}$ between ZrH$_x$, ZrH\textsubscript{2}, and ZrH evaluations of near or less than 150~pcm is observed more generally \cite{Svajger2023}. These findings are supported by critical mass calculations that illustrate small variations in critical mass for \textsuperscript{235}U loadings historically used in U-ZrH$_x$ fuel systems \cite{WORMALD2022}.

\subsubsection{Yttrium Hydride (YH\textsubscript{2})}
\label{sec:yh2}

TSL evaluations H and Y bound in YH\textsubscript{2} were generated with \FLASSH~\cite{Fleming2023} and AILD techniques as reevaluations of the ENDF/B-VIII.0 TSL File 7. Yttrium hydride (YH\textsubscript{2}) has a Fluorite crystal structure. Partial phonon spectra were generated from density functional theory and lattice dynamics, described in Refs. \cite{ZERKLE2017-EPJ} and \cite{ZERKLE2021-EPJ},  and are the same as those used for the ENDF/B-VIII.0 evaluations. The H(YH\textsubscript{2}) spectrum has been demonstrated to be in good agreement with the experiment documented in Ref.~\cite{ZERKLE2021-EPJ}.

TSL File 7 evaluations are available for H(YH\textsubscript{2}) and Y(YH\textsubscript{2}) at 293.6, 400, 500, 600, 700, 800, 1000, 1200, and 1600~K. The material TSL evaluation considers only \textsuperscript{1}H and \textsuperscript{89}Y. Both elements in the total scattering cross section and AWR were validated or updated to be consistent with the ENDF/B-VIII.1 nuclide evaluations. Coherent elastic scattering (MT2) for the yttrium sublattice, omitted in ENDF/B-VIII.0, has been added to the Y(YH\textsubscript{2}) evaluation using the experimental lattice parameter \cite{ZERKLE2021-EPJ}. The restriction of coherent elastic scattering to yttrium permits the present evaluation to be used with variable H stoichiometry defined by the phase diagram (e.g., Y-H system proposed by Fu \etal~\cite{FU2018}). Inelastic scattering (MT4) in all cases was reevaluated in the incoherent approximation with the phonon expansion and the automatic \FLASSH\ $\left(\alpha,\beta\right)$ gridding.

\begin{figure}
    \centering
    \includegraphics[width=\columnwidth]{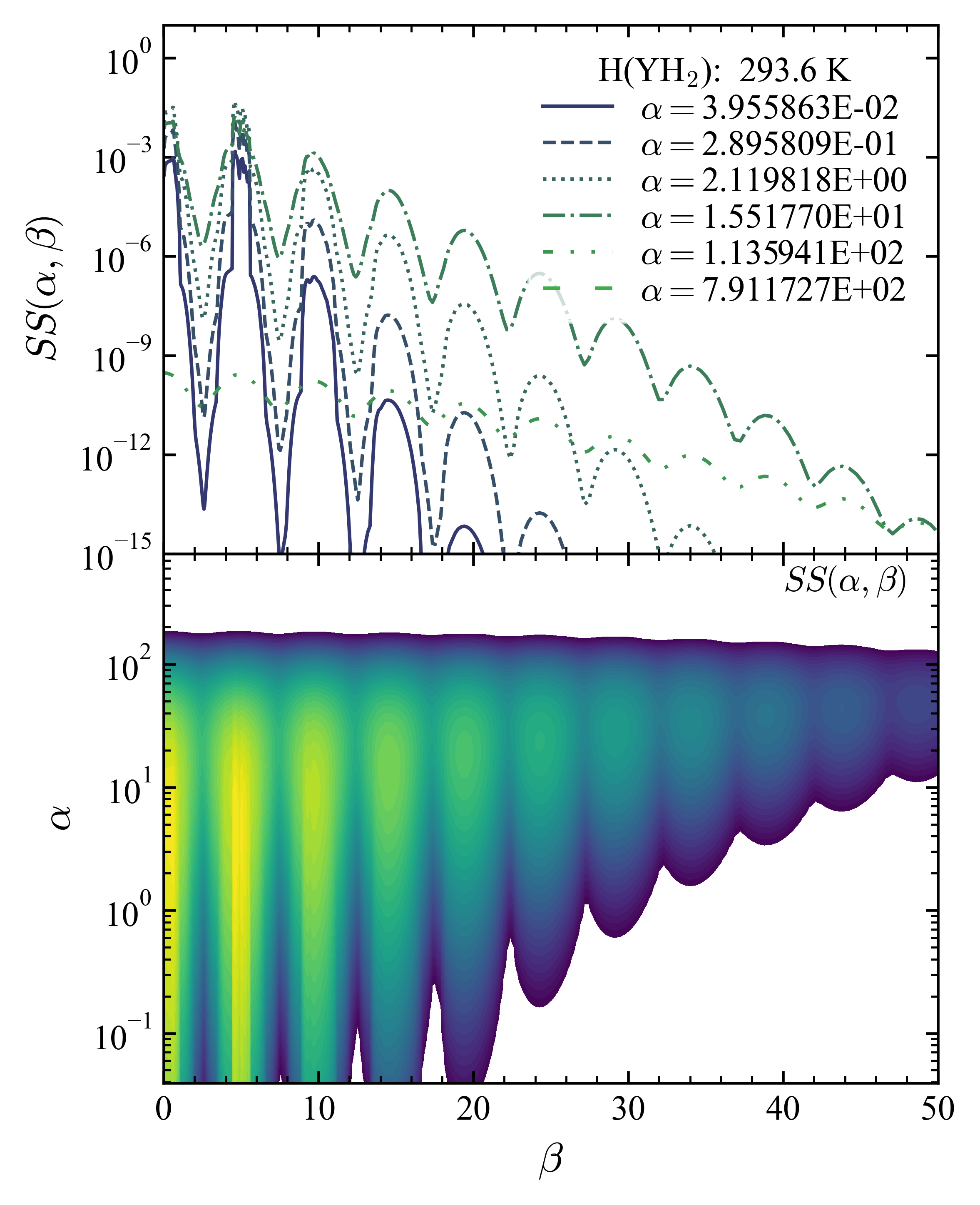}
    \caption{(Color online) H(YH\textsubscript{2}) symmetric TSL, $SS\left(\alpha,\beta\right)$, at 293.6~K. The $\left(\alpha,\beta\right)$ grid is referenced to ${k_B}T=0.0253$~eV. The contour plot illustrates only $SS\left(\alpha,\beta\right)>10^{-15}$; other non-zero values are not shown. Quantized oscillations in the TSL occur at integer values of $k_B T\beta\approx0.125$~eV.}
    \label{fig:TSL_YH2}
\end{figure}

\begin{figure}
    \centering
    \includegraphics[width=1.0\columnwidth,clip,trim=  0mm 5mm 0mm 0mm]{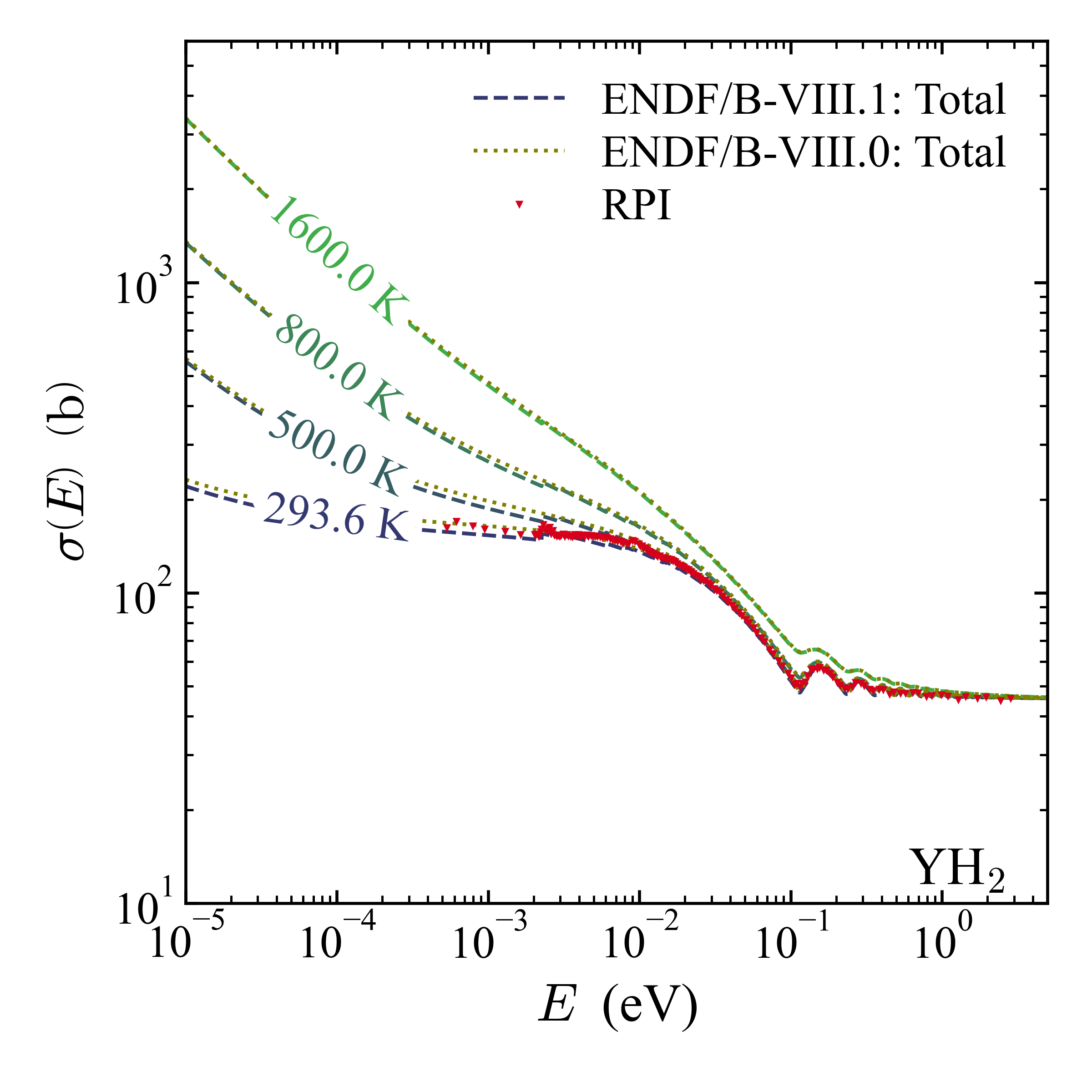}
    \caption{(Color online) Total scattering cross section of YH\textsubscript{2} ($x=1.85$) between 293.6--1200~K as compared to the ENDF/B-VIII.0 YH\textsubscript{2} material evaluation. Oscillations in the cross section correspond to the H oscillator energy of  $\approx0.125$~eV. Room temperature neutron transmission measurements at the RPI LINAC \cite{FRITZ2023109475} are also shown. LINAC cross sections have been adjusted for neutron capture using the ENDF/B-VIII.1 nuclide evaluations. The ENDF/B-VIII.1 YH\textsubscript{2} material evaluation captures the Bragg edges neglected in the ENDF/B-VIII.0 YH\textsubscript{2} material evaluation.}
    \label{fig:xsec_YH2}
\end{figure}

The symmetric TSL for H(YH\textsubscript{2}) at 293.6~K tabulated in File 7, MT4, is illustrated in Fig.~\ref{fig:TSL_YH2}. Quantized oscillations are present at fixed intervals of $k_B T\beta\approx0.125$~eV. The integrated total cross section for YH\textsubscript{2} computed with NDEX \cite{NDEX_TSL1,NDEX_TSL2} is illustrated in Fig.~\ref{fig:xsec_YH2} and compared to neutron transmission measurements at the RPI LINAC facility \cite{FRITZ2023109475} as well as the ENDF/B-VIII.0 YH\textsubscript{2} material evalution. The adaptive energy meshing in NDEX fully captures the oscillator behavior in the cross section. The revised ENDF/B-VIII.1 values are found to have good agreement with experiment and appropriately reproduce the Bragg edges. Oscillatory behavior above 0.1~eV due to inelastic contributions from H(YH\textsubscript{2}) are also consistent with experimental trends.

\subsubsection{Lithium-7 Hydride (\textsuperscript{7}LiH) and Deuteride (\textsuperscript{7}LiD)}
\label{sec:lih}

TSL evaluations for H and \textsuperscript{7}Li bound in lithium-7-enriched LiH as well as \textsuperscript{7}Li and D in lithium-7-enriched LiD have been generated with \FLASSH~\cite{Fleming2023} and AILD methods as new material evaluations in the ENDF/B database. LiH and LiD have a rock-salt crystal structure. The partial phonon spectra were generated in each case with density functional theory and lattice dynamics. In the AILD approach, the \ab\  force-field was the same for LiH and LiD; however, the mass of H and D differ in the lattice dynamics calculation.

TSL File 7 evaluations are available for H(\textsuperscript{7}LiH), \textsuperscript{7}Li(\textsuperscript{7}LiH), D(\textsuperscript{7}LiD), and \textsuperscript{7}Li(\textsuperscript{7}LiD) at 293.6, 400, 500, 600, 700 and 800~K. For both elements, the total scattering cross sections and AWR were extracted from the ENDF/B-VIII.1 nuclide evaluation; NIST cross sections were used to supplement incoherent elastic \cite{Sears1992}. The mixed elastic scattering format \cite{mixed-elastic,ENDF6-Format-2024} was used to capture both coherent and incoherent contributions to elastic scattering (MT2). Coherent elastic scattering was evaluated using the experimental lattice parameter at room temperature \cite{STARITZKY1956-lih}. Inelastic scattering (MT4) was evaluated in the incoherent approximation with the phonon expansion and automatic $\left(\alpha,\beta\right)$ gridding in \FLASSH.

\begin{figure}
    \centering
    \subfigure[~H($^{7}$LiH).]{\includegraphics[width=0.92\columnwidth]{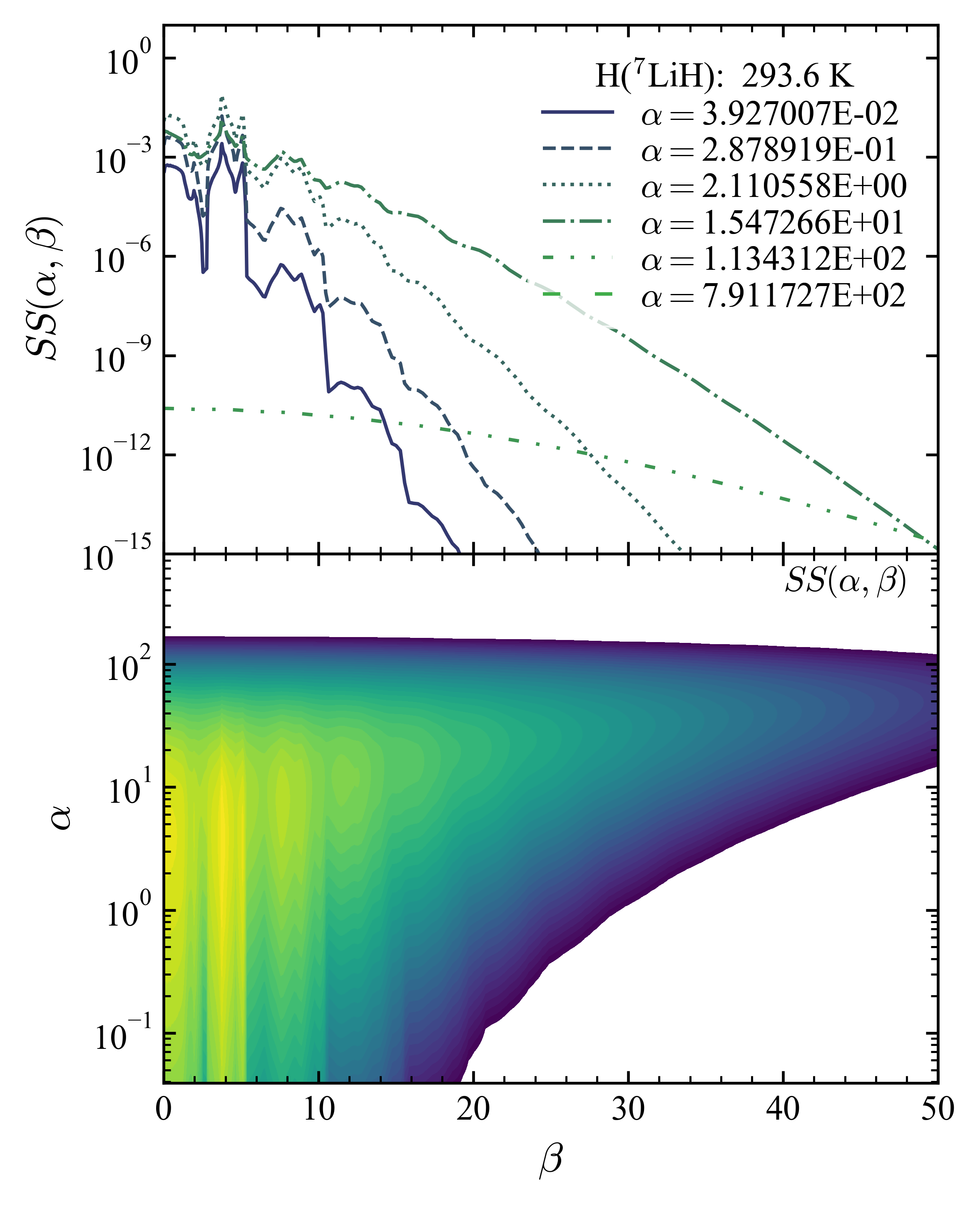}}
    \subfigure[~D($^{7}$LiD).]{\includegraphics[width=0.92\columnwidth]{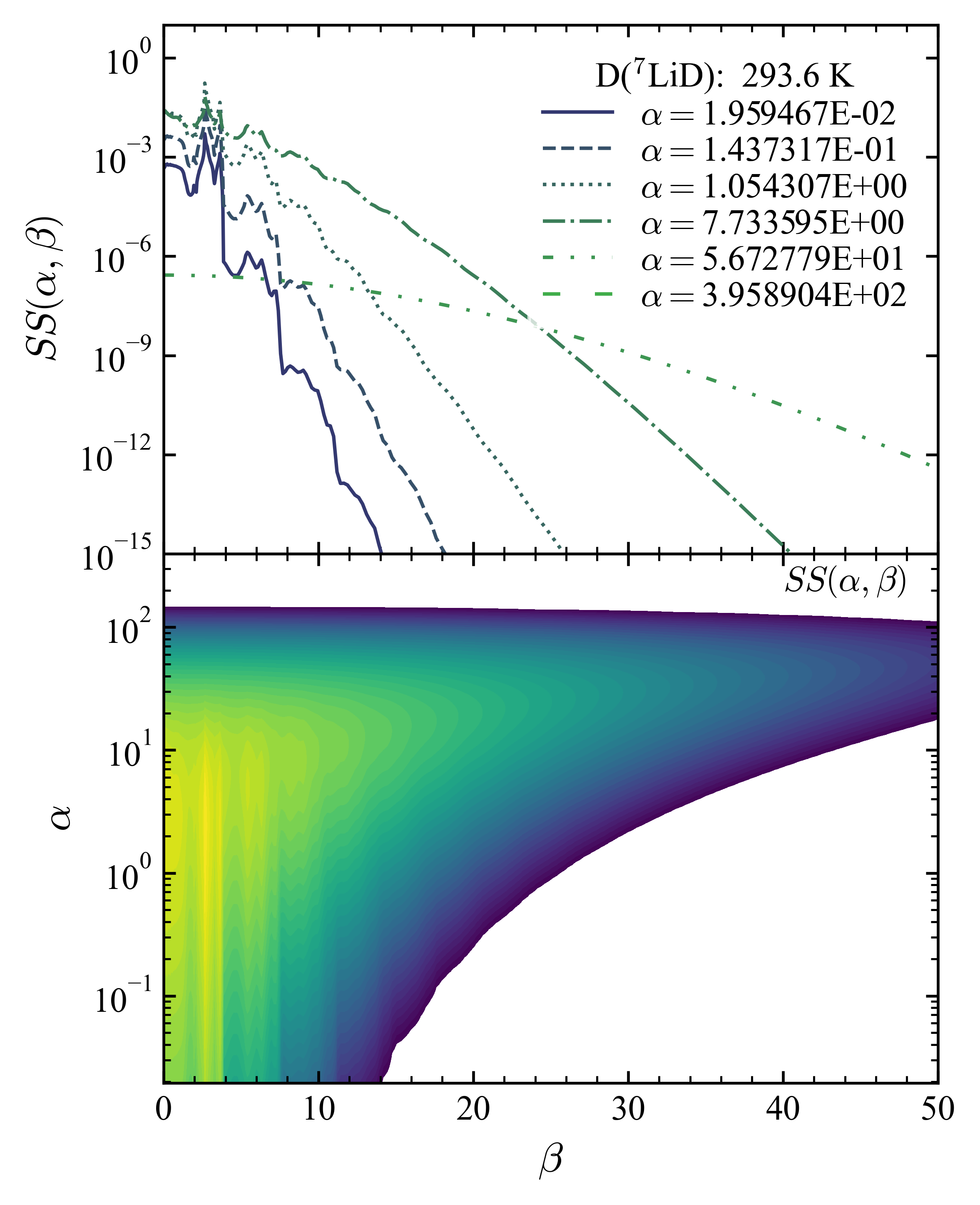}}
    \caption{(Color online) H(\textsuperscript{7}LiH) and D(\textsuperscript{7}LiD) symmetric TSL, $SS\left(\alpha,\beta\right)$, at 293.6~K. The $\left(\alpha,\beta\right)$ grid is referenced to ${k_B}T=0.0253$~eV. The contour plot illustrates only $SS\left(\alpha,\beta\right)>10^{-15}$: other non-zero values are not shown. Quantized oscillations occur at integer values of $k_B T\beta\approx0.1$~eV in H(\textsuperscript{7}LiH) and $k_B T\beta\approx0.07$~eV in D(\textsuperscript{7}LiD).}
    \label{fig:TSL_LiH}
\end{figure}

\begin{figure}
    \centering
    \subfigure[~H($^{7}$LiH).]{\includegraphics[width=0.96\columnwidth,clip,trim=  0mm 5mm 0mm 0mm]{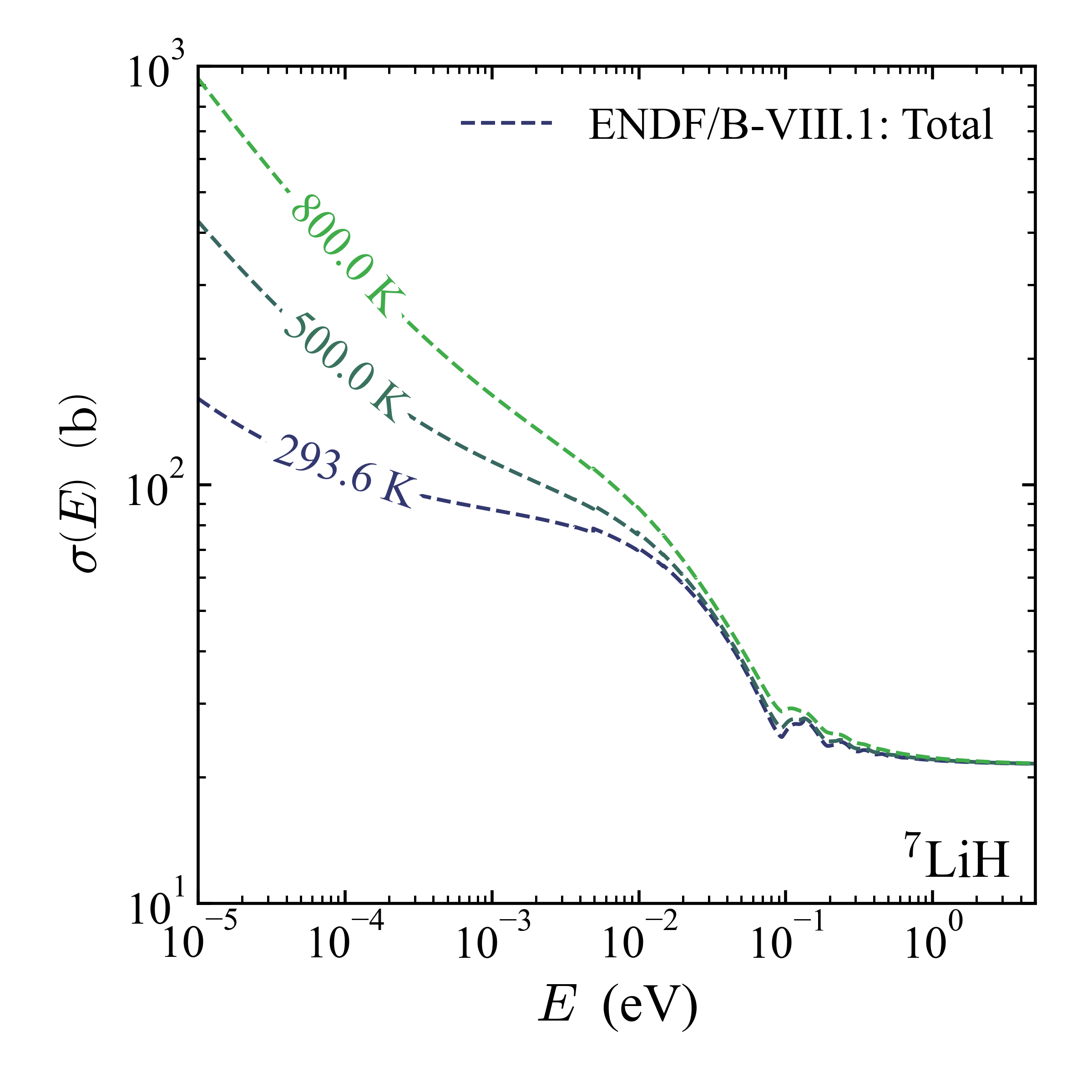}}
    \subfigure[~D($^{7}$LiD).]{\includegraphics[width=0.96\columnwidth,clip,trim=  0mm 5mm 0mm 0mm]{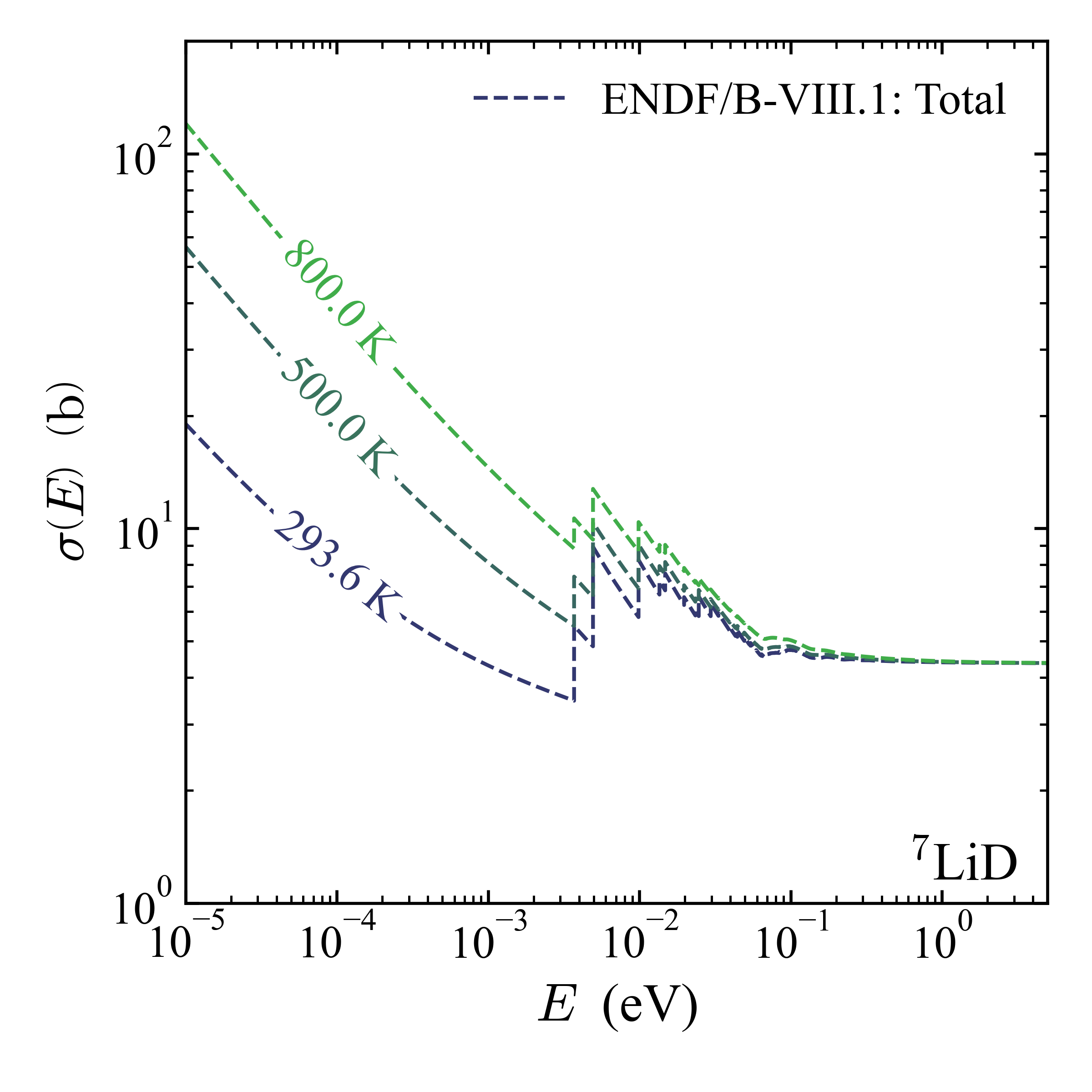}}
    \caption{(Color online) Total scattering cross section of  H(\textsuperscript{7}LiH) and D(\textsuperscript{7}LiD) between 293.6--800~K. Limited oscillations in the cross section H(\textsuperscript{7}LiH) correspond to the H oscillator energy of  $\approx0.1$~eV, whereas similar cross section behavior in D(\textsuperscript{7}LiD) correspond to the D oscillator energy of  $\approx0.07$~eV.}
    \label{fig:xsec_LiH}
\end{figure}

Symmetric TSLs for H(\textsuperscript{7}LiH) and D(\textsuperscript{7}LiD) at 293.6~K tabulated in File 7, MT4, are illustrated in Fig.~\ref{fig:TSL_LiH}. A limited number of quantized oscillations are present at fixed intervals of $k_B T\beta\approx0.1$~eV and $k_B T\beta\approx0.07$~eV for H(\textsuperscript{7}LiH) and D(\textsuperscript{7}LiD), respectively. The integrated scattering cross section for \textsuperscript{7}LiH and \textsuperscript{7}LiD computed with NDEX \cite{NDEX_TSL1,NDEX_TSL2} is illustrated in Fig.~\ref{fig:xsec_LiH}, respectively.

\subsection{Fuels}
\label{subsec:tsl:fuels}

\subsubsection{Plutonium Dioxide (\puo)}
\label{sec:puo}

\puo\ was evaluated using modern AILD techniques \AILD\ for its fluorite-type crystal structure. The minimized electronic structure was predicted with \texttt{VASP} \VASP\ DFT and accounts for spin-orbit coupling, semi-itinerant 5f electrons, Pu-5f / O-2p hybridization and AFM-1K magnetic ordering \cite{Crozier2023}. Hellmann-Feynman forces were used in PHONON \cite{Parlinski1997} to construct and solve the dynamical matrix for sampling partial PDOS of Pu(\puo) and O(\puo). The AILD lattice constant and predicted electronic DOS demonstrate good agreement with experimental data. Furthermore, the Lorentzian broadening of partial DOS using the resolution of the HERIX instrument compares well with inelastic X-ray scattering (IXS) measurements \cite{Manley2012}.

\begin{figure}
    \centering
    \includegraphics[width=1.0\columnwidth,clip,trim=  10mm 1mm 30mm 15mm]{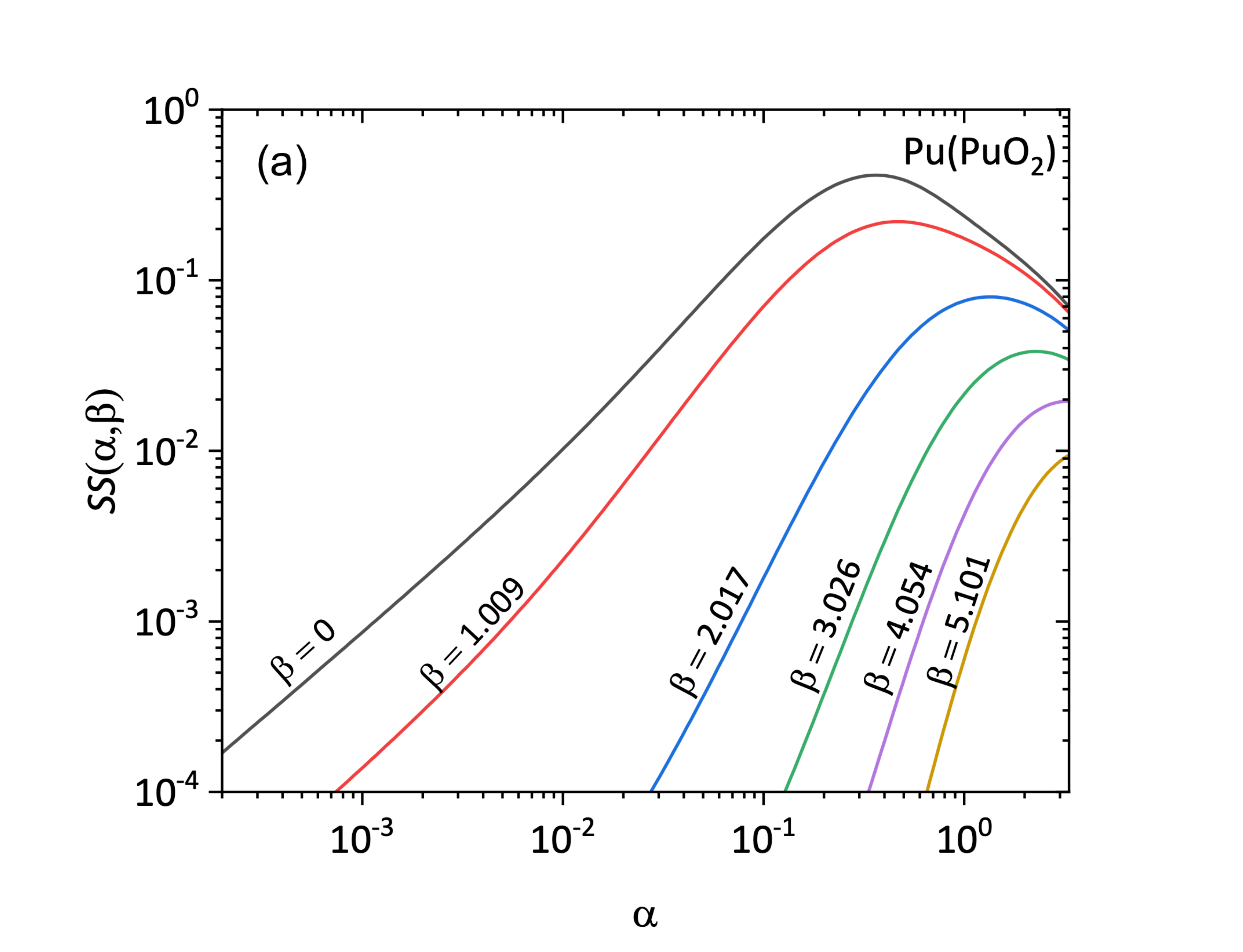}
    \includegraphics[width=1.0\columnwidth,clip,trim=  10mm 1mm 30mm 15mm]{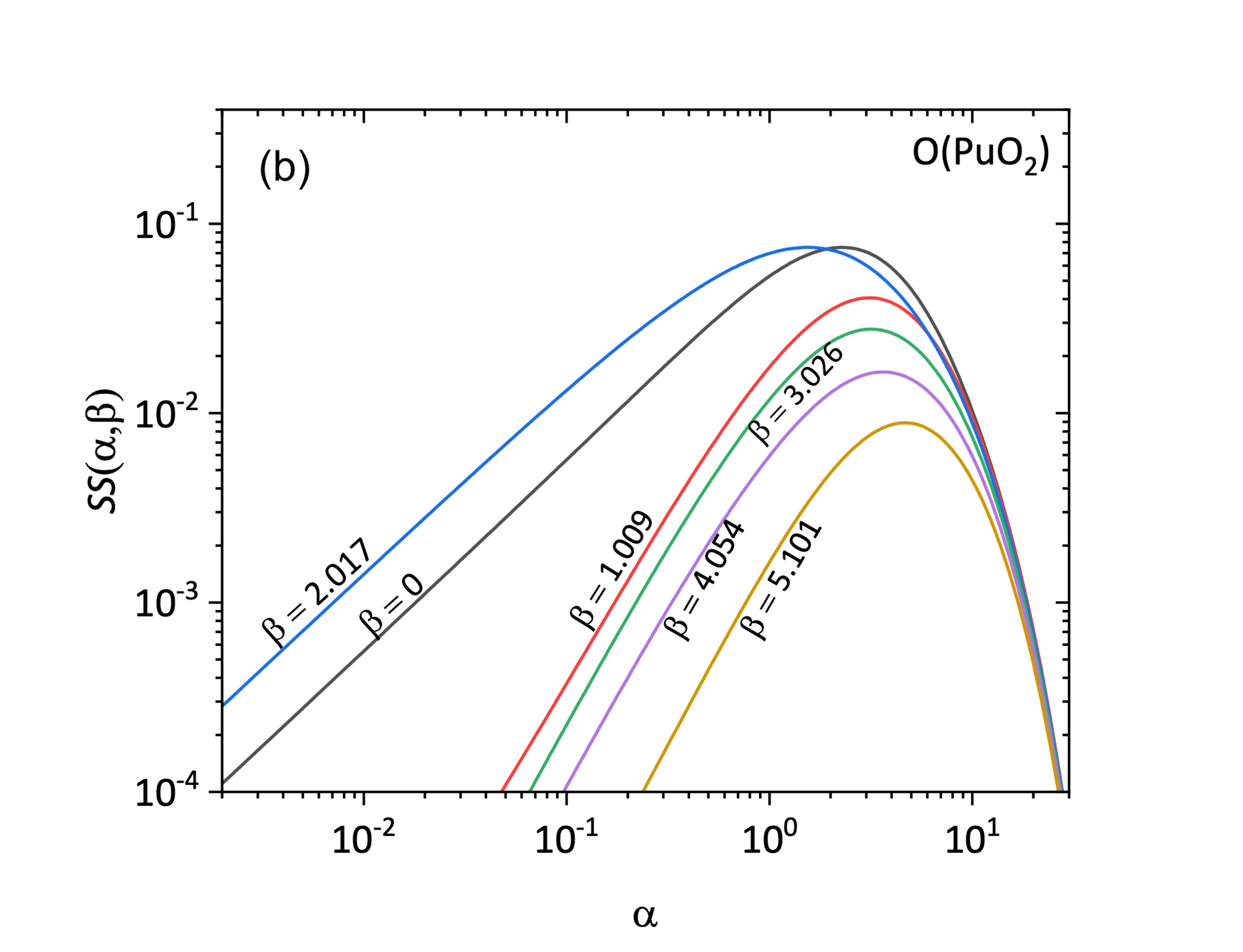}
    \caption{(Color online) The symmetric TSL of (a) Pu(\puo), and (b) O(\puo) as a function of momentum transfer, $\alpha$, for various values of $\beta$ at 296 K. $SS(\alpha,\beta)$ for each $\beta$ is labeled with the corresponding line.}
    \label{fig:TSL_PuO2}
\end{figure}

The TSLs for Pu(\puo) (File 7, MAT7200)  and O(\puo) (File 7, MAT7250) were evaluated in \FLASSH\ \cite{Fleming2023} with mass and free atom cross sections \cite{Brown2018} reflecting 100\% \ninePu\ and a natural abundance \cite{Sears1992} of oxygen at temperatures of 296, 400, 500, 600, 700, 800, 1000, 1200, 1600, 2000, 2400, 2800 K. \Ss\ (MT4) at 296 K for Pu(\puo) and O(\puo) are shown in Fig.~\ref{fig:TSL_PuO2}. 

\begin{figure}
    \centering
    \includegraphics[width=1.0\columnwidth,clip,trim=  20mm 0mm 30mm 15mm]{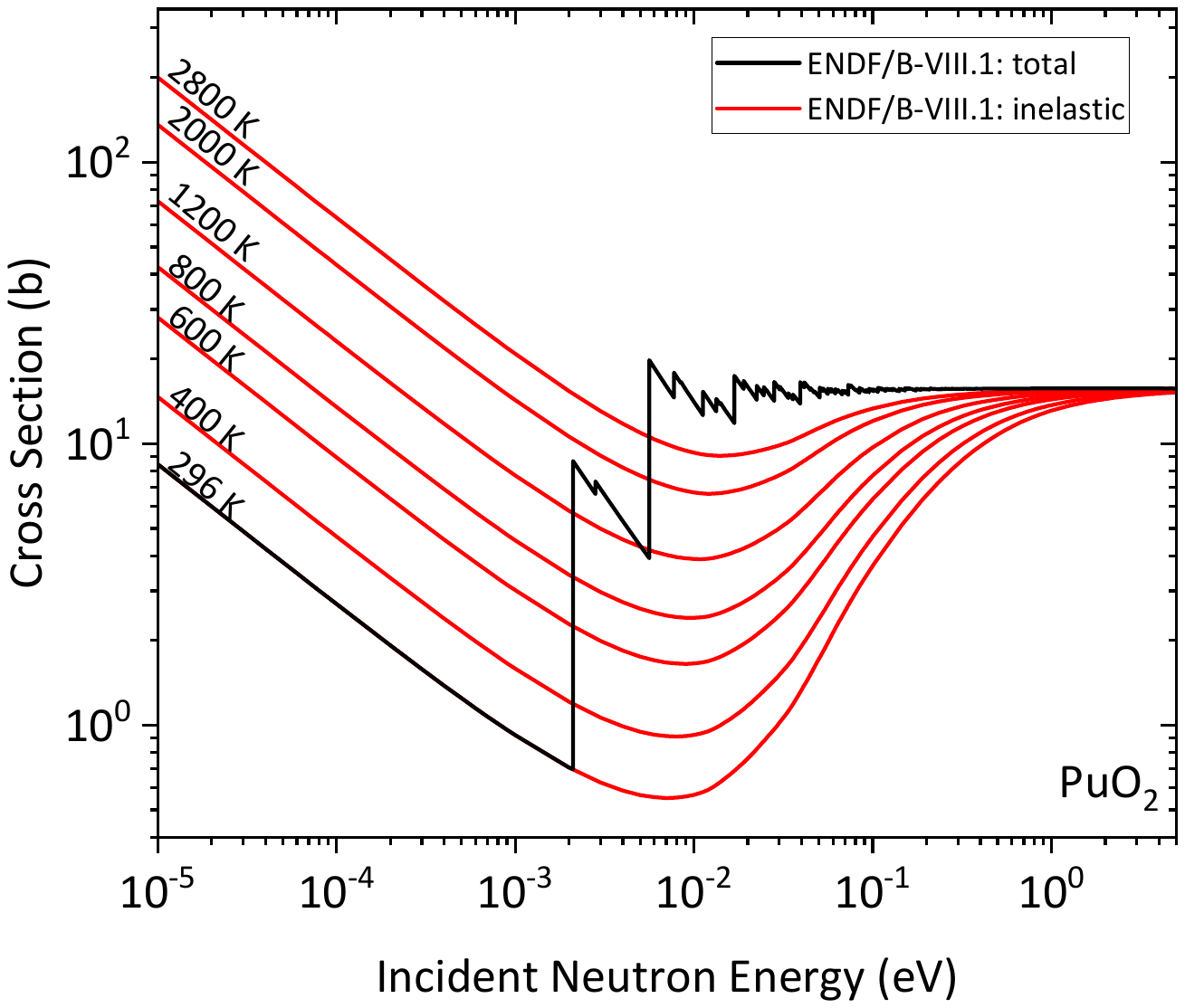}
    \caption{(Color online) The inelastic scattering cross section of \puo\ shown at selected temperatures. The total scattering cross sections of \puo\ is also shown at 296 K.}
    \label{fig:Xsec_puo}
\end{figure}

The \ab\ thermal scattering cross sections, with incoherent inelastic and coherent elastic data calculated in \FLASSH\, are shown in Fig.~\ref{fig:Xsec_puo}. These plutonium dioxide TSL evaluations are novel contributions to the \ENDF\ database. 

\subsubsection{Uranium Carbide (\uc)}
\label{sec:uc}

\uc\ was evaluated using modern AILD techniques \AILD\ for varying  \fiveU\ enrichment of its rock-salt crystal structure \cite{Crozier2022}. The minimized crystalline and electronic structures predicted within the \texttt{VASP} \VASP\ DFT framework account for spin-orbit-coupling, localized 5f interactions and approximate the paramagnetic ground state \cite{Wormald2021}. Averaged Hellmann-Feynman forces for AFM-1K orientations parallel and perpendicular to atomic displacements were used in PHONON \cite{Parlinski1997} to construct the dynamical matrix, relate phonon wave vectors to energy, and sample partial PDOS for U(\uc) and C(\uc). The predicted lattice parameter, electronic DOS, phonon dispersion relations, and total neutron-weighted PDOS demonstrate good agreement with measured data \cite{Crozier2022}. 

\begin{figure}
    \centering
    \includegraphics[width=1.0\columnwidth,clip,trim=  20mm 1mm 30mm 15mm]{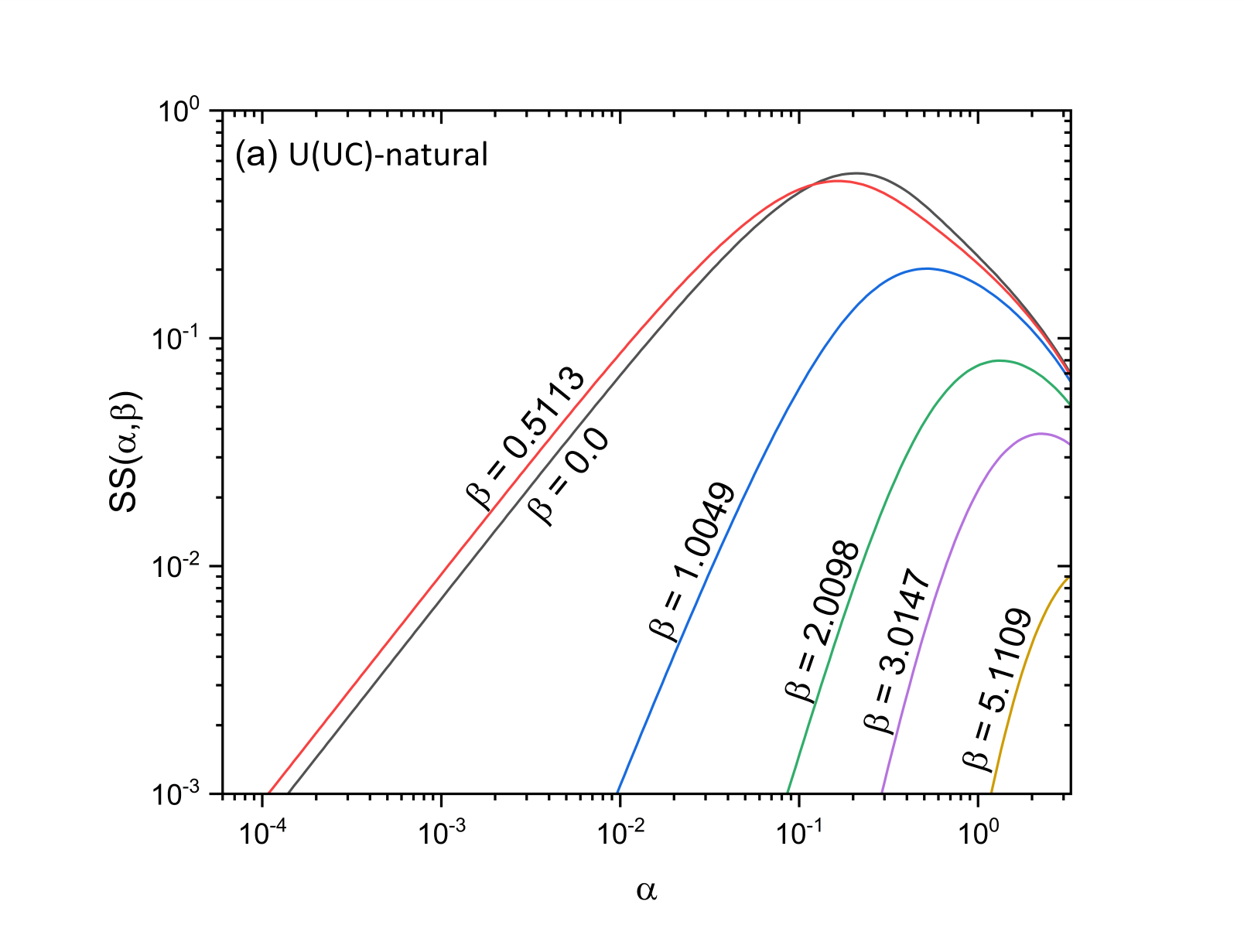}
    \includegraphics[width=1.0\columnwidth,clip,trim=  20mm 1mm 30mm 15mm]{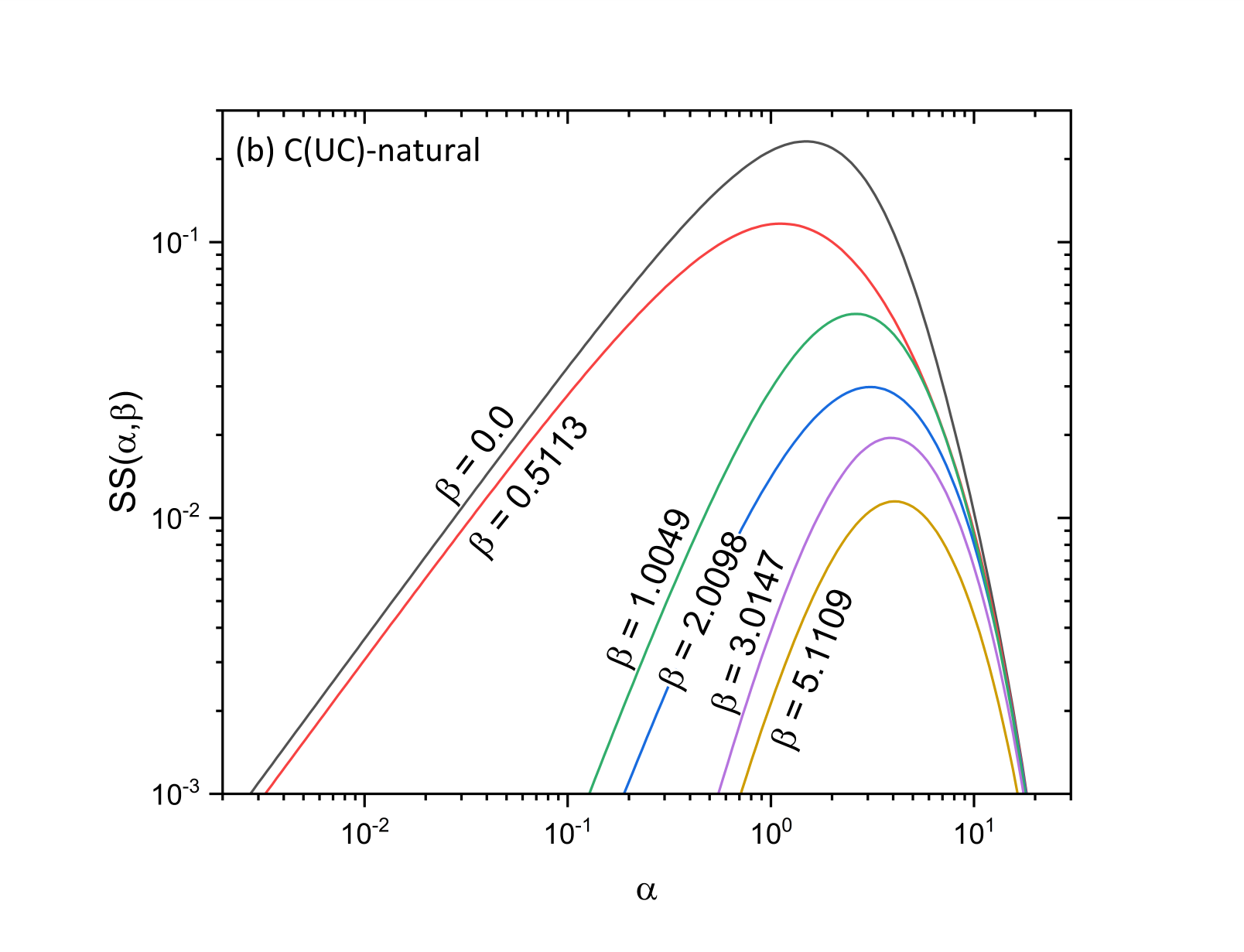}
    \caption{(Color online) The symmetric TSL of natural abundance (a) U(\uc) and (b) C(\uc) as a function of momentum transfer, $\alpha$, for various values of $\beta$ at 293.6 K. $SS(\alpha,\beta)$ for each $\beta$ is labeled with the corresponding line.}
    \label{fig:TSL_UC}
\end{figure}

With partial PDOS as fundamental input, the TSLs for U(\uc) and C(\uc) were evaluated at temperatures of 293.6, 400, 500, 600, 700, 800, 1000, 1200, 1600 and 2000 K. \Ss\ (MT4) at 293.6 K for natural U(\uc) and C(\uc) are shown in Fig.~\ref{fig:TSL_UC}. Enriched TSLs were evaluated using the characteristic AILD partial PDOS detailed above, and account for the impact of varying isotope dependent free-atom cross sections and masses \cite{Brown2018}. These evaluations reflect natural ($\approx$ 0.72\%) \cite{Sears1992}, 5\%, 10\%, 19.75\% (HALEU), 93\% (HEU), and 100\% \fiveU\ enrichments and capture the impact of isotope abundance on the total cross section. Evaluations were completed using the \FLASSH\ code for U(\uc) for naturally enriched, 5\%, 10\%, 19.75\% (HALEU), 93\% (HEU), and 100\%  \fiveU\ (File 7, MAT76, MAT8105, MAT8110, MAT8148, MAT8149, MAT8147, respectively) and for natural \cite{Berglund2011} C(\uc) paired with each \fiveU\ enrichment (File~7, MAT8150, MAT8155, MAT8160, MAT8198, MAT8199, MAT8197, respectively). Two files are required for each enrichment (a uranium and a carbon evaluation) because the tabulated compound coherent elastic will be impacted by the changing inputs associated with the uranium; MT4 for C(\uc) remains the same for each enrichment while MT2 varies. 

\begin{figure}
    \centering
    \includegraphics[width=1.0\columnwidth,clip,trim=  18mm 0mm 30mm 15mm]{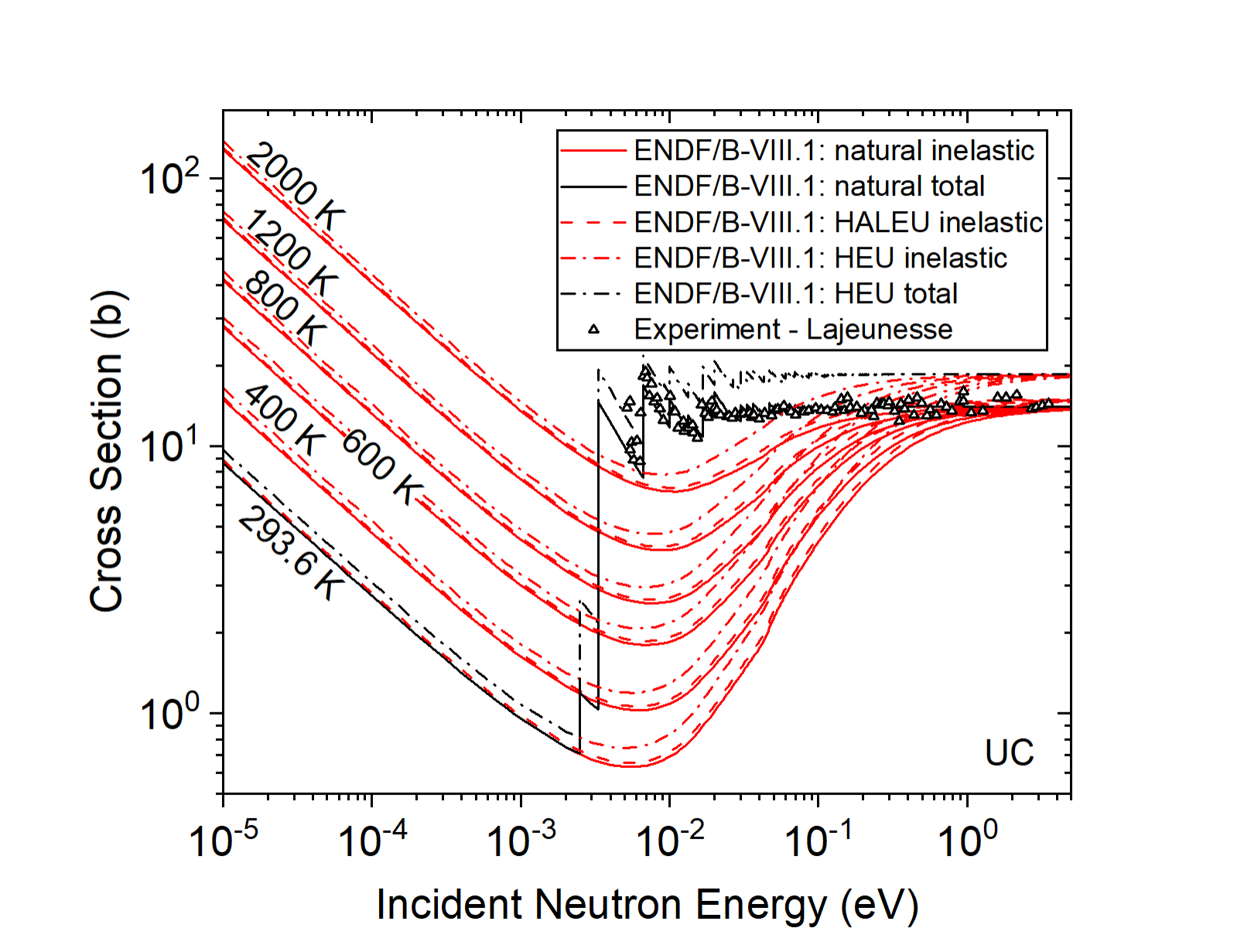}
    \caption{(Color online) The inelastic scattering cross section of \uc\ shown at selected temperatures and enrichments. The natural total scattering cross sections of \uc\ at 293.6 K and measurement \cite{Lajeunesse1972} are also shown.}
    \label{fig:Xsec_UC}
\end{figure}

The integrated total scattering cross sections containing incoherent inelastic (MT4) and coherent elastic (MT2) contributions are shown in Fig.~\ref{fig:Xsec_UC} for varying temperatures and enrichments. When accounting for neutron absorption effects, the \ENDF\ natural \uc\ cross sections demonstrate good agreement with neutron transmission measurements \cite{Lajeunesse1972}. These UC TSL evaluations are novel contributions to the ENDF/B database. 

\subsubsection{Uranium Metal (U-metal)}
\label{sec:um}

The thermal scattering law for U-metal was evaluated using modern molecular dynamics/ lattice dynamics (MDLD) techniques to capture the combined temperature and lattice effects \cite{Fleming2021}. The $\alpha$-phase of U-metal, which exists at room temperature, is a highly temperature-dependent structure and requires the use of MD techniques to capture the resulting temperature effects on the dynamical matrix. The MD model was composed of 8,000 atoms with a temperature of 296 K and zero pressure \cite{Fleming2021}. Perturbing the MD model, the resulting forces were used in the dynamical matrix method in phonopy \cite{Togo2015} to calculate the PDOS. The resulting phonon dispersion relations and DOS were found to be in good agreement with experimental data \cite{Fleming2021}. 

\begin{figure}
    \centering
    \includegraphics[width=1.0\columnwidth,clip,trim=  20mm 10mm 30mm 15mm]{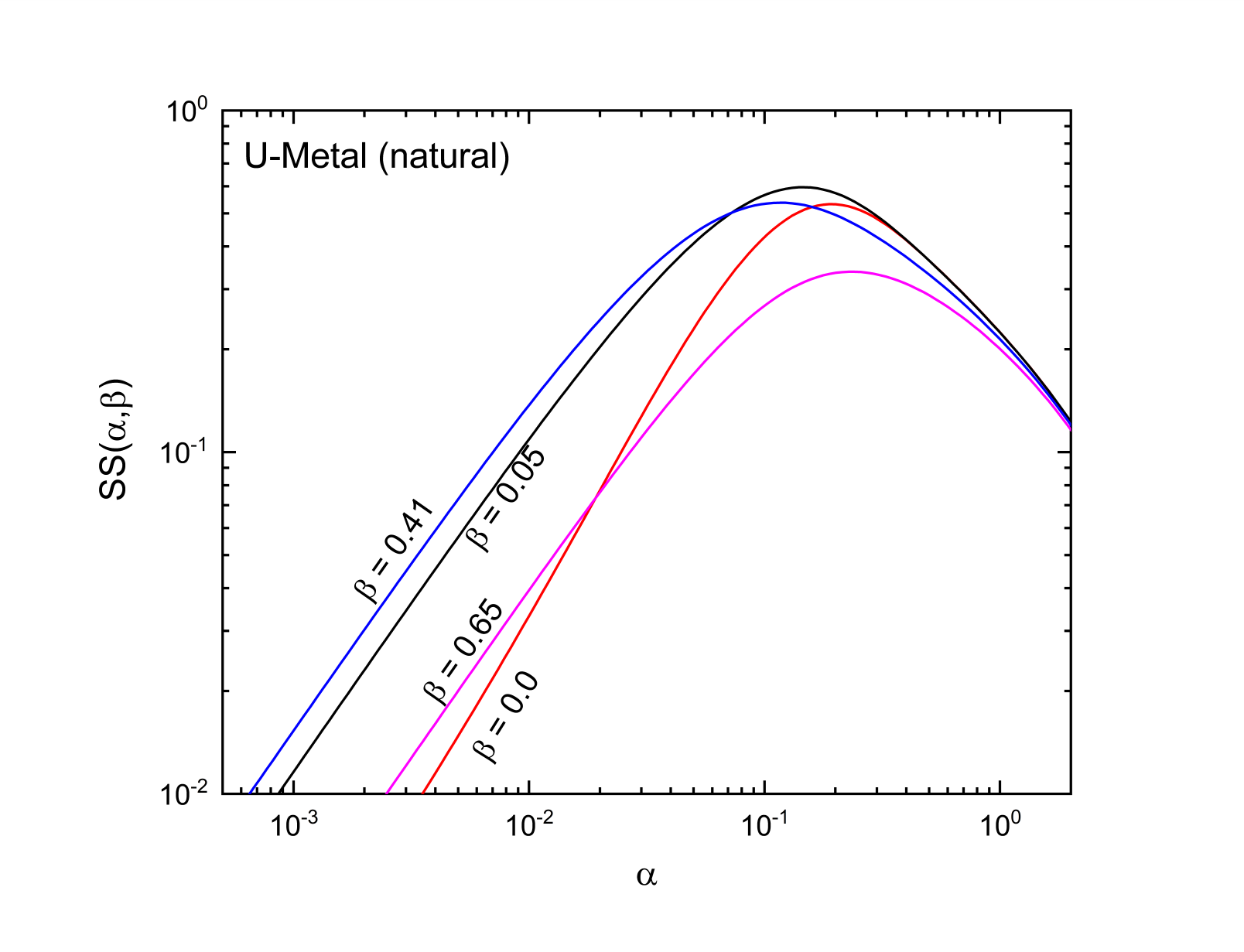}
    \caption{(Color online) The symmetric TSL for natural uranium metal at 293.6 K as a function of momentum transfer $\alpha$, for a range of neutron energy transfers, $\beta$. \Ss\ for each $\beta$ is labeled with the corresponding line.}
    \label{fig:TSL_U-metal}
\end{figure}

Using the PDOS as the fundamental input, the \FLASSH\ code \cite{Fleming2023} was used to evaluate the TSL for a natural enrichment U-metal (File 7, MAT8000) at temperatures of 293.6, 400, 500, 600, 700, 800, and 900 K, which correspond to the stable temperature region for U-metal. The symmetric TSL (MT4) at 293.6 K is shown in Fig. \ref{fig:TSL_U-metal}. 
 
\begin{figure}
    \centering
    \includegraphics[width=1.0\columnwidth,clip,trim=  15mm 10mm 30mm 15mm]{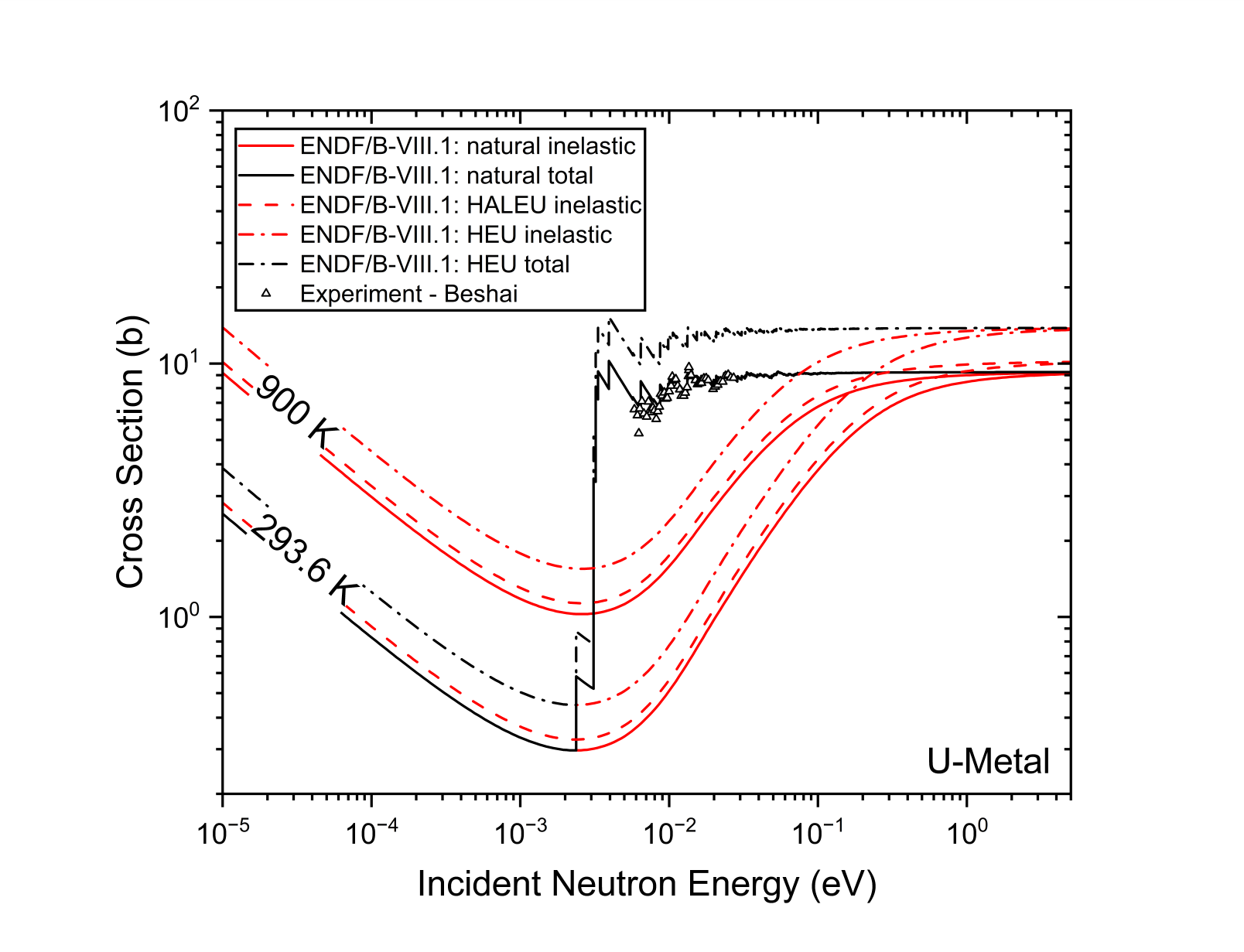}
    \caption{(Color online) The inelastic scattering cross section of U-metal shown at selected temperatures and enrichments. The natural and high enriched total scattering cross sections for U-metal at 293.6 K are also shown and compared to experimental data \cite{Beshai1966}.}
    \label{fig:Xsec_U-metal}
\end{figure}

Various enrichments of U-metal were evaluated using the PDOS described above. This DOS, which supports the structure of U-metal, was then combined with various mass and free atom cross sections \cite{Brown2018} to represent a natural \cite{Berglund2011}, 5\%, 10\%, 19.75\% (HALEU), 93\% (HEU), and 100\% enrichment of \fiveU. This then captures the impact of various compositions on the total scattering cross section. These evaluations were completed using the \FLASSH\ code for naturally enriched, 5\%, 10\%, 19.75\%, 93\%, and 100\%  \fiveU\ (File 7, MAT8000, MAT8005, MAT8010, MAT8048, MAT8099, and MAT8047, respectively). The integrated scattering cross sections as evaluated by \FLASSH\ are shown in Fig.~\ref{fig:Xsec_U-metal}. 

The use of enrichment-dependent evaluations provides cross sections which can be directly compared to experimental data. These evaluations have been preliminarily tested in criticality benchmarks which show impact due to the enrichment effects. U-metal and all the various enrichment-dependent evaluations represent new TSLs that are included for the first time in the ENDF/B database. 

\subsubsection{Uranium Nitride (\un)}
\label{sec:un}

\begin{figure}
    \centering
    \includegraphics[width=1.0\columnwidth,clip,trim=  20mm 10mm 30mm 15mm]{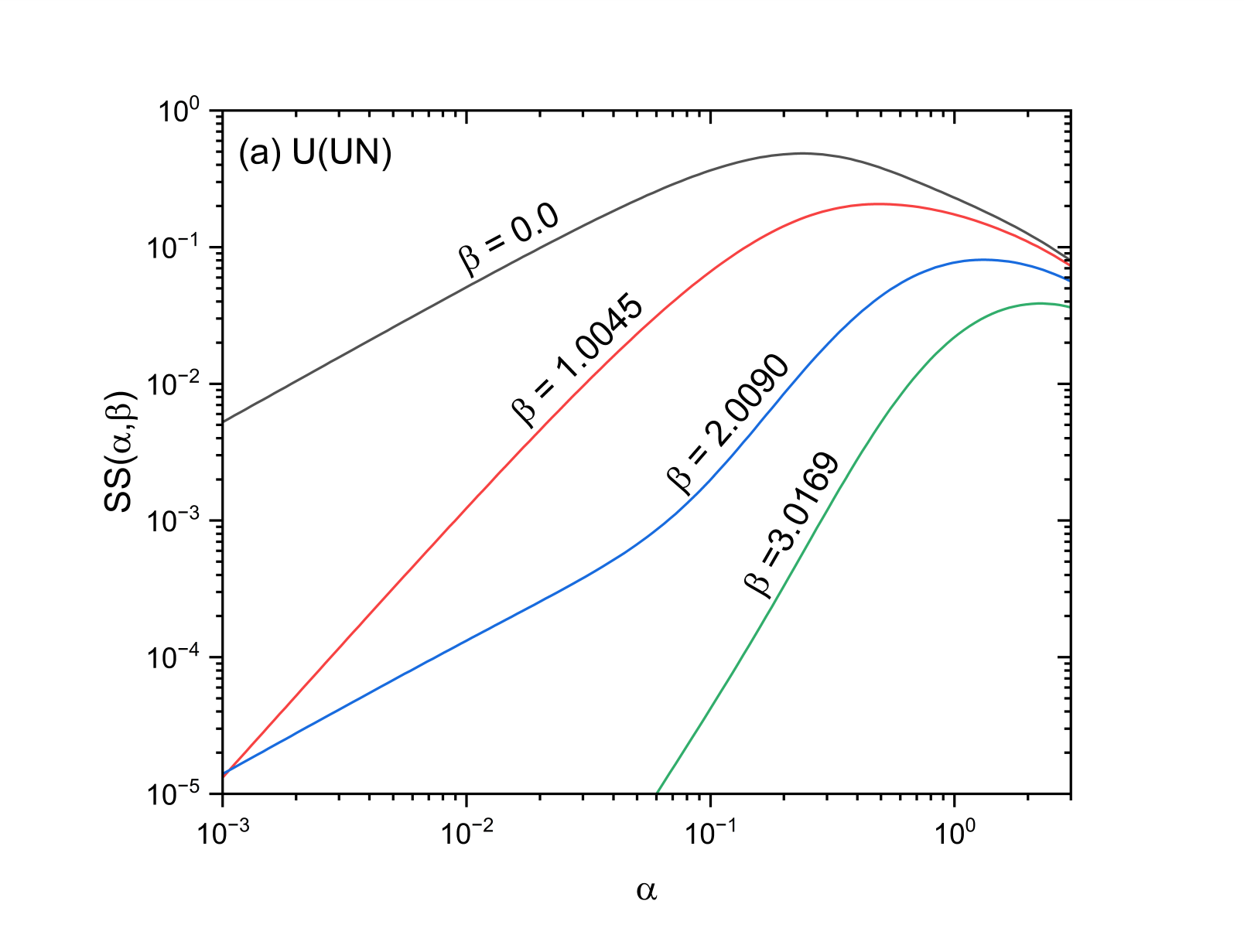}
    \includegraphics[width=1.0\columnwidth,clip,trim=  20mm 10mm 30mm 15mm]{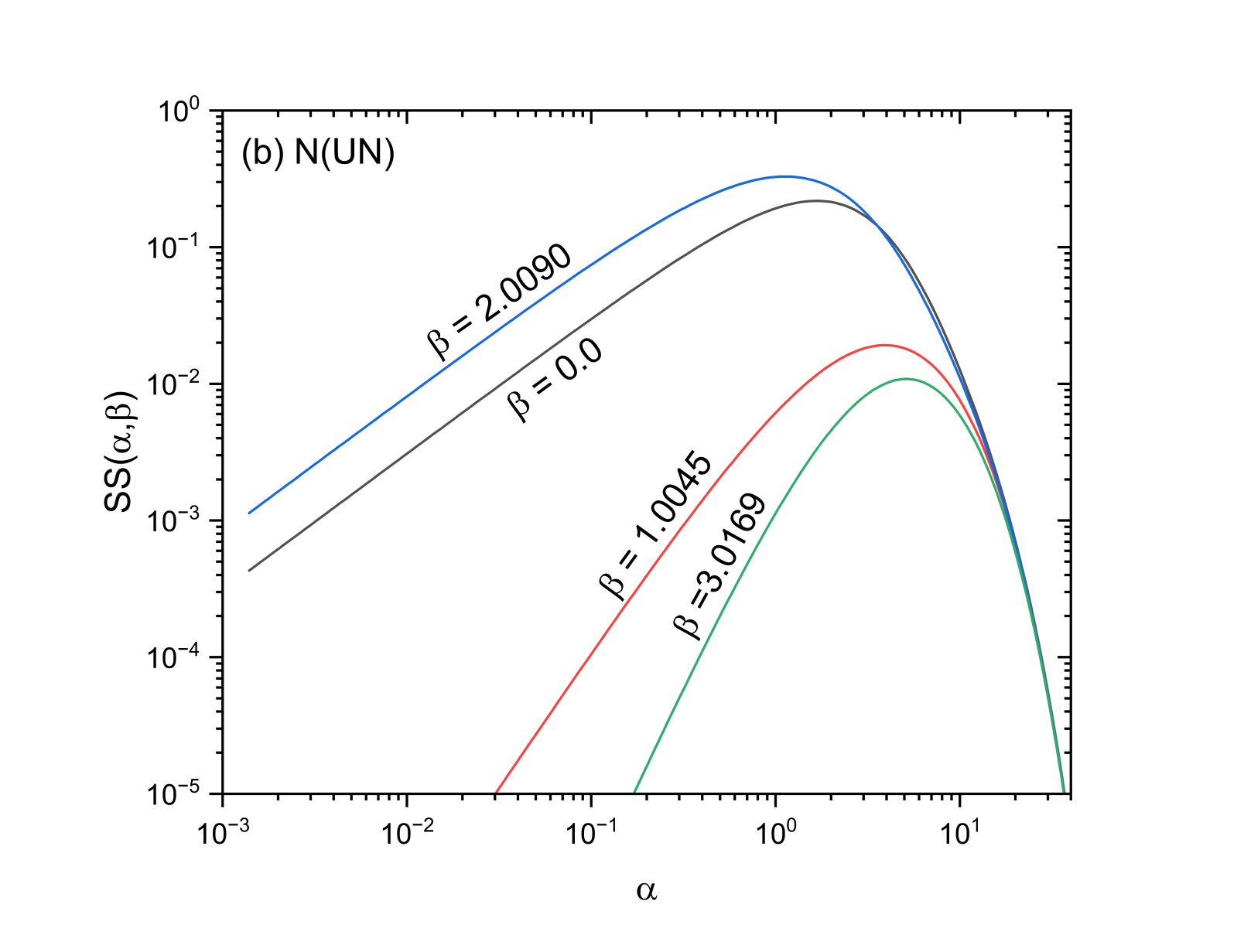}
    \caption{(Color online) The symmetric TSL for (a) U(\un)-natural and (b) N(\un)-natural at 296 K as a function of momentum transfer $\alpha$, for a range of neutron energy transfers, $\beta$. \Ss\ for each $\beta$ is labeled with the corresponding line.}
    \label{fig:TSL_UN}
\end{figure}

The evaluation for \un\ is based on the previous \prENDF\ evaluation with key updates to improve the TSL and resulting cross section. The original AILD simulations of the lattice structure and resulting PDOS were maintained \cite{Wormald2020_2}. The DOS was used as the primary input to the \FLASSH\ code \cite{Fleming2023} to evaluate both the TSL (MT4) as shown in Fig. \ref{fig:TSL_UN} and the incoherent and coherent elastic scattering (MT2) at 296, 400, 500, 600, 700, 800, 1000, and 1200 K. 

\begin{figure}
    \centering
    \includegraphics[width=1.0\columnwidth,clip,trim=  20mm 10mm 30mm 15mm]{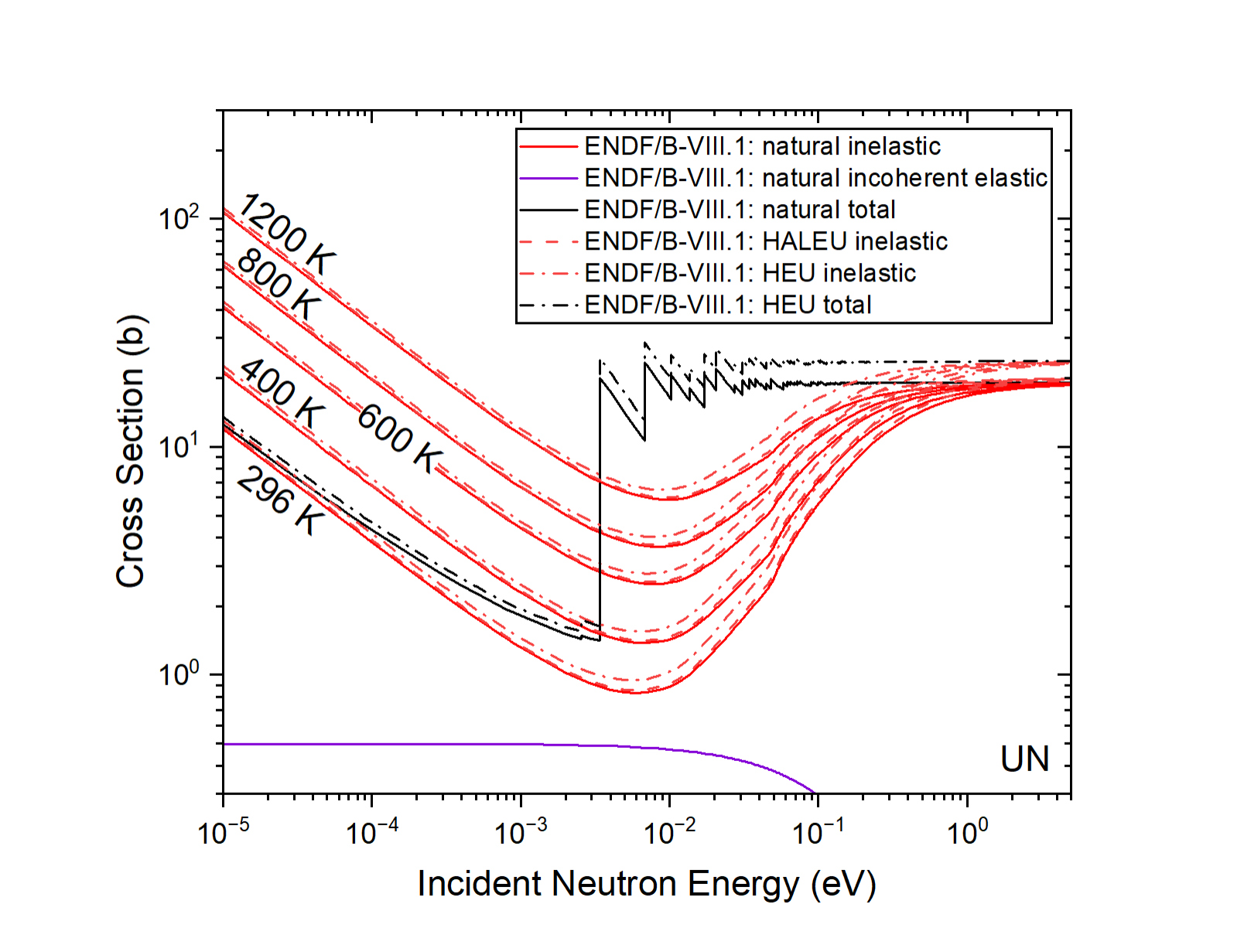}
    \caption{(Color online) The inelastic scattering cross section of \un\ shown at selected temperatures and enrichments. The natural and high enriched total scattering cross sections for \un\ at 296 K are also shown, in addition to the natural incoherent elastic scattering cross section.}
    \label{fig:Xsec_UN}
\end{figure}

Because of the significant contribution from both coherent and incoherent elastic effects, coherent elastic effects were tabulated for the compound \un. Half the coherent elastic is stored with the U(\un) evaluation with the other half stored using the mixed elastic ENDF formatting \cite{ENDF6-Format-2024} in the N(\un) evaluation. The incoherent elastic contributions, which arise solely from the nitrogen, are tabulated in MT2 of N(\un) using the mixed elastic formatting. The resulting total cross sections are shown in Fig. \ref{fig:Xsec_UN}. Various enrichments of uranium within \un\ were evaluated using the PDOS described above. This DOS, which supports the structure of \un, was combined with various mass and free atom cross section \cite{Brown2018} to represent natural \cite{Berglund2011}, 5\%, 10\%, 19.75\% (HALEU), 93\% (HEU), and 100\% enrichment of \fiveU\ with natural \cite{Sears1992} nitrogen. 

Evaluations were completed using the \FLASSH\ code for U(\un) for naturally enriched, 5\%, 10\%, 19.75\% (HALEU), 93\% (HEU), and 100\%  \fiveU\ (File 7, MAT72, MAT8305, MAT8310, MAT8348, MAT8349, and MAT8347, respectively) and for N(\un) paired with each \fiveU\ enrichment (File 7, MAT71, MAT8355, MAT8360, MAT8398, MAT8399, and MAT8397, respectively). Two files are required for each enrichment (a uranium and a nitrogen evaluation) because the tabulated compound coherent elastic will be impacted by the changing inputs associated with the uranium; MT4 for N(\un) remains the same for each enrichment while MT2 varies. The integrated scattering cross section as evaluated by \FLASSH\ are shown in Fig. \ref{fig:Xsec_UN}. 

The use of enrichment-dependent evaluations provides cross sections, which can be directly compared to experimental data. The enrichment-dependent \un\ evaluations represent novel TSLs that are included for the first time in the ENDF/B database.

\subsubsection{Uranium Dioxide (\uo)}
\label{sec:uo}

\begin{figure}
    \centering
    \includegraphics[width=1.0\columnwidth,clip,trim=  15mm 10mm 30mm 15mm]{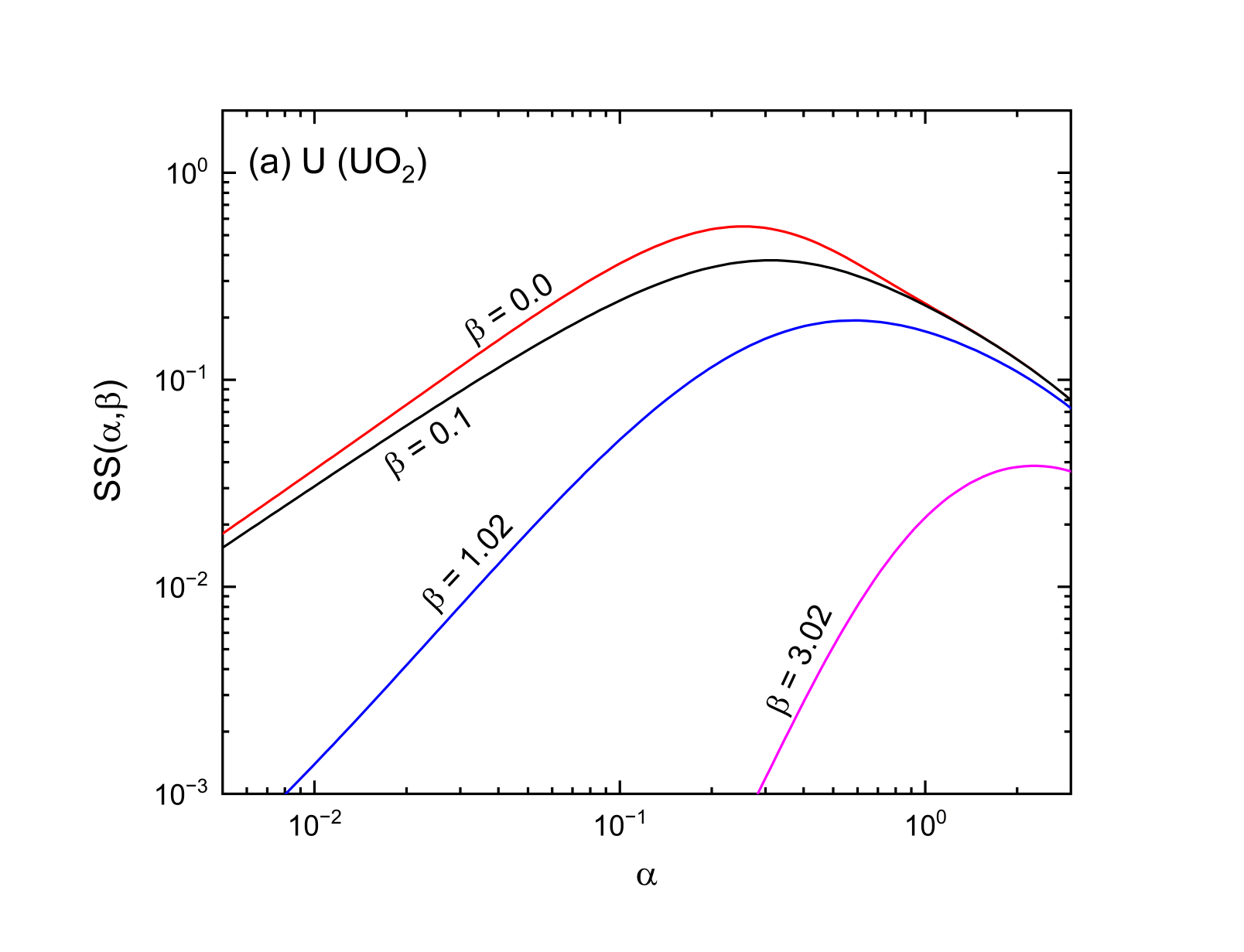}
    \includegraphics[width=1.0\columnwidth,clip,trim=  15mm 10mm 30mm 15mm]{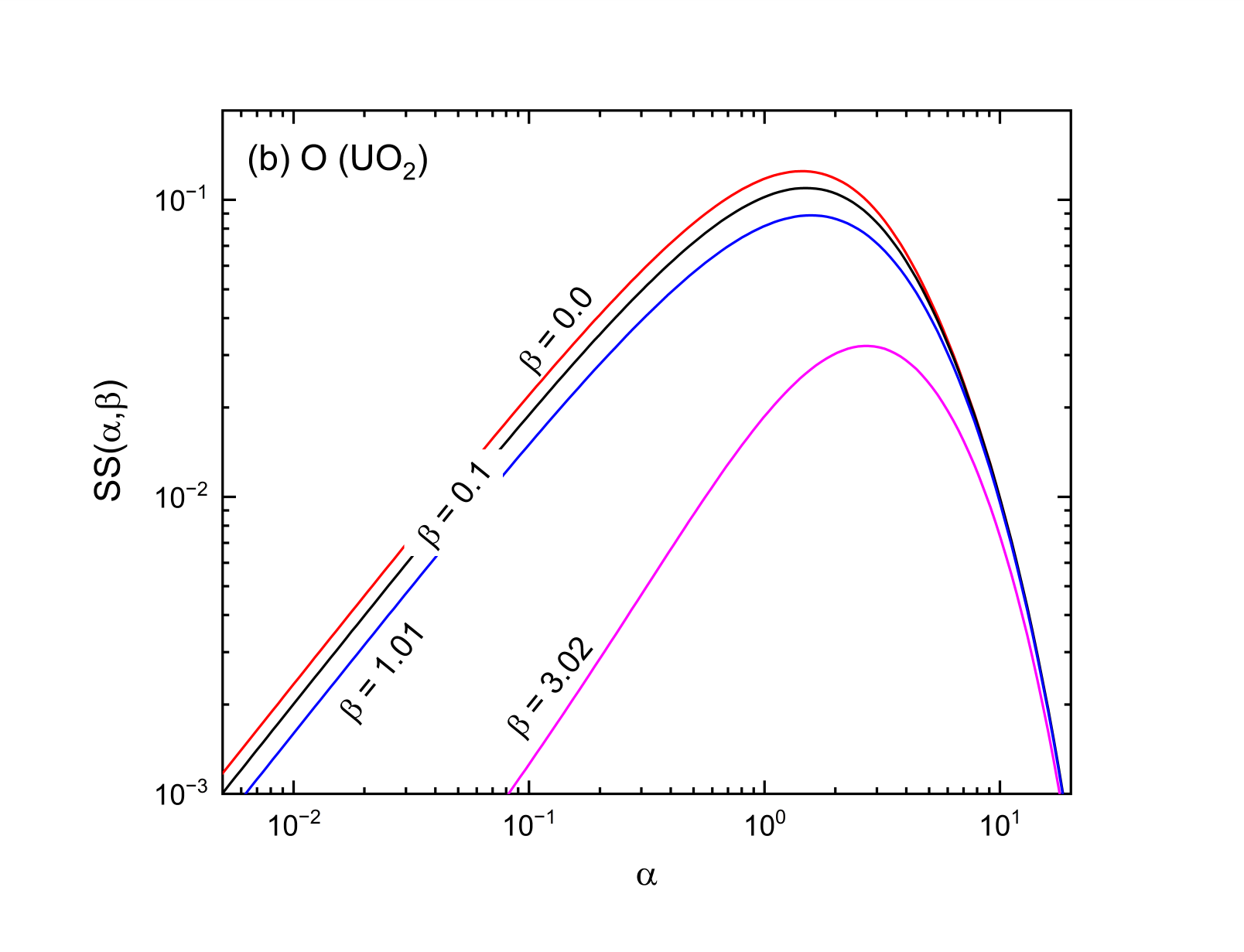}
    \caption{(Color online) The symmetric TSL for U(\uo)-natural and (b) O(\uo)-natural at 296 K as a function of momentum transfer $\alpha$, for a range of neutron energy transfers, $\beta$. \Ss\ for each $\beta$ is labeled with the corresponding line.}
    \label{fig:TSL_UO2}
\end{figure}

The evaluation for \uo\ is based on the previous \prENDF\ evaluation with key updates to improve the TSL and resulting cross section. The original AILD simulations of the lattice structure and resulting PDOS were maintained \cite{Wormald2021}. The DOS was used as the primary input to the \FLASSH\ code \cite{Fleming2023} to evaluate both the TSL (MT4) as shown in Fig. \ref{fig:TSL_UO2} and the coherent elastic scattering (MT2) at 296, 400, 500, 600, 700, 800, 1000, and 1200 K. 

\begin{figure}
    \centering
    \includegraphics[width=1.0\columnwidth,clip,trim=  20mm 10mm 30mm 15mm]{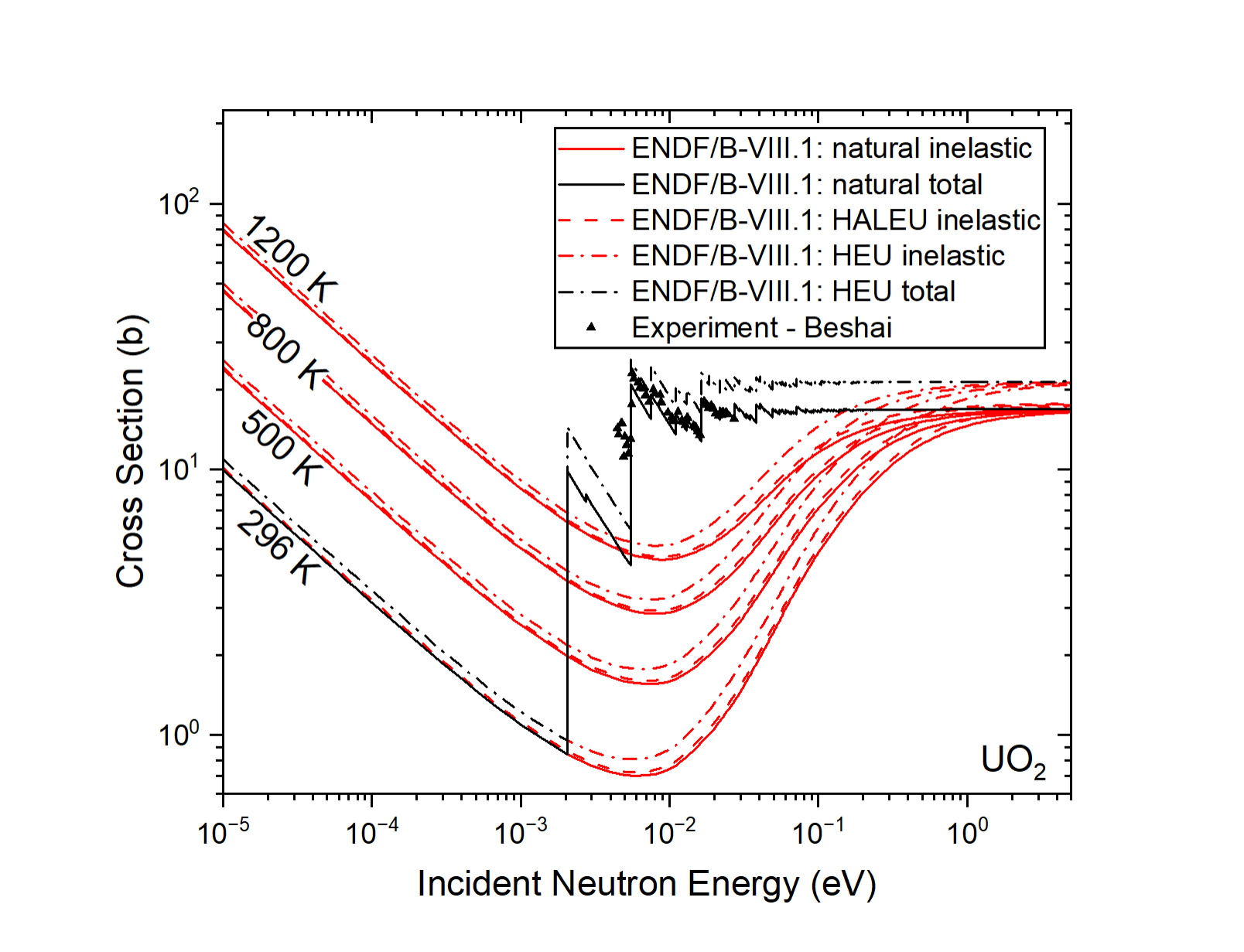}
    \caption{(Color online)  The inelastic scattering cross section of \uo\ shown at selected temperatures and enrichments. The natural and high enriched total scattering cross sections for \uo\ at 296 K are also shown, with the former compared to experimental data \cite{Beshai1966}.}
    \label{fig:Xsec_UO2}
\end{figure}

The coherent elastic was evaluated using the cubic approximation updated to reference experimental lattice parameters \cite{Leinders2015} with a third of the compound's elastic tabulated in MT2 in each of the U and O evaluations \cite{Wormald2021}. The resulting total scattering cross sections are shown in Fig. \ref{fig:Xsec_UO2}. Various enrichments of \uo\ were evaluated using the PDOS described above. This DOS, which supports the structure of \uo, was combined with different masses and free atom cross sections \cite{Brown2018} to represent \fiveU\ enrichments varying from natural \cite{Berglund2011} to 100\% to capture the impact of enrichment on the total scattering cross section. Evaluations were completed using the \FLASSH\ code for U(\uo) for naturally enriched, 5\%, 10\%, 19.75\% (HALEU), 93\% (HEU), and 100\%  \fiveU\ (File~7, MAT75, MAT8205, MAT8210, MAT8248, MAT8249, and MAT8247, respectively) and for O(\uo) paired with each \fiveU\ enrichment (File 7, MAT45, MAT8255, MAT8260, MAT8298, MAT8299, and MAT8297, respectively). Two files are required for each enrichment (a uranium and an oxygen evaluation) because the tabulated compound coherent elastic will be impacted by the changing inputs associated with the uranium; MT4 for O(\uo) remains the same for each enrichment while MT2 varies. The integrated scattering cross section as evaluated by \FLASSH\ are shown in Fig. \ref{fig:Xsec_UO2} and compared to experimental data, which have been corrected for neutron absorption \cite{Beshai1966}. 

The use of enrichment-dependent evaluations provides cross sections which can be directly compared to experimental data. The enrichment-dependent \uo\ evaluations represent novel TSLs that are included for the first time in the ENDF/B database. 

\subsubsection{Uranium Hydride (UH\textsubscript{3})}
\label{sec:uh3}

The TSL evaluation for H bound in UH\textsubscript{3} has been generated with \NJOY2012 \cite{NJOY2012} and AILD as a new contribution to the ENDF/B database. The partial phonon spectrum of H(UH\textsubscript{3}) was generated with Hubbard model corrected density functional theory and lattice dynamics described in Ref. \cite{ZERKLE2018-ANS}. The H(UH\textsubscript{3}) phonon spectrum is found to be in reasonable agreement with the experiment described in Ref.~\cite{ZERKLE2018-ANS}.

A TSL File 7 evaluation is available for H(UH\textsubscript{3}) at 293.6~K. Only \textsuperscript{1}H was considered in the evaluation. When using the H(UH\textsubscript{3}) evaluation, the uranium isotopes should be modeled with the free gas approximation. The total scattering cross sections and AWR were extracted from the ENDF/B-VIII.1 nuclide evaluation. Elastic scattering (MT2) was evaluated as incoherent. Inelastic scattering (MT4) was evaluated in the incoherent approximation with the phonon expansion method on a custom $\left(\alpha,\beta\right)$ grid.

\begin{figure}
    \centering
    \includegraphics[width=\columnwidth]{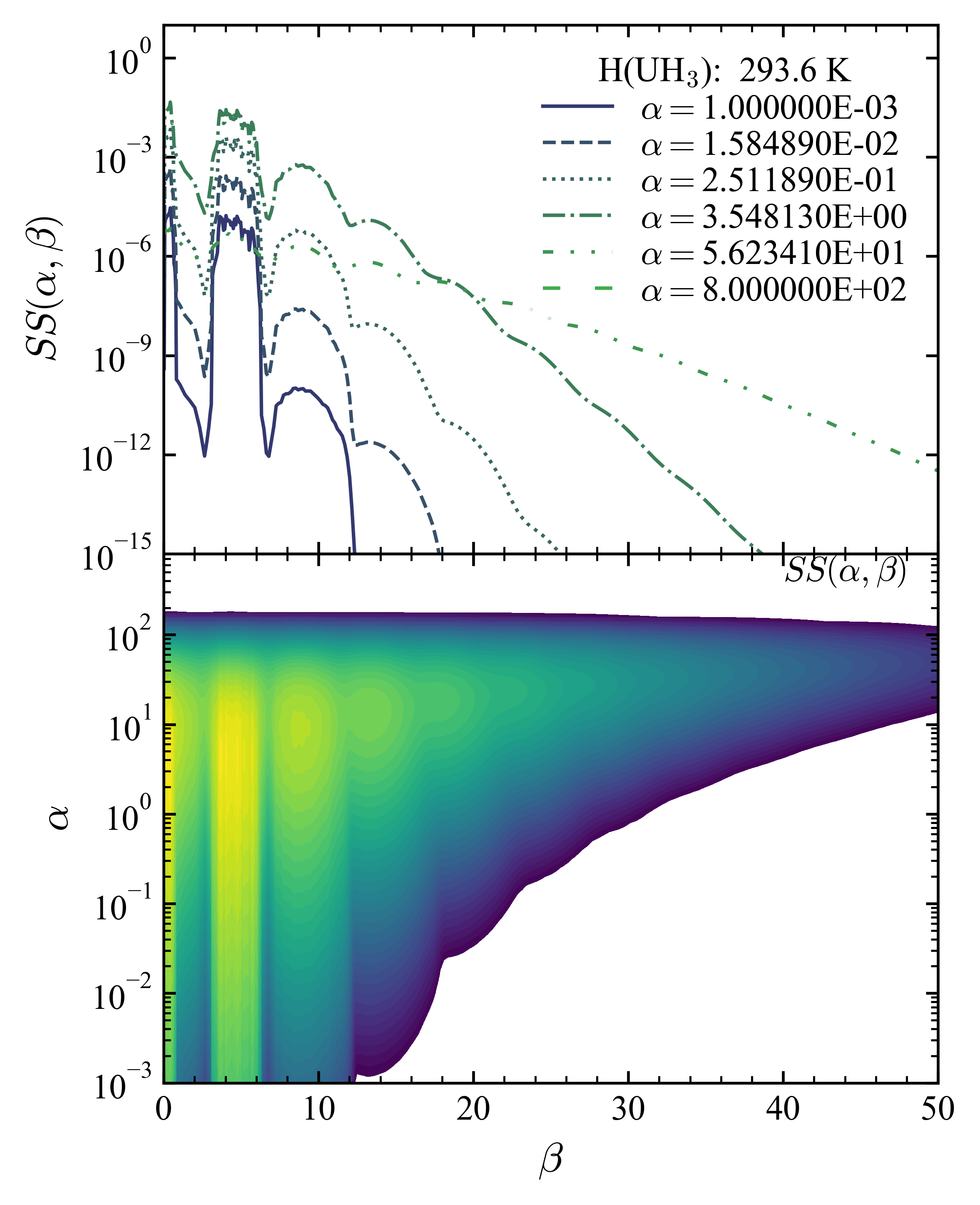}
    \caption{(Color online) H(UH\textsubscript{3}) symmetric TSL, $SS\left(\alpha,\beta\right)$, at 293.6~K. The $\left(\alpha,\beta\right)$ grid is referenced to ${k_B}T=0.0253$~eV. The contour plot illustrates only $SS\left(\alpha,\beta\right)>10^{-15}$; other non-zero values are not shown. Quantized oscillations in the TSL occur at integer values of $k_B T\beta\approx0.12$~eV.}
    \label{fig:TSL_UH3}
\end{figure}

\begin{figure}
    \centering
    \includegraphics[width=1.0\columnwidth,clip,trim=  0mm 5mm 0mm 0mm]{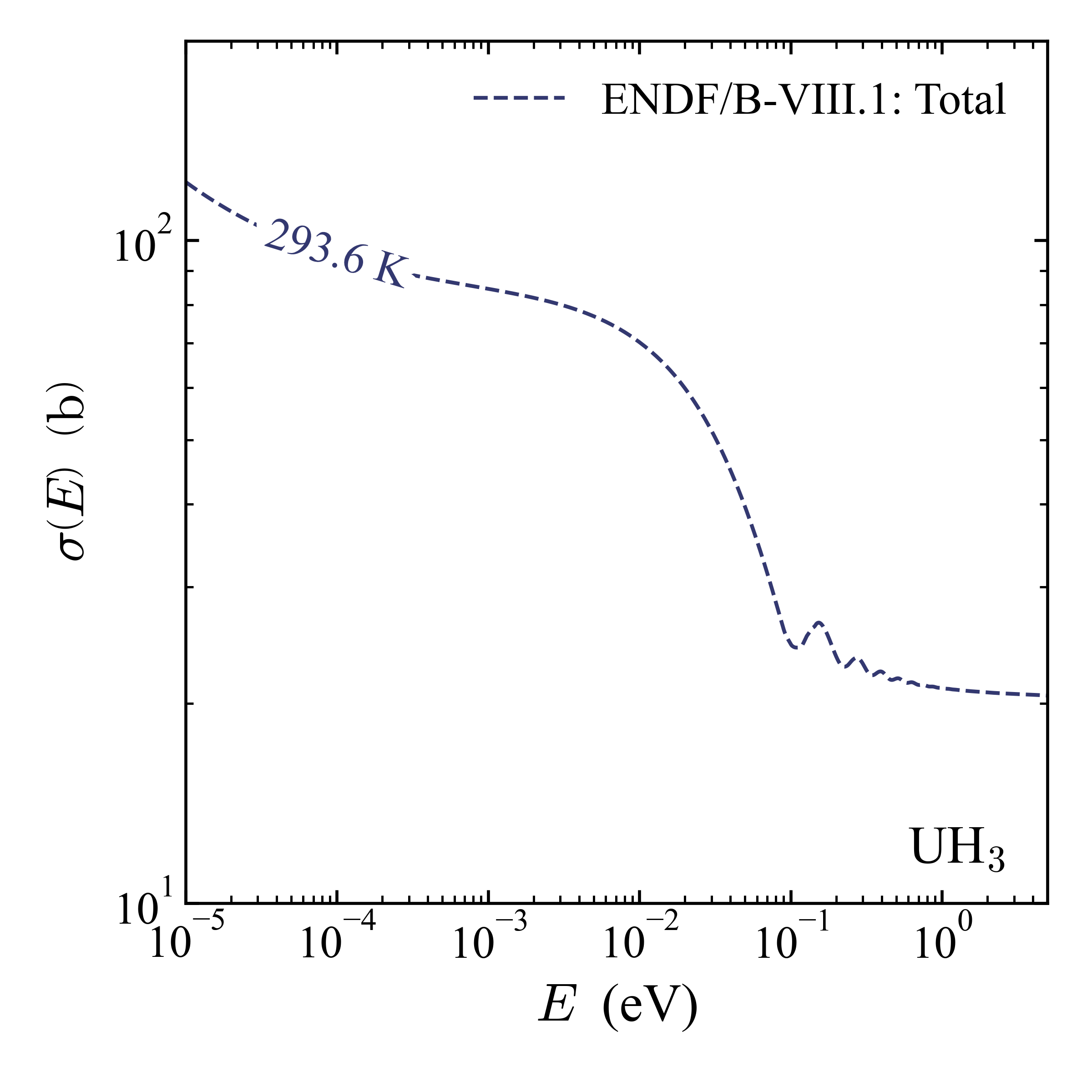}
    \caption{(Color online) Total scattering cross section of UH\textsubscript{3} between 293.6~K as compared to the ENDF/B-VIII.0 YH\textsubscript{2} material evaluation. Oscillations in the cross section correspond to the H oscillator energy of  $\approx0.12$~eV.}
    \label{fig:xsec_UH3}
\end{figure}

The symmetric TSL for H(UH\textsubscript{3}) at 293.6~K tabulated in File 7, MT4, is illustrated in Fig.~\ref{fig:TSL_UH3}. A limited number of quantized oscillations are present at fixed intervals of $k_B T\beta\approx0.12$~eV. The integrated inelastic and total cross sections computed with the NDEX nuclear data processing code \cite{NDEX_TSL1,NDEX_TSL2} is illustrated in Fig.~\ref{fig:xsec_UH3}. The adaptive energy meshing in NDEX is necessary to fully capture the oscillator behavior in the cross section.

\subsection{Special Purpose}
\label{subsec:tsl:special_purpose}

\subsubsection{Liquid hydrogen and deuterium (l-H$_2$, l-D$_2$)} 
\label{sec:lH2D2tsl}

\begin{figure}
    \centering
    \includegraphics[width=1.0\columnwidth]{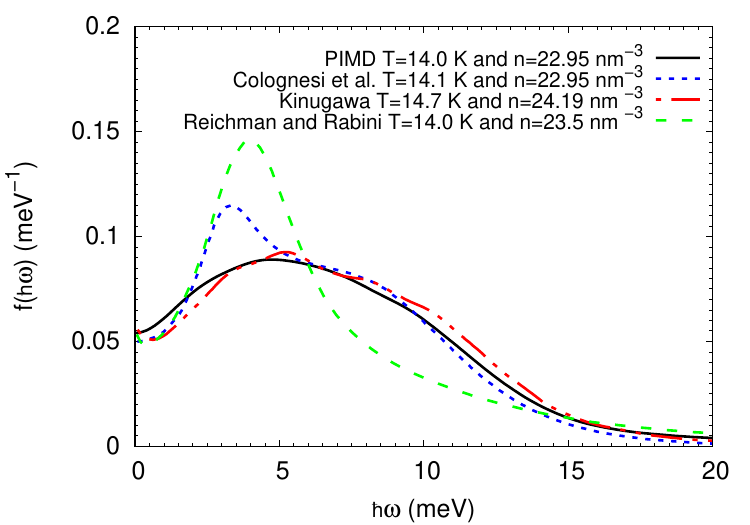}
    \caption{(Color online) Simulated frequency spectrum for liquid hydrogen at 14~K from the PIMD calculations compared to experimentally derived data from Colognesi \etal~\cite{colognesi2004microscopic} and simulations from Kinugawa \cite{kinugawa1998path} and from Reichman and Rabini \cite{reichman2001self}. Data is digitized from Ref.~\cite{colognesi2004microscopic}.}
    \label{fig:h2_RPMD}
\end{figure}

\begin{figure}
    \centering
    \includegraphics[width=1.0\columnwidth]{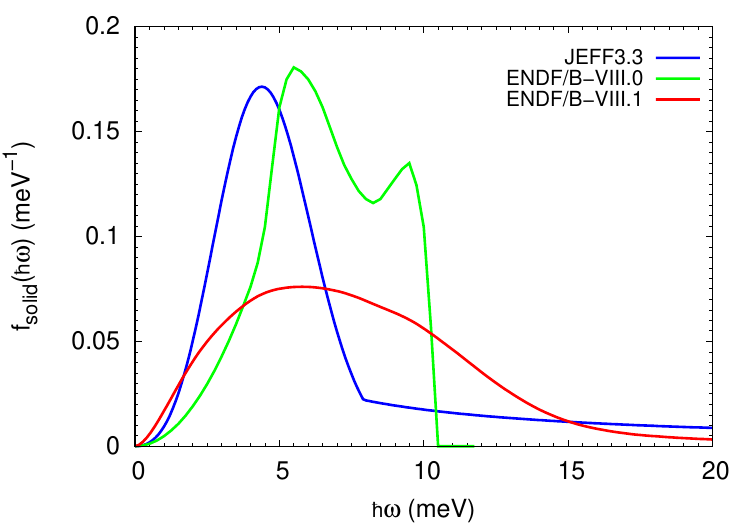}
    \caption{(Color online) Comparison of the solid-like frequency spectrum for liquid hydrogen at 14 K used in the ENDF/B-VIII.1 evaluation to the solid-like frequency spectra used in the ENDF/B-VIII.0 and JEFF-3.3 evaluations.}
    \label{fig:h2_frequency_spectrum}
\end{figure}

\begin{figure}
    \centering
    \includegraphics[width=1.0\columnwidth]{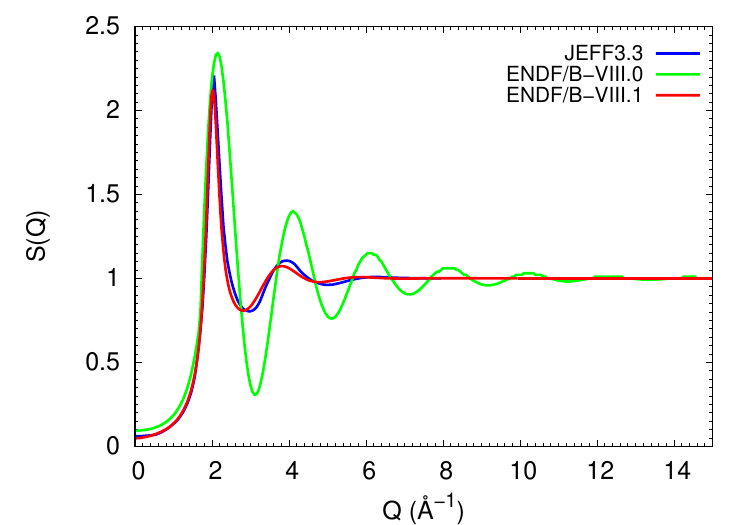}
    \caption{(Color online) Comparison of the static structure factor for liquid hydrogen at 14 K used in the ENDF/B-VIII.1 evaluation to the static-structure factors used in the ENDF/B-VIII.0 and JEFF-3.3 evaluations.}
    \label{fig:h2_static_structure}
\end{figure}

Compared to ENDF/B-VIII.0, the evaluations for liquid hydrogen and deuterium are now available at temperatures 
of 14, 15, 16, 17, 18, 19, and 20 K for para/ortho hydrogen and 19, 20, 21, 22, and 23 K for para/ortho deuterium. 
The evaluations are based on the approach described in Ref.~\cite{dijulio_2023}, where path-integral molecular dynamics (PIMD) 
techniques have been used to produce input data for the neutron scattering models in a modified \LEAPR\ module of \NJOY\ 
called \texttt{NJOY-H2D2}~\cite{report42_2020}. The main ingredients are the static structure factors and the frequency 
spectra, which have been generated from the radial distribution and velocity autocorrelation functions computed from the 
MD simulations, respectively. An example of the frequency spectrum for liquid hydrogen from the PIMD 
calculations at 14 K, compared to previous experimentally derived and simulated data, is shown in Fig.~\ref{fig:h2_RPMD}. 
For input into \LEAPR, the frequency spectra were decomposed into solid-like and diffusive parts, and the rotational and 
vibrational components were included using previously existing methods \cite{young_1964,granada_2004}. The solid-like part 
and the static structure factor at 14 K for liquid hydrogen are compared to data used for ENDF/B-VIII.0 and JEFF-3.3 evaluations 
in Figs.~\ref{fig:h2_frequency_spectrum} and~\ref{fig:h2_static_structure}, respectively. In the ENDF/B-VIII.0 evaluations, 
the coherent scattering contributions were computed using the Vineyard approximation \cite{vineyard_1958}. In the ENDF/B-VIII.1 
evaluations, the coherent scattering was instead computed using the Sk\"old approximation \cite{skold_1967}, which is 
included in \texttt{NJOY-H2D2}.

\begin{figure}
    \centering
    \includegraphics[width=1.0\columnwidth]{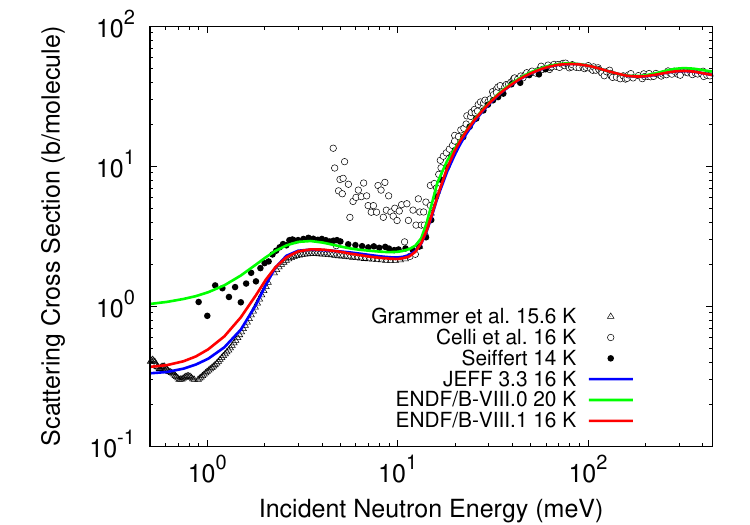}
    \caption{(Color online) Total scattering cross sections for liquid para-hydrogen at the indicated temperature from ENDF/B-VIII.0 
    and ENDF/B-VIII.1 compared to experimental data from Refs.~\cite{Seiffert_1970,Celli_1999,Grammer_2015} and JEFF-3.3 \cite{JEFF33}.}
    \label{fig:pH2}
\end{figure}

\begin{figure}
    \centering
    \includegraphics[width=1.0\columnwidth]{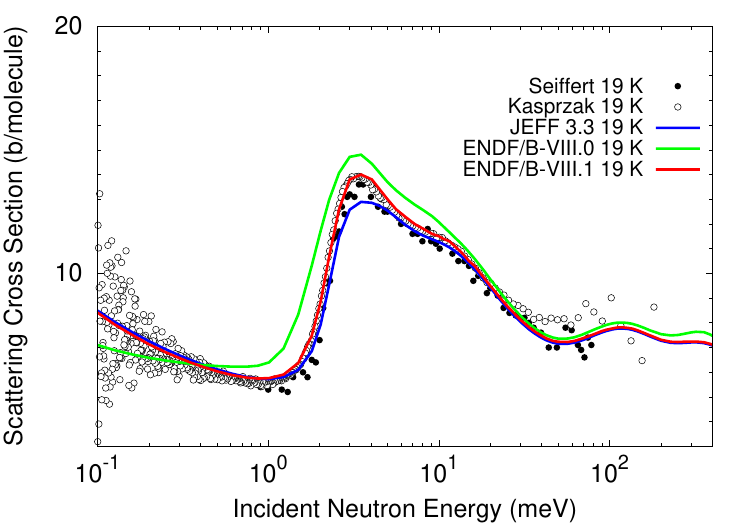}
    \caption{(Color online) Total scattering cross-sections for liquid ortho-deuterium at the indicated temperature from ENDF/B-VIII.0 
    and ENDF/B-VIII.1 compared to experimental data from Refs.~\cite{Seiffert_1970,Atchison_2005,Kasprzak_2008} and JEFF-3.3 \cite{JEFF33}.}
    \label{fig:oD2}
\end{figure}

Figs.~\ref{fig:pH2} and~\ref{fig:oD2} show a selection of the total scattering cross sections for the ENDF/B-VIII.1 evaluations for 
liquid para-hydrogen and ortho-deuterium at the indicated temperatures compared to available experimental data, the ENDF/B-VIII.0 
evaluations, and also the evaluations from JEFF-3.3 \cite{JEFF33}. For para-hydrogen, the ENDF/B-VIII.1 evaluation shows good agreement 
with the more recent measurements of Ref.~\cite{Grammer_2015} in the range of about 2--12 meV, which is also supported by a separate earlier 
study based also on path-integral MD techniques \cite{Guarini_2015}. In the case of ortho-deuterium, the ENDF/B-VIII.1 
evaluation shows improved agreement with experimental data compared to the ENDF/B-VIII.0 evaluation.

\subsubsection{Sapphire Single-Crystal Neutron Filter (\alo)} 
\label{sec:alo}

Sapphire (\alo) was evaluated using modern AILD techniques \AILD. Sapphire filters are single crystal materials, which consist of highly ordered structure to allow neutrons of select energies to pass through. As a filter material, sapphire is used in research reactors, which require the neutron beam to be filtered to transmit specific neutron energies and remove others \cite{Hawari2006}. The structure of sapphire was modeled using the \texttt{VASP} code \VASP\ to predict the Hellmann-Feynman forces. The resulting forces were used in the dynamical matrix approach within the PHONON code \cite{Parlinski1997} to calculate the phonon dispersion relations which were found to be in good agreement with experimental data \cite{Bialas1975, Schober1993}.

\begin{figure}
    \centering
    \includegraphics[width=1.0\columnwidth,clip,trim=  20mm 10mm 30mm 15mm]{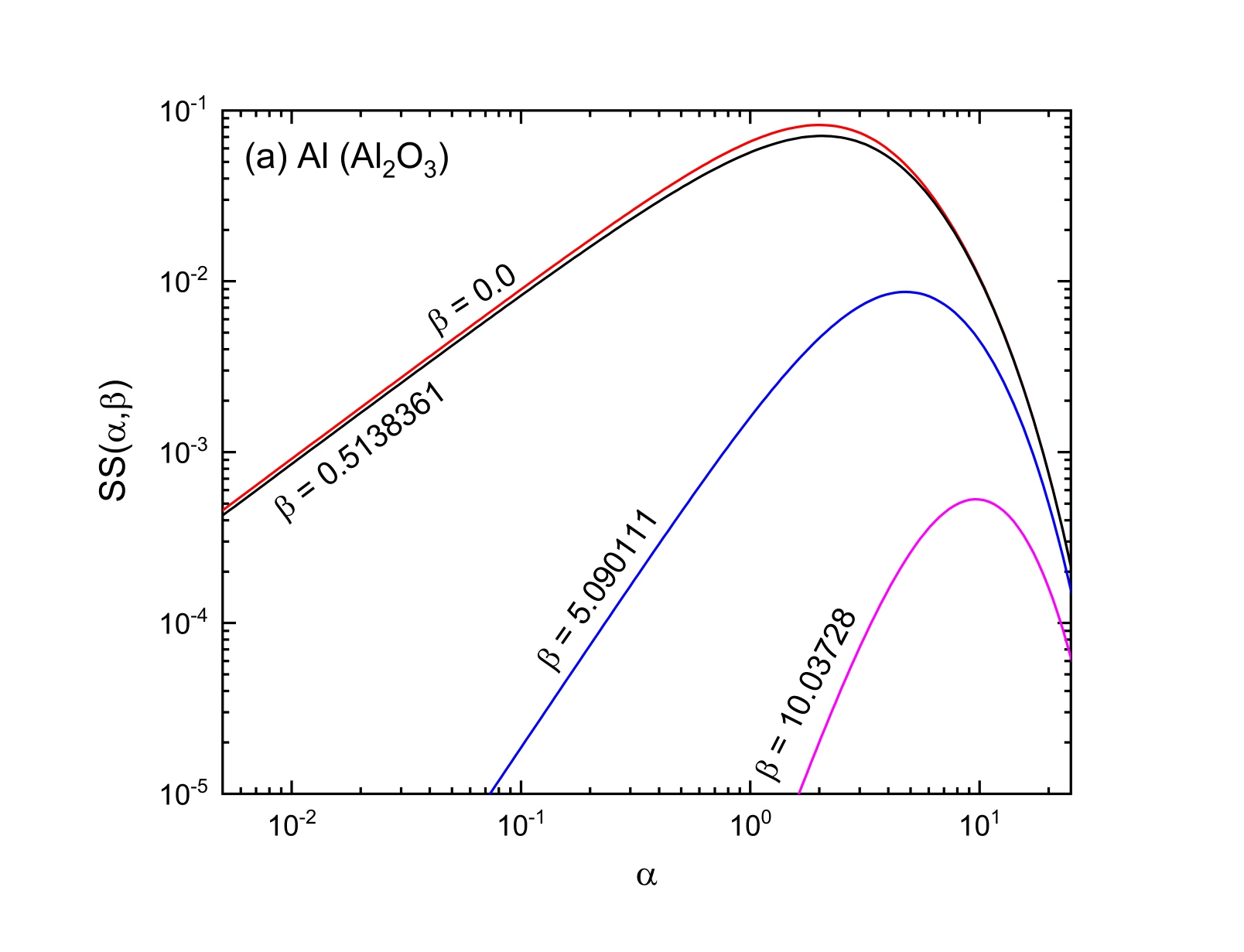}
    \includegraphics[width=1.0\columnwidth,clip,trim=  20mm 10mm 30mm 15mm]{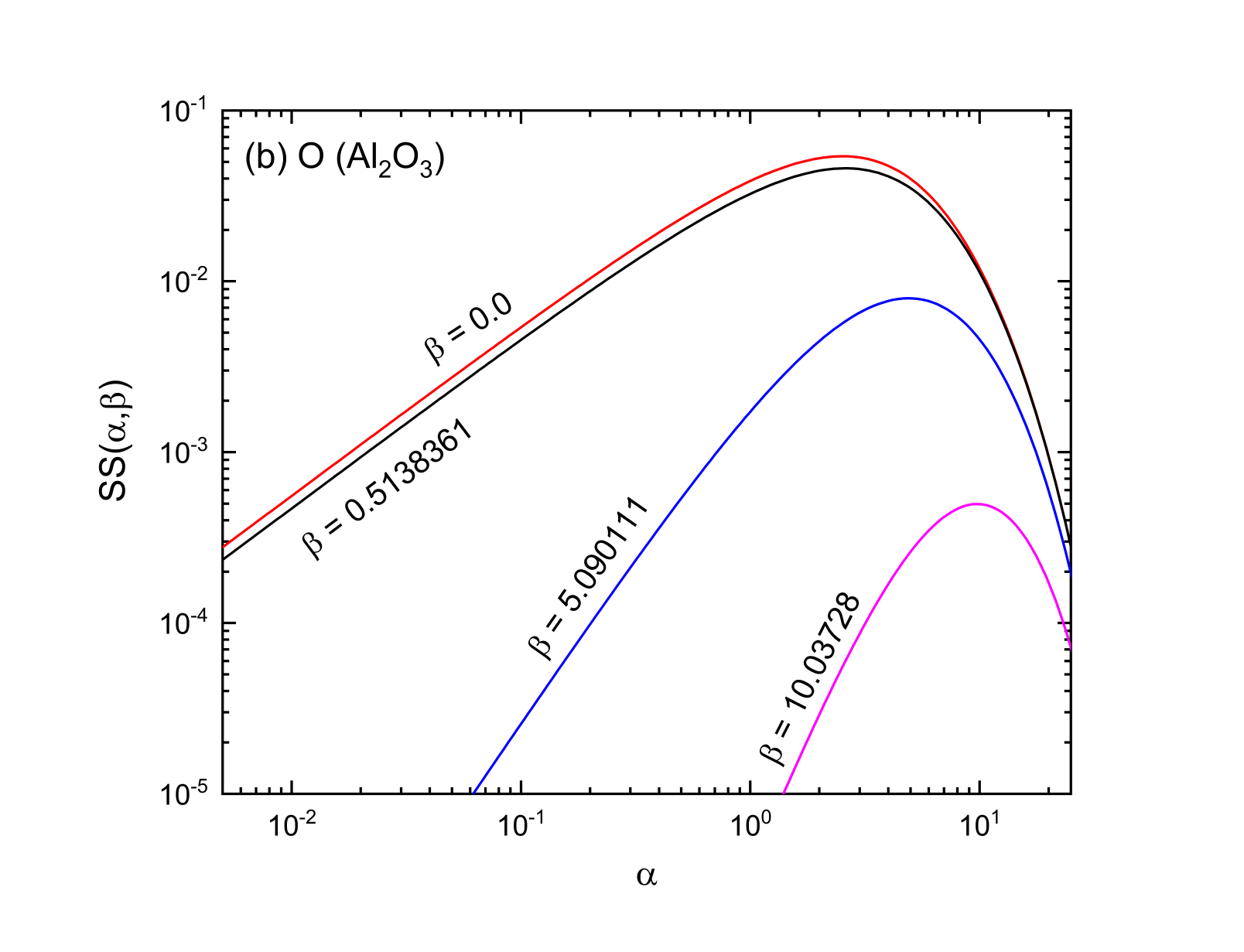}
    \caption{(Color online) The symmetric TSL for (a) Al(\alo) and (b) O(\alo) at 296 K as a function of momentum transfer $\alpha$, for a range of neutron energy transfers, $\beta$. \Ss\ for each $\beta$ is labeled with the corresponding line. }
    \label{fig:TSL_Al2O3}
\end{figure}

Using the PDOS as the fundamental input, the \FLASSH\ code \cite{Fleming2023} was used to evaluate the TSL for both Al(\alo) (File 7, MAT3052) and O(\alo) (File 7, MAT3053) with mass and free atom cross sections \cite{Brown2018} for $^{27}$Al and $^{16}$O at temperatures of 77, 80, 100, 200, 293.6, 296, and 300 K. This temperature grid supports common single crystal filter applications. The symmetric TSL (MT4) at 296 K is shown in Fig. \ref{fig:TSL_Al2O3}. 

\begin{figure}
    \centering
    \includegraphics[width=1.0\columnwidth,clip,trim=  20mm 10mm 30mm 15mm]{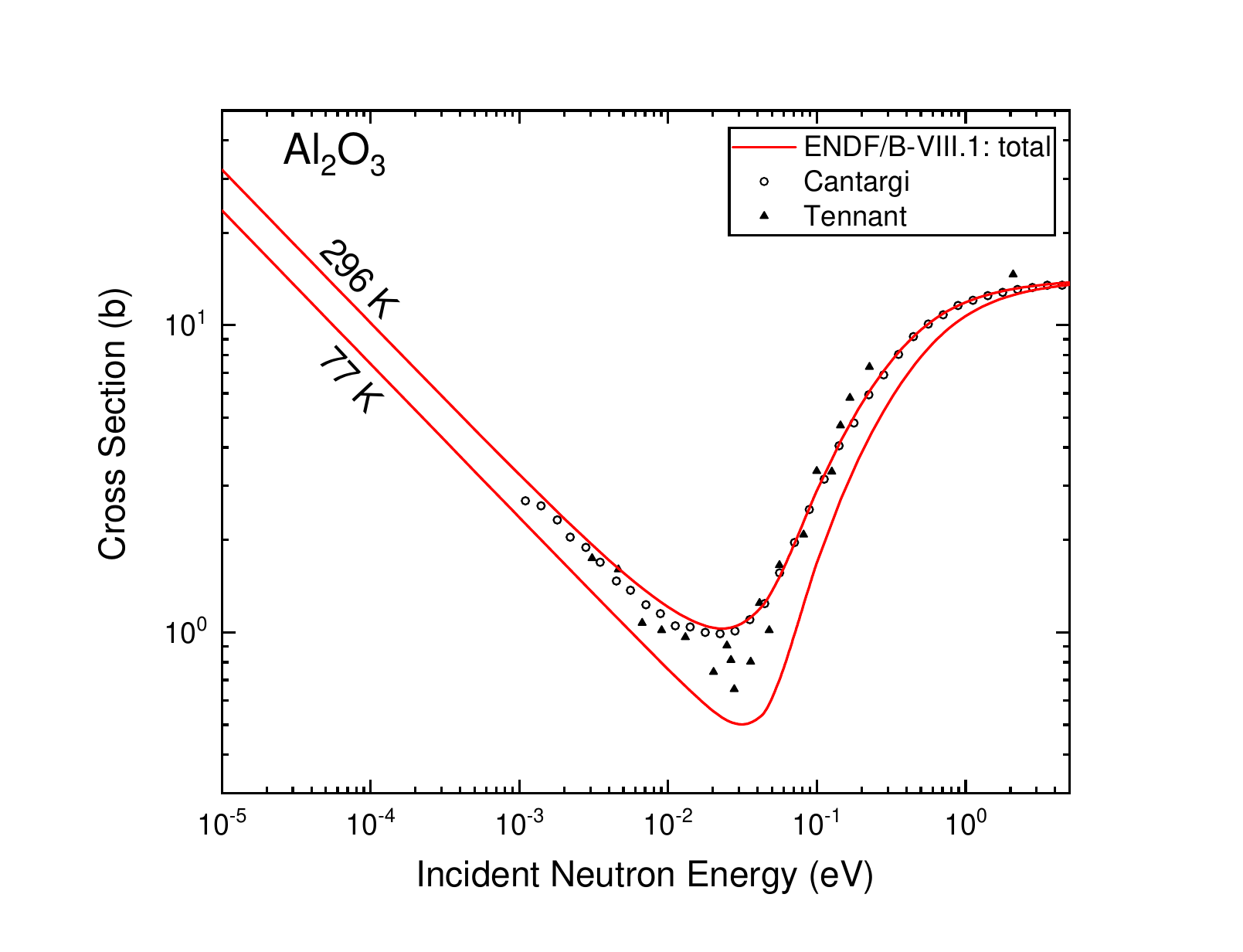}
    \caption{(Color online) Total cross section for \ENDF\ sapphire (\alo) compared to measured data at room temperature \cite{Tennant1988, Cantargi2015}. No elastic components are included in the evaluation.}
    \label{fig:Xsec_Al2O3}
\end{figure}

In order to serve as a filter material, the coherent elastic impacts must be removed from the evaluation. As such, both the Al(\alo) and O(\alo) TSL evaluations provide information only with respect to inelastic scattering (MT4), and the MT2 card has been removed. The inelastic cross sections integrated from the MT4 TSL were processed using the \FLASSH\ \cite{Fleming2023} cross section routines and are shown in Fig. \ref{fig:Xsec_Al2O3} in comparison to experimental data \cite{Tennant1988, Cantargi2015}.

\subsubsection{Magnesium Oxide Neutron Filter (MgO)} 
\label{sec:mgofil}

The objective of this evaluation is to utilize MgO as a neutron filter, a critical component for selectively transmitting thermal neutrons while attenuating higher-order neutrons and gamma rays, thereby reducing radiation background. The vibrational properties of MgO were calculated utilizing density functional perturbation theory employing linear response calculations with the ABINIT code \cite{GONZE2002478}. This approach efficiently determines phonon excitations at any wavevectors by examining the system's response to perturbations, including atomic displacements and constant electric fields. These perturbations facilitate the computation of interatomic force constants, dielectric constants, and Born effective charges, crucial for accurately describing phenomena like the longitudinal optical-transverse optical (LO-TO) splitting in polar materials, such as MgO.

\begin{figure}
    \centering
    \includegraphics[width=1.0\columnwidth]{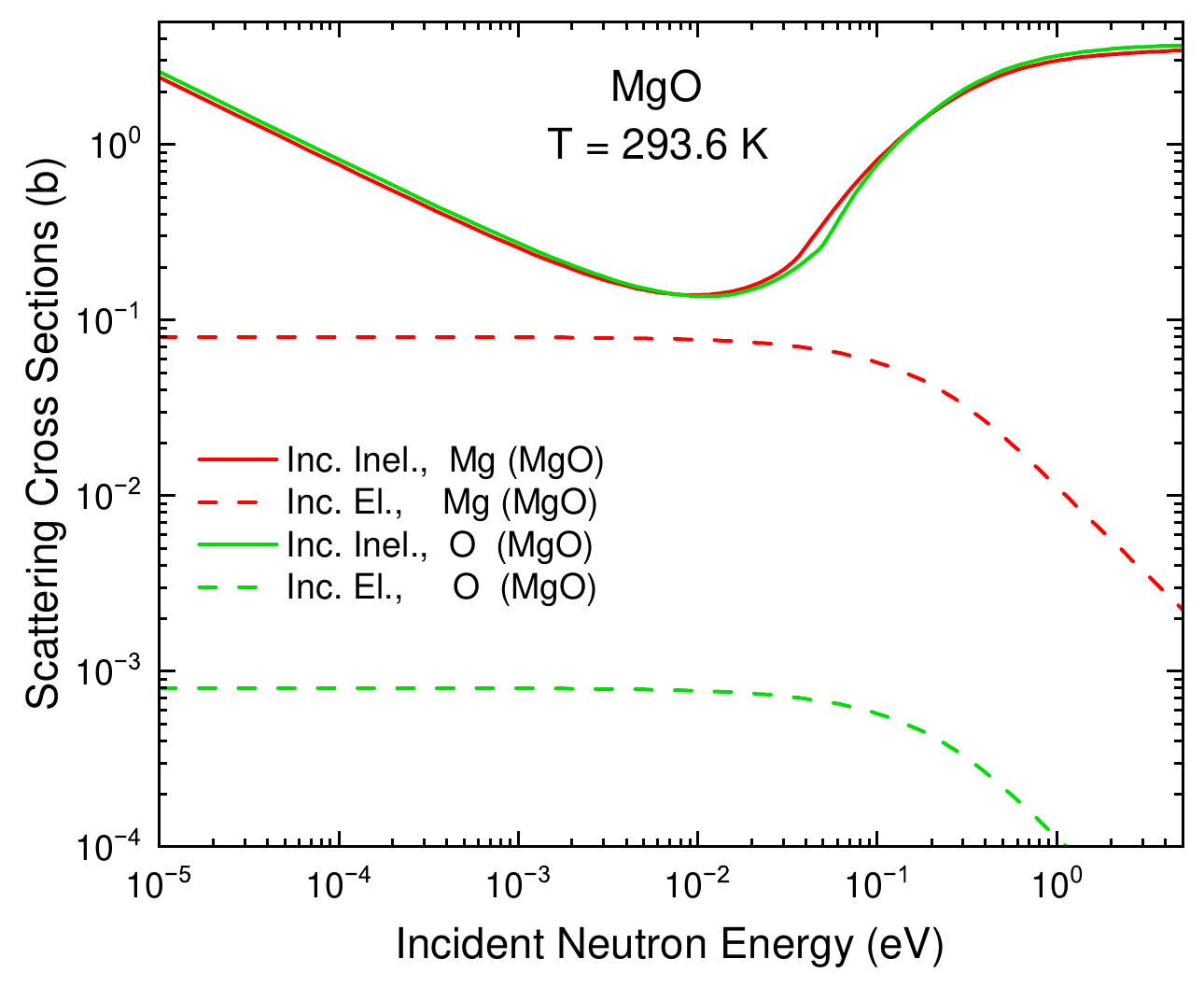}
    \caption{(Color online) Calculated scattering cross section (solid curves) Incoherent inelastic of Mg (red) and O (green) in MgO, and (dashed curves) incoherent elastic of Mg (red) and O (green) in MgO.}
    \label{fig:Xsec_MgO}
\end{figure}

MgO has a cubic symmetry and crystallizes in the rock-salt structure. It has two atoms per primitive unit cell. The calculated phonon dispersion relations exhibit excellent agreement with experimental data, validating the accuracy of the calculated PDOS. The partial PDOSs are then employed to calculate the TSL (File 7) for Mg(MAT3146) and O(MAT3147) in MgO, utilizing the incoherent approximation at 293.6 K as well as the incoherent elastic scattering cross section utilizing the \LEAPR\ module of \NJOY~\cite{NJOY} code. The free atom cross sections and AWR for both oxygen (O) and magnesium (Mg) were obtained from the nuclide evaluations in ENDF/B-VIII.0 \cite{Brown2018}. Furthermore, the incoherent bound scattering cross sections and the natural abundances of isotopes were defined according to the recommendations of the NIST database \cite{Sears1992}. For more details about the first-principles calculations and thermal neutron scattering analysis of MgO as a neutron filter, refer to Ref \cite{ALQASIR2016242}. Fig. \ref{fig:Xsec_MgO} shows the room temperature incoherent inelastic and incoherent elastic scattering cross sections of Mg and O in MgO.

\subsubsection{Magnesium Fluoride Neutron Filter (\mgft)} 
\label{sec:mgffil}

Neutron transmission through a single crystal of \mgft\ at liquid-nitrogen and room temperatures was first measured by Barker \etal~\cite{Barker:ks5194} to explore its potential as a neutron filter. In this work, the lattice vibrations of \mgft\ were computed using first-principles calculations. The calculated phonon dispersion relations showed excellent agreement with available experimental data, confirming the accuracy and reliability of the calculated PDOS. This was further verified by investigating the heat capacity and comparing it with experimental data. The obtained PDOS was then used to calculate the incoherent inelastic and incoherent elastic thermal neutron scattering cross sections at room temperature, resulting in excellent agreement between the calculated total cross section and the measured transmission data.

At ambient pressure and temperature, \mgft\ crystallizes into the rutile structure. It possesses a body-centered tetragonal symmetry and belongs to the $P4_2$/mnm space group. The hexagonal primitive unit cell of \mgft\ contains six atoms. The structural and vibrational characteristics of \mgft\ were investigated through a density functional perturbation theory as implemented in the ABINIT code \cite{GONZE2002478}, utilizing the local density approximation (LDA) for the exchange-correlation potential. The pseudopotentials for Mg and F were generated using the Troullier \& Martins \cite{PhysRevB.43.1993} optimization scheme with the FHI98PP code \cite{FUCHS199967}.

\begin{figure}
    \centering
    \includegraphics[width=1.0\columnwidth]{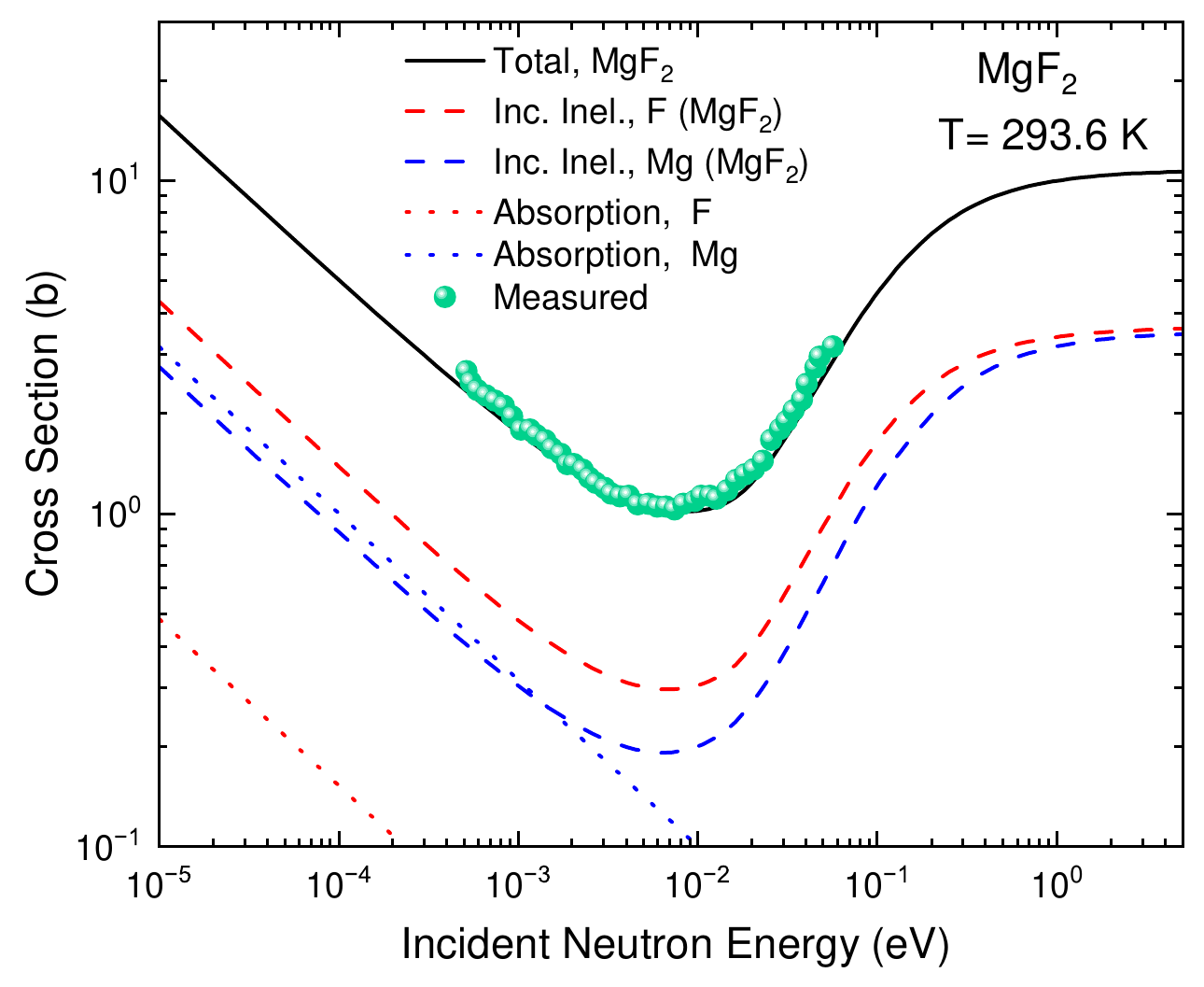}
    \caption{(Color online) \mgft\ Calculated (black solid curve) total cross section, (dashed curves) incoherent inelastic scattering cross sections, (dotted lines) absorption cross sections, and (spheres) measured total cross section \cite{Barker:ks5194}.}
    \label{fig:Xsec_MgF2}
\end{figure}

The TSL for Mg (MAT3142) and F (MAT3141) in \mgft\ were computed using partial PDOSs of Mg and F. This calculation was conducted at 293.6 K, employing the incoherent approximation as implemented in the \LEAPR\ module of \NJOY\ code \cite{NJOY}. Additionally, incoherent elastic scattering cross section was determined. The cross sections for free atoms and AWR for Mg and F were obtained from nuclide evaluations in ENDF/B-VIII.0 \cite{Brown2018}. Moreover, incoherent bound scattering cross sections for MF=7 and MT=2, as well as natural abundances of isotopes, were extracted from the NIST database \cite{Sears1992}. For further insights into the first-principles calculations and thermal neutron scattering analysis of \mgft\ as a neutron filter, readers are encouraged to refer to Ref.~\cite{Al-Qasir:aj5286}. Fig. \ref{fig:Xsec_MgF2} compares the calculated total cross section at room temperature with measured data \cite{Barker:ks5194}. It also displays the incoherent inelastic scattering cross sections of Mg and F in \mgft\ as well as their absorption cross sections. The incoherent elastic cross sections are excluded for clarity, as they are several orders of magnitude smaller than the incoherent inelastic cross sections.

\subsubsection{Beryllium Fluoride Neutron Filter (\beft)} 
\label{sec:beffil}

The Be nucleus has a smaller absorption and higher bound scattering cross section than that of Mg, which suggests that \beft\ is also a promising neutron filter. The \beft\ crystal structure has a trigonal symmetry and P3$_1$21 space group. It has a tetrahedral network where each Be atom is bonded to four F atoms, and each F atom is bonded to two Be atoms. The hexagonal primitive unit cell has nine atoms. A 2 x 2 x 2 supercell containing 72 atoms was constructed to determine the vibrational properties using the direct method. First-principles projector augmented wave method as implemented in the \texttt{VASP} \VASP\ was used to calculate the Hellmann-Feynman forces. The calculations were conducted at the GGA level, employing the PBE functional \cite{PhysRevB.54.16533}. The Phonopy software \cite{Togo2015} was utilized to calculate the interatomic force constants and obtain the partial PDOSs for Be and F in BeF$_2$.

\begin{figure}
    \centering
    \includegraphics[width=1.0\columnwidth]{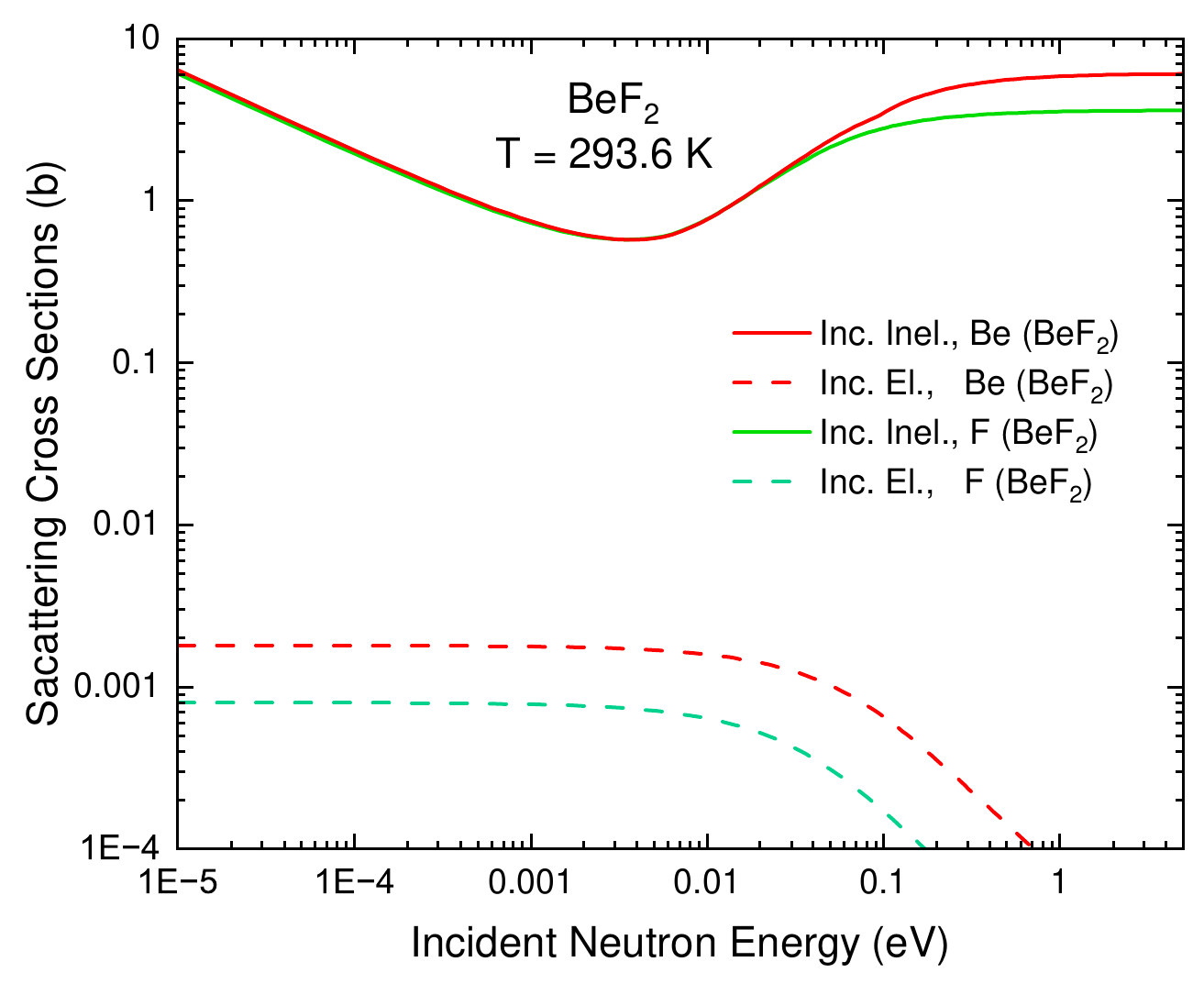}
    \caption{(Color online) Calculated scattering cross sections (solid curves) Incoherent inelastic of Be (red) and F (green) in \beft\, and (dashed curves) incoherent elastic of Be (red) and F (green) in BeF$_2$.}
    \label{fig:Xsec_BeF2}
\end{figure}

The \LEAPR\ module of the \NJOY\ code \cite{NJOY} was used to compute the TSL for Be (MAT3137) and F (MAT3136) in \beft\, utilizing the partial PDOSs for Be and F at 293.6~K under the incoherent approximation. Additionally, the incoherent elastic scattering cross section was calculated. Cross sections for free atoms and AWR for Be and F were taken from nuclide evaluations in ENDF/B-VIII.0 \cite{Brown2018}. Incoherent bound scattering cross sections for MT=2 and MF=7, along with natural isotope abundances, were sourced from the NIST database \cite{Sears1992}. For more details of the first-principles calculations and thermal neutron scattering analysis of \beft\ as a neutron filter, readers are directed to Ref \cite{Al-Qasir:aj5286}. Fig. \ref{fig:Xsec_BeF2} shows the room temperature incoherent inelastic and incoherent elastic scattering cross sections of Be and F in BeF$_2$.

\section{PHOTONUCLEAR SUBLIBRARY}
\label{sec:photonuclear}



In the late 1990s, the first IAEA CRP on photonuclear data was chaired by Chadwick~\cite{IAEAPhoto1999}. During this period, LANL developed evaluations for various isotopes of hydrogen (D),  carbon, nitrogen, oxygen, aluminum, silicon, calcium, copper, tantalum, tungsten, lead, uranium, plutonium, neptunium, and americium. The actinides were developed through a LANL-Saclay (CEA) collaboration. Some of these evaluations were incorporated into the IAEA database. The results of the LANL-CEA collaboration were documented in three Nuclear Science and Engineering journal papers~\cite{Chadwick:2003,White:2003,Giacri-Mauborgne:2006} and the comprehensive ENDF/B-VII.0 publication in 2007. These evaluations were also released alongside over 100 others in the IAEA's 1999 library -- IAEA1999~\cite{IAEAPhoto1999}.

To meet the increasing demand for precise and dependable photonuclear data, the IAEA conducted a new CRP entitled  \ql Updating the IAEA Photonuclear Data Library and generating a Reference Database for Photon Strength Functions\qr from 2016 to 2019 \cite{Goriely2019,Kawano2020}. The component dedicated to updating the IAEA1999 photonuclear data library was chaired by Kawano. This project resulted in the creation of the IAEA2019 Photonuclear Data Library, detailed in Ref.~\cite{Kawano2020}, available on the IAEA website at \url{https://www-nds.iaea.org/photonuclear}. The library encompasses photon absorption data, total and partial photoneutron reaction cross-sections, and neutron spectra for 220 isotopes.
The 220 isotopes included in the IAEA2019 Photonuclear Data Library \cite{Kawano2020} fall into five categories:
\begin{itemize}
\item Structural, shielding, and bremsstrahlung target materials: Be, Al, Si, Ti, V, Cr, Fe, Co, Ni, Cu, Zn, Zr, Mo, Sn, Ta, W, and Pb;
\item Biological materials: C, N, O, Na, S, P, Cl, and Ca;
\item Fissionable materials: Th, U, Np, and Pu;
\item Other materials: H, K, Ge, Sr, Nb, Pd, Ag, Cd, Sb, Te, I, Cs, Sm, and Tb;
\item Potential diagnostic and therapeutic radionuclides.
\end{itemize}

With the emergence of new facilities generating intense photon beams, the photo-production of certain significant radionuclides could emerge as a competitive alternative to traditional methods, utilizing neutrons from highly enriched Uranium reactors or charged-particle beams \cite{Habs2011, Nichols2014, INDC0596, INDC0717}. The identified radionuclides for potential diagnostic and therapeutic applications in nuclear medicine suitable for production via photonuclear reactions include:\nuc{168,170}{Er} for the production of \nuc{167,169}{Er}, \nuc{187}{Re} for the production of \nuc{186}{Re},\nuc{226}{Ra} for the production of \nuc{225}{Ra} which further decays to \nuc{225}{Ac}, \nuc{98}{Ru} for the production of \nuc{97}{Ru}, \nuc{194}{Pt} for the production of \nuc{193}{Pt}, \nuc{132}{Xe} for the production of \nuc{131}{I}, \nuc{162}{Dy} for the production of \nuc{161}{Tb}, and \nuc{178}{Hf} for the production of \nuc{177}{Lu}.

In brief, the 164 isotopes from the IAEA1999 library \cite{IAEAPhoto1999} underwent comprehensive reevaluation, incorporating new data when available, results from experimental-based assessments by Varlamov,
updated GDR parameters from the recently revised Atlas of GDR parameters \cite{Plujko2018}, and enhanced reaction models. Additionally, evaluations were conducted for 37 isotopes, not included in IAEA1999, with available experimental data, along with the nine isotopes identified for medical applications. Where experimental data was lacking, model predictions were
provided for evaluation. A total of 220 isotopes were assessed, encompassing all available experimental data on photo-absorption cross sections, photo-neutron production cross sections
and yields, as well as partial photo-neutron and photo-charged-particle cross sections from the EXFOR database \cite{EXFOR} up to April 2019.

Particular attention was given to re-assessing the partial photoneutron cross sections from multiplicity sorting measurements performed at LLNL and Saclay between 1960 and 1990. By applying a new approach using physical criteria, as detailed in Ref.~\cite{Kawano2020}, Varlamov identified issues in the $1n$, $2n$, and $3n$ partial photoneutron cross sections of several Livermore and Saclay measurements. These criteria were consistent with the new measurements obtained using the neutron-multiplicity sorting technique \cite{Utsunomiya2017}. In total, photonuclear data for 34 nuclides were assessed, and new recommendations were made within the CRP. While many of these recommendations were adopted by the CRP evaluators, they were not implemented in the JENDL and JAEA evaluations included in IAEA2019. Lastly, all the new measurements obtained through the direct neutron-multiplicity sorting technique \cite{Utsunomiya2017}, which were fully developed and implemented for the CRP, were incorporated into the evaluations. The assessments were extended to energies of 200 MeV to support accelerator-driven transmutation technologies, complementing the ongoing development of high-energy neutron and proton libraries for radiation transport simulation codes.

To summarize, significant progress was made in this CRP, benefiting from many new photonuclear measurements conducted since the first CRP. However, no new evaluations authored by the US were included in the resulting IAEA2019 file. For our new ENDF-VIII.1 evaluation of photon-induced reactions, we have adopted the majority of the new IAEA CRP evaluations. However, we have retained almost all the previous US (LANL) ENDF evaluations, only replacing them when new experimental data support the new IAEA work over our previous US evaluations. The previous ENDF-VIII.0=ENDF-VII.0 library is widely used in DOE photonuclear applications today, and these applications depend on high-quality photonuclear data. While the IAEA2019 documentation presents comparison plots against IAEA1999 data, these are not particularly useful for ENDF evaluations, as IAEA1999 does not equate to ENDF-VIII.0=ENDF-VII.0 in many cases. Therefore, comparisons should focus on ENDF versus IAEA2019 \cite{Kawano2020}, and indeed it was such comparisons that guided our evaluation decisions for ENDF-B/VIII.1. Our detailed principles for the choices we made were discussed in Sec.~\ref{sec:Principles}.

Photonuclear measurement discrepancies have long posed challenges, particularly regarding cross section magnitudes and multiplicity sorting assessments (e.g., LLNL versus Saclay versus Russian data). Resolving these discrepancies requires care and study. Conclusions drawn in 1999, such as normalization techniques for LLNL 1980 (Caldwell-Berman) data \cite{Fultz1962,Bramblett1963,Bramblett1964,Berman1969b,Alvarez1979,Berman1987} versus Saclay (Veyssiere) \cite{Bergere1968,Veyssiere1970} and Russian bremsstrahlung (Gurevich) data \cite{Gurevich1976}, are provisional and need continuous scrutiny. To date, we are not aware of peer-reviewed papers that have contested the 1999 reasoning by Varlamov concerning the magnitude of $^{235}$U photonuclear data \cite{Varlamov1987,IAEAPhoto1999}, as adopted by ENDF/B-VIII.0=ENDF/B-VII.0, IAEA1999, and Obninsk. Varlamov analysis was based on a least-square fit of available data as discussed in EXFOR entry M030 \cite{Varlamov1987} and recently reviewed in Ref.~\cite{Varlamov2007}. New ratio data from the collaboration between the Triangle Universities Nuclear Laboratory (TUNL) and LLNL \cite{Krishichayan2018} do not support the latest IAEA evaluation adopted from JAEA evaluations (JENDL/PD-2016 and JENDL(u)) over the earlier ENDF data for the $^{235}$U/$^{239}$Pu ratio.

\subsection{Evaluation choices}
For $^{239}$Pu, the existing ENDF evaluation appears superior versus the new IAEA evaluation. ENDF/B-VIII.0 provides a better fit to the LLNL Berman data. In the range of 7-17 MeV, the integrated ENDF/measured-data ratio is 1.01, while IAEA2019/measured-data is 1.025. Both evaluations aimed to model the LLNL Berman data, but ENDF/B-VIII.0 shows a superior fit for ($\gamma$,F), ($\gamma$,1n), ($\gamma$,2n), and ($\gamma$,xn). Hence, ENDF/B-VIII.0=ENDF/B-VII.0 is retained for ENDF-VIII.1.

Regarding $^{235}$U, many of the various measured data sets are discrepant, and ENDF and IAEA2019 (JAEA) made different evaluation decisions. New ratio data from TUNL/LLNL  at the Duke University Higgs facility \cite{Krishichayan2018} do not support IAEA2019's evaluations of the $^{235}$U/$^{239}$Pu ratio. As noted earlier, the original ENDF normalization choice was based on Varlamov analysis \cite{Varlamov1987}; we are not aware of a peer-reviewed paper that challenges Varlamov logic \cite{Varlamov1987,Varlamov2007}, so for now ENDF-VIII.0=ENDF-VII.0 is retained for ENDF-VIII.1. It is prudent to await  new measurements  before making changes to ENDF here.

For many of the other LANL photonuclear evaluations in ENDF/B-VIII.0=ENDF/B-VII.0, no new accurate photonuclear measurements have been made. Consequently, the LANL ENDF/B-VIII.0=ENDF/B-VII.0 evaluations will be retained for ENDF/B-VIII.1 for these cases, as are listed below. For $^{181}$Ta, new IAEA2019 evaluation for $^{181}$Ta follows new NewSUBARU measured data, which appear reasonable; therefore, we adopt IAEA2019 evaluation for this isotope.

Below we list the sources of the evaluations used in \ENDF:

\begin{itemize}

\item Nuclei with evaluated files taken from the IAEA CRP (IAEA-PD2019)~ \cite{Kawano2020} without any changes: 
\nuc{3}{He}, 
\nuc{6}{Li}, 
\nuc{7}{Li}, 
\nuc{9}{Be}, 
\nuc{13}{C}, 
\nuc{14}{C}, 
\nuc{15}{N}, 
\nuc{17}{O}, 
\nuc{18}{O}, 
\nuc{19}{F}, 
\nuc{23}{Na}, 
\nuc{24}{Mg}, 
\nuc{25}{Mg}, 
\nuc{26}{Mg}, 
\nuc{27}{Si}, 
\nuc{29}{Si}, 
\nuc{30}{Si}, 
\nuc{32}{S}, 
\nuc{33}{S}, 
\nuc{34}{S}, 
\nuc{36}{S}, 
\nuc{35}{Cl}, 
\nuc{37}{Cl}, 
\nuc{36}{Ar}, 
\nuc{38}{Ar}, 
\nuc{40}{Ar}, 
\nuc{39}{K}, 
\nuc{40}{K}, 
\nuc{41}{K}, 
\nuc{42}{Ca}, 
\nuc{43}{Ca}, 
\nuc{44}{Ca}, 
\nuc{46}{Ca}, 
\nuc{48}{Ca}, 
\nuc{45}{Sc}, 
\nuc{46}{Ti}, 
\nuc{47}{Ti}, 
\nuc{48}{Ti}, 
\nuc{49}{Ti}, 
\nuc{50}{Ti}, 
\nuc{50}{V}, 
\nuc{51}{V}, 
\nuc{50}{Cr}, 
\nuc{52}{Cr}, 
\nuc{53}{Cr}, 
\nuc{54}{Cr}, 
\nuc{55}{Mn}, 
\nuc{56}{Fe}, 
\nuc{57}{Fe}, 
\nuc{58}{Fe}, 
\nuc{59}{Co}, 
\nuc{58}{Ni}, 
\nuc{60}{Ni}, 
\nuc{61}{Ni}, 
\nuc{62}{Ni}, 
\nuc{64}{Ni}, 
\nuc{65}{Cu}, 
\nuc{64}{Zn}, 
\nuc{66}{Zn}, 
\nuc{67}{Zn}, 
\nuc{68}{Zn}, 
\nuc{70}{Zn}, 
\nuc{70}{Ge}, 
\nuc{72}{Ge}, 
\nuc{73}{Ge}, 
\nuc{74}{Ge}, 
\nuc{76}{Ge}, 
\nuc{75}{As}, 
\nuc{76}{Se}, 
\nuc{78}{Se}, 
\nuc{80}{Se}, 
\nuc{82}{Se}, 
\nuc{84}{Sr}, 
\nuc{86}{Sr}, 
\nuc{87}{Sr}, 
\nuc{88}{Sr}, 
\nuc{90}{Sr}, 
\nuc{89}{Y}, 
\nuc{90}{Zr}, 
\nuc{91}{Zr}, 
\nuc{92}{Zr}, 
\nuc{93}{Zr}, 
\nuc{94}{Zr}, 
\nuc{96}{Zr}, 
\nuc{93}{Nb}, 
\nuc{94}{Nb}, 
\nuc{92}{Mo}, 
\nuc{94}{Mo}, 
\nuc{95}{Mo}, 
\nuc{96}{Mo}, 
\nuc{97}{Mo}, 
\nuc{98}{Mo}, 
\nuc{100}{Mo}, 
\nuc{98}{Ru}, 
\nuc{102}{Pd}, 
\nuc{104}{Pd}, 
\nuc{105}{Pd}, 
\nuc{106}{Pd}, 
\nuc{107}{Pd}, 
\nuc{108}{Pd}, 
\nuc{110}{Pd}, 
\nuc{107}{Ag}, 
\nuc{108}{Ag}, 
\nuc{109}{Ag}, 
\nuc{106}{Cd}, 
\nuc{108}{Cd}, 
\nuc{110}{Cd}, 
\nuc{111}{Cd}, 
\nuc{112}{Cd}, 
\nuc{113}{Cd}, 
\nuc{114}{Cd}, 
\nuc{116}{Cd}, 
\nuc{115}{In}, 
\nuc{112}{Sn}, 
\nuc{114}{Sn}, 
\nuc{115}{Sn}, 
\nuc{116}{Sn}, 
\nuc{117}{Sn}, 
\nuc{118}{Sn}, 
\nuc{119}{Sn}, 
\nuc{120}{Sn}, 
\nuc{122}{Sn}, 
\nuc{124}{Sn}, 
\nuc{121}{Sb}, 
\nuc{123}{Sb}, 
\nuc{120}{Te}, 
\nuc{122}{Te}, 
\nuc{123}{Te}, 
\nuc{124}{Te}, 
\nuc{125}{Te}, 
\nuc{126}{Te}, 
\nuc{128}{Te}, 
\nuc{130}{Te}, 
\nuc{127}{I}, 
\nuc{129}{I}, 
\nuc{132}{Xe}, 
\nuc{133}{Cs}, 
\nuc{135}{Cs}, 
\nuc{137}{Cs}, 
\nuc{138}{Ba}, 
\nuc{139}{La}, 
\nuc{140}{Ce}, 
\nuc{142}{Ce}, 
\nuc{141}{Pr}, 
\nuc{142}{Nd}, 
\nuc{143}{Nd}, 
\nuc{144}{Nd}, 
\nuc{145}{Nd}, 
\nuc{146}{Nd}, 
\nuc{148}{Nd}, 
\nuc{150}{Nd}, 
\nuc{144}{Sm}, 
\nuc{147}{Sm}, 
\nuc{148}{Sm}, 
\nuc{149}{Sm}, 
\nuc{150}{Sm}, 
\nuc{151}{Sm}, 
\nuc{152}{Sm}, 
\nuc{154}{Sm}, 
\nuc{153}{Eu}, 
\nuc{156}{Gd}, 
\nuc{160}{Gd}, 
\nuc{158}{Tb}, 
\nuc{162}{Dy}, 
\nuc{163}{Dy}, 
\nuc{166}{Er}, 
\nuc{170}{Er}, 
\nuc{169}{Tm}, 
\nuc{175}{Lu}, 
\nuc{174}{Hf}, 
\nuc{176}{Hf}, 
\nuc{177}{Hf}, 
\nuc{178}{Hf}, 
\nuc{179}{Hf}, 
\nuc{180}{Hf}, 
\nuc{186}{Os}, 
\nuc{188}{Os}, 
\nuc{189}{Os}, 
\nuc{190}{Os}, 
\nuc{192}{Os}, 
\nuc{209}{Bi}.

\item Nuclei with evaluated files taken from the IAEA CRP (IAEA-PD2019)~ \cite{Kawano2020} after fixes to level index on isomeric production were implemented: 
\nuc{103}{Rh}, 
\nuc{159}{Tb}, 
\nuc{165}{Ho}, 
\nuc{181}{Ta}.

\item Nuclei with evaluated files adopted from the IAEA CRP (IAEA-PD2019)~ \cite{Kawano2020} after minor fixes were implemented:
\nuc{185}{Re}, 
\nuc{187}{Re}, 
\nuc{194}{Pt}, 
\nuc{197}{Au}, 
\nuc{226}{Ra}, 
\nuc{232}{Th}, 
\nuc{233}{U}, 
\nuc{234}{U}, 
\nuc{236}{U}, 
\nuc{238}{Pu}, 
\nuc{241}{Pu}.

\item Nuclei with evaluated files kept from \prENDF\ without changes: 
\nuc{2}{H}, 
\nuc{12}{C}, 
\nuc{14}{N}, 
\nuc{16}{O}, 
\nuc{27}{Al}, 
\nuc{28}{Si}, 
\nuc{40}{Ca}, 
\nuc{63}{Cu}, 
\nuc{180}{W}, 
\nuc{182}{W}, 
\nuc{183}{W}, 
\nuc{184}{W}, 
\nuc{186}{W}, 
\nuc{206}{Pb}, 
\nuc{207}{Pb}, 
\nuc{208}{Pb}, 
\nuc{235}{U}, 
\nuc{238}{U}, 
\nuc{237}{Np}, 
\nuc{239}{Pu}.

\item Nuclei with evaluated files adopted from \prENDF\ after fixes were implemented: 
\nuc{40}{Ca} (Fix to discrepant masses), 
\nuc{14}{N} (added missing photon distributions to MT=102 with energies and branching ratios taken from ENSDF).

\item Nuclei with evaluated files kept from \prENDF\ after minor fixes were implemented: 
\nuc{240}{Pu}, 
\nuc{241}{Am}.

\item New evaluated file adopted from JENDL-5~\cite{jendl5}: \nuc{242}{Pu}.

\end{itemize}

\section{CHARGED PARTICLE REACTION SUBLIBRARIES}
\label{sec:chargedparticles-sublib}


The changes from \ENDF\ relative to \prENDF\ for the charged particle sublibraries are described in the following sections.

\subsection{Alphas sublibrary}
\label{subsec:cp:alphas}


The incident alpha-particle sublibrary expanded in the ENDF/B-VIII.1 release to include a
\nuc{6}{Li} evalution from the LLNL ECPL  as well as 
new evaluations for \nuc{9}{Be}, \nuc{17}{O} and \nuc{18}{O} produced by NNL.

\begin{itemize}

	\item \nuc{4}{He}: The evaluation is nearly identical to ENDF/B-VIII.0. Minor changes were made to the target mass (adding the electron mass), and an invalid interpolation flag was fixed for Coulomb elastic outgoing product distributions.
	\item \nuc{6}{Li}: The evaluation from the LLNL ECPL~\cite{white:1991ecp} is adopted. In addition to elastic scattering, the evaluation includes ($\alpha, p$) and ($\alpha, d$) reactions.
\item \nuc{9}{Be}: The NNL adopted the JENDL/AN-2005 evaluation but made additional modifications to improve modeling of neutron spectra produced by Am-Be neutron sources. The ENDF/B-VIII.1 evaluation produces a better match to the experimental neutron spectrum, especially for outgoing neutron energies below 1.5 MeV.
\item \nuc{17, 18}{O}: The NNL produced a modified version of the JENDL/AN-2005 evaluations for two Oxygen isotopes. The modifications focused on improving modeling of outgoing neutron spectra from ($\alpha, n$) reactions in actinide oxide fuels.

\end{itemize}

\subsubsection{\nuc{9}{Be}}

For incident $\alpha$ particles above a threshold near 4.5 MeV, the measured low-energy neutron yield cannot be adequately explained by the $^9$Be($\alpha$,n)$^{12}$C reaction alone, as the $^9$Be($\alpha,\alpha'$)$^{9}$Be* $\rightarrow$ $^{8}$Be+n ``breakup'' reaction becomes significant~\cite{romain1962low,obst1972reaction,geiger1975radioactive}. Cross sections for the $^9$Be($\alpha,n$)$^{12}$C and breakup reaction are given by {MT=4} and {MT=22}, respectively.

Am-Be sources are an application where the neutron yield and energy spectrum resulting from ($\alpha,n$)-type reactions with $^{9}$Be are of interest.  In this case, $^{241}$Am is the primary source for $\alpha$ particles, with $\alpha$ decay energies of about 5.5~MeV. At this maximum $\alpha$ energy, the breakup reaction accounts for about 30\% of total neutron yield~\cite{geiger1976evaluation}, with almost all below 1.5 MeV~\cite{obst1972reaction,geiger1975radioactive}. A similar situation exists for Pu-Be sources, where $^{239}$Pu has $\alpha$ decay energies of about 5.15~MeV~\cite{anderson1972neutron}. The NNL Monte Carlo radiation transport code MC21 was used to simulate the neutron energy spectrum for an Am--Be source using the JENDL/AN-2005 (NMOD=2) evaluation for $^{9}$Be. A continuous-slowing-down model is used for $\alpha$ particles in MC21 without transport or use of MT=2 elastic cross sections~\cite{griesheimer2017line}. Fission neutrons were accounted for but were statistically insignificant. The results generally compared well to experimental data, except for neutron energies below about 1.5~MeV. In this low-energy region, results were substantially low compared with Marsh \etal~\cite{marsh1995high}, one of the few publications reporting high-quality Am--Be neutron energy spectrum measurements including energies from near-zero to 1.5~MeV.

The JENDL/AN-2005 evaluation (NMOD=2) uses experimental data from Gibbons \etal~\cite{gibbons} in the evaluation of the MT=201 total cross sections. MT=22 cross sections are based on experimental data from Obst \etal~\cite{obst1972reaction}.  As a result, above the breakup threshold, the MT=4 cross sections (MT4 = MT201 – MT22) depend on the difference between two separately normalized components derived from completely independent experiments.  MT=50-52 were informed by measurements from Van Der Zwan \etal~\cite{van19709be}, and the MT=50-52,91 partials were set to maintain consistency with MT=4.

Geiger \etal~\cite{geiger1976evaluation} measured the MT=50-53 partial cross sections (physically comprising all of MT=4 for the $\alpha$ energies studied) and MT=201 total cross sections, allowing MT=22/MT=201 ratios to be determined directly in a single experiment.  Based on Ref.~\cite{geiger1976evaluation}, the JENDL/AN-2005 (NMOD=2) breakup reaction fraction appears to be highly underestimated for $\alpha$ energies from the minimum threshold for MT=22 up to about 6.0~MeV. Above about 7.0~MeV, the JENDL fraction is generally consistent with Ref.~\cite{geiger1976evaluation}.

The following modifications were made to $^{9}$Be MF=3 data by NNL. Within the $\alpha$ energy range 4.64~MeV (the JENDL/AN-2005 minimum threshold for MT=22) to 7.90~MeV (the maximum energy studied in Ref.~\cite{geiger1976evaluation}), MT=4 and MT=22 were adjusted to match the experimentally observed MT=4/MT=22 ratios from Ref.~\cite{geiger1976evaluation}. The MT=201 total cross sections were unchanged, and the sum of MT=4 + MT=22 remains equal to the original MT=201 total cross sections. The internal ratios for MT=50-52,91 were also unchanged. However, MT=50-52,91 were rescaled to be consistent with the modified MT=4. The original energy grids were retained in all cases. No changes were made to MF=6 data by NNL. IAEA added elastic scattering MT=2 data from TENDL-2019 to MF=3 and MF=6 to complete the evaluation.  

The modified NNL evaluation for $^{9}$Be was retested in MC21. The resulting Am--Be neutron energy spectrum agreed well with Ref.~\cite{marsh1995high} and showed reasonable agreement in the low-energy region below 1.5~MeV,  as shown in Ref.~\cite[Fig. 8]{griesheimer2017line}. The MC21 results and improvement in the predicted neutron energy spectrum are consistent with those of Shores \etal~\cite{shores2003new}, where the SOURCES-4C code is specifically modified, based on Ref.~\cite{geiger1976evaluation}, to better account for the $^{9}$Be breakup reaction.

\subsubsection{\nuc{{17},{18}}{O}}

The original JENDL/AN-2005 evaluations (NMOD=2) use Kalbach-Mann systematics (LAW=1, LANG=2) for MF=6 neutron energy/angle distributions for MT=4,22 for $^{17}$O (MT=4,16,22 for $^{18}$O).  No neutron energy/angle distributions are given for the MT=50-53,91 partial cross sections for $^{17}$O (MT=50-54,91 for $^{18}$O), and the MF=6 Kalbach-Mann distributions for MT=4 are decoupled from the constituent partial cross sections and associated residual energy states.

The NNL Monte Carlo radiation transport code MC21 was used to simulate the neutron energy spectrum for thick $^{238}$UO$_2$ and $^{238}$PuO$_2$ using the original JENDL/AN-2005 (NMOD=2) evaluations for $^{17}$O and $^{18}$O.  A continuous-slowing-down model is used for $\alpha$ particles in MC21 without transport or use of MT=2 elastic cross sections~\cite{griesheimer2017line}. Fission neutrons were accounted for but were of lesser significance. While neutron yields were predicted very well, the resulting neutron energy spectrum compared very poorly to published experimental measurements~\cite{jacobs1983energy,herold1968neutron,anderson1967neutron,anderson1980neutron}.

Kalbach-Mann systematics may provide a reasonable approximation for ($\alpha,n$) neutron energy/angle distributions for high-incident-energy $\alpha$ particles in the tens to hundreds of MeV range, where higher-level residual excitations dominate or where a continuum excitation approximation may be appropriate. Kalbach-Mann systematics may also be an acceptable option for ($\alpha,n$)-type reactions with multiple particle emission or when no partial cross section information is available for discrete residual states.

For the incident $\alpha$ energy spectrum ranging from actinide decay energies (4-8~MeV) down to below the minimum threshold for ($\alpha,n$) reactions, the total ($\alpha,n$) neutron yield will be largely due to reactions where the residual is left in the ground state or in low-level excited states. These low-energy residual states have widely separated energy levels that strongly influence the resulting neutron energy spectrum.  Therefore, the use of Kalbach-Mann systematics is inappropriate for characterizing ($\alpha,n$) neutron energy/angle distributions for applications dependent on $\alpha$ particle emission via actinide decay.

The following modifications were made to $^{17}$O and $^{18}$O MF=6 data by NNL. MT=4 was removed and replaced with MT=50-53,91 for $^{17}$O (MT=50-54,91 for $^{18}$O). MT=50-54 use LAW=3 (isotropic two-body kinematics) for neutron emission based on AWR, AWP, and QI from the respective MF=3 MT=50-54 sections. MT=91 uses the same Kalbach-Mann systematics neutron distribution data previously used for MT=4. No changes were made to MF=3 data by NNL. IAEA added elastic scattering MT=2 data from TENDL-2019 to MF=3 and MF=6 to complete the evaluation.  

\begin{figure}
\centering
\includegraphics[scale=.35,clip, trim = 0mm 0mm 0mm 0mm]{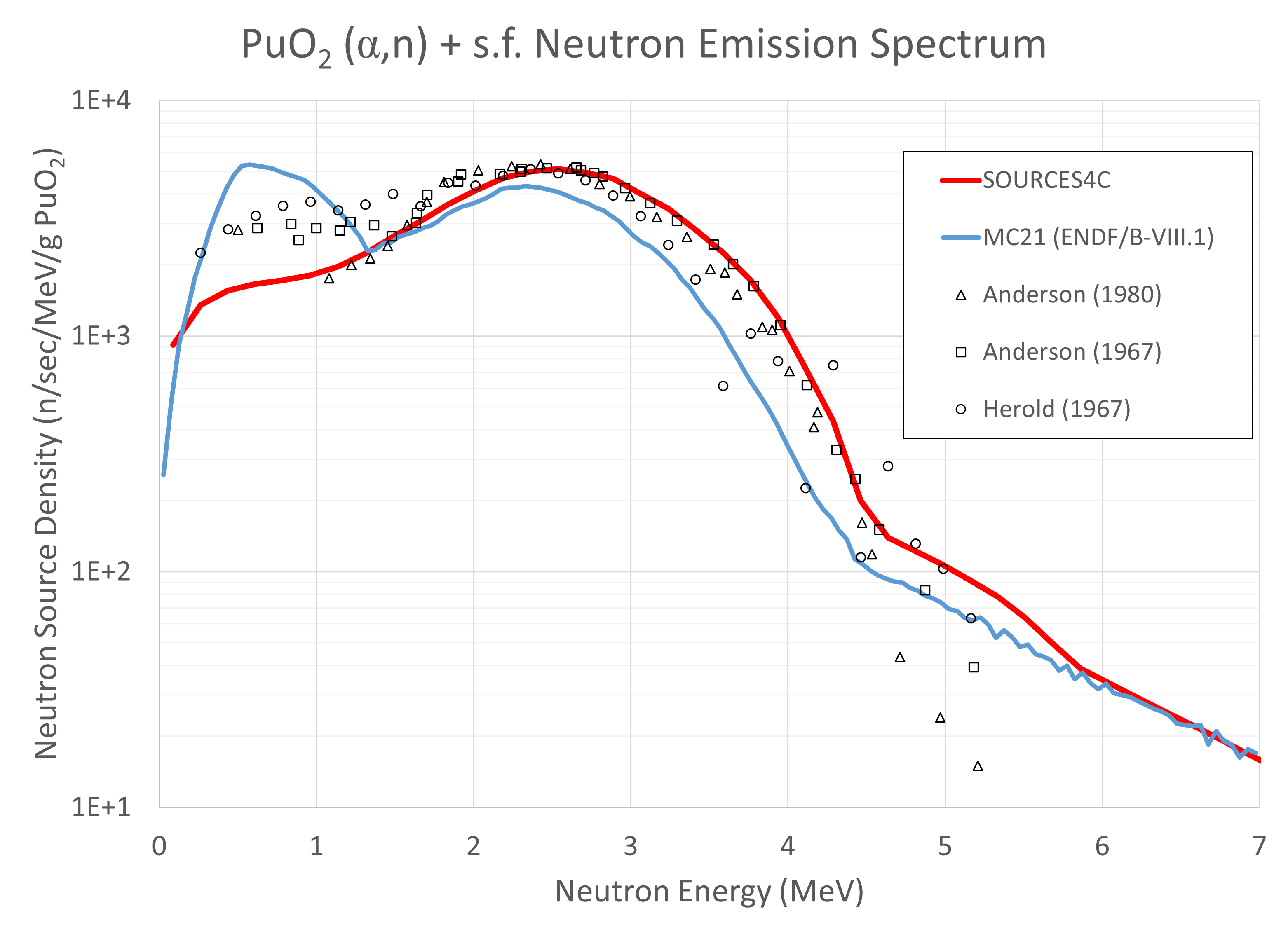} 
\caption{Comparison of MC21 neutron emission energy spectra for PuO$_2$ using ENDF/B-VIII.1 to SOURCES-4C and to experimental data Refs.~\cite{anderson1967neutron,anderson1980neutron,griesheimer2017line}.}
\label{fig:alphas_PuO2}
\end{figure}

The modified NNL evaluations for $^{17}$O and $^{18}$O were retested in MC21. The resulting neutron energy spectra for thick $^{238}$UO$_2$ and $^{238}$PuO$_2$ compared well to Refs.~\cite{jacobs1983energy,herold1968neutron,anderson1967neutron,anderson1980neutron,griesheimer2017line}, 
as shown in Ref.~\cite[Fig.~3]{griesheimer2017line}
and Fig.~\ref{fig:alphas_PuO2}, respectively. Below 1.5 MeV in Fig.~\ref{fig:alphas_PuO2} results could be improved further by explicit inclusion of MT=53
with two-body kinematics instead of including it as part of MT=91
with  Kalbach-Mann systematics.

\subsection{Helions sublibrary}
\label{subsec:cp:helions}


For the incident  \nuc{3}{He} sublibrary, the \nuc{3}{He} + \nuc{3}{He} evaluated file from \prENDF\ was kept.

For \nuc{3}{He} + \nuc{4}{He}, a new evaluation from the INDEN light elements collaboration was adopted.
The new evaluation is based on an R-Matrix fit using the Brune parameterization, including experimental data from multiple experiments that populated the same Be7 system.
Some data renormalization was required, including revising estimated uncertainties to take discretization error into account.
The fit included 14 resonances including some broad 3/2+ and 5/2+ poles contributing to the background.
Details about the evaluation were presented as part of the 2022 CSEWG meeting~\cite{CSEWG2022-Thompson}.
The R-Matrix resonance parameters are included in the evaluation, but since resonance reconstruction for incident charged particles
has not been fully tested in processing codes, the reconstructed cross sections are also provided for MTs 2 and 600,
and the LRP=2 flag is set to indicate that resonance reconstruction is not needed.

The \nuc{6}{Li} \ENDF\ evaluation replaced the (h,2n) and (h,p) cross sections and distributions from \prENDF\ with
values from the LLNL ECPL, while keeping the Coulomb elastic scattering from \prENDF.

A new evaluation of \nuc{7}{Li} was adopted from the LLNL ECPL. This evaluation includes (h,p), (h,d), (h,t), (h,n+p) reactions
as well as (h,a) reactions to both the ground state and second excited state in \nuc{6}{Li}.

\subsection{Deuterons sublibrary}
\label{subsec:cp:deuterons}


Updates to the deuteron sublibrary are being provided for target materials \nuc{3}{H}, \nuc{3}{He}, and \nuc{6}{Li}. Minor
modifications and corrections were made for tritium, which is based on an R-matrix evaluation~\cite{Hale:1987zz} and described for the
ENDF/B-VII.0 library~\cite{ENDFVII.0_2006}. Significant revisions to the $d+$\nuc{3}{He} reaction were made on the basis of an updated
R-matrix evaluation for the \nuc{5}{Li} system~\cite{osti_1770083} (which also results in an updated $p+$\nuc{4}{He} described below in
Section~\ref{sssec:p4he}).

\subsubsection{\nuc{3}{H}}


The previous evaluation of the $d+t$ reactions is summarized in Section VII.B.1 of the paper describing ENDF/B-VII.0
\cite[Section VII.B.1]{ENDFVII.0_2006}.  A few additions and corrections were made to this evaluation for ENDF/B-VIII.1.  The evaluation is based
fundamentally on a LANL R-matrix analysis of reactions in the $^5$He system at deuteron energies up to 10 MeV, which includes data for all
the reactions possible between $d+t$ and $n+^4$He channels, as well as branching-ratio measurements for T$(d,\gamma)^5$He/T$(d,n)^4$He.  The
addition of the branching-ratio data allowed us to include gamma-production cross sections from the first and second excited states of
$^5$He for the T$(d,\gamma)^5$He reaction in MF=3, MT=91.  Also, the integrated cross section for the inelastic process to the first excited
state of $^4$He, T$(d,n_1)^4$He$^*$ (MT=51), was extended to 20 MeV in MF=3 and MF=6 using extrapolations calculated from the $^5$He
R-matrix analysis.  The LR flag for this reaction was changed to LR=0 to reflect that it is a two-body reaction, with subsequent decay of
the $^4$He$^*$ to $p+t$.  The Q-values for the T$(d,n_0)$ and T$(d,n_1)$ reactions (MT=50,51) were corrected in MF=3.  And finally, the
integrated cross sections and angular distributions for the T$(d,n_0)^4$He reaction (MF=3,6 for MT=50) were extended to 40 MeV, using newer
results from the experimentally determined coefficients provided by M. Drosg (U. Vienna) \cite{Drosg15}.

\subsubsection{\nuc{3}{He}}
\label{subsec:d:3He}

The evaluation update for $d+$\nuc{3}{He} is derived from the multichannel unitary R-matrix evaluation of the \nuc{5}{Li} compound system
described in some detail in Ref.~\cite{osti_1770083}.  The top portion of Table~\ref{tab:5li-config} shows the channels along with the
channel-radius parameters $a_c$ of the R-matrix parametrization and the maximum value of the orbital angular momentum $\ell_{\text{max}}$
for the channel. The scattering and reaction processes (collectively ``Reactions'') of the \nuc{5}{Li} compound system are shown in the left
column of the lower portion of the table.  Energy ranges, numbers of data points and types of observables are shown in the remaining columns
of the lower portion.

\begin{table}[bhtp]
   \begin{center}
      \caption[width=0.3\textwidth]{\label{tab:5li-config}Channel configuration (top) and data summary (bottom) for the \nuc{5}{Li} system analysis.}
      \begin{tabular}{lcc}
\toprule \toprule
         Channel                           & $a_c$ (fm)       &  $\ell_{\text{max}}$      \\ \midrule
         $d$ +\nuc{3}{He}($\tfrac{1}{2}^+$)   &        4.8       &      4          \\
         $p$ +\nuc{4}{He}($0^+$     )         &        2.9       &      4          \\
         $p$ +\nuc{4}{He}$^*$($0^+$; 20.2 MeV)&        3.4       &      2          \\
         $d_0$+\nuc{3}{He}($\tfrac{1}{2}^+$)  &        5.1       &      0          \\
         \bottomrule
      \end{tabular}\\
      \begin{tabular}{ccrc}
         \toprule
         \multirow{2}{*}{Reaction}                         & $E_{\mathrm{proj}}$ range      & \# Data    & \multirow{2}{*}{Observables}                   \\
                                           &    (MeV)              &    Points  &                               \\
         \midrule
         \nuc{3}{He}($d,d$)\nuc{3}{He}     &  $0.32-10.0 $     & 2,229      &\begin{tabular}[c]{@{}c@{}}$\sigma(\theta),A_i, A_{ii}$, $C_{i,j}$,\\ $C_{ij,k}, K_{i,j'k'}, K_{ij,k'l'}$\end{tabular} \\
         \nuc{3}{He}($d,p$)\nuc{4}{He}     &  $0.13-10.0 $     & 3,839      &\begin{tabular}[c]{@{}c@{}}$\sigma(E),\sigma(\theta),A_i, A_{ii}$,\\ $C_{i,j}, K_{ij,k'}$              \end{tabular} \\
         \nuc{3}{He}($d,p$)\nuc{4}{He}$^*$ &  $3.70- 6.70$     &    28      & $\sigma(\theta)$ \\
         \nuc{4}{He}($p,p$)\nuc{4}{He}     &  $0.92-34.3 $     &   867      &$\sigma(E), \sigma(\theta), A_y, P_y$ \\
         \midrule
   Total:         &              & 6963       & \\
         \bottomrule \bottomrule
         \end{tabular}
   \end{center}
\end{table}

The column labeled ``Observables'' indicates the following data types: $\sigma(E)$, integrated cross section; $\sigma(\theta)$,
unpolarized angular distributions (energy-dependence suppressed); $A$, initial-state analyzing power; $P$, final-state polarization; $C$,
spin correlation coefficients; $K$, polarization transfer coefficients. (We have suppressed the indices $i,j,\ldots$ which take on
values $x,y,z$ for spins/polarization directions in configuration space.) All polarization and spin distributions are angular
distributions, which depend on the angle of the outgoing particle.  The $\chi$-squared per degree of freedom for the analysis is
$\chi^2/\text{dof}\simeq 2.7$ over 7,178 data points.

In the current evaluation, the differential cross section angular distributions in the center-of-mass frame
and the experimentally observed data for \nuc{3}{He}($d,d$)\nuc{3}{He} elastic scattering are fit to 
data up to $E_d \simeq 10$ MeV. (A previous, unpublished, LANL-internal evaluation, ``CP2011''
covered deuteron incident laboratory energies from $0 \le E_d \le 1.4$ MeV.)
New data has been added to both the existing region from
Refs.~\cite{brown:1954dh,brolley:1960hh,tombrello:1967de,jarmie:1974vc,jenny:1979ap}. The evaluated data has been encoded in the following
files and sections: MF=3 and MF=6, MT=600 for \nuc{3}{He}$(d,p)$\nuc{4}{He} and MF=6 for MT=2, \nuc{3}{He}$(d,d)$\nuc{3}{He}.



\subsubsection{\nuc{6}{Li}}


The evaluation of \nuc{6}{Li} is based on an R-matrix analysis of reactions in the $^8$Be system done at LANL in 2004 at
incident deuteron energies up to 5 MeV.  It includes data for $d+^6$Li elastic scattering in the 3 - 5 MeV range, for $^6$Li$(d,n)^7$Be in
the 0.1 - 3.7 MeV range, for $^6$Li$(d,p)^7$Li in the 0.1 - 5.0 MeV range, and for $^6$Li$(d,\alpha)^4$He in the 0.0 - 5 MeV range.  Smooth
extensions of these cross sections to 20 MeV based on experimental data were added in 2020.

\begin{figure}[tb]
   \centering
   \vspace{-2mm}
   \includegraphics[width=\columnwidth]{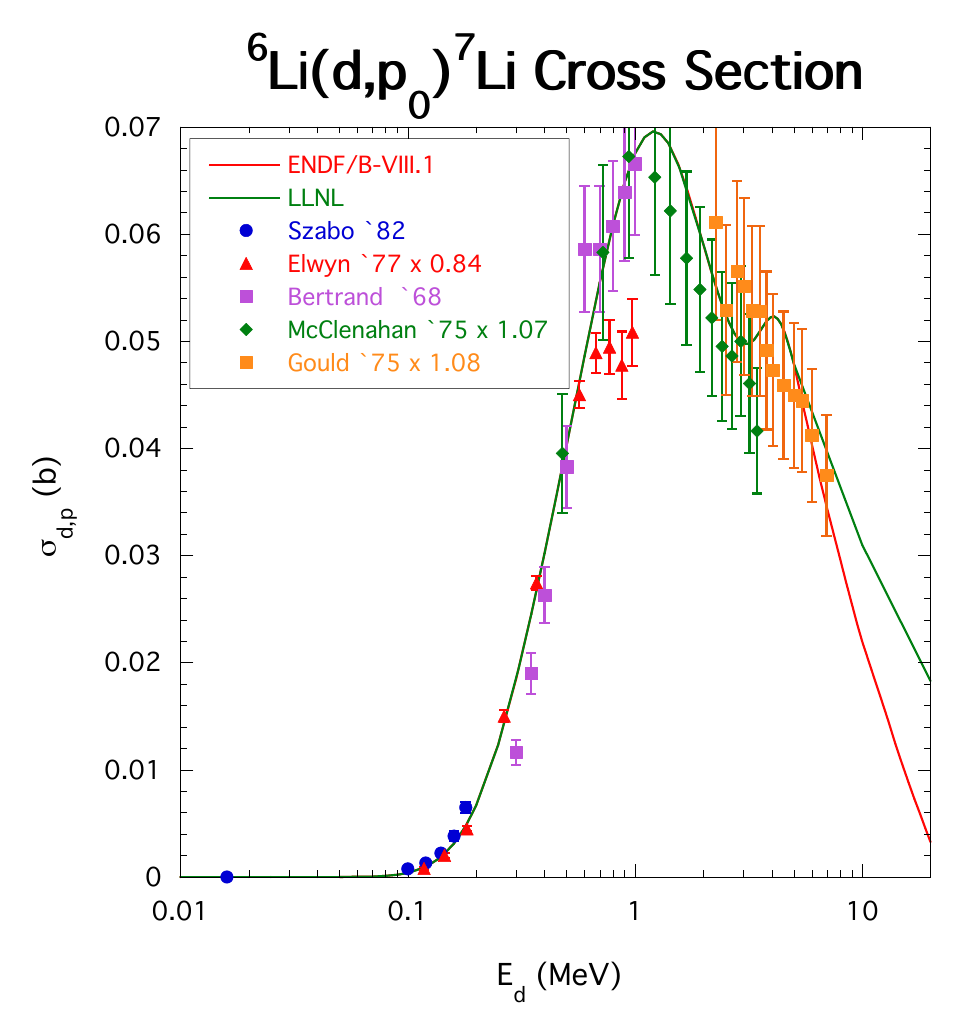}
   \vspace{-2mm}
   \caption{Measurements of the \nuc{6}{Li}(d,p)\nuc{7}{Li} cross section compared to the (LANL) ENDF/B-VIII.1 and LLNL evaluations at incident
   neutron energies between 0.01 and 20 MeV.  The experimental data are from
   Refs.~\cite{Body79,Elwyn77,Bertrand68,McClen75,Gould75}.}
   \label{6Lidpxs}
\end{figure}

\begin{figure}[tb]
   \centering
   \vspace{-2mm}
   \includegraphics[width=\columnwidth]{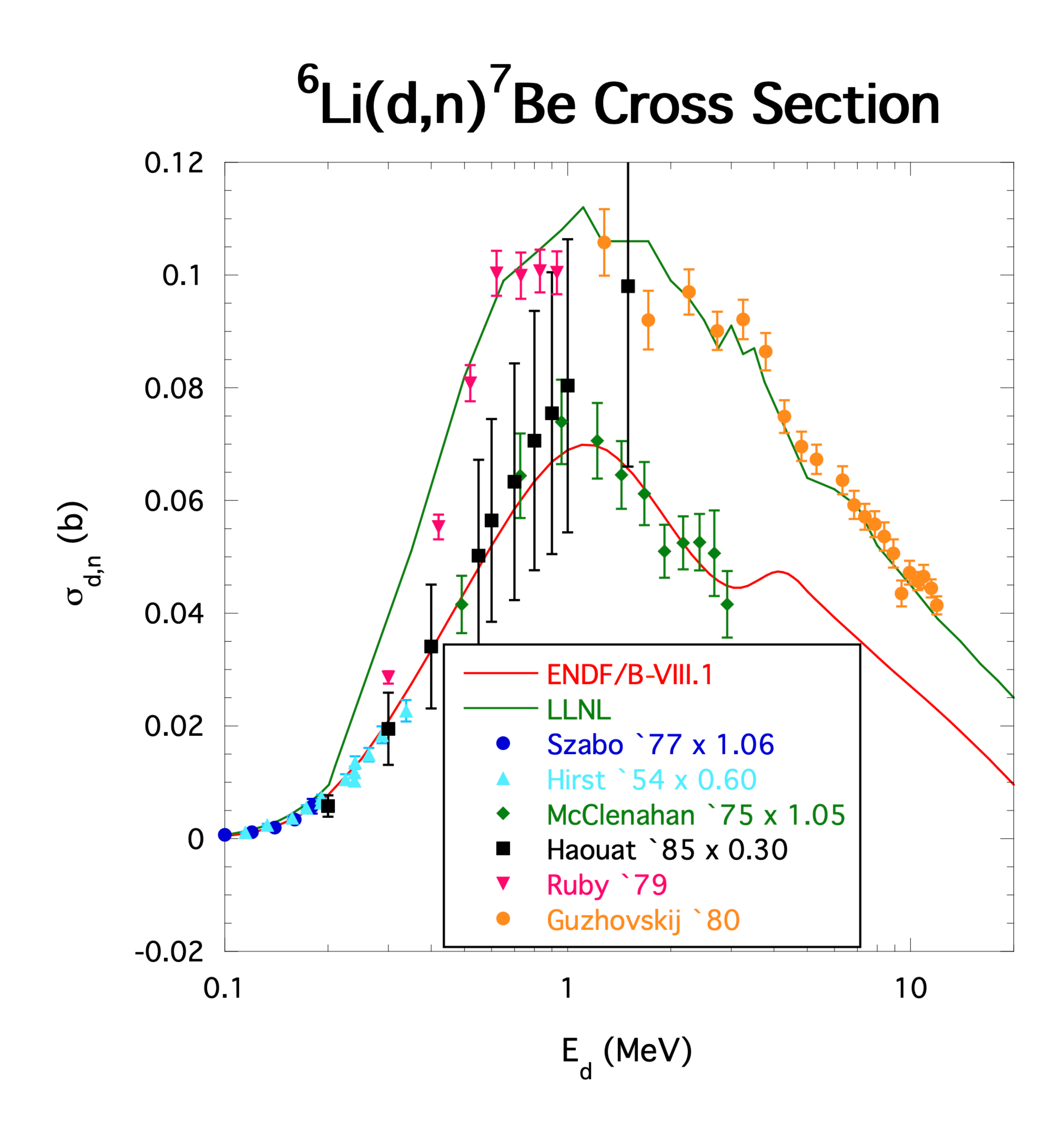}
   \vspace{-2mm}
   \caption{Measurements of the \nuc{6}{Li}(d,n)\nuc{7}{Be} cross section compared to the (LANL) ENDF/B-VIII.1 and LLNL evaluations at incident
   neutron energies between 0.1 and 20 MeV.  The experimental data are from
   Refs.~\cite{Szabo77,Hirst54,McClen75,Haouat85,Ruby79}, and \cite{Guzh80}. The data of Ruby, Guzhovskij, and
   possibly those of Haouat, contain contributions from excited-state neutrons, which are included in the LLNL curve.}
   \label{6Lidnxs}
\end{figure}

The analysis treats $p+^7$Li and $n+^7$Be as charge-conjugate channels, which imposes isospin ($T$) constraints on the reduced widths in the
$T=0$ and $T=1$ levels and relates the cross sections for reactions involving those channels.  Therefore, the cross sections for
$^6$Li$(d,n)^7$Be and $^6$Li$(d,p)^7$Li, for example, are expected to be similar.  This is illustrated in Figs.~\ref{6Lidpxs} and
\ref{6Lidnxs}, in which the evaluation is compared with some of the experimental data, and with an evaluation done at LLNL.  One sees that
the LLNL evaluation is same as LANL's for the $^6$Li$(d,p)^7$Li cross section at energies below 5 MeV but has a different extension to 20
MeV.  For the $^6$Li$(d,n)^7$Be reaction, however, the LLNL evaluation is substantially different from the LANL one at energies above 200
keV, following the higher data of Ruby \cite{Ruby79} and Guzhovskij \cite{Guzh80}.  The LANL evaluation is more consistent with the data of
McClenahan~\cite{McClen75}, which are roughly comparable in magnitude for the two reactions, in agreement with the charge-conjugate R-matrix
calculation.  The reason for this difference is that the measurements of Ruby and of Guzhovskij on which the LLNL evaluation is based
include neutrons recoiling from the ground- and first-excited states of $^7$Be, whereas the McClenahan measurement separates the two neutron
groups, and only the ground-state cross section is shown in Fig.~\ref{6Lidnxs}.  Presumably, the difference between the two evaluated curves
represents the contribution from the $^6$Li$(d,n_1)^7$Be$^*$ reaction.

Work is continuing on the $^8$Be system R-matrix analysis, in collaboration with S. Paneru, who is using the AZURE-2 R-matrix code to
cross-check the results of the LANL code EDA, and extend the results to higher energies~\cite{msrx-3fjr}.  This involves adding excited-state channels of
$n+^7$Be$^*$ and $p+^7$Li$^*$, and of course, more experimental data at higher energies~\cite{PhysRevC.110.044603}.  We expect improved evaluations will result from
this combined effort, not only for the $d+^6$Li reactions, but also for $n+^7$Be, $p+^7$Li, and $\alpha+^4$He.

\subsection{Protons sublibrary}
\label{subsec:cp:protons}


The proton sublibrary has been updated to include the proton-\nuc{4}{He} elastic scattering angular distributions. This evaluation has been
derived from an update of a previous LANL EDA-code based evaluation, not previously published, and is described in the following section.

\subsubsection{\nuc{4}{He}}
\label{sssec:p4he}
The present evaluation for ENDF/B-VIII.1 subsumes the previous ENDF/B-VIII.0 file, which was derived from the evaluations of
Refs.~\cite{perkins:1981enp,white:1991ecp}. The present evaluation is preferred since it is based on the multichannel unitary R-matrix
evaluation of \nuc{5}{Li} discussed in Section~\ref{subsec:d:3He}.

The current evaluation includes reaction data up to $E_d < 10$ MeV, to just above the three-body break-up threshold at $E_d
\simeq 8.9$ MeV. The resulting extension of the elastic p-induced data on \nuc{4}{He}, part of the same \nuc{5}{Li} compound system, to $E_p
\lesssim 34.3$ MeV, permits application codes testing between the ENDF/B-VIII.0 and the present libraries to determine requirements on upper energy
limits for charged particles, which is a point of increasing concern that is not completely or consistently discussed in the ENDF-6 format
manual.

\subsection{Tritons sublibrary}
\label{subsec:cp:tritons}


The triton sublibrary has been updated for \nuc{4}{He}. The new evaluation is based on the \nuc{7}{Li} evaluation, which couples to
$t+$\nuc{4}{He}, which was described in Section~\ref{subsec:n:6Li}.

\subsubsection{\nuc{4}{He}}
The $t$+\nuc{4}{He} entrance channel for the \nuc{7}{Li} compound system has been significantly updated since the previous
evaluation~\cite{White:2011ace,White:2011ndi}. The R-matrix analysis, described in detail in the Section~\ref{subsec:n:6Li}, includes elastic
\nuc{4}{He}($t,t$)\nuc{4}{He} 
differential scattering cross sections and analyzing powers ($A_y$) for $E_t \le 17$ MeV.  
These data extend beyond the three-body break-up threshold at $E_t \approx 11$ MeV, the \nuc{4}{He}($t,n$)\nuc{6}{Li} reaction data,
for $8.5\text{ MeV} < E_t < 14.4\text{MeV}$ and \nuc{4}{He}($t,n_1$)\nuc{6}{Li}($3^+$;0) for $E_t = 12.9$ MeV.

This information has been encoded in the file $\texttt{t-002\_He\_004.endf}$ in files MF=3, MF=4 and MF=6. The MT=50 section corresponds to
observations of neutron groups recoiling against the ground state nucleus \nuc{6}{Li} that give the \nuc{4}{He}(t,n)\nuc{6}{Li} reaction
cross section.  The observed data for this reaction is mainly from the inverse reaction, \nuc{6}{Li}$(n,t)$\nuc{4}{He} (as in Table
\ref{7Lisumm}), including the measurements of integrated and differential cross sections at neutron energies up to incident neutron energy of
$E_n = 8.0$ MeV. The MT=51, \nuc{4}{He}$(t,n)$\nuc{6}{Li*} reaction cross section has been determined
primarily by the t-alpha total reaction cross section, plus a direct measurement at one energy of the differential cross section. The MT=52,
600, and
650 sections for \nuc{4}{He}$(t,n)$\nuc{6}{Li**}, \nuc{4}{He}$(t,p)$\nuc{6}{He}, and \nuc{4}{He}$(t,d)$\nuc{5}{He} reaction cross sections predicted 
from the \nuc{7}{Li} R-matrix analysis of Section~\ref{subsec:n:6Li}. 
                                                                   
The MT=2 sections encode the nuclear cross sections for $t+\alpha$ elastic scattering. 
The data set included measurements of the differential cross
section and triton analyzing power, including precise measurements from LANL~\cite{Devlin09}. Legendre coefficients are calculated from the R-matrix fit
using the exact (LAW=5) nuclear amplitude representation. Nuclear partial waves through $H$-waves were allowed in the analysis
giving Legendre orders $L=0$ to $10$ in the nuclear cross section and $L=0$ to $5$ in the complex nuclear amplitude that interferes with
the Coulomb amplitude. 

The MT=50, \nuc{4}{He}$(t,n)$\nuc{6}{Li} reaction, is represented by Legendre coefficients (for MF=4) through $L=6$, which  are given for the calculated angular distributions. Data
included differential cross section measurements for both the forward and reverse reactions at triton energies up to 14 MeV
and at neutron energies up to 8.0 MeV. 

The MT=51, \nuc{4}{He}$(t,n)$\nuc{6}{Li*} reaction is given in terms of Legendre coefficients through $L=2$ for the calculated angular distributions. Only
one measurement exists, at $E_t = 12.888$ MeV. 
                                                                   

%
%
%

\section{FISSION YIELDS SUBLIBRARIES}
\label{sec:fpy}



In this updated release of the fission product yields sublibraries, we address an issue related to anomalously large uncertainties of nuclides close to stability, as well as a nonphysical discontinuity in the thermal neutron-induced fission yields of \nuc{241}{Pu}, which has affected the ENDF/B library since the release VI.2. 

\subsection{Correction to the uncertainties of cumulative yields near stability}

The update of the uncertainties affects the cumulative fission yields (MT=459) of 17 nuclides: \nuc{90g}{Y}, \nuc{90g}{Zr}, \nuc{91g}{Y}, \nuc{91m}{Y}, \nuc{93g}{Y}, \nuc{93m}{Y}, \nuc{109g}{Ru}, \nuc{109g}{Rh}, \nuc{109g}{Pd}, \nuc{109g}{Ag}, \nuc{109m}{Ag}, \nuc{132g}{I}, \nuc{133g}{I}, \nuc{133g}{Xe}, \nuc{135g}{Cs}, \nuc{135g}{Ba}, \nuc{148g}{Pr}.

In addition, 4 independent yields (MT=454) were modified to account for the fact that isomers for \nuc{109}{Ru} and \nuc{109}{Rh} have not been confirmed.

A brief description of the changes and the methodology will follow. 
For more details and a complete list of the updated values, the reader is referred to Ref.~\cite{BNL-220804-2021-INRE}.
The changes have been applied to all NFY and SFY files, except for the cumulative yields of \nuc{135g}{Ba} in the neutron-induced fission of \nuc{239}{Pu} at 500~keV, 2~MeV and 14~MeV incident energies, which were recently revised contextually with the release of ENDF/B-VII.1 \cite{ENDF-VII.1} and were not modified further.

\begin{figure}[tbp]
\centering
\includegraphics[width=0.5\textwidth]{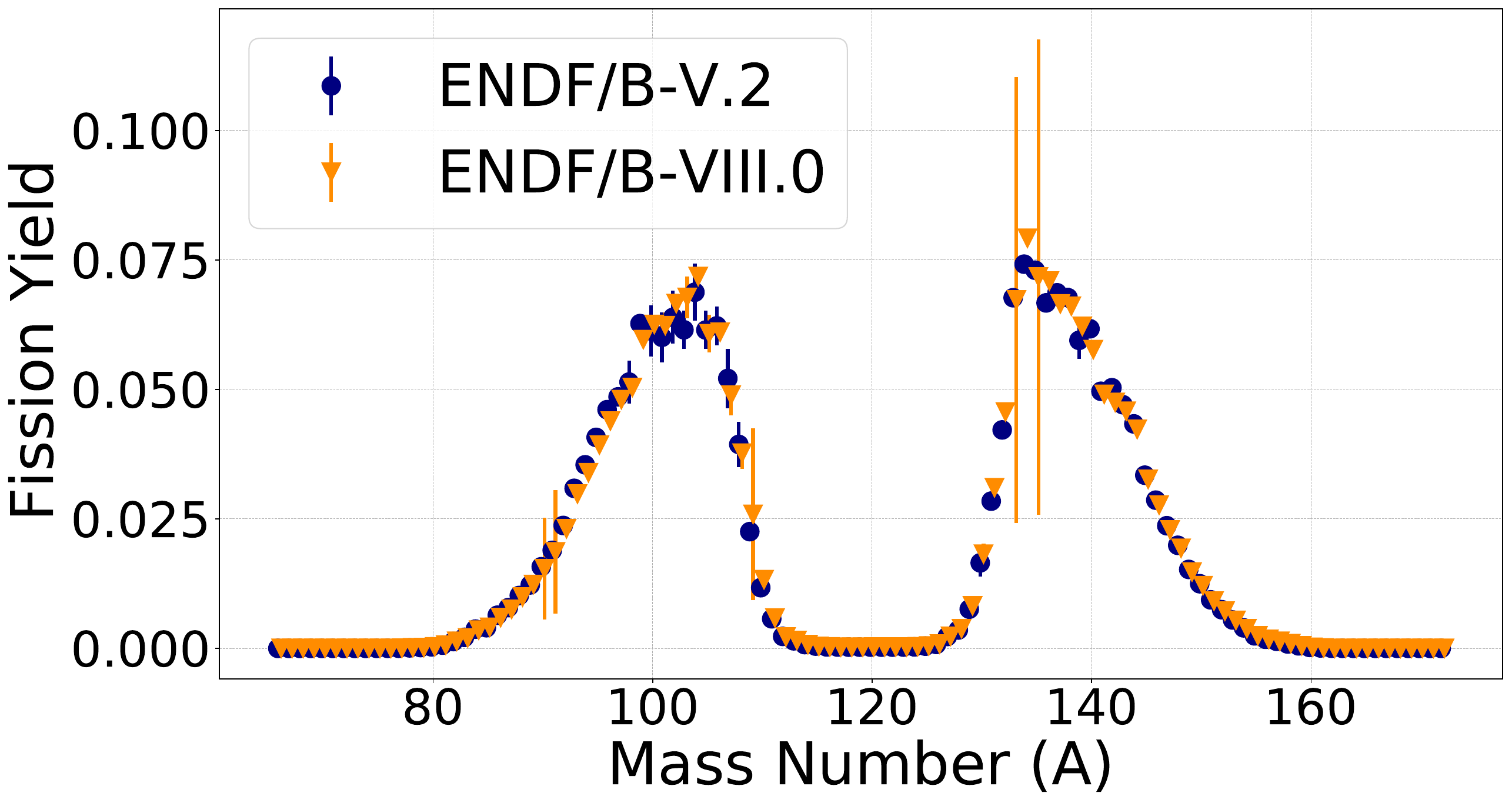}
\vspace{-2mm}
\caption{Chain yields for thermal-neutron induced fission of \nuc{241}{Pu} for ENDF/B-V.2 (blue circles) and ENDF/B-VIII.0 (orange triangles).}
\label{fig:FPYcompare_B8-B6}
\vspace{-2mm}
\end{figure}


The yields included in the ENDF/B-VIII.1 NFY (31 materials) and SFY sublibraries (9 materials) are based on the seminal work of T.~R. England and B.~F. Rider~\cite{ENDF349}, published for the first time in the ENDF/B-VI.1 release and subsequently updated for ENDF/B-VI.3. 
Yields were then inherited from one minor release of ENDF/B-VI to the next, and finally incorporated in ENDF/B-VII and -VIII. 
The only documented change to the yields was introduced in ENDF/B-VII.1, where NFYs of \nuc{239}{Pu} at 500~keV, 2~MeV and 14~MeV incident energies were revised \cite{chadwick2010fission,ENDF-VII.1}.

Upon careful analysis of ENDF/B-VIII.0, we noticed abnormally high uncertainties of the cumulative yields of nuclides in proximity to the valley of stability.
The chain yields for thermal-neutron-induced fission of \nuc{241}{Pu} from ENDF/B-VIII.0 are compared to those of an earlier evaluation (ENDF/B-V.2) in Fig.~\ref{fig:FPYcompare_B8-B6}, revealing unusually large error bars for masses 133 and 135. 
These anomalous uncertainties appear to have been introduced with the ENDF/B-VI.1 release, and they are not consistent with the expected trends of the uncertainties of cumulative yields for end-of-chain fission products nor are they supported by any other experimental evidence.

\begin{figure}[htbp]
\centering
\includegraphics[width=0.5\textwidth]{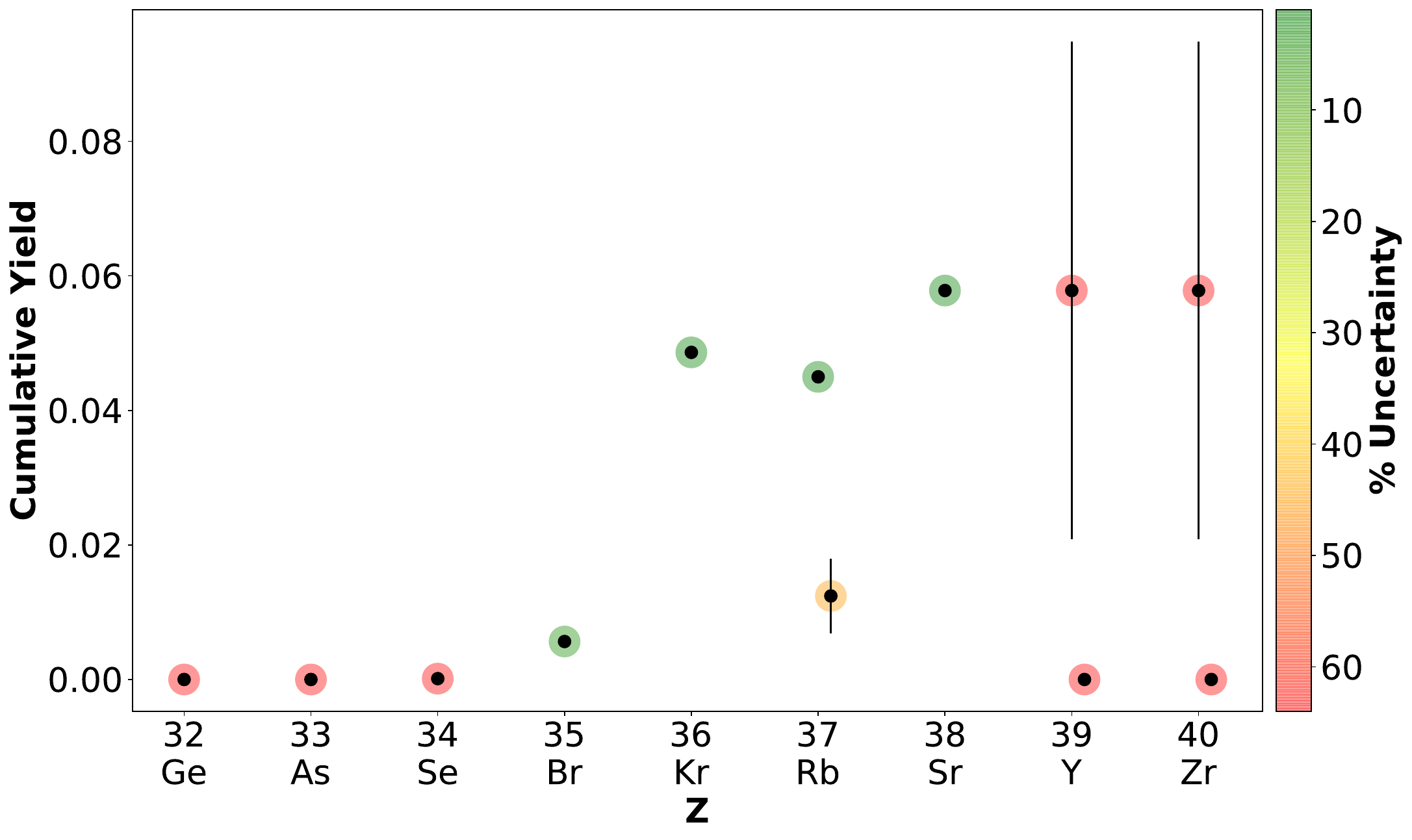}
\vspace{-2mm}
\caption{Cumulative yields and relative uncertainties of fission products with mass A=90 for thermal-neutron-induced fission of \nuc{235}{U}. The large uncertainties for \nuc{90}{Y} and \nuc{90}{Zr} are not consistent with the expected trends of cumulative yields for end-of-chain fission products and are not supported by experimental evidence.}
\label{fig:FPYA90}
\vspace{-2mm}
\end{figure}

As an example, one can look at the cumulative yields of the A=90 mass chain, shown in Fig.~\ref{fig:FPYA90}. 
Here the anomalously large uncertainties are evident for \nuc{90g}{Y} and \nuc{90g}{Zr}.
As an example, the uncertainty of \nuc{90g}{Y} was re-calculated as:

$$ \Delta^2CY(^\mathrm{90g}\mathrm{Y}) = \Delta^2CY(^\mathrm{90g}\mathrm{Sr}) + \Delta^2IY(^\mathrm{90m}\mathrm{Y}) + \Delta^2IY(^\mathrm{90g}\mathrm{Y})  $$

These corrections lead to a reduction of the uncertainty that, depending on the fissioning system, can be up to two orders of magnitude.

In addition to corrections such as the one discussed above for \nuc{90g}{Y}, which were applied to 17 fission products in all NFY and SFY materials (with the exception noted for \nuc{239}{Pu}), two isomers were removed from the decay chain in mass A=109.
Long-lived isomers for \nuc{109}{Ru} and \nuc{109}{Rh} are assumed in the ENDF/B decay data sublibrary but have not been confirmed by nuclear structure or FY measurements \cite{KUMAR20161}. 

For these two isotopes, the independent yield of the isomer was set to 0, and the independent yield of the ground state was defined as:
$$
    IY'(g.s.) = IY(g.s.) + IY(m.s.),
$$

where IY'(g.s.) is the new value for the ground state yield, and IY(g.s.) (IY(m.s.)) is the yield of the ground (isomeric) state included in ENDF/B-VIII.0.

\subsection{Corrections to the thermal-neutron-induced fission yields of \nuc{241}{Pu}}

\begin{figure}[htb]
\centering
\includegraphics[width=.95\columnwidth]{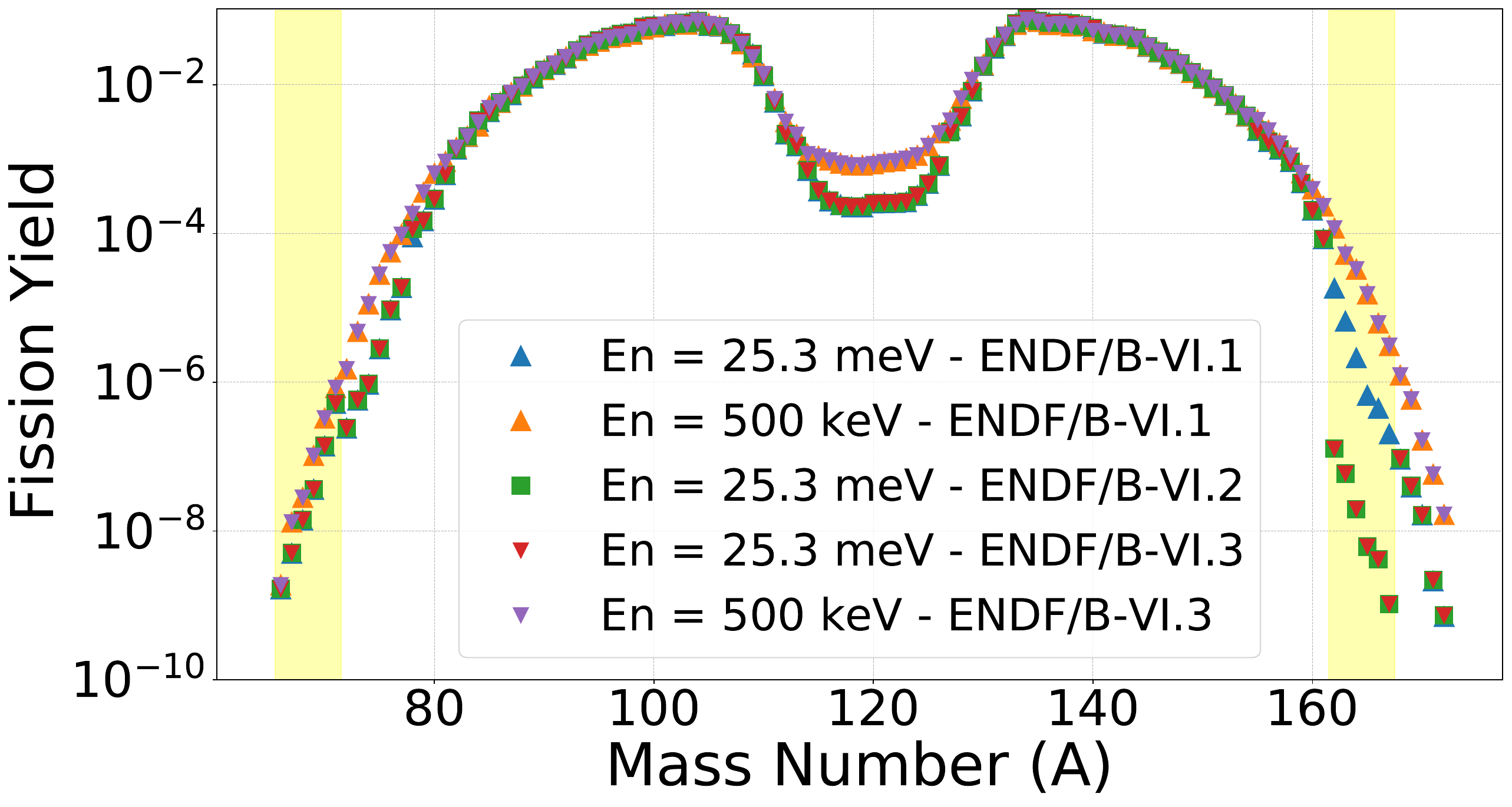}
\vspace{-2mm}
\caption{\footnotesize Mass-chain yields for thermal- and fast-neutron-induced fission of \nuc{241}{Pu} in the releases from ENDF/B-VI.1 to ENDF/B-VI.3. Notice that ENDF/B-VI.2 mistakenly omitted the fast fission energy point. The FYs included in ENDF/B-VIII.0 have been substantially inherited from ENDF/B-VI.3. FYs in the shaded regions are those for which the yields were changed in the current ENDF/B-VIII.1 release.}
\label{fig:241Pu-anomaly}
\vspace{-2mm}
\end{figure}

Another long-standing issue addressed in this release affected the thermal-neutron-induced fission yields of \nuc{241}{Pu} showing a nonphysical discontinuity in the yields of fission products in 13 mass chains. 
The anomalous yields appeared after the ENDF/B-VI.2 release and was left unchanged in all releases since, as shown in Fig.~\ref{fig:241Pu-anomaly}.
For a detailed list of all the changes, the reader is referred to Ref.~\cite{BNL-225087-2023-INRE}.

The extensive documentation that accompanies the evaluation work of England and Rider in Ref.~\cite{ENDF349} allowed us to take a close look at the possible origin of the anomalous yields.
For mass chains between 162 and 168, no experimental data are available, and the cumulative yields of the end-of-chain isotopes are obtained from  Ref.~\cite[Table~VIII]{1987RIZT}, where the chain yields of every fissioning system included in the evaluation are estimated and tabulated.
Comparing the values in the table with those included in Ref.~\cite{ENDF349}, it seems that the data used for the evaluation were mistakenly obtained from the chain yields of \nuc{233}{U}(n\textsubscript{th},f).
With a deeper dive into the England and Rider documentation in Ref.~\cite{ENDF349}, we found additional mass chains at the edge of the distribution (A=66-71), where no experimental data are available, that are also affected by this very same issue. 

In order to address the anomaly, we introduced correction factors ($CF$) for the independent yields of masses 66 to 71 and 162 to 168 (reported in Table~\ref{tab:correctionFactors}), based on the ratio between the current mass chain yields and the ones tabulated in Ref.~\cite{1987RIZT}. 

\begin{table}[!htph]
\centering
\caption{Correction factors proposed for masses 66-71 and 162-168, based on the ratio between the mass chain yields from ENDF/B-VI.3 and those tabulated in Ref.~\cite{1987RIZT}. All other masses were corrected with a CF=0.99998 defined in Eq.~\ref{eqn:normalize}.}  
    \label{tab:correctionFactors}  
    \begin{tabular}{cc@{\hskip 15pt}cc}
    \toprule \toprule
    \multicolumn{2}{c}{Light Fragments} & \multicolumn{2}{c}{Heavy Fragments}\\ 
    A & Correction Factor & A & Correction Factor \\
    \midrule
    &  & 162 & $2.057 \times 10^{2}$ \\
    66 & $8.02 \times 10^{-1}$ & 163 & $1.597 \times 10^{2}$ \\
    67 & $4.93 \times 10^{-1}$ & 164 & $1.533 \times 10^{2}$ \\
    68 & $4.08 \times 10^{-1}$ & 165 & $1.535 \times 10^{2}$ \\
    69 & $3.43 \times 10^{-1}$ & 166 & $1.529 \times 10^{2}$ \\
    70 & $3.21 \times 10^{-1}$ & 167 & $2.750 \times 10^{2}$ \\
    71 & $1.28 \times 10^{-1}$ & 168 & $1.379$ \\
    \bottomrule \bottomrule
    \end{tabular}
\end{table}

Despite the fact that the affected yields only represent a small fraction of the entire mass distribution, all independent FYs had to be corrected to normalize the sum to 200\%.

The $CF$ was obtained as:
\begin{equation}
CF = \frac{200\%}{\sum_i IY'_{i}} = 0.99998,
\label{eqn:normalize}
\end{equation}
where $IY'_{i}$ are the independent FYs after the factors in Table~\ref{tab:correctionFactors} were applied to masses 66-71 and 162-168 and are summed over the full mass distribution. 
$CF$ was uniformly applied to all independent FYs and, in all cases, represents only a fraction of the uncertainty on the mass yield, that -- for \nuc{241}{Pu} -- are estimated to be known at best at 1.4\% accuracy even for fission products in the peaks of the distribution \cite{1987RIZT}.

Cumulative FYs could have been recalculated from independent yields using the decay data sublibrary, but to ensure consistency with the remaining body of FYs, we decided to apply the same correction factors to the cumulative FYs.
This is possible since there is no transfer of yields via $\beta$-delayed neutron decay between two mass chains with different correction factors, and with this method, we introduce only negligible changes to those mass chains not affected by the anomaly.

Uncertainties of both independent and cumulative yields were also rescaled by the same factor, to preserve the percentage uncertainty estimated by England and Rider in their original work.

A number of tests were made on the corrected FY data, in terms of atom, neutron, proton balances; total sum of chain yields in the heavy and mass peaks, and over the entire distribution.
Results are summarized in Tables~\ref{tab:testSum} and~\ref{tab:testBalance}, expressed as total sum of independent yields, as well as residuals, i.e., difference between $1$ ($2$) and the sum of the yields of the light/heavy peaks (full distribution).
Despite a slight worsening of the sum of the light/heavy peaks, the difference (0.0023\%) is still an order of magnitude lower than the accepted differences for a number of other nuclides.

\begin{table}[!tbhp]
\centering
\vspace{-2mm}
\caption{Comparison of the sum of independent yields between ENDF/B-VIII.0 and ENDF/B-VIII.1. The sum was performed in the light/heavy peaks, in the full mass distribution. The residuals, defined as the difference between $1$ ($2$) and the sum of the yields of the light/heavy peaks (full distribution), are also shown.}  
    \label{tab:testSum}  
    \begin{tabular}{cccc}
    \toprule \toprule
    \multicolumn{4}{c}{ENDF/B-VIII.0}\\
    \midrule
    & Light & Heavy & Full Dist. \\
    Sum & 0.99999741 & 1.00000278 & 2.00000019 \\
    Residual & 2.587e-06 & -2.776e-06 & -1.882e-07 \\
    \midrule
    & Zlow & Zhigh & Full Dist. \\
    Sum &1.00015263 & 0.99984756 & 2.00000019 \\
    \bottomrule
    \toprule
    \multicolumn{4}{c}{ENDF/B-VIII.1}\\
    \midrule
    & Light & Heavy & Full Dist. \\
    Sum & 0.99997721 & 1.00002336 & 2.00000057 \\
    Residual &   2.279e-05 & -2.336e-05 & -5.740e-07 \\
    \midrule
    & Zlow & Zhigh & Full Dist. \\
    Sum & 1.00013243 & 0.99986814 & 2.00000057 \\    
    \bottomrule \bottomrule
    \end{tabular}
\vspace{-3mm}
\end{table}

Additional tests included calculating the average mass of the heavy and light fission products, the average number of emitted neutrons ($\bar{\nu}$), and the sum of the yield-weighted atomic number (Z).
Also in this case, the values after the implemented corrections are comparable with those in the original ENDF/B-VIII.0 evaluation, and always within the errors accepted for other fissioning systems in the library.

\begin{table*}[tb]
    \centering
    \caption{Comparison of the average atomic and mass numbers, and apparent average neutron multiplicity for ENDF/B-VIII.0 and ENDF/B-VIII.1.}  
    \label{tab:testBalance}  
    \begin{tabular}{c@{\hskip 15pt}cc@{\hskip 30pt}cc@{\hskip 30pt}c}
    \toprule \toprule
    & Sum of & Atomic & Avg. Light & Avg. Heavy & Apparent  \\
    & Z*Yields & Number & Mass & Mass & $\bar{\nu}$  \\
    \midrule
    ENDF/B-VIII.0 & 94.00 & 94 & 100.2828 & 138.7657 & 2.95  \\
    ENDF/B-VIII.1 & 94.00 & 94 & 100.2808	& 138.7695 & 2.95  \\
    \bottomrule \bottomrule
    \end{tabular}
\end{table*}





\section{DECAY SUBLIBRARY}

Although some preliminary decay evaluations were developed since \prENDF, they did not reach the stage of maturity necessary to be incorporated into the final \ENDF. Therefore, there were no significant updates or major evaluations incorporated into the \ENDF\ decay sublibrary. For the following library release, expected to be ENDF/B-IX.0, we anticipate a major evaluation effort for the decay sublibrary.

Compared to \prENDF, only 10 decay files  were updated; namely, \nuc{91,92}{Rb}, \nuc{97}{Y}, \nuc{101}{Nb}, \nuc{105}{Mo}, \nuc{104-107}{Tc}, and \nuc{135}{Sb}. The reason for these updates was simply to accommodate updates from their corresponding ENSDF evaluations from May/June 2012~\cite{ENSDF-June-2012} and improved neutrino spectra, and they are not expected to have any significant impact in applications. Although in some cases there is an improvement, compared to \prENDF, in accounting for missing energy (\nuc{92}{Rb}, \nuc{97}{Y}), or the missing energy is preserved (\nuc{135}{Sb}), in most cases an increase in missing energy is observed. This indicates that these nuclides should also be considered for reevaluation for a future release.

\section{ATOMIC SUBLIBRARIES}
\label{sec:atomic}

For the atomic sublibraries in \ENDF, we adopted the updated files from the Electron-Photon Interaction Cross Sections, 2023 version (EPICS2023), described in Ref.~\cite{EPICS-2023}. The nomenclature correspondence between \ENDF\ and EPICS is as follows: the EADL, EEDL, and EPDL in EPICS correspond to the atomic relaxation, electrons, and photo-atomic sublibraries in ENDF/B, respectively. More information about EPICS can be found in the EPICS website~\cite{EPICS-website}.

As stated in Ref.~\cite{EPICS-2023}, EPICS2023 includes elemental, cold, neutral, isolated atomic data, in the absence of electrical or magnetic fields. 
EPICS is designed to handle ENDF applications involving photons and
electrons in the keV and MeV energy ranges, which can accurately be described by simple elemental data with atomic number ranging from 1 to 100. However, EPICS is not designed to handle density or molecular and other binding
effects that are important at low energies. Nor is it designed to handle nuclear effects that are important
at higher energy. The atomic sublibraries in \ENDF\ inherit from EPICS its atomic self-consistency as all elements were obtaining simultaneously using the same methods, preserving important atomic number dependencies, such as sub-shell binding energies. Additionally, for the same reason, all three atomic sublibraries use the same sub-shell binding energies, in order to
conserve energy when performing coupled electron-photon calculations. 

The changes in \ENDF\ from \prENDF\ for the atomic sublibraries are thus the same as the changes in EPICS2023 compared to its earlier version (EPICS2017), that is:

\begin{itemize} 

\item Corrected electrons (EEDL) bremsstrahlung ZAP (particle emitted), to 0 (photons), which in earlier
versions was incorrectly defined as 11 (electrons); this caused the bremsstrahlung to appear to emit
only electrons and no photons.
\item Deleted extraneous repeated energies in earlier versions of the photo-atomic data (EPDL).
\item Ensured that all thresholds start with a zero cross section, e.g., all photoionization
sub-shells); earlier versions potentially confused users regarding interpolating or extrapolating data.

\end{itemize}

\section{NEUTRON DATA STANDARDS}
\label{sec:standards}

There were no changes made in the standards for ENDF/B-VIII.1. A new evaluation of the standards leads to a new version of ENDF/B. It is understood that the standards should not change within a given version of ENDF/B. 
The energy ranges of validity of neutron cross section standards including the Thermal Neutron Constants, high energy reference cross
sections, prompt $\gamma$-ray reference cross sections, and prompt fission fission spectra are given
in Table~\ref{tab:standards}. The Thermal Neutron Constants included in the Standards are reviewed in Section~\ref{sec:TNC} and listed in Table~\ref{tab:standards_thermal} below.


We would like to remark on two important points regarding the dissemination of Neutron Data Standards:
\begin{enumerate}
  \item In the IAEA Neutron Data Standards webpage (\url{https://nds.iaea.org/standards/}), the $^{239}$Pu(n,f) and the $^{238}$U(n,$\gamma$) cross sections are listed as additional cross sections used within the standard GMA fit.
   This is due to the fact that those cross sections were fitted within the Neutron Standard cross sections, but the quality and consistency of available data was not enough to declare them standard cross sections.    
  \item The ENDF/B-VIII.0  neutron standards sublibrary (NSUB-19) includes evaluated data both inside and outside the energy range of the standards as well as non-standard reactions, without distinction. The correct energy range for the neutron data standards reactions is shown in Table~\ref{tab:standards}.
 
\end{enumerate}

\begin{table}
\centering
\caption{\label{tab:standards} Neutron Data Standards.}
\begin{tabular}{l|l}
\toprule \toprule
\multicolumn{2}{c}{Neutron cross section standards}\\
\midrule
Reaction & Standards incident neutron energy range \\
\midrule
H(n,n) & 1~keV to 20~MeV \\
\nuc{3}{He}(n,p) & 0.0253~eV to 50~keV \\
\nuc{6}{Li}(n,t) &   0.0253~eV to 1.0~MeV \\
\nuc{10}{B}(n,$\alpha$) &   0.0253~eV to 1~MeV \\
\nuc{10}{B}(n,$\alpha_1 \gamma$) &   0.0253~eV to 1~MeV \\
\nuc{}{C}(n,n) &   10~eV to 1.8~MeV\footnote{The angular distributions at these incident energies are declared standard} \\
\nuc{}{Au}(n,$\gamma$) &   0.0253~eV, 0.2 to 2.5~MeV, 30~keV MACS \\
\nuc{235}{U}(n,f) & 8--11eV (integral) \\
                  &   0.15 to 200~MeV \\
\nuc{238}{U}(n,f) &   2~MeV to 200~MeV \\
\toprule
\multicolumn{2}{c}{High energy reference fission cross sections}\\
\midrule
Reaction & Reference incident neutron energy range \\
\midrule
\nuc{Nat}{Pb}(n,f) & $\thickapprox$ 34~MeV up to 1~GeV\\
\nuc{209}{Bi}(n,f) & $\thickapprox$ 34~MeV up to 1~GeV\\
\nuc{235}{U}(n,f) & 200~MeV to 1~GeV\\
\nuc{238}{U}(n,f) & 200~MeV to 1~GeV\\
\nuc{239}{Pu}(n,f) & 200~MeV to 1~GeV\\
\toprule
\multicolumn{2}{c}{Prompt $\gamma$-ray production reference cross sections}\\
\midrule
Reaction & Reference incident neutron energy range \\
\midrule
\nuc{10}{B}(n,$\alpha_1 \gamma$) &   0.0253~eV to 1~MeV \\
\nuc{7}{Li}(n,n$^{\prime}\gamma$) &   0.9~MeV to 8~MeV \\
\nuc{48}{Ti}(n,n$^{\prime}\gamma$) &   2.8~MeV to 16~MeV \\
\toprule
\multicolumn{2}{c}{Thermal neutron constants at E=0.0253~eV (2200 m/s)}\\
\toprule
\multicolumn{2}{c}{Prompt fission neutron spectra (PFNS)}\\
\midrule
Reaction & Reference outgoing neutron energy range \\
\midrule
\nuc{235}{U}(n$_{\mathrm{th}}$,f) &  0.00001~eV to 30~MeV \\
\nuc{252}{Cf}(sf) &  0.00001~eV to 30~MeV \\
\bottomrule \bottomrule
\end{tabular}
\end{table}

\subsection{Thermal Neutron Constants (TNC)}
\label{sec:TNC}
These data historically have been evaluated using least-square techniques by the Standards community since the 60's. The fission, capture and elastic thermal cross sections at the thermal point (2200 m/s of neutron velocity) play an especially important role for the normalization of time-of-flight measurements for R-matrix evaluations. 

These data are particularly important in the determination of the neutron economy in thermal reactors. Since thermal data are included in the Standards'  evaluation, the thermal constants will have an impact on the results of the evaluation \cite{carlson2018a}. It should be noted, however, that most of the TNC were never directly used in the evaluated data files, mainly due to constraints in the R-matrix evaluations used in the RRR of heavy actinides. The TNC in Standards 2017~\cite{carlson2018} are listed in Table~\ref{tab:standards_thermal} and compared with values used in ENDF/B-VIII.0 (in {\it italics}) and ENDF/B-VIII.1 evaluations (in \textbf{bold}). It is strongly suggested that any changes in the thermal constants by evaluators be within the uncertainties shown here.
The consistency for $^{235}$U evaluated cross sections and $\bar{\nu}_{\mathrm{total}}$ with TNC is excellent. 
However, the situation is different for other fissile actinides. For example, the $^{233}$U thermal-capture cross section of 42.3 barn used in ENDF/B-VIII.0 was almost three sigmas lower than the recommended TNC value of 44.9(9) barn. Similarly, the $^{239}$Pu thermal fission cross section of 747.4 barn used in ENDF/B-VIII.0  was much lower than the recommended TNC value of 752.4(22) barn. 

Evaluated values used in the ENDF/B-VIII.1 evaluations are fully consistent with the TNC, a clear progress from the ENDF/B-VIII.0 situation. 

\begin{table}
\centering
\caption{\label{tab:standards_thermal} The thermal neutron constants in Standards 2017~\cite{carlson2018}. The constants in \textit{italics} are those obtained from the \prENDF\ evaluations, while those in \textbf{bold} are the values in the \ENDF\ neutron sublibrary files.}
\begin{tabular}{l|cccc}
\toprule \toprule
Quantity & \nuc{233}{U} & \nuc{235}{U} & \nuc{239}{Pu} & \nuc{241}{Pu} \\
\midrule
\multirow{3}{*}{$\sigma_{\mathrm{(n,f)}}$(b)} & 533.0(2.2) & 587.3(1.4) & 752.4(2.2) & 1024(11) \\
                                                                    &\textbf{533.4} & \textbf{586.1 } & \textbf{751.1} & \textbf{1024} \\
                                                                    &\textit{534.1} & \textit{586.7} & \textit{747.4} & \textit{1012} \\
\midrule
\multirow{3}{*}{$\sigma_{\mathrm{(n,\gamma)}}$(b)} & 44.9(9) & 99.5(1.3) & 269.8(2.5) & 362.3(6.1) \\
                                                                    &\textbf{44.1} & \textbf{99.4} & \textbf{270.4} & \textbf{363.8} \\
                                                                    &\textit{42.3} & \textit{99.4} & \textit{270.1} & \textit{363.1} \\
\midrule
\multirow{3}{*}{$\sigma_{\mathrm{(n,n)}}$(b)} & 12.2(7) & 14.09(22) & 7.8(1.0) & 11.9(2.6) \\
                                                                    &\textbf{12.1} & \textbf{14.07} & \textbf{8.1} & \textbf{12.0} \\
                                                                    &\textit{12.2} & \textit{14.11} & \textit{8.1} & \textit{11.3} \\
\midrule
\multirow{3}{*}{$\bar{\nu}_{\mathrm{total}}$} & 2.487(11) & 2.425(11) & 2.878(13) & 2.940(13) \\
                                                                    &\textbf{2.4869} & \textbf{2.4299} & \textbf{2.8614} & \textbf{2.9453} \\
                                                                    &\textit{2.4852} & \textit{2.4298} & \textit{2.8769} & \textit{2.9453} \\
\bottomrule \bottomrule
\end{tabular}
\end{table}

\section{SUMMARY OF FORMAT CHANGES SINCE \prENDF}
\label{sec:formats}





In the early stage of work on the ENDF/B-VII.1 library release, CSEWG made the practical and important decision to use the same ENDF-6 format \cite{ENDF6-Format-2012} for the next few ENDF releases.
This decision was adopted even though it was argued that it would be timely to modernize the formats as the Generalised Nuclear Database Structure (GNDS) was now available.
After careful deliberation, CSEWG concluded that actual GNDS implementation requires considerable resources to modify processing codes and to
guarantee high quality of the files processed by these codes. 
As the transition to GNDS is now underway, CSEWG determined that both ENDF/B-VIII libraries will be released in both the ENDF-6
format and in the GNDS format.  

For both ENDF/B-VIII.0 and ENDF/B-VIII.1 releases, CSEWG adopted minor format changes to correct the ENDF-6 format and deal with emerging needs.  
ENDF-6 format changes for ENDF/B-VIII.0 are described in the 2018 ENDF-102 report \cite{ENDF6-Format-2018}.
ENDF-6 changes for the current ENDF/B-VIII.1 are summarized here and detailed in the 2024 ENDF-102 report \cite{ENDF6-Format-2024}.
The GNDS-2.0 format changes are summarized here as well and are detailed in the GNDS specifications \cite{GNDS2.0}.

\subsection{ENDF-6 Format Changes}
For the ENDF/B-VIII.1 release, significant changes were made to both the TSL data and to the covariance data.  
In addition, there are a number of important changes made to resonance, decay data and gamma data formats.
Finally, there were a number of minor and sometimes cosmetic changes.

Significant changes to the thermal-neutron scattering formats include:
\begin{itemize}
\item Adding a generalized information file for MF7 (thermal-neutron scattering) files. 
      This format allows the evaluator to explicitly state the isotopic distribution of atoms used 
      to generate the MF7 data.  This, in turn, enables the linkage to corresponding neutron 
      sublibrary evaluations during the processing step (Chapter 7).
\item Added a mixed elastic scattering format to File 7 to improve thermal 
      scattering physics for nuclides with substantial coherent 
      and incoherent scattering cross sections (Chapter 7).
\item As the number of new thermal-neutron scattering evaluations submitted to ENDF/B-VIII.1 far exceeded the previous MAT number availablility, CSEWG adopted an entirely different scheme for thermal-neutron scattering evaluation MAT (Appendix C).
\end{itemize}

Significant changes to covariance formats include:
\begin{itemize}
\item Clarifying the procedure for collapsing File 35 covariance data to match those in Files 31 and 33 (Chapter 35).
\item Changing one of the LTY=0 sub-subsection rules to disallow overlapping NI-type subsections (Chapter 33).  The previous rule potentially allowed for recursive definitions of covariance matrices.  Although this never occurred in practice, processing codes have no easy way to determine whether covariance matrices were recursively defined.  Therefore, it was decided that allowing for overlapping NI-type subsections was unsafe.
\end{itemize}

Other significant changes:
\begin{itemize}
\item In order to be both consistent with resonance physics and to meet evaluator needs, various background R-matrix options were adopted (Chapter 2).  
\item Added an option to denote fission as a ``decay product'' in MF=10 (Chapter 10).  
\item Equations describing the energies of primary gammas were adjusted to include nuclear recoil (Chapters 12 and 13).  This change improves energy balance for capture reactions when primary gammas are given.
\end{itemize}

Other less important changes:
\begin{itemize}
\item Various references to the $^3$He nuclide were replaced with the word ``helion'' or the abbreviation ``h''.
\item The IAEA Photonuclear Data Library was assigned a library identifier (NLIB).
\item ENDF-6 sum rules are now illustrated with graphics (Sections 0.4.3.11 and 23.2).
\item Various other clarifications and minor revisions.
\end{itemize}

\subsection{GNDS Format Changes}

The GNDS underwent two sizeable revisions (GNDS 1.9 \cite{GNDS1.9} and GNDS 2.0 \cite{GNDS2.0}) since the previous ENDF release.  As of the 2.0 release, GNDS supports all ENDF-6 formats and meets all of the requirements laid out by WPEC Subgroup 38.   It should be mentioned that GNDS, unlike the ENDF-6 format, is managed by the WPEC Expert Group (EG) on the Generalised Nuclear Database Structure (EG-GNDS). In all, EG-GNDS made 149 formal merge requests and 29 of the merge requests were substantive format proposal requiring peer review from the EG and domain experts from the larger nuclear data community.  The changes made to GNDS centered on nine themes, summarized below:

\begin{enumerate}
	\item \textbf{Atomic data:}
		\begin{itemize}
		    \item Support for electron sub-shells as given in the ENDF-6 format.
		\end{itemize}

	\item \textbf{Covariances:}
		\begin{itemize}
		    \item Support for ``sandwich product'' covariance, allowing for storage of sensitivity matrices;
		    \item Add support for multi-dimensional covariances, such as encountered in TSL covariance data; 
		    \item Simplify the covariance section naming.
		\end{itemize}

	\item \textbf{Documentation:}
		\begin{itemize}
		    \item Change the GNDS date formats to comply with international standards; 
		    \item Improve documentation markup, bringing it into compliance with the requirements document;
		    \item Align the GNDS \texttt{documentation} node and decedants with DataCite.org's schema \cite{DataCiteSchema};
		    \item Make many small corrections, clarifications, simplifications and renamings to improve the usability of the documentation markups.
		\end{itemize}

	\item \textbf{File management:}
		\begin{itemize}
		    \item Add an \texttt{externalFiles} markup to denote external resources that may need to be read before processing interaction;
		    \item Add support for file checksums;
		    \item Add support for \texttt{map} files, the GNDS analog of \texttt{MCNP}'s \texttt{xsdir} file \cite{MCNP6.2}.
		\end{itemize}

	\item \textbf{Gamma data:}
		\begin{itemize}
		    \item Add a \texttt{finalState} attribute to primary gamma data markup so that enough information is available to perform a gamma cascade initiated by a primary gamma transition.
		\end{itemize}

	\item \textbf{Resonances:}
		\begin{itemize}
		    \item Consolidate the ENDF-6 ``scattering radius'' and ``hard sphere radius'' concepts into one R-matrix radius;
		    \item Expand options for the boundary condition of the R-matrix;
		    \item Add support for the various background R-matrix treatments recently added to the ENDF-6 format as noted above;
		    \item Add URR probability tables.
		\end{itemize}

	\item \textbf{SG-38 Compliance:}
		\begin{itemize}
		    \item Improve multi-group data styles;
		    \item Consolidate and clean up fission data support; 
		    \item Add appendices, including a change log and mapping between ENDF and GNDS.
		\end{itemize}

	\item \textbf{Thermal neutron scattering:}
		\begin{itemize}
		    \item Improve thermal-neutron scattering markup by bring it into compliance with the requirements document, including:
		    	\begin{itemize}
		    		\item Added a mixed elastic scattering format to improve thermal scattering physics for nuclides with substantial coherent and incoherent scattering cross sections;
		    		\item Reorganize the TNSL transport data to mimic the organization of other transport data;
				\item Add support for storing isotopic abundances (approved for GNDS-2.1).
					This section is equivalent to the new MF7 MT451 general information section in the ENDF-6 format;
		    	\end{itemize}
		    \item TNSL erroneously required a special \texttt{interaction} attribute in the main \texttt{map} container.
		\end{itemize}

	\item \textbf{Usability:}
		\begin{itemize}
		    \item Resolve possible discrepancy between branchings and PoPs;
		    \item Denote what kind of reaction evaluation is being considered using the \texttt{interaction} tag;
		    \item Fix inconsistency in orphan product organization;
		    \item Add functional container organization to low-level containers;
		    \item Rename `value' attribute for clarity in multi-dimensional containers;
		    \item Add \texttt{uncertainty} as allowed child node of \texttt{shell} in the PoPs database; 
		    \item Improve element names, making them comply with conventions elsewhere in the specifications.
		\end{itemize}

\end{enumerate}






\section{PROCESSING UPDATES}
\label{sec:processing}



Evaluated nuclear data files are formatted in a compact way that is not the most suitable for transport calculations. Therefore, the files in \ENDF\ should be properly processed to ensure the users will access the data in a more standardized way. Also, processing codes are stringent test of the proper formatting of the evaluated files.

The ENDF-6 files in this ENDF/B-VIII.1 release were successfully processed using \NJOY-16.76 and \FUDGE-6.7 for all sublibraries and \AMPX\ (\SCALE-7.0b08) for neutrons and photoatomic sublibraries.  \AMPX\ processing for the thermal-neutron scattering sublibrary used features to be made available in a future \SCALE-7.0 beta. The sections below describe the updates done in those three processing codes to process correctly \ENDF.

\subsection{NJOY}

Each release of the ENDF/B nuclear data library pushes the limits of the processing codes through the
introduction of new data and new data formats. Processing codes must be adapted to ensure that they
are capable of using these new formats, or at the very least understand them. \NJOY~\cite{NJOY},
the nuclear data processing code developed at LANL is no exception to this.
As \NJOY2016 is still the main production code for nuclear data libraries at LANL, we will continue to
make the necessary changes in this version of \NJOY\ to ensure full compatibility with the ENDF/B-VIII.1
nuclear data library.

The ENDF/B-VIII.1 nuclear data library added the following capabilities to the ENDF-6 format:
\begin{itemize}
    \item  The addition of the mixed elastic scattering option (LTHR = 2) for elastic thermal scattering (found in MF7 MT2).
    \item  The addition of general information concerning moderator materials in thermal scattering (found in MF7 MT451).
    \item  The correction of the background R-matrix elements for resonance parameters (found in MF2 MT151).
\end{itemize}

\begin{figure}[tbp]
    \centering
    \includegraphics[width=\columnwidth]{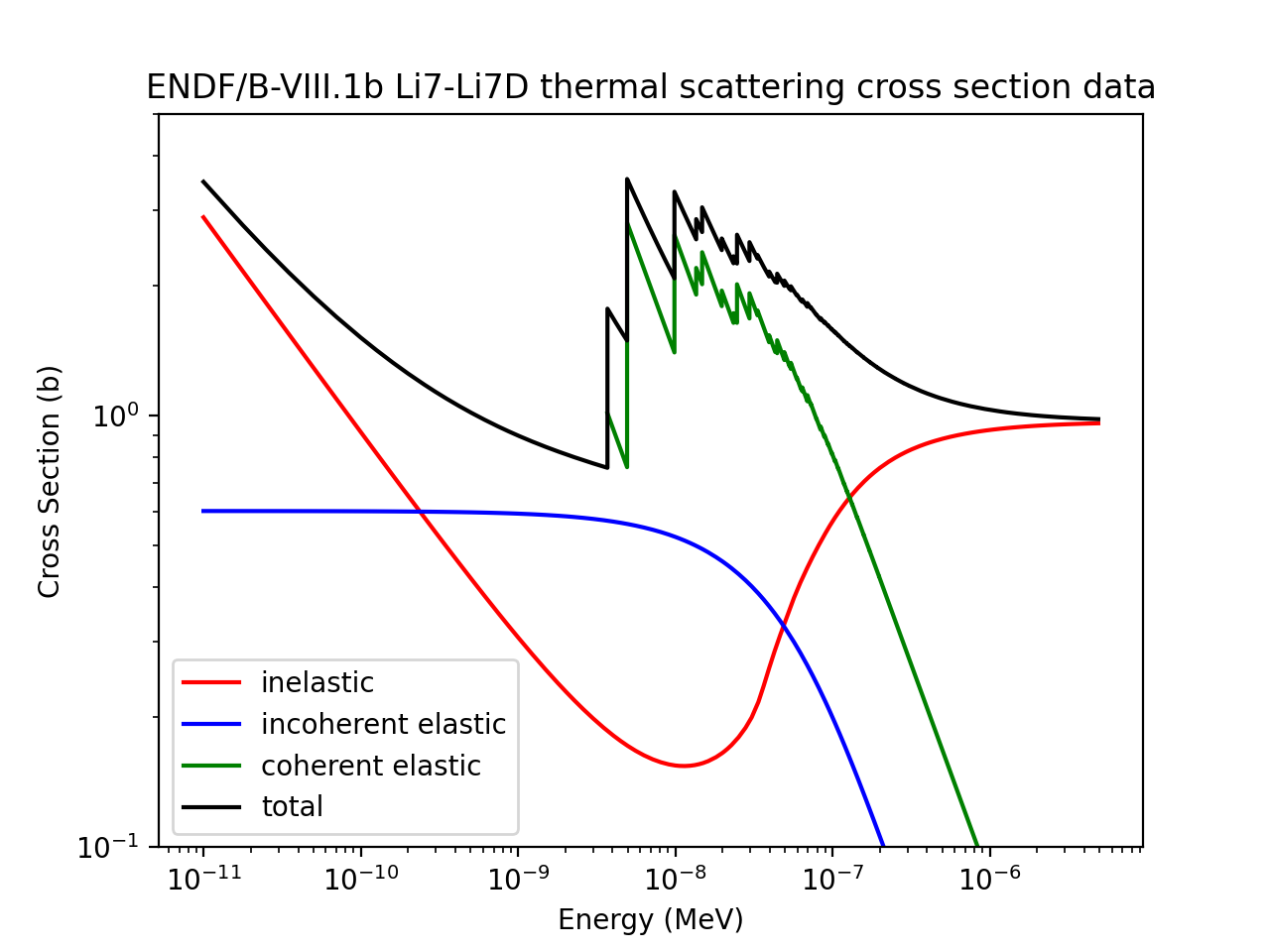}
    \vspace{-.1in}
    \caption{Mixed mode thermal scattering in D-7LiD at 400~K processed by \NJOY2016.65.}\label{fig:njoytsl}
\end{figure}

The addition of the mixed elastic scattering option (LTHR = 2) for elastic thermal scattering is one of the most
anticipated changes for the ENDF/B-VIII.1 nuclear data library since it lifts the limitation of using either
coherent or incoherent elastic scattering for moderators. For \NJOY2016, this new format has required changes
in various modules, the most important ones being THERMR (which processes the thermal scattering data) and ACER
(which formats the data into an ACE file for use in application codes like the MCNP code~\cite{MCNP63}).
Fig.~\ref{fig:njoytsl} gives an example of what this mixed mode data looks like for D in 7LiD after processing
through ACER in \NJOY2016. To enable the use of this new type of thermal scattering data, \NJOY\ users have to use
\NJOY2016.65 or above.

Because the coherent and incoherent data needs to be treated separately by \texttt{MCNP}6, the mixed mode scattering
also necessitated an update to the ACE format itself by adding an additional elastic scattering data block~\cite{aceformat},
and a subsequent modification to \texttt{MCNP} so that the mixed mode elastic data can be used during a Monte Carlo
simulation. This new format is fully supported in \texttt{MCNP}, Version 6.3.0 and above~\cite{MCNP63}.

The second change to thermal scattering that was introduced in the ENDF/B-VIII.1 library is the general
information concerning moderator materials in thermal scattering (found in MF7 MT451). This section was
added to the ENDF-6 format to have a machine-readable description of some of the moderator characteristics
(such as the uranium enrichment for the U in UO2 thermal scattering law). None of this information is
required for processing the thermal scattering data. As a result, only MODER in \NJOY2016 had to be modified to
account for the presence of the MF7 MT451 general information section in the ENDF-6 format.

The third ENDF format change that impacts the processing of ENDF/B-VIII.1 is the addition of R-matrix background
elements for resonance channels in the R-matrix limited format (LRF = 7, see Section 2.2.1.6 and D.1.7.5 of
the ENDF-6 format manual~\cite{ENDF6-Format-2024}). The ENDF-6 format has had the option for R-matrix background elements for
quite some time, but the formats that were laid out in the format manual were broken and the use of this format
option would have led to a malformatted ENDF file. These format issues were corrected in 2021 and were subsequently
used in the ENDF/B-VIII.1 evaluation for \nuc{88}{Sr}. This evaluation is the only evaluation in the ENDF/B-VIII.1
nuclear data library that uses these options.

The background R-matrix elements modify the diagonal of the R-matrix for a particular spin group through an
external function (the background elements are, therefore, defined per channel). The corrected ENDF-6 format
for the background R-matrix elements allow for four different options:
\begin{itemize}
    \item  no background element (the R-matrix for the spin group is thus not modified),
    \item  a tabulated complex function of energy (this would be the most general case possible),
    \item  a real statistical parameterization of the form available in \SAMMY~\cite{SAMMY},
    \item  a complex statistical parameterization of the forms described by Frohner~\cite{frohner1978, frohner1981}.
\end{itemize}

The new \nuc{88}{Sr} evaluation in ENDF/B-VIII.1 uses the \SAMMY\ parametrization option.

Contrary to the mixed mode elastic scattering where changes to the ACE format and, therefore, \texttt{MCNP} were required,
only the data processing itself is impacted by these new options. The changes in \NJOY2016 have, therefore,
been limited to MODER (to account for the change in the ENDF-6 format in MF2 MT151), RECONR and ERRORR.
Although only the \SAMMY\ parametrization has been used in the ENDF/B-VIII.1 \nuc{88}{Sr} evaluation, RECONR was modified
to be able to use all background options. The \SAMMY\ parametrization option was tested against the \nuc{88}{Sr} evaluation
in collaboration with ORNL. ERRORR was also updated to allow for covariance data with respect to these background
R-matrix options, although currently only the \SAMMY\ parametrization is supported. Since the new \nuc{88}{Sr} evaluation does
not include covariance data for the resonance parameters (in MF32), these updates could not be validated. \NJOY2016.73
or higher is required to process evaluations that use these background R-matrix element options.

In addition to these format changes that have had an impact on the nuclear data processing software, some processing
limitations in \NJOY2016 that were identified in the past also have been lifted to process ENDF/B-VIII.1 evaluations.
Examples here would be MF34 covariance processing with multiple subsections in ERRORR or improved photonuclear data
processing. For a detailed overview of all changes to \NJOY2016, we refer to the release notes on the \NJOY2016
repository on GitHub\footnote{See https://github.com/njoy/NJOY2016/releases}.

\subsection{FUDGE}\label{sec:fudge-processing}

LLNL maintains a nuclear data infrastructure code called \FUDGE\
(For Updating Data and Generating Evaluations)~\cite{FUDGE2023}.
\FUDGE\ is written primarily in Python, with computationally intensive tasks implemented in C and C++.
\FUDGE\ supports interacting with GNDS-formatted nuclear data via the Python interpreter,
including visualizing, checking, modifying and processing GNDS nuclear data. It is designed primarily
as a toolkit to enable users to write their own workflow for handling nuclear data,
but it also includes many useful Python scripts for common workflows, such as visualization, checking and processing.
\FUDGE\ is an open source code and is available for download at \url{https://github.com/LLNL/fudge}.

\FUDGE\ does not directly process a file in the ENDF-6 format, but it includes tools for translating
an evaluated ENDF-6 file into the GNDS/XML format and then processing the resulting GNDS data.
The ENDF-6 to GNDS translator strictly interprets the ENDF/B format manual~\cite{ENDF6-Format-2024}
and raises warnings or errors when data and/or format issues are encountered,
making the translator a useful tool for uncovering problems in an ENDF-6 evaluation.
The translator was used to detect and resolve many issues before the release of several recent ENDF/B library releases,
starting with the ENDF/B-VII.1 release in 2011~\cite{ENDF-VII.1}.
The same translator was also used to prepare the GNDS-2.0 version of the final ENDF/B-VIII.1 library,
available from the NNDC at ~\url{https://www.nndc.bnl.gov/endf-library/B-VIII.1/GNDS/}.
\FUDGE\ also comes packaged with a physics checking tool called `checkGNDS.py' that was used to help uncover
a variety of problems in ENDF/B-VIII.1 candidate evaluations and beta releases.

Several recent improvements have been made to processing and physics checking capabilities in \FUDGE, including:
\begin{itemize}
 \item support for external R-Matrix contributions as seen in the new n + Sr-88 evaluation,
 \item improved support for generating cross-section probability tables in the URR,
 \item improvements in processing thermal-neutron scattering evaluations, to better support the many new
   TNSL evaluations introduced in ENDF/B-VIII.1,
 \item many enhancements to the physics checking capabilities in `checkGNDS.py', including better options for
   filtering out minor issues to focus on the most important problems, and
 \item improved capabilities for exporting continuous-energy processed libraries to ACE format~\cite{aceformat}
   to support transport codes that depend on ACE-formatted data.
\end{itemize}

Using \FUDGE, LLNL processed each ENDF/B-VIII.1 beta release into both multi-group and continuous-energy libraries.
The resulting processed libraries are also stored in the GNDS format, using a hybrid combination of XML and HDF5 files
to reduce disk space and improve data load times.
Processed libraries were then tested with the LLNL Metis Validation and Verification platform~\cite{Metis2017}
against a wide range of benchmarks, including critical assemblies, pulsed spheres, and reaction ratios.
Results were compared to those obtained with the ENDF/B-VIII.0 library in GNDS format.
Metis supports running several radiation transport codes,
including the LLNL Monte Carlo code Mercury~\cite{Mercury5.40} and deterministic code Ardra~\cite{ARDRA}.
Both Mercury and Ardra use the C++ \GIDI\ API (also developed at LLNL) to access the GNDS data~\cite{GIDIPLUS2023}.
Additionally, Mercury uses the GPU-enabled \GIDI/\texttt{MCGIDI} API for cross-section lookup and to sample product distributions.
\GIDI\ is available for download from \url{https://github.com/LLNL/gidiplus}.

\subsection{AMPX}\label{sec:ampx-processing}

The \AMPX\ code system \cite{AMPX} is the nuclear data processing code for the \SCALE\ code system \cite{scale_631}.  
\AMPX\ is developed and distributed with the \SCALE\ code system.  
Additionally, an open source version of \AMPX\ is now available \cite{scale_public}.  
Modernization and development efforts in \AMPX\ are funded through the US 
NCSP.  

Key processing capabilities of \AMPX\ include: 
\begin{itemize}
    \item generation of temperature-dependent continuous-energy cross-section data; 
    \item resonance self-shielding for resolved and unresolved resonance ranges; 
    \item generation of energy and angle distributions for secondary particles; 
    \item processing of thermal scattering law data, $S(\alpha,\beta)$, for thermal moderators; 
    \item processing of particle-yield and decay data; 
    \item production of cross-section covariance data files for sensitivity/uncertainty 
    analyses; 
    \item performance of multi-group averaging operations; and 
    \item generation of continuous-energy weighting spectra.
\end{itemize}

The detailed thermal moderator evaluations present in the ENDF/B-VIII.1, especially 
those with large numbers of Bragg edges, spurred development in \AMPX\ and \SCALE\ 
in order to prepare for the library's release.  \AMPX\ and the \SCALE\ transport codes 
have historically adopted the philosophy to represent continuous energy data uniformly, 
with all reactions represented as fully double-differential cross-section 
cumulative probability distributions.  
While that approach facilitates sampling in the transport codes, it comes with the cost 
of very large CE libraries.  The ENDF/B-VIII.0 CE library, for example, is 28~GB; 
the ENDF/B-VIII.1 CE library in double-differential representation would more than double 
that size.  
Some of the thermal moderator files proved to be too prohibitive in run-time memory to process in 
double-differential representation.  

This issue is being addressed by ongoing development in \AMPX\ and \SCALE\ to 
update the CE library format to accommodate a simplified representation for Bragg 
edge data.  This development is targeting testing and release with \SCALE\ 7.0.0.  

Another improvement to the processing of thermal moderator materials has been developed 
in the XSPROC problem-specific cross-section processing module of \SCALE~\cite{kim_xsproc_improvement_2023}.  In particular, it proved necessary to improve the interpolation 
within the pointwise slowing down calculations for materials that have 
bound thermal scattering data with high forward peaks, such as the light 
water evaluation in ENDF/B-VIII.0.  

Another significant development for \AMPX\ is the capability to process the photonuclear 
sublibrary \cite{alexander_photonuclear_2024}, 
funded by a collaboration between National Nuclear Security Administration
(NNSA) and industry partnerships.  
The photonuclear sublibrary processing is available in the open source version of \AMPX, 
and will be released generally with the corresponding transport capabilities in 
\SCALE\ 7.0.0.  

The ENDF/B-VIII.1 library was processed into multigroup and continuous-energy libraries.  
The library was subsequently tested with a suite derived from 
multiple applications of interest: 
criticality safety benchmarks (VALID), reactor physics benchmarks, and 
nuclide inventory benchmarks.  

\section{INTEGRAL DATA TESTING SUMMARY}
\label{sec:integral}


The new nuclear data evaluations in the ENDF/B-VIII.1 nuclear data library were validated with experimentally measured values of the neutron multiplication factor in criticality benchmark experiments, delayed neutron benchmark experiments, reaction rate ratios in criticality benchmark experiments, reactor depletion benchmarks,  and
 neutron leakage spectra in LLNL 14--MeV pulsed-sphere experiments. An emphasis is placed on the validation of the neutron multiplication factor with measured values from recently performed criticality safety benchmark experiments. The ``modern criticality benchmark experiment suite'' has been organized to showcase this validation. The significant effort to support new criticality safety experiments and reevaluation of important nuclides has successfully led to the production of reduced nuclear data uncertainties and more accurate nuclear data mean values. Information about thermal-neutron cross sections, Westcott factors, resonance integrals, MACS, astrophysical reaction rates, and solar system $r$-process abundances obtained through  calculations  using the \ENDF\ data can be found in  Ref.~\cite{Boris}.

Additionally, for the first time ever, the new ENDF/B-VIII.1 library has been validated in depletion prediction and reactivity calculations in power reactors, which is crucial for economics of operating commercial reactors. This validation addresses deficiencies found in the ENDF/B-VIII.0 library.

\subsection{Criticality Benchmark Experiment Results}
\label{sec:criticality}

\subsubsection{The Mosteller Suite}

\begin{figure}
\vspace{-2mm}
\centering
\includegraphics[width=\columnwidth]{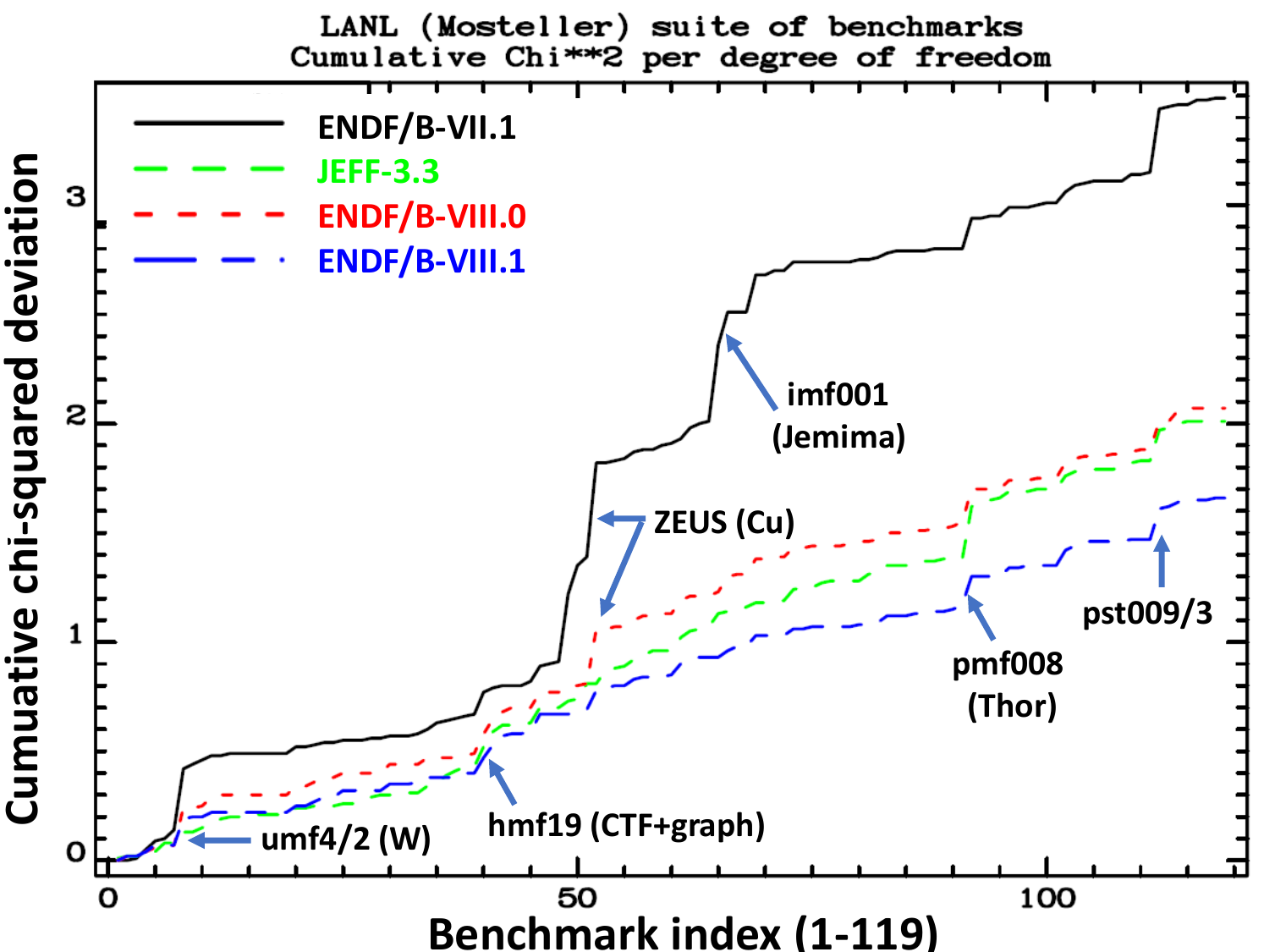}
\vspace{-2mm}
\caption{The cumulative chi-squared deviation for the new ENDF/B-VIII.1 evaluation, versus ENDF/B-VII.1, ENDF/B-VIII.0, and JEFF-3.3 libraries for a suite of 119 assemblies defined by Mosteller \etal \cite{Mosteller2011}.}
\label{fig:Mosteller}
\vspace{-2mm}
\end{figure}

An example of the progress made by the current ENDF/BVIII.1 is shown in Fig.~\ref{fig:Mosteller}, from the analysis by A. Trkov. Over the Mosteller suite of 119 benchmarks that have been systematically modeled by LANL for many decades~\cite{Mosteller2011}, and compared with reference (measured) \keff\ values, the cumulative Chi-squared per degree of freedom value is seen to have been reduced compared to both JEFF-3.3 and ENDF/B-VIII.0 values. A more detailed investigation is still needed for discrepant benchmarks bearing \nuc{233}{U} with a tungsten reflector (umf4/2), \nuc{235}{U} with a graphite or a copper reflector (hmf19, ZEUS), \nuc{239}{Pu} with a \nuc{232}{Th} reflector (Thor) and \nuc{239}{Pu} nitrate solutions (pst900). Some of the discrepancies could be caused by materials other than the primary actinides. We note that the Chi-squared per degree of freedom values shown in Fig.~\ref{fig:Mosteller} follow the usual definition where the squared (C-E) differences are divided by the squared experimental uncertainty in \keff\  for each critical assembly. This differs from the definition shown in the next subsection.

\subsubsection{Complete Criticality Testing}
\label{sec:complete_criticality}

\begin{figure}
\vspace{-2mm}
\centering
\includegraphics[width=0.45\textwidth]{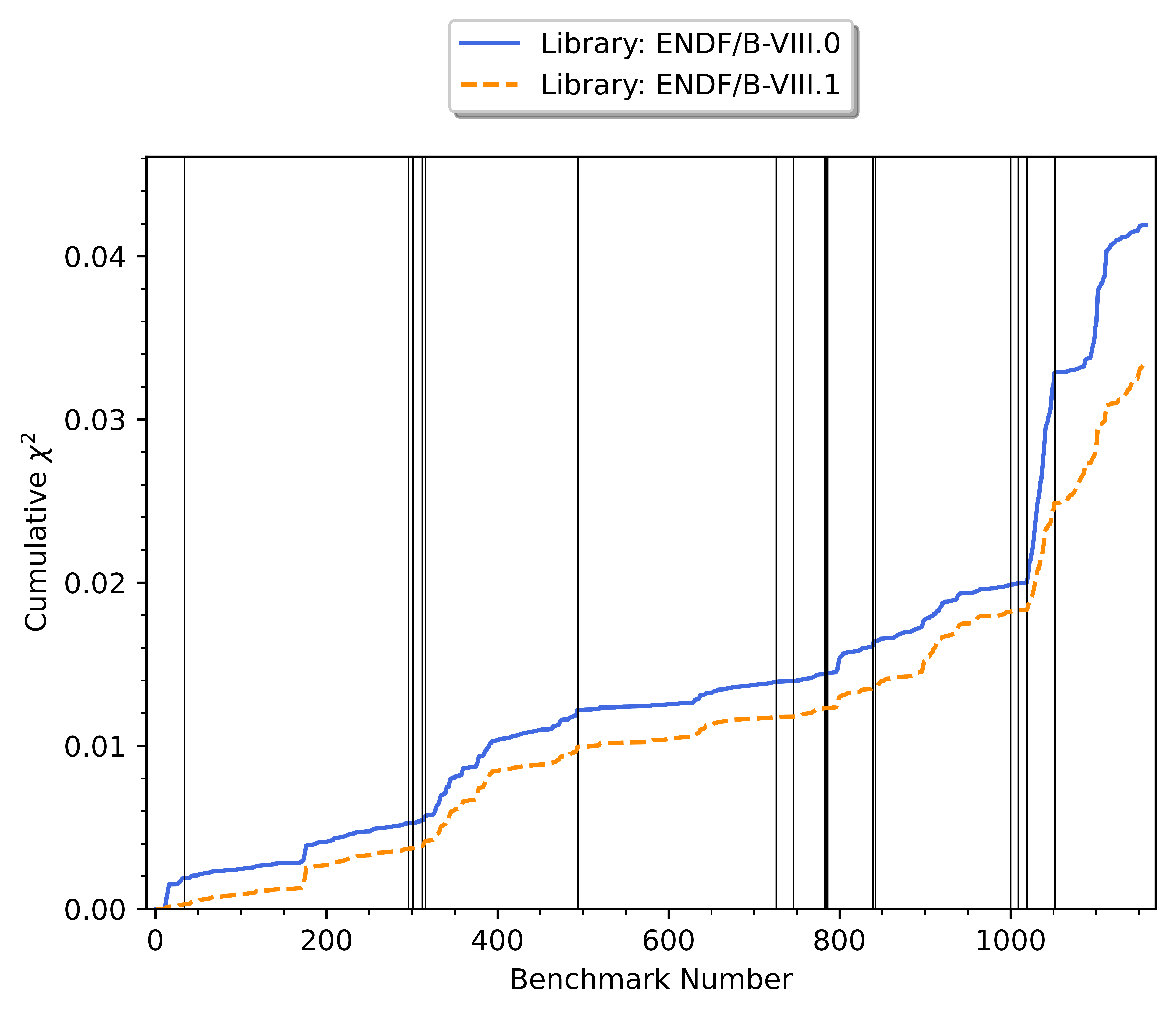}
\vspace{-2mm}
\caption{The value of $\chi^{2}$ plotted cumulatively as a function of criticality benchmark experiment number. Each benchmark category is separated with solid black vertical lines.}
\label{fig:crit_overview_chiSq}
\vspace{-2mm}
\end{figure}

The criticality benchmark experiments used for validation in this work are documented in the ICSBEP Handbook \cite{ICSBEP}. The categories of criticality benchmark experiments used in this work include HEU-MET-FAST, HEU-MET-INTER, HEU-MET-MIXED, HEU-MET-THERM, HEU-SOL-THERM, LEU-COMP-THERM, LEU-SOL-THERM, MIX-MET-FAST, MIX-MET-INTER, MIX-MET-MIXED, PU-MET-FAST, PU-MET-INTER, PU-SOL-THERM, U233-COMP-THERM, U233-MET-FAST, U233-SOL-INTER, and U233-SOL-THERM. The categories have three abbreviations which denote the (1) fissile material, (2) physical form of the fissile material, and (3) energy range of a majority of neutrons causing fission. All benchmark input files were run with the radiation transport code, Monte Carlo N-Particle\textsuperscript{\textregistered} Code Version 6.3\footnote{MCNP\textsuperscript{\textregistered} and Monte Carlo N-Particle\textsuperscript{\textregistered} are registered trademarks owned by Triad National Security, LLC, manager and operator of Los Alamos National Laboratory. Any third party use of such registered marks should be properly attributed to Triad National Security, LLC, including the use of the designation as appropriate. For the purposes of visual clarity, the registered trademark symbol is assumed for all references to MCNP within the remainder of this paper.} (\texttt{MCNP}). Two metrics, $\chi^2$ and mean absolute bias, were used to compare the ENDF/B-VIII.1 nuclear data library to the previously released ENDF/B-VIII.0 nuclear data library. The $\chi^2$ metric can be written mathematically as
\begin{equation}
    \chi^2 = \sum_{i}^{N}\frac{(C_i - E_i)^2}{E_i},
\label{eq:chi_sq}
\end{equation}
where $N$ is the number of criticality benchmark experiments, $C$ is the calculated neutron multiplication factor, and $E$ is the experimentally measured neutron multiplication factor. This definition of chi-squared in the above equation differs from that used in the previous subsection and does not include experimental uncertainties in the denominator; hence, the y-values in Fig.~\ref{fig:crit_overview_chiSq} are very different in scale from those in Fig.~\ref{fig:Mosteller}.
Additionally, mean absolute bias can be calculated using
\begin{equation}
    \frac{\sum_{i}^{N}\lvert C_i - E_i \rvert}{N},
\end{equation}
where $N$, $C$, and $E$ have the same definition as in Equation \ref{eq:chi_sq}. The mean absolute bias is reported in units of pcm (percent mille, 10$^{-5}$ $k_{\textrm{eff}}$). Two figures were produced for all benchmarks to compare ENDF/B-VIII.0 and ENDF/B-VIII.1. The first figure, Fig.~\ref{fig:crit_overview_chiSq}, shows the cumulative $\chi^2$ value as a function of criticality benchmark experiment number for both nuclear data libraries. The solid black vertical lines in this figure represent a division of benchmark categories. From this figure, it can be deduced that the calculated values of criticality using the ENDF/B-VIII.1 nuclear data library are closer to experimentally measured values than those calculated using the ENDF/B-VIII.0 nuclear data library as evident by the lower calculated cumulative $\chi^2$ value. Although the final cumulative $\chi^2$ value for ENDF/B-VIII.1 is lower than ENDF/B-VIII.0, the $\chi^2$ value for individual benchmarks for ENDF/B-VIII.1 is not always lower than ENDF/B-VIII.0.

\begin{figure}
\vspace{-2mm}
\centering
\includegraphics[width=0.45\textwidth]{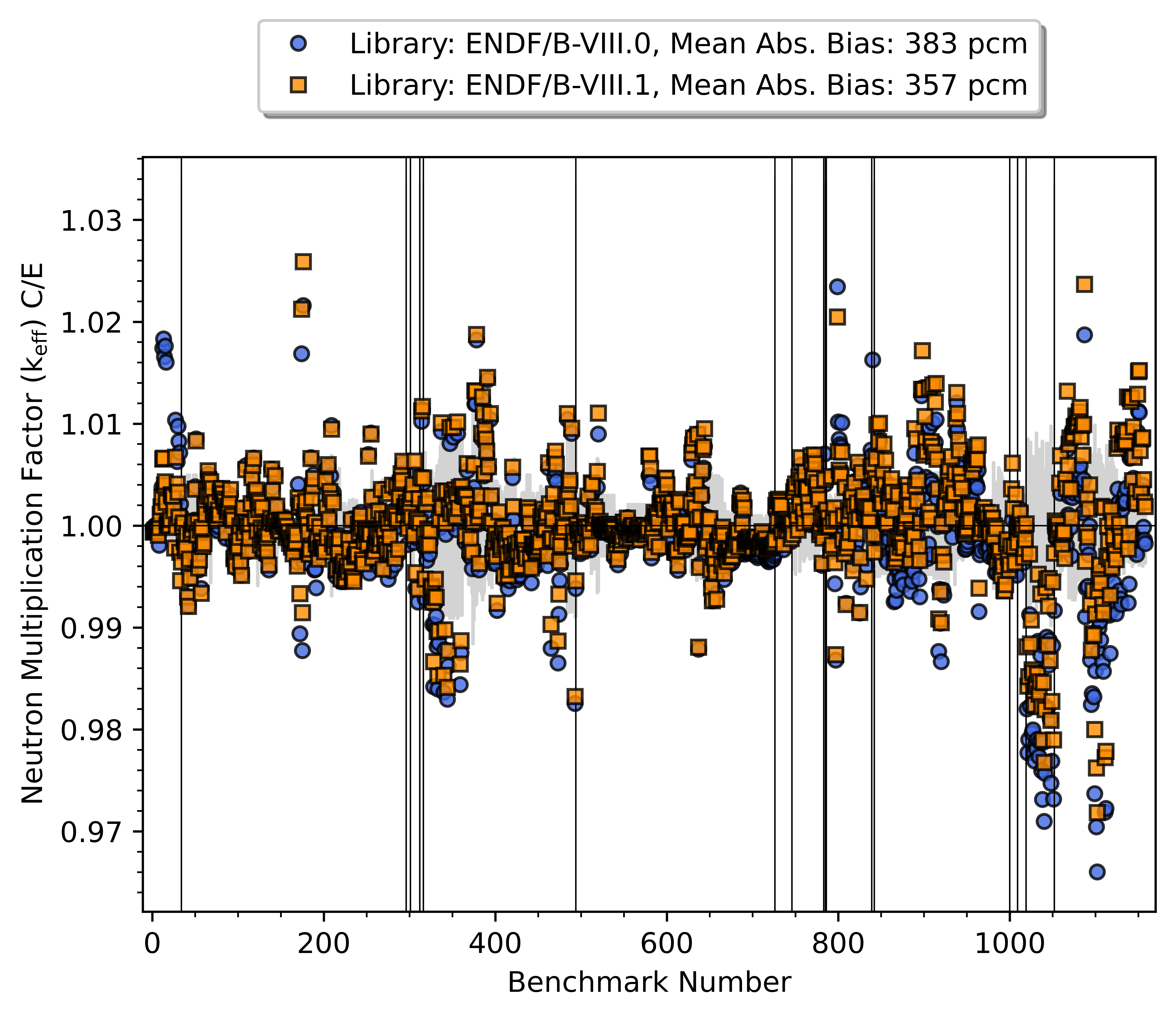}
\vspace{-2mm}
\caption{The value of the neutron multiplication factor ($k_{\textrm{eff}}$) plotted as a function of criticality benchmark experiment number. The grey shaded region outlines the experimental one-$\sigma$ uncertainty.}
\label{fig:crit_overview_keff}
\vspace{-2mm}
\end{figure}

The second figure, Fig.~\ref{fig:crit_overview_keff}, shows how well the calculated values of the neutron multiplication factor match experimentally measured values on a benchmark-by-benchmark basis. Each benchmark was run for a sufficient amount of time such that the calculation uncertainty is negligble; therefore, the grey shading outlines the experimental uncertainty. The mean absolute bias calculated for results using ENDF/B-VIII.1 is lower than the mean absolute bias calculated for results using ENDF/B-VIII.0. An improvement in C/E values is especially evident in the last benchmark categories, which contain $^{233}$U fissile material. This figure gives a brief overview of the entire library as a whole; however, it is important to investigate each category of benchmarks. After analyzing legacy and modern suites of criticality experiment benchmarks, Fig.~\ref{fig:crit_overview_keff} will be divided by fissile material into plots to provide more insightful analysis.

\begin{figure}
\vspace{-2mm}
\centering
\includegraphics[width=0.45\textwidth]{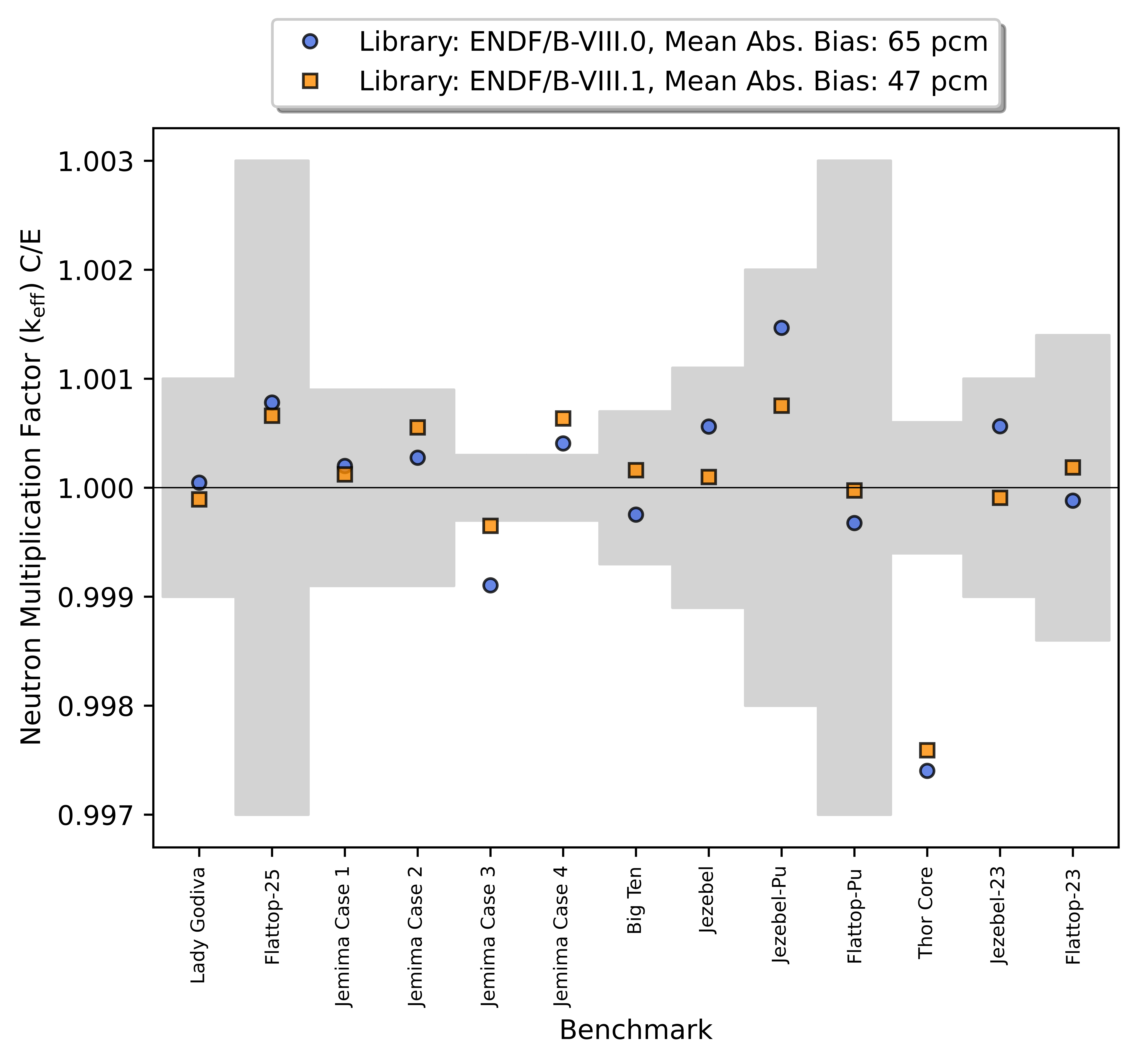}
\vspace{-2mm}
\caption{The value of the neutron multiplication factor ($k_{\textrm{eff}}$) plotted as a function of criticality benchmark experiment for the legacy criticality experiment benchmark suite. The grey shaded region outlines the experimental one-$\sigma$ uncertainty.}
\label{fig:crit_legacy}
\vspace{-2mm}
\end{figure}

The legacy criticality experiment benchmark suite has been used for many years to provide validation of intermediate (i.e., 1~eV--100~keV) and fast (i.e., $>$ 100~keV) energy cross-section values for $^{235,238}$U, $^{239}$Pu, $^{233}$U, and $^{232}$Th. The benchmarks and associated ICSBEP Handbook designations included in the legacy suite: (1) Lady Godiva, HEU-MET-FAST-001 Revision 2, (2) Flattop with HEU core (Flattop-25), HEU-MET-FAST-028 Revision 2 (3) Jemima, IEU-MET-FAST-001 Revision 0, (4) Bigten, IEU-MET-FAST-007 Revision 1, (5) Jezebel, PU-MET-FAST-001 Revision 5, (6) $^{240}$Pu Jezebel (Jezebel-Pu), PU-MET-FAST-002 Revision 1, (7) Flattop with Pu core (Flattop-Pu), PU-MET-FAST-006 Revision 1, (8) Thor Core, PU-MET-FAST-008, (9) U-233 Jezebel (Jezebel-23), U233-MET-FAST-001 Revision 3, and (10) Flattop with U-233 core (Flattop-23), U233-MET-FAST-006 Revision 1. As shown in Fig.~\ref{fig:crit_legacy} by the calculated results for plutonium benchmarks Jezebel, Jezebel-Pu, and Flattop-Pu, the updated evaluation for $^{239}$Pu has contributed to a much better agreement of calculated results with ENDF/B-VIII.1 to experimentally measured values compared to the calculated results of ENDF/B-VIII.0. Evaluation changes for $^{238}$U and $^{233}$U have also contributed to substantial improvements in calculating $k_{\textrm{eff}}$ for Jemima Case 3, Big Ten, and Jezebel-23. Overall, there is very good agreement between calculated and experimental values for fast plutonium and uranium bare systems (i.e., critical assemblies without reflectors). The results for the legacy criticality experiment benchmark suite are supplemented by results in the modern criticality experiment benchmark suite. 
As stated earlier, in Fig.~\ref{fig:crit_legacy}, for Jezebel, ENDF/B-VIII.1 matches the data better than ENDF/B-VIII.0 simply because CSEWG made a decision to adopt the recently reassessed (Jeff Favorite) rev-5 measured benchmark value, and calibrated ENDF/B-VIII.1 to that value.

\begin{figure}
\vspace{-2mm}
\centering
\includegraphics[width=0.45\textwidth]{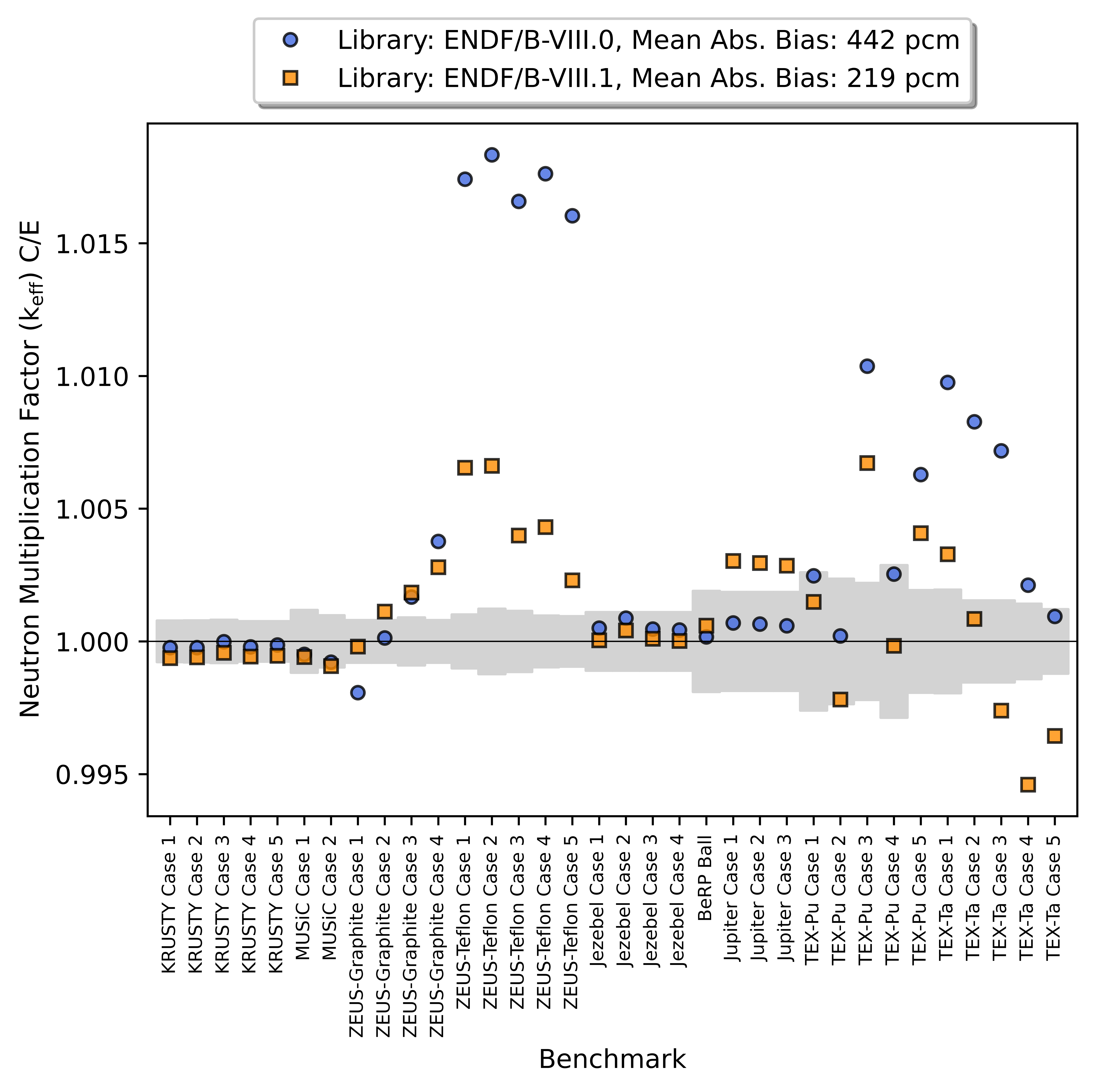}
\vspace{-2mm}
\caption{The value of the neutron multiplication factor ($k_{\textrm{eff}}$) plotted as a function of criticality benchmark experiment for the modern criticality experiment benchmark suite. The grey shaded region outlines the experimental one-$\sigma$ uncertainty.}
\label{fig:crit_modern}
\vspace{-2mm}
\end{figure}

The modern criticality experiment benchmark suite was put together to showcase recently performed criticality safety experiments. The benchmark evaluations for these criticality safety experiments are extremely well documented. The benchmarks and associated ICSBEP Handbook designations included in the modern suite: (1) Kilowatt Reactor Using Stirling Technology (KRUSTY), HEU-MET-FAST-101 (Detailed Cases 1-5), (2) Measurement of Uranium Subcritical and Critical (MUSiC), HEU-MET-FAST-104 (Detailed Cases 1-2), (3) ZEUS-Graphite, HEU-MET-INTER-006 (Detailed Cases 1-4), (4) ZEUS-Teflon, also known as Critical Unresolved Region Integral Experiment (CURIE), HEU-MET-INTER-011 (Detailed Cases 1-5), (5) Jezebel Revision 5, PU-MET-FAST-001 (Detailed Cases 1-4), (6) Beryllium-reflected Plutonium (BeRP) Ball, PU-MET-FAST-038 (Detailed Case 1), (7) Jupiter, PU-MET-FAST-047 (Detailed Cases 1-3), (8) Plutonium Thermal/Epithermal eXperiments (TEX), PU-MET-MIXED-002 (Detailed Cases 1-5), and (9) Tantalum TEX, PU-MET-MIXED-003 (Detailed Cases 1-5). This suite of criticality experiment benchmarks provides validation for thermal (i.e., $<$ 1 eV), intermediate, and fast cross sections of various fissile materials, moderator materials, and reflector materials. Fig.~\ref{fig:crit_modern} shows the calculated over experimental values of $k_{\textrm{eff}}$ for all benchmarks in this suite. A significant reduction in bias can be seen for many benchmarks in Fig.~\ref{fig:crit_modern}. The significant reduction in bias for ZEUS-Teflon and specific cases of TEX-Ta are directly related to evaluation updates made to $^{19}$F and $^{181}$Ta. An important note: the reduction of bias in Jezebel was mainly due to the model being updated from Revision~4 to Revision~5. In the future, it will be important to investigate evaluations of $^{9}$Be, Pb isotopes, and $^{181}$Ta to better understand the undesired calculated value changes to KRUSTY, BeRP Ball, Jupiter, and cases of TEX-Ta with higher percentages of thermal neutrons causing fission. One of the major evaluation changes of this new nuclear data library for actinides was $^{239}$Pu.

\begin{figure}
\vspace{-2mm}
\centering
\includegraphics[width=0.45\textwidth]{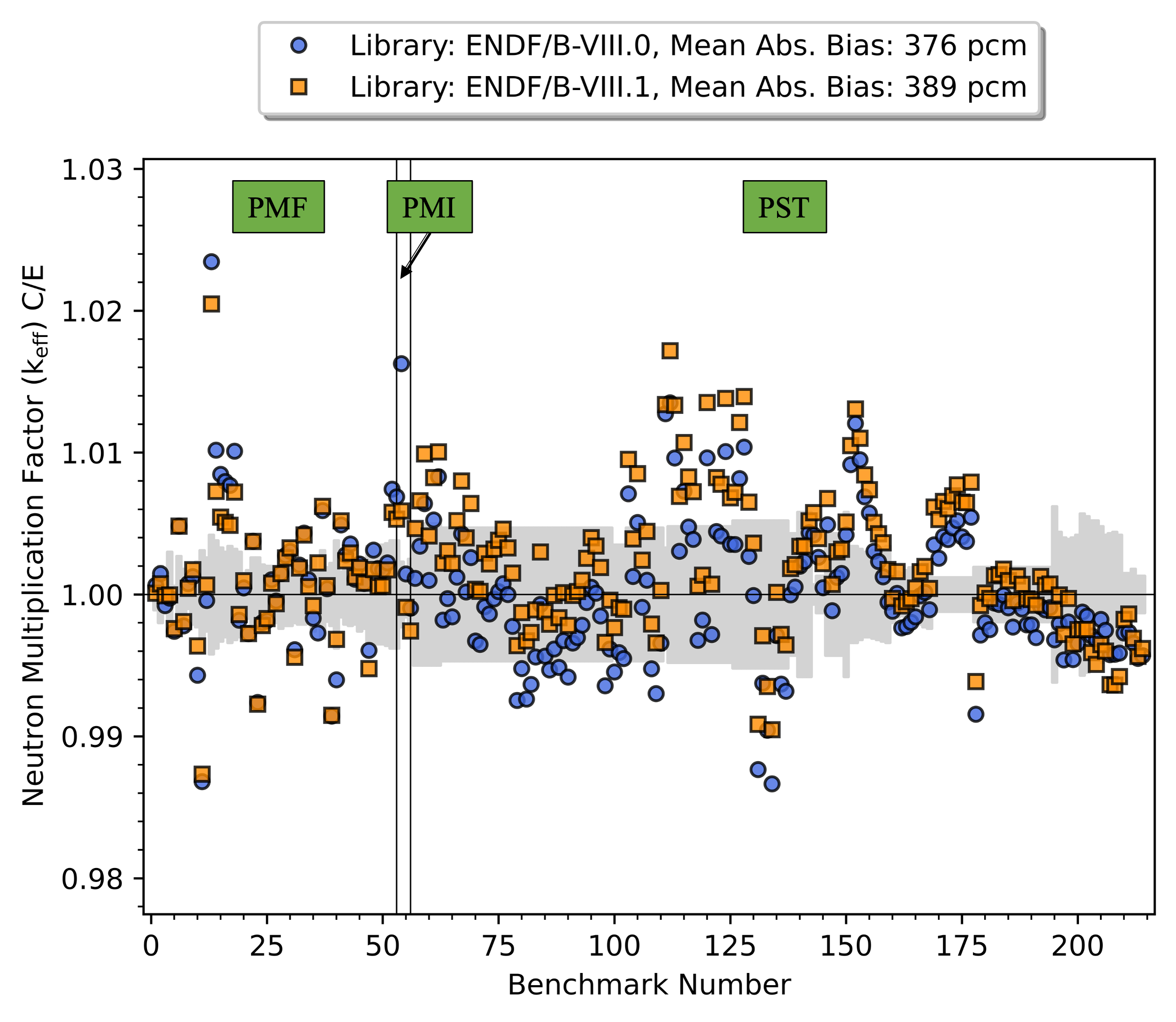}
\vspace{-2mm}
\caption{The value of the neutron multiplication factor ($k_{\textrm{eff}}$) plotted as a function of criticality benchmark experiment for the plutonium criticality experiment benchmark suite. The grey shaded region outlines the experimental one-$\sigma$ uncertainty.}
\label{fig:crit_pu}
\vspace{-2mm}
\end{figure}

\begin{figure}
\vspace{-2mm}
\centering
\includegraphics[width=\columnwidth]{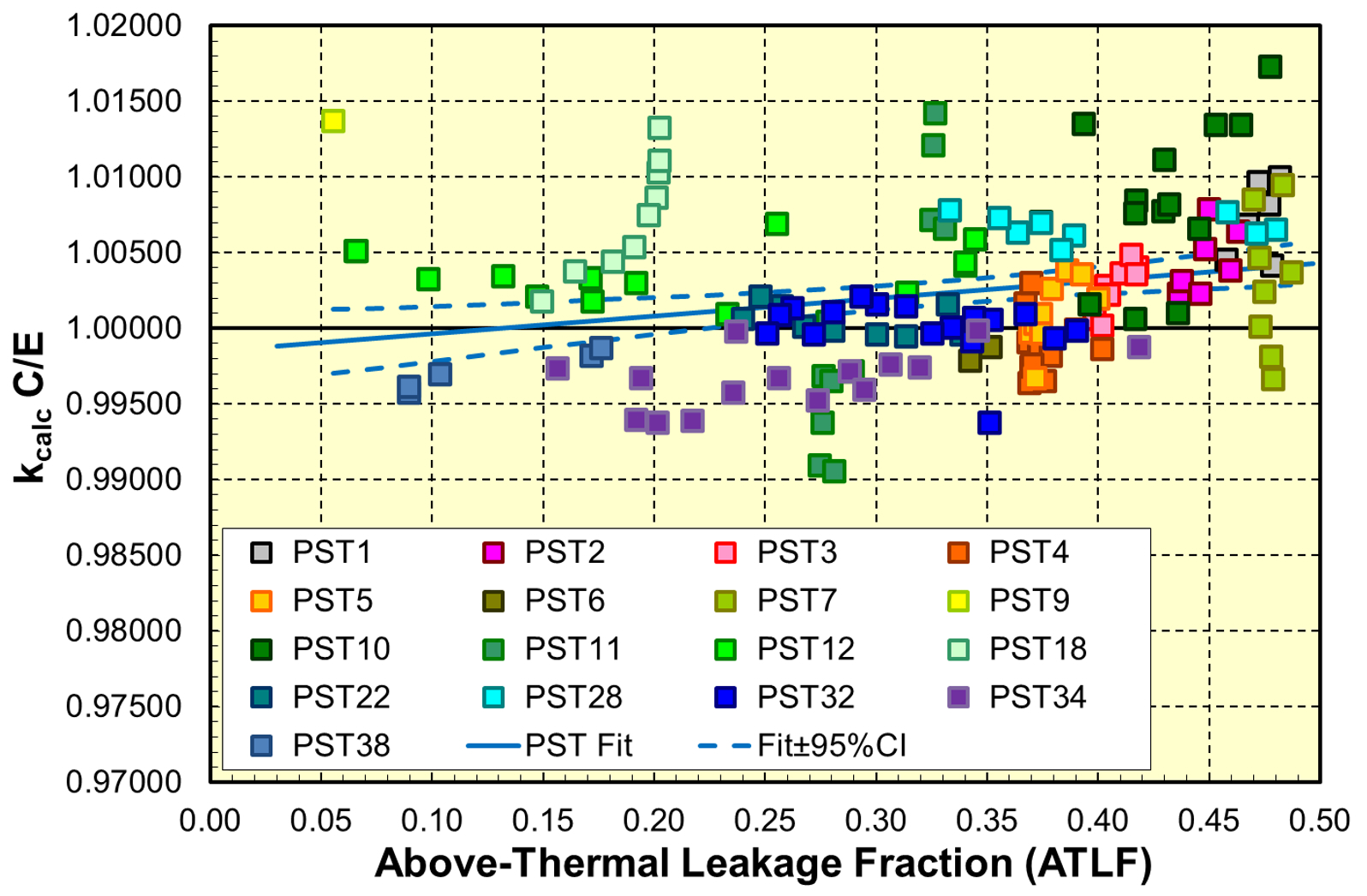}
\vspace{-2mm}
\caption{The calculated over experimental value of the neutron multiplication factor ($k_{\textrm{eff}}$) plotted as a function of above-thermal leakage fraction (ATLF) for PU-SOL-THERM (PST) benchmarks. A regression fit is shown for ENDF/B-VIII.1. The ideal regression fit would have an intercept of 1 and slope of zero.}
\label{fig:pst_atlf}
\vspace{-2mm}
\end{figure}

The benchmarks containing $^{239}$Pu fissile material are shown in Fig.~\ref{fig:crit_pu}. In this figure, benchmarks with a majority of fissions being caused by higher energy neutrons have similar calculated values using both ENDF/B-VIII.0 and ENDF/B-VIII.1. The $^{239}$Pu evaluation was specifically adjusted such that the calculated value of $k_{\textrm{eff}}$ for the latest revision of the Jezebel benchmark (Revision 5) exactly matched the experimental value\footnote{A similar adjustment was made for the ENDF/B-VIII.0 library using Revision 4 of the Jezebel model.}. Therefore, calculated values of plutonium benchmarks with higher energy neutron spectra using ENDF/B-VIII.1 are slightly closer to experimentally measured values than those calculated with ENDF/B-VIII.0. The opposite is true for most plutonium solution benchmarks that have thermal-energy neutron spectra. The exception to this statement includes the case of plutonium solution benchmarks with gadolinium (PU-SOL-THERM-034) C/E values that are now closer to unity. The average energy of the thermal PFNS for the latest $^{239}$Pu evaluation is 30 keV lower than the one used in the ENDF/B-VIII.0 library, which has resulted in increased criticality of high-leakage critical assemblies. The increase in criticality breaks the nearly flat trend for PST benchmark C/E values as a function of above-thermal leakage fraction (ATLF) observed for the ENDF/B-VIII.0 library. The new positively sloped trend line for C/E values as a function of ATLF for the ENDF/B-VIII.1 library is plotted in Fig.~\ref{fig:pst_atlf}. Due to the updated changes in thermal PFNS, the $^{239}$Pu cross sections and $\overline{\nu}$ need to be adjusted to reduce criticality. Unfortunately, such reduction is in contradiction with the required criticality increase to satisfy depletion metrics and reactivity temperature coefficients (e.g., the CEA MISTRAL benchmark).  The success of the ENDF/B-VIII.0 $^{239}$Pu evaluation to reduce the magnitude of the trend line as a function of the ATLF should be revisited in future releases considering the recommended $^{239}$Pu thermal PFNS. Measurements of PFNS at the thermal point and fluctuations of neutron multiplicity below 5~eV need to be prioritized to improve the PST performance. Uranium isotope RRR evaluations did not have as significant of performance changes as the updated $^{239}$Pu evaluation showing the challenge of this reevaluation.

\begin{figure}
\vspace{-2mm}
\centering
\includegraphics[width=0.45\textwidth]{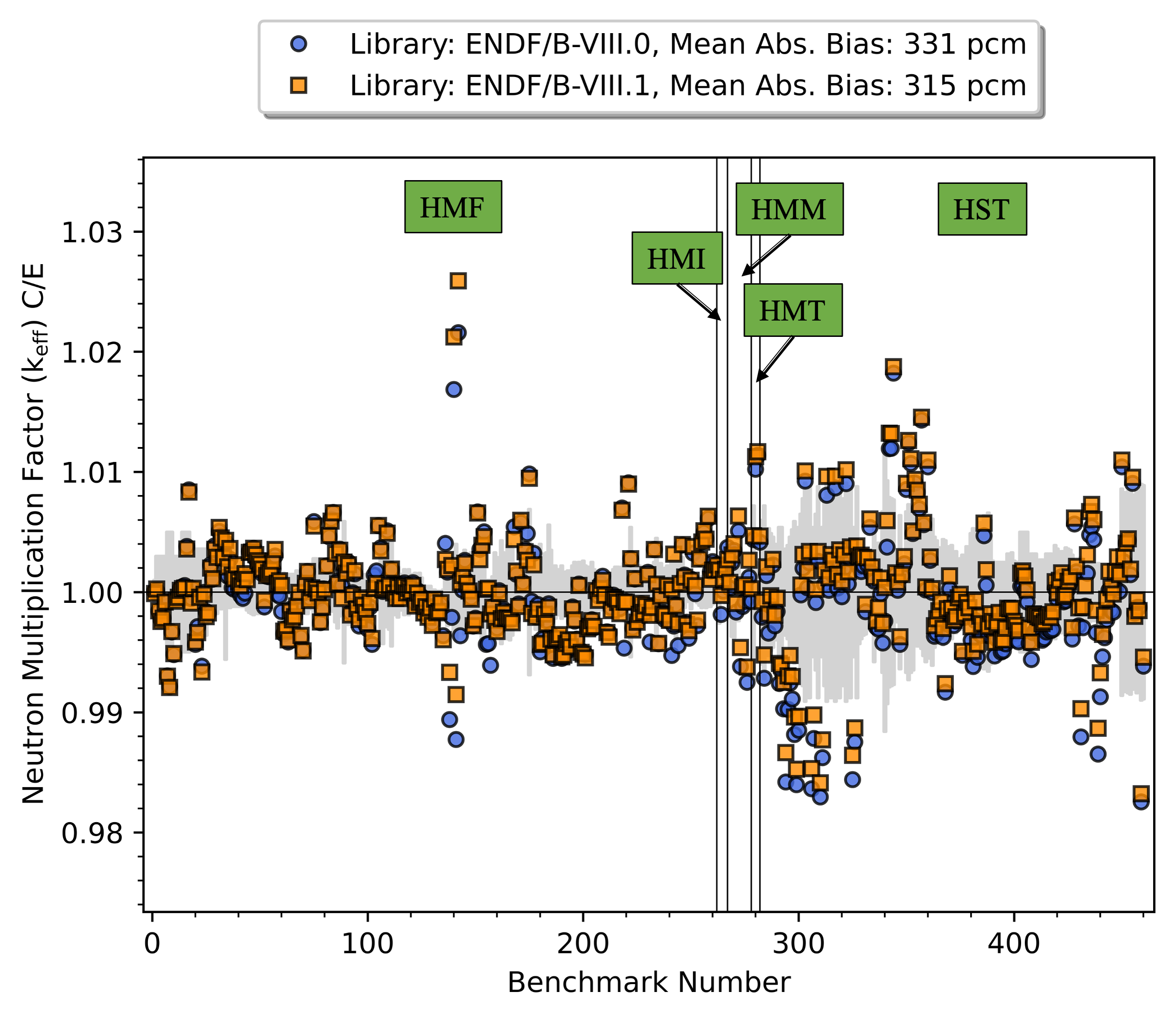}
\vspace{-2mm}
\caption{The value of the neutron multiplication factor ($k_{\textrm{eff}}$) plotted as a function of criticality benchmark experiment for the highly enriched uranium (HEU) criticality experiment benchmark suite. The grey shaded region outlines the experimental one-$\sigma$ uncertainty.}
\label{fig:crit_heu}
\vspace{-2mm}
\end{figure}

\begin{figure}
\vspace{-2mm}
\centering
\includegraphics[width=0.45\textwidth]{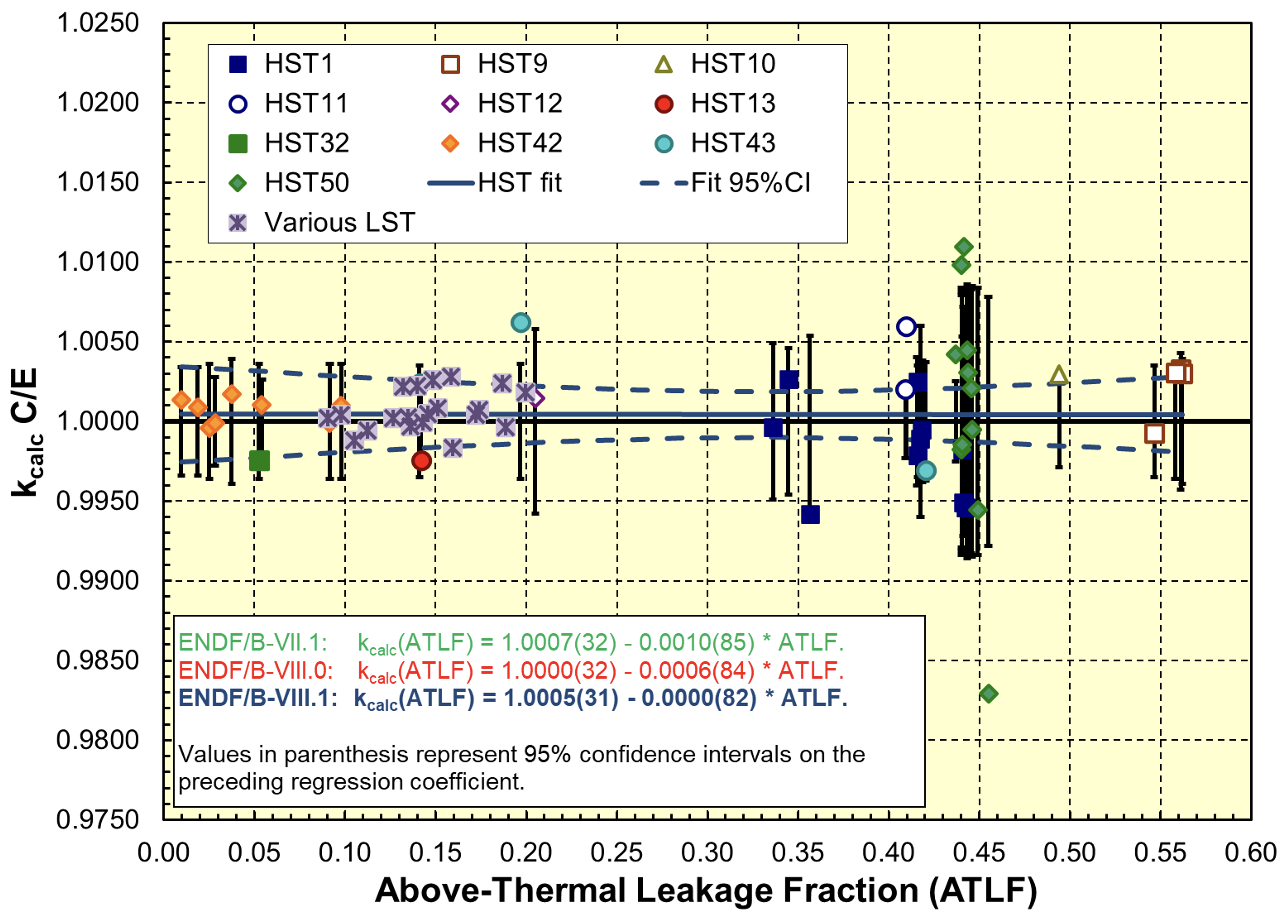}
\vspace{-2mm}
\caption{The calculated over experimental value of the neutron multiplication factor ($k_{\textrm{eff}}$) plotted as a function of above-thermal leakage fraction (ATLF) for HEU-SOL-THERM (HST) and LEU-SOL-THERM (LST) benchmarks. A regression fit is shown for ENDF/B-VII.1, ENDF/B-VIII.0, and ENDF/B-VIII.1. The ideal regression fit would have an intercept of 1 and slope of zero.}
\label{fig:crit_hst}
\vspace{-2mm}
\end{figure}

The changes in $^{235}$U and $^{238}$U did not have a significant effect on the calculated $k_{\textrm{eff}}$ values of HEU systems as shown in Fig.~\ref{fig:crit_heu}. The calculated values for the HEU benchmark series are in excellent agreement with experimentally measured values. Even with more benchmarks, the mean absolute bias shown calculated for ENDF/B-VIII.1 in Fig.~\ref{fig:crit_heu} is lower than that calculated for ENDF/B-VIII.1 in Fig.~\ref{fig:crit_pu}. Normally, the HEU solution benchmarks from Fig.~\ref{fig:crit_heu} are isolated to test thermal $^{235}$U nuclear data. The dependence of $k_{\textrm{eff}}$ as a function of ATLF is calculated for ENDF/B-VII.1, ENDF/B-VIII.0, and ENDF/B-VIII.1 in Fig.~\ref{fig:crit_hst}. The LEU-SOL-THERM (LST) benchmarks included in Fig.~\ref{fig:crit_hst} were not considered in the calculation but rather are plotted for comparison. The regression fit for ENDF/B-VIII.1 has an intercept slightly higher than unity, which is undesired, but has a very small slope, which is desired. The results are very consistent with ENDF/B-VIII.0 and do not require considerable attention. The calculated results for benchmarks containing LEU instead of HEU are also comparable for both nuclear data libraries, with an slight +70~pcm increase in LEU-COMP-THERM (LCT) criticality of the ENDF/B-VIII.1 results compared to \prENDF{}
, which makes the new ENDF/B-VIII.1 evaluation looks much closer to the reference ENDF/B-VII.1 LCT criticality, which is important for reactor applications.

\begin{figure}
\vspace{-2mm}
\centering
\includegraphics[width=0.45\textwidth]{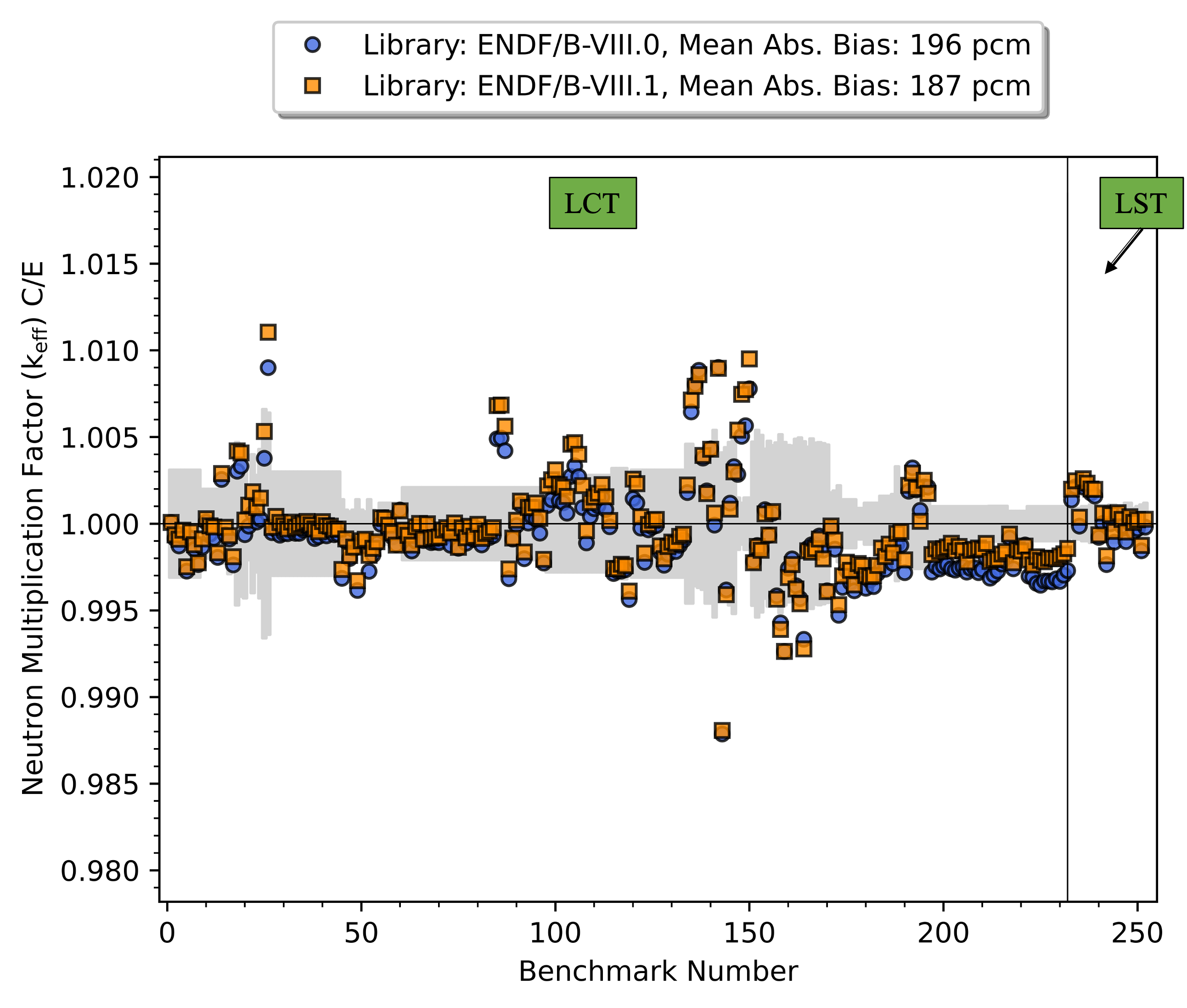}
\vspace{-2mm}
\caption{The value of the neutron multiplication factor ($k_{\textrm{eff}}$) plotted as a function of criticality benchmark experiment for the low enriched uranium (LEU) criticality experiment benchmark suite. The grey shaded region outlines the experimental one-$\sigma$ uncertainty.}
\label{fig:crit_leu}
\vspace{-2mm}
\end{figure}

Calculated results for benchmarks containing LEU are shown in Fig.~\ref{fig:crit_leu}. Conclusive statements about the results shown in this figure are very similar to those previously stated for benchmarks containing HEU. One important trend to take note of in Fig.~\ref{fig:crit_leu} is the slight increase in reactivity calculated for almost all criticality benchmark experiments using ENDF/B-VIII.1. The LCT set of benchmarks is important for reactor lattice calculations. The low mean absolute bias calculated for LCT using ENDF/B-VIII.1 is a great sign that these reactor validation benchmarks are well characterized and useful for future validation. 

\begin{figure}
\vspace{-2mm}
\centering
\includegraphics[width=0.44\textwidth]{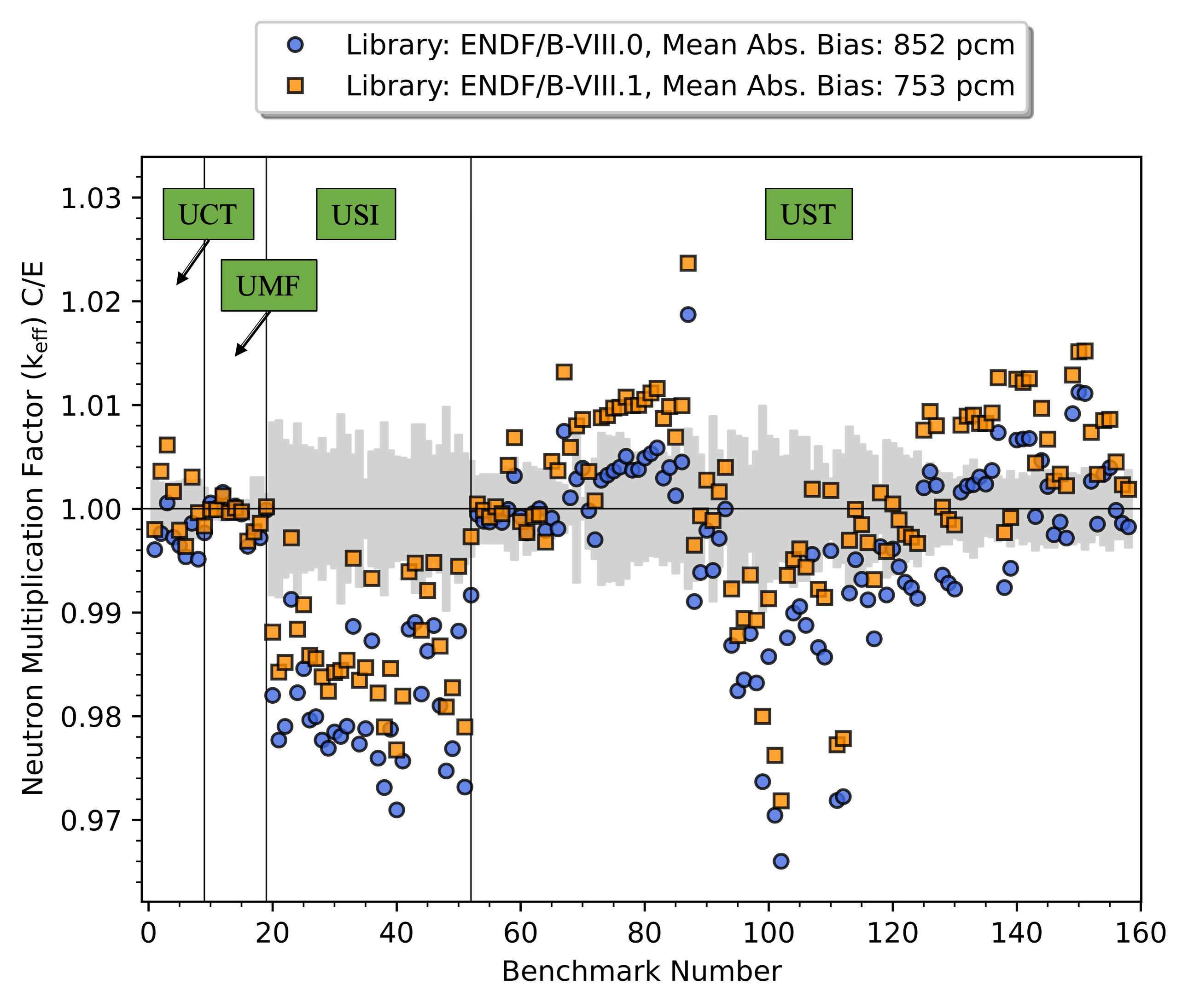}
\vspace{-2mm}
\caption{The value of the neutron multiplication factor ($k_{\textrm{eff}}$) plotted as a function of criticality benchmark experiment for the $^{233}$U criticality experiment benchmark suite. The grey shaded region outlines the experimental one-$\sigma$ uncertainty.}
\label{fig:crit_u233}
\vspace{-2mm}
\end{figure}

\begin{figure}
\vspace{-2mm}
    \centering
    \subfigure[ENDF/B-VIII.0]{\includegraphics[width=0.44\textwidth]{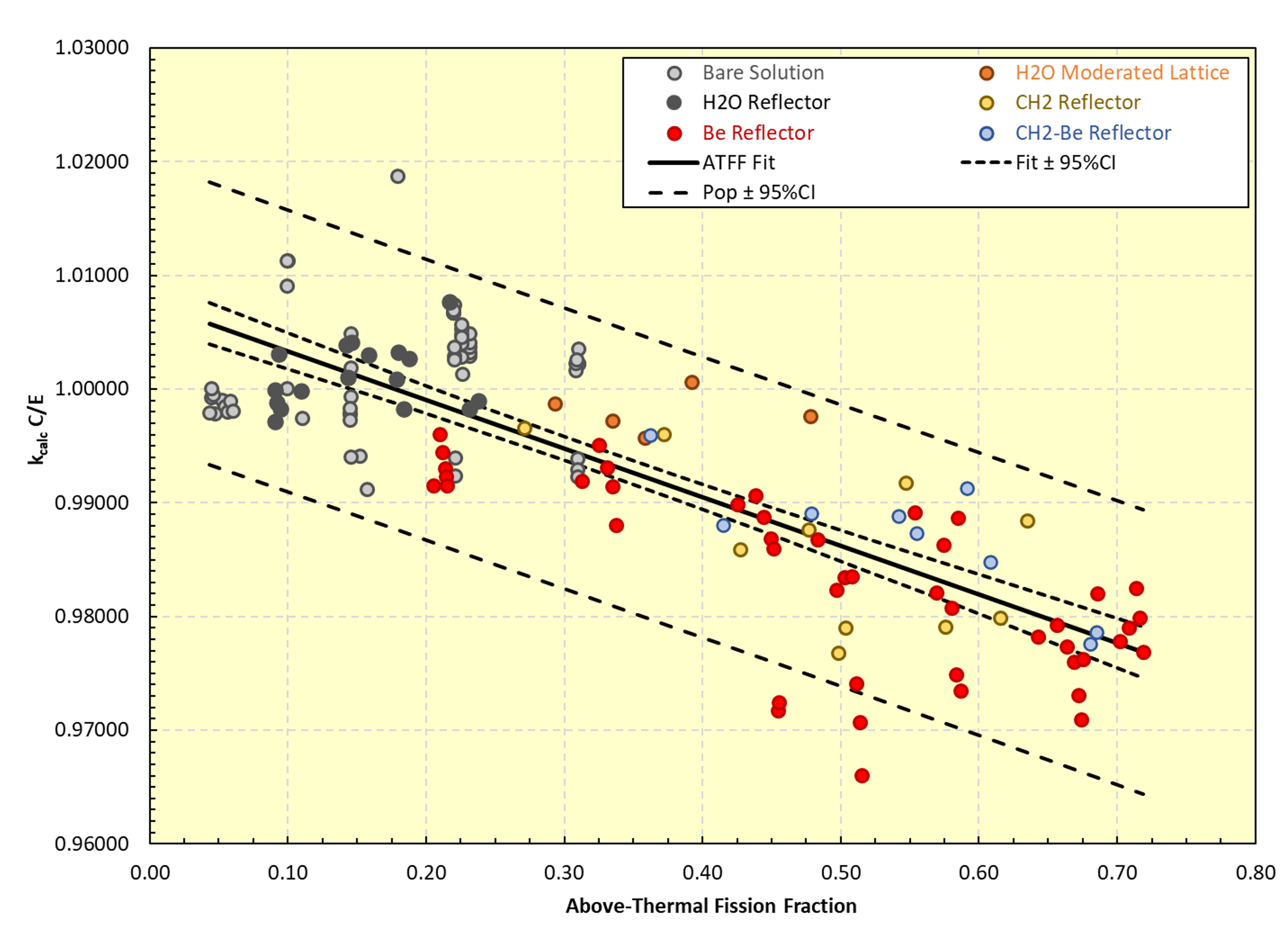}} 
    \subfigure[ENDF/B-VIII.1]{\includegraphics[width=0.44\textwidth]{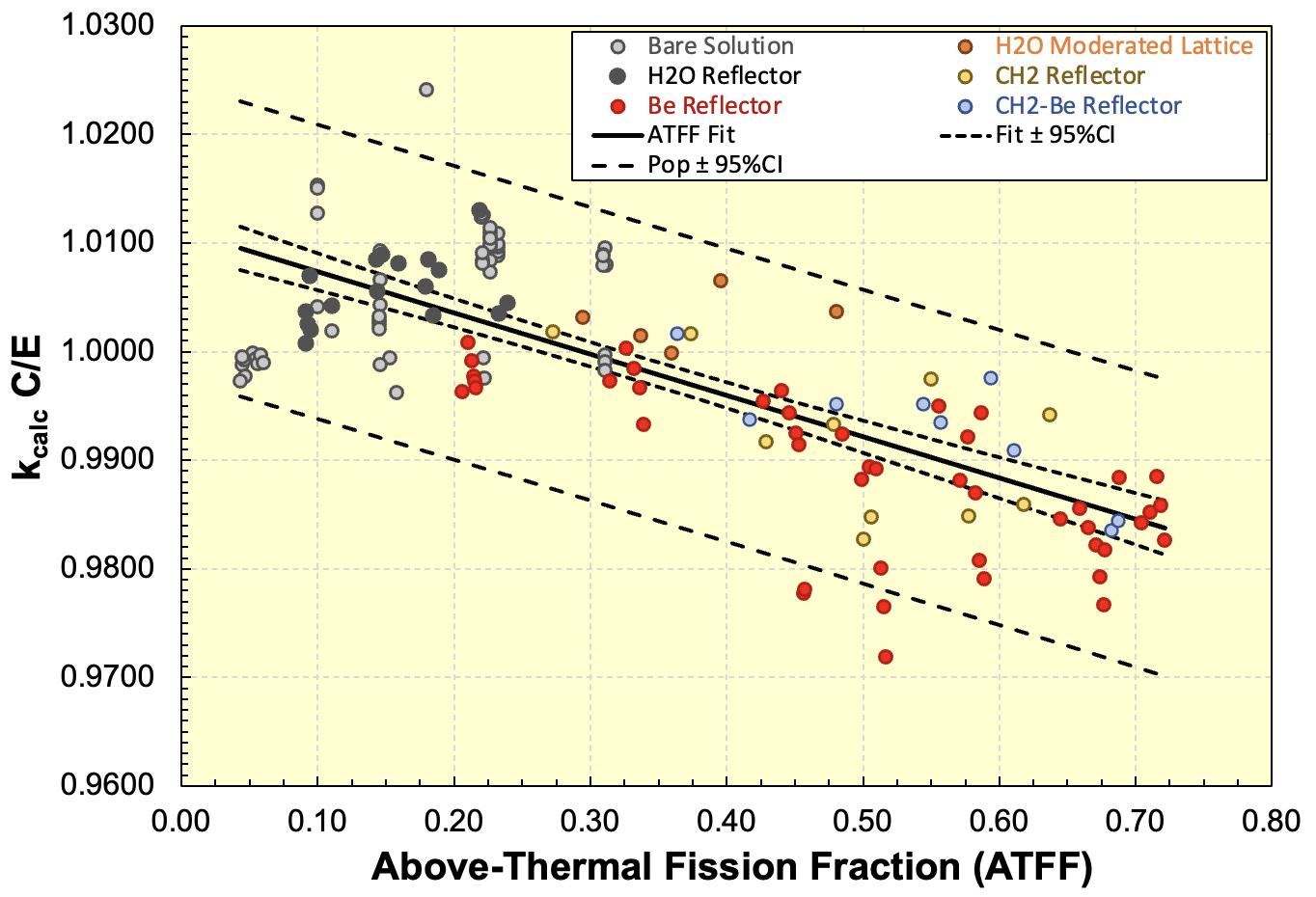}} 
\vspace{-2mm}
    \caption{The calculated over experimental value of $k_{\textrm{eff}}$ plotted as a function of above-thermal fission fraction for (a) ENDF/B-VIII.0 and (b) ENDF/B-VIII.1. The regression fit and confidence intervals are plotted with the bare and reflected criticality benchmark experiments.}
    \label{fig:crit_ust}
\vspace{-2mm}
\end{figure}

The last set of benchmarks used for validation contained $^{233}$U.
The changes of the $^{233}$U evaluation in the resonance region created significant performance changes in calculated values of $k_{\textrm{eff}}$ using ENDF/B-VIII.1 compared to ENDF/B-VIII.0 as shown in Fig.~\ref{fig:crit_u233}. The mean absolute bias calculated for benchmarks using ENDF/B-VIII.1 was substantially reduced compared to the mean absolute bias calculated for benchmarks using ENDF/B-VIII.0. Although there was a significant reduction in mean absolute bias with the new $^{233}$U evaluation changes, the calculated over experimental values for many U233-SOL-INTER and U233-SOL-THERM are far from unity. A more detailed analysis of the calculated over experimental value trend for U233-SOL-THERM benchmarks is presented in Fig.~\ref{fig:crit_ust}. Eigenvalue calculations for $^{233}$U thermal and intermediate energy benchmarks have exhibited a strong, negative trend with increasing energy for decades. It is shown in Fig.~\ref{fig:crit_ust}(b) that the trend for ENDF/B-VIII.1 is very much the same as previously calculated. At higher energies, beryllium and Be-CH2 reflected systems are now calculated about 1000 pcm higher, which is a better result compared to \prENDF{}. However, the average calculated $k_{\textrm{eff}}$ is still low. At lower energies (i.e., above-thermal fission fraction 0.1--0.3), calculated $k_{\textrm{eff}}$ values for bare solution and water-reflected systems are now too large and show a positive trend compared to \prENDF{}, which is bad. Overall, the trend line for $^{233}$U criticality benchmark experiments is flattening, but careful attention will be needed for bare solutions and water-reflected systems. An in-depth analysis of $^{233}$U benchmark quality is long due, especially for benchmarks having above-thermal fission fraction larger than 0.35. In summary, the evaluation of $^{233}$U in the resonance region is still challenging; additional efforts will be needed in future libraries.


\subsection{Delayed Neutron Experiment Results}

Although the delayed neutron data have not been changed from their ENDF/B-VIII.0 values, they were tested against measurements of effective delayed neutron fraction $\beta_\mathrm{eff}$ in critical configurations. Unlike the situation for $k_\mathrm{eff}$, only a handful of measurements of $\beta_\mathrm{eff}$ have been reported in open literature with sufficiently detailed information.  Twenty-one measurements originally used in Ref.~\cite{van2006benchmarking} were supplemented with two ORSphere measurements. We avoid the term ``benchmark'' for these cases, because good benchmark descriptions, comparable to those given in the ICSBEP Handbook, are generally not available. 

\begin{figure}[!htb]
    \centering
    \includegraphics[width=1.0\linewidth]{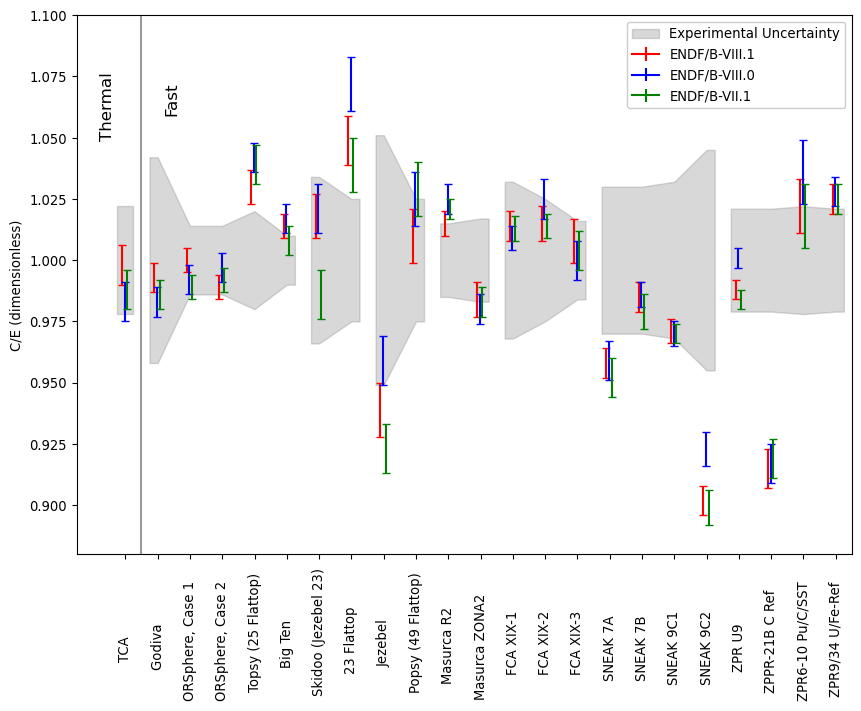}
    \caption{C/E values for delayed neutron tests. The systems are roughly ordered by spectrum and fissile nuclide driving the assembly, thermal (left) and fast (right) then \nuc{235}{U}-driven, \nuc{233}{U}-driven, \nuc{239}{Pu}-driven, and fast reactor mockup assemblies. The grey shaded region outlines the experimental one-$\sigma$ uncertainty.}
    \label{fig:delayed_neutron_exp}
\end{figure}

The delayed neutron experiment measurement data were tested using the MC21 correlated sampling method~\cite{griesheimer2019simplified}. The $\beta_\mathrm{eff}$ measurement data include assemblies with $^{233}$U, $^{235}$U, $^{238}$U, and $^{239}$Pu. There is one thermal spectrum assembly fueled by low enriched $^{235}$U (TCA). The majority of the measurement data is for fast spectrum assemblies, including $^{233}$U-driven assemblies (Skidoo and 23 Flattop), $^{235}$U-driven assemblies (Big Ten, Topsy, Godiva, Masurca, FCA, SNEAK, ZPR, ORSphere), $^{239}$Pu-driven assemblies (Popsy, Jezebel, Masurca, FCA, SNEAK, ZPR) and some mixed assemblies (Masurca, FCA, SNEAK, ZPPR). Table~\ref{tab:delayed_neutron_c_e_comparison} provides the MC21 C/E values for the $\beta_\mathrm{eff}$ experiments, which are also plotted in Fig.~\ref{fig:delayed_neutron_exp}. C/E values are provided for ENDF/B-VIII.1, ENDF/B-VIII.0, and ENDF/B-VII.1 along with the standard deviation of the experimental uncertainty of the $\beta_\mathrm{eff}$ measurements. Good agreement with the delayed neutron experiments is observed.  One should note that the tests performed are only sensitive to the total delayed neutron yields. The delayed neutron yields per group are not tested, nor are the values for the decay constant per group.  

\begin{table*}[!htb]
    \centering
    \caption{MC21 C/E values for the $\beta_\mathrm{eff}$ calculations. The uncertainties quoted for the C/E values include only the statistical uncertainty of the calculation. All cases are fast spectrum except TCA.}
    \begin{tabular}{ccccc}
    \toprule \toprule
    & Experiment & ENDF/B-VIII.1 & ENDF/B-VIII.0 & ENDF/B-VII.1 \\
    & $\beta_\mathrm{eff}$ (pcm) & C/E  & C/E & C/E \\\midrule
TCA                 & 771  (2.2\%)  &0.998$\pm$0.008	& 0.983$\pm$0.008	& 0.988$\pm$0.008 \\
Big Ten	            & 720 (1.0\%)   &1.014$\pm$0.005	& 1.017$\pm$0.006	& 1.008$\pm$0.006 \\
Topsy (25 Flattop)	& 665 (2.0\%)   &1.030$\pm$0.007	& 1.042$\pm$0.006	& 1.039$\pm$0.008 \\
Godiva	            & 659 (4.2\%)   &0.993$\pm$0.006	& 0.983$\pm$0.006	& 0.986$\pm$0.006 \\
23 Flattop	        & 360 (2.5\%)   &1.049$\pm$0.010	& 1.072$\pm$0.011	& 1.039$\pm$0.011 \\
Skidoo (Jezebel 23)	& 290 (3.4\%)   &1.018$\pm$0.009	& 1.021$\pm$0.010	& 0.986$\pm$0.010 \\
Popsy (49 Flattop)	& 276 (2.5\%)   &1.010$\pm$0.011	& 1.025$\pm$0.011	& 1.029$\pm$0.011 \\
Jezebel	            & 195 (5.1\%)   &0.939$\pm$0.011	& 0.959$\pm$0.010	& 0.923$\pm$0.010 \\
Masurca R2	        & 721 (1.5\%)   &1.015$\pm$0.005	& 1.025$\pm$0.006	& 1.021$\pm$0.004 \\
Masurca ZONA2 & 349 (1.7\%)   &0.984$\pm$0.007	& 0.980$\pm$0.006	& 0.983$\pm$0.006 \\
FCA XIX-1	        & 742 (3.2\%)   &1.014$\pm$0.006	& 1.009$\pm$0.005	& 1.013$\pm$0.005 \\
FCA XIX-2	        & 364 (2.5\%)   &1.015$\pm$0.007	& 1.025$\pm$0.008	& 1.014$\pm$0.005 \\
FCA XIX-3	        & 251 (1.6\%)   &1.008$\pm$0.009	& 1.000$\pm$0.008	& 1.004$\pm$0.008 \\
SNEAK 7A	        & 395 (3.0\%)   &0.958$\pm$0.006	& 0.959$\pm$0.008	& 0.952$\pm$0.008 \\
SNEAK 7B	        & 429 (3.0\%)   &0.985$\pm$0.006	& 0.986$\pm$0.005	& 0.979$\pm$0.007 \\
SNEAK 9C1	        & 758 (3.2\%)   &0.971$\pm$0.005	& 0.970$\pm$0.005	& 0.970$\pm$0.004 \\
SNEAK 9C2	        & 426 (4.5\%)   &0.902$\pm$0.006	& 0.923$\pm$0.007	& 0.899$\pm$0.007 \\
ZPR U9	            & 731 (2.1\%)   &0.988$\pm$0.004	& 1.001$\pm$0.004	& 0.984$\pm$0.004 \\
ZPPR-21B C Ref	    & 384 (2.1\%)   &0.915$\pm$0.008	& 0.917$\pm$0.008	& 0.919$\pm$0.008 \\
ZPR6-10 Pu/C/SST	& 223 (2.2\%)   &1.022$\pm$0.011	& 1.036$\pm$0.013	& 1.018$\pm$0.013 \\
ZPR9/34 U/Fe-Ref	& 671 (2.1\%)	&1.025$\pm$0.006	& 1.028$\pm$0.006	& 1.025$\pm$0.006 \\
ORSphere, Case 1	& 657 (1.4\%)	&1.000$\pm$0.005	& 0.992$\pm$0.006	& 0.989$\pm$0.005 \\
ORSphere, Case 2	& 657 (1.4\%)	&0.989$\pm$0.005	& 0.997$\pm$0.006	& 0.992$\pm$0.005 \\
\bottomrule \bottomrule 
    \end{tabular}
    \label{tab:delayed_neutron_c_e_comparison}
\end{table*}

\subsection{Reaction Rate Ratio Results}

Reaction rate ratios were measured in the legacy criticality benchmark experiments. Typically, foils made of certain materials were placed inside a central cavity or glory hole, irradiated, and then subsequently placed near/inside a detection system. The radiation detection system would count radiation emitted from the foils to measure the activity. Later, the measurement of activity would be used to calculate a reaction rate ratio. The measurement of these ratios was well established; however, the procedure for measuring reaction rate ratios is less common today. Reaction rate ratios were calculated using ENDF/B-VIII.0 and ENDF/B-VIII.1 for legacy criticality benchmark experiments in Table \ref{tab:rxnRate1}. The naming convention used in Section 5.1 for these legacy criticality benchmark experiments is replicated here. Calculation of these reaction rate ratios was performed in MCNP by applying a detector tally in the center of the fissile material of each benchmark. The neutron flux was calculated in a sphere with radius equal to 1.3335 cm (0.525 inches), which is the typical size of a foil measured for reaction rate ratios. The values calculated using ENDF/B-VIII.0 and ENDF/B-VIII.1 are very consistent. There are no significant differences that need to be highlighted, addressed, or investigated.   

\begin{table*}[!tbp]
    \centering
    \caption{Reaction rate ratio values for legacy criticality benchmark experiments. The calculated ratios were rounded to the same significant figures as the measured values. The four ratios that do not have associated measured values are included for historical comparison. These LANL-calculated values can be compared to LLNL GNDS/Mercury calculated values in Table~\ref{tab:appendix:GNDS_fission_ratios}. They are similar.}
    \begin{tabular}{cccc}
    \toprule \toprule
      & & \multicolumn{2}{c}{Calculated Value} \\
        \cmidrule{3-4}
        Experiment, Ratio & Measured Value & ENDF/B-VIII.0 & ENDF/B-VIII.1 \\ 
\midrule
        Lady Godiva, $\frac{^{233}\textrm{U(n,f)}}{^{235}\textrm{U(n,f)}}$ & 1.59(3) & 1.58 & 1.58 \\
        Lady Godiva, $\frac{^{237}\textrm{Np(n,f)}}{^{235}\textrm{U(n,f)}}$ & 0.8516(120) & 0.8311 & 0.8307 \\
        Lady Godiva, $\frac{^{238}\textrm{U(n,f)}}{^{235}\textrm{U(n,f)}}$ & 0.1643(18) & 0.1582 & 0.1580 \\
        Lady Godiva, $\frac{^{239}\textrm{Pu(n,f)}}{^{235}\textrm{U(n,f)}}$ & 1.4152(140) & 1.3844 & 1.3832 \\
\midrule
        Big Ten, $\frac{^{238}\textrm{U(n,f)}}{^{235}\textrm{U(n,f)}}$ & 0.0375(9) & 0.0357 & 0.0362 \\
        Big Ten, $\frac{^{239}\textrm{Pu(n,f)}}{^{235}\textrm{U(n,f)}}$ & 1.198(28) & 1.170 & 1.170 \\
\midrule
        Jezebel, $\frac{^{233}\textrm{U(n,f)}}{^{235}\textrm{U(n,f)}}$ & 1.578(27) & 1.566 & 1.566 \\
        Jezebel, $\frac{^{237}\textrm{Np(n,f)}}{^{235}\textrm{U(n,f)}}$ & 0.9835(14) & 0.9768 & 0.9710 \\
        Jezebel, $\frac{^{238}\textrm{U(n,f)}}{^{235}\textrm{U(n,f)}}$ & 0.2133(23) & 0.2119 & 0.2106 \\
        Jezebel, $\frac{^{239}\textrm{Pu(n,f)}}{^{235}\textrm{U(n,f)}}$ & 1.4609(130) & 1.4273 & 1.4242 \\
        Jezebel, $\frac{^{239}\textrm{Pu(n,2n)}}{^{239}\textrm{Pu(n,f)}}$ & None & 0.0023 & 0.0022 \\
        Jezebel, $\frac{^{239}\textrm{Pu(n,}\gamma\textrm{)}}{^{239}\textrm{Pu(n,f)}}$ & None & 0.0345 & 0.0359 \\
\midrule
        Jezebel-23, $\frac{^{237}\textrm{Np(n,f)}}{^{235}\textrm{U(n,f)}}$ & 0.997(15) & 0.984 & 0.984 \\
        Jezebel-23, $\frac{^{238}\textrm{U(n,f)}}{^{235}\textrm{U(n,f)}}$ & 0.2131(26) & 0.2116 & 0.2110 \\
\midrule
        Flattop-Pu, $\frac{^{237}\textrm{Np(n,f)}}{^{235}\textrm{U(n,f)}}$ & 0.8561(120) & 0.8569 & 0.8513 \\
        Flattop-Pu, $\frac{^{238}\textrm{U(n,f)}}{^{235}\textrm{U(n,f)}}$ & 0.1799(20) & 0.1793 & 0.1779 \\
        Flattop-Pu, $\frac{^{239}\textrm{Pu(n,2n)}}{^{239}\textrm{Pu(n,f)}}$ & None & 0.0020 & 0.0019 \\
        Flattop-Pu, $\frac{^{239}\textrm{Pu(n,}\gamma\textrm{)}}{^{239}\textrm{Pu(n,f)}}$ & None & 0.0458 & 0.0468 \\
\midrule
        Flattop-23, $\frac{^{237}\textrm{Np(n,f)}}{^{235}\textrm{U(n,f)}}$ & 0.910(13) & 0.900 & 0.898 \\
        Flattop-23, $\frac{^{238}\textrm{U(n,f)}}{^{235}\textrm{U(n,f)}}$ & 0.1916(21) & 0.1882 & 0.1869 \\
\midrule
        Flattop-25, $\frac{^{233}\textrm{U(n,f)}}{^{235}\textrm{U(n,f)}}$ & 1.608(30) & 1.578 & 1.578 \\
        Flattop-25, $\frac{^{237}\textrm{Np(n,f)}}{^{235}\textrm{U(n,f)}}$ & 0.7804(100) & 0.7716 & 0.7712 \\
        Flattop-25, $\frac{^{238}\textrm{U(n,f)}}{^{235}\textrm{U(n,f)}}$ & 0.1492(16) & 0.1445 & 0.1444 \\
        Flattop-25, $\frac{^{239}\textrm{Pu(n,f)}}{^{235}\textrm{U(n,f)}}$ & 1.3847(120) & 1.3615 & 1.3602 \\

     \bottomrule \bottomrule
    \end{tabular}
    \label{tab:rxnRate1}
\end{table*}

We show graphically the variation of various reaction rate spectral indexes at different locations in LANL's fast critical assemblies, and for a variety of assemblies, that cover a range of neutron spectra. These plots represent updates of previous similar plots we have shown in publications documenting earlier versions of the ENDF libraries, such as in Ref.~\cite{Brown2018}.
 
\begin{figure}[tbph]
    \centering
    \includegraphics[width=1.0\linewidth]{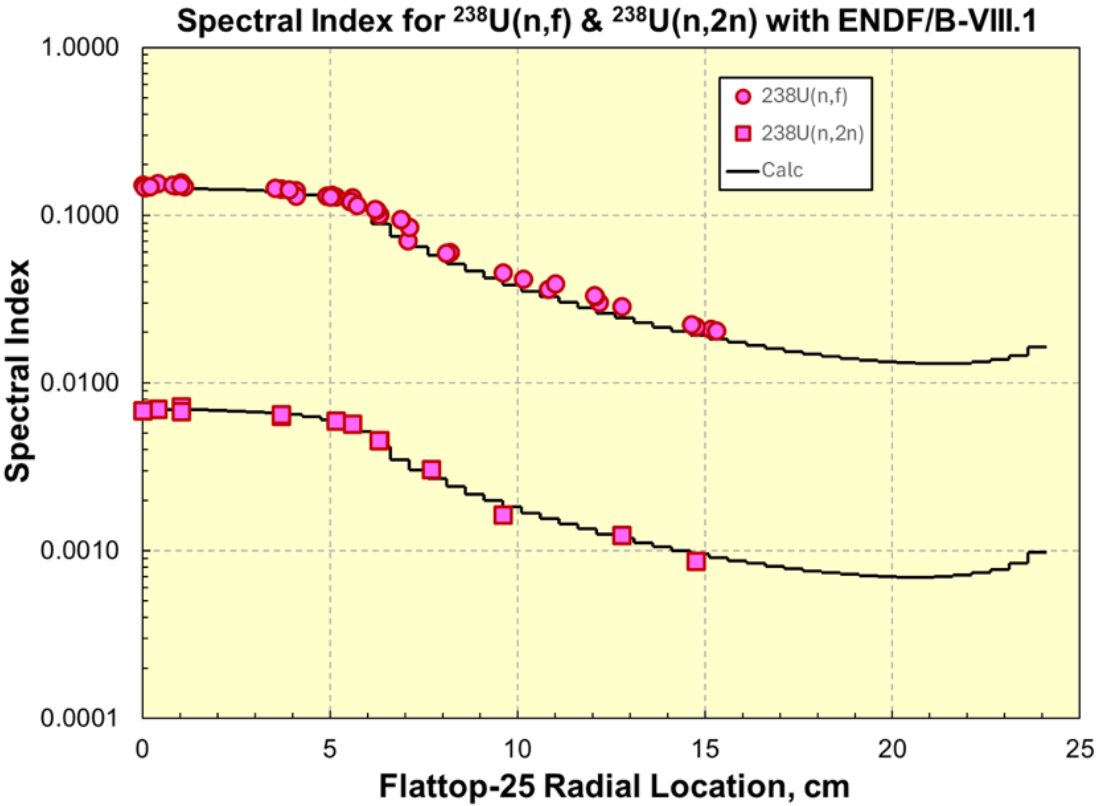}
    \caption{MCNP-calculated and measured spectral indices for \nuc{238}{U}(n,f)/\nuc{235}{U}(n,f) and \nuc{238}{U}(n,2n)/\nuc{235}{U}(n,f), as a function of the radial location, in the Flattop-25 critical assembly.}
    \label{fig:p17a}
\end{figure}

Fig.~\ref{fig:p17a} shows the \nuc{238}{U} fission and (n,2n) reaction rates (divided by \nuc{235}{U}(n,f)) as a function of the radial distance in the Flattop-25 assembly. 
As expected, the spectral indices decrease with increasing distance from the center of the assembly. For larger radial locations (larger x-axis values), the neutron spectrum becomes softer and, therefore, threshold reactions (both \nuc{238}{U}(n,f) and (n,2n)) become relatively smaller compared to \nuc{235}{U}(n,f), so the ratio decreases.
Agreement between calculation and measurement is good, similar to the previous \prENDF\ evaluation.
 
\begin{figure}[tbph]
    \centering
    \includegraphics[width=1.0\linewidth]{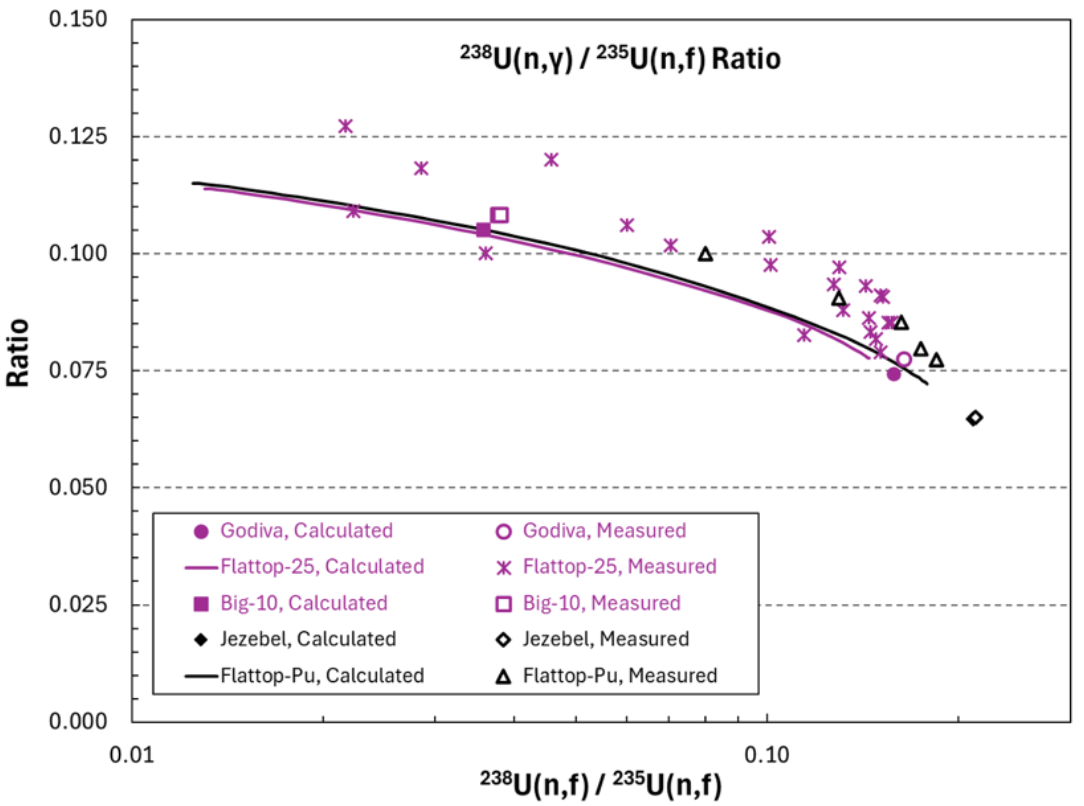}
    \caption{MCNP-calculated and measured rates for \nuc{238}{U}(n,$\gamma$)/\nuc{235}{U}(n,f), as a function of the hardness of the neutron energy spectrum (as characterized by the spectral index \nuc{238}{U}(n,f)/\nuc{235}{U}(n,f)), in a variety of fast Los Alamos critical assemblies.}
    \label{fig:p1ba}
\end{figure}

Fig.~\ref{fig:p1ba} shows the \nuc{238}{U}(n,g)/\nuc{235}{U}(n,f) rate as a function of the \nuc{238}{U}(n,f)/\nuc{235}{U}(n,f) spectral index measured at the same location, allowing us to plot many critical assembly measurements on the same plot. Again, agreement between calculation and measurement is fairly good, similar to the previous \prENDF\ evaluation. In some assemblies, the trend of the calculation is below the measured data; this reflects the fact  that the evaluation for \nuc{238}{U}(n,g) was based on fundamental lab cross section-measurements and not on these integral data.
 
\begin{figure}[tbph]
    \centering
    \includegraphics[width=1.0\linewidth]{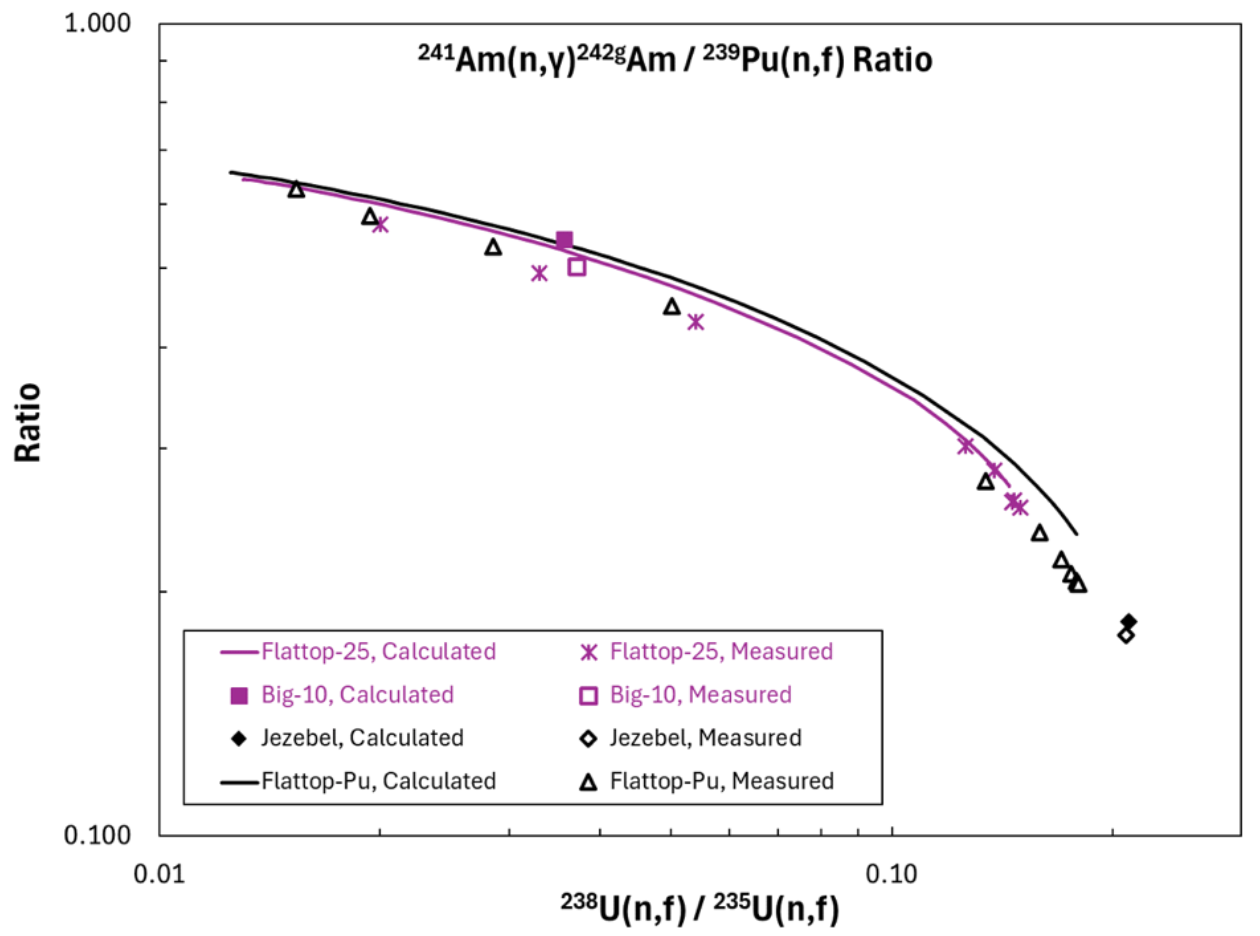}
    \caption{MCNP-calculated and measured rates for \nuc{241}{Am}(n,$\gamma$)/\nuc{239}{Pu}(n,f), as a function of the hardness of the neutron energy spectrum (as characterized by the spectral index \nuc{238}{U}(n,f)/\nuc{235}{U}(n,f)), in a variety of fast Los Alamos critical assemblies.}
    \label{fig:p18a}
\end{figure}

Fig.~\ref{fig:p18a} is similar to the previous plot, but this time for \nuc{241}{Am}(n,g) (relative to \nuc{239}{Pu}(n,f)). Agreement between calculation and measurement is fairly good, similar to the previous \prENDF\ evaluation.

\subsection{Reactor Depletion}
\label{sec:depletion}

The accuracy of predicting the depletion and subsequent loss of reactivity in power reactors is crucial for fuel management and particularly for the determination of the cycle length in operating commercial reactors, which has significant economic implications. The industry was reasonably happy with the performance of the ENDF/B-VII.1 library\footnote{and similarly with the JEFF-3.1.1 library performance in Europe.} in this respect. Although the ENDF/B-VIII.0 library contained better physics, it was clearly pointed out by ORNL (Kim and Wieselquist) \cite{kim2021-318} that the loss of reactivity with burnup is considerably larger with the ENDF/B-VIII.0 cross-section data (a swing of about -500~pcm compared to ENDF/B-VII.1), which means that the industry could not accept the ENDF/B-VIII.0 library for nuclear power applications without expensive data adjustment. It is also worth to point out that an even larger problem at high burnup was identified in the JEFF-3.3 library. This led to several studies, both at CSEWG and JEFF committees, made to pin down the root cause of the unfavorable loss of reactivity with burnup in ENDF/B-VIII.0 and JEFF-3.3 evaluations and to restore the ENDF/B-VII.1 performance in the new ENDF/B-VIII.1 and JEFF-4 libraries. The relevant ENDF/B-VIII.0 changes will be summarized below. Achieved consistent performance for the new ENDF/B-VIII.1 will be reviewed with calculations presented by four groups using different depletion benchmarks.

\subsubsection{Depletion calculations using \SERPENT}
\label{Sec:depletion_serpent}


The calculations carried out to test the performance of the ENDF/B-VIII.1 library during depletion calculations are divided in two main parts. In the first part, a code-to-code comparison is carried out while in the second step, the Duke1 benchmark from IRPhE is analyzed. 

\paragraph{Three Mile Island Pincell depletion calculations\newline}

The pincell configuration utilized for testing the performance of the ENDF/B-VIII.1 library with respect to its older declinations (ENDF/B-VII.1 \cite{ENDF-VII.1} and ENDF/B-VIII.0 \cite{Brown2018}) is modeled after the TMI-1 PWR unit cell setup specified in the OECD/NEA Uncertainty Analysis in Modeling (UAM) benchmark, Exercise I-1b \cite{Ivanov2016}. The fuel is made of 4.85 wt.\% enriched uranium with a boron free moderator density of 0.7484 g/cm$^3$. 

The \Serpent2 code \cite{Leppanen2015} (Version 2.2.0) is used with ACE-formatted libraries based on ENDF/B-VII.1, ENDF/B-VIII.0, ENDF/B-VIII.1 distributed by LANL. Consistent fission yield and decay data information is used as well. The number of neutron histories considered is $25\times 10^6$ (500 active batches of $10^4$ neutrons each).  The closest temperature available in the library is used for the $S(\alpha,\beta)$ tables (consistent temperatures are used for all libraries). The energy released per fission model takes into account the facts that the neutrons deposit energy along their history in various reactions; the energy deposition due to reactions other than fission is calculated using KERMA coefficients \cite{Tuominen2019a}. The resonance upscattering -- Doppler broadened rejection correction (DBRC) treatment \cite{Becker2009} -- is considered only for one special case as described below. A standard predictor/corrector approach is used. 

\begin{figure}[!thbp] 
    \vspace{-2mm}
	\centering
	\includegraphics[width=0.5\textwidth]{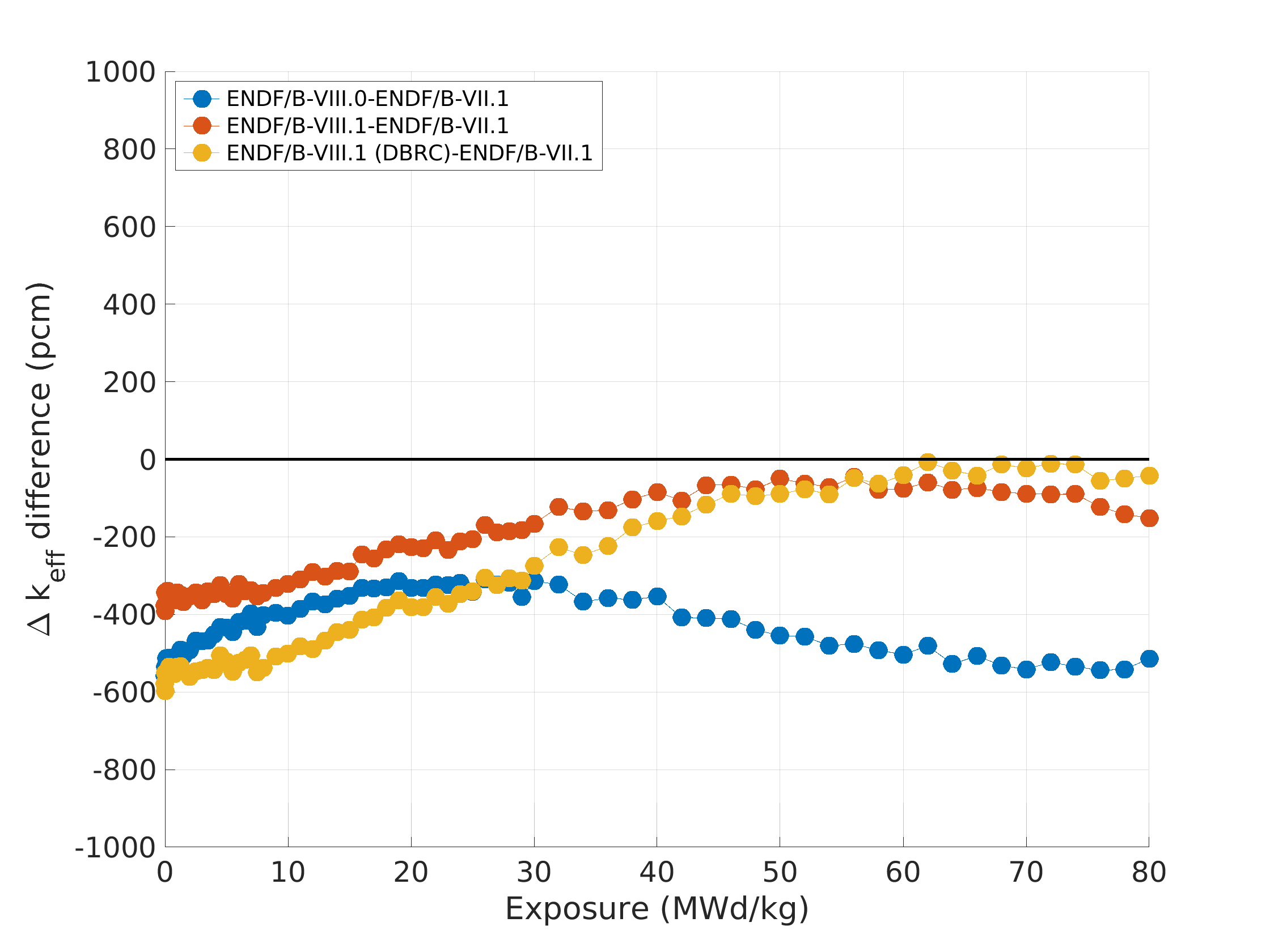}
    \vspace{-2mm}
    \caption{The $k_{\mathrm{eff}}$ obtained with the ENDF/B-VIII.0 and ENDF/B-VIII.1 libraries (red and blue curves) is compared to ENDF/B-VII.1 by subtracting the VII.1 $k_{\mathrm{eff}}$ value. The yellow curve represents the reactivity loss change due to the resonance upscattering model (DBRC) obtained with ENDF/B-VIII.1.} 
	\label{fig:post}
    \vspace{-2mm}
\end{figure}

The depletion benchmark results are plotted in Fig.~\ref{fig:post}. Each library was used to calculate the effective multiplication coefficient of the reactor, $k_{\mathrm{eff}}$, which was compared to the ENDF/B-VII.1 reference as a function of burnup. In the figure, the black solid line at 0~pcm defines the ENDF/B-VII.1 performance, and the blue and red dots represent the difference in $k_{\mathrm{eff}}$, $\Delta k_{\mathrm{eff}}$, for ENDF/B-VIII.0 and ENDF/B-VIII.1, respectively. Also shown by the yellow markers is the depletion calculation with ENDF/B-VIII.1 using the DBRC. ENDF/B-VII.1 is the reference as it is used in the industry for routine reactor applications with good performance~\cite{Georgieva:2024}.

For this computational benchmark, the ENDF/B-VIII.0 library shows decreased criticality of about -500~pcm relative to ENDF/B-VII.1. The $\Delta k_{\mathrm{eff}}$ is fairly constant with exposure within 200~pcm. The ENDF/B-VIII.1 library shows a lower criticality (about -300~pcm) at BOL compared to the ENDF/B-VII.1. However, the ENDF/B-VIII.1 criticality already at 30 MWd/kg exposure becomes quite similar to the ENDF/B-VII.1 showing a significant improvement compared to the ENDF/B-VIII.0 performance.
The DBRC correction reduces the criticality at the BOL by 200~pcm; a reduction which decreases in magnitude with exposure.

\begin{figure}[!thbp] 
    \vspace{-2mm}
	\centering
	\includegraphics[width=0.5\textwidth]{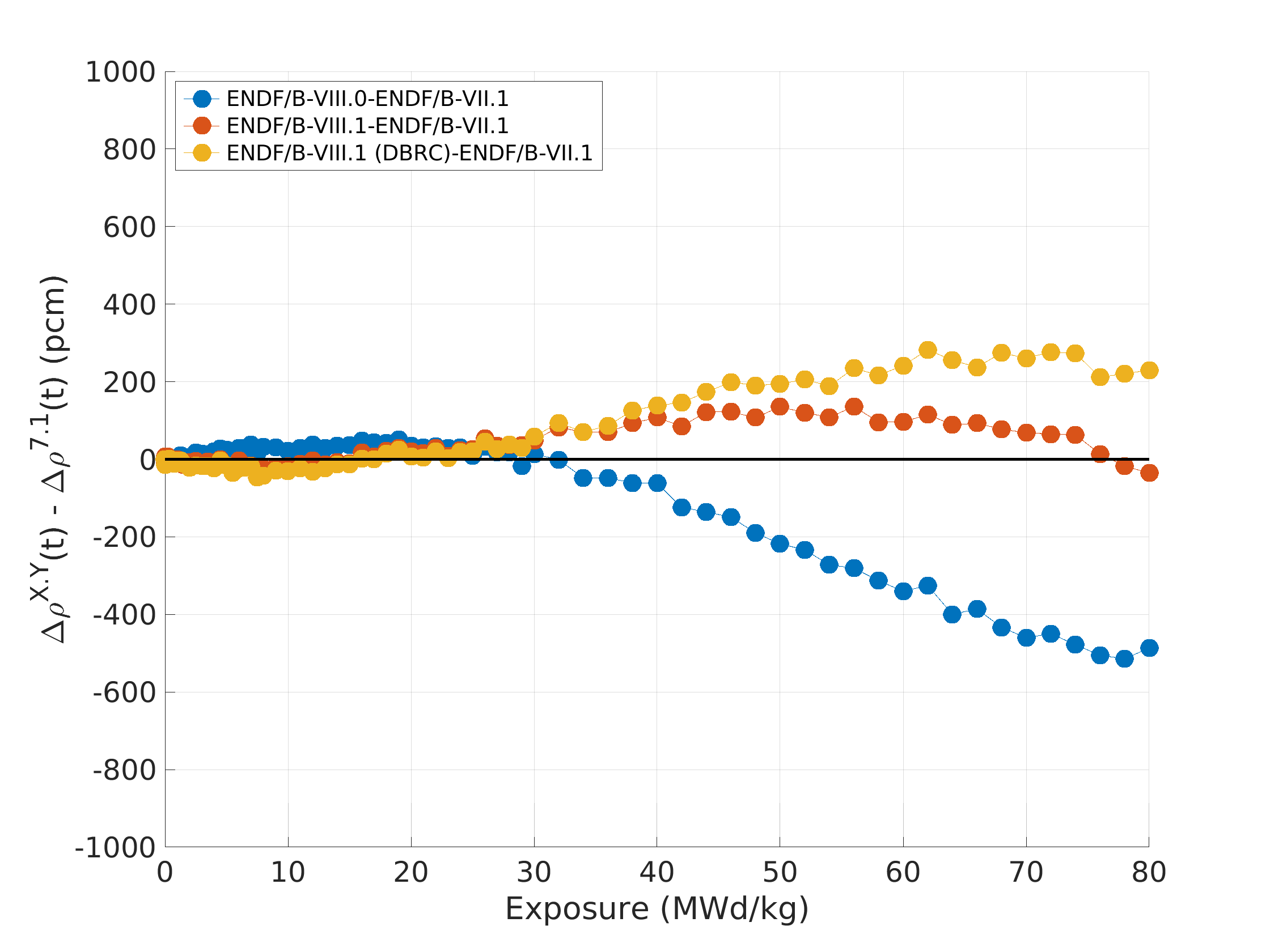}
    \vspace{-2mm}
    \caption{The reactivity loss difference $\Delta \rho^{XY}(t)$ obtained with the ENDF/B-VIII.0 and ENDF/B-VIII.1 libraries (red and blue curves) is compared to ENDF/B-VII.1 by subtracting the $\Delta \rho^{B71}(t)$ value. The yellow curve represents the reactivity loss change due to the resonance upscattering model (DBRC) obtained with ENDF/B-VIII.1.} 
	\label{fig:post-rho}
    \vspace{-2mm}
\end{figure}

In power reactor operation, the cycle length depends on the loss of reactivity $\Delta \rho(t)$ and not on $\Delta k_{eff}$.  
The reactivity loss $\Delta \rho^{XY}(t)$ with exposure $t$  for a given nuclear data library labelled \textit{XY} is defined in Eq.~\eqref{eq:sens}: 
\begin{equation} 
	\label{eq:sens}
	\Delta\rho ^{XY} (t) = \left[\frac{1}{k^{XY}_{eff}(t)} - \frac{1}{k^{XY}_{eff}(0)}\right] \times 10^5\text{~pcm}  
\end{equation}
The reactivity loss with exposure difference $\Delta \rho^{XY}(t)$ between the various libraries labeled $XY$ relative to the one calculated for the ENDF/B-VII.1 library $\Delta \rho^{B71}(t)$ is shown in Fig.~\ref{fig:post-rho}. 
A clear negative swing of reactivity of about -500~pcm is observable when using the ENDF/B-VIII.0 library. This issue is fixed when using ENDF/B-VIII.1 as the reactivity difference with exposure is now fairly constant. 
The DBRC correction induces a reactivity gain of about 200 pcm at the EOL for the ENDF/B-VIII.1 library. Further research on the impact of the DBRC on reactivity is needed. 

\paragraph{Modeling of the IRPhE/DUKE benchmark\newline}
The IRPhE/DUKE \cite{IRPHE} experimental benchmark provides  valuable information on the integral prediction of the reactivity decrement during burnup that can be used to assess the depletion reactivity change in LWRs.  This has been inferred from an extensive study of differences between 680 core-wide fission rate distribution measurements and calculated neutron flux during 44 cycles PWR operation at McGuire and Catawba NPPs. This magnitude has been demonstrated to be essentially code independent.

The fuel assembly configuration is a typical PWR $17\times 17$. The fuel assembly contains 160 rods UO$_2$ (4.25 wt\%), Zircaloy-4 cladding and water with 990~ppm boron as moderator. There are 104 rods with ZrB$_2$ coating. The burnup is modeled at constant power level of 104.5~W/cm$^3$, with an average temperature of 900~K in the fuel, 618.4~K in the cladding and 580~K in the moderator.

The \Serpent2 code (Version 2.2.1) is used with a number of neutron histories of $12.8\times 10^6$ (400 active batches of 32,000 neutrons each), which gives a statistical uncertainty in $k_\infty$ of around 15 pcm. To model the burnup, similar options to TMI pincell were used; except for the time integration method, only predictor method with constant extrapolation is used in this calculation. DBRC treatment for $^{238}$U is also included in the calculation to show the impact on the burnup reactivity loss.  The Serpent2 calculations are performed with ACE-formatted libraries based on ENDF/B-VII.1 distributed by SERPENT, ENDF/B-VIII.0 distributed by LANL and ENDF/B-VIII.1 processed with \NJOY2016.76. Consistent fission yield and decay data information is used as well.

\begin{figure}[!thbp]
    \vspace{-2mm}
	\centering
	\includegraphics[width=0.5\textwidth]{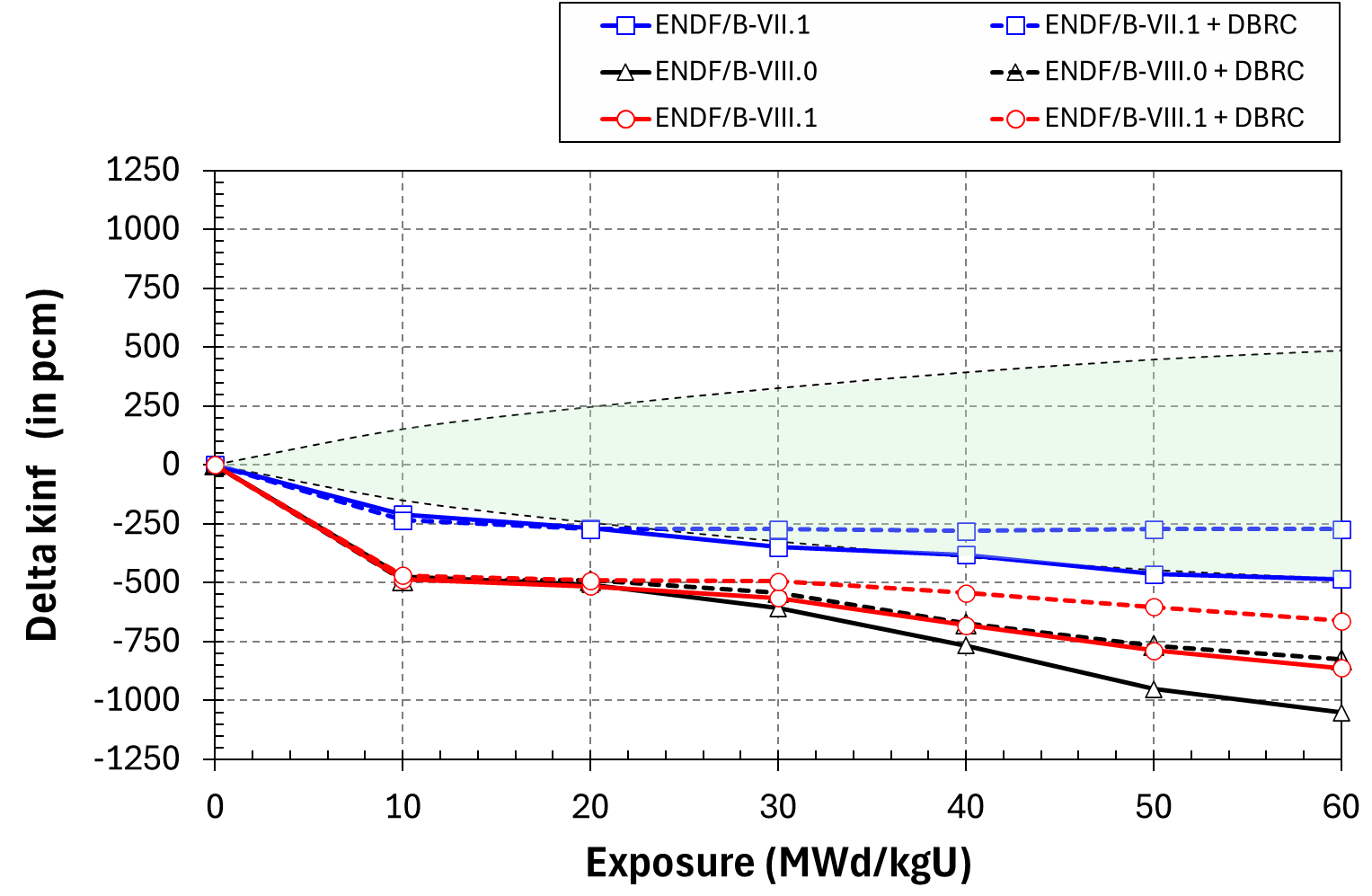}
    \vspace{-2mm}
	\caption{PWR Fuel Assembly Depletion Criticality Bias  (in pcm) of the DUKE-PWR-POWER-001-REAC Benchmark of IRPhE.}
	\label{fig:duke}
    \vspace{-2mm}
\end{figure}

For this benchmark, the benchmark depletion criticality difference 
($\Delta k_{\mathrm{eff}}^{\mathrm{exp}}$) is defined as the $\Delta k_\infty$ from the initial fresh fuel assembly condition. The fresh fuel is assumed to be at hot full-power (HFP) conditions with burnable absorbers and no Xe, so depletion criticality difference 
includes the worth of \nuc{135}{Xe} as well as the worth of partially depleted burnable absorbers. The PWR depletion 
benchmark is for one HFP lattice depletion at six specific burnups: 10, 20, 30, 40, 50, and 60 GWd/MTU.
The SERPENT 2 code is used to predict the calculated $k_\infty$  for fresh fuel at HFP and no burnup, and the  $k_\infty$ for fuel with the specified burnup at the specified temperature. Then, the calculated depletion 
criticality difference ($\Delta k_{\mathrm{eff}}^{\mathrm{C}}$) is predicted as the $\Delta k_\infty$ from fresh to the specified burnup. These values are compared with the $\Delta k_{\mathrm{eff}}^{\mathrm{exp}}$ for that burnup.  The depletion criticality difference 
 bias ($\Delta k_{\mathrm{eff}}^{\mathrm{C}} - \Delta k_{\mathrm{eff}}^{\mathrm{exp}}$) is shown in Fig.~\ref{fig:duke}.

The 1-sigma total uncertainty of the benchmark depletion criticality difference 
($\Delta k_{\mathrm{eff}}^{\mathrm{exp}}$) as a function of the exposure is shown in the shadow area. The ENDF/B-VII.1 is close to the 1-sigma total uncertainty while ENDFB/B-VIII.0 and ENDF/B-VIII.1 are within 2-sigmas. It can be noted a better performance of ENDF/B-VIII.1 at high burnup.

The DBRC correction increases the criticality at the largest calculated 60 MWd/kgU burnup by about 200 pcm for this benchmark.

\subsubsection{Depletion calculations using \Shift}
\label{Sec:depletion_shift}

Shortly after the ENDF/B-VIII.0 release~\cite{Brown2018}, a depletion analysis comparing ENDF/B-VII.1~\cite{ENDF-VII.1} and ENDF/B-VIII.0 was published by Kim \etal~\cite{kim2021-318}; the results showed that reactor criticality using ENDF/B-VIII.0 was underestimated by about -600~pcm at high burnups. A similar method of analysis was performed in this work using the same benchmark problems and the ENDF/B-VIII.1 library. This analysis used the new capability in the \Shift\ code~\cite{DAVIDSON2018} for massively parallel continuous-energy Monte Carlo calculations with depletion. The total run time for the current analysis was approximately 1.6 CPU-years. The model was a single PWR pin with a reflecting boundary condition based on Virtual Environment for Reactor Applications (VERA) PWR depletion benchmark problem 1C, as described in Kim and Wieselquist~\cite{kim2021-318}. The modeled fuel temperature was 900~K, and all other materials were 600~K. Fresh fuel enrichment was 3.1\% $^{235}$U. Performance of the full assembly was found to be nearly identical to the single pin problem in Ref.~\cite{kim2021-318}, so no full assembly problems were tested. The same fission product yields and decay data as those validated by Ilas \etal~\cite{ilas_2022} were used for the present analysis.

\begin{figure}[!thb]
    \vspace{-3mm}
    \centering
    \includegraphics[width=\columnwidth]{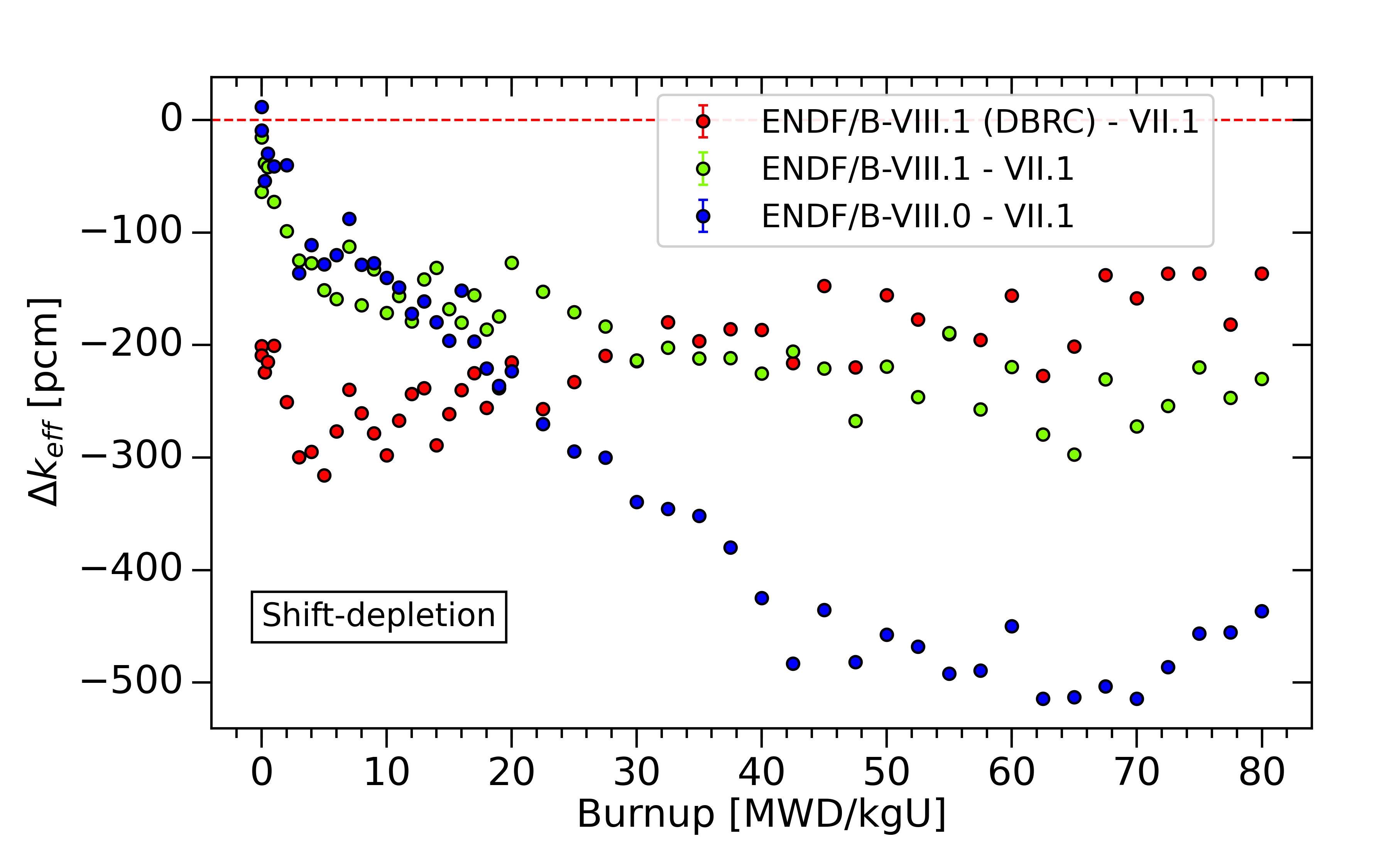}
    \vspace{-4mm}
    \caption{The depletion calculation for each library is compared to ENDF/B-VII.1 by subtracting the ENDF/B-VII.1 $k_\mathrm{eff}$ value. The changes made to the neutron sublibrary have a significant impact on reactor criticality as a function of burnup, bringing the performance much closer to the values using ENDF/B-VII.1.} 
    \label{fig:shift-depl}
    \vspace{-3mm}
\end{figure}

The depletion benchmark results are plotted in Fig.~\ref{fig:shift-depl}. Following the method of Kim \etal, each library version used to calculate the effective multiplication of the reactor, $k_\mathrm{eff}$, was compared to ENDF/B-VII.1 as a function of burnup. In the figure, the red dashed line at 0~pcm defines the ENDF/B-VII.1 performance, and the blue and green markers represent the difference in $k_\mathrm{eff}$, $\Delta k_\mathrm{eff}$, for ENDF/B-VIII.0 and ENDF/B-VIII.1, respectively. Also shown by the red markers is the depletion calculation with ENDF/B-VIII.1 using the DBRC. The DBRC is calculated on-the-fly during transport \cite{hart-2013} using the 0~K cross sections from \AMPX\ processing (see Section~\ref{sec:ampx-processing}) and is made available for isotopes of the following elements: Hg, Tl, Pb, Bi, Po, Ra, Ac, Th, Pa, U, Np, Pu, Am, Cm, Bk, Cf, Es, and Fm. The depletion benchmark results indicate that the ENDF/B-VIII.1 library is a significant improvement compared to ENDF/B-VIII.0 and that performance is closer to ENDF/B-VII.1 for high-burnup criticality. 
The $\Delta k$ trends shown in Figs.~\ref{fig:post}  and~\ref{fig:shift-depl} are slightly different due to the different enrichments and boron concentrations in the two models. The effects of DBRC, however, appear to have a similar magnitude.

\subsubsection{Depletion calculations using \CASMO}
\label{Sec:depletion_casmo}

\begin{figure}[!tbh]
    \vspace{-2mm}
    \centering
    \includegraphics[width=0.5\textwidth]{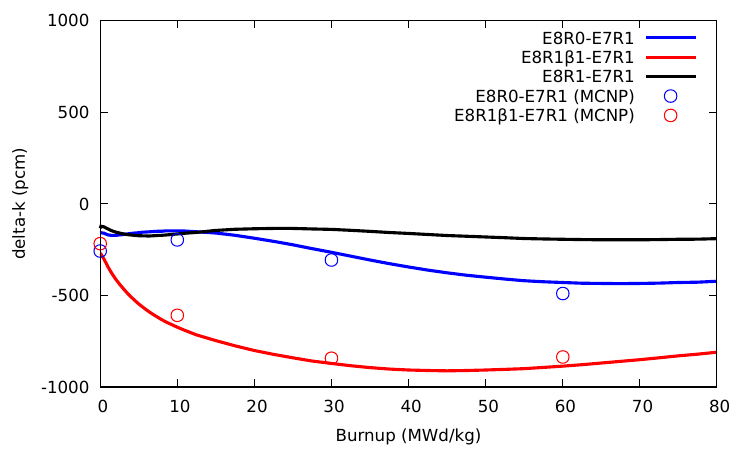}
    \vspace{-2mm}
    \caption{Comparison of \CASMO\ pincell depletion results for ENDF/B-VIII.0 (E8R0), ENDF/B-VIII.1$\beta$1 (E8R1$\beta$1), and ENDF/B-VIII.1 (E8R1) with ENDF/B-VII.1 (E7R1).} 
    \label{fig:casmo-depl1}
    \vspace{-2mm}
\end{figure}

\begin{figure}[!hbt]
    \vspace{-2mm}
    \centering
    \includegraphics[width=0.5\textwidth]{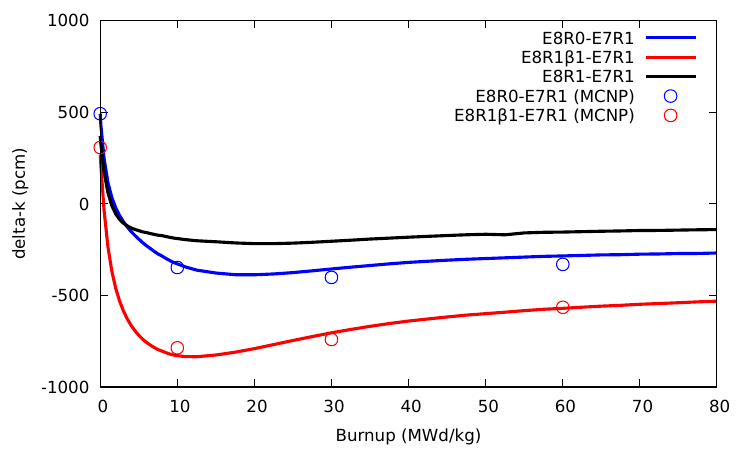}
    \vspace{-2mm}
    \caption{Comparison of \CASMO\ BWR fuel assembly depletion results for ENDF/B-VIII.0 (E8R0), ENDF/B-VIII.1$\beta$1 (E8R1$\beta$1), and ENDF/B-VIII.1 (E8R1) with ENDF/B-VII.1 (E7R1).} 
    \label{fig:casmo-depl2}
    \vspace{-2mm}
\end{figure}

\begin{figure}[!tbh]
    \vspace{-2mm}
    \centering
    \includegraphics[width=0.5\textwidth]{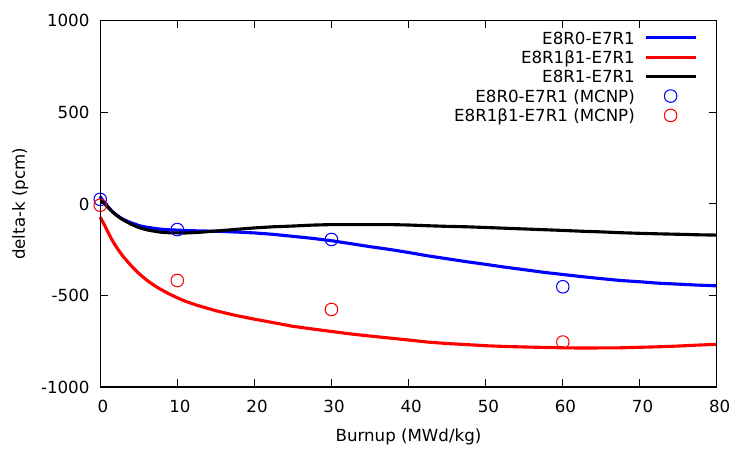}
    \vspace{-2mm}
    \caption{Comparison of \CASMO\ PWR fuel assembly depletion results for ENDF/B-VIII.0 (E8R0), ENDF/B-VIII.1$\beta$1 (E8R1$\beta$1), and ENDF/B-VIII.1 (E8R1) with ENDF/B-VII.1 (E7R1).} 
    \label{fig:casmo-depl3}
    \vspace{-2mm}
\end{figure}

Initial testing of the ENDF/B-VIII.1$\beta$1 and of the final ENDF/B-VIII.1 evaluations was performed by Studsvik using the \CASMO\ advanced lattice physics code \cite{CASMO5}, which solves fixed-source and eigenvalue problems using the Method of Characteristics (MOC). The burnup calculation uses a predictor-corrector scheme and the Chebyshev Rational Approximation Method (CRAM). Each ENDF/B evaluation has been processed with \NJOY\ (\NJOY2016~\cite{NJOY}) and the resulting \CASMO\ multi-group nuclear data library is denoted by the letter ``E'' followed by the release,  revision number and $\beta$ release if applicable, e.g., the ENDF/B-VIII.1-based \CASMO\ multi-group library is denoted as E8R1. Depletion calculations using \CASMO\ demonstrated a decrease in the ENDF/B-VIII.1$\beta$1-predicted $k_\infty$ as a function of burnup relative to previous libraries \cite{Ferrer2024}. This trend was observed for depletions of a typical single pin cell (Fig.~\ref{fig:casmo-depl1}), boiling water reactor bundle (Fig.~\ref{fig:casmo-depl2}), and PWR assembly (Fig.~\ref{fig:casmo-depl3}), all involving low-enrichment uranium fuel. For instance, the ENDF/B-VIII.1$\beta$1-predicted $k_\infty$  was observed to be as much as 600 pcm lower than the ENDF/B-VII.1-based result in the range of 0 to 20 MWd/kg for the PWR assembly depletion case. This trend was verified by performing independent MCNP (MCNPv6.2 \cite{MCNP6.2}) calculations at selected exposure points using the same details as used in the \CASMO\ models. The $^{239}$Pu ENDF/B-VIII.1$\beta$1 evaluation was identified as the source of most of the discrepancy. The initial ENDF/B-VIII.1$\beta$1 had focused mostly on criticality benchmarks and not on depletion results. Although there were many changes between ENDF/B-VIII.1$\beta$1 and the final release, the most impactful change was found to be $^{239}$Pu \cite{Ferrer2024}. In the end, ENDF/B-VIII.1 was a compromise between optimal performance in criticality and depletion tests, with the latter practically recovering the ENDF/B-VII.1 performance, which is considered here as the reference in reactor applications. Following the release of ENDF/B-VIII.1, which included the revised $^{239}$Pu evaluation relative to ENDF/B-VIII.1$\beta$1, the \CASMO\ depletions were performed again. Improved agreement was observed between the ENDF/B-VIII.1 results relative to the ENDF/B-VII.1 $k_\infty$ values calculated by \CASMO\ (Figs.~\ref{fig:casmo-depl1}--\ref{fig:casmo-depl3}). This agreement is a desirable improvement given that the ENDF/B-VII.1 evaluation, as the basis for the \CASMO\ E7R1 multi-group library, is routinely used for in-core fuel management calculations and found to provide the best agreement with operating power reactor measurements~\cite{Georgieva:2024}.

\subsubsection{Fuel cycle calculations using CORD-2 and WIMSD-5B}
\label{Sec:depletion_CORD-2}

The CORD-2 package~\cite{cord2} is used at the Jo\v{z}ef Stefan Institute in Ljubljana, Slovenia for core design calculations of the Kr\v{s}ko NPP. The package uses the WIMSD-5B code~\cite{askew-general-1966, WIMSD5} for lattice cell calculations. The WIMSD-5B is a 1D reactor lattice transport solver that uses a 69-group library in WIMS-D format~\cite{wimslib,WLUP,LopezAldama2014} and features lattice cell homogenization, group condensation and depletion calculations. 

The WIMSD-5B code was verified by comparing the isotopic composition and reactivity loss with burnup to \Serpent\ and \OpenMC\ Monte Carlo calculations. The test case was the TMI-1 pin-cell benchmark. For code comparison, the libraries based on ENDF/B-VIII.0 data were used. In the process, the following tests were done:
\begin{itemize}
  \item The WIMS-D libraries from different evaluated data files were generated with \NJOY2016.75 according to the procedures developed within the WLUP project of the IAEA~\cite{WLUP,LopezAldama2014}.
  \item A few explicitly treated fission products were added to the WIMS-D libraries to achieve agreement with \Serpent\ and \OpenMC\ results. The updated WIMS-D libraries are available from the IAEA web site \cite{wimslib}.
  \item The $^{135}$Xe production by the decay of $^{135}$I is treated explicitly in WIMS. Small changes (within the uncertainty) of the decay constant of $^{135}$I has negligible effect on reactivity as a function of burnup. Changes to the fission yield of $^{135}$I (for example, adopting the value from JEFF-3.3) produces a positive shift in reactivity at HFP conditions, but has negligible effect on the reactivity gradient with burnup.
  \item Taking all fission yield and decay data of the major actinides from JEFF-3.3 increases the reactivity gradient in an unfavourable way.
  \item The strongest impact on the reactivity gradient comes from the assumptions about the energy released per fission. The option of fixing the energy released per fission of $^{235}$U to $E^*=202.23$~MeV, defining the ratio $R=E^*/E_{U235}$, where $E_{U235}$ is the fission Q-value in the ENDF file, and normalizing the fission Q-values of all fissile isotopes by this ratio produces a burnup reactivity gradient marginally steeper than predicted by WIMSD-5B. Using a similar assumption in \texttt{OpenMC} with $E^{*}=205.0$~ MeV the same reactivity gradient is reproduced.
\end{itemize}

\begin{figure}[!htbp]
\vspace{-2mm}
	\centering
	\includegraphics[width=0.5\textwidth]{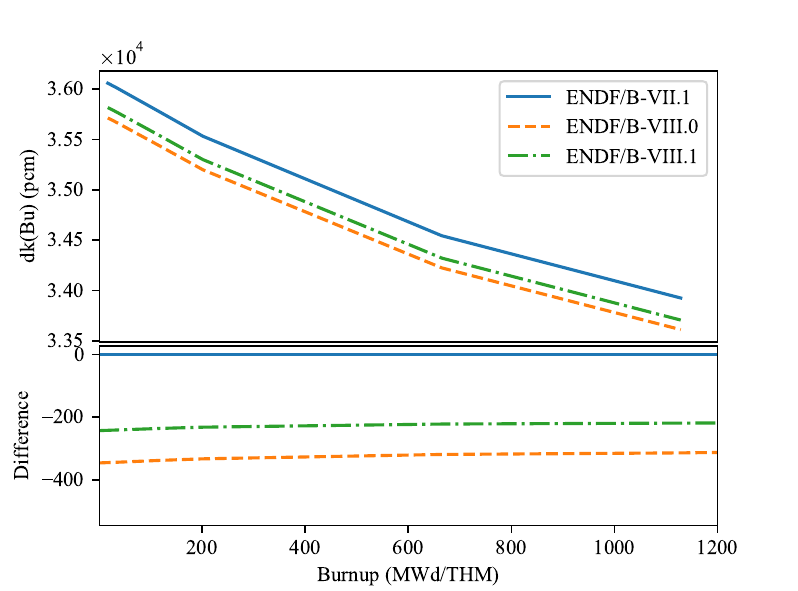}
\vspace{-4mm}
	\caption{Differences in the multiplication factor $\Delta k$ relative to ENDF/B-VII.1 with various version of the ENDF/B libraries at short burnups for the TMI-1 pincell benchmark.}
	\label{fig:pin-short_dk}
\end{figure}

Differences in the calculated multiplication factor $\Delta k$ relative to ENDF/B-VII.1 for short burnups are shown in Fig.~\ref{fig:pin-short_dk}. As soon as xenon reaches equilibrium, the reactivity is lower by about 300~pcm with ENDF/B-VIII.0 data. About 100~pcm is recovered when ENDF/B-VIII.1 data are used.

\begin{figure}[!htbp]
\vspace{-3mm}
	\centering
	\includegraphics[clip, trim = 0mm 0mm 0mm 0mm, width=0.5\textwidth]{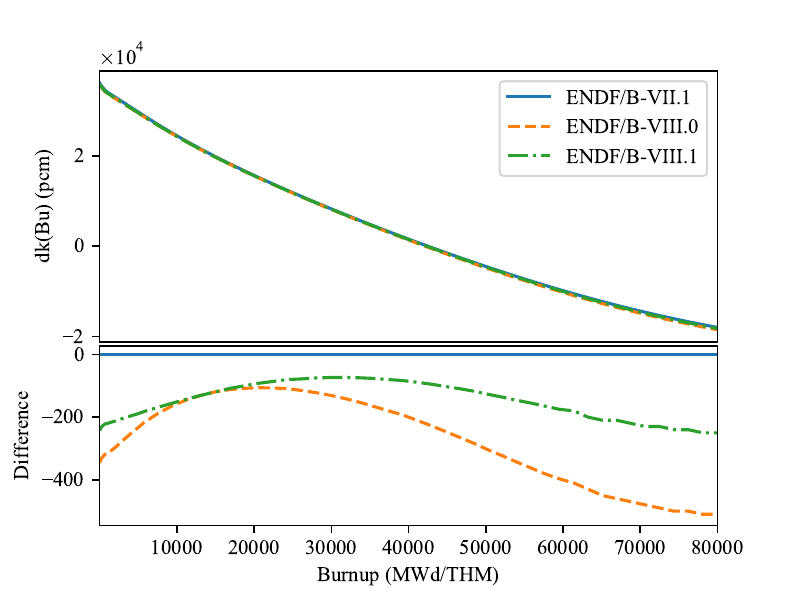}
	\caption{Differences in the multiplication factor $\Delta k$ relative to ENDF/B-VII.1 with various version of the ENDF/B libraries at long burnups for the TMI-1 pincell benchmark.}
	\label{fig:pin-long_dk}
\end{figure}

The trends for long burnups are shown in Fig.~\ref{fig:pin-long_dk}. The plot of $\Delta k$ shows a similar trend like in Fig.~\ref{fig:post}, but there are some differences, mainly because the absolute $k_{\mbox{eff}}$ calculated with WIMSD-5B has an inherent bias due to the simplified transport model. It has been checked independently that displaying the trends in terms of reactivity loss the agreement is good.

\begin{figure}[!htbp]
\vspace{-2mm}
	\centering
	\includegraphics[width=0.5\textwidth]{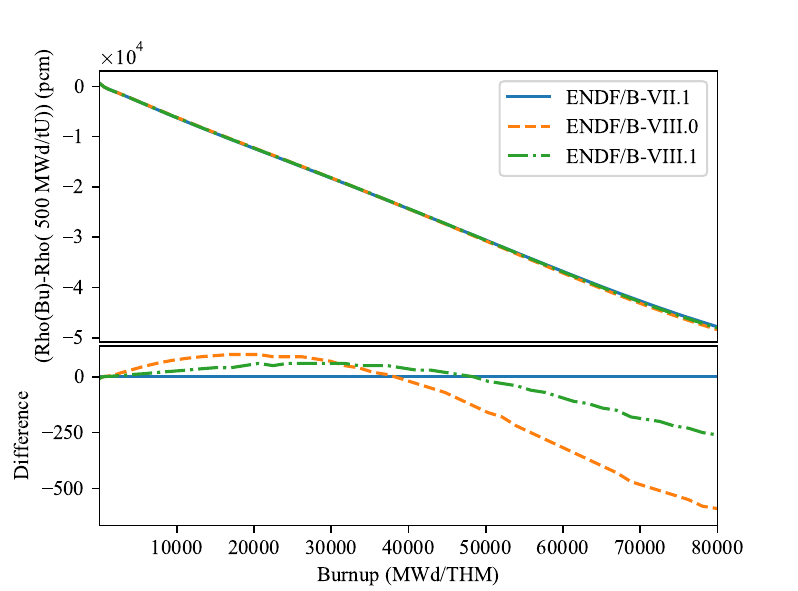}
\vspace{-4mm}
	\caption{Reactivity differences relative to ENDF/B-VII.1 with various version of the ENDF/B libraries at long burnups for the TMI-1 pincell benchmark.}
	\label{fig:pin-long}
\end{figure}

In power reactor operation, the cycle length depends on the loss of reactivity and not on $\Delta k$.  The reactivity loss as a function of burnup is shown in Fig.~\ref{fig:pin-long}. To separate out the initial transient when short-lived fission products approach equilibrium, a bias was added to each curve so that reactivities with all libraries coincide at 500~MWd/tU. The plot shows that with ENDF/B-VIII.0 data compared to ENDF/B-VII.1, the reactivity increases by up to 80~pcm below 20~GWd/tU but then decreases sharply, reaching about -600~pcm near 80 GWd/tU. On the contrary, with ENDF/B-VIII.1 data, the increase of reactivity above 500~MWd/tU is smaller, but the total loss of reactivity at 60~GWd/tU compared to ENDF/B-VII.1 stays below 100~pcm.

Having verified that WIMSD-5B~\cite{WIMSD5} calculations are representative of the detailed Monte Carlo calculations in terms of reactivity, the full core design calculations were launched for 32 cycles of the Kr\v{s}ko NPP using WIMS-D 69-group libraries based on the following data:
\begin{itemize}
  \item ENDF/B-VII.1
  \item ENDF/B-VIII.0
  \item ENDF/B-VIII.1
\end{itemize}

The results were compared against data measured at the plant during startup physics tests and boron concentration monitoring during each cycle.
\begin{figure}[!tbhp]
\vspace{-3mm}
	\centering
	\includegraphics[scale=0.32, clip, trim = 1mm 2mm 2mm 1mm]{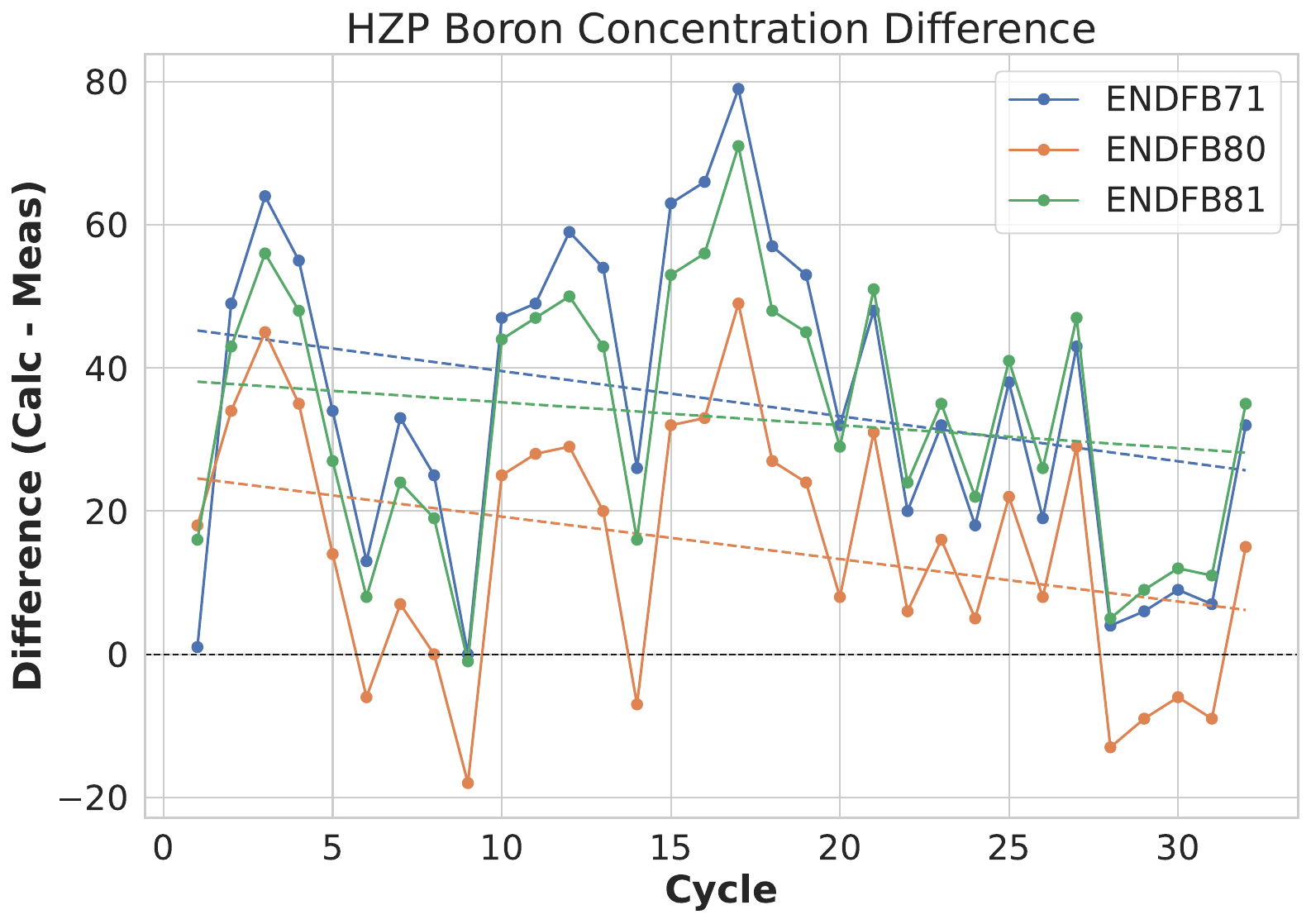}
\vspace{-4mm}
	\caption{Critical boron concentration differences with various version of the ENDF/B libraries at HZP conditions of the Kr\v{s}ko NPP.}
	\label{fig:HZP}
\vspace{-2mm}
\end{figure}

In Fig.~\ref{fig:HZP}, the hot-zero-power (HZP) boron concentrations  are compared to measured values. The simple 1-D transport cell calculations in WIMSD-5B are less accurate than Monte Carlo calculations, so a global bias in terms of a buckling is applied, which normalizes the calculations. In this particular case, a fixed buckling was used so that different libraries could be compared. The scattering is partly due to the simplified computational model, but also partly due to detailed plant specifications, which are not captured in the computational models and can change from cycle to cycle. The review criterion for predicting the critical boron concentration at HZP conditions is 50~ppm. 

In spite of the scattering in the calculated values, the relative differences between calculations with different libraries are representative of the differences in the data.  We can see that there is some decrease of reactivity with ENDF/B-VIII.0 data, which is partly compensated with ENDF/B-VIII.1 data, which is consistent with the pincell results reported above. There is a slight negative trend with cycle number, but over the last 40 years of operation, the enrichment steadily increased, the low-leakage loading pattern was introduced, burnable absorber rods were replaced by integral fuel burnable absorber rods, and the power rating was increased from 1876~MW to 1994~MW thermal power; thus some differences are not unexpected.

\begin{figure}[!htbp]
\vspace{-3mm}
	\centering
	\includegraphics[scale=0.32, clip, trim = 1mm 2mm 2mm 1mm]{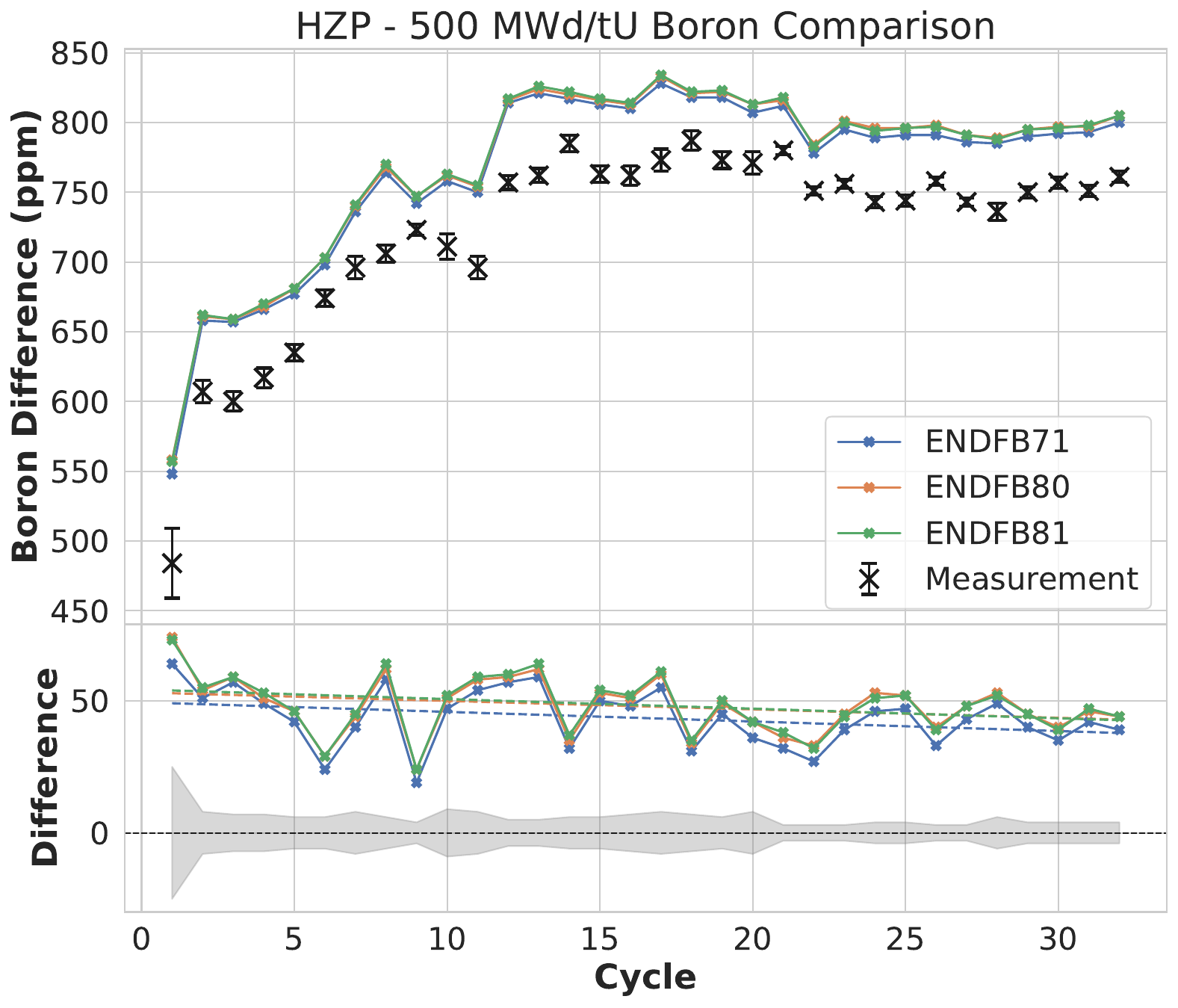}
\vspace{-4mm}
	\caption{Differences in the critical boron concentration at HZP and at 500~MWd/tU compared to measured values of the Kr\v{s}ko NPP.}
	\label{fig:HFP-EqXe}
\end{figure}

In Fig.~\ref{fig:HFP-EqXe} the differences in boron concentrations at HZP and at 500~MWd/tU are compared to measured values. This is an indication of the power and xenon defect. We can see that prediction of this defect is too strong. As shown with the testing of the libraries at the cell level, the $^{135}$I fission yield in JEFF-3.3 is lower and helps to reduce the power defect. This could be a hint to review the fission yields of $^{135}$I, its precursors, and other short-lived fission products in the next library, although a contribution from the simplifications in the thermal hydraulics models are also partly responsible for the differences.

\begin{figure}[!htbp]
\vspace{-3mm}
	\centering
	\includegraphics[scale=0.32, clip, trim = 1mm 2mm 2mm 1mm]{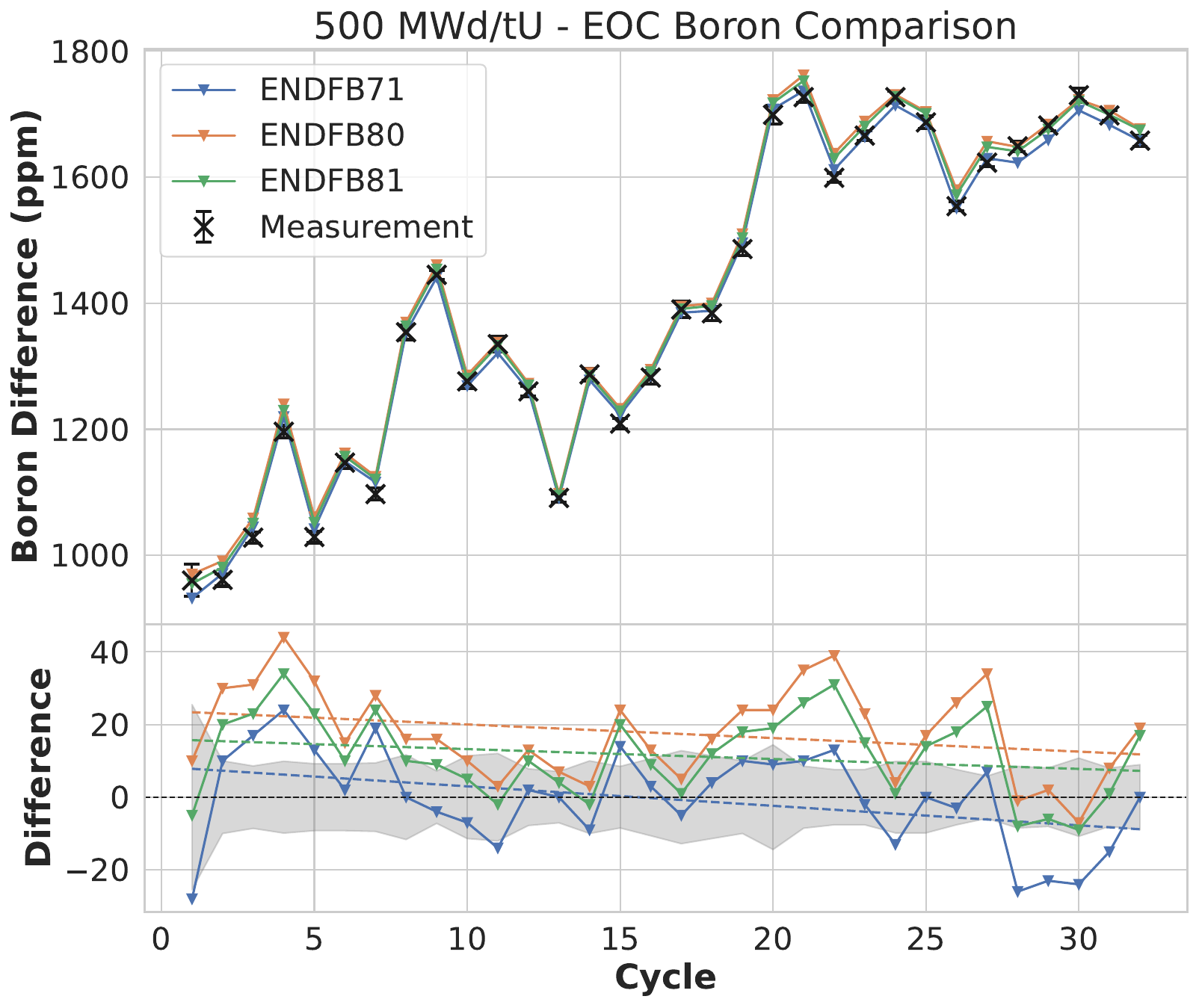}
\vspace{-4mm}
	\caption{Differences in the critical boron concentration at 500~MWd/tU and at end-of-cycle compared to measured values of the Kr\v{s}ko NPP.}
	\label{fig:CycleLength}
\end{figure}

The difference in boron concentration at 500 MWd/tU and at end-of-cycle is closely related to the cycle length. Calculated differences in boron concentration are compared to the measured differences in Fig.~\ref{fig:CycleLength}.
From the figure, it is evident that overall the cycle length prediction improves with ENDF/B-VIII.1 data and that the excessive reactivity loss in terms of critical boron concentration is only about 10~ppm on average over 32 cycles of operation.

The overall conclusion from the validation exercise with CORD-2 is that the calculations with WIMSD-5B~\cite{WIMSD5} at the cell level are representative of the more accurate Monte Carlo calculations. The full core design calculations show some over-prediction of the power and xenon defect, which could partly be attributed to the fission yield of $^{135}$I and its precursors. The cycle length prediction is improved with ENDF/B-VIII.1 data compared to previous versions of the library.

\subsection{Pulsed Spheres Results}

LLNL pulsed-spheres measurements provide neutron-leakage spectra over time-of-flight pulsed by a 14-MeV neutron source in the center of the sphere~\cite{Wong1972}.
They allow us to validate nuclear data from 15 MeV down to several MeV dependent on the sphere's thickness and the level scheme of the nuclei at hand~\cite{Neudecker:2021_PS,EUCLIDSensLibrary}.
Below, validation results with ENDF/B-VIII.0 and ENDF/B-VIII.1$\beta$3 are given.
These simulated values are virtually the same as for $\beta$4 except for Pb where simulated values for $\beta$4 are shown.
Also, hints are provided how predictions of neutron-leakage spectra can be improved by nuclear data changes based on sensitivity profiles computed as part of the LANL EUCLID LDRD-DR project~\cite{Neudecker:2021_PS,EUCLIDSensLibrary}.

\begin{figure}[!htb]
\vspace{-2mm}
\centering
\includegraphics[width=0.45\textwidth]{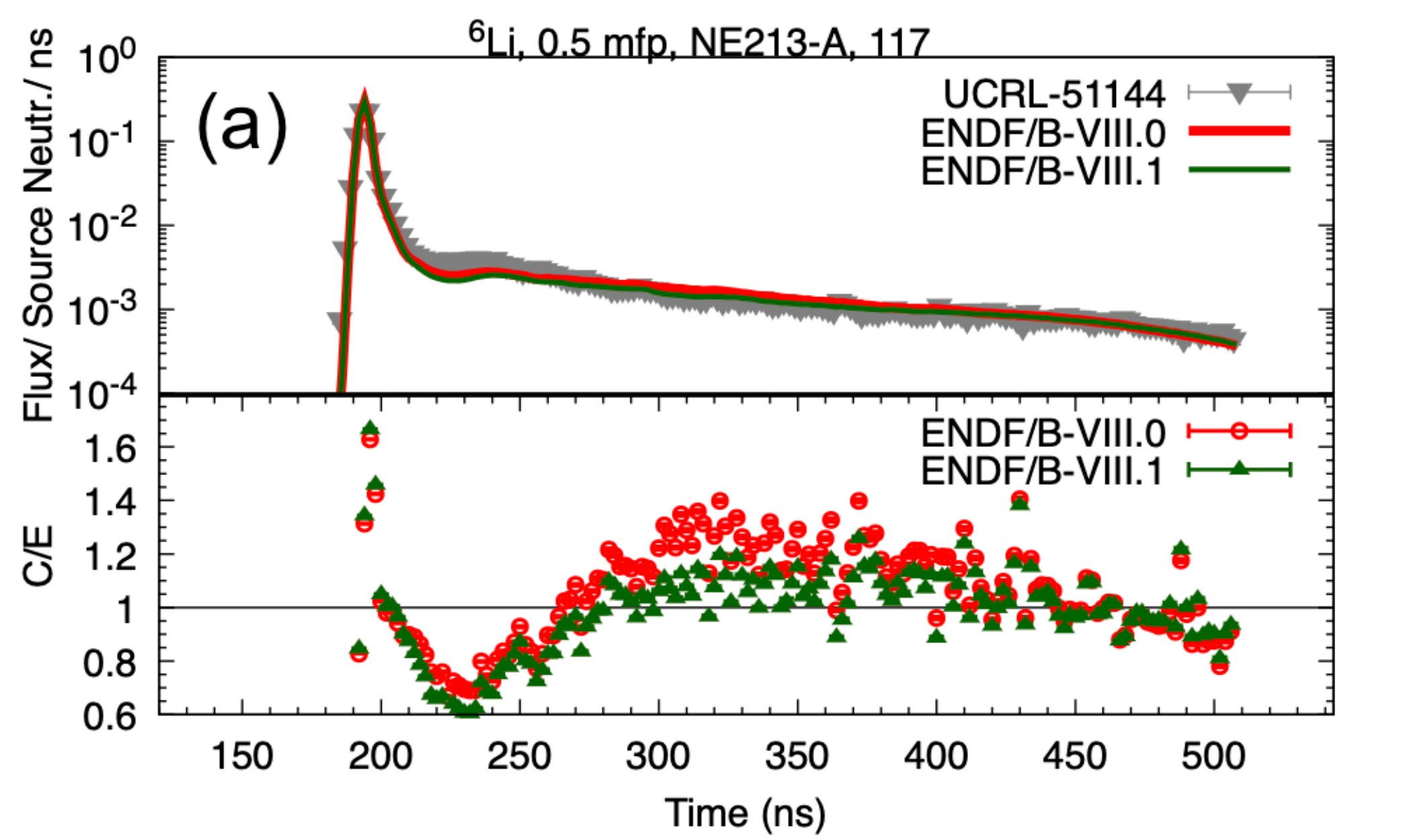}
\includegraphics[width=0.45\textwidth]{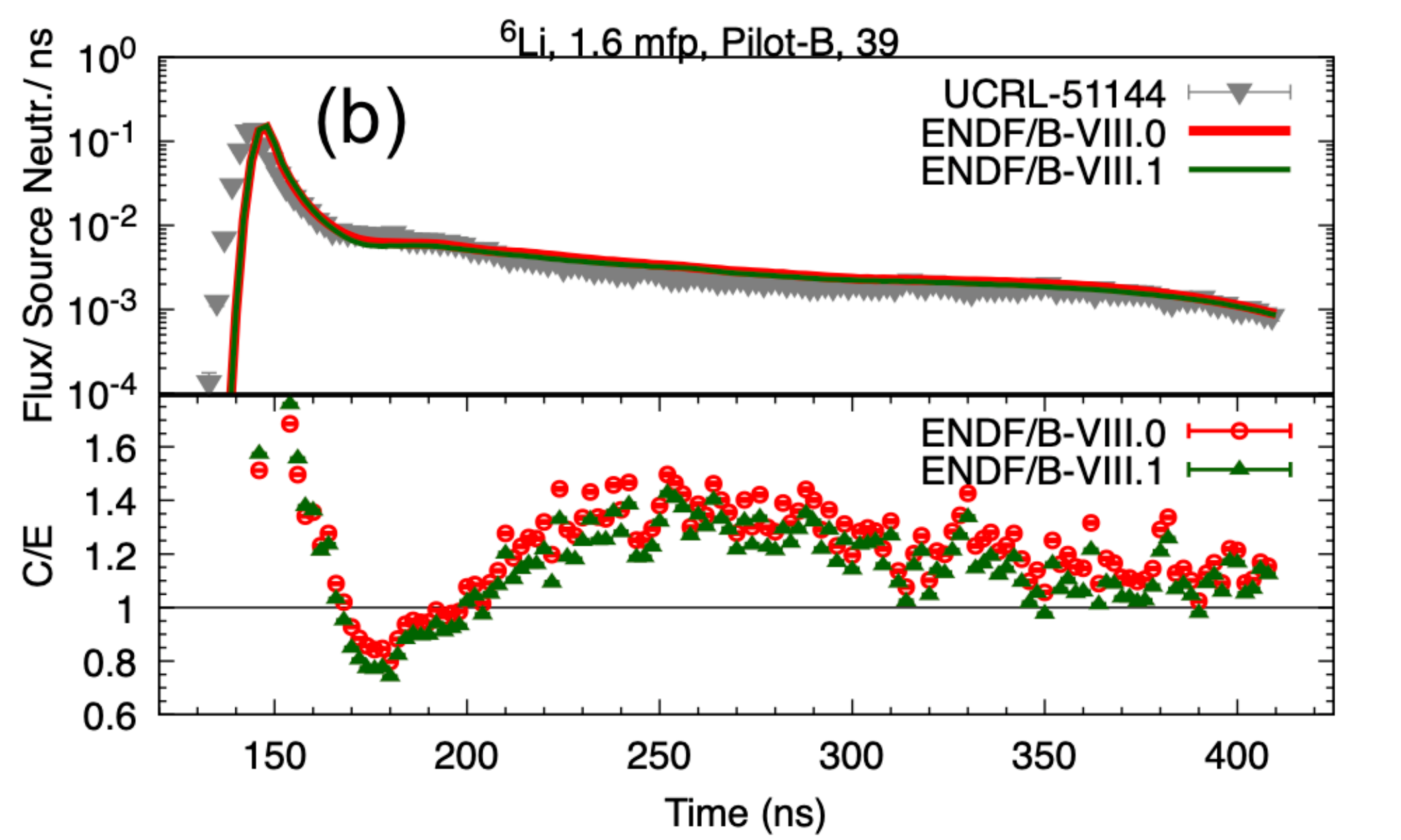}\\
\includegraphics[width=0.45\textwidth]{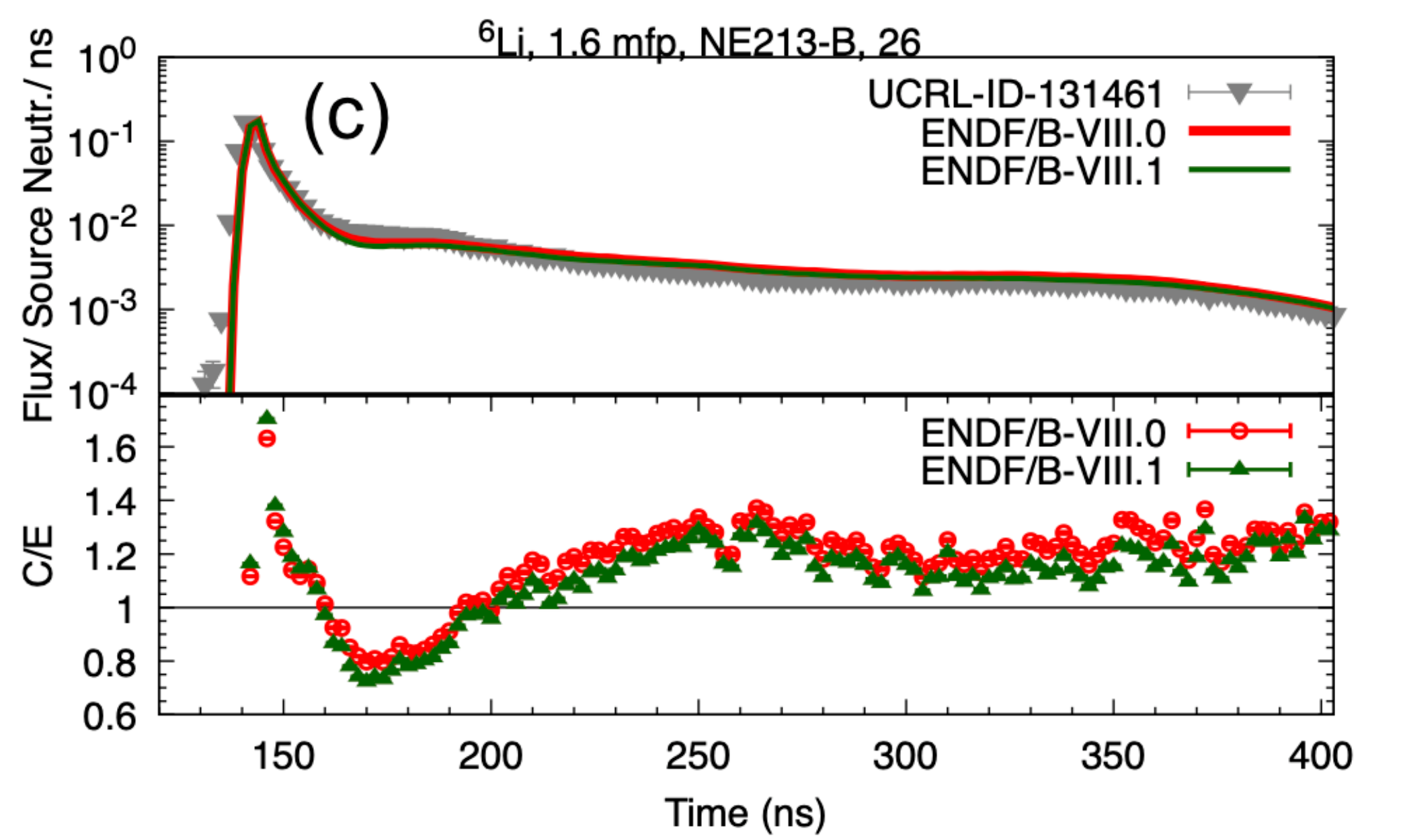}
\caption{LLNL pulsed-sphere neutron-leakage spectra simulated with ENDF/B-VIII.0 and ENDF/B-VIII.1 for $^{6}$Li compared to experimental data.}
\label{fig:LLNLpulsedspheres6Li}
\vspace{-2mm}
\end{figure}
$^6$Li LLNL pulsed-sphere neutron-leakage spectra change distinctly in Fig.~\ref{fig:LLNLpulsedspheres6Li} when simulated with ENDF/B-VIII.0 and ENDF/B-VIII.1.
The spheres are thin, with mfp 0.5 and 1.6; little multiple scattering of neutrons will happen and most neutrons will be induced in the sphere by the incident-neutron beam with energies of 12--15 MeV.
In this energy range, one major change happened for the $^6$Li(n,n') cross sections; the reaction was changed from the laboratory to the center-of-mass frame.
To be clear, both evaluations are leading to large biases in simulating pulsed-sphere spectra.
While the simulations with the new evaluation get closer at later times, it is slightly worse in the peak region and right after.
One could study changing MF=$\{3,4\}$ MT=53 for better predicting $^6$Li LLNL pulsed-sphere neutron-leakage spectra.

\begin{figure}[!htb]
\vspace{-2mm}
\centering
\includegraphics[width=0.45\textwidth]{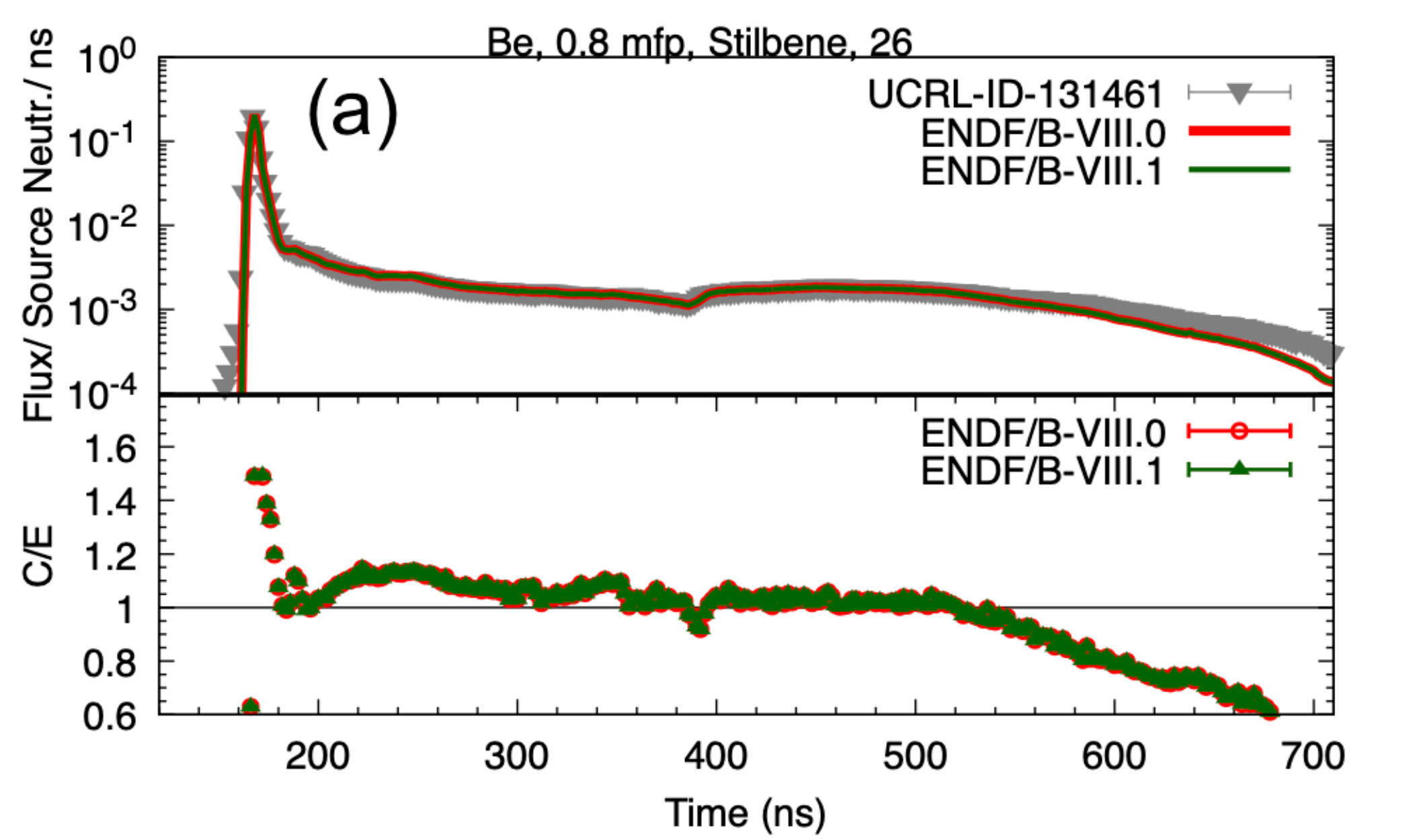}
\includegraphics[width=0.45\textwidth]{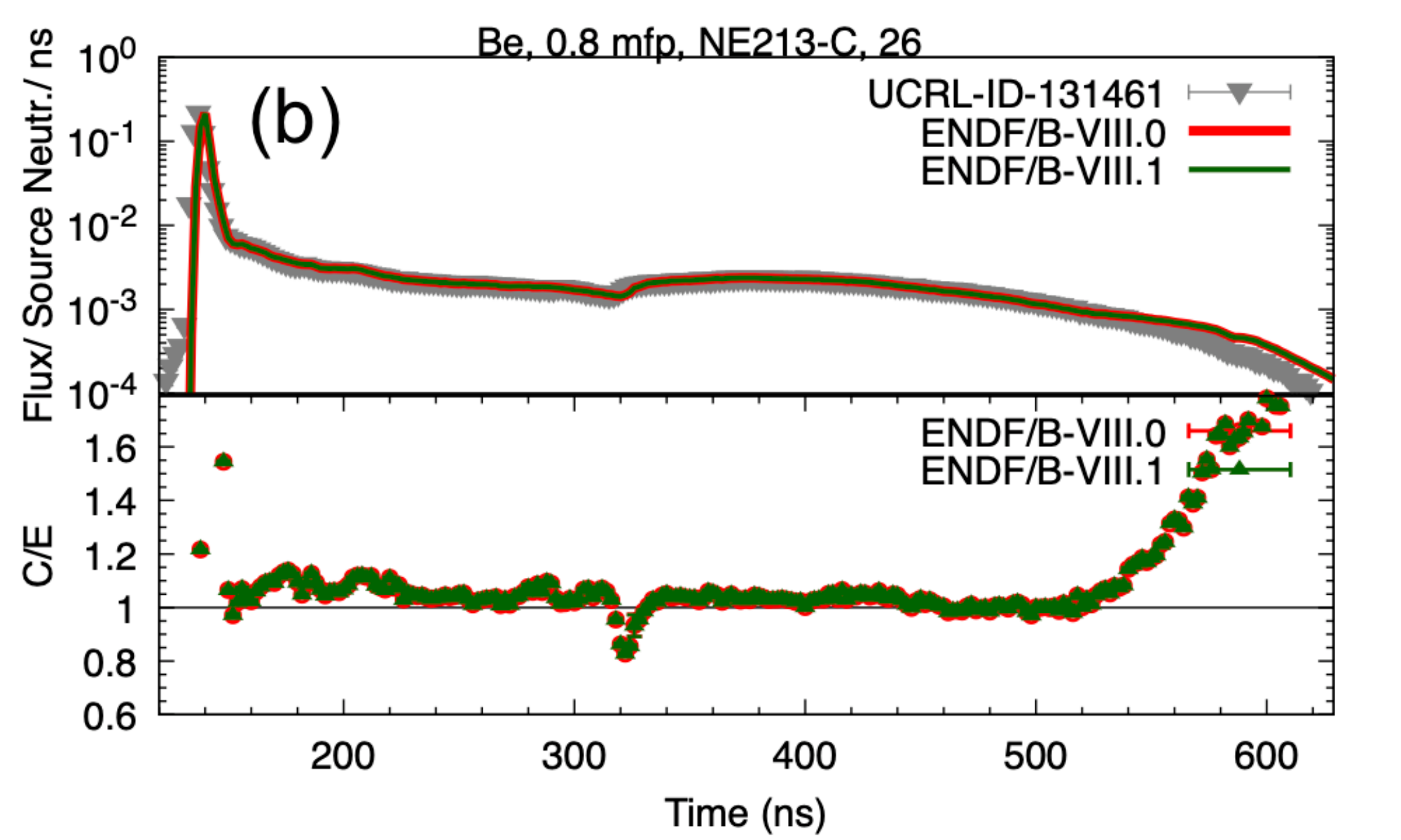}
\caption{LLNL pulsed-sphere neutron-leakage spectra simulated with ENDF/B-VIII.0 and ENDF/B-VIII.1 for $^{8}$Be compared to experimental data.}
\label{fig:LLNLpulsedspheresBe}
\vspace{-2mm}
\end{figure}
The $^9$Be LLNL pulsed-sphere neutron-leakage spectra in Fig.~\ref{fig:LLNLpulsedspheresBe} are described similarly in ENDF/B-VIII.1 to ENDF/B-VIII.0.
This similarity was expected given that mostly the capture cross section changed, which would minimally impact the neutron-leakage spectra.
In general, the agreement between simulated and experimental data is within the knowledge we have of the pulsed spheres.

\begin{figure}[!thb]
\vspace{-2mm}
\centering
\includegraphics[width=0.45\textwidth]{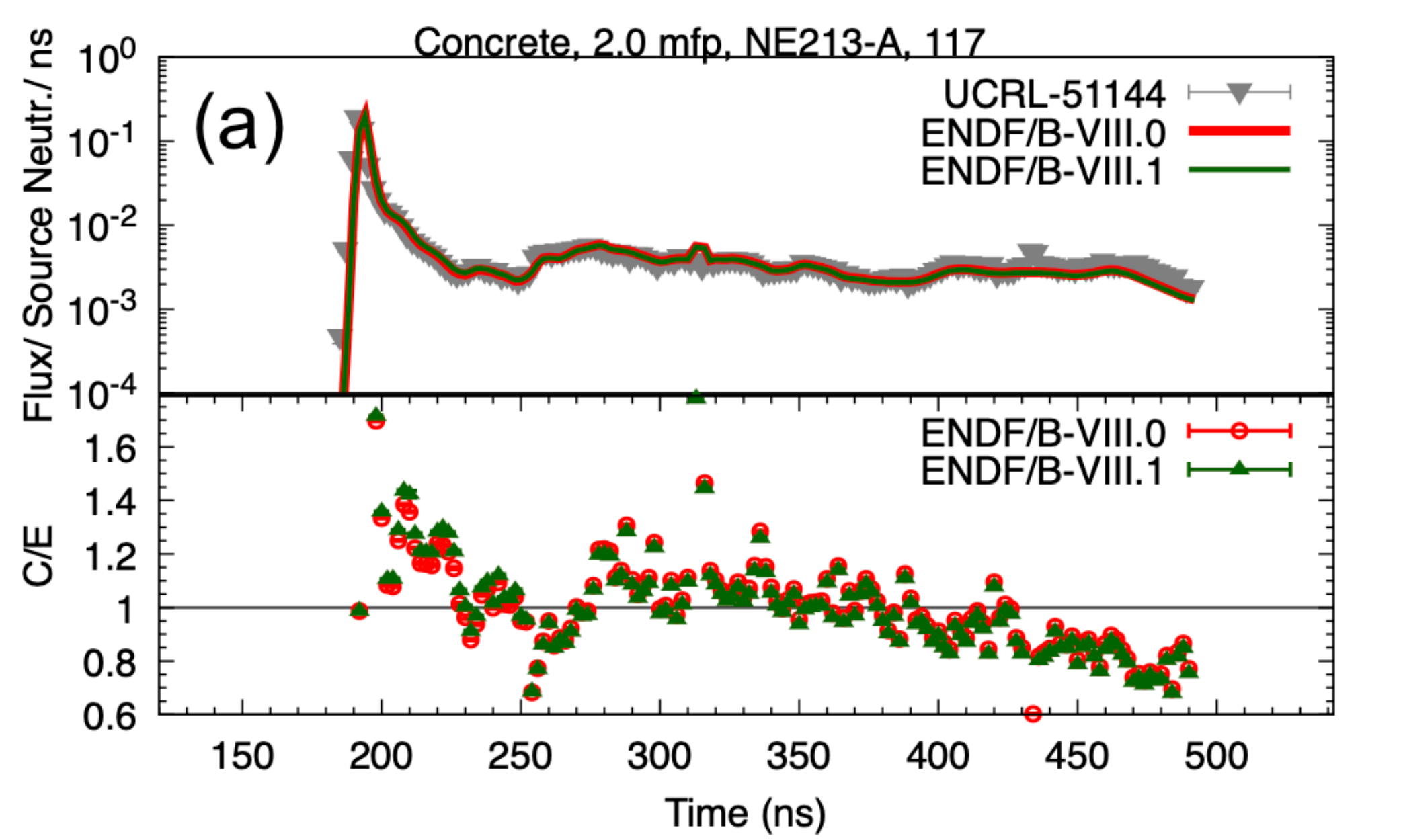}
\includegraphics[width=0.45\textwidth]{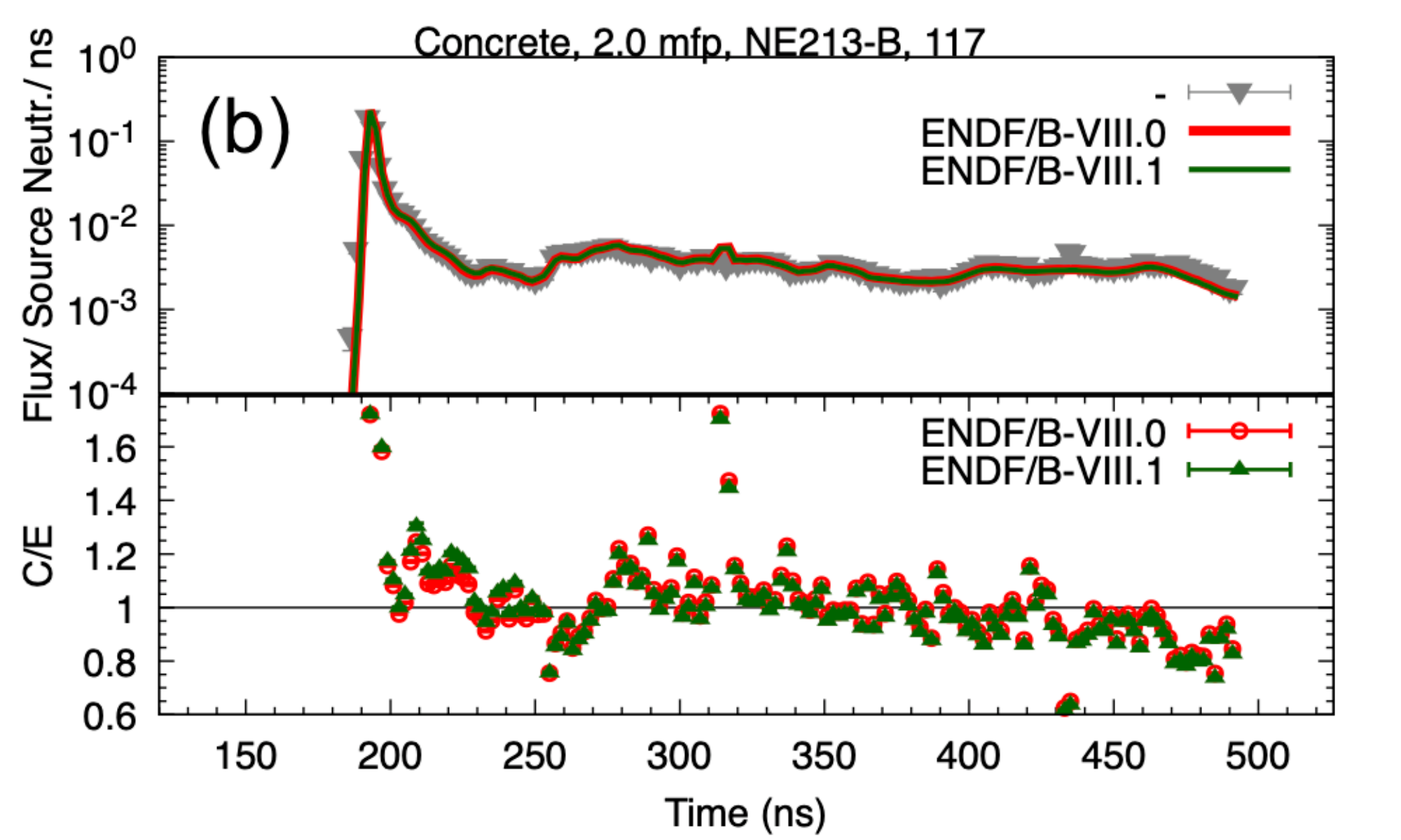}
\includegraphics[width=0.45\textwidth]{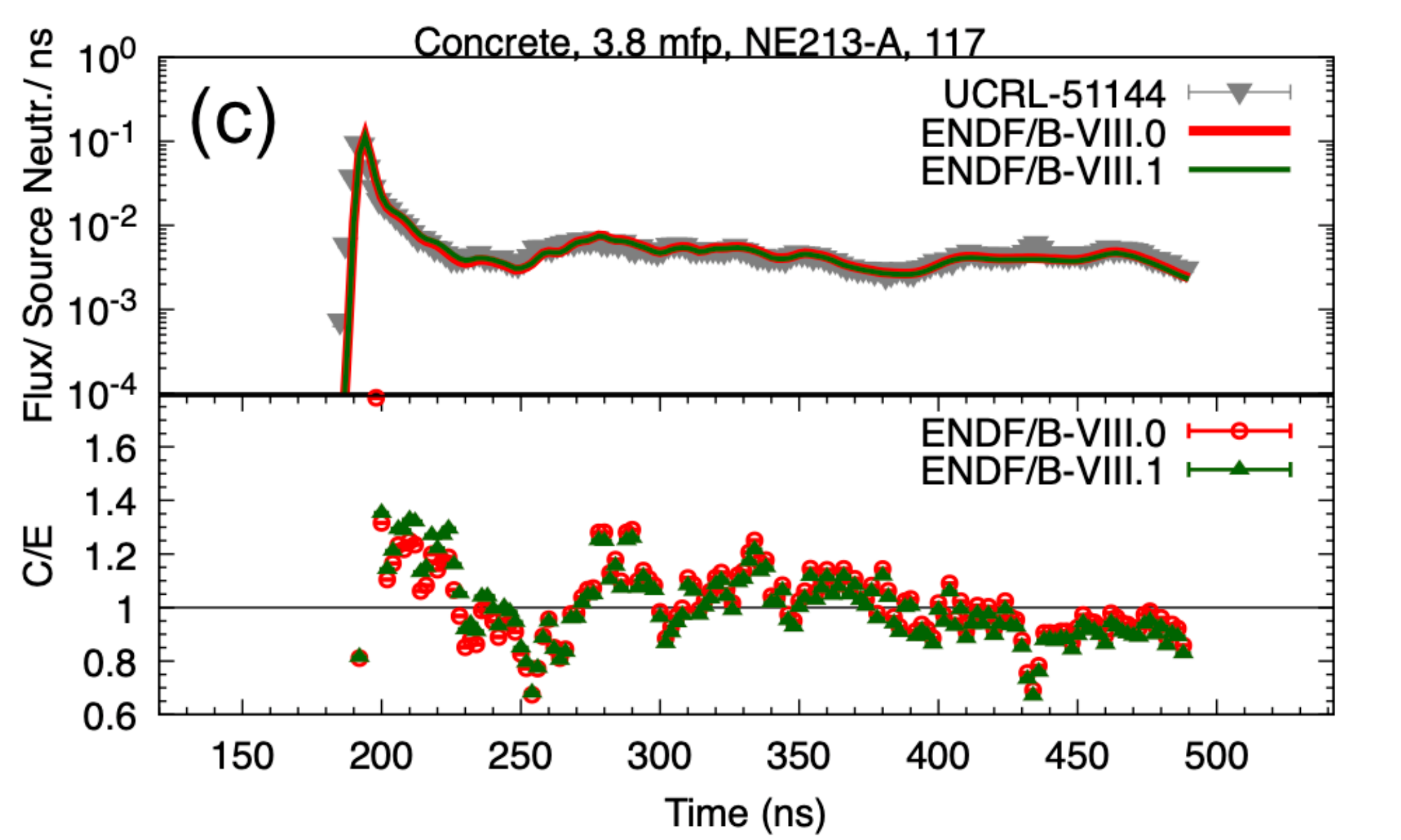}
\includegraphics[width=0.45\textwidth]{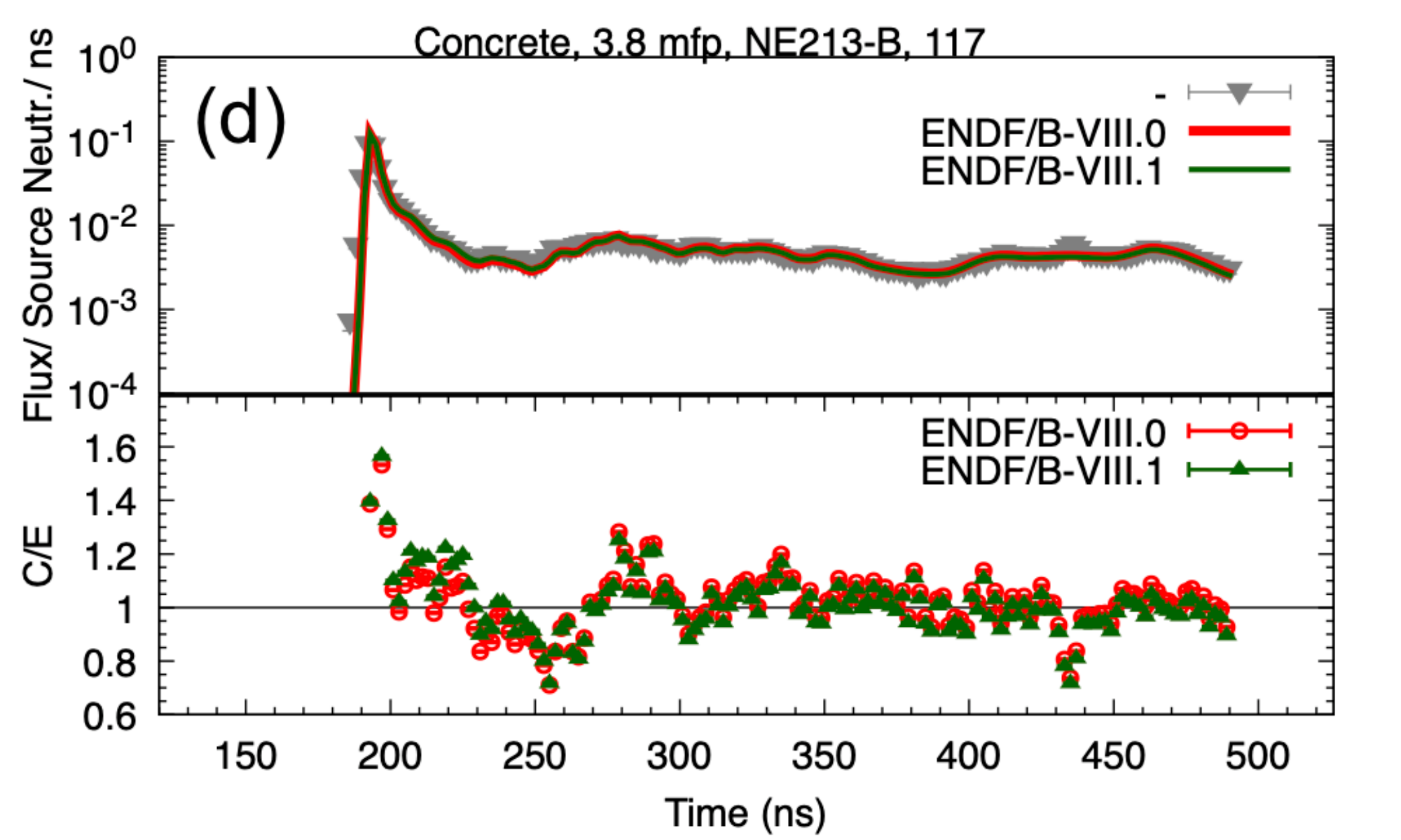}
\vspace{-2mm}
\caption{LLNL pulsed-sphere neutron-leakage spectra simulated with ENDF/B-VIII.0 and ENDF/B-VIII.1 for concrete compared to experimental data.}
\label{fig:LLNLpulsedspheresConcrete}
\vspace{-2mm}
\end{figure}

\begin{figure}[!thb]
\vspace{-2mm}
\centering
\includegraphics[width=0.45\textwidth]{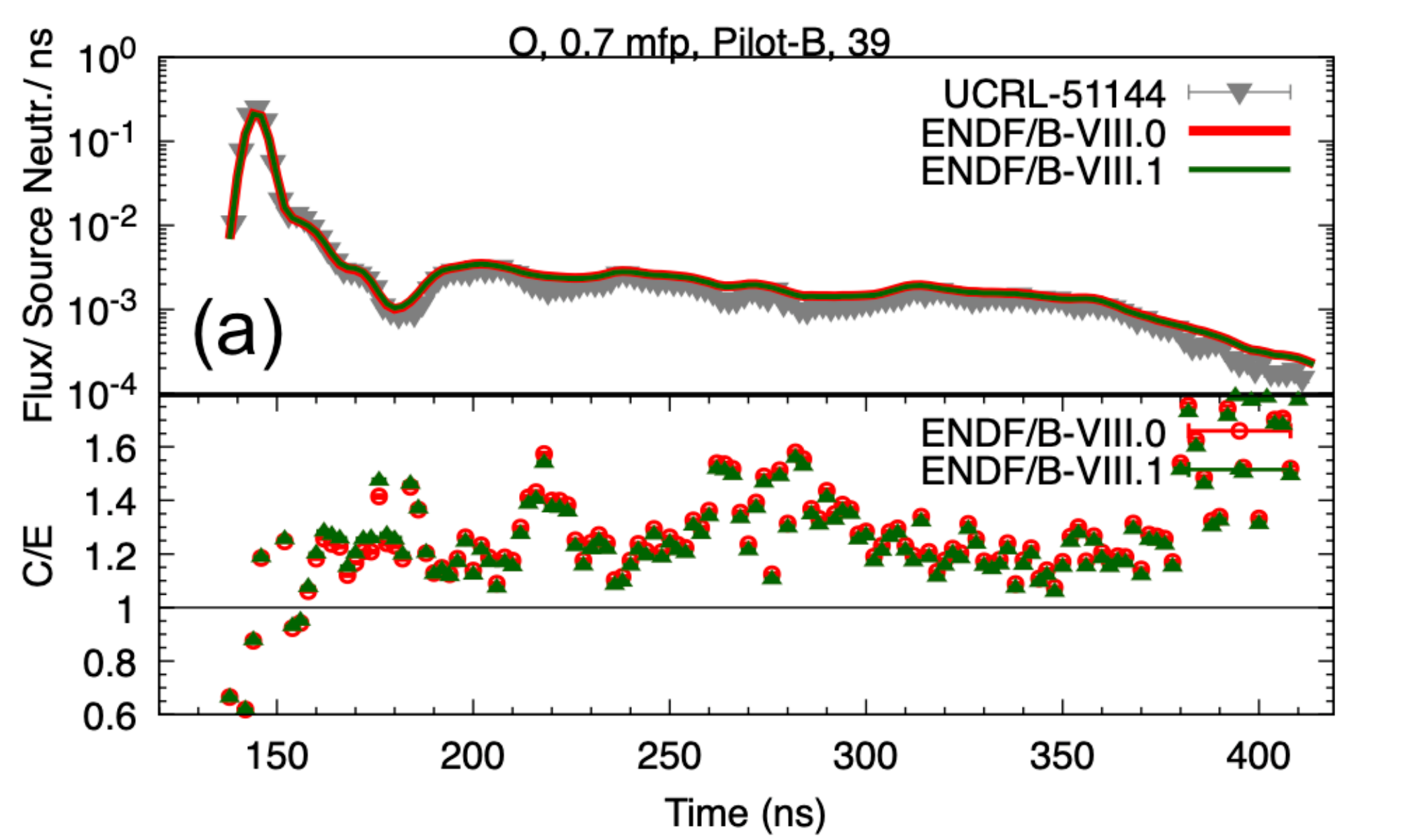}
\includegraphics[width=0.45\textwidth]{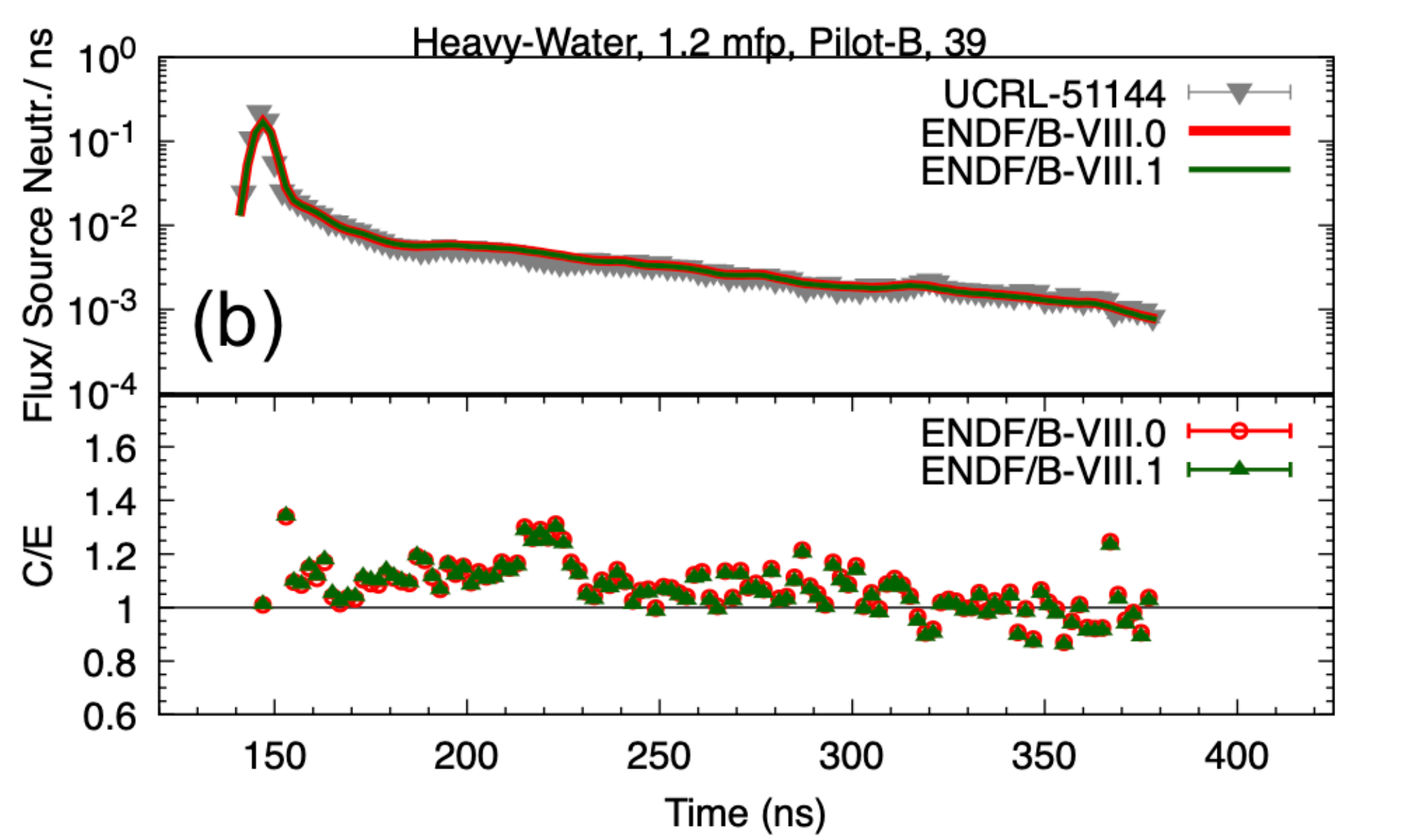}
\includegraphics[width=0.45\textwidth]{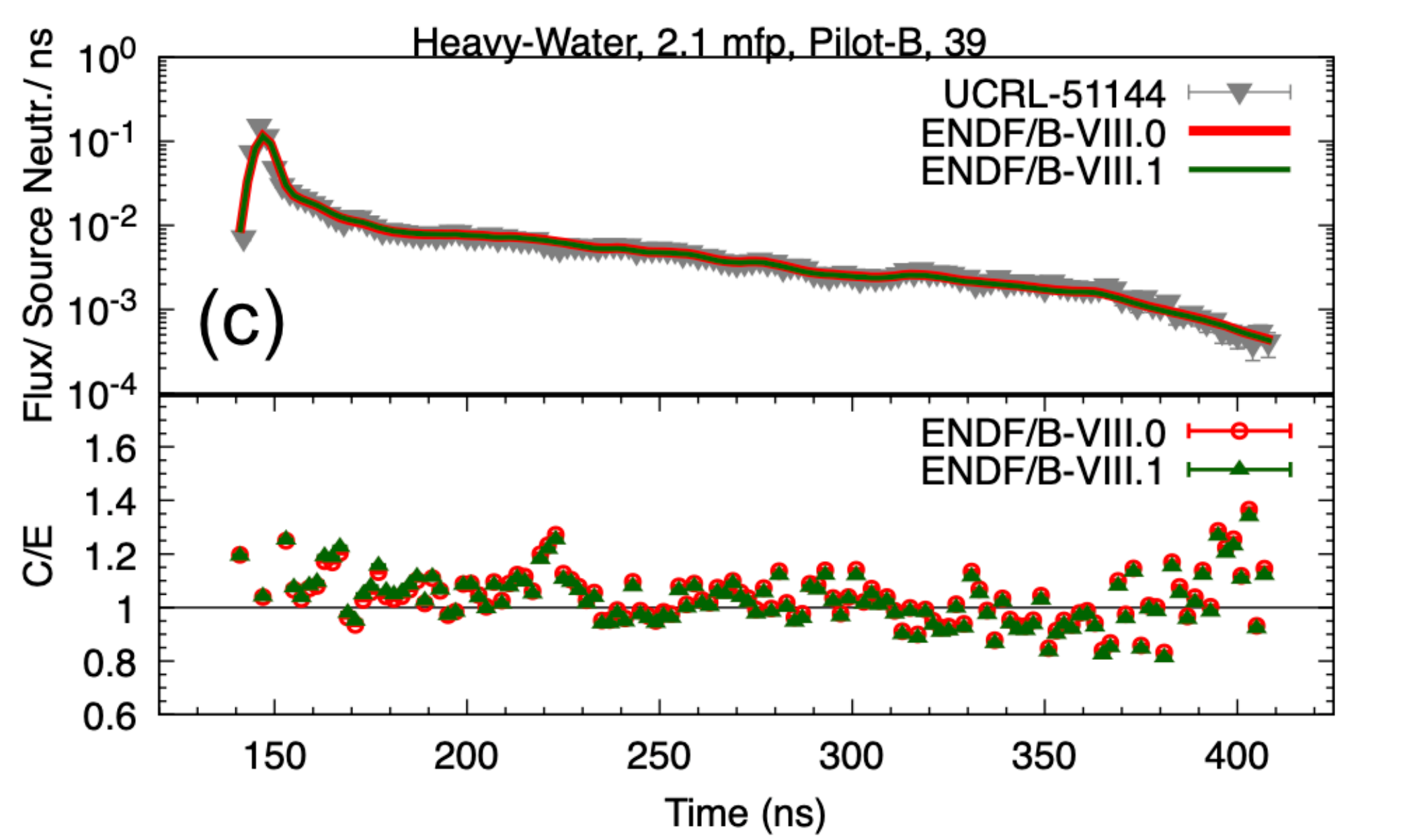}
\vspace{-2mm}
\caption{LLNL pulsed-sphere neutron-leakage spectra simulated with ENDF/B-VIII.0 and ENDF/B-VIII.1 for $^{16}$O and heavy water compared to experimental data.}
\label{fig:LLNLpulsedspheresO16hwt}
\end{figure}

\begin{figure}[!thbp]
\centering
\includegraphics[width=0.45\textwidth]{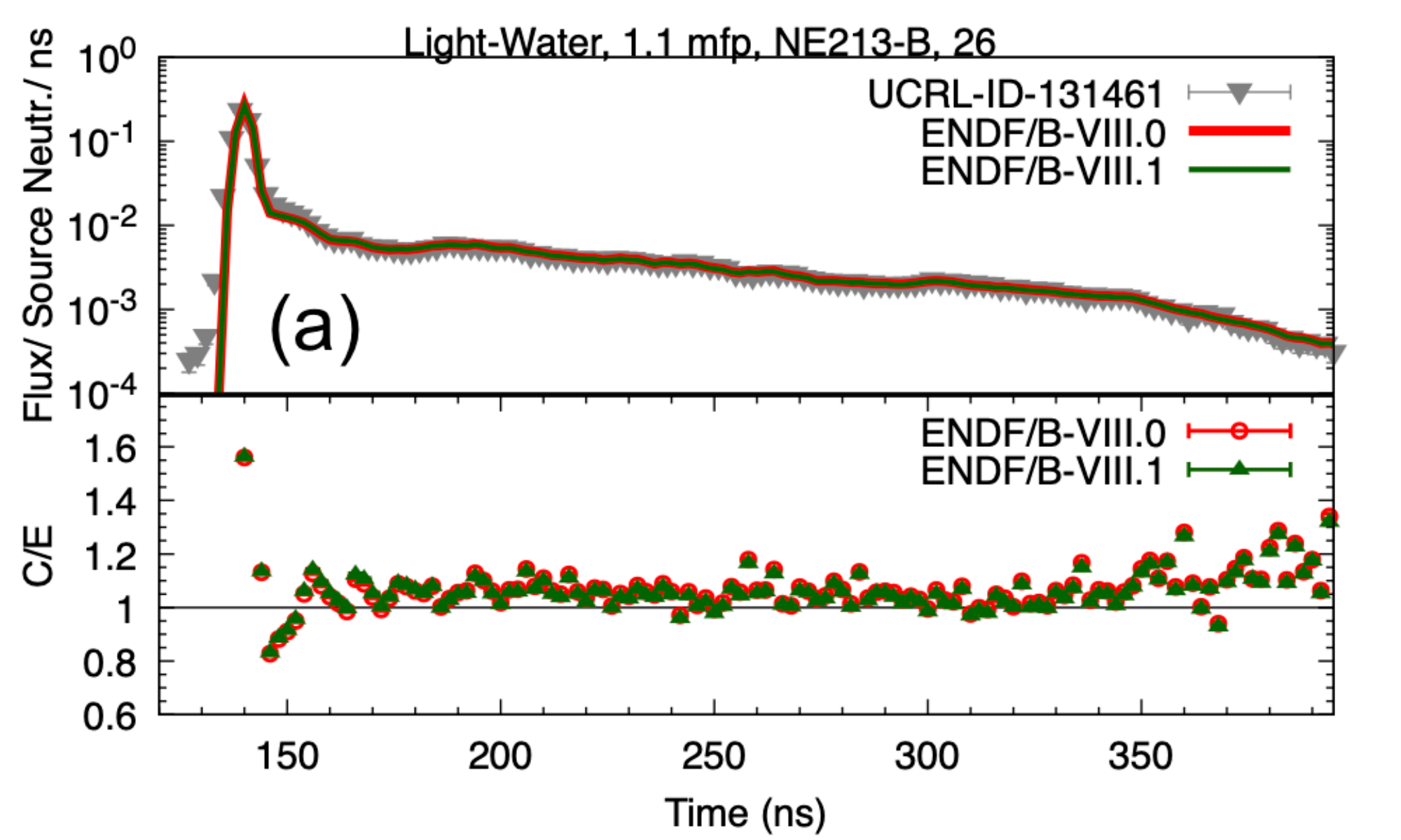}
\includegraphics[width=0.45\textwidth]{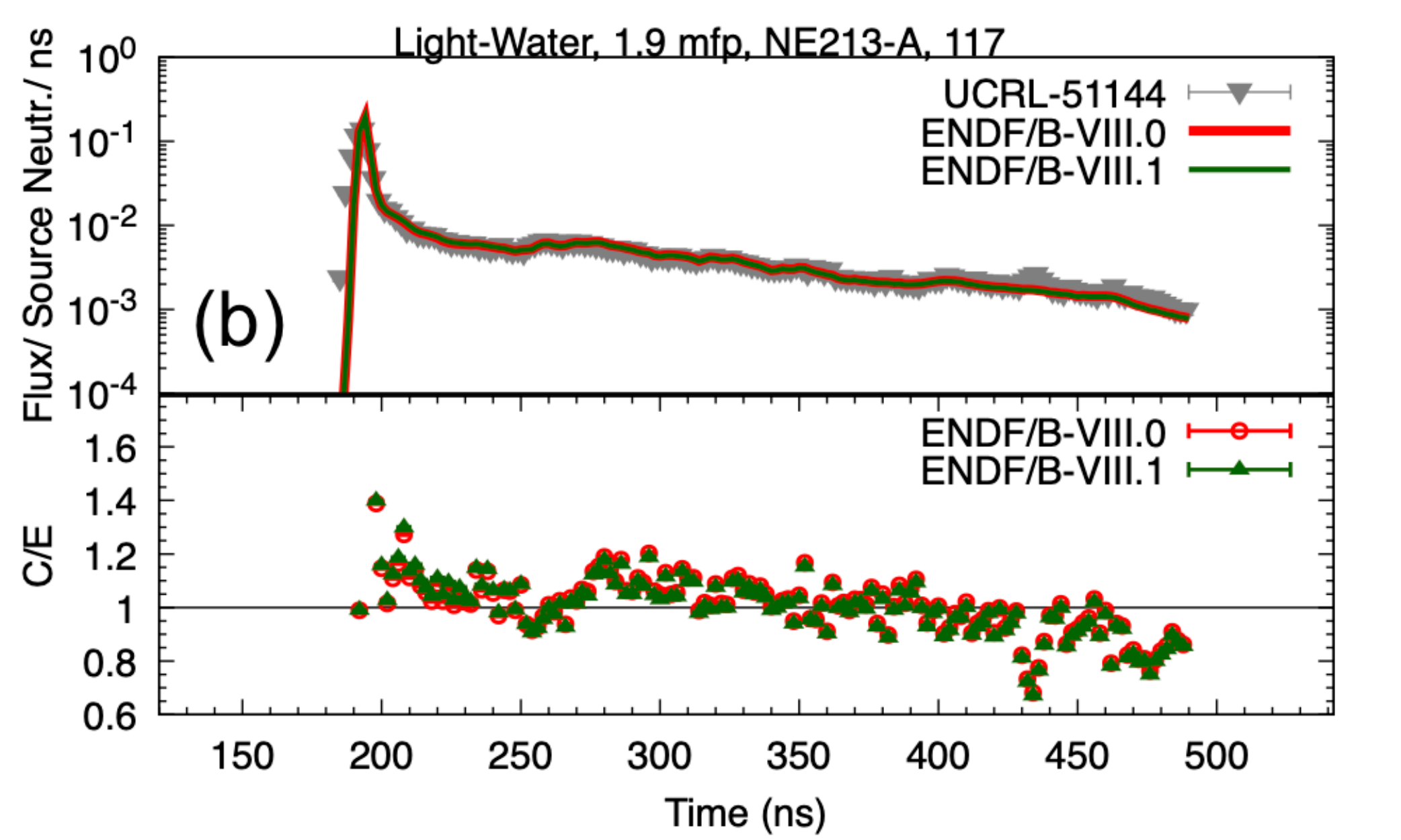}
\includegraphics[width=0.45\textwidth]{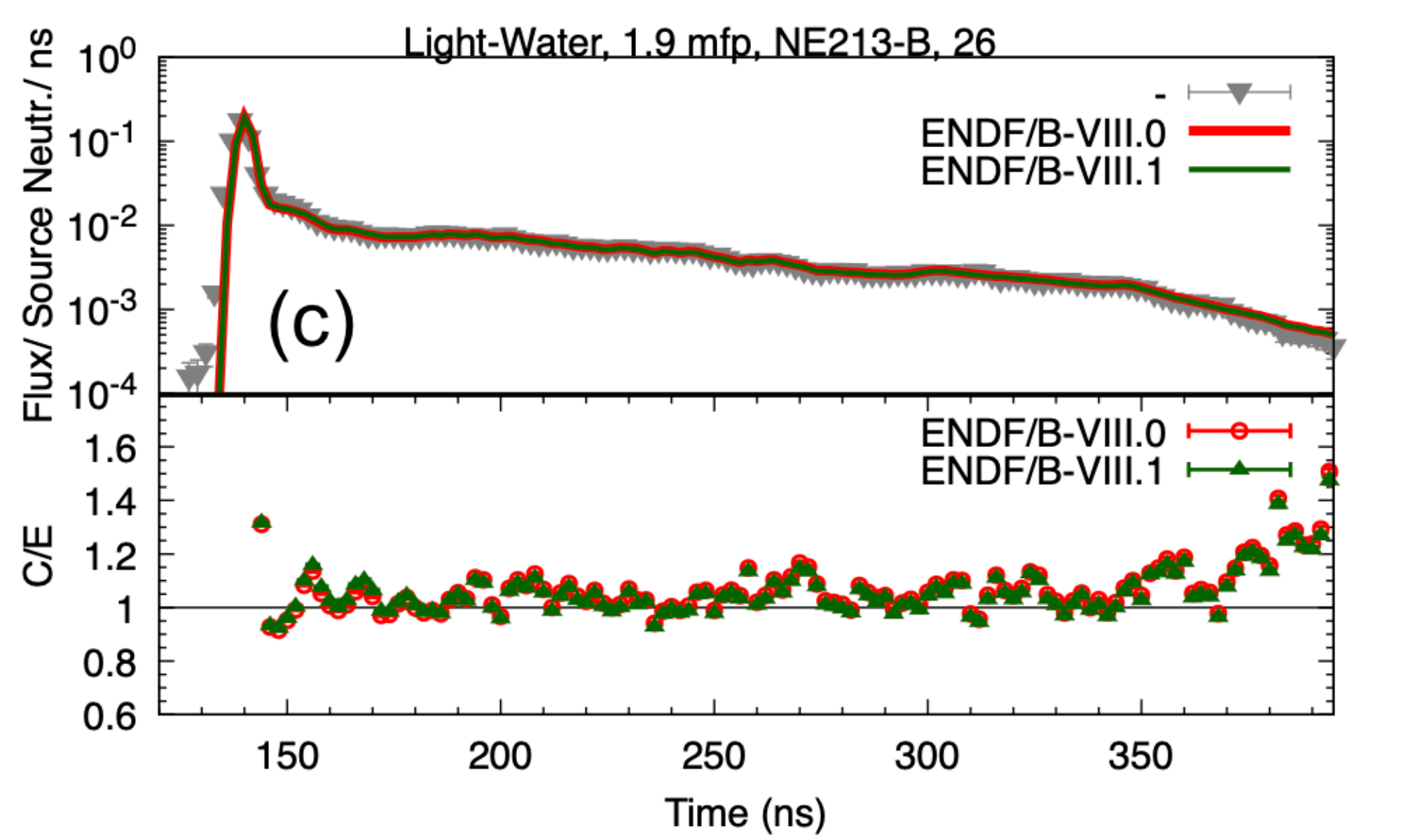}
\includegraphics[width=0.45\textwidth]{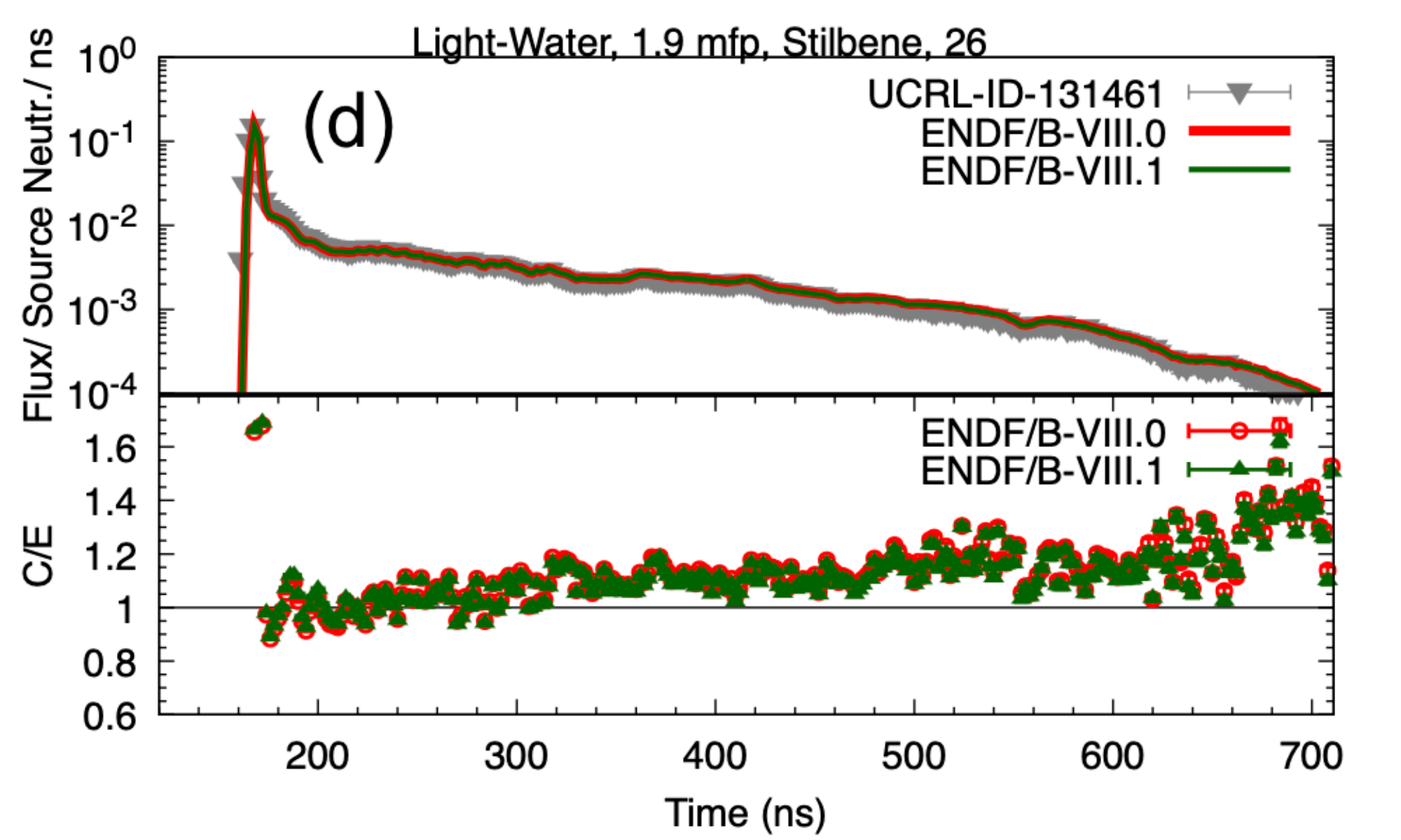}
\caption{LLNL pulsed-sphere neutron-leakage spectra simulated with ENDF/B-VIII.0 and ENDF/B-VIII.1 for light water compared to experimental data.}
\label{fig:LLNLpulsedsphereslwt}
\end{figure}

\begin{figure}[!thbp]
\centering
\includegraphics[width=0.45\textwidth]{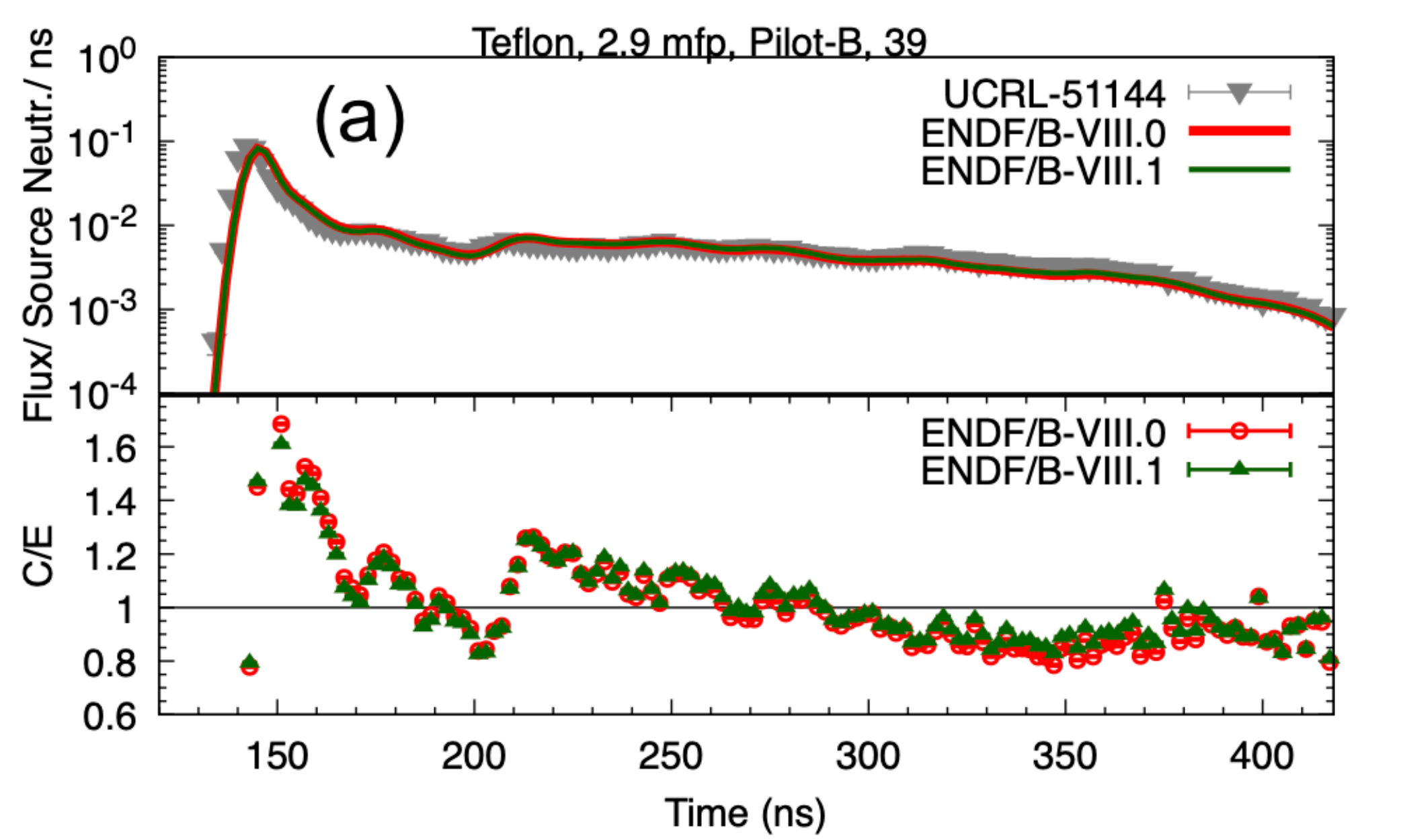}
\includegraphics[width=0.45\textwidth]{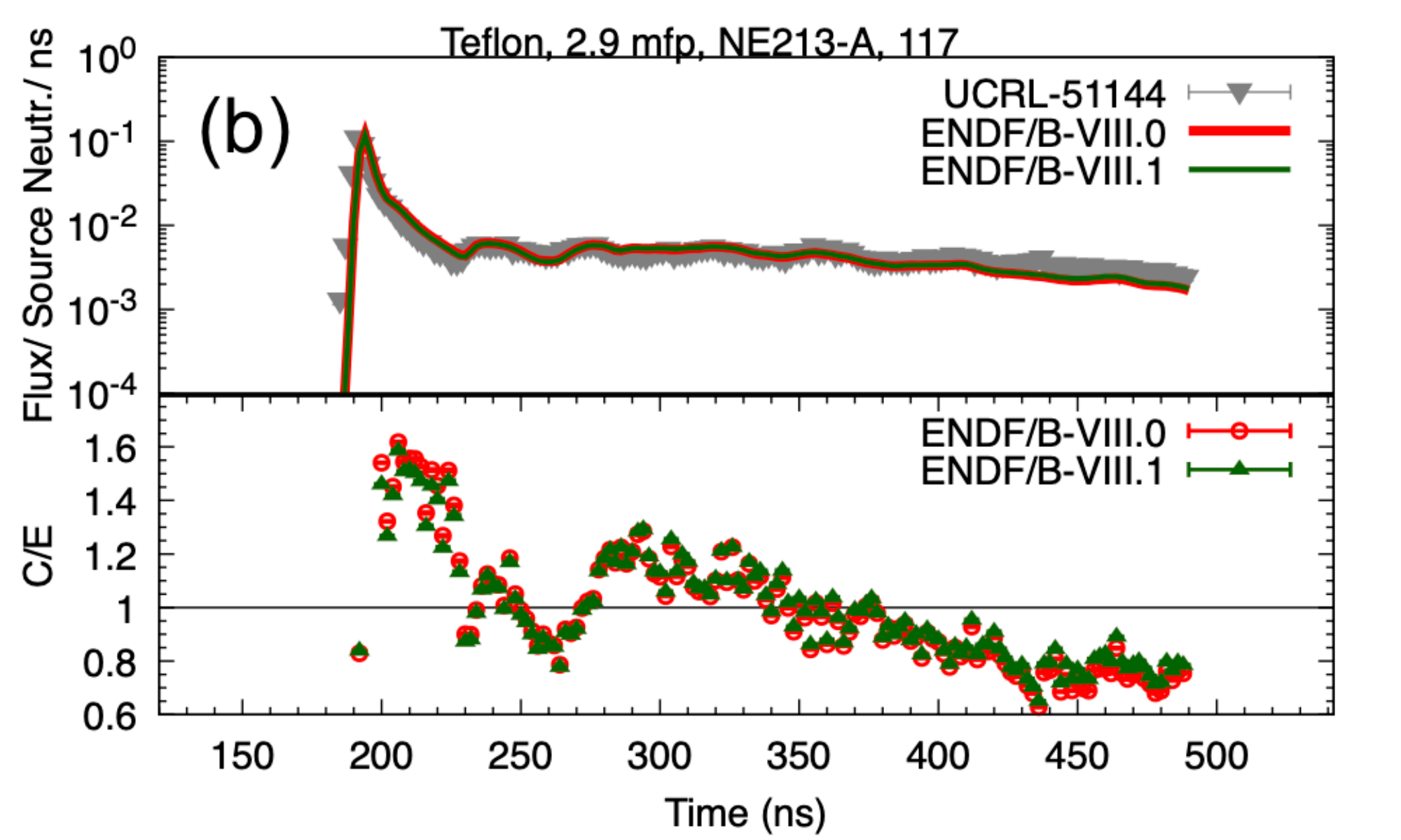}
\caption{LLNL pulsed-sphere neutron-leakage spectra simulated with ENDF/B-VIII.0 and ENDF/B-VIII.1 for teflon compared to experimental data.}
\label{fig:LLNLpulsedspheresteflon}
\end{figure}

\begin{figure}[!tphb]
\vspace{-2mm}
\centering
\includegraphics[width=0.45\textwidth]{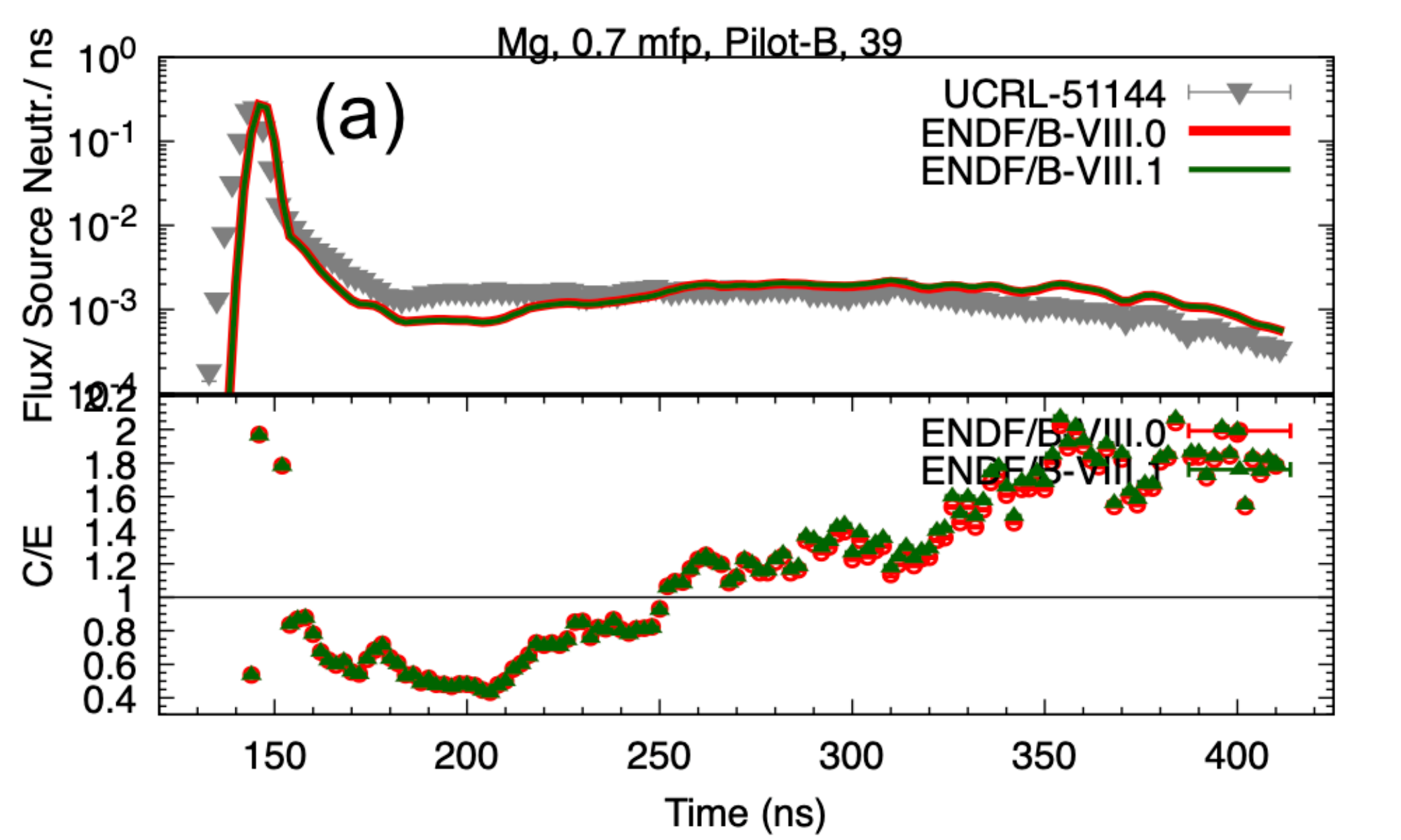}
\includegraphics[width=0.45\textwidth]{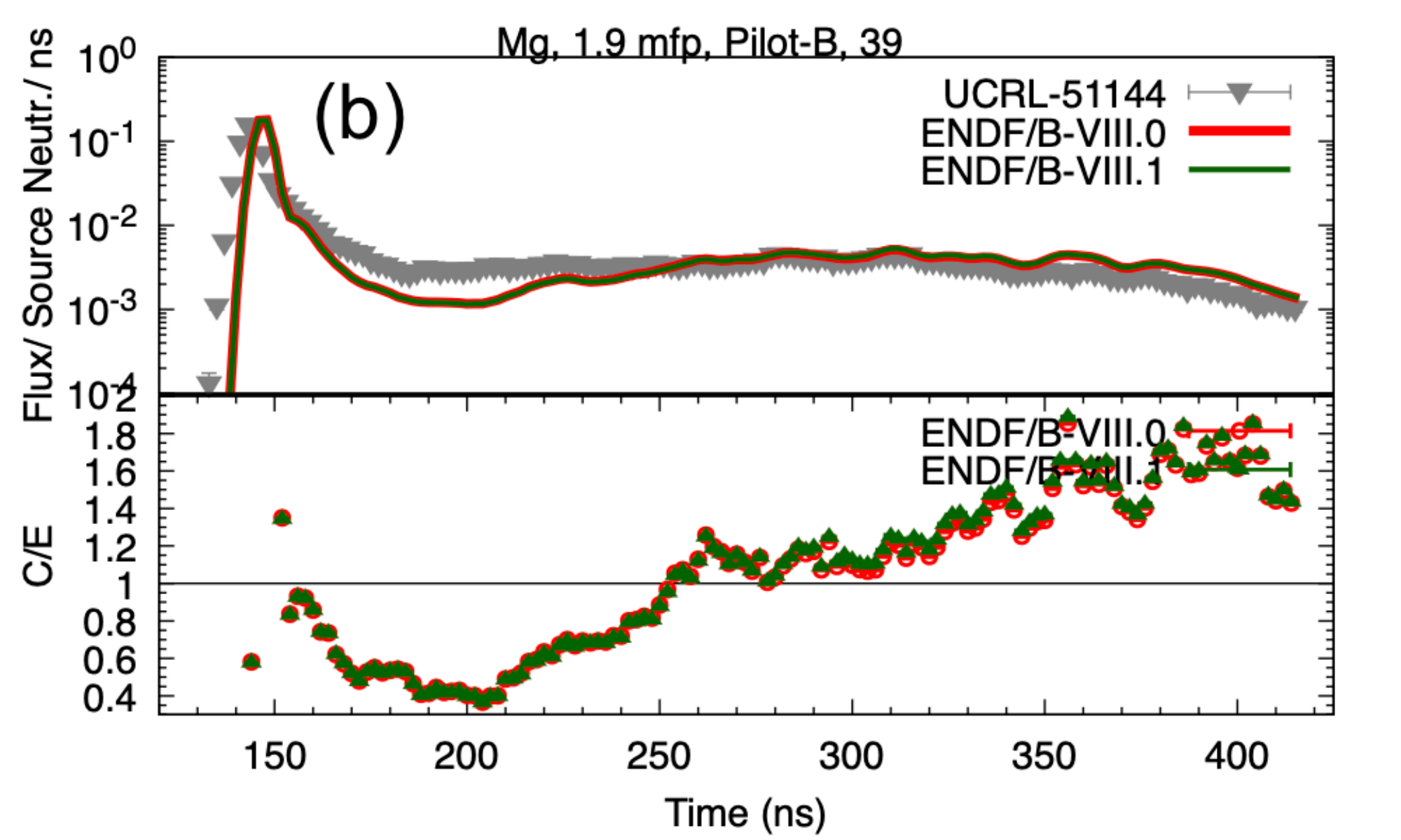}
\caption{LLNL pulsed-sphere neutron-leakage spectra simulated with ENDF/B-VIII.0 and ENDF/B-VIII.1 for Mg compared to experimental data.}
\label{fig:LLNLpulsedspheresmg}
\end{figure}

\begin{figure}[!tphb]
\vspace{-2mm}
\centering
\includegraphics[width=0.45\textwidth]{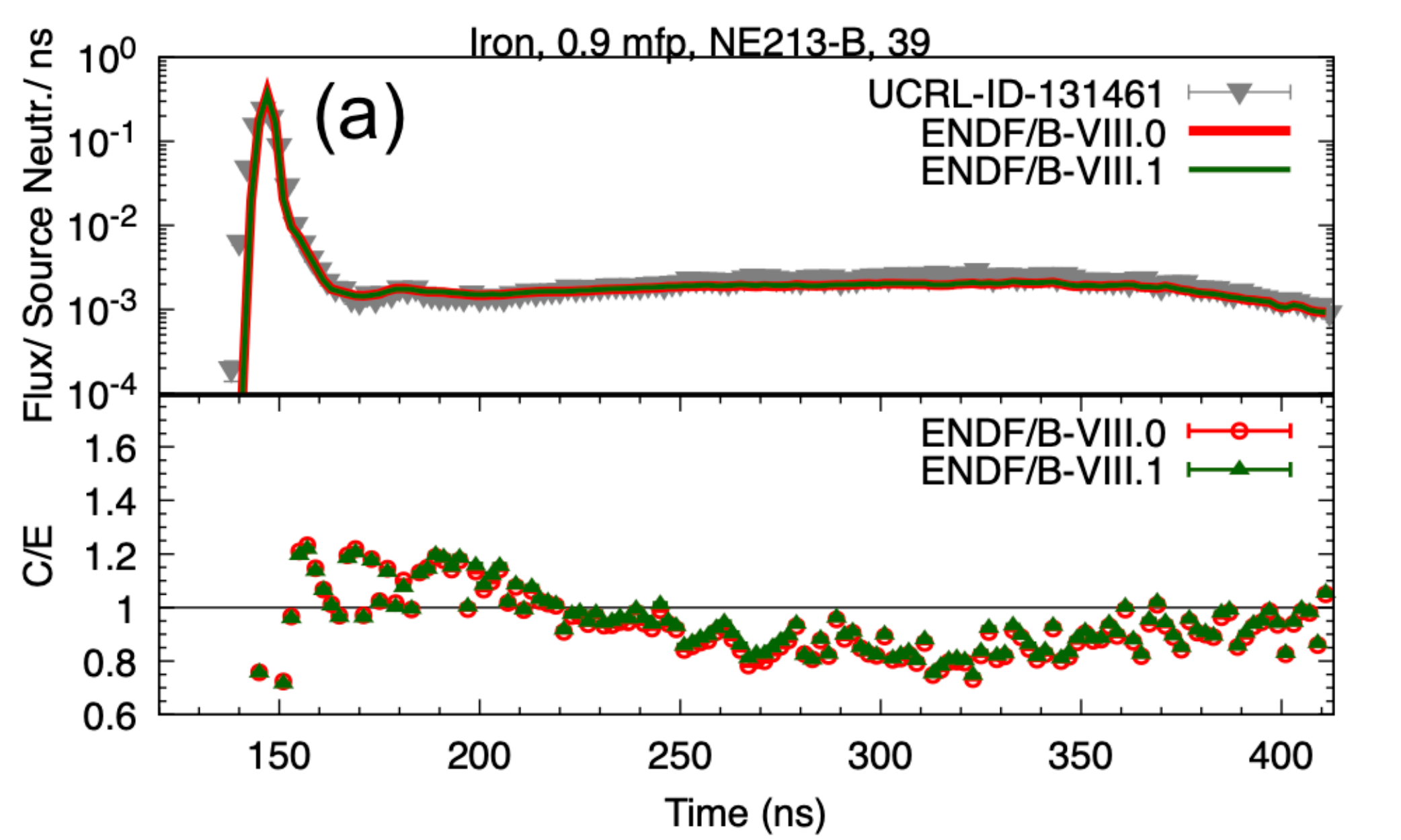}
\includegraphics[width=0.45\textwidth]{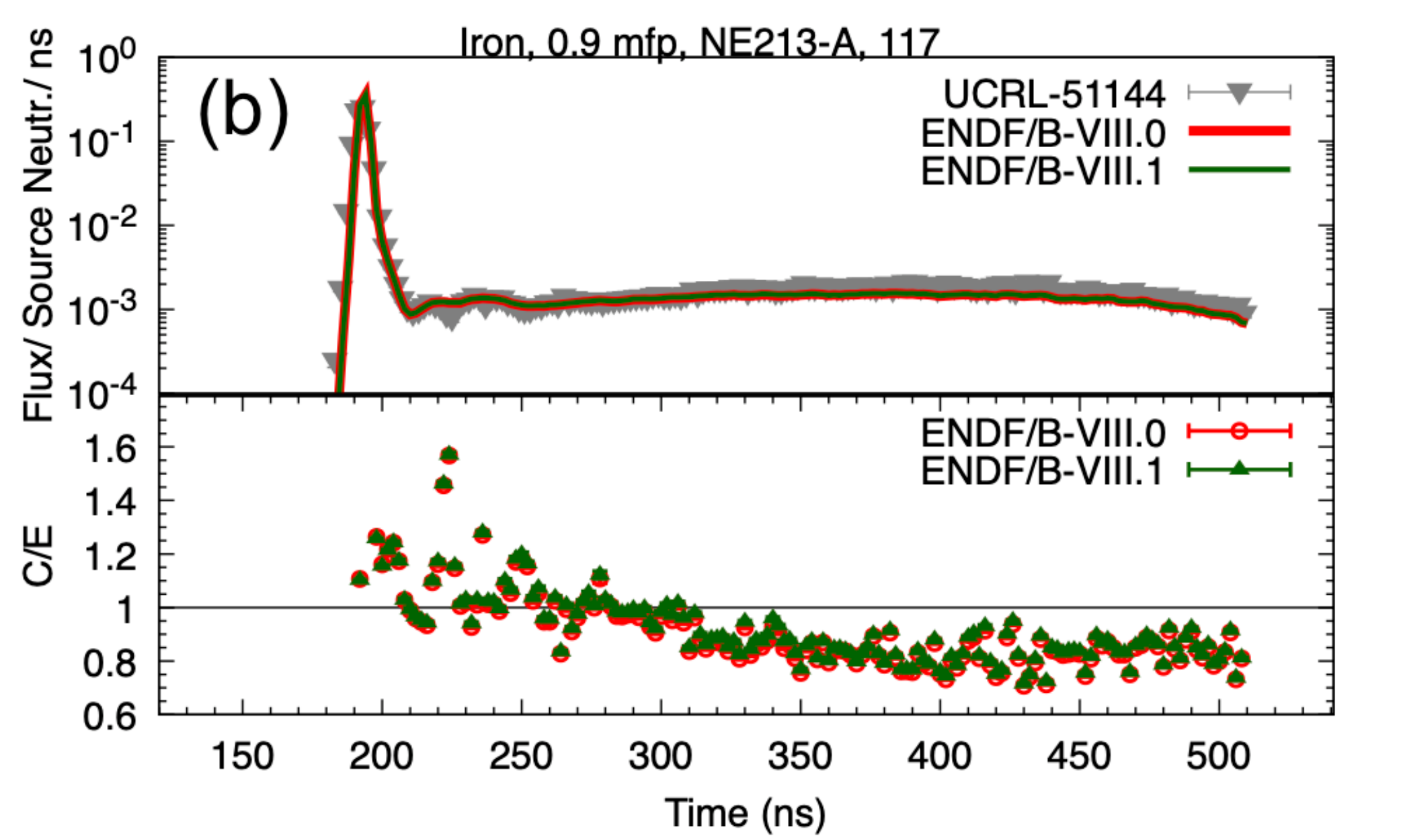}
\includegraphics[width=0.45\textwidth]{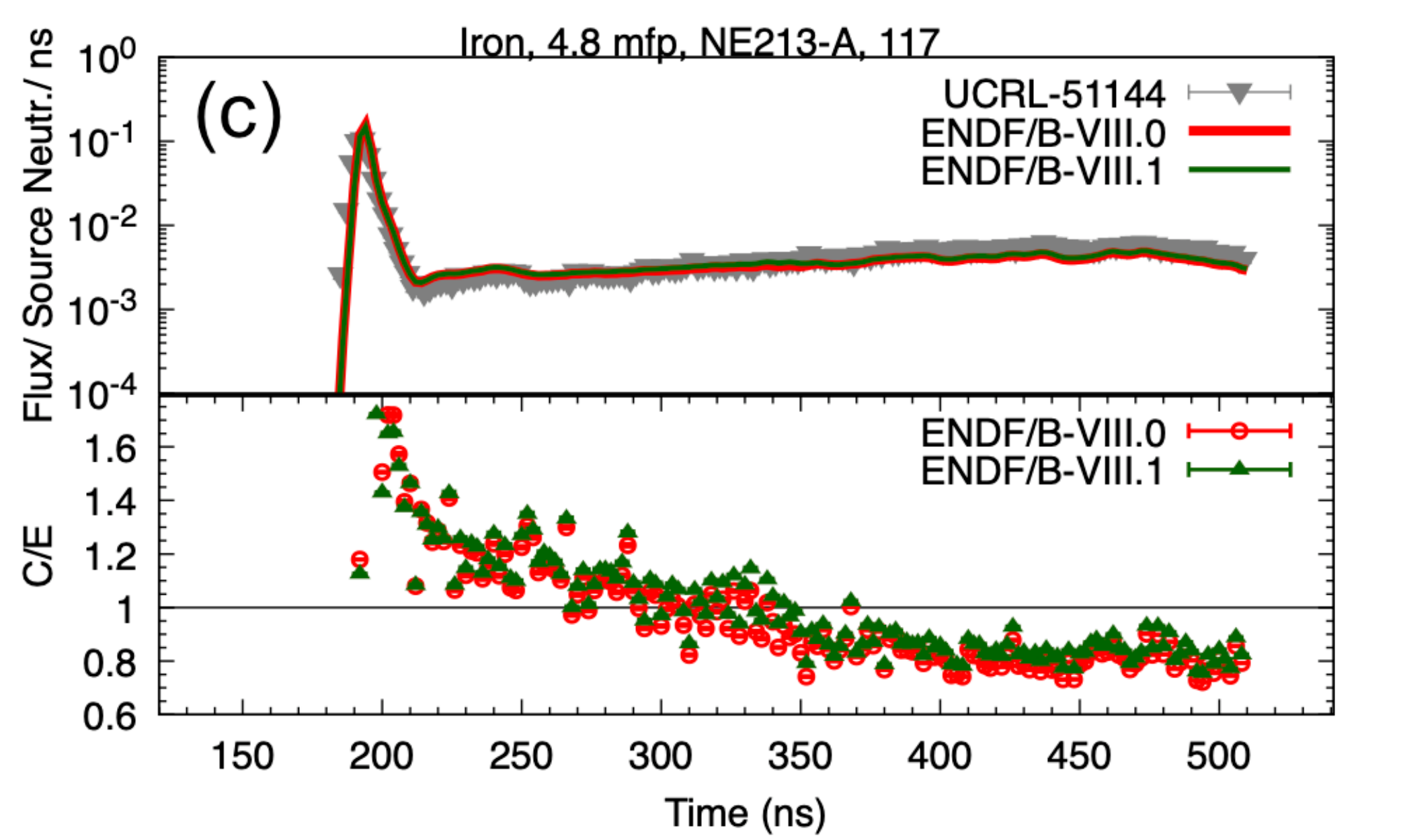}
\includegraphics[width=0.45\textwidth]{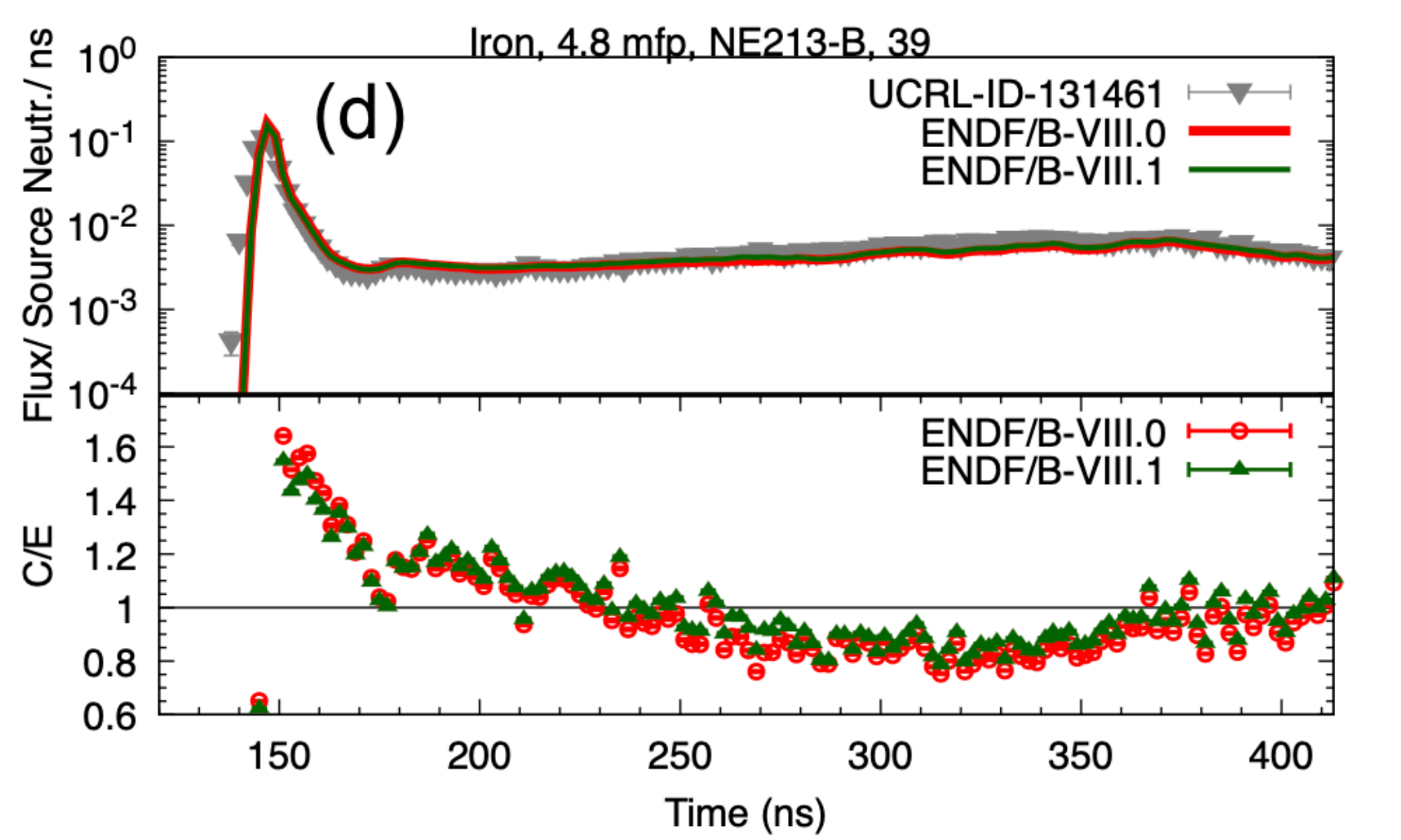}
\caption{LLNL pulsed-sphere neutron-leakage spectra simulated with ENDF/B-VIII.0 and ENDF/B-VIII.1 for iron compared to experimental data.}
\label{fig:LLNLpulsedspheresfe}
\end{figure}

\begin{figure}[!tph]
\vspace{-2mm}
    \centering
    \includegraphics[width=0.45\textwidth]{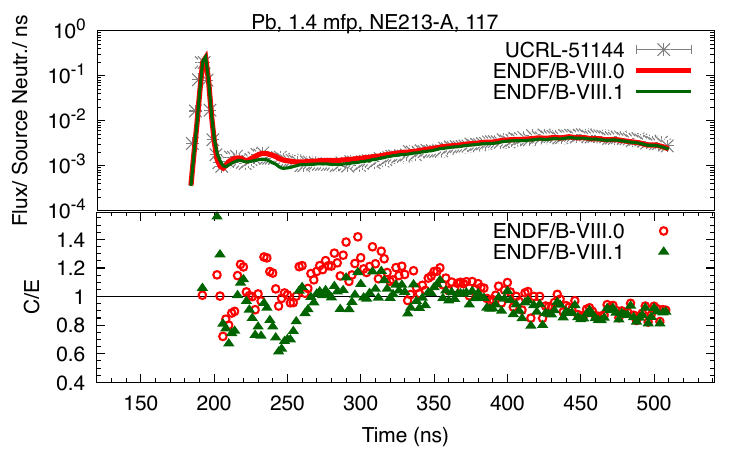}
\vspace{-2mm}
    \caption{LLNL pulsed-sphere neutron-leakage spectra simulated with ENDF/B-VIII.0 and ENDF/B-VIII.1 for Pb compared to experimental data.}
    \label{fig:LLNLpulsedspherespb}
    \end{figure}

\begin{figure}[!tph]
\vspace{-2mm}
\centering
\includegraphics[width=0.4\textwidth]{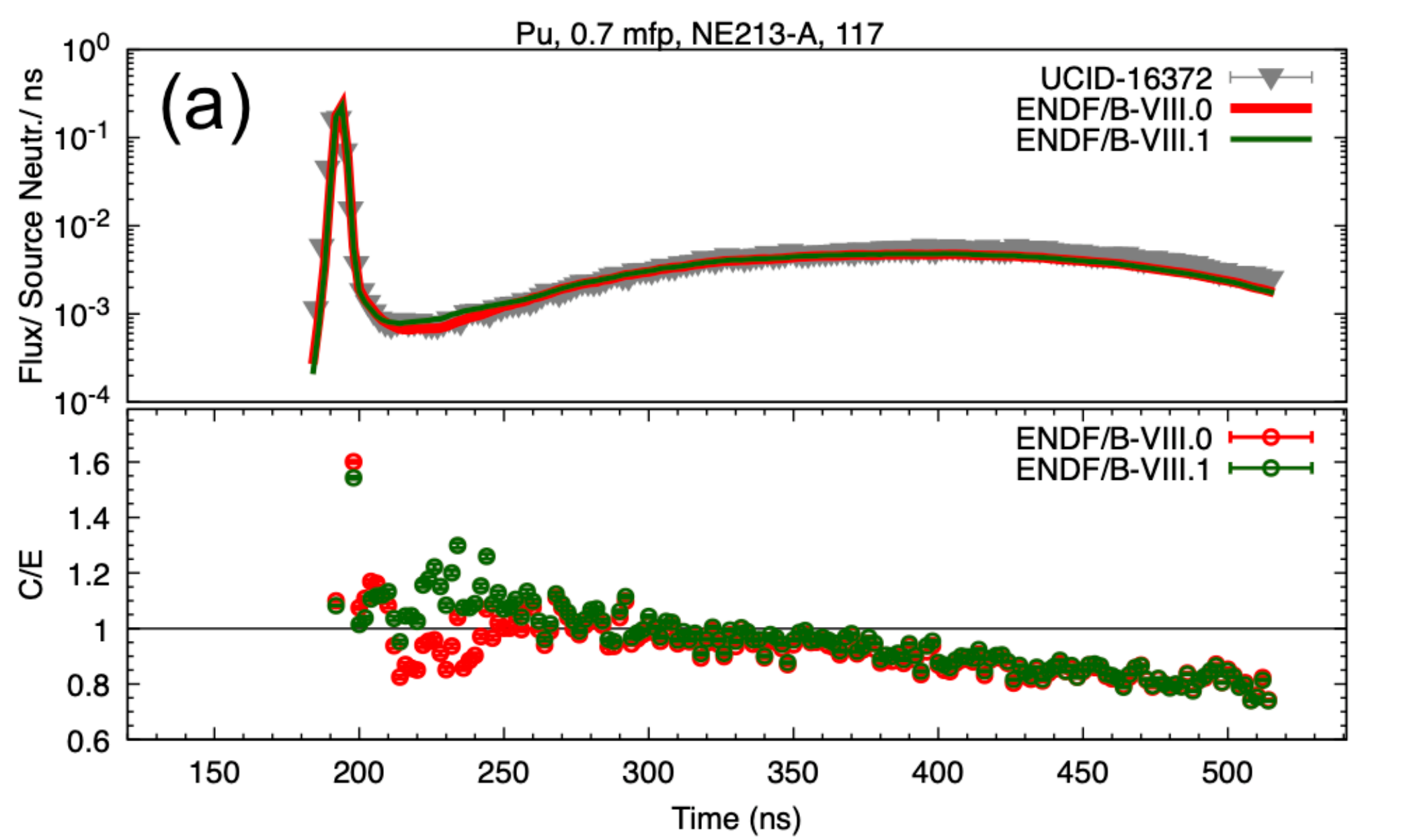}
\includegraphics[width=0.4\textwidth]{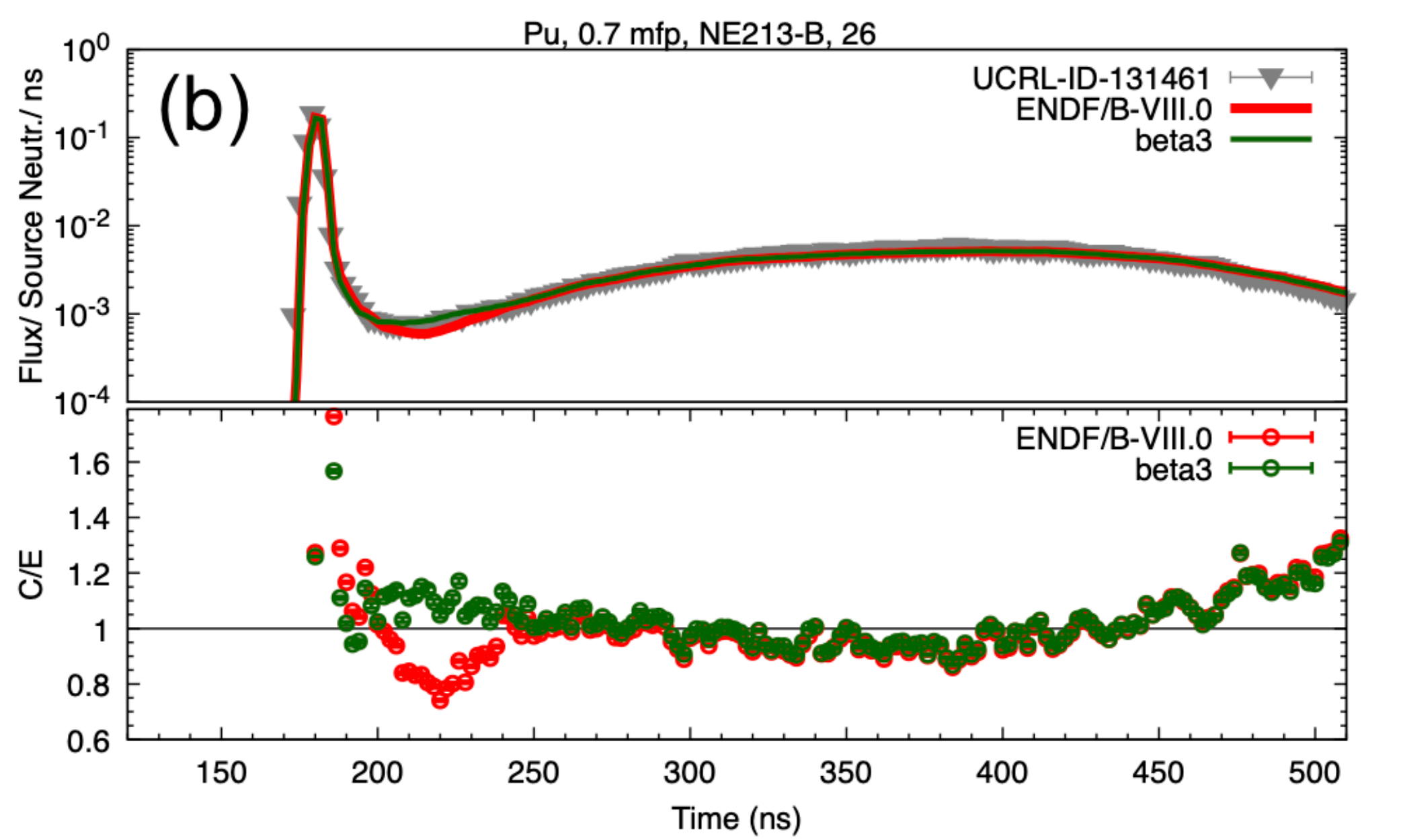}
\vspace{-2mm}
\caption{LLNL pulsed-sphere neutron-leakage spectra simulated with ENDF/B-VIII.0 and ENDF/B-VIII.1 for Pu compared to experimental data.}
\label{fig:LLNLpulsedspheresPu}
\end{figure}

\begin{figure}[!tph]
\vspace{-2mm}
\centering
\includegraphics[width=0.4\textwidth]{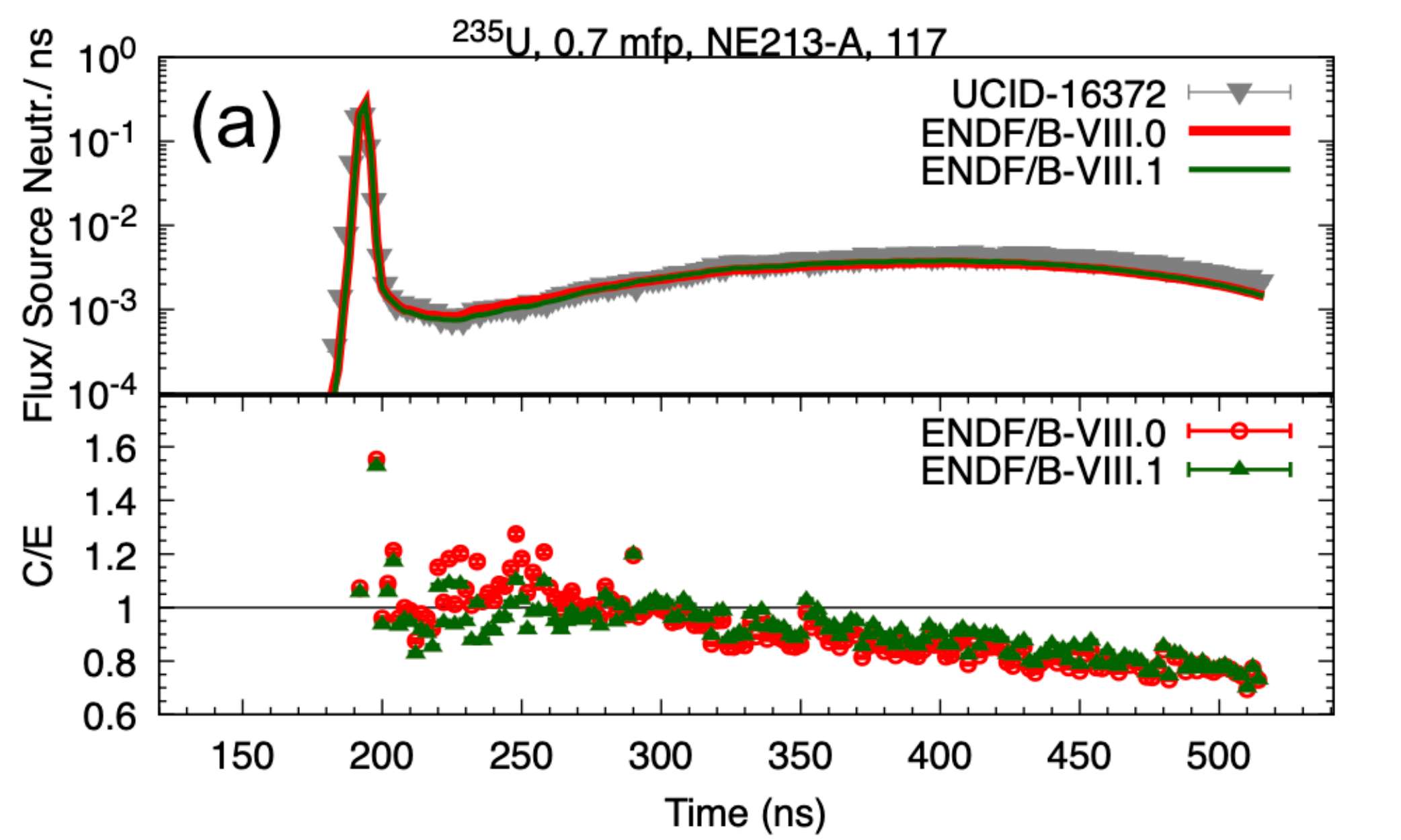}
\includegraphics[width=0.4\textwidth]{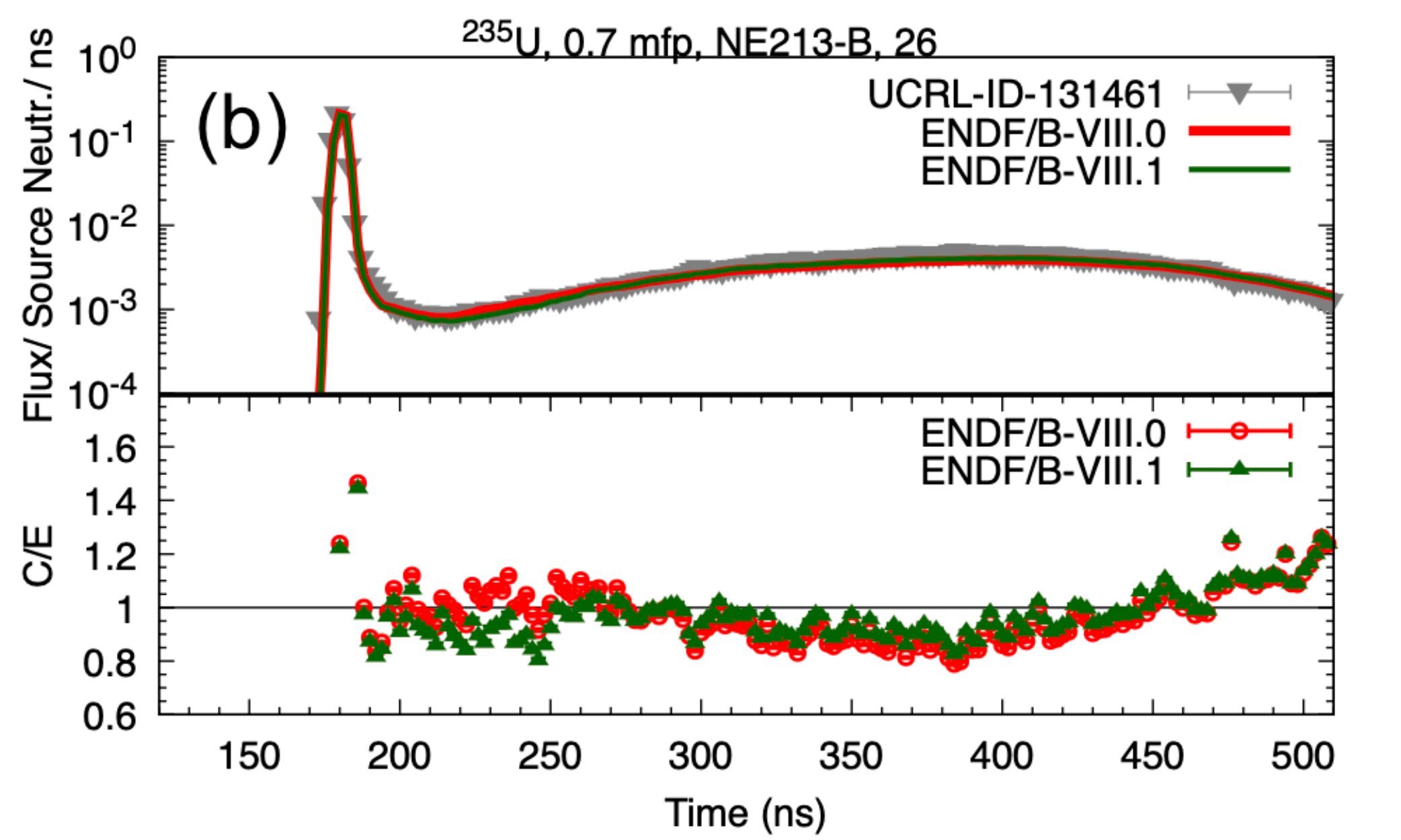}
\includegraphics[width=0.4\textwidth]{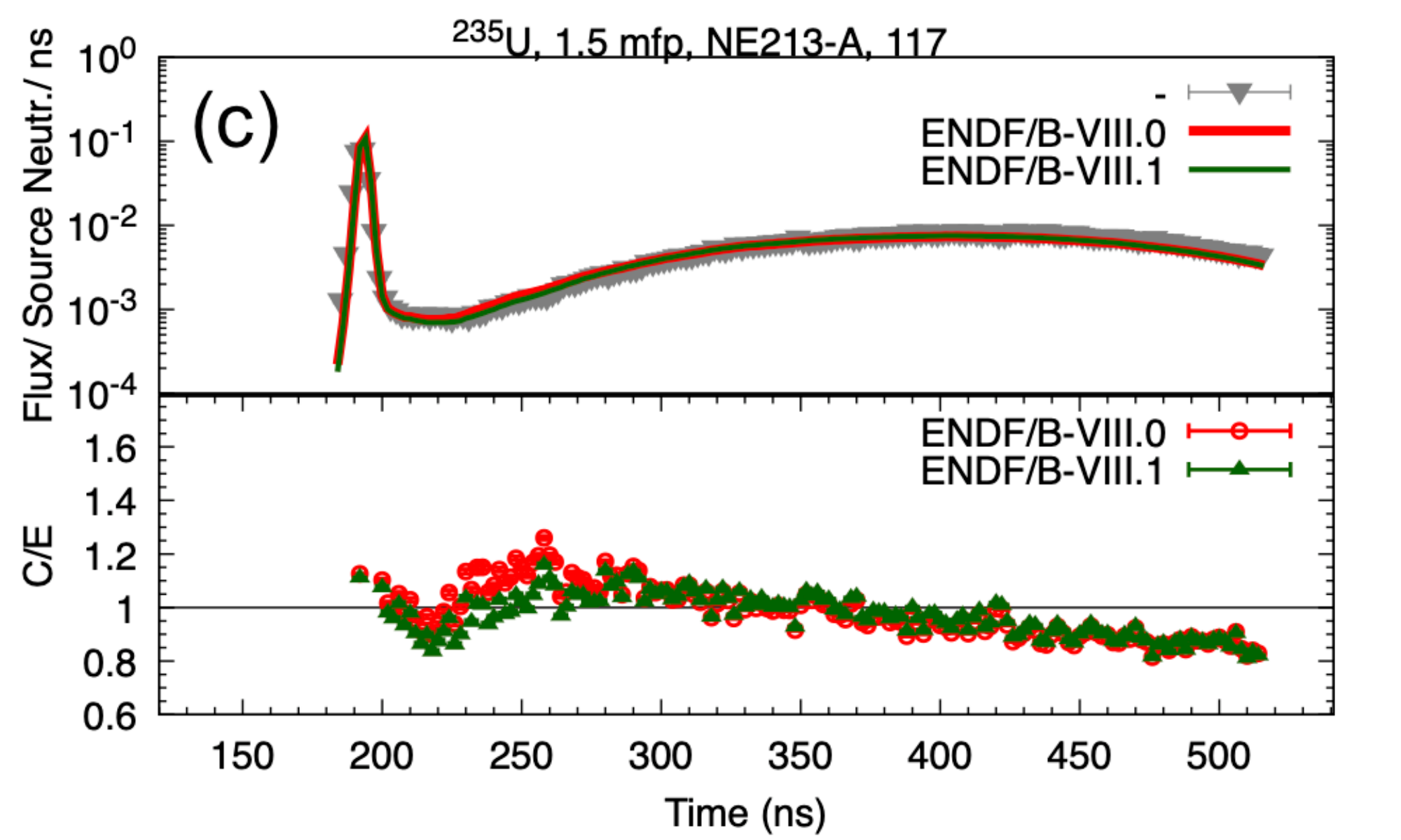}
\includegraphics[width=0.4\textwidth]{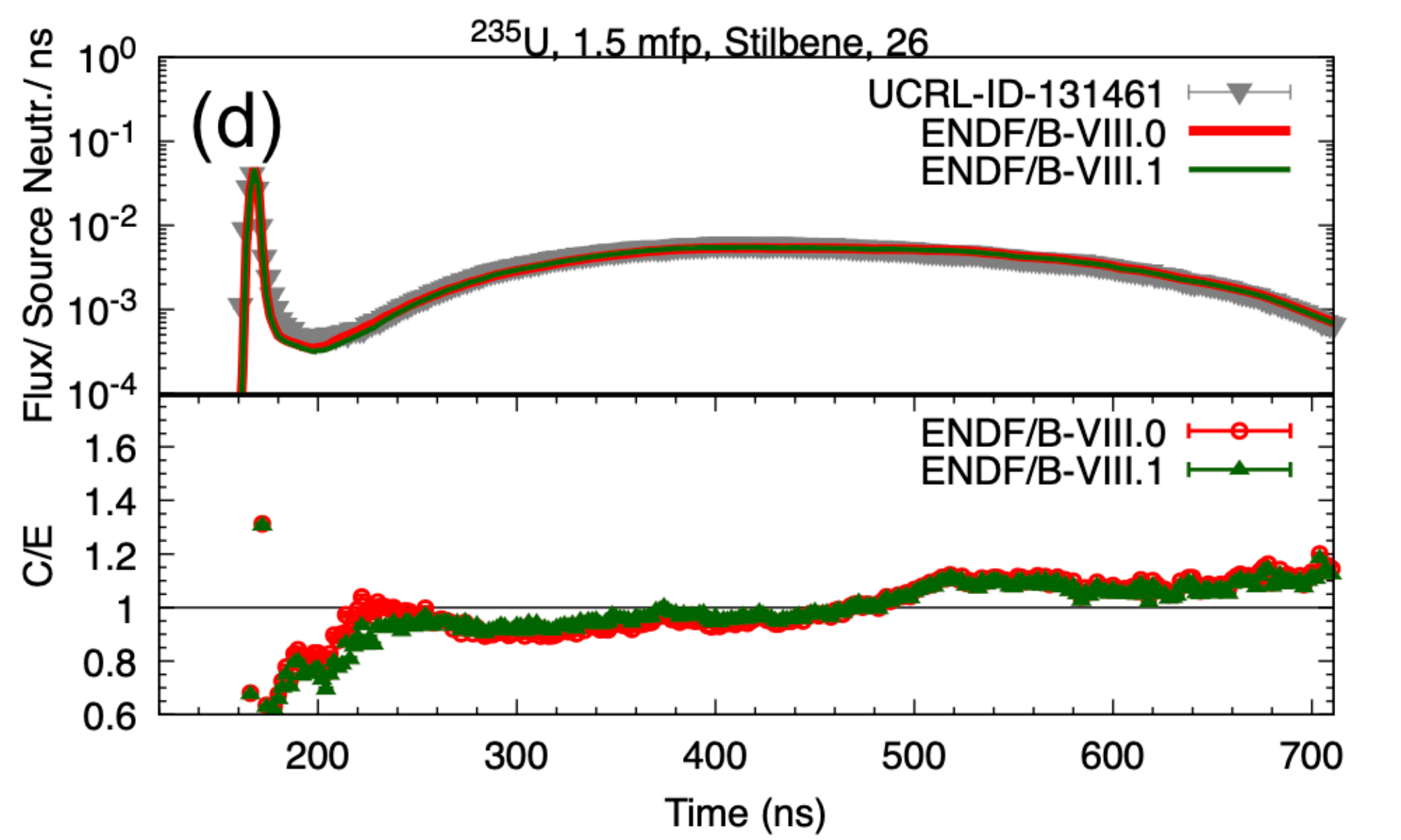}
\includegraphics[width=0.4\textwidth]{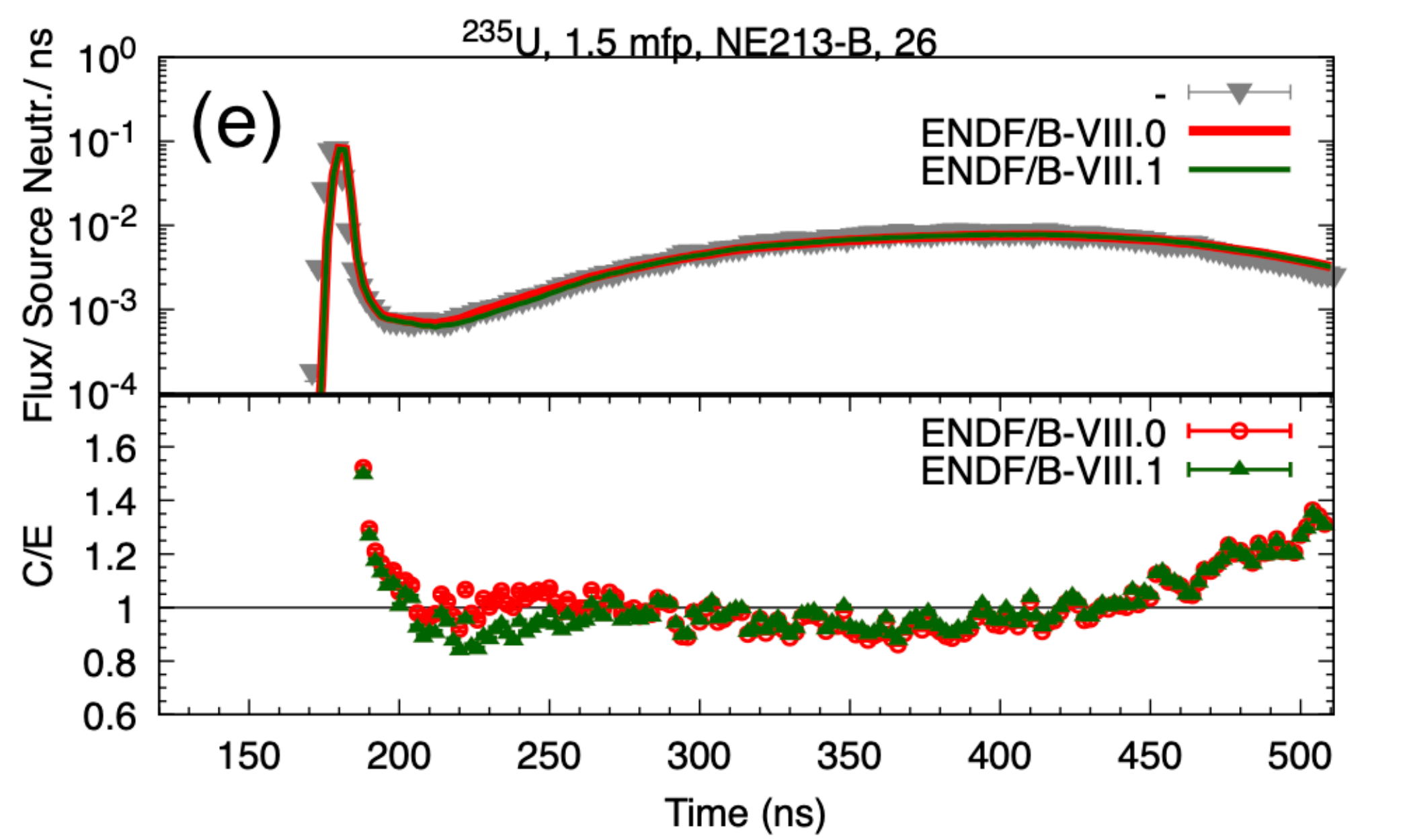}
\vspace{-2mm}
\caption{LLNL pulsed-sphere neutron-leakage spectra simulated with ENDF/B-VIII.0 and ENDF/B-VIII.1 for $^{235}$U compared to experimental data.}
\label{fig:LLNLpulsedspheres235U}
\end{figure}

There are only small changes observed in the simulations of pulsed spheres containing $^{16}$O to a non-negligible amount (light water in Fig.~\ref{fig:LLNLpulsedsphereslwt}, ``pure'' $^{16}$O and heavy water in Fig.~\ref{fig:LLNLpulsedspheresO16hwt} and concrete in Fig.~\ref{fig:LLNLpulsedspheresConcrete}) if one uses ENDF/B-VIII.0 and ENDF/B-VIII.1 nuclear data.
If one would like to improve agreement of simulations right after the peak, one could investigate MF=4, MT=2, and MF=$\{3,4\}$, MT=52, as these $^{16}$O observables matter most in the simulations of the spheres shown.
Slightly larger changes are observed between 
ENDF/B-VIII.0 and ENDF/B-VIII.1 in concrete spheres (Fig.~\ref{fig:LLNLpulsedspheresConcrete}) than in the $^{16}$O, light and heavy water spheres.
The concrete spheres are made up to close to 15\% of $^1$H and Si (mostly $^{28}$Si).
While $^1$H nuclear data changed in ENDF/B-VIII.0, the change's impact on light-water spheres is minimal.
Hence, the $^{28}$Si nuclear data changes can be seen in concrete-sphere spectra but are well within experimental uncertainties.
Small improvements can be observed in simulations of Teflon LLNL pulsed-sphere neutron-leakage spectra in Fig.~\ref{fig:LLNLpulsedspheresteflon} if ENDF/B-VIII.1 nuclear data are used.
The overall agreement is poor, which could be either due to carbon or $^{19}$F nuclear data given the fact that the predictions of carbon pulsed spheres is also poor.
If one would like to try to improve predictions of Teflon pulsed-sphere spectra via $^{19}$F nuclear data, one could investigate MF=4, MT=2 and MF=$\{3,4\}$, MT=51,55 nuclear data.
The updates in Mg ENDF/B-VIII.1 nuclear data compared to ENDF/B-VIII.0 led to only small changes in the simulated Mg LLNL pulsed-sphere neutron-leakage spectra (Fig.~\ref{fig:LLNLpulsedspheresmg}). 
These differences are only at late times and are within experimental uncertainties.
In general, the nuclear data describes the spectrum poorly; 
79\% of the sphere is made up of $^{28}$Mg.
One could investigate MF=4, MT=2 and MF=$\{3,4,6\}$, MT=$\{2, 51, 91\}$ nuclear data.
The updates for iron nuclear data in  ENDF/B-VIII.1  compared to ENDF/B-VIII.0 led to only small changes in the simulated iron LLNL pulsed-sphere neutron-leakage spectra. 
It can be seen in Fig.~\ref{fig:LLNLpulsedspheresfe} that the changes are a bit larger for thicker spheres than thinner ones.
The thicker the sphere, the more multiple scattering of neutrons happens, and one uses lower-energy nuclear data (4--10 MeV) for the spectra's simulations.
Hence, changes in inelastic data (MF=$\{3,4,6\}$) are likely an important contributor
to differences in simulated spectra.
If one would like to improve agreement between simulations and experiments at later times, one could investigate at MF=6, MT=91 iron nuclear data.
The new ENDF/B-VIII.1 Pb nuclear data led to a distinct change in simulated Pb LLNL pulsed-sphere spectra in Fig.~\ref{fig:LLNLpulsedspherespb} compared to ENDF/B-VIII.0.
The agreement is worse from 240--260~ns but better at later times. 
Future research could study elastic, inelastic and (n,2) for cross sections and angular distributions.
Large changes can be observed for $^{239}$Pu LLNL pulsed-sphere neutron-leakage spectra in Fig.~\ref{fig:LLNLpulsedspheresPu}.
The main difference between the ENDF/B-VIII.0 and ENDF/B-VIII.1 simulation is caused by (n,inl) and (n,el) cross sections and angular distributions.
The $^{235}$U LLNL pulsed-sphere neutron-leakage spectra in Fig.~\ref{fig:LLNLpulsedspheres235U} are noticeably better described with ENDF/B-VIII.1 data than ENDF/B-VIII.0 in the valley after the elastic peak.
This improved agreement was attributed to the new $^{235}$U PFNS evaluation that changed significantly from 12--15 MeV due to inclusion of high-precision Chi-Nu experimental data~\cite{Neudecker:2022U235PFNS}.
Negligible changes were observed between $^7$Li, $^{12,13}$C, $^{27}$Al, Ti, and $^{238}$U LLNL pulsed-sphere neutron-leakage spectra simulated with  ENDF/B-VIII.0 and ENDF/B-VIII.1$\beta$1 despite changes in the nuclear data.


%
%
%
%
%
%

\section{CONCLUSIONS \& FUTURE WORK}
\label{sec:conclusion}

The ENDF/B-VIII.1 library constitutes a substantive advance in the development of evaluated nuclear data, representing an improvement on its predecessor ENDF/B-VIII.0. Through the integration of new evaluations (informed by new experimental data, theory and evaluation methods), adoption of international collaborative efforts, and implementation of improved validation and peer-review protocols, this library provides enhanced fidelity and robustness for applications across the nuclear sciences.  Throughout we strove to adhere to these key principles:
\begin{itemize}
\item{Do no harm}
\item{Be aware of historical and previous evaluation decisions}
\item{Ensure proposed changes are demonstrably better}
\item{Anticipate potential problems and conduct validation testing}
\end{itemize}

The main achievements of ENDF/B-VIII.1 may be summarized as follows:

\begin{itemize}

\item Extensive updates to important actinide and structural material evaluations, notably isotopes of uranium, plutonium, fluorine, silicon, chromium, iron, copper, tantalum, and lead.
\item Incorporation of high-quality evaluations from international collaborations, such as INDEN and IRDFF-II, which significantly improve neutron dosimetry standards and reactor benchmark performance.
\item Introduction of new and reevaluated thermal-neutron scattering kernels, thereby broadening the coverage and accuracy of low-energy neutron interaction data.
\item Resolution of deficiencies in high burnup reactor applications through improved resonance region data for key actinides, thereby enhancing the performance of nuclear data libraries for power reactor simulations.
\item Adoption of modern covariance methodologies and improved uncertainty quantification standards, providing more rigorous support for uncertainty propagation in nuclear applications.
\item Implementation of a fully online, transparent peer-review and validation workflow, ensuring reproducibility, traceability, and community-wide engagement in the evaluation process.
\item Library release in the ENDF-6 and GNDS format.
\item Marked improvement in the agreement with fundamental cross-section data, together with demonstrably improved predictive performance for integral critical assembly experimental benchmarks.

\end{itemize}

Collectively, these advancements underscore the technical maturity and expanded scope of the ENDF/B-VIII.1 release. By improving both differential cross-section fidelity and integral benchmark predictive capability, the library is a cornerstone for modern nuclear data applications. This release not only sustains confidence in current engineering and scientific uses but also establishes a robust foundation for addressing the demands of next-generation applications.

ENDF/B-VIII.1 is most likely the last release of the ENDF/B-VIII series as the next ENDF/B release is expected to contain a new Nuclear Data Standards set of evaluations.  ENDF/B-VIII.1 is also expected to be the last ENDF/B library to be released primarily in the legacy ENDF-6 format as ENDF/B-IX.0 is expected to be released in the GNDS format.  

In addition to these major changes, we can expect the usual battery of smaller changes, focusing on not just library quality improvements, but on adding features addressing emerging nuclear data needs.  These needs often are hard to predict five to six years before another release, but we do expect new generation of nuclear reactors, SMR, fusion and space physics to drive new evaluation efforts.  We also hope that the increasing reliance on AI/ML techniques will improve the quality and reach of the library.  We also expect these techniques to enable new ways to prioritize evaluations and identify new work that can deliver the greatest impacts to users.


\section*{Acknowledgments}

Work at Brookhaven National Laboratory was sponsored by the Office of Nuclear Physics, Office of Science of the U.S. Department of Energy under Contract No. DE-SC0012704 with Brookhaven Science Associates, LLC.
Work at Lawrence Livermore National Laboratory was performed under Contract DE-AC52-07NA27344.
Work at Los Alamos National Laboratory, operated by Triad National Security, LLC, was carried out under the auspices of the National Nuclear Security Administration of the U.S. Department of Energy under Contract No. 89233218CNA000001.
Work at Oak Ridge National Laboratory was authored by UT-Battelle, LLC under Contract No. DE-AC05-00OR22725 with the U.S. Department of Energy.
Work at Argonne National Laboratory was supported by the U.S. Department of Energy, Office of Science, under Contract DE-AC02-06CH11357.
Work at Naval Nuclear Laboratory, operated by Fluor Marine Propulsion, LLC, was performed under Contract No. 89233018CNR000004 with the U.S. Department of Energy.
This work was supported by the Naval Nuclear Propulsion Program and Nuclear Criticality Safety Program, funded and managed by the National Nuclear Security Administration for the U.S. Department of Energy.
This work received funding support from the National Nuclear Security Administration, Office of Defense Nuclear Nonproliferation R\&D.

The U.S. Government retains right to use the published form of this manuscript, or allow others to do so, for U. S. Government purposes.




\bibliography{endf-VIII-1}


\clearpage
\appendix

\renewcommand{\arraystretch}{0.5}

\section{Provenance of evaluations in \ENDF}

To facilitate future reference, we document the provenance and update information of each individual file in the \ENDF\ release for the following sublibraries: neutrons (Table~\ref{tab:appendix:provenance_neutrons}), thermal-neutron scattering law (Table~\ref{tab:appendix:provenance_tsl}), photonuclear (Table~\ref{tab:appendix:provenance_gammas}), 
neutron-induced fission yields (Table~\ref{tab:appendix:provenance_nfy}), 
spontaneous fission yields (Table~\ref{tab:appendix:provenance_sfy}), 
protons (Table~\ref{tab:appendix:provenance_protons}), deuterons (Table~\ref{tab:appendix:provenance_deuterons}), tritons (Table~\ref{tab:appendix:provenance_tritons}), \nuc{3}{He} (Table~\ref{tab:appendix:provenance_helions}), and alphas (Table~\ref{tab:appendix:provenance_alphas}).  

As indicated in the main text, both the neutron-induced and spontaneous fission-yield libraries were updated.  The nature of the fission yield changes imply that {\em every} fission yield evaluation was  impacted as reflected in Tables~\ref{tab:appendix:provenance_nfy} and~\ref{tab:appendix:provenance_sfy}.



\begin{longtable*}{@{}>{\scriptsize}l>{\scriptsize}c>{\scriptsize}c>{\scriptsize}c>{\scriptsize}c>{\scriptsize}c>{\scriptsize}c>{\scriptsize}c>{\scriptsize}c>{\scriptsize}c>{\scriptsize}c>{\scriptsize}c@{}}
\caption{\footnotesize Summary of the origins of and recent changes to the neutrons 
sublibrary. For each file we list the library source, the year when the last significant evaluation effort or impactful modification was performed, and the evaluation source and year for each incident-neutron energy range: Resolved Resonance Range (RRR), Unresolved Resonance Range (URR), fast-neutron range below 20~MeV (Fast), and High Energy above 20~MeV (HE). We also provide additional comments, where relevant.  Evaluations updated in  \ENDF\ are in bold.}
\label{tab:appendix:provenance_neutrons}\\
\toprule
\toprule
     \multirow{3}{*}{File name (.endf)}         & \multirow{3}{*}{\centering Source}   &  \multirow{3}{*}{\parbox{0.88cm}{\centering Last \\ updated}}    & \multicolumn{2}{c}{\scriptsize RRR}       & \multicolumn{2}{c}{\scriptsize URR}   & \multicolumn{2}{c}{\scriptsize Fast} & \multicolumn{2}{c}{\scriptsize HE}   &  \multirow{3}{*}{\centering Comments}       \\
\cmidrule(lr){4-5}\cmidrule(lr){6-7}\cmidrule(lr){8-9}\cmidrule(lr){10-11}
                                   &                                               &           &     Inst.        &      Year        &     Inst.        &      Year    &   Inst.        &      Year     &      Inst.        &      Year  &                                           \\
\midrule
\endfirsthead
\caption{\footnotesize (continued) Provenance for the neutrons 
sublibrary.}\\
\toprule
     \multirow{3}{*}{File name (.endf)}         & \multirow{3}{*}{\centering Source}   &  \multirow{3}{*}{\parbox{0.88cm}{\centering Last \\ updated}}    & \multicolumn{2}{c}{\scriptsize RRR}       & \multicolumn{2}{c}{\scriptsize URR}   & \multicolumn{2}{c}{\scriptsize Fast} & \multicolumn{2}{c}{\scriptsize HE}   &  \multirow{3}{*}{\centering Comments}       \\
\cmidrule(lr){4-5}\cmidrule(lr){6-7}\cmidrule(lr){8-9}\cmidrule(lr){10-11}
                                   &                                               &           &     Inst.        &      Year        &     Inst.        &      Year    &   Inst.        &      Year     &      Inst.        &      Year  &                                           \\
\midrule
\endhead
\bottomrule\addlinespace[\belowrulesep]
\multicolumn{12}{r}{{Continued on the next page}}\\
\endfoot
\bottomrule \bottomrule
\endlastfoot
 n-000{\_}n{\_}001           & ENDF/B-VIII.0                                                                            & 2024          &                                                             &               &                                                   &               & LANL                                                          & 2016          &                                                              &               &                                                                                                                                                                         \\
\midrule                                                                                                                                                                                                                                                                                                                                                                                                                                 
n-001{\_}H{\_}001            & \parbox[t]{2.4cm}{\centering ENDF/B-VIII.0; CIELO 2017; Standards 2017 \vspace{1mm}}     & 2023          &                                                             &               &                                                   &               & LANL                                                          & 2016          &                                                              &               & \parbox[t]{4.3cm}{R-matrix analysis; COMMARA-2.0 covariances added in 2011 }                                                                                                              \\
\end{longtable*}


\begin{longtable*}{@{}>{\scriptsize}l>{\scriptsize}c>{\scriptsize}c>{\scriptsize}c>{\scriptsize}c}
\caption{\footnotesize Summary of the origins of and recent changes to the thermal-neutron scattering law 
sublibrary. 
For each file we list the library source, the year when the last significant evaluation effort or impactful modification was performed, the institution responsible for its most recent evaluation and the corresponding year said evaluation was completed. 
Evaluations updated in  \ENDF\ are in bold.}
\label{tab:appendix:provenance_tsl}\\
\toprule
\toprule
     \multirow{2}{*}{File name (.endf)}         & \multirow{2}{*}{Source}        & Last          & \multirow{2}{*}{Institution}   & \multirow{2}{*}{Year}                       \\
                                             &                                             &  updated   &                                            &                                                                                           \\
\midrule
\endfirsthead
\caption{\footnotesize (continued) Provenance for the thermal-neutron scattering law 
sublibrary.}\\
\toprule
     \multirow{2}{*}{File name (.endf)}         & \multirow{2}{*}{Source}        & Last          & \multirow{2}{*}{Institution}   & \multirow{2}{*}{Year}                                      \\
                                             &                                             &  updated   &                                            &                                                                                               \\
\midrule
\endhead
\bottomrule\addlinespace[\belowrulesep]
\multicolumn{5}{r}{{Continued on the next page}}\\
\endfoot
\bottomrule \bottomrule
\endlastfoot
 tsl-013\_Al\_027                    & ENDF/B-VII.0           & 2024          & LANL              & 2005         \\
\midrule 
 \textbf{tsl-AlinAl2O3}            & \textbf{ENDF/B-VIII.1} & \textbf{2024} & \textbf{LEIP}     & \textbf{2020}  \\
 \textbf{tsl-OinAl2O3}             & \textbf{ENDF/B-VIII.1} & \textbf{2023} & \textbf{LEIP}     & \textbf{2020}  \\
\midrule 
 \textbf{tsl-Be-metal}             & \textbf{ENDF/B-VIII.1} & \textbf{2024} & \textbf{LEIP}     & \textbf{2021}  \\
 \textbf{tsl-Be-metal+Sd}          & \textbf{ENDF/B-VIII.1} & \textbf{2024} & \textbf{LEIP}     & \textbf{2021}  \\
\midrule 
 \textbf{tsl-BeinBe2C}             & \textbf{ENDF/B-VIII.1} & \textbf{2023} & \textbf{NNL}      & \textbf{2023}  \\
 \textbf{tsl-CinBe2C}              & \textbf{ENDF/B-VIII.1} & \textbf{2023} & \textbf{NNL}      & \textbf{2023}  \\
\midrule 
 \textbf{tsl-BeinBeF2}             & \textbf{ENDF/B-VIII.1} & \textbf{2023} & \textbf{UoS}      & \textbf{2021}  \\
\end{longtable*}



\begin{longtable*}{@{}>{\scriptsize}l>{\scriptsize}c>{\scriptsize}c>{\scriptsize}c>{\scriptsize}c>{\scriptsize}c@{}}
\caption{\footnotesize Summary of the origins of and recent changes to the photonuclear 
sublibrary. For each file we list the library source, the year when the last significant evaluation effort or impactful modification was performed, the institution responsible for its most recent evaluation and the corresponding year said evaluation was completed. We also provide additional comments, where relevant.  Evaluations updated in  \ENDF\ are in bold.}
\label{tab:appendix:provenance_gammas}\\
\toprule
\toprule
     \multirow{2}{*}{File name (.endf)}         & \multirow{2}{*}{Source}        & Last          & \multirow{2}{*}{Institution}   & \multirow{2}{*}{Year}       & \multirow{2}{*}{Comments}                                  \\
                                             &                                             &  updated   &                                            &                                         &                                                                             \\
\midrule
\endfirsthead
\caption{\footnotesize (continued) Provenance for the photonuclear 
sublibrary.}\\
\toprule
     \multirow{2}{*}{File name (.endf)}         & \multirow{2}{*}{Source}        & Last          & \multirow{2}{*}{Institution}   & \multirow{2}{*}{Year}       & \multirow{2}{*}{Comments}                                  \\
                                             &                                             &  updated   &                                            &                                         &                                                                             \\
\midrule
\endhead
\bottomrule\addlinespace[\belowrulesep]
\multicolumn{6}{r}{{Continued on the next page}}\\
\endfoot
\bottomrule \bottomrule
\endlastfoot
 g-001\_H\_002           & ENDF/B-VII.0                   & 2023          & LANL               & 2005          & \parbox[t]{6.0cm}{2019 IAEA Photonuclear CRP                                              }\\
\midrule
 \textbf{g-002\_He\_003} & \textbf{JENDL-5}               & \textbf{2023} & \textbf{JAEA}      & \textbf{2003} & \parbox[t]{6.0cm}{\textbf{2019 IAEA Photonuclear CRP}                                     }\\
\midrule
 \textbf{g-003\_Li\_006} & \textbf{JENDL-5}               & \textbf{2023} & \textbf{JAEA}      & \textbf{2013} & \parbox[t]{6.0cm}{\textbf{2019 IAEA Photonuclear CRP}                                     }\\
 \textbf{g-003\_Li\_007} & \textbf{JENDL-5}               & \textbf{2023} & \textbf{JAEA}      & \textbf{2013} & \parbox[t]{6.0cm}{\textbf{2019 IAEA Photonuclear CRP}                                     }\\
\midrule
 \textbf{g-004\_Be\_009} & \textbf{CENDL-3}               & \textbf{2023} & \textbf{CNDC}      & \textbf{2018} & \parbox[t]{6.0cm}{\textbf{2019 IAEA Photonuclear CRP}                                     }\\
\midrule
 g-006\_C\_012           & LANL150u                       & 2023          & LANL               & 1999          & \parbox[t]{6.0cm}{2019 IAEA Photonuclear CRP                                              }\\
 g-006\_C\_013           & KAERI/G-1                      & 2023          & KAERI              & 1999          & \parbox[t]{6.0cm}{2019 IAEA Photonuclear CRP                                              }\\
 \textbf{g-006\_C\_014}  & \textbf{TENDL2012}             & \textbf{2023} & \textbf{KAERI/NRG} & \textbf{2019} & \parbox[t]{6.0cm}{\textbf{2019 IAEA Photonuclear CRP}                                     }\\
\end{longtable*}


\begin{longtable*}{@{}>{\scriptsize}l>{\scriptsize}c>{\scriptsize}c>{\scriptsize}c>{\scriptsize}c>{\scriptsize}l@{}}
\caption{\footnotesize Summary of the origins of and recent changes to the neutron-induced fission yields 
sublibrary. For each file we list the library source, the year when the last significant evaluation effort or impactful modification was performed, the institution responsible for its most recent evaluation and the corresponding year said evaluation was completed. We also provide additional comments, where relevant.  Evaluations updated in  \ENDF\ are in bold.} \label{tab:appendix:provenance_nfy} \\
\toprule \toprule
     \multirow{2}{*}{File name (.endf)}         & \multirow{2}{*}{Source}        & Last          & \multirow{2}{*}{Institution}   & \multirow{2}{*}{Year}       & \multirow{2}{*}{Comments}                                  \\
                                             &                                             &  updated   &                                            &                                         &                                                                             \\
\midrule
\endfirsthead
\caption{\footnotesize (continued) Provenance for the  neutron-induced fission yields 
sublibrary.}\\
\toprule
     \multirow{2}{*}{File name (.endf)}         & \multirow{2}{*}{Source}        & Last          & \multirow{2}{*}{Institution}   & \multirow{2}{*}{Year}       & \multirow{2}{*}{Comments}                                  \\
                                             &                                             &  updated   &                                            &                                         &                                                                             \\
\midrule
\endhead
\bottomrule\addlinespace[\belowrulesep]
\multicolumn{6}{r}{{Continued on the next page}}\\
\endfoot
\bottomrule \bottomrule
\endlastfoot
 \textbf{nfy-090\_Th\_227}   & \textbf{ENDF/B-VI.0}  & \textbf{2023} & \textbf{LANL}  & \textbf{1989} & \textbf{England Rider evaluation (ENDF-349); Corrected uncertainties} \\
 \textbf{nfy-090\_Th\_229}   & \textbf{ENDF/B-VI.0}  & \textbf{2023} & \textbf{LANL}  & \textbf{1989} & \textbf{England Rider evaluation (ENDF-349); Corrected uncertainties} \\
 \textbf{nfy-090\_Th\_232}   & \textbf{ENDF/B-VI.0}  & \textbf{2023} & \textbf{LANL}  & \textbf{1989} & \textbf{England Rider evaluation (ENDF-349); Corrected uncertainties} \\
\midrule
 \textbf{nfy-091\_Pa\_231}   & \textbf{ENDF/B-VI.0}  & \textbf{2023} & \textbf{LANL}  & \textbf{1989} & \textbf{England Rider evaluation (ENDF-349); Corrected uncertainties} \\
\midrule
 \textbf{nfy-092\_U\_232}    & \textbf{ENDF/B-VI.0}  & \textbf{2023} & \textbf{LANL}  & \textbf{1989} & \textbf{England Rider evaluation (ENDF-349); Corrected uncertainties} \\
 \textbf{nfy-092\_U\_233}    & \textbf{ENDF/B-VI.0}  & \textbf{2023} & \textbf{LANL}  & \textbf{1989} & \textbf{England Rider evaluation (ENDF-349); Corrected uncertainties} \\
 \textbf{nfy-092\_U\_234}    & \textbf{ENDF/B-VI.0}  & \textbf{2023} & \textbf{LANL}  & \textbf{1989} & \textbf{England Rider evaluation (ENDF-349); Corrected uncertainties} \\
 \textbf{nfy-092\_U\_235}    & \textbf{ENDF/B-VI.0}  & \textbf{2023} & \textbf{LANL}  & \textbf{1989} & \textbf{England Rider evaluation (ENDF-349); Corrected uncertainties} \\
\end{longtable*}

\begin{longtable*}{@{}>{\scriptsize}l>{\scriptsize}c>{\scriptsize}c>{\scriptsize}c>{\scriptsize}c>{\scriptsize}l@{}}
\caption{\footnotesize Summary of the origins of and recent changes to the spontaneous fission yields 
sublibrary. For each file we list the library source, the year when the last significant evaluation effort or impactful modification was performed, the institution responsible for its most recent evaluation and the corresponding year said evaluation was completed. We also provide additional comments, where relevant.  Evaluations updated in  \ENDF\ are in bold.} \label{tab:appendix:provenance_sfy}\\
\toprule \toprule
     \multirow{2}{*}{File name (.endf)}         & \multirow{2}{*}{Source}        & Last          & \multirow{2}{*}{Institution}   & \multirow{2}{*}{Year}       & \multirow{2}{*}{Comments}                                  \\
                                             &                                             &  updated   &                                            &                                         &                                                                             \\
\midrule
\endfirsthead
\caption{\footnotesize (continued) Provenance for the  spontaneous fission yields 
sublibrary.}\\
\toprule
     \multirow{2}{*}{File name (.endf)}         & \multirow{2}{*}{Source}        & Last          & \multirow{2}{*}{Institution}   & \multirow{2}{*}{Year}       & \multirow{2}{*}{Comments}                                  \\
                                             &                                             &  updated   &                                            &                                         &                                                                             \\
\midrule
\endhead
\bottomrule\addlinespace[\belowrulesep]
\multicolumn{6}{r}{{Continued on the next page}}\\
\endfoot
\bottomrule \bottomrule
\endlastfoot
 \textbf{sfy-092\_U\_238}  & \textbf{ENDF/B-VI.0} & \textbf{2023} & \textbf{LANL}  & \textbf{1989} & \textbf{England Rider evaluation (ENDF-349); Corrected uncertainties} \\
\midrule
 \textbf{sfy-096\_Cm\_244} & \textbf{ENDF/B-VI.0} & \textbf{2023} & \textbf{LANL}  & \textbf{1989} & \textbf{England Rider evaluation (ENDF-349); Corrected uncertainties} \\
 \textbf{sfy-096\_Cm\_246} & \textbf{ENDF/B-VI.0} & \textbf{2023} & \textbf{LANL}  & \textbf{1989} & \textbf{England Rider evaluation (ENDF-349); Corrected uncertainties} \\
 \textbf{sfy-096\_Cm\_248} & \textbf{ENDF/B-VI.0} & \textbf{2023} & \textbf{LANL}  & \textbf{1989} & \textbf{England Rider evaluation (ENDF-349); Corrected uncertainties} \\
\midrule
 \textbf{sfy-098\_Cf\_250} & \textbf{ENDF/B-VI.0} & \textbf{2023} & \textbf{LANL}  & \textbf{1989} & \textbf{England Rider evaluation (ENDF-349); Corrected uncertainties} \\
 \textbf{sfy-098\_Cf\_252} & \textbf{ENDF/B-VI.0} & \textbf{2023} & \textbf{LANL}  & \textbf{1989} & \textbf{England Rider evaluation (ENDF-349); Corrected uncertainties} \\
\midrule
 \textbf{sfy-099\_Es\_253} & \textbf{ENDF/B-VI.0} & \textbf{2023} & \textbf{LANL}  & \textbf{1989} & \textbf{England Rider evaluation (ENDF-349); Corrected uncertainties} \\
\midrule
 \textbf{sfy-100\_Fm\_254} & \textbf{ENDF/B-VI.0} & \textbf{2023} & \textbf{LANL}  & \textbf{1989} & \textbf{England Rider evaluation (ENDF-349); Corrected uncertainties} \\
\end{longtable*}



\begin{longtable*}{@{}>{\scriptsize}l>{\scriptsize}c>{\scriptsize}c>{\scriptsize}c>{\scriptsize}c>{\scriptsize}l@{}}
\caption{\footnotesize Summary of the origins of and recent changes to the protons 
sublibrary. For each file we list the library source, the year when the last significant evaluation effort or impactful modification was performed, the institution responsible for its most recent evaluation and the corresponding year said evaluation was completed. We also provide additional comments, where relevant.  Evaluations updated in  \ENDF\ are in bold.} \label{tab:appendix:provenance_protons}\\
\toprule \toprule
     \multirow{2}{*}{File name (.endf)}         & \multirow{2}{*}{Source}        & Last          & \multirow{2}{*}{Institution}   & \multirow{2}{*}{Year}       & \multirow{2}{*}{Comments}                                  \\
                                             &                                             &  updated   &                                            &                                         &                                                                             \\
\midrule
\endfirsthead
\caption{\footnotesize (continued) Provenance for the  protons 
sublibrary.}\\
\toprule
     \multirow{2}{*}{File name (.endf)}         & \multirow{2}{*}{Source}        & Last          & \multirow{2}{*}{Institution}   & \multirow{2}{*}{Year}       & \multirow{2}{*}{Comments}                                  \\
                                             &                                             &  updated   &                                            &                                         &                                                                             \\
\midrule
\endhead
\bottomrule\addlinespace[\belowrulesep]
\multicolumn{6}{r}{{Continued on the next page}}\\
\endfoot
\bottomrule \bottomrule
\endlastfoot
 p-001\_H\_001           & ENDF/B-VI.6           & 2006          & LANL           & 1998          & R-matrix analysis                              \\
 p-001\_H\_002           & ENDF/B-VI.6           & 2006          & LANL/ECN       & 1997          & R-matrix analysis; primary gamma fix           \\
 p-001\_H\_003           & ENDF/B-VII.0          & 2006          & LANL           & 2001          & R-matrix analysis                              \\
\midrule 
 p-002\_He\_003          & ENDF/B-VII.0          & 2006          & LANL           & 1983          & R-matrix analysis                              \\
 \textbf{p-002\_He\_004} & \textbf{ECPL}         & \textbf{2024} & \textbf{LLNL}  & \textbf{1999} & \textbf{Minor corrections}                     \\
\midrule 
 p-003\_Li\_006          & ENDF/B-VII.0          & 2006          & LANL           & 2001          & R-matrix analysis                              \\
 p-003\_Li\_007          & ENDF/B-VII.0          & 2016          & LANL/LLNL      & 2010          & R-matrix analysis; modifications from ENDL2009 \\
\midrule 
 p-004\_Be\_009          & ENDF/B-VI.0           & 2006          & LANL           & 1988          & LA-11753-MS                                    \\
\end{longtable*}

\begin{table*}
\begin{center}
\caption{\footnotesize Summary of the origins of and recent changes to the deuterons 
sublibrary. For each file we list the library source, the year when the last significant evaluation effort or impactful modification was performed, the institution responsible for its most recent evaluation and the corresponding year said evaluation was completed. We also provide additional comments, where relevant.  Evaluations updated in  \ENDF\ are in bold.}  \label{tab:appendix:provenance_deuterons}
\begin{scriptsize}
\begin{tabular}{lccccl}
\toprule \toprule
     \multirow{2}{*}{File name (.endf)}         & \multirow{2}{*}{Source}        & Last          & \multirow{2}{*}{Institution}   & \multirow{2}{*}{Year}       & \multirow{2}{*}{Comments}                                  \\
                                             &                                             &  updated   &                                            &                                         &                                                                             \\
\midrule
 d-001\_H\_002           & ENDF/B-VII.0           & 2024          & LLNL/LANL      & 1992          & Updated ECPL               \\
 \textbf{d-001\_H\_003}  & \textbf{ENDF/B-VIII.1} & \textbf{2023} & \textbf{LANL}  & \textbf{2023} & \textbf{R-matrix analysis} \\
\midrule 
 \textbf{d-002\_He\_003} & \textbf{ENDF/B-VIII.1} & \textbf{2024} & \textbf{LANL}  & \textbf{2020} & \textbf{R-matrix analysis} \\
 d-002\_He\_004          & ENDF/B-VII.0           & 2024          & LANL           & 2001          & R-matrix analysis          \\
\midrule 
 \textbf{d-003\_Li\_006} & \textbf{ENDF/B-VIII.1} & \textbf{2024} & \textbf{LANL}  & \textbf{2020} & \textbf{R-matrix analysis} \\
 d-003\_Li\_007          & ENDL2010               & 2024          & LLNL           & 2010          & Updated ECPL               \\
\bottomrule \bottomrule
\end{tabular}
\end{scriptsize}
\end{center}
\end{table*}

\begin{table*}
\begin{center}
\caption{\footnotesize Summary of the origins of and recent changes to the tritons 
sublibrary. For each file we list the library source, the year when the last significant evaluation effort or impactful modification was performed, the institution responsible for its most recent evaluation and the corresponding year said evaluation was completed. We also provide additional comments, where relevant.  Evaluations updated in  \ENDF\ are in bold.} \label{tab:appendix:provenance_tritons}
\begin{scriptsize}
\begin{tabular}{lccccl}
\toprule \toprule
     \multirow{2}{*}{File name (.endf)}         & \multirow{2}{*}{Source}        & Last          & \multirow{2}{*}{Institution}   & \multirow{2}{*}{Year}       & \multirow{2}{*}{Comments}                                  \\
                                             &                                             &  updated   &                                            &                                         &                                                                             \\
\midrule
 t-001\_H\_003           & ENDF/B-VII.0          & 2006          & LANL           & 2001          & R-matrix analysis          \\
\midrule 
 \textbf{t-002\_He\_003} & \textbf{ENDF/B-VII.0} & \textbf{2023} & \textbf{LANL}  & \textbf{2001} & \textbf{R-matrix analysis} \\
 \textbf{t-002\_He\_004} & \textbf{ECPL}         & \textbf{2023} & \textbf{LLNL}  & \textbf{1999} & \textbf{}                  \\
\midrule 
 \textbf{t-003\_Li\_006} & \textbf{ENDF/B-VII.0} & \textbf{2023} & \textbf{LANL}  & \textbf{2001} & \textbf{R-matrix analysis} \\
 t-003\_Li\_007          & ENDL2011              & 2016          & LLNL           & 2016          & Updated ECPL               \\
\bottomrule \bottomrule
\end{tabular}
\end{scriptsize}
\end{center}
\end{table*}

\begin{table*}
\begin{center}
\caption{\footnotesize Summary of the origins of and recent changes to the helions (\nuc{3}{He}) 
sublibrary. For each file we list the library source, the year when the last significant evaluation effort or impactful modification was performed, the institution responsible for its most recent evaluation and the corresponding year said evaluation was completed. We also provide additional comments, where relevant.  Evaluations updated in  \ENDF\ are in bold.} \label{tab:appendix:provenance_helions}
\begin{scriptsize}
\begin{tabular}{lccccl}
\toprule \toprule
     \multirow{2}{*}{File name (.endf)}         & \multirow{2}{*}{Source}        & Last          & \multirow{2}{*}{Institution}   & \multirow{2}{*}{Year}       & \multirow{2}{*}{Comments}                                  \\
                                             &                                             &  updated   &                                            &                                         &                                                                             \\
\midrule
 h-002\_He\_003          & ENDL2011                   & 2016          & LLNL/LANL           & 2010          & Updated ENDF/B-VII.0       \\
 \textbf{h-002\_He\_004} & \textbf{ENDF/B-VIII.1}     & \textbf{2024} & \textbf{LLNL/INDEN} & \textbf{2022} & \textbf{R-matrix analysis} \\
\midrule 
 \textbf{h-003\_Li\_006} & \textbf{ENDF/B-VII.0+ECPL} & \textbf{2024} & \textbf{LANL/LLNL}  & \textbf{2002} & \textbf{R-matrix analysis} \\
 \textbf{h-003\_Li\_007} & \textbf{ECPL}              & \textbf{2024} & \textbf{LLNL}       & \textbf{1985} & \textbf{}                  \\
\bottomrule \bottomrule
\end{tabular}
\end{scriptsize}
\end{center}
\end{table*}

\begin{table*}
\begin{center}
\caption{\footnotesize Summary of the origins of and recent changes to the alphas (\nuc{4}{He})   
sublibrary. For each file we list the library source, the year when the last significant evaluation effort or impactful modification was performed, the institution responsible for its most recent evaluation and the corresponding year said evaluation was completed. We also provide additional comments, where relevant.  Evaluations updated in  \ENDF\ are in bold.}\label{tab:appendix:provenance_alphas}
\begin{scriptsize}
\begin{tabular}{lccccl}
\toprule \toprule
     \multirow{2}{*}{File name (.endf)}         & \multirow{2}{*}{Source}        & Last          & \multirow{2}{*}{Institution}   & \multirow{2}{*}{Year}       & \multirow{2}{*}{Comments}                                  \\
                                             &                                             &  updated   &                                            &                                         &                                                                             \\
\midrule
 \textbf{a-002\_He\_004} & \textbf{ENDF/B-VII.1}  & \textbf{2024} & \textbf{LANL/LLNL} & \textbf{2010} & \textbf{R-matrix analysis}      \\
\midrule 
 a-003\_Li\_006          & ECPL                   & 2024          & LLNL               & 1985          & Minor revisions                 \\
\midrule 
 a-004\_Be\_009          & ENDF/B-VIII.1          & 2024          & NNL                & 2018          & Based on JENDL/AN-2005          \\
\midrule 
 a-008\_O\_017           & ENDF/B-VIII.1          & 2024          & NNL                & 2018          & Based on JENDL/AN-2005          \\
 \textbf{a-008\_O\_018}  & \textbf{ENDF/B-VIII.1} & \textbf{2024} & \textbf{NNL}       & \textbf{2018} & \textbf{Based on JENDL/AN-2005} \\
\bottomrule \bottomrule
\end{tabular}
\end{scriptsize}
\end{center}
\end{table*}

\clearpage

\section{Summary of criticality $k_{\textrm{eff}}$ C/E Changes Between ENDF/B-VIII.0 and ENDF/B-VIII.1 validation results}

The calculated results for multiple benchmark suites presented in Section \ref{sec:complete_criticality} are listed in Table~\ref{tab:appendix:ENDF_criticality}. The benchmark name listed in this table is the name used in the International Criticality Safety Benchmark Evaluation Project (ICSBEP) Handbook \cite{ICSBEP}. The results include the use of the following scattering kernels: (1) water (h-h2o), (2) heavy water (d-d2o), (3) polyethylene (h-poly), (4) beryllium metal (be-met), (5) beryllium-oxide (be-beo, o-beo), and (6) reactor graphite (grph, grph10). The ``$\textrm{S}_{\textrm{d}}$'' versions of beryllium metal and reactor graphite scattering kernels were used. In the table, benchmark names include an ``s'' for simplified models and ``d'' for detailed models (e.g., the Jezebel benchmark PU-MET-FAST-001 includes simplified ``PU-MET-FAST-001-001s'' and detailed ``PU-MET-FAST-001-001-d-rev5'' models).

For the convenience of potential users of the data in Table~\ref{tab:appendix:ENDF_criticality}, we provide the corresponding data in a comma-separated-value (CSV) formatted file in the \ENDF\ release page hosted by the NNDC~\cite{ENDF-release-page}.

\begin{longtable*}{lcccccccc}
\caption{\footnotesize Complete list of validation results of \ENDF\ in criticality benchmarks performed by Los Alamos National Laboratory.}
\label{tab:appendix:ENDF_criticality}\\
\toprule
\toprule
 \multirow{3}{*}{Benchmark}  & \multirow{3}{*}{E}   & \multirow{3}{*}{unc.$_\mathrm{E}$}  & \multicolumn{3}{c}{ENDF/B-VIII.1} & \multicolumn{3}{c}{ENDF/B-VIII.0} \\
\cmidrule(lr){4-6}\cmidrule(lr){7-9}
                                    &                      &                        & C & unc.$_\mathrm{C}$ & C/E                    &  C & unc.$_\mathrm{C}$ & C/E  \\
\midrule 
\endfirsthead
\caption{(continued) ENDF-6 ENDF/B-VIII.1 criticality validation.}\\
\toprule
 \multirow{3}{*}{Benchmark}  & \multirow{3}{*}{E}   & \multirow{3}{*}{unc.$_\mathrm{E}$}  & \multicolumn{3}{c}{ENDF/B-VIII.1} & \multicolumn{3}{c}{ENDF/B-VIII.0} \\
\cmidrule(lr){4-6}\cmidrule(lr){7-9}
                                    &                      &                        & C & unc.$_\mathrm{C}$ & C/E                    &  C & unc.$_\mathrm{C}$ & C/E  \\
\midrule 
\endhead
\bottomrule\addlinespace[\belowrulesep]
\multicolumn{9}{r}{{Continued on the next page}}\\
\endfoot
\bottomrule \bottomrule
\endlastfoot
 HEU-MET-FAST-101-001        & 1.00065 & 0.00078 & 1.00002         & 0.00002           & 0.99937           & 1.00041         & 0.00002           & 0.99976           \\
 HEU-MET-FAST-101-002        & 1.00345 & 0.00079 & 1.00285         & 0.00002           & 0.99940           & 1.00321         & 0.00002           & 0.99976           \\
 HEU-MET-FAST-101-003        & 1.00017 & 0.00081 & 0.99974         & 0.00002           & 0.99957           & 1.00016         & 0.00002           & 0.99999           \\
 HEU-MET-FAST-101-004        & 1.00048 & 0.00076 & 0.99991         & 0.00002           & 0.99943           & 1.00027         & 0.00002           & 0.99979           \\
 HEU-MET-FAST-101-005        & 1.00189 & 0.00076 & 1.00135         & 0.00002           & 0.99946           & 1.00175         & 0.00002           & 0.99987           \\
\midrule
 HEU-MET-FAST-104-001-d      & 1.00072 & 0.00118 & 1.00013         & 0.00001           & 0.99941           & 1.00023         & 0.00002           & 0.99951           \\
 HEU-MET-FAST-104-002-d      & 1.00091 & 0.00098 & 0.99998         & 0.00002           & 0.99907           & 1.00013         & 0.00002           & 0.99922           \\
\midrule
 HEU-MET-INTER-006-001-rev4  & 0.99770 & 0.00080 & 0.99750         & 0.00001           & 0.99980           & 0.99578         & 0.00001           & 0.99807           \\
\end{longtable*}

\section{Complete validation results from GNDS files}

LLNL tested ENDF/B-VIII.1 in the GNDS format against a wide range of fast criticality and fission ratio benchmarks from the ICSBEP~\cite{ICSBEP}
using the Metis Validation and Verification platform~\cite{Metis2017} and the Monte Carlo transport code Mercury~\cite{Mercury5.40}.
These tests were performed by first translating the evaluations into GNDS-2.0 format, processing with \FUDGE~\cite{FUDGE2023},
and using the \GIDI\ API~\cite{GIDIPLUS2023} within Mercury to read and sample data.
See section~\ref{sec:fudge-processing} for more detail on \FUDGE, \GIDI\ and Metis.
Mercury tests of 17 LLNL pulsed spheres available in Metis were consistent with the MCNP6 results presented in the text.

For the convenience of potential users of the data in Tables~\ref{tab:appendix:GNDS_criticality} and~\ref{tab:appendix:GNDS_fission_ratios}, we provide the corresponding data in comma-separated-value (CSV) formatted files in the \ENDF\ release page hosted by the NNDC~\cite{ENDF-release-page}.

\begin{longtable*}{l|cc|cc|cc}
\caption{\footnotesize Validation of the ENDF/B-VIII.1 nuclear data library in GNDS-2.0 format against effective multiplication factors (keff)
	of criticality benchmarks published in the ICSBEP handbook~\cite{ICSBEP}. Simulation results were obtained with the LLNL Monte Carlo code
	Mercury version 5.42.0 with \GIDI\ API version 3.31.0. Results for the released version of ENDF/B-VIII.0 in GNDS-2.0  format are also provided.
	The  ENDF/B-VIII.0 and ENDF/B-VIII.1 libraries were processed with the \FUDGE\ processing code at a temperature of 293.6K,
	and include thermal neutron scattering laws and unresolved region probability tables. }
\label{tab:appendix:GNDS_criticality}\\
\toprule
\toprule
 GNDS-2.0 library     & \multicolumn{2}{c|}{ENDF/B-VIII.1 at 293.6K}             & \multicolumn{2}{c|}{ENDF/B-VIII.0 at 293.6K}             & \multicolumn{2}{c}{Experimental Benchmark}        \\
 Case Name            & Calculated keff         & uncertainty & Calculated keff         & uncertainty & keff                   & uncertainty  \\
\midrule
\endfirsthead
\caption{(continued) GNDS ENDF/B-VIII.1 criticality validation}\\
\toprule
 GNDS-2.0 library     & \multicolumn{2}{c|}{ENDF/B-VIII.1 at 293.6K}             & \multicolumn{2}{c|}{ENDF/B-VIII.0 at 293.6K}             & \multicolumn{2}{c}{Experimental Benchmark}        \\
 Case Name            & Calculated keff         & uncertainty & Calculated keff         & uncertainty & keff                   & uncertainty  \\
\midrule
\endhead
\bottomrule\addlinespace[\belowrulesep]
\multicolumn{7}{r}{{Continued on the next page}}\\
\endfoot
\bottomrule \bottomrule
\endlastfoot
 HEU-MET-FAST-001     & 0.99995                 & 0.00010     & 1.00007                 & 0.00010     & 1.00000                & 0.00100     \\
\midrule
 HEU-MET-FAST-002     & 0.99951                 & 0.00010     & 0.99961                 & 0.00010     & 1.00000                & 0.00300     \\
 HEU-MET-FAST-002-002 & 0.99991                 & 0.00014     & 0.99975                 & 0.00014     & 1.00000                & 0.00300     \\
\midrule
 HEU-MET-FAST-003     & 0.99893                 & 0.00010     & 0.99906                 & 0.00010     & 1.00000                & 0.00300     \\
 HEU-MET-FAST-003-001 & 0.99294                 & 0.00013     & 0.99297                 & 0.00013     & 1.00000                & 0.00500     \\
 HEU-MET-FAST-003-010 & 1.00362                 & 0.00013     & 1.00398                 & 0.00013     & 1.00000                & 0.00500     \\
 HEU-MET-FAST-003-011 & 1.00864                 & 0.00013     & 1.00886                 & 0.00013     & 1.00000                & 0.00500     \\
 HEU-MET-FAST-003-002 & 0.99196                 & 0.00013     & 0.99235                 & 0.00013     & 1.00000                & 0.00500     \\
\end{longtable*}

\begin{table*}
\begin{center}
\caption{\footnotesize Comparison of calculated fission ratios for the ENDF/B-VIII.1 nuclear data library in GNDS-2.0 format with measured values in the center of critical assemblies,
where fission ratio is defined as the fission rate for an actinide normalized by the \nuc{235}{U} fission rate.
Calculated fission ratios were obtained from fission rates simulated with LLNL's Mercury Monte Carlo transport code.
Experimental fission ratios for Exp-A and Exp-B are taken from Ref.~\cite[Table XXXIX]{Brown2018}.
Note that Exp-A and Exp-B refer to the same experimental data. Values for Exp-A are found in the CSEWG Fast Reactor Benchmark Compilation, BNL 19302 (June 1973)~\cite{BNL19302}.
These measurements were later reevaluated and transmitted to R. Mc Farlane in 1984 (personal communication): the reevaluated fission ratios are named Exp-B.
Calc/Exp Ratio were calculated with Exp-B values when available. Experimental values for IMF007 are from Ref.~\cite{SELBY20102891}. These LLNL-calculated values can be compared to LANL MCNP calculated values in Table~\ref{tab:rxnRate1}. They are similar.}

\begin{footnotesize}
\label{tab:appendix:GNDS_fission_ratios}
\begin{tabular}{l|l|cc|cc|cc|cc}
\toprule
\toprule
 \multirow{ 2}{*}{Assembly}   & \multirow{ 2}{*}{Quantity}   & \multicolumn{2}{c|}{\nuc{238}{U}(n,f)/\nuc{235}{U}(n,f)} & \multicolumn{2}{c|}{\nuc{237}{Np}(n,f)/\nuc{235}{U}(n,f)} & \multicolumn{2}{c|}{\nuc{233}{U}(n,f)/\nuc{235}{U}(n,f)} & \multicolumn{2}{c}{\nuc{239}{Pu}(n,f)/\nuc{235}{U}(n,f)} \\
            &            & value               & unc.                     & value                & unc.                      & value               & unc.                     & value                & unc.                      \\ 
\midrule
 \multirow{4}{*}{Godiva (HMF001)}     & Calculated & 0.1581              &                          & 0.8307               &                           & 1.5794              &                          & 1.383                &                           \\
           & Exp-B      & 0.1643              & 0.0018                   & 0.8516               & 0.012                     &                     &                          & 1.4152               & 0.014                     \\
            & Exp-A      & 0.1642              & 0.0018                   & 0.837                & 0.013                     & 1.59                & 0.03                     & 1.402                & 0.025                     \\
            & Calc/Exp   & 0.9623              &                          & 0.9755               &                           & 0.9933              &                          & 0.9772               &                           \\
\midrule 
 \multirow{4}{*}{Jezebel (PMF001)}    & Calculated & 0.2109              &                          & 0.9714               &                           & 1.5663              &                          & 1.4244               &                           \\
          & Exp-B      & 0.2133              & 0.0023                   & 0.9835               & 0.014                     &                     &                          & 1.4609               & 0.013                     \\
            & Exp-A      & 0.2137              & 0.0023                   & 0.962                & 0.016                     & 1.578               & 0.027                    & 1.448                & 0.029                     \\
            & Calc/Exp   & 0.9887              &                          & 0.9877               &                           & 0.9926              &                          & 0.975                &                           \\
\midrule 
 Jezebel-23 & Calculated & 0.2112              &                          & 0.9842               &                           &                     &                          &                      &                           \\
 (UMF001)   & Exp-B      & 0.2131              & 0.0026                   & 0.997                & 0.015                     &                     &                          &                      &                           \\
            & Exp-A      & 0.2131              & 0.0023                   & 0.977                & 0.016                     &                     &                          &                      &                           \\
            & Calc/Exp   & 0.9912              &                          & 0.9872               &                           &                     &                          &                      &                           \\
\midrule 
 \multirow{4}{*}{Flattop-25 (HMF028)} & Calculated & 0.1447              &                          & 0.7719               &                           & 1.5779              &                          & 1.3603               &                           \\
          & Exp-B      & 0.1492              & 0.0016                   & 0.7804               & 0.01                      & 1.608               & 0.003                    & 1.3847               & 0.012                     \\
            & Exp-A      & 0.149               & 0.002                    & 0.76                 & 0.01                      & 1.6                 & 0.003                    & 1.37                 & 0.02                      \\
            & Calc/Exp   & 0.9699              &                          & 0.9892               &                           & 0.9813              &                          & 0.9824               &                           \\
\midrule 
 \multirow{4}{*}{Flattop-Pu (PMF006)} & Calculated & 0.1779              &                          & 0.8508               &                           &                     &                          &                      &                           \\
          & Exp-B      & 0.1799              & 0.002                    & 0.8561               & 0.012                     &                     &                          &                      &                           \\
            & Exp-A      & 0.18                & 0.003                    & 0.84                 & 0.01                      &                     &                          &                      &                           \\
            & Calc/Exp   & 0.9889              &                          & 0.9938               &                           &                     &                          &                      &                           \\
\midrule 
 \multirow{4}{*}{Flattop-23 (UMF006)} & Calculated & 0.1872              &                          & 0.8992               &                           &                     &                          &                      &                           \\
          & Exp-B      & 0.1916              & 0.0021                   & 0.9103               & 0.013                     &                     &                          &                      &                           \\
            & Exp-A      & 0.191               & 0.003                    & 0.89                 & 0.01                      &                     &                          &                      &                           \\
            & Calc/Exp   & 0.9771              &                          & 0.9878               &                           &                     &                          &                      &                           \\
\bottomrule
\bottomrule
\end{tabular}
\end{footnotesize}
\end{center}
\end{table*}

\end{document}